\documentclass{aa}  

\usepackage{graphicx}
\usepackage{hyperref}
\usepackage{txfonts}
\usepackage{booktabs}
\usepackage{multirow}
\usepackage{longtable}
\usepackage{lscape}
\usepackage{url}

\usepackage{subfigure}
\usepackage{caption}
\usepackage{romannum}
\usepackage{siunitx}
\usepackage{longtable}
\usepackage{bibentry}
\nobibliography*

\begin{document}

   \title{The relation between DCN/DNC ratios and gas temperature in late-stage massive star-forming regions using Yebes observations
   }

\author{
Yuqiang Li\inst{1,2}\thanks{E-mail: lyq@kasi.re.kr}
\and
Junzhi Wang\inst{3}
\and
Aran Lyo\inst{1}
\and
Juan Li\inst{2,4}
\and
Siqi Zheng\inst{2,4,5}
\and
Woojin Kwon\inst{6,7,8}
\and
Seokho Lee\inst{1}
\and
Jongsoo Kim\inst{1}
\and\\
Huangxue Feng\inst{9}
\and
Miguel Santander-Garc\'ia\inst{10,11}
\and
Jos\'e Jairo D\'iaz-Luis\inst{10,11}
\and
Shaoyuan Lu\inst{3}
\and
Fu Mo\inst{3}
         }

 \institute{
Korea Astronomy and Space Science Institute, No. 776, Daedeok-daero, Yuseong-gu, Daejeon, Republic of Korea
\and
Shanghai Astronomical Observatory, Chinese Academy of Sciences, 80 Nandan Road, Shanghai 200030, People’s Republic of China
\and
Guangxi Key Laboratory for Relativistic Astrophysics, Department of Physics, Guangxi University, Nanning 530004, People’s Republic of China
\and 
State Key Laboratory of Radio Astronomy and Technology, A20 Datun Road, Chaoyang District, Beijing 100101, People’s Republic of China
\and 
University of Chinese Academy of Sciences, 19A Yuquanlu, Beijing 100049, People’s Republic of China
\and 
Department of Earth Science Education, Seoul National University, 1 Gwanak-ro, Gwanak-gu, Seoul 08826, Republic of Korea
\and 
SNU Astronomy Research Center, Seoul National University, Gwanak-ro 1, Gwanak-gu, Seoul, 08826, Republic of Korea
\and
The Center for Educational Research, Seoul National University, 1 Gwanak-ro, Gwanak-gu, Seoul 08826, Republic of Korea
\and
School of Physics and Astronomy, Sun Yat-sen University, Zhuhai 519082, People’s Republic of China
\and
Observatorio Astron\'omico Nacional (OAN, IGN), C/ Alfonso XII 3, 28014 Madrid, Spain
\and
Observatorio de Yebes (IGN). Cerro de la Palera s/n, 19141 Yebes, Guadalajara, Spain
          }

   \date{Received XXX; accepted YYY}

 
  \abstract{
Deuterated molecules are important probes of the physical and chemical processes that regulate molecular cloud evolution and star formation. Their abundances are affected by the surrounding environment, such as temperature, density, and evolutionary stage. DCN and DNC are relatively abundant deuterated isomers that provide an opportunity to investigate temperature-dependent deuterium chemistry. Although the abundance ratio between their hydrogenated counterparts, HCN and HNC, has often been thought to increase in warmer gas, the temperature dependence of the abundance ratio [DCN]/[DNC] remains less well constrained. To investigate how [DCN]/[DNC] depends on the gas kinetic temperature, we observed the $J$=1-0 transitions of DCN, DNC, H$^{13}$CN, HN$^{13}$C, HC$^{15}$N, and H$^{15}$NC toward 101 late-stage massive star-forming regions using the Yebes 40m telescope. Gas kinetic temperatures were estimated from CH$_3$CCH 5-4, whose rotation temperature is closely related to the kinetic temperature of dense gas. These temperatures were adopted in the column density calculations and in the analysis of temperature-dependent abundance ratios. We find a moderate positive correlation between the abundance ratio [DCN]/[DNC] and kinetic temperature, which is statistically significant and indicates that DCN becomes progressively more abundant relative to DNC in warmer gas. A weak negative correlation with temperature is observed for $D_{\rm frac}({\rm HNC})$, whereas no statistically significant correlation is found for $D_{\rm frac}({\rm HCN})$. The [HCN]/[HNC] ratios show substantial scatter and no statistically significant correlation with temperature, possibly due in part to optical-depth effects in the $^{13}$C isotopologues and the limited sensitivity of the $^{15}$N isotopologue observations. These results suggest that [DCN]/[DNC] may be governed not only by HCN/HNC isomeric chemistry but also by temperature-dependent deuterium fractionation pathways and the chemical evolution of dense gas. 
  }


\titlerunning{DCN/DNC ratio and gas temperature}
\authorrunning{Li et al.}

   \maketitle
   \nolinenumbers
%

\section{Introduction}


Deuterated molecules are important probes of interstellar chemistry and provide valuable insights into the physical and chemical evolution of molecular clouds and star-forming regions \citep[e.g., ][]{2012A&ARv..20...56C,2014prpl.conf..859C}. Their abundances and deuterium fractionation ratios are controlled by a combination of environmental and chemical factors, such as evolutionary stage \citep[e.g., ][]{2005ApJ...619..379C,2010A&A...517L...6B,2010ApJ...713L..50C,2015A&A...575A..87F,2015A&A...579A..80G}, temperature \citep[e.g., ][]{2009A&A...493...89E,2012ApJ...747..140S,2022ARep...66.1302P}, density \citep[e.g., ][]{2000A&A...364..780R,2000A&A...361..388R,2007A&A...470..221C,2007A&A...467..179P}, and cosmic ray ionization rate \citep[e.g., ][]{2008A&A...492..703C,2010A&A...509A..98S}. Therefore, studies of deuterated molecules provide insight into the chemical processes that regulate molecular deuteration during star formation. Because of the slightly lower zero-point energy for deuterated species \citep{2003A&A...406..383R,2013JPCA..117.9959R}, they form more easily than their non-deuterated counterparts \citep{laidler1987chemical,atkins2010molecular,levine2013quantum}. As a result, the abundance ratios between deuterated molecules and their hydrogenated counterparts can be orders of magnitude higher than the elemental deuterium-to-hydrogen ratio in the interstellar medium (ISM), D/H $\approx 1.5\times10^{-5}$ \citep{1995ApJ...451..335L,2003ApJ...587..235O}. This enhancement is commonly quantified by the deuterium fractionation, $D_{\rm frac}$, defined as the abundance ratio between a deuterated molecule and its hydrogenated counterpart.

DCN and DNC were first detected in the ISM in the 1970s \citep{1973ApJ...179L..57J,1973ApJ...179L.107W,1977MNRAS.180P..83G,1977ApJ...216L.111S}. Since then, these two deuterated species have been widely detected in different interstellar environments, from starless cores \citep[e.g., ][]{2020ApJS..249...33K,2024ApJS..270...35Y,2025A&A...696A.140M} to hot molecular cores \citep[e.g., ][]{2015A&A...579A..80G,2023MNRAS.526.3673P,2024A&A...688A.215C} and ultracompact H\Romannum{2} regions \citep[e.g., ][]{2014MNRAS.440..448F,2014A&A...569A..19T,2015A&A...579A..80G}. In particular, DCN is often observed to be relatively abundant in warm and hot environments, while DNC is typically enhanced in colder environments \citep{2015A&A...579A..80G}. This contrast suggests that the relative abundance of DCN and DNC may be sensitive to temperature. Such a temperature dependence may arise either from differences in the deuterium fractionation of HCN and HNC or from the thermal evolution of the parent HCN/HNC isomeric system. 

As commonly used dense gas tracers, HCN and HNC -- including their $^{13}$C and $^{15}$N isotopologues -- play a special role due to their isomeric nature. The abundance ratio, [HCN]/[HNC], has long been considered temperature-dependent, generally increasing with temperature \citep[e.g., ][]{1986ApJ...310..383G,2020A&A...635A...4H}. 
At low temperatures, [DCN]/[DNC] has been found to show only a weak dependence on temperature in TMC-1 \citep{2021A&A...653A..15N}. In warmer and more evolved environments, however, \citet{2020ARep...64..244T} suggest that [DCN]/[DNC] increases with gas temperature based on eight sources with reliable [DCN]/[DNC] measurements, and \citet{2024A&A...688A.215C} find that the velocity-integrated intensity ratio of DCN and DNC increases with dust temperature in DR21. 
These results suggest that warm, evolved star-forming regions may provide more favorable conditions than cold dark-cloud environments for detecting a temperature dependence of [DCN]/[DNC].
Late-stage massive star-forming regions are therefore particularly suitable for testing this trend, because they are already heated by embedded massive young stellar objects and span a relatively broad range of gas temperatures \citep[e.g.,][]{2017A&A...603A..33G,2018A&A...611A...6T}. However, previous studies were based on limited samples or intensity ratios rather than abundance ratios derived from column densities. A systematic investigation of the temperature dependence of [DCN]/[DNC] in a large sample of late-stage massive star-forming regions is therefore still lacking.

In this work, we investigate the relationship between [DCN]/[DNC] and gas temperature. We observed the $J$=1-0 transitions of DCN, DNC, H$^{13}$CN, HN$^{13}$C, HC$^{15}$N, and H$^{15}$NC toward a large sample of 101 late-stage star-forming regions using Yebes 40m observations. 
These observations allow us to derive [DCN]/[DNC] and [HCN]/[HNC] as well as the deuterium fractionation $D_{\rm frac}$(HCN) and $D_{\rm frac}$(HNC).
Gas kinetic temperatures were adopted from CH$_3$CCH 5-4 observations reported by \citet{2026ApJ...998..182L}. By comparing these ratios as a function of gas kinetic temperature, we examine whether the observed [DCN]/[DNC] trend can be interpreted solely in terms of the parent HCN/HNC isomeric chemistry or whether the different deuterium fractionation behavior of HCN and HNC also plays a role.
The observations are described in Sect. \ref{observations}, our main results are reported in Sect. \ref{results}, a discussion is given in Sect. \ref{discussion}, and a brief summary is given in Sect. \ref{summary}.


\section{Observations}
\label{observations}

Our sample consists of 101 late-stage massive star-forming regions with class \Romannum{2} 6.7 GHz CH$_3$OH masers selected from \citet{2014ApJ...783..130R}. Two additional targets in that study, G348.70-01.04 and G351.44+00.65, were excluded because their maximum elevations are below $\ang{20}$. 
The properties of the representative sources are listed in Table \ref{source}. The complete source list is shown in the online material. The observations were performed with the Yebes 40m radio telescope from May to June 2023 (Project ID: 23A013). The position-switch mode was used, with the OFF positions located \ang{;30;} away from the target positions in azimuth. Calibration was performed using the chopper-wheel method, with hot (ambient-temperature) and cold loads, together with observations of the sky near each source to calibrate the temperature scale and the atmospheric opacity using standard atmosphere models (ATMs). 
 The frequency coverage was about 18.5 GHz, spanning from 72 to 90.5 GHz. The Fast Fourier Transform Spectrometer (XFFTS) was used with 65536 channels, providing a frequency resolution of 38 kHz, corresponding to a velocity resolution of 0.16 km s$^{-1}$ and 0.13 km s$^{-1}$ at about 72 GHz and 90 GHz, respectively. The system temperatures ranged from 110 to 390 K. The beam sizes of the Yebes 40 m telescope are approximately $\ang{;;24.3}$ and $\ang{;;19.9}$ at 72.5 and 88.5 GHz, respectively.

\begin{table*}[!htbp]
\centering
\caption{Source properties.}
\label{source}
\begin{tabular}{llccrccc}
\hline\hline
Source name & Alias & R.A. (J2000) & Decl. (J2000) &
$v_{\rm LSR}$ &
$D_{\sun}$ &
$D_{\rm gc}$ &
$T_{\rm kin}$\\
 & & (hh:mm:ss) & (dd:mm:ss) &
(km s$^{-1}$) &
(kpc) & (kpc) & (K)\\
\hline

G005.88-00.39       	&		&	18:00:30.31    	&	-24:04:04.50   	&	9	$\pm$	3	&	3.0	&	5.2	& 48.5$\pm$0.4	\\
G014.33-00.64       	&		&	18:18:54.67    	&	-16:47:50.30   	&	22	$\pm$	5	&	1.1	&	7.1	& 31.4$\pm$0.3 	\\
G035.19-00.74       	&		&	18:58:13.05    	&	+01:40:35.70   	&	30	$\pm$	7	&	2.2	&	6.5	& 34.9$\pm$0.3 	\\
G081.75+00.59       	&	DR 21	&	20:39:01.99    	&	+42:24:59.30   	&	-3	$\pm$	3	&	1.5	&	8.1	& 31.3$\pm$0.3 	\\
G109.87+02.11       	&	Cep A	&	22:56:18.10    	&	+62:01:49.50   	&	-7	$\pm$	5	&	0.7	&	8.4	& 36.1$\pm$0.4 	\\
G209.00-19.38       	&	Orion Nebula	&	05:35:15.80    	&	-05:23:14.10   	&	3	$\pm$	5	&	0.4	&	8.5	& 38.1$\pm$0.7	\\
\hline
\end{tabular}
\tablefoot{
1. The $v_{\rm LSR}$ values are from \citet{2014ApJ...783..130R}. \\
2. The $D_{\sun}$ values were derived from the trigonometric parallax reported in \citet{2014ApJ...783..130R}, and the $D_{\rm gc}$ values were calculated from those parallaxes. \\
3. The $T_{\rm kin}$ values were adopted from \citet{2026ApJ...998..182L}.\\
4. The complete table for all 101 sources is available online.
}
\end{table*}

The main beam temperature was obtained from $T_{\rm mb}=T_{\rm A}^*/\eta_{\rm mb}$, where $\eta_{\rm mb}$ is 0.30 and 0.21 at 72.5 GHz and 88.5 GHz, respectively\footnote{\url{https://rt40m.oan.es/rt40m_en.php}}. 
Data reduction was conducted using the CLASS/GILDAS software\footnote{\url{http://www.iram.fr/IRAMFR/GILDAS}}. First-order baselines were used for all spectra.
In the following analysis, we assume a beam-filling factor of unity. The spectra were smoothed to a velocity resolution of approximately 0.6 km s$^{-1}$, yielding a typical rms noise level of approximately 15 mK ($T_{\rm A}^*$ scale).

\section{Results}
\label{results}
\subsection{The detected lines}
\label{The detected lines}

\begin{figure*}[!htbp]
\centering

{\scriptsize
\makebox[0.155\textwidth][c]{DCN 1-0}\hfill
\makebox[0.155\textwidth][c]{DNC 1-0}\hfill
\makebox[0.155\textwidth][c]{H$^{13}$CN 1-0}\hfill
\makebox[0.155\textwidth][c]{HN$^{13}$C 1-0}\hfill
\makebox[0.155\textwidth][c]{HC$^{15}$N 1-0}\hfill
\makebox[0.155\textwidth][c]{H$^{15}$NC 1-0}
\par
}
\vspace{0.2em}

\includegraphics[width=0.155\textwidth]{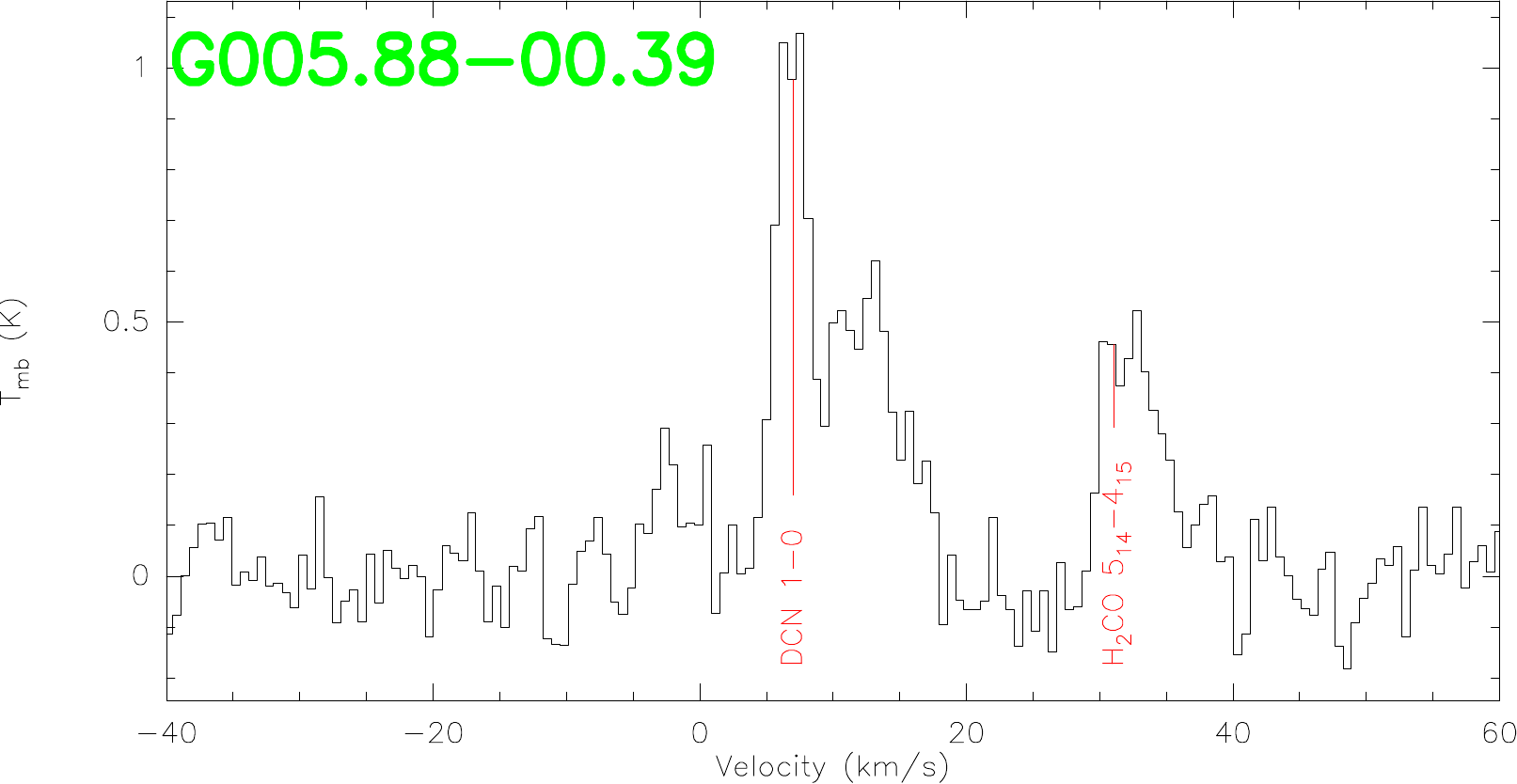}\hfill
\includegraphics[width=0.155\textwidth]{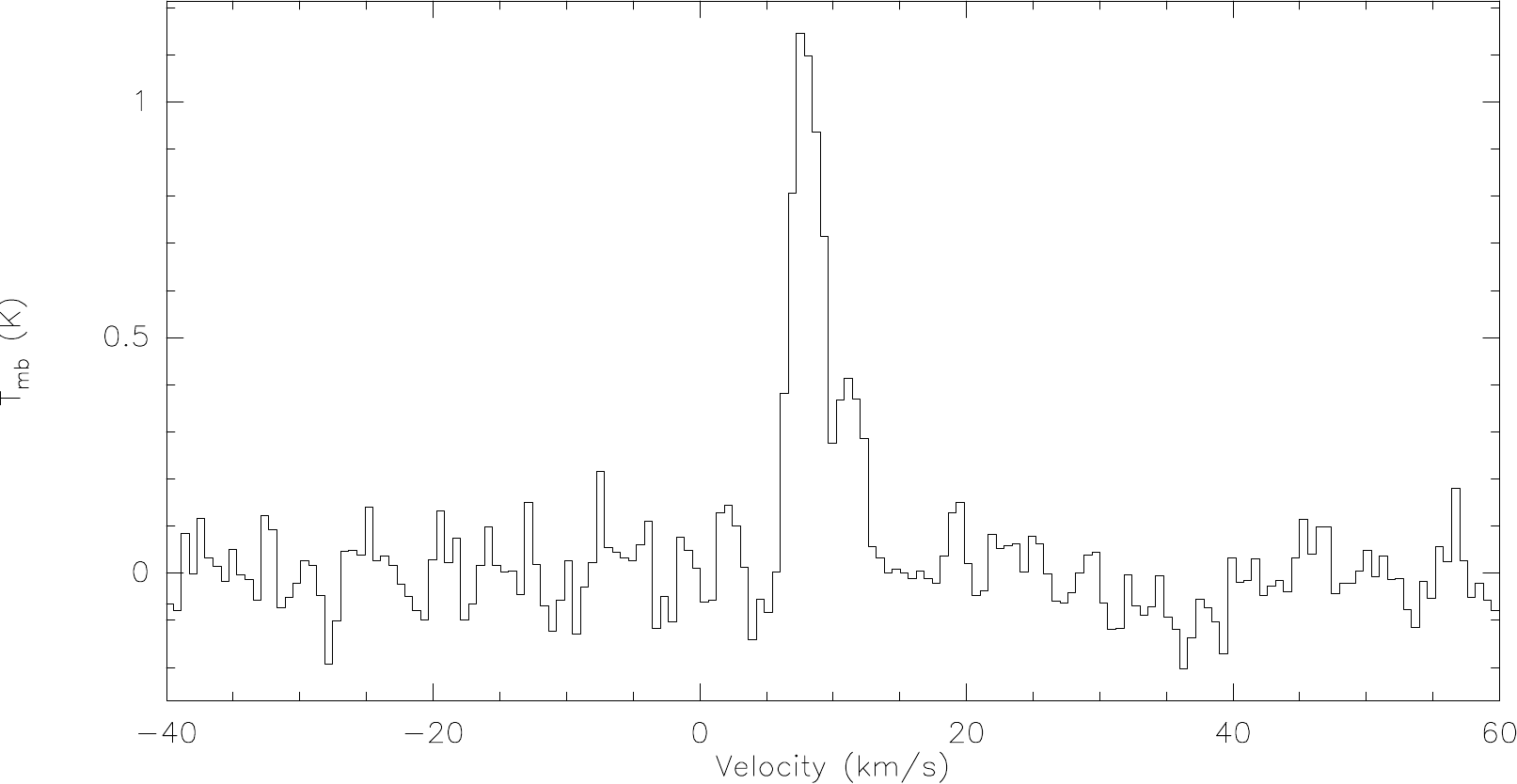}\hfill
\includegraphics[width=0.155\textwidth]{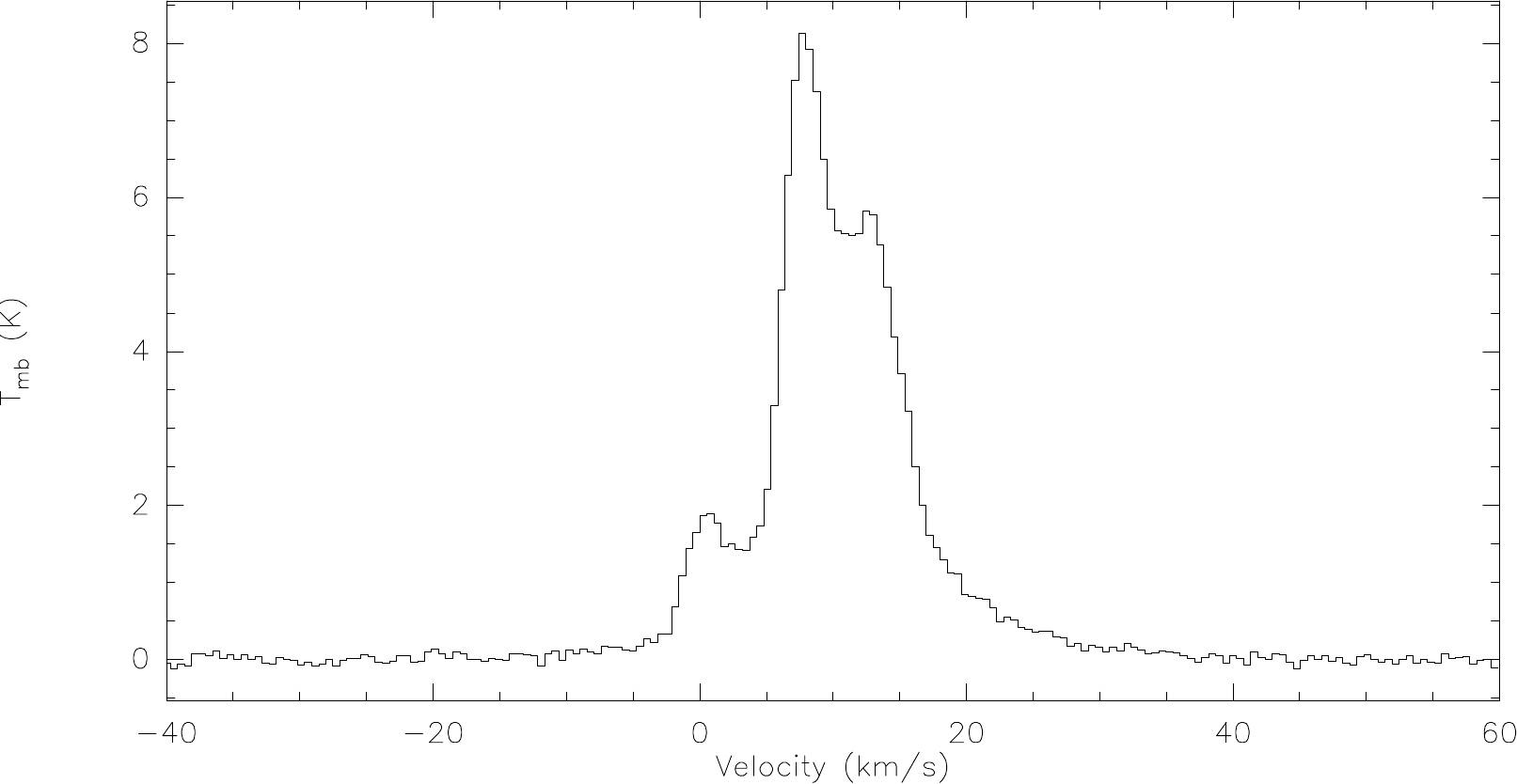}\hfill
\includegraphics[width=0.155\textwidth]{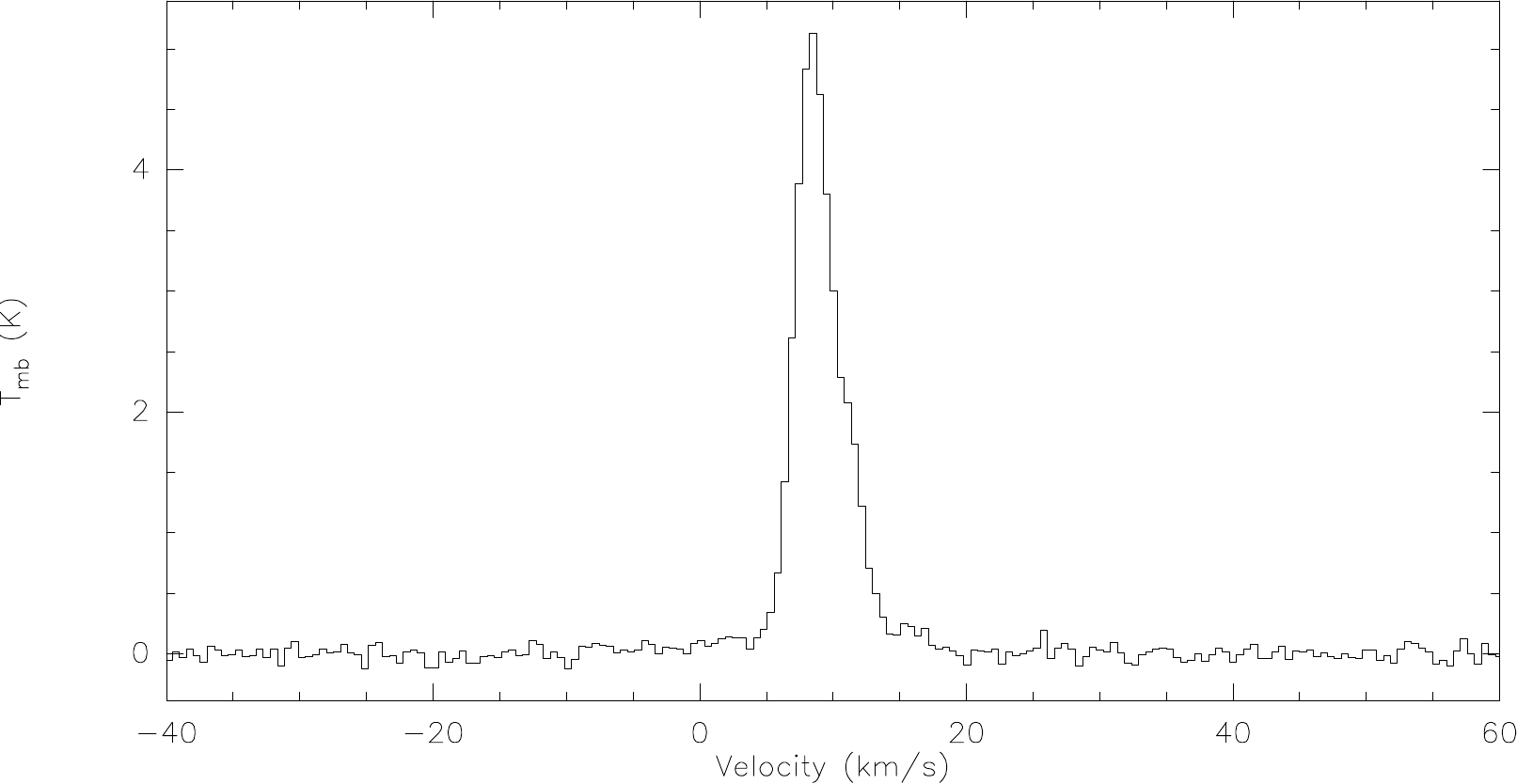}\hfill
\includegraphics[width=0.155\textwidth]{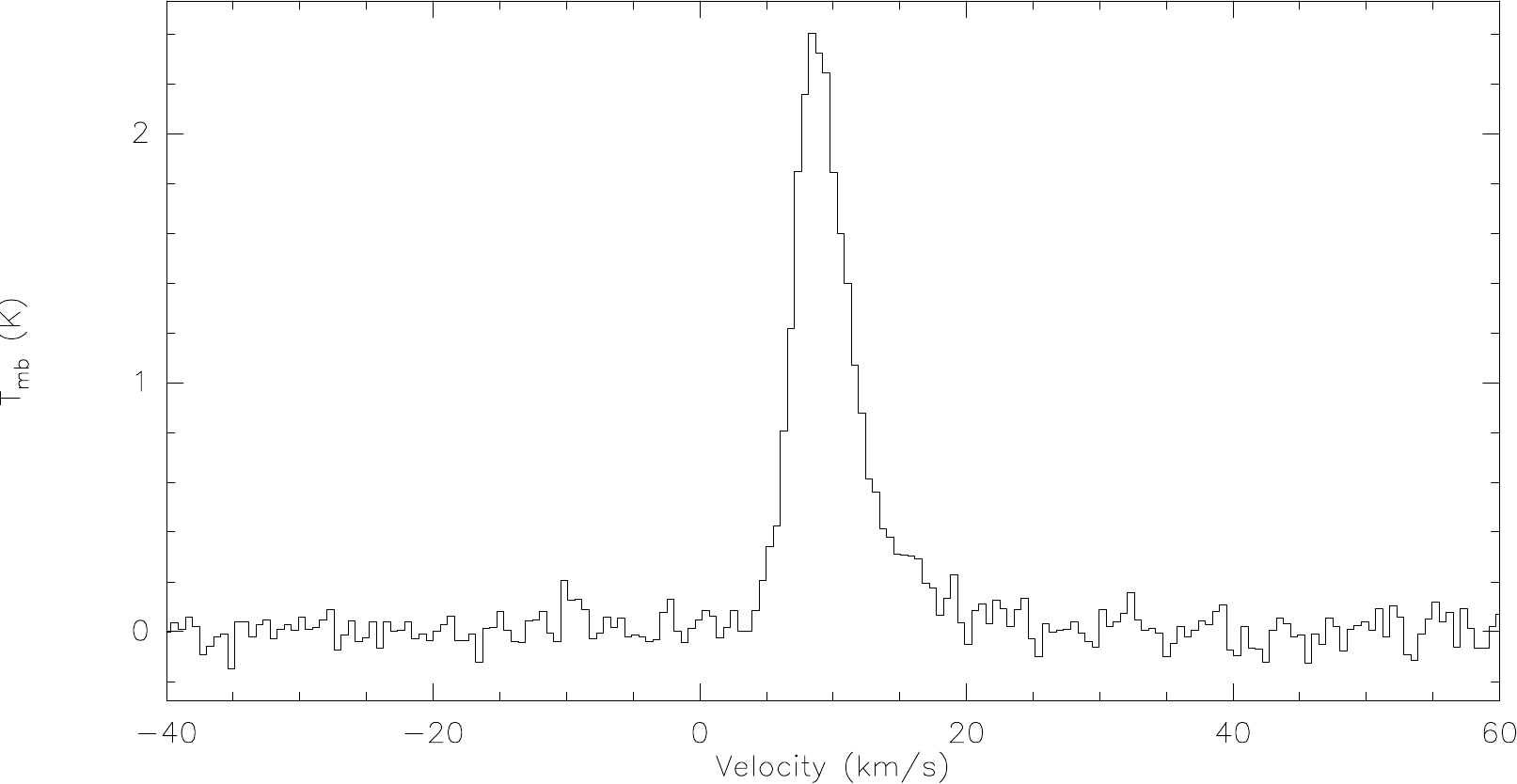}\hfill
\includegraphics[width=0.155\textwidth]{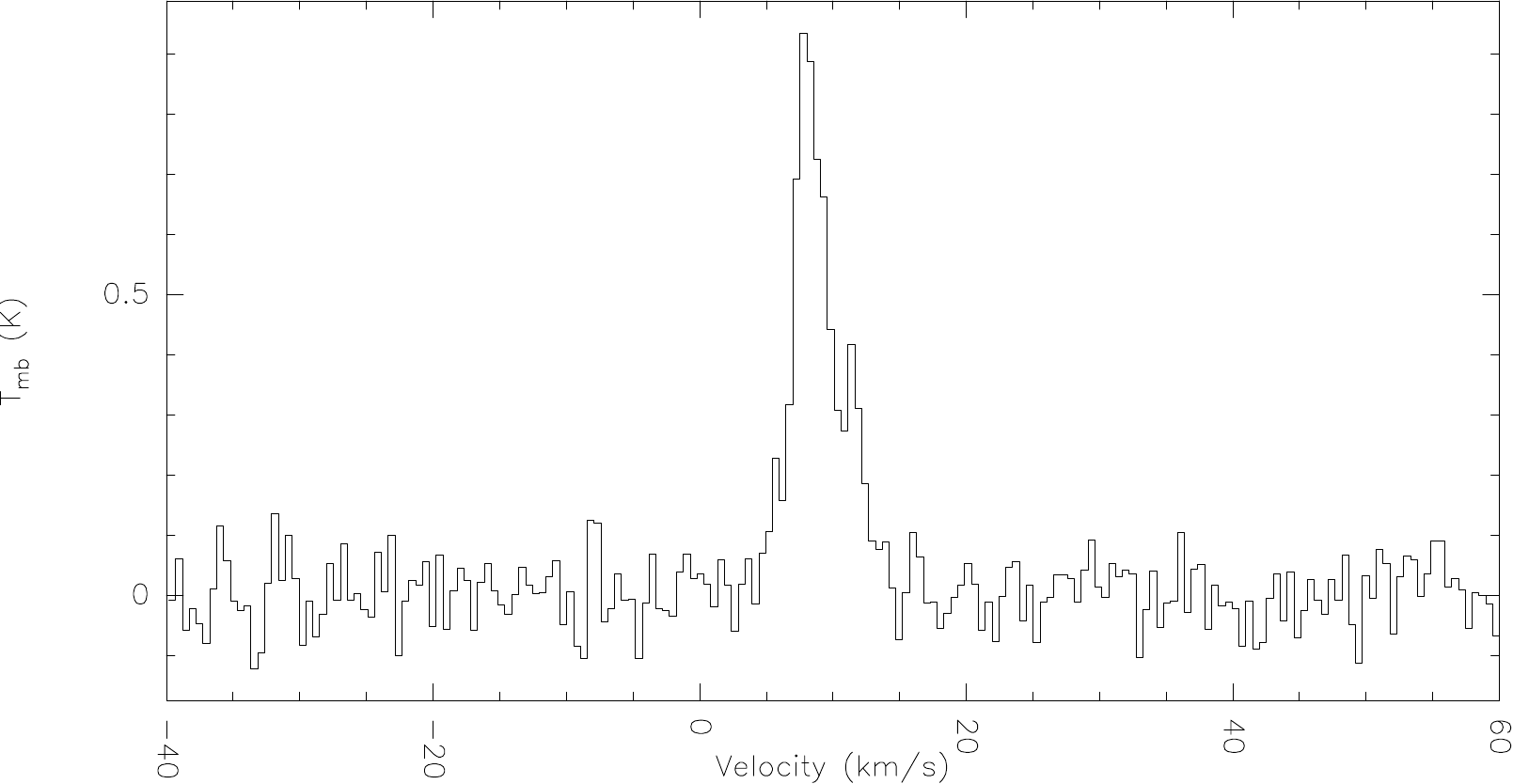}
\par

\includegraphics[width=0.155\textwidth]{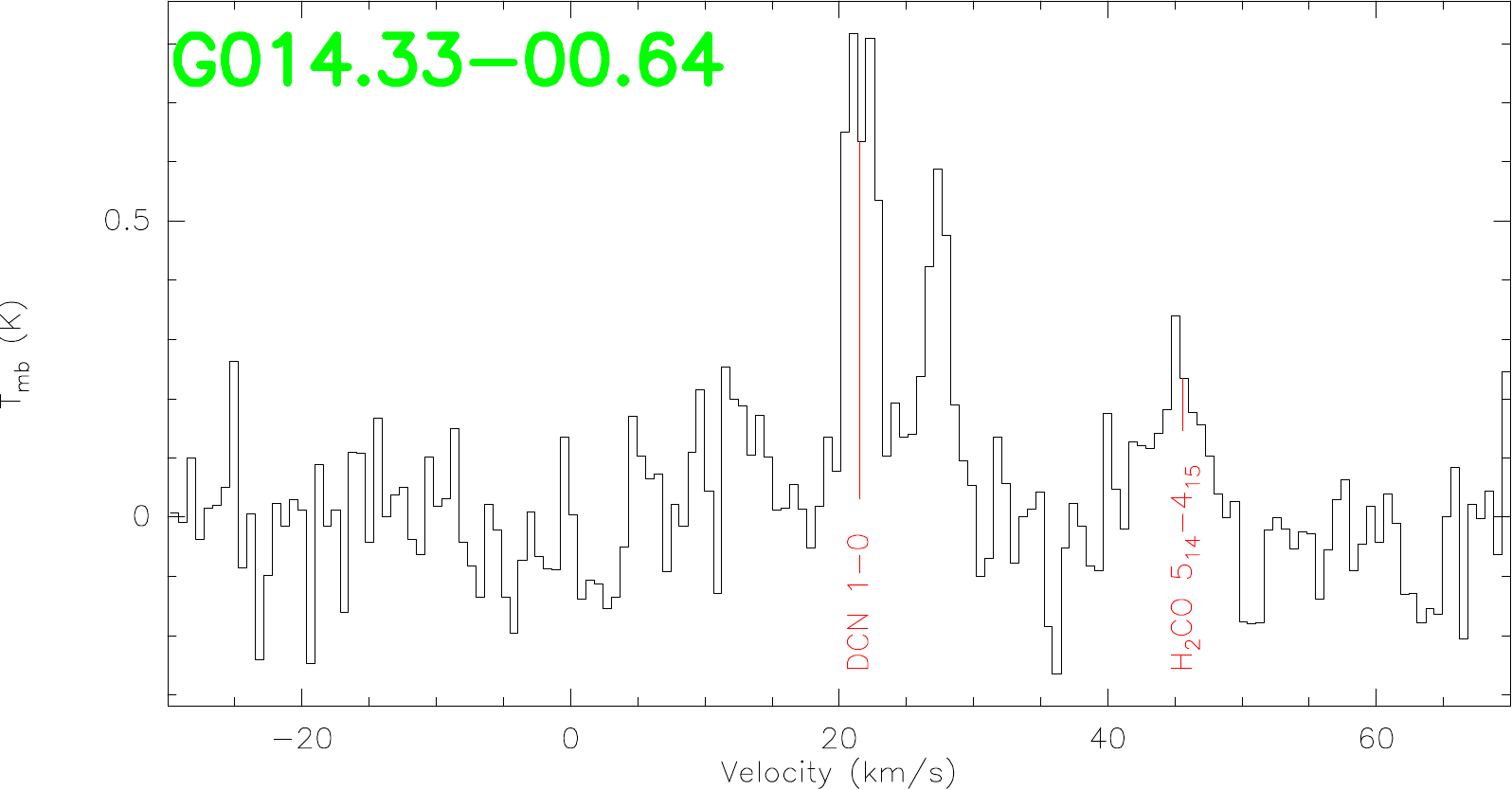}\hfill
\includegraphics[width=0.155\textwidth]{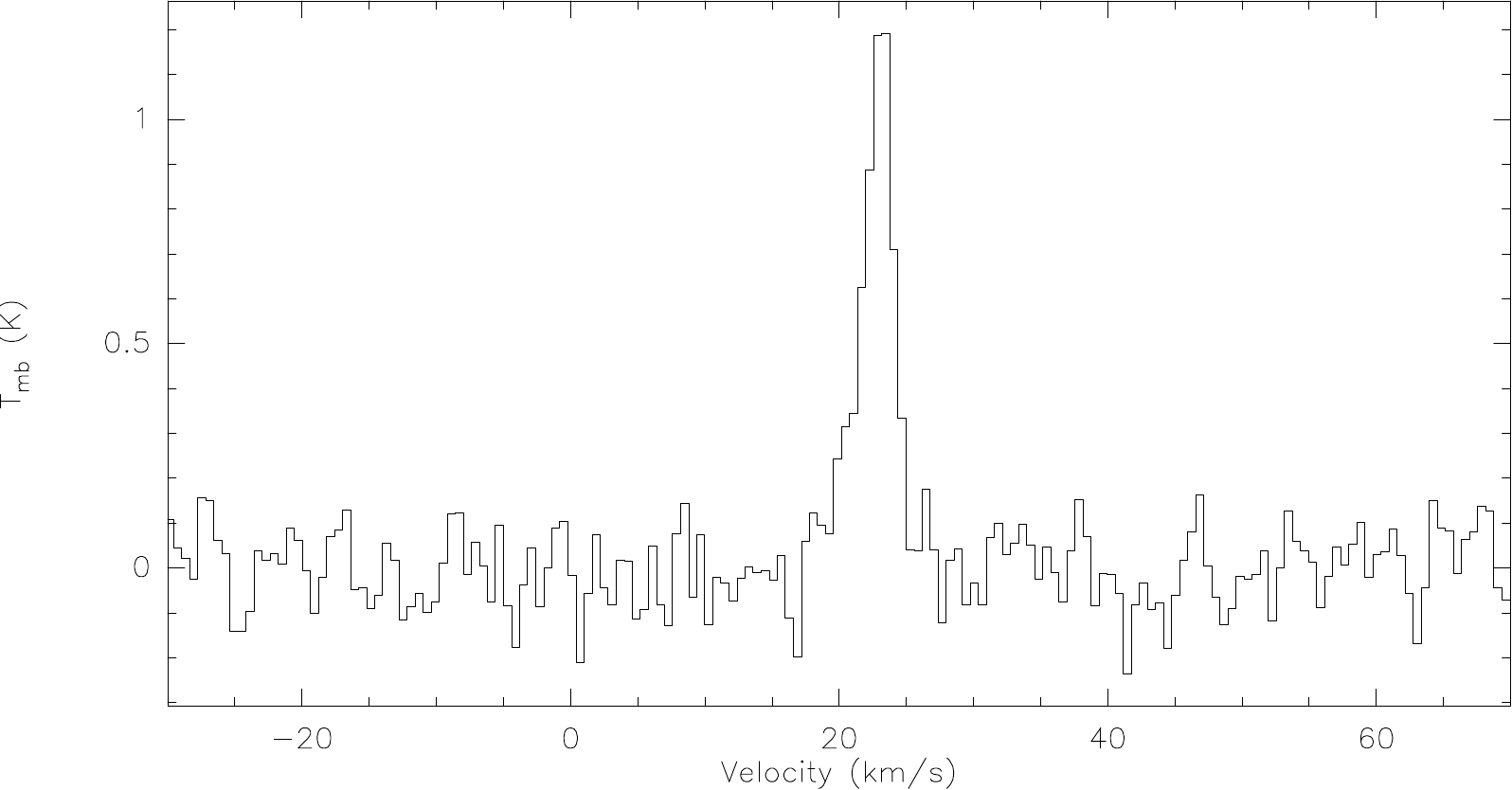}\hfill
\includegraphics[width=0.155\textwidth]{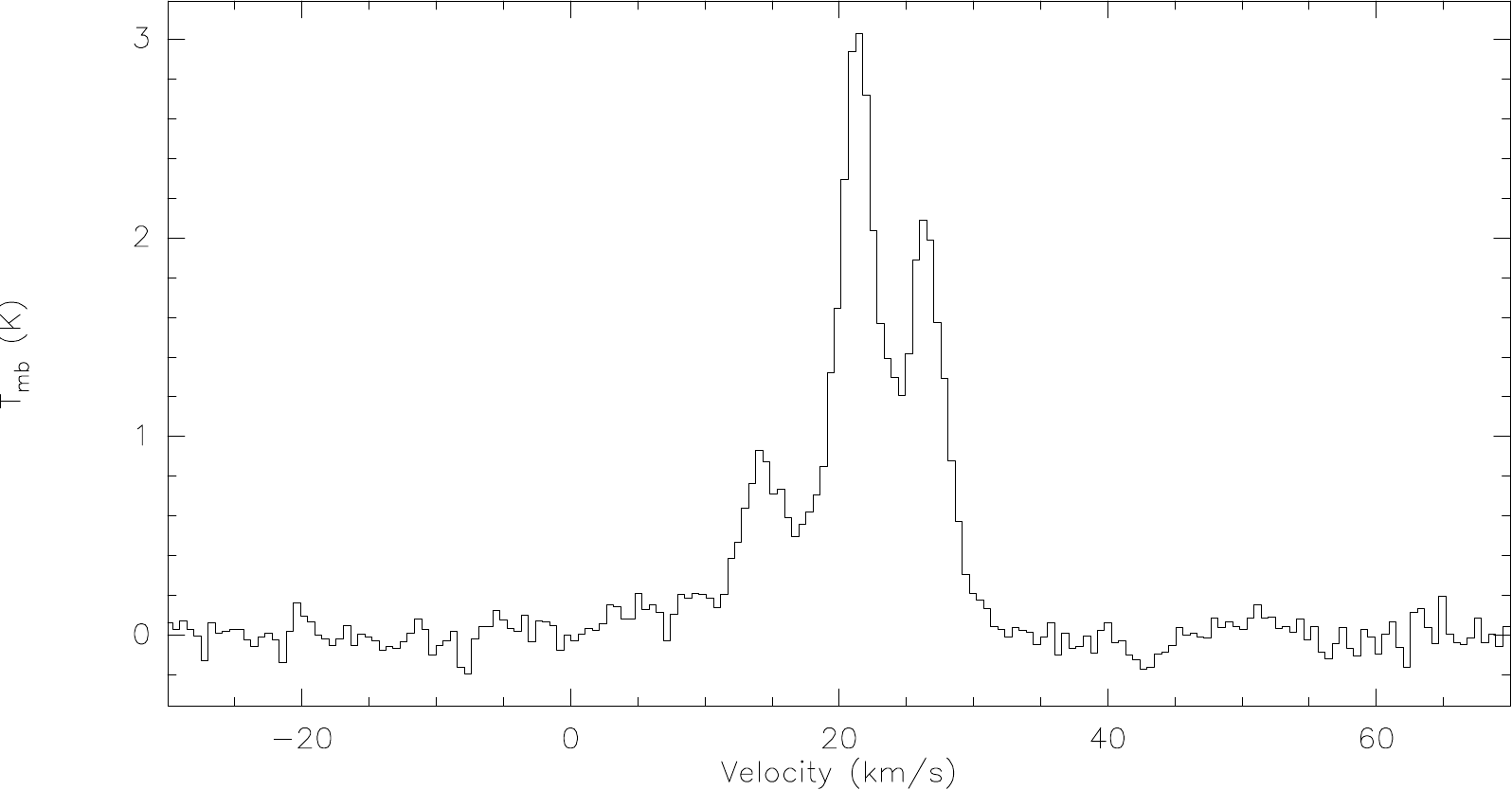}\hfill
\includegraphics[width=0.155\textwidth]{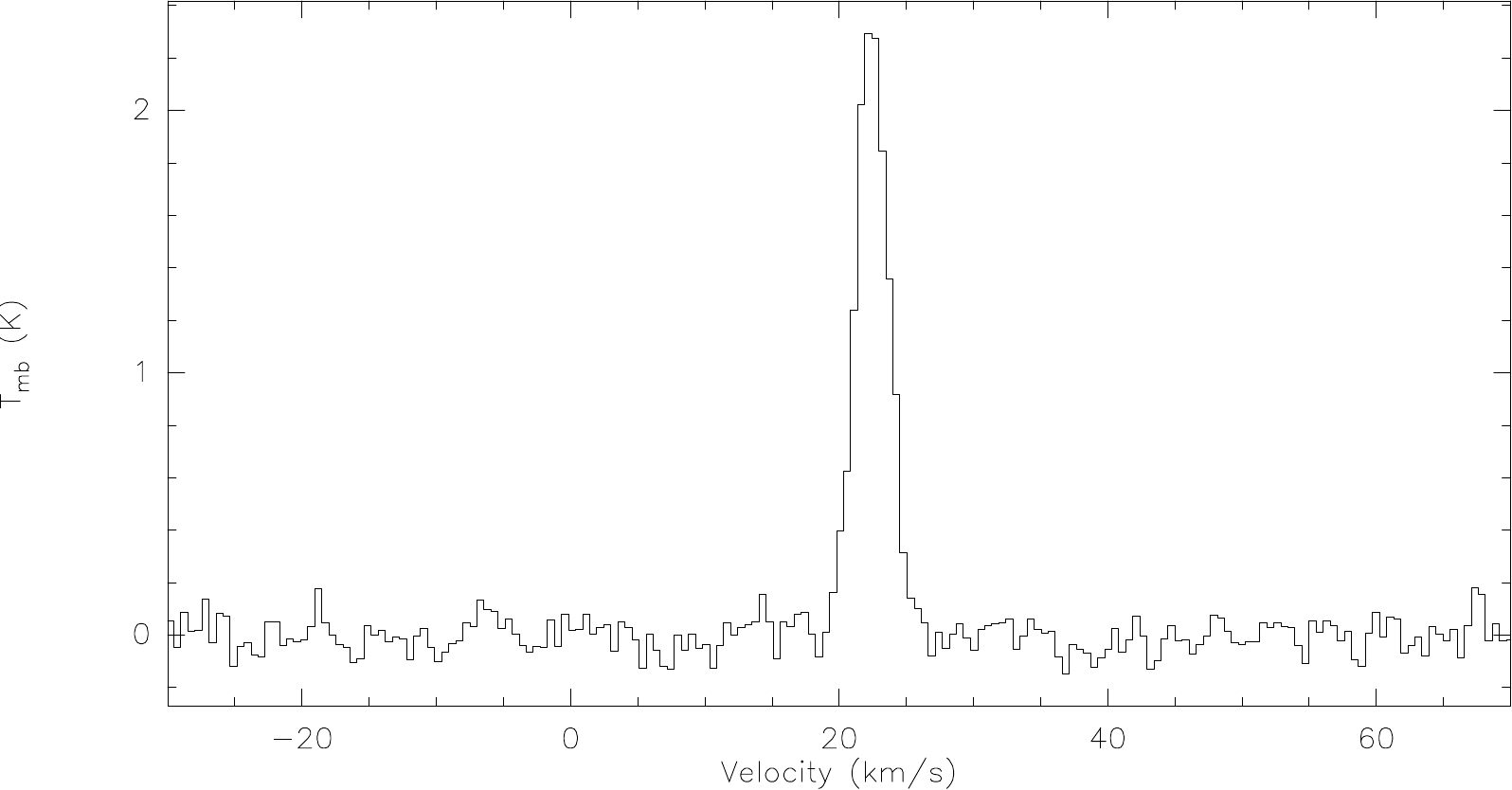}\hfill
\includegraphics[width=0.155\textwidth]{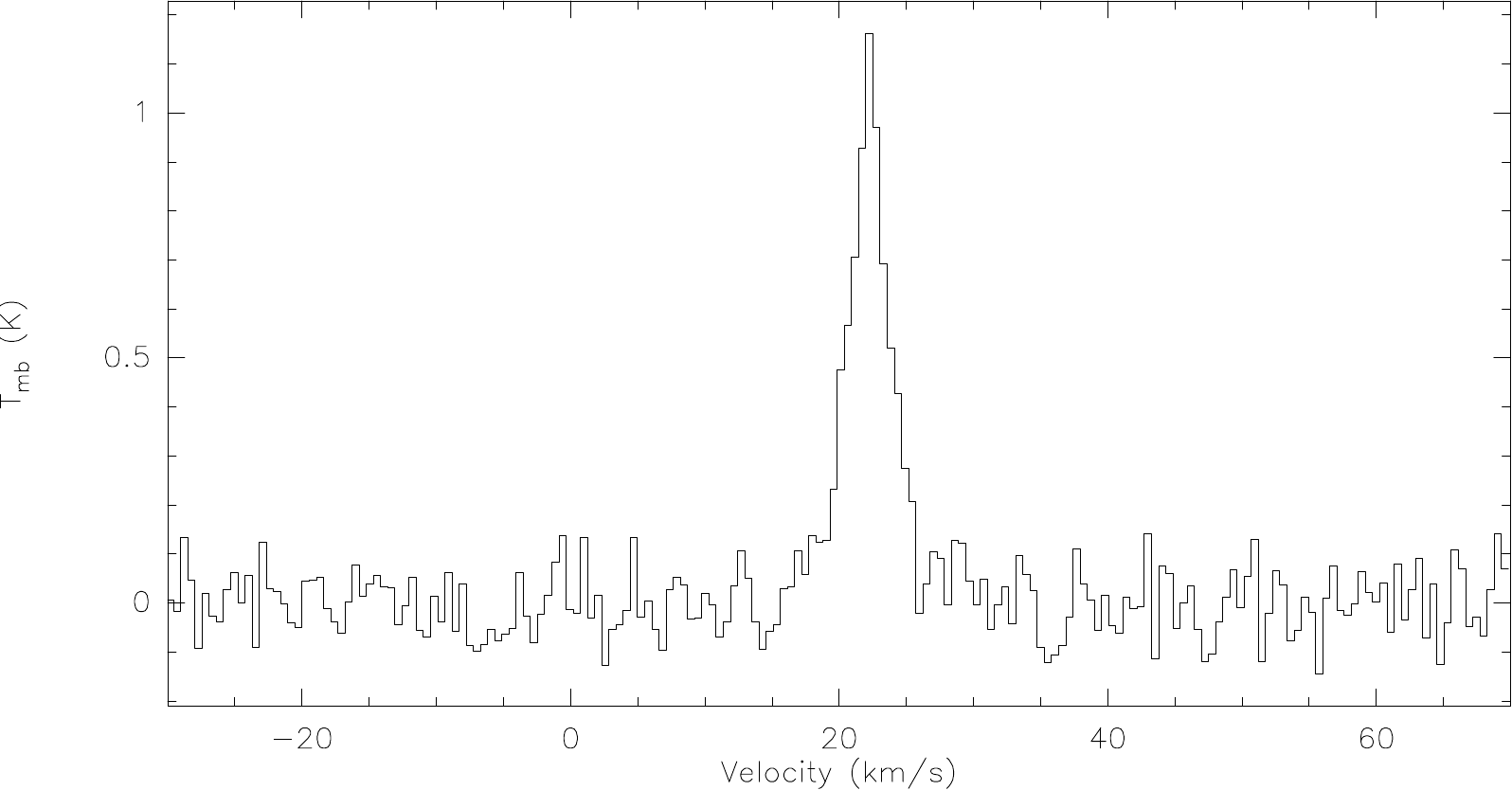}\hfill
\includegraphics[width=0.155\textwidth]{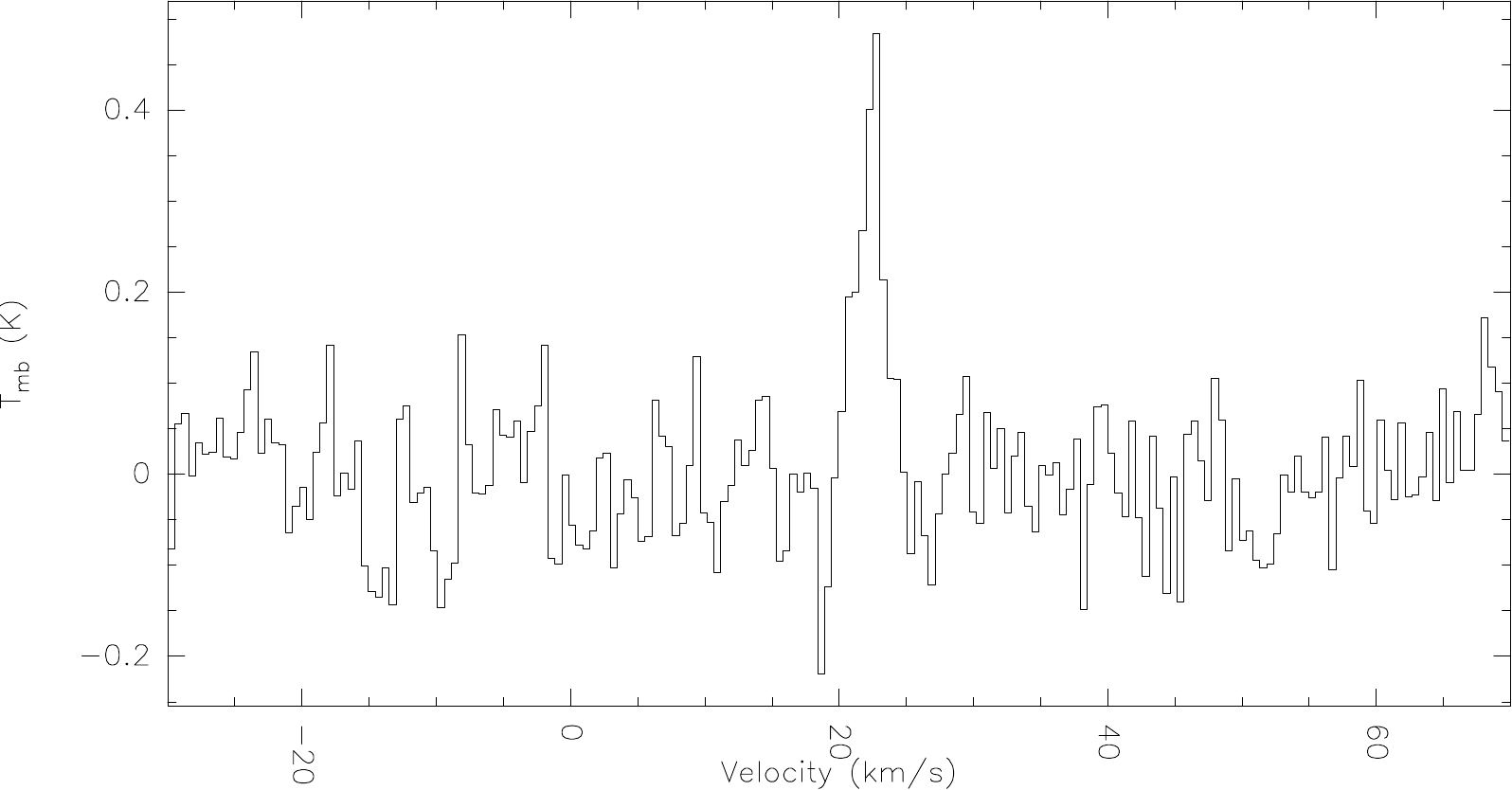}
\par

\includegraphics[width=0.155\textwidth]{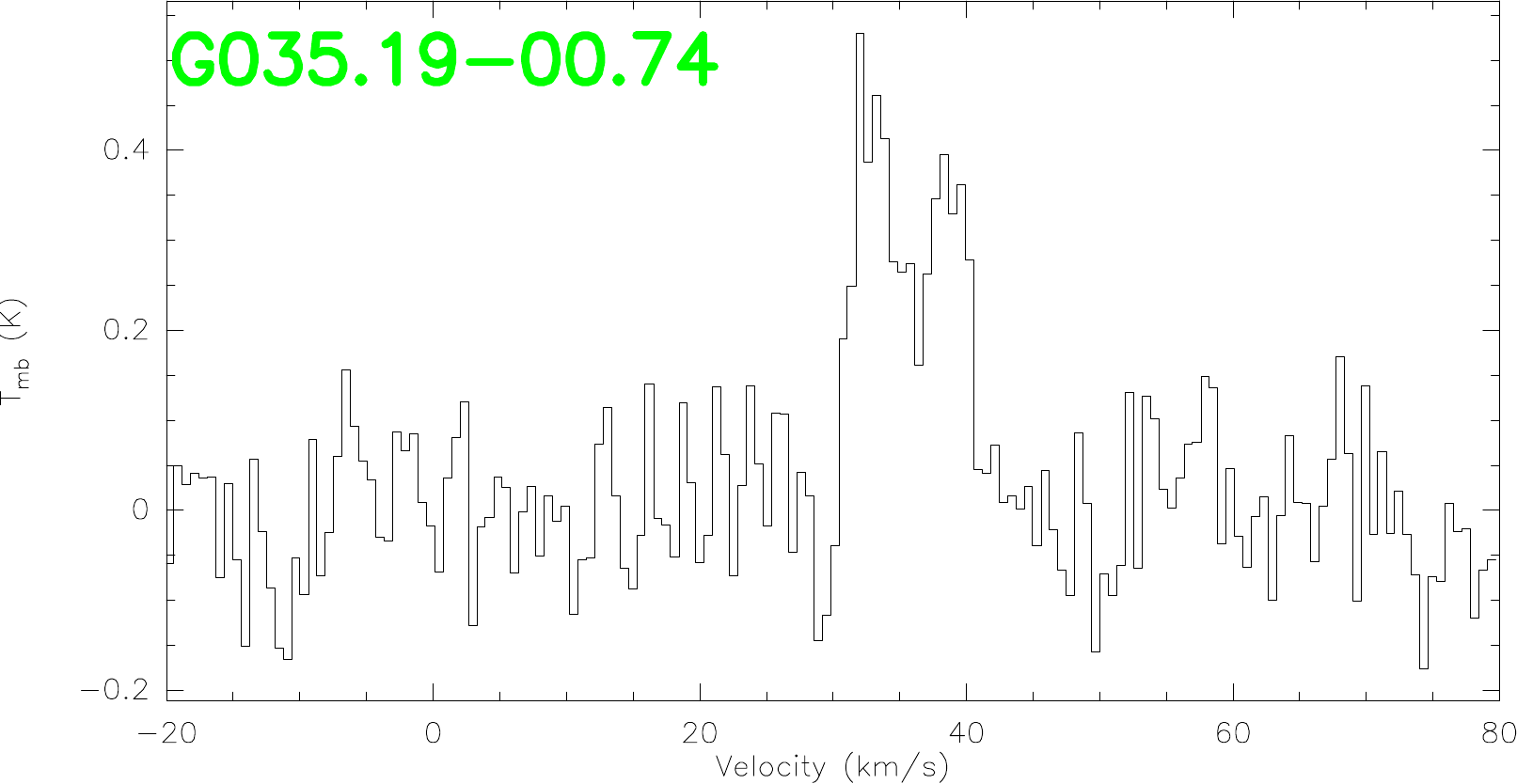}\hfill
\includegraphics[width=0.155\textwidth]{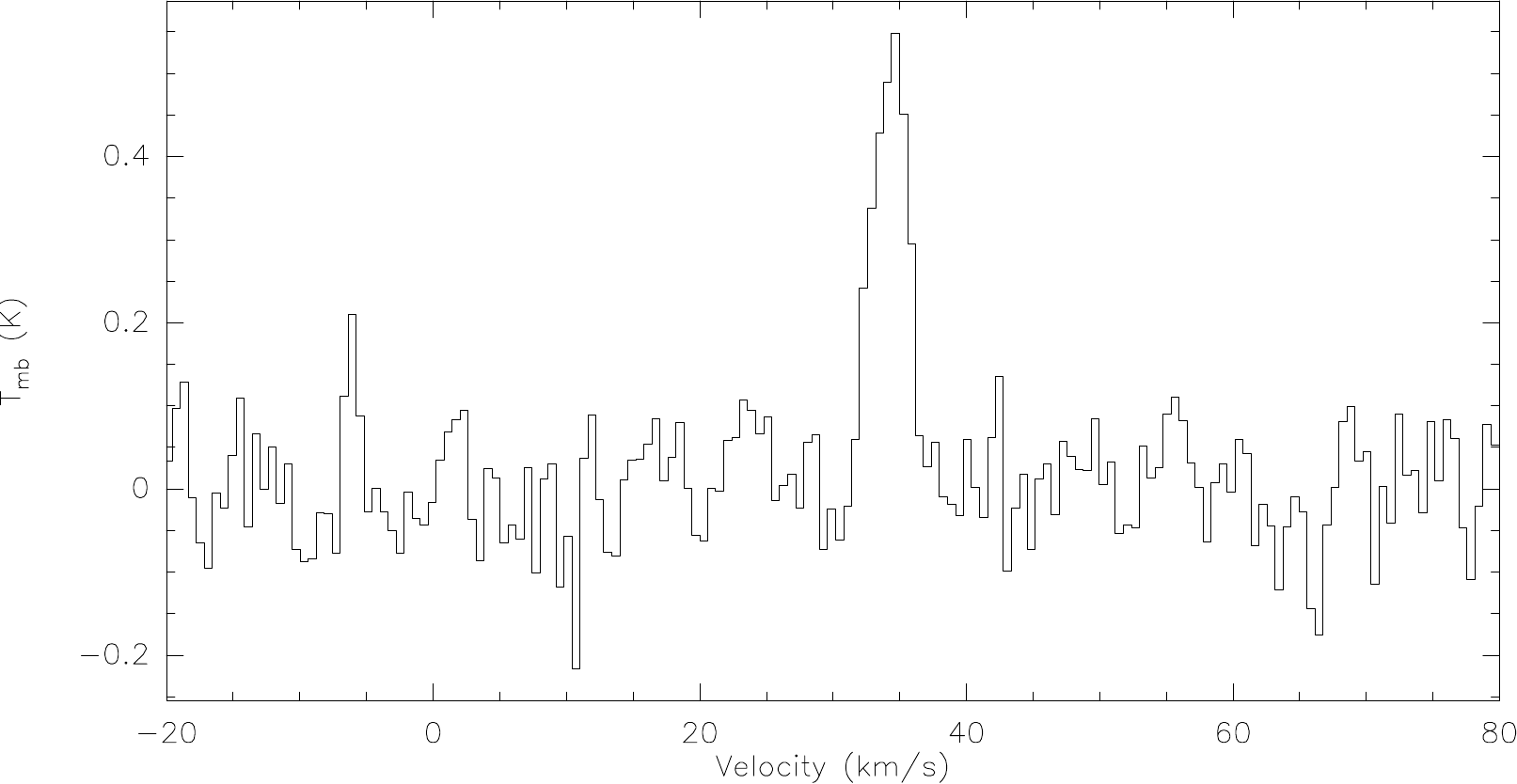}\hfill
\includegraphics[width=0.155\textwidth]{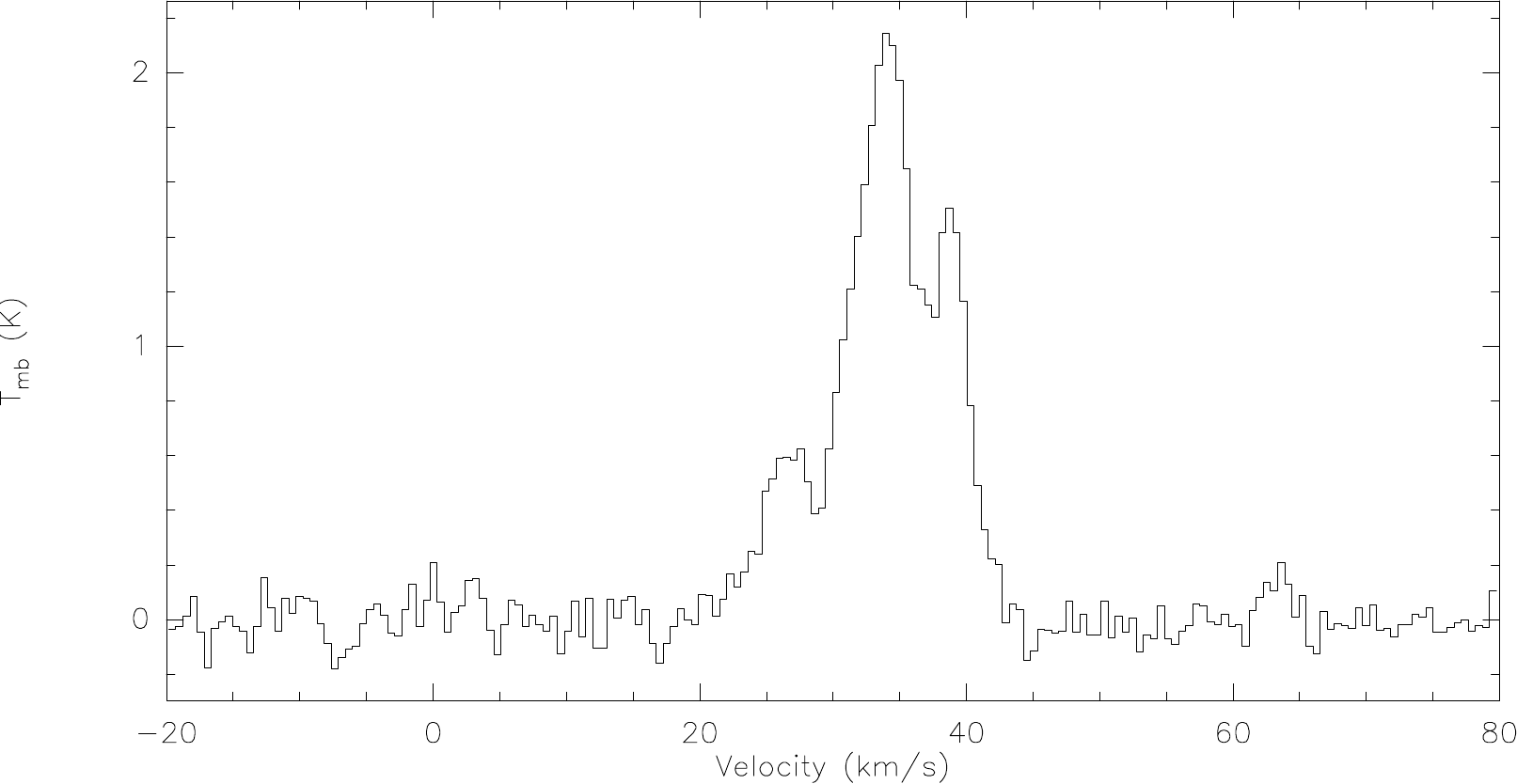}\hfill
\includegraphics[width=0.155\textwidth]{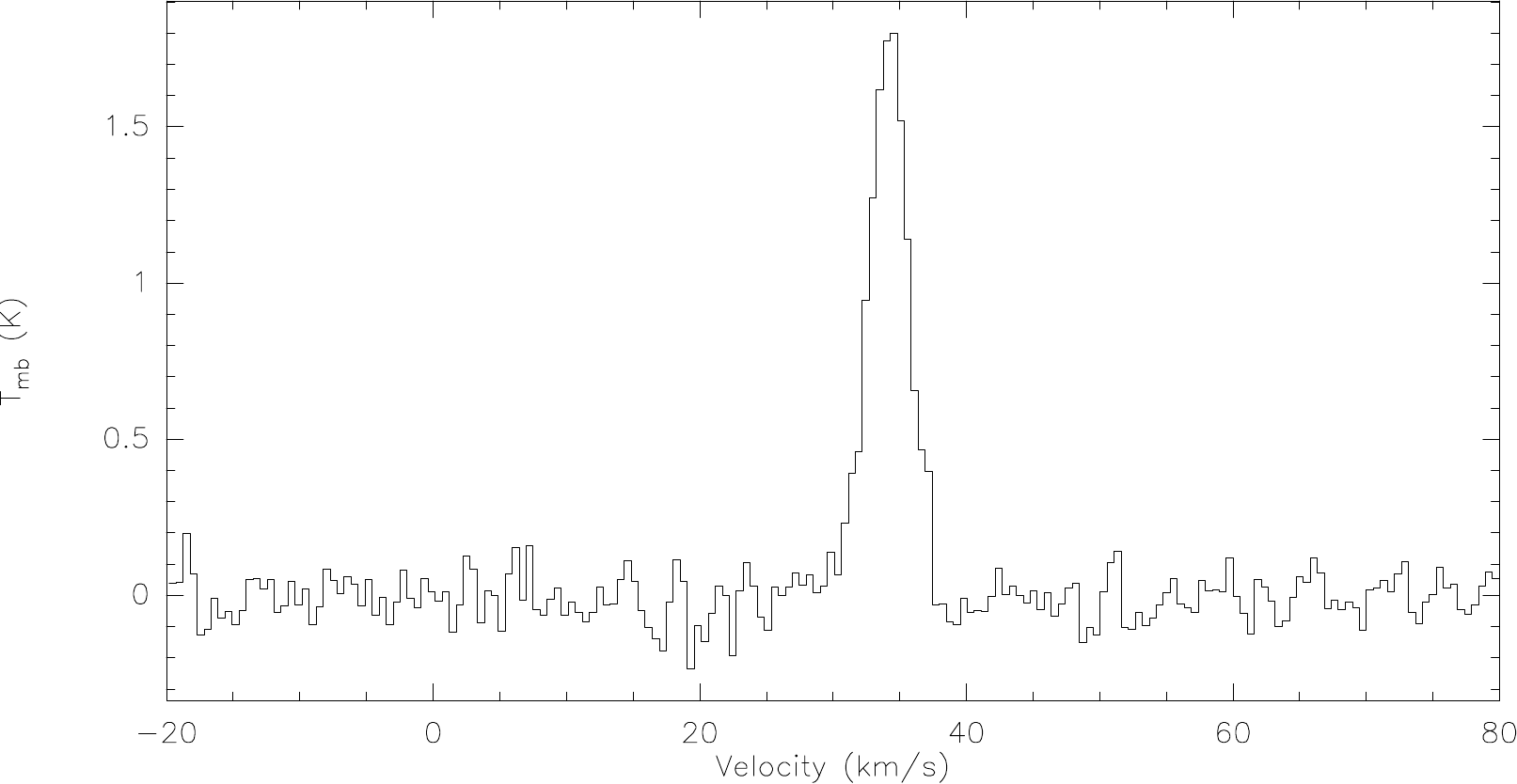}\hfill
\includegraphics[width=0.155\textwidth]{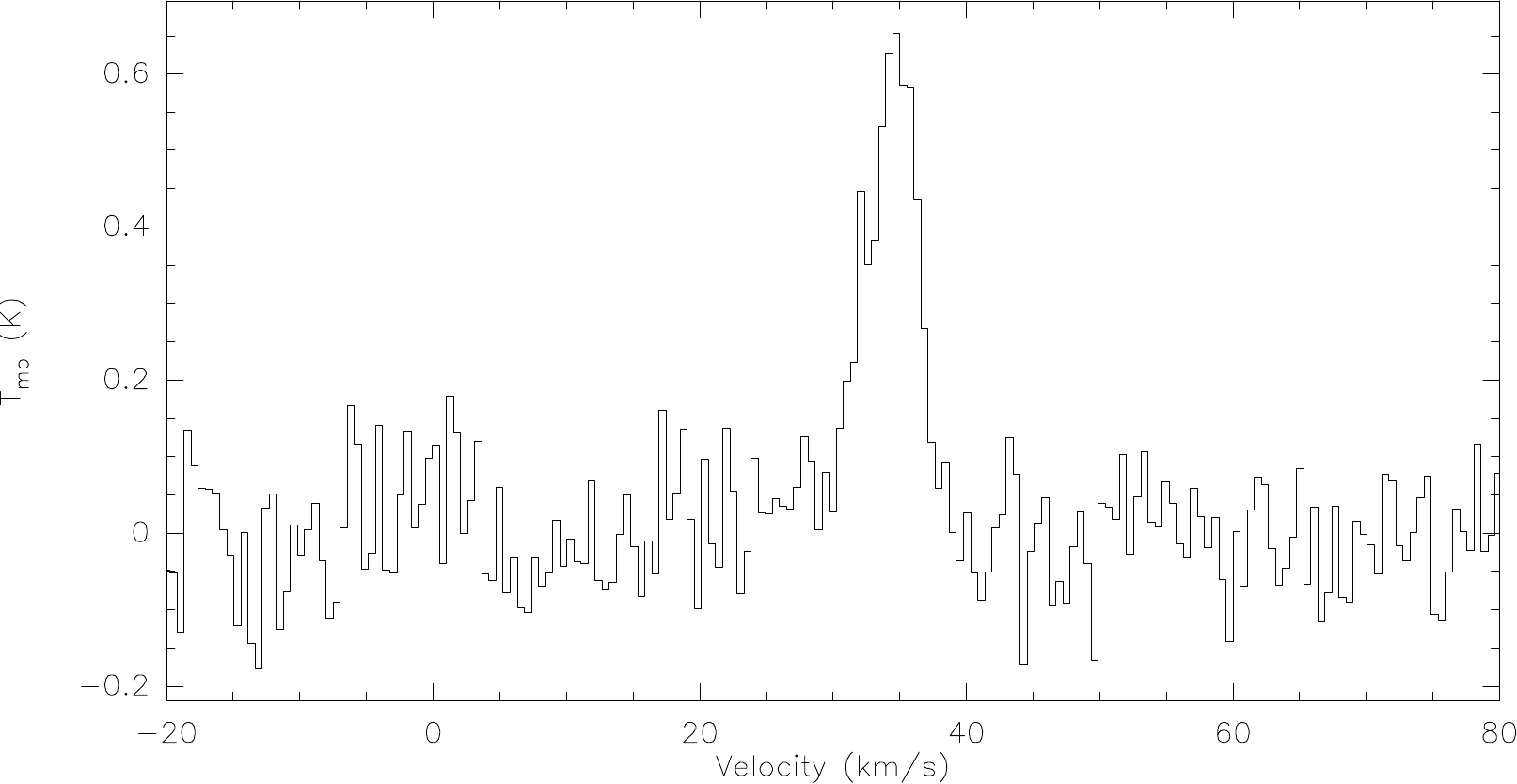}\hfill
\includegraphics[width=0.155\textwidth]{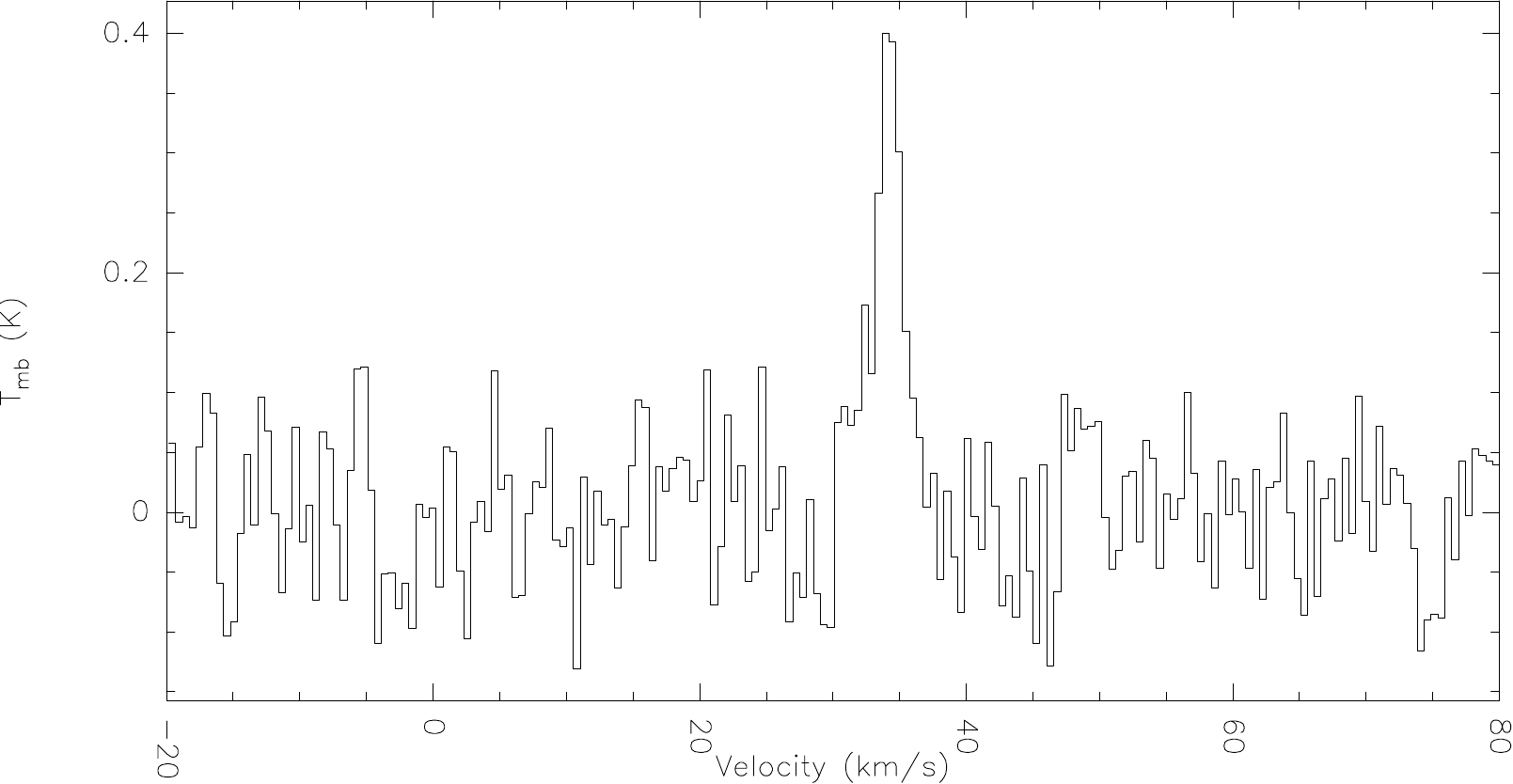}
\par

\includegraphics[width=0.155\textwidth]{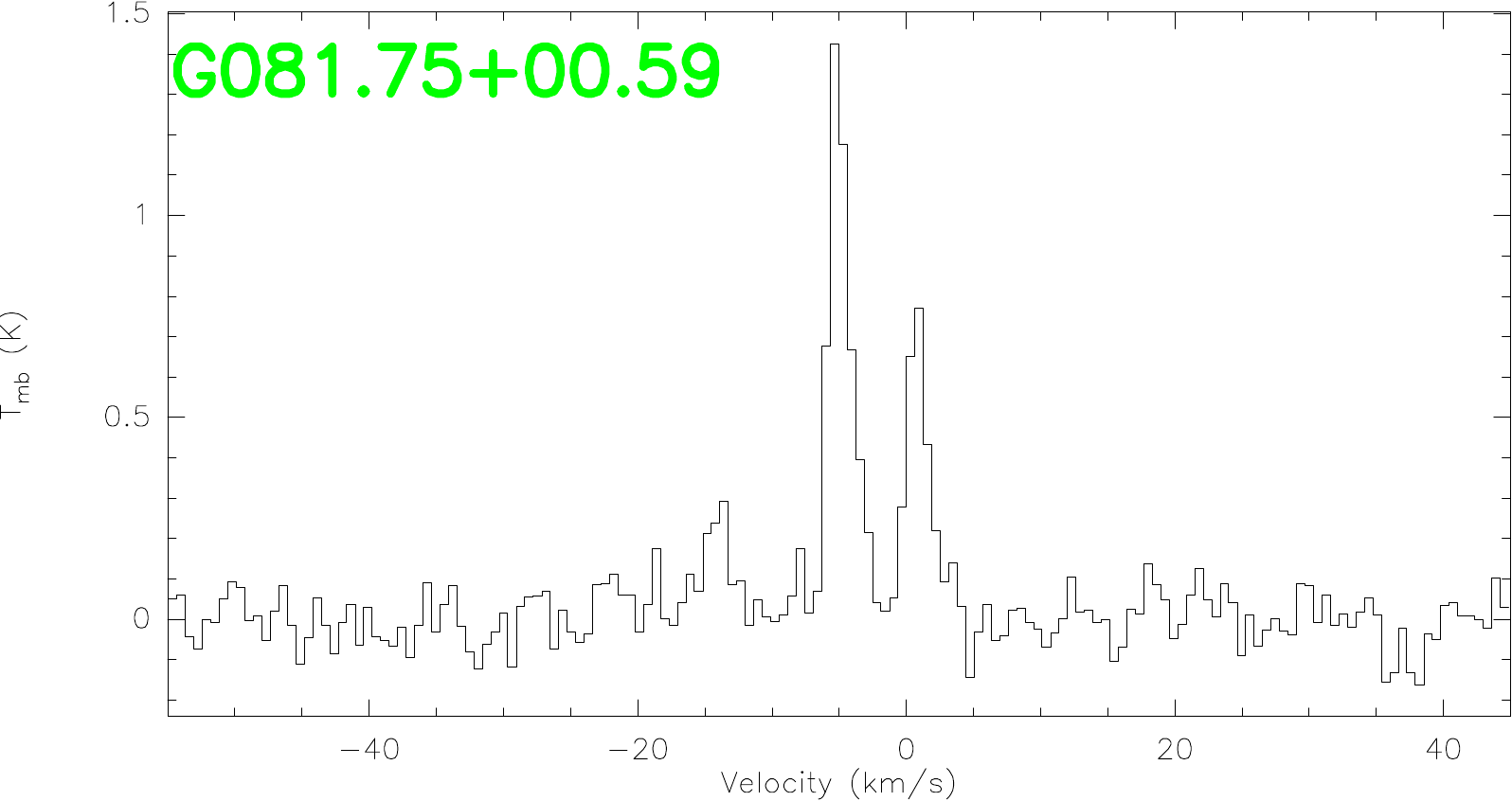}\hfill
\includegraphics[width=0.155\textwidth]{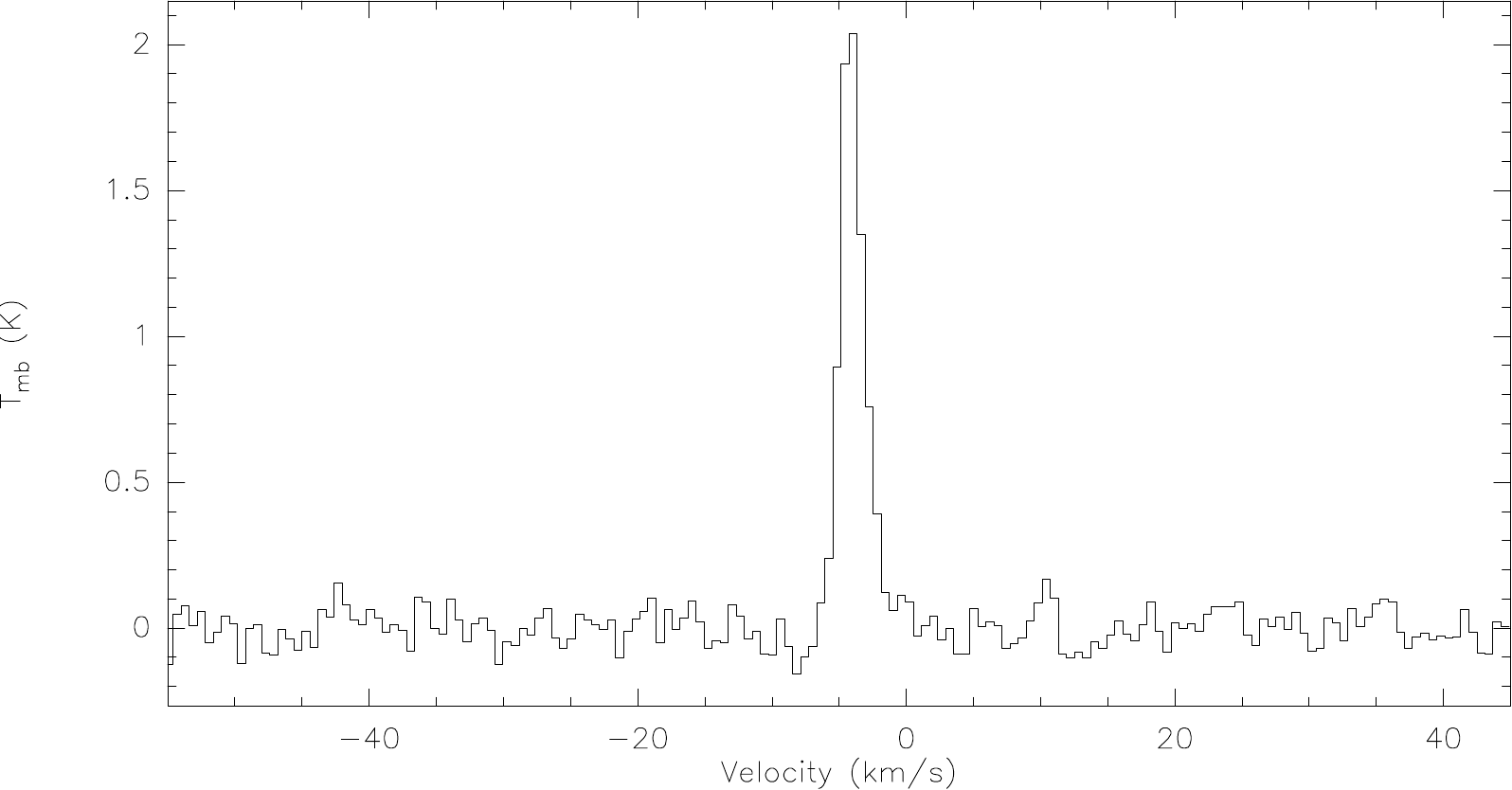}\hfill
\includegraphics[width=0.155\textwidth]{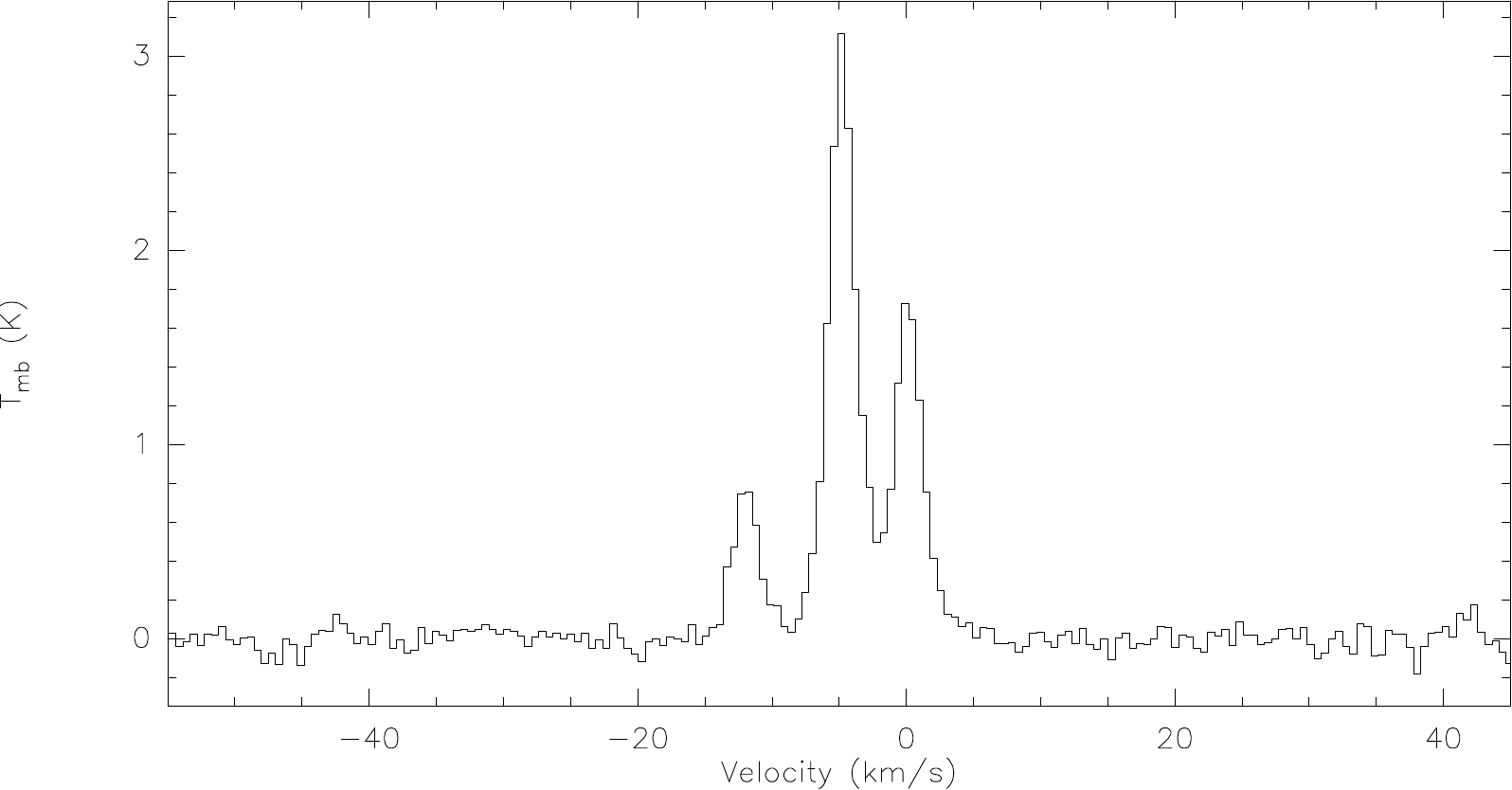}\hfill
\includegraphics[width=0.155\textwidth]{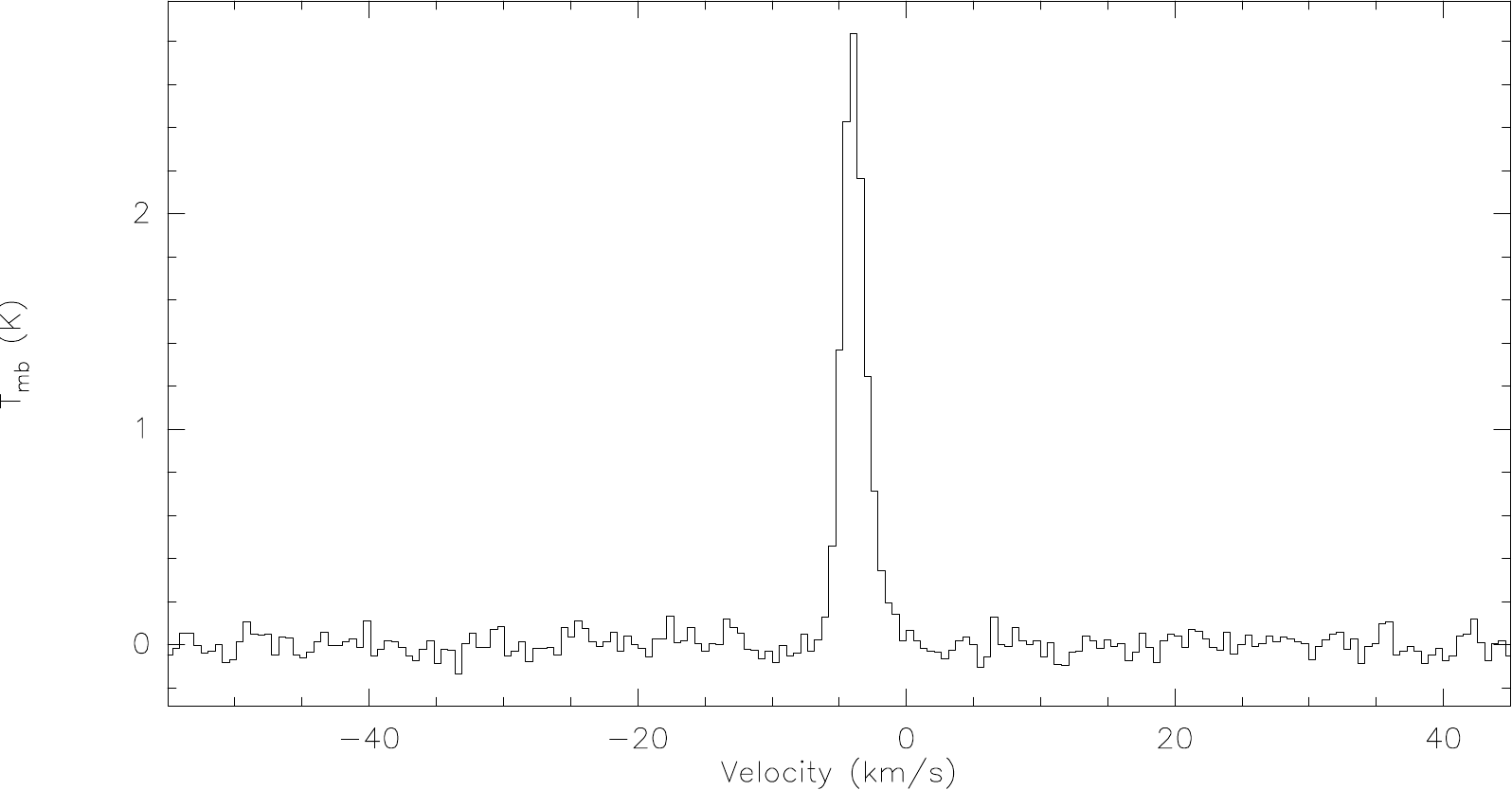}\hfill
\includegraphics[width=0.155\textwidth]{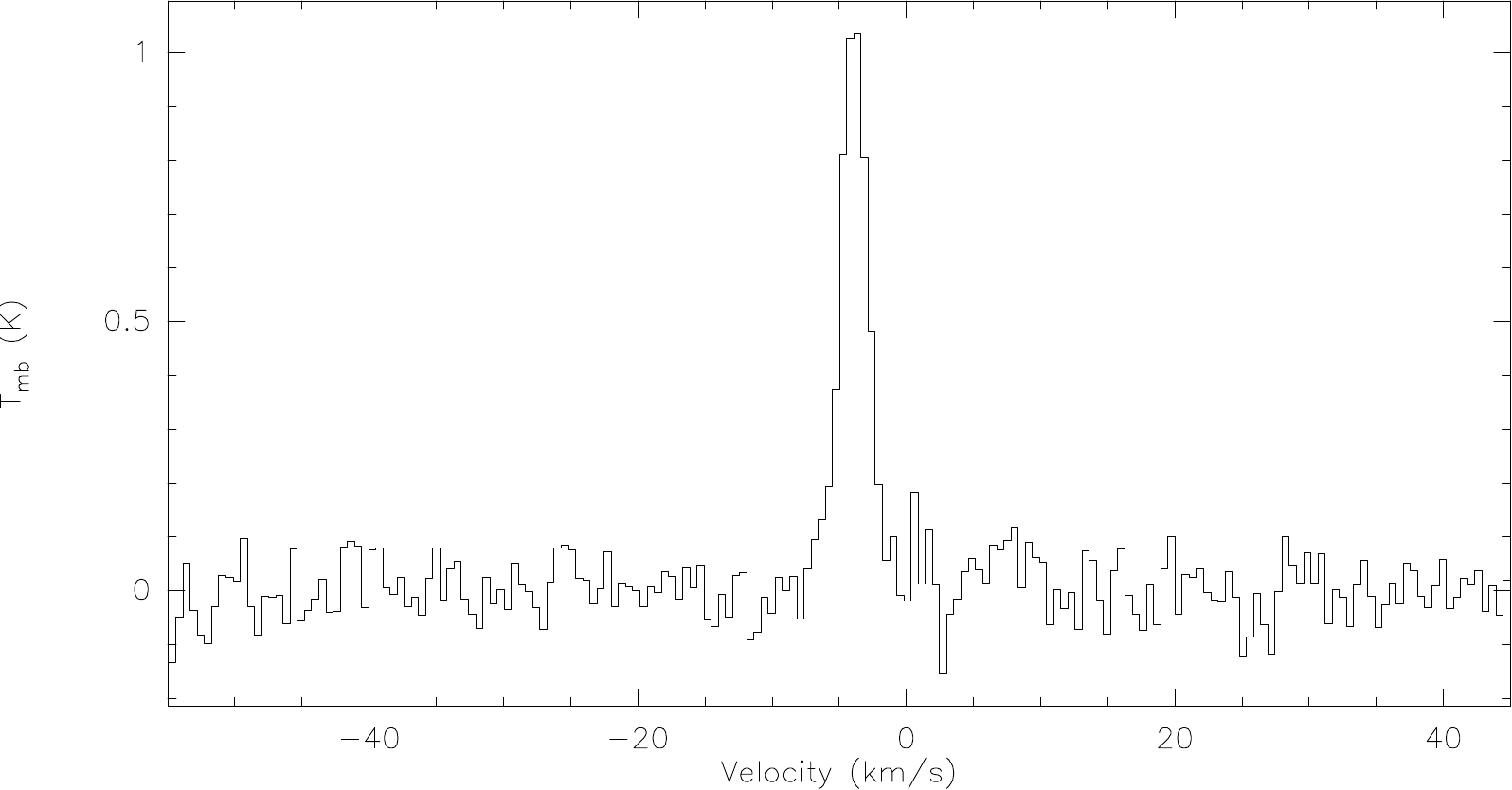}\hfill
\includegraphics[width=0.155\textwidth]{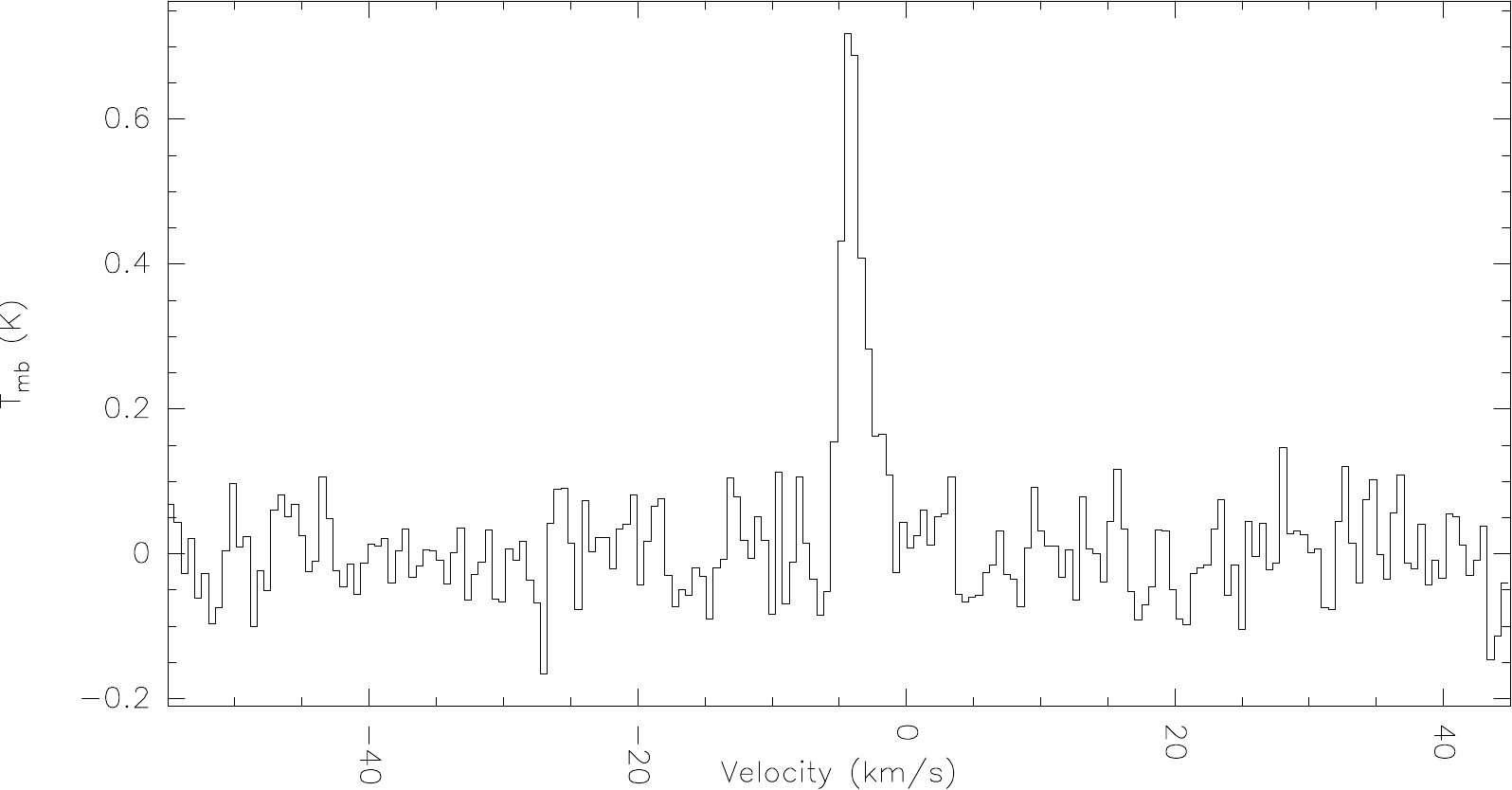}
\par

\includegraphics[width=0.155\textwidth]{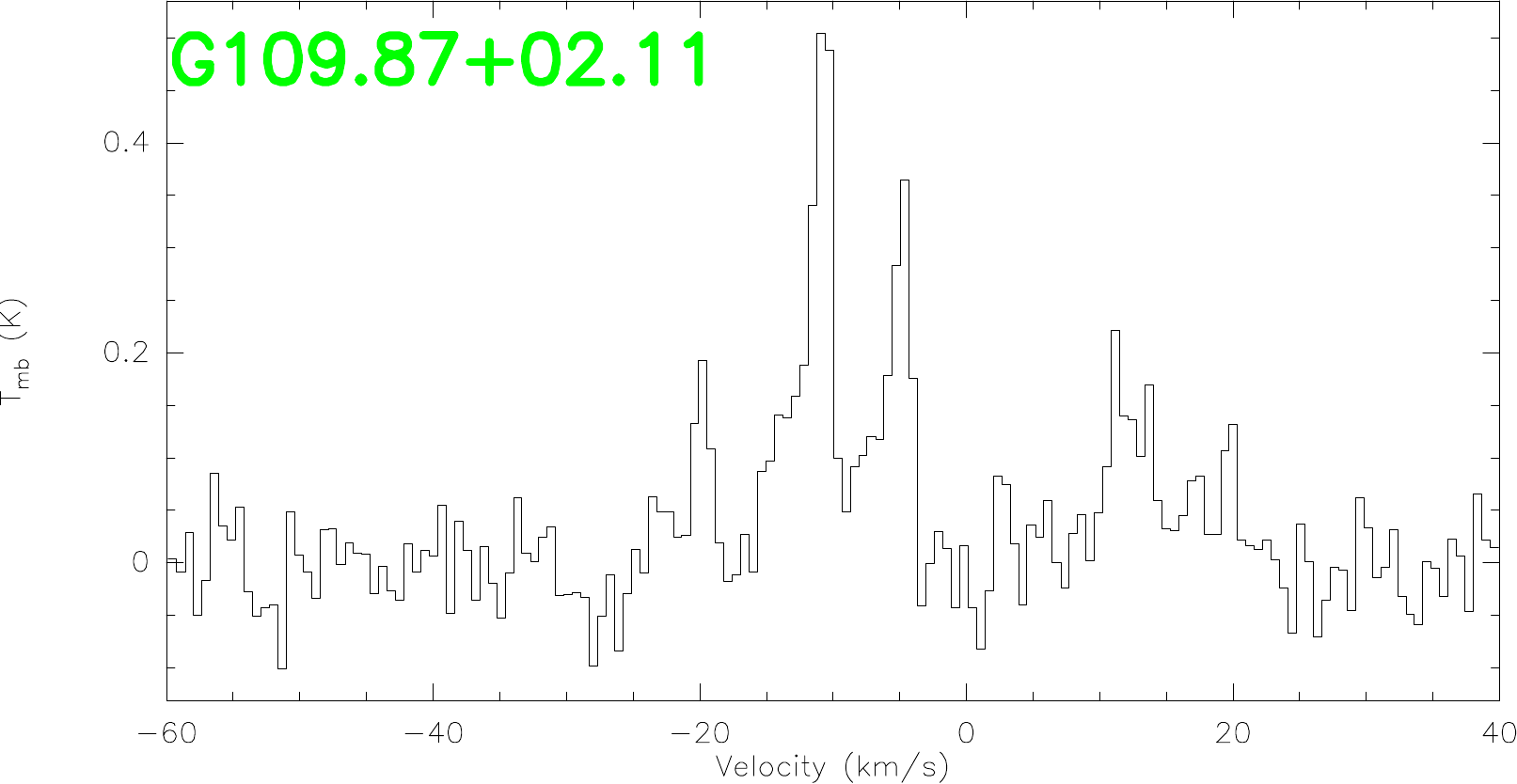}\hfill
\includegraphics[width=0.155\textwidth]{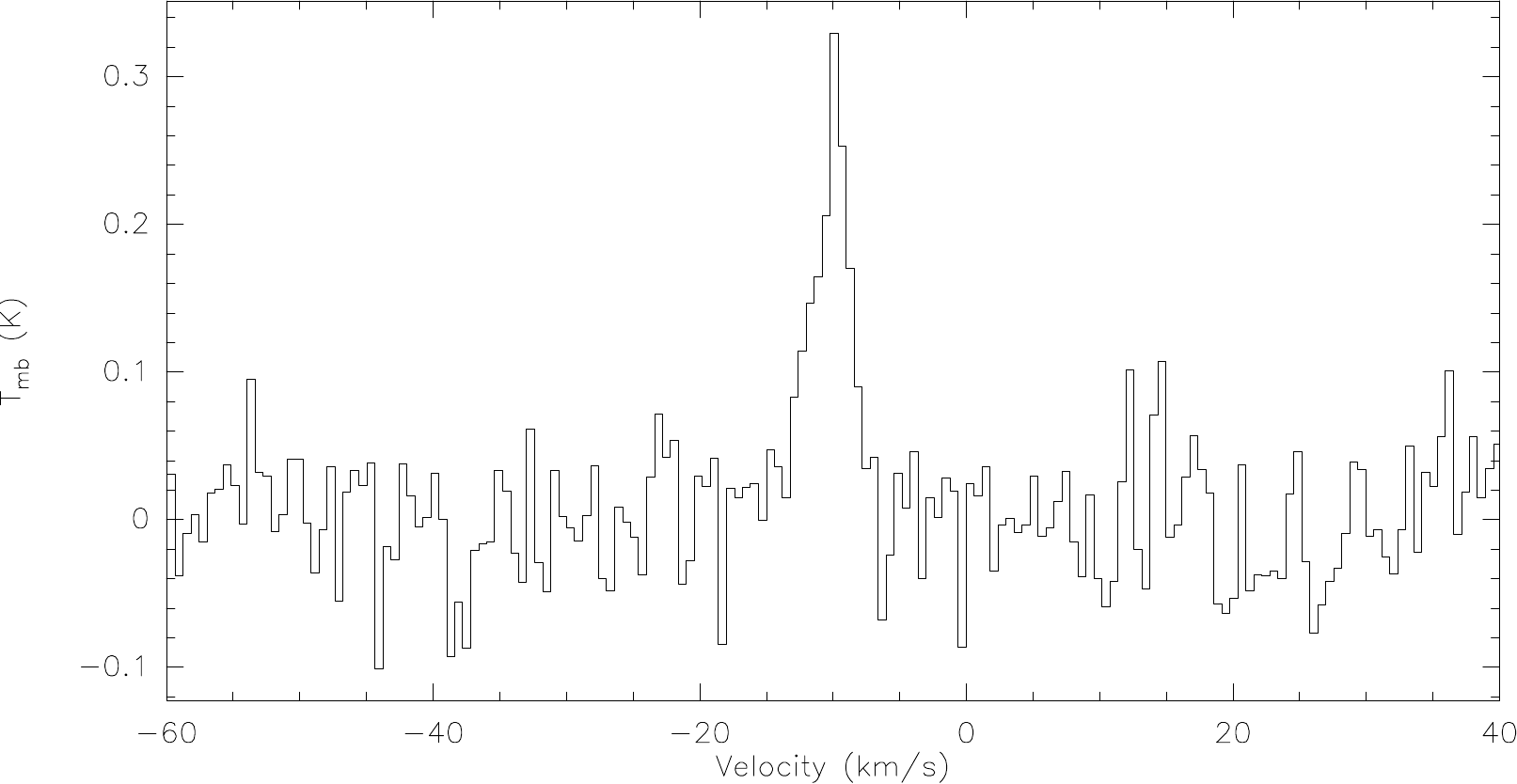}\hfill
\includegraphics[width=0.155\textwidth]{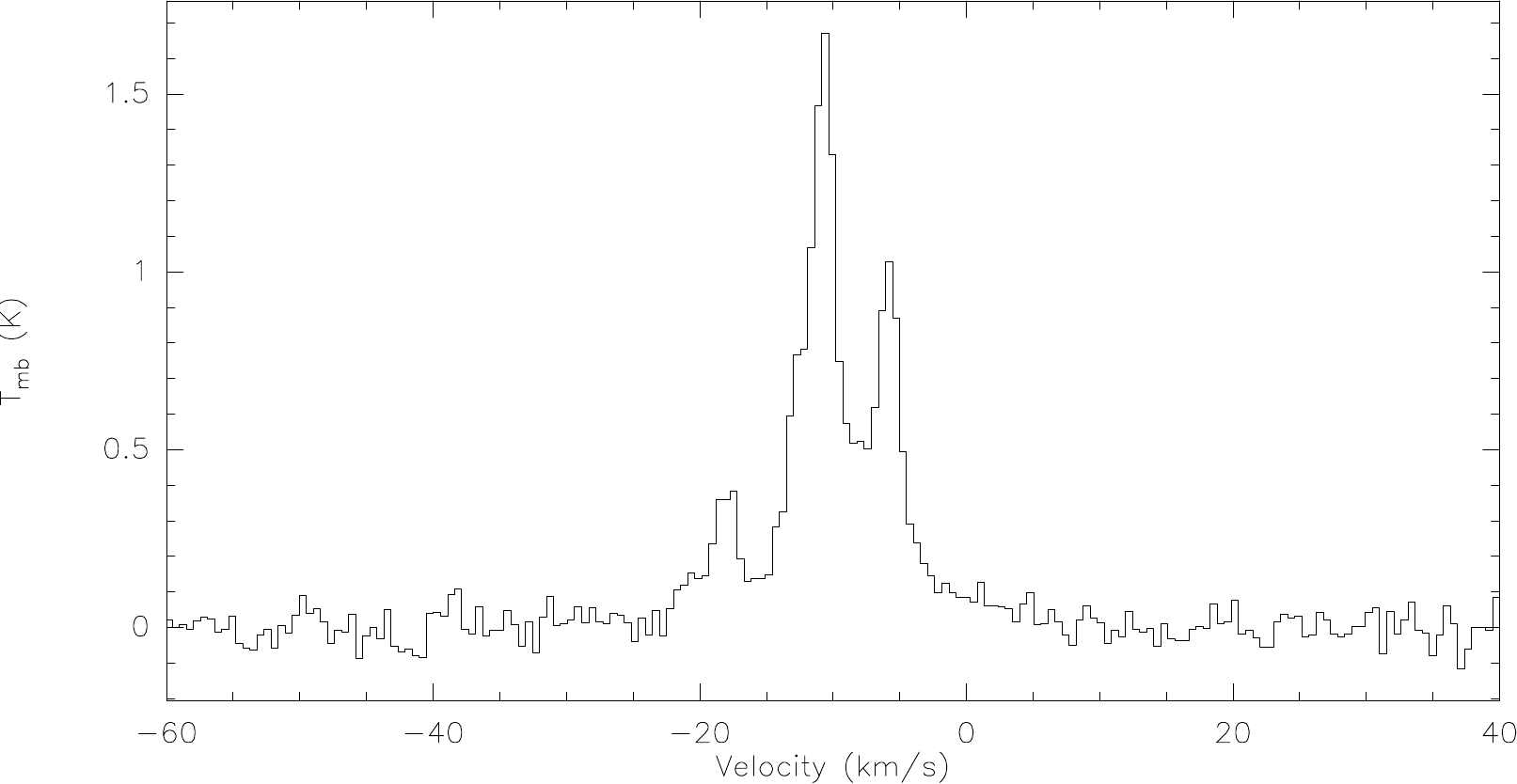}\hfill
\includegraphics[width=0.155\textwidth]{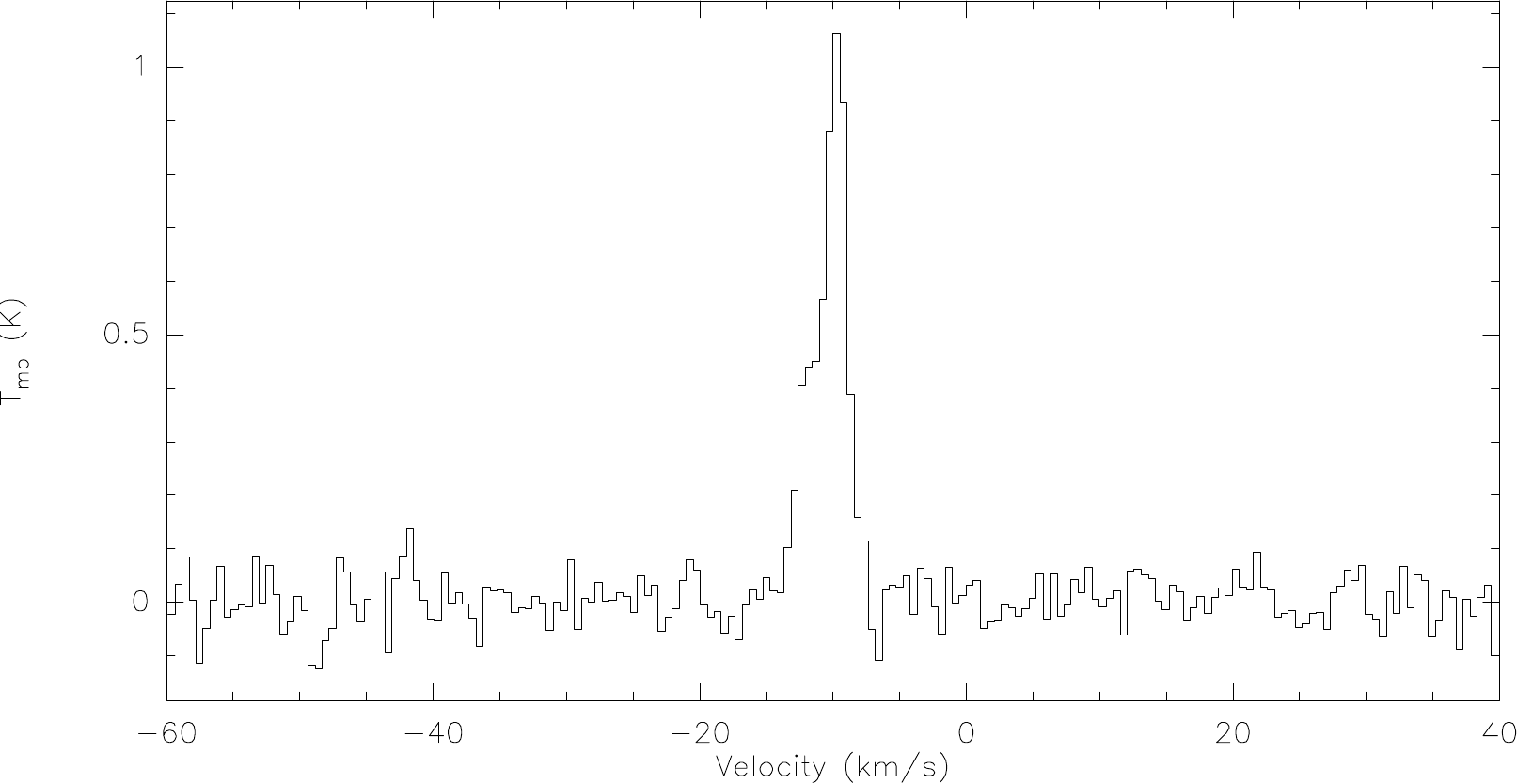}\hfill
\includegraphics[width=0.155\textwidth]{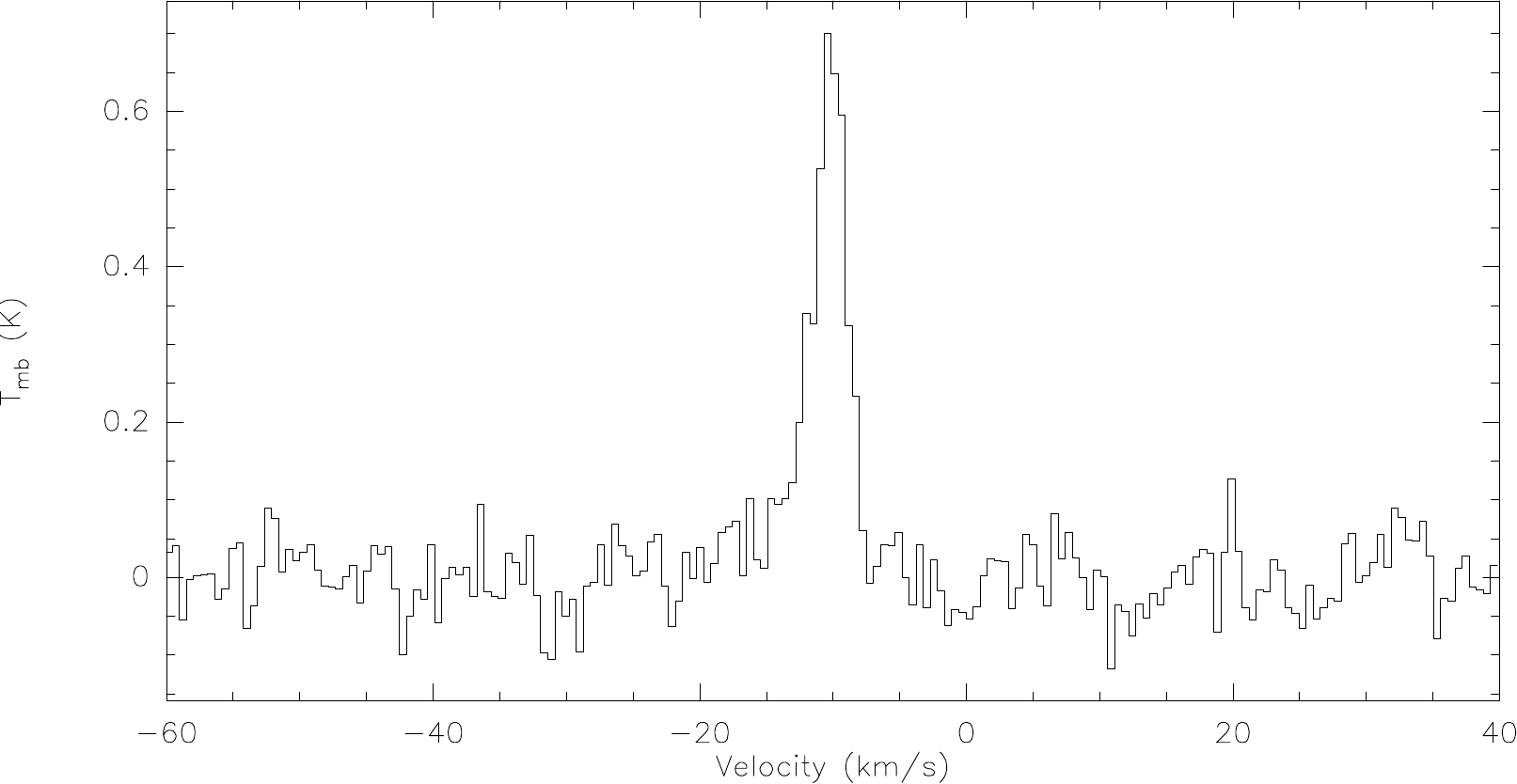}\hfill
\includegraphics[width=0.155\textwidth]{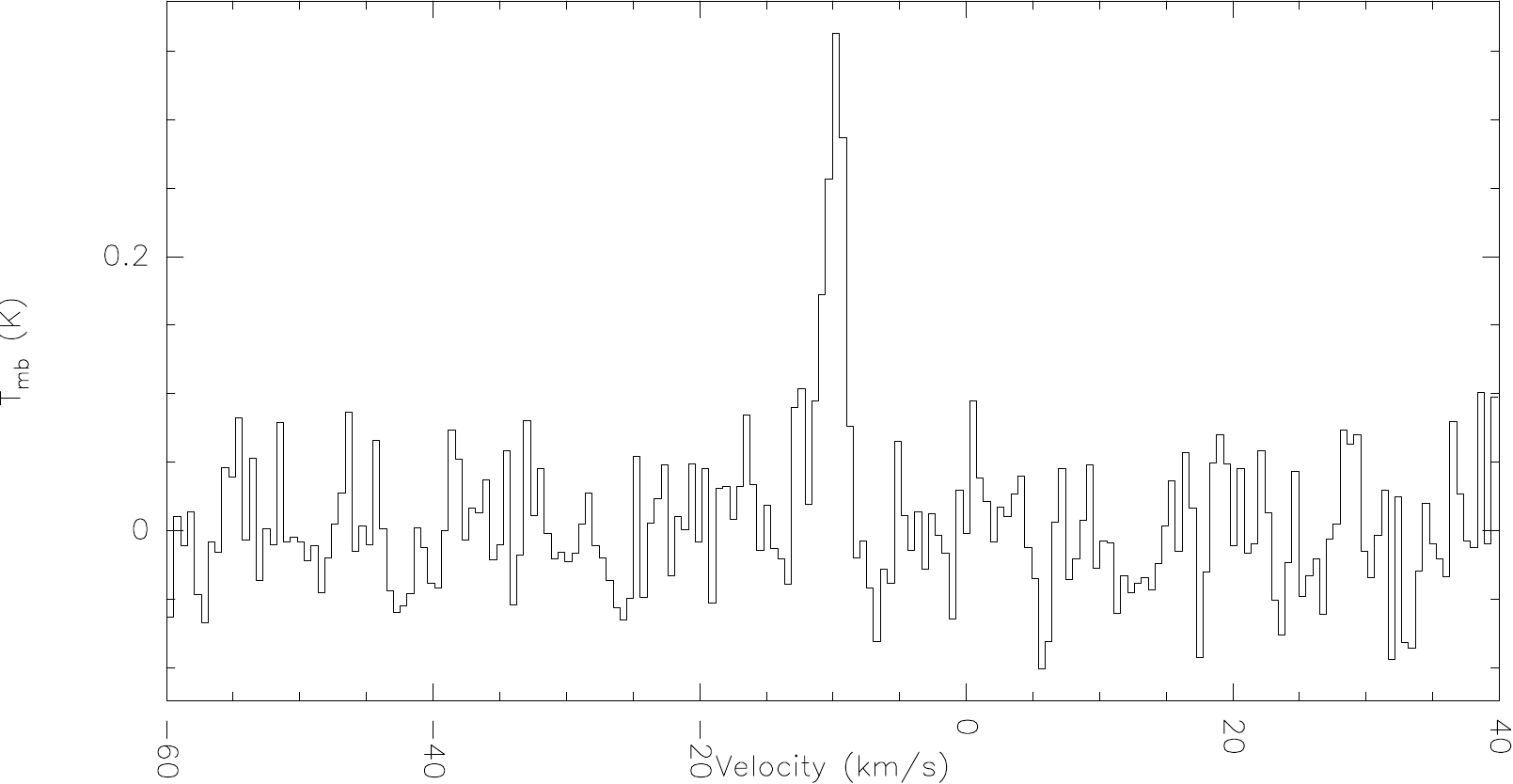}
\par

\includegraphics[width=0.155\textwidth]{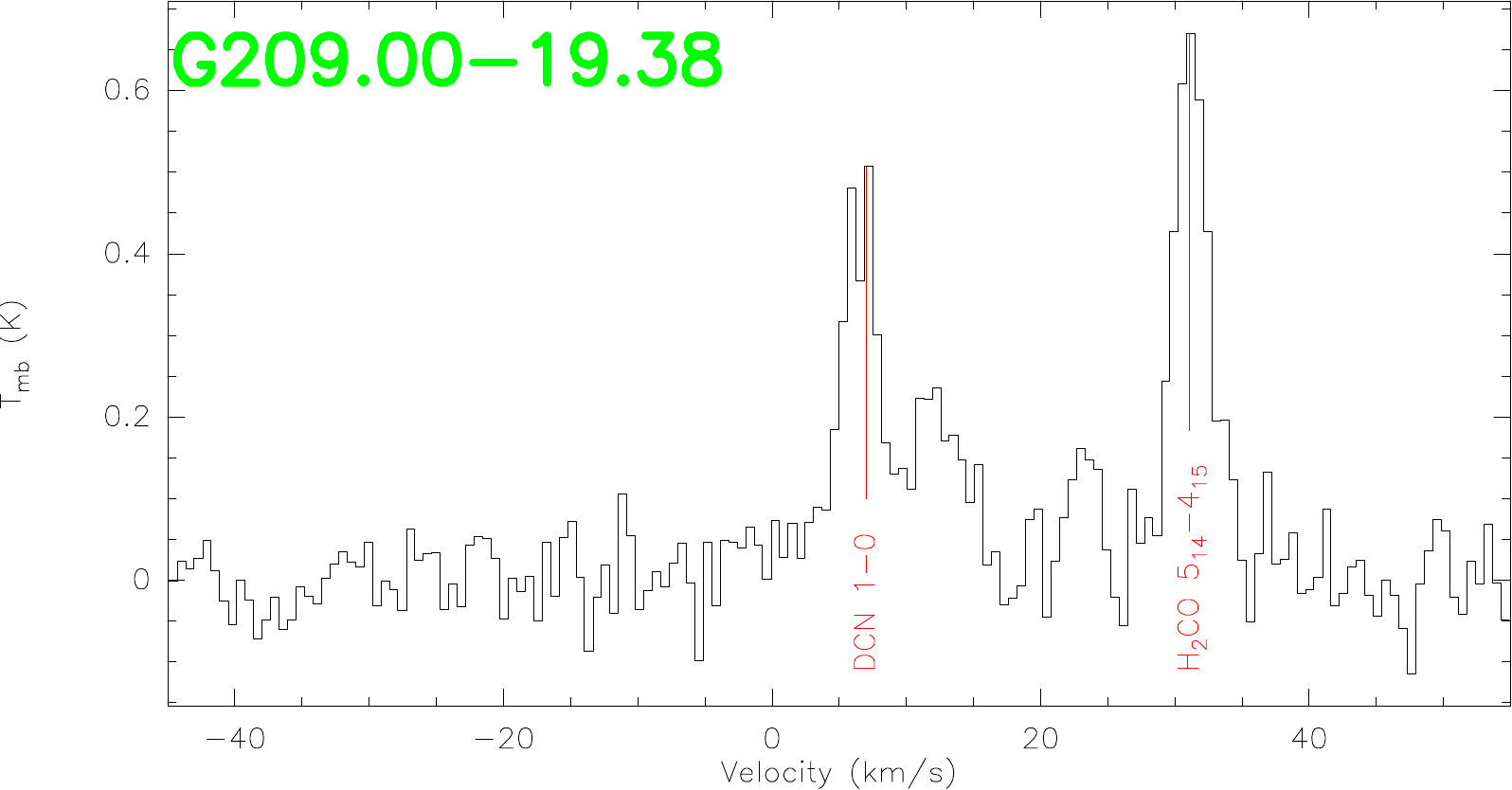}\hfill
\includegraphics[width=0.155\textwidth]{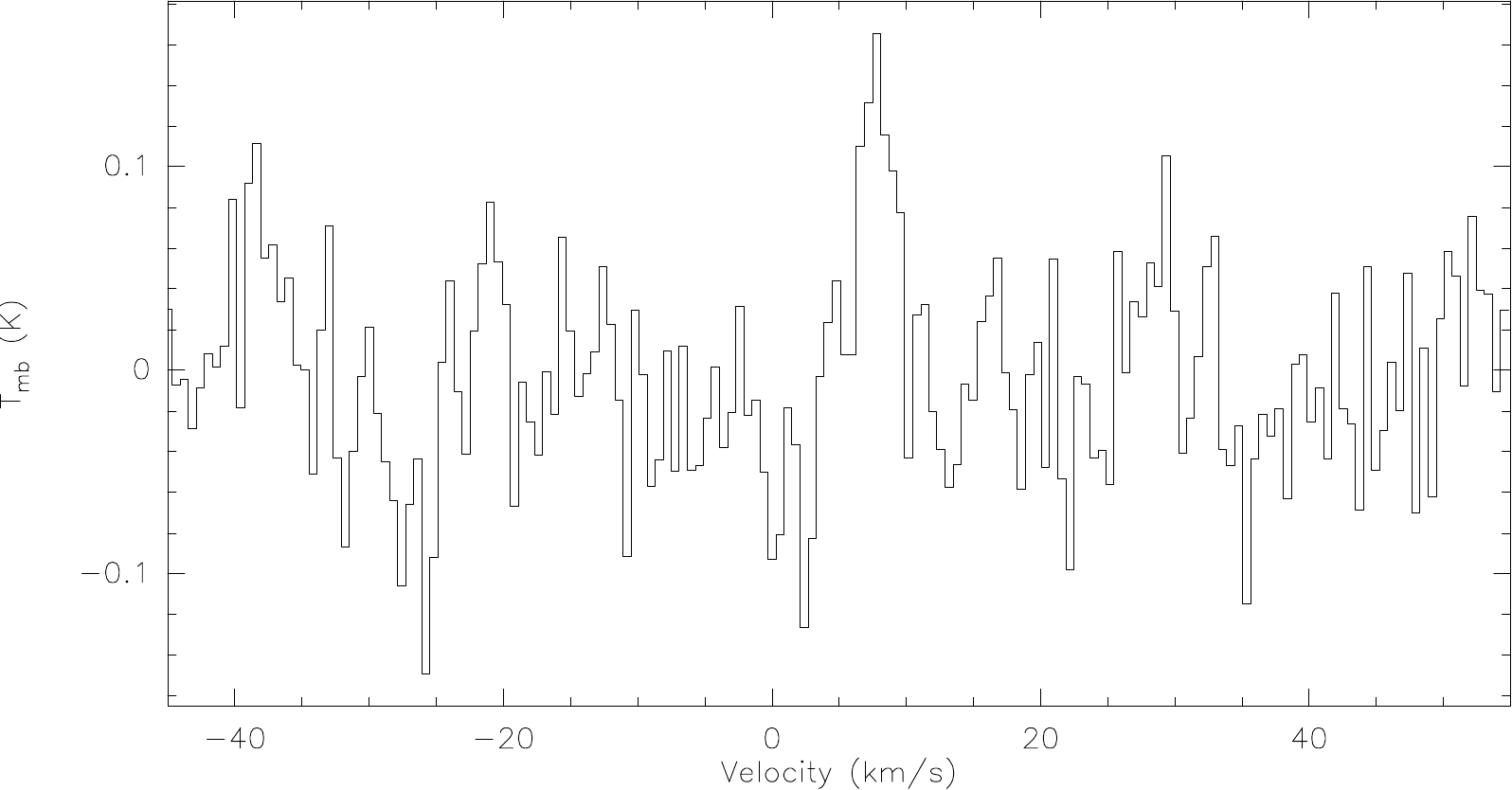}\hfill
\includegraphics[width=0.155\textwidth]{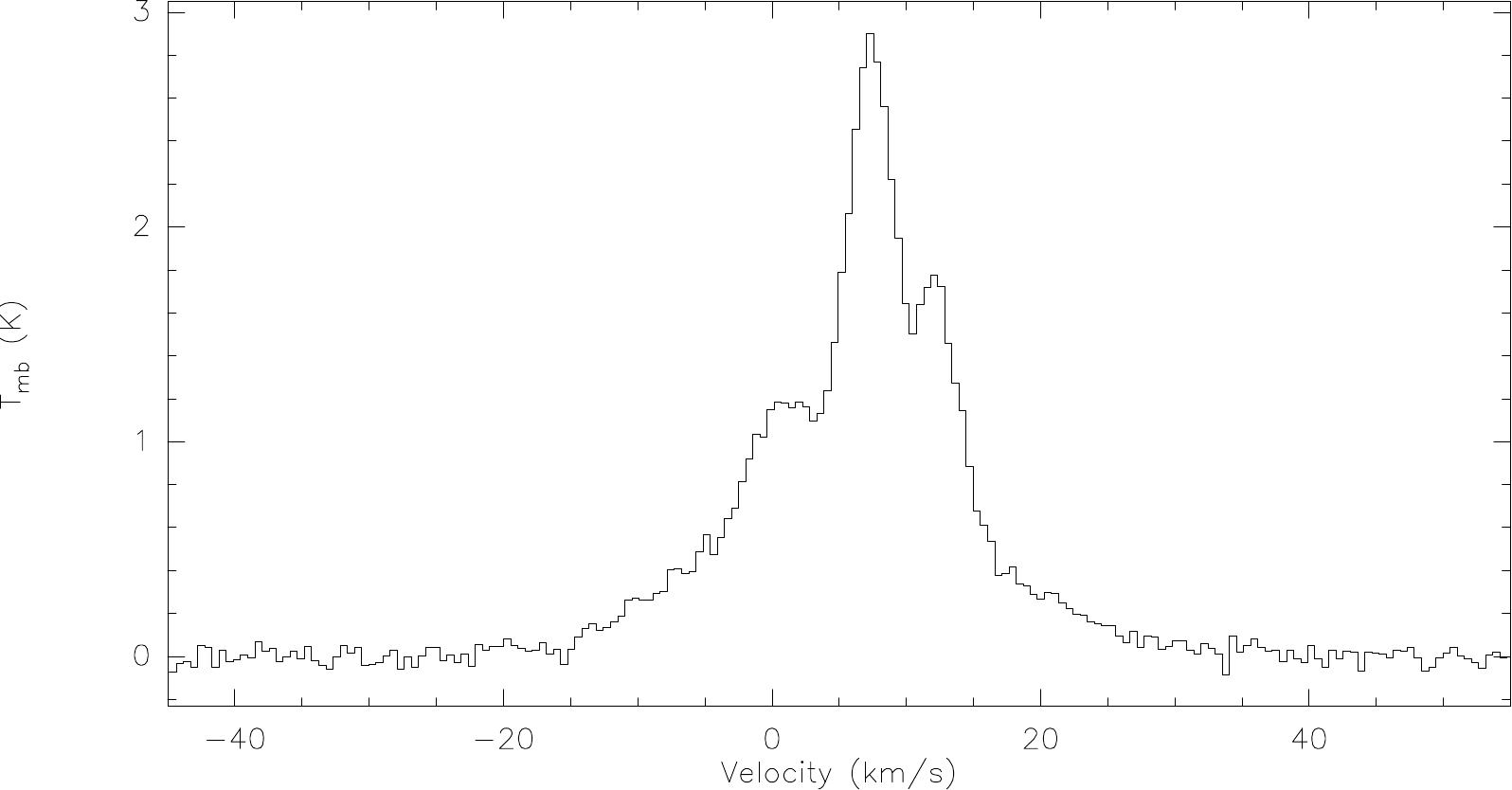}\hfill
\includegraphics[width=0.155\textwidth]{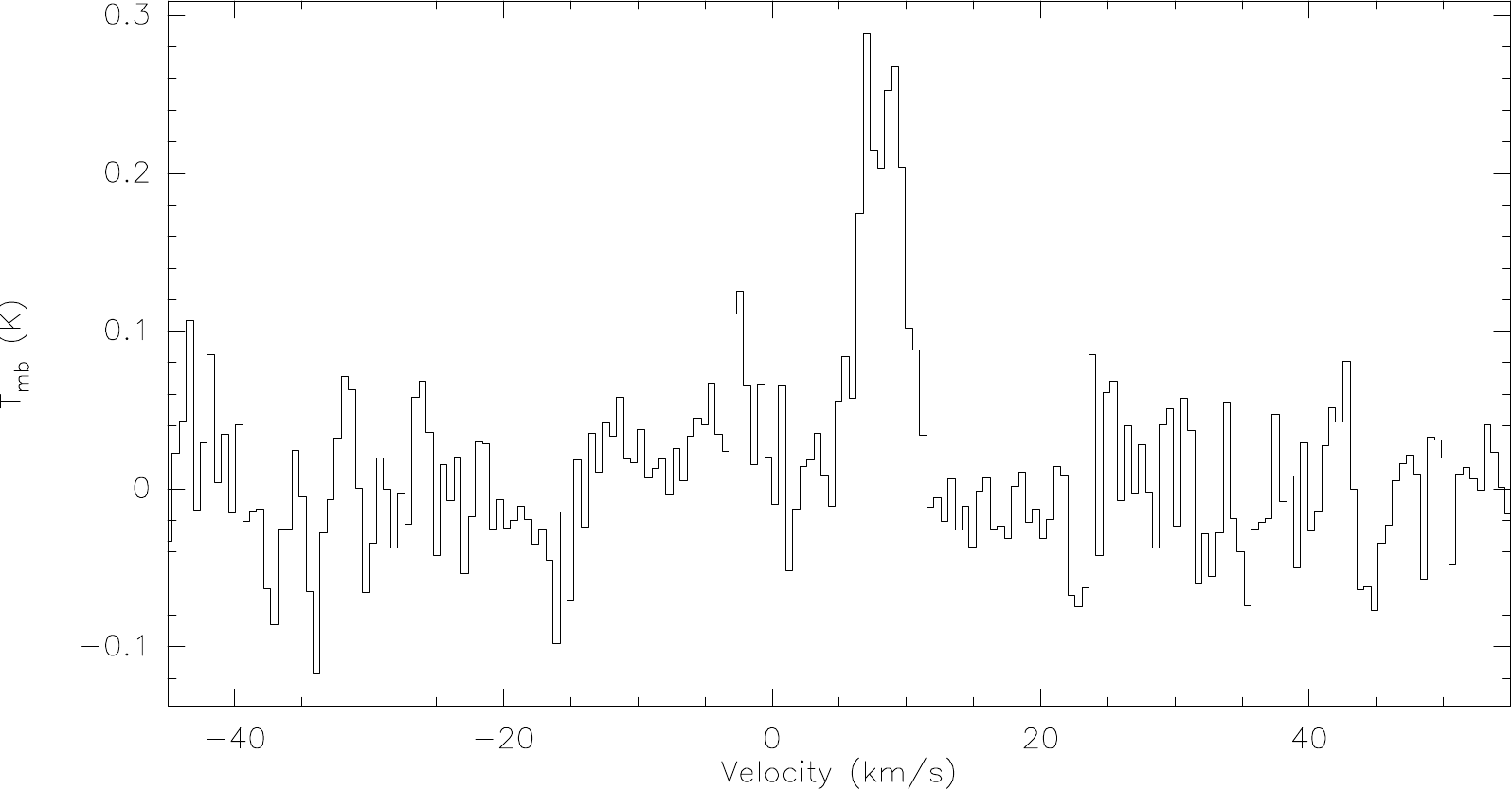}\hfill
\includegraphics[width=0.155\textwidth]{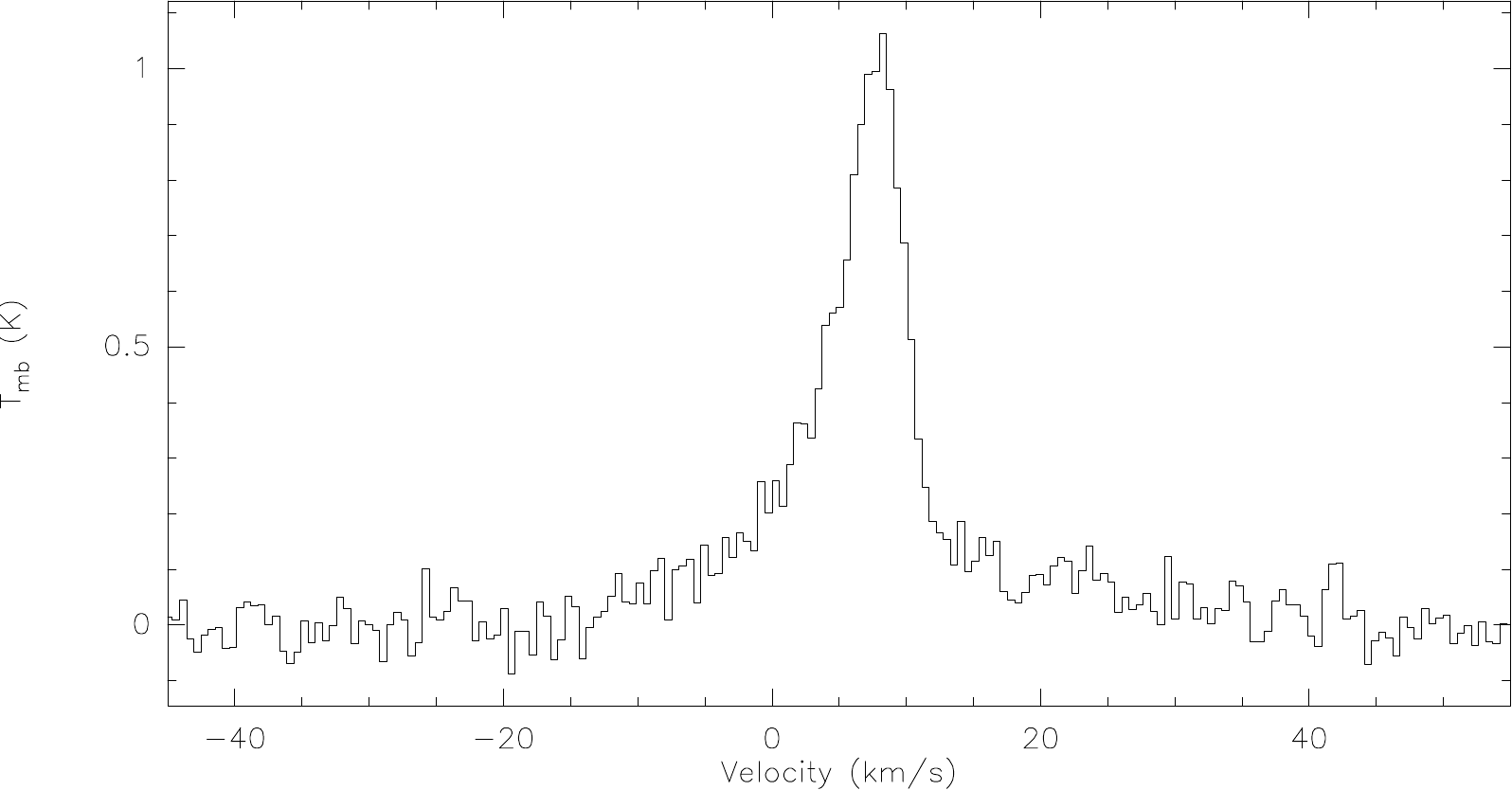}\hfill
\includegraphics[width=0.155\textwidth]{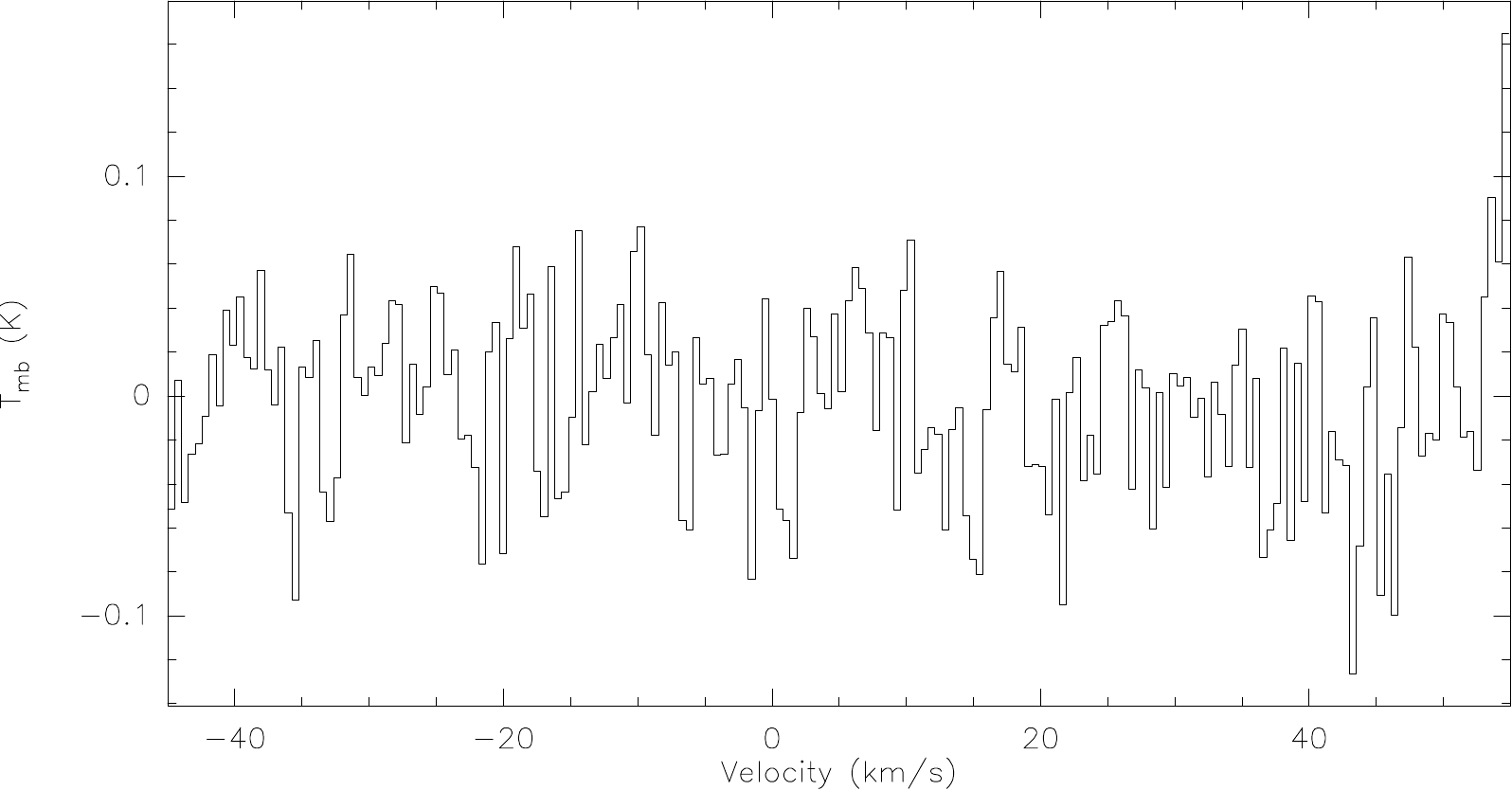}
\par

\caption{Spectra of the six observed molecular transitions toward the selected representative sources. From left to right: DCN 1-0, DNC 1-0, H$^{13}$CN 1-0, HN$^{13}$C 1-0, HC$^{15}$N 1-0, and H$^{15}$NC 1-0. The complete spectra for all 101 sources are provided online.}
\label{fig: spec}
\end{figure*}

DCN 1-0, DNC 1-0, H$^{13}$CN 1-0, HN$^{13}$C 1-0, HC$^{15}$N 1-0, and H$^{15}$NC 1-0 were detected in 62, 60, 92, 78, 79, and 46 sources, respectively, corresponding to detection rates of approximately 61\%, 59\%, 91\%, 77\%, 78\%, and 46\%, respectively. For the abundance ratio analysis, both transitions involved in a given ratio are required to be detected to obtain the ratio measurement. Representative spectra of the six observed transitions are shown in Fig. \ref{fig: spec}. The complete spectra for all 101 sources are provided as online material. 

The velocity-integrated intensities of DNC 1-0, HN$^{13}$C 1-0, HC$^{15}$N 1-0, and H$^{15}$NC 1-0 were obtained from single-component Gaussian fitting. Because DCN 1-0 and H$^{13}$CN 1-0 exhibit resolved or partially blended hyperfine structure (HFS), their velocity-integrated intensities were instead calculated by directly integrating the emission over the velocity range covering all hyperfine components. 
For sources in which the target line was not detected, the 3$\sigma$ upper limit of the integrated intensity, $\int$$T_{\rm mb}\rm dv$, was estimated as
\begin{equation}
    \int T_{\rm mb} {\rm d}{\rm v} = 3\,{\rm rms}\sqrt{\delta {\rm v} \Delta {\rm v}} \quad ({\rm K\; km\; s^{-1}}),
\end{equation}
where $\delta {\rm v}$  denotes channel spacing in velocity and $\Delta {\rm v}$ is the assumed line width. For most species, $\Delta {\rm v}$ was taken to be equal to the full width at half maximum (FWHM) of the detected HN$^{13}$C 1-0 line. If HN$^{13}$C 1-0 was not detected, the median FWHM (=3.1 km s$^{-1}$) of the detected HN$^{13}$C 1-0 lines in the sample was adopted. For DCN 1-0 and H$^{13}$CN 1-0, 
which exhibit HFS, the upper limits were calculated over velocity ranges that include both the line width of each hyperfine component and the velocity spread among the hyperfine components. The hyperfine splitting itself spans 14.6 and 12.2 km s$^{-1}$ for DCN 1-0 and H$^{13}$CN 1-0, respectively. Since the lines in our sample are relatively broad, with a median HN$^{13}$C 1-0 FWHM of 3.1 km s$^{-1}$, the hyperfine components are expected to be partially blended. Therefore, this treatment captures the full expected emission without using an unnecessarily large velocity window. The velocity-integrated intensities for representative sources are presented in Table \ref{tab: line flux}, while the full set of velocity-integrated intensities is provided online.

\begin{table*}[!htbp]
\centering
\caption{Velocity-integrated intensities of the observed molecular transitions.}
\label{tab: line flux}

\scriptsize
\begin{tabular}{lcccccccccccc}
\hline
\hline
\multirow{3}{*}{Source} 
& \multicolumn{2}{c}{DCN 1-0} 
& \multicolumn{2}{c}{DNC 1-0} 
& \multicolumn{2}{c}{H$^{13}$CN 1-0} 
& \multicolumn{2}{c}{HN$^{13}$C 1-0} 
& \multicolumn{2}{c}{HC$^{15}$N 1-0} 
& \multicolumn{2}{c}{H$^{15}$NC 1-0} \\

\cmidrule(lr){2-3}
\cmidrule(lr){4-5}
\cmidrule(lr){6-7}
\cmidrule(lr){8-9}
\cmidrule(lr){10-11}
\cmidrule(lr){12-13}

& rms & $\int$$T_{\rm mb}{\rm dv}$
& rms & $\int$$T_{\rm mb}{\rm dv}$
& rms & $\int$$T_{\rm mb}{\rm dv}$
& rms & $\int$$T_{\rm mb}{\rm dv}$
& rms & $\int$$T_{\rm mb}{\rm dv}$
& rms & $\int$$T_{\rm mb}{\rm dv}$ \\

& (K) & (K $\cdot$ km\; s$^{-1}$)
& (K) & (K $\cdot$ km\; s$^{-1}$)
& (K) & (K $\cdot$ km\; s$^{-1}$)
& (K) & (K $\cdot$ km\; s$^{-1}$)
& (K) & (K $\cdot$ km\; s$^{-1}$)
& (K) & (K $\cdot$ km\; s$^{-1}$) \\

\hline

G005.88-00.39  & 0.13 & 6.85$\pm$0.51 & 0.02 & 3.70$\pm$0.18 & 0.05 & 84.38$\pm$0.29 & 0.01 & 20.32$\pm$0.15 & 0.01 & 12.06$\pm$0.19 & 0.01 & 3.38$\pm$0.13 \\
G014.33-00.64  & 0.11 & 4.55$\pm$0.41 & 0.02 & 3.49$\pm$0.18 & 0.07 & 24.24$\pm$0.33 & 0.01 & 7.23$\pm$0.12 & 0.01 & 4.00$\pm$0.14 & 0.02 & 1.04$\pm$0.11 \\ 
G035.19-00.74  & 0.08 & 3.30$\pm$0.20 & 0.02 & 1.78$\pm$0.13 & 0.07 & 19.00$\pm$0.29 & 0.02 & 6.77$\pm$0.15 & 0.02 & 3.15$\pm$0.18 & 0.01 & 1.04$\pm$0.11 \\ 
G081.75+00.59  & 0.06 & 5.58$\pm$0.24 & 0.02 & 4.58$\pm$0.11 & 0.06 & 15.13$\pm$0.21 & 0.01 & 6.10$\pm$0.08 & 0.01 & 2.71$\pm$0.09 & 0.01 & 1.48$\pm$0.09 \\  
G109.87+02.11  & 0.05 & 2.56$\pm$0.19 & 0.01 & 0.97$\pm$0.09 & 0.04 & 10.47$\pm$0.18 & 0.01 & 2.87$\pm$0.09 & 0.01 & 2.23$\pm$0.09 & 0.01 & 0.66$\pm$0.06 \\
G209.00-19.38  & 0.12 & 2.05$\pm$0.42 & 0.01 & 0.51$\pm$0.10 & 0.04 & 35.68$\pm$0.19 & 0.01 & 1.11$\pm$0.09 & 0.01 & 7.41$\pm$0.19 & 0.01 & $\leq$0.01 \\ 

\hline
\end{tabular}
\tablefoot{
The complete table for all 101 sources is shown in the online material.
}
\end{table*}


\subsection{Comparison of molecular line widths}
\begin{figure}[!htbp]
\centering
\includegraphics[width=0.5\textwidth]{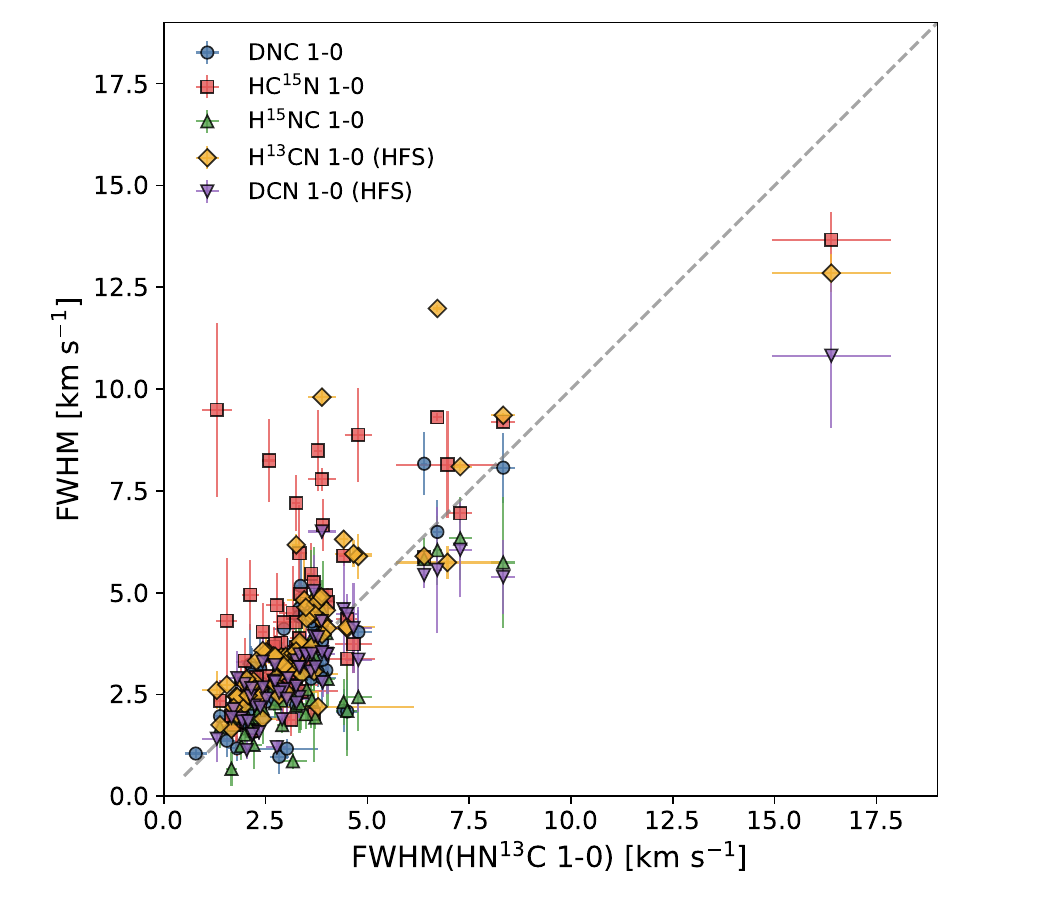}
\caption{Comparison of the FWHM of HN$^{13}$C 1-0 with those of the other five molecular transitions. The FWHMs of HN$^{13}$C 1-0, DNC 1-0, HC$^{15}$N 1-0, and H$^{15}$NC 1-0 were derived from Gaussian fitting, whereas those of H$^{13}$CN 1-0 and DCN 1-0 were derived from HFS fitting.}
\label{FWHM}
\end{figure}

The FWHMs of H$^{13}$CN 1-0 and DCN 1-0 were derived using HFS fitting in the CLASS/GILDAS package, while those of the other four transitions were obtained from single-component Gaussian fitting.
The FWHM of HN$^{13}$C 1-0 is comparable to those of the other molecular lines in most sources (see Fig. \ref{FWHM}). No clear systematic excess or deficit in FWHM is observed for H$^{13}$CN 1-0, H$^{15}$NC 1-0, DCN 1-0, and DNC 1-0. In contrast, HC$^{15}$N 1-0 exhibits noticeably broader line widths in several sources, such as G012.68-00.18, G016.58-00.05, and G023.00-00.41. These cases are associated with relatively low signal-to-noise HC$^{15}$N 1-0 spectra, for which the fitted line widths are less well constrained. Overall, except for a few sources with poorly constrained line widths due to low signal-to-noise ratios, HC$^{15}$N 1-0 shows FWHMs broadly similar to those of HN$^{13}$C 1-0. The overall similarity in line widths indicates that the different transitions trace gas with broadly similar kinematic properties. Therefore, there is no evidence that the observed temperature-dependent variations in the molecular abundance ratios are primarily caused by systematic differences in the gas components traced by the different species in our targets. The FWHM values of the six molecular lines are provided in the online material.

\subsection{Column density and deuterium fractionation}
\label{sect: column density}

Assuming optically thin emission and local thermodynamic equilibrium (LTE), we calculated the molecular column densities using 
\begin{equation}
\label{eq: column density}
	N_{\rm tot}=\frac{8\pi k\nu^2}{hc^3A_{ul}}\frac{Q(T_{\rm ex})}{\rm g_u}e^{E_u/kT_{\rm ex}}\int T_{\rm mb}\rm dv \quad (\rm cm^{-2}),
\end{equation} 
where $k=1.38\times10^{-16}$ erg/K is the Boltzmann constant, $\nu$ is the frequency of the transition, $h=6.626\times10^{-27} $erg$\cdot$s is the Planck constant, $c=2.998\times10^{10}$ cm $\rm s^{-1}$ is the speed of light, $A_{\rm ul}$ is the Einstein emission coefficient, $\rm g_u$ is the upper level degeneracy, and $E_u$ is the energy of the upper level above the ground state. The frequencies, $\rm g_u$, $A_{\rm ul}$, and $E_u$ for the six emission lines were taken from the Cologne Database for Molecular Spectroscopy (CDMS)\footnote{\url{https://cdms.astro.uni-koeln.de/classic/predictions/catalog/partition_function.html}} \citep{2001A&A...370L..49M,2005JMoSt.742..215M}. The parameters of the transitions used in this work are listed in Table \ref{parameters}. The value $Q(T_{\rm ex})$ is the partition function, calculated by summing the contributions from all energy levels obtained from CDMS using 
\begin{equation}
	Q(T_{\rm ex})=\sum_{i} g_i \exp\left(-\frac{E_i}{k_B T_{\rm ex}}\right).
\end{equation} 
Gas kinetic temperatures ($T_{\rm{kin}}$) derived from the CH$_3$CCH 5-4 for 54 targets in our sample were reported by \citet{2026ApJ...998..182L}. We adopted $T_{\rm{kin}}$ as the excitation temperature ($T_{\rm{ex}}$) under the LTE assumption. In the following analysis, $T_{\rm kin}$ therefore denotes the adopted gas temperature. Because of non-detections or weak emission, a reliable $T_{\rm{kin}}$ could not be derived from CH$_3$CCH 5-4 in some sources, so we adopted the median $T_{\rm{kin}}$ (=33.1 K) value of the successfully measured sample as their $T_{\rm{ex}}$. This assumption may introduce additional uncertainties in the 
absolute column densities. However, the abundance ratios discussed in this work are much less sensitive to the adopted $T_{\rm ex}$, because the temperature-dependent terms in the column density calculation largely cancel out when ratios are taken between molecular column densities derived from transitions with similar upper level energies. 
We verified that, within the temperature range of our sample, the temperature-dependent factors in the column density formula (Eq. \ref{eq: column density}), excluding the integrated intensity, introduce variations of less than $\sim$10\% in the derived column density ratios. Therefore, the temperature dependence introduced by the column density calculation itself has only a minor effect on the abundance ratios and is unlikely to dominate the observed trends. Besides, sources without kinetic temperatures derived from CH$_3$CCH 5-4 were used only for the column density and abundance ratio estimates and were excluded from the abundance ratio versus $T_{\rm kin}$ analysis.

\begingroup
\catcode`\&=12
\gdef\CDMSkey{2001A&A...370L..49M,2005JMoSt.742..215M}
\endgroup

\newcommand{\CDMScites}{\citep{\CDMSkey}}

\begin{table}[!htbp]
\centering
\captionof{table}{Spectroscopic parameters of the observed transitions.}
\label{parameters}
\begin{tabular}{lcccc}
\hline
\hline
Transition & $\nu$  & $A_{ul}$ & $g_{\rm up}$ & $E_u$ \\
            & (MHz) & (s$^{-1}$)        &              & (K)  \\
\hline
DCN 1-0               & 72414.6936 & 1.32$\times$10$^{-5}$ & 9 & 3.48 \\
DNC 1-0               & 76305.6995 & 1.60$\times$10$^{-5}$ & 3 & 3.67 \\
HC$^{15}$N 1-0   & 86054.9664 & 2.20$\times$10$^{-5}$ & 3 & 4.13 \\
H$^{13}$CN 1-0   & 86339.9214 & 2.23$\times$10$^{-5}$ & 9 & 4.14 \\
HN$^{13}$C 1-0   & 87090.8252 & 2.38$\times$10$^{-5}$ & 3 & 4.18 \\
H$^{15}$NC 1-0   & 88865.7150 & 1.98$\times$10$^{-5}$ & 3 & 4.26 \\
\hline
\end{tabular}
\tablefoot{
These parameters were taken from CDMS \citep{2001A&A...370L..49M,2005JMoSt.742..215M}.
}
\end{table}

The derived column densities of  DCN, DNC, H$^{13}$CN, HN$^{13}$C, HC$^{15}$N, and H$^{15}$NC for representative sources are listed in Table \ref{tab: column densities}. The complete column density table for the full sample is provided online. The derived column densities range from 0.27$\times$10$^{13}$ to 5.50$\times$10$^{13}$ cm$^{-2}$ for DCN, from 0.13$\times$10$^{13}$ to 2.52$\times$10$^{13}$ cm$^{-2}$ for DNC, from 0.22$\times$10$^{13}$ to 47.34$\times$10$^{13}$ cm$^{-2}$ for H$^{13}$CN, from 0.11$\times$10$^{13}$ to 10.75$\times$10$^{13}$ cm$^{-2}$ for HN$^{13}$C, from 0.11$\times$10$^{13}$ to 10.27$\times$10$^{13}$ cm$^{-2}$ for HC$^{15}$N, and from 0.09$\times$10$^{13}$ to 2.20$\times$10$^{13}$ cm$^{-2}$ for H$^{15}$NC. 
G000.67-00.03 (Sgr B2) is included with the online material for completeness, but it is excluded from the follow-up analysis because of its complex chemical environment \citep[e.g., ][]{2013A&A...559A..47B} and multiple velocity components \citep[e.g., ][]{2019A&A...623A..68T}.

\begin{table*}[!htbp]
\centering
\caption{Molecular column densities.}
\label{tab: column densities}
\begin{tabular}{lcccccc}
\hline\hline
Source 
& $N_{\rm tot}$(DCN) 
& $N_{\rm tot}$(DNC) 
& $N_{\rm tot}$(H$^{13}$CN) 
& $N_{\rm tot}$(HN$^{13}$C) 
& $N_{\rm tot}$(HC$^{15}$N) 
& $N_{\rm tot}$(H$^{15}$NC) \\
 & ($\times$10$^{13}$ cm$^{-2}$) & ($\times$10$^{13}$ cm$^{-2}$) & ($\times$10$^{13}$ cm$^{-2}$) &
($\times$10$^{13}$ cm$^{-2}$) & ($\times$10$^{13}$ cm$^{-2}$) & ($\times$10$^{13}$ cm$^{-2}$) \\
\hline
G005.88-00.39  & 5.35$\pm$0.40 & 2.52$\pm$0.12 & 47.34$\pm$0.17 & 10.75$\pm$0.08 & 6.81$\pm$0.10 & 2.19$\pm$0.08 \\ 
G014.33-00.64  & 2.41$\pm$0.22 & 1.61$\pm$0.08 & 9.29$\pm$0.13 & 2.61$\pm$0.04 & 1.54$\pm$0.06 & 0.46$\pm$0.05 \\ 
G035.19-00.74  & 1.92$\pm$0.12 & 0.90$\pm$0.07 & 7.97$\pm$0.12 & 2.68$\pm$0.06 & 1.33$\pm$0.07 & 0.51$\pm$0.05 \\ 
G081.75+00.59  & 2.95$\pm$0.13 & 2.11$\pm$0.05 & 5.79$\pm$0.08 & 2.20$\pm$0.03 & 1.04$\pm$0.03 & 0.66$\pm$0.04 \\ 
G109.87+02.11  & 1.53$\pm$0.11 & 0.51$\pm$0.05 & 4.61$\pm$0.06 & 1.17$\pm$0.04 & 0.97$\pm$0.04 & 0.33$\pm$0.03 \\ 
G209.00-19.38  & 1.29$\pm$0.27 & 0.28$\pm$0.05 & 16.16$\pm$0.09 & 0.47$\pm$0.04 & 3.38$\pm$0.09 & $\leq$0.09 \\ 
\hline
\end{tabular}
\tablefoot{
The complete table for all 101 sources is provided online.
}
\end{table*}

\subsection{Abundance ratio and gas temperature}
\label{sect: abundance ratio}
The deuterium fractionation, [DCN]/[HCN] and [DNC]/[HNC] (hereafter denoted as $D_{\rm frac}$(HCN) and $D_{\rm frac}$(HNC), respectively) and the abundance ratios [DCN]/[DNC] and [HCN]/[HNC] for our representative sources are listed in Table \ref{tab: ratios}. The full table, including upper and lower limits, is provided online. The HCN and HNC column densities were first derived from H$^{13}$CN and HN$^{13}$C using the $^{12}$C/$^{13}$C ratio as a function of galactocentric distance, $D_{\rm gc}$, given by $^{12}$C/$^{13}$C=4.08$D_{\rm gc}$+18.8 \citep{2024MNRAS.527.8151S}. 
These ratios are listed in Table \ref{tab: ratios}.
In addition, we independently derived the HCN and HNC column densities from HC$^{15}$N and H$^{15}$NC using the $^{14}$N/$^{15}$N ratio as a function of galactocentric distance, given by $^{14}$N/$^{15}$N=10.8$D_{\rm gc}$+141.1 \citep{2025A&A...695L...6R}. These column densities derived from the $^{15}$N isotopologues were then used to calculate an independent set of $D_{\rm frac}$(HCN), $D_{\rm frac}$(HNC), and [HCN]/[HNC]. 
These ratios are also listed in Table \ref{tab: ratios} and provide a complementary check on the ratios derived from the $^{13}$C isotopologues. 
The quoted uncertainties in the column densities and abundance ratios include the errors propagated from the line measurements and column density calculations but do not include the systematic uncertainties associated with the adopted $^{12}$C/$^{13}$C and $^{14}$N/$^{15}$N isotopic conversion factors.

\begin{table*}[!htbp]
\fontsize{7.5pt}{9pt}\selectfont
\centering
\caption{Abundance ratios and deuterium fractionation.}
\label{tab: ratios}
\begin{tabular}{lccccccc}
\hline
\hline
\multirow{2}{*}{Source}
& \multirow{2}{*}{[DCN]/[DNC]}
& \multicolumn{3}{c}{Ratios from $^{13}$C isotopologues}
& \multicolumn{3}{c}{Ratios from $^{15}$N isotopologues} \\

\cmidrule(lr){3-5} \cmidrule(lr){6-8}

&
& [HCN]/[HNC]
& $D_{\rm frac}({\rm HCN})$
& $D_{\rm frac}({\rm HNC})$
& [HCN]/[HNC]
& $D_{\rm frac}({\rm HCN})$
& $D_{\rm frac}({\rm HNC})$ \\

\hline

G005.88-00.39 & 2.12$\pm$0.19 & 4.40$\pm$0.04 & 0.0028$\pm$0.0002 & 0.0059$\pm$0.0003 & 3.10$\pm$0.13 & 0.0034$\pm$0.0003 & 0.0050$\pm$0.0003 \\
G014.33-00.64 & 1.50$\pm$0.15 & 3.56$\pm$0.08 & 0.0054$\pm$0.0005 & 0.013$\pm$0.001 & 3.35$\pm$0.38 & 0.0072$\pm$0.0007 & 0.016$\pm$0.002 \\  
G035.19-00.74 & 2.13$\pm$0.20 & 2.98$\pm$0.08 & 0.0053$\pm$0.0003 & 0.0074$\pm$0.0006 & 2.62$\pm$0.32 & 0.0073$\pm$0.0006 & 0.0090$\pm$0.0012 \\
G081.75+00.59 & 1.40$\pm$0.07 & 2.63$\pm$0.05 & 0.0098$\pm$0.0004 & 0.0186$\pm$0.0005 & 1.59$\pm$0.11 & 0.011$\pm$0.001 & 0.012$\pm$0.001 \\ 
G109.87+02.11 & 3.03$\pm$0.35 & 3.93$\pm$0.13 & 0.0063$\pm$0.0005 & 0.0081$\pm$0.0008 & 2.94$\pm$0.30 & 0.0097$\pm$0.0008 & 0.0094$\pm$0.0012 \\  
G209.00-19.38 & 4.63$\pm$1.32 & 34.25$\pm$2.69 & 0.0015$\pm$0.0003 & 0.011$\pm$0.002 & $\geq$37.83  & 0.0015$\pm$0.0003 & $\geq$0.012  \\  

\hline
\end{tabular}
\tablefoot{
The complete table for all 101 sources is included with the online material.
}
\end{table*}

\begin{figure*}[!htbp]
\sidecaption
\begin{minipage}[b]{12cm}
\centering
\includegraphics[width=0.49\linewidth]{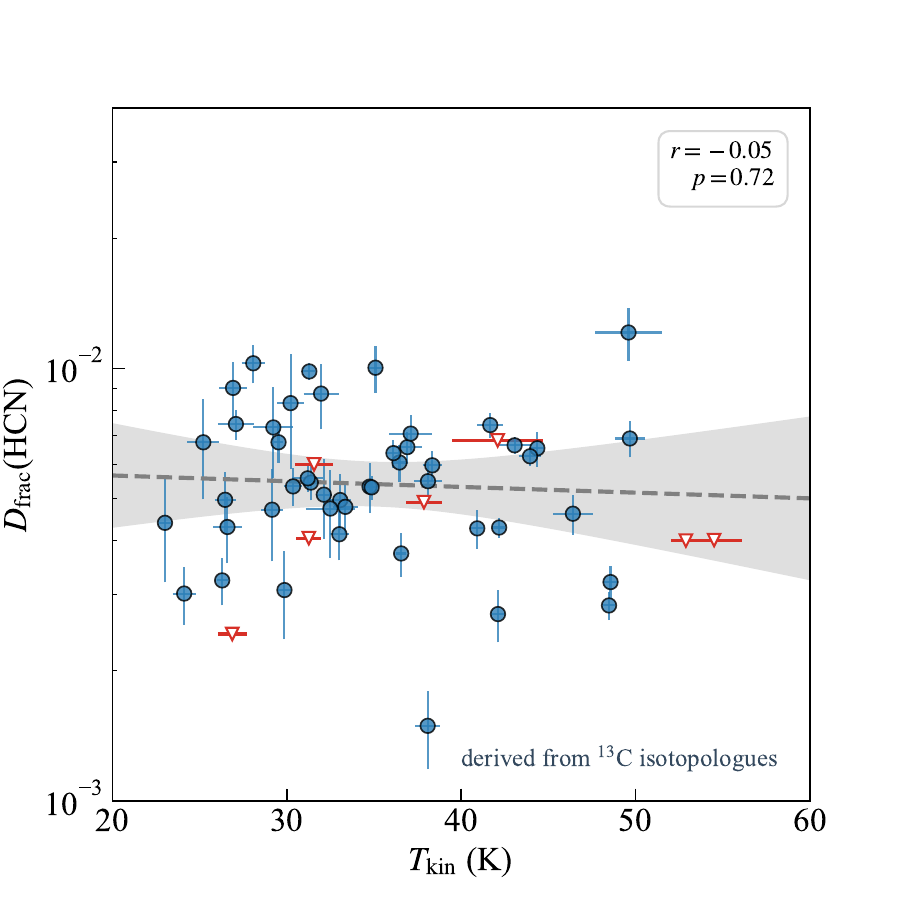}\hfill
\includegraphics[width=0.49\linewidth]{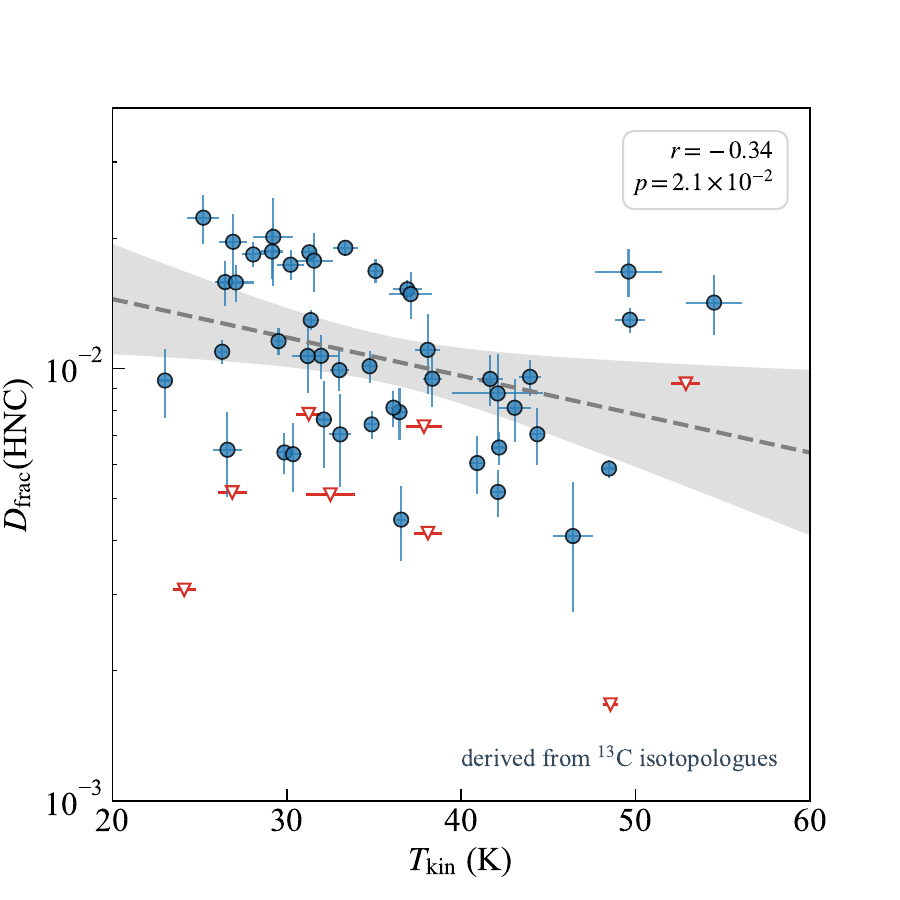}\\[0.3em]
\includegraphics[width=0.49\linewidth]{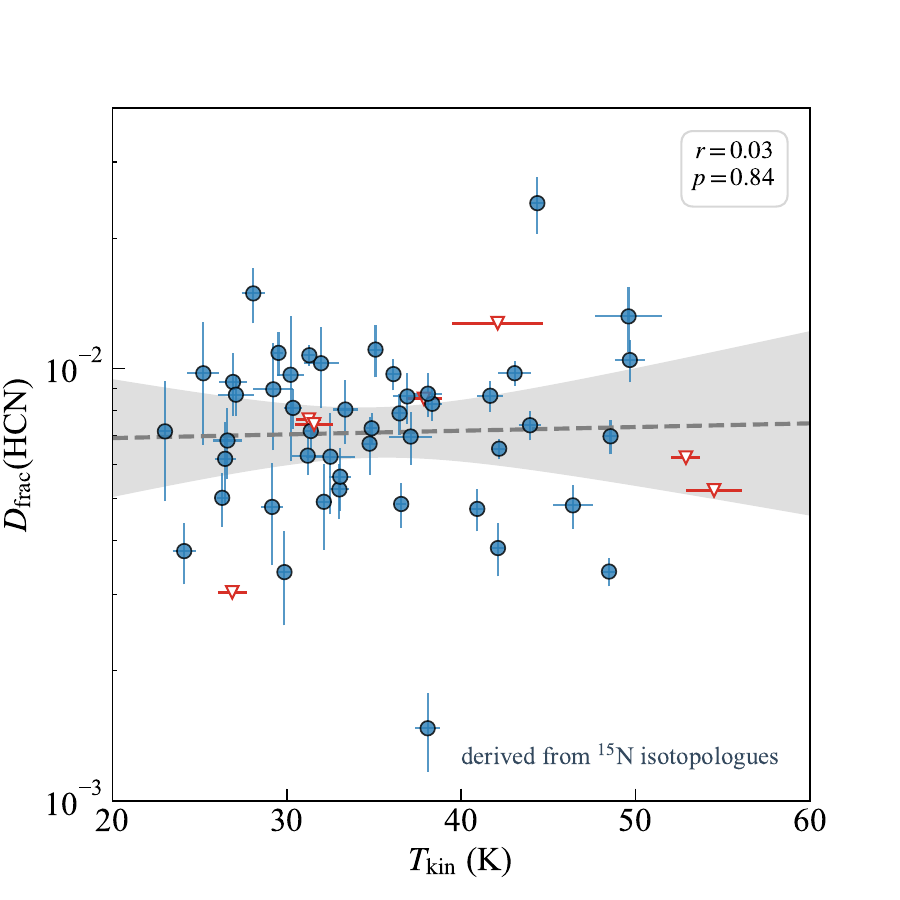}\hfill
\includegraphics[width=0.49\linewidth]{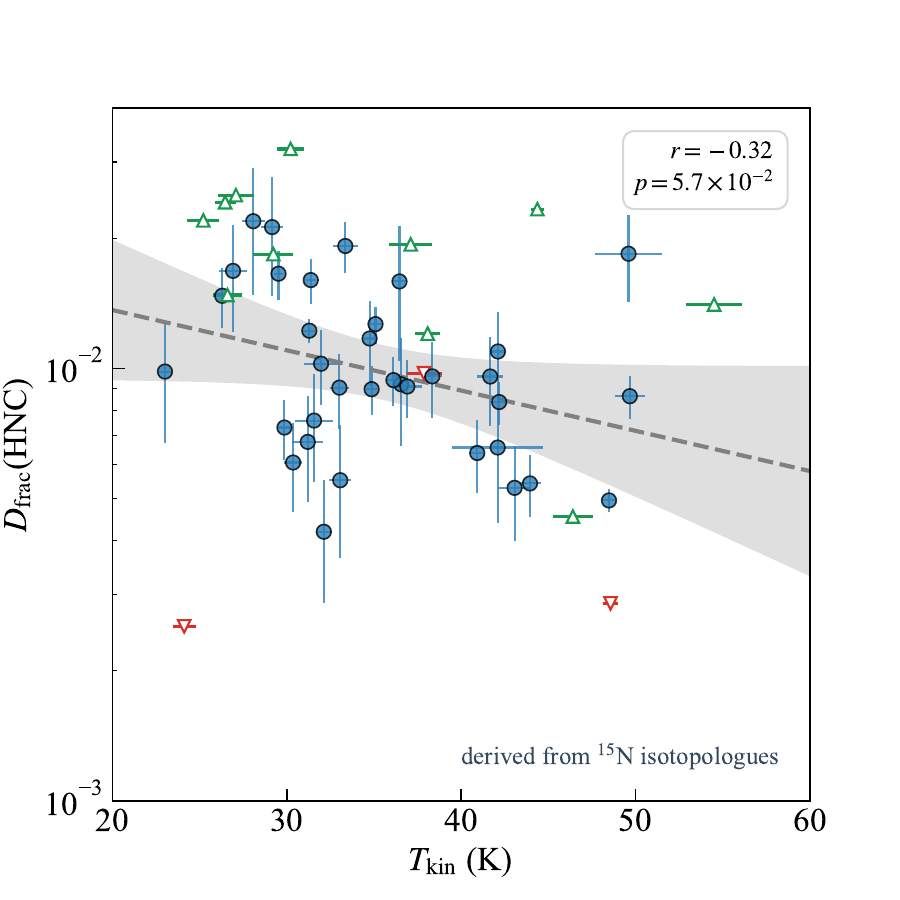}
\end{minipage}
\caption{Relationships between $D_{\rm frac}$(HCN) and $T_{\rm{kin}}$ (left panels) and between $D_{\rm frac}$(HNC) and $T_{\rm kin}$ (right panels). Top panels: $D_{\rm frac}$ (HCN) and $D_{\rm frac}$ (HNC) calculated using HCN and HNC column densities derived from their $^{13}$C isotopologues, H$^{13}$CN and HN$^{13}$C, respectively. Bottom panels: $D_{\rm frac}$(HCN) and $D_{\rm frac}$(HNC) calculated using HCN and HNC column densities derived from their $^{15}$N isotopologues, HC$^{15}$N and H$^{15}$NC, respectively. The blue points indicate ratios for which both transitions are detected, the downward red triangles indicate upper limits, and the upward green triangles indicate lower limits. The correlation coefficients were calculated using only the blue points with measured CH$_3$CCH temperatures. The upper and lower limits are shown for reference but are excluded from the Pearson correlation analysis. The dashed gray line shows the linear least-squares fit to the detections in log space, and the shaded gray region indicates the 95\% confidence interval of the fit. The Pearson correlation coefficients are reported in each panel.}
\label{Dfrac}
\end{figure*}

\begin{figure*}[!htbp]
\centering
\includegraphics[width=0.33\textwidth]{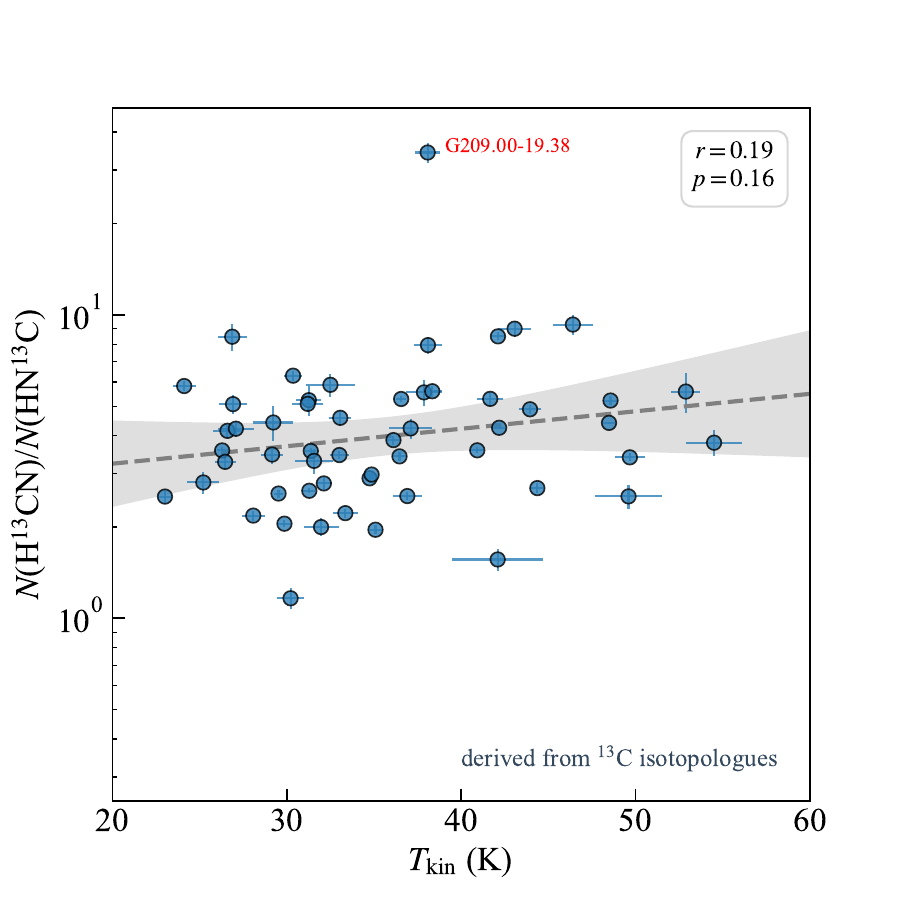}\includegraphics[width=0.33\textwidth]{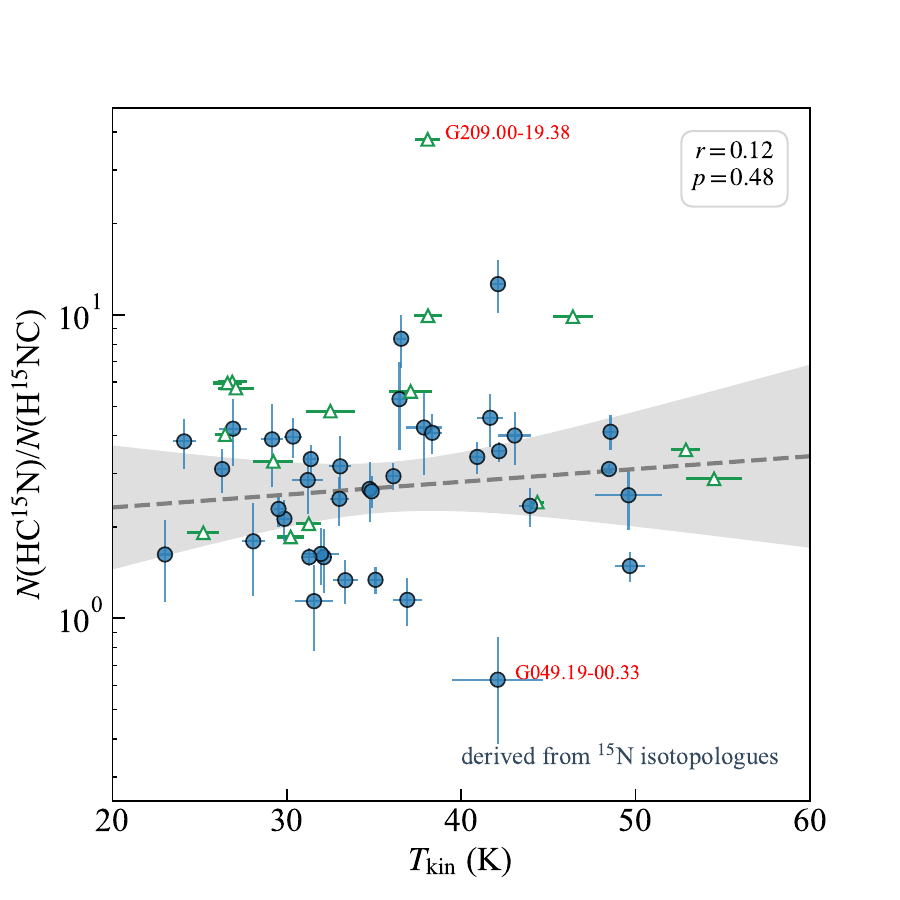}\includegraphics[width=0.33\textwidth]{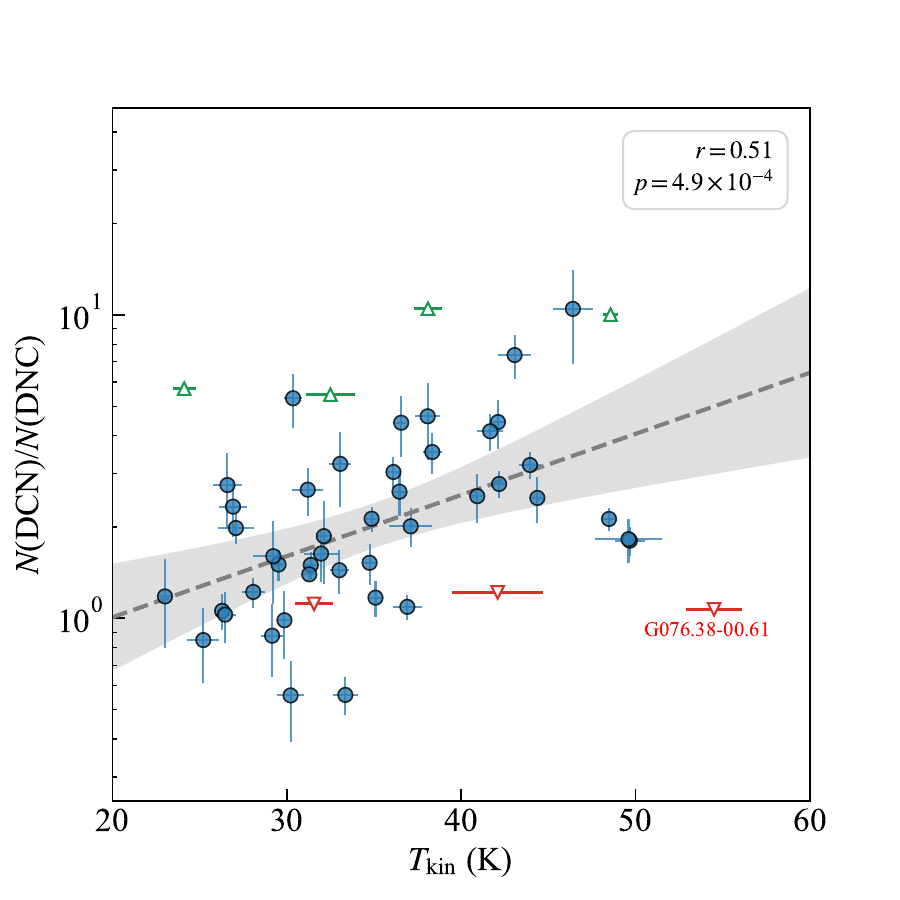}
\caption{Relationships between the isomeric abundance ratios and $T_{\rm{kin}}$. Left: Relationship between [HCN]/[HNC] derived from $^{13}$C isotopologues and $T_{\rm{kin}}$.
Middle: Relationship between [HCN]/[HNC] derived from $^{15}$N isotopologues and $T_{\rm{kin}}$.
Right: Relationship between [DCN]/[DNC] and $T_{\rm{kin}}$. The blue points indicate ratios for which both transitions are detected, the downward red triangles indicate upper limits, and the upward green triangles indicate lower limits. The correlation coefficients were calculated using only the blue points with measured CH$_3$CCH temperatures. Upper and lower limits are shown for reference but are excluded from the Pearson correlation analysis. The dashed gray line shows the linear least-squares fit to the detections in log space, and the shaded gray region indicates the 95\% confidence interval of the fit. The Pearson correlation coefficients are reported in each panel.
}
\label{DCN_DNC}
\end{figure*}

In the following analysis, only sources with $T_{\rm kin}$ measured from CH$_3$CCH 5-4 were included in the Pearson correlation analysis. Sources for which the median value of $T_{\rm kin}$=33.1 K was adopted as $T_{\rm ex}$  were excluded from the Pearson correlation analysis. Upper and lower limits are shown in Figs. \ref{Dfrac} and \ref{DCN_DNC} but were not included in the calculation of the correlation coefficients. Since the abundance ratios span more than one order of magnitude, the Pearson correlation coefficients were calculated in logarithmic space. Specifically, we evaluated the correlations between $T_{\rm kin}$ and $\log_{10}$ of each abundance ratio. Because the conventional Pearson correlation analysis cannot directly account for censored data, we additionally performed a generalized Kendall correlation analysis  \citep{1986ApJ...306..490I,2020ApJ...893..149P}, which explicitly incorporates censored data, including upper and lower limits. The results are presented in Appendix \ref{app: correlation coefficient}.

For sources with $T_{\rm{kin}}$ derived from the CH$_3$CCH 5-4, we examined the relationship between deuterium fractionation and $T_{\rm{kin}}$. As the temperature increases, the $D_{\rm frac}$(HCN) derived from the $^{13}$C isotopologues shows no statistically significant correlation with temperature, while $D_{\rm frac}$(HNC) derived from the $^{13}$C isotopologues exhibits a decreasing trend (see Fig. \ref{Dfrac}). Three sources -- G059.78+00.06, G076.38-00.61, and G176.51+00.20 -- show significant deviations from the general trend between $D_{\rm frac}$(HNC) derived from the $^{13}$C isotopologues and $T_{\rm{kin}}$. The $T_{\rm{kin}}$ values of these sources are relatively high compared with the $E_{\rm u}$ of the $K$=0, 1, and 2 components used in the temperature determination. As a result, the derived temperatures may be less tightly constrained and more sensitive to uncertainties in the spectral fitting. In addition, the $K$=0 and/or $K$=1 components of CH$_3$CCH may be optically thick in these sources, leading to an overestimate of $T_{\rm kin}$. However, we lack independent evidence demonstrating that these temperatures are incorrect. We therefore retain these sources in the correlation analysis. Their deviations may reflect either uncertainties in the $T_{\rm{kin}}$ derived from CH$_3$CCH 5-4 or genuine source-to-source variations in the physical and chemical conditions.
Including these sources, the Pearson correlation coefficients between $T_{\rm kin}$ and $D_{\rm frac}$(HCN) and between $T_{\rm kin}$ and $D_{\rm frac}$(HNC), both derived from the $^{13}$C isotopologues, are -0.05 and -0.34, respectively.

The deuterium fractionation was also independently derived from the $^{15}$N isotopologues, HC$^{15}$N and H$^{15}$NC. The ratios derived from the $^{15}$N isotopologues show trends broadly consistent with those derived from the $^{13}$C isotopologues, although the scatter may differ because of the uncertainties in the adopted isotopic ratios (see Fig. \ref{Dfrac}). 
$D_{\rm frac}({\rm HNC})$ shows a weak negative trend with $T_{\rm kin}$, whereas $D_{\rm frac}({\rm HCN})$ shows no statistically significant correlation. With the same analysis, the Pearson correlation coefficients between $T_{\rm kin}$ and $D_{\rm frac}$(HCN) and between $T_{\rm kin}$ and $D_{\rm frac}$(HNC), both derived from the $^{15}$N isotopologues, are 0.03 and -0.32, respectively.

The abundance ratio of the deuterated isomers, [DCN]/[DNC], shows a moderate increasing trend with increasing $T_{\rm kin}$ (see Fig. \ref{DCN_DNC}). For the detections, the Pearson correlation coefficient is r=0.51 with $p$=4.9$\times$10$^{-4}$. When the censored data are also taken into account using the generalized Kendall correlation analysis, the positive correlation remains statistically significant, with $\tau$=0.25 and $p$=1.1$\times$10$^{-2}$. 
This positive correlation is also statistically significant, indicating that the inferred positive temperature dependence is not driven solely by the exclusion of upper or lower limits in the Pearson analysis. 

In contrast, [HCN]/[HNC] derived from either the $^{13}$C or $^{15}$N isotopologues shows only a weak, statistically insignificant positive correlation with $T_{\rm kin}$, with substantial scatter (see Fig. \ref{DCN_DNC}). Two individual sources should be treated with caution. 
G209.00$-$19.38 (Orion Nebula) appears as an outlier in [HCN]/[HNC] derived from both the $^{13}$C and $^{15}$N isotopologues. This source shows strong radio recombination line emission \citep[e.g., ][]{2022ApJS..263...13L} and is nearby at a distance of 388 pc \citep{2017ApJ...834..142K}, suggesting that strong free-free continuum emission may affect the observed isotopologue line intensities. Therefore, [HCN]/[HNC] derived from both $^{13}$C and $^{15}$N isotopologues for this source may be unreliable. 
In addition, G049.19$-$00.33 deviates from the general trend in [HCN]/[HNC] derived from the $^{15}$N isotopologues. 
This deviation may be related to the weak detections of both $^{15}$N isotopologue lines, which introduce a large uncertainty in the derived ratio.
The Pearson correlation coefficients between [HCN]/[HNC] and $T_{\rm kin}$ are 0.19 for the estimates derived from the $^{13}$C isotopologues and 0.12 for those derived from the $^{15}$N isotopologues.

As discussed in Sect. \ref{sect: column density}, the abundance ratios are only weakly affected by the adopted $T_{\rm ex}$, and the temperature-dependent terms in the column density calculation introduce variations of less than $\sim$10\% in the derived ratios. Therefore, the observed trends are unlikely to arise from adopting the gas kinetic temperatures as the excitation temperatures.

\section{Discussion}
\label{discussion}
\subsection{[DCN]/[DNC] in star-forming regions}

In this work, we find a moderate, statistically significant, positive correlation between the abundance ratio [DCN]/[DNC] and the gas kinetic temperature in late-stage massive star-forming regions. For sources with detections of both species, the Pearson correlation coefficient between $T_{\rm kin}$ and $\log_{10}$([{DCN}]/[{DNC}]) is r=0.51 with $p$=4.9$\times$10$^{-4}$. This positive correlation is also supported by the generalized Kendall analysis of the censored data (Sect. \ref{sect: abundance ratio} and Appendix \ref{app: correlation coefficient}). It indicates that DCN becomes increasingly abundant relative to DNC as the gas temperature rises. Compared with previous studies, our analysis combines a larger number of sources with simultaneous DCN and DNC detections with gas kinetic temperatures derived from CH$_3$CCH 5-4. This provides a more direct test of the gas temperature dependence of DCN/DNC chemistry in dense gas.
However, the scatter around this relation, together with the presence of one upper limit (G076.38-00.61) below the general distribution, suggests that [DCN]/[DNC] may also be affected by other local physical or chemical conditions in addition to temperature.

This trend is broadly consistent with previous observational results in warm or hot star-forming regions. \citet{2020ARep...64..244T} report an increasing trend of [DCN]/[DNC] with kinetic temperature in massive star-forming regions and obtain a linear regression coefficient of 0.059 with the Onsala Space Observatory 20m telescope observations, but reliable [DCN]/[DNC] measurements are available for only eight sources, with a temperature range of 17-42 K. Using IRAM 30m, \citet{2024A&A...688A.215C} also observe a moderate positive relation between the DCN/DNC velocity-integrated intensity ratio and dust temperature in DR21, with a Pearson correlation coefficient of 0.58. Their analysis, however, is limited to one case study. It relies on dust temperature rather than kinetic temperature and covers a relatively narrow temperature range of $\sim$15-32 K. 
In comparison, our survey includes 51 sources with simultaneous DCN 1-0 and DNC 1-0 detections, among which 43 sources have gas kinetic temperatures derived from CH$_3$CCH 5-4 and are used in the correlation analysis. These 43 sources cover a broader gas-temperature range of $\sim$20–60 K. Therefore, our result provides a more systematic confirmation that the DCN/DNC chemistry is sensitive to temperature in late-stage massive star-forming regions.

The behavior in warm massive star-forming regions appears to be different from that in cold dark-cloud environments. In TMC-1, where the gas temperature is below about 10 K, \citet{2021A&A...653A..15N} observe only a weak dependence of [DCN]/[DNC] on kinetic temperature. This contrast suggests that the temperature dependence of [DCN]/[DNC] becomes more evident if the gas is heated by processes associated with embedded star formation. In cold gas, [DCN]/[DNC] may be more strongly regulated by cold deuterium chemistry, whereas in warmer regions the temperature dependence of this ratio may become more apparent as the relative formation and destruction efficiencies of DCN and DNC change with the physical conditions.

Nevertheless, the scatter in the observed [DCN]/[DNC]-$T_{\rm kin}$ relation indicates that the ratio is not determined by gas temperature alone, because the measured abundance ratio may also depend on the local chemical conditions. Therefore, the positive relation observed in this work should be interpreted as evidence that temperature is an important factor affecting [DCN]/[DNC] rather than as a purely temperature-controlled relation. In the following sections, we further discuss this behavior in the context of the parent HCN/HNC isomeric chemistry and the different temperature dependences of the deuterium fractionation of HCN and HNC.

\subsection{Possible optical-depth effects on [HCN]/[HNC]}
\label{optical-depth effects}

[HCN]/[HNC] in this work was estimated from the isotopologues rather than from the HCN and HNC lines, because the main isotopologue lines are expected to be optically thick in dense molecular clouds \citep[e.g., ][]{2023MNRAS.525.4761L}. However, even the isotopologue lines may not always be completely optically thin, especially toward hot cores with high column densities. Therefore, possible optical-depth effects should still be considered when interpreting the temperature dependence of [HCN]/[HNC].
In particular, [HCN]/[HNC] derived from the $^{13}$C isotopologues may still be affected by optical-depth effects. Since our sample mainly consists of hot cores and the beam size of the Yebes observations is relatively small ($\sim$20$\arcsec$), H$^{13}$CN 1-0 and HN$^{13}$C 1-0 may become optically thick in some sources. If this is the case, saturation effects could weaken the apparent relation between [HCN]/[HNC] derived from $^{13}$C isotopologues and $T_{\rm kin}$.

As shown in Fig. \ref{DCN_DNC}, [HCN]/[HNC] derived from either the $^{13}$C or $^{15}$N isotopologues exhibits a weak, statistically insignificant positive correlation with $T_{\rm kin}$ and substantial scatter. Optical-depth effects remain a possible source of uncertainty for the ratios derived from H$^{13}$CN and HN$^{13}$C $J$=1-0 lines, but their contribution cannot be established from this comparison.
Although the H$^{13}$CN 1-0 HFS was fitted to derive the FWHM, no optical-depth correction was applied because the broad line widths and partial blending of the hyperfine components make the fitted optical depths uncertain. The $^{13}$C isotopologues may nevertheless be more affected by optical-depth saturation than the $^{15}$N-bearing species, since their lines are typically about four times stronger \citep[e.g.,][]{2024MNRAS.527.8151S}. Such optical-depth effects may also affect the $D_{\rm frac}$(HCN) and $D_{\rm frac}$(HNC) derived from the $^{13}$C isotopologues and contribute to the scatter in their relations with $T_{\rm{kin}}$.  
In contrast, HC$^{15}$N and H$^{15}$NC $J$=1-0 lines are expected to be less affected by optical-depth effects. However, [HCN]/[HNC] derived from $^{15}$N-bearing species also has significant uncertainties because the $^{15}$N-bearing lines are generally weak and H$^{15}$NC 1-0 has a relatively low detection rate. Higher-sensitivity observations are required to better constrain the intrinsic relationship between [HCN]/[HNC] and $T_{\rm kin}$ using optically thin tracers.

Overall, the isotopologue analysis suggests that [HCN]/[HNC] may still have an intrinsic temperature dependence, but the observed relation with $T_{\rm{kin}}$ can be weakened or obscured by optical-depth effects in the $^{13}$C isotopologues and by the limited signal-to-noise ratio and detection rate of the $^{15}$N isotopologues. 
However, the temperature dependence masked by observational effects cannot be ruled out, nor can we confirm it on this basis.
Additional physical or chemical factors, such as evolutionary stage, may also contribute to the scatter, as discussed in the following subsection. These limitations should be considered when comparing the temperature behavior of [HCN]/[HNC] with that of [DCN]/[DNC].

\subsection{Origin of the temperature dependence of [DCN]/[DNC]}

\begin{figure*}[!htbp]
\sidecaption
\begin{minipage}[b]{12cm}
\centering
\includegraphics[width=0.49\linewidth]{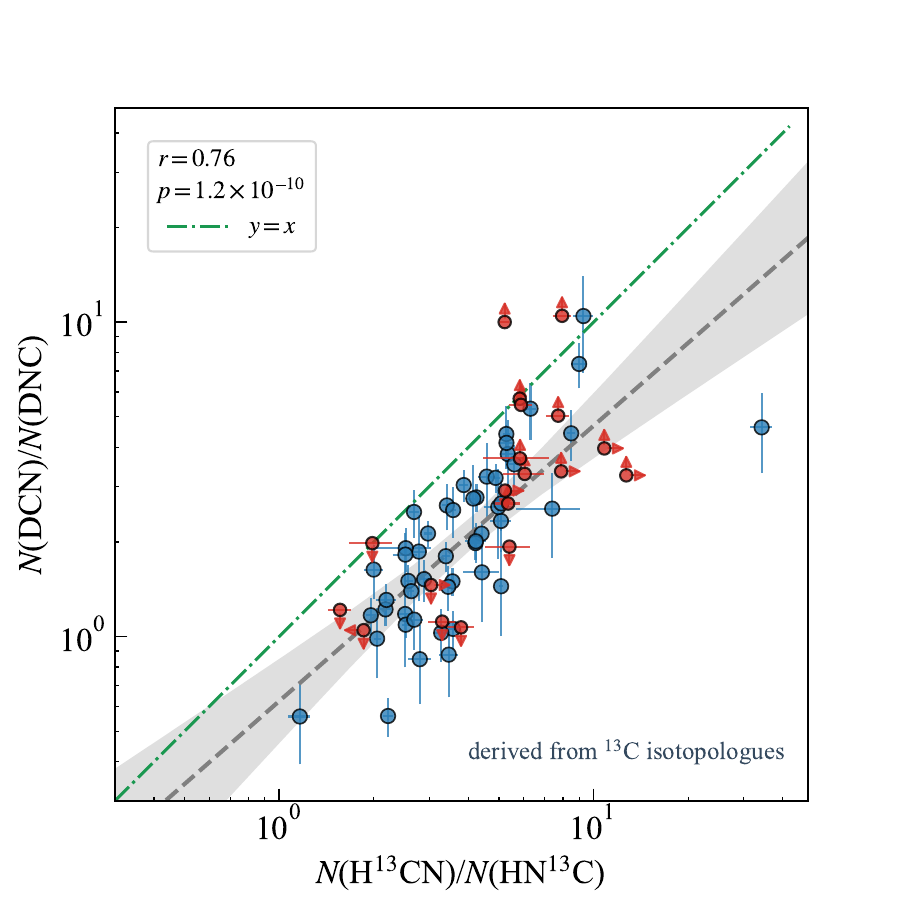}\hfill
\includegraphics[width=0.49\linewidth]{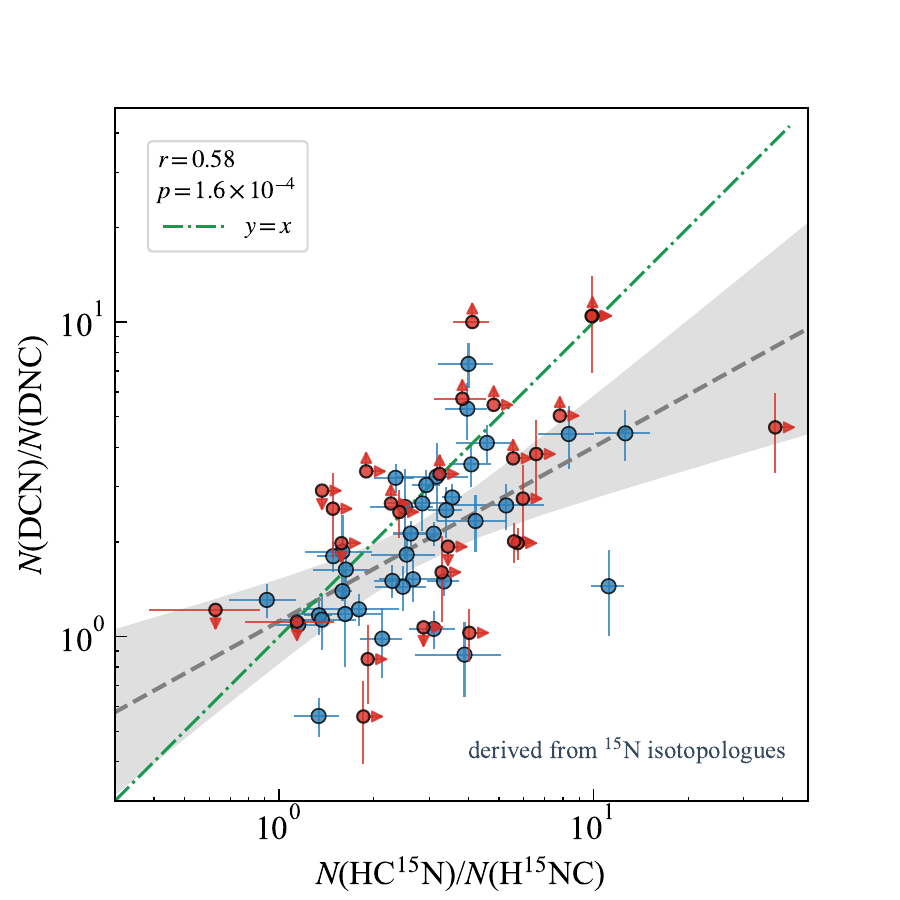}\\[0.3em]
\includegraphics[width=0.49\linewidth]{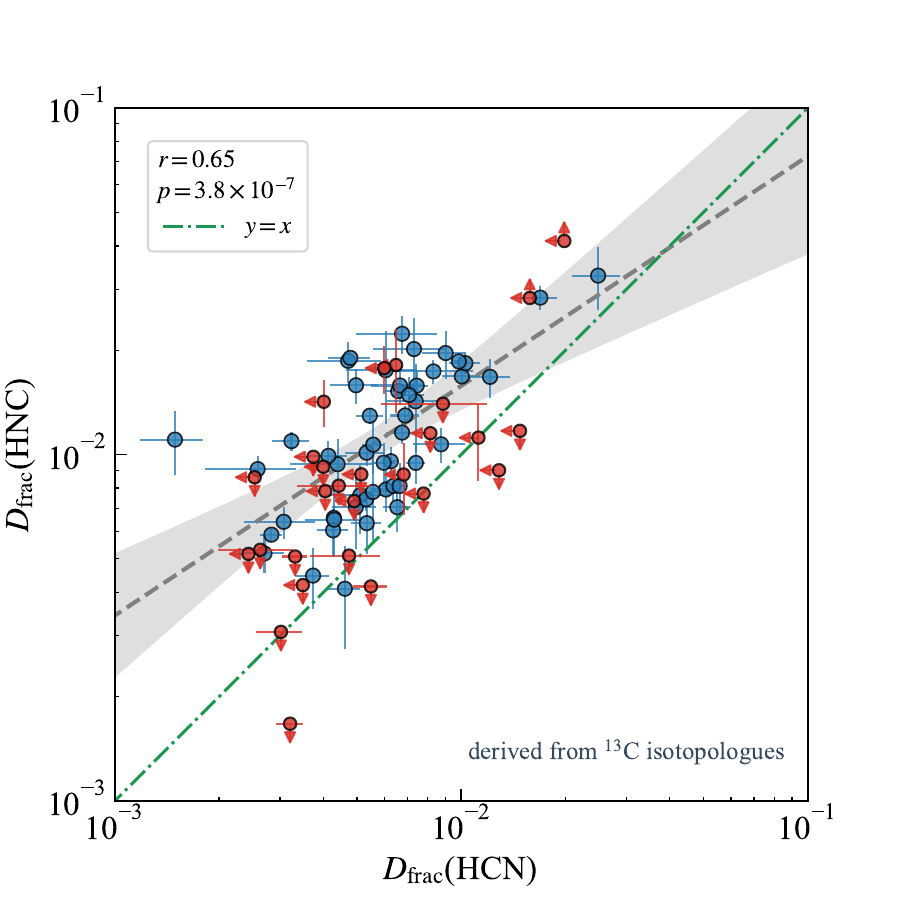}\hfill
\includegraphics[width=0.49\linewidth]{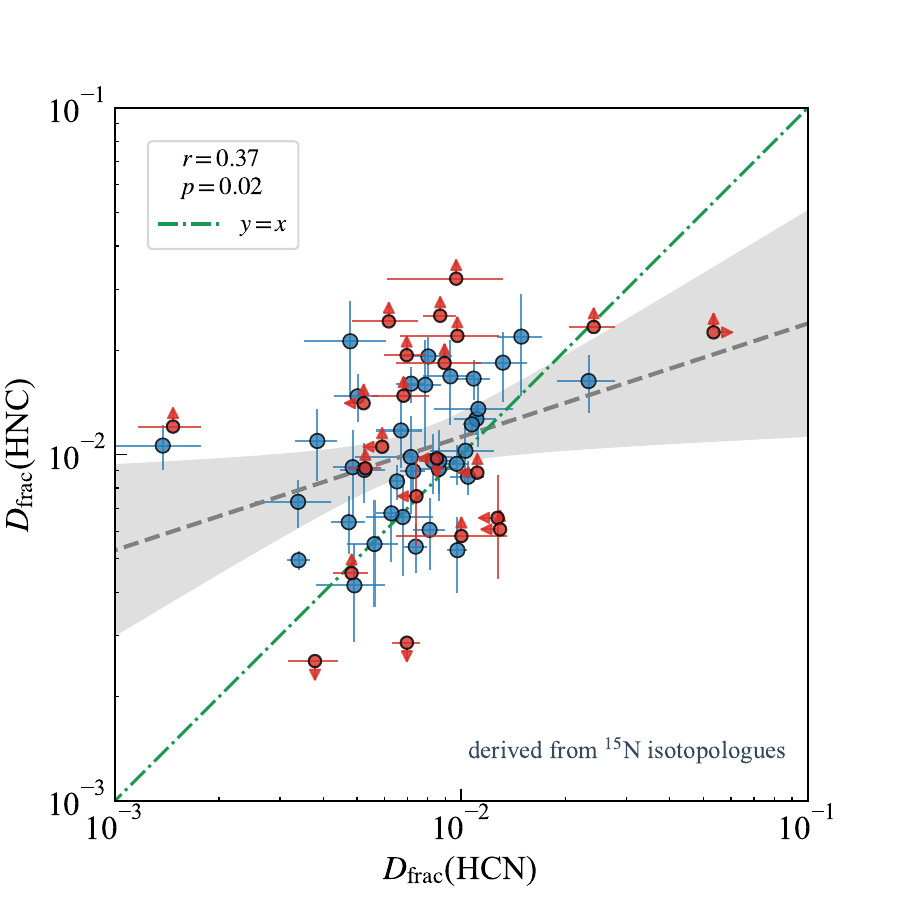}
\end{minipage}
\caption{Relationships between the [HCN]/[HNC] and [DCN]/[DNC] abundance ratios (upper panels) and between $D_{\rm frac}$(HCN) and $D_{\rm frac}$(HNC) (lower panels). The HCN/HNC ratios and deuterium fractions were derived using the $^{13}$C isotopologues in the left panels and the $^{15}$N isotopologues in the right panels. The blue markers denote sources detected in all four relevant transitions. The red markers denote sources for which one or both plotted quantities are constrained only by upper or lower limits owing to non-detections in the relevant transitions. The red arrows point toward the allowed values, with leftward and downward arrows indicating upper limits and rightward and upward arrows indicating lower limits on the corresponding axes. Sources are omitted from a given panel if both transitions used to derive either plotted ratio are undetected, so that neither a measurement nor an upper or lower limit can be obtained for that ratio. The dashed gray lines show linear least-squares fits to the detections in logarithmic space, and the shaded gray regions represent the 95\% confidence intervals of the fits. The Pearson correlation coefficients and $p$ values, calculated in logarithmic space using detections only, are reported in each panel. The dash-dotted green lines indicate the one-to-one relation (y=x).}
\label{ratio to ratio}
\end{figure*}

As shown in Fig. \ref{DCN_DNC}, [DCN]/[DNC] shows a moderate positive trend with $T_{\rm{kin}}$, while the relationship between [HCN]/[HNC] and $T_{\rm{kin}}$ is weaker and more scattered. In addition, [DCN]/[DNC] exhibits a relatively large dynamic range and reaches values below 1 in some colder sources, while [HCN]/[HNC] is generally larger than 1 in our sample, especially when derived from the $^{13}$C isotopologues. 
To examine the relationship between the deuterated and parent isomeric ratios directly, the upper panels of Fig. \ref{ratio to ratio} compare [DCN]/[DNC] with [HCN]/[HNC] derived from $^{13}$C and $^{15}$N isotopologues. Both comparisons show positive correlations with Pearson coefficients of $r$=0.76 and $r$=0.58 for $^{13}$C and $^{15}$N isotopologues, respectively. 
These positive correlations suggest that the chemistry of the deuterated isomers is linked to that of their hydrogenated counterparts and that the parent HCN/HNC chemistry may contribute to the observed source-to-source variations in [DCN]/[DNC]. However, these correlations do not establish whether the temperature dependence of [DCN]/[DNC] can be fully explained by the parent HCN/HNC chemistry.

[HCN]/[HNC] has long been recognized as sensitive to temperature, generally increasing with temperature \citep[e.g., ][]{2015ApJS..219....2J,2020A&A...635A...4H,2022ARep...66.1302P}. HCN and HNC both efficiently form through the dissociative recombination of HCNH$^+$ \citep{2014MNRAS.443..398L}:
\begin{equation*}
\mathrm{HCNH^+ + e^- \longrightarrow}
\left\{
\begin{aligned}
    &\mathrm{HCN + H} \\
    &\mathrm{HNC + H}
\end{aligned}
\right..
\end{equation*} 
Moreover, HNC can be converted into HCN through the following reaction \citep{2014ApJ...787...74G,2014MNRAS.443..398L}:
\begin{equation*}
	\mathrm{HNC + H \longrightarrow HCN + H}.
\end{equation*}
Theoretical calculations predict an activation barrier of about 2000 K for this reaction \citep{1996CPL...263..385T}. However, astrochemical models adopting such a high barrier have difficulty reproducing the observed temperature dependence of [HCN]/[HNC], and substantially lower effective barriers, typically of a few hundred kelvin, are required \citep[e.g.,][]{2014ApJ...787...74G,2020A&A...635A...4H,1998ApJ...503..717H}. The physical origin of this discrepancy remains unclear.
In contrast to the temperature dependence reproduced by these models, [HCN]/[HNC] shows only weak, statistically insignificant correlations with $T_{\rm kin}$ in our sample.
This may reflect a combination of optical-depth effects in the $^{13}$C isotopologues, the limited sensitivity and lower detection rate of the $^{15}$N isotopologues, and source-to-source variations in density, evolutionary stage, and UV radiation \citep[e.g., ][]{2014A&A...563A..97G,2014ApJ...787...74G,2018MNRAS.481.4662R,2021arXiv211102074T,2023A&A...679A...4S}. 
In addition, shock chemistry may affect [HCN]/[HNC] and contribute to the observed source-to-source variations \citep{2021MNRAS.507.1034L}.

As in the HCN/HNC system, DCN and DNC can form through the dissociative recombination of DCNH$^+$ \citep{2013ApJS..207...27A}: 
\begin{equation*}
\mathrm{DCNH^+ + e^- \longrightarrow}
\left\{
\begin{aligned}
    &\mathrm{DCN + H} \\
    &\mathrm{DNC + H}
\end{aligned}
\right..
\end{equation*}
These analogous formation routes suggest that the chemistry of DCN and DNC may share some features with that of HCN and HNC.
Direct isomeric conversion may also contribute to the temperature dependence of [DCN]/[DNC]. The deuterated isomeric pair may undergo a conversion reaction analogous to HNC $\rightarrow$ HCN:
\begin{equation*}
    \mathrm{DNC + H \longrightarrow DCN + H}.
\end{equation*}
This reaction is included in the KInetic Database for Astrochemistry (KIDA; \citealt{2015ApJS..217...20W})\footnote{\url{https://kida.astrochem-tools.org}} network associated with \cite{2017MNRAS.466.4470M}, where an activation barrier of $\sim$2000 K is adopted in the modified Arrhenius rate expression. However, given the uncertainty in the effective barrier of the analogous HNC+H reaction, the effective barrier for DNC+H may also be uncertain. Therefore, the contribution of this reaction to the observed [DCN]/[DNC] trend should be regarded as tentative. 
In addition, the different trends shown by [DCN]/[DNC] and [HCN]/[HNC] suggest that additional processes may be required. A natural explanation is the different temperature dependences of the deuterium fractionation of HCN and HNC.

The lower panels of Fig. \ref{ratio to ratio} compare $D_{\rm frac}({\rm HCN})$ and $D_{\rm frac}({\rm HNC})$. The two quantities are positively correlated, with r=0.65 and r=0.37 for estimates derived from the $^{13}$C and $^{15}$N isotopologues, respectively. Therefore, sources with higher $D_{\rm frac}({\rm HCN})$ also tend to have higher $D_{\rm frac}({\rm HNC})$. Most sources lie above the one-to-one relation, indicating greater deuterium fractionation in HNC than in HCN.
Their relative temperature dependence is examined more directly in Figs. \ref{Dfrac} and \ref{Dfrac ratio and Tkin}. While $D_{\rm frac}({\rm HCN})$ shows no clear trend with temperature, $D_{\rm frac}({\rm HNC})$ decreases. The ratio $D_{\rm frac}({\rm HCN})/D_{\rm frac}({\rm HNC})$ shows positive trends with $T_{\rm kin}$, with Pearson coefficients of r=0.39 ($p$=9.7$\times$10$^{-3}$) and r=0.33 ($p$=0.06) for the $^{13}$C and $^{15}$N isotopologues, respectively. These results support a possible contribution from differential deuterium fractionation to the observed increase in [DCN]/[DNC] with temperature.

A possible chemical explanation for these trends is that DNC is more closely associated with cold deuteration pathways, whereas DCN can continue to form through CH$_2$D$^+$-related pathways in relatively warm gas  \citep{2000A&A...364..780R,2000A&A...361..388R}. In colder gas, efficient enhancement of DNC leads to lower [DCN]/[DNC]. As the gas warms, cold deuteration becomes less efficient, causing $D_{\rm frac}$(HNC) to decrease. In contrast, DCN may either survive over longer timescales or continue to form through warmer deuteration pathways, resulting in higher [DCN]/[DNC] in more evolved sources. 
The continued formation of DCN through CH$_2$D$^+$-related pathways in relatively warm gas may help explain the weak temperature dependence of $D_{\rm frac}$(HCN). Together with the observed decrease in $D_{\rm frac}$(HNC), this provides a qualitative explanation for the increasing ratio $D_{\rm frac}$(HCN)/$D_{\rm frac}$(HNC).
Thus, the different temperature dependences of the two deuterium fractions can increase $D_{\rm frac}$(HCN)/$D_{\rm frac}$(HNC), contributing to the observed increase in [DCN]/[DNC], without requiring the deuterium fraction of HCN to exceed that of HNC.

Temperature and evolutionary stage are closely coupled in star-forming regions \citep[e.g., ][]{2017A&A...603A..33G,2017A&A...599A.139K,2018MNRAS.473.1059U}. Therefore, the observed relation between [DCN]/[DNC] and temperature may partly reflect evolutionary effects, since both gas heating and time-dependent chemistry proceed as protostellar objects evolve \citep{2009ARA&A..47..427H}. Nevertheless, the stronger temperature dependence of [DCN]/[DNC] compared with that of [HCN]/[HNC] suggests that deuterated isomer chemistry is more sensitive to the physical and evolutionary conditions of dense gas. Further progress will require spatially resolved observations and dedicated chemical modeling to determine the relative importance of temperature-dependent deuterium fractionation, isomeric conversion, and evolutionary effects.

Overall, the temperature dependence of [DCN]/[DNC] is best interpreted as the result of both the chemistry of the parent molecules and chemistry specific to the deuterated species. The decrease in $D_{\rm frac}$(HNC) may contribute to the observed trend, while possible DNC-to-DCN conversion and changes in [HCN]/[HNC] may provide additional contributions. This may help explain why [DCN]/[DNC] shows a clearer temperature dependence than [HCN]/[HNC] in our sample.

\section{Conclusions}
\label{summary}
We investigated the temperature dependence of [DCN]/[DNC] toward a sample of 101 late-stage massive star-forming regions. The $J$=1-0 transitions of DCN, DNC, H$^{13}$CN, HN$^{13}$C, HC$^{15}$N, and H$^{15}$NC were observed with the Yebes 40m telescope, and gas kinetic temperatures were derived from CH$_3$CCH 5-4. Our main conclusions are summarized as follows:

\begin{enumerate}
    \item The $J$=1-0 transitions of DCN, DNC, H$^{13}$CN, HN$^{13}$C, HC$^{15}$N, and H$^{15}$NC are widely detected in our sample, with detection rates of about 61\%, 59\%, 91\%, 77\%, 78\%, and 46\%, respectively. 
    \item $D_{\rm frac}({\rm HNC})$ shows a weak negative trend with temperature, whereas $D_{\rm frac}({\rm HCN})$ shows no clear trend with temperature over the observed range. Both estimates derived from the $^{13}$C and $^{15}$N isotopologues support this behavior.
    \item In our sample, [DCN]/[DNC] exhibits a statistically significant positive correlation with $T_{\rm kin}$, whereas no statistically significant correlation is found for [HCN]/[HNC] derived from either the $^{13}$C or $^{15}$N isotopologues. The latter ratios also show substantial scatter. Possible optical-depth effects in the $^{13}$C isotopologues and uncertainties associated with the weaker $^{15}$N lines may obscure an underlying temperature dependence.
    \item The abundance ratio [DCN]/[DNC] shows a moderate, statistically significant positive correlation with gas kinetic temperature, with a Pearson correlation coefficient of r=0.51 and $p$=4.9$\times$10$^{-4}$, while the generalized Kendall analysis including censored data gives $\tau$=0.25 and $p$=1.1$\times$10$^{-2}$. This trend is consistent with temperature-dependent deuterium fractionation, in which DNC is more efficiently suppressed than DCN as the gas temperature increases. 
This suggests that DCN/DNC chemistry is not simply a deuterated analogue of the HCN/HNC system but is additionally influenced by temperature-dependent deuterium fractionation and source evolution.
\end{enumerate}

\section{Data availability}
The spectral data used in this study and the complete Tables 1-5 are available on Zenodo at \url{https://doi.org/10.5281/zenodo.24005181}.

\begin{acknowledgements}
This research was supported by the Korea Astronomy and Space Science Institute under the R\&D program (Project No. 2026-1-844-00) supervised by the Korea AeroSpace Administration. The 40m radio telescope at Yebes Observatory is operated by the Spanish Geographic Institute (IGN; Ministerio de Transportes y Movilidad Sostenible). 
\end{acknowledgements}

%
%
\bibpunct{(}{)}{;}{a}{}{,}
\bibliographystyle{aa}
\bibliography{bibtex.bib}

\clearpage
\onecolumn

\begin{appendix}

\section{Check using Kendall correlation coefficients}
\label{app: correlation coefficient}

In Figs. \ref{Dfrac}-\ref{DCN_DNC}, we show the correlations between several abundance ratios and $T_{\rm kin}$, together with the Pearson correlation coefficients. The Pearson correlation coefficient was used to assess whether these abundance ratios exhibit linear correlations with $T_{\rm kin}$. To further assess the relationships between the abundance ratios and $T_{\rm kin}$ while accounting for upper and lower limits, we calculated Kendall correlation coefficients to test for monotonic correlations between the abundance ratios and $T_{\rm kin}$. The generalized Kendall correlation coefficients were calculated using the pymccorrelation package \citep{2020ApJ...893..149P}, which implements correlation analysis for censored data \citep{1986ApJ...306..490I}. The comparison between the Pearson and Kendall correlation coefficients is listed in Table \ref{tab:corr_compare}. 

The generalized Kendall correlation coefficients indicate that $D_{\rm frac}$(HNC) shows a weak but statistically significant negative monotonic correlation with $T_{\rm kin}$,  with $p$ values below 0.05 for both estimates based on the $^{13}$C and $^{15}$N isotopologues. 
In contrast, $D_{\rm frac}$(HCN) shows only a very weak monotonic dependence on $T_{\rm kin}$, with both Pearson and Kendall $p$ values well above 0.05, indicating that the correlation is not statistically significant. Overall, these results suggest that $D_{\rm frac}$(HNC) is more sensitive to $T_{\rm kin}$ than $D_{\rm frac}$(HCN).

For [DCN]/[DNC], the generalized Kendall analysis confirms a statistically significant positive correlation with $T_{\rm kin}$. In contrast, no statistically significant monotonic correlation with $T_{\rm kin}$ is found for [HCN]/[HNC] derived from either the $^{13}$C or $^{15}$N isotopologues. These results are consistent with the qualitative conclusions obtained from the Pearson analysis and indicate that the positive correlation between [DCN]/[DNC] and $T_{\rm kin}$ remains significant when upper and lower limits are included.

We also applied the generalized Kendall analysis to the direct comparisons between the isomeric abundance ratios and between the deuterium fractions shown in Fig. \ref{ratio to ratio}. The corresponding statistics, together with the Pearson results for detections, are listed in Table \ref{tab:corr_compare_ratio}.

\begin{table}[h]
\centering
\caption{Comparison of Pearson and generalized Kendall correlation coefficients for correlations with $T_{\rm kin}$.}
\label{tab:corr_compare}
\begin{tabular}{ccccccccc}
\hline
\hline
\multirow{2}{*}{Ratio} 
& \multirow{2}{*}{Transitions required} 
& \multicolumn{3}{c}{Pearson} 
& \multicolumn{3}{c}{Kendall} \\
\cmidrule(lr){3-5}
\cmidrule(lr){6-8}
& 
& $N_{\rm P}$ & $r$ & $p$ value 
& $N_{\rm K}$ & $\tau$ & $p$ value \\
\hline

[DCN]/[DNC] & DCN 1-0 and DNC 1-0 & 43 & 0.51 & 4.9$\times$10$^{-4}$ & 50 & 0.25 & 1.1$\times$10$^{-2}$ \\

[HCN]/[HNC] from $^{13}$C & H$^{13}$CN 1-0 and HN$^{13}$C 1-0 & 54 & 0.19 & 0.16 & 54 & 0.15 & 0.10 \\

[HCN]/[HNC] from $^{15}$N & HC$^{15}$N 1-0 and H$^{15}$NC 1-0  & 38 & 0.12 & 0.48 & 53 & 0.03 & 0.79 \\

$D_{\rm frac}({\rm HCN})$ from $^{13}$C & DCN 1-0 and H$^{13}$CN 1-0 & 47 & -0.05 & 0.72 & 54 & -0.05 & 0.60 \\

$D_{\rm frac}({\rm HNC})$ from $^{13}$C & DNC 1-0 and HN$^{13}$C 1-0  & 46 & -0.34 & 2.1$\times$10$^{-2}$ & 54 & -0.22 & 1.7$\times$10$^{-2}$ \\

$D_{\rm frac}({\rm HCN})$ from $^{15}$N & DCN 1-0 and HC$^{15}$N 1-0 & 47 & 0.03 & 0.84 & 54 & -0.01 & 0.94 \\

$D_{\rm frac}({\rm HNC})$ from $^{15}$N & DNC 1-0 and H$^{15}$NC 1-0 & 35 & -0.32 & 5.7$\times$10$^{-2}$ & 49 & -0.24 & 1.7$\times$10$^{-2}$ \\
\hline
\multicolumn{6}{@{}l}{\footnotesize Notes.}\\
\multicolumn{6}{@{}l}{
\makebox[0pt][l]{
\parbox{0.75\linewidth}{
\footnotesize 1. $N_{\rm P}$ is the number of detections used for the Pearson analysis. 
}}}\\
\multicolumn{8}{@{}l}{
\makebox[0pt][l]{
\parbox{0.8\linewidth}{
\footnotesize 2. $N_{\rm K}$ is the total number of sources included in the generalized Kendall analysis, including censored data.
}}}\\
\end{tabular}
\end{table}

\begin{table}[h]
\centering
\caption{Pearson and generalized Kendall correlation statistics for isomeric abundance ratios and deuterium fractions.}
\label{tab:corr_compare_ratio}
\begin{tabular}{ccccccccc}
\hline
\hline
\multirow{2}{*}{Ratio pair} &
\multirow{2}{*}{Isotopologues} 
& \multicolumn{3}{c}{Pearson} 
& \multicolumn{3}{c}{Kendall} \\
\cmidrule(lr){3-5}
\cmidrule(lr){6-8}
& 
& $N_{\rm P}$ & $r$ & $p$ value 
& $N_{\rm K}$ & $\tau$ & $p$ value \\
\hline

[HCN]/[HNC] and [DCN]/[DNC] & $^{13}$C & 50 & 0.76 & 1.2$\times$10$^{-10}$ & 69 & 0.60 & 3.3$\times$10$^{-13}$ \\

[HCN]/[HNC] and [DCN]/[DNC] & $^{15}$N & 37 & 0.58 & 1.6$\times$10$^{-4}$ & 64 & 0.42 & 1.1$\times$10$^{-6}$ \\

$D_{\rm frac}({\rm HCN})$ and $D_{\rm frac}({\rm HNC})$ & $^{13}$C & 50 & 0.65 & 3.8$\times$10$^{-7}$ & 77 & 0.46 & 2.3$\times$10$^{-9}$ \\

$D_{\rm frac}({\rm HCN})$ and $D_{\rm frac}({\rm HNC})$ & $^{15}$N & 37 & 0.37 & 0.02 & 59 & 0.24 & 6.6$\times$10$^{-3}$ \\

\hline
\multicolumn{8}{@{}l}{\footnotesize Notes.}\\
\multicolumn{8}{@{}l}{
\makebox[0pt][l]{
\parbox{0.75\linewidth}{
\footnotesize 1. The labels $^{13}$C and $^{15}$N indicate the isotopologues used to derive [HCN]/[HNC] and the deuterium fractions.
}}}\\
\multicolumn{8}{@{}l}{
\makebox[0pt][l]{
\parbox{0.75\linewidth}{
\footnotesize 2. $N_{\rm P}$ is the number of sources with detections of all four relevant transitions used in the Pearson correlation analysis.
}}}\\
\multicolumn{8}{@{}l}{
\makebox[0pt][l]{
\parbox{0.8\linewidth}{
\footnotesize 3. $N_{\rm K}$ is the total number of sources included in the generalized Kendall analysis, including censored data.
}}}\\
\end{tabular}
\end{table}

\section{$D_{\rm frac}({\rm HCN})$/$D_{\rm frac}({\rm HNC})$ versus $T_{\rm kin}$}
\label{app: Dfrac ratio and Tkin}

\begin{figure}[h]
\centering
\includegraphics[width=0.45\textwidth]{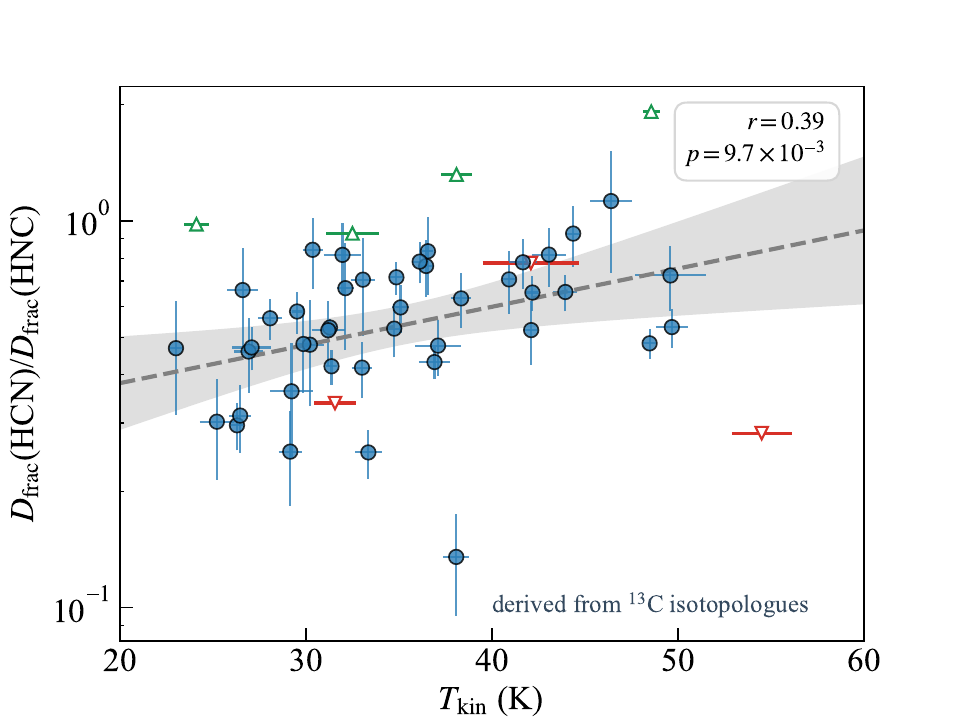}\includegraphics[width=0.45\textwidth]{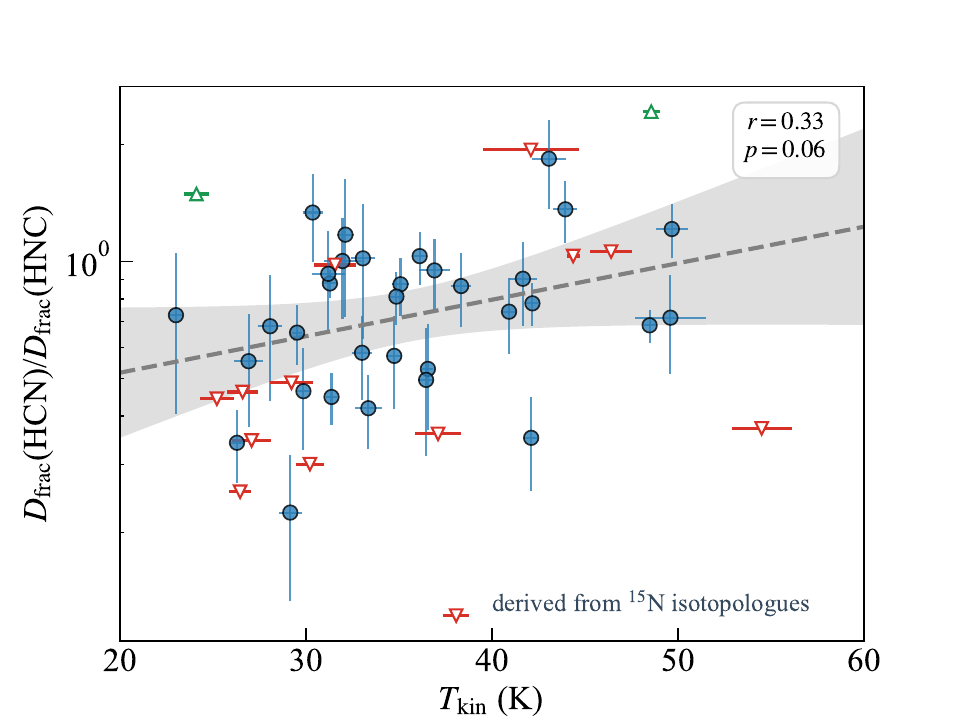}
\caption{Relationship between $D_{\rm frac}({\rm HCN})$/$D_{\rm frac}({\rm HNC})$ and $T_{\rm{kin}}$. Left: Relationship between $D_{\rm frac}({\rm HCN})$/$D_{\rm frac}({\rm HNC})$ derived from $^{13}$C isotopologues and $T_{\rm{kin}}$.
Right: Relationship between $D_{\rm frac}({\rm HCN})$/$D_{\rm frac}({\rm HNC})$ derived from $^{15}$N isotopologues and $T_{\rm{kin}}$. The blue points indicate ratios for which all four relevant transitions are detected, the downward red triangles indicate upper limits, and the upward green triangles indicate lower limits. 
Sources for which both deuterium fractions are constrained only by upper limits or both by lower limits are not shown, because no meaningful upper or lower limit can be derived for their ratio. 
The Pearson correlation coefficients were calculated using only the blue points with measured CH$_3$CCH temperatures. Upper and lower limits are shown for reference but are excluded from the Pearson correlation analysis. The dashed gray line shows the linear least-squares fit to the detections in log space, and the shaded gray region indicates the 95\% confidence interval of the fit. The Pearson correlation coefficients are reported in each panel.}
\label{Dfrac ratio and Tkin}
\end{figure}

Figure \ref{Dfrac ratio and Tkin} shows $D_{\rm frac}({\rm HCN})/D_{\rm frac}({\rm HNC})$ as a function of $T_{\rm kin}$, with estimates derived from the $^{13}$C and $^{15}$N isotopologues. Including upper and lower limits, the generalized Kendall analysis yields $\tau=0.25$ ($p$=1.1$\times$10$^{-2}$) and $\tau=0.20$ ($p$=4.6$\times$10$^{-2}$), respectively. These weak but statistically significant positive correlations support a possible contribution from differential deuterium fractionation to the increase in [DCN]/[DNC] with temperature. A detailed chemical interpretation will be presented in our follow-up study (Feng et al., in prep.).

\end{appendix}

\clearpage

\begin{center}
    {\Huge\bfseries \underline{Online material}}
  \end{center}

\setcounter{section}{0}
\setcounter{figure}{0}
\setcounter{table}{0}

\renewcommand{\thesection}{\arabic{section}}
\renewcommand{\thefigure}{\arabic{figure}}
\renewcommand{\thetable}{\arabic{table}}

\section{Source information}
\label{app: sources}

Table \ref{app_tab: source} lists the basic properties of the full source sample. 

\begin{longtable}{llccrccc}
\caption{Source list}\label{app_tab: source}\\
\hline\hline
Source name & Alias & R.A. (J2000) & Decl. (J2000) &
$v_{\rm LSR}$ &
$D_{\sun}$ &
$D_{\rm gc}$ &
$T_{\rm kin}$\\
 & & (hh:mm:ss) & (dd:mm:ss) &
(km s$^{-1}$) &
(kpc) & (kpc) & (K)\\
\hline
\endfirsthead

\caption{continued.}\\
\hline\hline
Source name & Alias & R.A. (J2000) & Decl. (J2000) &
$v_{\rm LSR}$ &
$D_{\sun}$ &
$D_{\rm gc}$ &
$T_{\rm kin}$\\
 & & (hh:mm:ss) & (dd:mm:ss) &
(km s$^{-1}$) &
(kpc) & (kpc) & (K)\\
\hline
\endhead

\hline
\multicolumn{8}{r}{\emph{Continued on next page}}\\
\endfoot

\hline
\multicolumn{8}{@{}l}{\footnotesize Notes.}\\
\multicolumn{8}{@{}l}{
\makebox[0pt][l]{
\parbox{0.88\linewidth}{
\footnotesize 1. The $v_{\rm LSR}$ values are from \citet{2014ApJ...783..130R}.
}}}\\
\multicolumn{8}{@{}l}{
\makebox[0pt][l]{
\parbox{0.88\linewidth}{
\footnotesize 2. The $D_{\sun}$ values are derived from the trigonometric parallax reported in \citet{2014ApJ...783..130R} and the $D_{\rm gc}$ values are calculated from those parallaxes.
}}}\\
\multicolumn{8}{@{}l}{
\makebox[0pt][l]{
\parbox{0.88\linewidth}{
\footnotesize
3. The $T_{\rm kin}$ values are adopted from \citet{2026ApJ...998..182L}, while the ellipses indicate that no $T_{\rm kin}$ value is available from CH$_3$CCH 5-4 observations.
}}}\\
\endlastfoot
G000.67-00.03       	&	Sgr B2	&	17:47:20.00    	&	-28:22:40.00   	&	62	$\pm$	5	&	7.8	&	0.4	& ...	\\
G005.88-00.39       	&		&	18:00:30.31    	&	-24:04:04.50   	&	9	$\pm$	3	&	3.0	&	5.2	& 48.5$\pm$0.4	\\
G009.62+00.19       	&		&	18:06:14.66    	&	-20:31:31.70   	&	2	$\pm$	3	&	5.2	&	3.2	& 36.6$\pm$0.3 	\\
G010.47+00.02       	&		&	18:08:38.23    	&	-19:51:50.30   	&	69	$\pm$	5	&	8.5	&	1.6	& ...	\\
G010.62-00.38       	&	W 31	&	18:10:28.55    	&	-19:55:48.60   	&	-3	$\pm$	5	&	5.0	&	3.4	& 42.2$\pm$0.3 	\\
G011.49-01.48       	&		&	18:16:22.13    	&	-19:41:27.20   	&	11	$\pm$	3	&	1.3	&	6.9	& 26.9$\pm$0.8 	\\
G011.91-00.61       	&		&	18:13:58.12    	&	-18:54:20.30   	&	37	$\pm$	5	&	3.4	&	4.9	& 26.3$\pm$0.3 	\\
G012.02-00.03       	&		&	18:12:01.84    	&	-18:31:55.80   	&	108	$\pm$	5	&	9.4	&	2.2	& ...	\\
G012.68-00.18       	&		&	18:13:54.75    	&	-18:01:46.60   	&	58	$\pm$	10	&	2.4	&	5.8	& ...	\\
G012.80-00.20       	&		&	18:14:14.23    	&	-17:55:40.50   	&	34	$\pm$	5	&	2.9	&	5.3	& 44.4$\pm$0.4 	\\
G012.88+00.48       	&	IRAS 18089-1732	&	18:11:51.42    	&	-17:31:29.00   	&	31	$\pm$	7	&	2.5	&	5.7	& 36.5$\pm$0.4	\\
G012.90-00.24       	&		&	18:14:34.42    	&	-17:51:51.90   	&	36	$\pm$	10	&	2.5	&	5.8	& 30.2$\pm$0.8	\\
G012.90-00.26       	&		&	18:14:39.57    	&	-17:52:00.40   	&	39	$\pm$	10	&	2.5	&	5.7	& 34.7$\pm$0.4 	\\
G013.87+00.28       	&		&	18:14:35.83    	&	-16:45:35.90   	&	48	$\pm$	10	&	3.9	&	4.4	& 31.3$\pm$0.7 	\\
G014.33-00.64       	&		&	18:18:54.67    	&	-16:47:50.30   	&	22	$\pm$	5	&	1.1	&	7.1	& 31.4$\pm$0.3 	\\
G014.63-00.57       	&		&	18:19:15.54    	&	-16:29:45.80   	&	19	$\pm$	5	&	1.8	&	6.4	& 35.1$\pm$0.5	\\
G015.03-00.67       	&	M 17	&	18:20:24.81    	&	-16:11:35.30   	&	22	$\pm$	3	&	2.0	&	6.3	& 52.9$\pm$0.8 	\\
G016.58-00.05       	&		&	18:21:09.08    	&	-14:31:48.80   	&	60	$\pm$	5	&	3.6	&	4.8	& 26.5$\pm$0.6 	\\
G023.00-00.41       	&		&	18:34:40.20    	&	-09:00:37.00   	&	80	$\pm$	3	&	4.6	&	4.3	& 29.9$\pm$0.4 	\\
G023.43-00.18       	&		&	18:34:39.29    	&	-08:31:25.40   	&	97	$\pm$	3	&	5.9	&	3.6	& 29.2$\pm$0.6 	\\
G023.65-00.12       	&		&	18:34:51.59    	&	-08:18:21.40   	&	83	$\pm$	3	&	3.2	&	5.4	& ...	\\
G023.70-00.19       	&		&	18:35:12.36    	&	-08:17:39.50   	&	73	$\pm$	5	&	6.2	&	3.5	& ... 	\\
G025.70+00.04       	&		&	18:38:03.14    	&	-06:24:15.50   	&	93	$\pm$	5	&	10.2	&	4.5	& 23.0$\pm$0.4 	\\
G027.36-00.16       	&		&	18:41:51.06    	&	-05:01:43.40   	&	92	$\pm$	3	&	8.0	&	3.8	& 28.1$\pm$0.7 	\\
G028.86+00.06       	&		&	18:43:46.22    	&	-03:35:29.60   	&	100	$\pm$	10	&	7.4	&	3.9	& 37.9$\pm$1.0	\\
G029.86-00.04       	&		&	18:45:59.57    	&	-02:45:06.70   	&	100	$\pm$	3	&	6.2	&	4.1	& ...	\\
G029.95-00.01       	&	W 43S	&	18:46:03.74    	&	-02:39:22.30   	&	98	$\pm$	3	&	5.3	&	4.4	& 24.1$\pm$0.7 	\\
G031.28+00.06       	&		&	18:48:12.39    	&	-01:26:30.70   	&	109	$\pm$	3	&	4.3	&	5.0	& 40.9$\pm$0.4 	\\
G031.58+00.07       	&	W 43Main	&	18:48:41.68    	&	-01:09:59.00   	&	96	$\pm$	5	&	4.9	&	4.7	& 32.1$\pm$0.5 	\\
G032.04+00.05       	&		&	18:49:36.58    	&	-00:45:46.90   	&	97	$\pm$	5	&	5.2	&	4.7	& 33.0$\pm$0.6 	\\
G033.64-00.22       	&		&	18:53:32.56    	&	+00:31:39.10   	&	60	$\pm$	3	&	6.5	&	4.5	& ...	\\
G034.39+00.22       	&		&	18:53:19.00    	&	+01:24:08.80   	&	57	$\pm$	5	&	1.6	&	6.9	& 36.9$\pm$0.8 	\\
G035.02+00.34       	&		&	18:54:00.67    	&	+02:01:19.20   	&	52	$\pm$	5	&	2.3	&	6.4	& 33.1$\pm$0.6 	\\
G035.19-00.74       	&		&	18:58:13.05    	&	+01:40:35.70   	&	30	$\pm$	7	&	2.2	&	6.5	& 34.9$\pm$0.3 	\\
G035.20-01.73       	&		&	19:01:45.54    	&	+01:13:32.50   	&	42	$\pm$	3	&	3.3	&	5.8	& ...	\\
G037.43+01.51       	&		&	18:54:14.35    	&	+04:41:41.70   	&	41	$\pm$	3	&	1.9	&	6.8	& ...	\\
G043.16+00.01       	&	W 49N	&	19:10:13.41    	&	+09:06:12.80   	&	10	$\pm$	5	&	11.1	&	7.6	& ...	\\
G043.79-00.12       	&	OH 43.8-0.1	&	19:11:53.99    	&	+09:35:50.30   	&	44	$\pm$	10	&	6.0	&	5.6	& ...	\\
G043.89-00.78       	&		&	19:14:26.39    	&	+09:22:36.50   	&	54	$\pm$	5	&	8.3	&	6.1	& ...	\\
G045.07+00.13       	&		&	19:13:22.04    	&	+10:50:53.30   	&	59	$\pm$	5	&	8.0	&	6.2	& ...	\\
G045.45+00.05       	&		&	19:14:21.27    	&	+11:09:15.90   	&	55	$\pm$	7	&	8.4	&	6.4	& ...	\\
G048.60+00.02       	&		&	19:20:31.18    	&	+13:55:25.20   	&	18	$\pm$	5	&	10.8	&	8.1	& ...	\\
G049.19-00.33       	&		&	19:22:57.77    	&	+14:16:10.00   	&	67	$\pm$	5	&	5.3	&	6.2	& 42.1$\pm$2.6	\\
G049.48-00.36       	&	W 51 IRS2	&	19:23:39.82    	&	+14:31:05.00   	&	56	$\pm$	3	&	5.1	&	6.2	& 42.1$\pm$0.4 	\\
G049.48-00.38       	&	W 51M	&	19:23:43.87    	&	+14:30:29.50   	&	58	$\pm$	4	&	5.4	&	6.2	& 48.6$\pm$0.4 	\\
G052.10+01.04       	&	IRAS 19213+1723	&	19:23:37.32    	&	+17:29:10.50   	&	42	$\pm$	5	&	4.0	&	6.5	& ...	\\
G059.78+00.06       	&		&	19:43:11.25    	&	+23:44:03.30   	&	25	$\pm$	3	&	2.2	&	7.3	& 49.7$\pm$0.9 	\\
G069.54-00.97       	&	ON 1	&	20:10:09.07    	&	+31:31:36.00   	&	12	$\pm$	5	&	2.5	&	7.6	& 29.5$\pm$0.3 	\\
G074.03-01.71       	&		&	20:25:07.11    	&	+34:49:57.60   	&	5	$\pm$	5	&	1.6	&	7.9	& 33.4$\pm$0.7	\\
G075.29+01.32       	&		&	20:16:16.01    	&	+37:35:45.80   	&	-58	$\pm$	5	&	9.3	&	10.7	& ...	\\
G075.76+00.33       	&		&	20:21:41.09    	&	+37:25:29.30   	&	-9	$\pm$	9	&	3.5	&	8.0	& 43.9$\pm$0.6	\\
G075.78+00.34       	&	ON 2N	&	20:21:44.01    	&	+37:26:37.50   	&	1	$\pm$	5	&	3.8	&	8.1	& 41.7$\pm$0.7	\\
G076.38-00.61       	&		&	20:27:25.48    	&	+37:22:48.50   	&	-2	$\pm$	5	&	1.3	&	7.9	& 54.5$\pm$1.6 	\\
G078.12+03.63       	&	IRAS 20126+4104	&	20:14:26.07    	&	+41:13:32.70   	&	-4	$\pm$	5	&	1.6	&	8.0	& 32.0$\pm$1.0 	\\
G078.88+00.70       	&	AFGL 2591	&	20:29:24.82    	&	+40:11:19.60   	&	-6	$\pm$	7	&	3.3	&	8.2	& 26.9$\pm$0.8	\\
G079.73+00.99       	&	IRAS 20290+4052	&	20:30:50.67    	&	+41:02:27.50   	&	-3	$\pm$	5	&	1.4	&	8.0	& 29.2$\pm$1.1	\\
G079.87+01.17       	&		&	20:30:29.14    	&	+41:15:53.60   	&	-5	$\pm$	10	&	1.6	&	8.0	& 26.6$\pm$0.8 	\\
G080.79-01.92       	&	NML Cyg	&	20:46:25.54    	&	+40:06:59.40   	&	-3	$\pm$	3	&	1.6	&	8.1	& ...	\\
G080.86+00.38       	&	DR 20	&	20:37:00.96    	&	+41:34:55.70   	&	-3	$\pm$	5	&	1.5	&	8.0	& 32.5$\pm$1.4	\\
G081.75+00.59       	&	DR 21	&	20:39:01.99    	&	+42:24:59.30   	&	-3	$\pm$	3	&	1.5	&	8.1	& 31.3$\pm$0.3 	\\
G081.87+00.78       	&	W 75N	&	20:38:36.43    	&	+42:37:34.80   	&	7	$\pm$	3	&	1.3	&	8.1	& 30.4$\pm$0.5 	\\
G090.21+02.32       	&		&	21:02:22.70    	&	+50:03:08.30   	&	-3	$\pm$	5	&	0.7	&	8.2	& ...	\\
G092.67+03.07       	&		&	21:09:21.73    	&	+52:22:37.10   	&	-5	$\pm$	10	&	1.6	&	8.4	& 38.3$\pm$0.6 	\\
G094.60-01.79       	&	AFGL 2789	&	21:39:58.27    	&	+50:14:21.00   	&	-46	$\pm$	5	&	3.6	&	9.2	& ...	\\
G095.29-00.93       	&		&	21:39:40.51    	&	+51:20:32.80   	&	-38	$\pm$	5	&	4.9	&	9.9	& ...	\\
G097.53+03.18       	&		&	21:32:12.43    	&	+55:53:49.70   	&	-73	$\pm$	5	&	7.5	&	11.8	& ...	\\
G100.37-03.57       	&		&	22:16:10.37    	&	+52:21:34.10   	&	-37	$\pm$	10	&	3.4	&	9.4	& ...	\\
G105.41+09.87       	&		&	21:43:06.48    	&	+66:06:55.30   	&	-10	$\pm$	5	&	0.9	&	8.4	& ...	\\
G107.29+05.63       	&	IRAS 22198+6336	&	22:21:26.73    	&	+63:51:37.90   	&	-11	$\pm$	5	&	0.8	&	8.4	& ...	\\
G108.18+05.51       	&	L 1206	&	22:28:51.41    	&	+64:13:41.30   	&	-11	$\pm$	3	&	0.8	&	8.4	& ...	\\
G108.20+00.58       	&		&	22:49:31.48    	&	+59:55:42.00   	&	-49	$\pm$	5	&	4.4	&	10.4	& ...	\\
G108.47-02.81       	&		&	23:02:32.08    	&	+56:57:51.40   	&	-54	$\pm$	5	&	3.2	&	9.7	& ...	\\
G108.59+00.49       	&		&	22:52:38.30    	&	+60:00:52.00   	&	-52	$\pm$	5	&	2.5	&	9.3	& ...	\\
G109.87+02.11       	&	Cep A	&	22:56:18.10    	&	+62:01:49.50   	&	-7	$\pm$	5	&	0.7	&	8.4	& 36.1$\pm$0.4 	\\
G111.23-01.23       	&		&	23:17:20.79    	&	+59:28:47.00   	&	-53	$\pm$	10	&	3.5	&	9.9	& ...	\\
G111.25-00.76       	&		&	23:16:10.36    	&	+59:55:28.50   	&	-43	$\pm$	5	&	3.4	&	9.9	& ...	\\
G111.54+00.77       	&	NGC 7538	&	23:13:45.36    	&	+61:28:10.60   	&	-57	$\pm$	5	&	2.6	&	9.4	& 43.1$\pm$0.9 	\\
G121.29+00.65       	&	L 1287	&	00:36:47.35    	&	+63:29:02.20   	&	-23	$\pm$	5	&	0.9	&	8.7	& 25.2$\pm$0.9 	\\
G122.01-07.08       	&	IRAS 00420+5530	&	00:44:58.40    	&	+55:46:47.60   	&	-50	$\pm$	5	&	2.2	&	9.5	& ...	\\
G123.06-06.30       	&	NGC 281	&	00:52:24.70    	&	+56:33:50.50   	&	-30	$\pm$	5	&	2.8	&	10.0	& 27.1$\pm$1.0 	\\
G123.06-06.30       	&	NGC 281W	&	00:52:24.20    	&	+56:33:43.20   	&	-29	$\pm$	3	&	2.4	&	9.7	& 37.1$\pm$1.2 	\\
G133.94+01.06       	&	W 3OH	&	02:27:03.82    	&	+61:52:25.20   	&	-47	$\pm$	3	&	2.0	&	9.6	& 38.1$\pm$0.8 	\\
G134.62-02.19       	&	S Per	&	02:22:51.71    	&	+58:35:11.40   	&	-39	$\pm$	5	&	2.4	&	10.0	& ...	\\
G135.27+02.79       	&	WB 89-437	&	02:43:28.57    	&	+62:57:08.40   	&	-72	$\pm$	3	&	6.0	&	13.1	& ...	\\
G160.14+03.15       	&		&	05:01:40.24    	&	+47:07:19.00   	&	-18	$\pm$	5	&	4.1	&	12.1	& ...	\\
G168.06+00.82       	&	IRAS 05137+3919	&	05:17:13.74    	&	+39:22:19.90   	&	-27	$\pm$	5	&	7.7	&	15.8	& ...	\\
G176.51+00.20       	&		&	05:37:52.14    	&	+32:00:03.90   	&	-17	$\pm$	5	&	1.0	&	9.1	& 49.6$\pm$1.9 	\\
G182.67-03.26       	&		&	05:39:28.42    	&	+24:56:32.10   	&	-7	$\pm$	10	&	6.7	&	14.9	& ...	\\
G183.72-03.66       	&		&	05:40:24.23    	&	+23:50:54.70   	&	3	$\pm$	5	&	1.8	&	9.9	& 31.6$\pm$1.1	\\
G188.79+01.03       	&	IRAS 06061+2151	&	06:09:06.97    	&	+21:50:41.40   	&	-5	$\pm$	5	&	2.0	&	10.1	& ...	\\
G188.94+00.88       	&	S 252	&	06:08:53.35    	&	+21:38:28.70   	&	8	$\pm$	5	&	2.1	&	10.2	& 31.2$\pm$0.9 	\\
G192.16-03.81       	&		&	05:58:13.53    	&	+16:31:58.90   	&	5	$\pm$	5	&	1.5	&	9.6	& ...	\\
G192.60-00.04       	&	S 255	&	06:12:54.02    	&	+17:59:23.30   	&	6	$\pm$	5	&	1.6	&	9.7	& 46.4$\pm$1.1 	\\
G196.45-01.67       	&	S 269	&	06:14:37.08    	&	+13:49:36.70   	&	19	$\pm$	5	&	5.3	&	13.3	& ...	\\
G209.00-19.38       	&	Orion Nebula	&	05:35:15.80    	&	-05:23:14.10   	&	3	$\pm$	5	&	0.4	&	8.5	& 38.1$\pm$0.7	\\
G211.59+01.05       	&		&	06:52:45.32    	&	+01:40:23.10   	&	45	$\pm$	5	&	4.4	&	12.1	& ...	\\
G229.57+00.15       	&		&	07:23:01.84    	&	-14:41:32.80   	&	47	$\pm$	10	&	4.5	&	11.6	& ...	\\
G232.62+00.99       	&		&	07:32:09.78    	&	-16:58:12.80   	&	21	$\pm$	3	&	1.7	&	9.3	& ...	\\
G236.81+01.98       	&		&	07:44:28.24    	&	-20:08:30.20   	&	43	$\pm$	7	&	3.4	&	10.4	& ...	\\
G239.35-05.06       	&	VY CMa	&	07:22:58.33    	&	-25:46:03.10   	&	20	$\pm$	3	&	1.2	&	8.8	& ...	\\
G240.31+00.07       	&		&	07:44:51.92    	&	-24:07:41.50   	&	67	$\pm$	5	&	4.7	&	11.3	& ...	\\
\end{longtable}

\section{Spectra of the full sample}
\label{app: spectra}

We present the spectra of DCN 1-0, DNC 1-0, H$^{13}$CN 1-0, HN$^{13}$C 1-0, HC$^{15}$N 1-0 and H$^{15}$NC 1-0.

\newcommand{\colw}{0.155\textwidth}

\begin{figure}[p]
\centering

\makebox[\colw][c]{DCN 1-0}
\makebox[\colw][c]{DNC 1-0}
\makebox[\colw][c]{H$^{13}$CN 1-0}
\makebox[\colw][c]{HN$^{13}$C 1-0}
\makebox[\colw][c]{HC$^{15}$N 1-0}
\makebox[\colw][c]{H$^{15}$NC 1-0}\\[0.2em]

\includegraphics[width=0.155\textwidth]{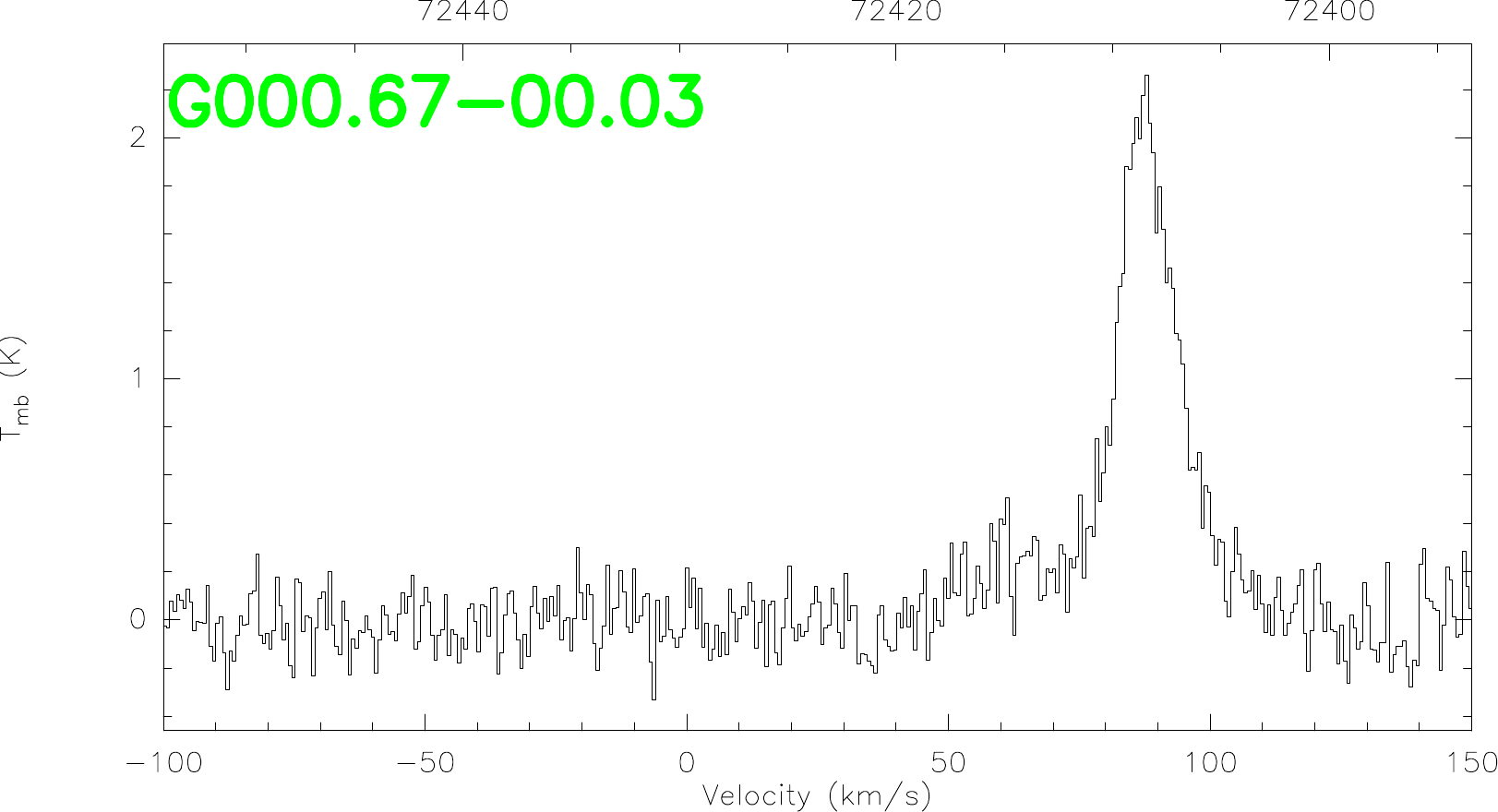}
\includegraphics[width=0.155\textwidth]{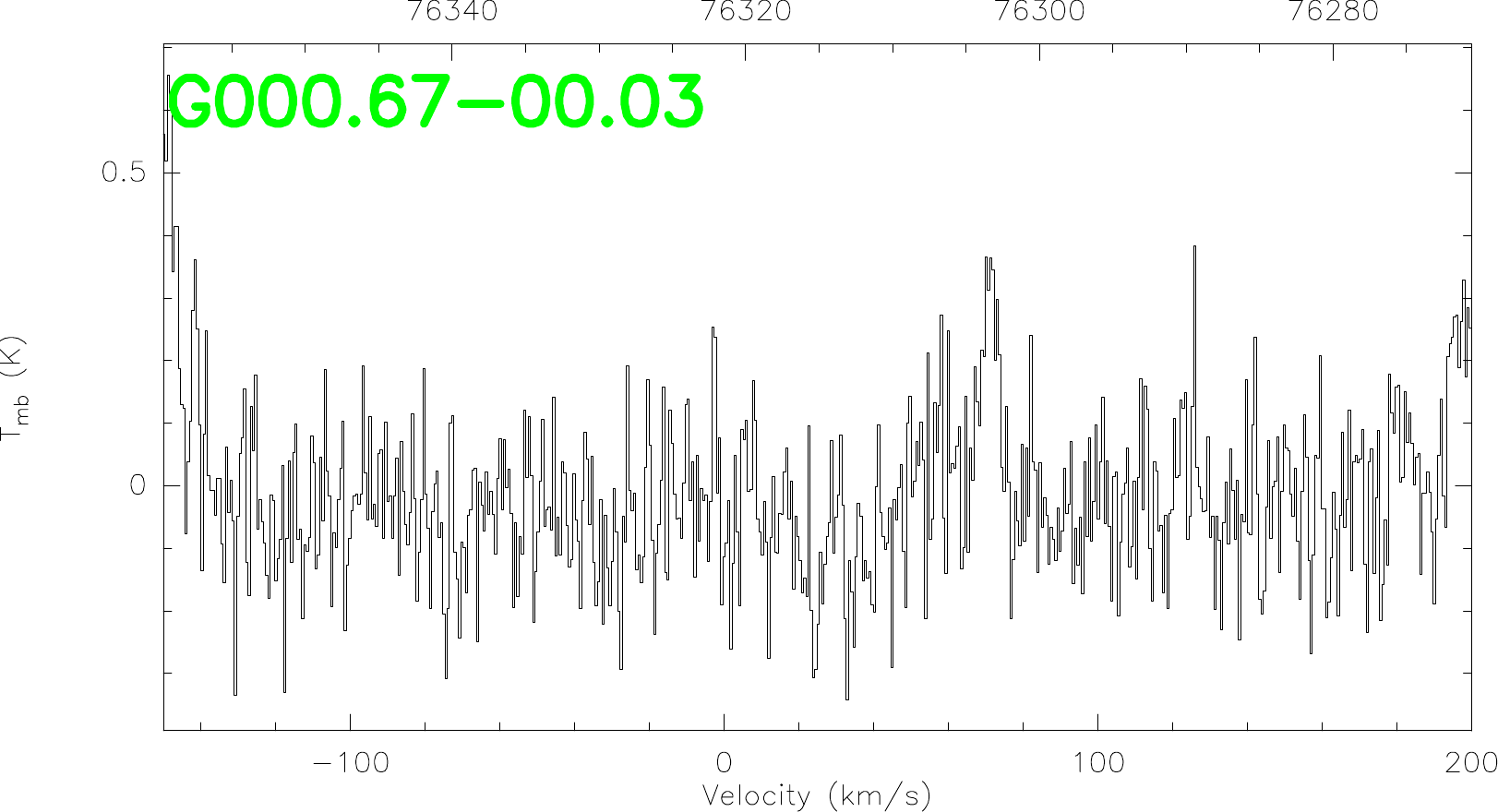}
\includegraphics[width=0.155\textwidth]{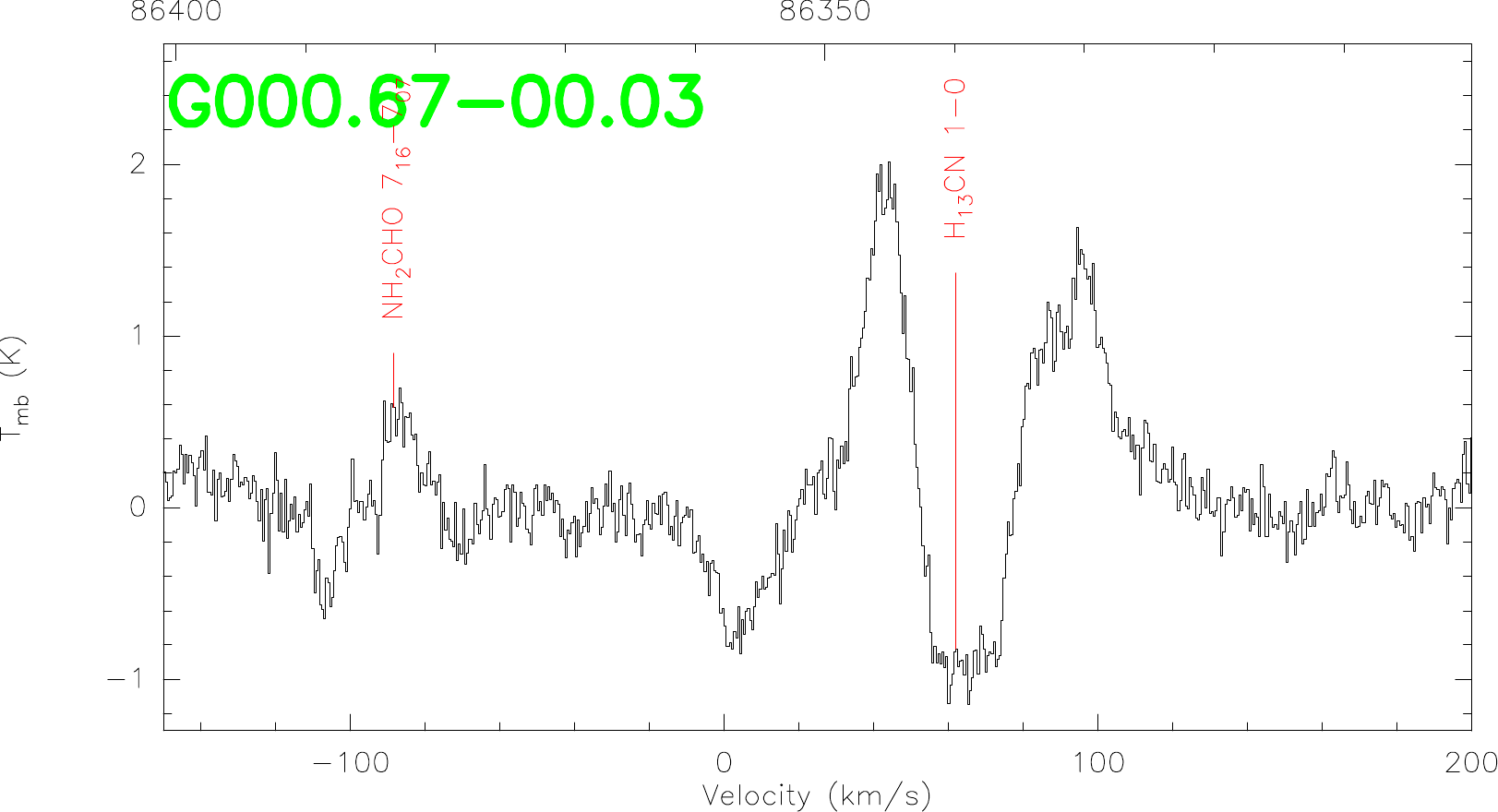}
\includegraphics[width=0.155\textwidth]{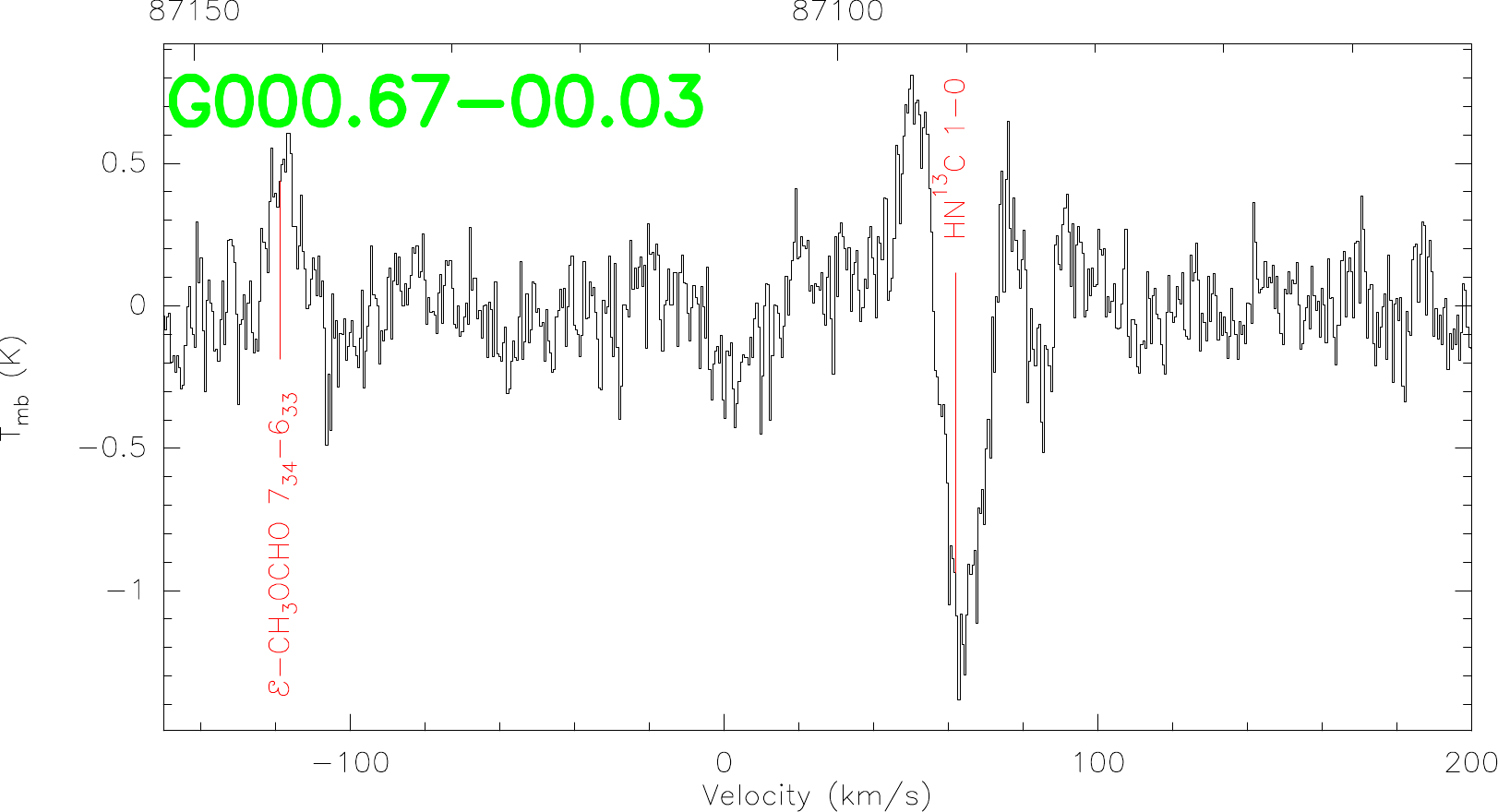}
\includegraphics[width=0.155\textwidth]{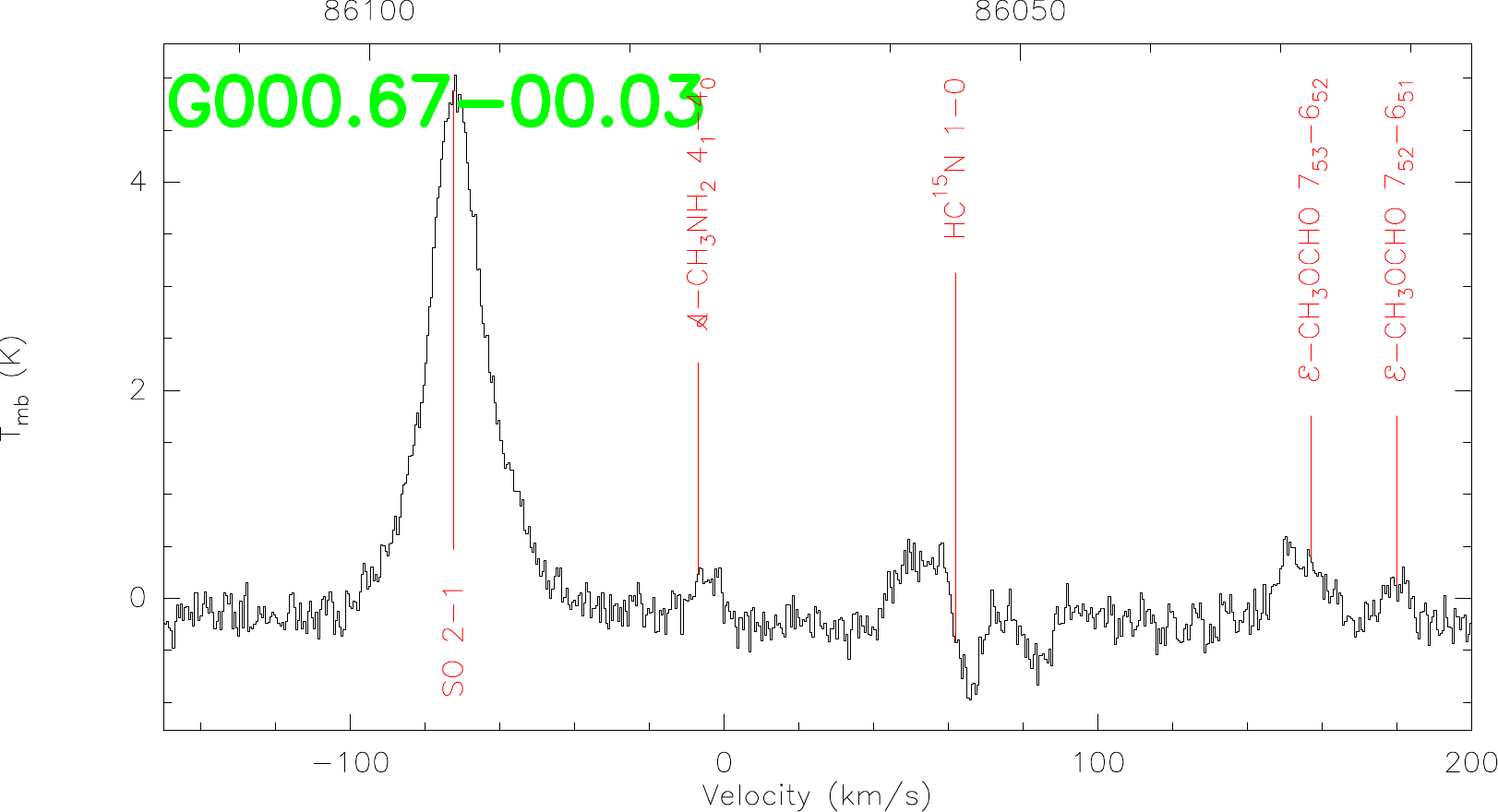}
\includegraphics[width=0.155\textwidth]{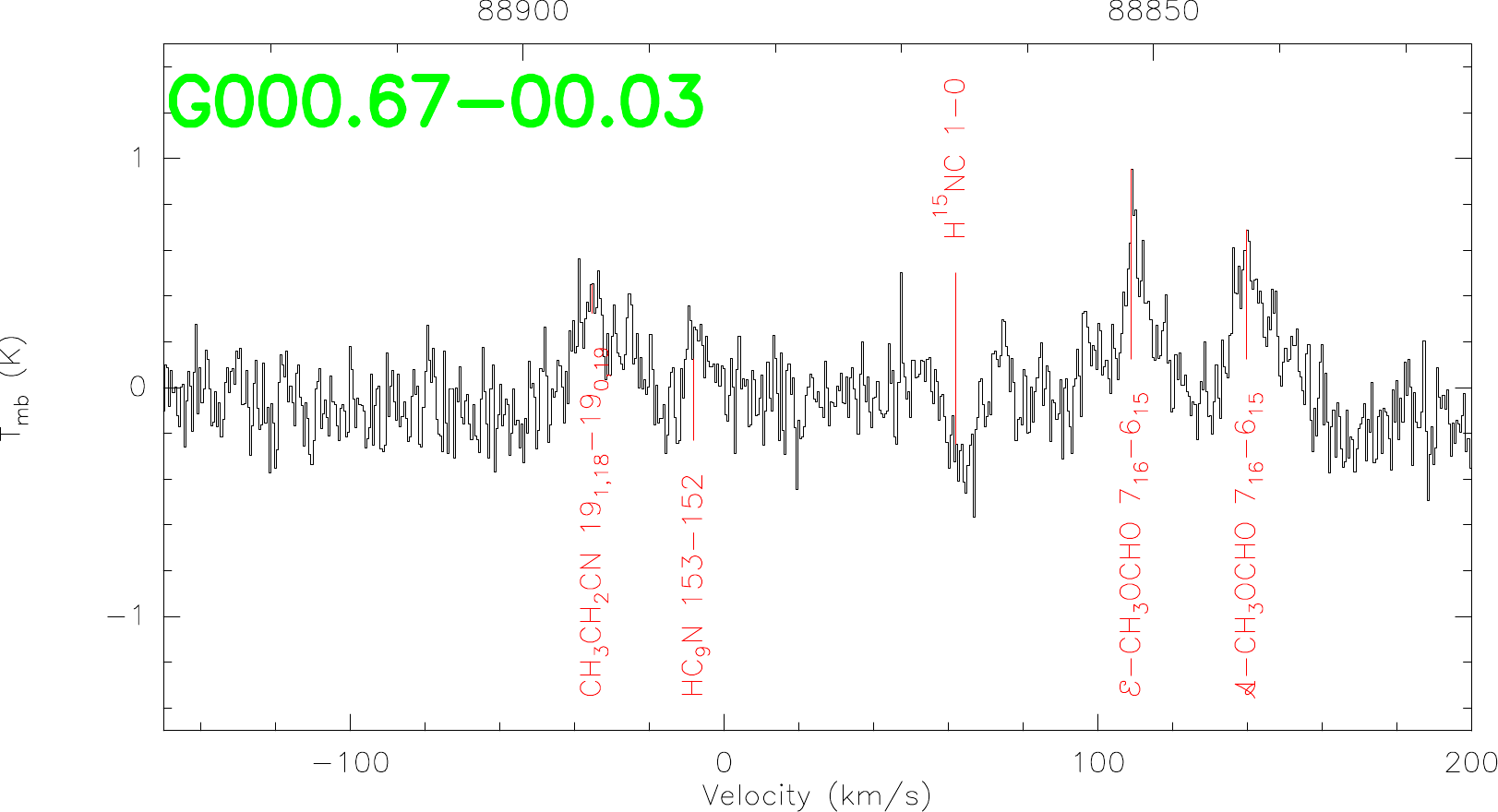}\\

\includegraphics[width=0.155\textwidth]{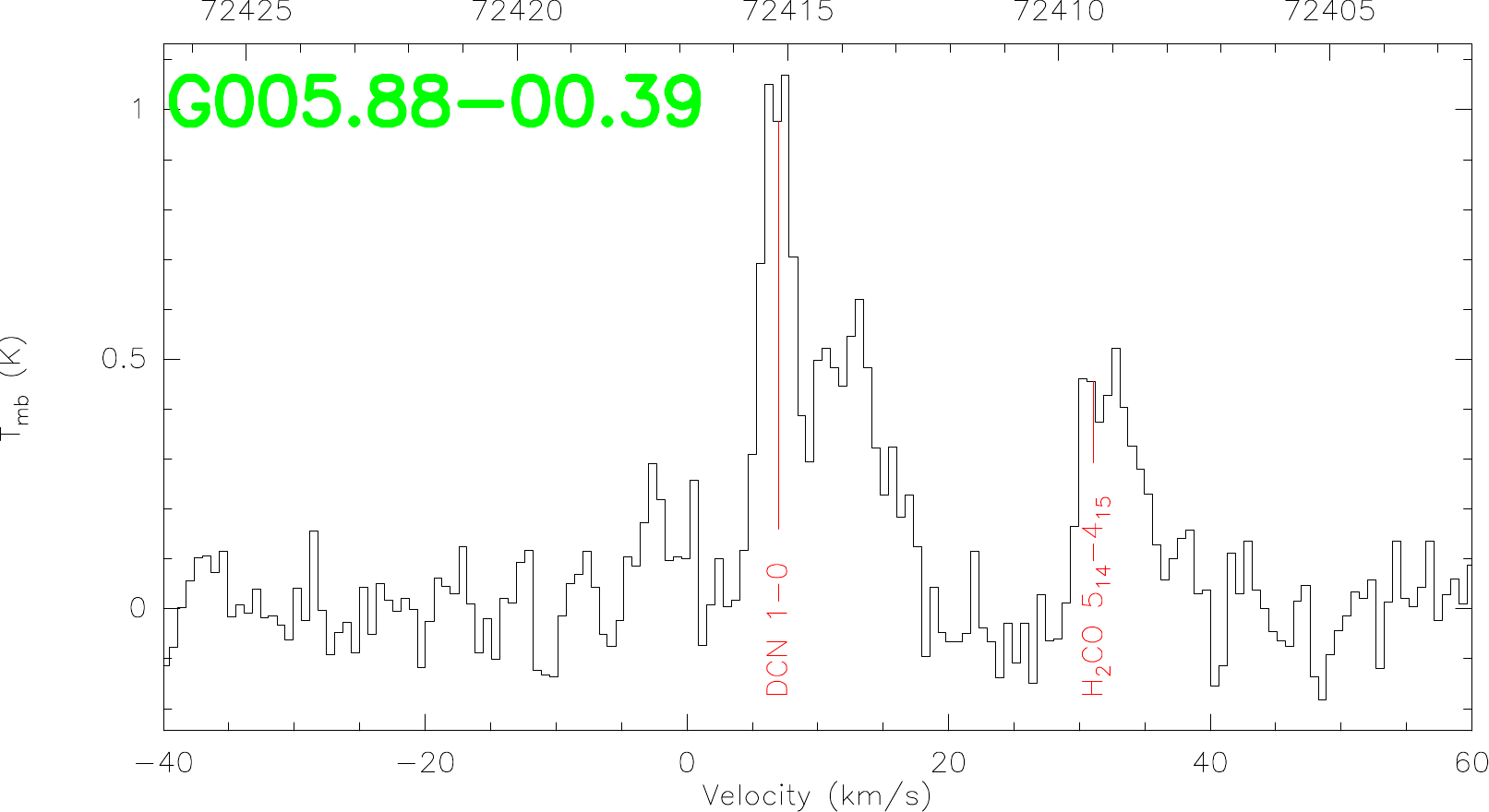}
\includegraphics[width=0.155\textwidth]{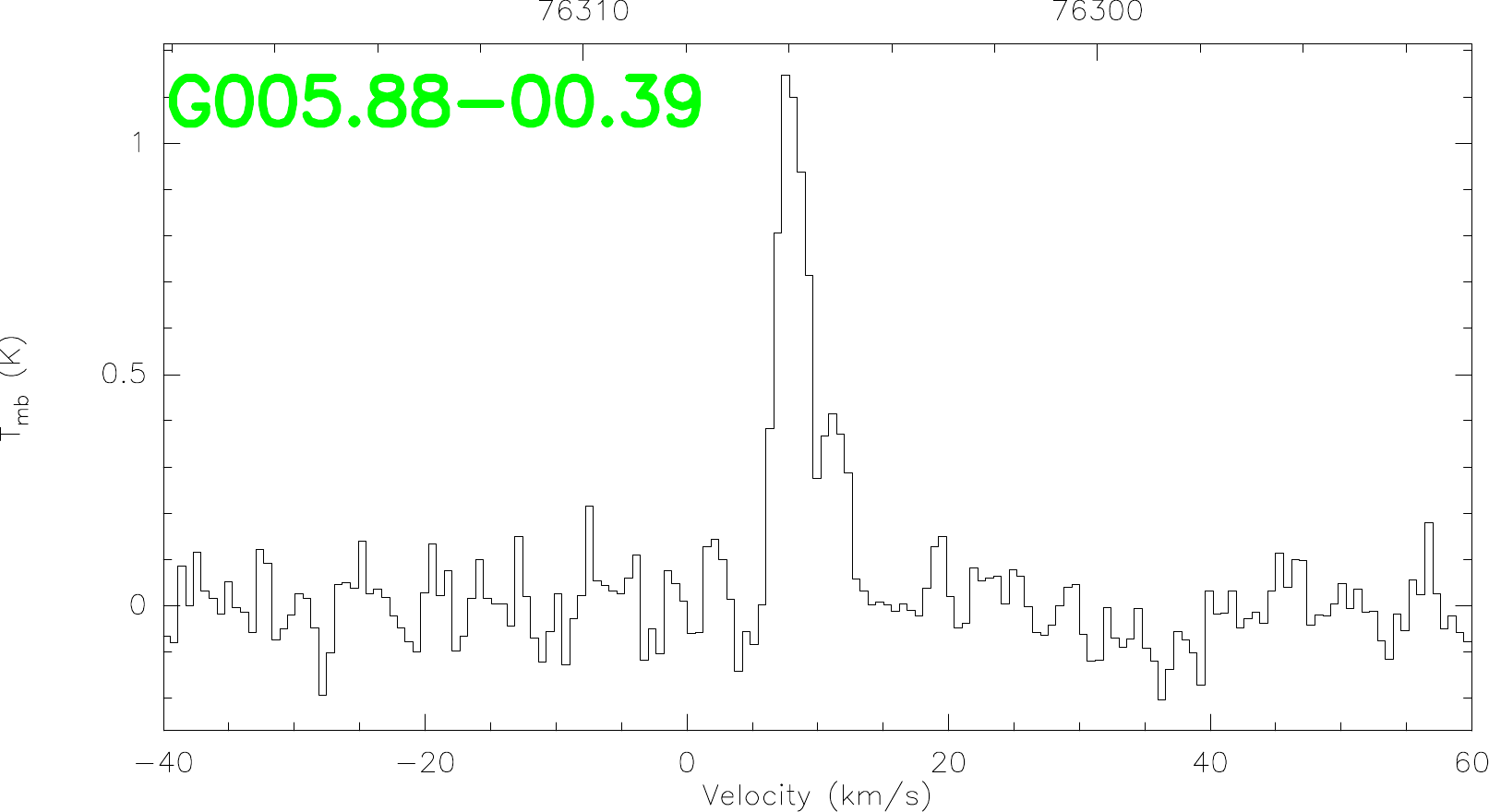}
\includegraphics[width=0.155\textwidth]{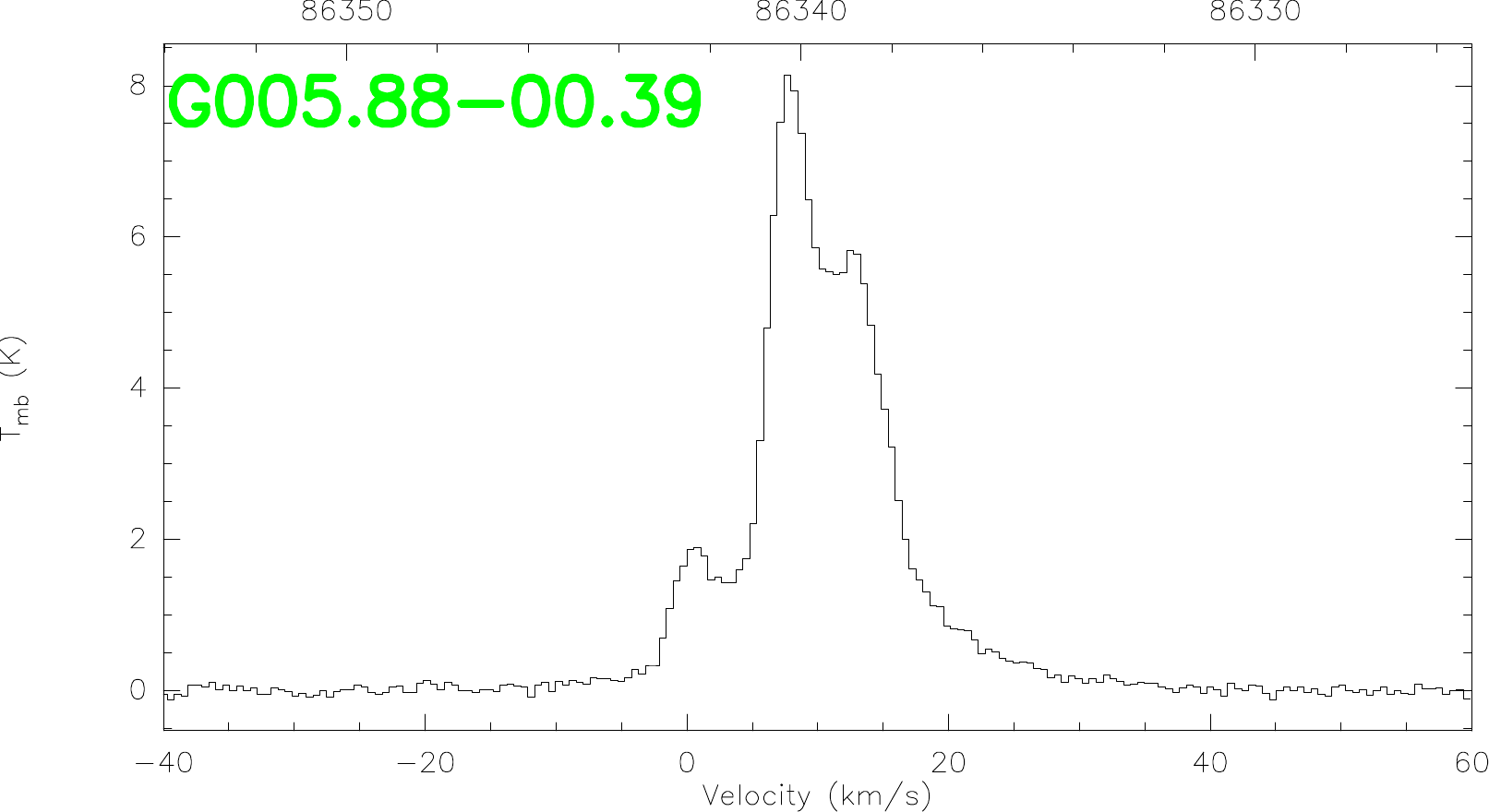}
\includegraphics[width=0.155\textwidth]{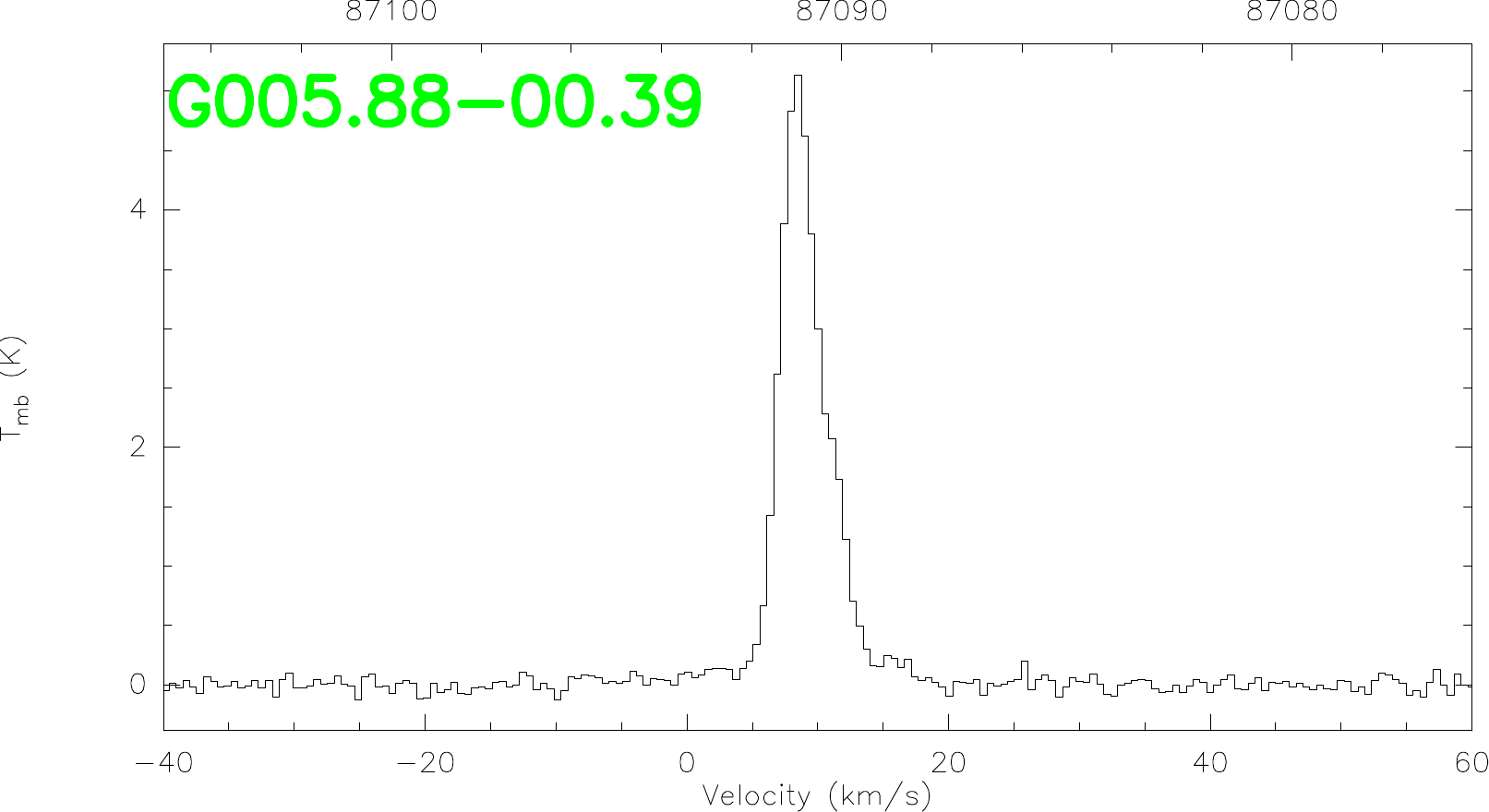}
\includegraphics[width=0.155\textwidth]{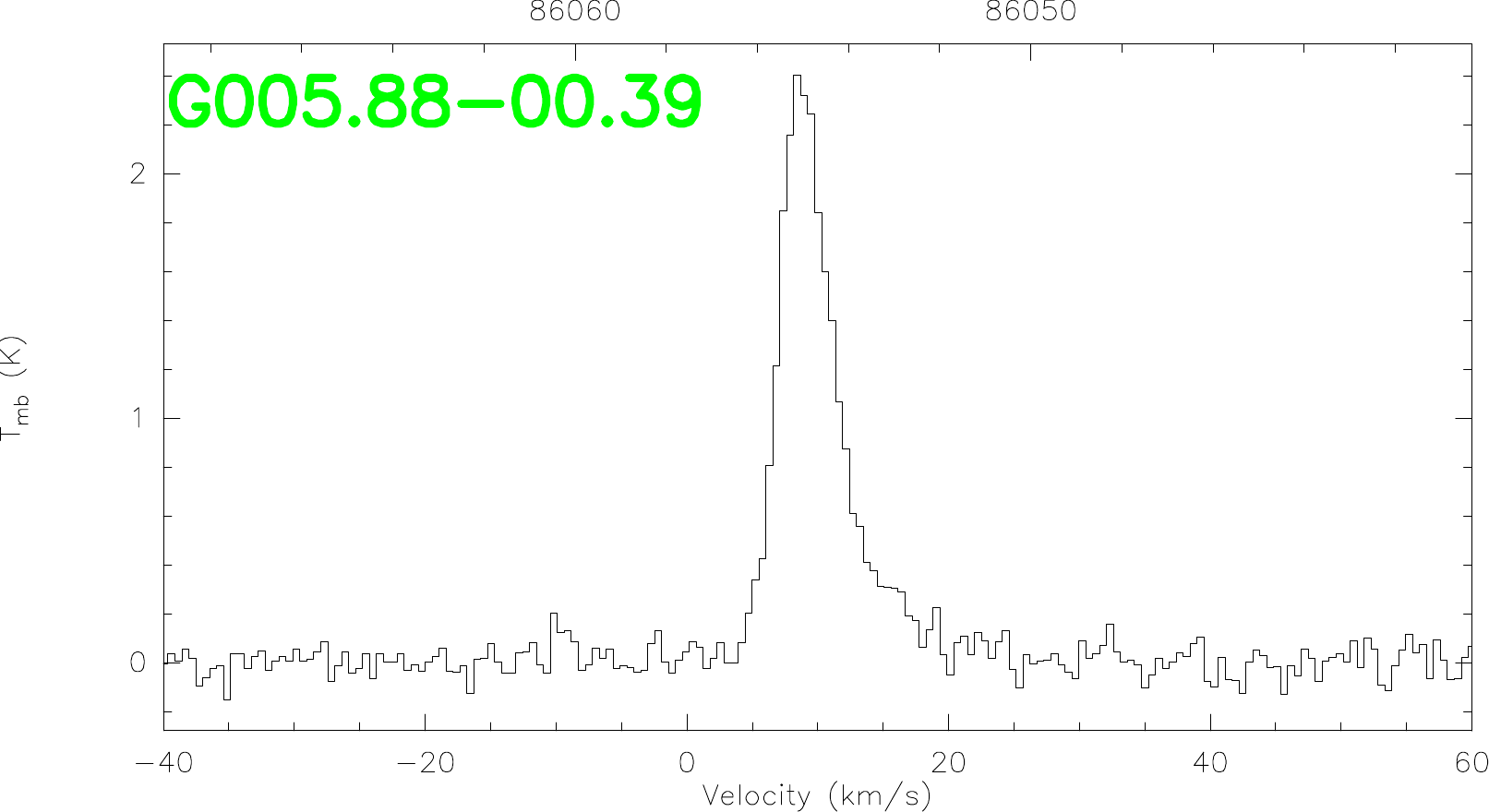}
\includegraphics[width=0.155\textwidth]{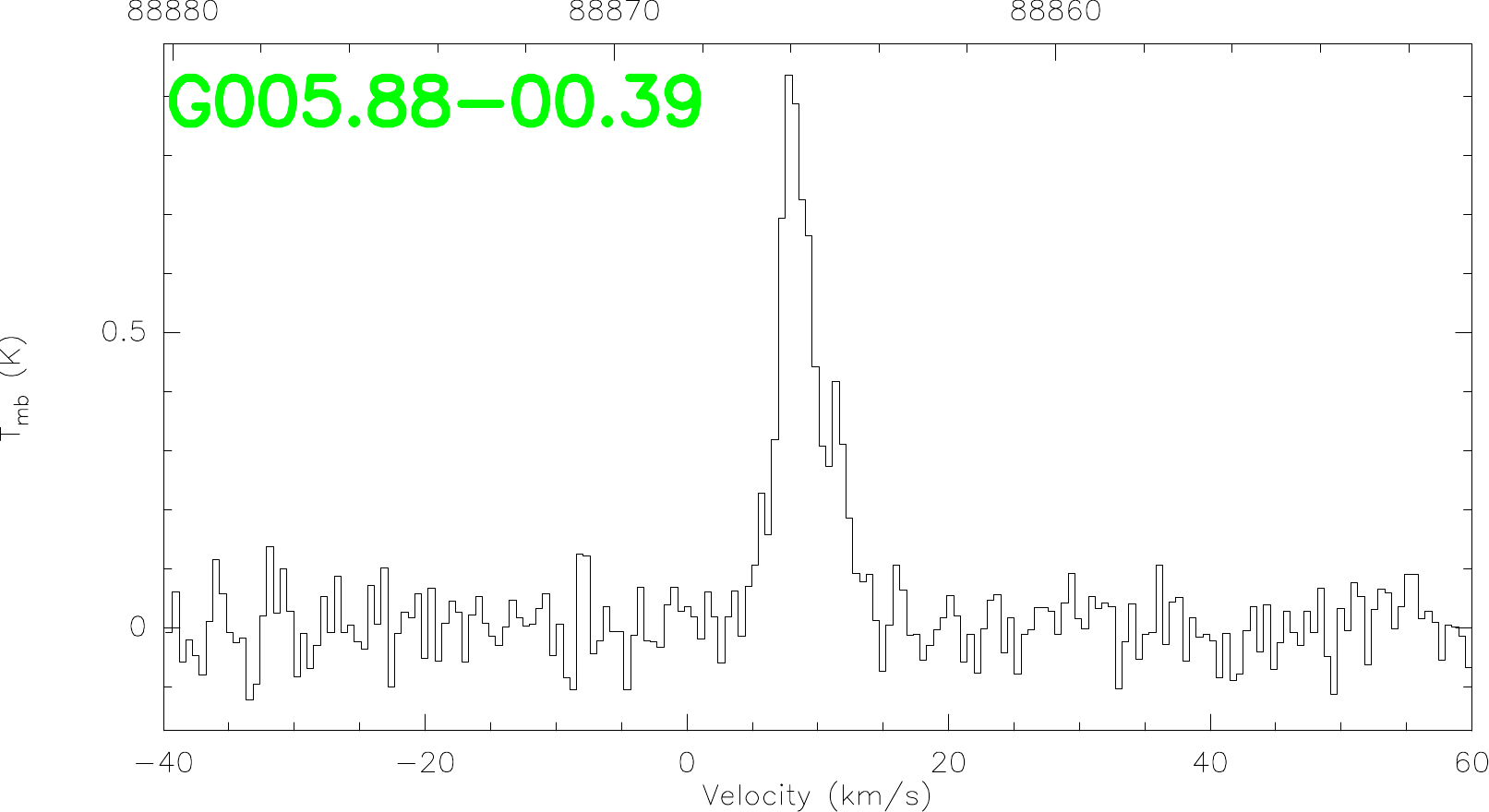}\\

\includegraphics[width=0.155\textwidth]{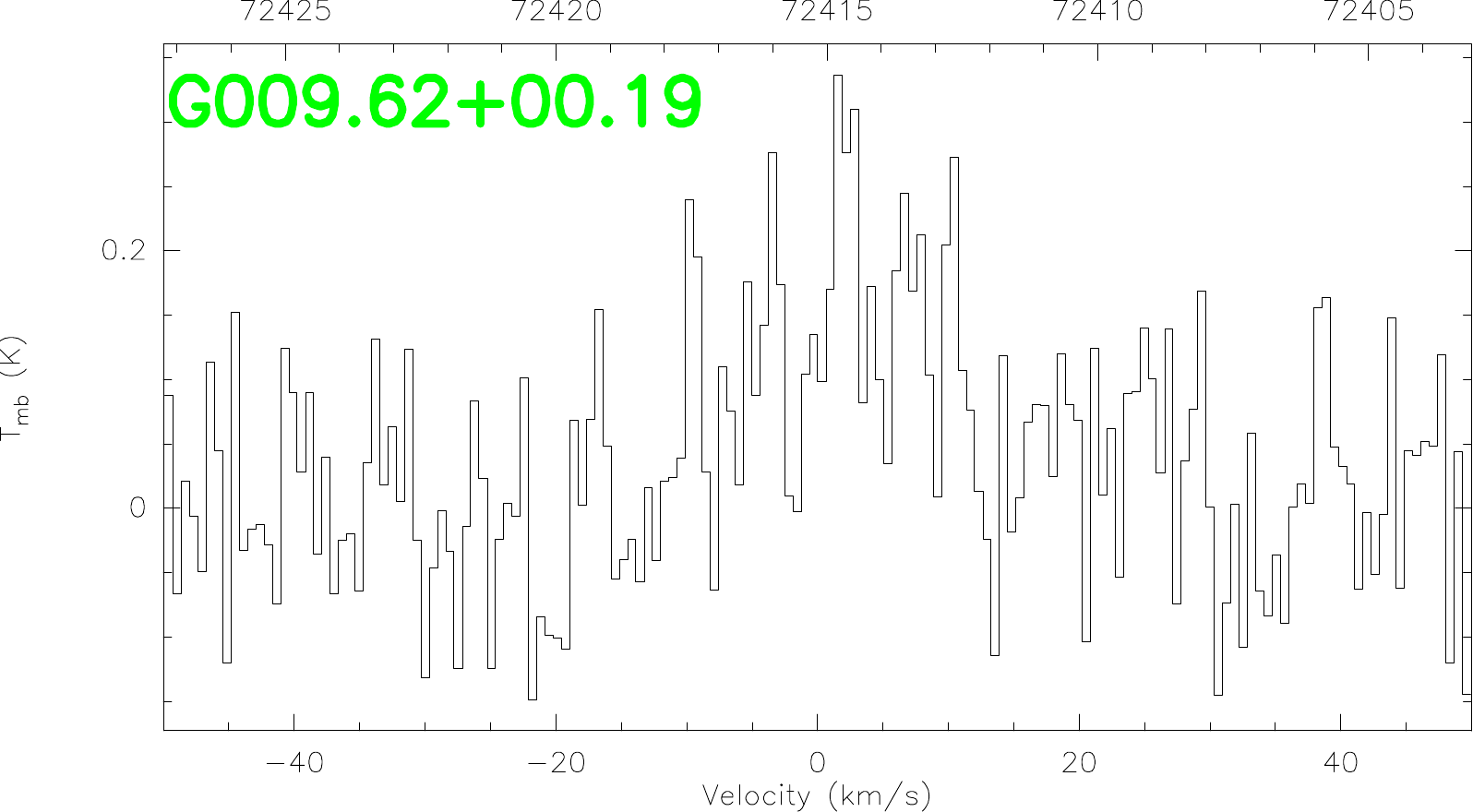}
\includegraphics[width=0.155\textwidth]{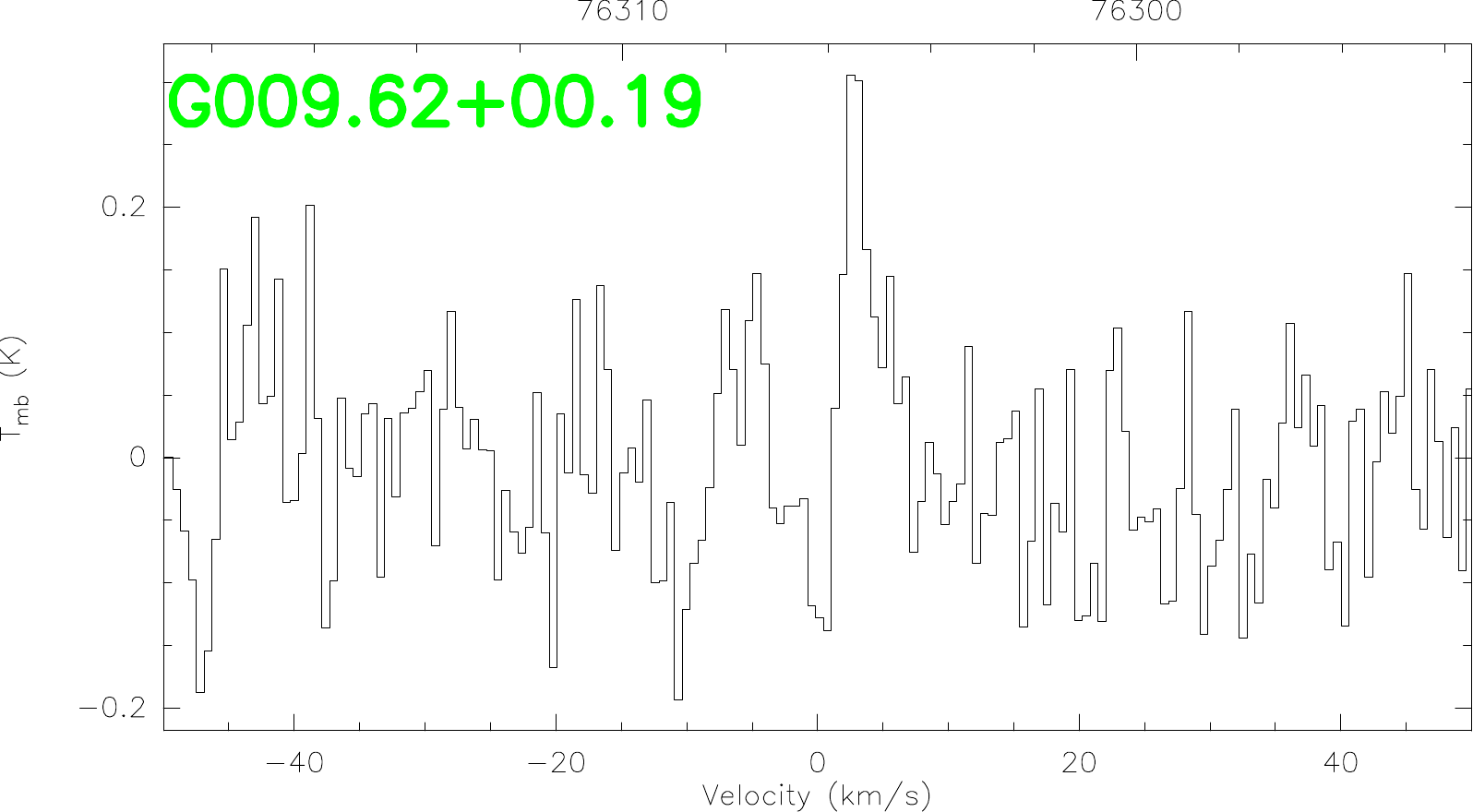}
\includegraphics[width=0.155\textwidth]{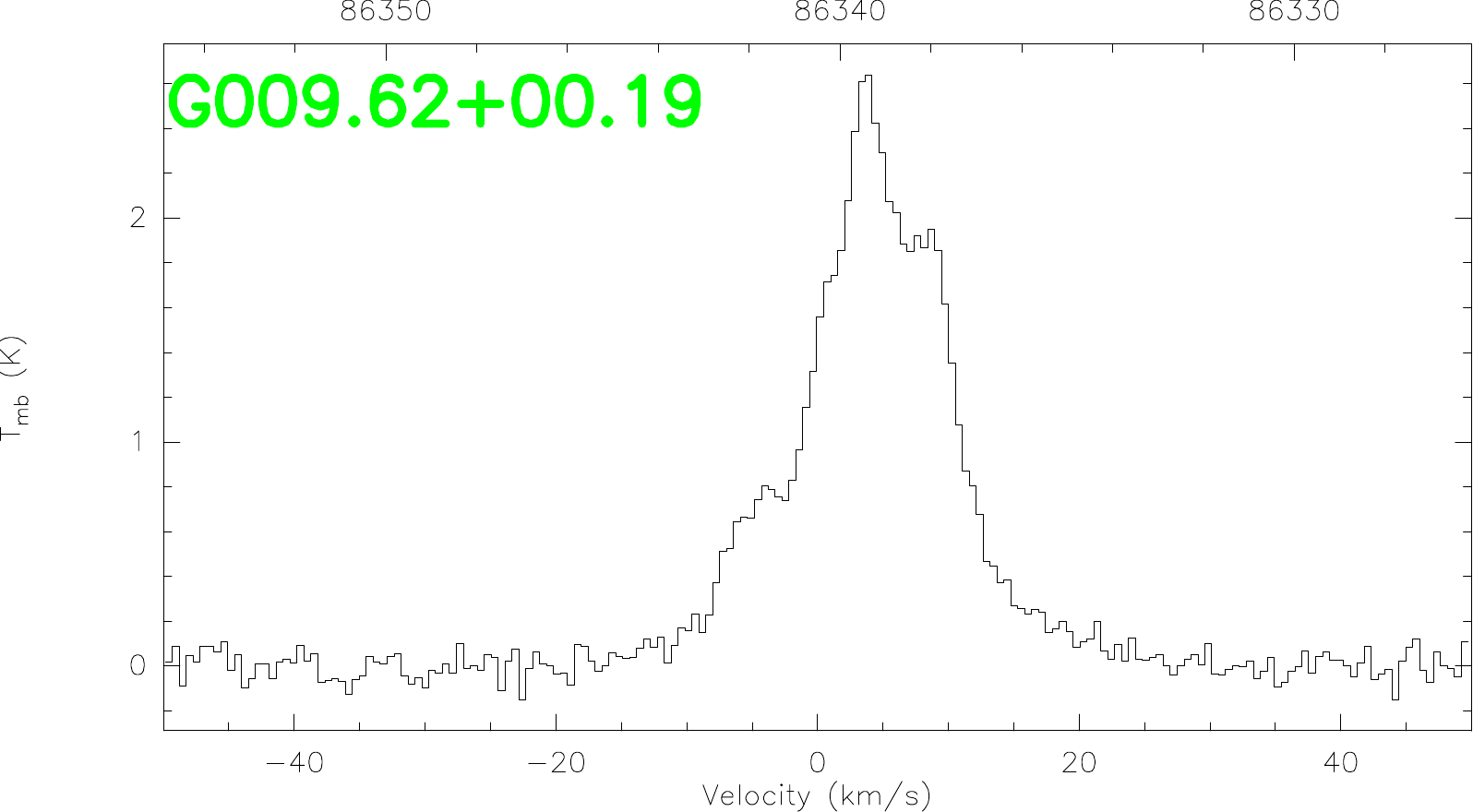}
\includegraphics[width=0.155\textwidth]{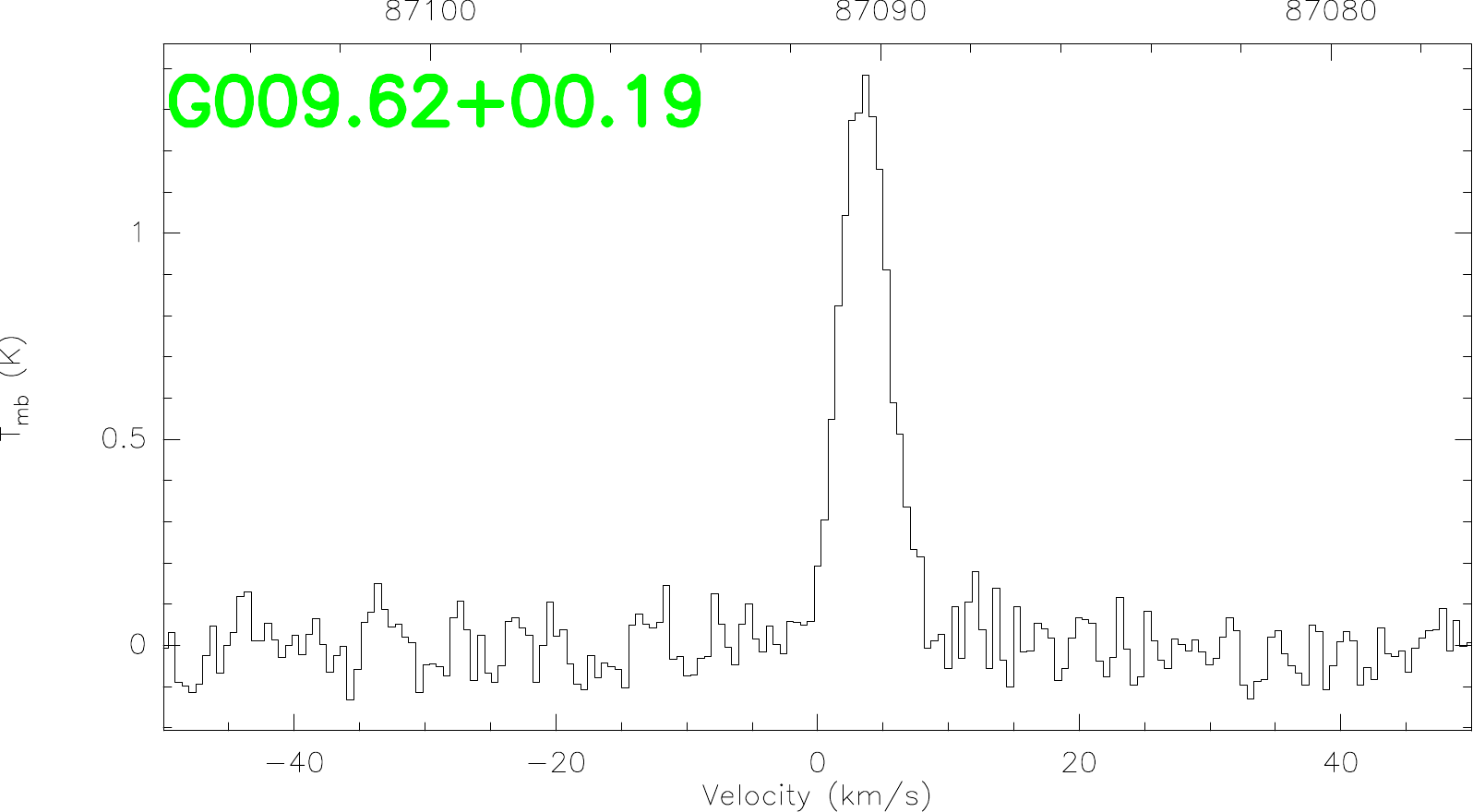}
\includegraphics[width=0.155\textwidth]{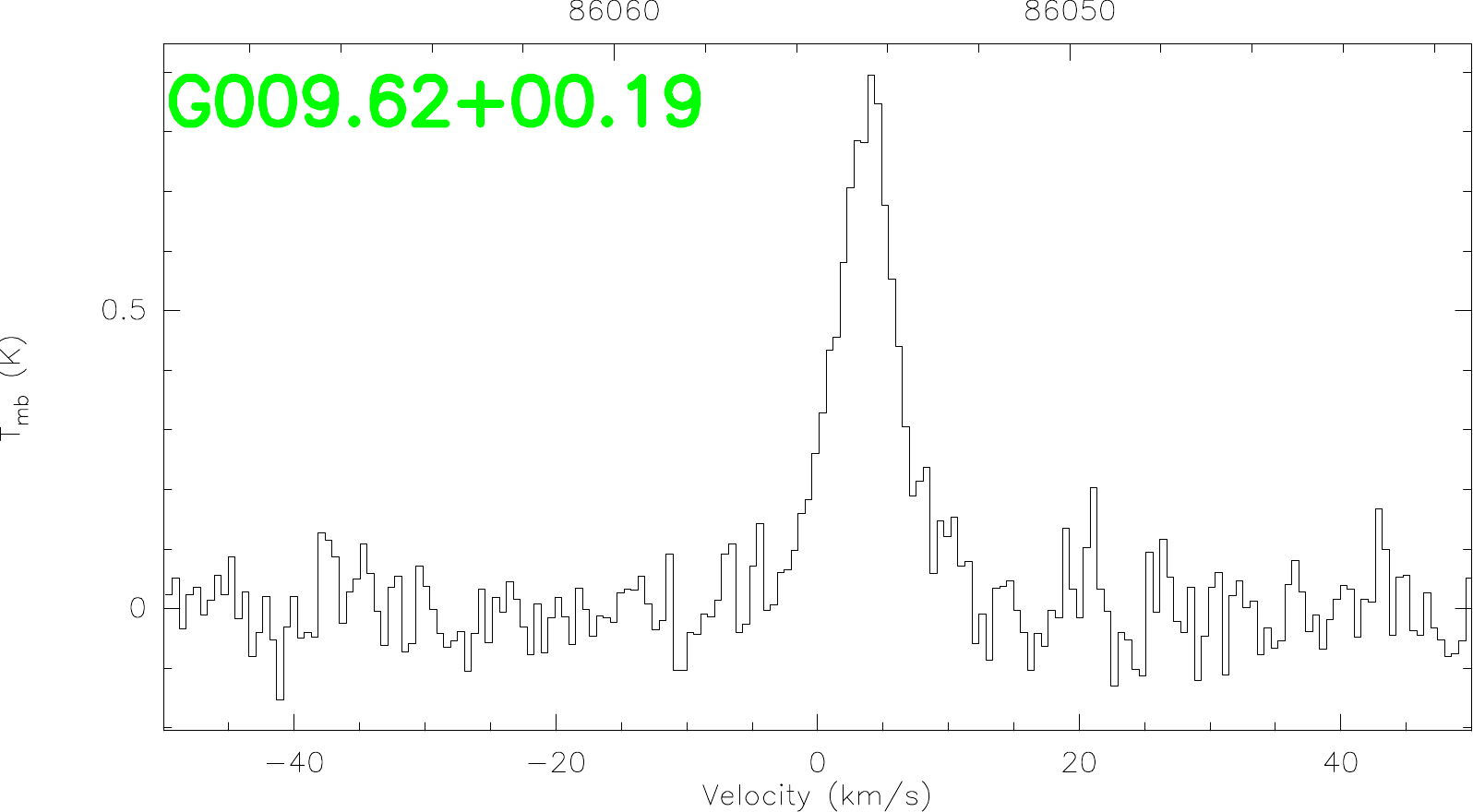}
\includegraphics[width=0.155\textwidth]{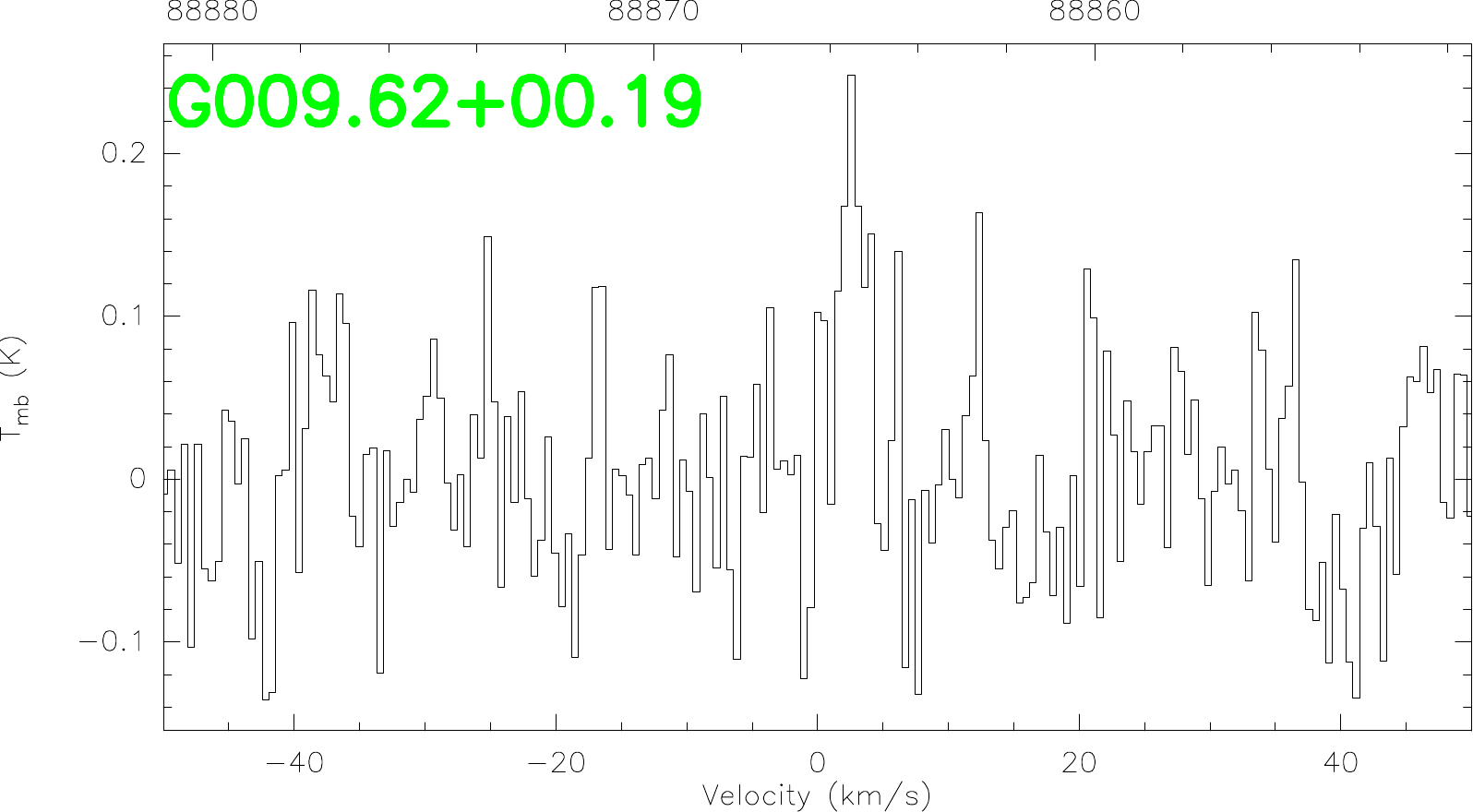}\\

\includegraphics[width=0.155\textwidth]{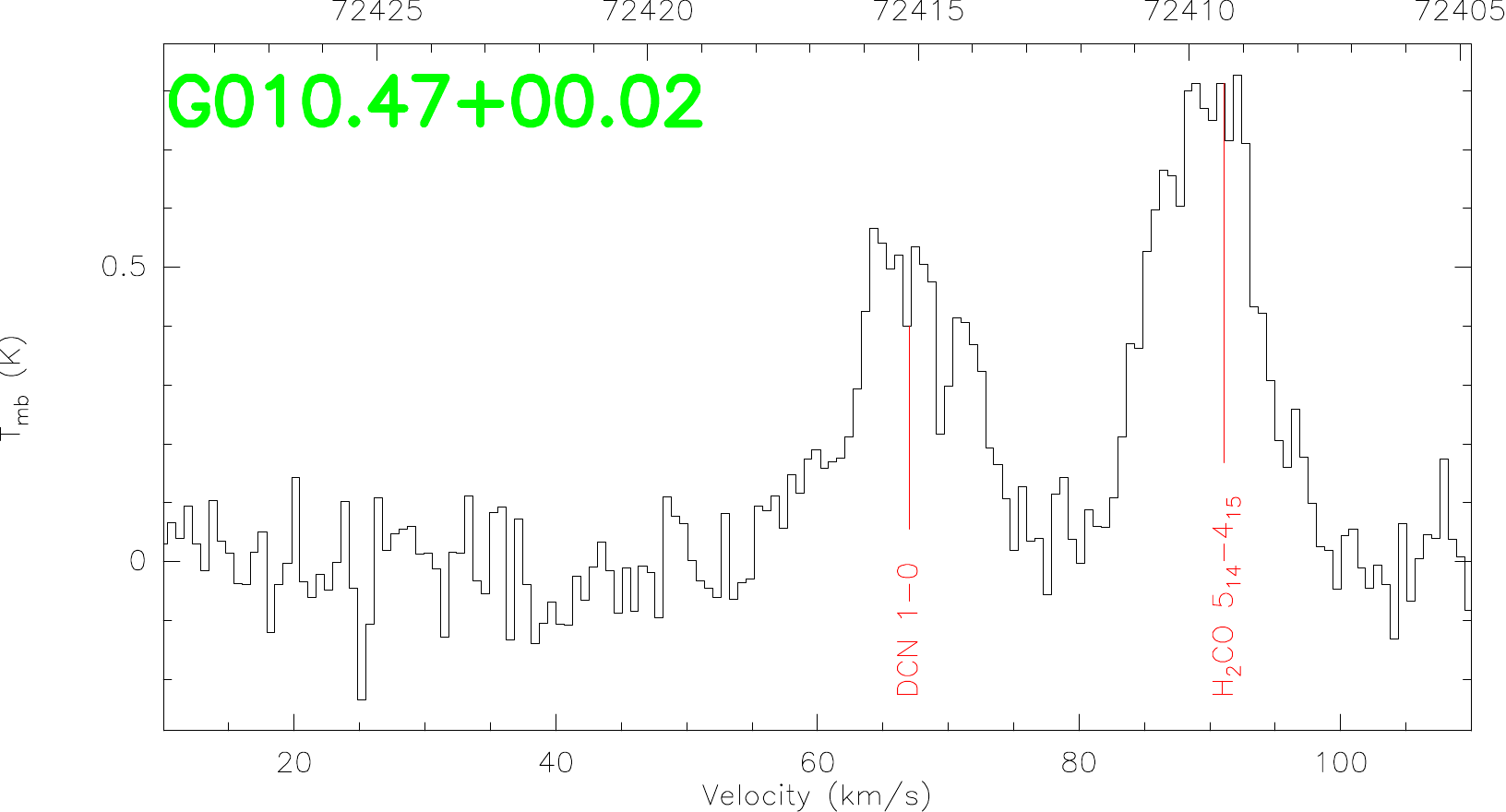}
\includegraphics[width=0.155\textwidth]{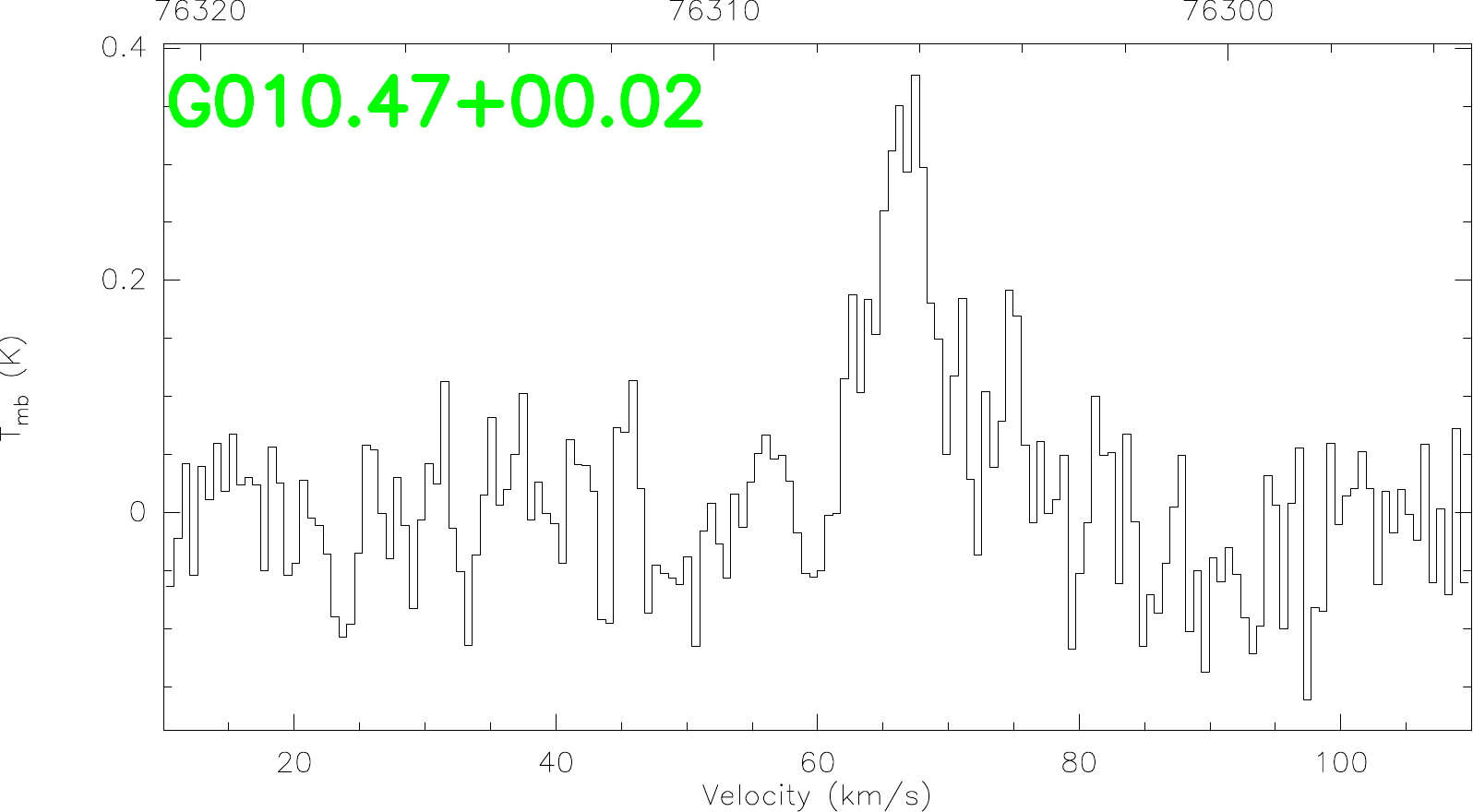}
\includegraphics[width=0.155\textwidth]{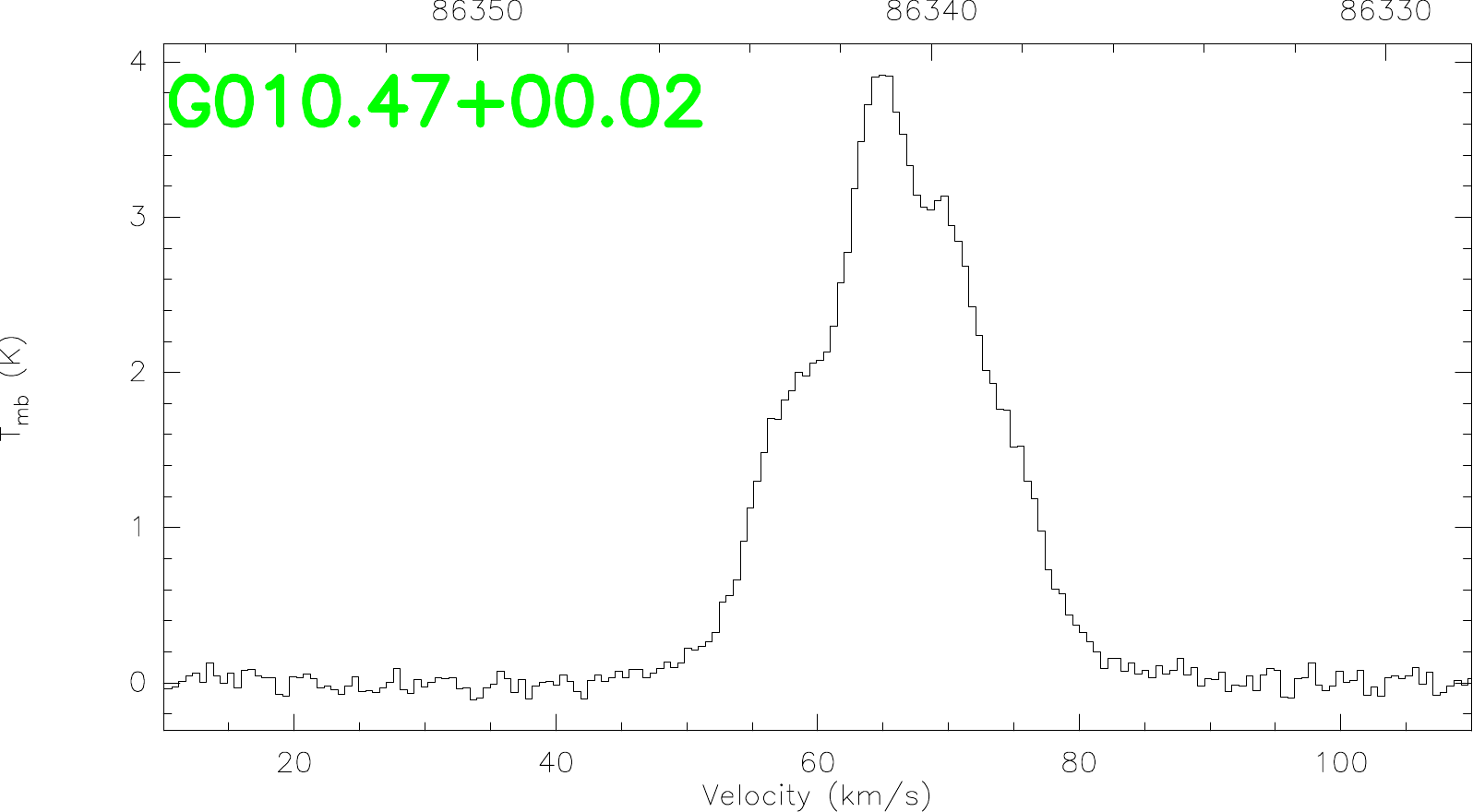}
\includegraphics[width=0.155\textwidth]{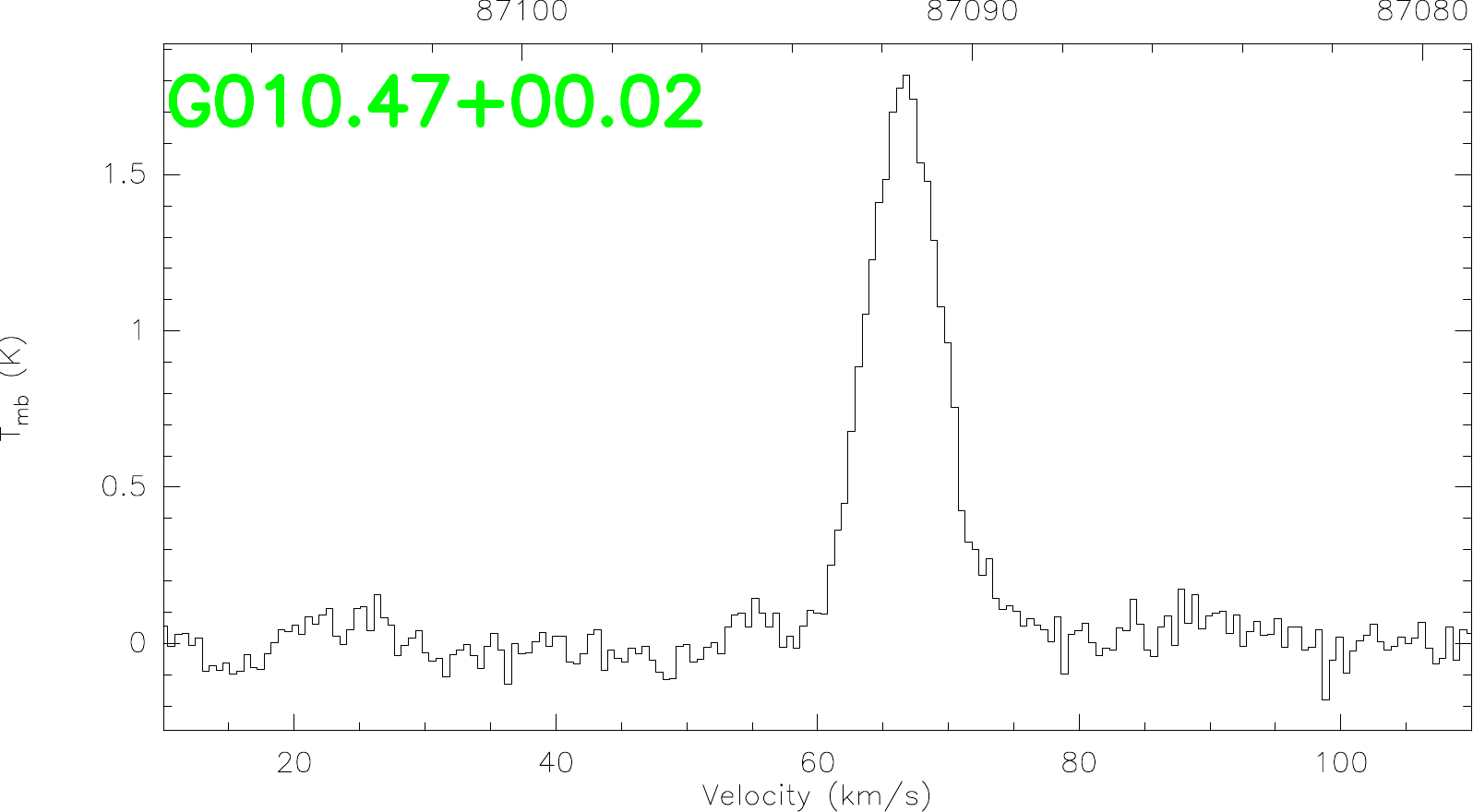}
\includegraphics[width=0.155\textwidth]{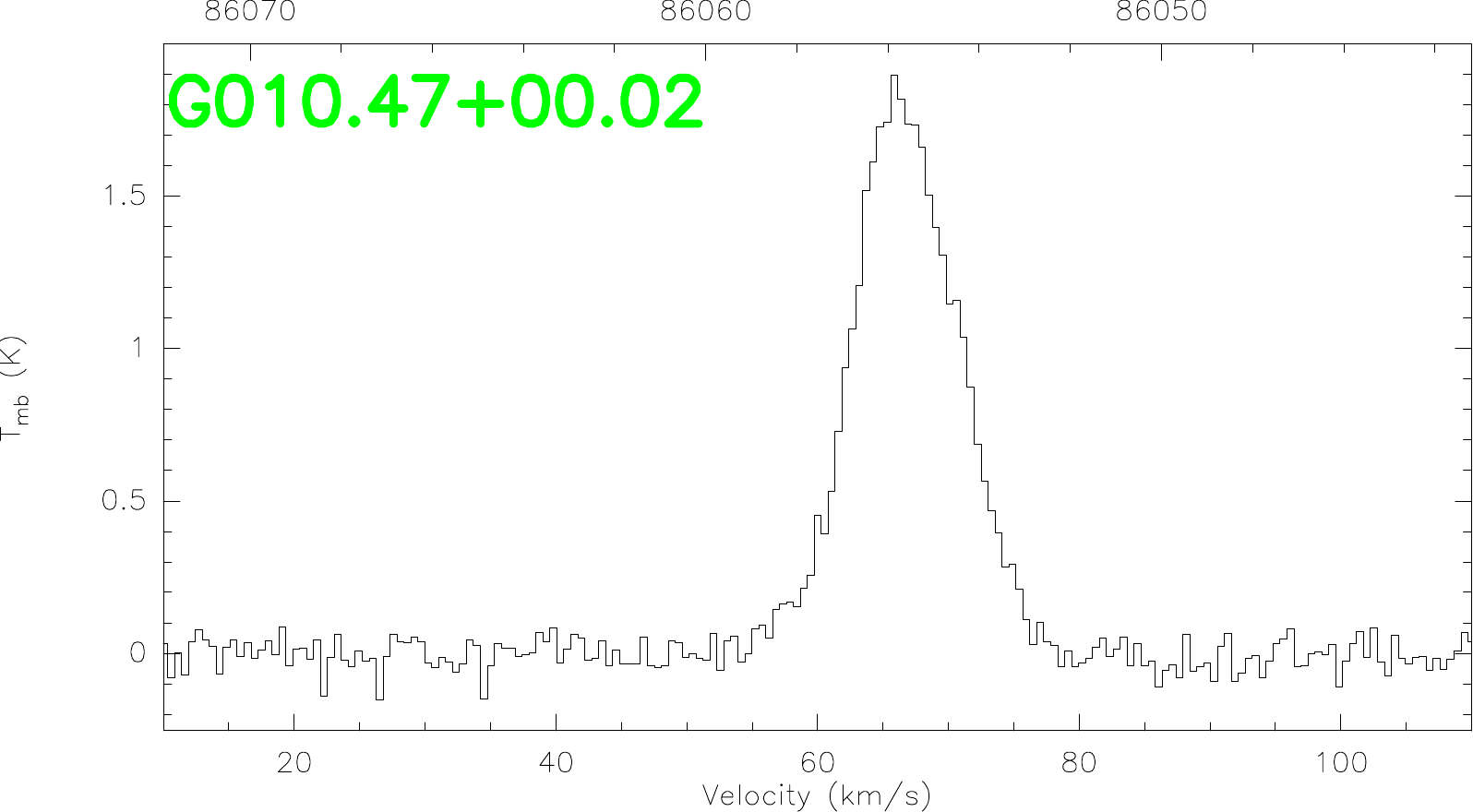}
\includegraphics[width=0.155\textwidth]{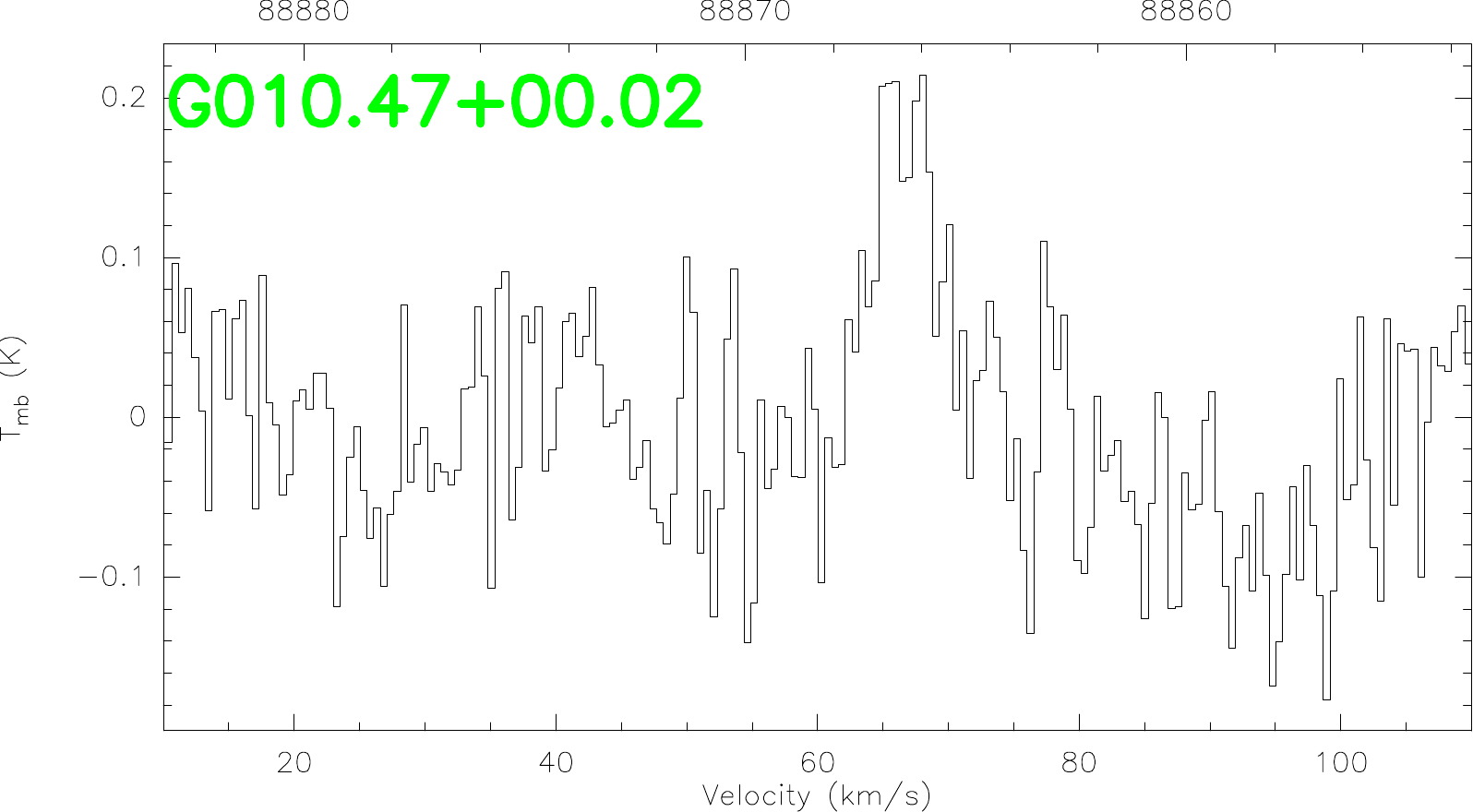}\\

\includegraphics[width=0.155\textwidth]{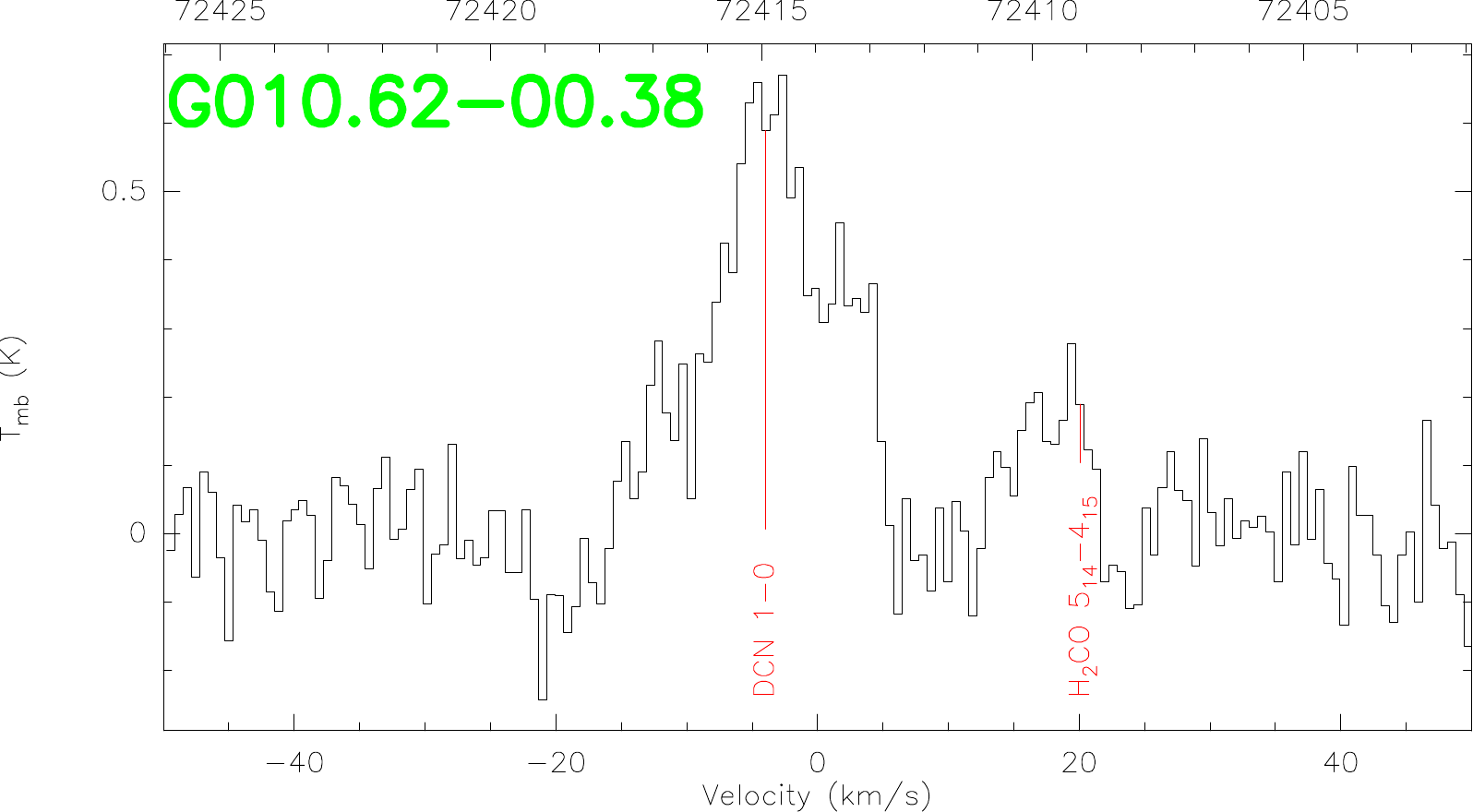}
\includegraphics[width=0.155\textwidth]{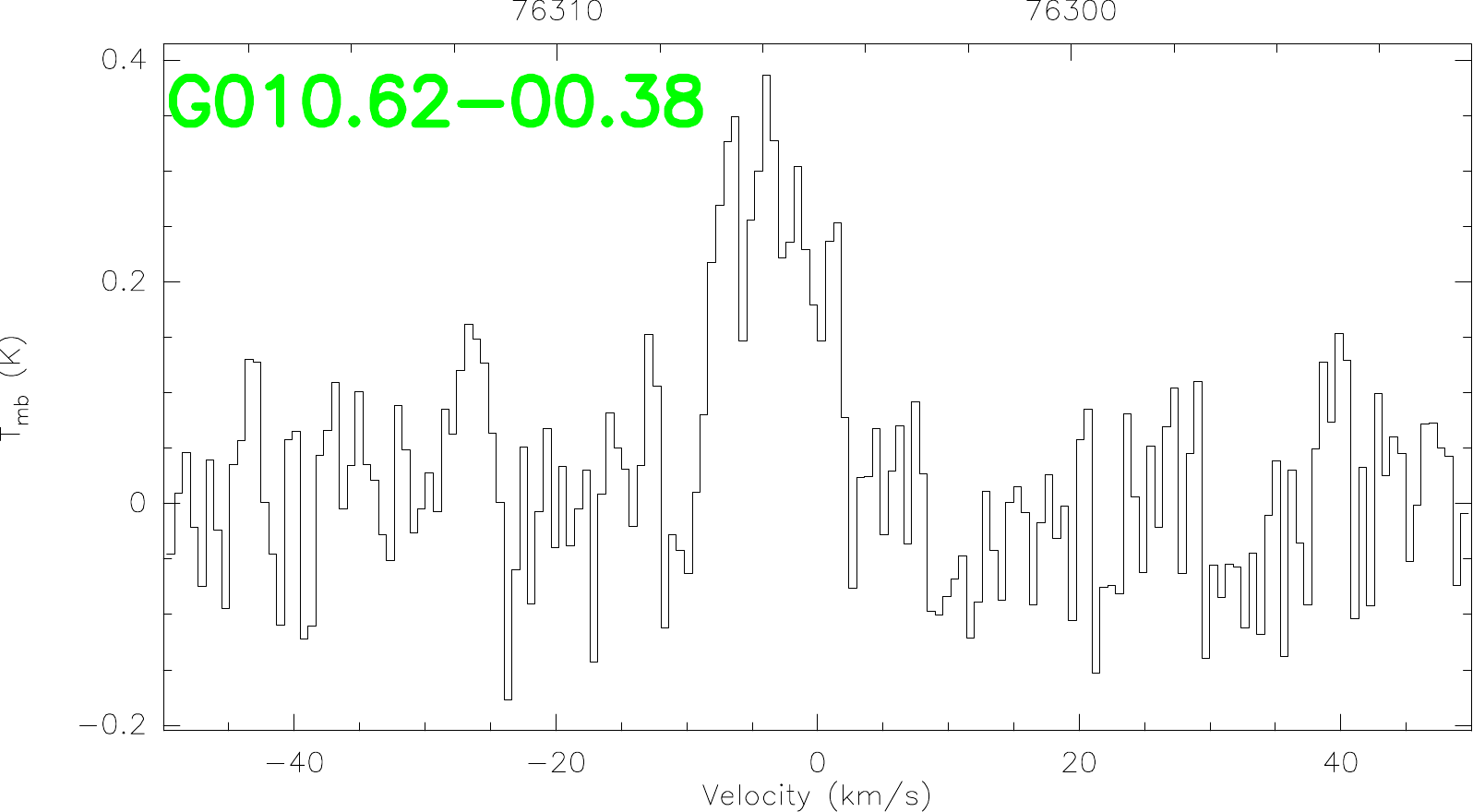}
\includegraphics[width=0.155\textwidth]{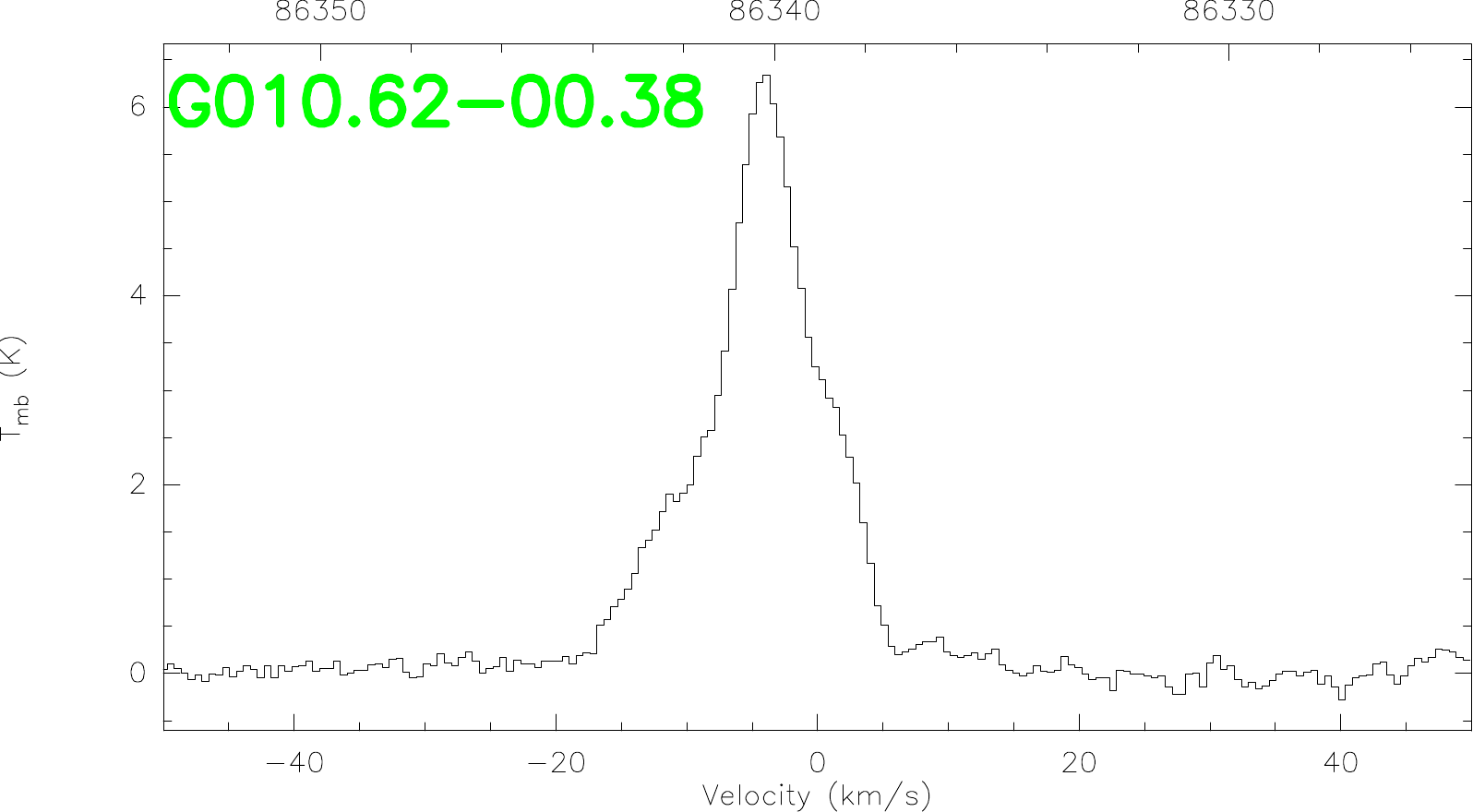}
\includegraphics[width=0.155\textwidth]{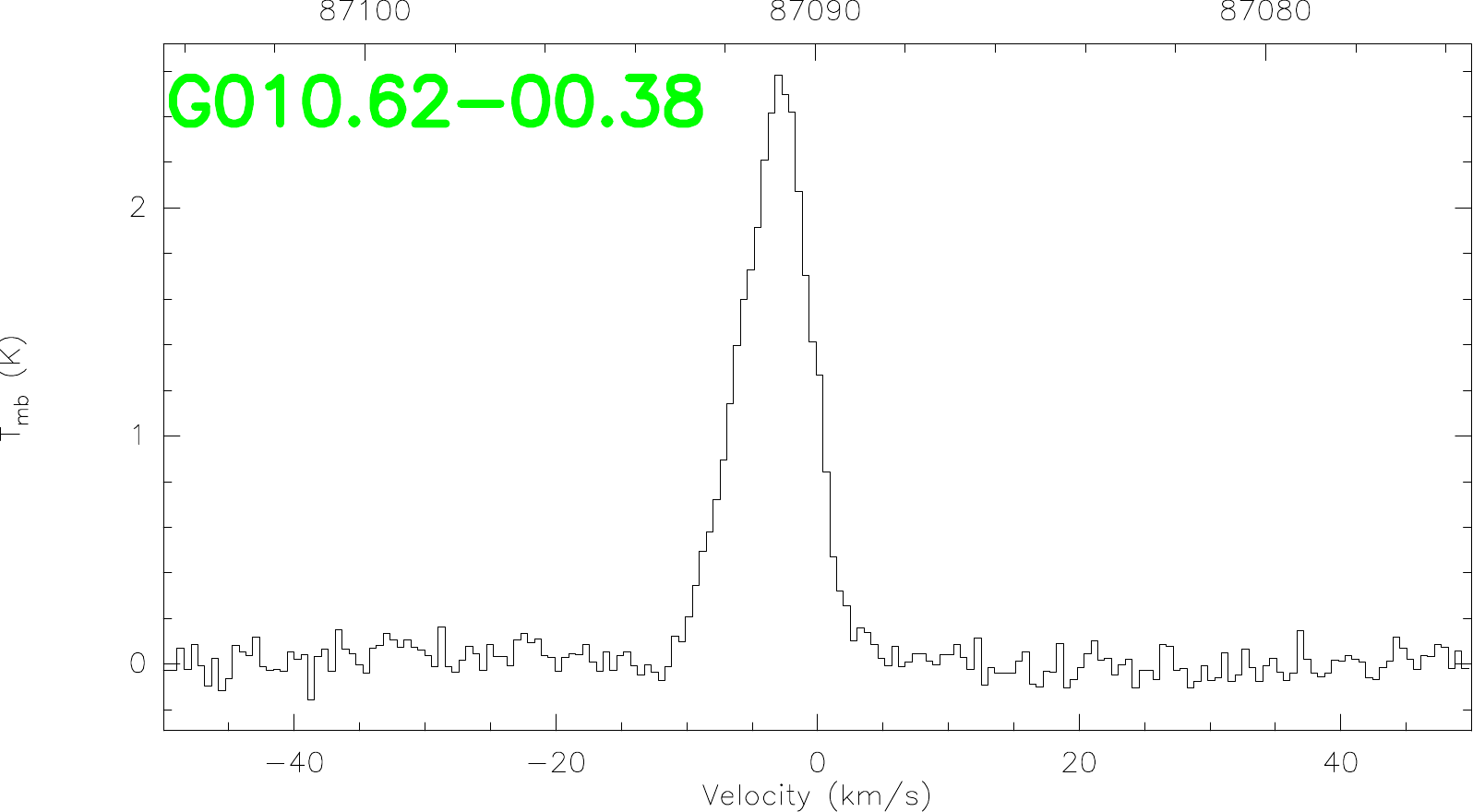}
\includegraphics[width=0.155\textwidth]{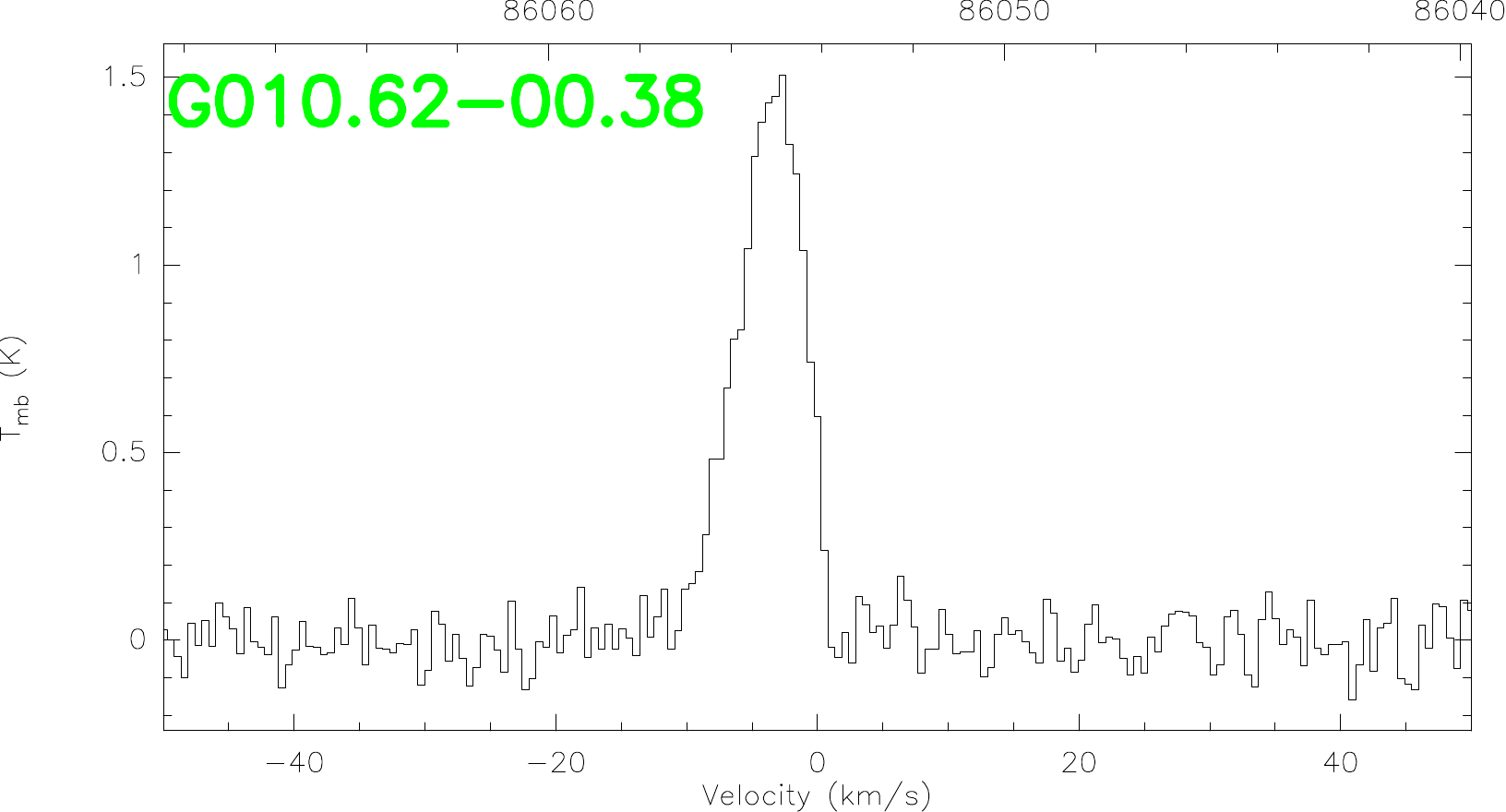}
\includegraphics[width=0.155\textwidth]{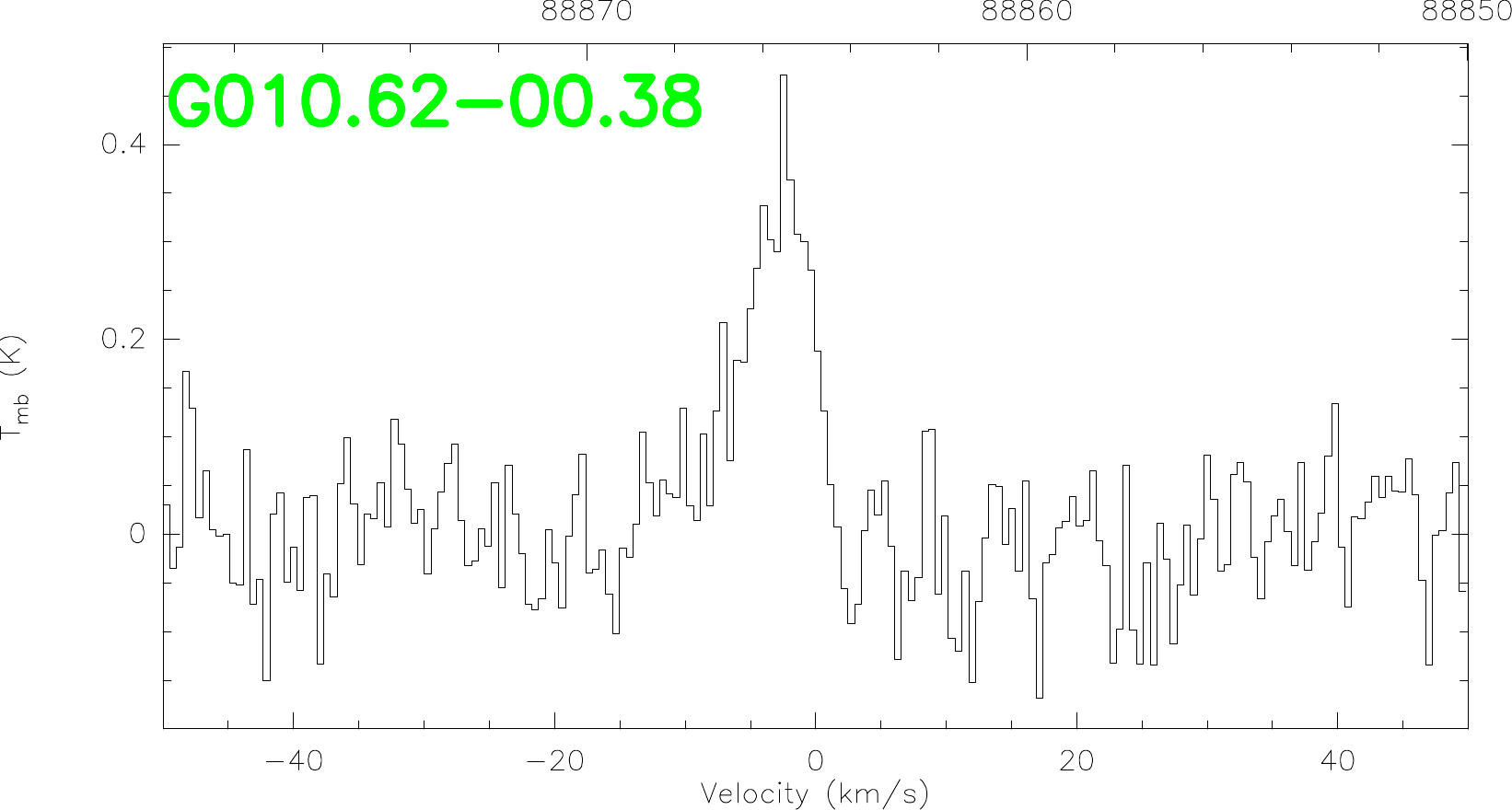}\\

\includegraphics[width=0.155\textwidth]{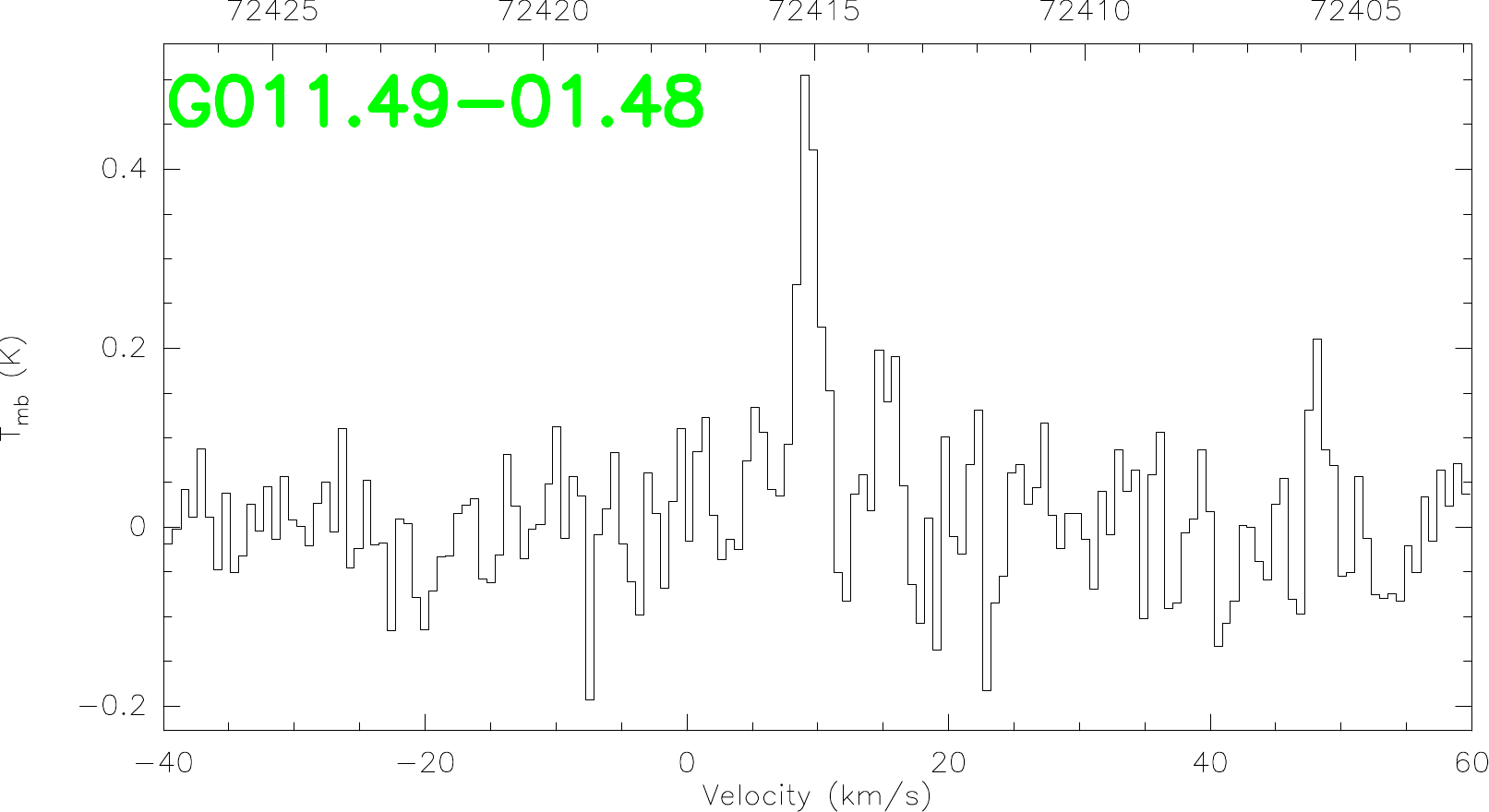}
\includegraphics[width=0.155\textwidth]{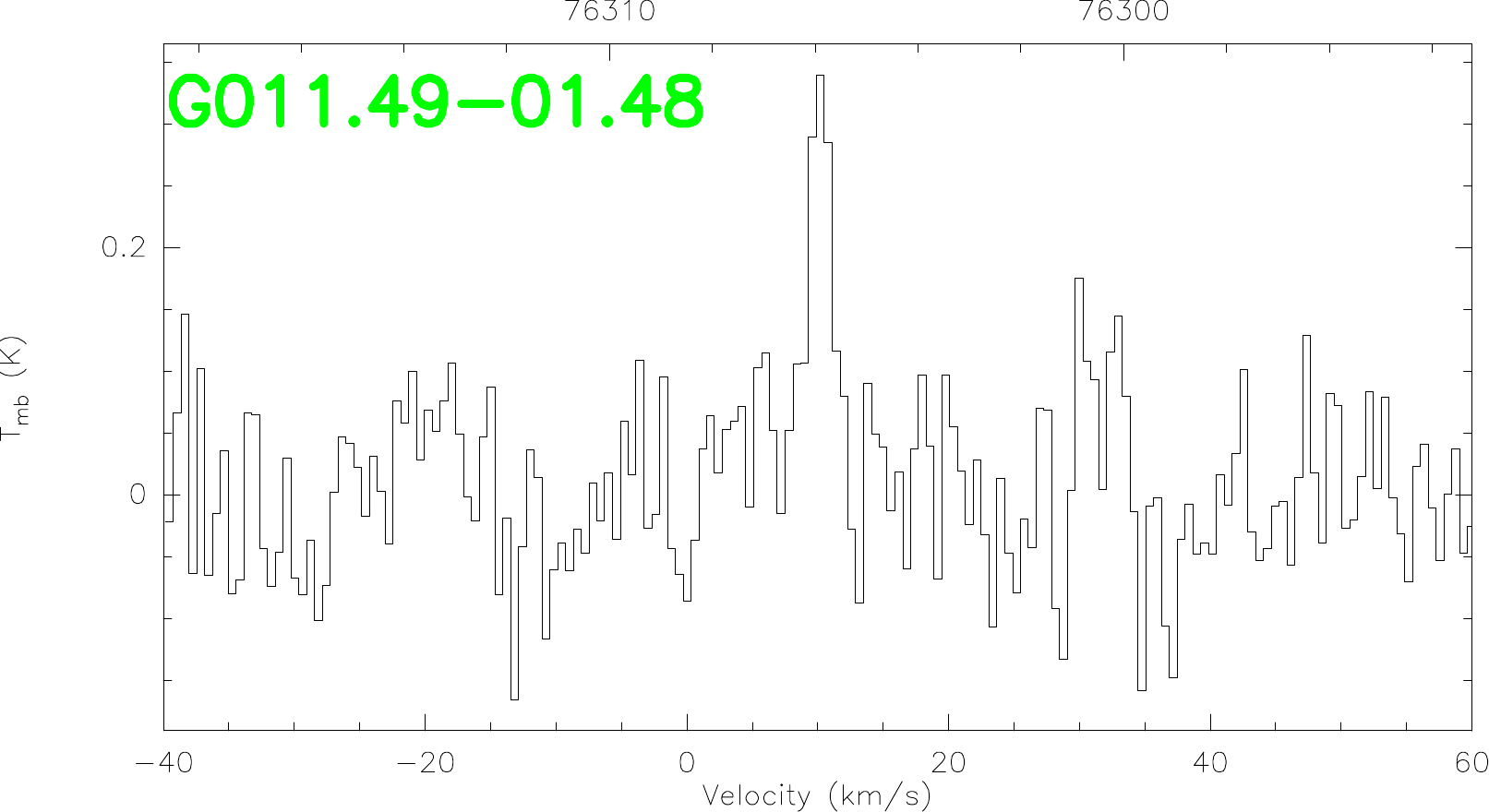}
\includegraphics[width=0.155\textwidth]{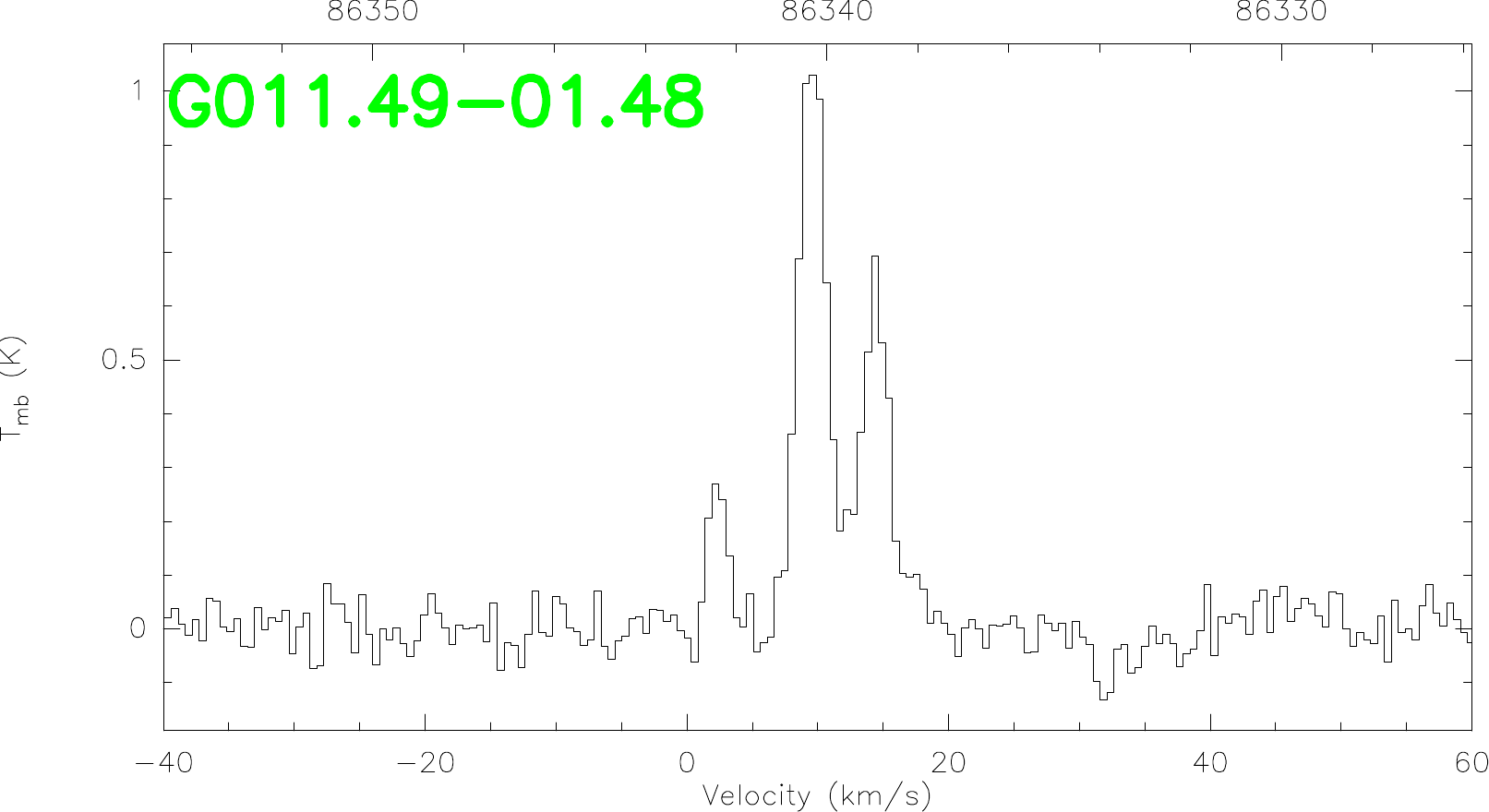}
\includegraphics[width=0.155\textwidth]{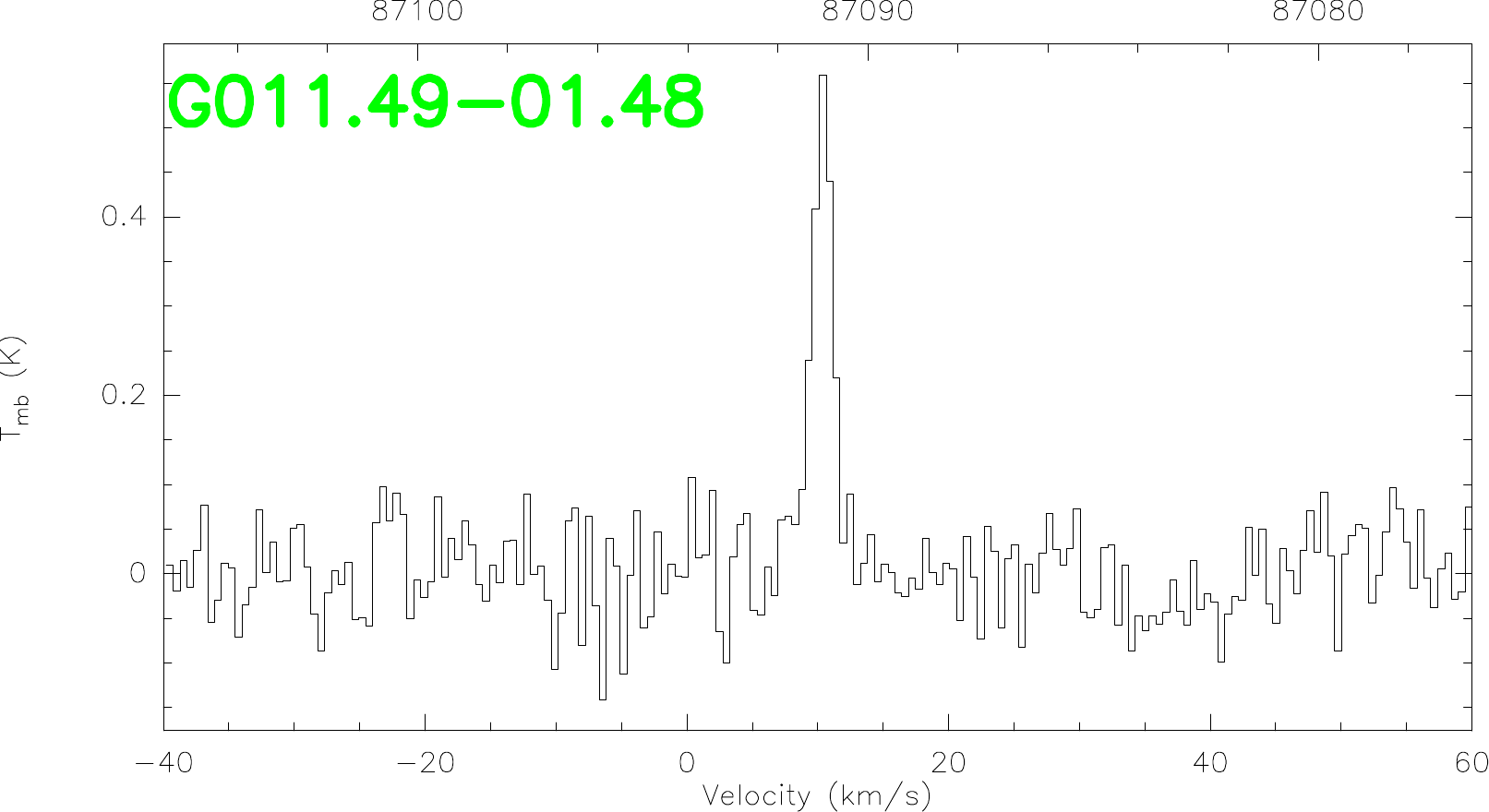}
\includegraphics[width=0.155\textwidth]{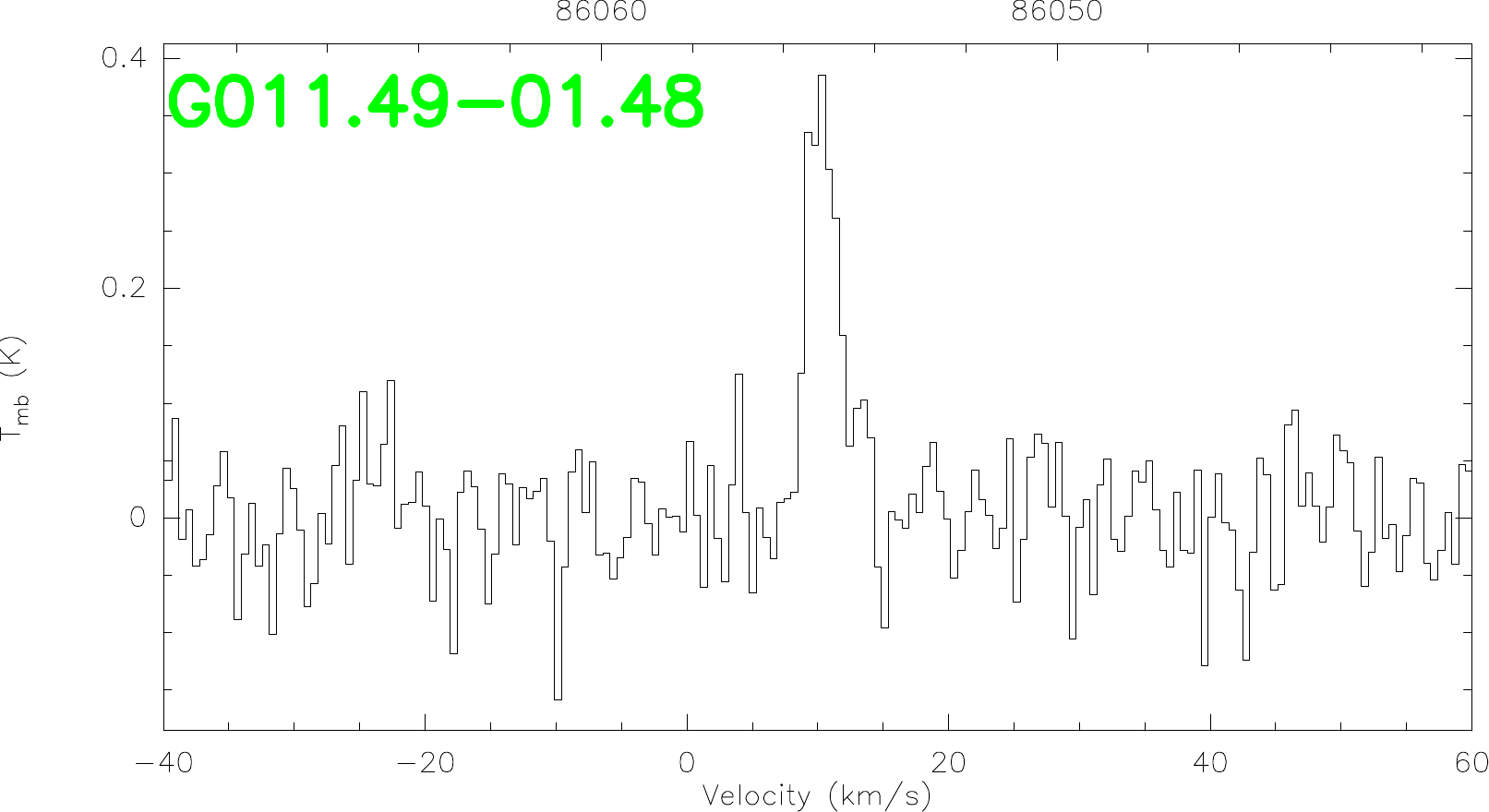}
\includegraphics[width=0.155\textwidth]{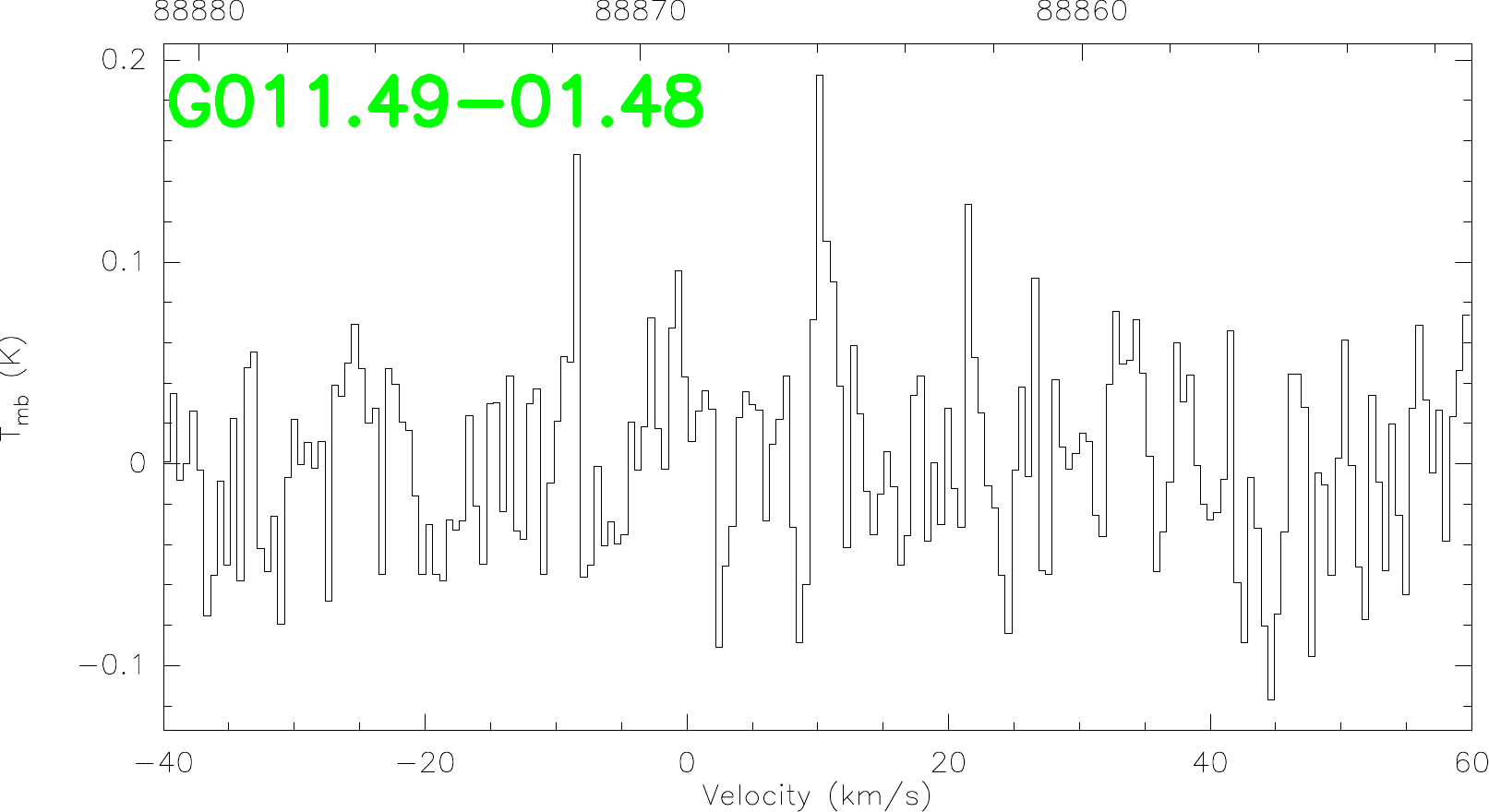}\\

\includegraphics[width=0.155\textwidth]{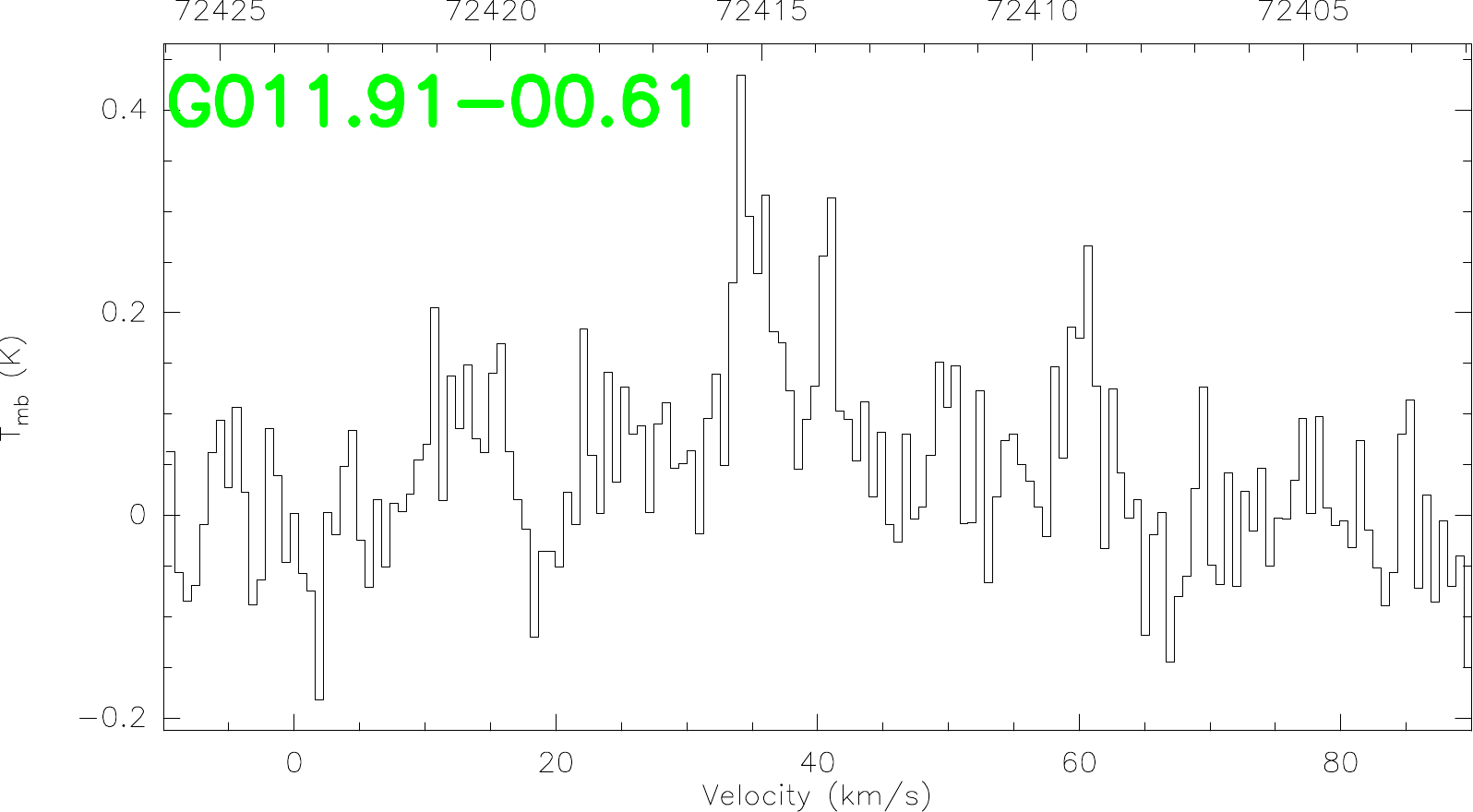}
\includegraphics[width=0.155\textwidth]{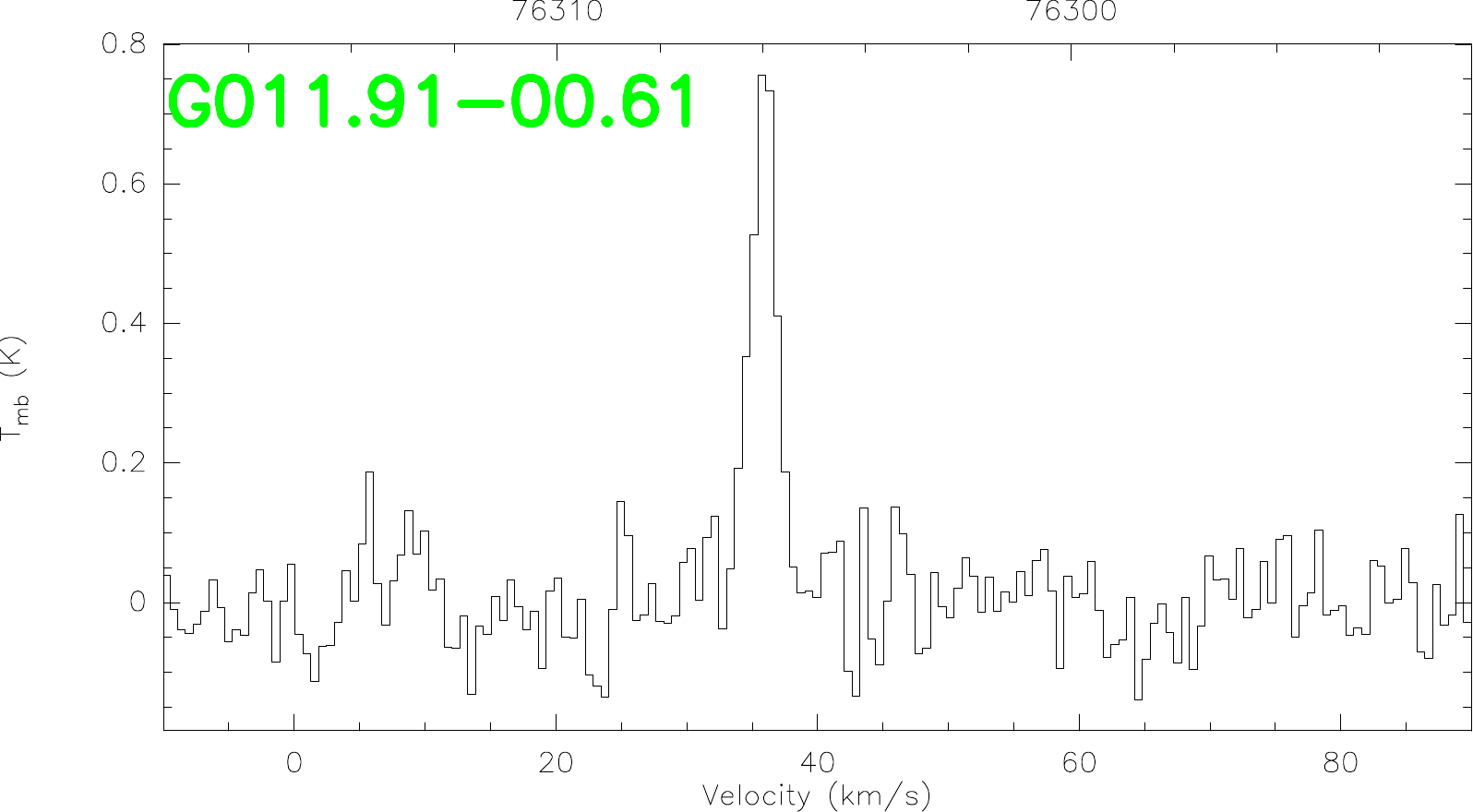}
\includegraphics[width=0.155\textwidth]{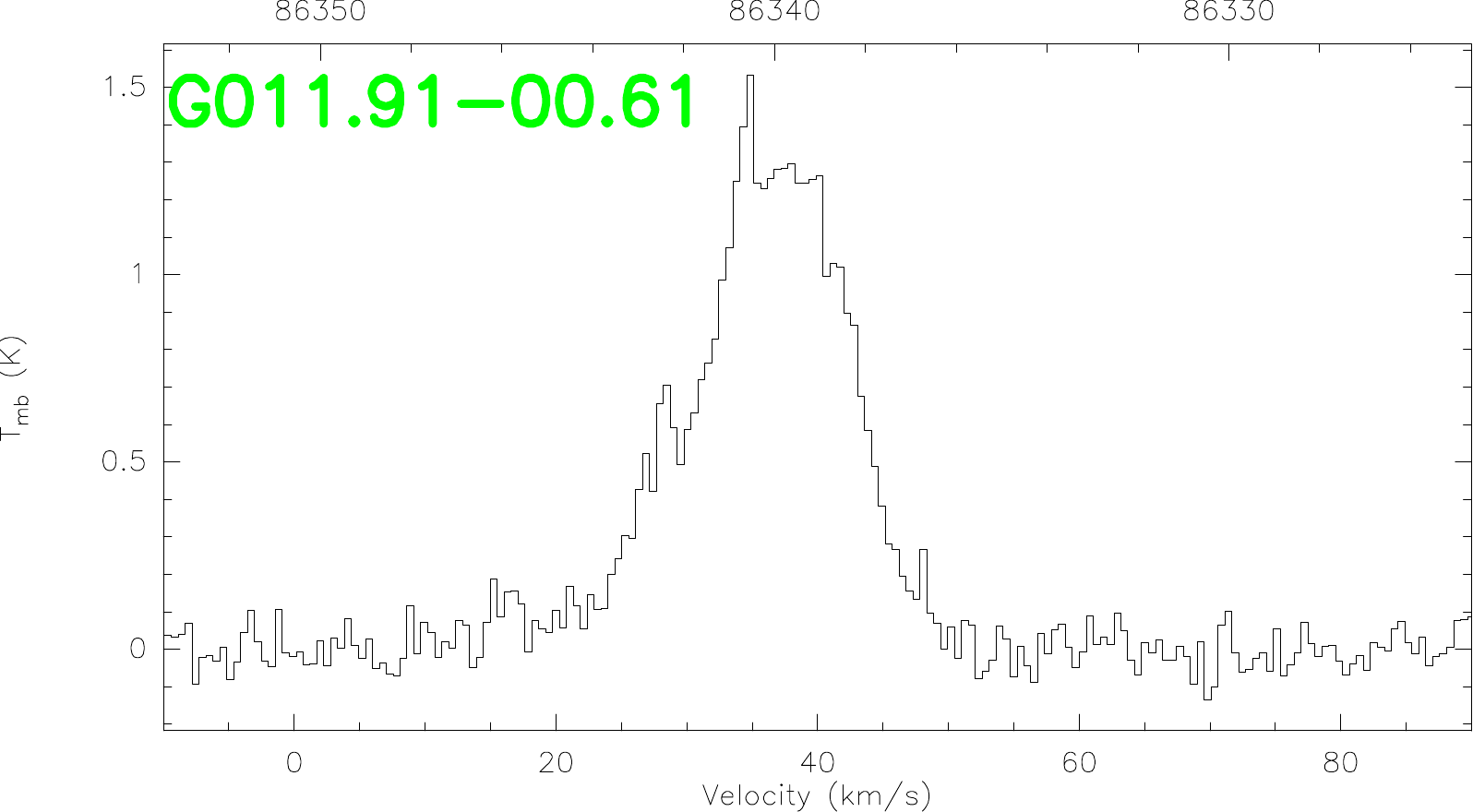}
\includegraphics[width=0.155\textwidth]{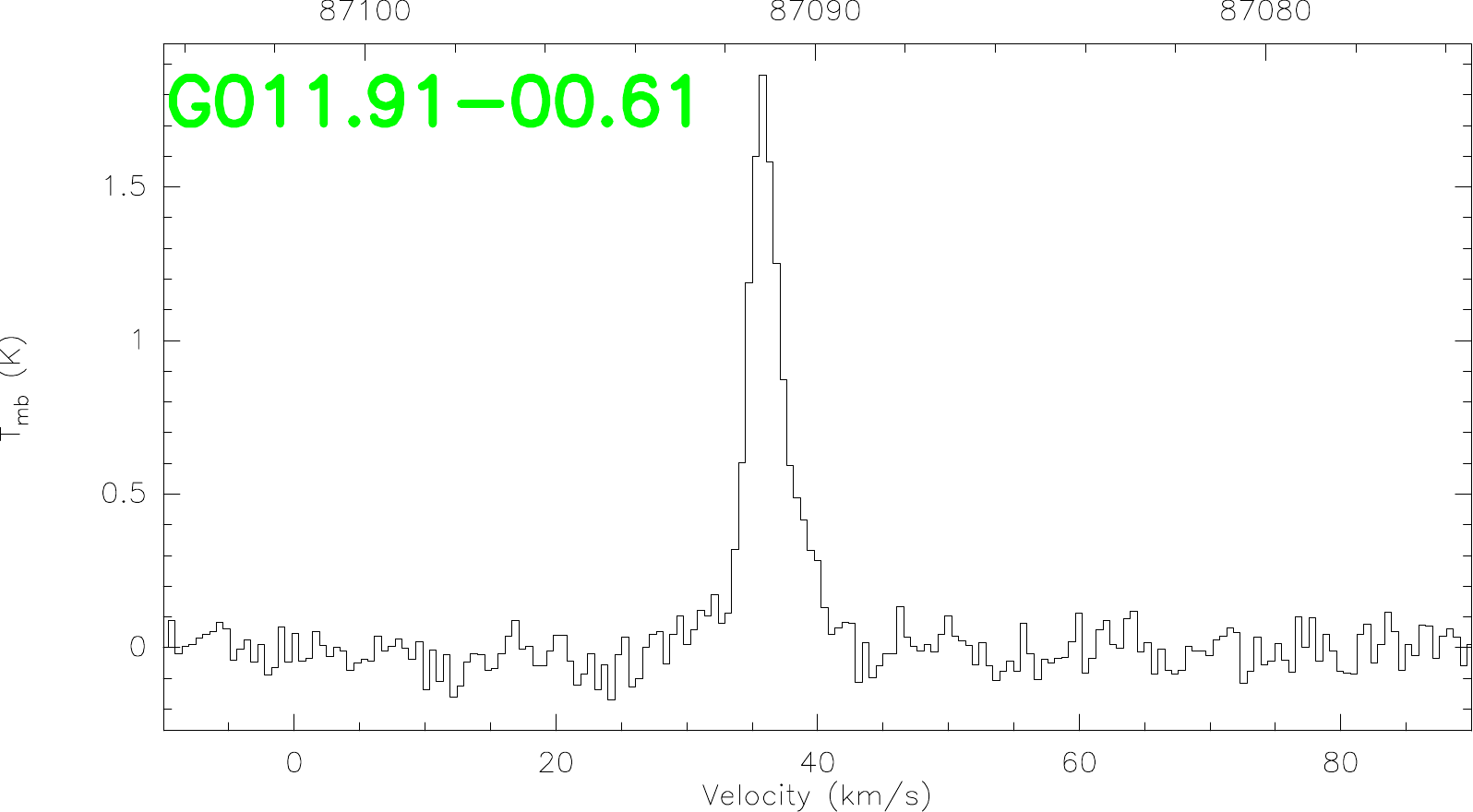}
\includegraphics[width=0.155\textwidth]{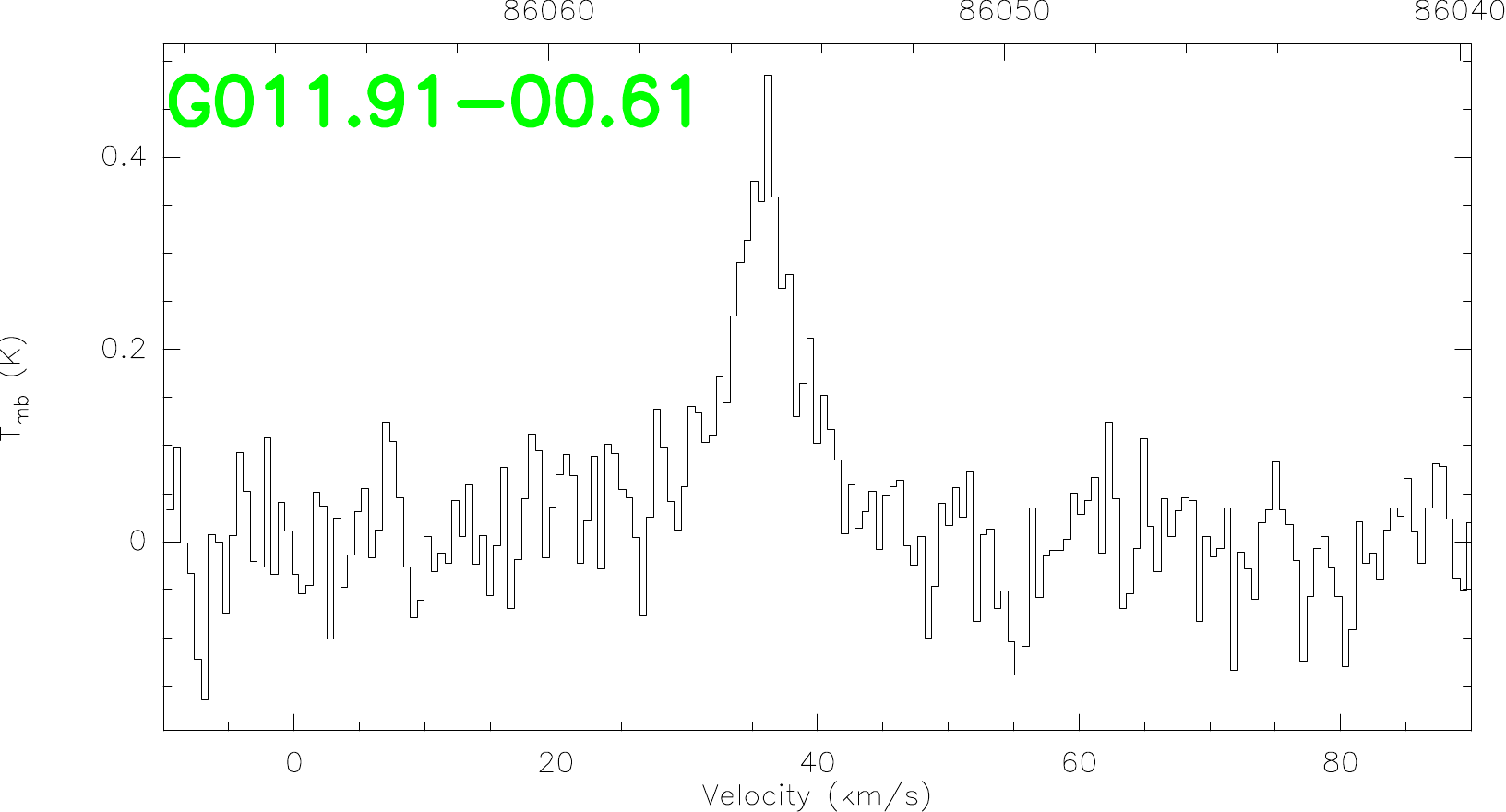}
\includegraphics[width=0.155\textwidth]{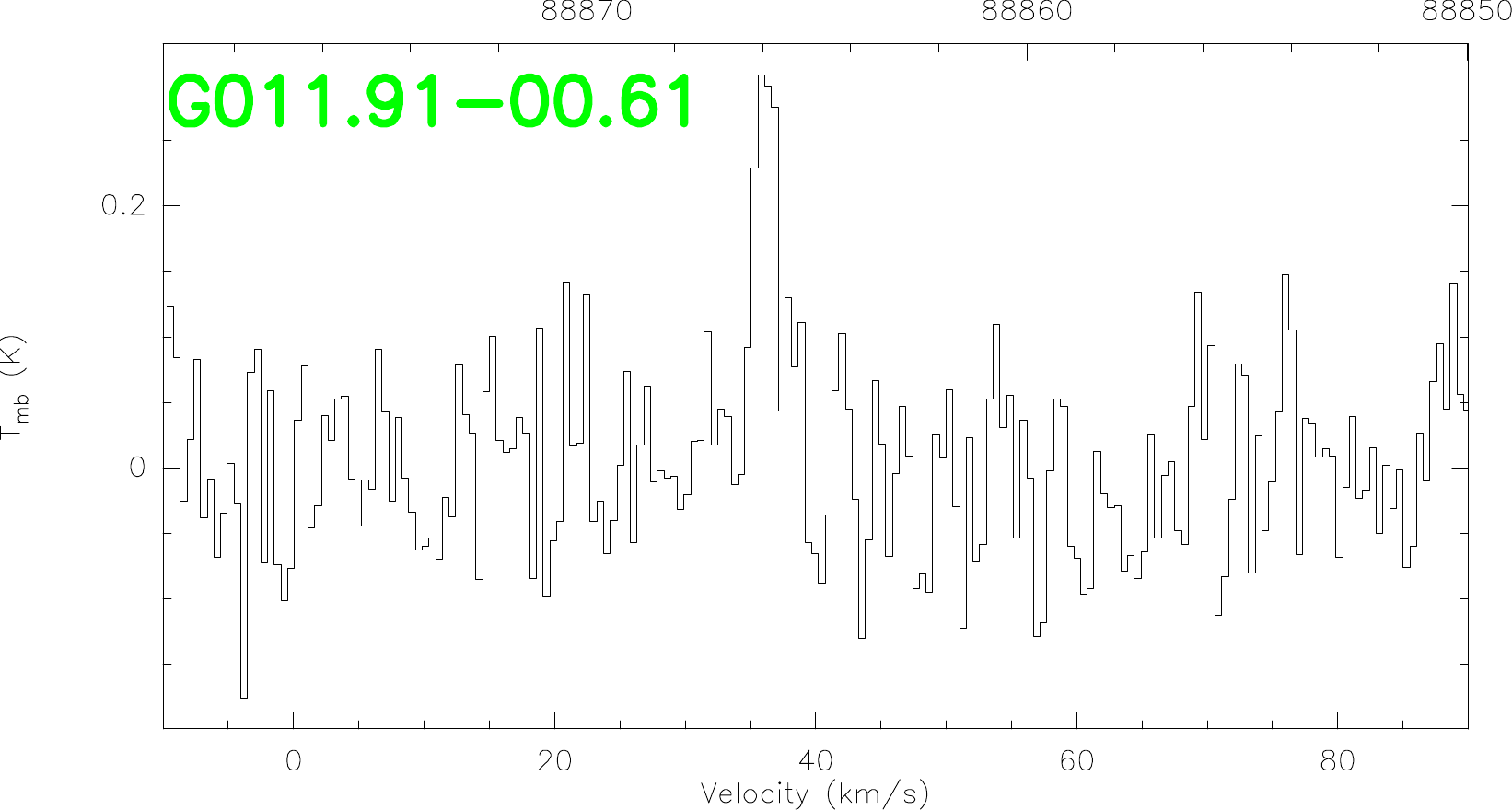}\\

\includegraphics[width=0.155\textwidth]{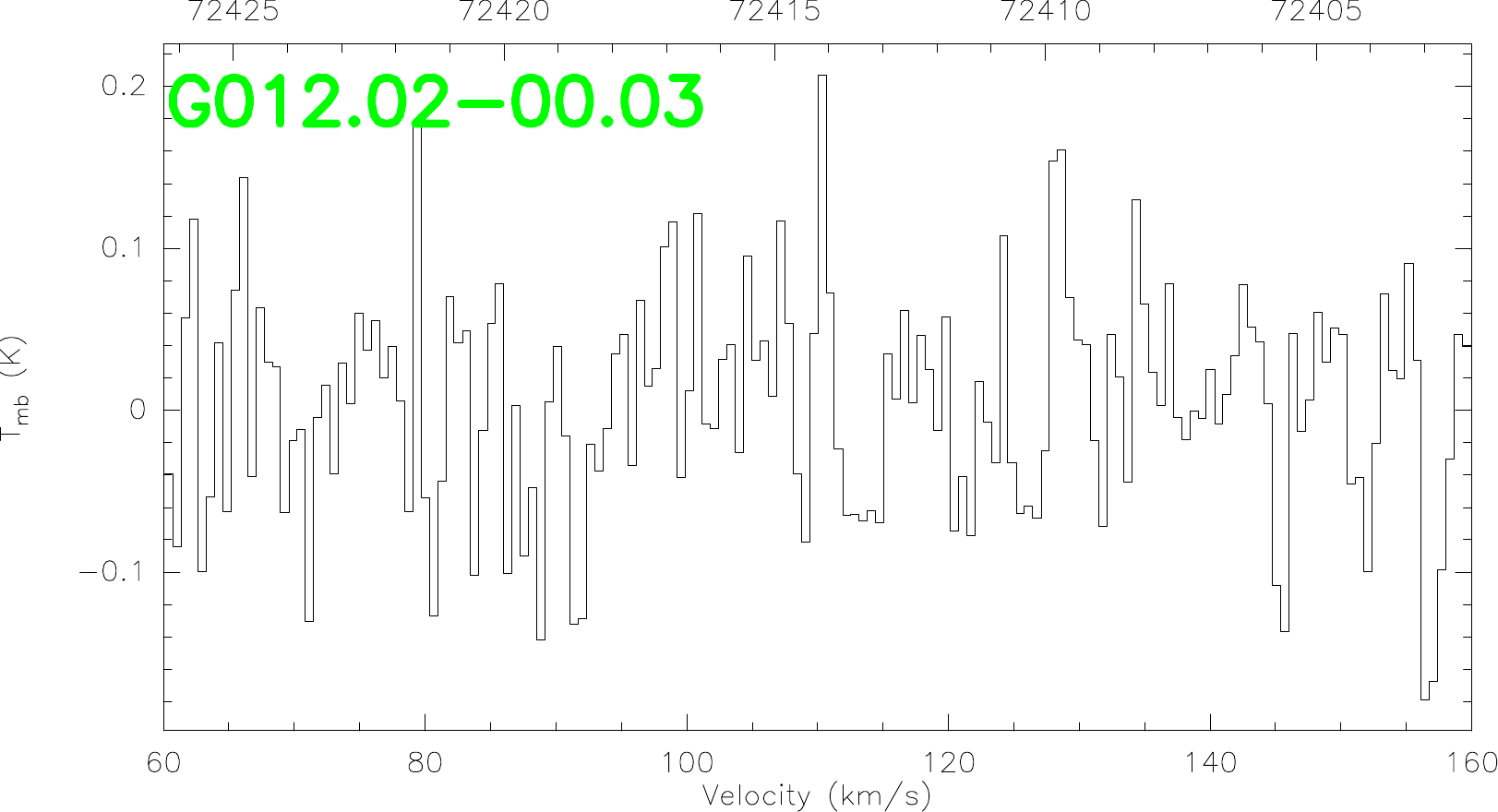}
\includegraphics[width=0.155\textwidth]{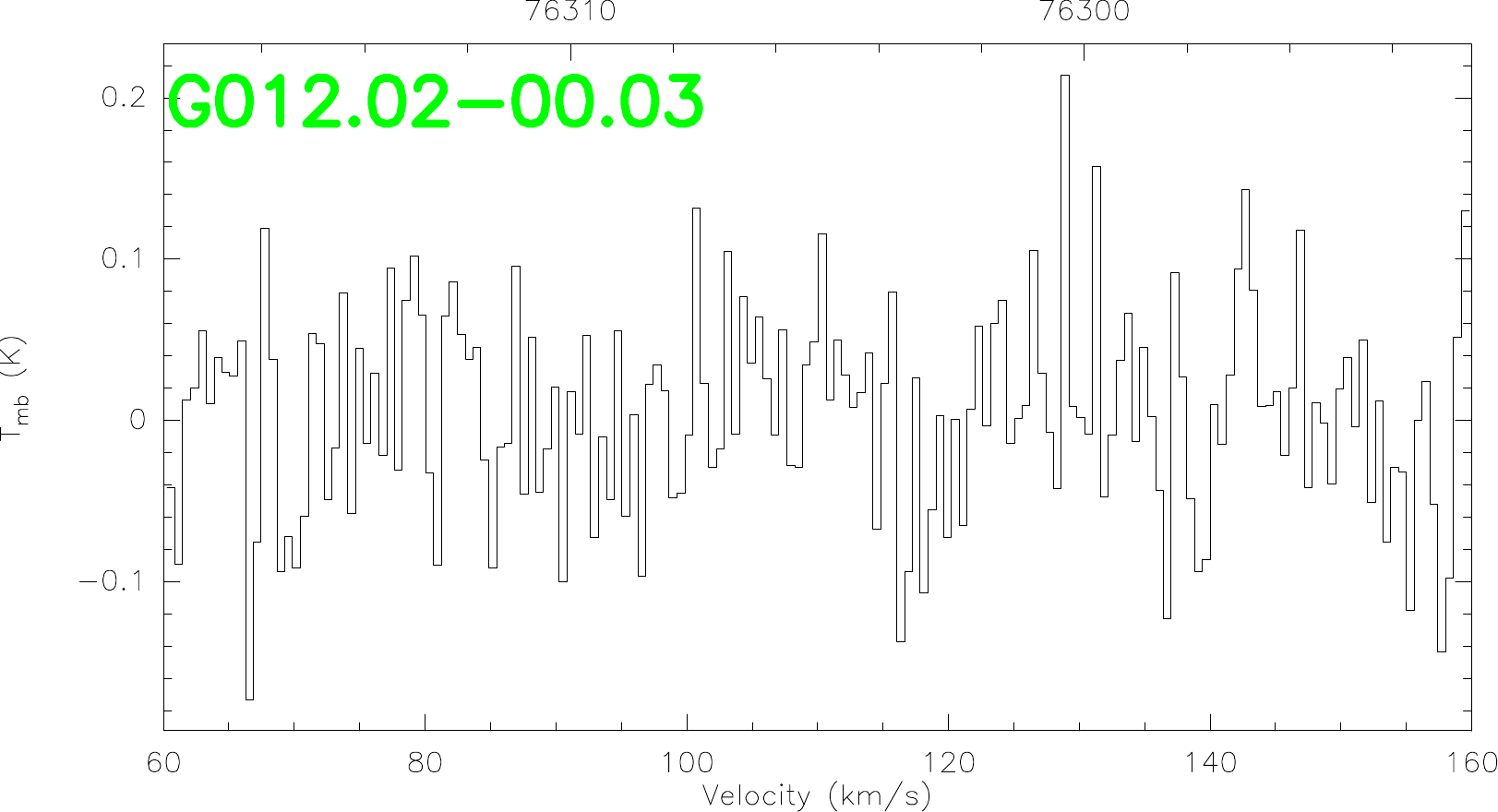}
\includegraphics[width=0.155\textwidth]{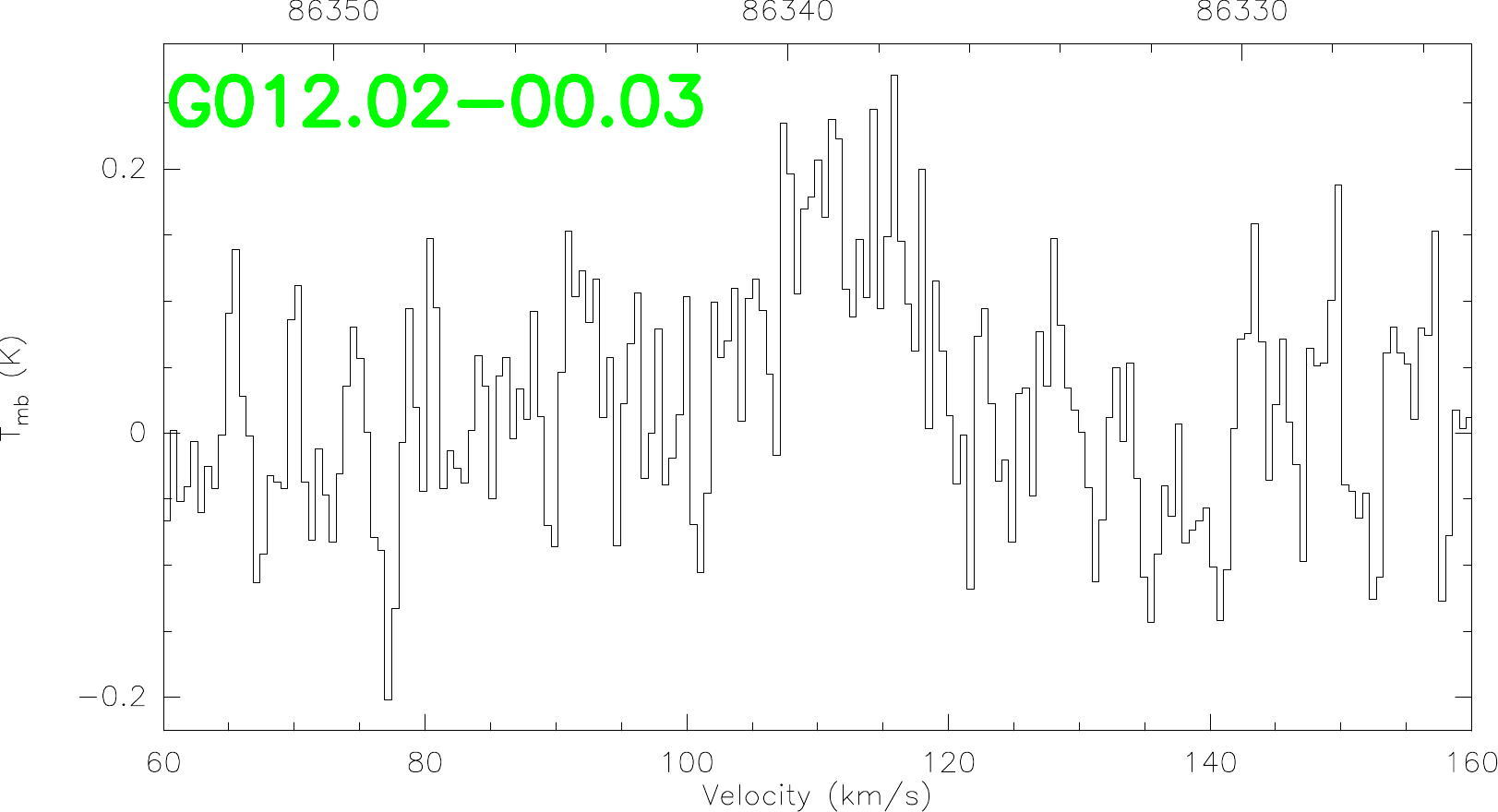}
\includegraphics[width=0.155\textwidth]{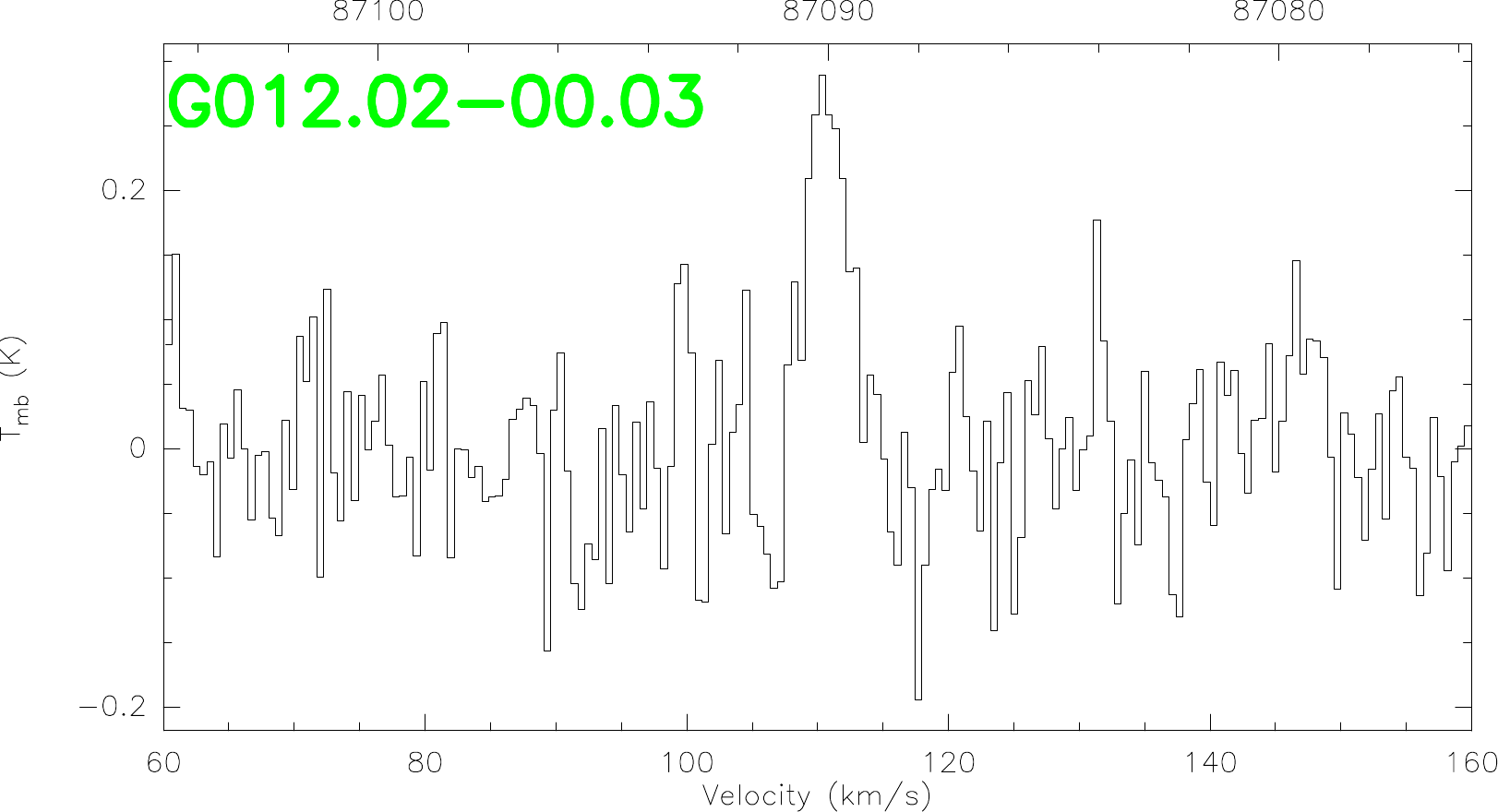}
\includegraphics[width=0.155\textwidth]{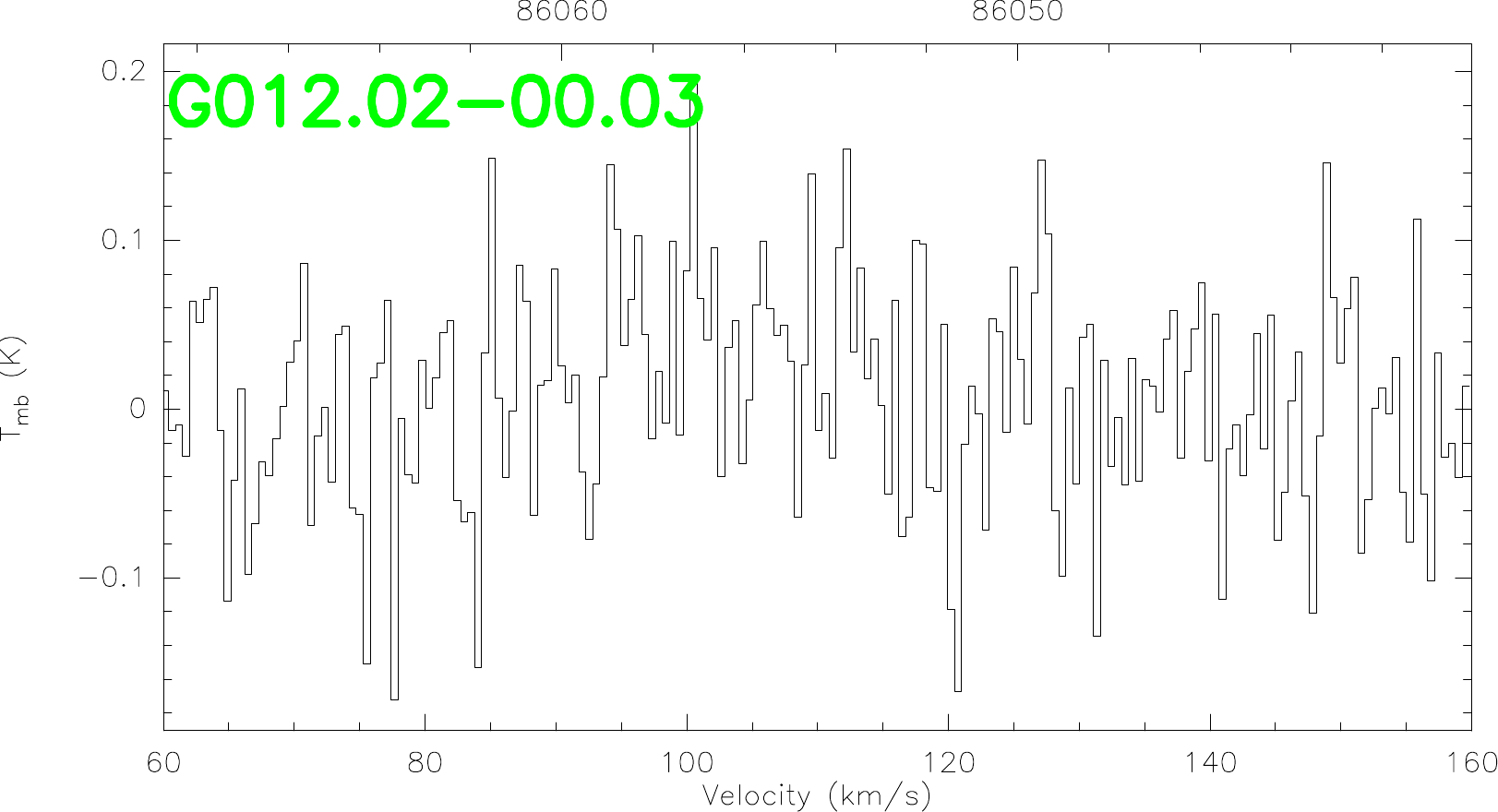}
\includegraphics[width=0.155\textwidth]{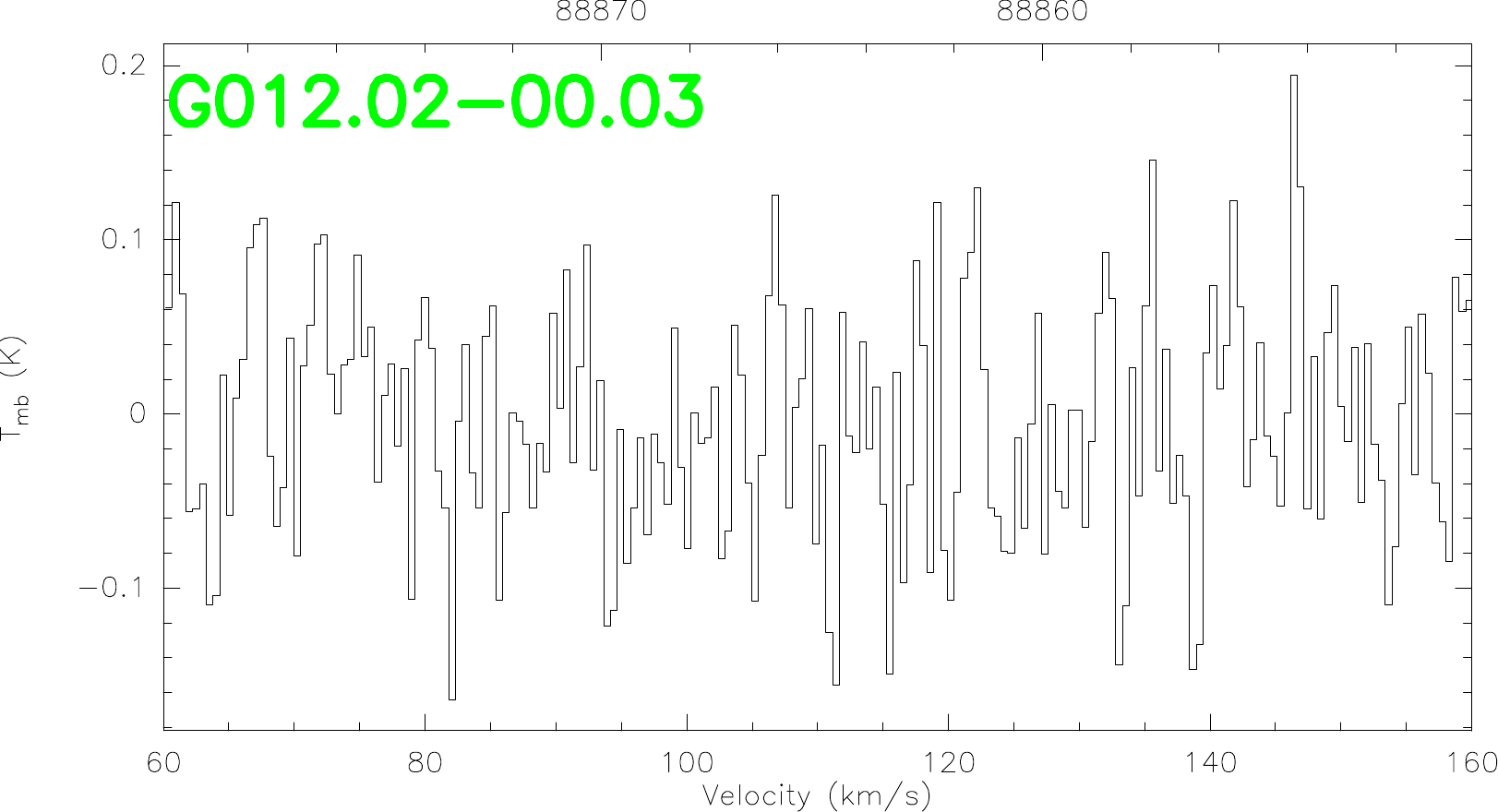}\\

\includegraphics[width=0.155\textwidth]{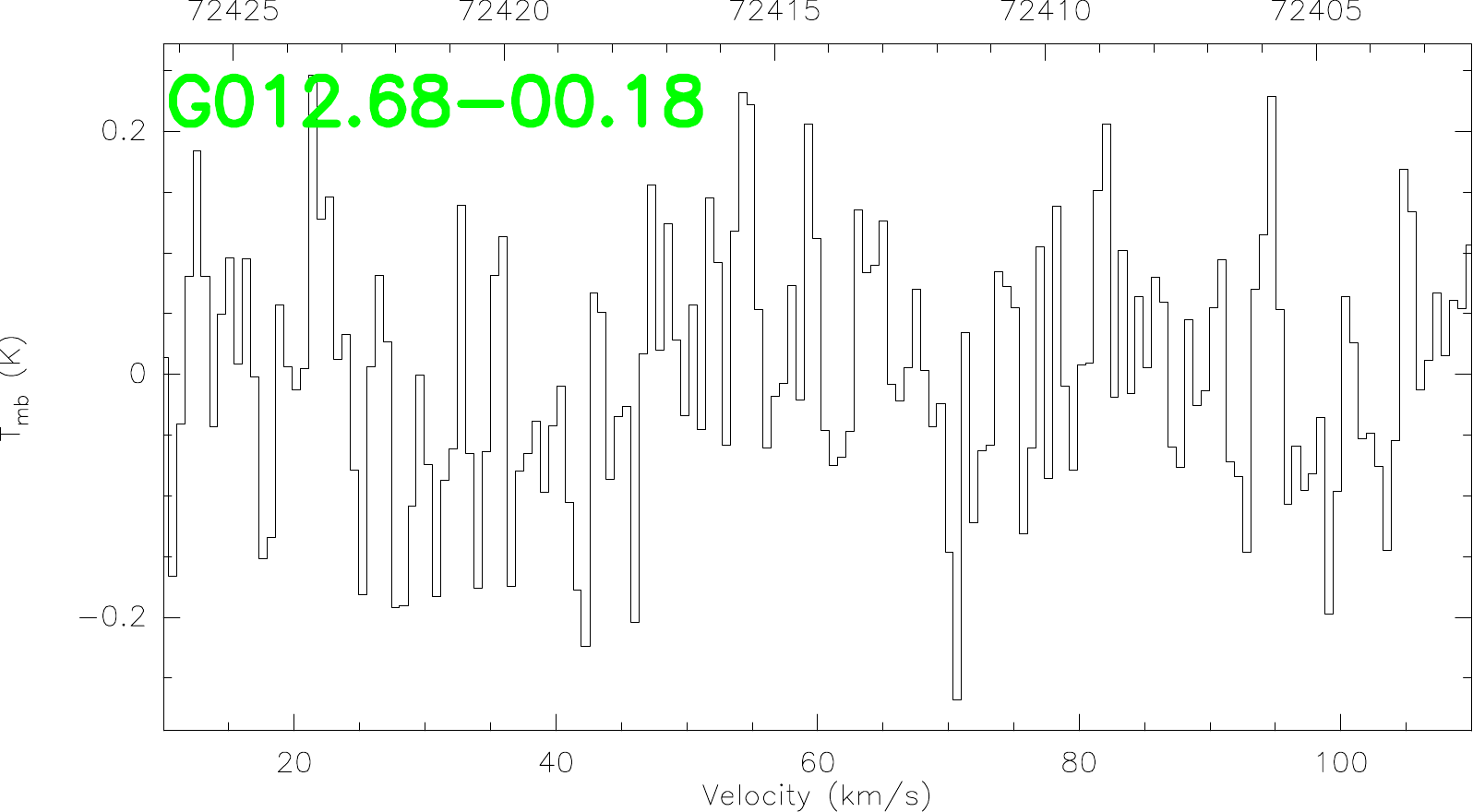}
\includegraphics[width=0.155\textwidth]{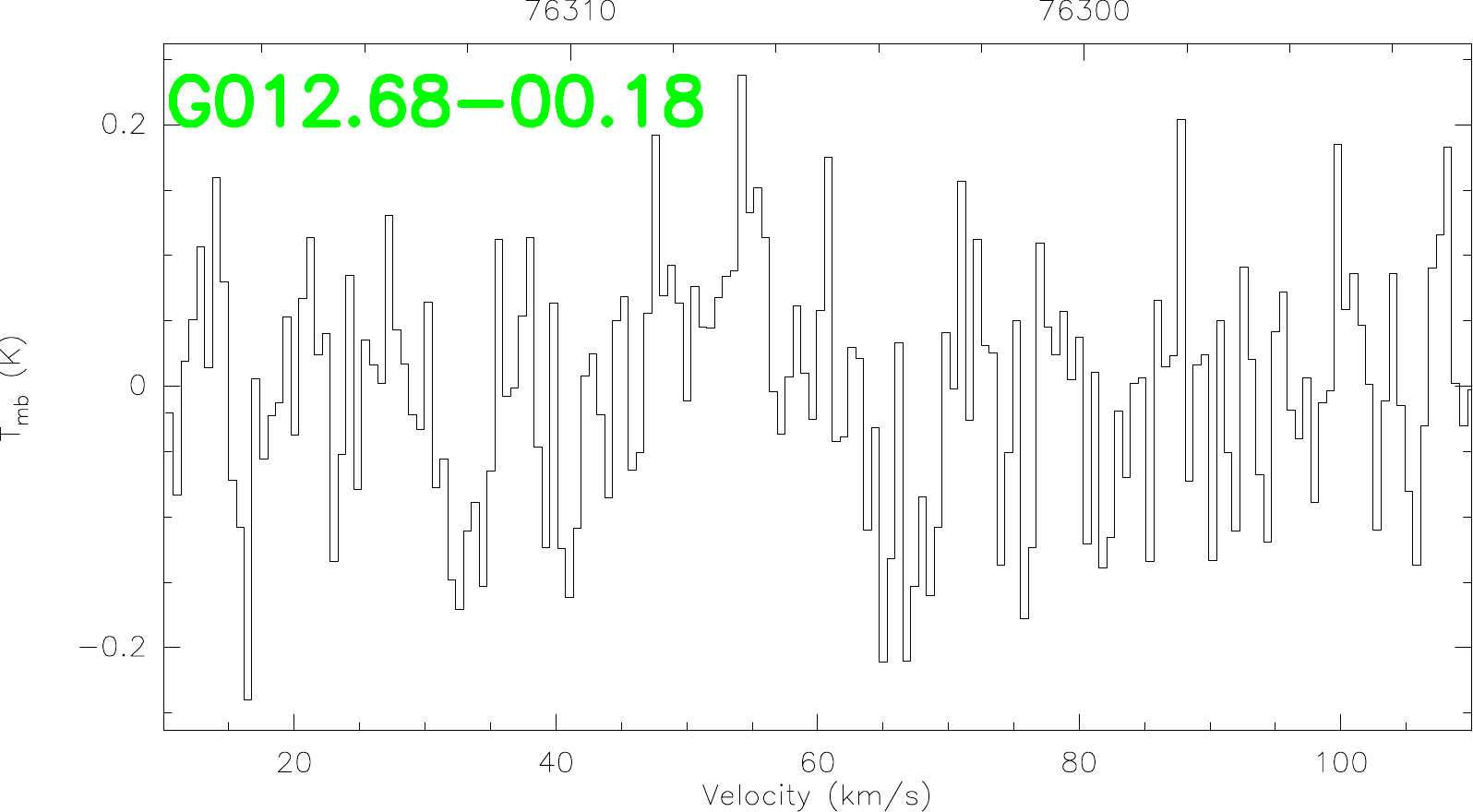}
\includegraphics[width=0.155\textwidth]{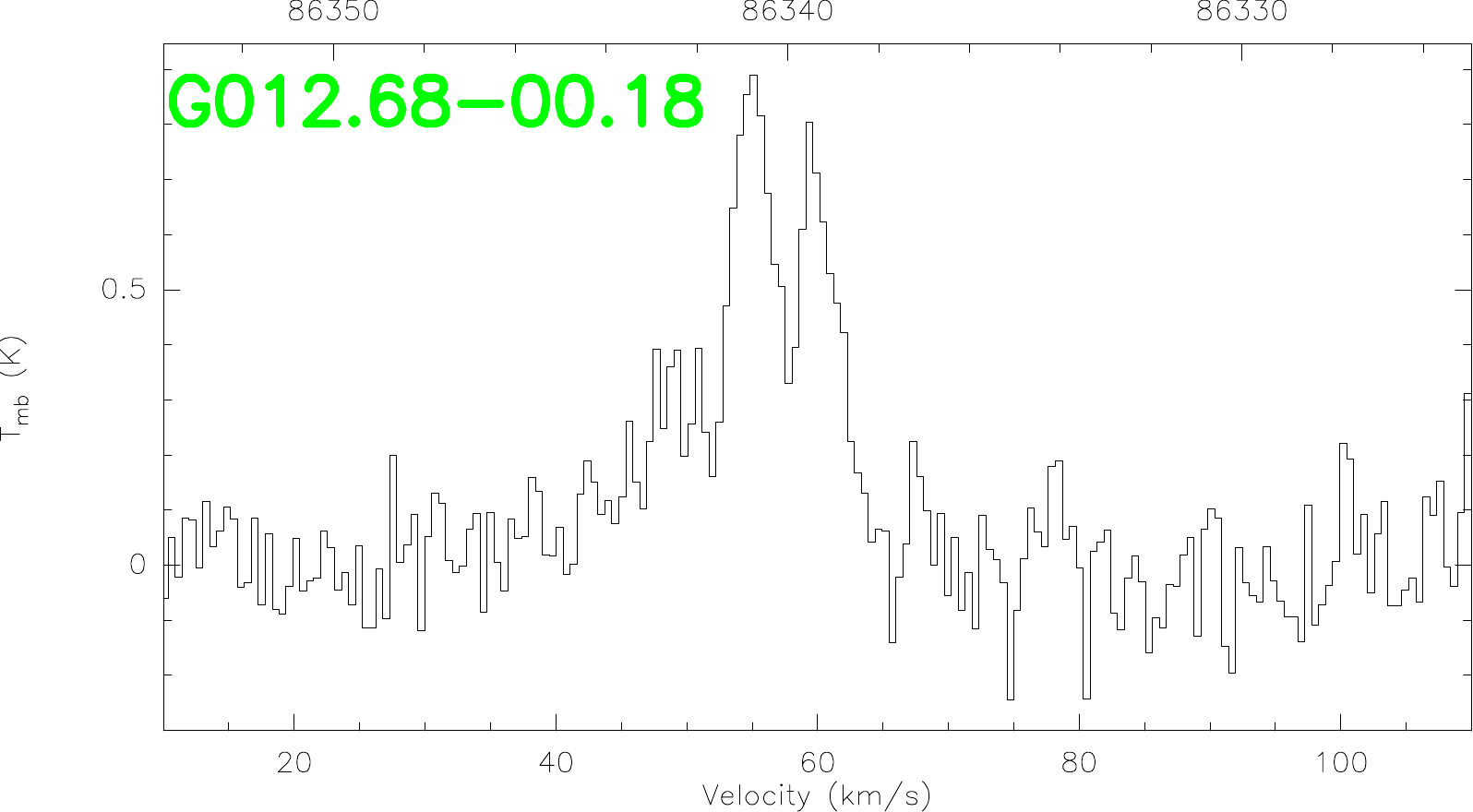}
\includegraphics[width=0.155\textwidth]{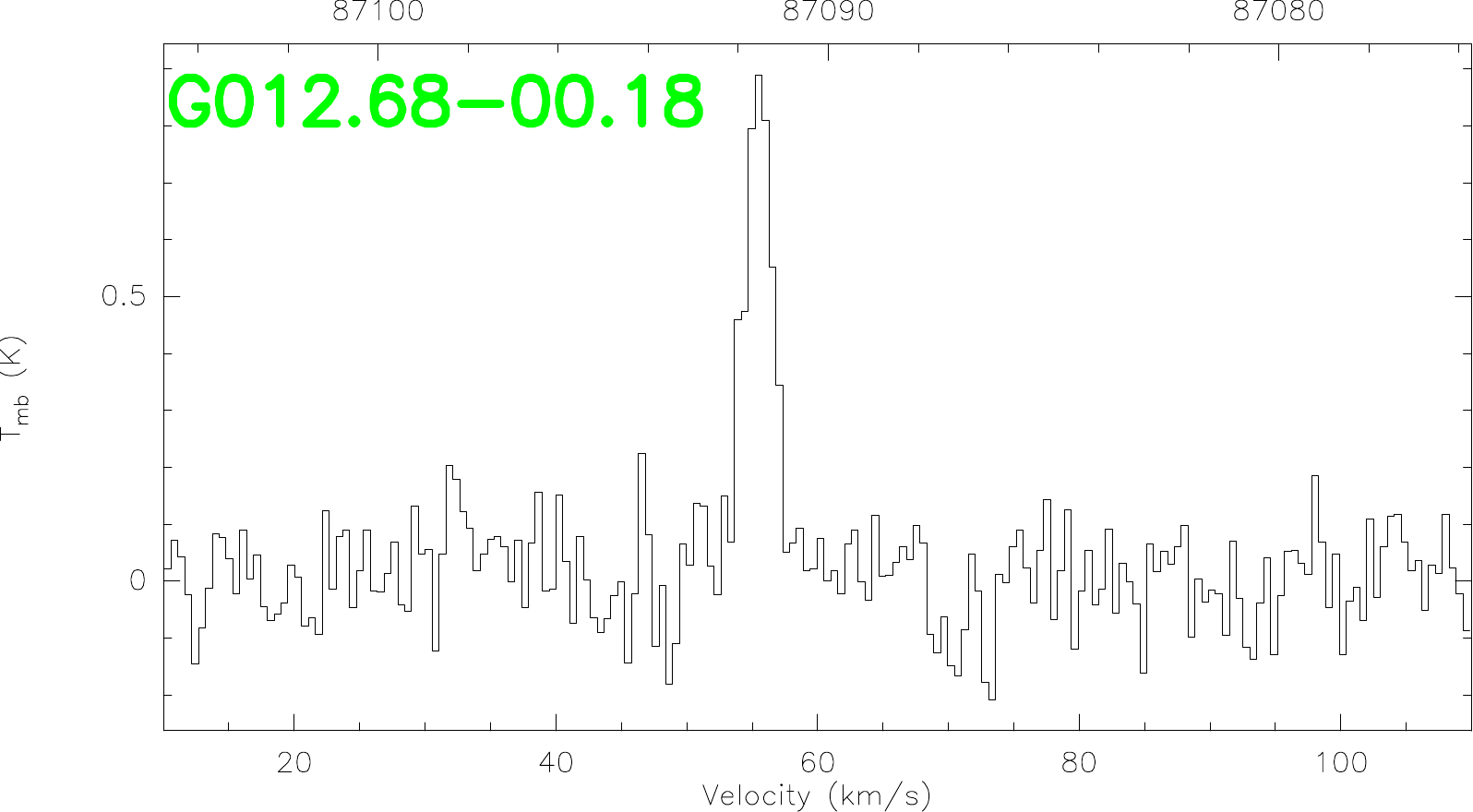}
\includegraphics[width=0.155\textwidth]{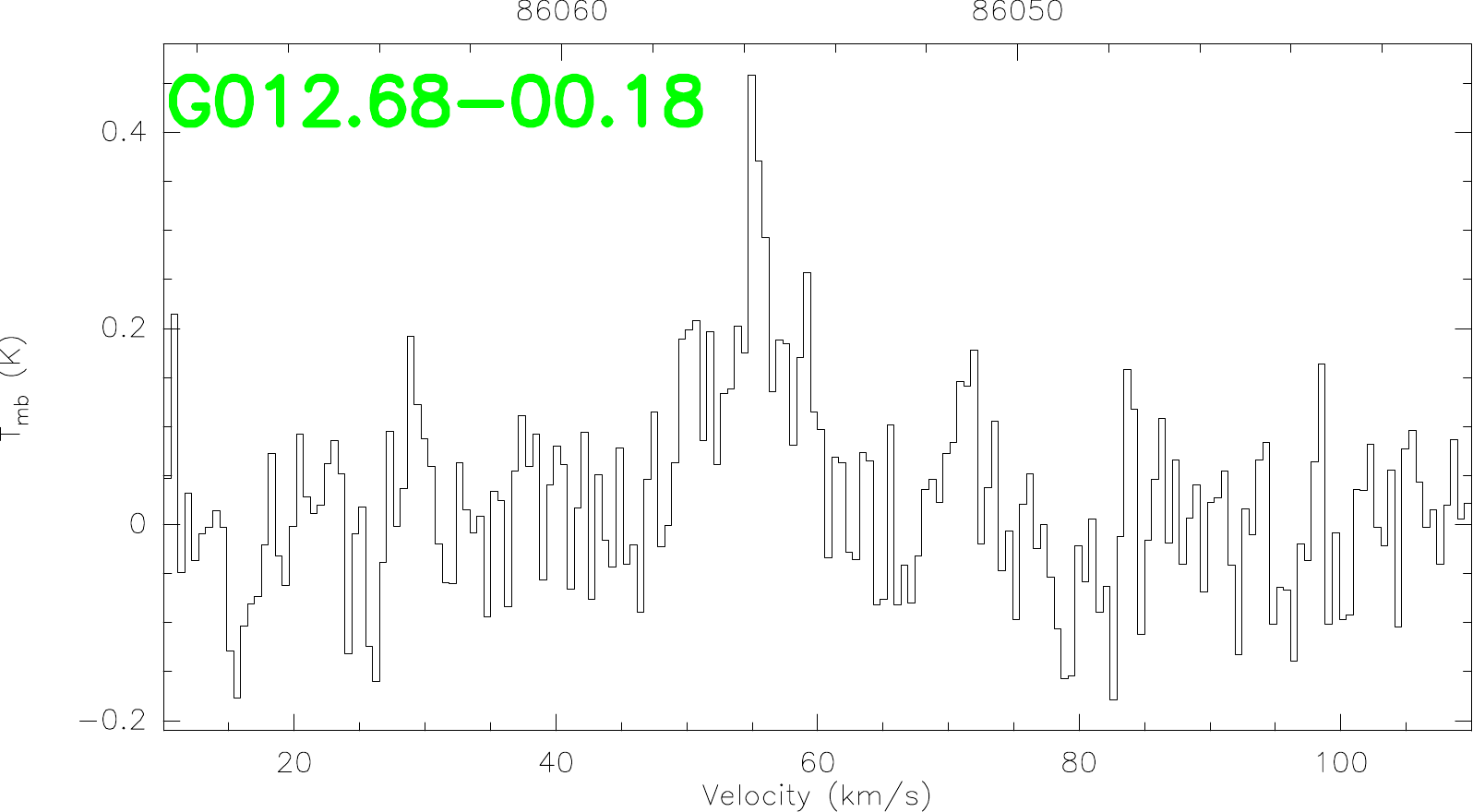}
\includegraphics[width=0.155\textwidth]{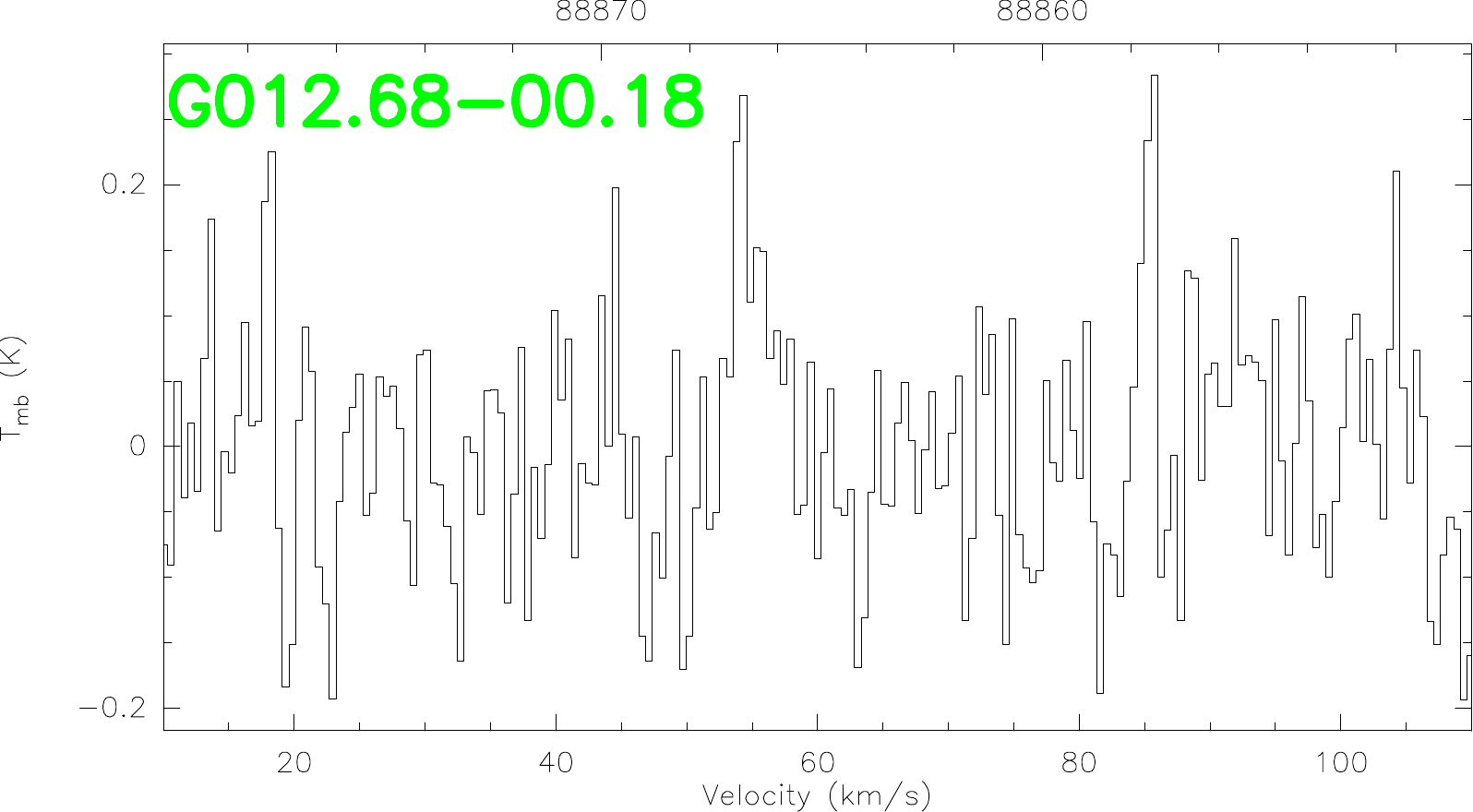}\\

\includegraphics[width=0.155\textwidth]{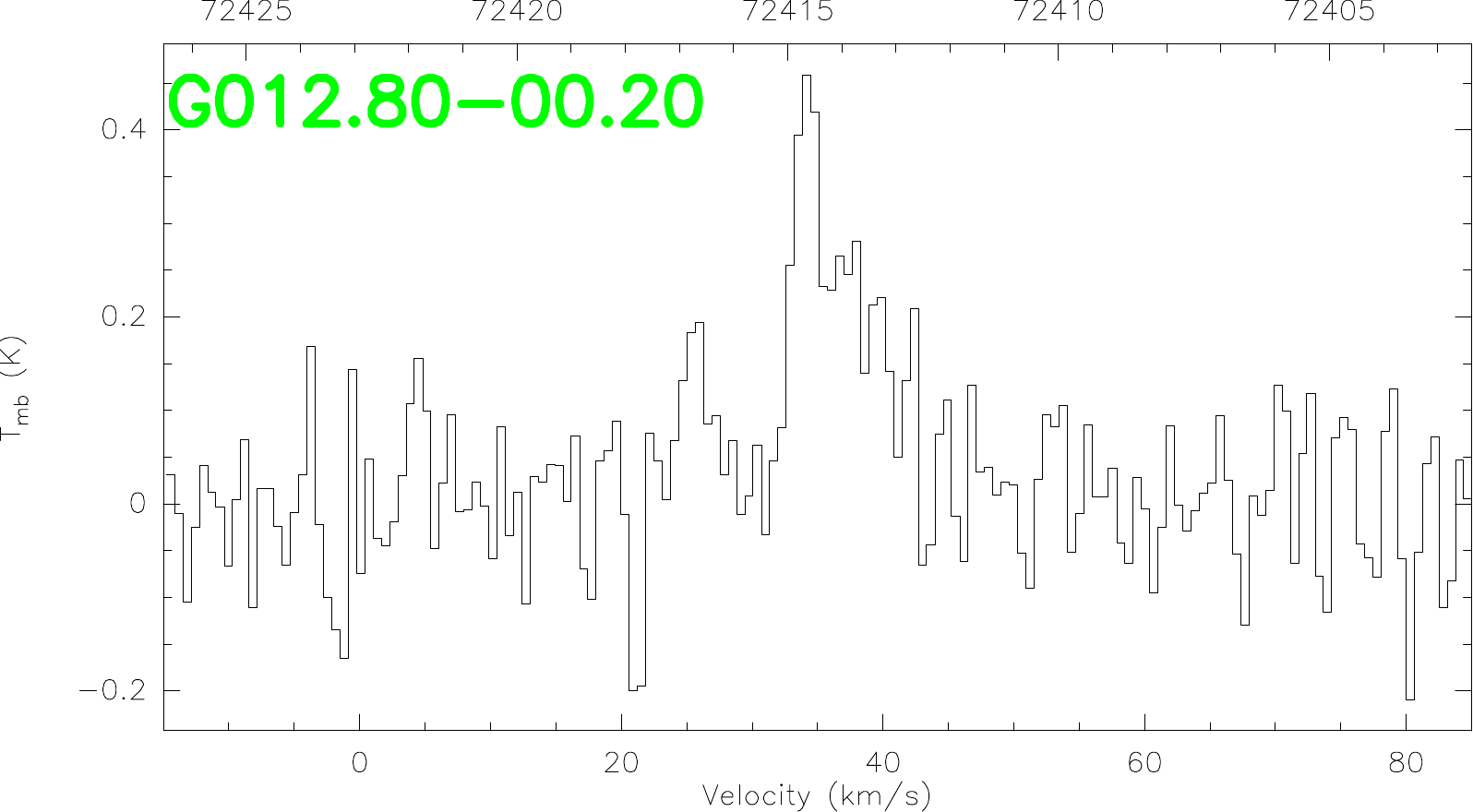}
\includegraphics[width=0.155\textwidth]{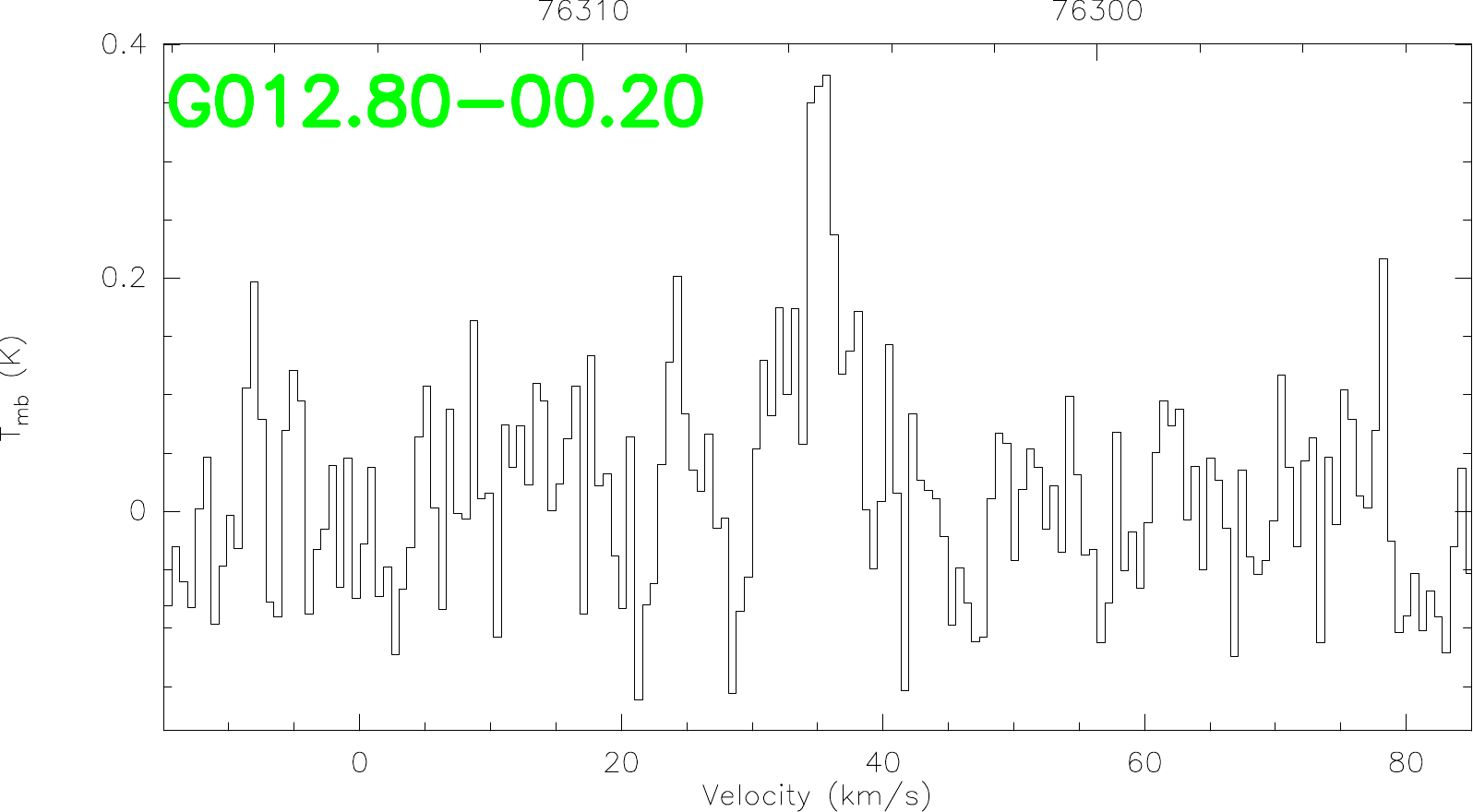}
\includegraphics[width=0.155\textwidth]{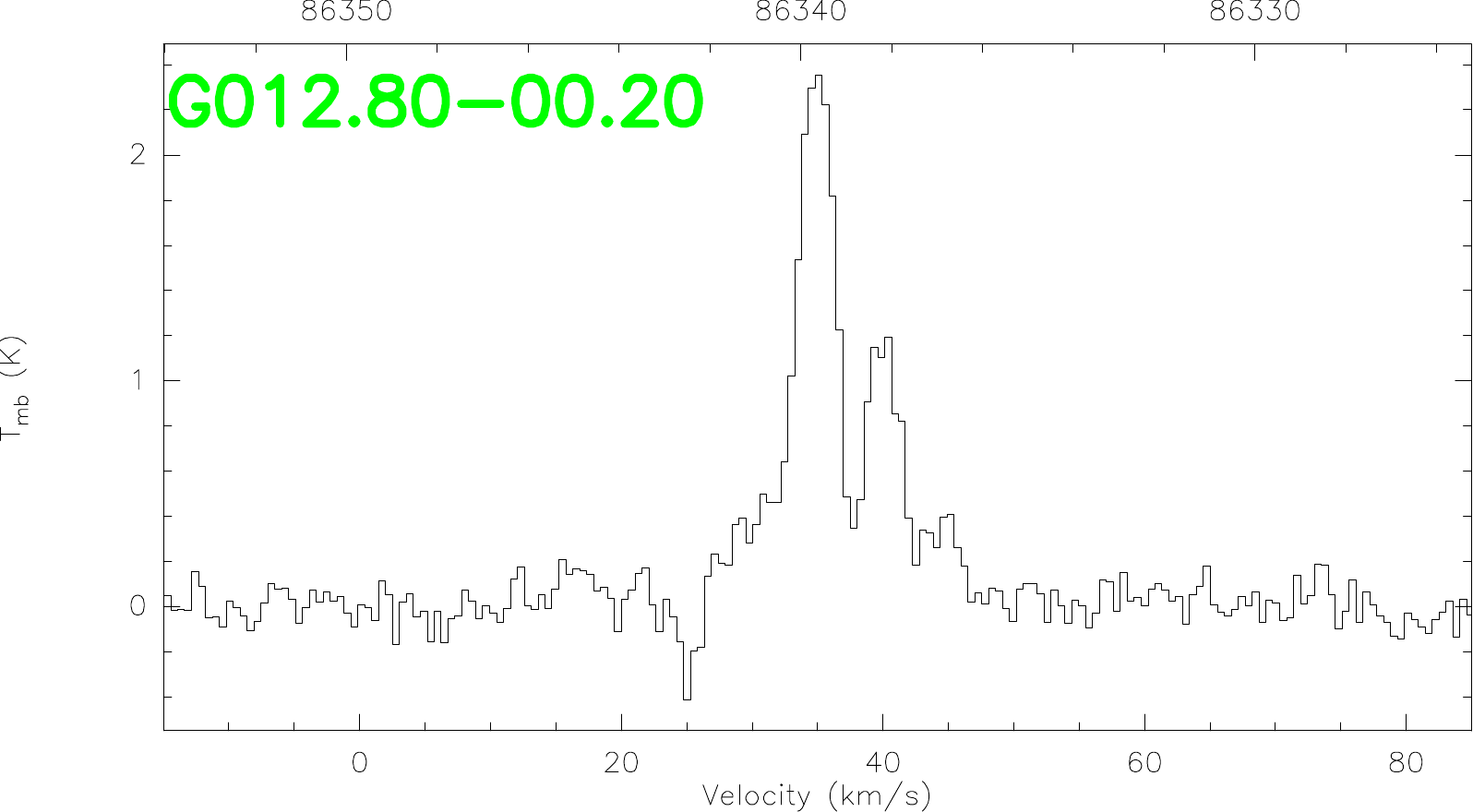}
\includegraphics[width=0.155\textwidth]{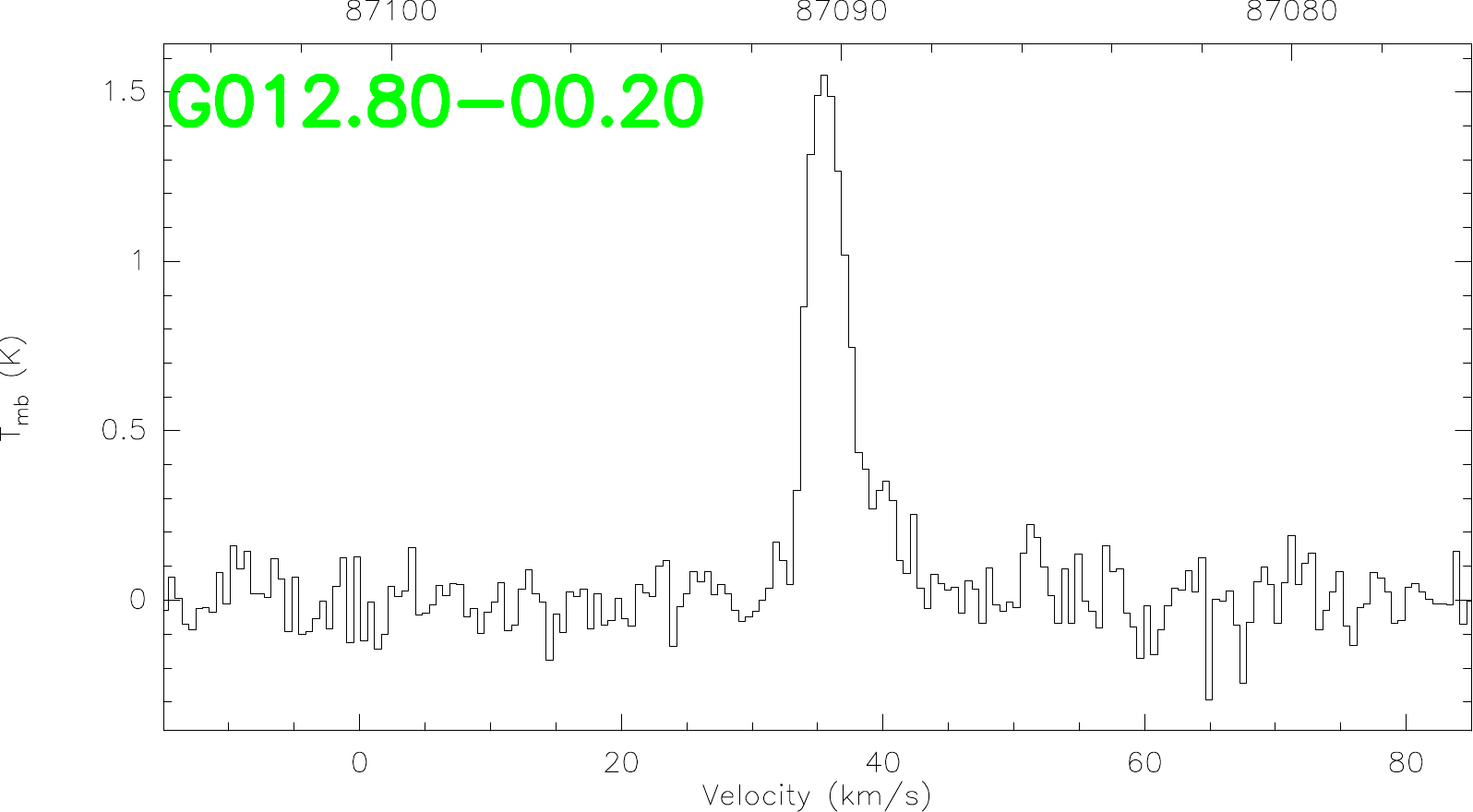}
\includegraphics[width=0.155\textwidth]{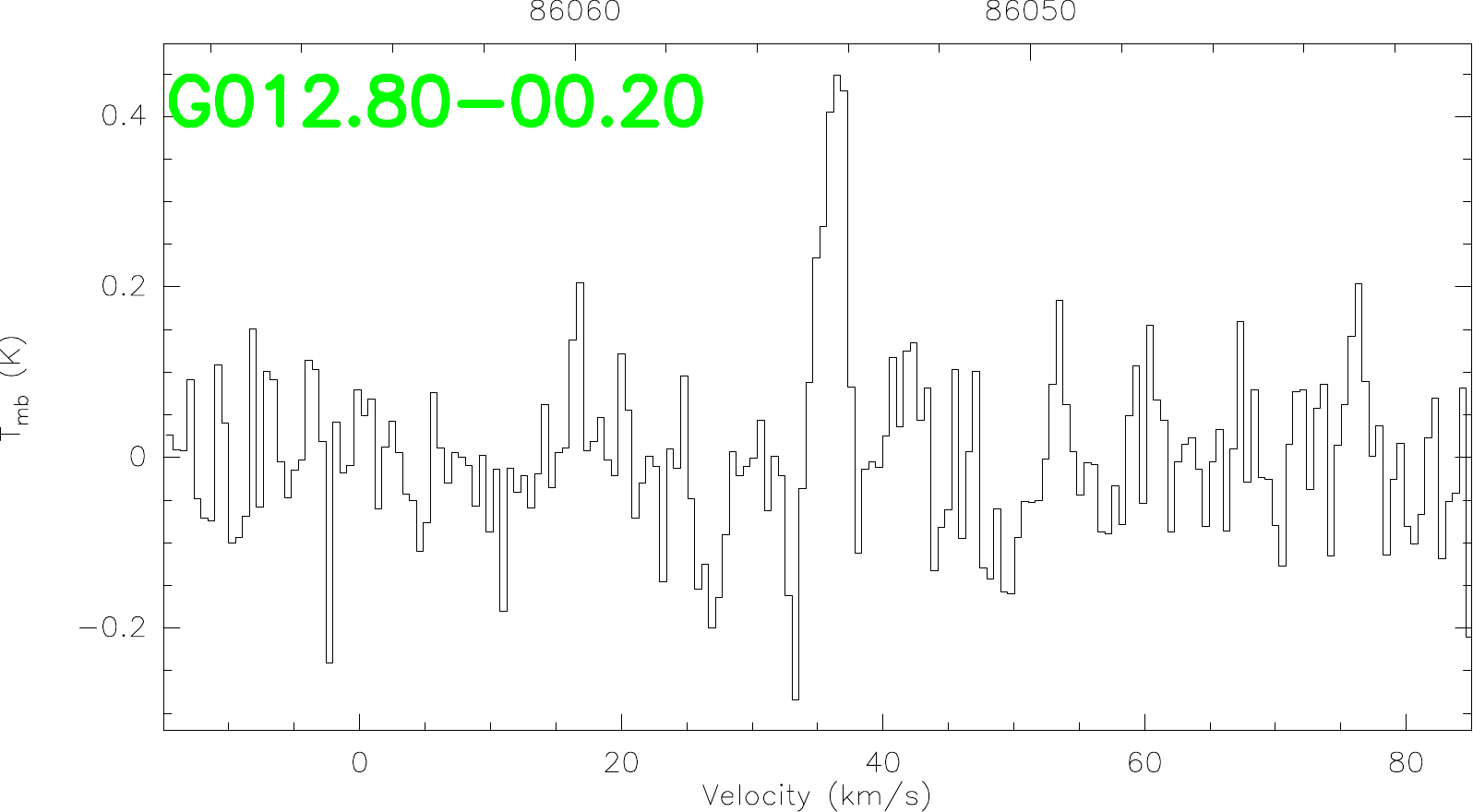}
\includegraphics[width=0.155\textwidth]{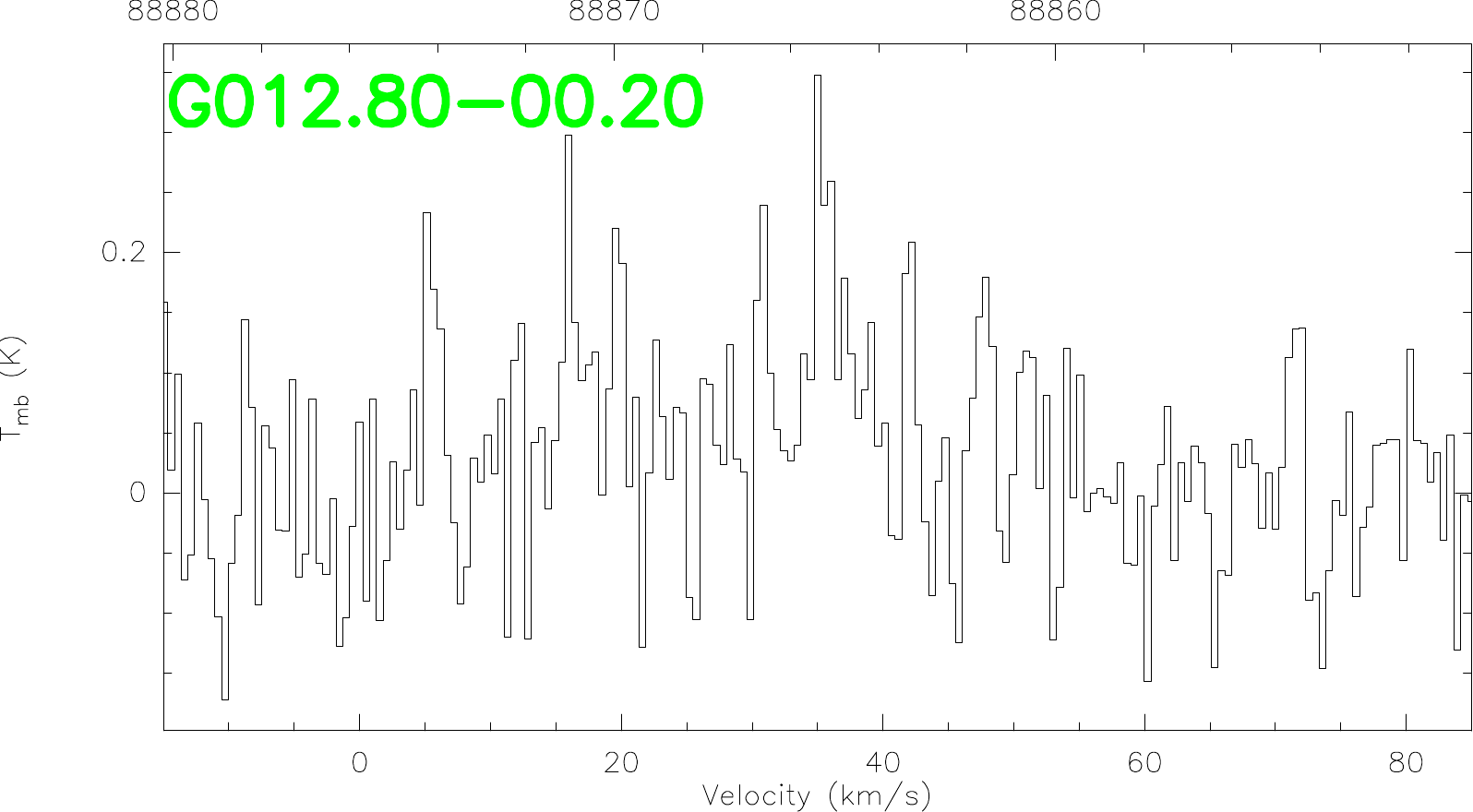}\\

\includegraphics[width=0.155\textwidth]{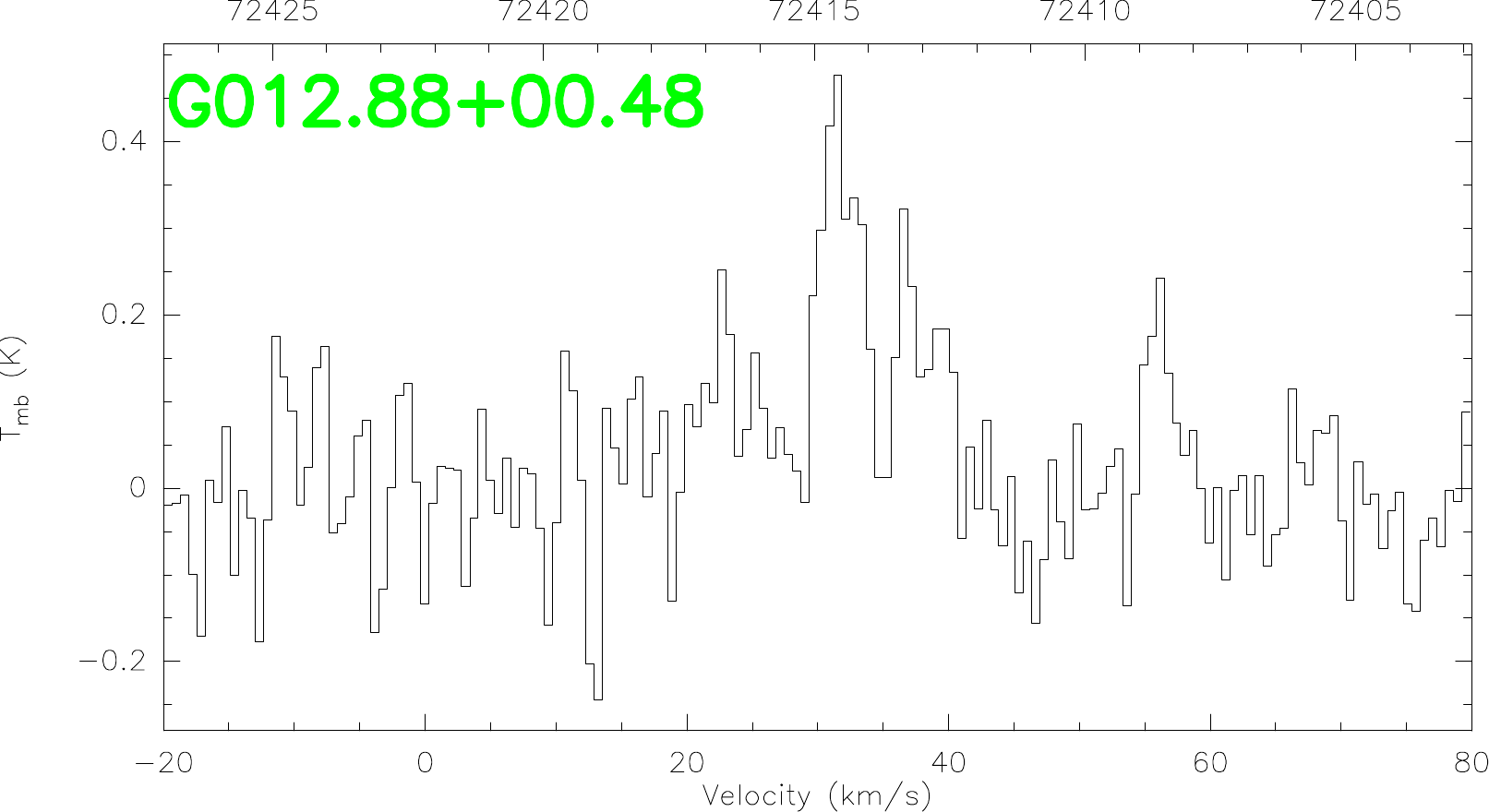}
\includegraphics[width=0.155\textwidth]{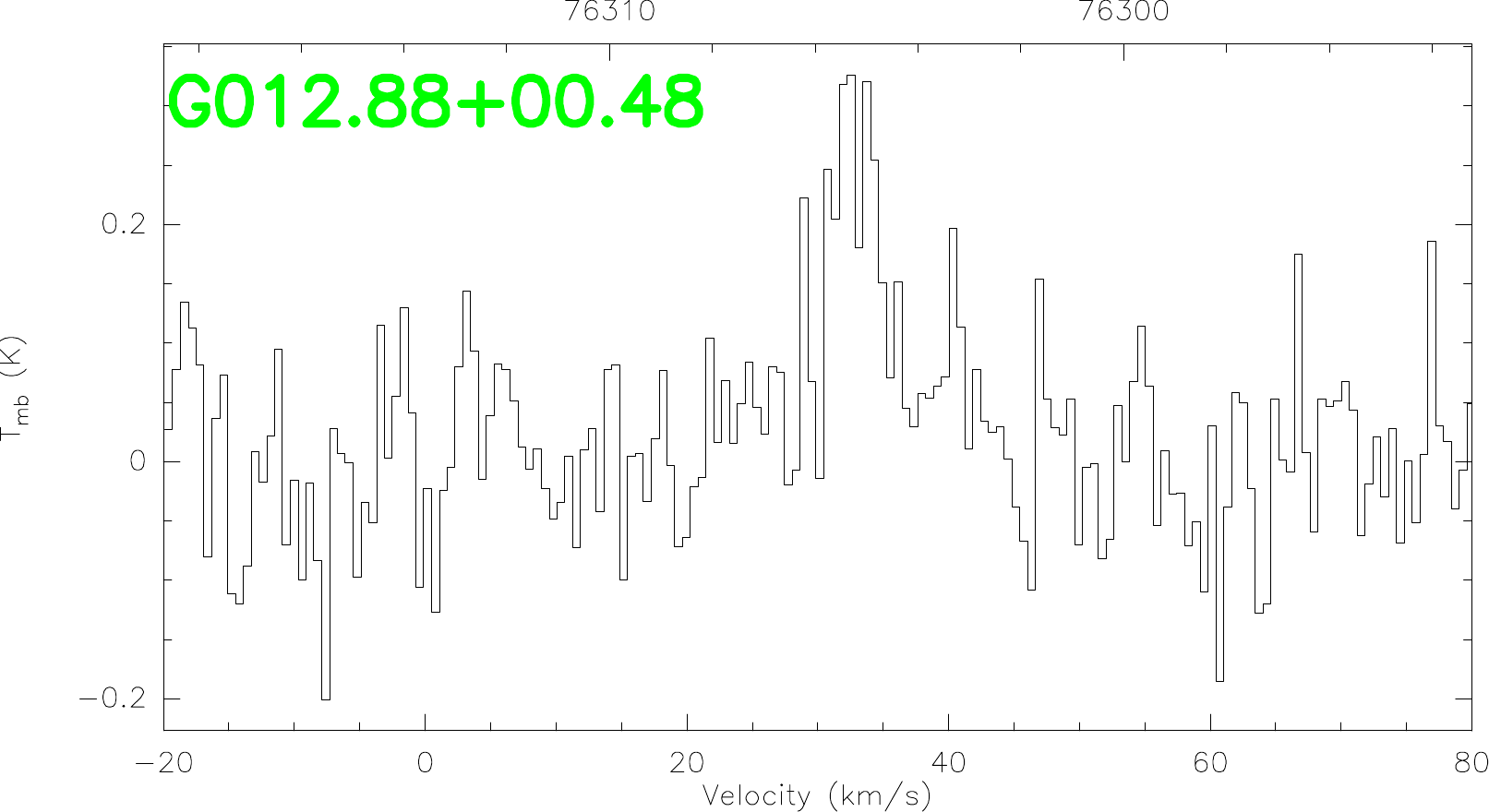}
\includegraphics[width=0.155\textwidth]{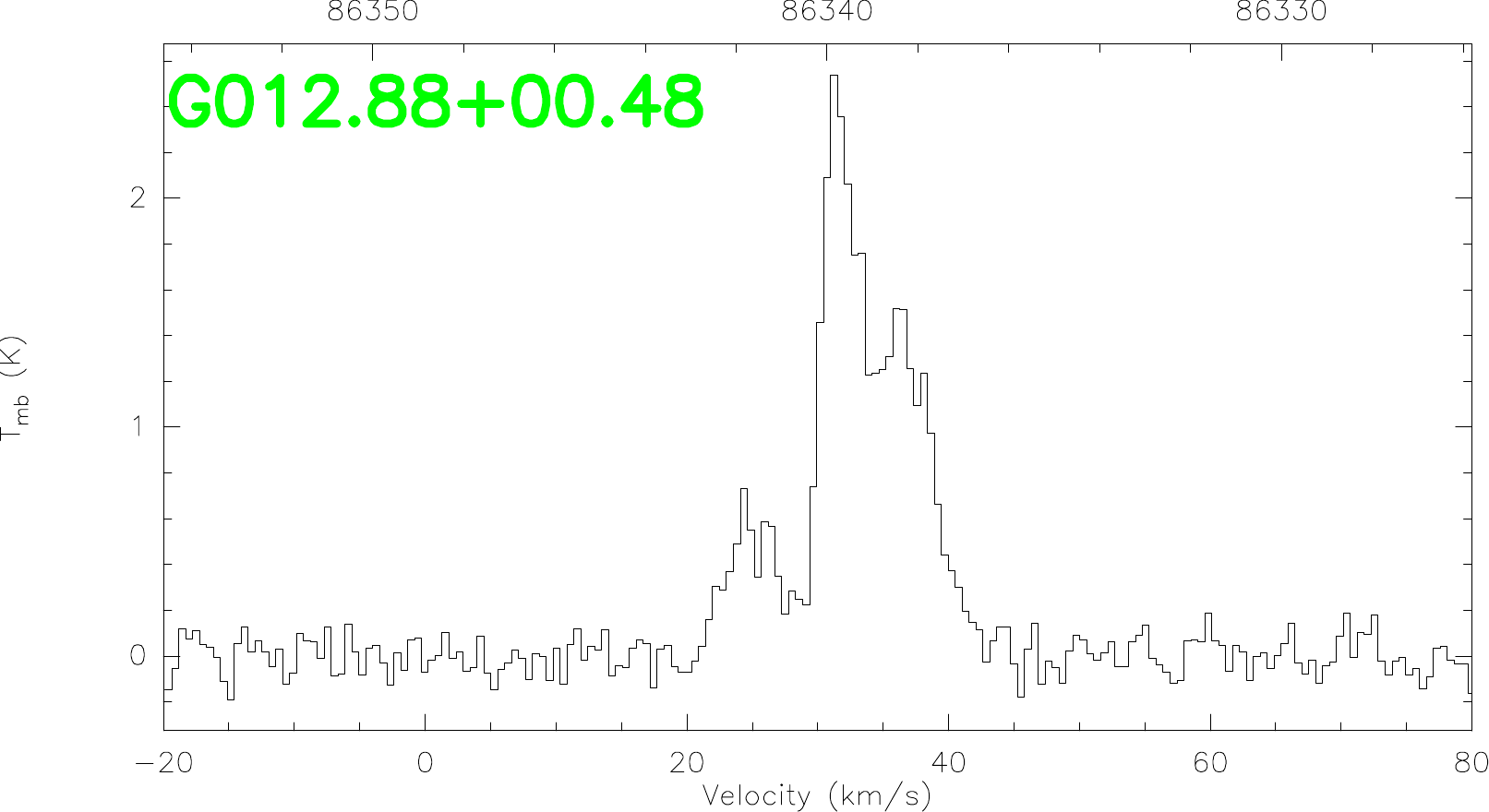}
\includegraphics[width=0.155\textwidth]{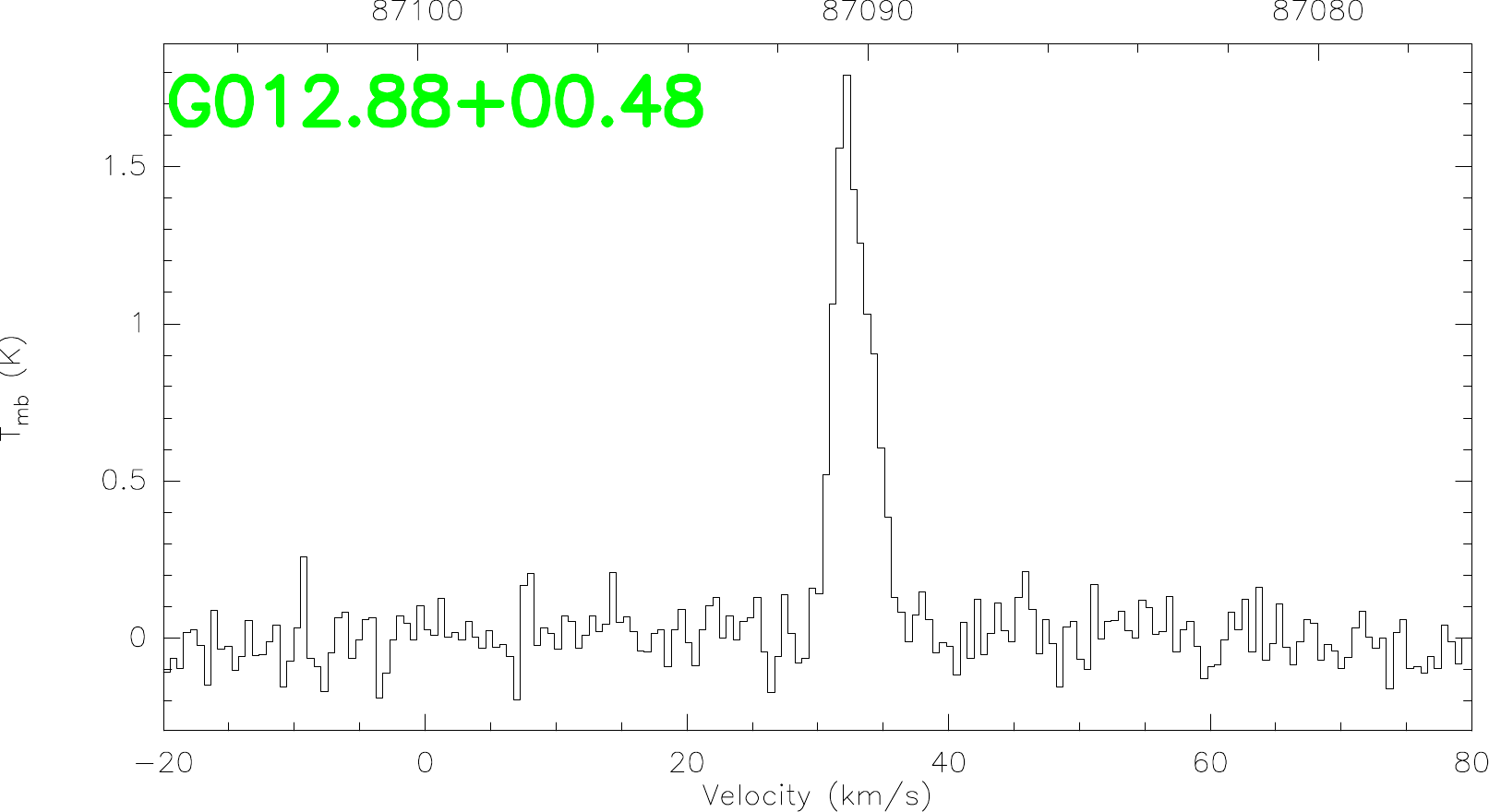}\includegraphics[width=0.155\textwidth]{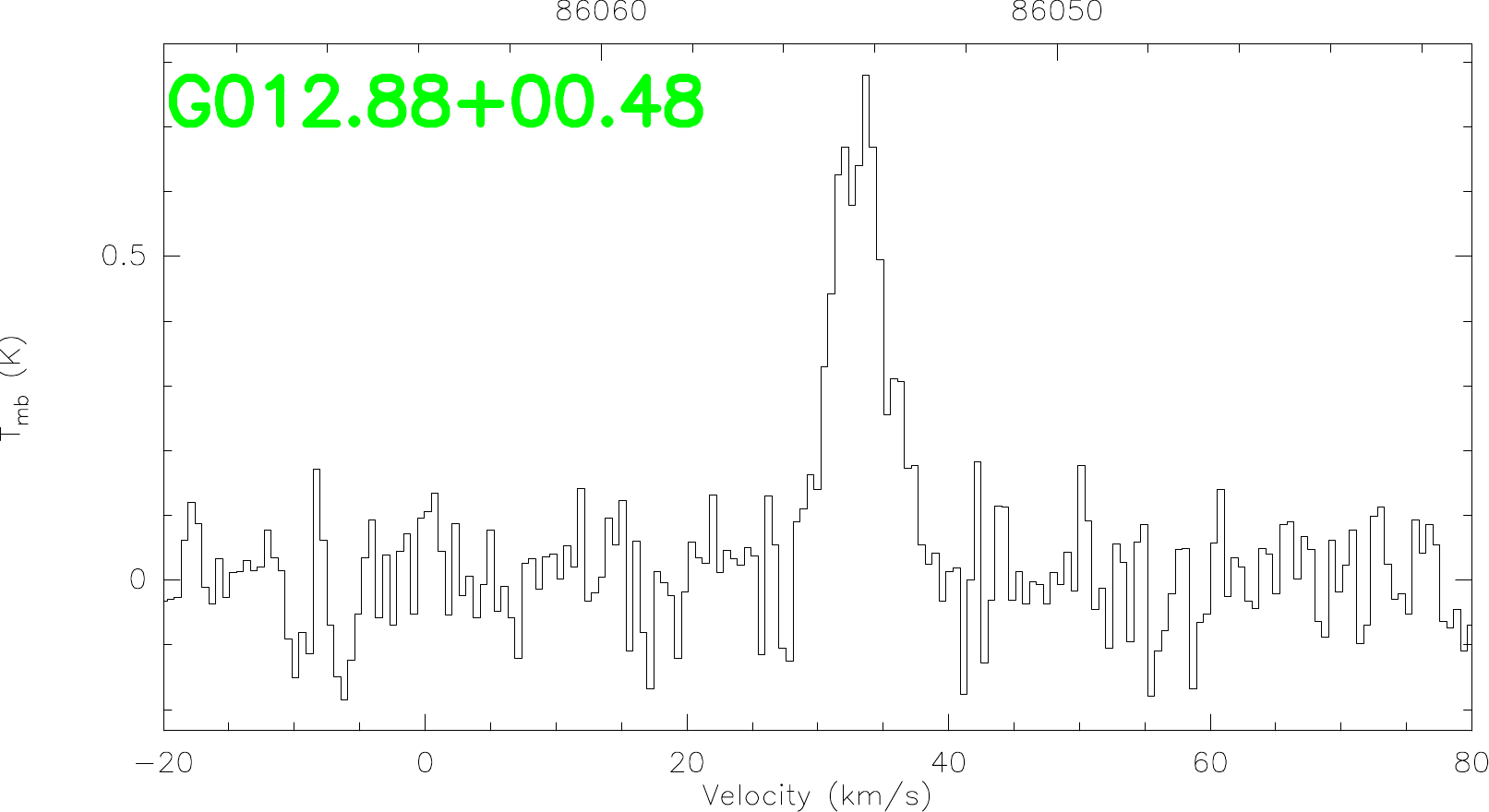}
\includegraphics[width=0.155\textwidth]{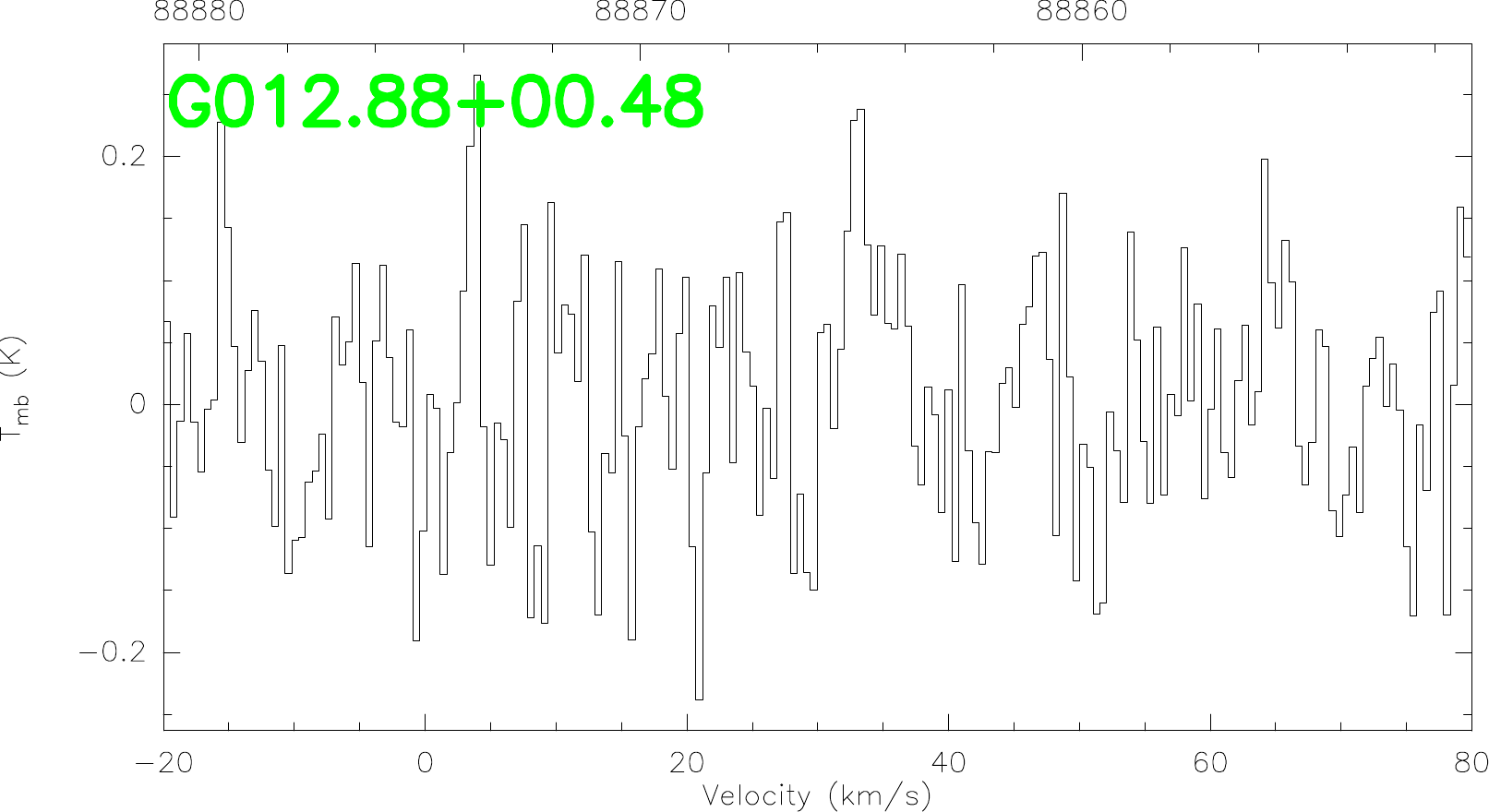}\\

\includegraphics[width=0.155\textwidth]{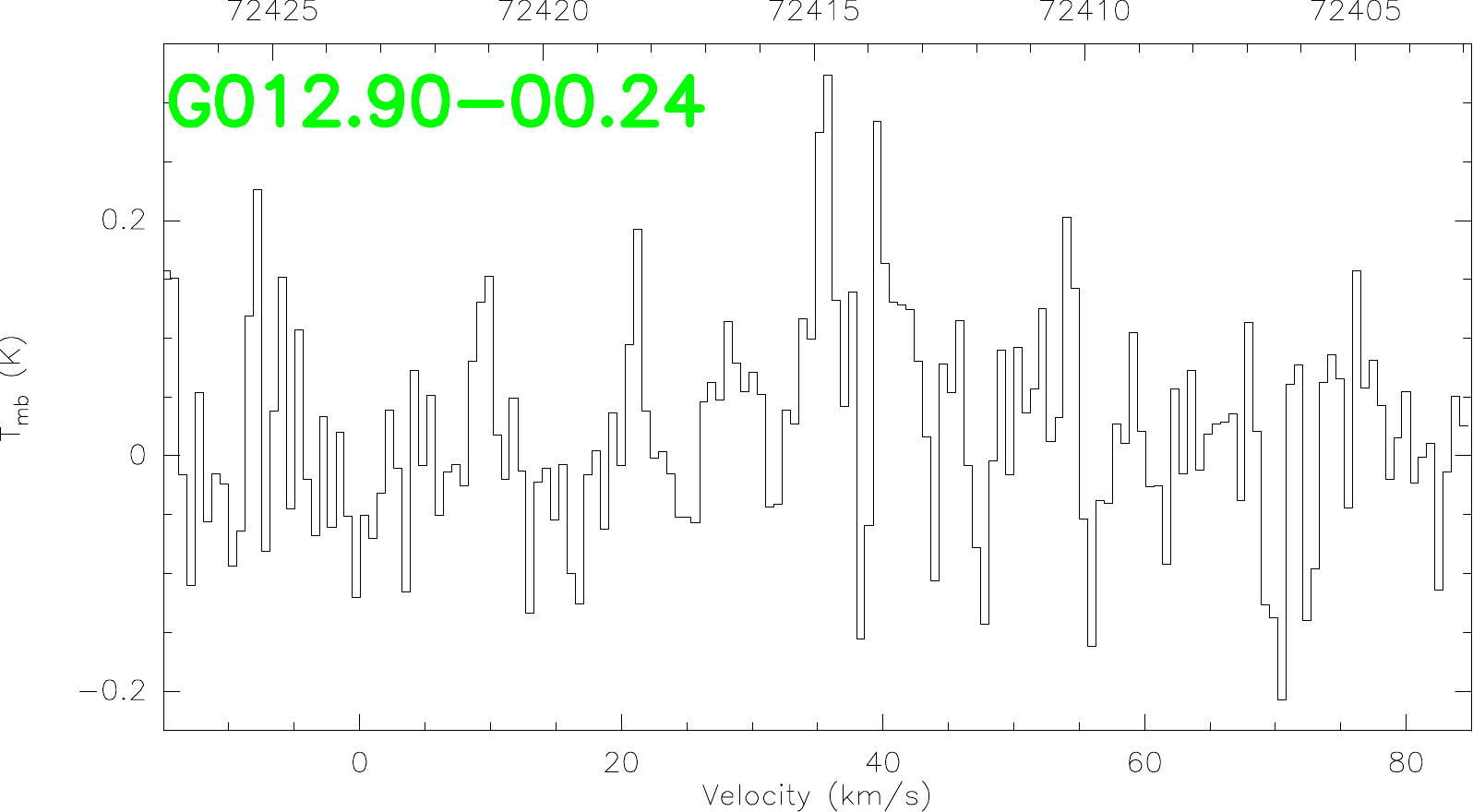}
\includegraphics[width=0.155\textwidth]{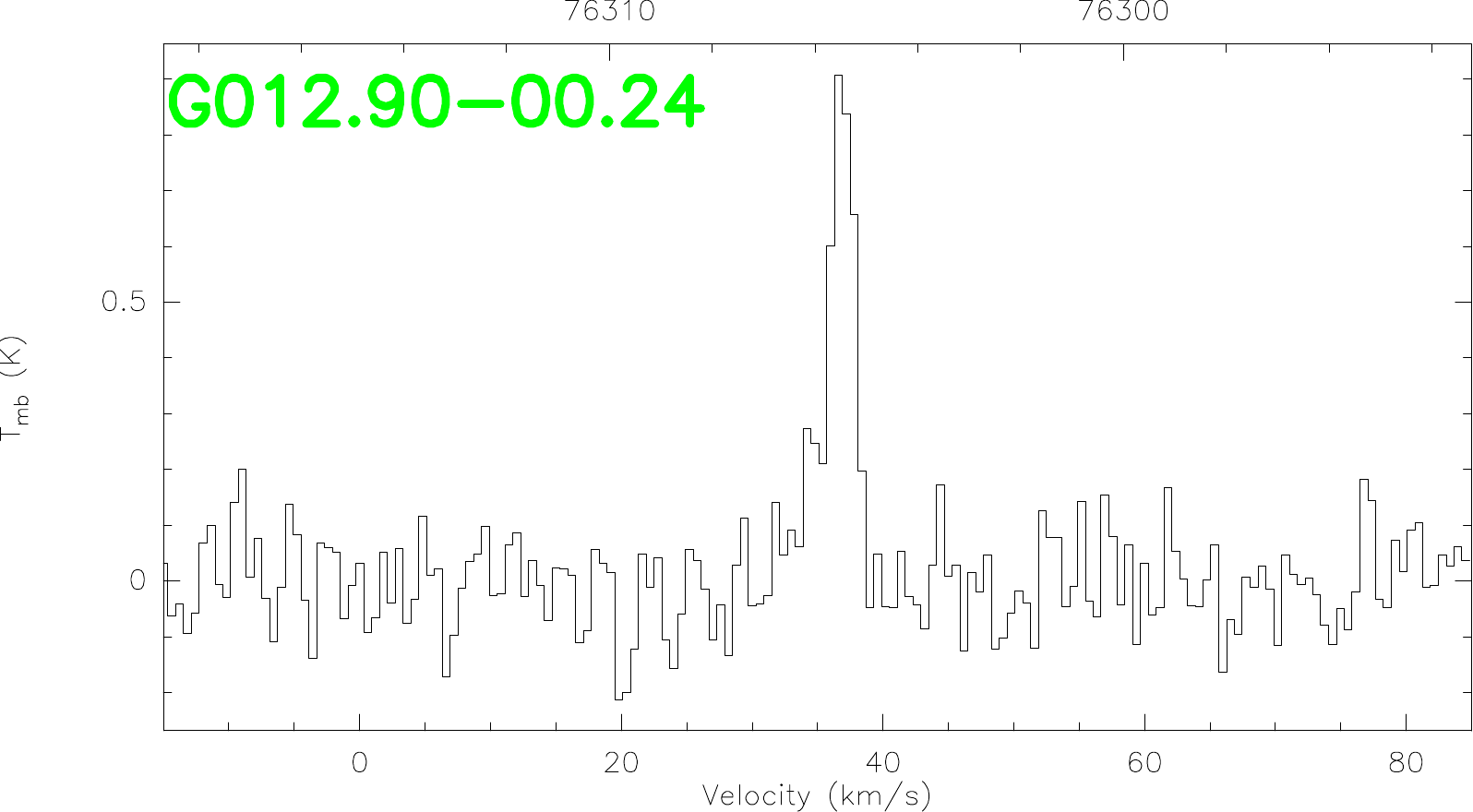}
\includegraphics[width=0.155\textwidth]{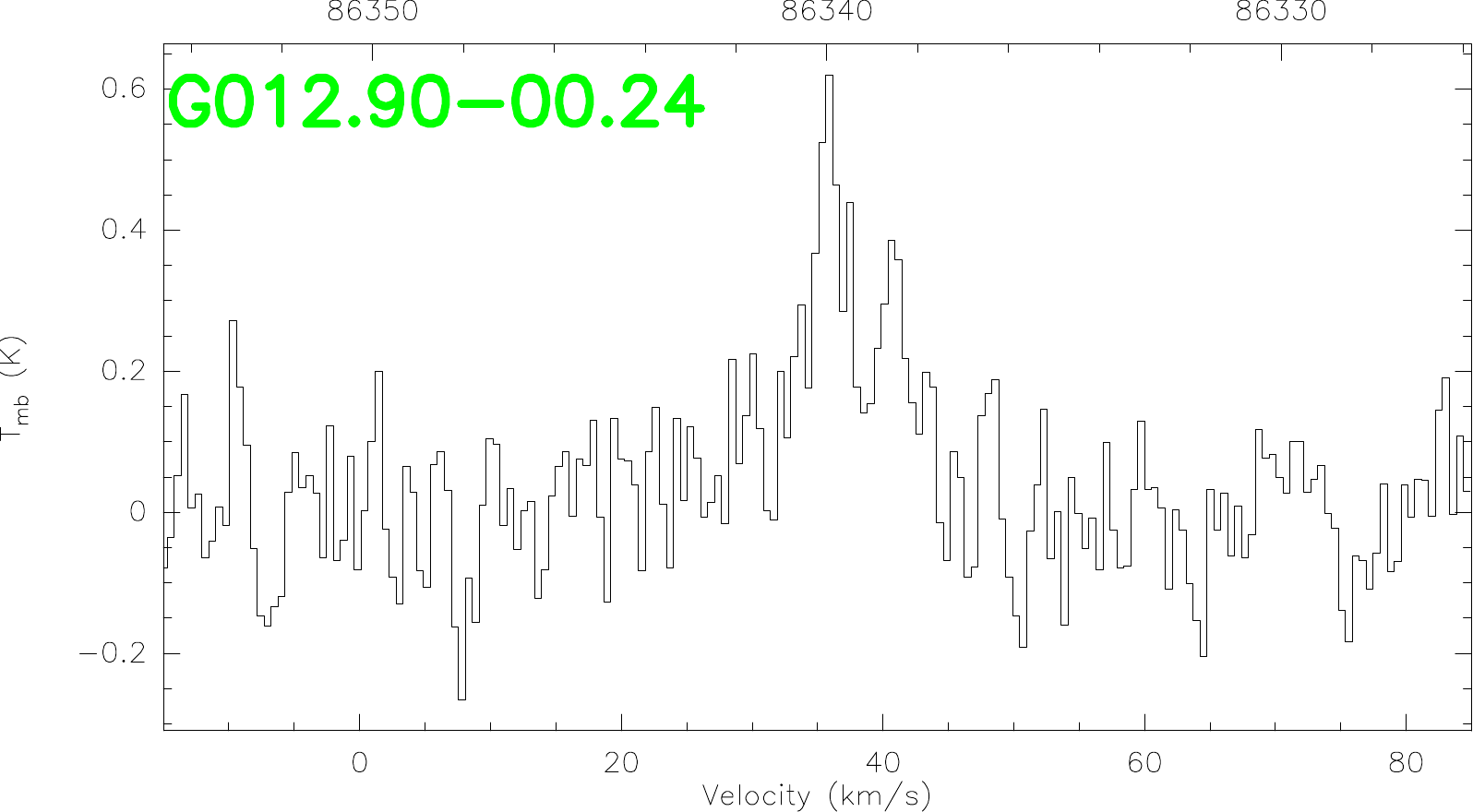}
\includegraphics[width=0.155\textwidth]{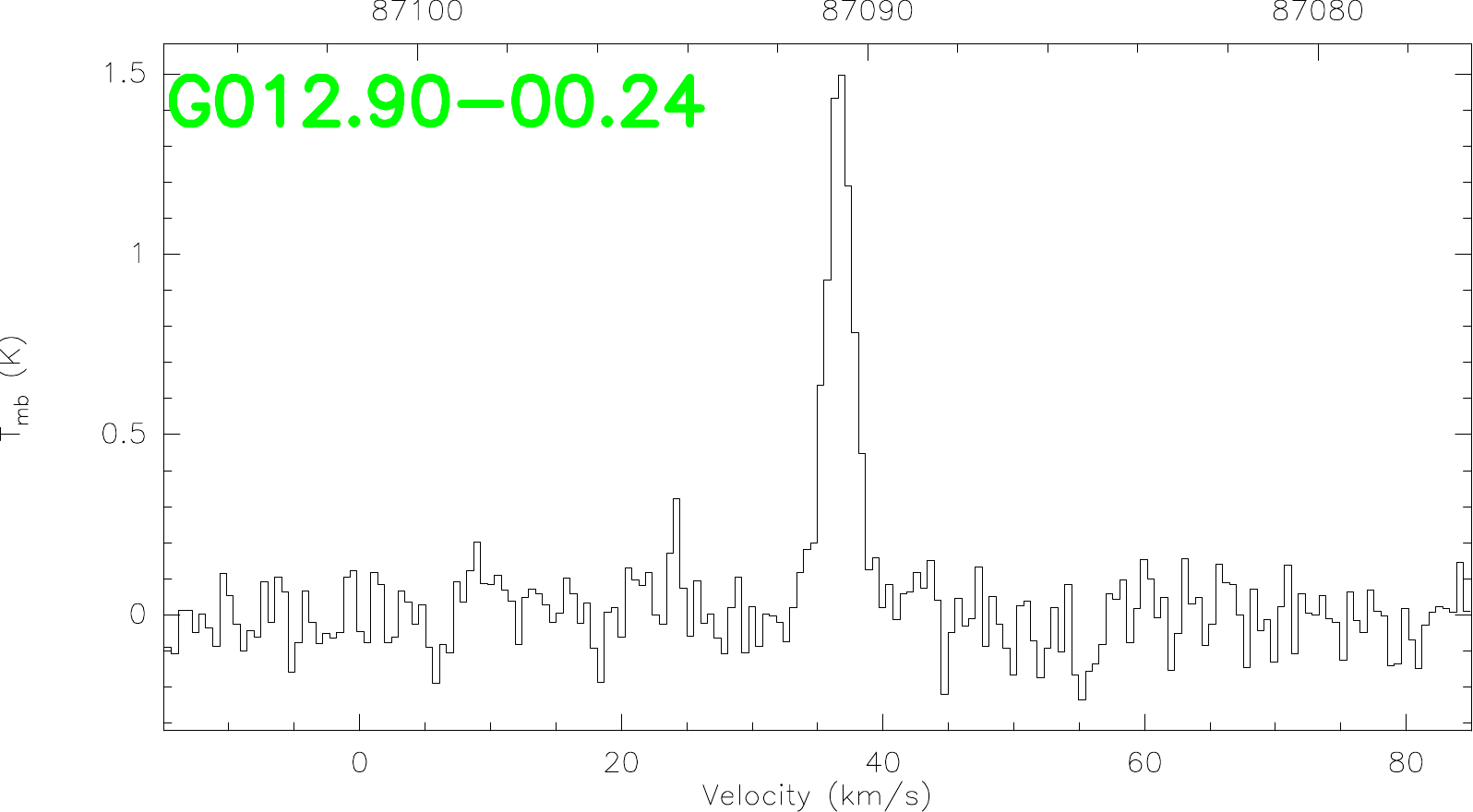}
\includegraphics[width=0.155\textwidth]{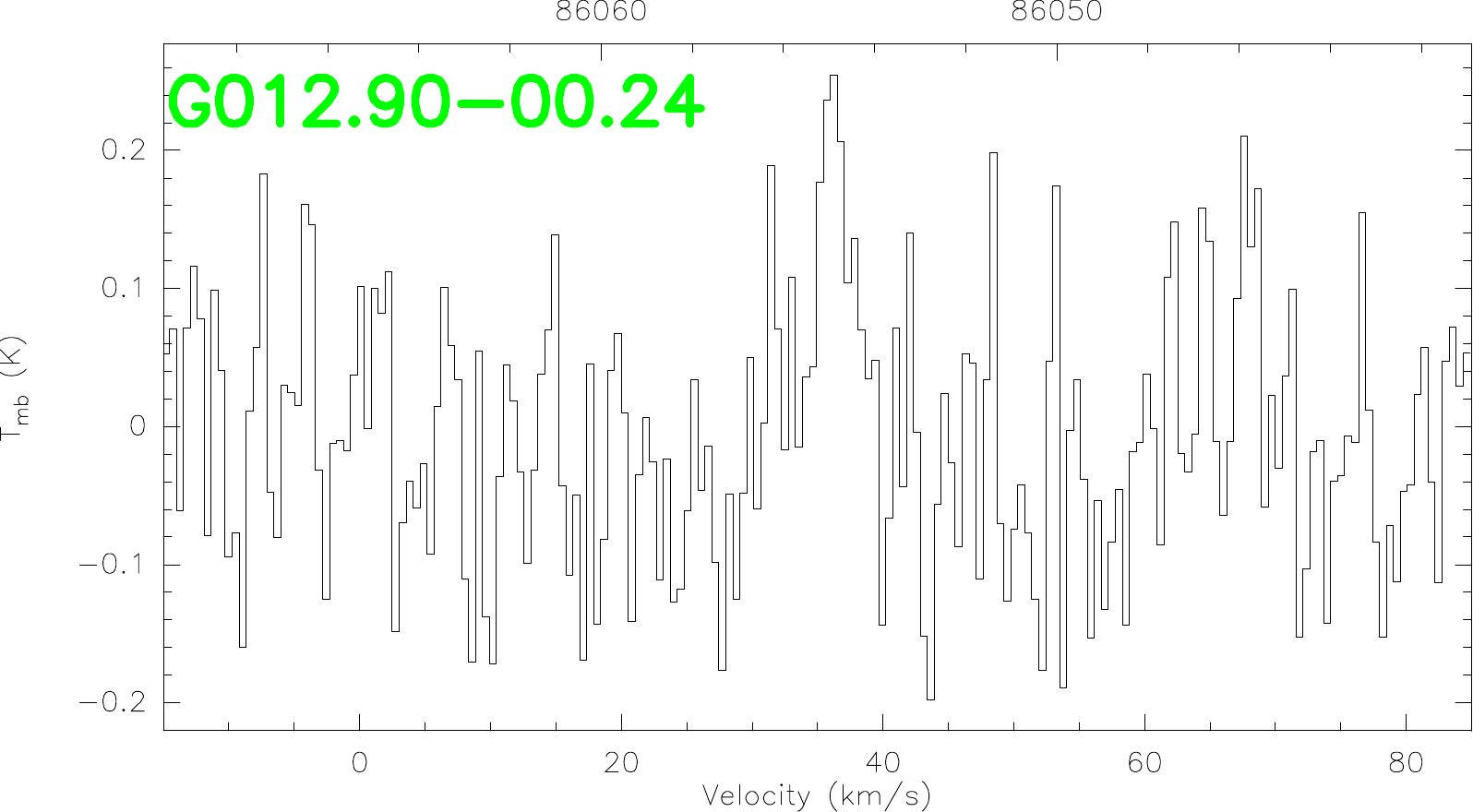}
\includegraphics[width=0.155\textwidth]{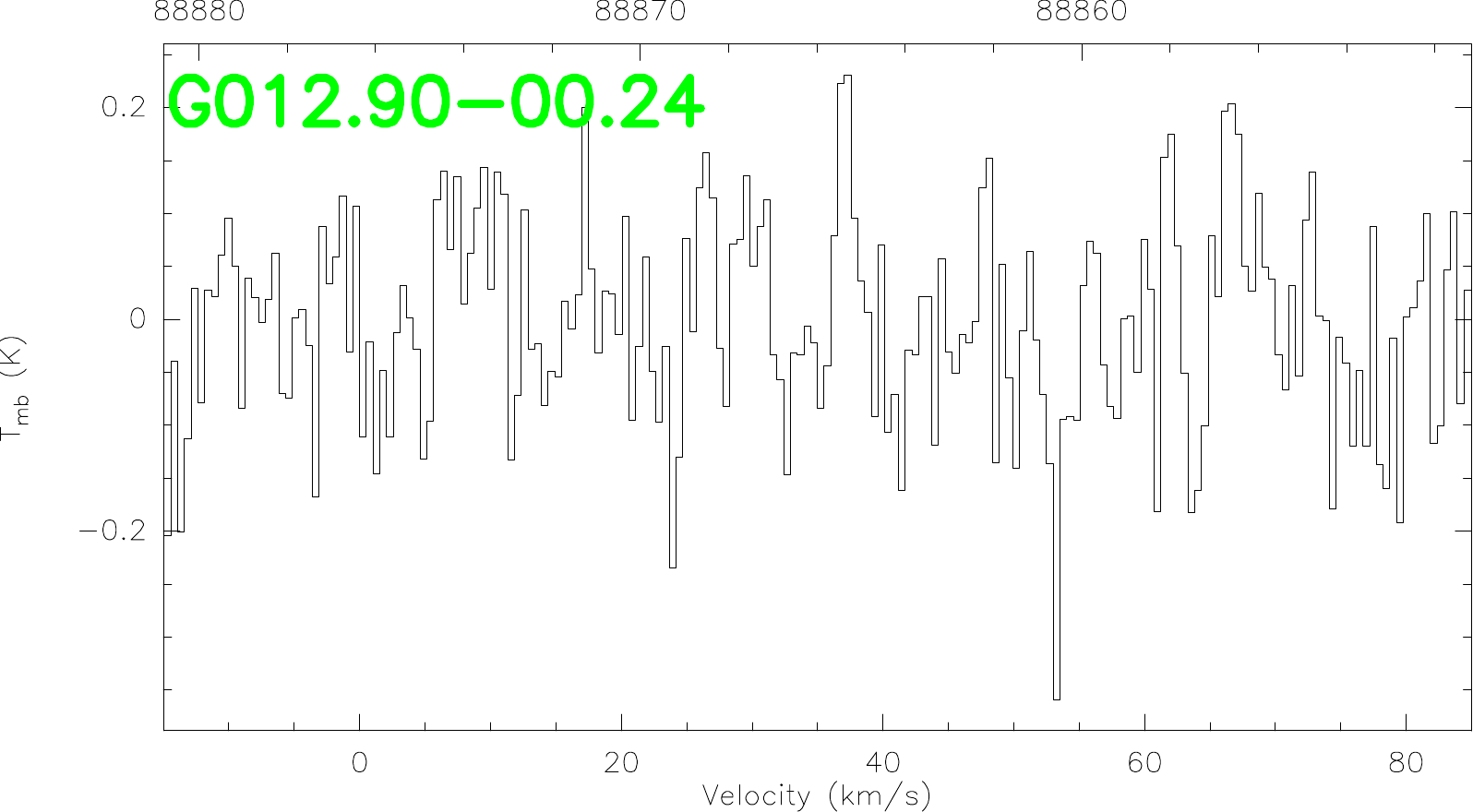}\\

\includegraphics[width=0.155\textwidth]{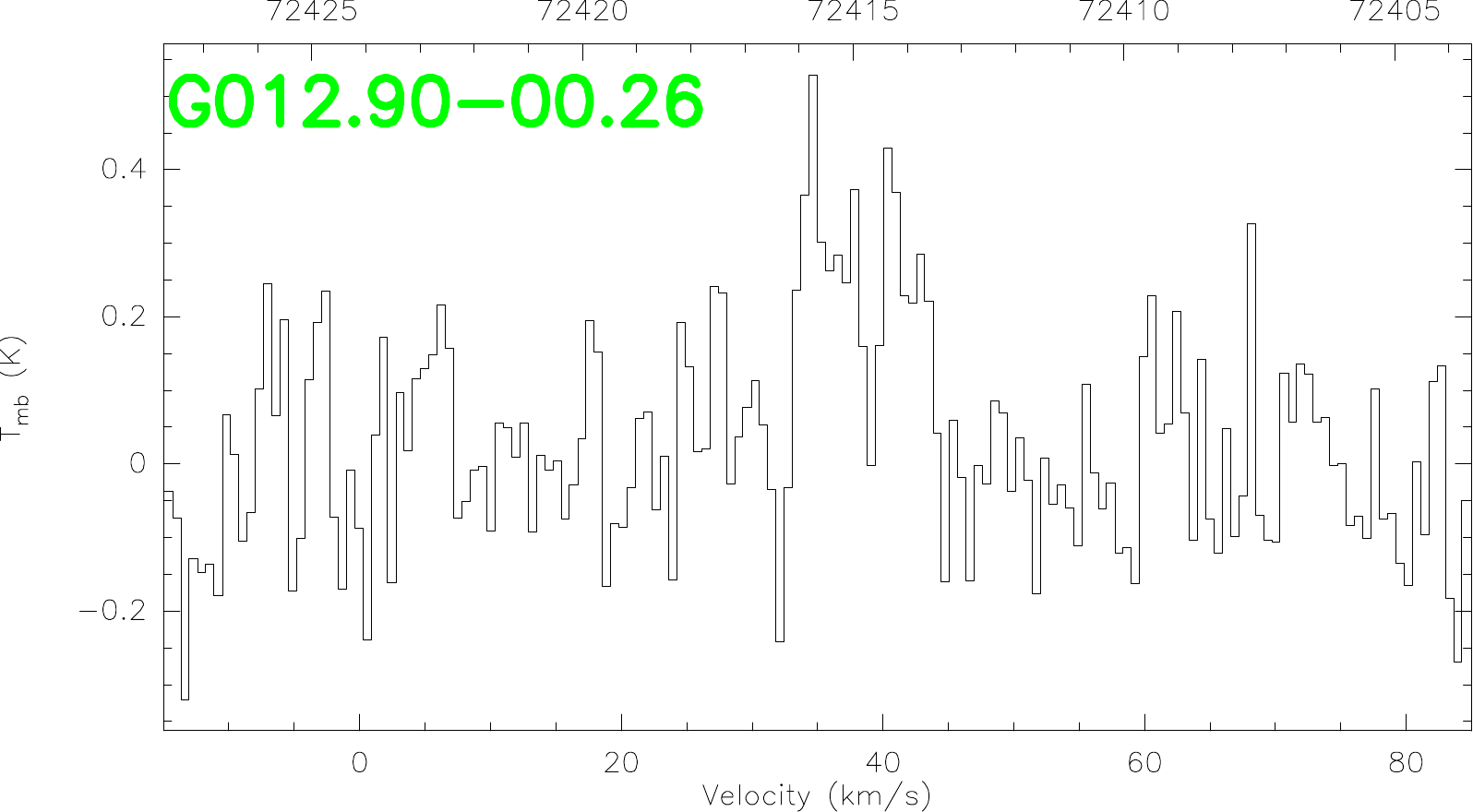}
\includegraphics[width=0.155\textwidth]{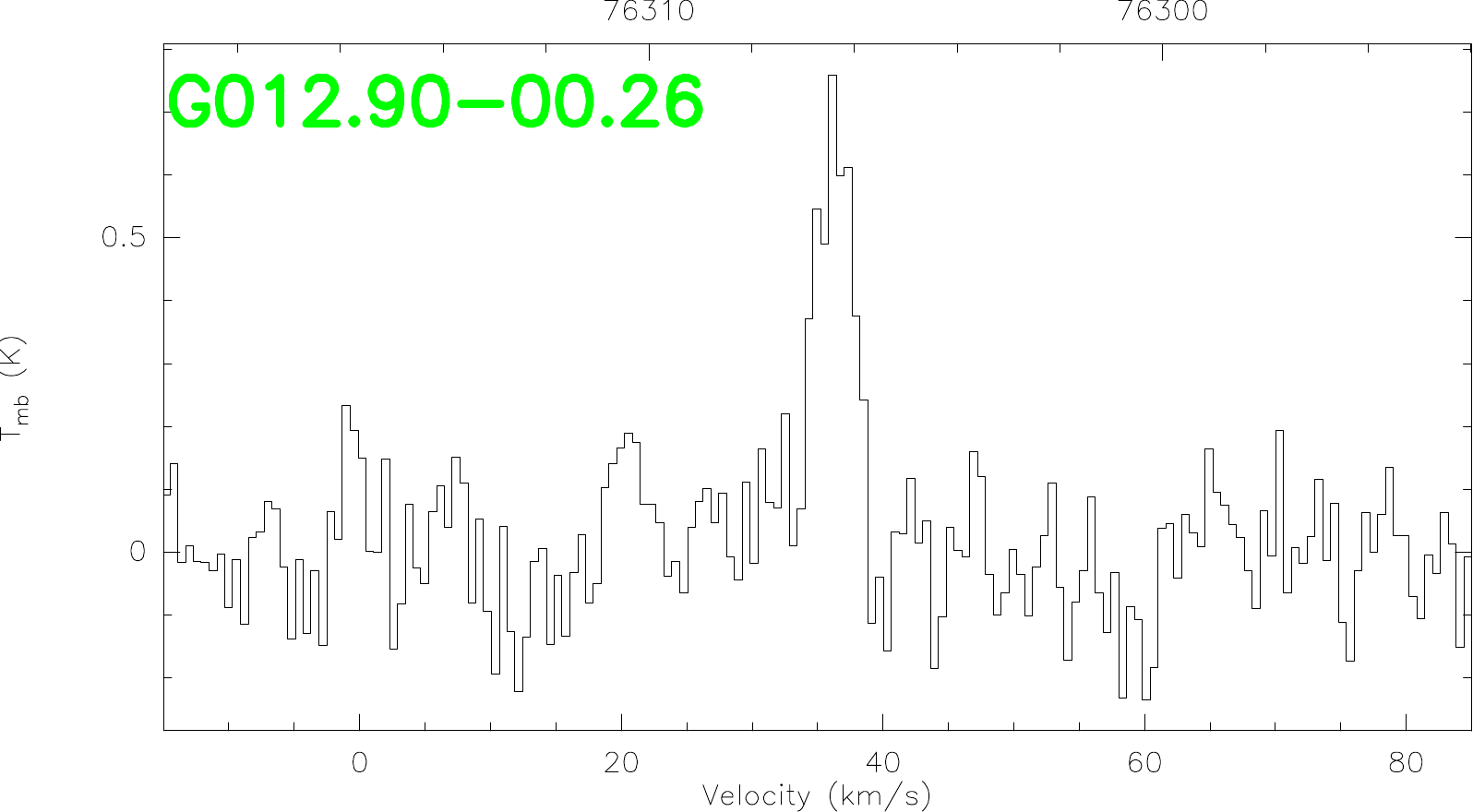}
\includegraphics[width=0.155\textwidth]{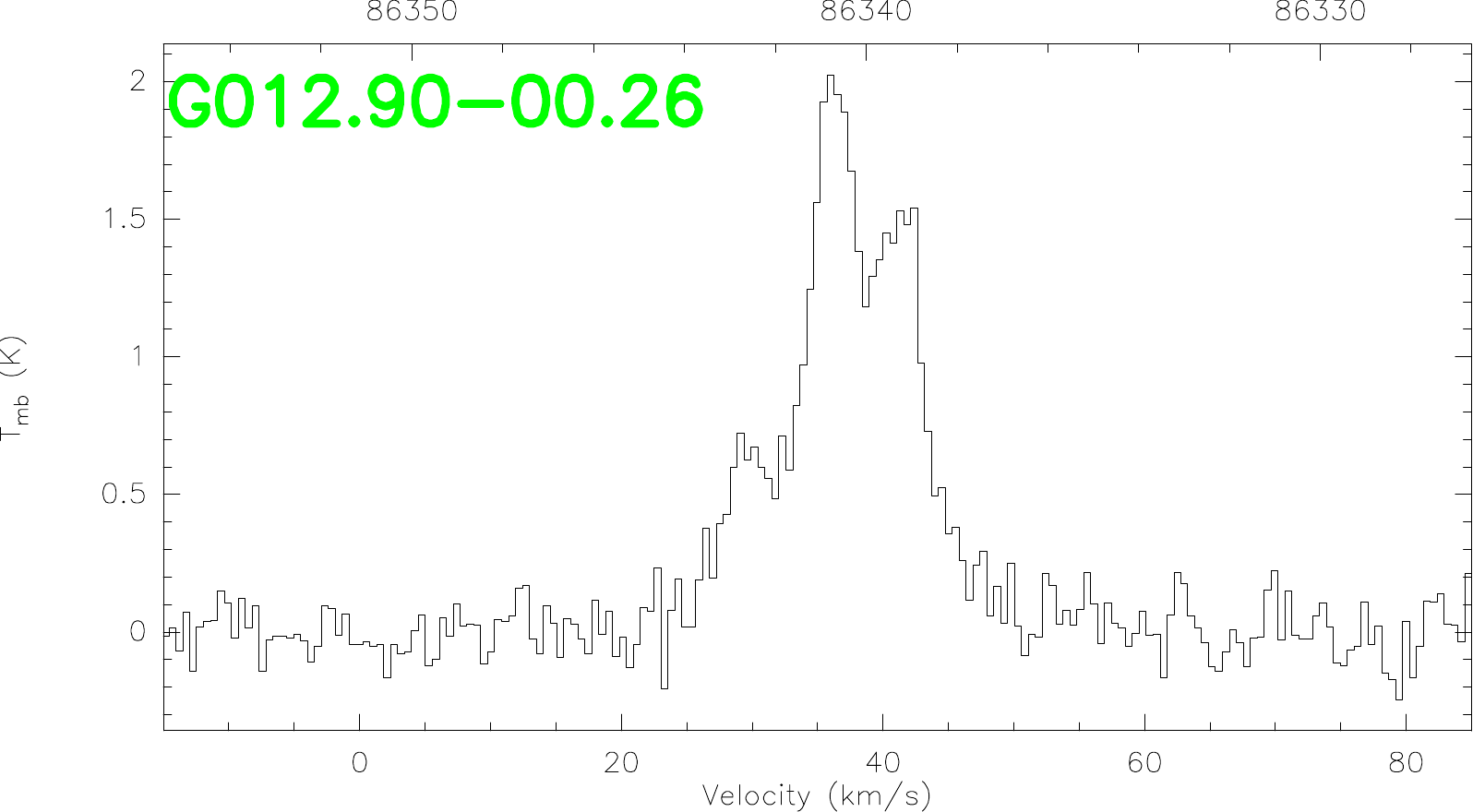}
\includegraphics[width=0.155\textwidth]{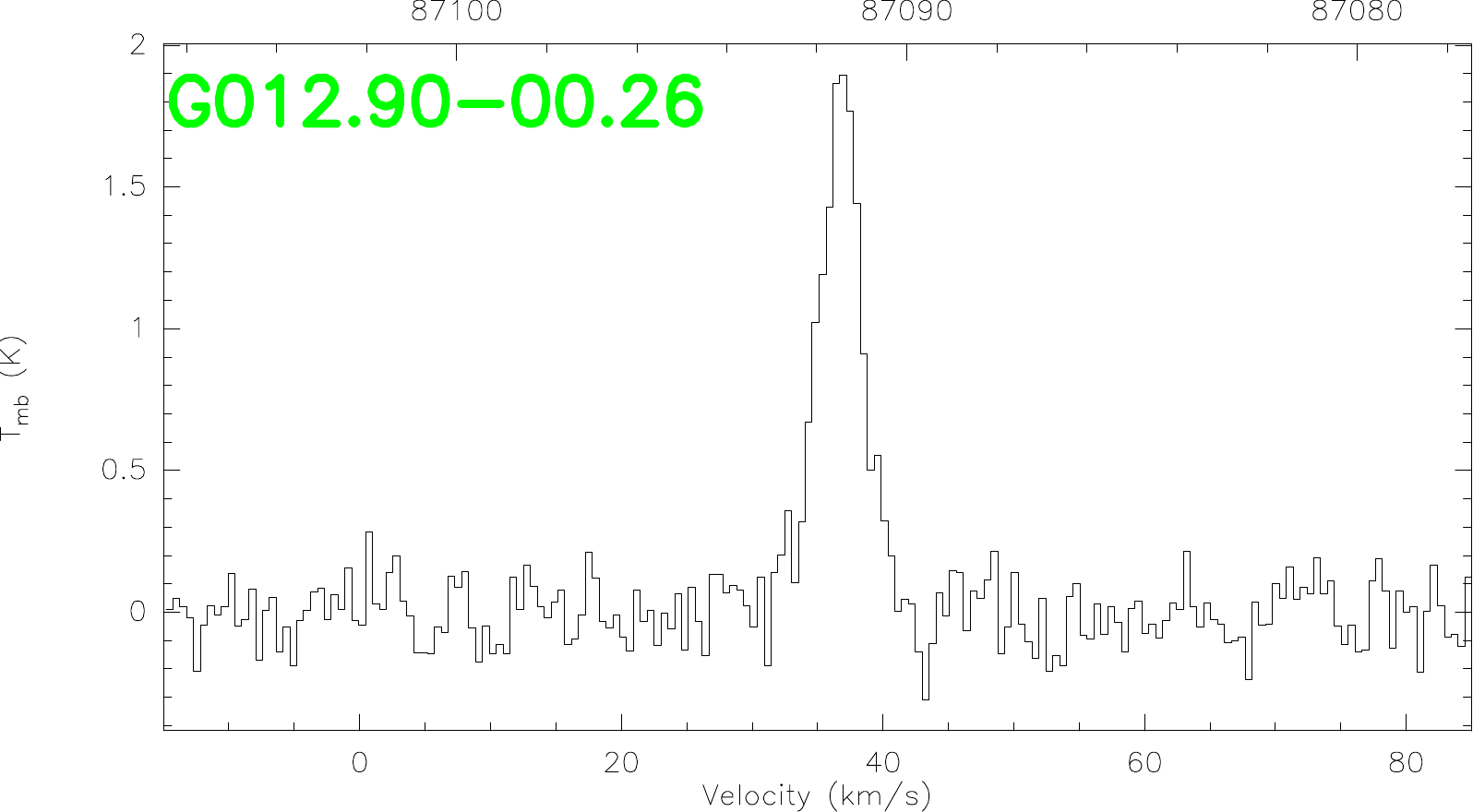}
\includegraphics[width=0.155\textwidth]{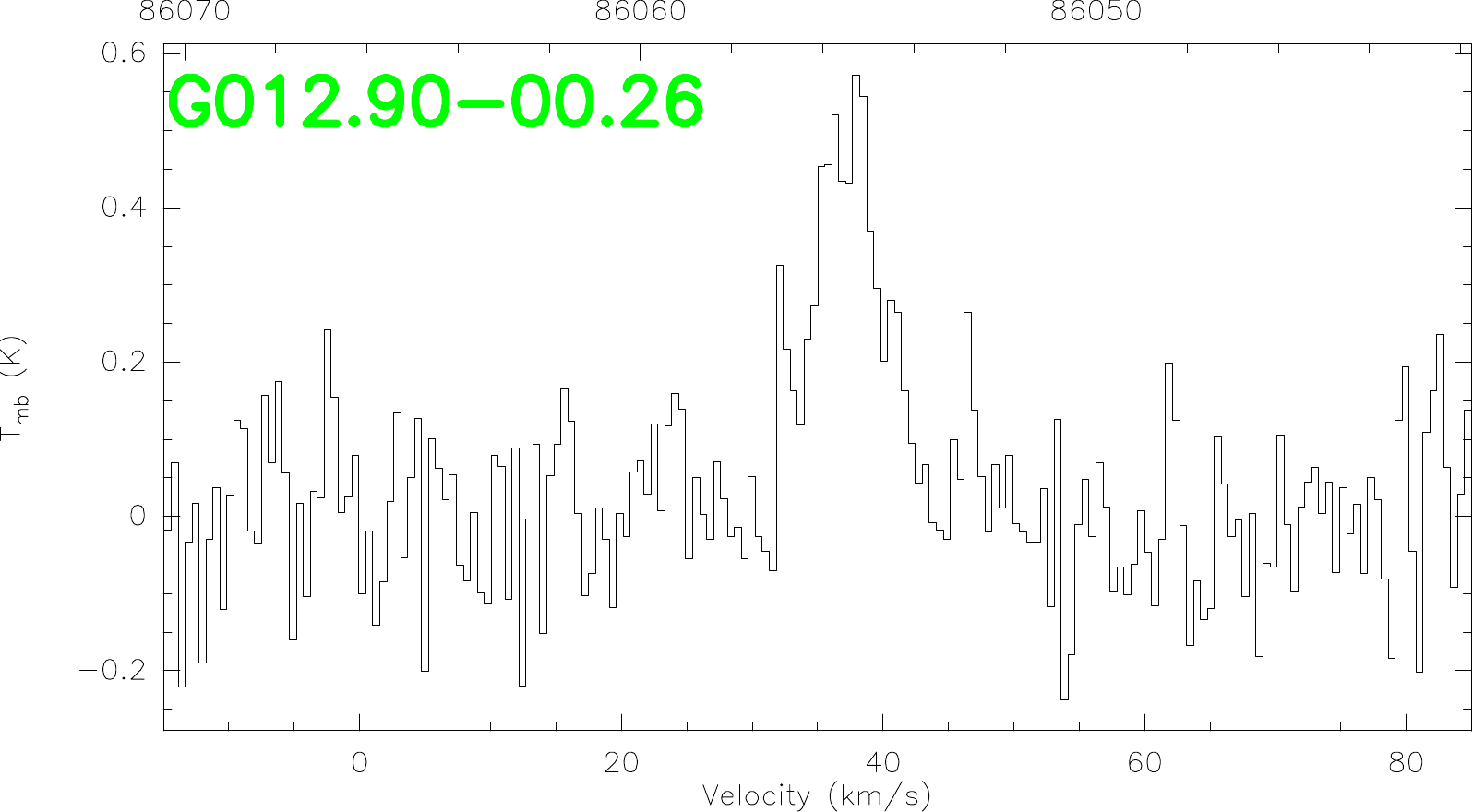}
\includegraphics[width=0.155\textwidth]{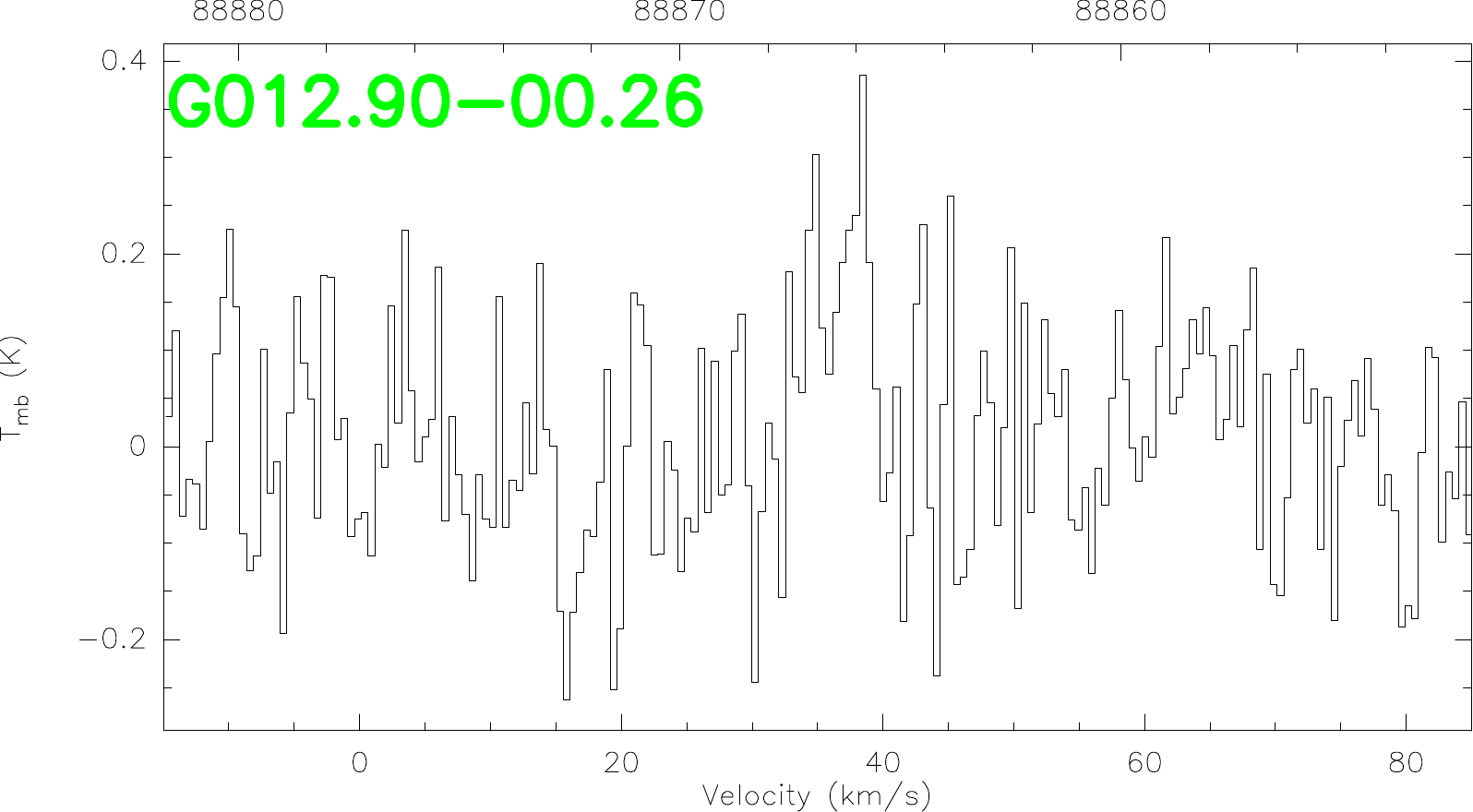}\\

\includegraphics[width=0.155\textwidth]{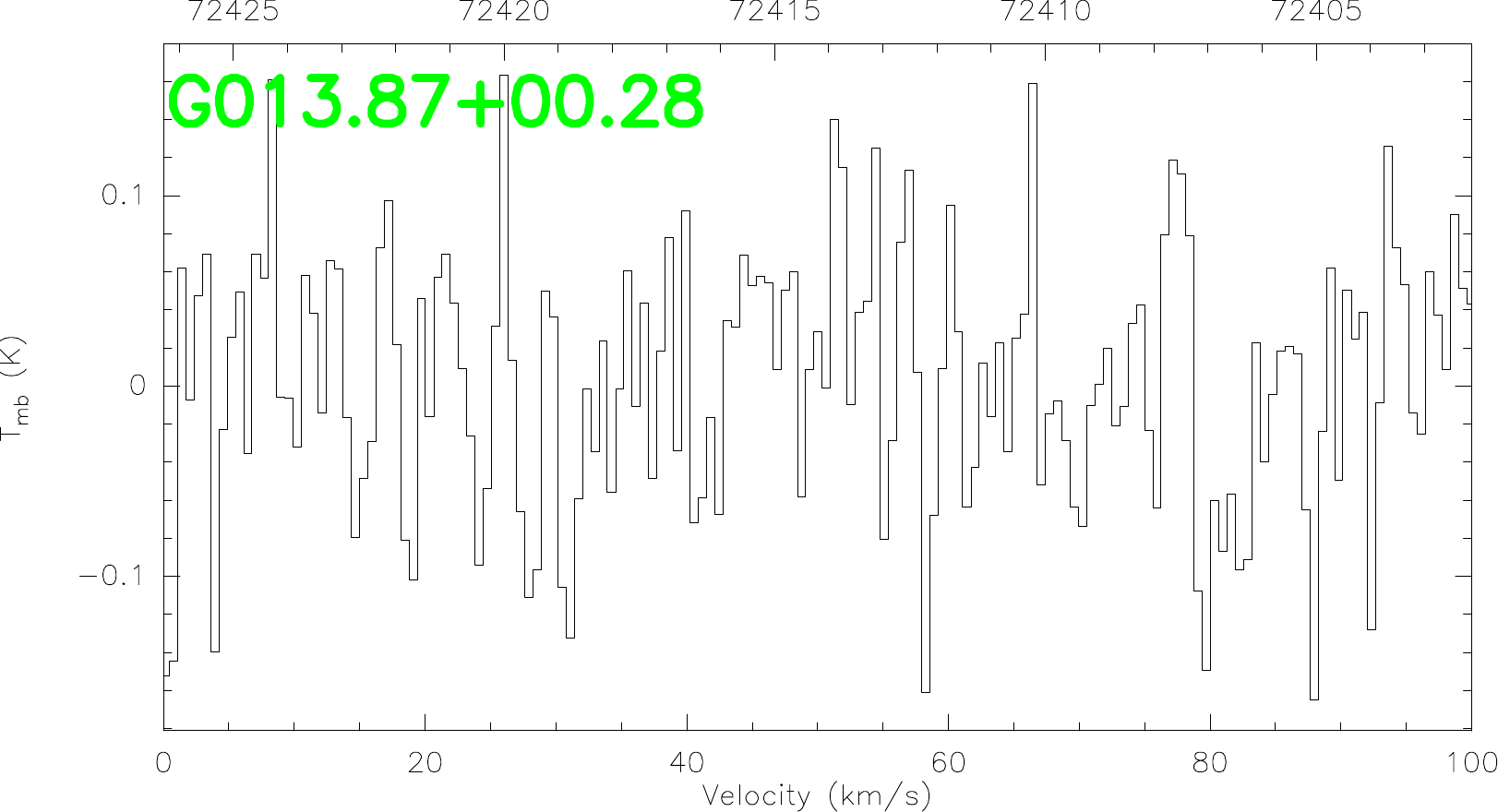}
\includegraphics[width=0.155\textwidth]{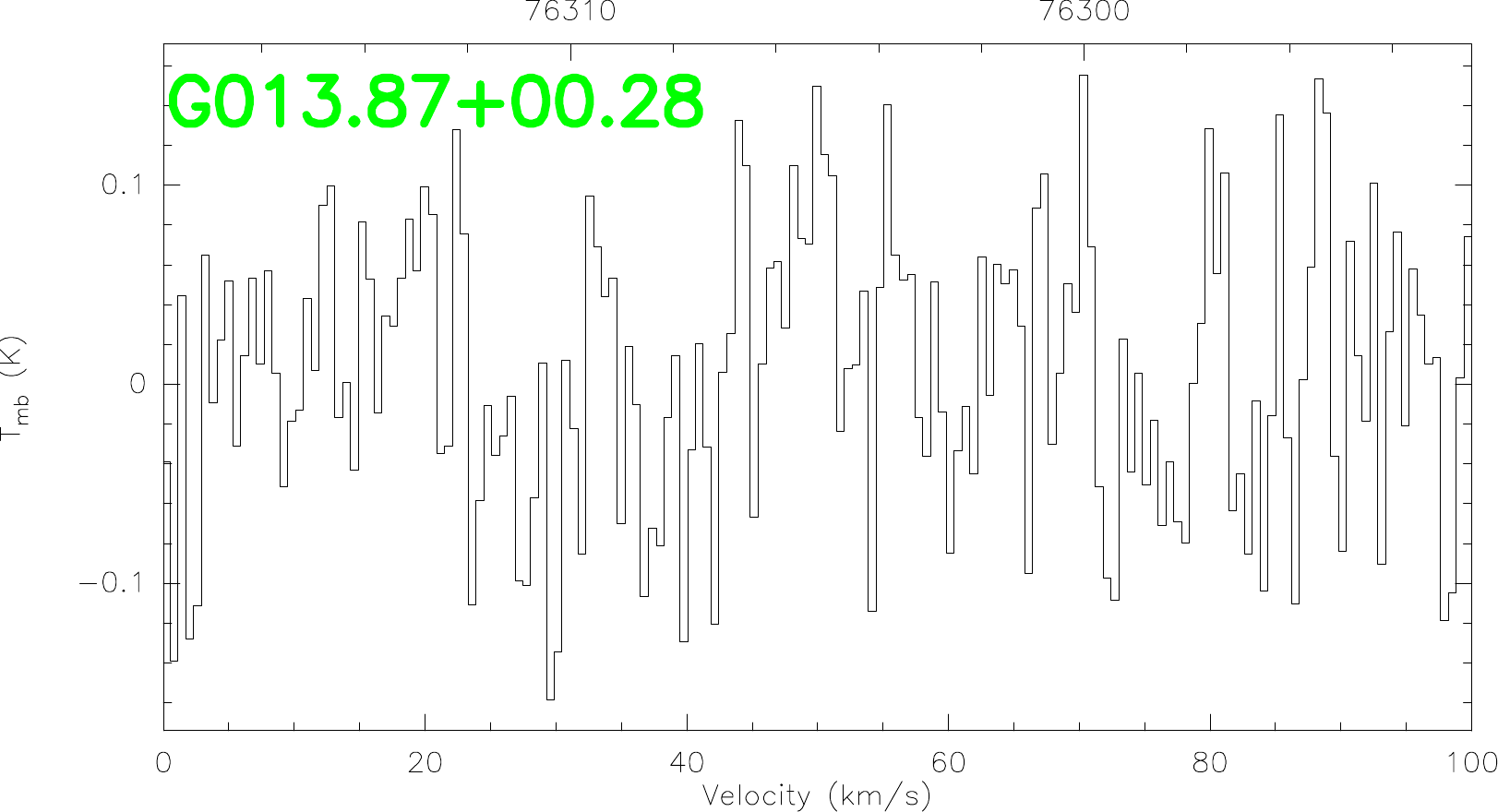}
\includegraphics[width=0.155\textwidth]{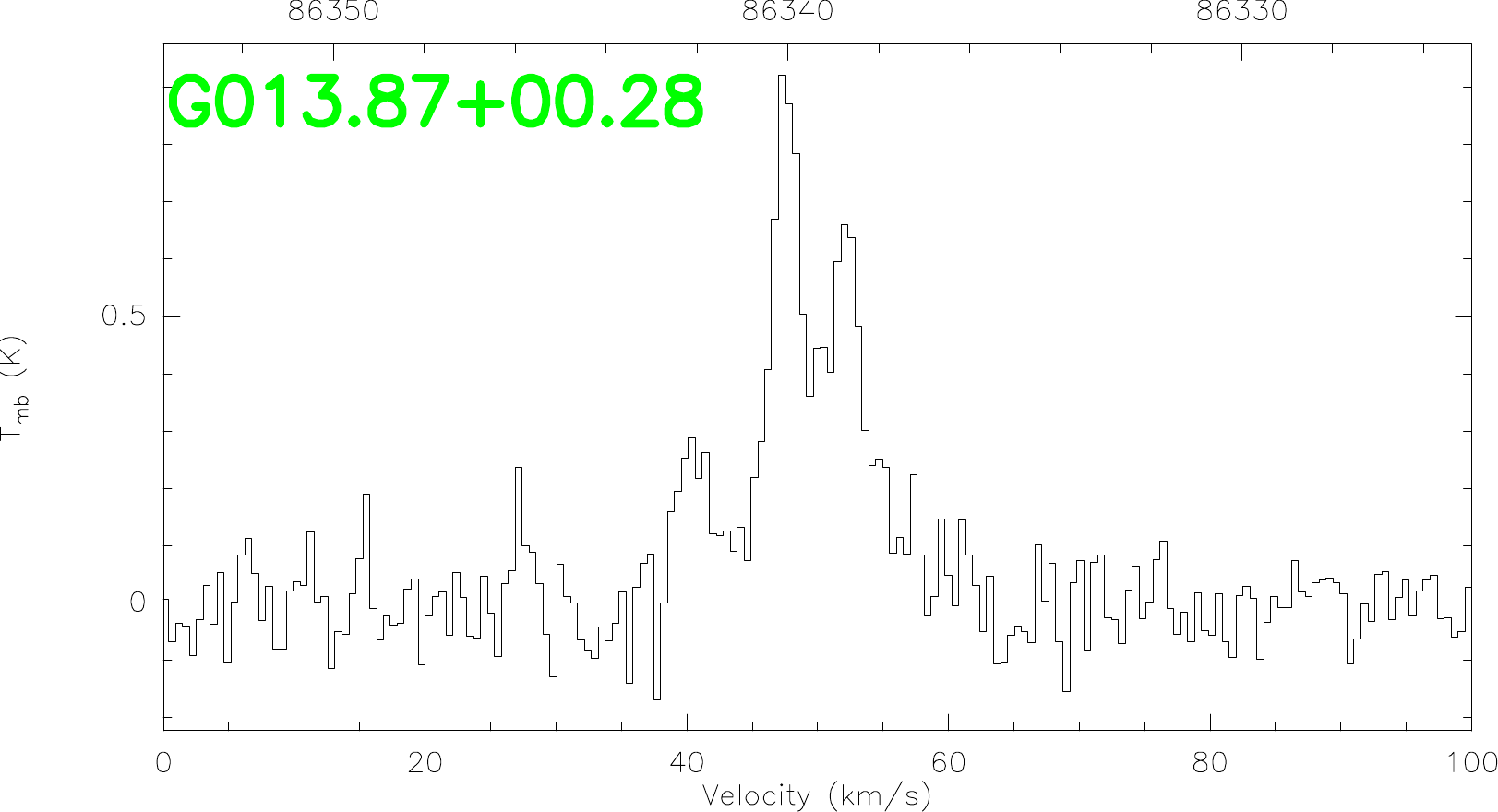}
\includegraphics[width=0.155\textwidth]{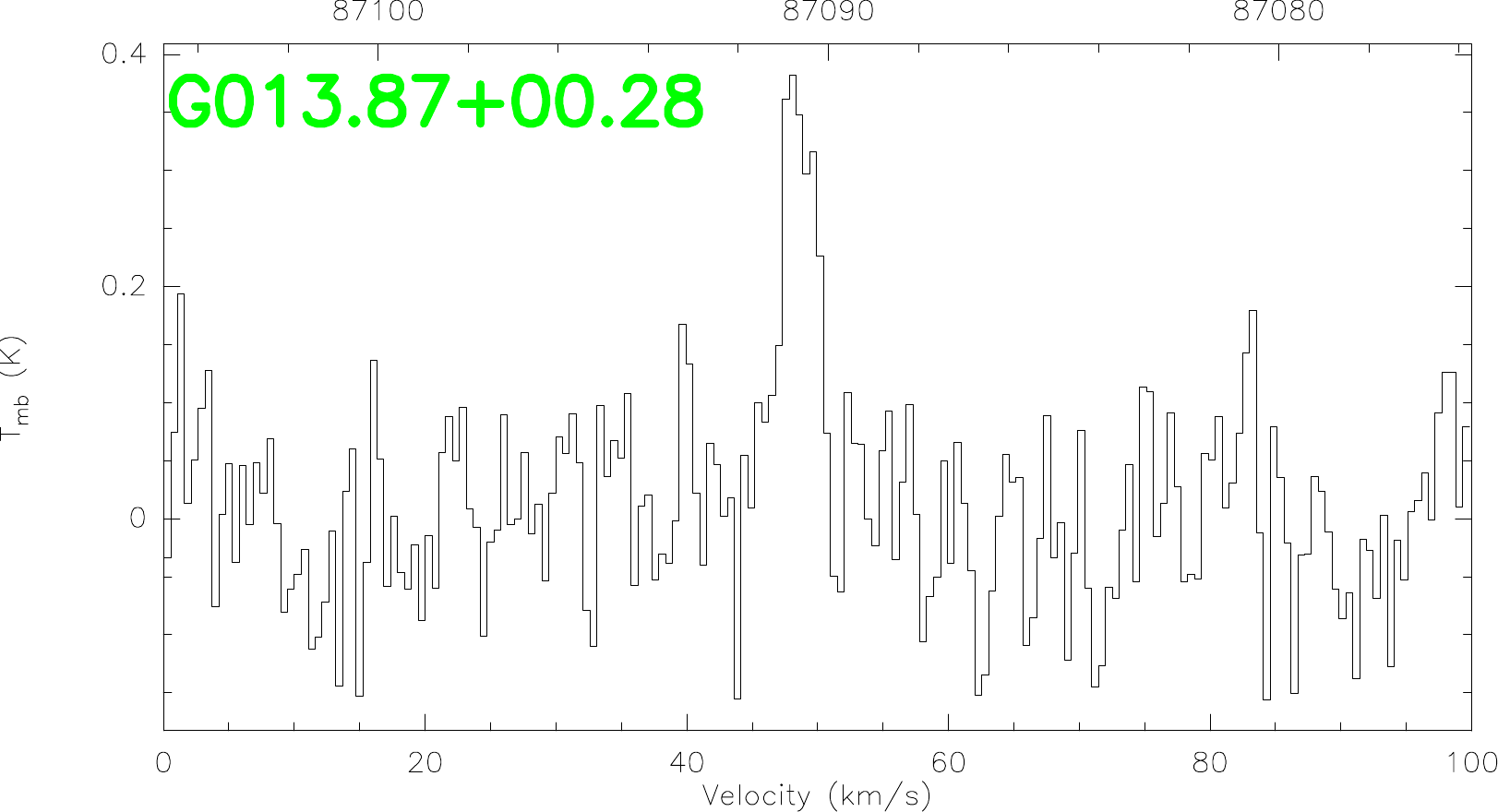}
\includegraphics[width=0.155\textwidth]{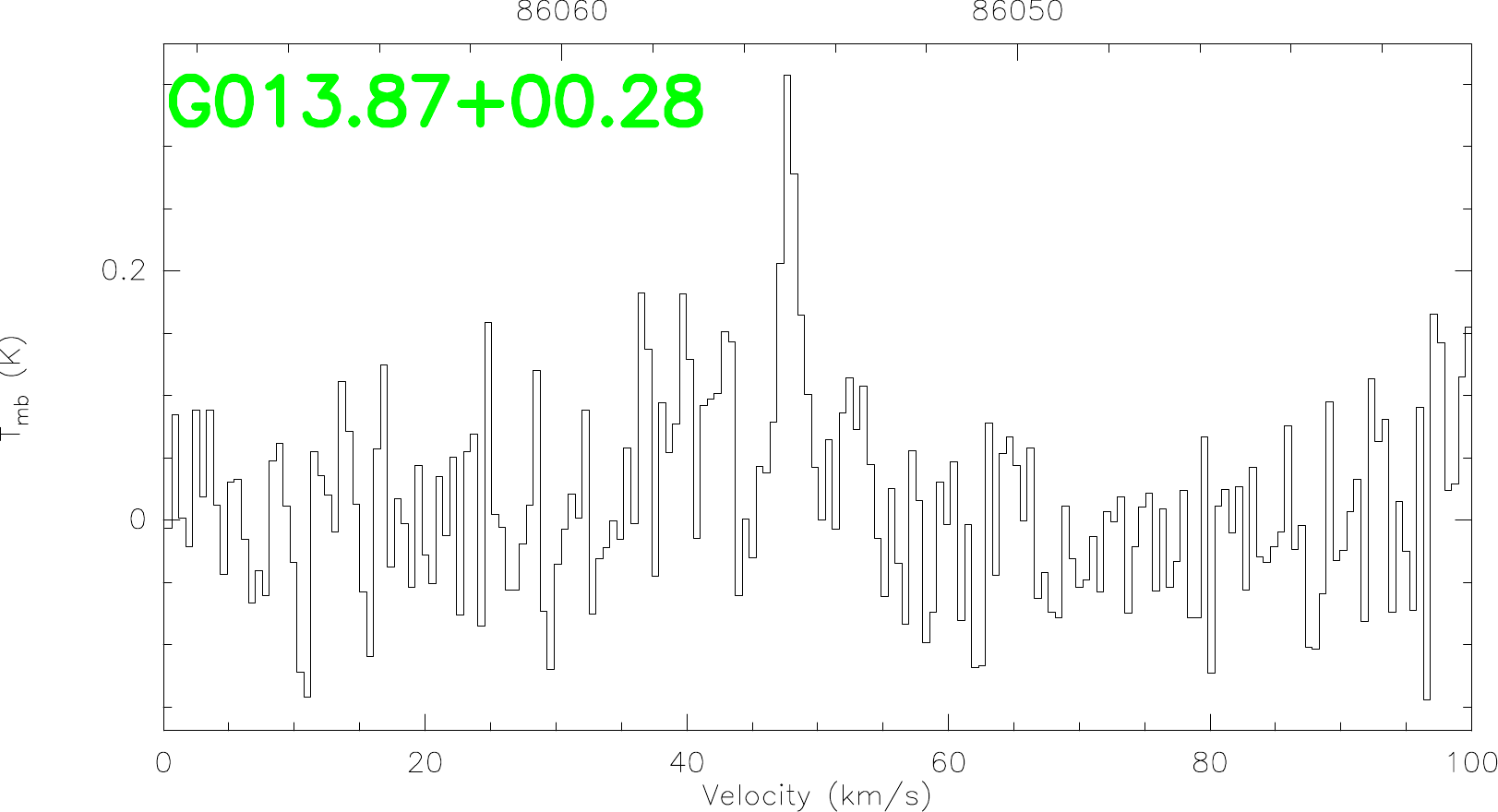}
\includegraphics[width=0.155\textwidth]{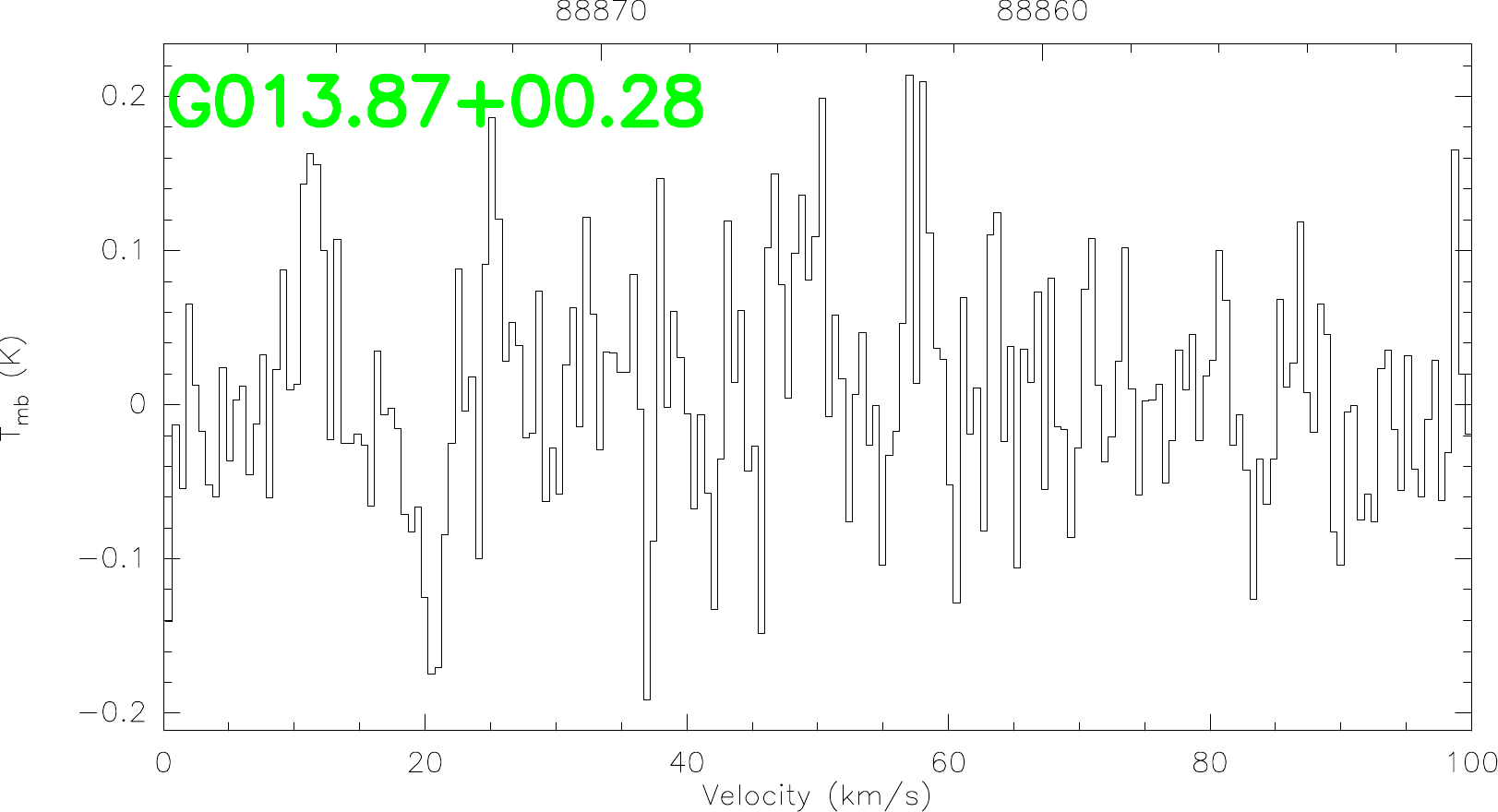}\\

\includegraphics[width=0.155\textwidth]{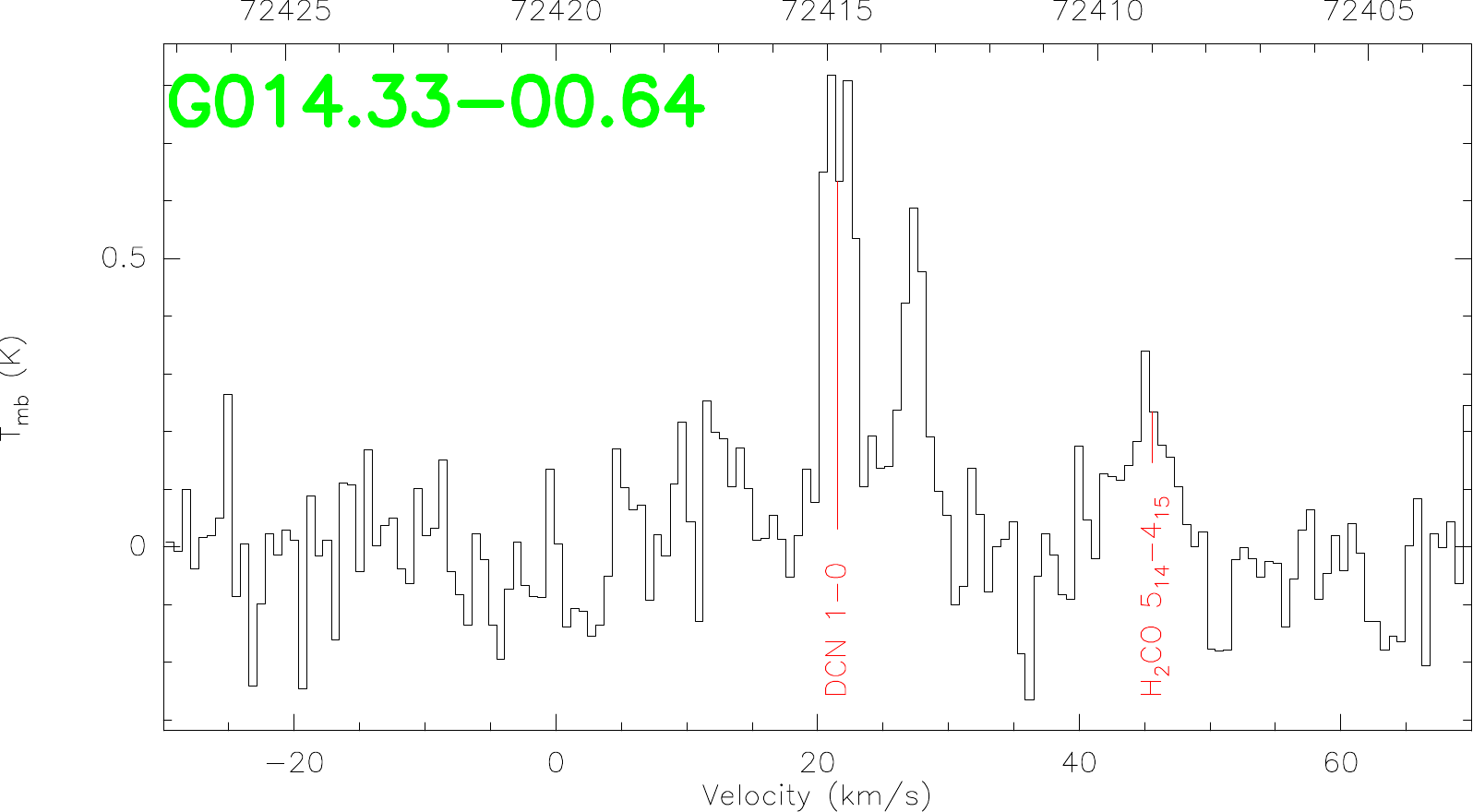}
\includegraphics[width=0.155\textwidth]{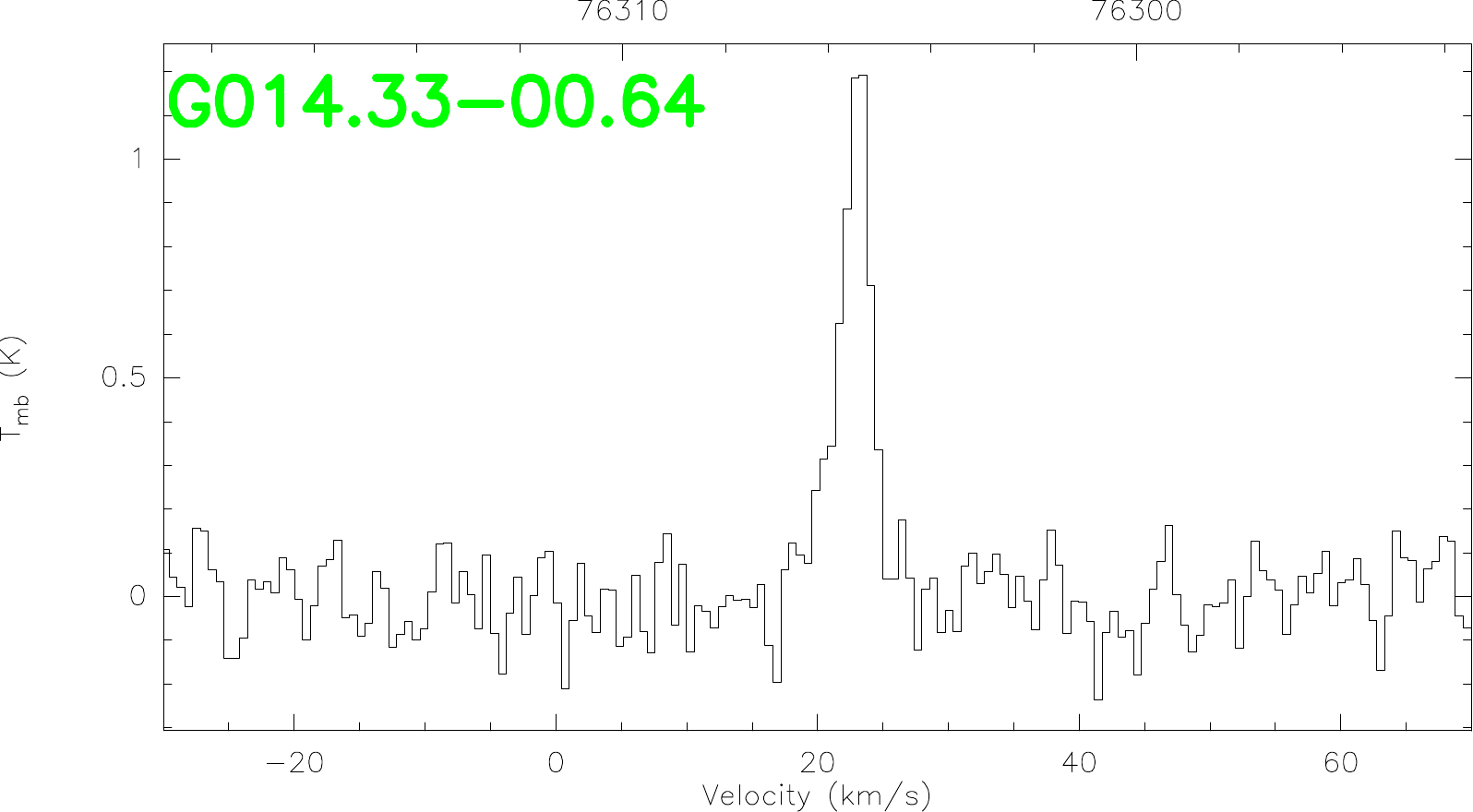}
\includegraphics[width=0.155\textwidth]{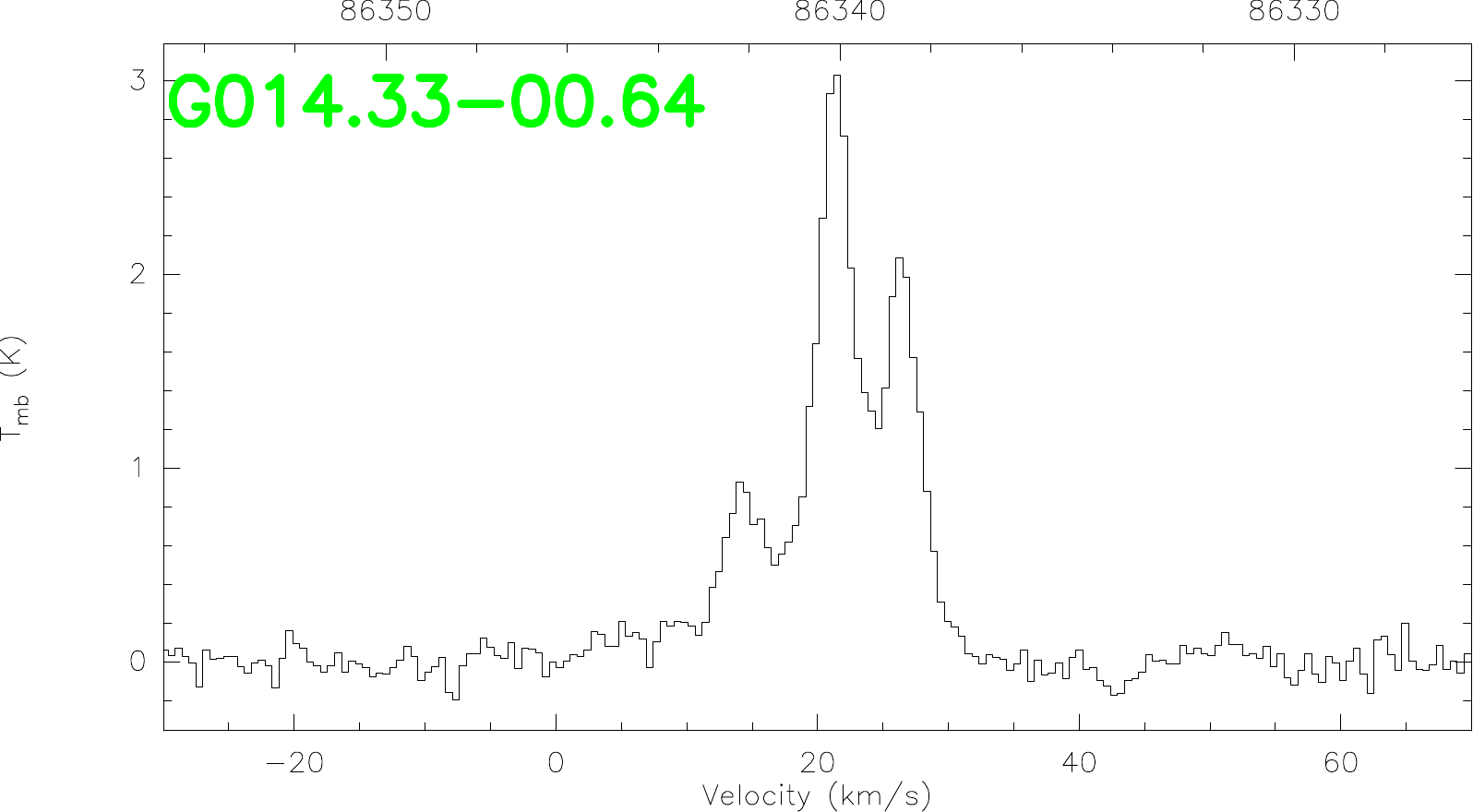}
\includegraphics[width=0.155\textwidth]{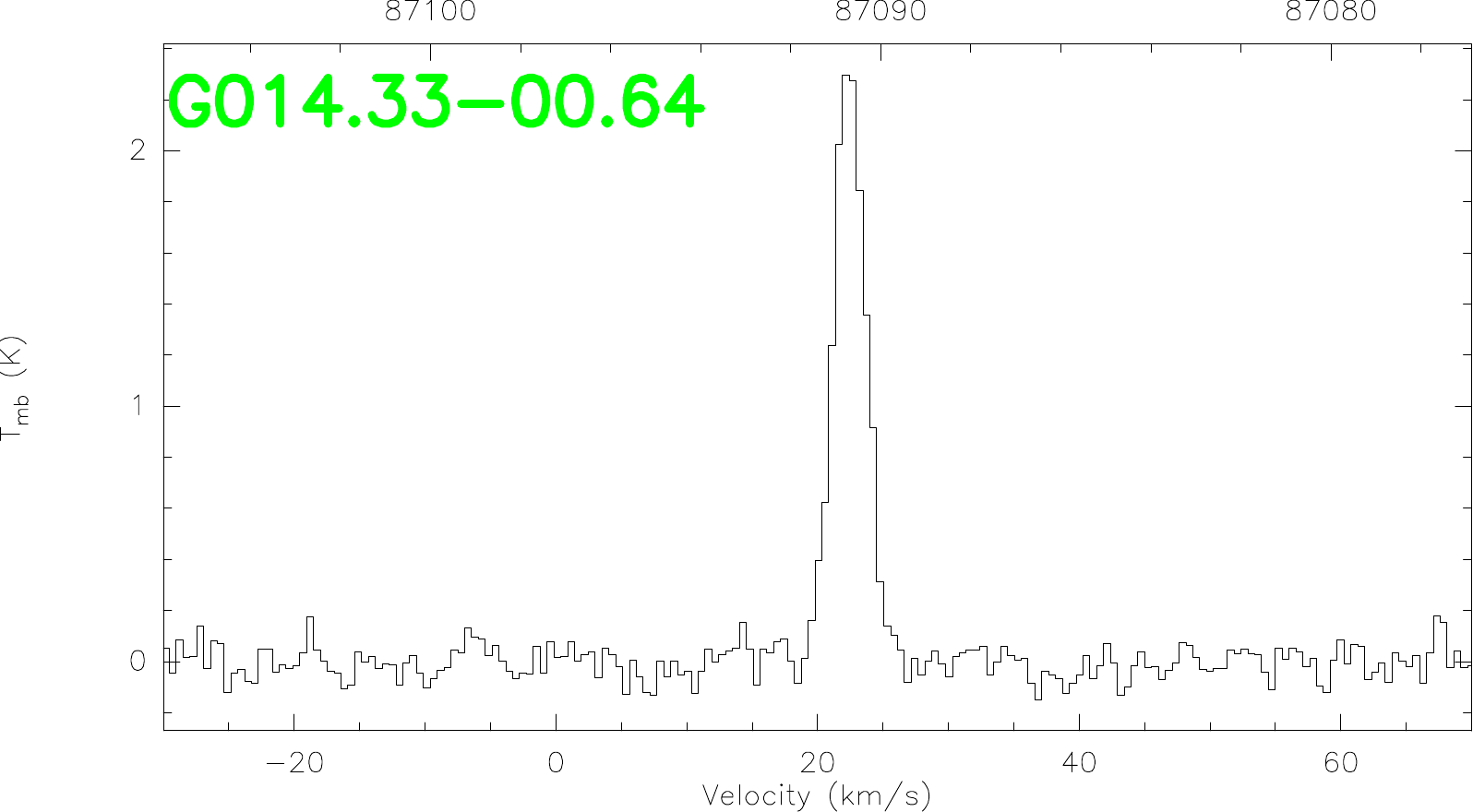}
\includegraphics[width=0.155\textwidth]{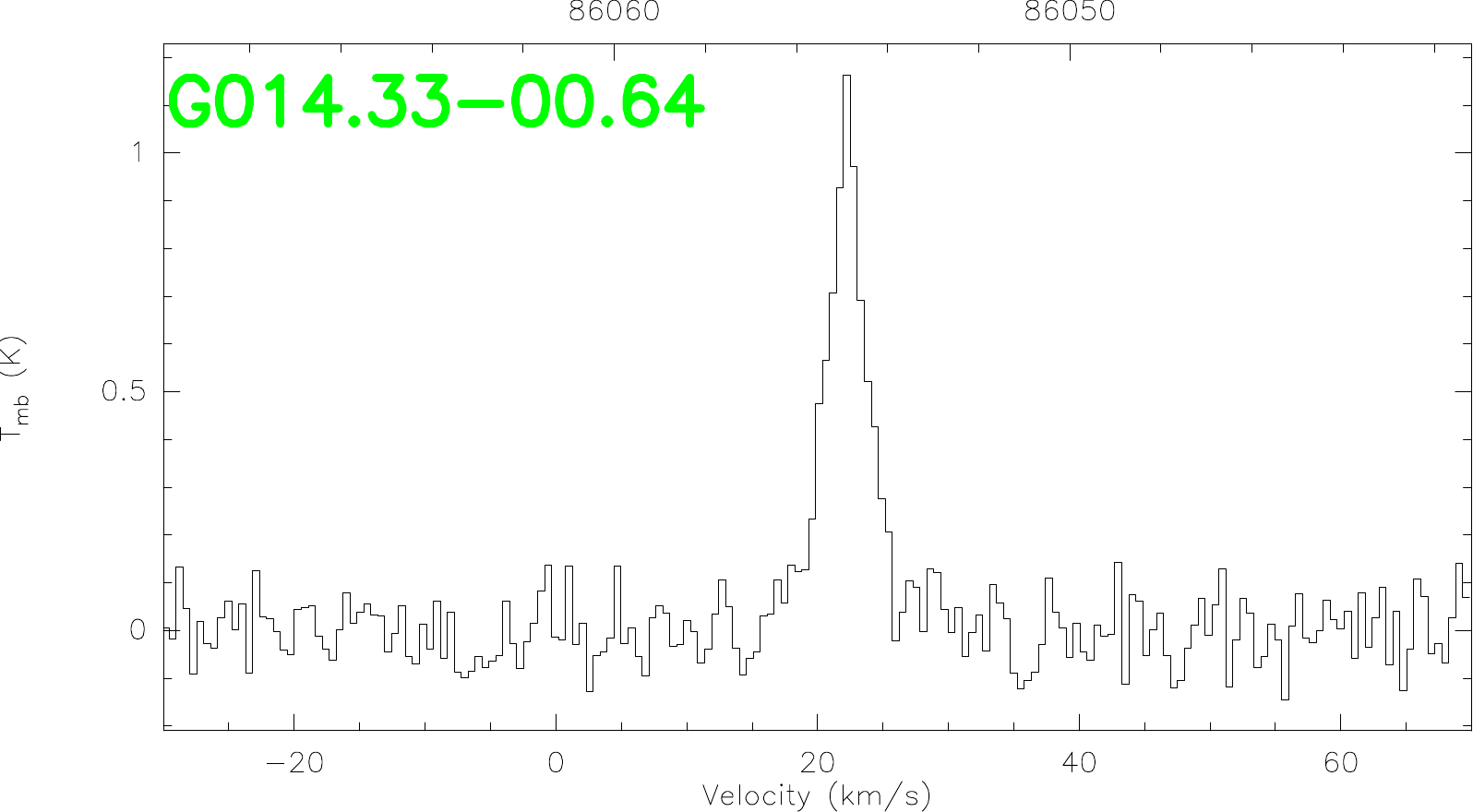}
\includegraphics[width=0.155\textwidth]{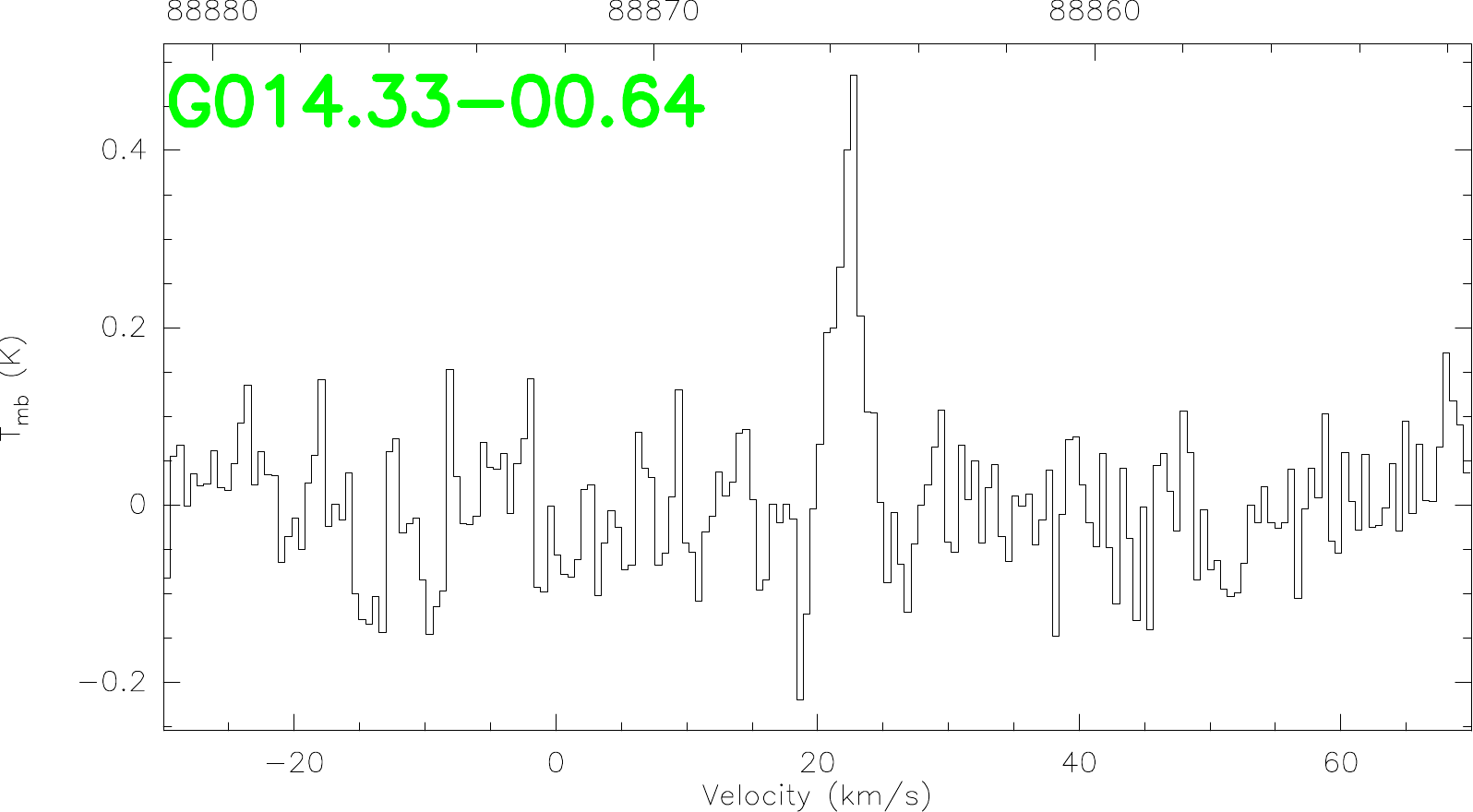}\\

\caption{Spectra of different molecular lines. From left to right: 1-0 transition of DCN, DNC, H$^{13}$CN, HN$^{13}$C, HC$^{15}$N and H$^{15}$NC.}
\label{fig:multi_lines_page1}
\end{figure}

\begin{figure}[h]
\centering

\makebox[\colw][c]{DCN 1-0}
\makebox[\colw][c]{DNC 1-0}
\makebox[\colw][c]{H$^{13}$CN 1-0}
\makebox[\colw][c]{HN$^{13}$C 1-0}
\makebox[\colw][c]{HC$^{15}$N 1-0}
\makebox[\colw][c]{H$^{15}$NC 1-0}\\[0.2em]

\includegraphics[width=0.155\textwidth]{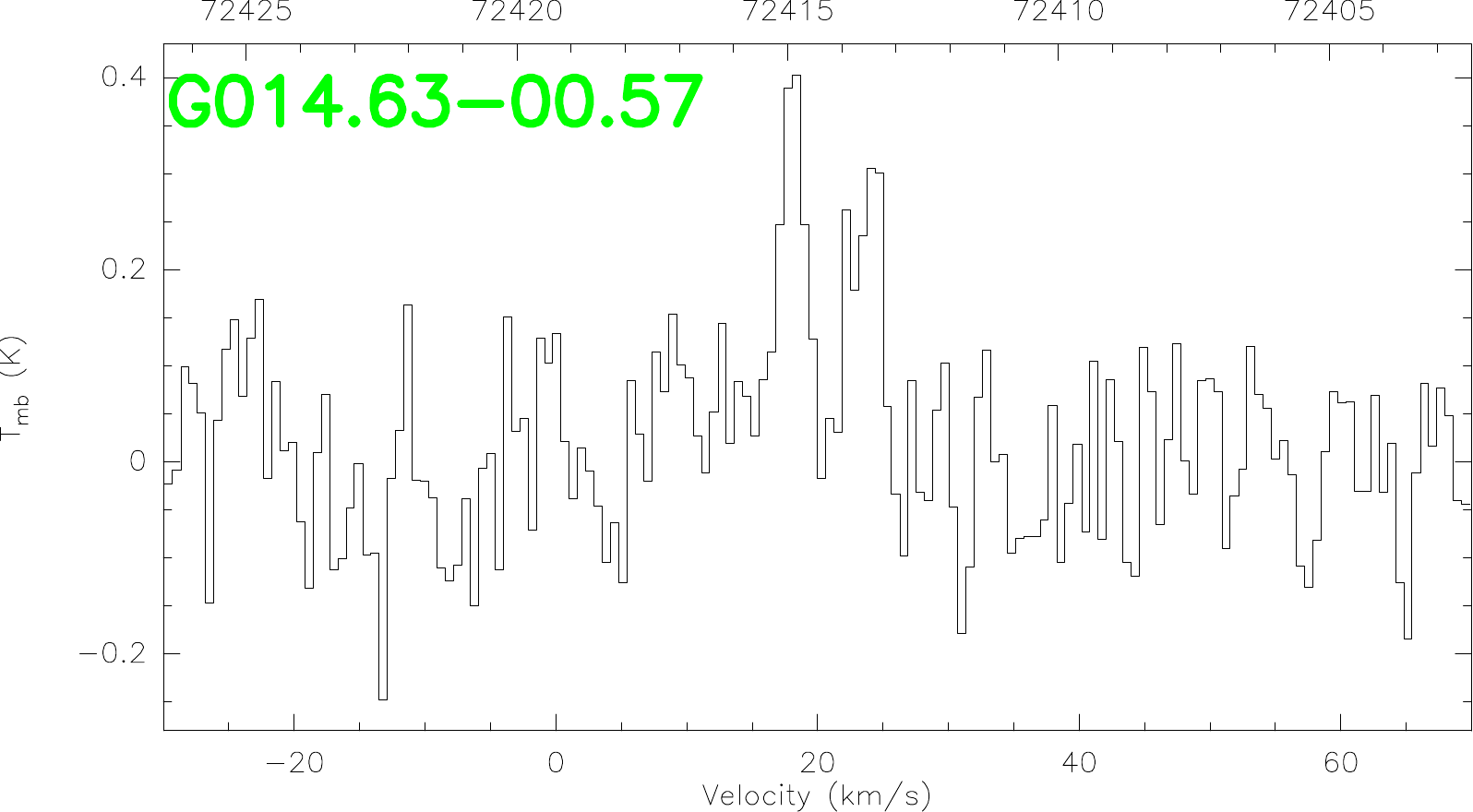}
\includegraphics[width=0.155\textwidth]{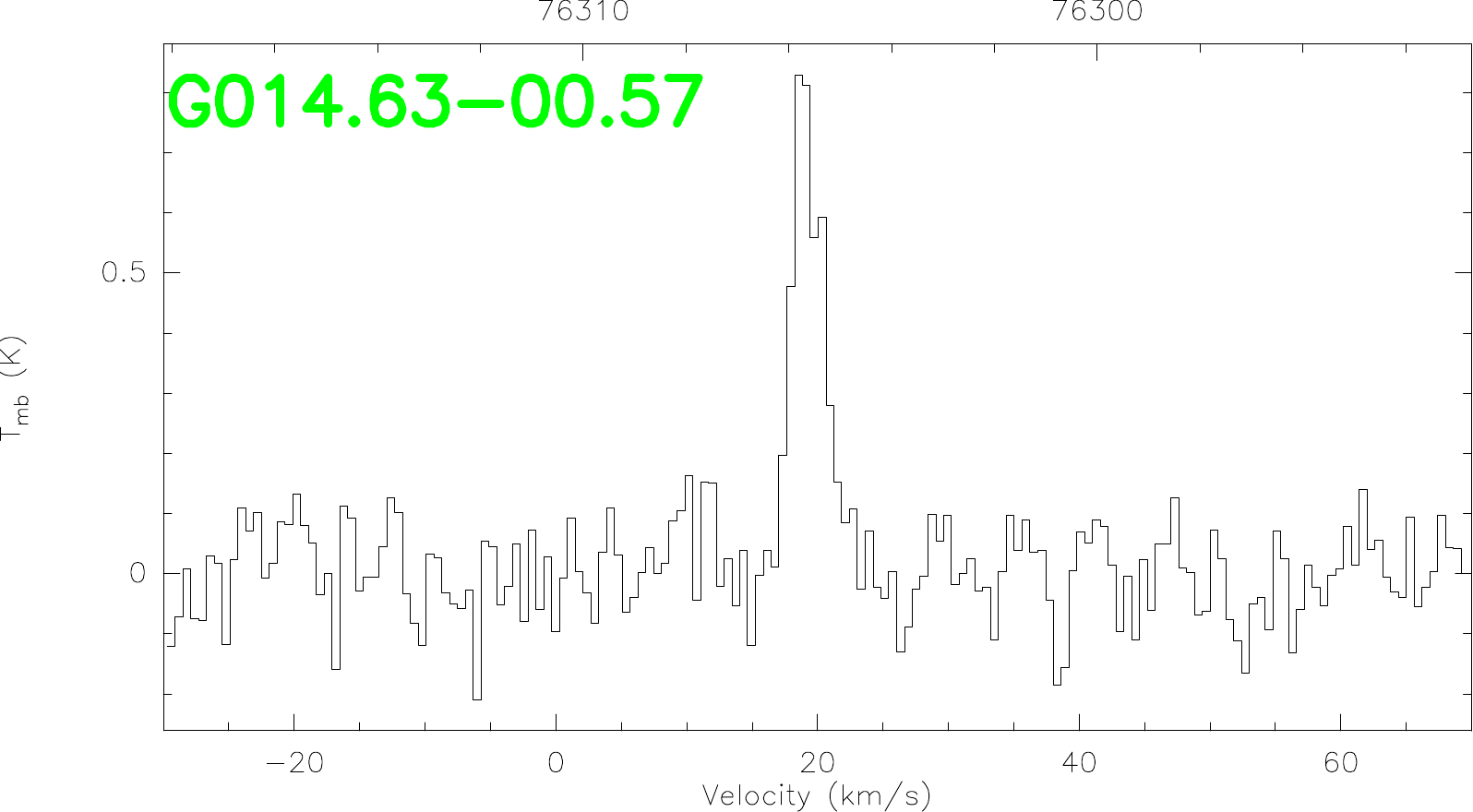}
\includegraphics[width=0.155\textwidth]{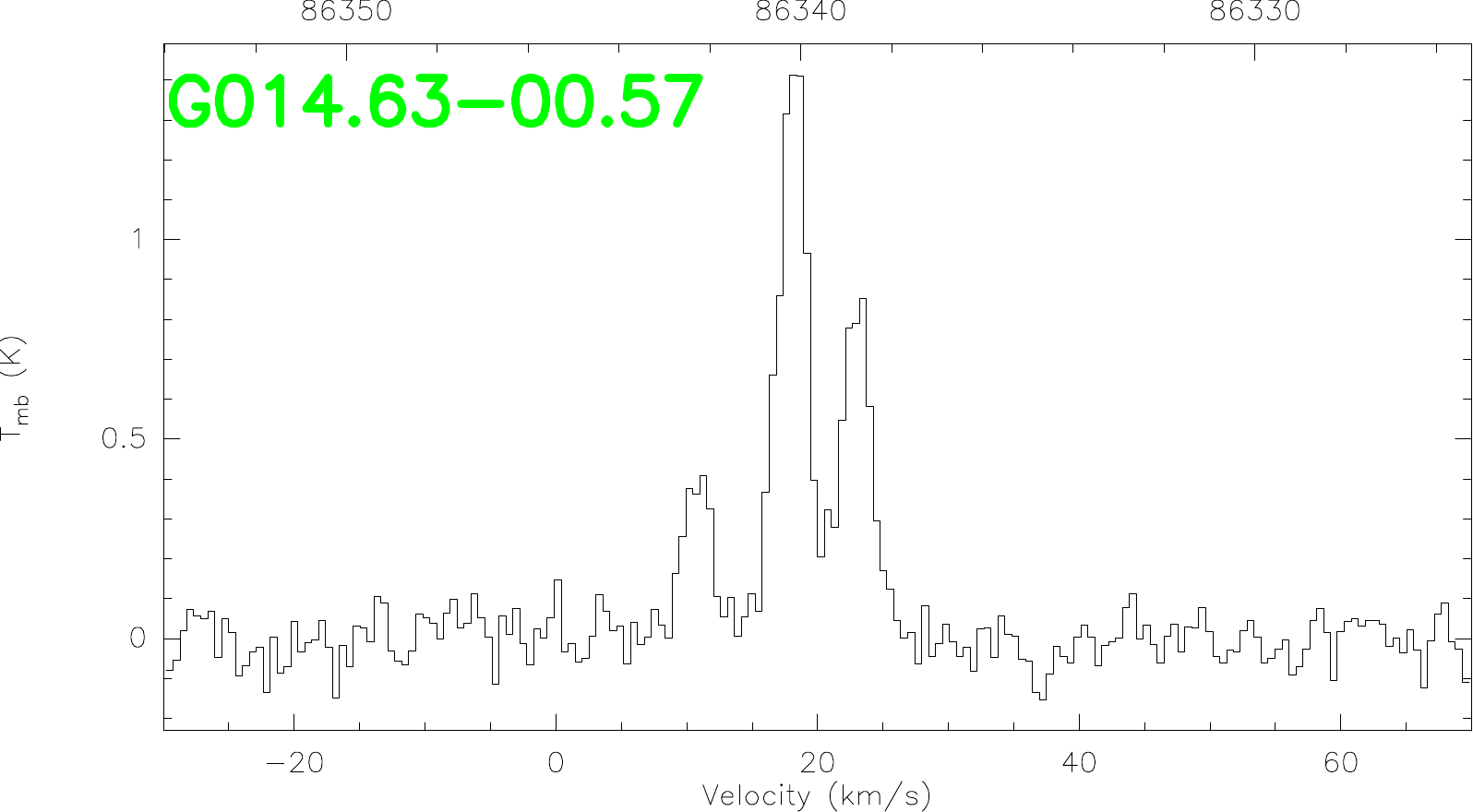}
\includegraphics[width=0.155\textwidth]{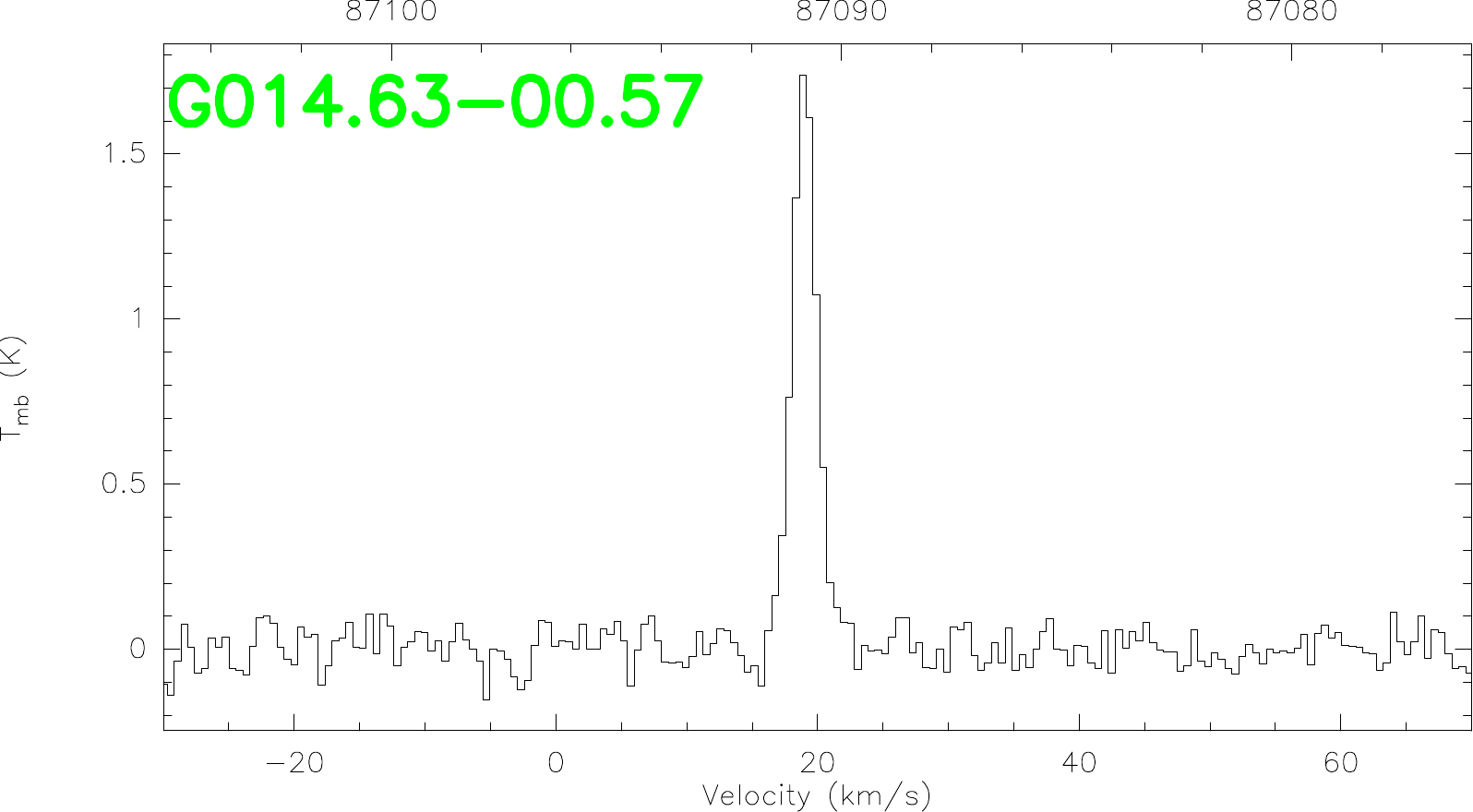}
\includegraphics[width=0.155\textwidth]{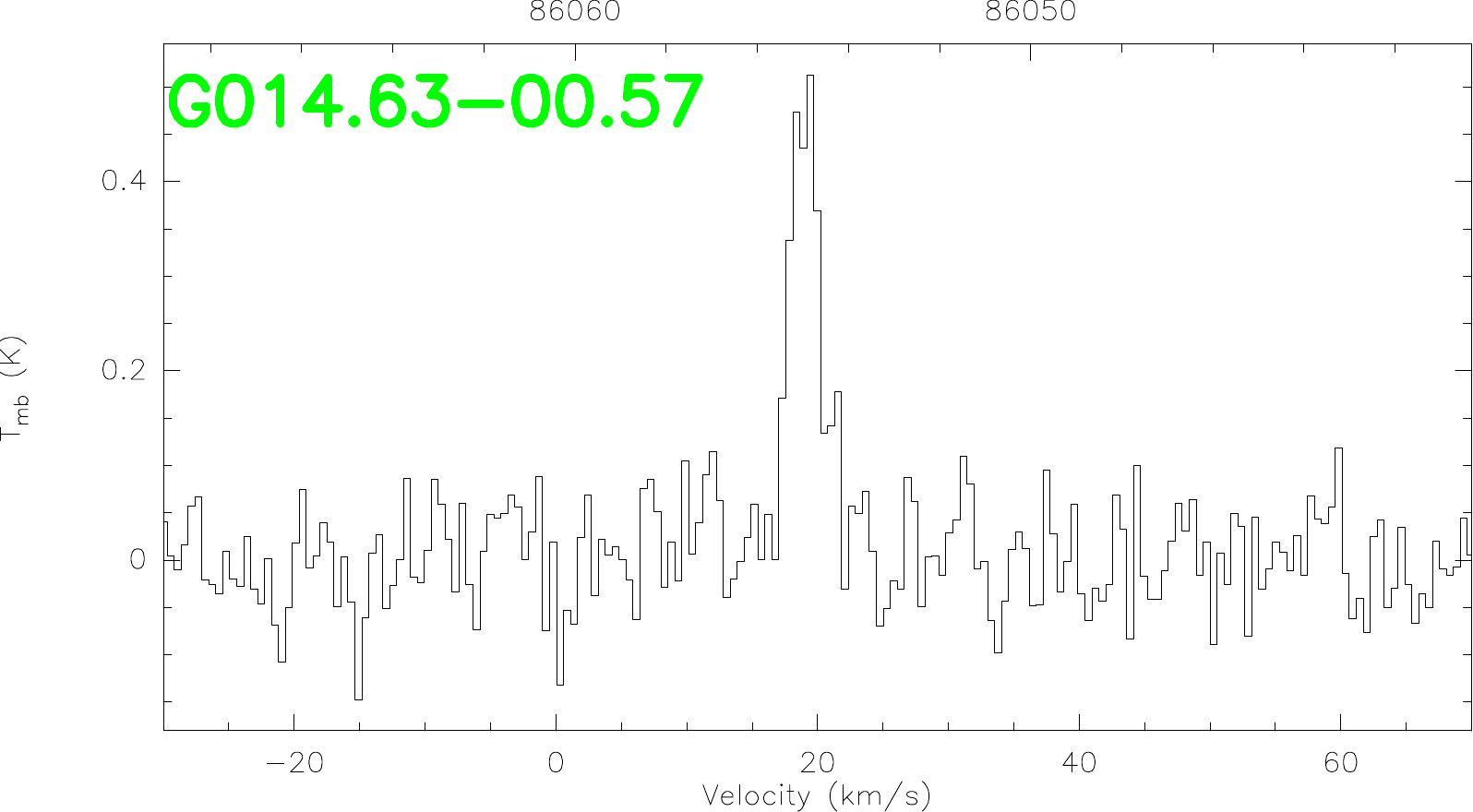}
\includegraphics[width=0.155\textwidth]{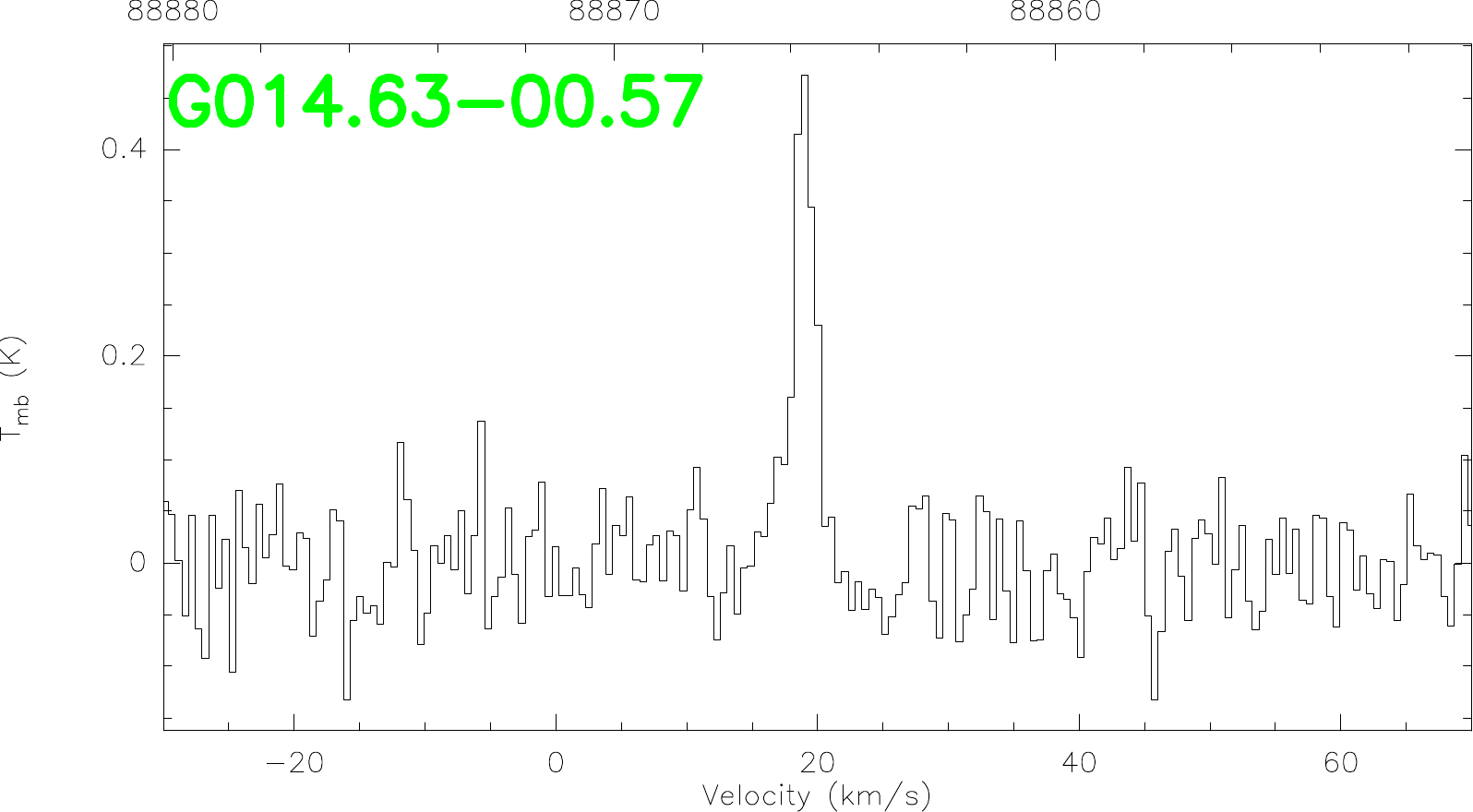}\\

\includegraphics[width=0.155\textwidth]{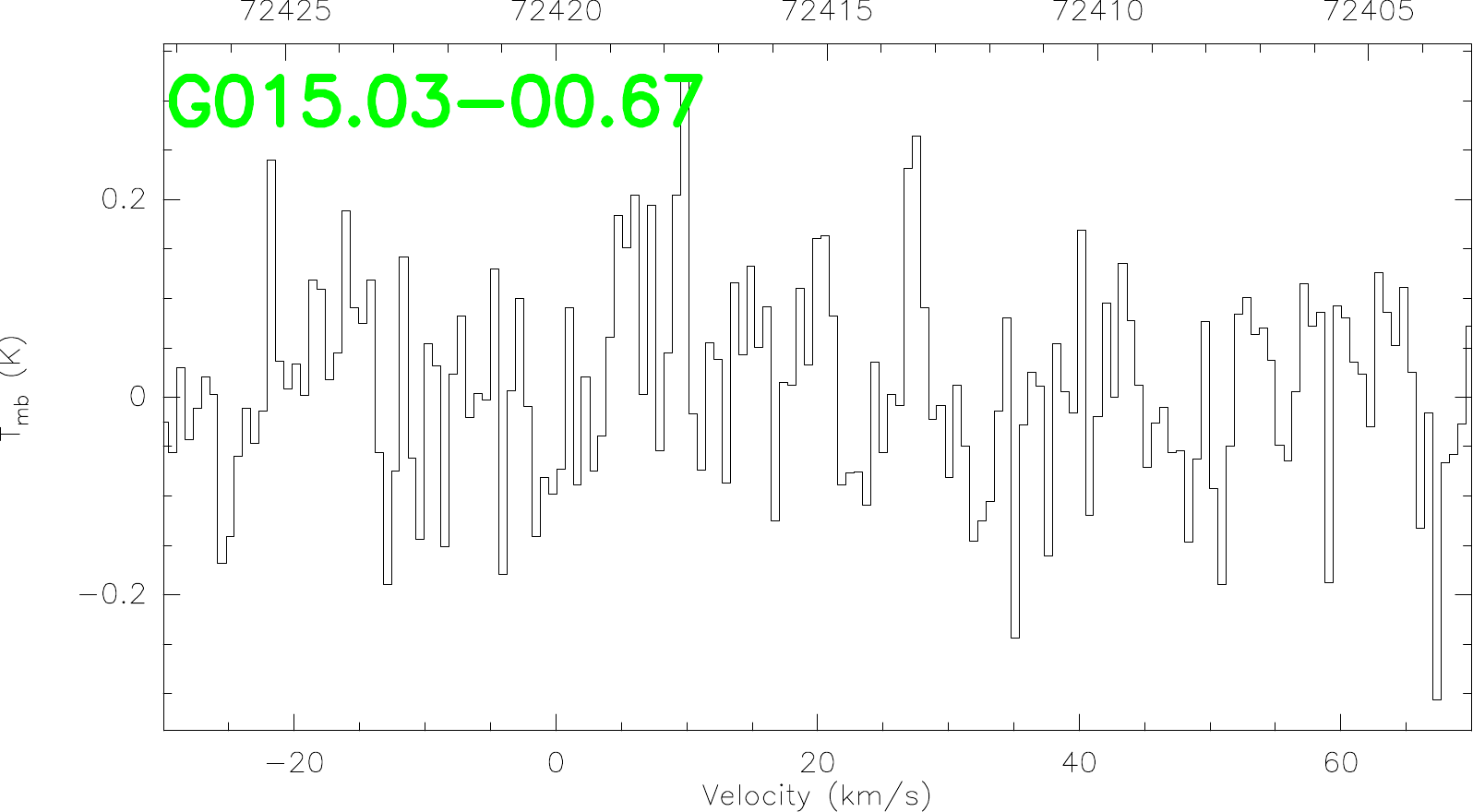}
\includegraphics[width=0.155\textwidth]{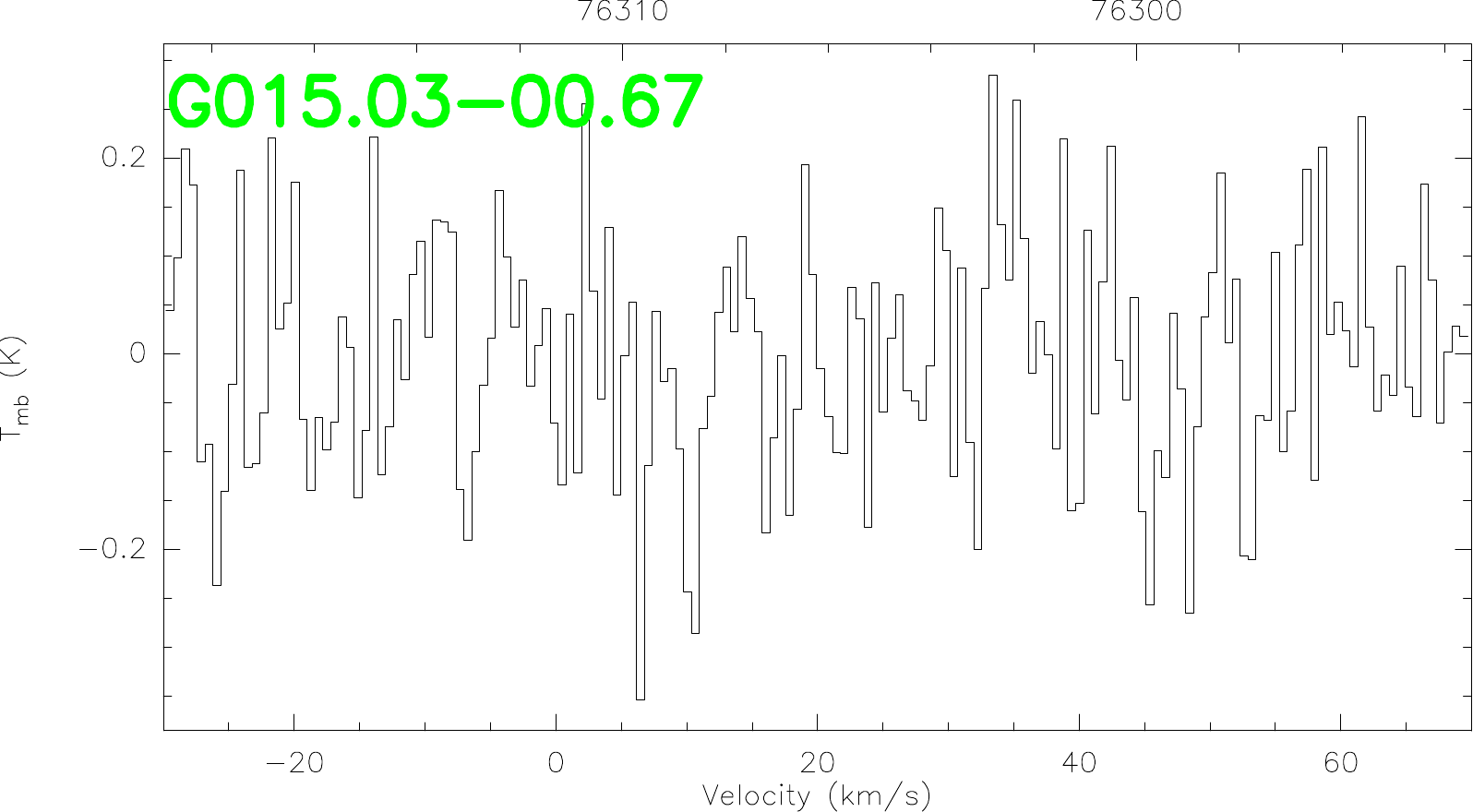}
\includegraphics[width=0.155\textwidth]{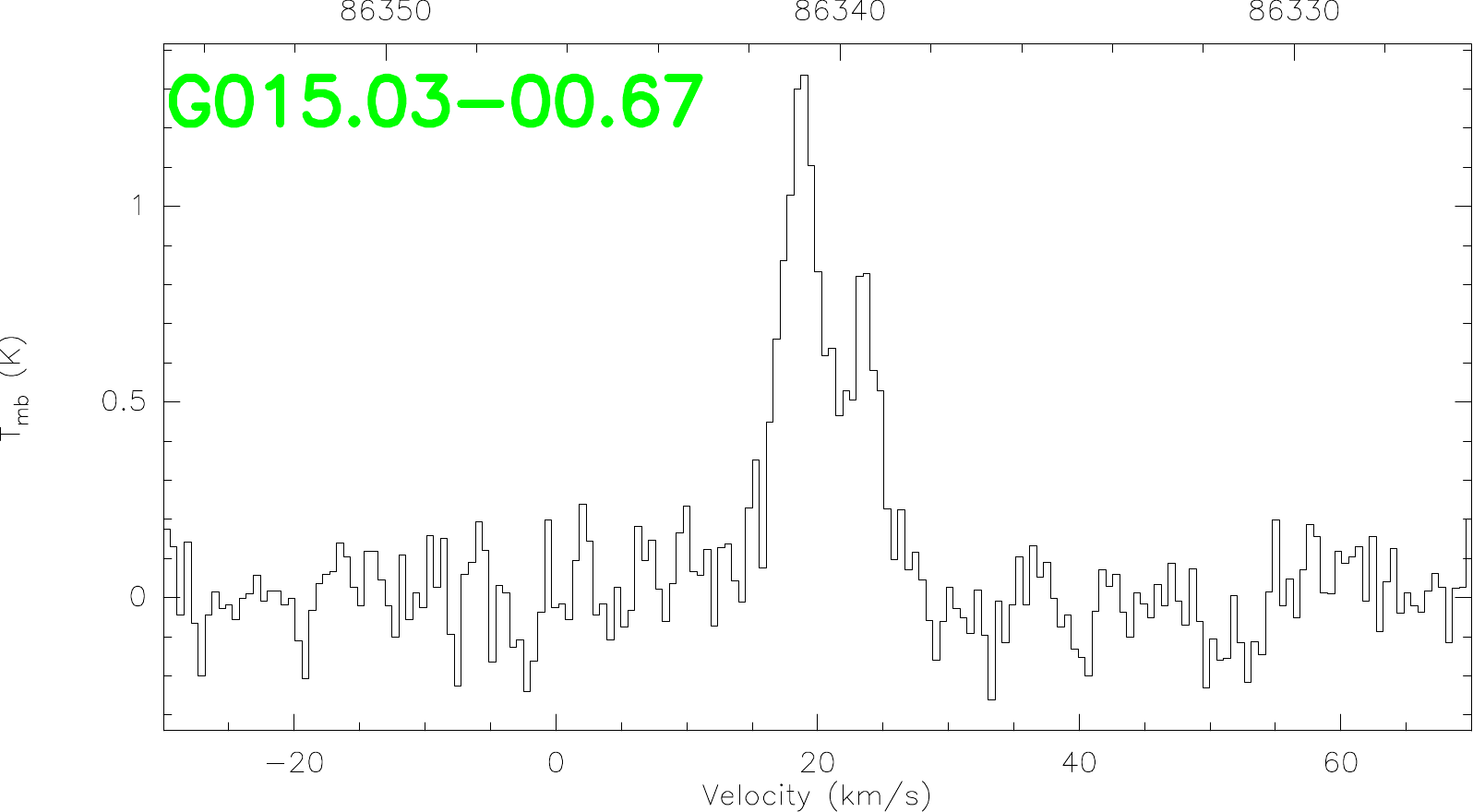}
\includegraphics[width=0.155\textwidth]{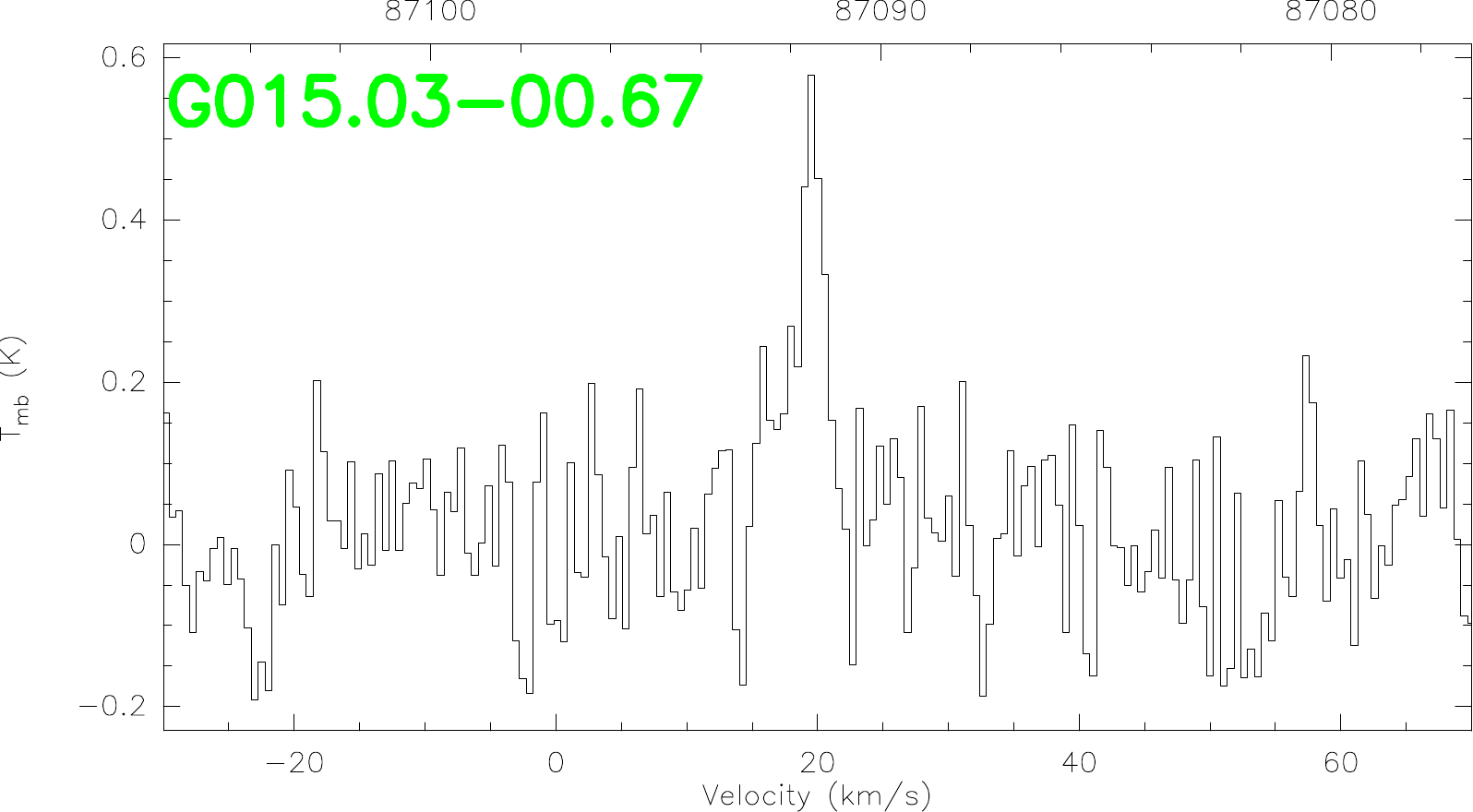}
\includegraphics[width=0.155\textwidth]{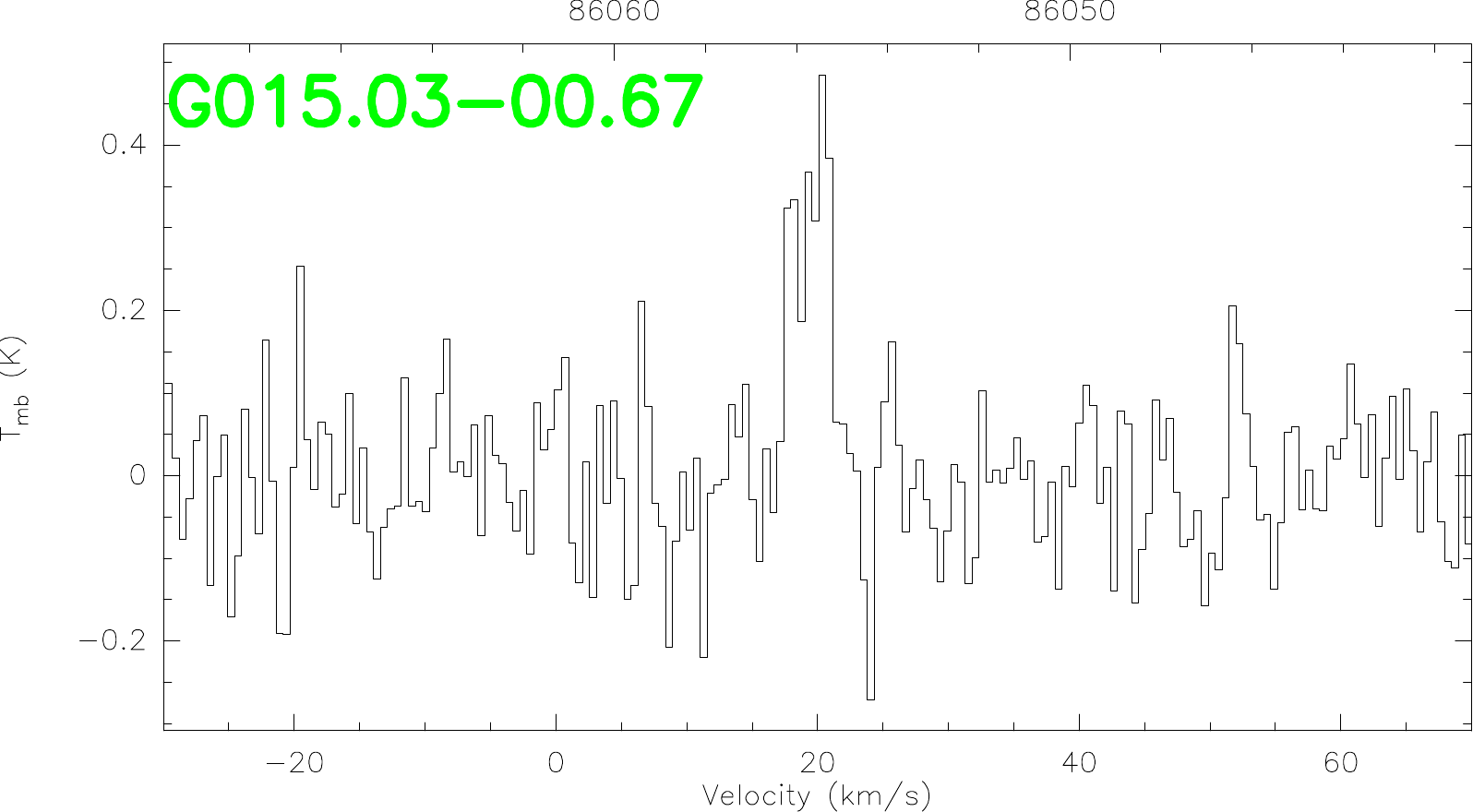}
\includegraphics[width=0.155\textwidth]{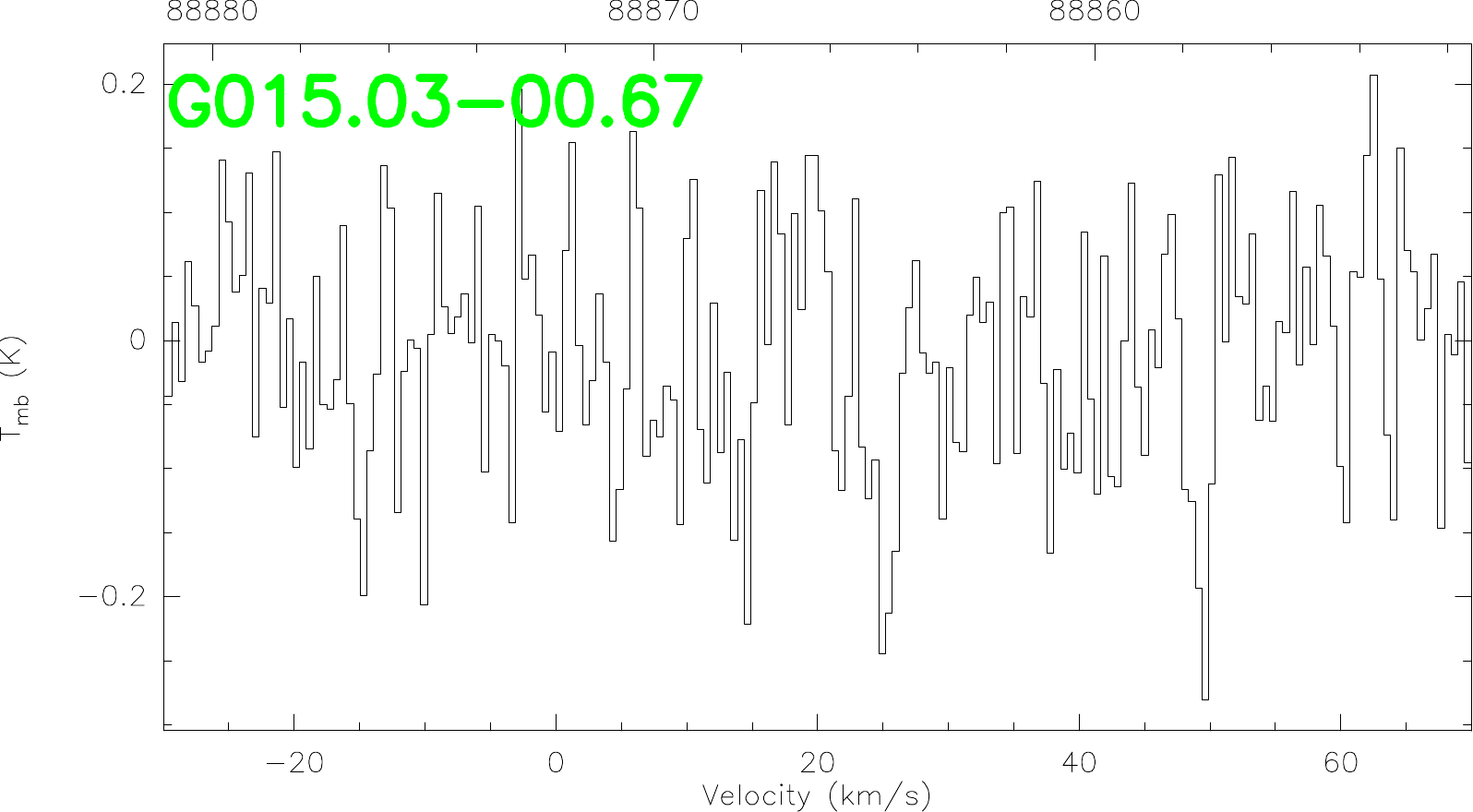}\\

\includegraphics[width=0.155\textwidth]{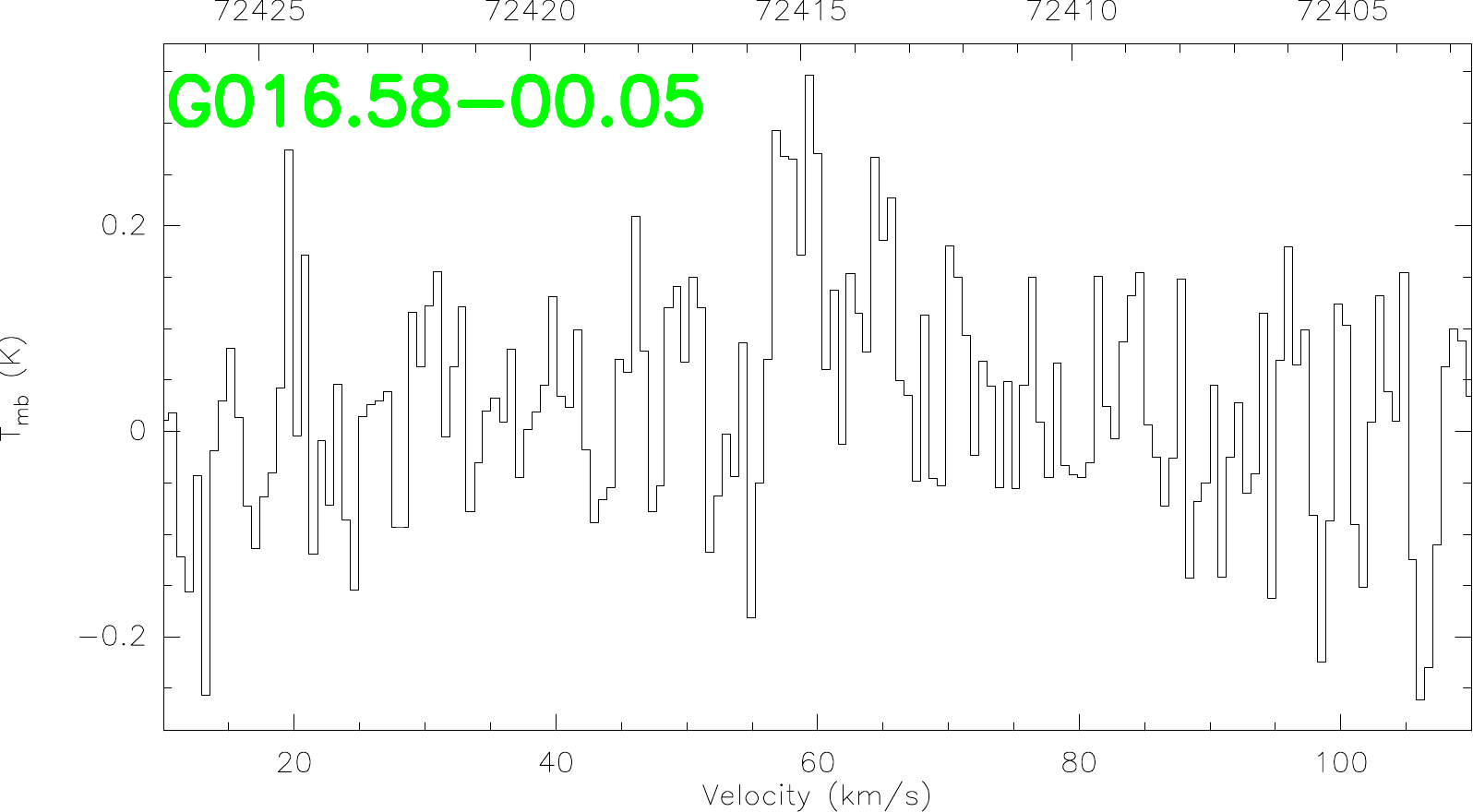}
\includegraphics[width=0.155\textwidth]{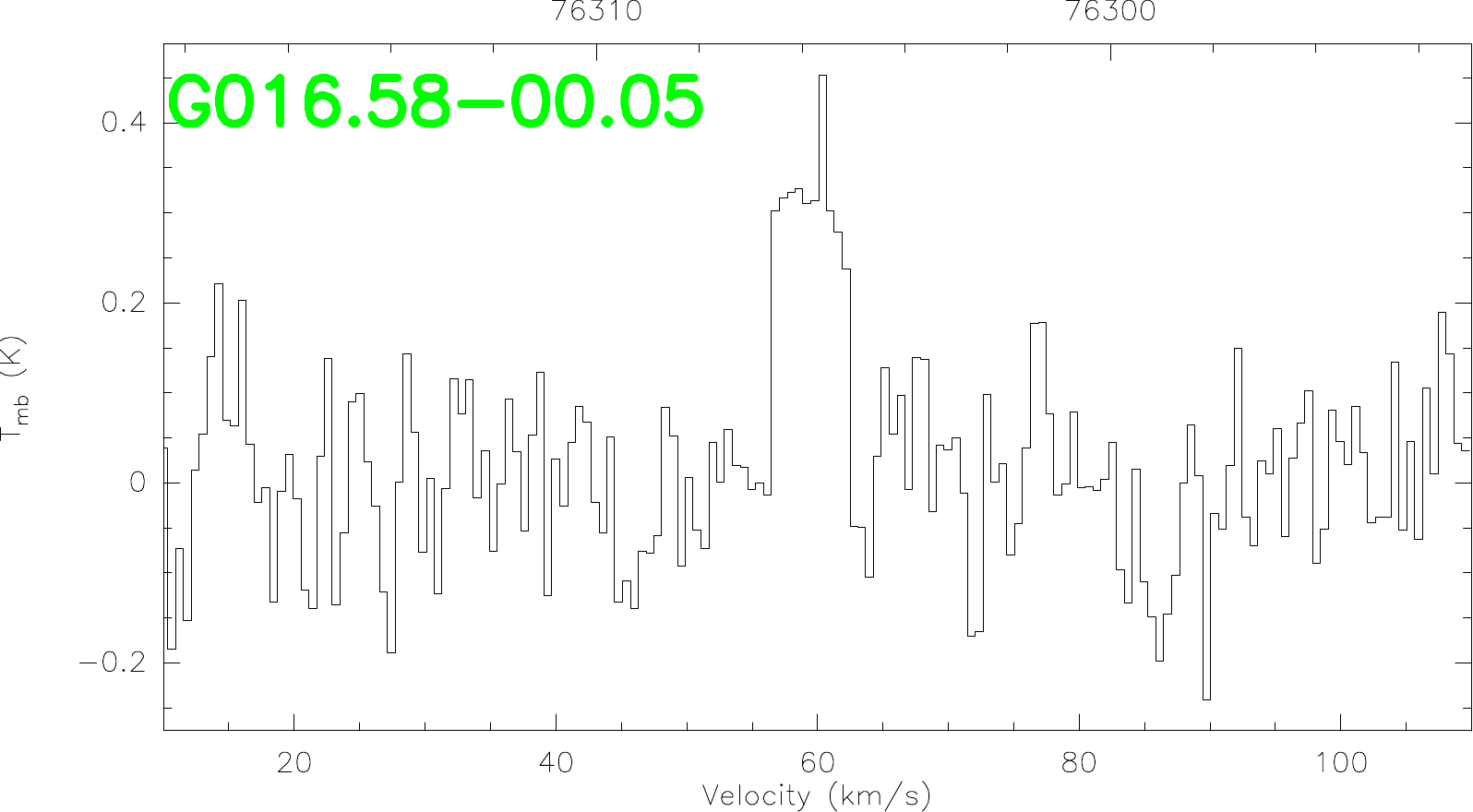}
\includegraphics[width=0.155\textwidth]{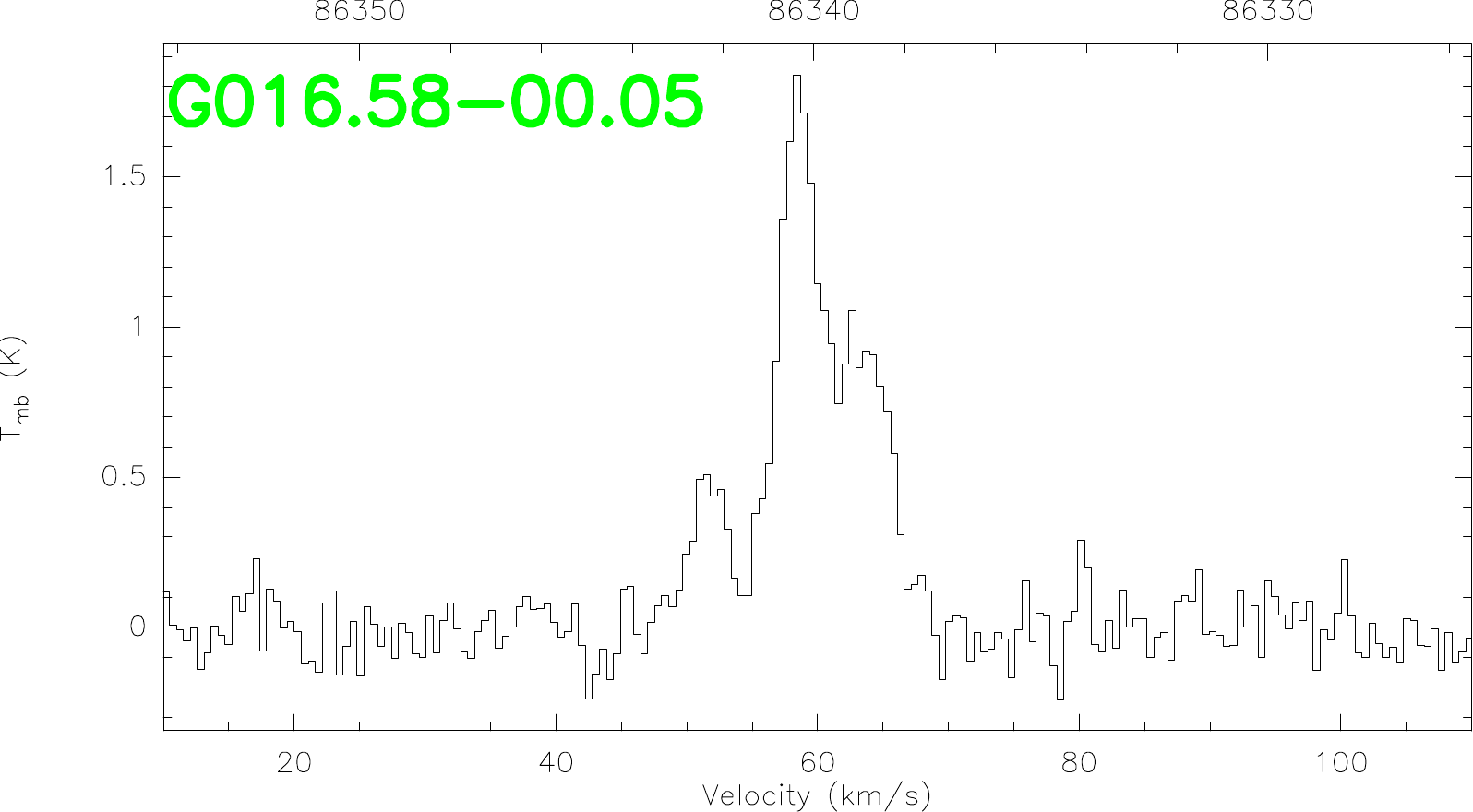}
\includegraphics[width=0.155\textwidth]{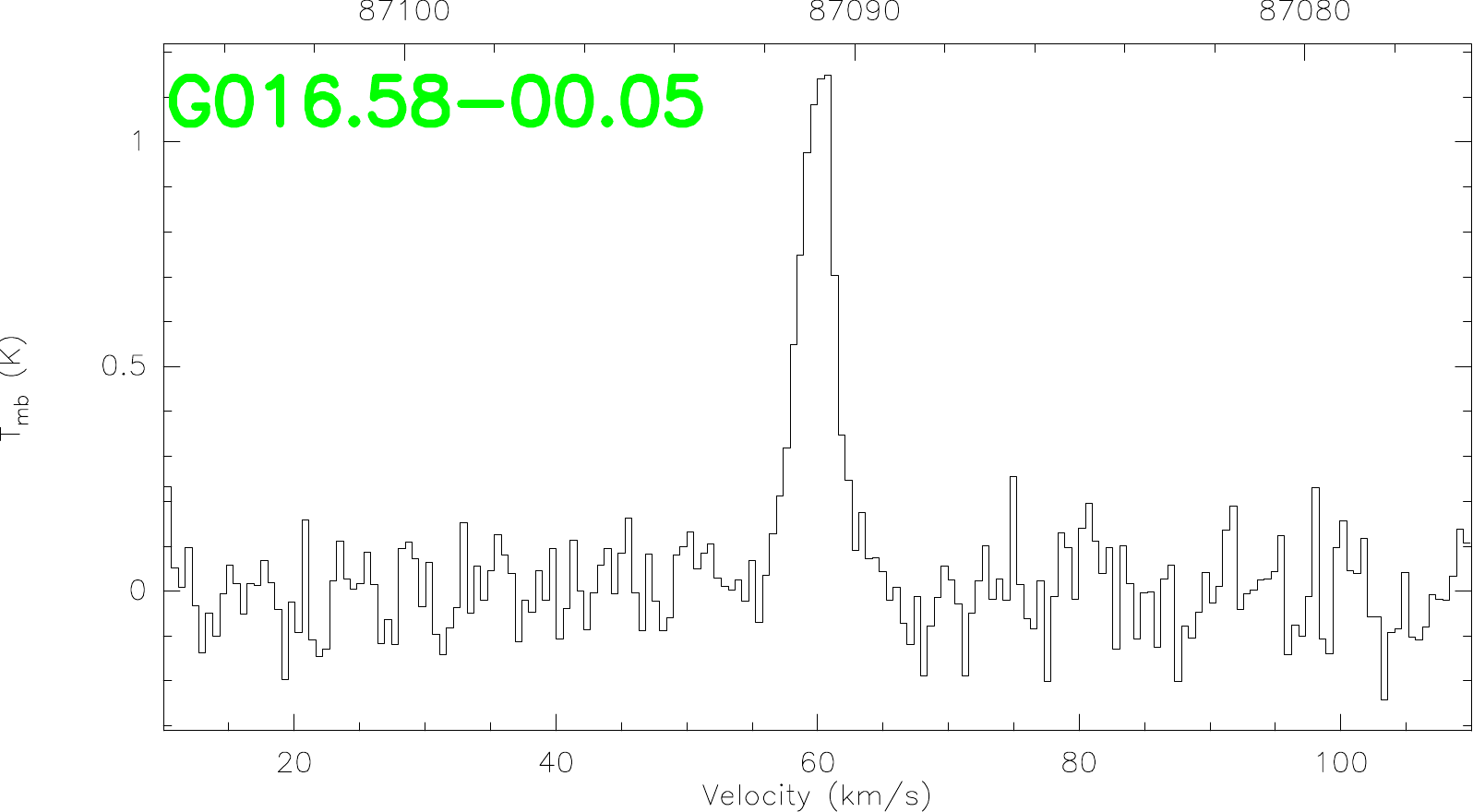}
\includegraphics[width=0.155\textwidth]{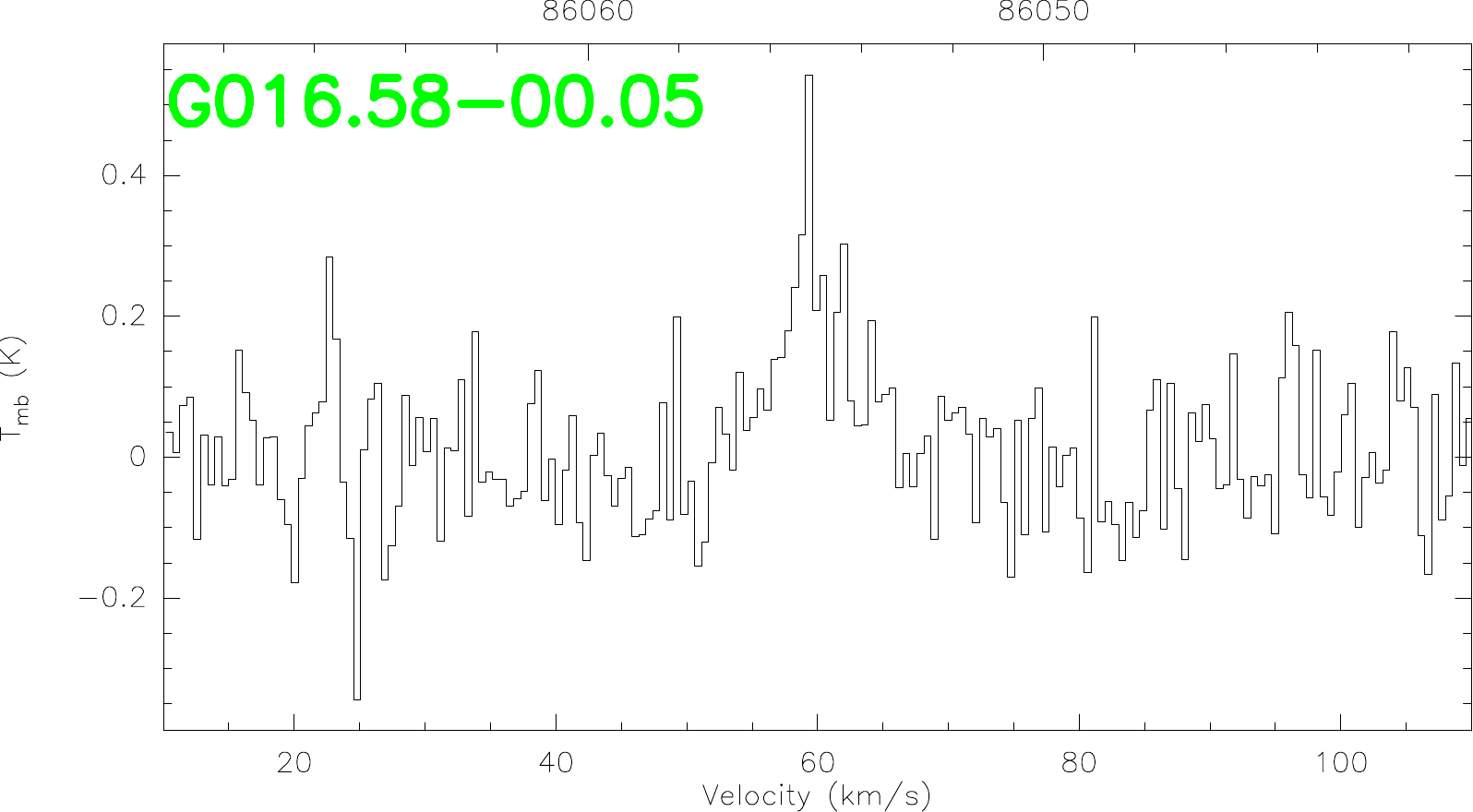}
\includegraphics[width=0.155\textwidth]{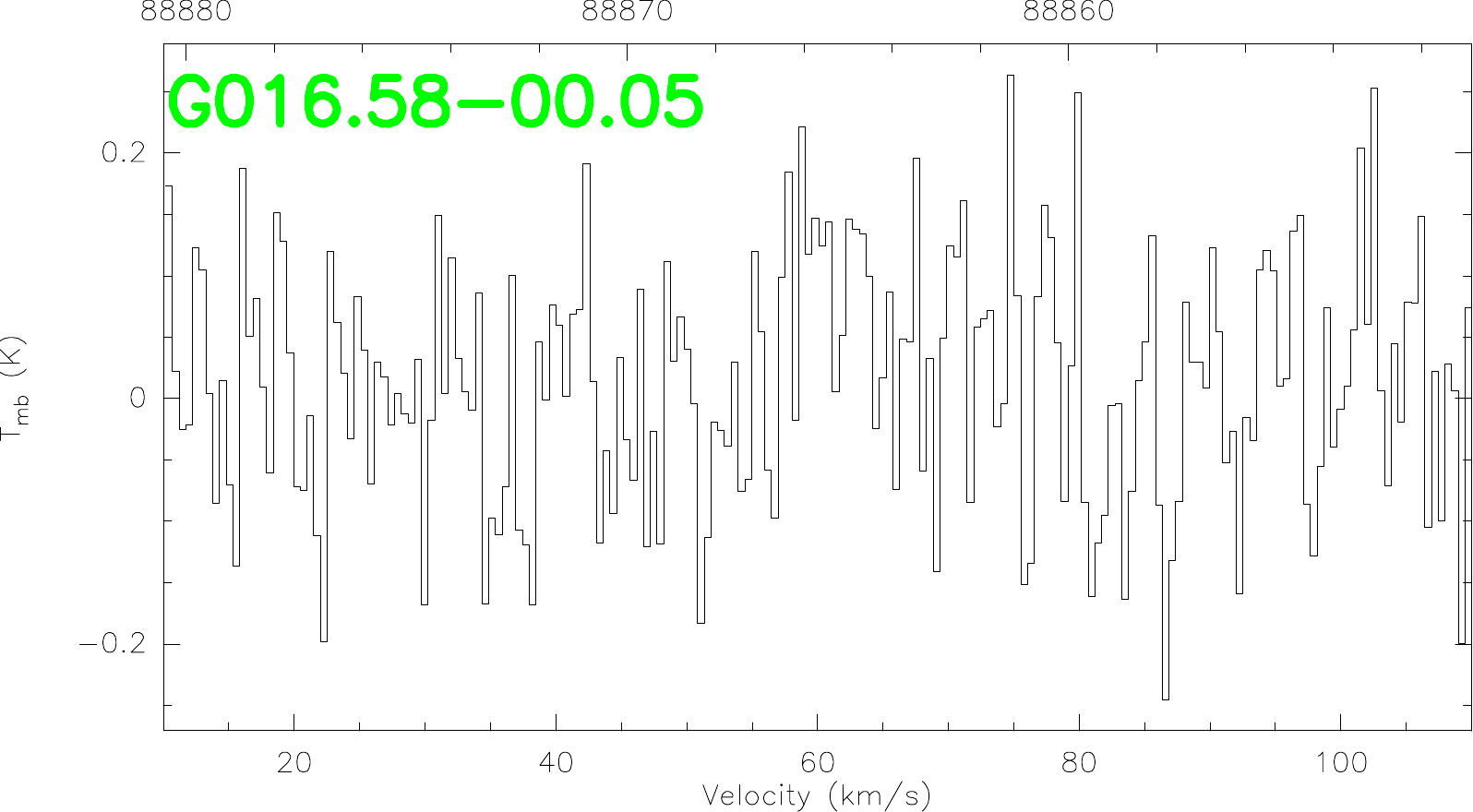}\\

\includegraphics[width=0.155\textwidth]{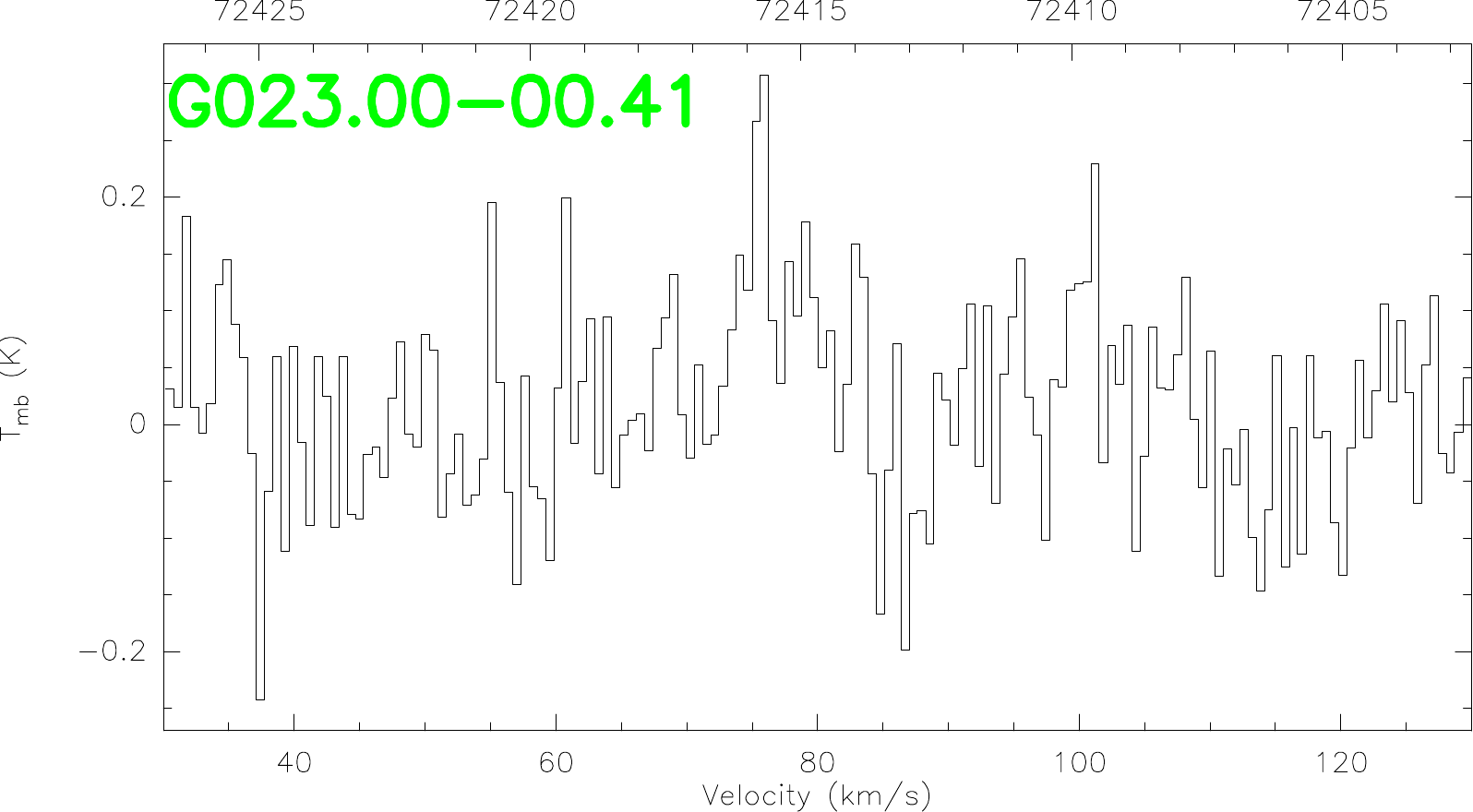}
\includegraphics[width=0.155\textwidth]{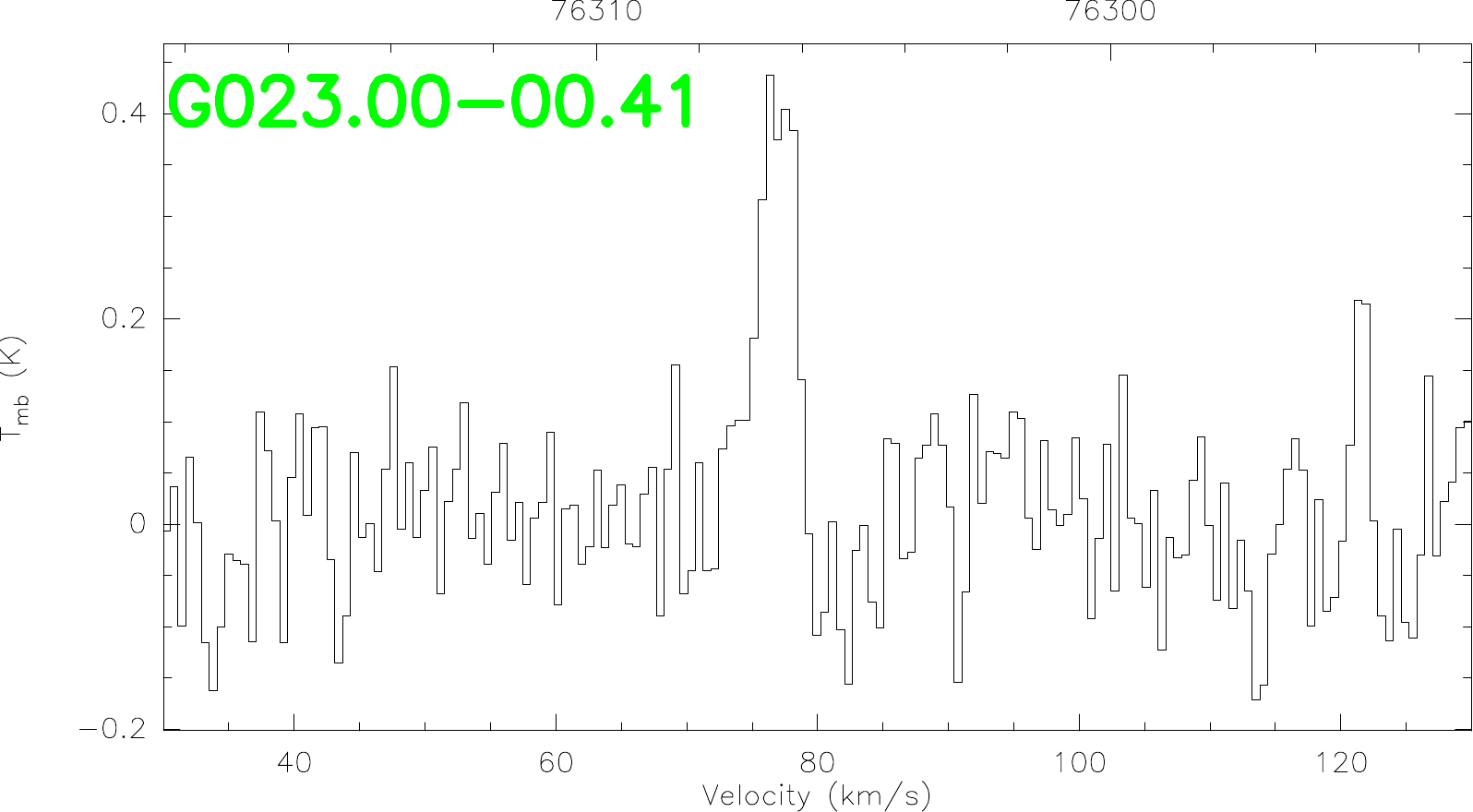}
\includegraphics[width=0.155\textwidth]{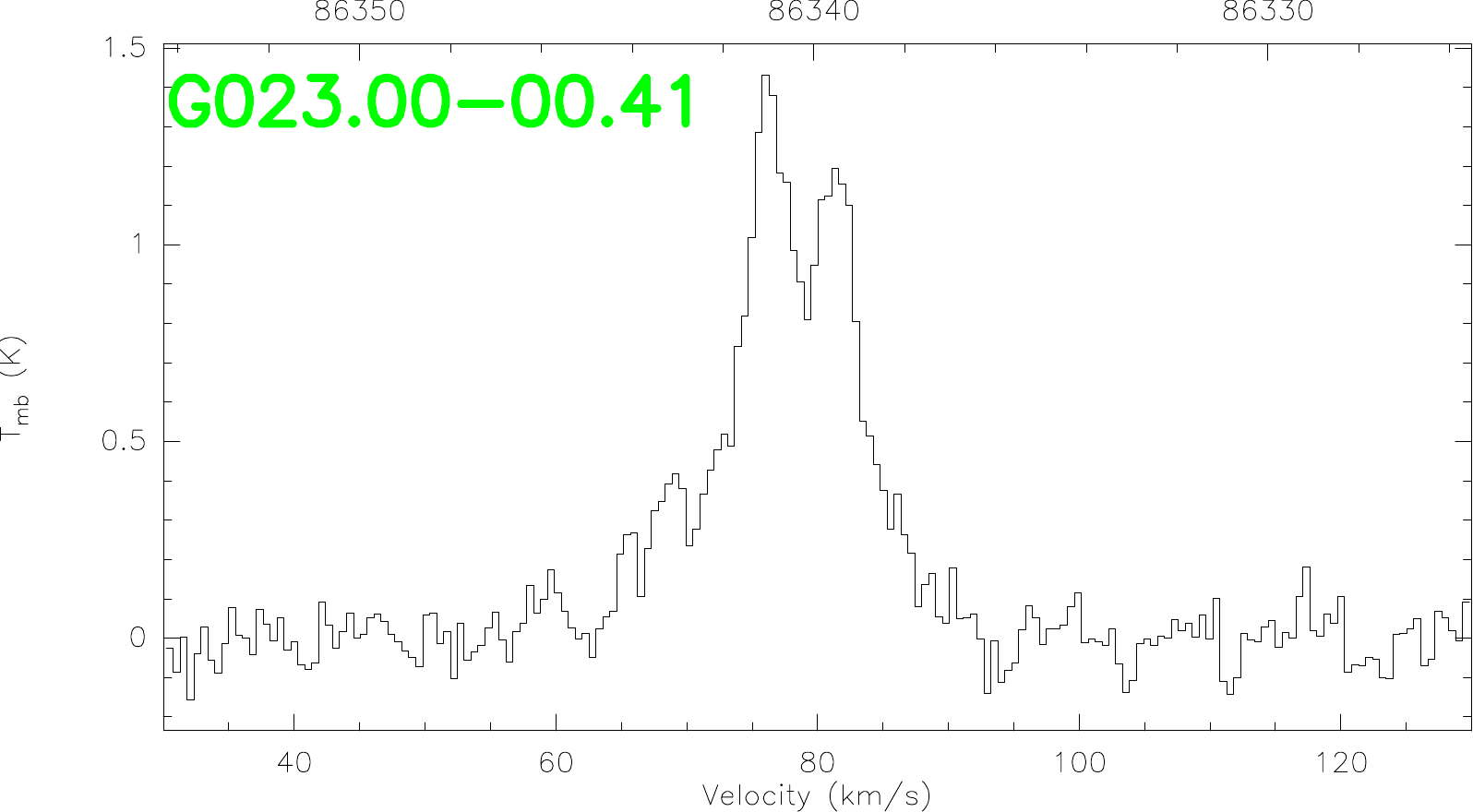}
\includegraphics[width=0.155\textwidth]{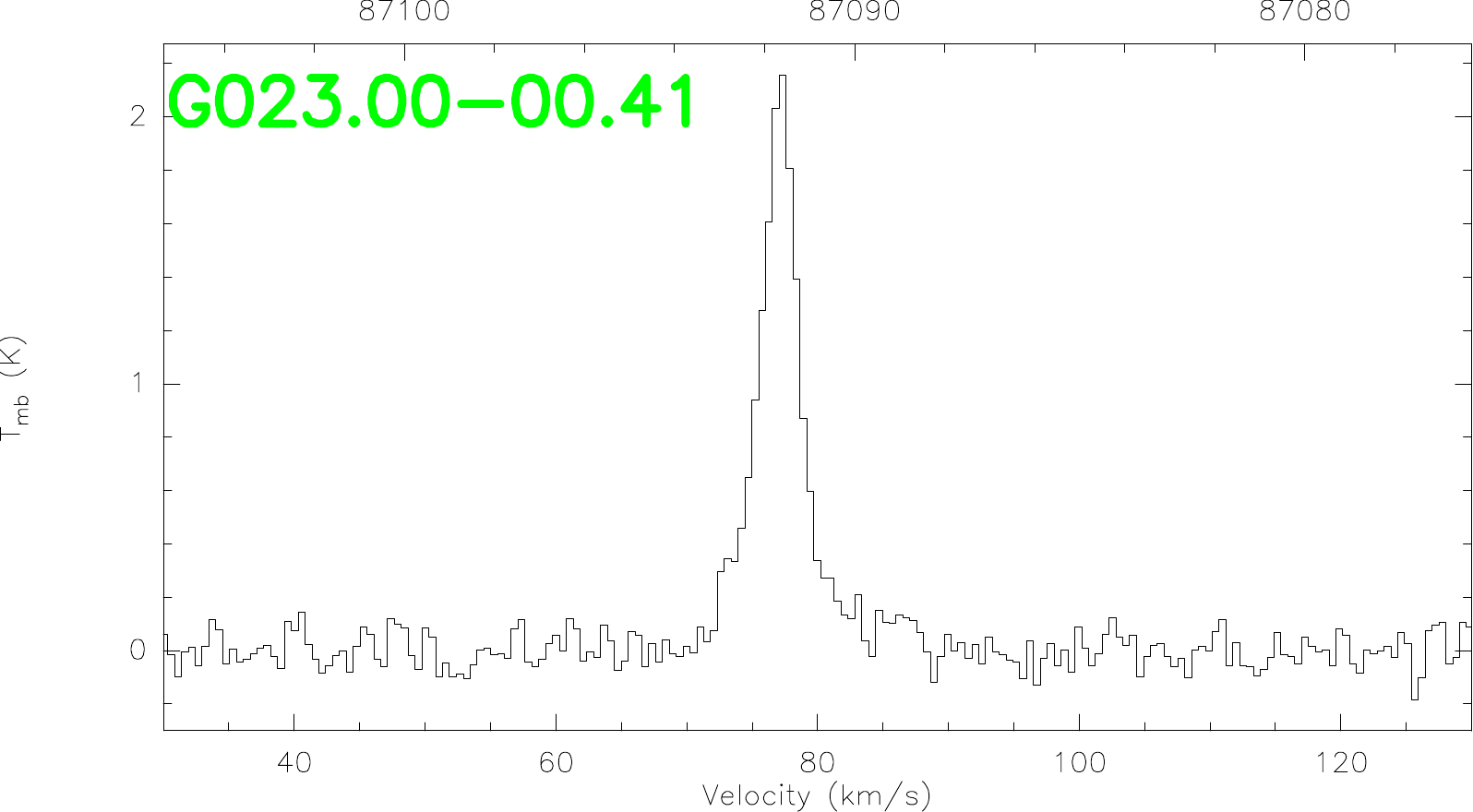}
\includegraphics[width=0.155\textwidth]{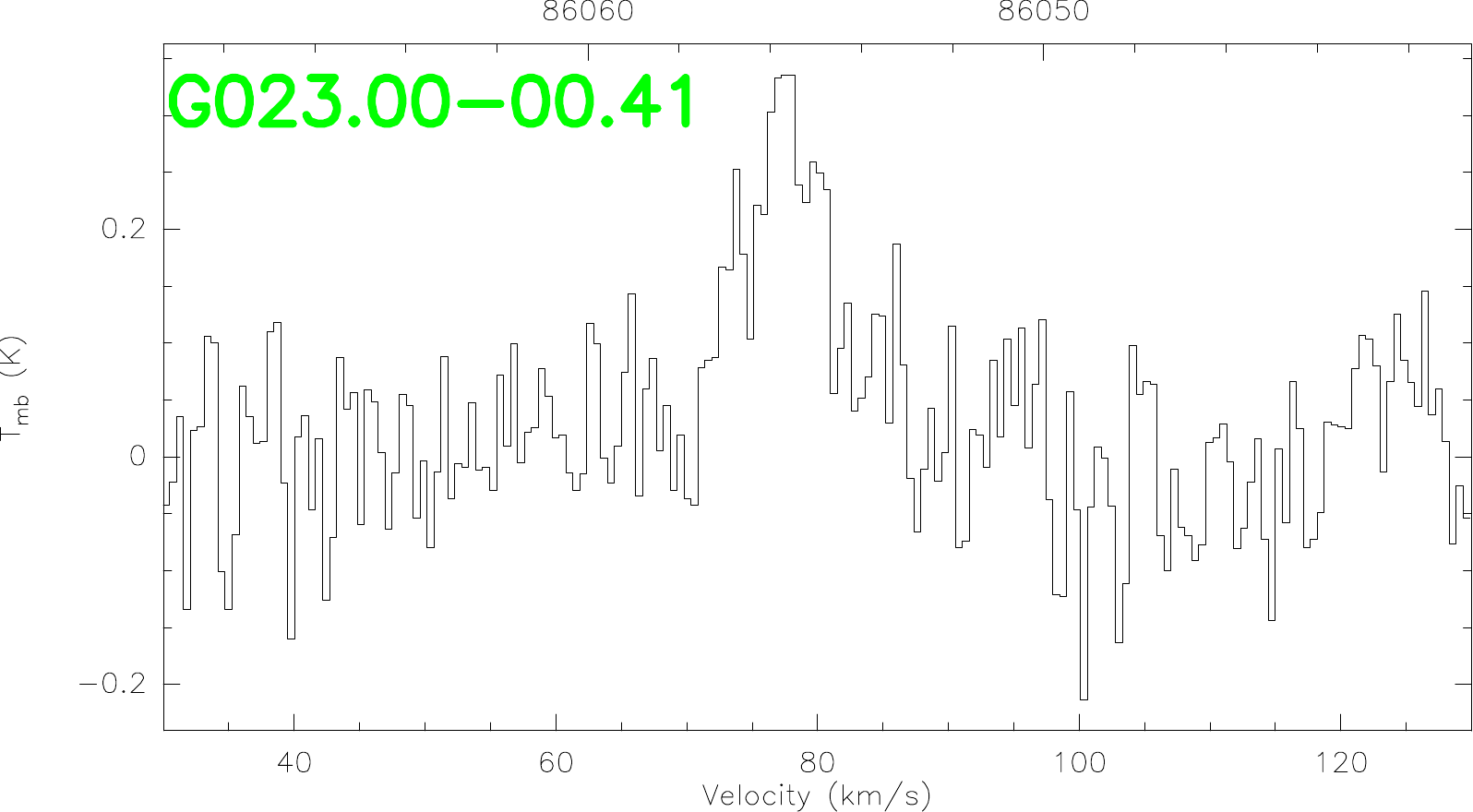}
\includegraphics[width=0.155\textwidth]{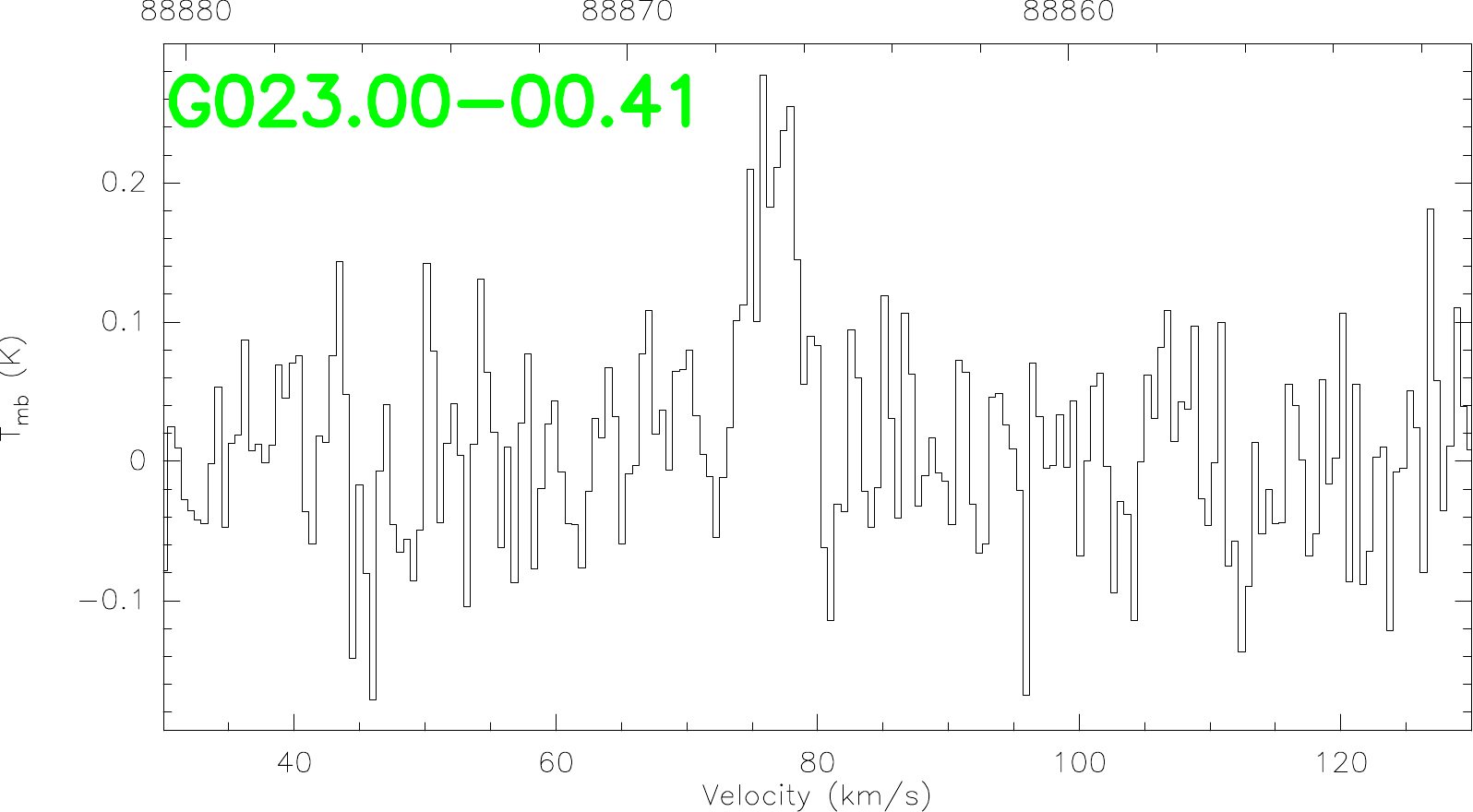}\\

\includegraphics[width=0.155\textwidth]{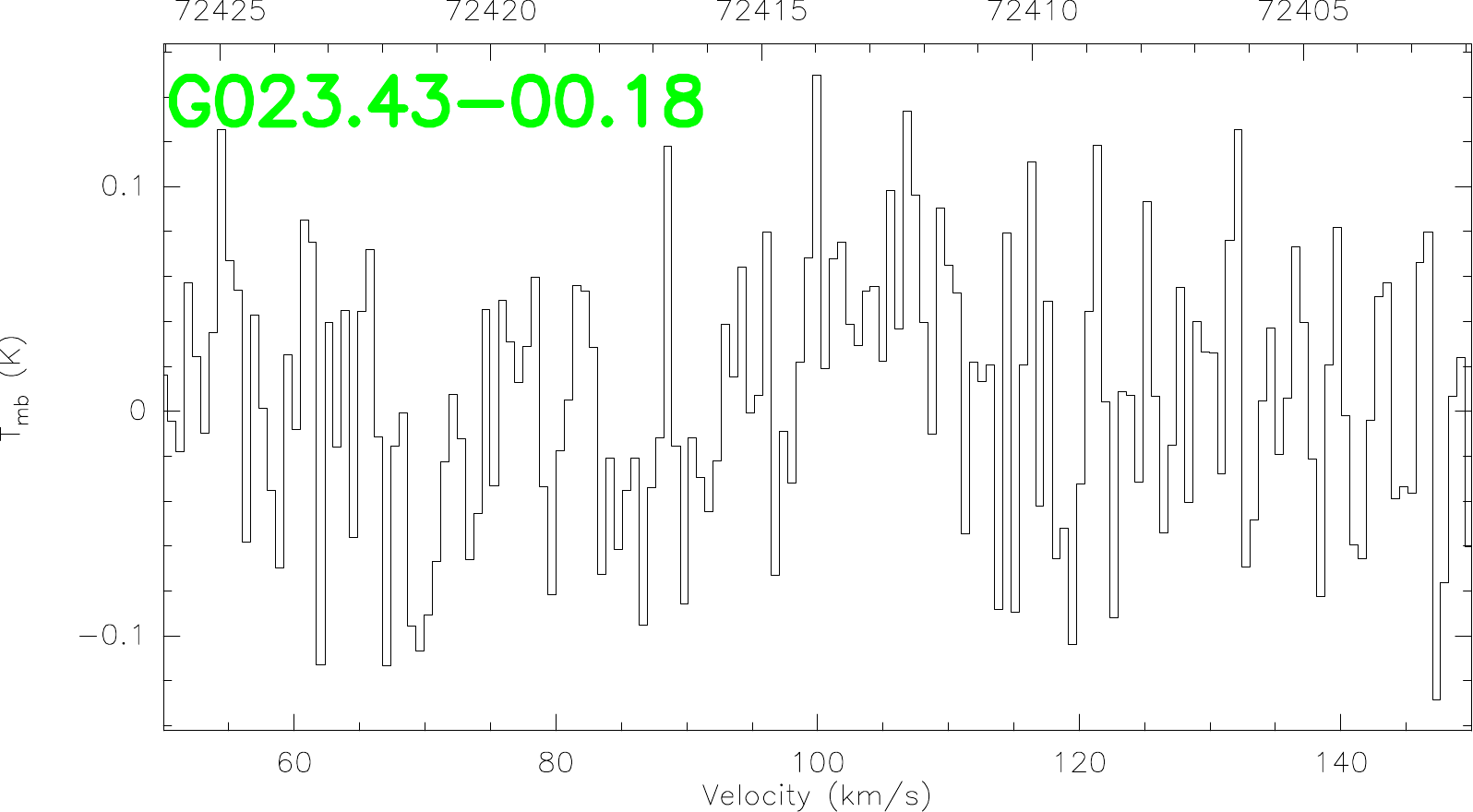}
\includegraphics[width=0.155\textwidth]{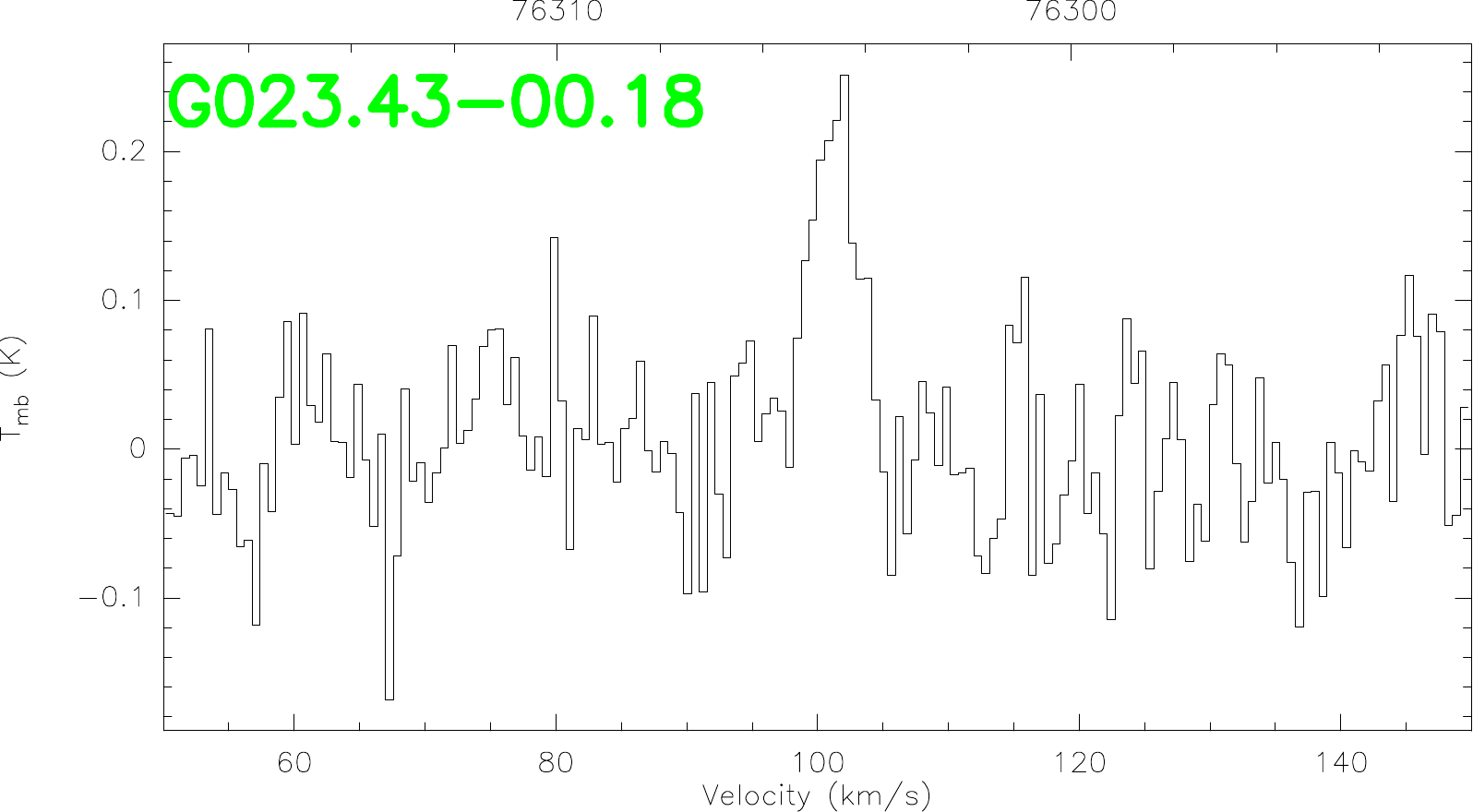}
\includegraphics[width=0.155\textwidth]{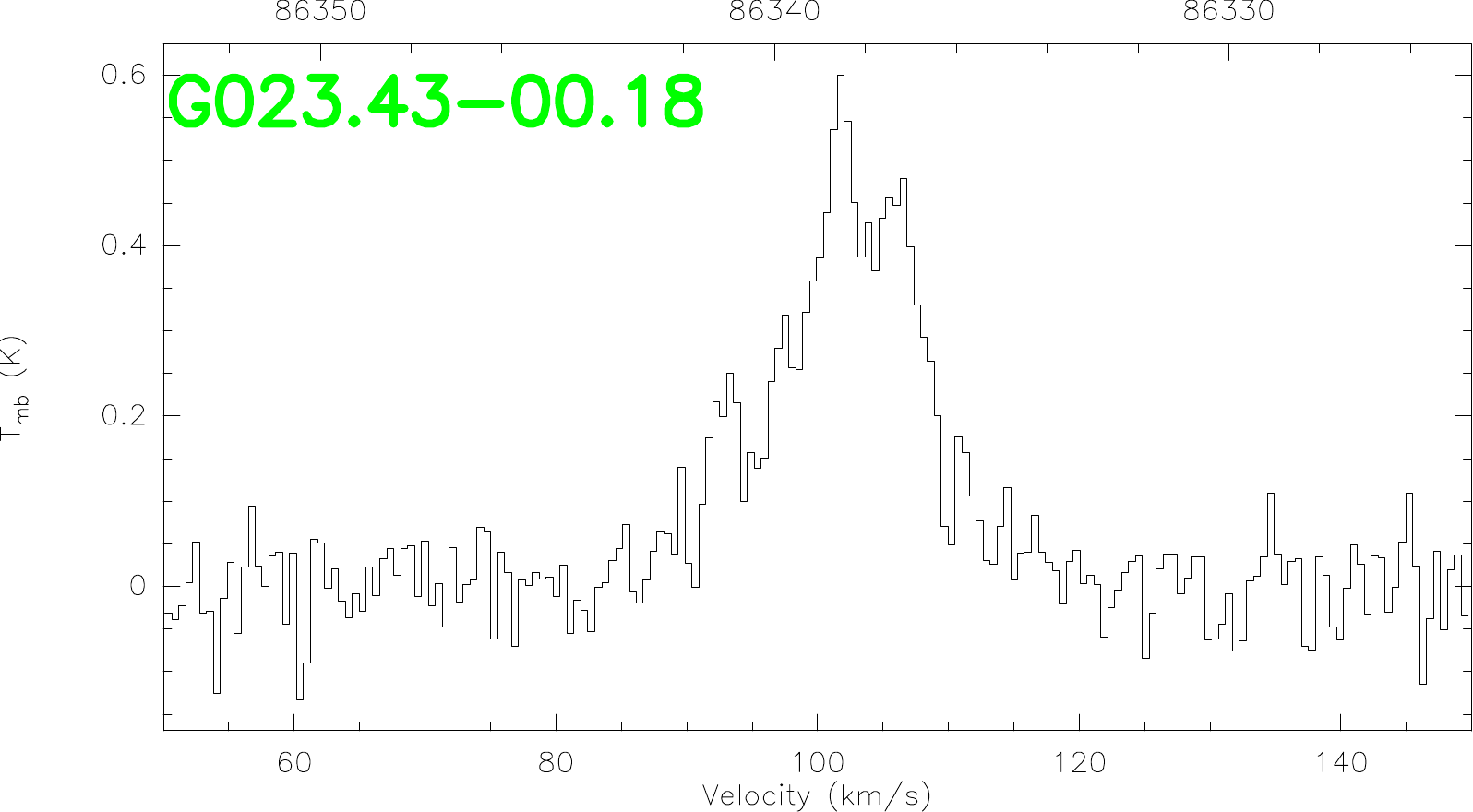}
\includegraphics[width=0.155\textwidth]{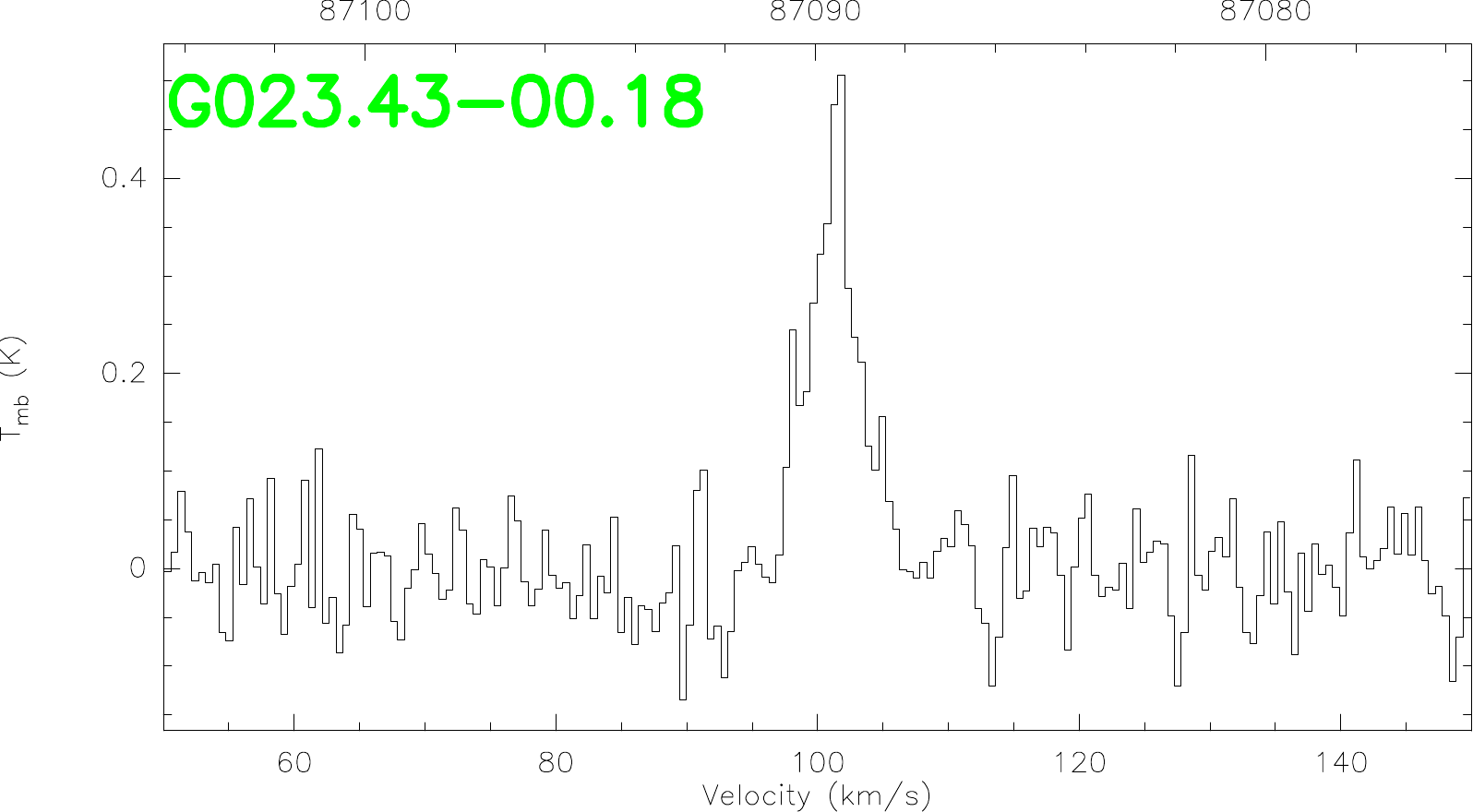}
\includegraphics[width=0.155\textwidth]{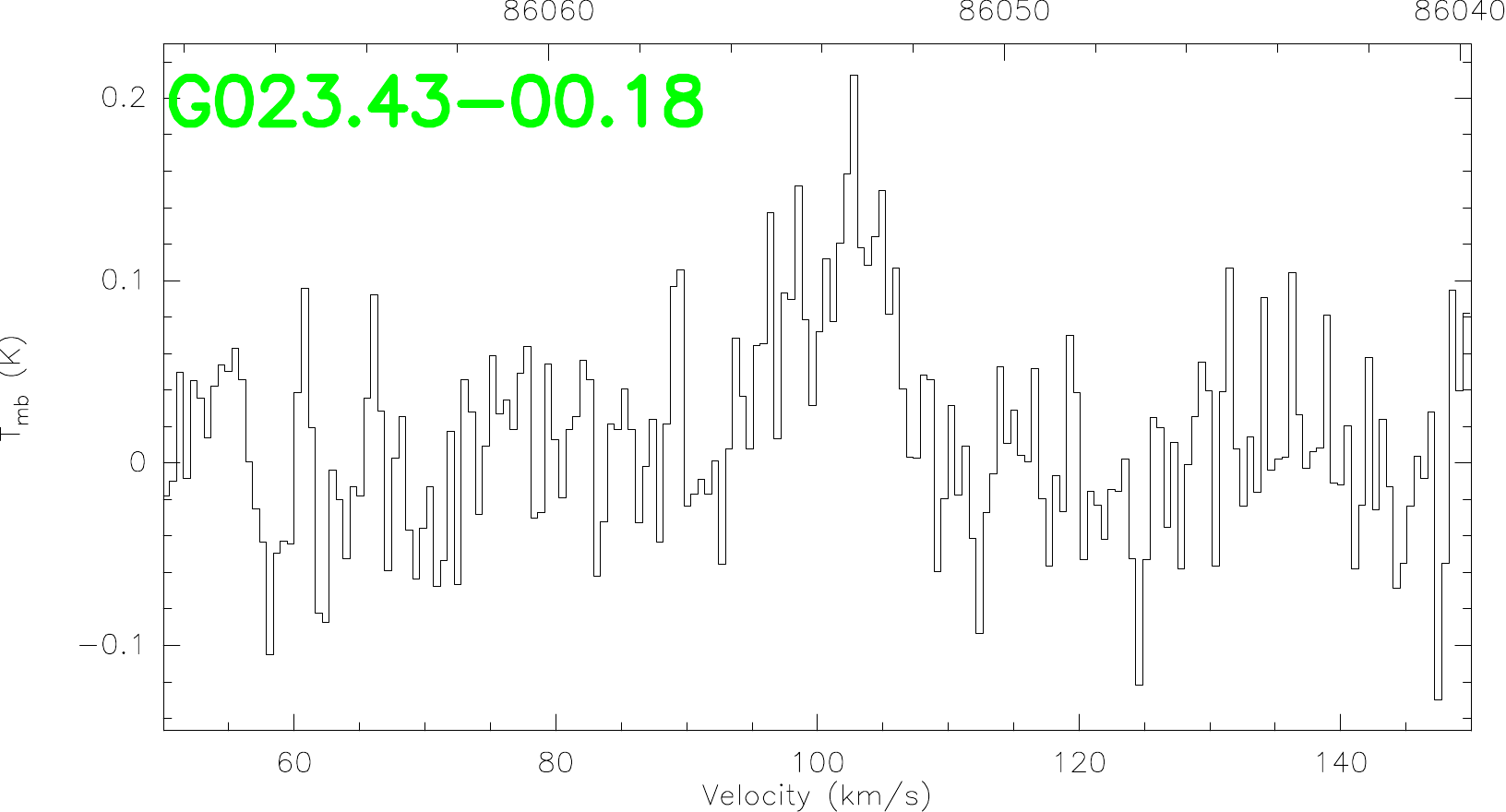}
\includegraphics[width=0.155\textwidth]{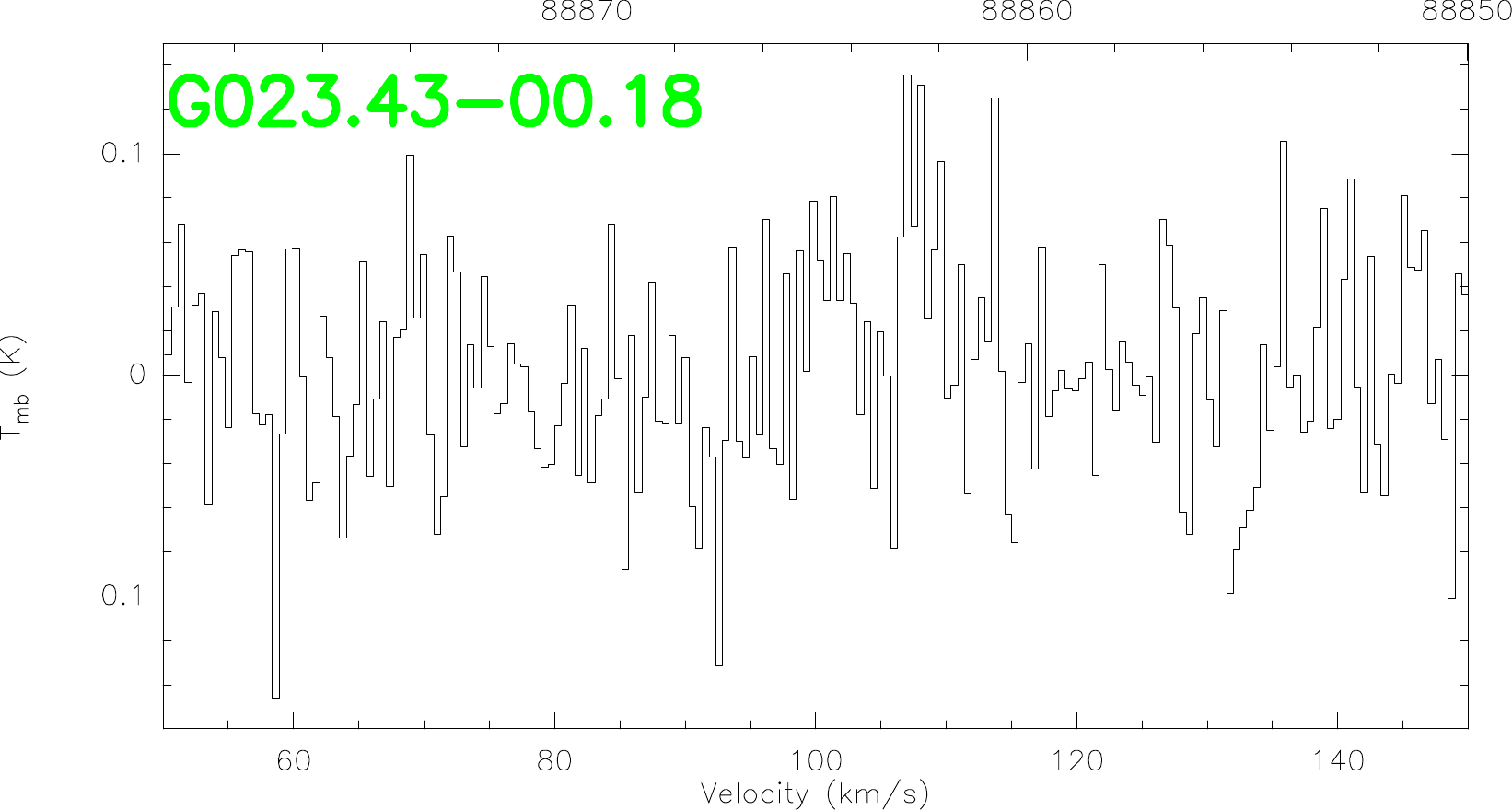}\\

\includegraphics[width=0.155\textwidth]{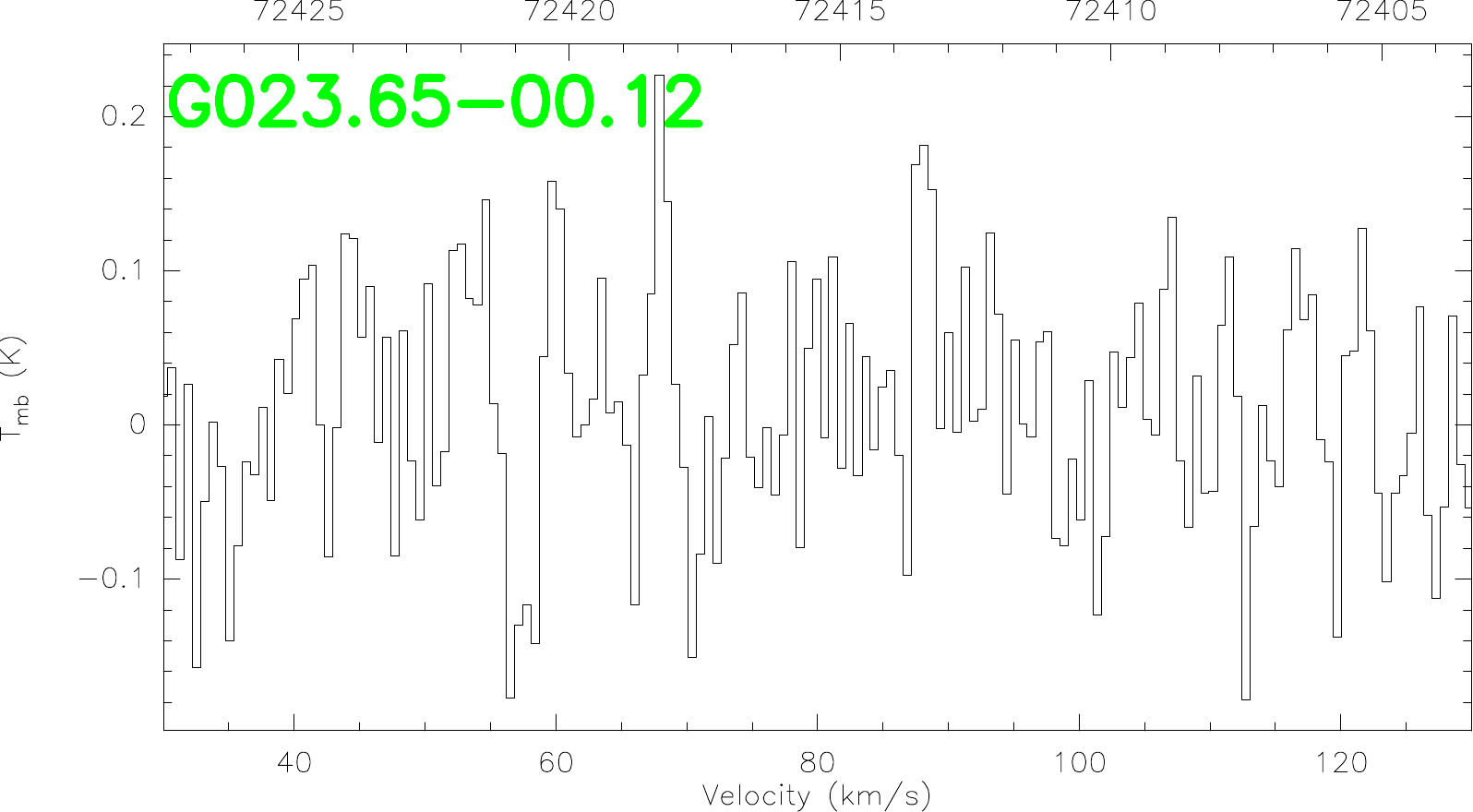}
\includegraphics[width=0.155\textwidth]{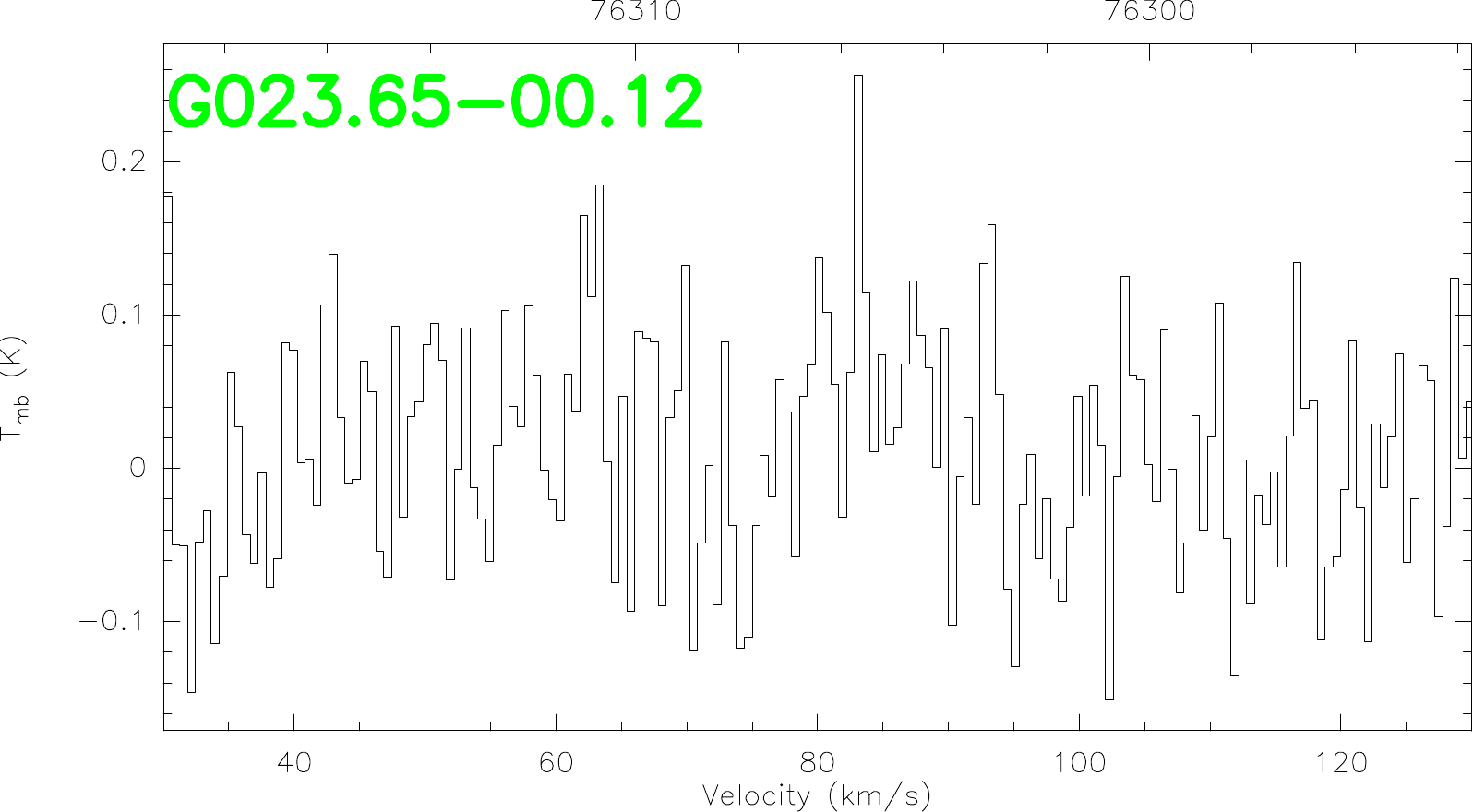}
\includegraphics[width=0.155\textwidth]{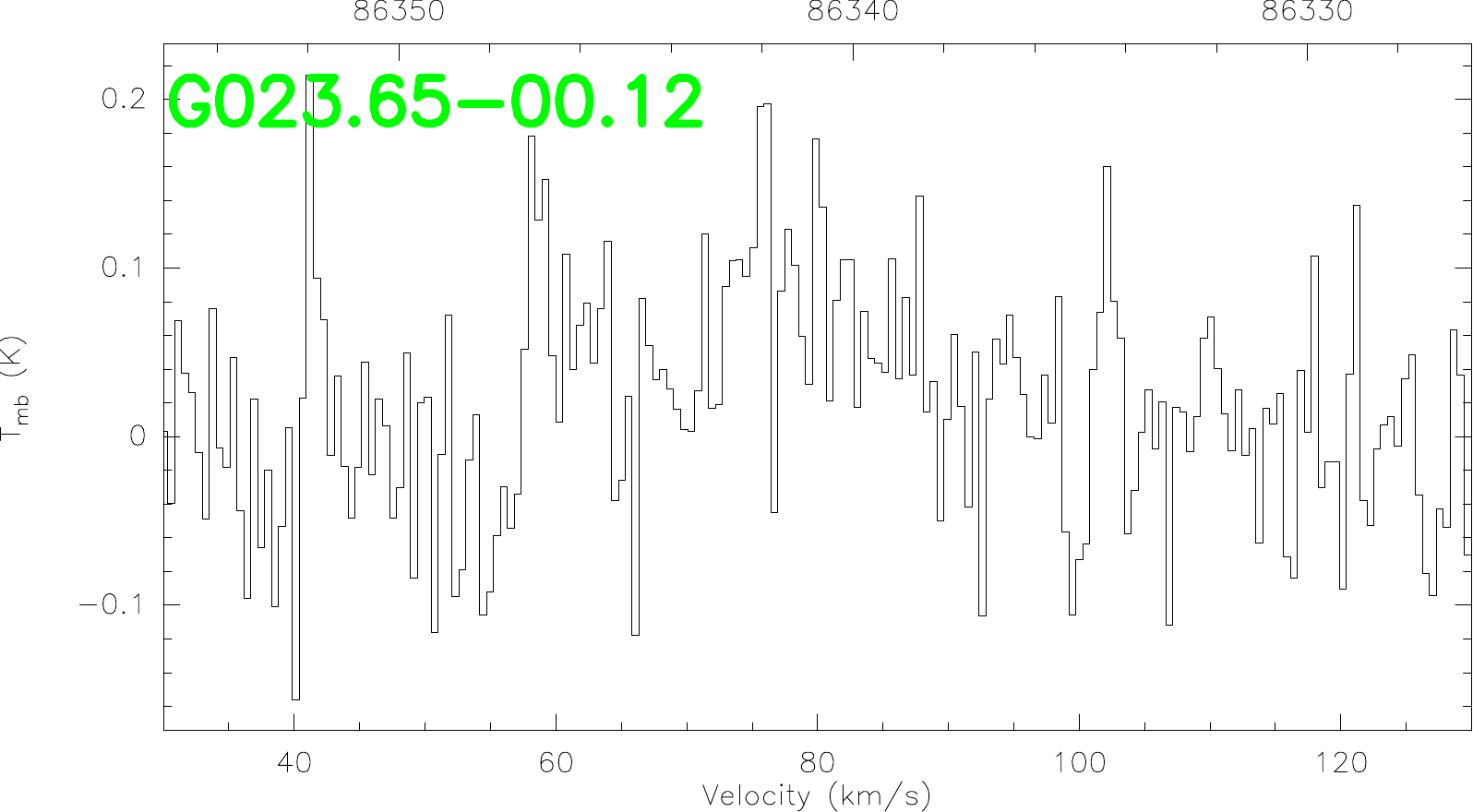}
\includegraphics[width=0.155\textwidth]{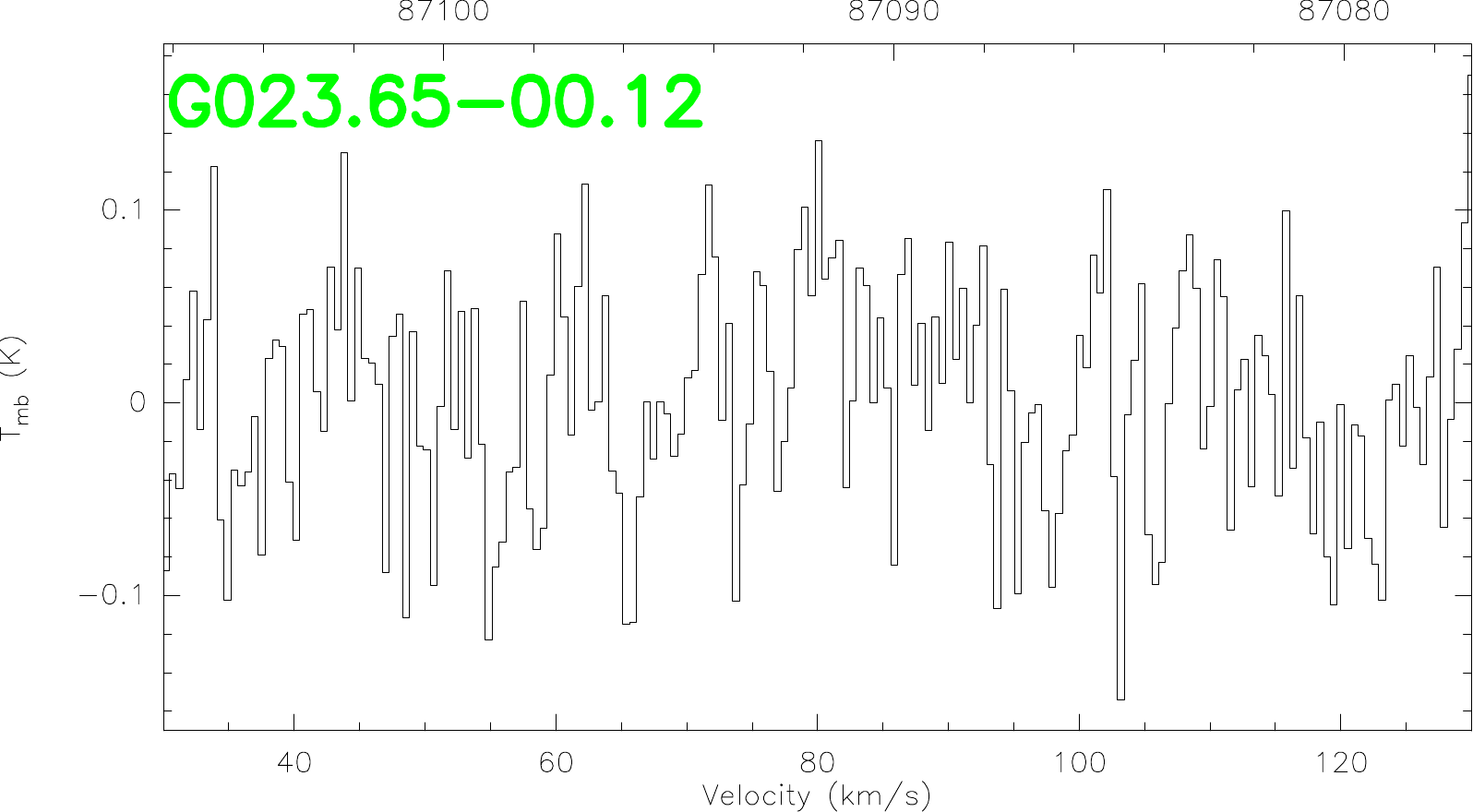}
\includegraphics[width=0.155\textwidth]{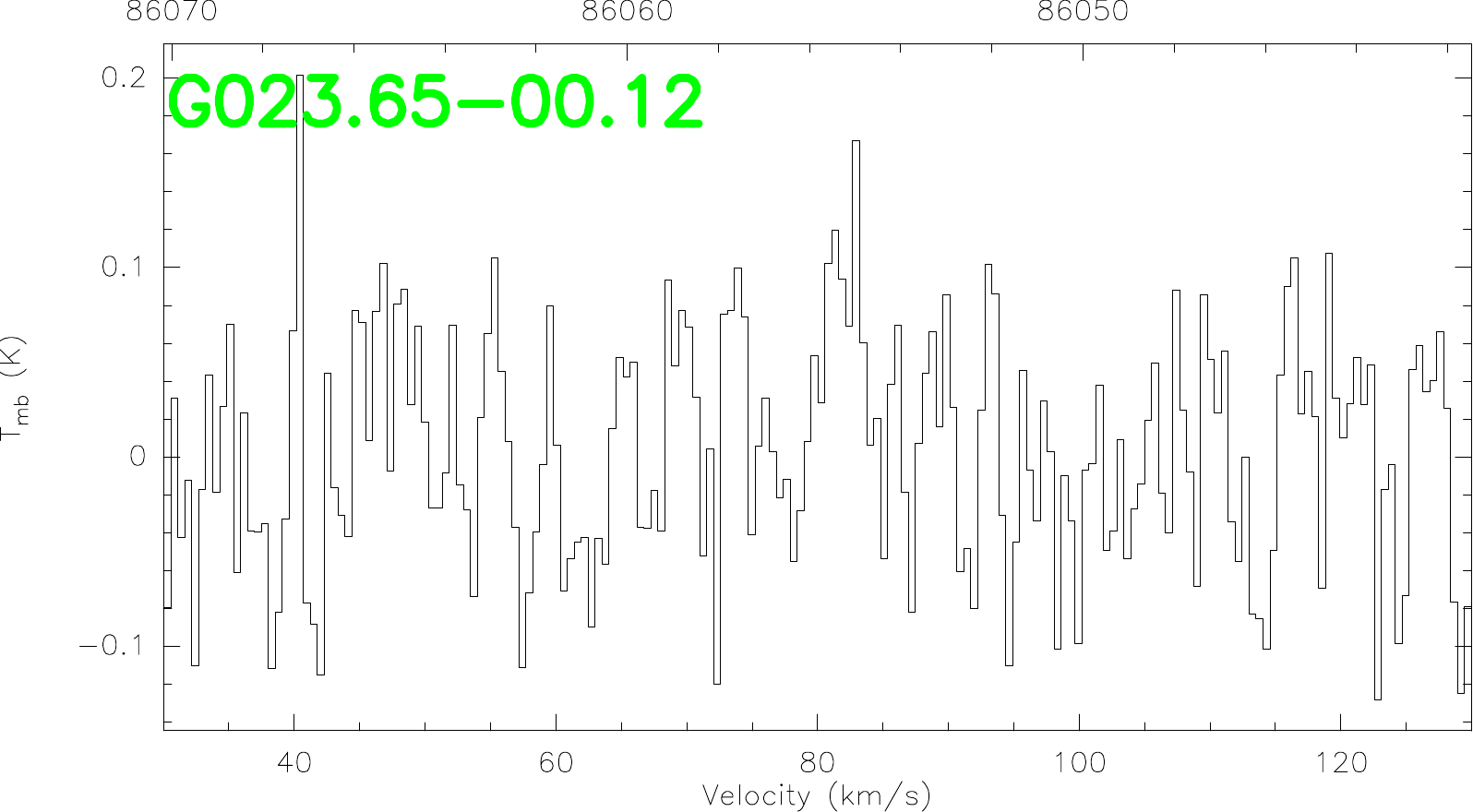}
\includegraphics[width=0.155\textwidth]{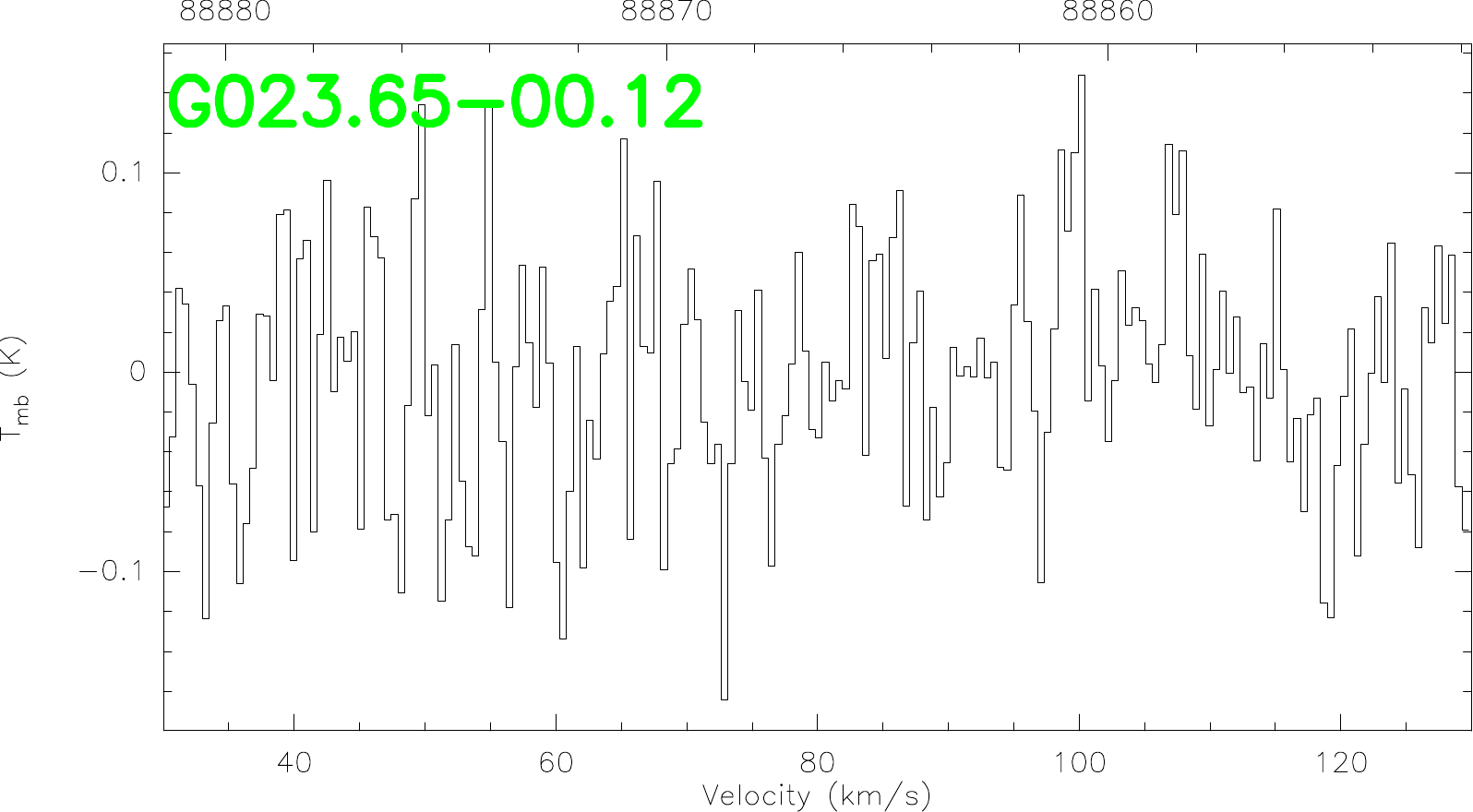}\\

\includegraphics[width=0.155\textwidth]{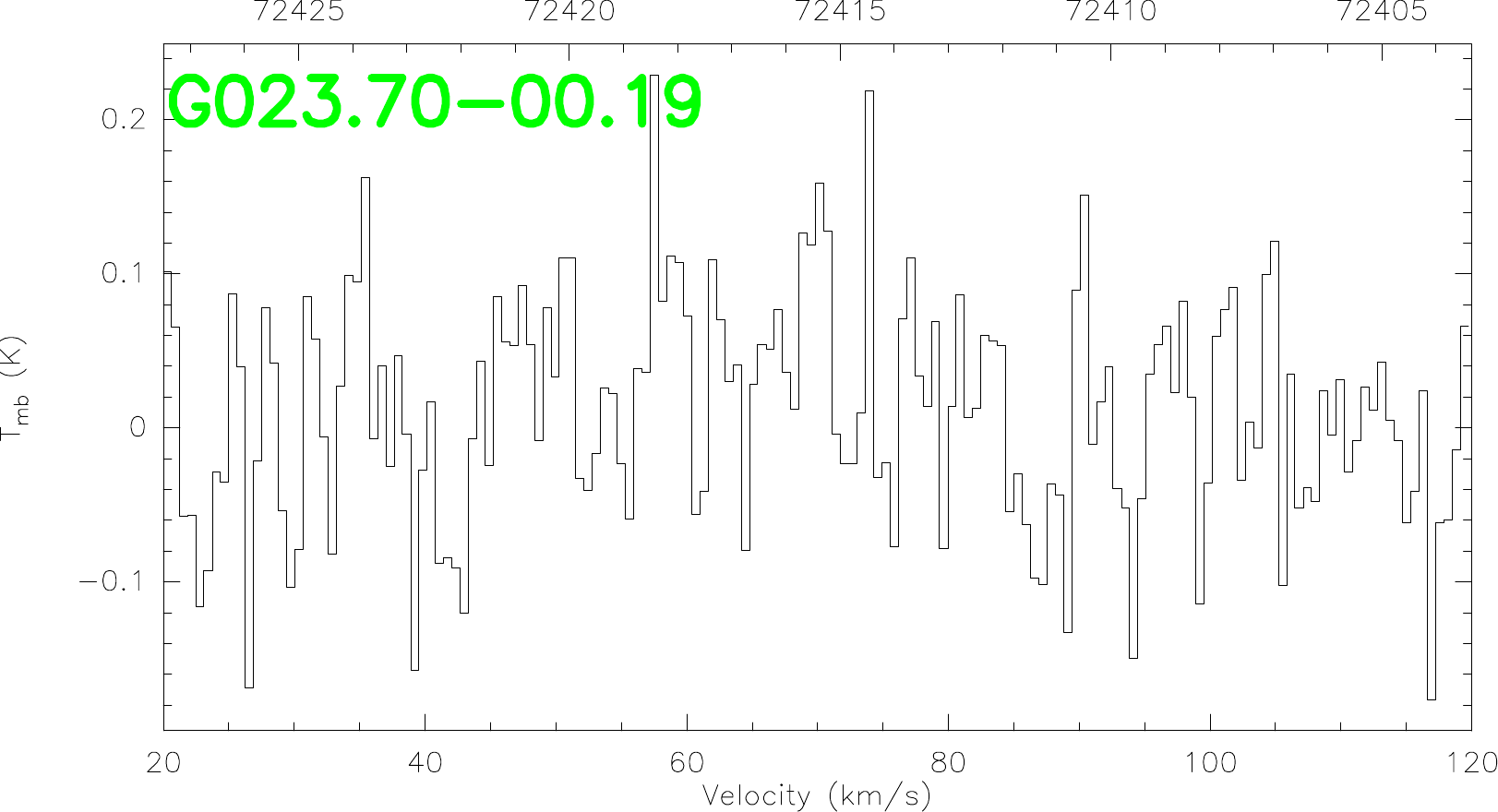}
\includegraphics[width=0.155\textwidth]{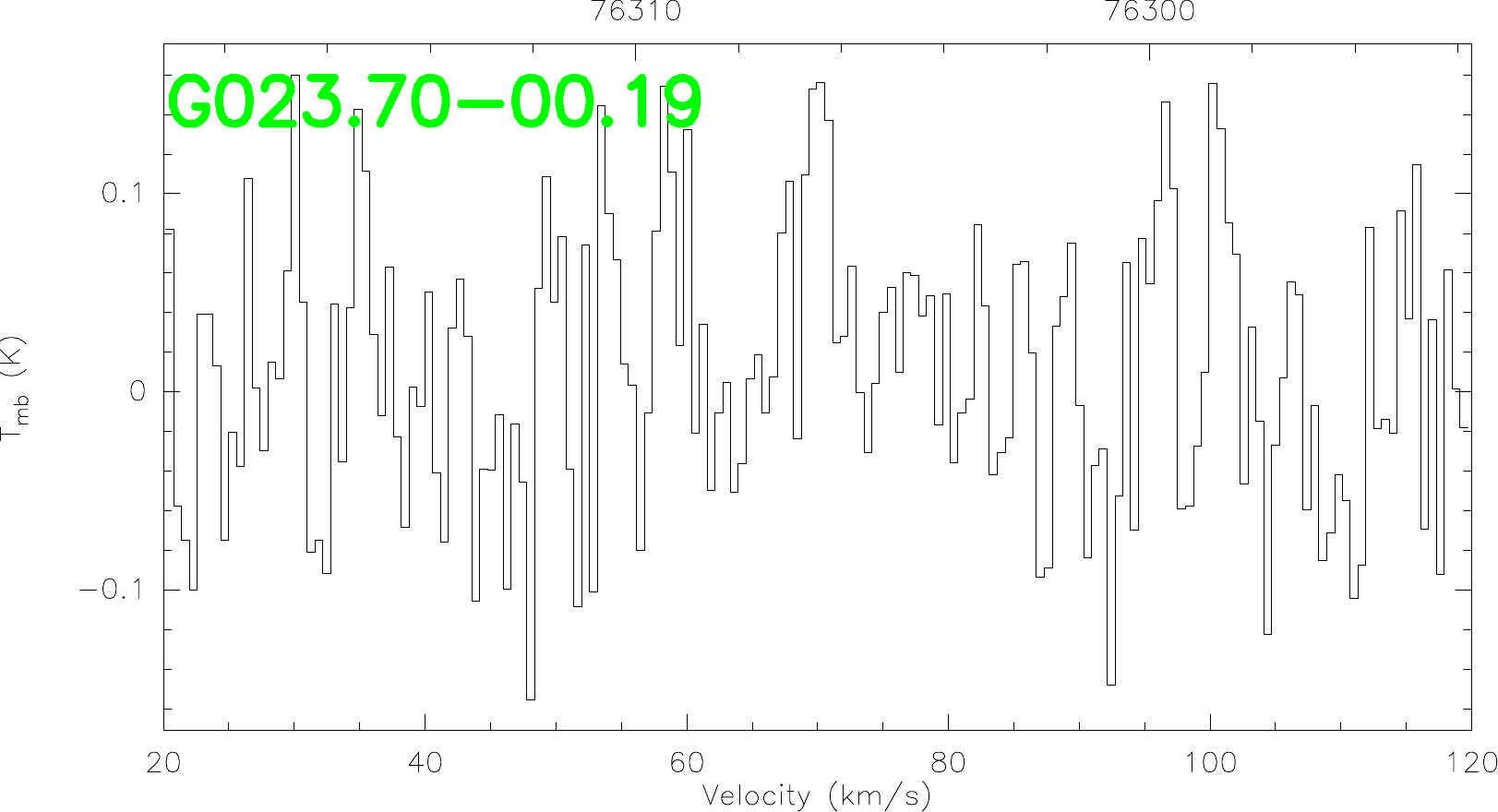}
\includegraphics[width=0.155\textwidth]{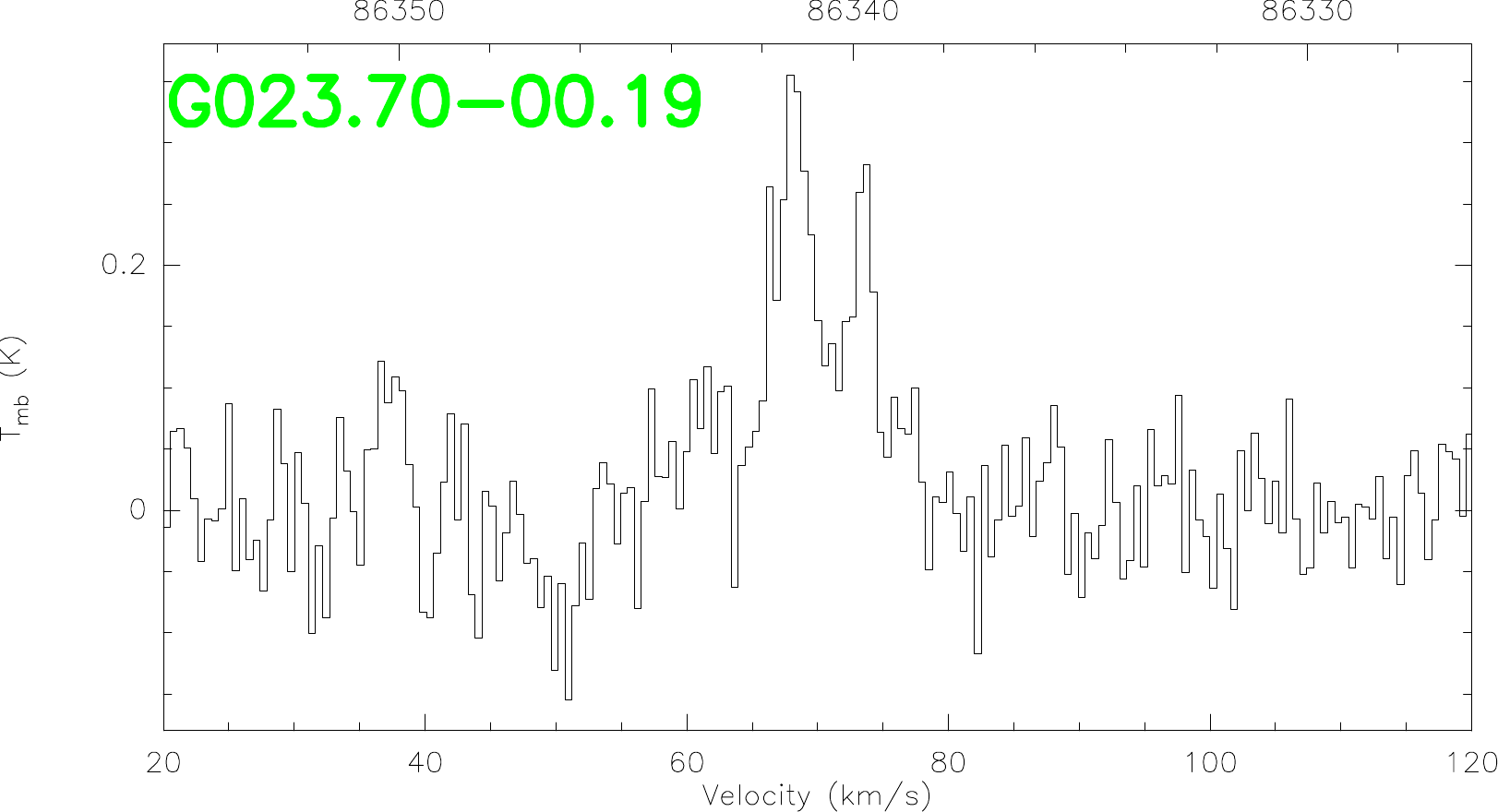}
\includegraphics[width=0.155\textwidth]{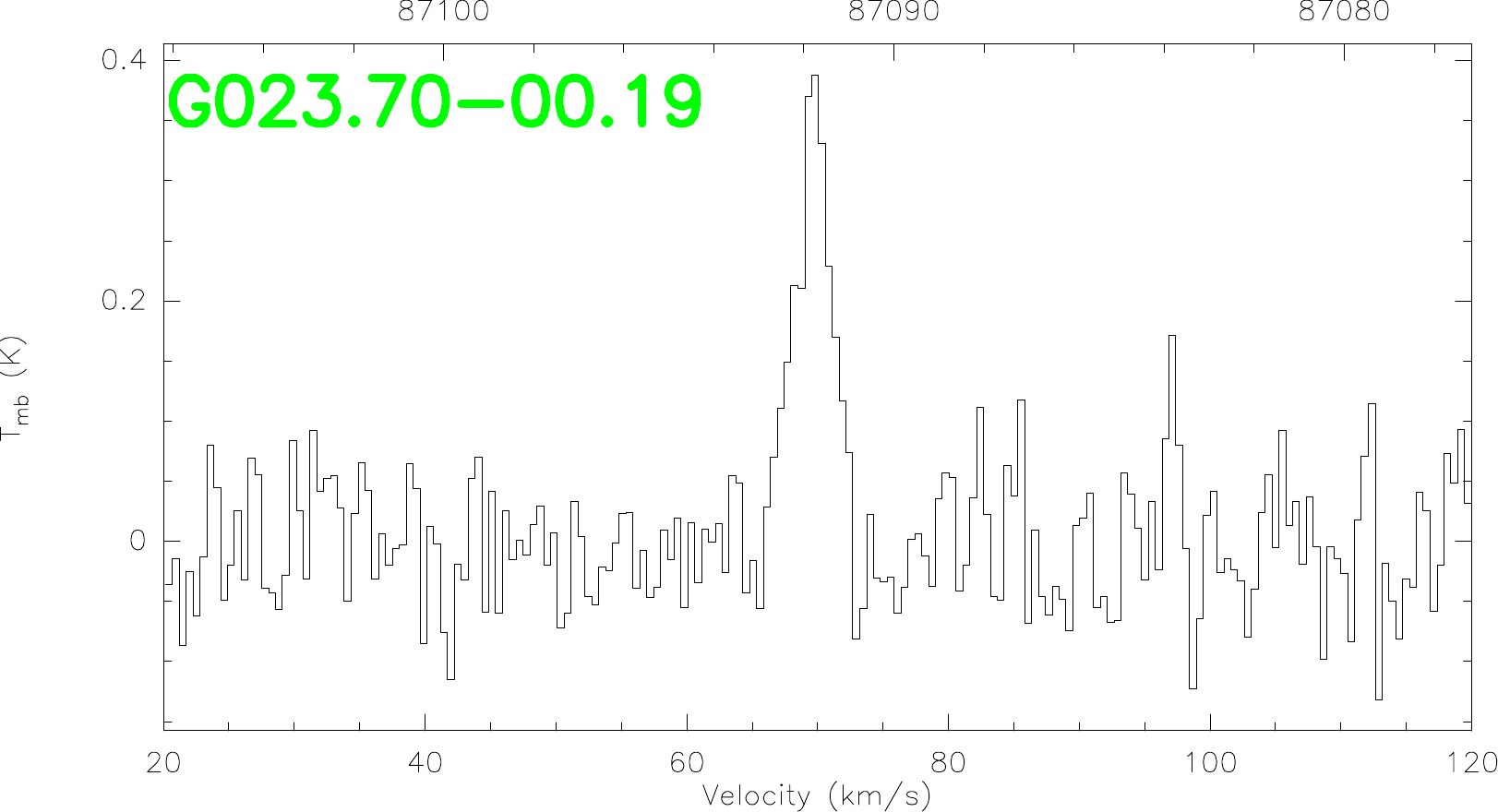}
\includegraphics[width=0.155\textwidth]{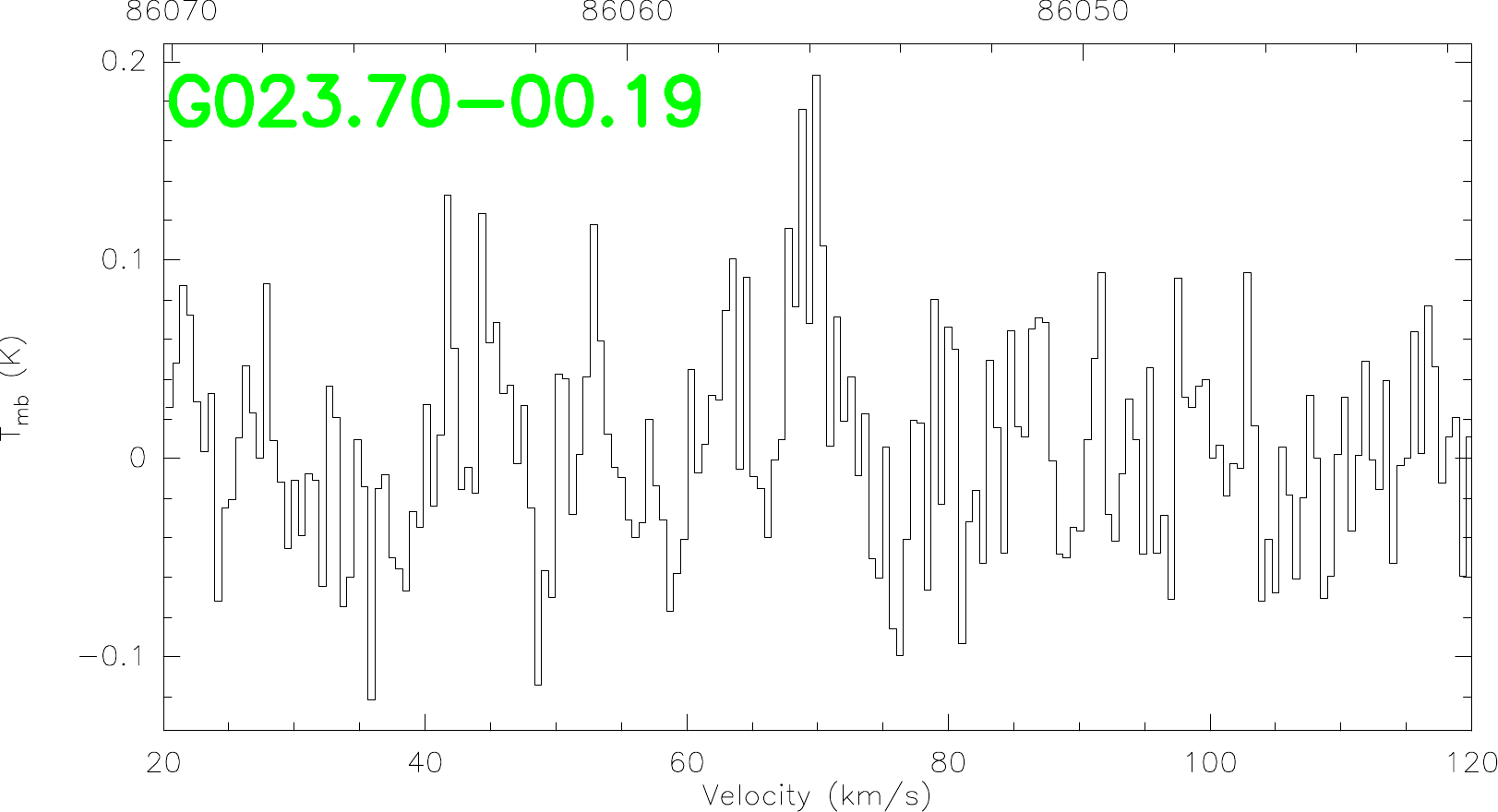}
\includegraphics[width=0.155\textwidth]{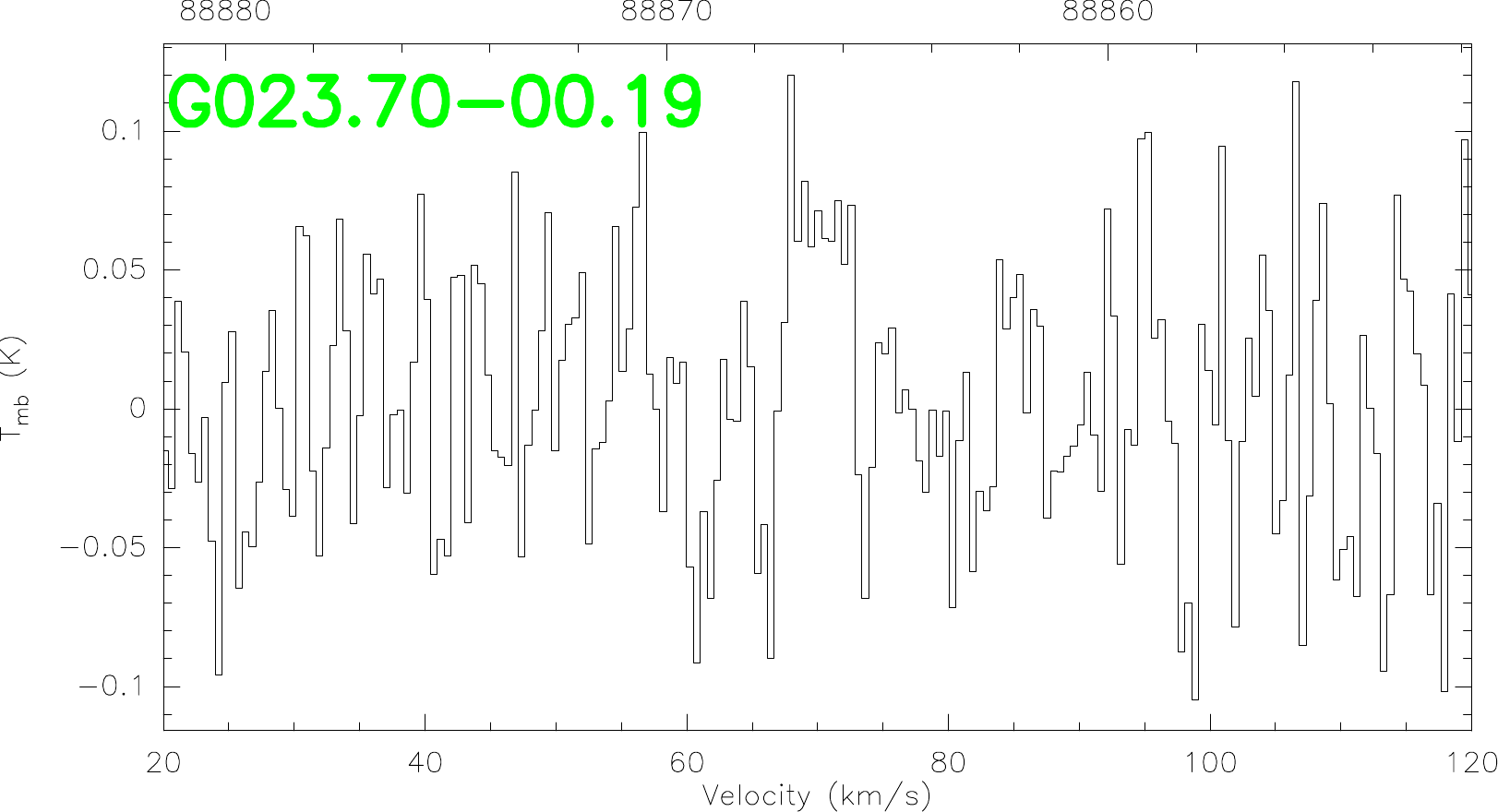}\\

\includegraphics[width=0.155\textwidth]{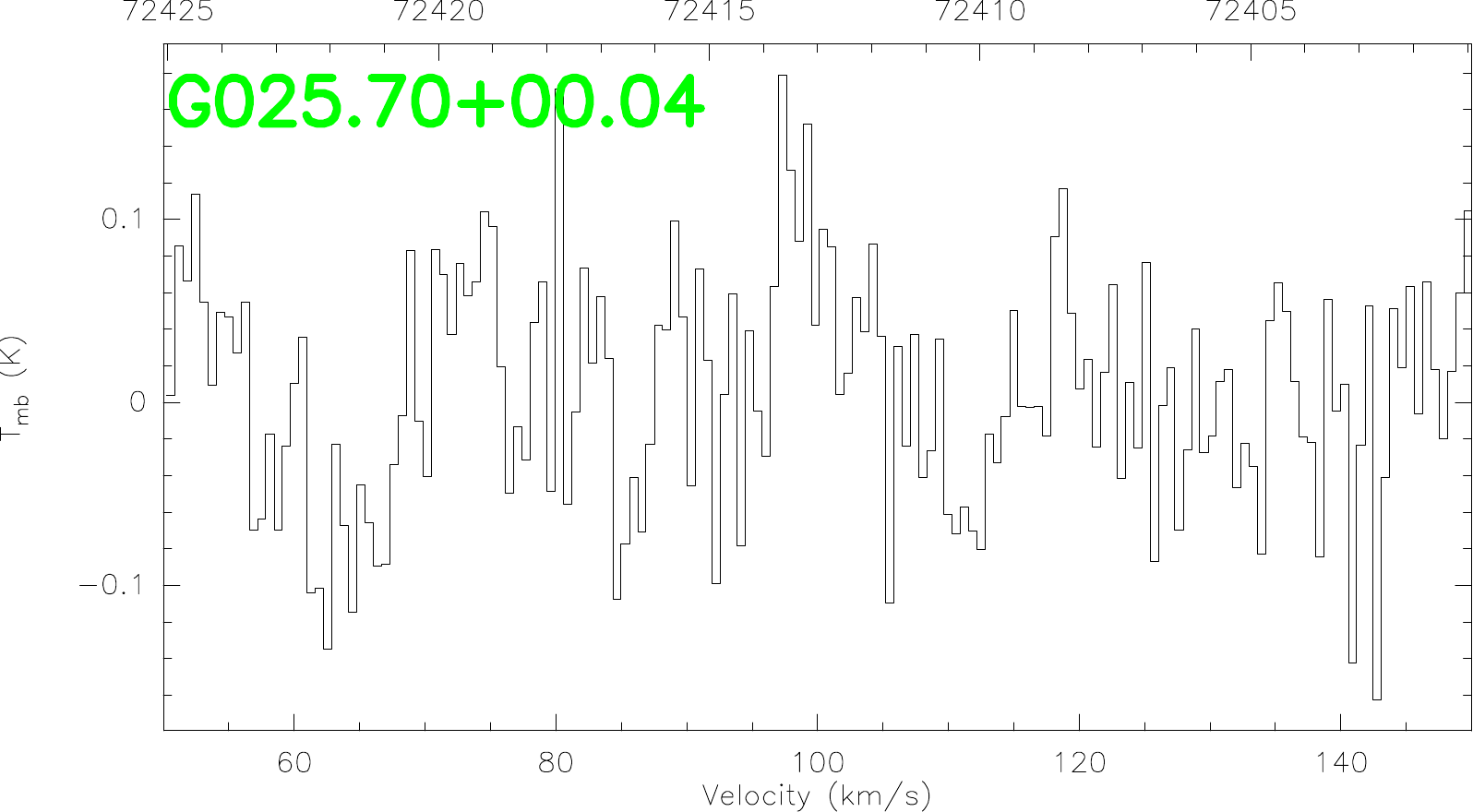}
\includegraphics[width=0.155\textwidth]{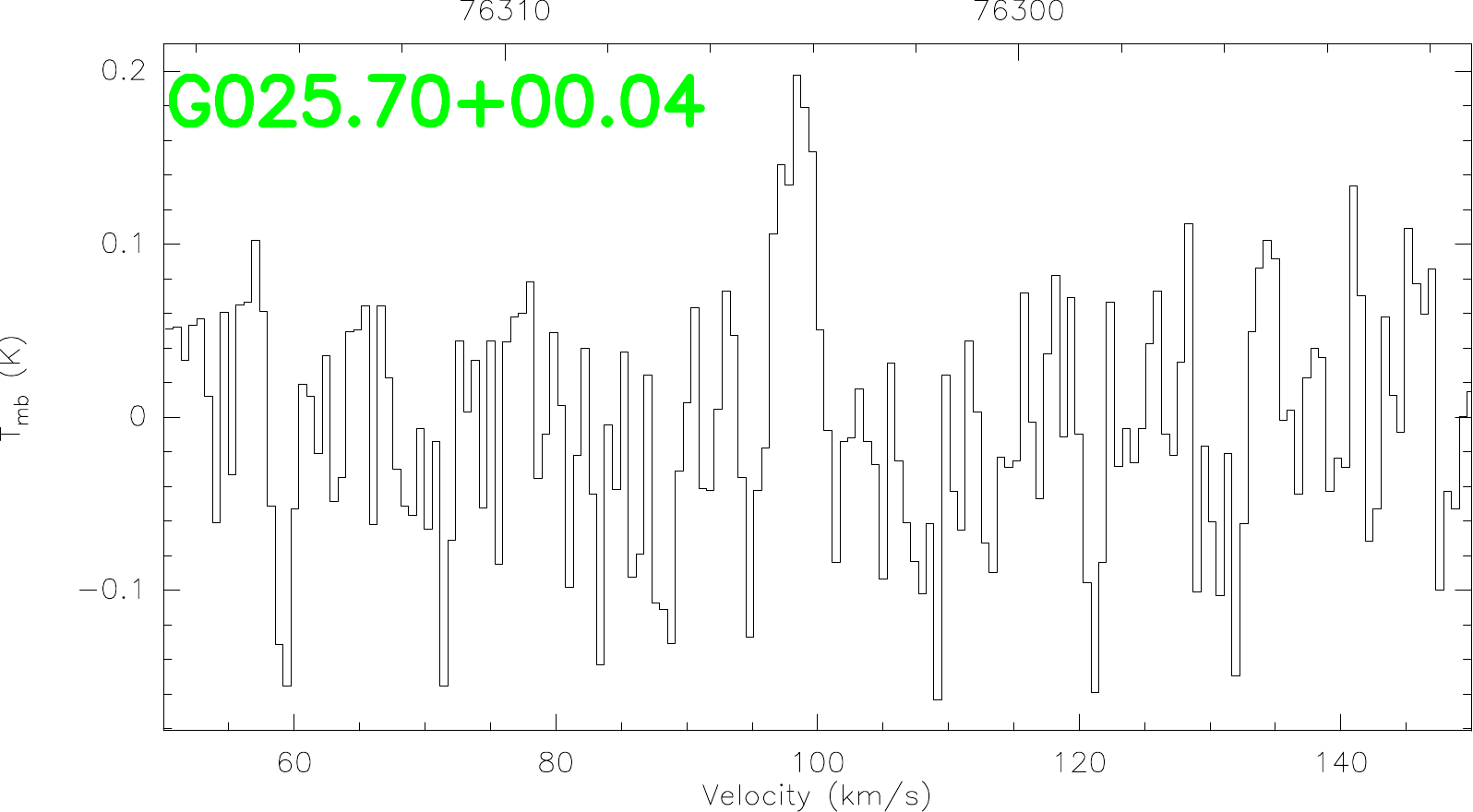}
\includegraphics[width=0.155\textwidth]{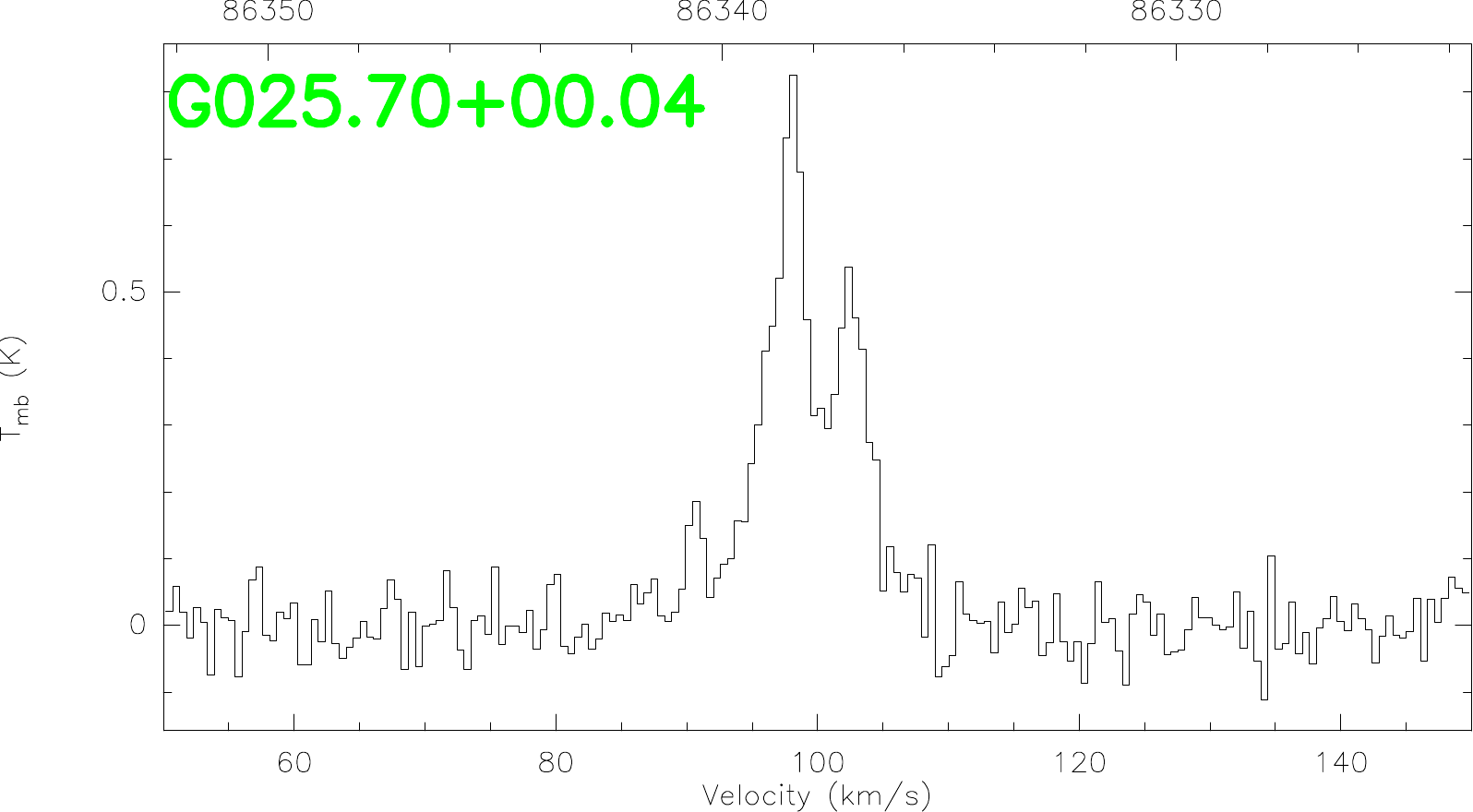}
\includegraphics[width=0.155\textwidth]{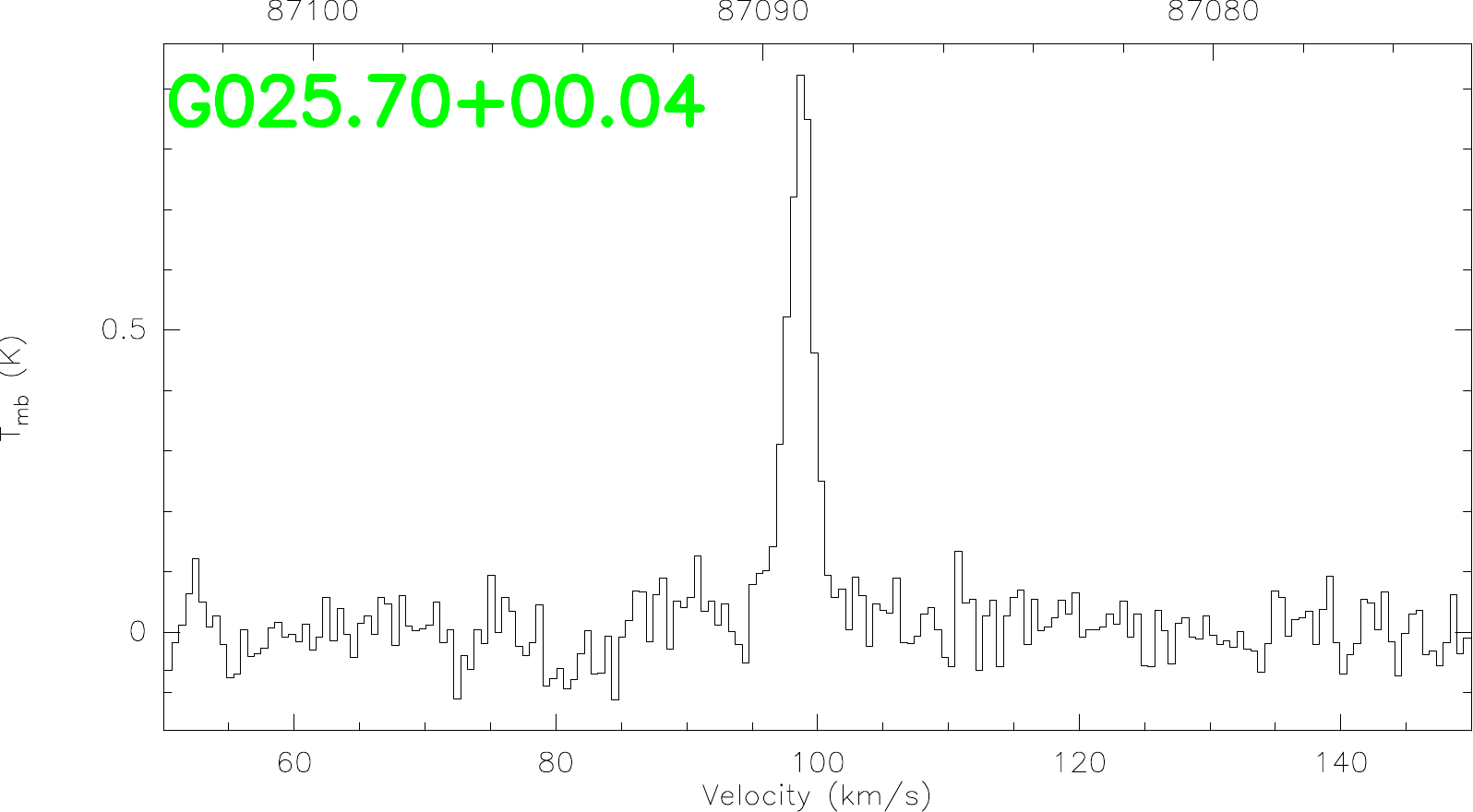}
\includegraphics[width=0.155\textwidth]{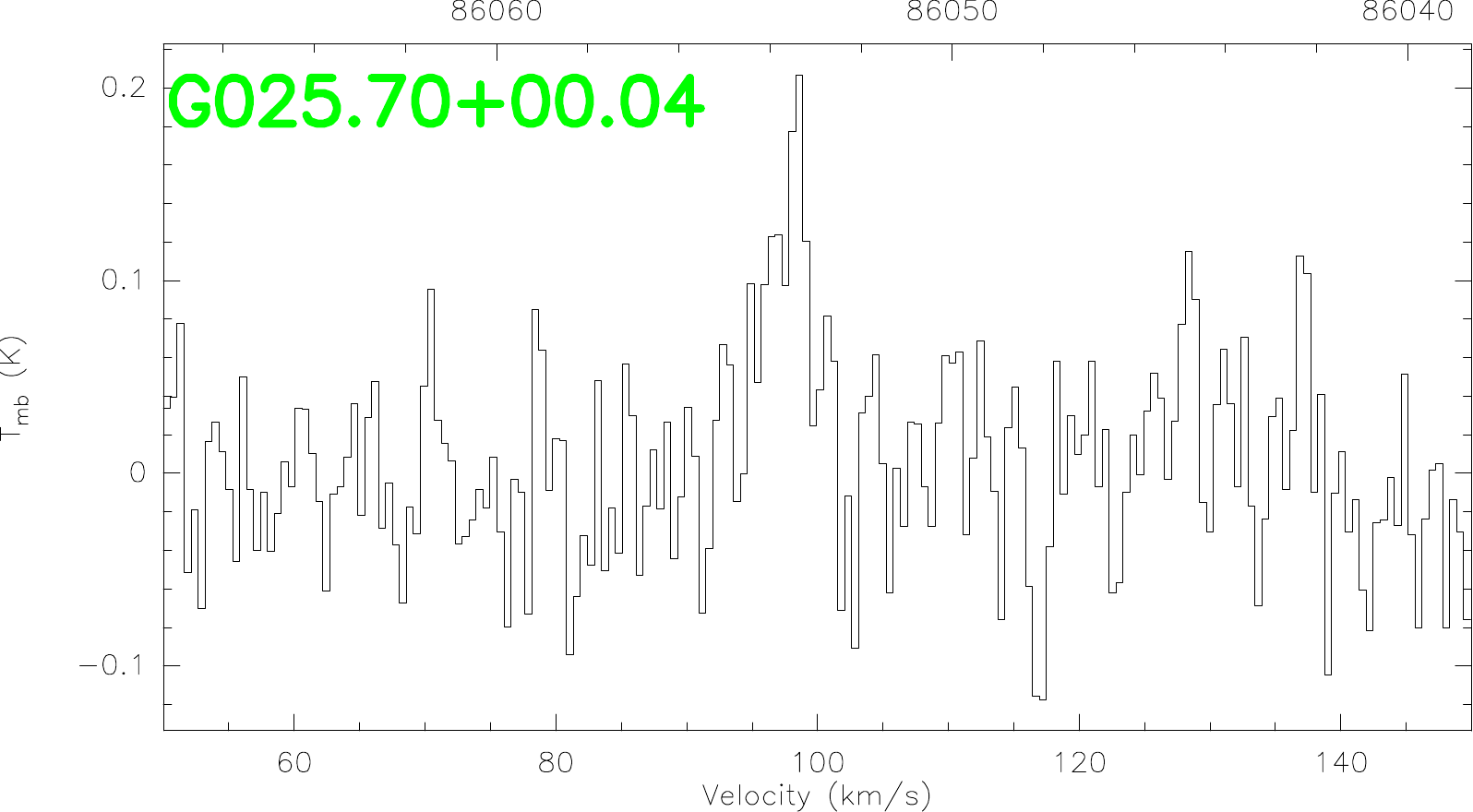}
\includegraphics[width=0.155\textwidth]{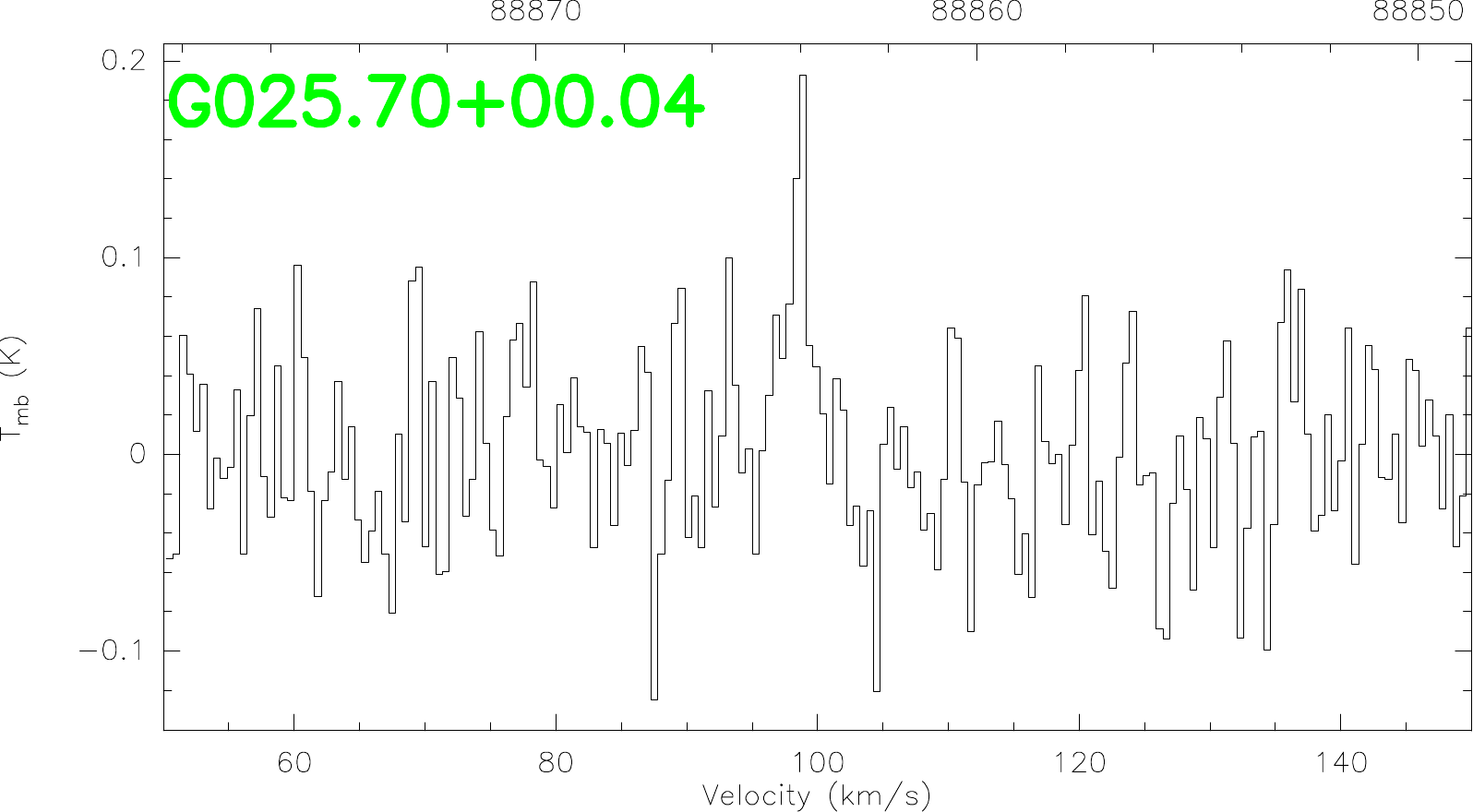}\\

\includegraphics[width=0.155\textwidth]{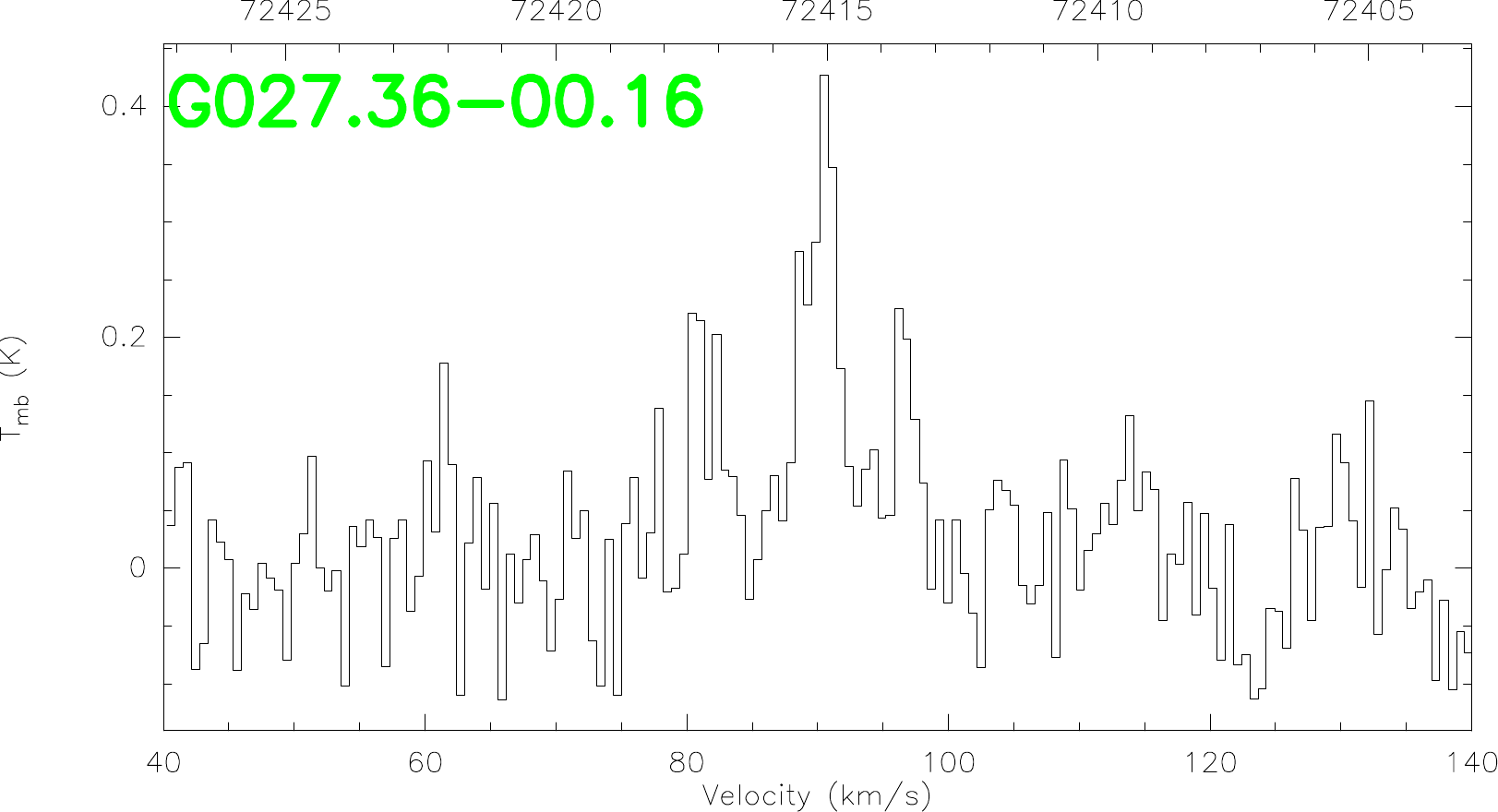}
\includegraphics[width=0.155\textwidth]{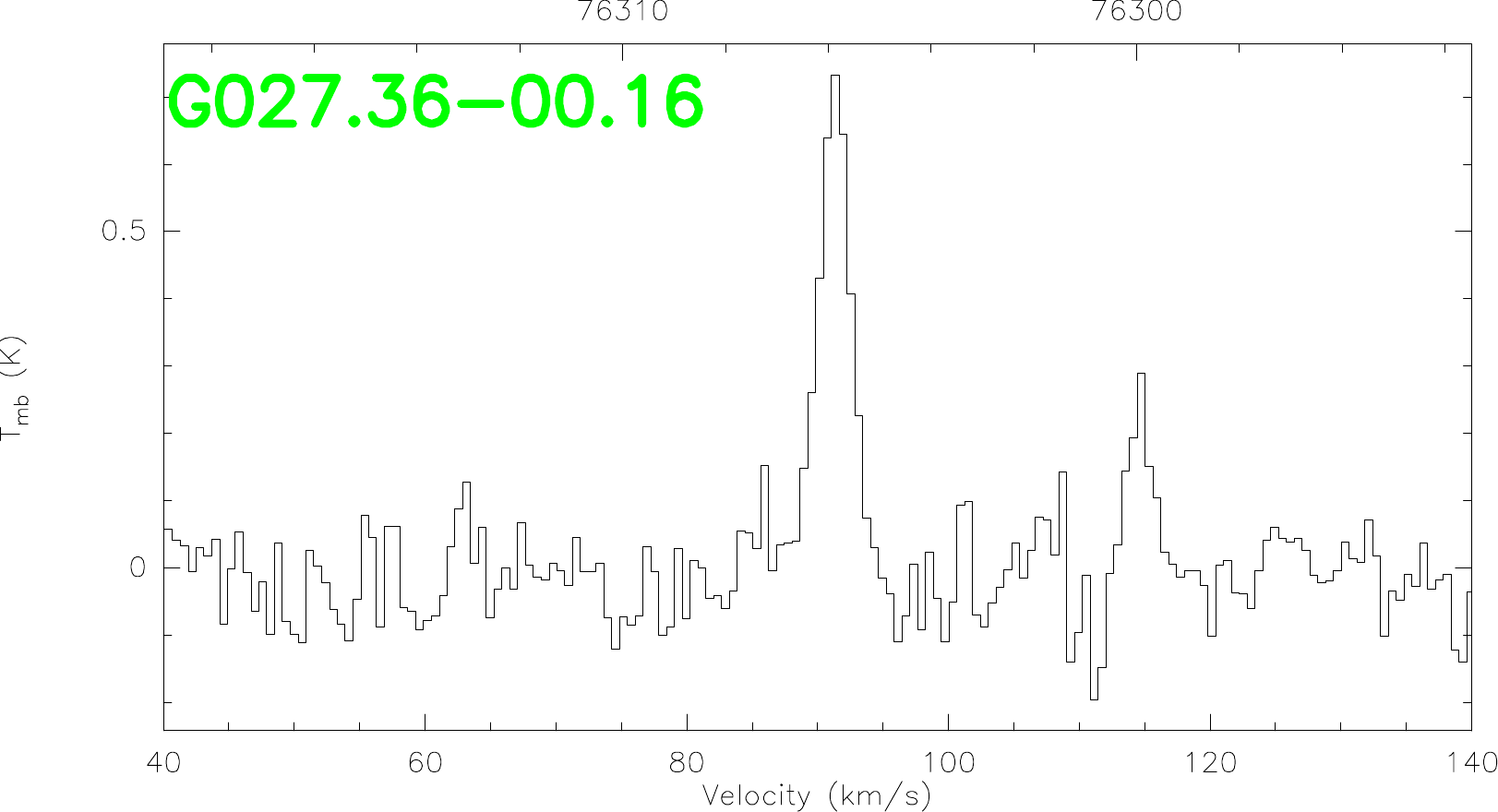}
\includegraphics[width=0.155\textwidth]{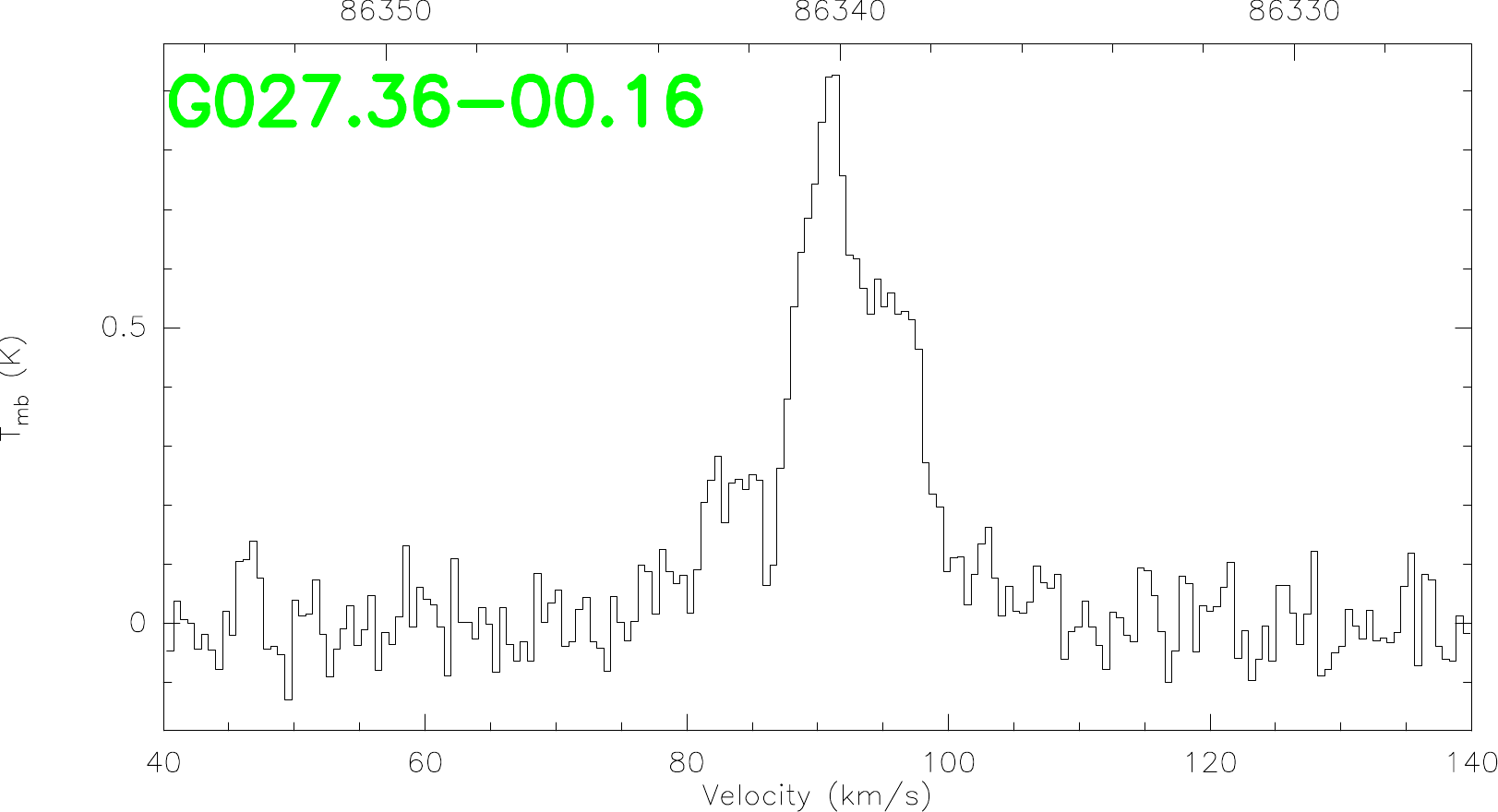}
\includegraphics[width=0.155\textwidth]{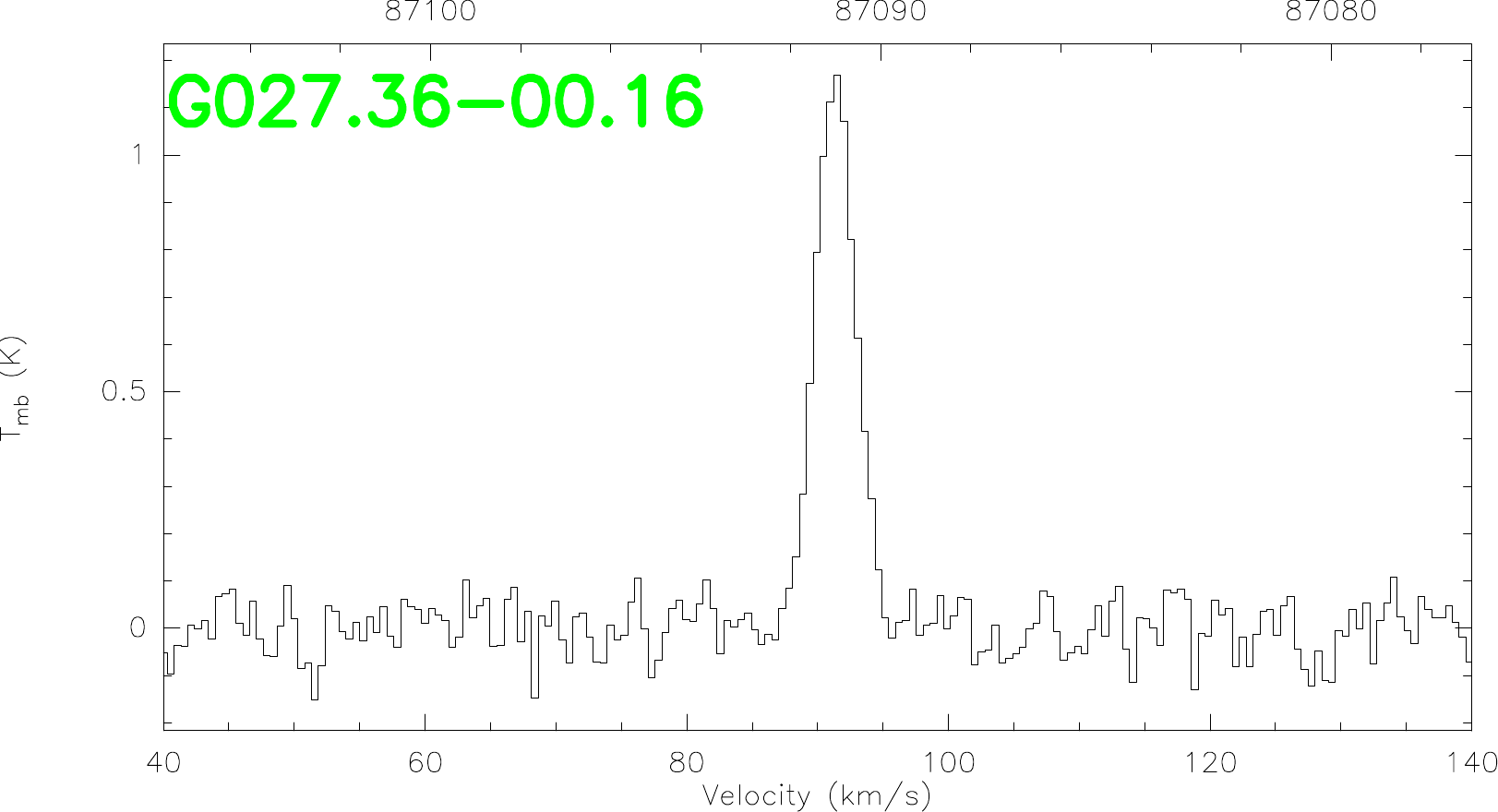}
\includegraphics[width=0.155\textwidth]{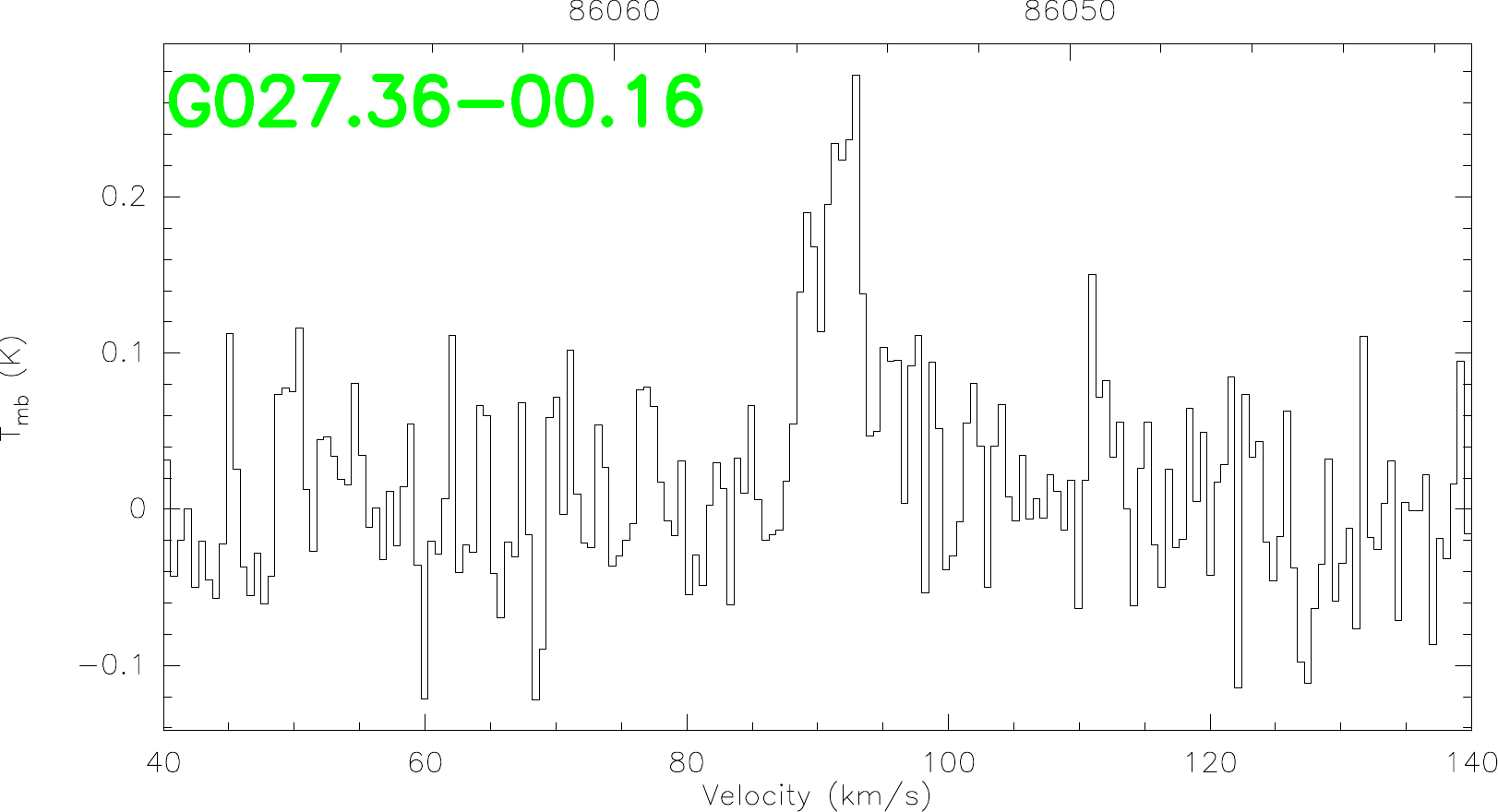}
\includegraphics[width=0.155\textwidth]{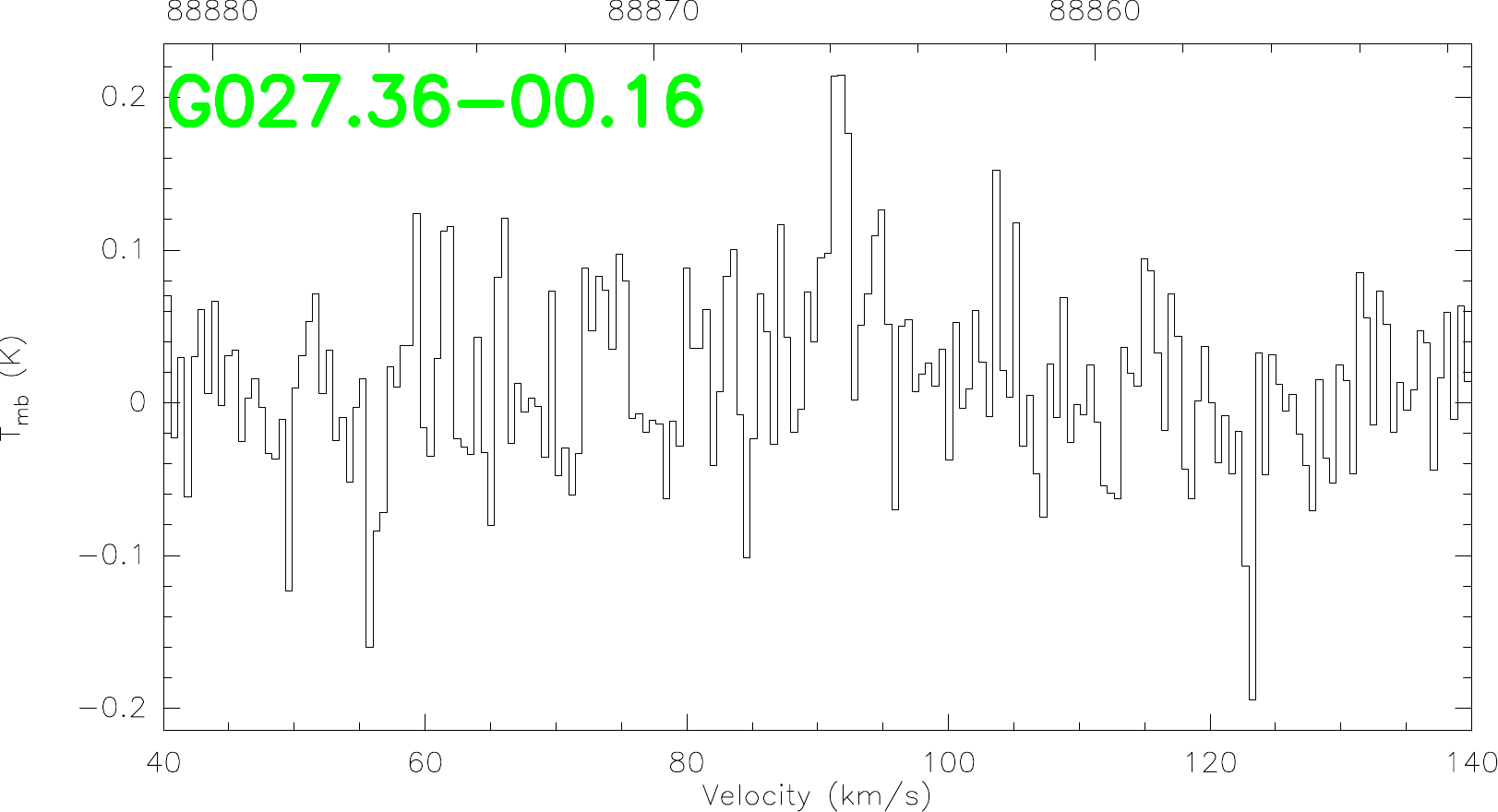}\\

\includegraphics[width=0.155\textwidth]{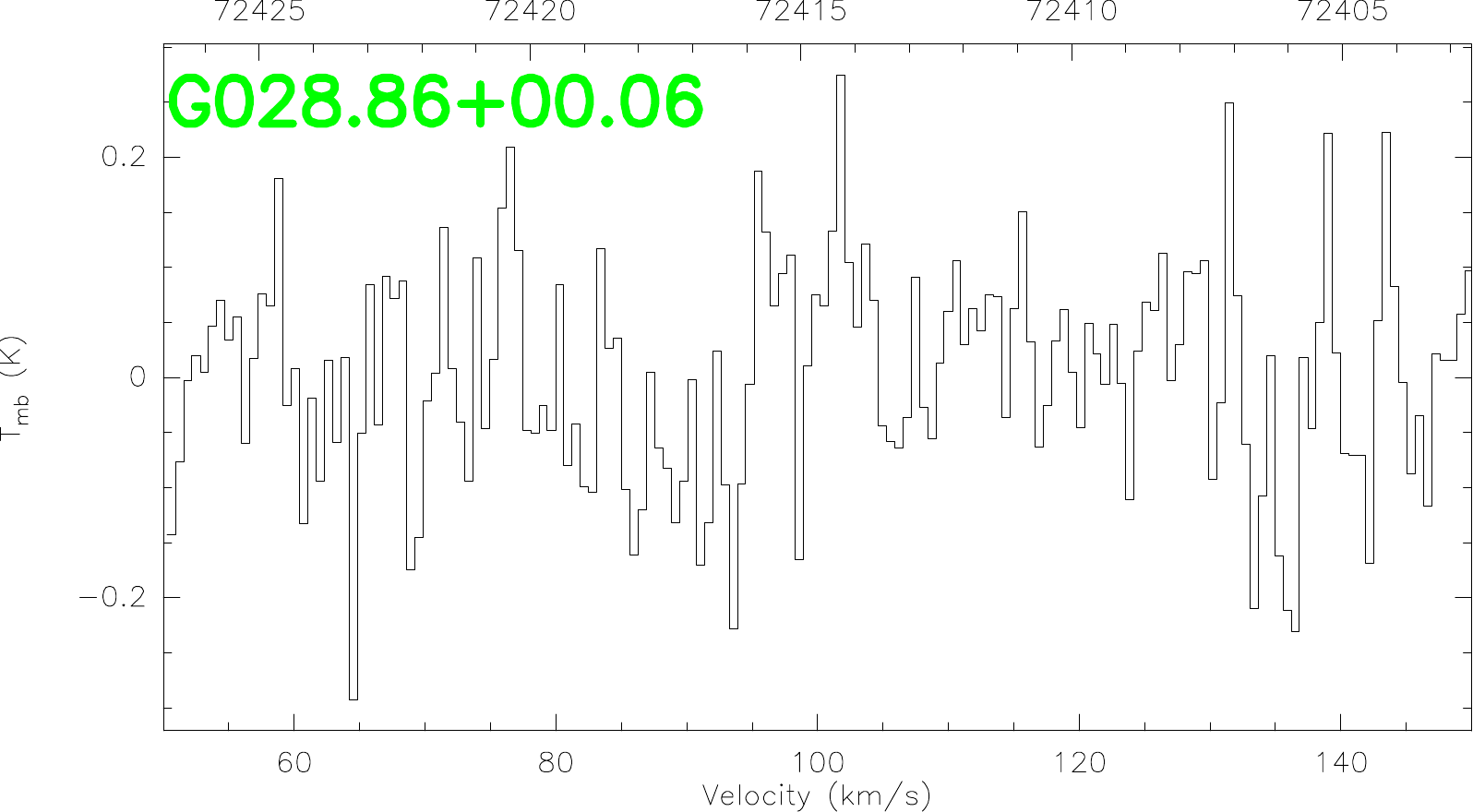}
\includegraphics[width=0.155\textwidth]{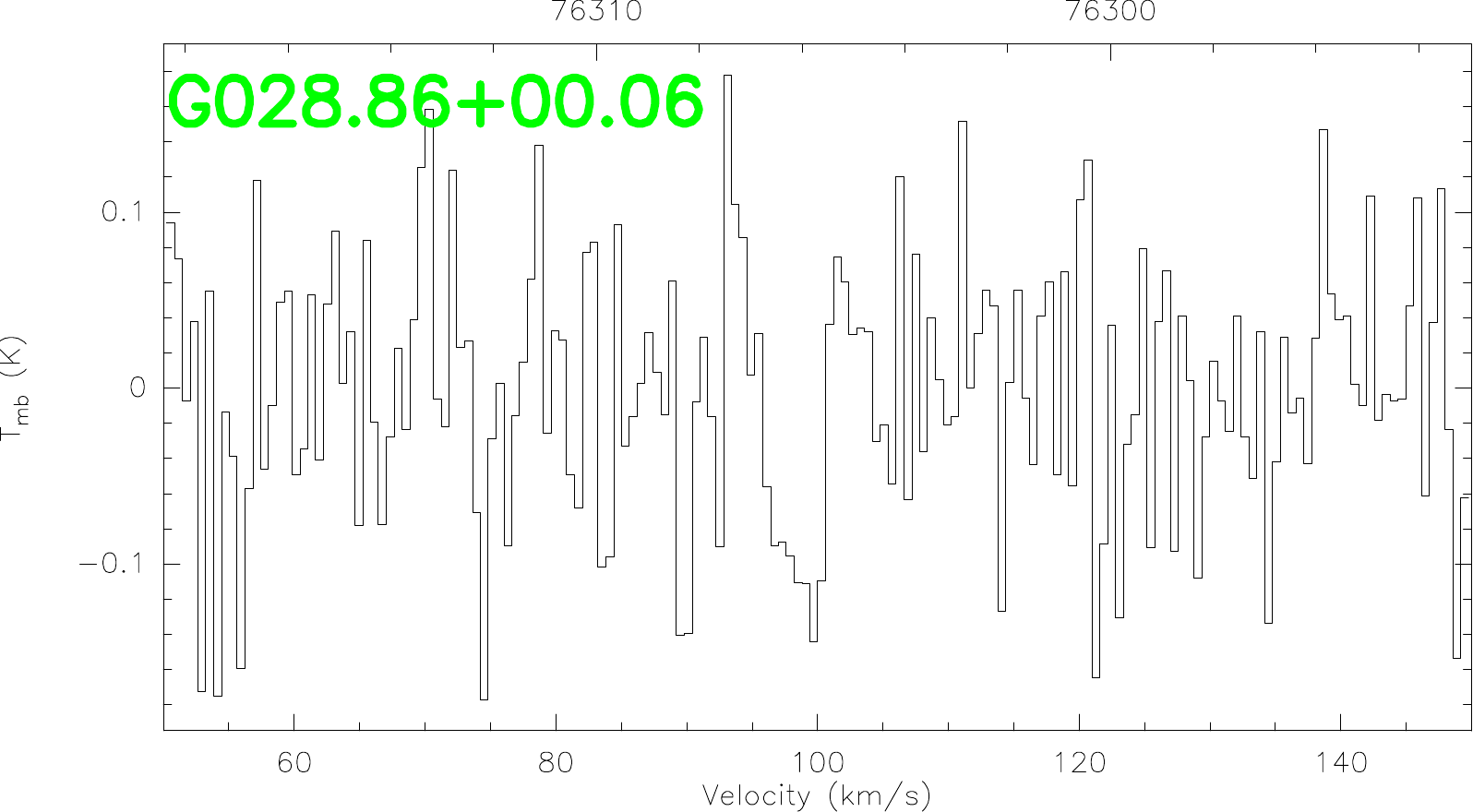}
\includegraphics[width=0.155\textwidth]{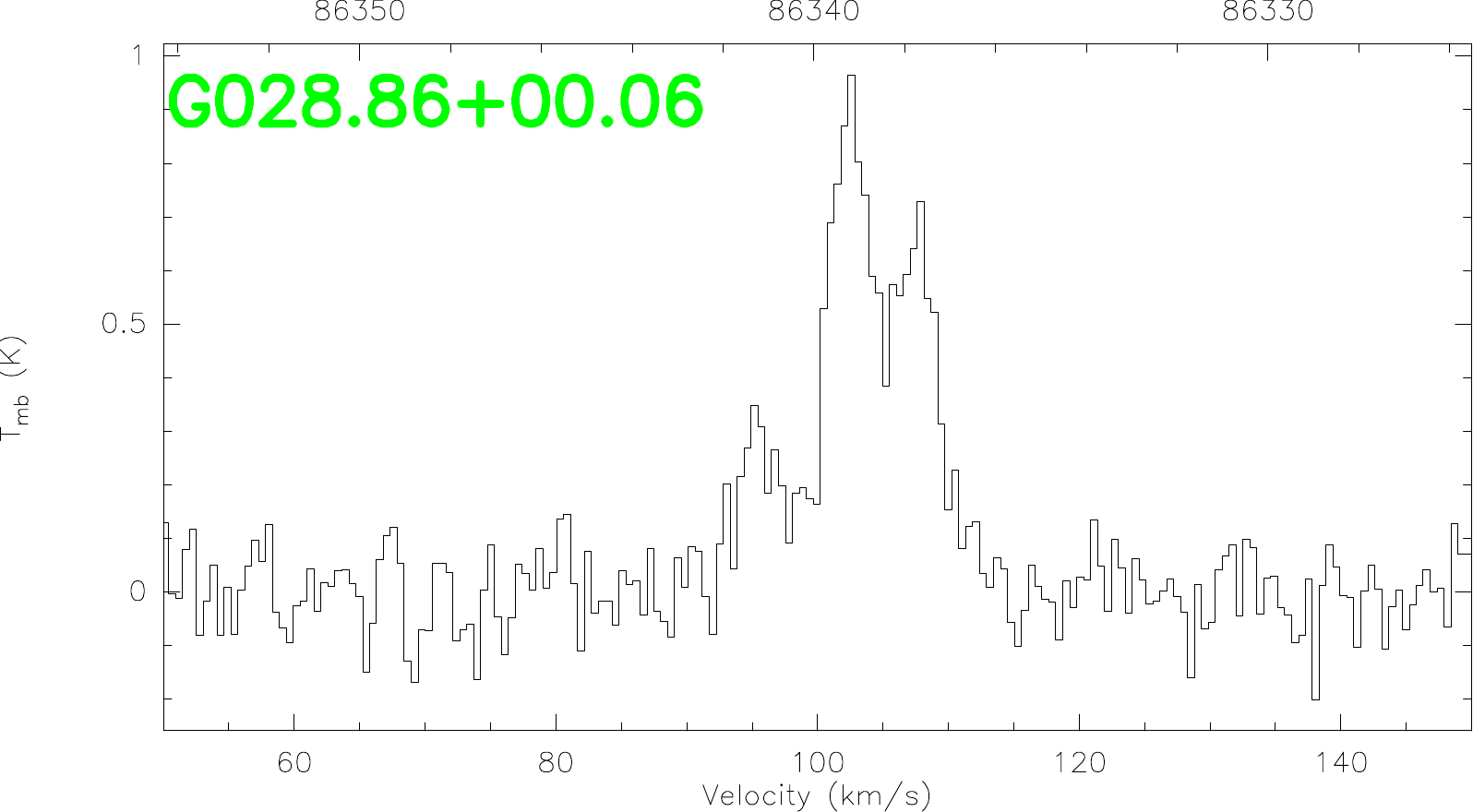}
\includegraphics[width=0.155\textwidth]{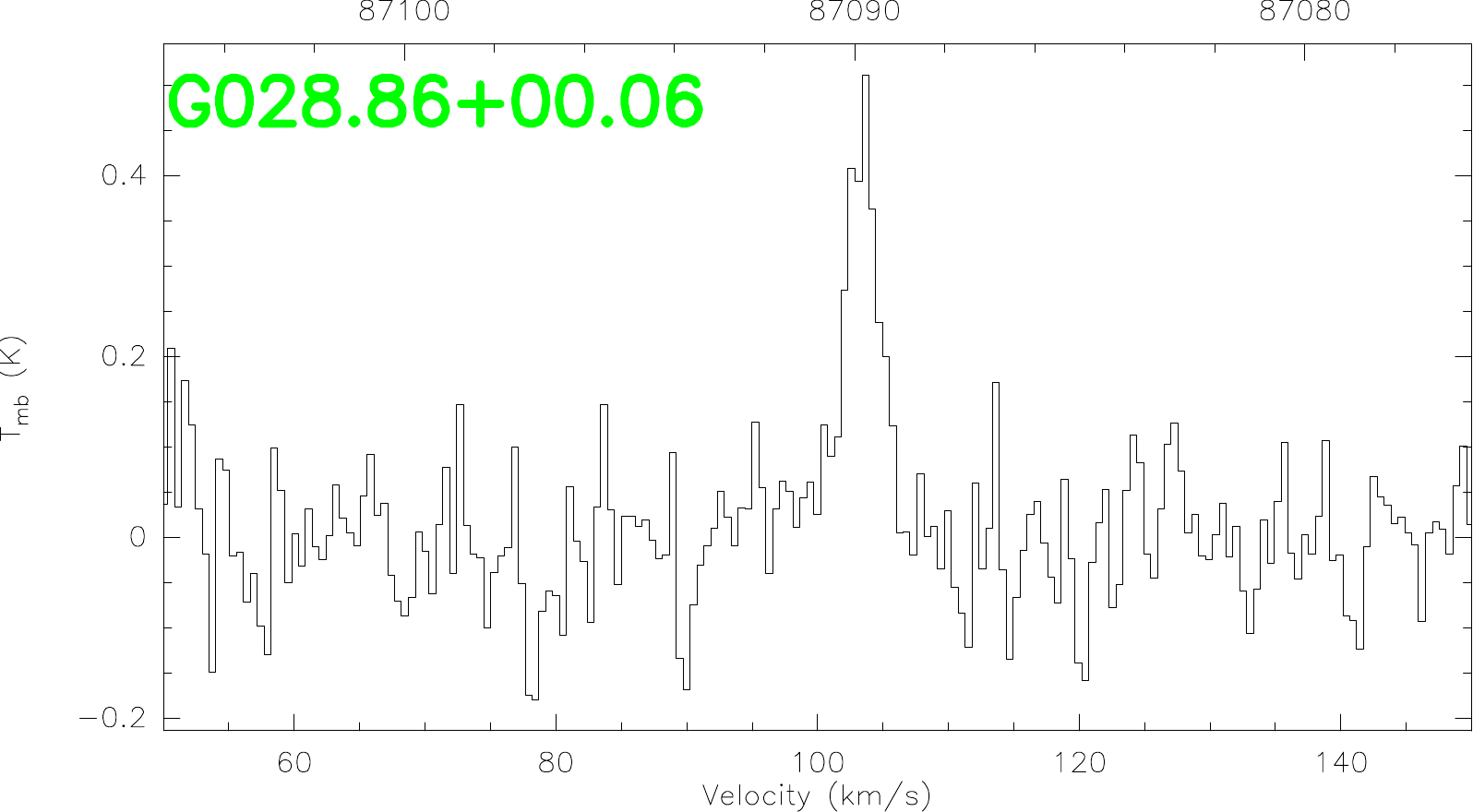}
\includegraphics[width=0.155\textwidth]{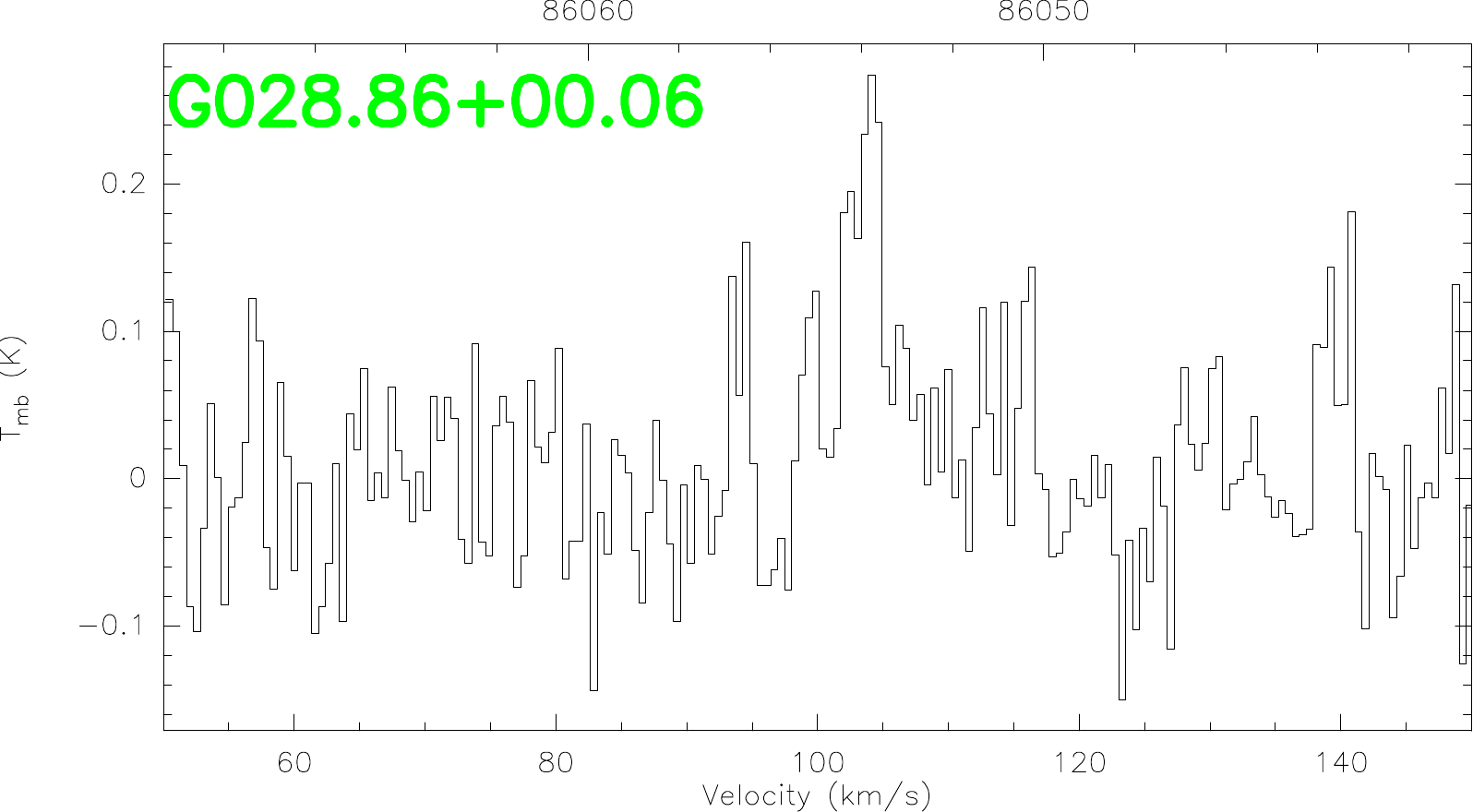}
\includegraphics[width=0.155\textwidth]{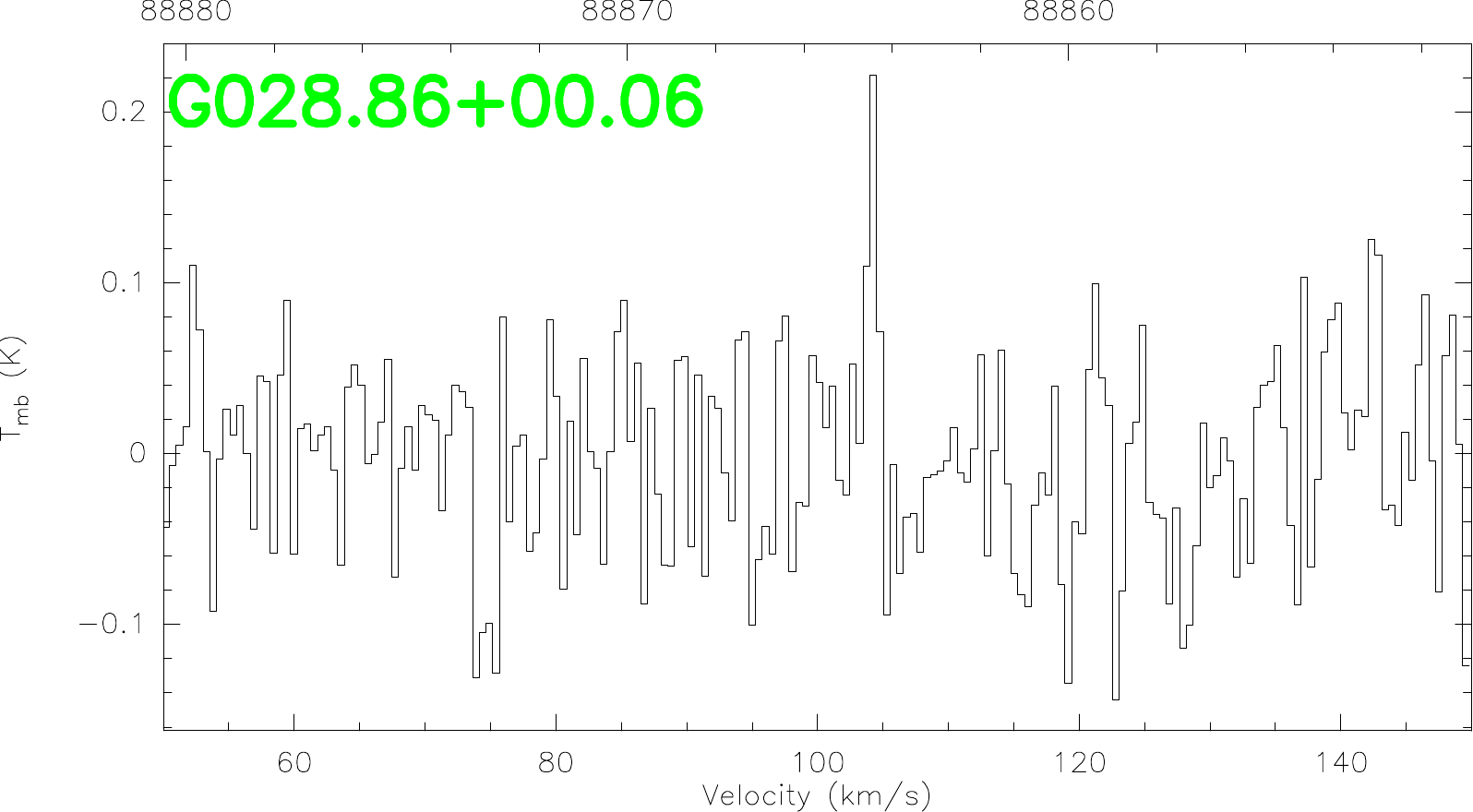}\\

\includegraphics[width=0.155\textwidth]{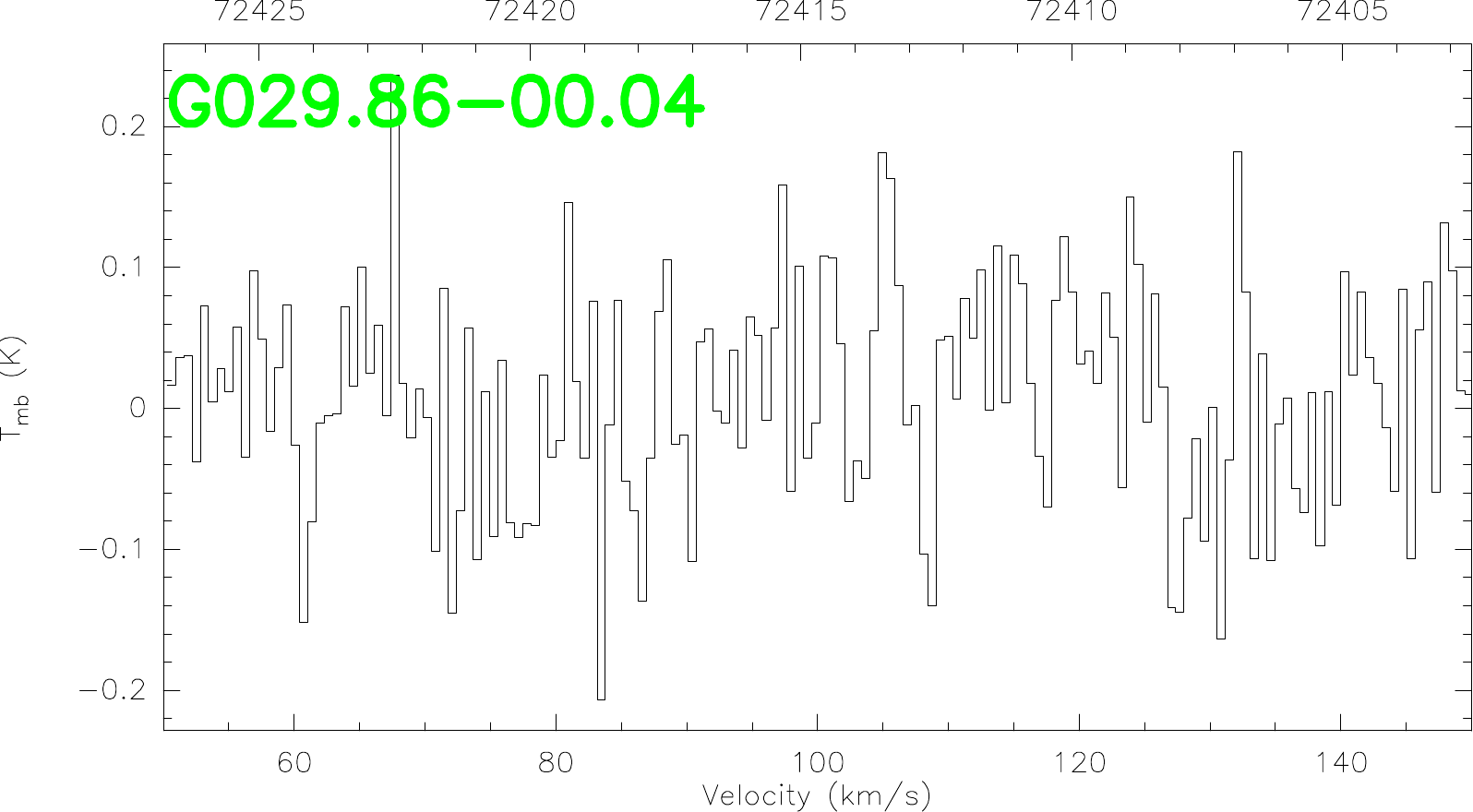}
\includegraphics[width=0.155\textwidth]{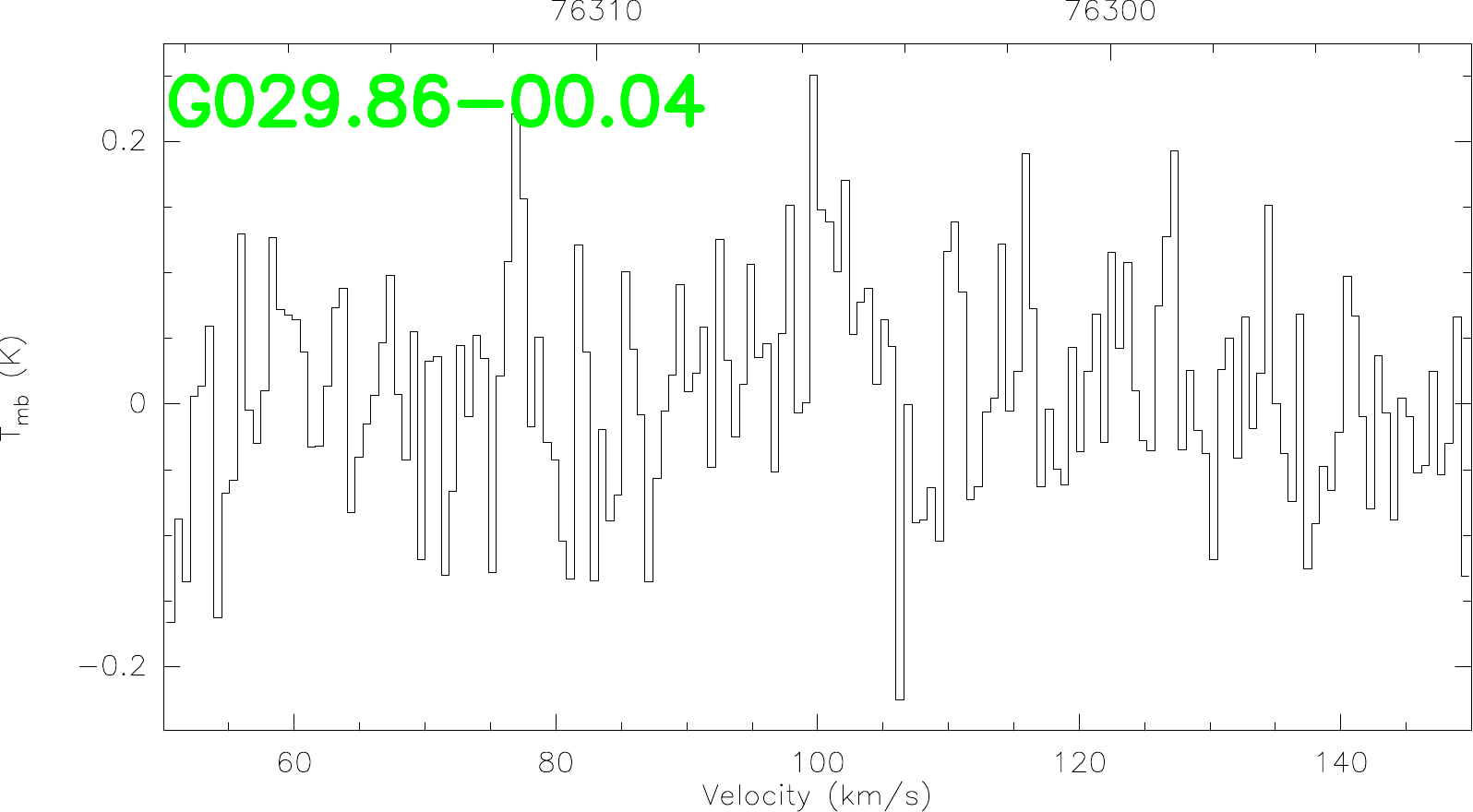}
\includegraphics[width=0.155\textwidth]{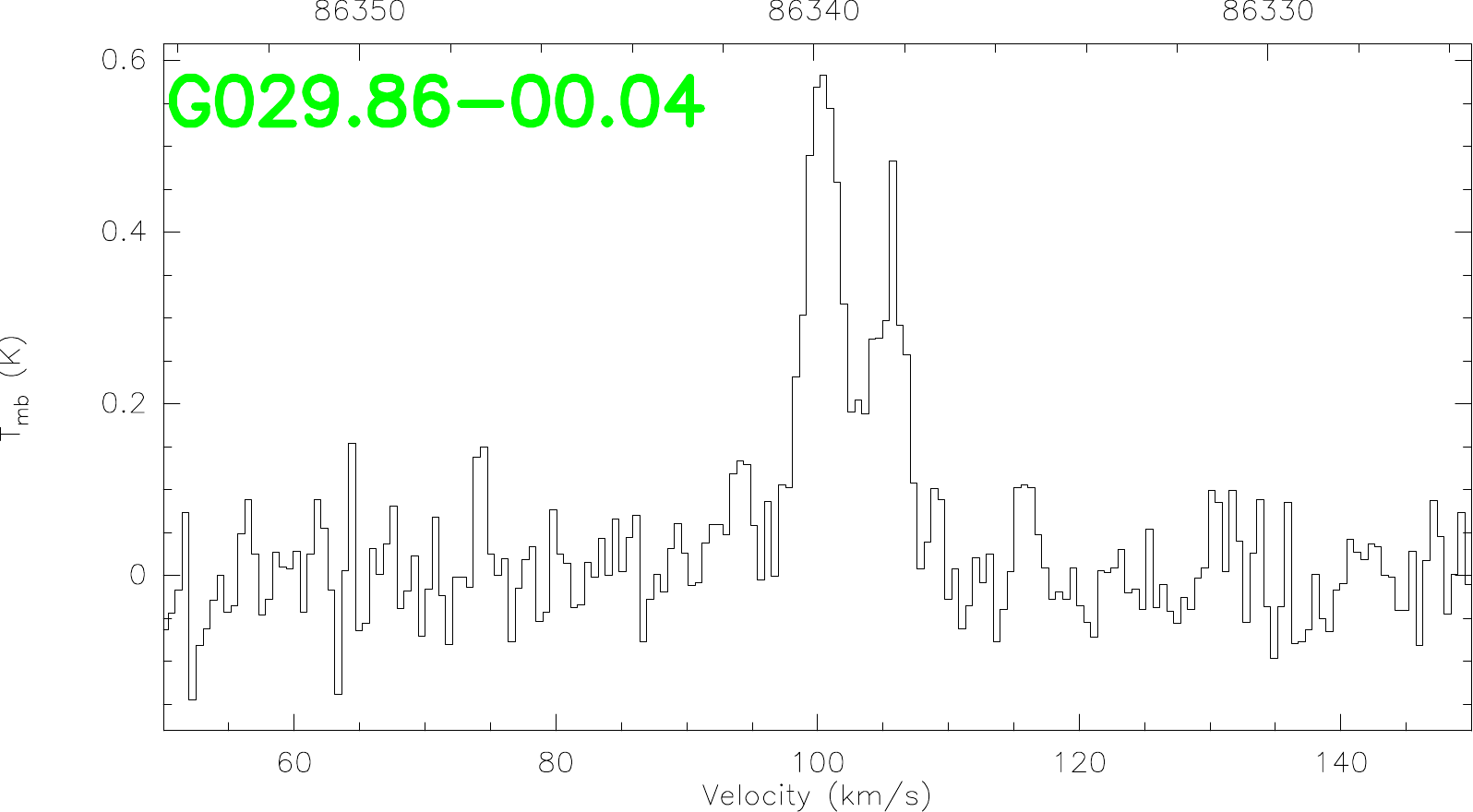}
\includegraphics[width=0.155\textwidth]{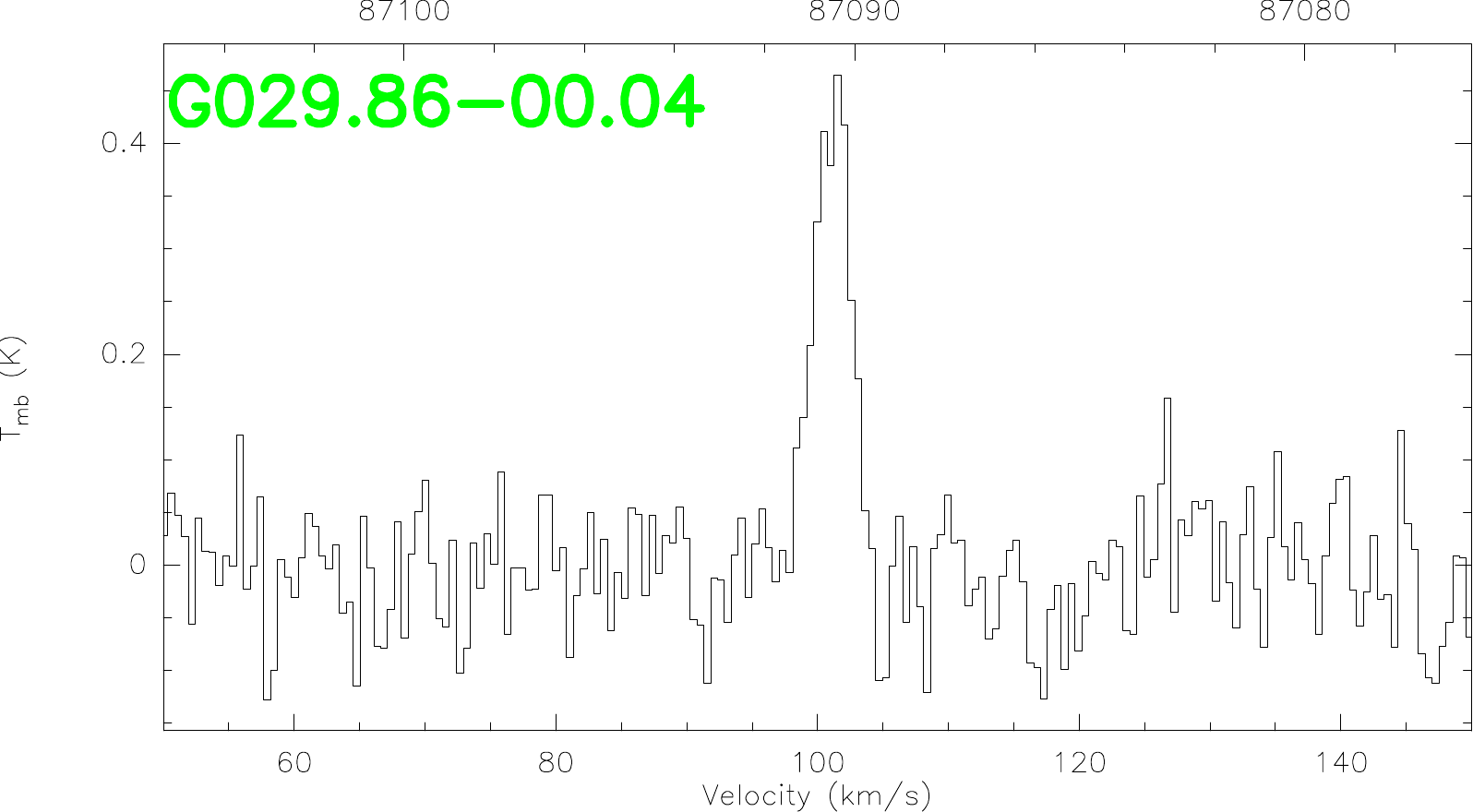}
\includegraphics[width=0.155\textwidth]{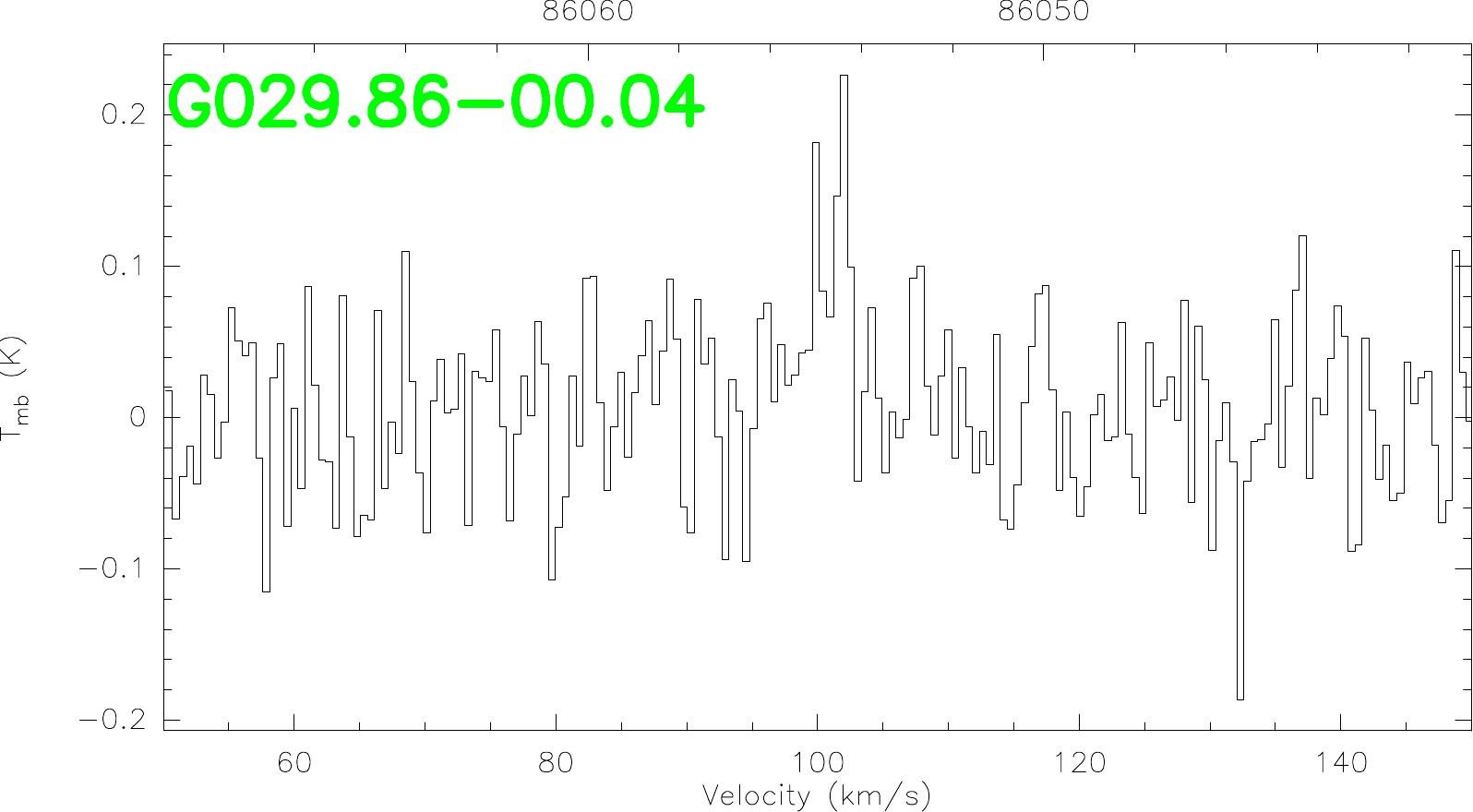}
\includegraphics[width=0.155\textwidth]{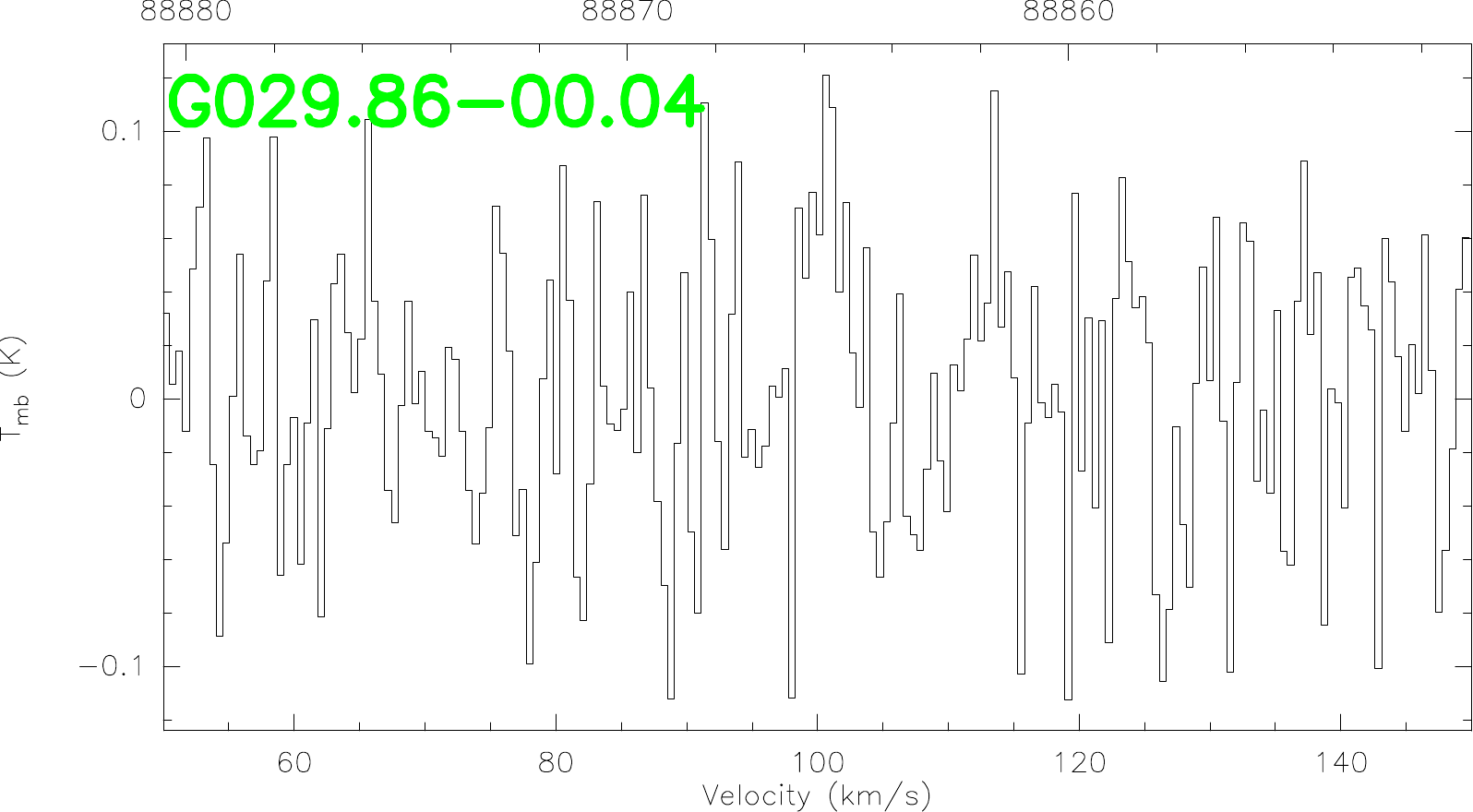}\\

\includegraphics[width=0.155\textwidth]{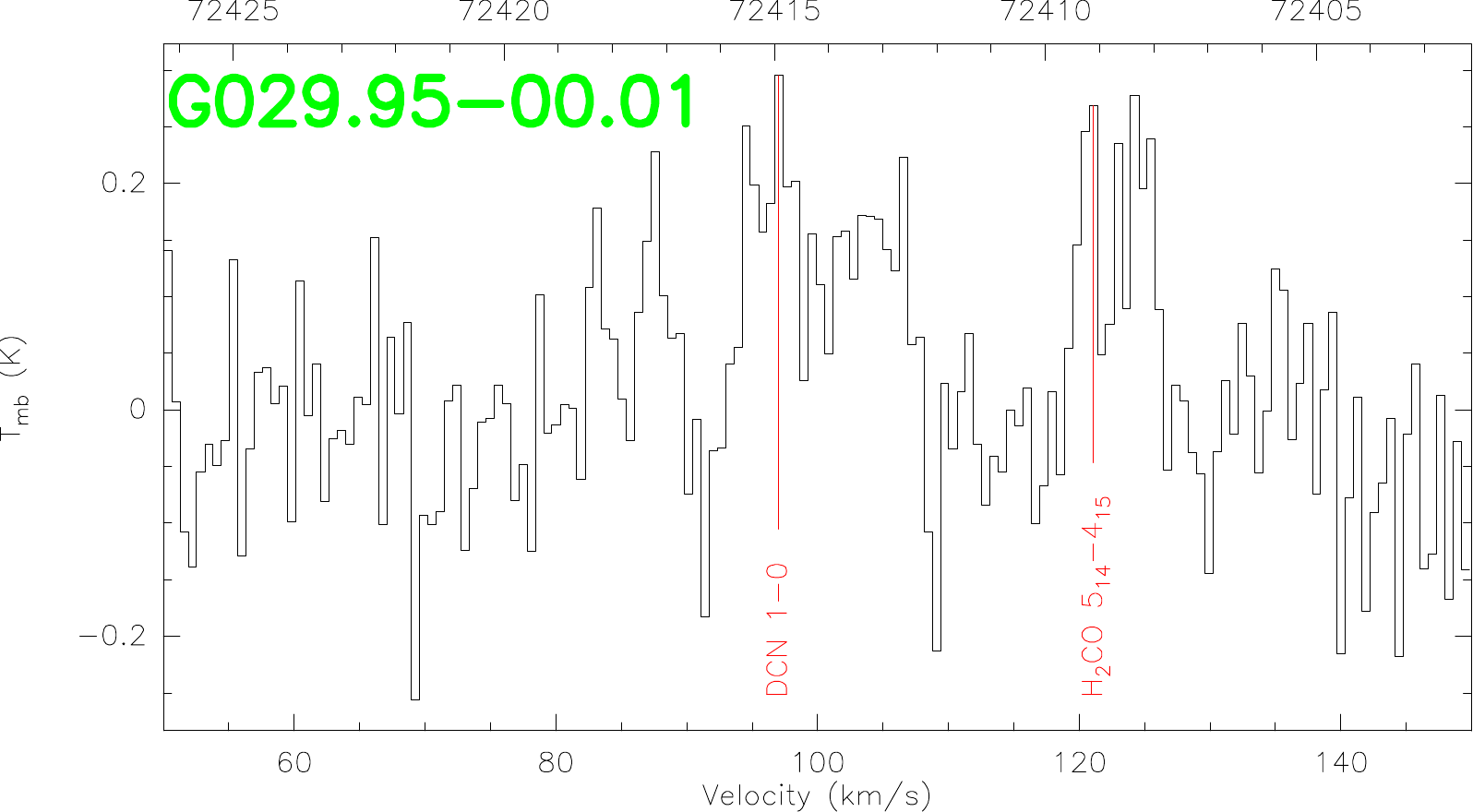}
\includegraphics[width=0.155\textwidth]{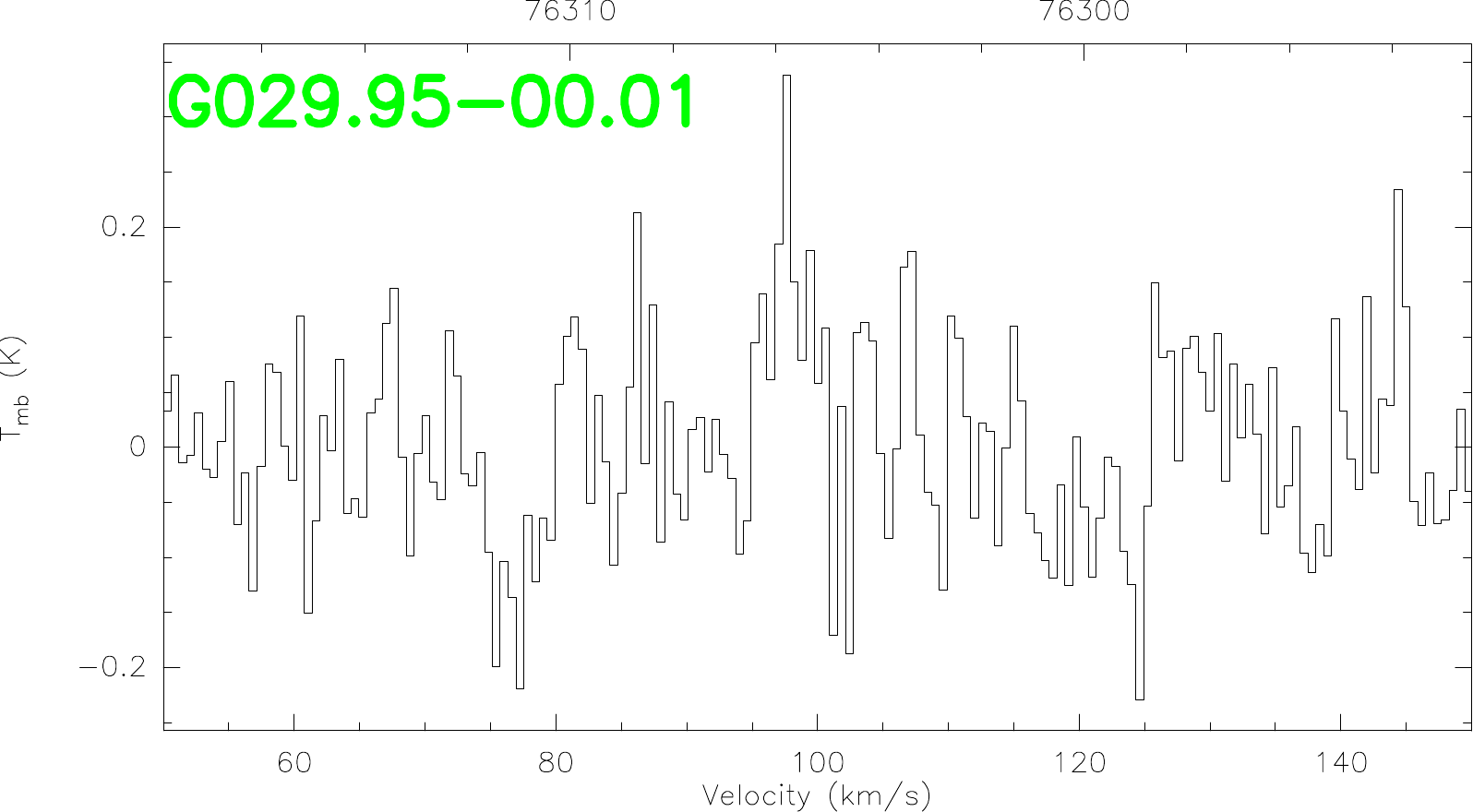}
\includegraphics[width=0.155\textwidth]{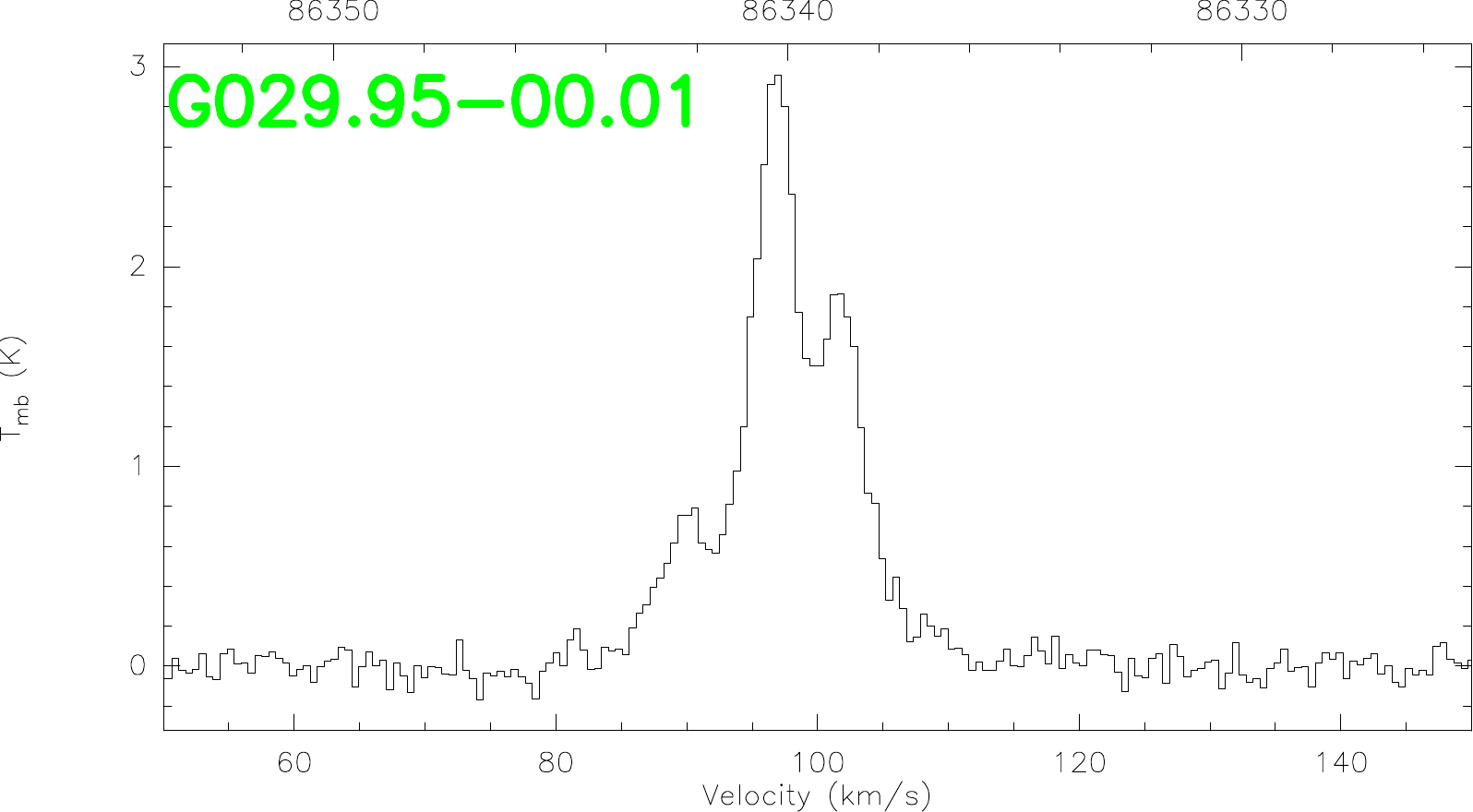}
\includegraphics[width=0.155\textwidth]{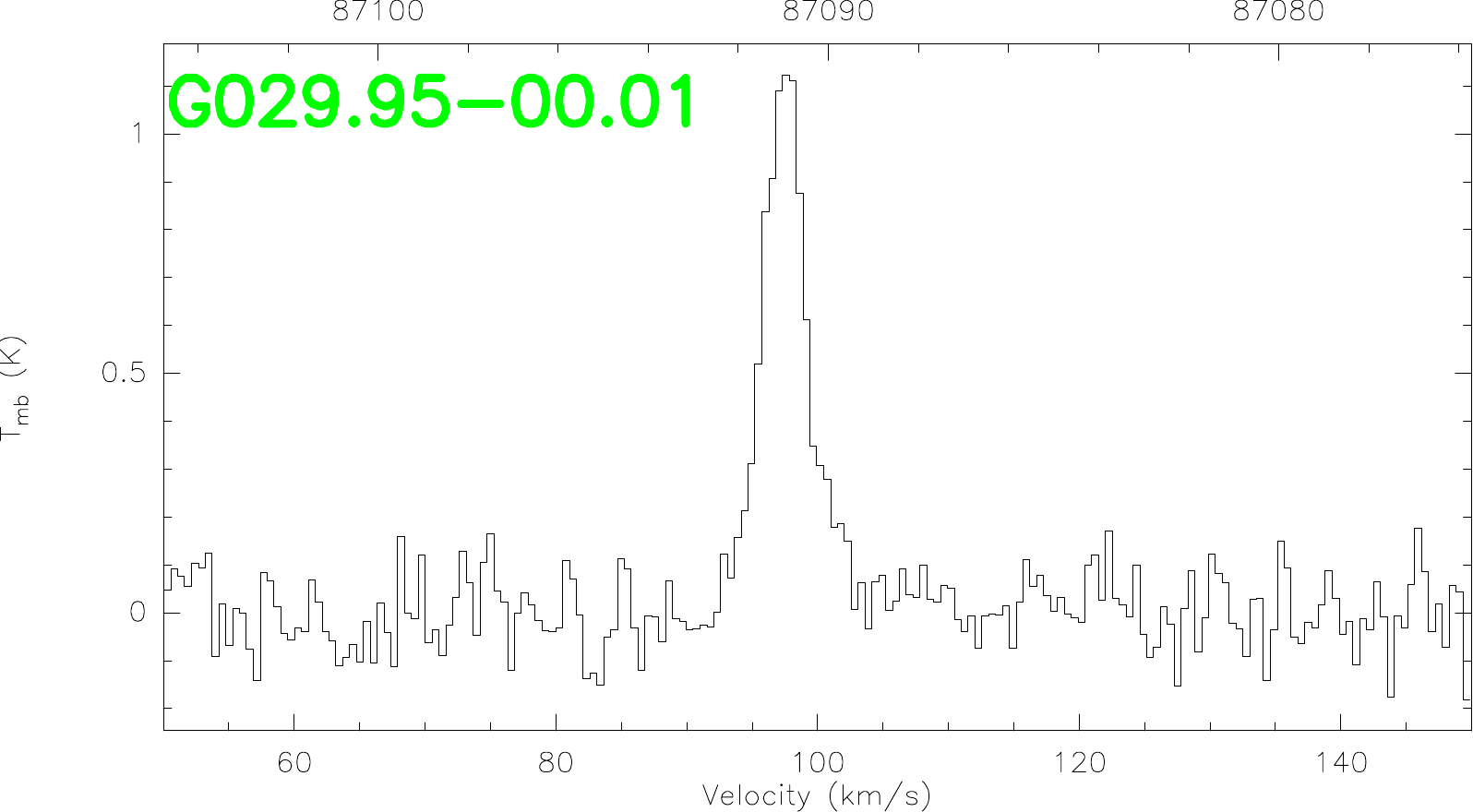}
\includegraphics[width=0.155\textwidth]{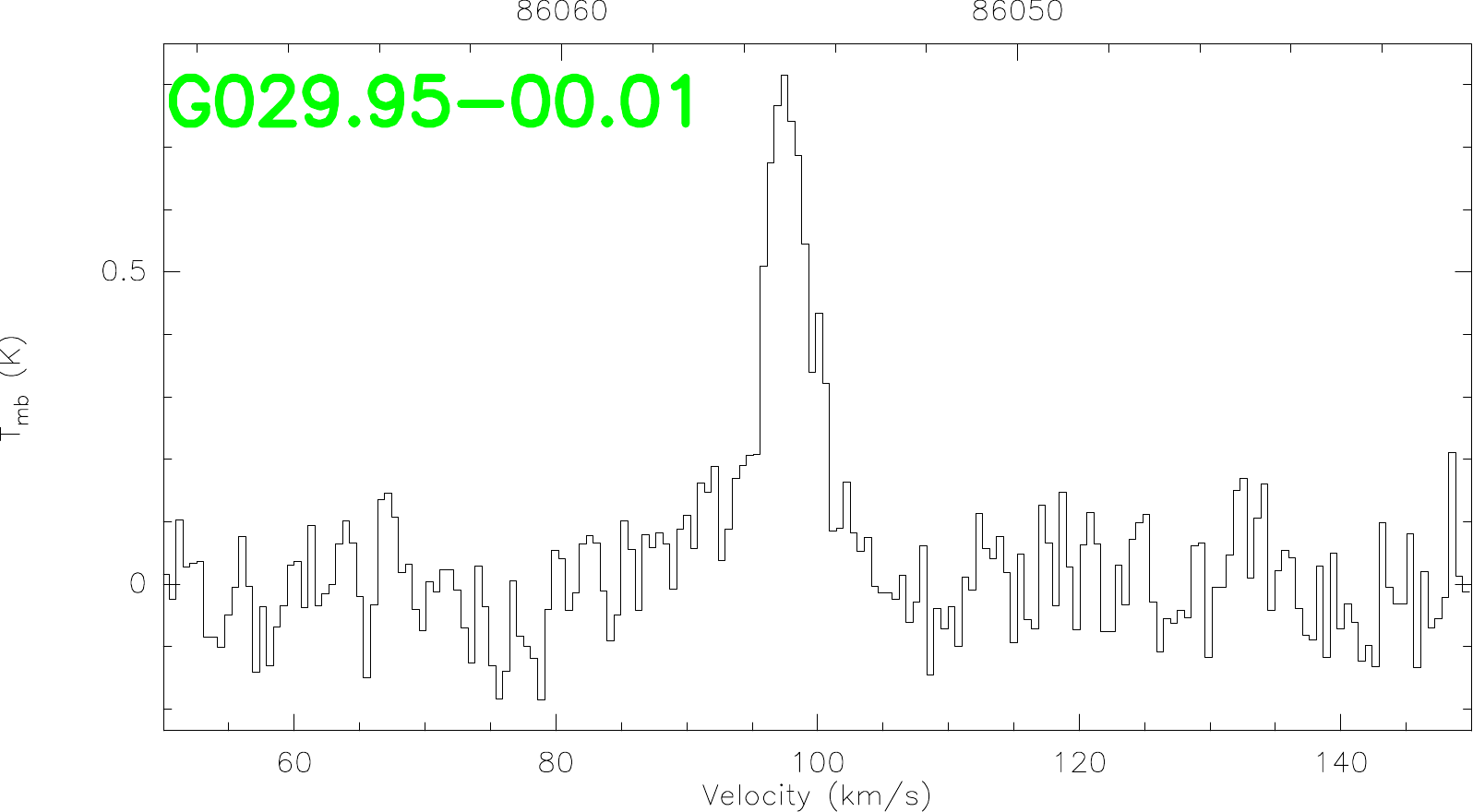}
\includegraphics[width=0.155\textwidth]{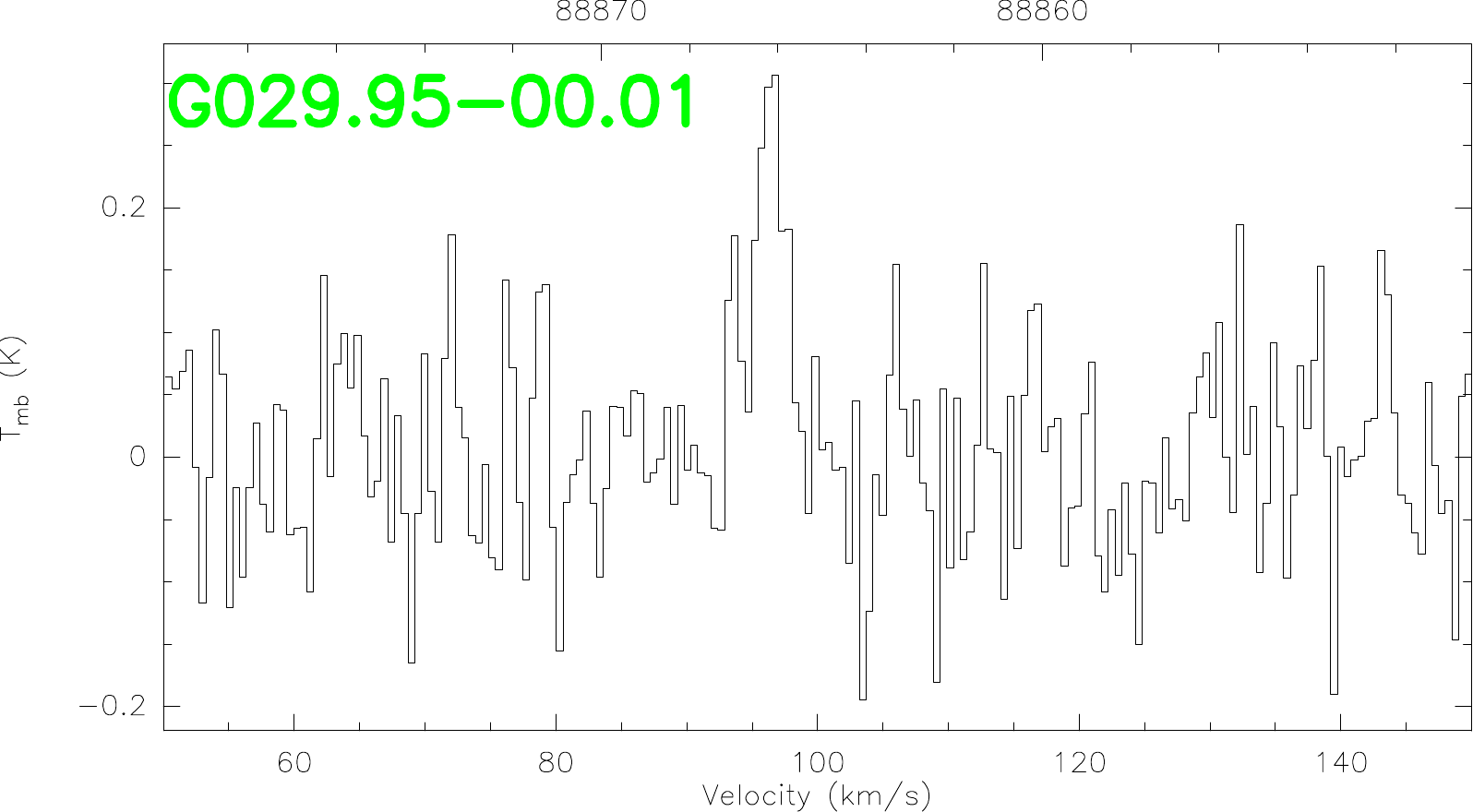}\\

\includegraphics[width=0.155\textwidth]{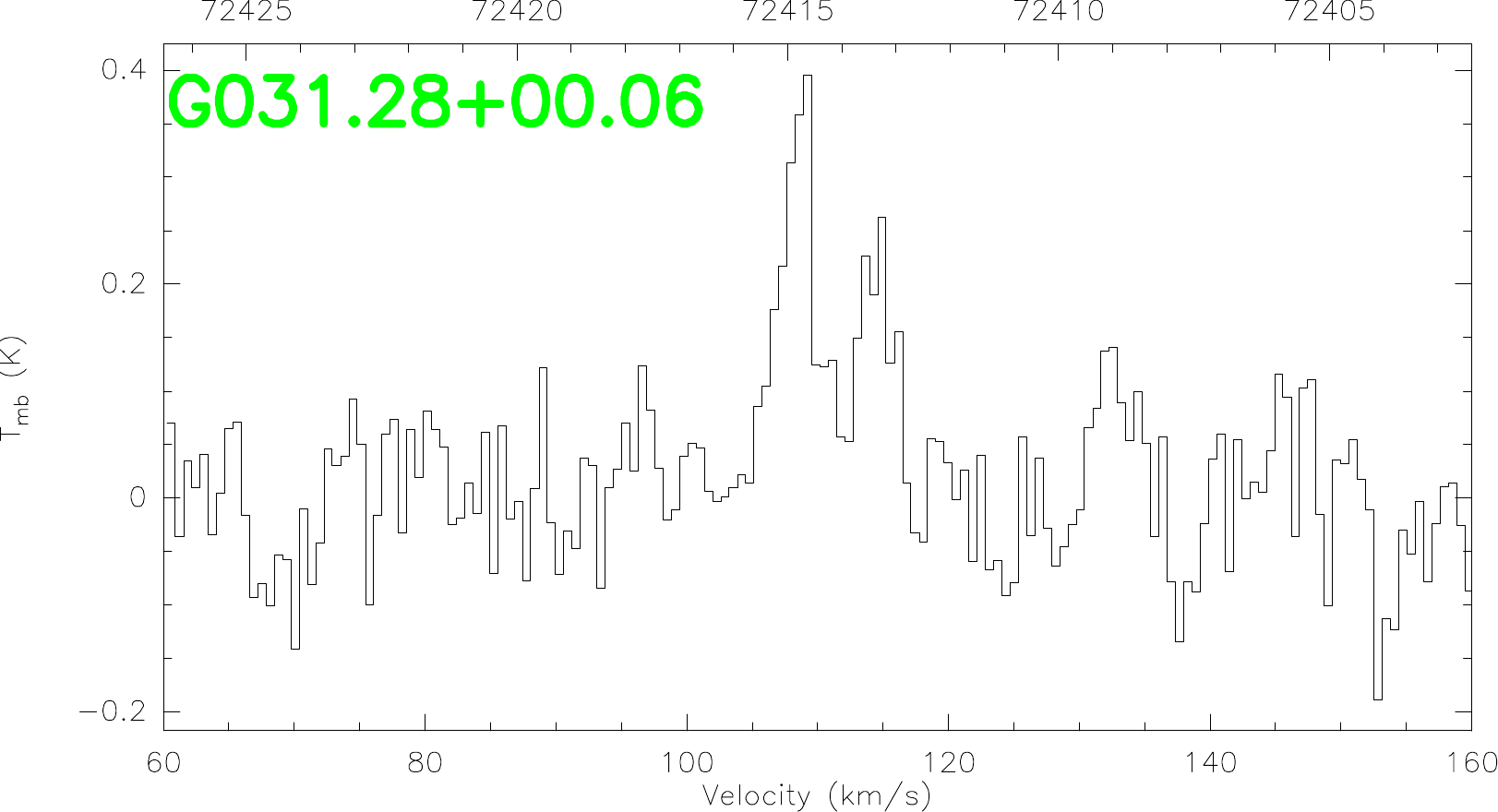}
\includegraphics[width=0.155\textwidth]{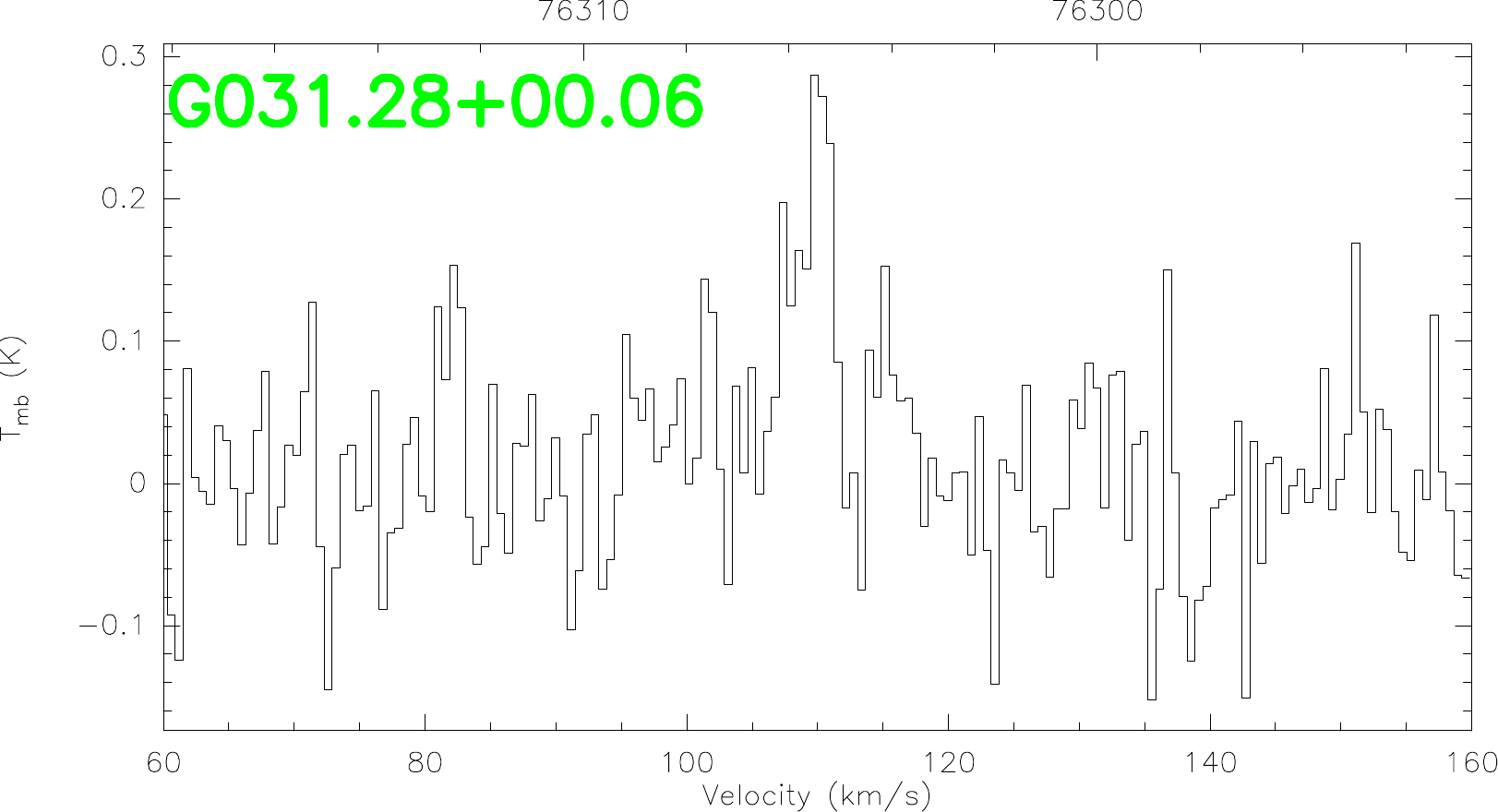}
\includegraphics[width=0.155\textwidth]{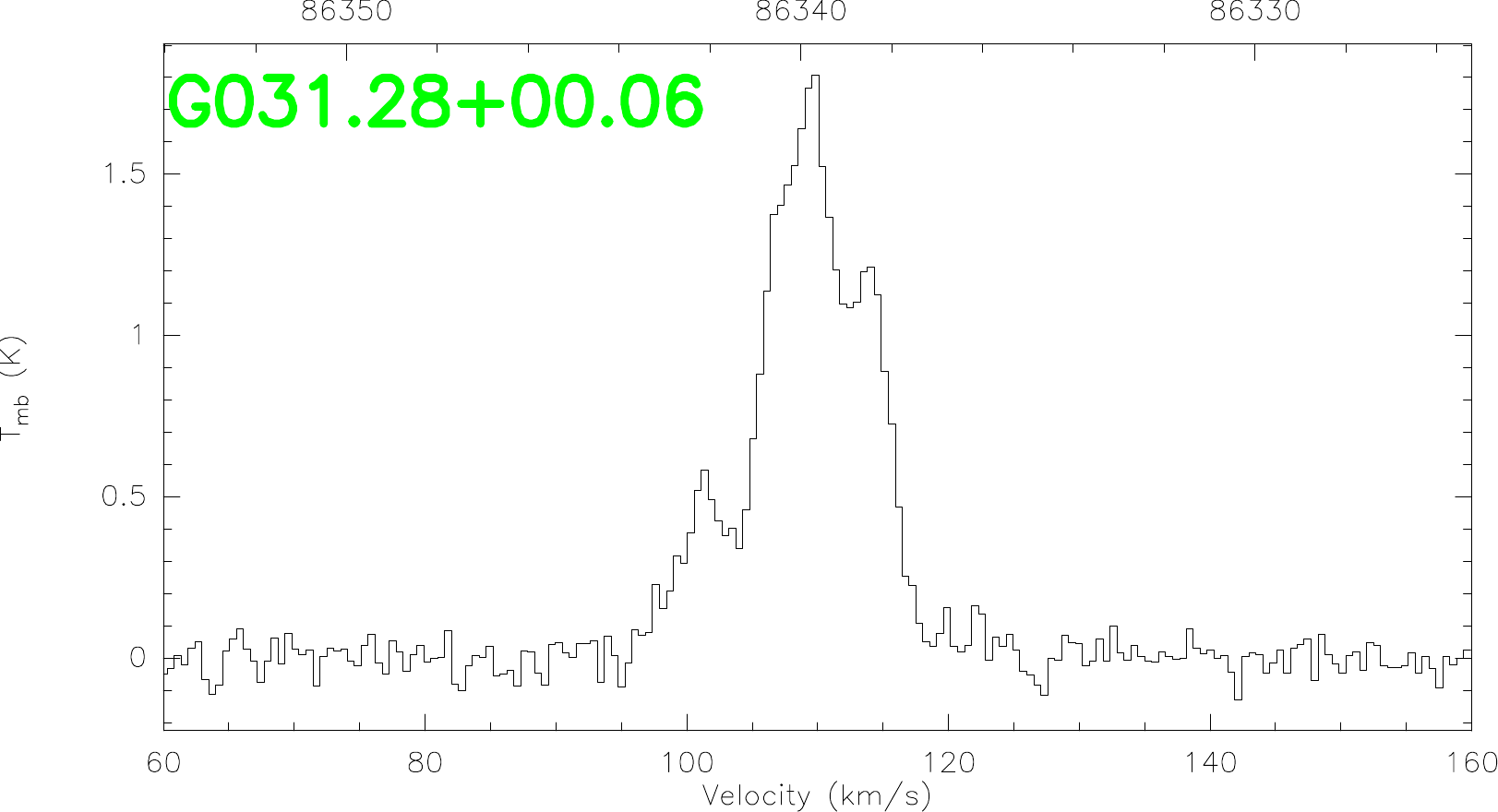}
\includegraphics[width=0.155\textwidth]{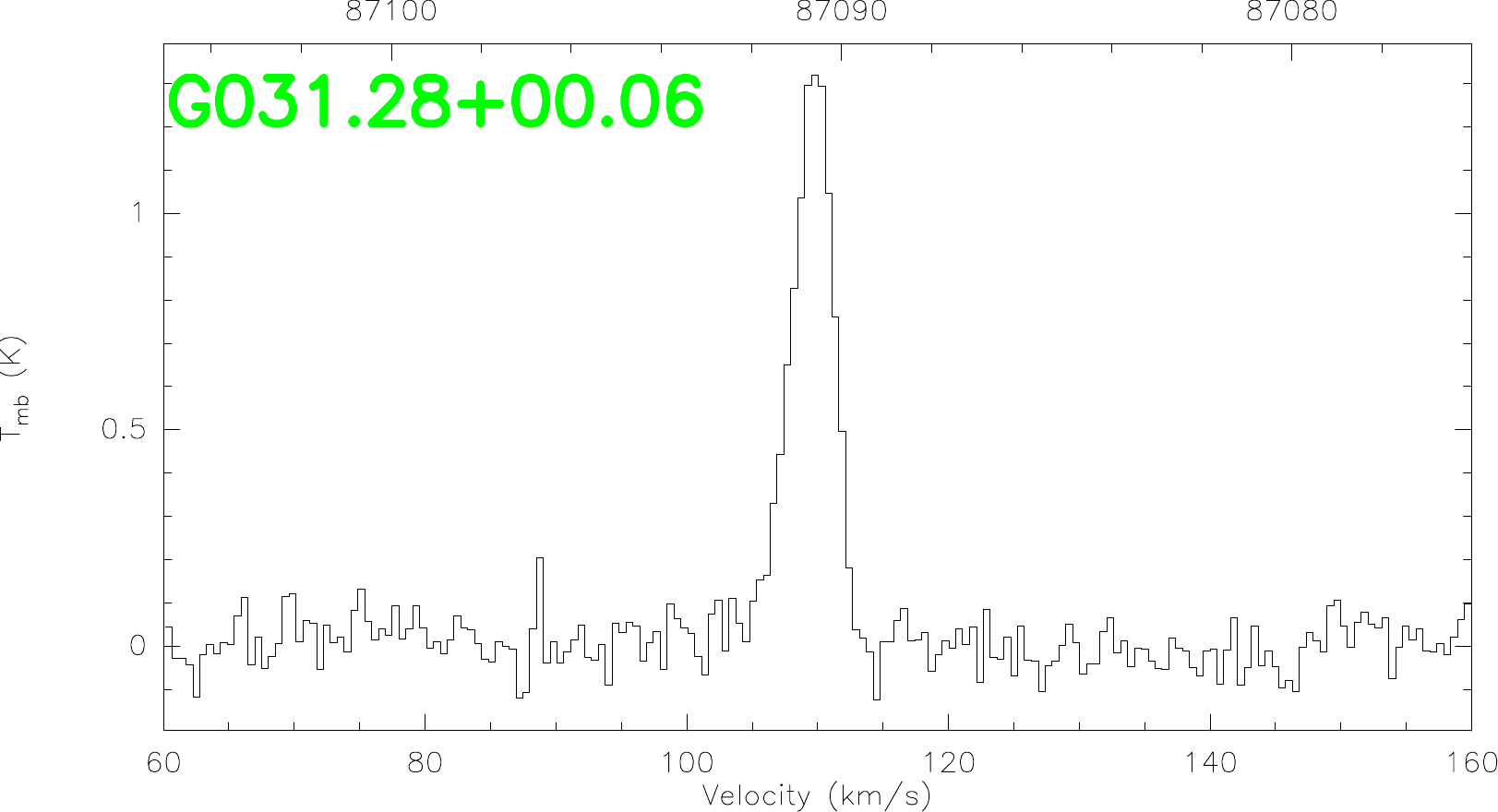}
\includegraphics[width=0.155\textwidth]{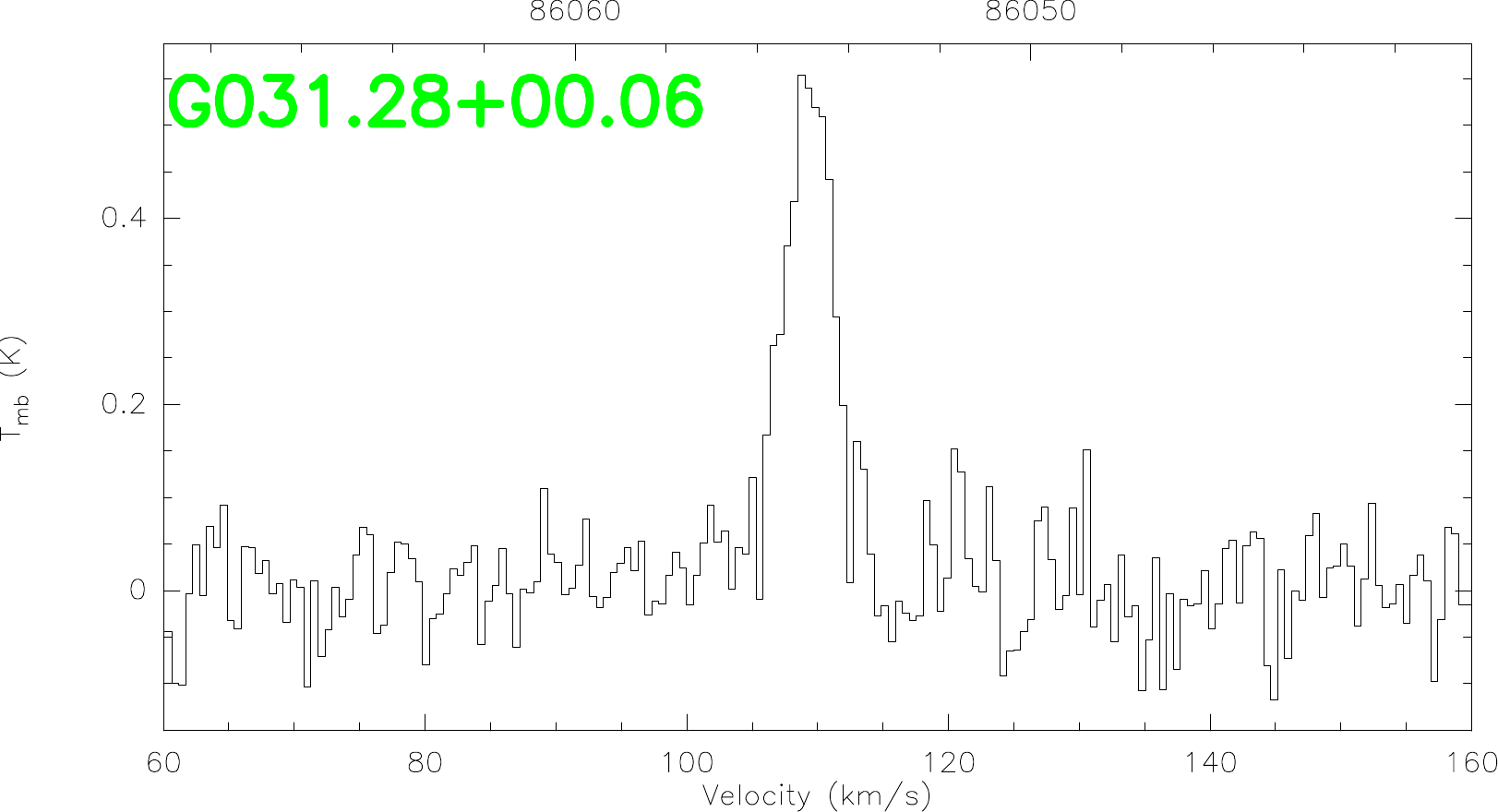}
\includegraphics[width=0.155\textwidth]{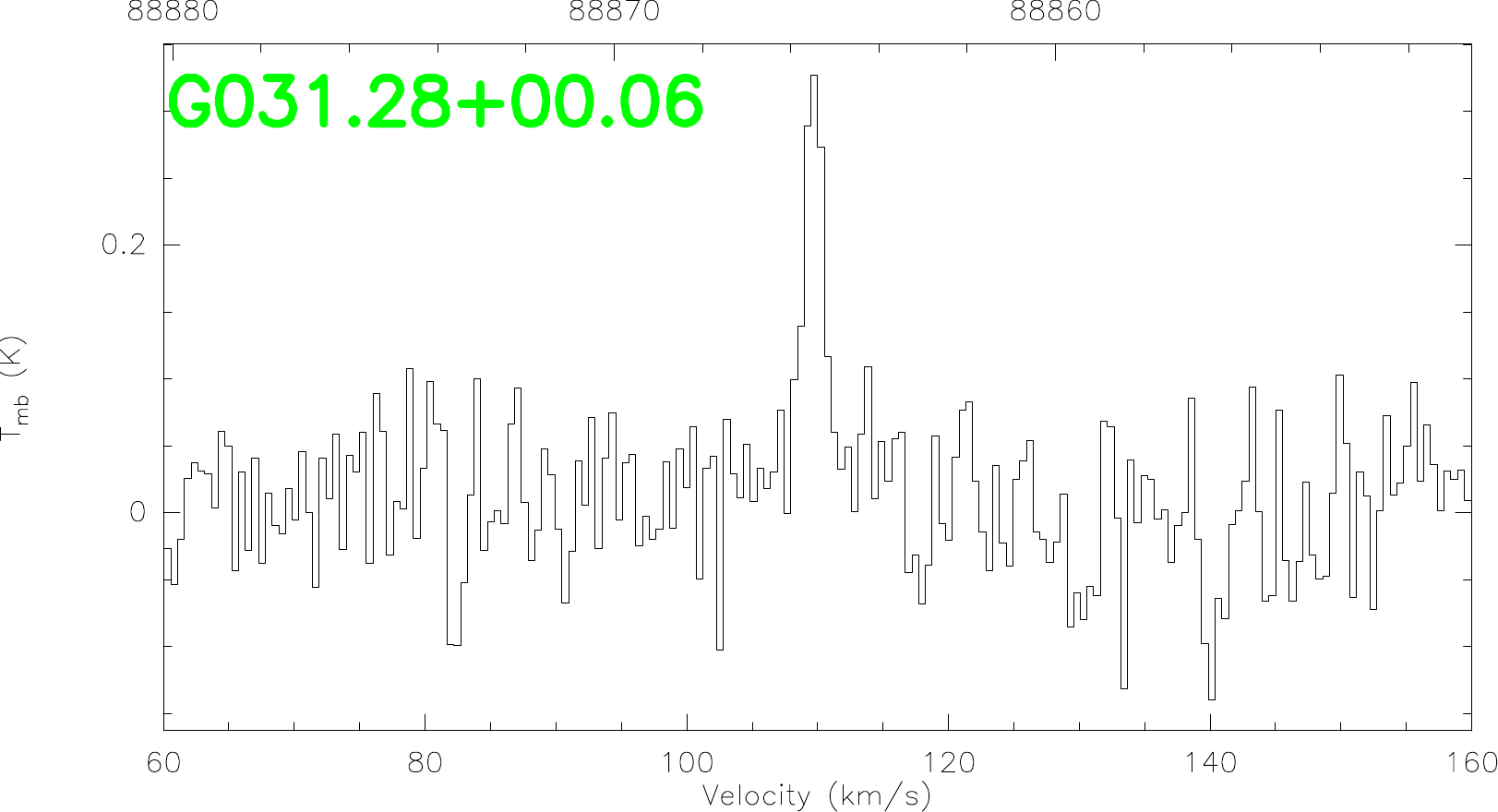}\\

\includegraphics[width=0.155\textwidth]{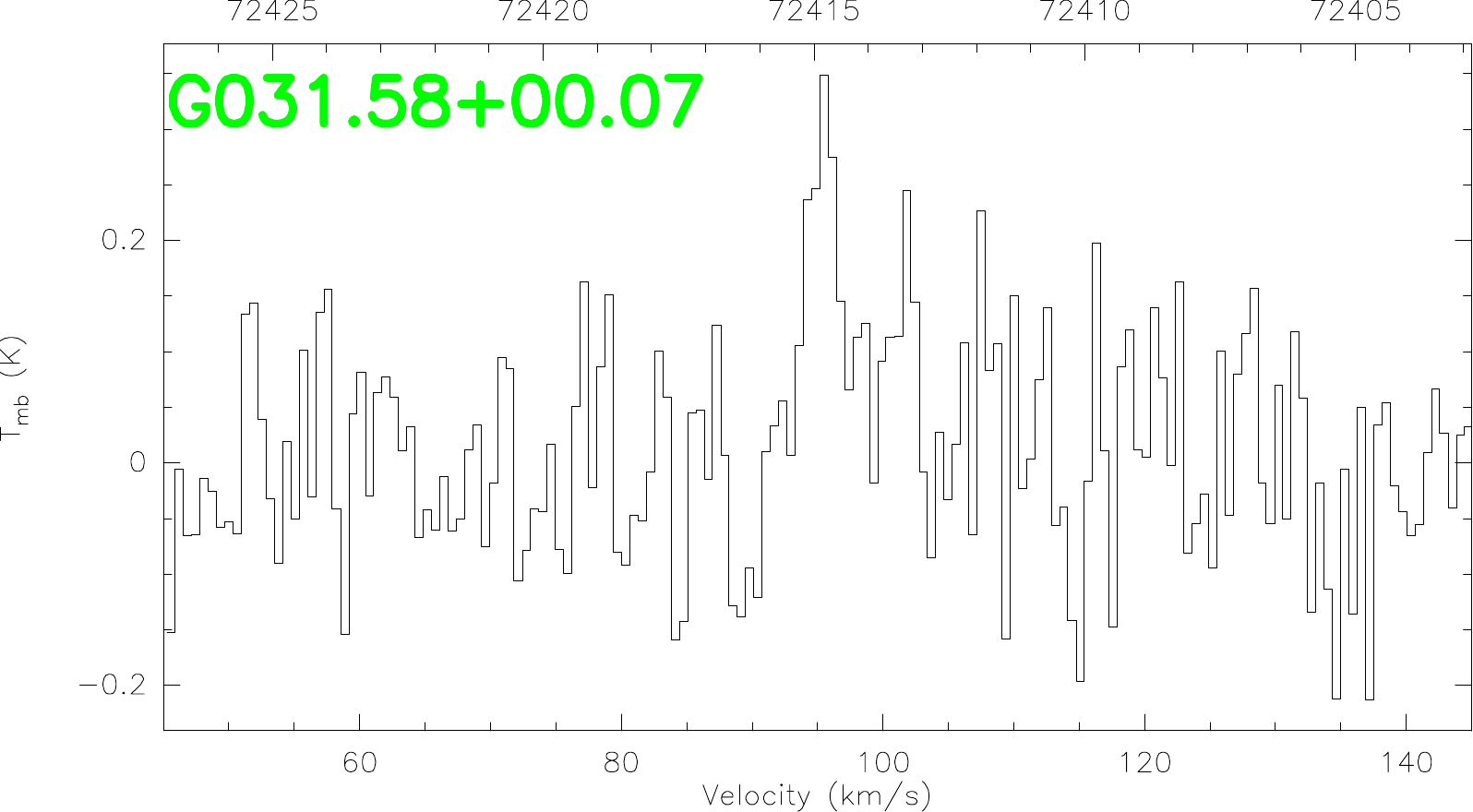}
\includegraphics[width=0.155\textwidth]{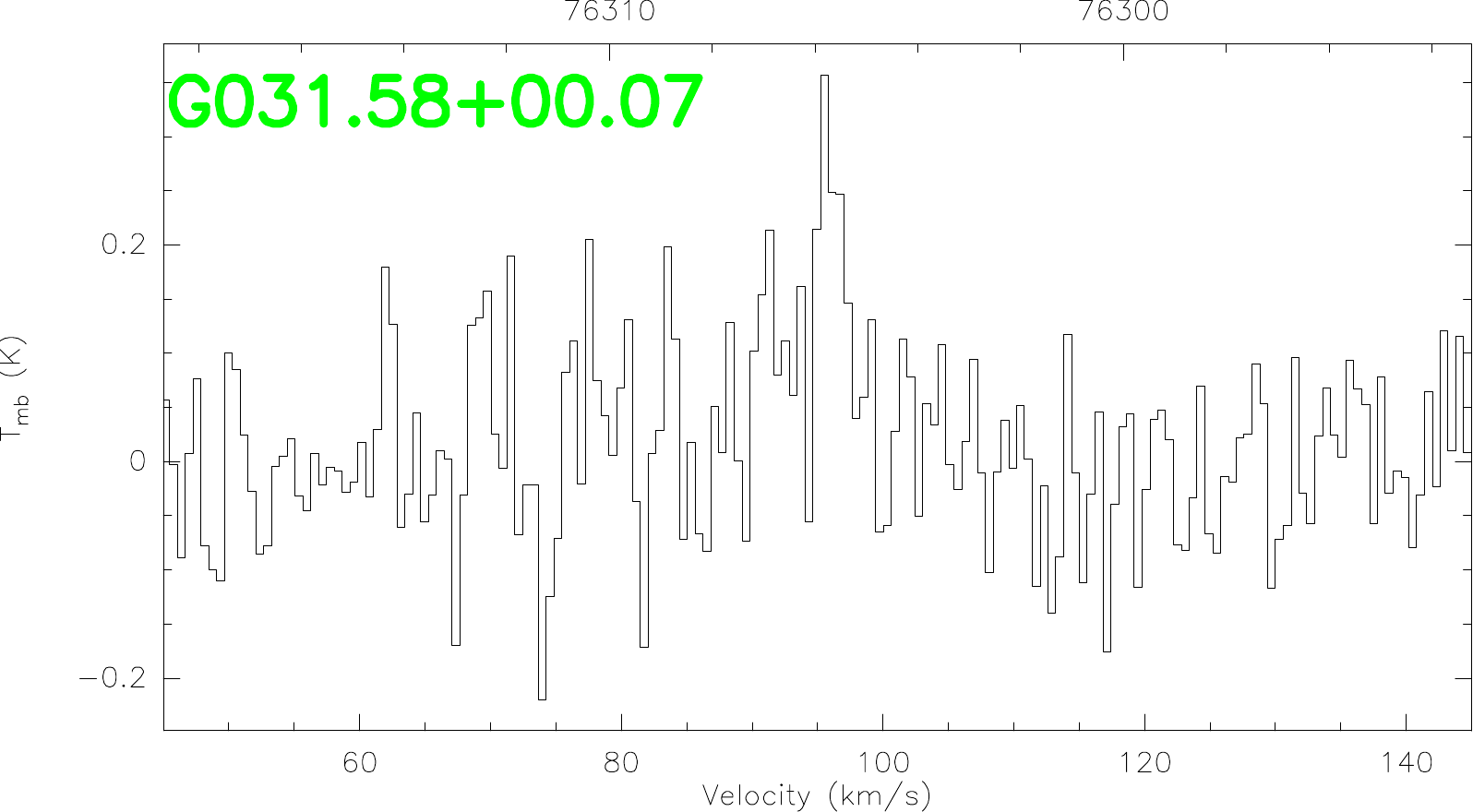}
\includegraphics[width=0.155\textwidth]{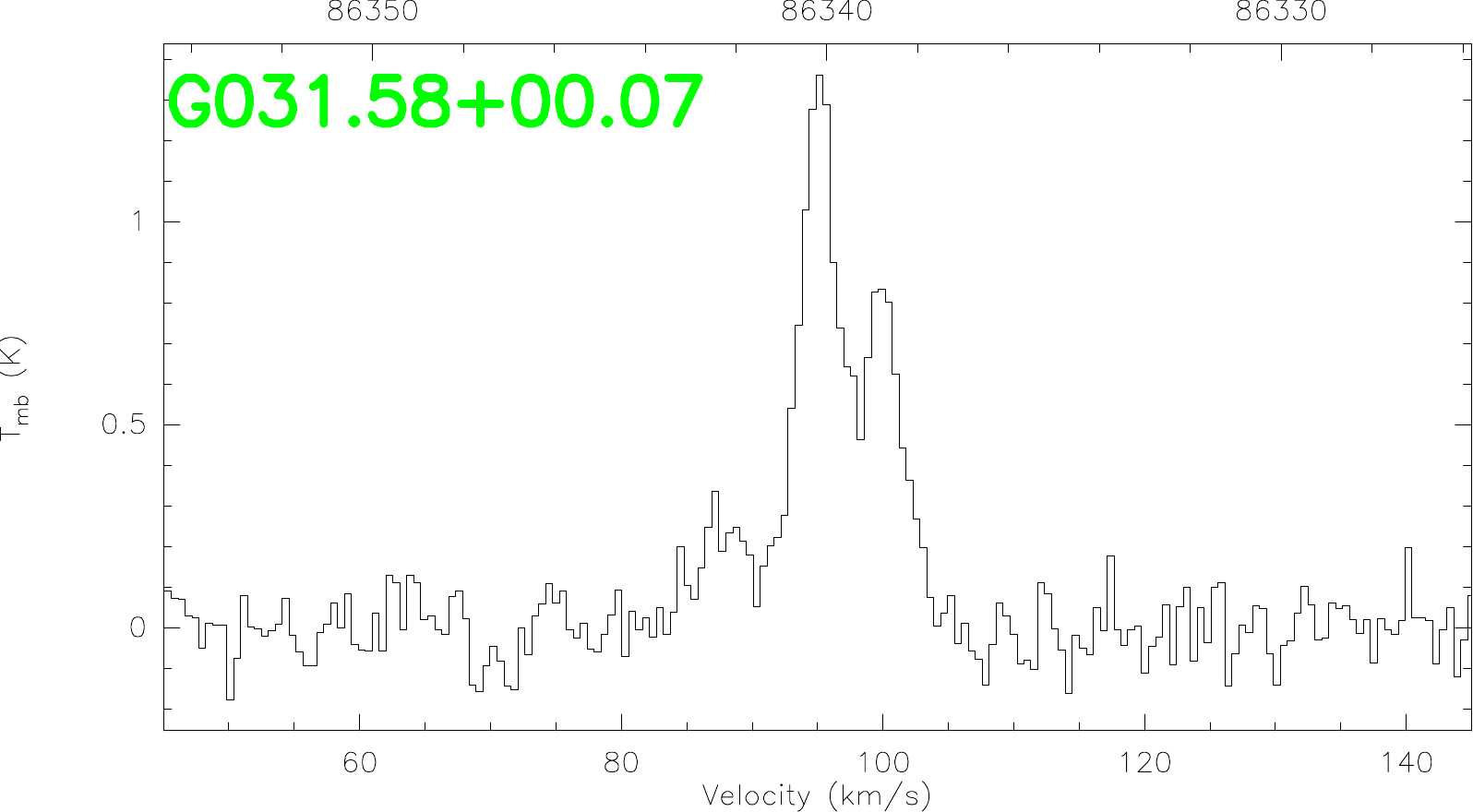}
\includegraphics[width=0.155\textwidth]{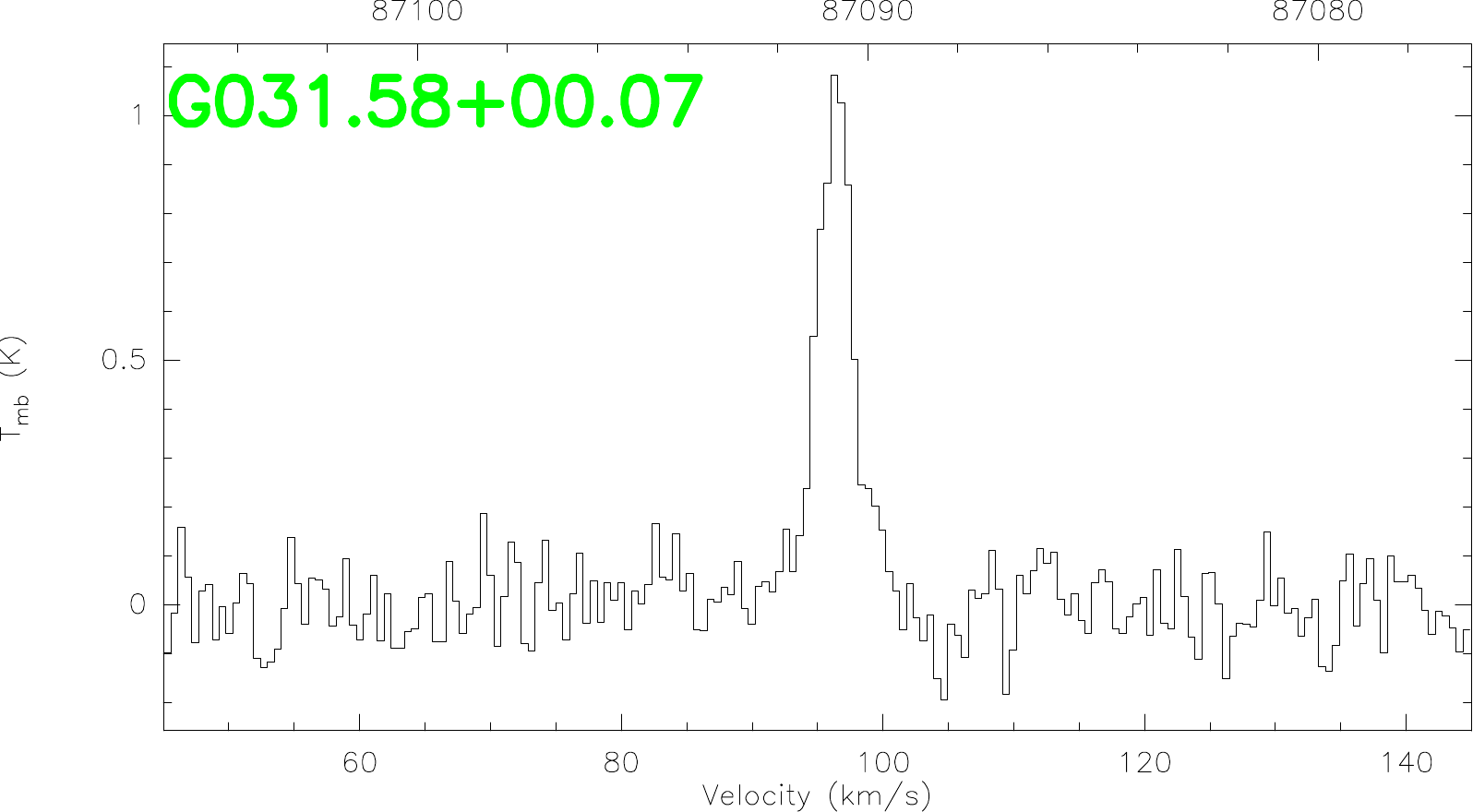}
\includegraphics[width=0.155\textwidth]{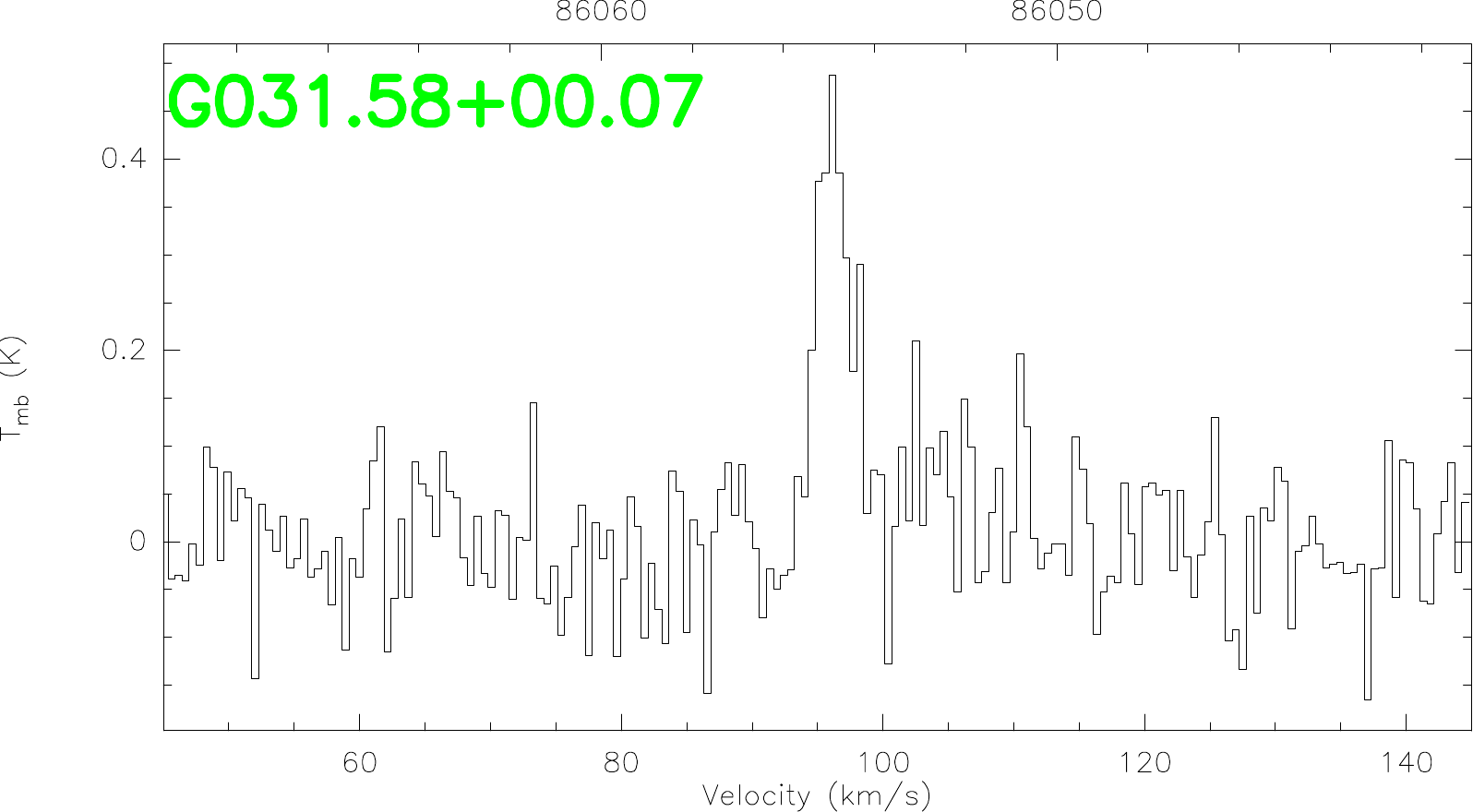}
\includegraphics[width=0.155\textwidth]{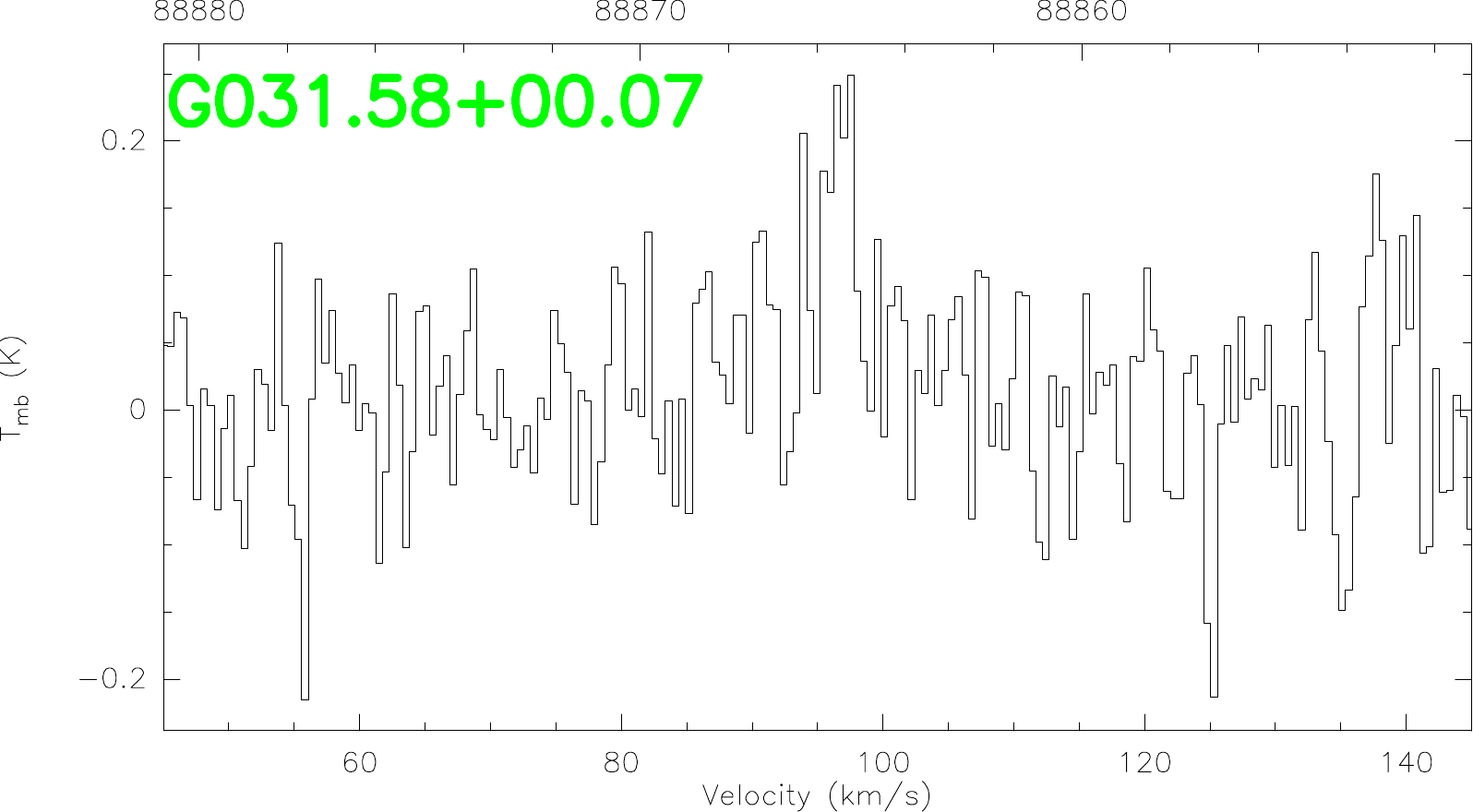}\\

\includegraphics[width=0.155\textwidth]{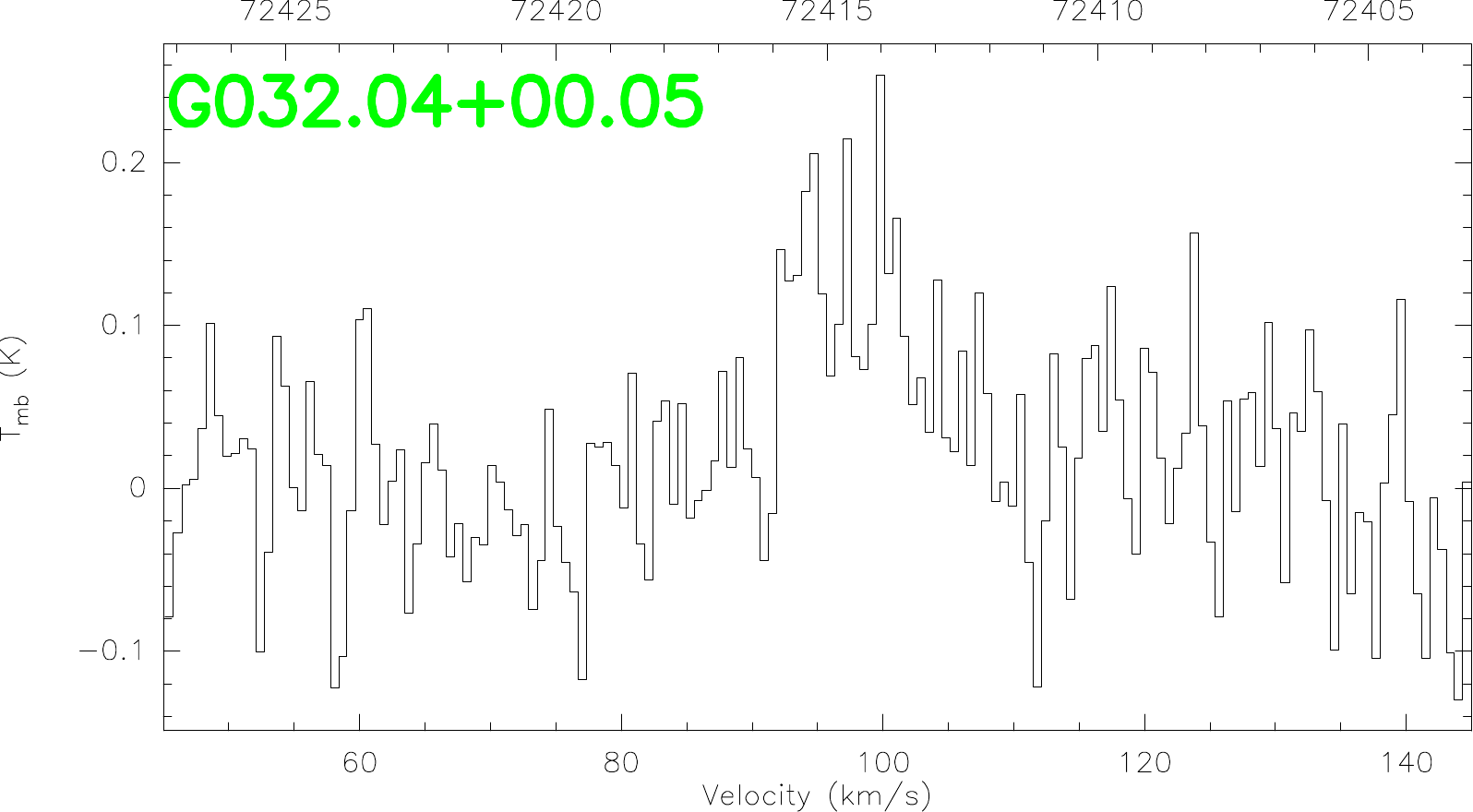}
\includegraphics[width=0.155\textwidth]{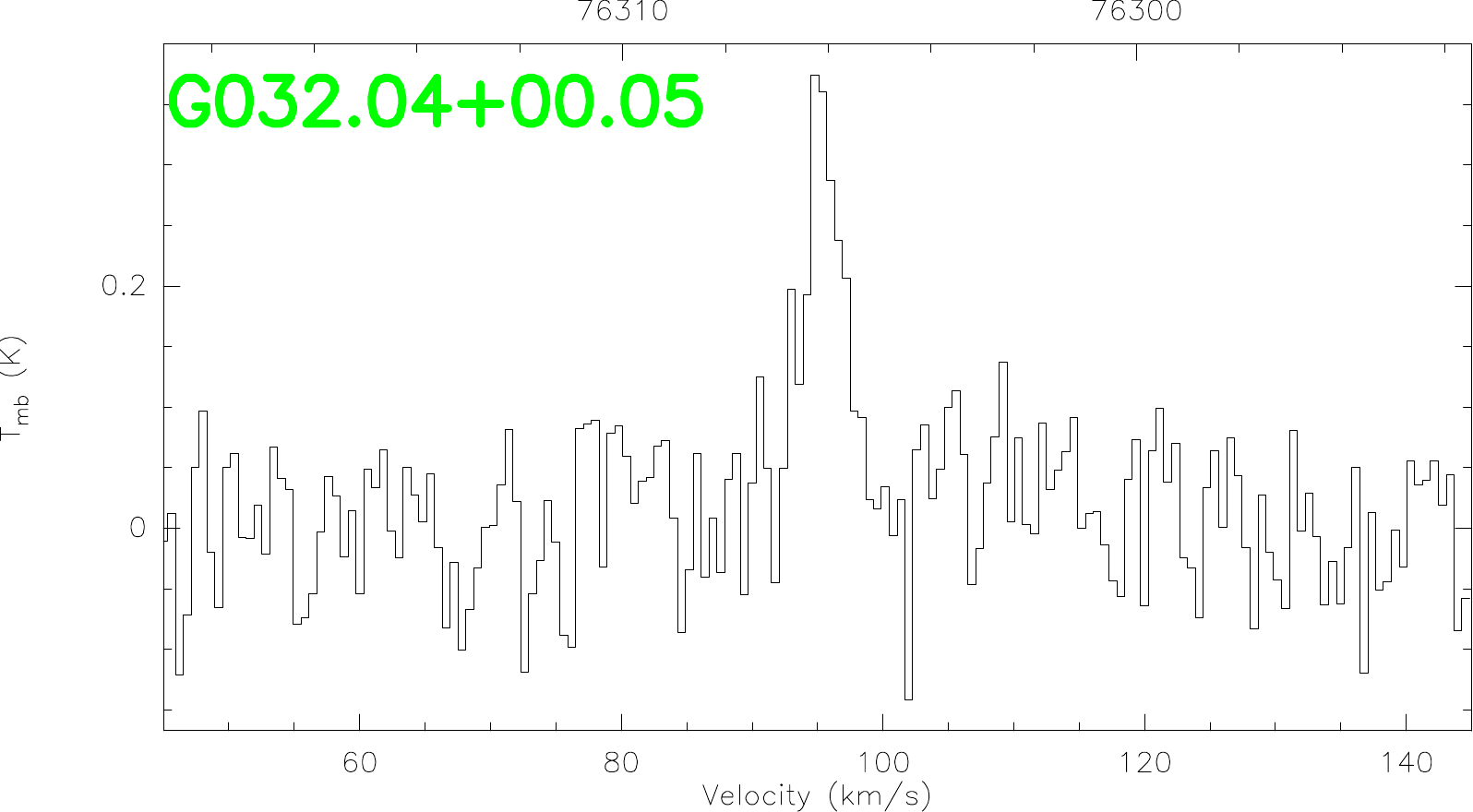}
\includegraphics[width=0.155\textwidth]{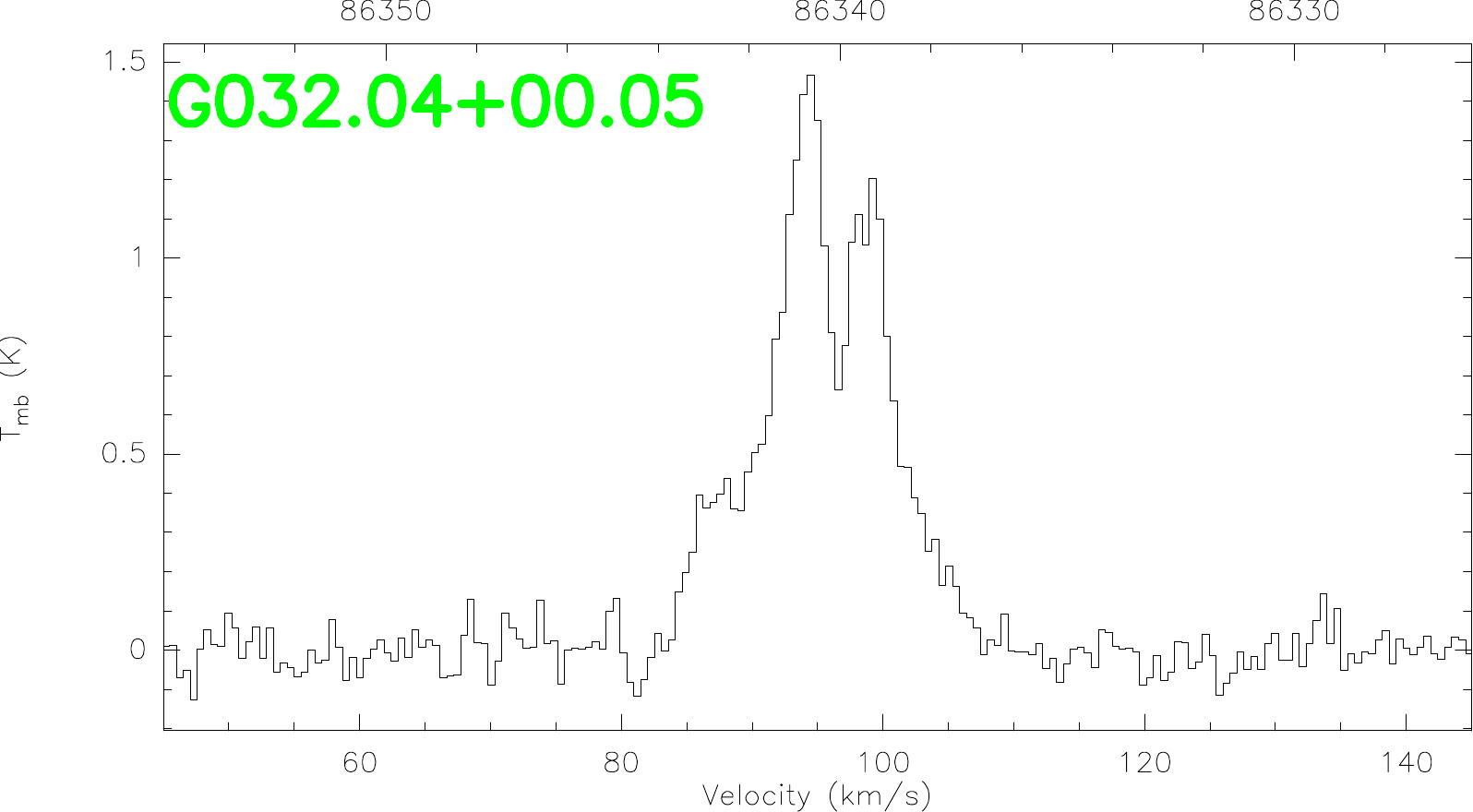}
\includegraphics[width=0.155\textwidth]{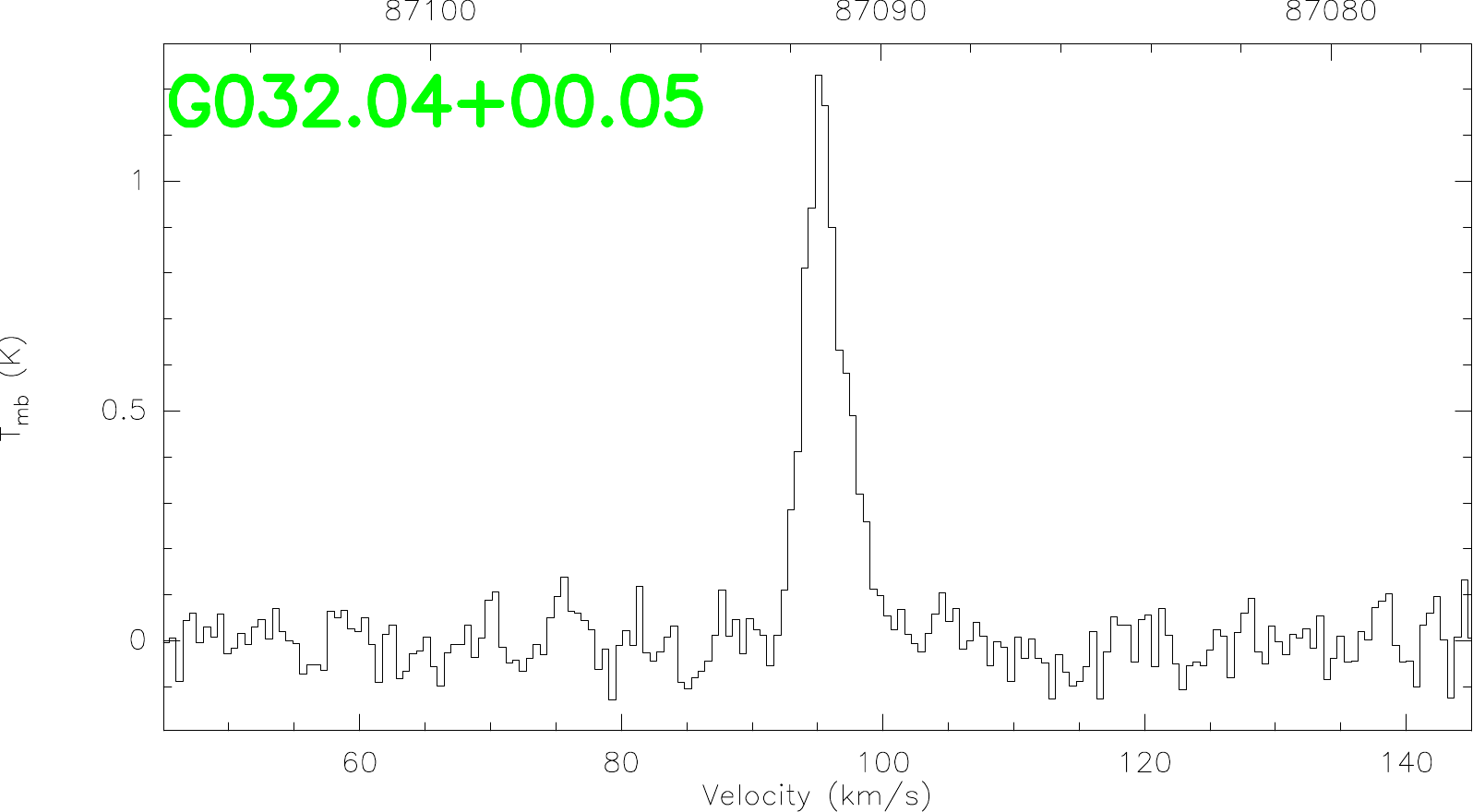}
\includegraphics[width=0.155\textwidth]{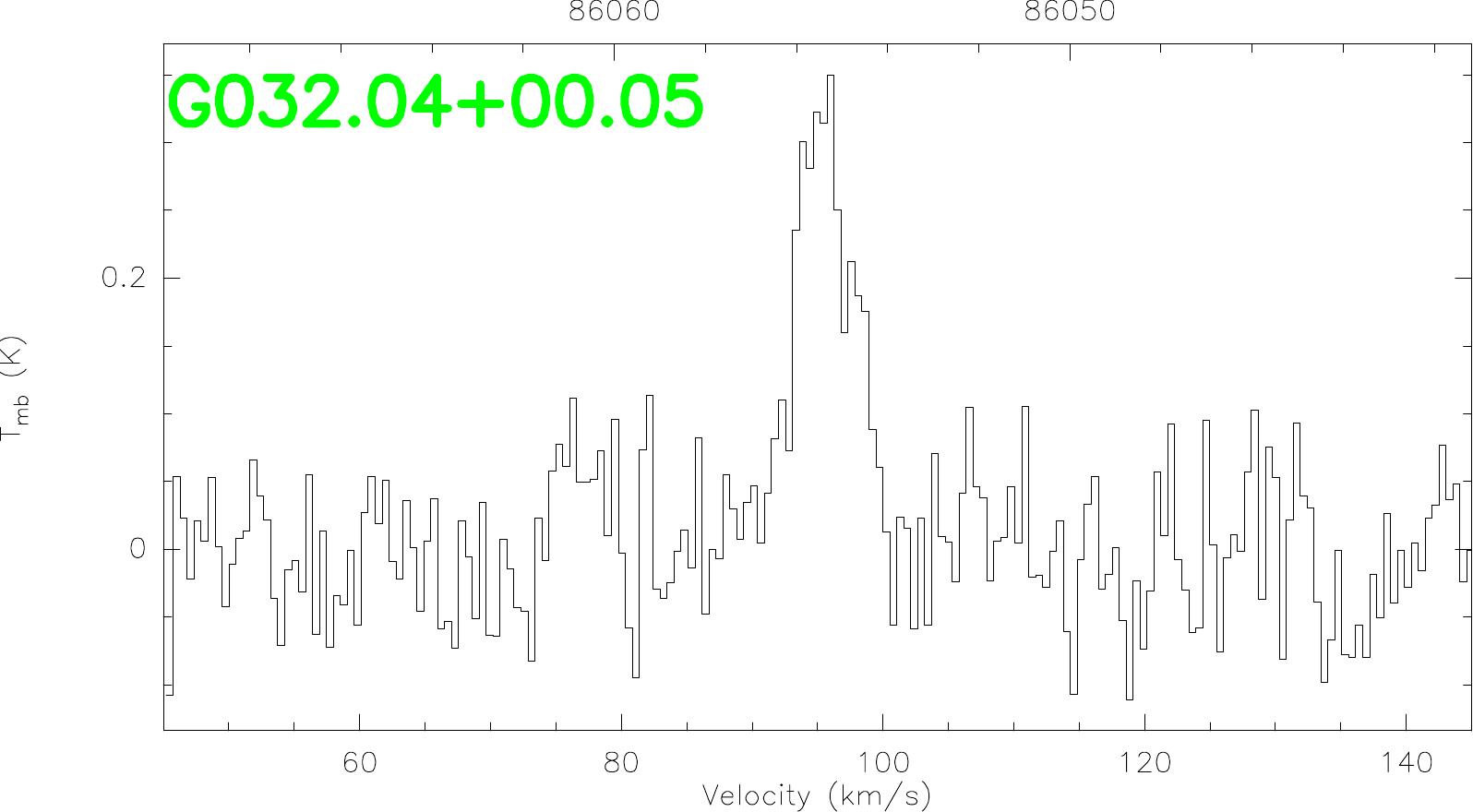}
\includegraphics[width=0.155\textwidth]{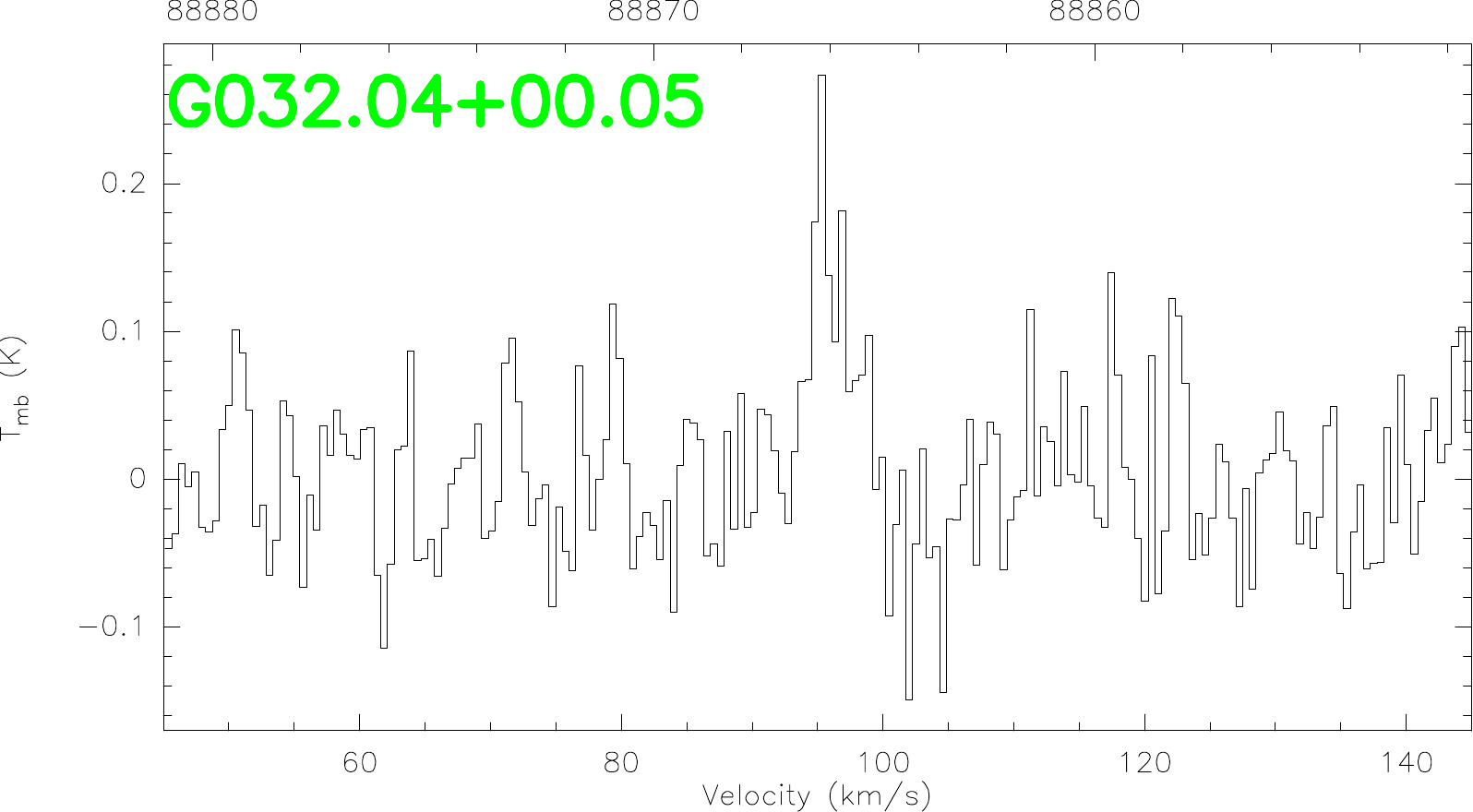}\\

\caption{Continued. Spectra of different molecular lines.}
\label{fig:multi_lines_page2}
\end{figure}

\begin{figure}[h]
\centering

\makebox[\colw][c]{DCN 1-0}
\makebox[\colw][c]{DNC 1-0}
\makebox[\colw][c]{H$^{13}$CN 1-0}
\makebox[\colw][c]{HN$^{13}$C 1-0}
\makebox[\colw][c]{HC$^{15}$N 1-0}
\makebox[\colw][c]{H$^{15}$NC 1-0}\\[0.2em]

\includegraphics[width=0.155\textwidth]{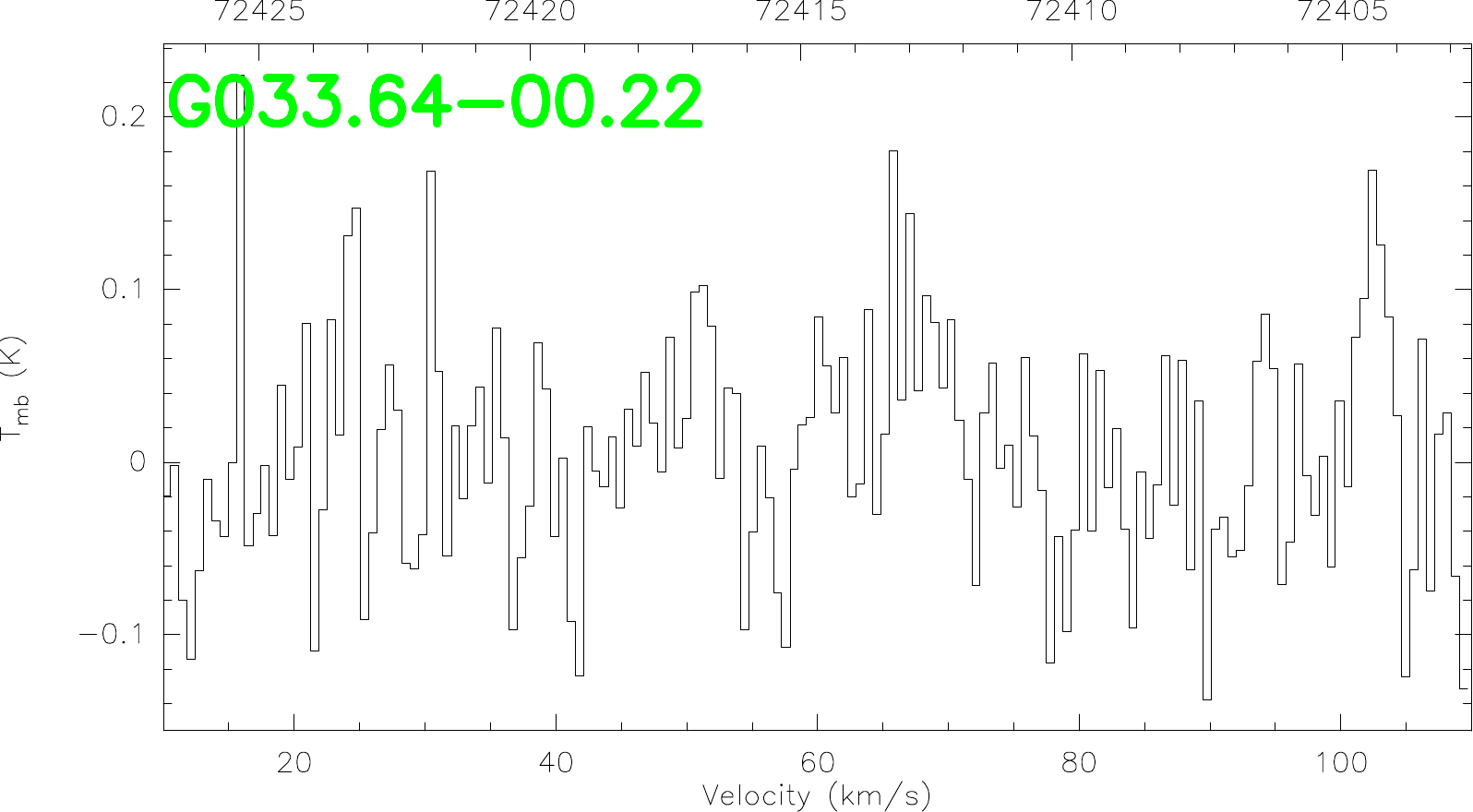}
\includegraphics[width=0.155\textwidth]{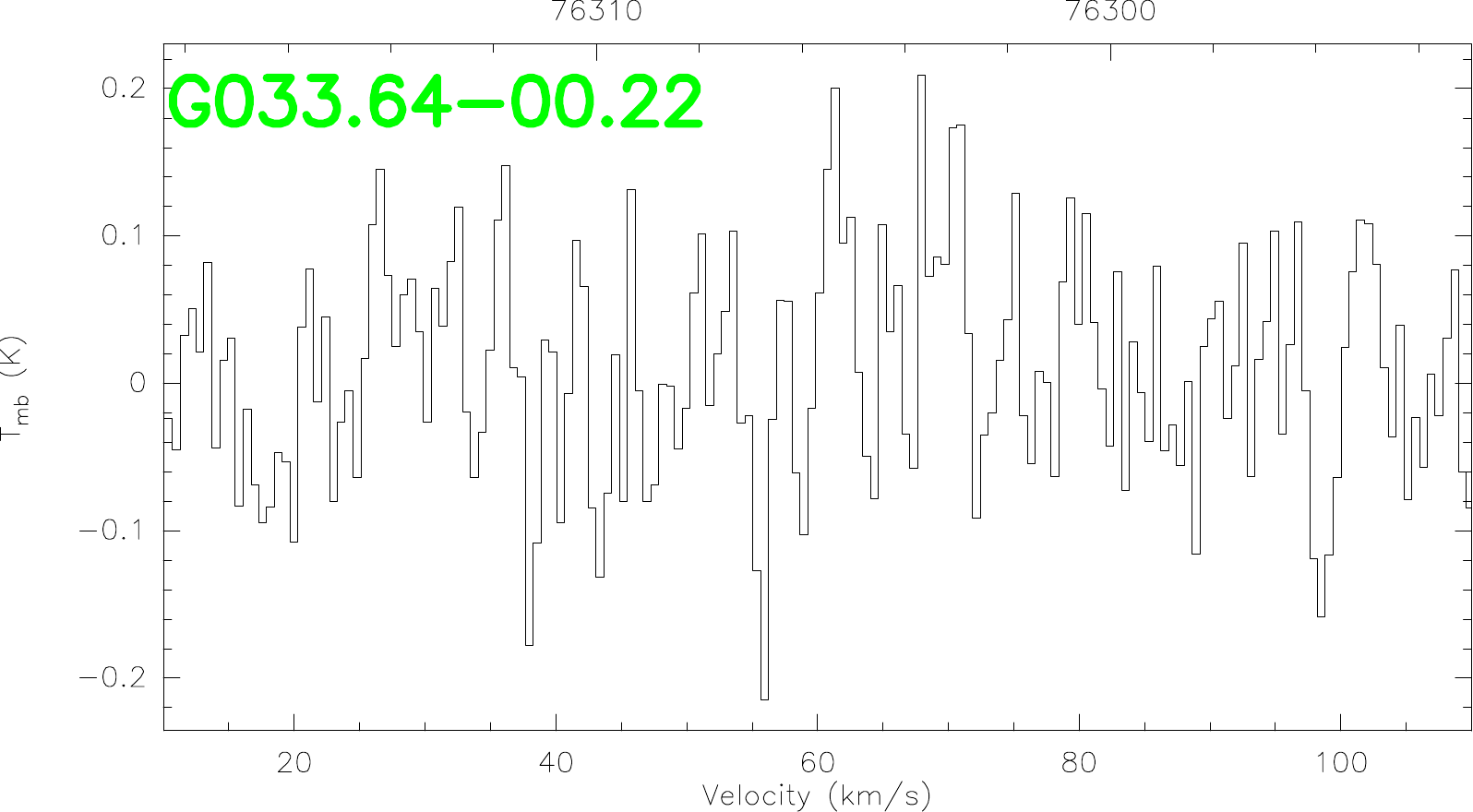}
\includegraphics[width=0.155\textwidth]{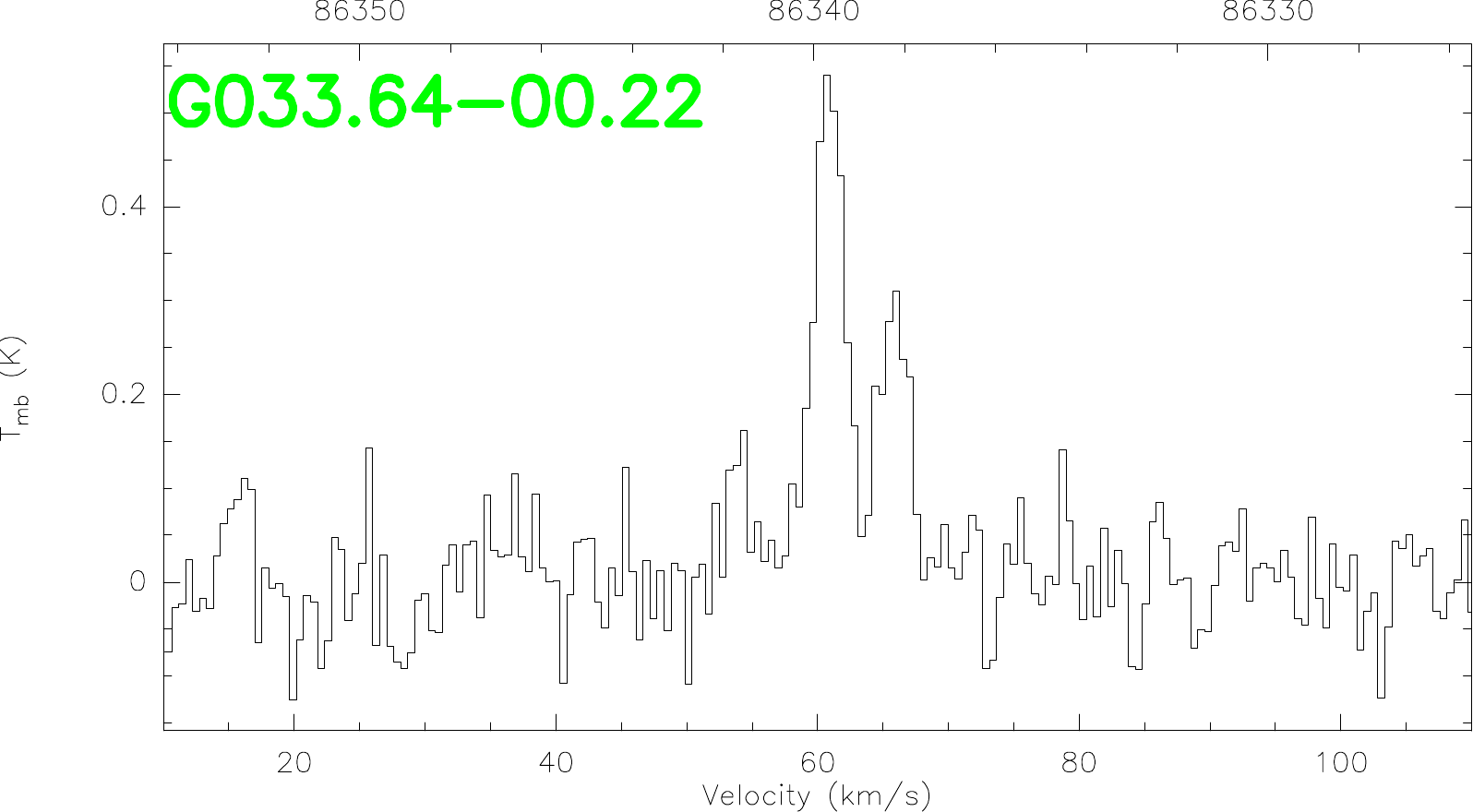}
\includegraphics[width=0.155\textwidth]{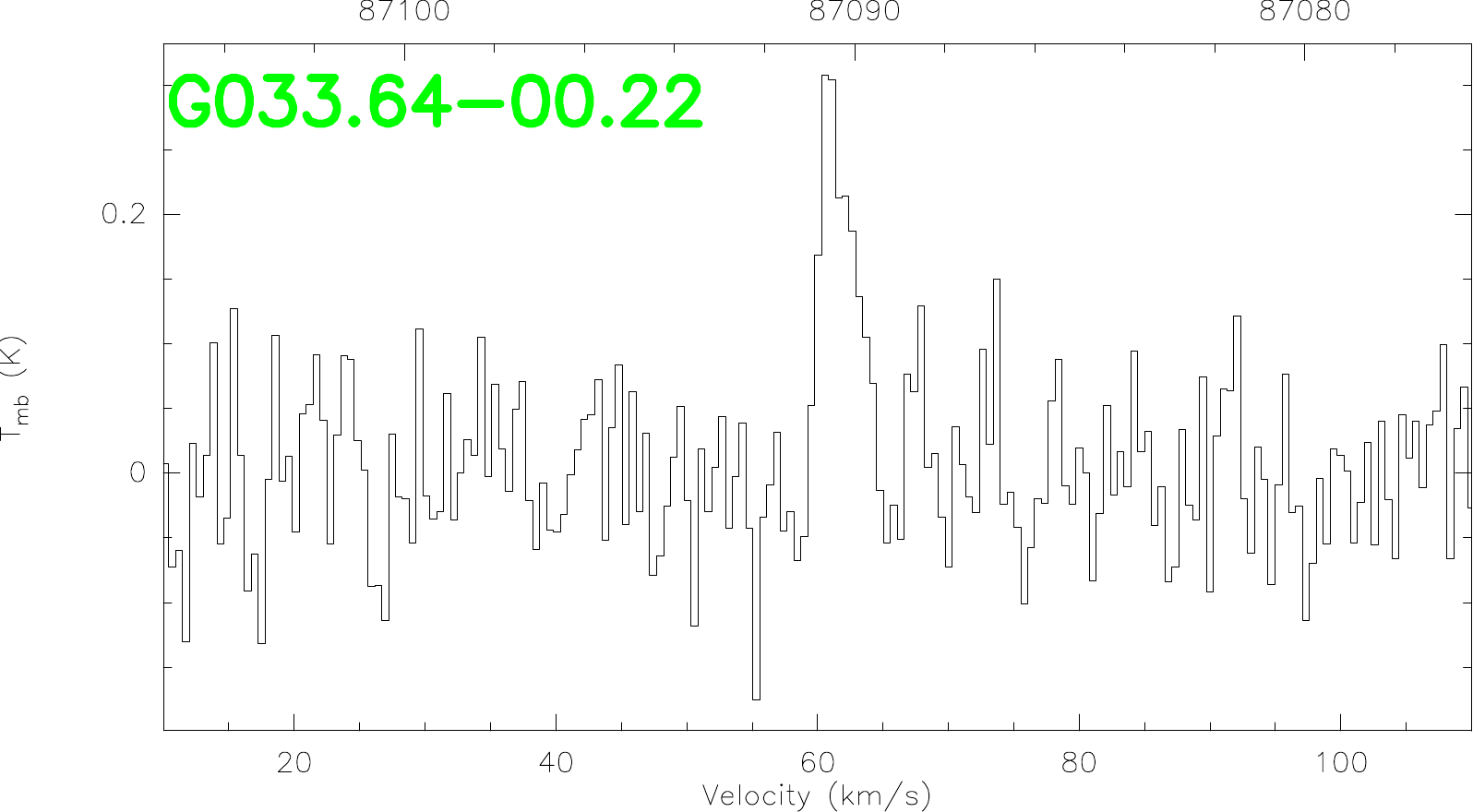}
\includegraphics[width=0.155\textwidth]{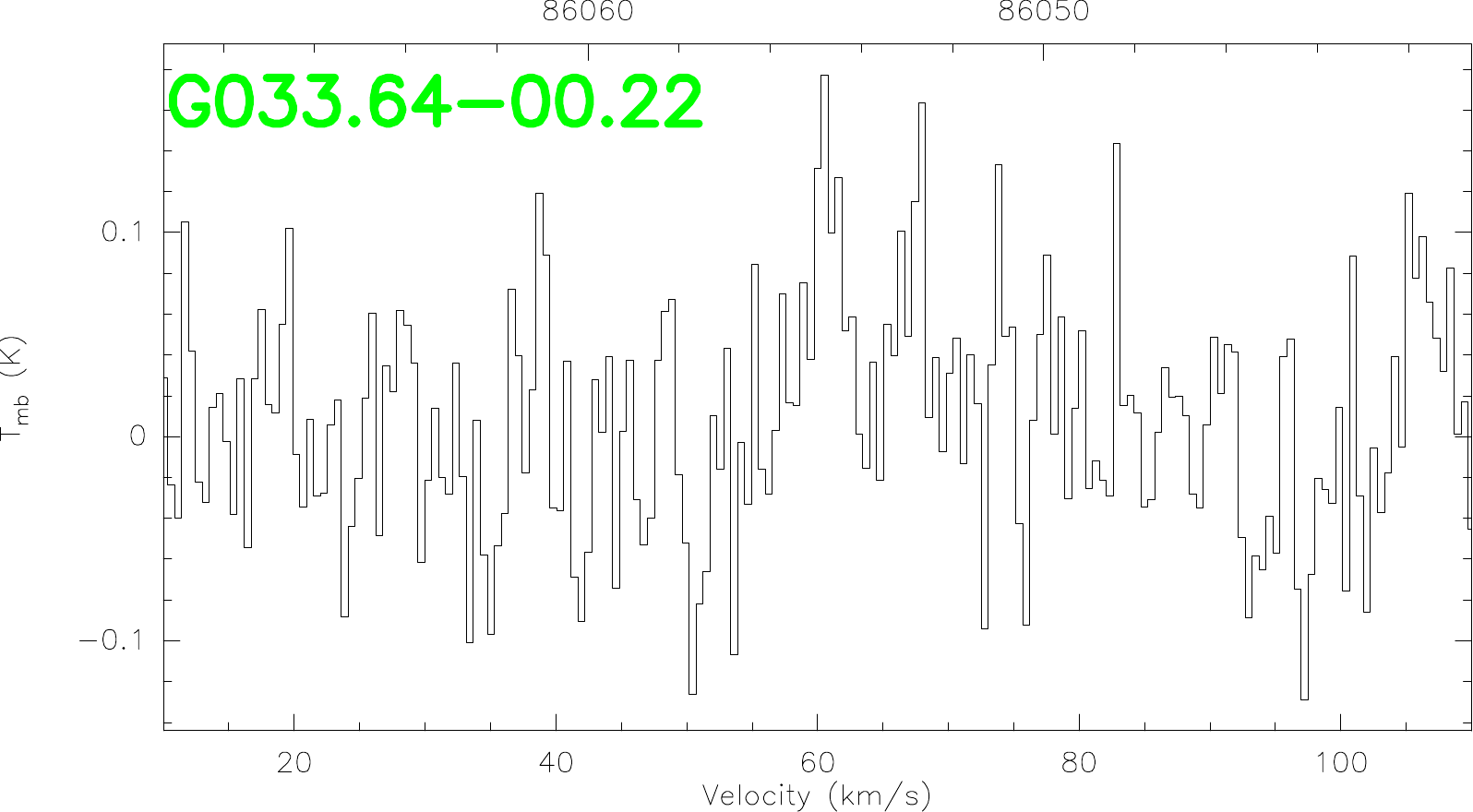}
\includegraphics[width=0.155\textwidth]{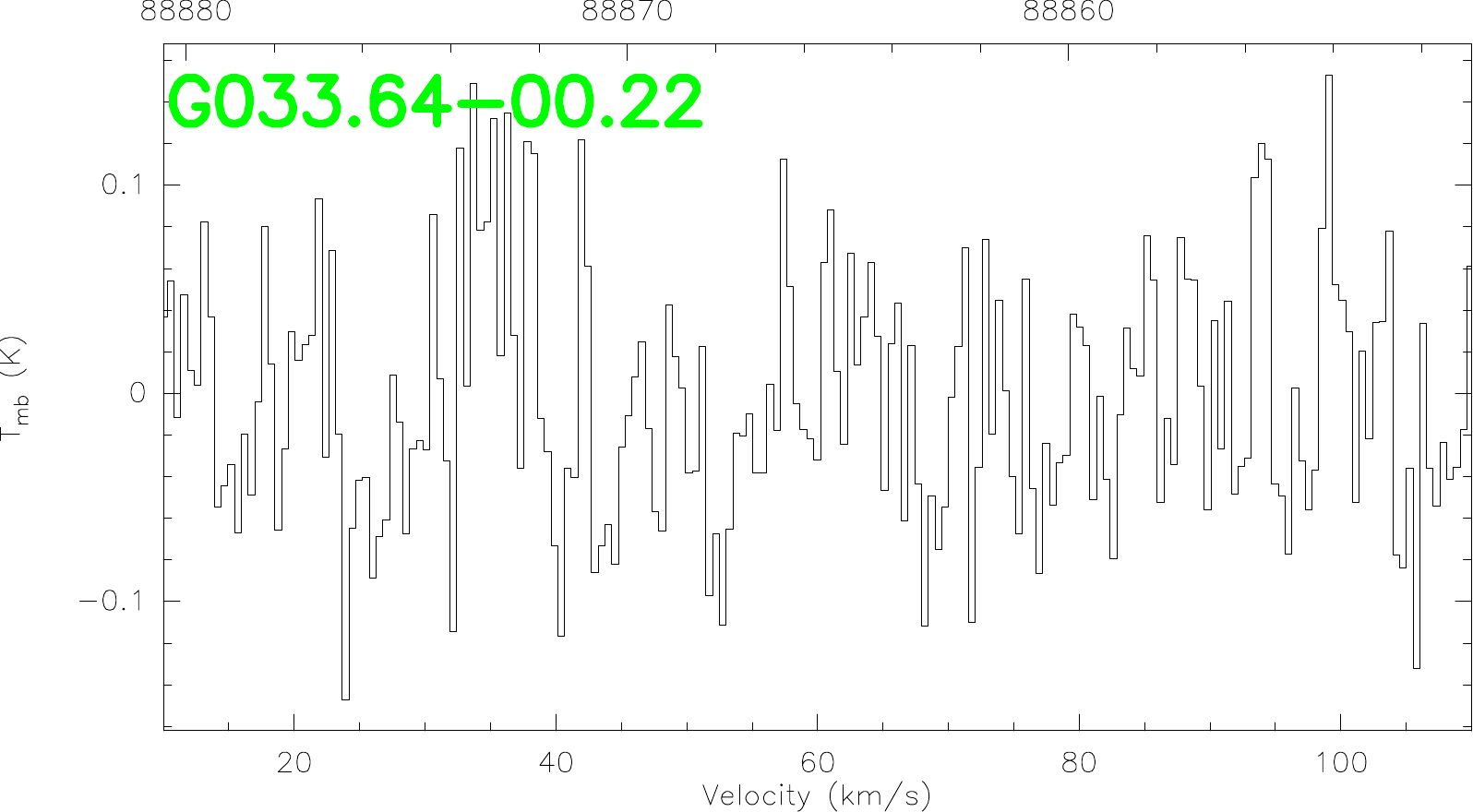}\\

\includegraphics[width=0.155\textwidth]{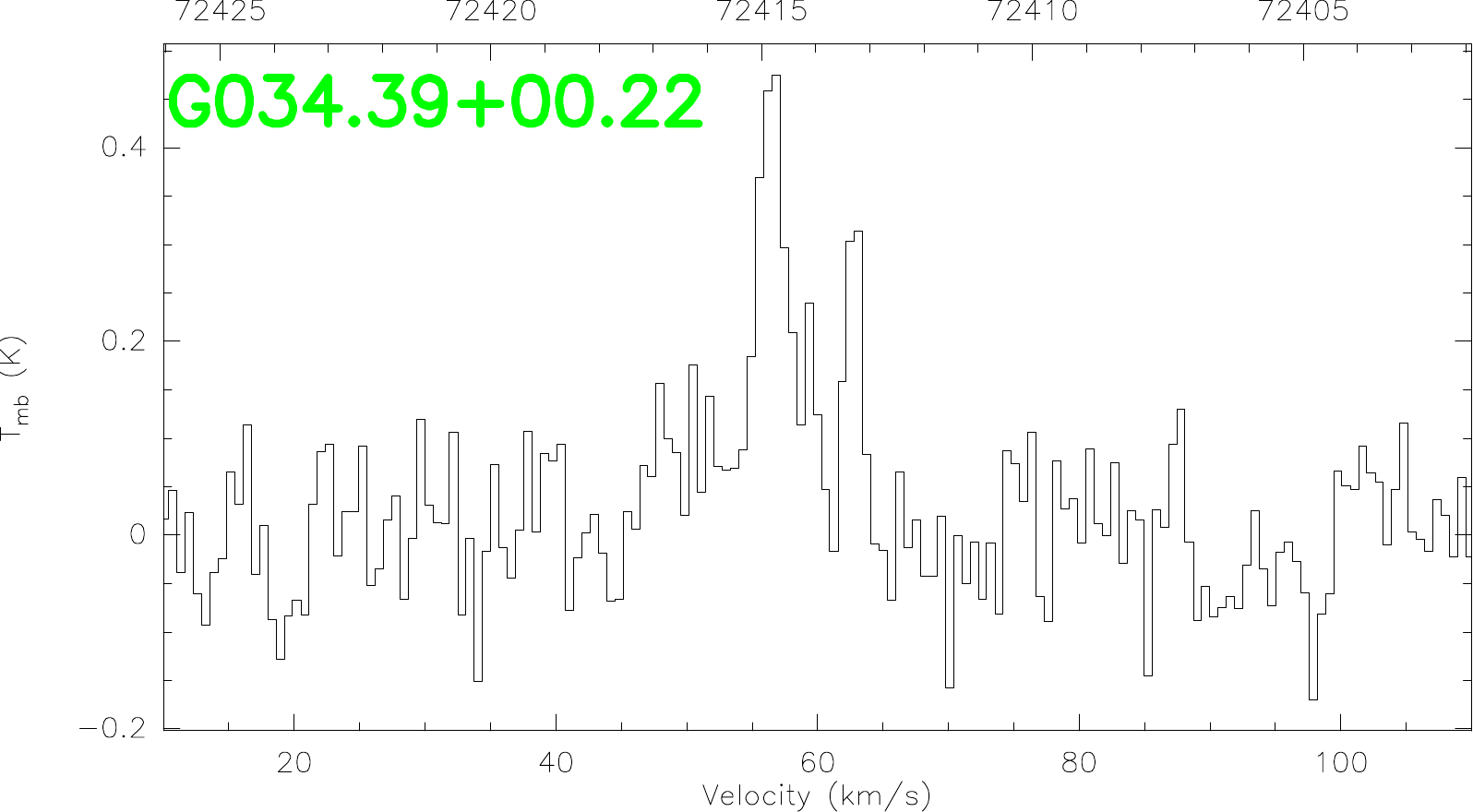}
\includegraphics[width=0.155\textwidth]{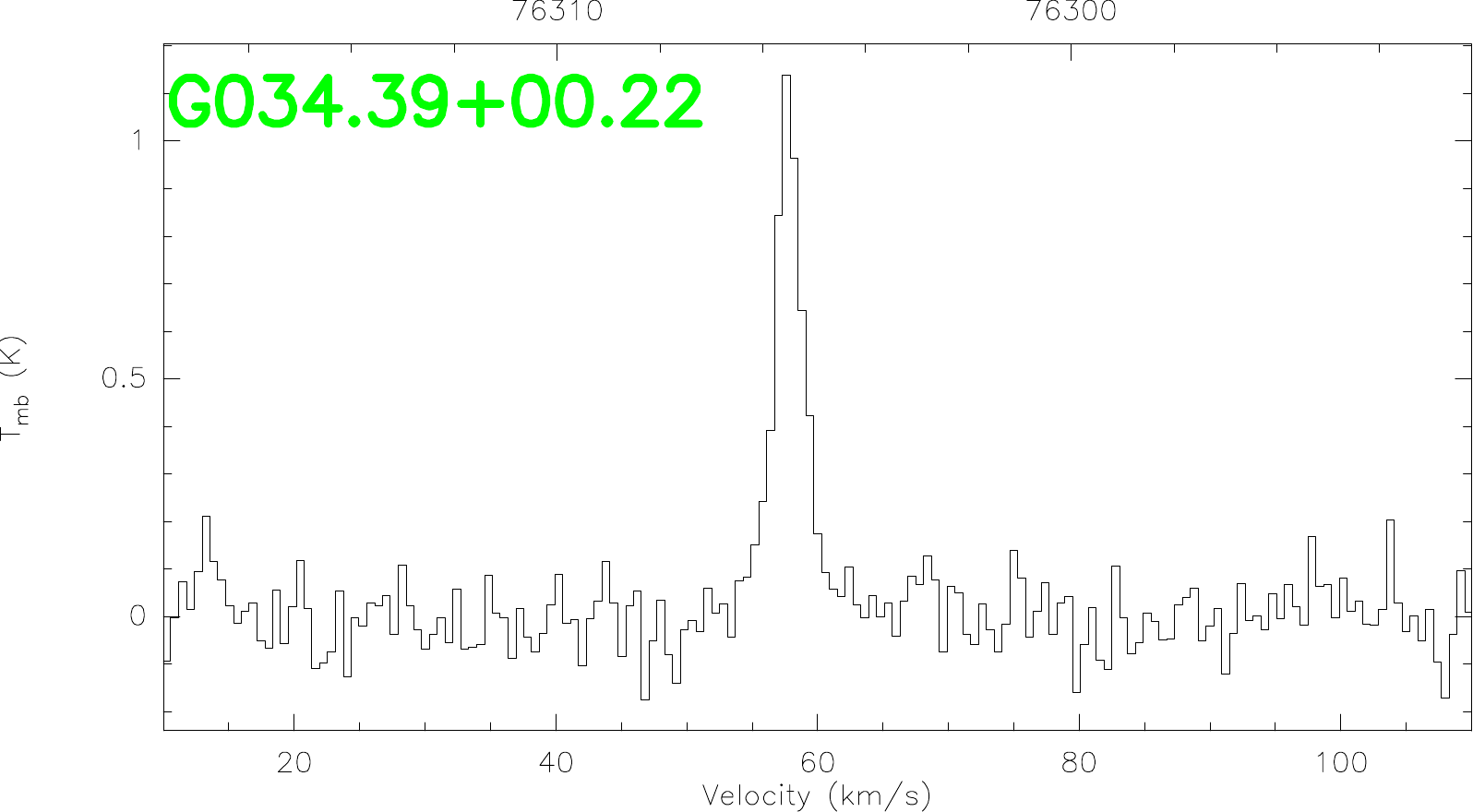}
\includegraphics[width=0.155\textwidth]{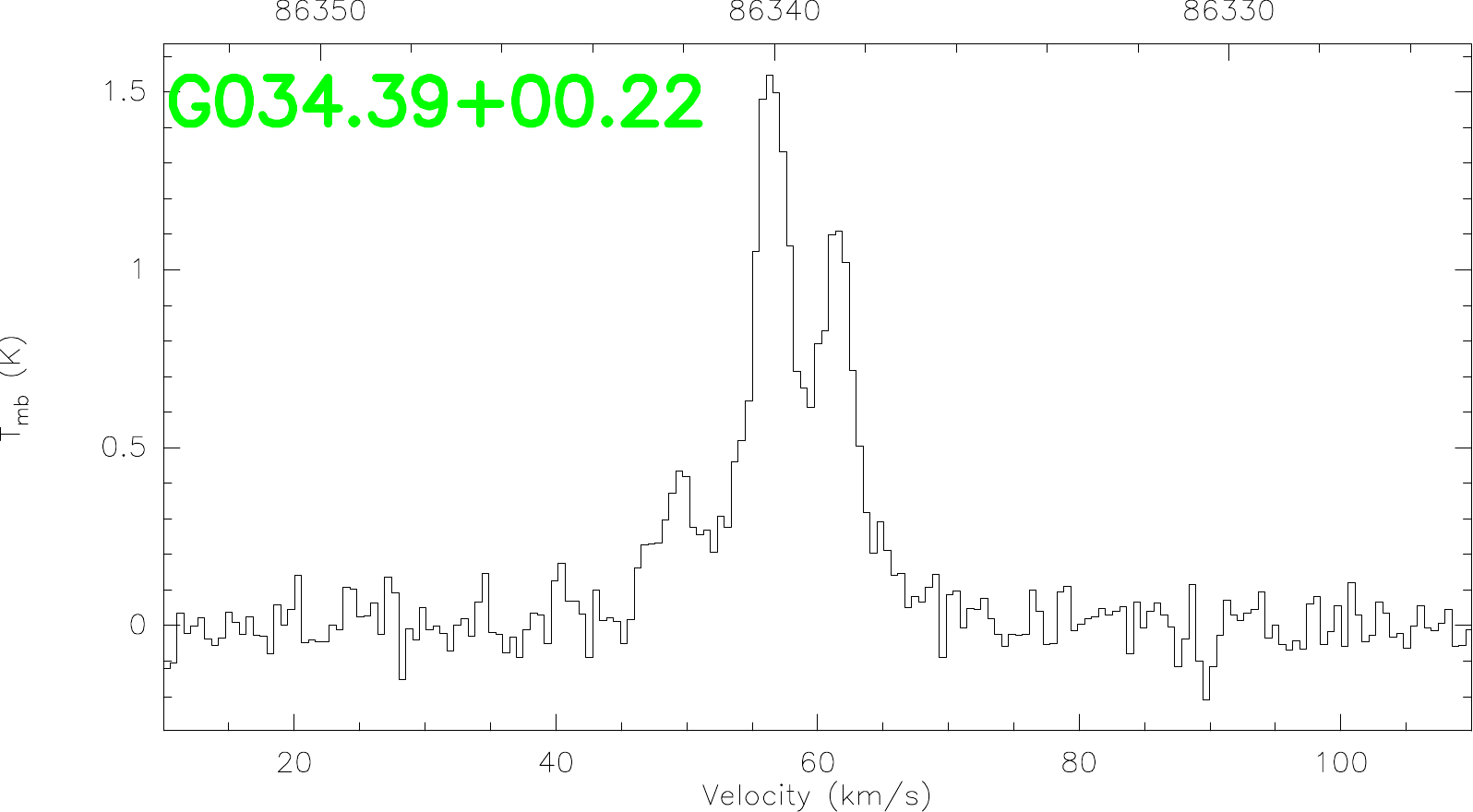}
\includegraphics[width=0.155\textwidth]{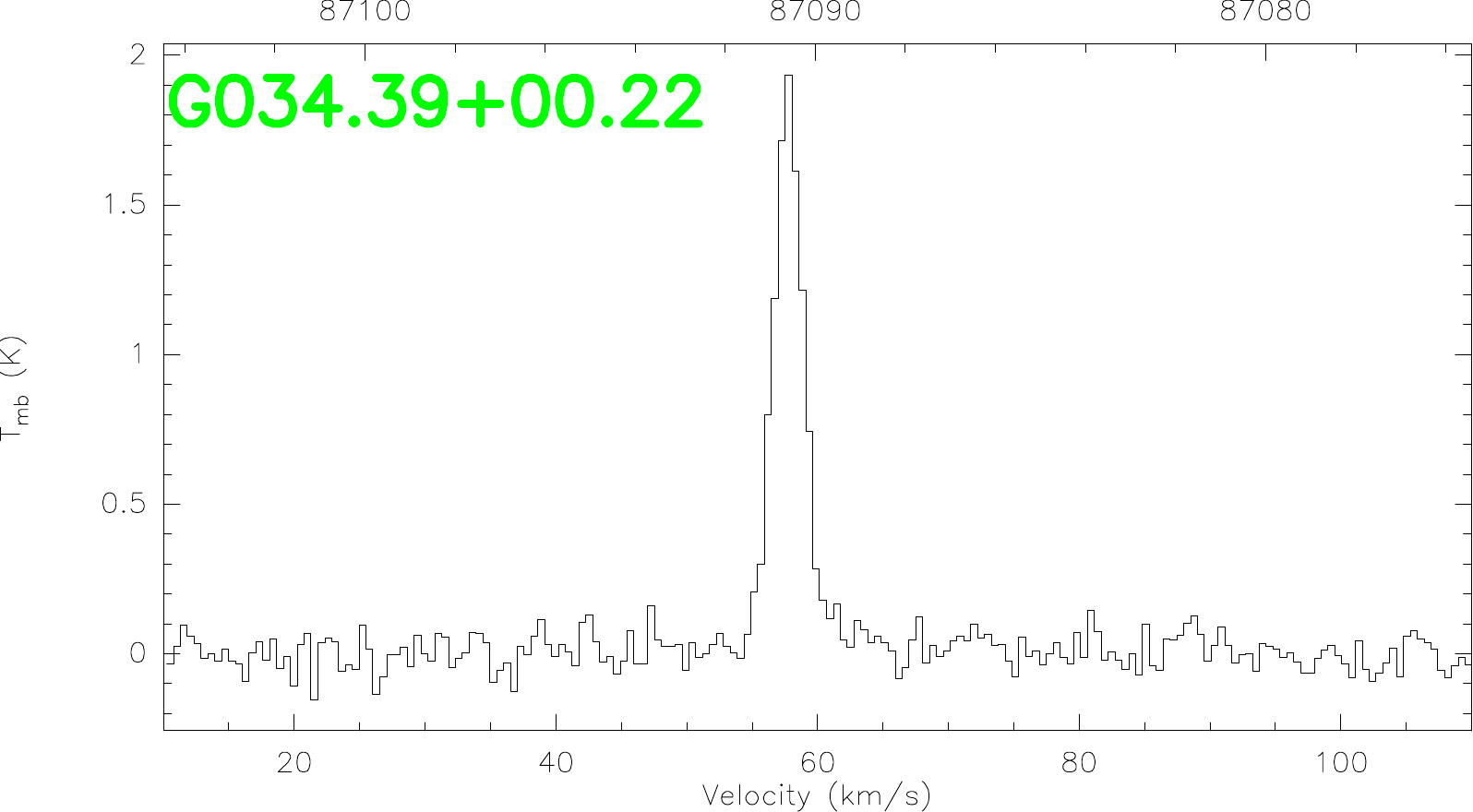}
\includegraphics[width=0.155\textwidth]{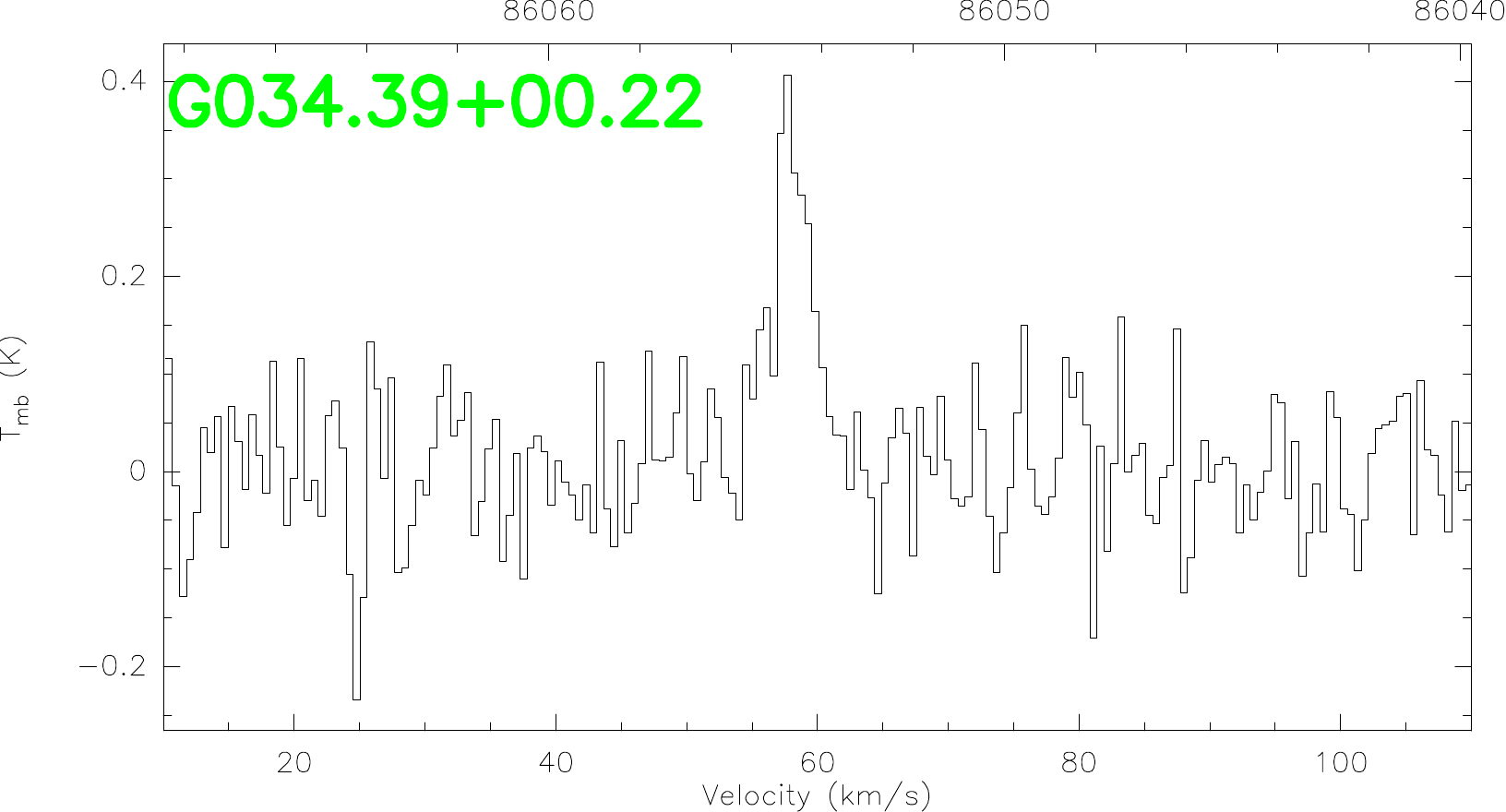}
\includegraphics[width=0.155\textwidth]{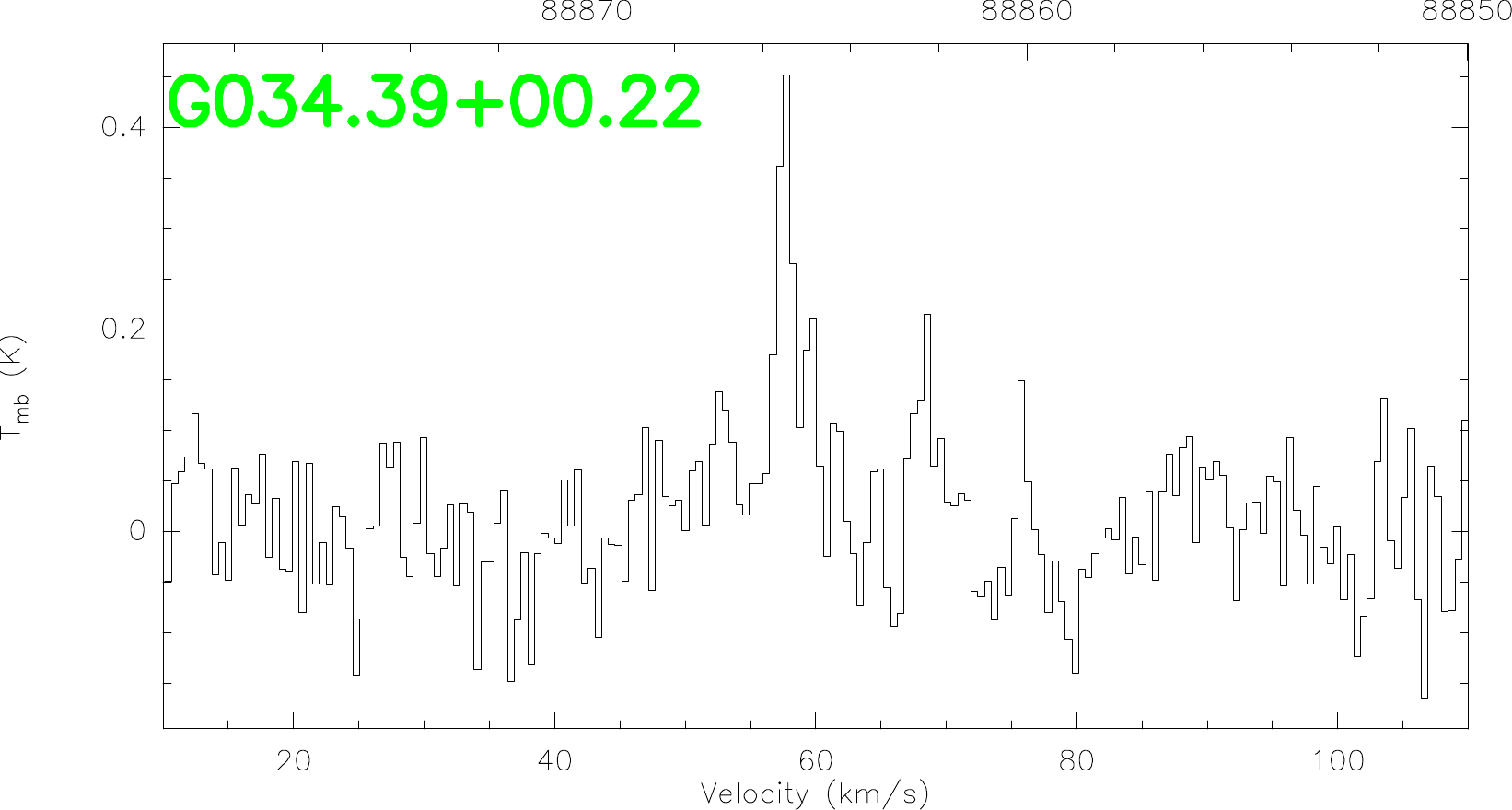}\\

\includegraphics[width=0.155\textwidth]{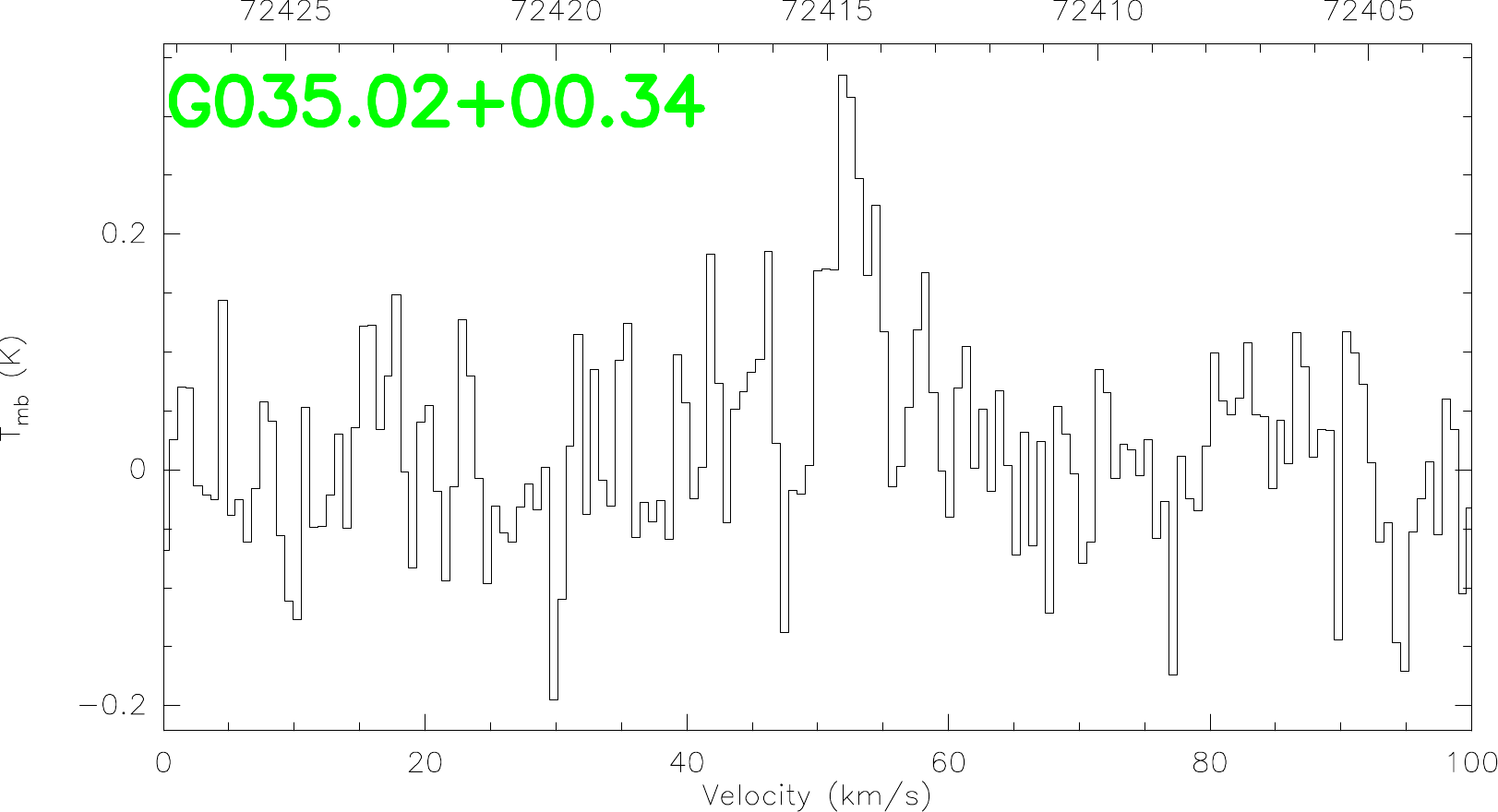}
\includegraphics[width=0.155\textwidth]{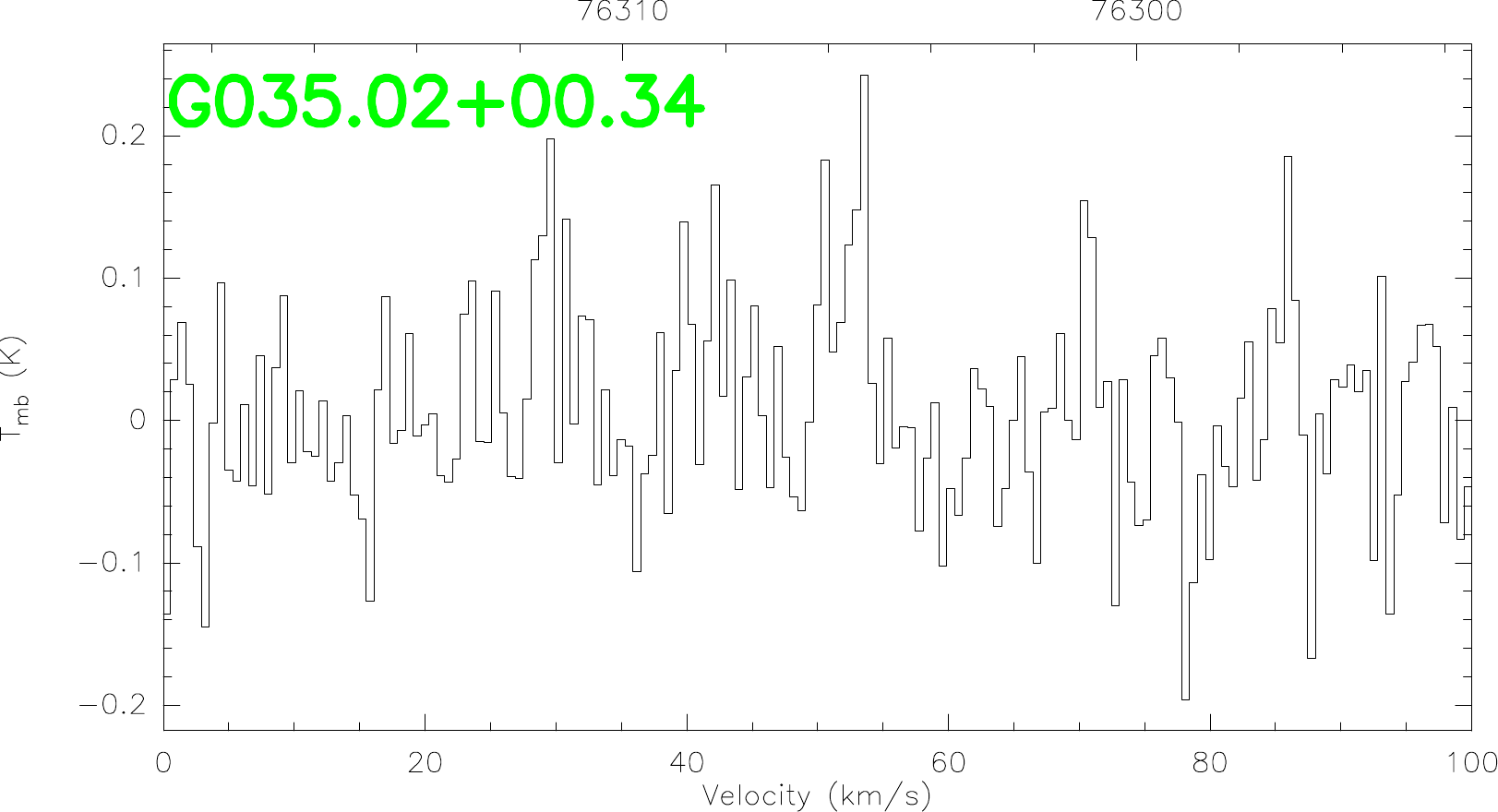}
\includegraphics[width=0.155\textwidth]{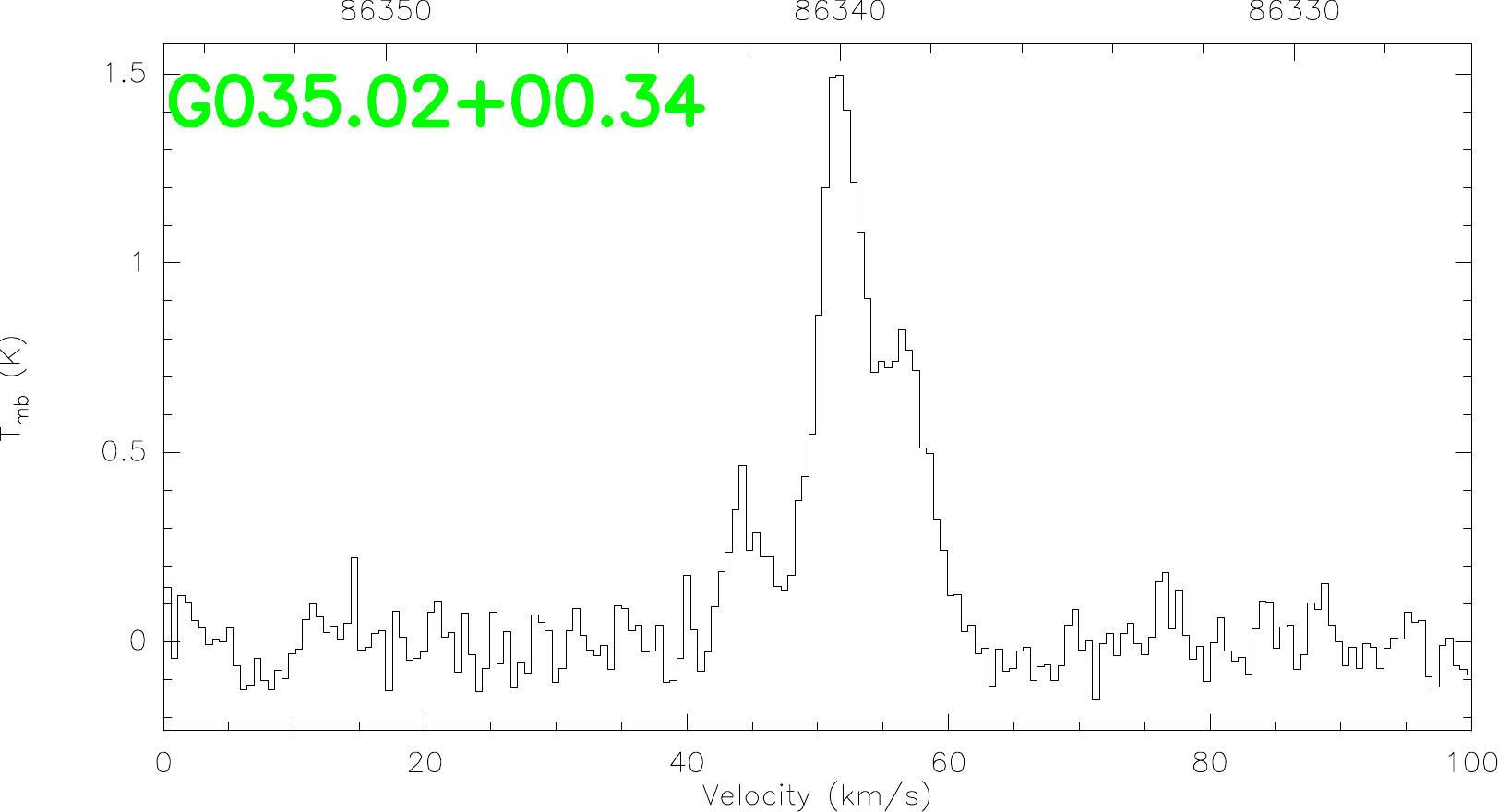}
\includegraphics[width=0.155\textwidth]{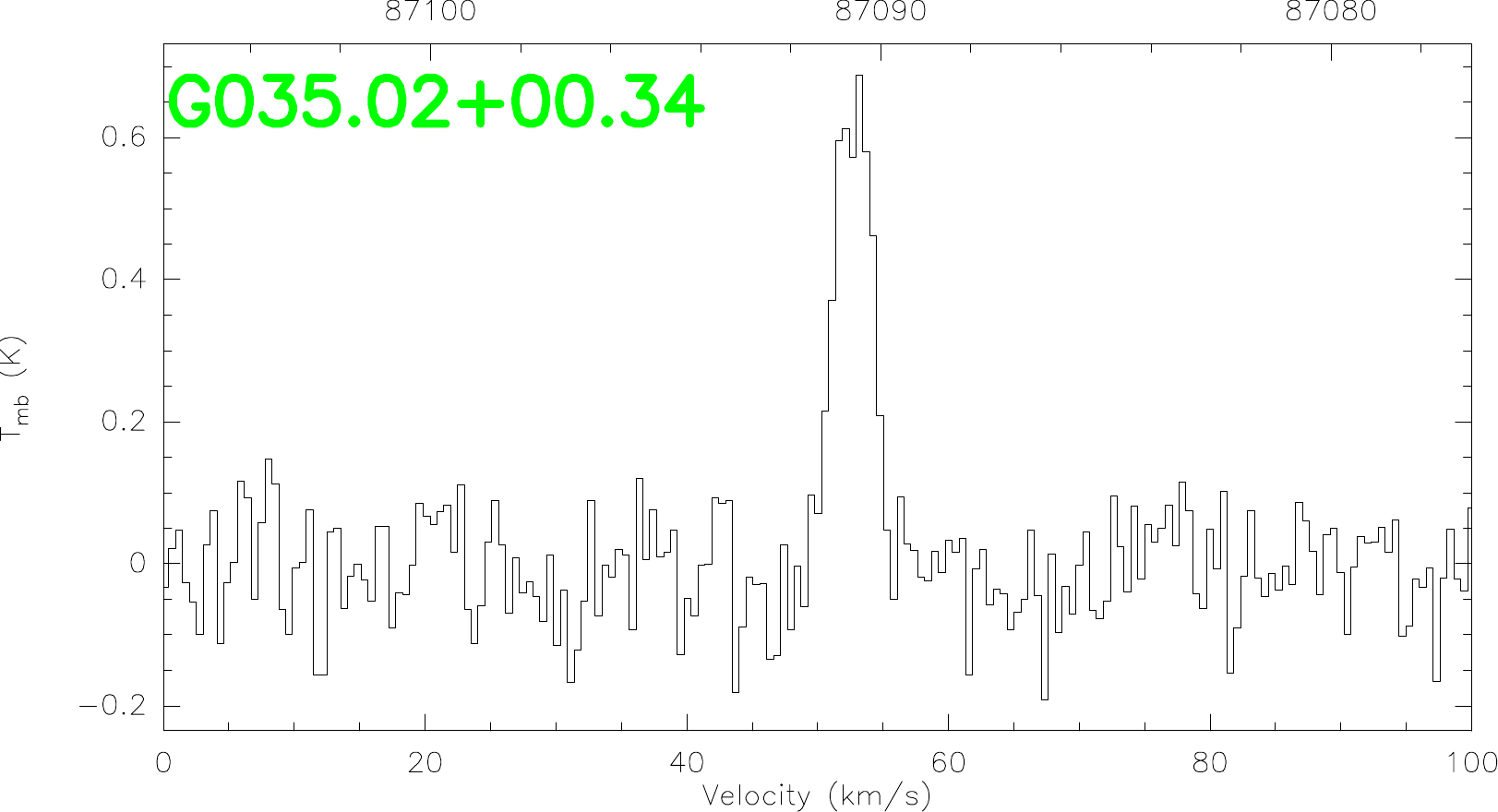}
\includegraphics[width=0.155\textwidth]{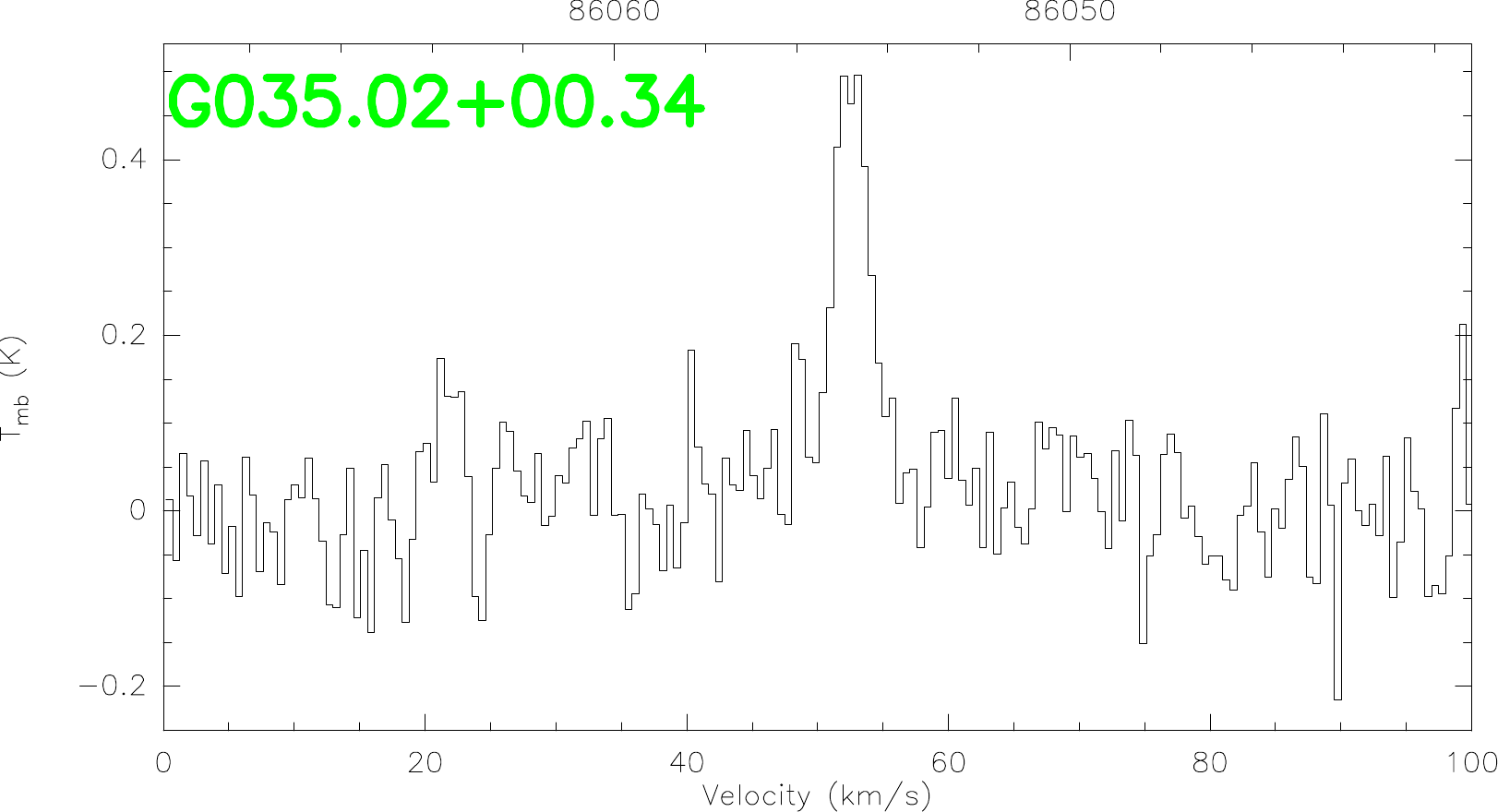}
\includegraphics[width=0.155\textwidth]{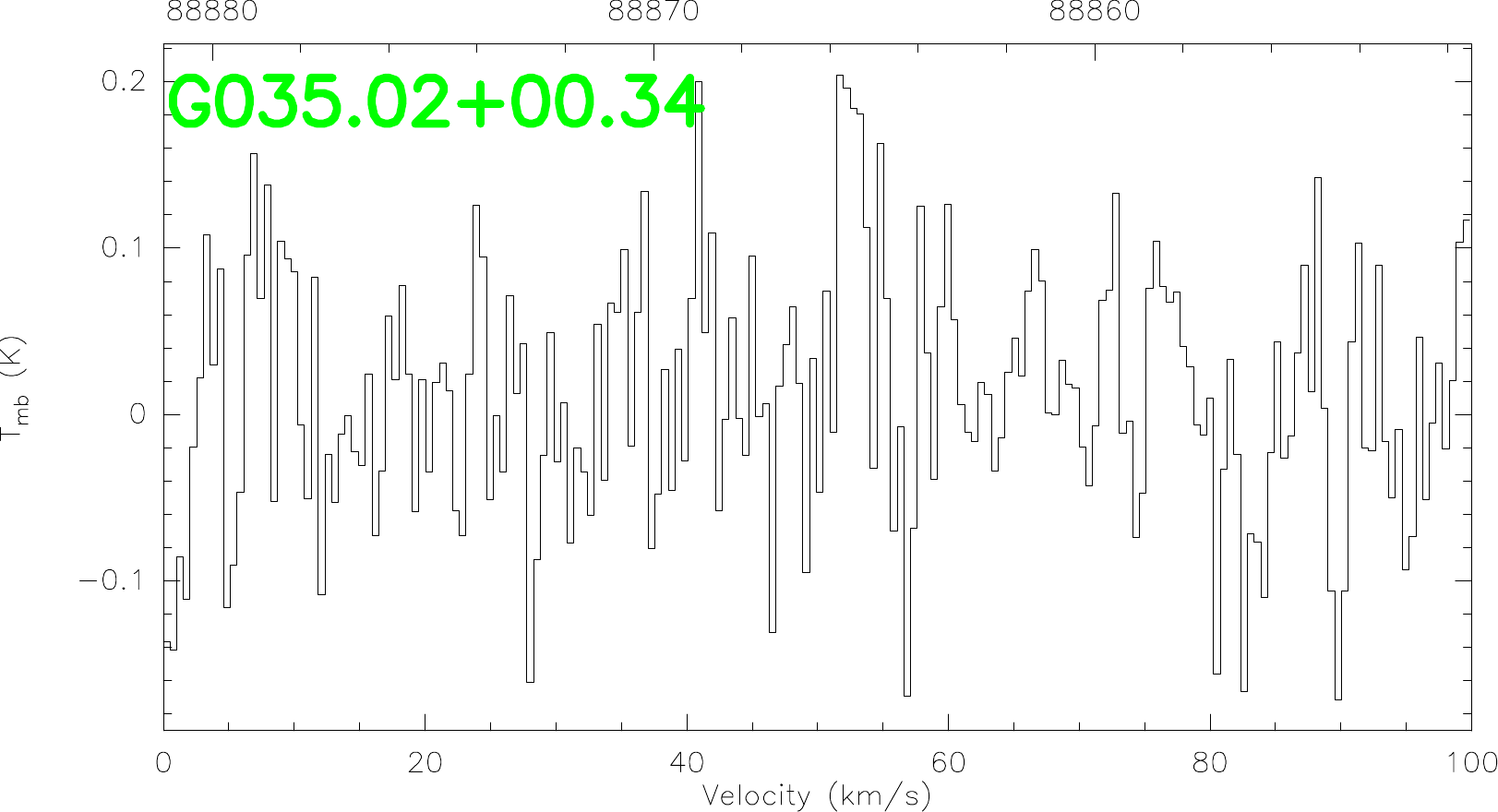}\\

\includegraphics[width=0.155\textwidth]{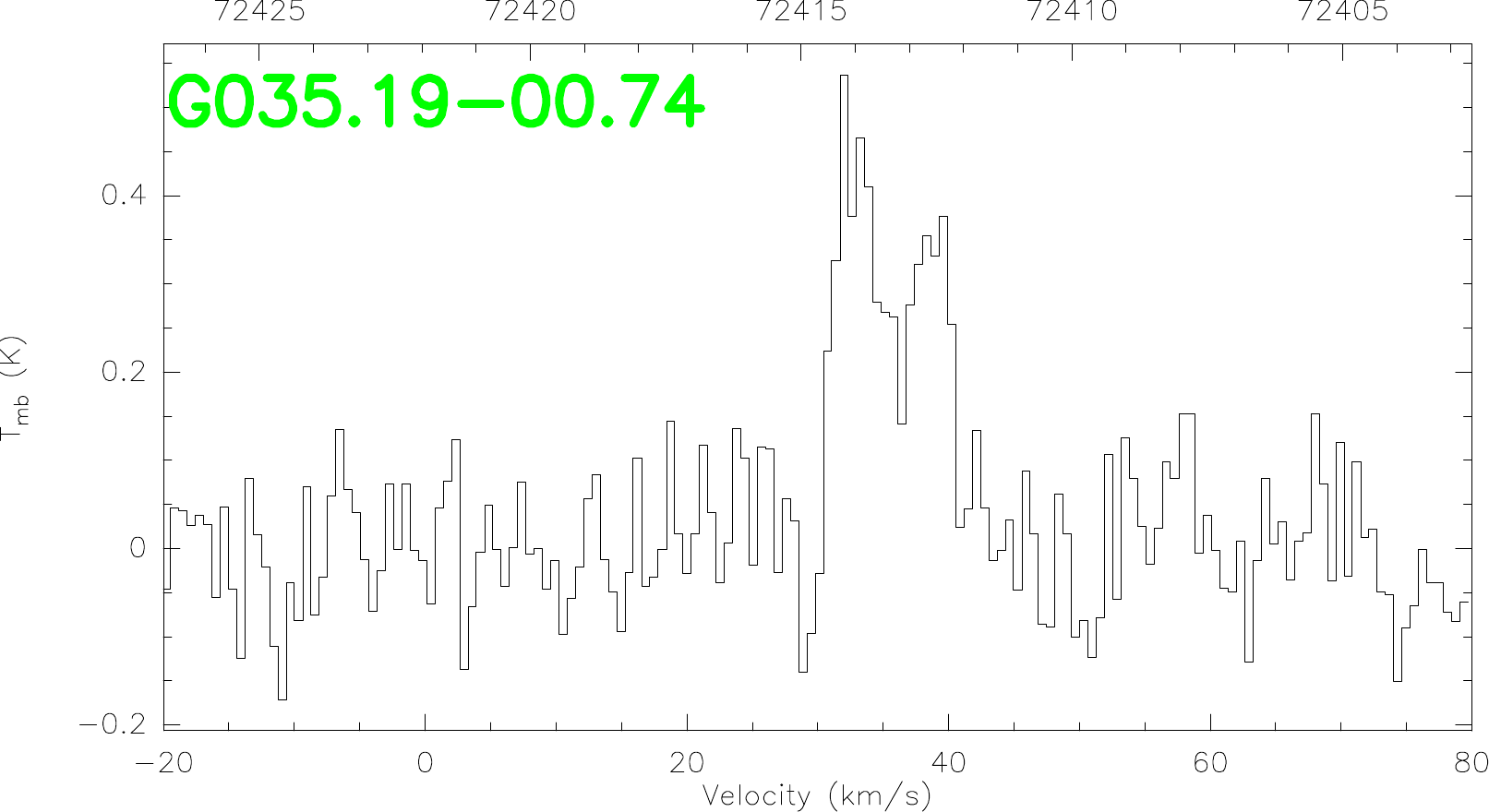}
\includegraphics[width=0.155\textwidth]{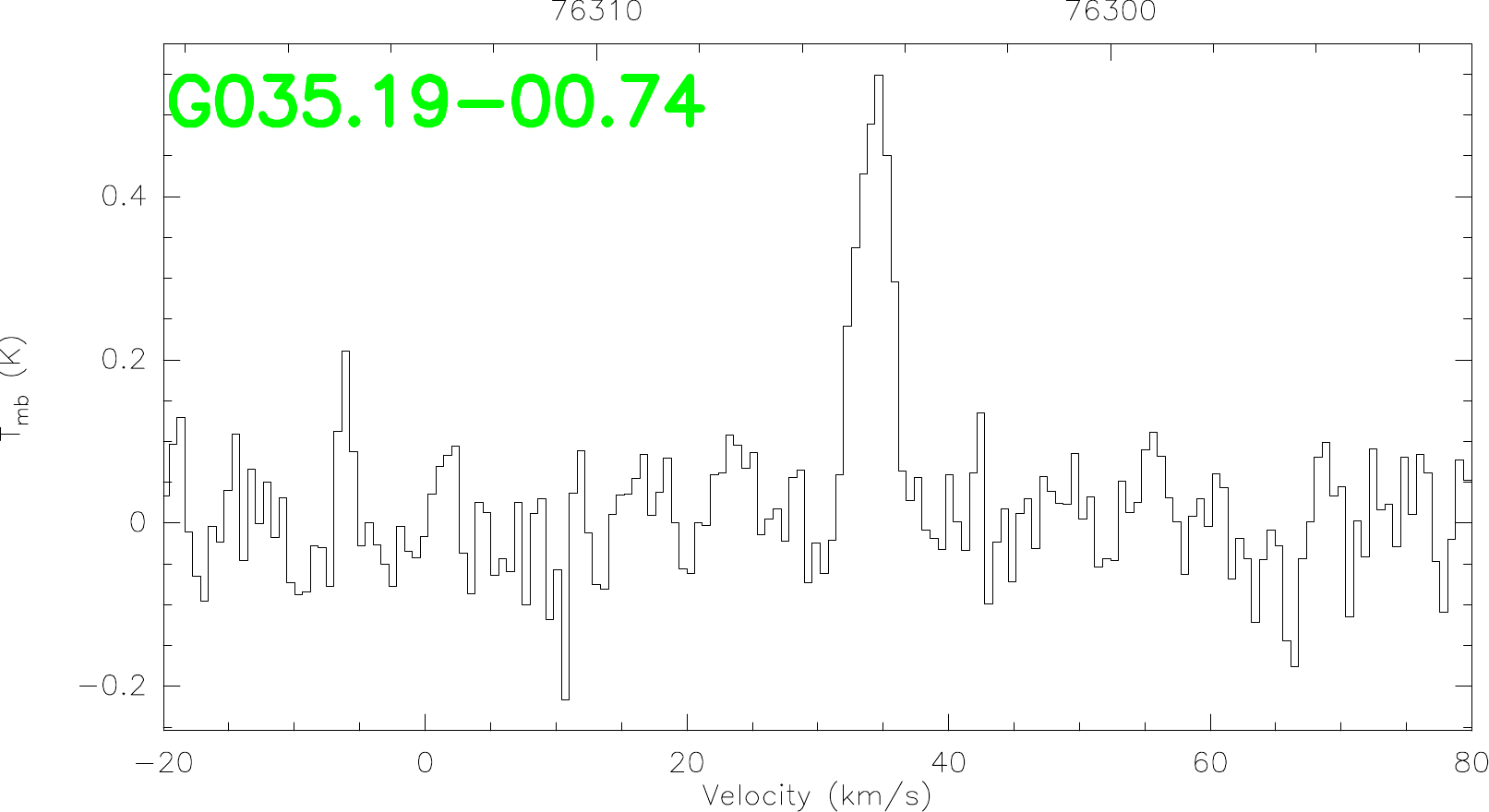}
\includegraphics[width=0.155\textwidth]{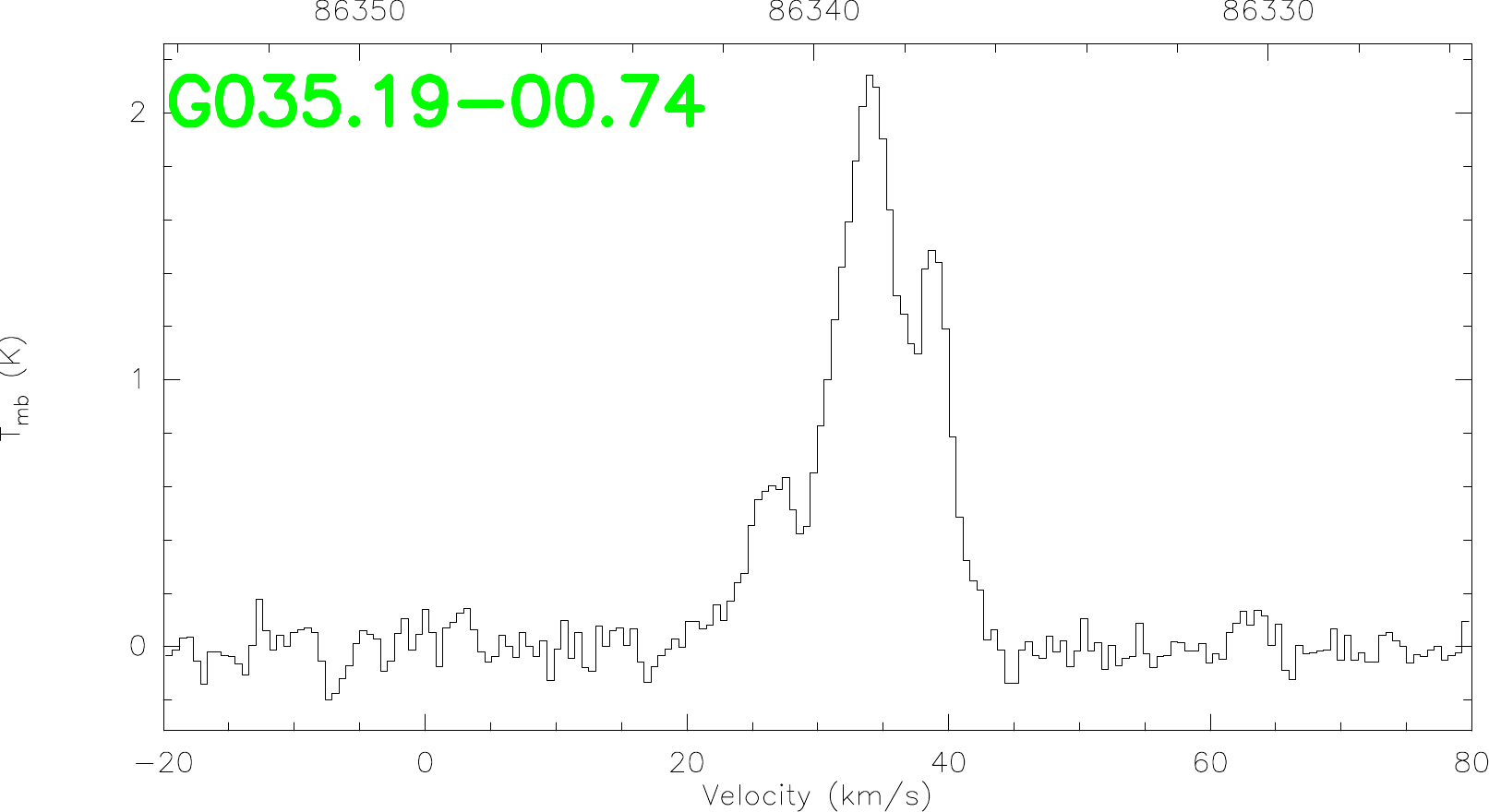}
\includegraphics[width=0.155\textwidth]{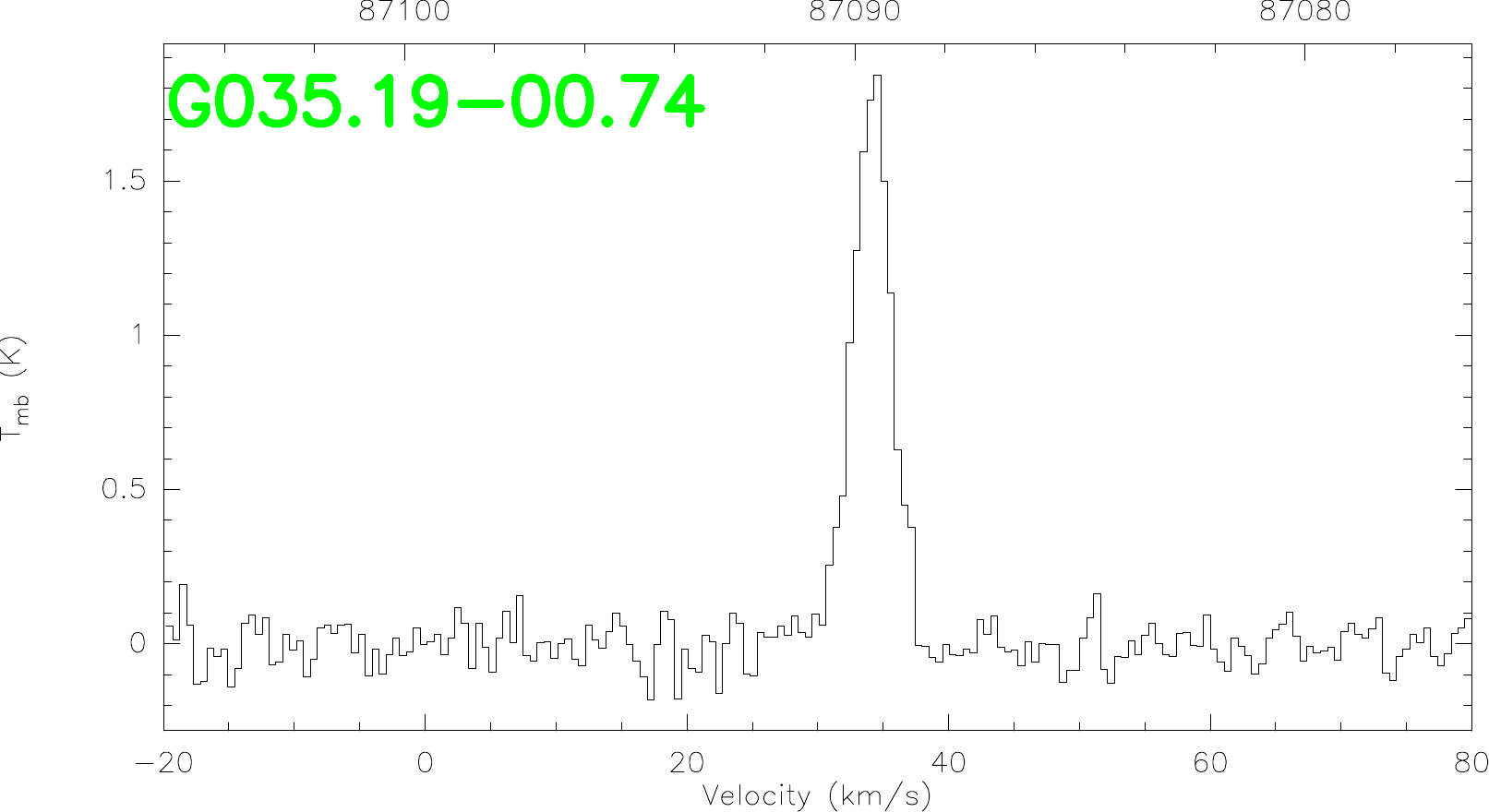}
\includegraphics[width=0.155\textwidth]{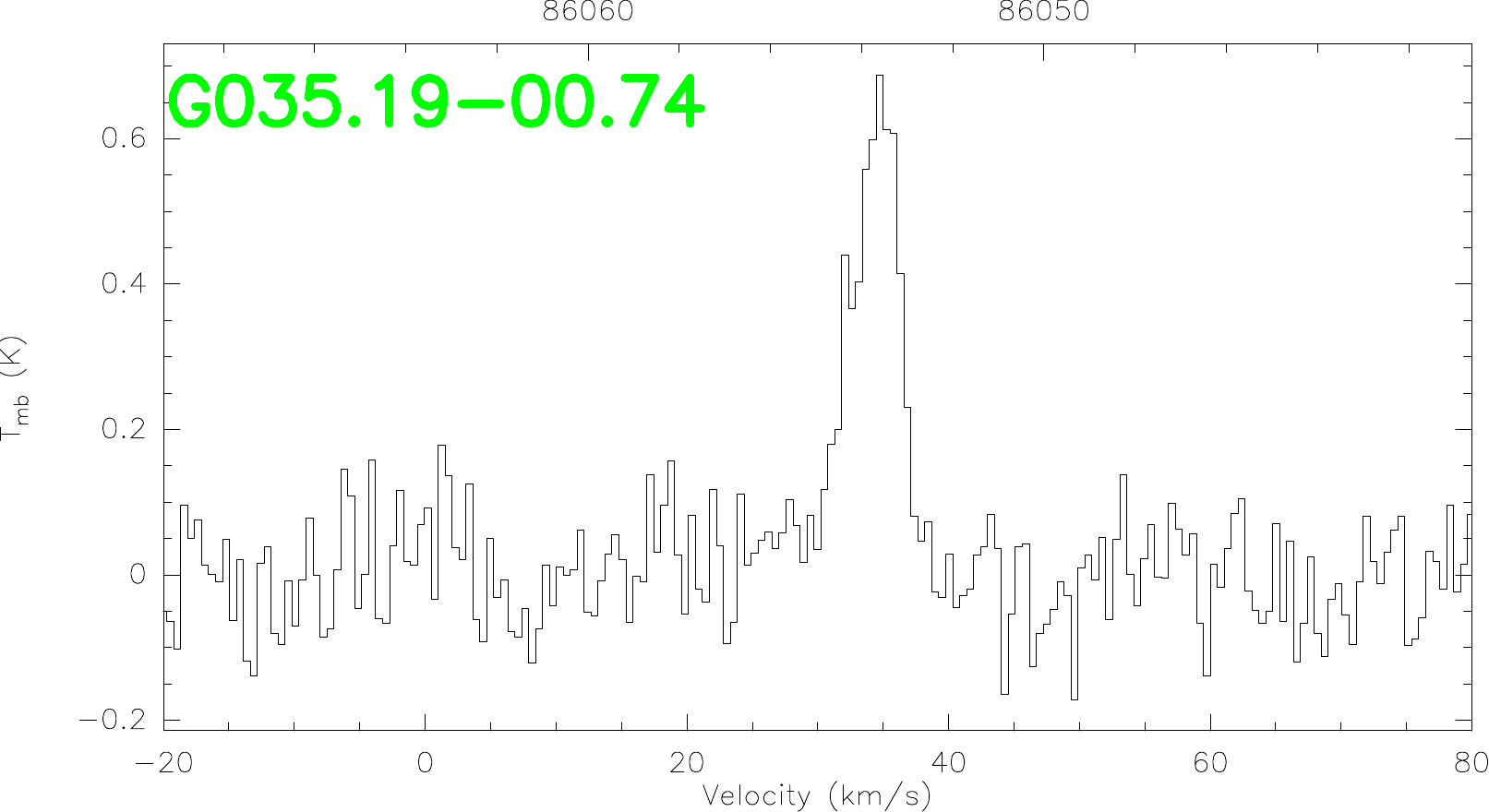}
\includegraphics[width=0.155\textwidth]{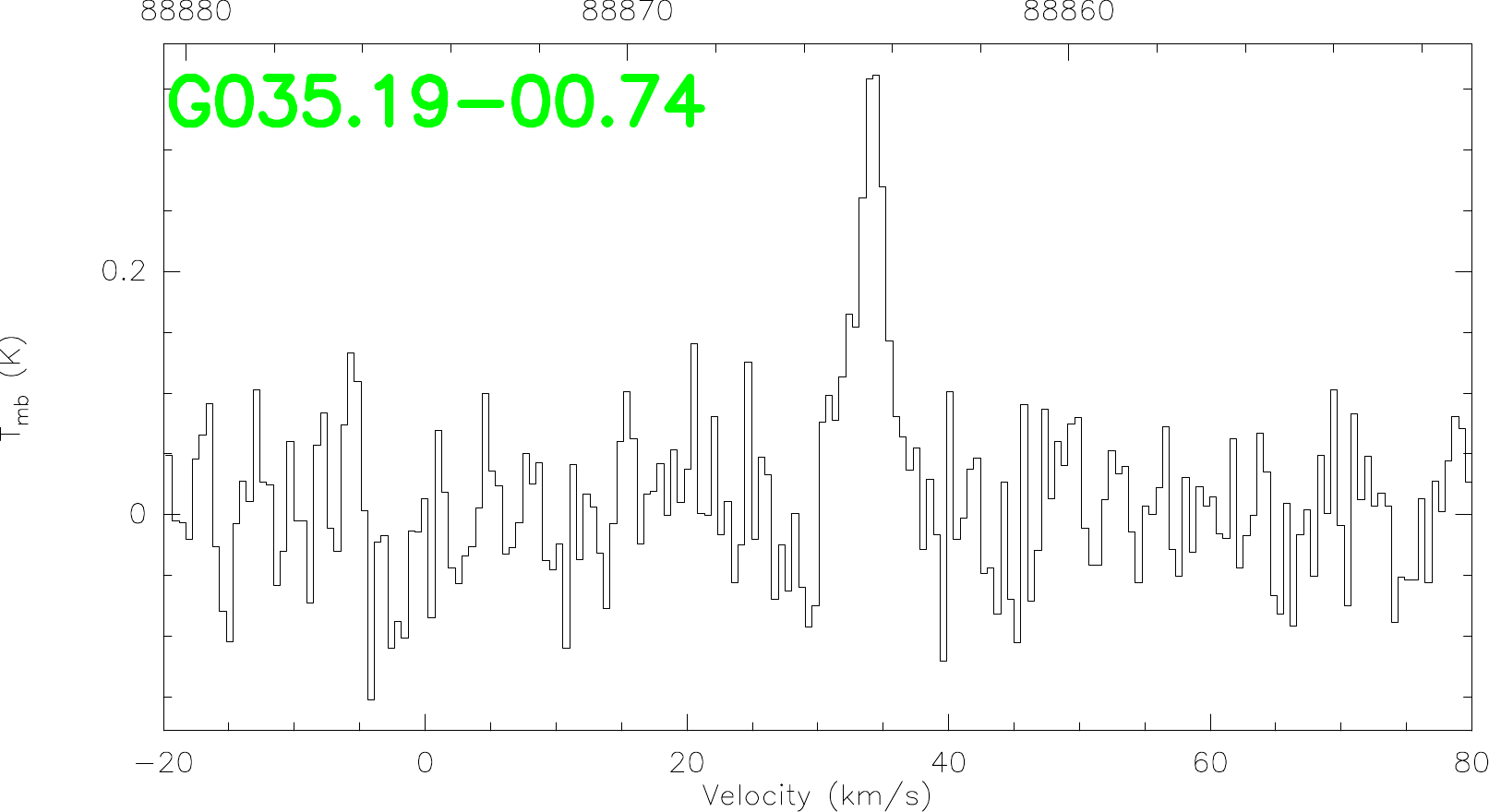}\\

\includegraphics[width=0.155\textwidth]{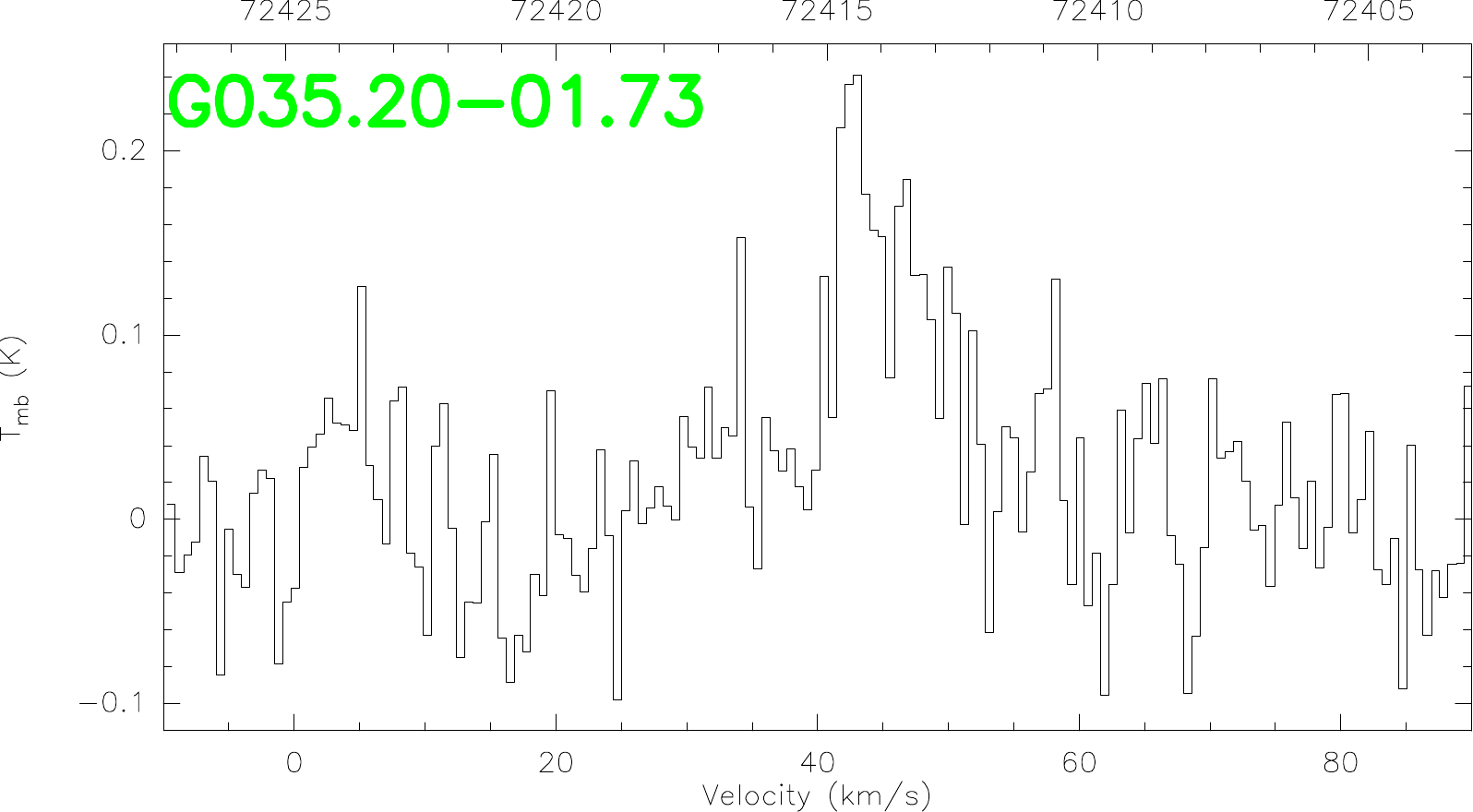}
\includegraphics[width=0.155\textwidth]{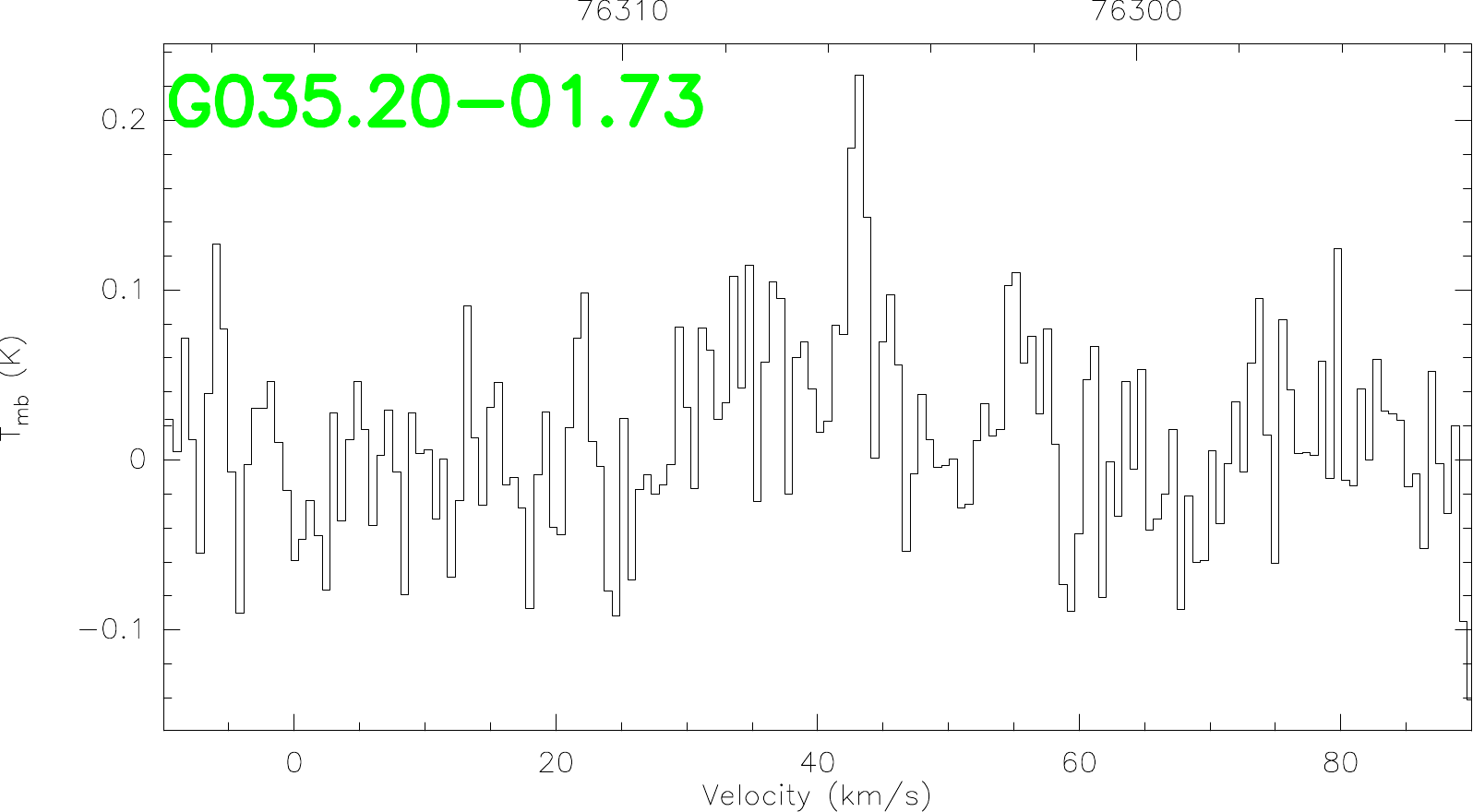}
\includegraphics[width=0.155\textwidth]{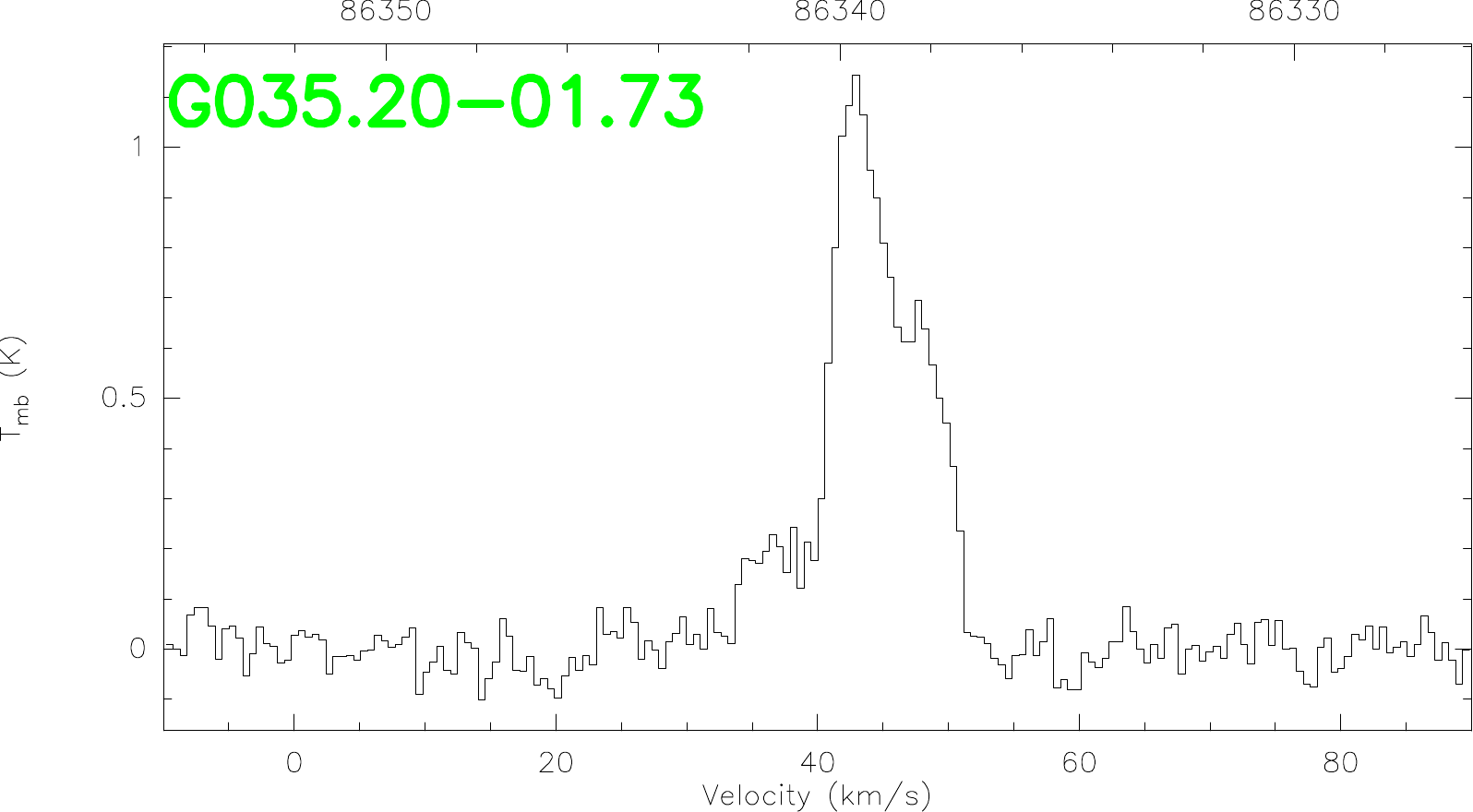}
\includegraphics[width=0.155\textwidth]{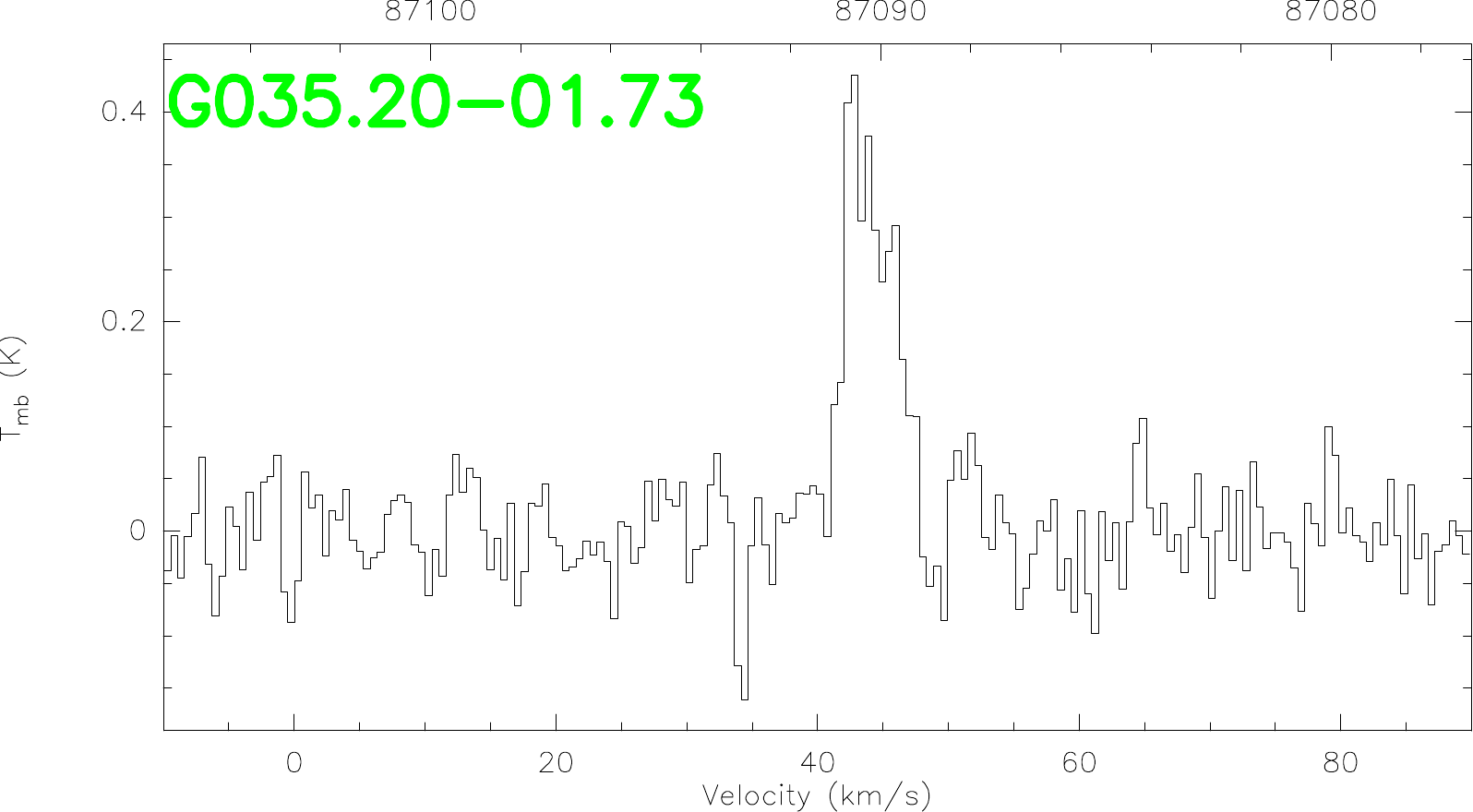}
\includegraphics[width=0.155\textwidth]{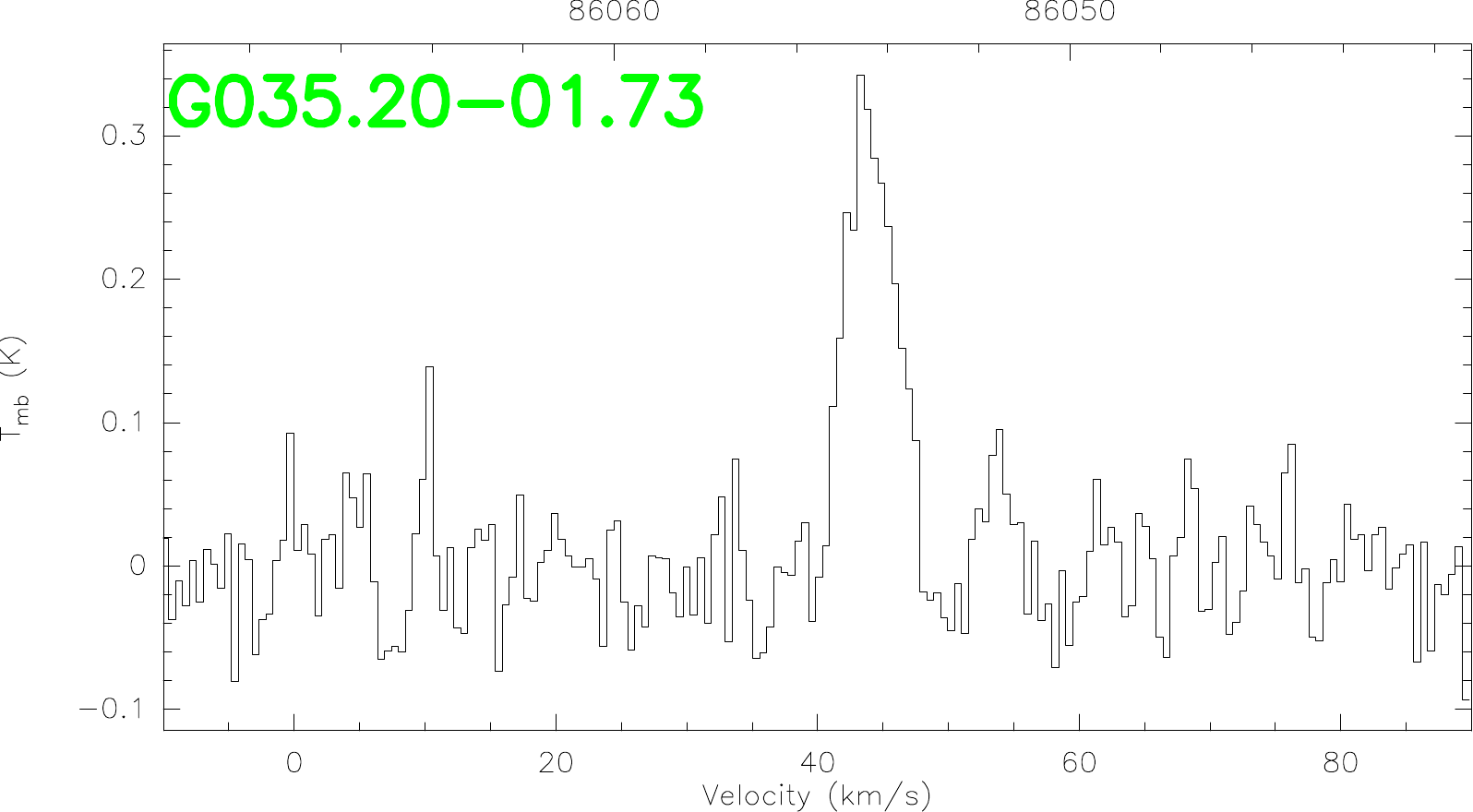}
\includegraphics[width=0.155\textwidth]{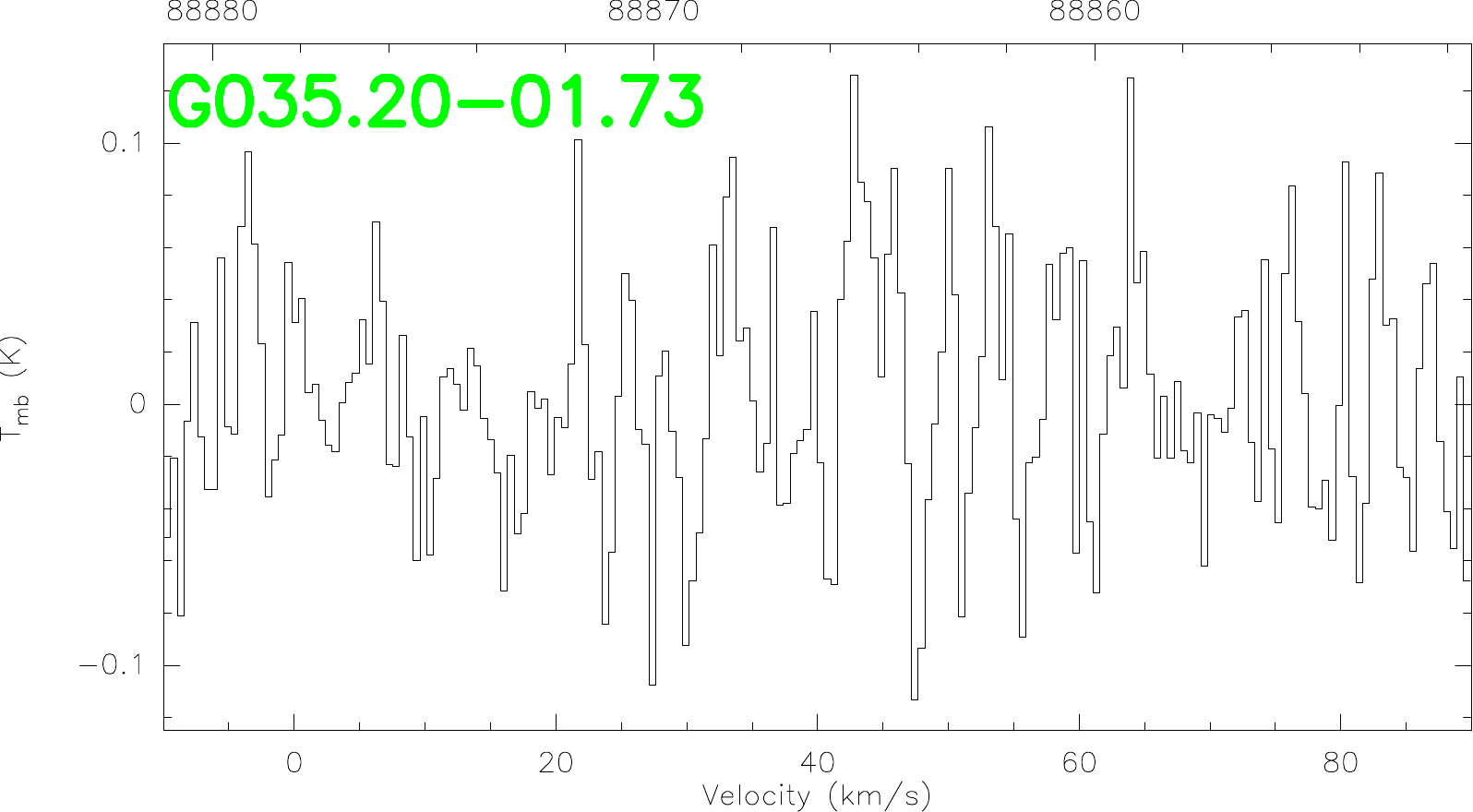}\\

\includegraphics[width=0.155\textwidth]{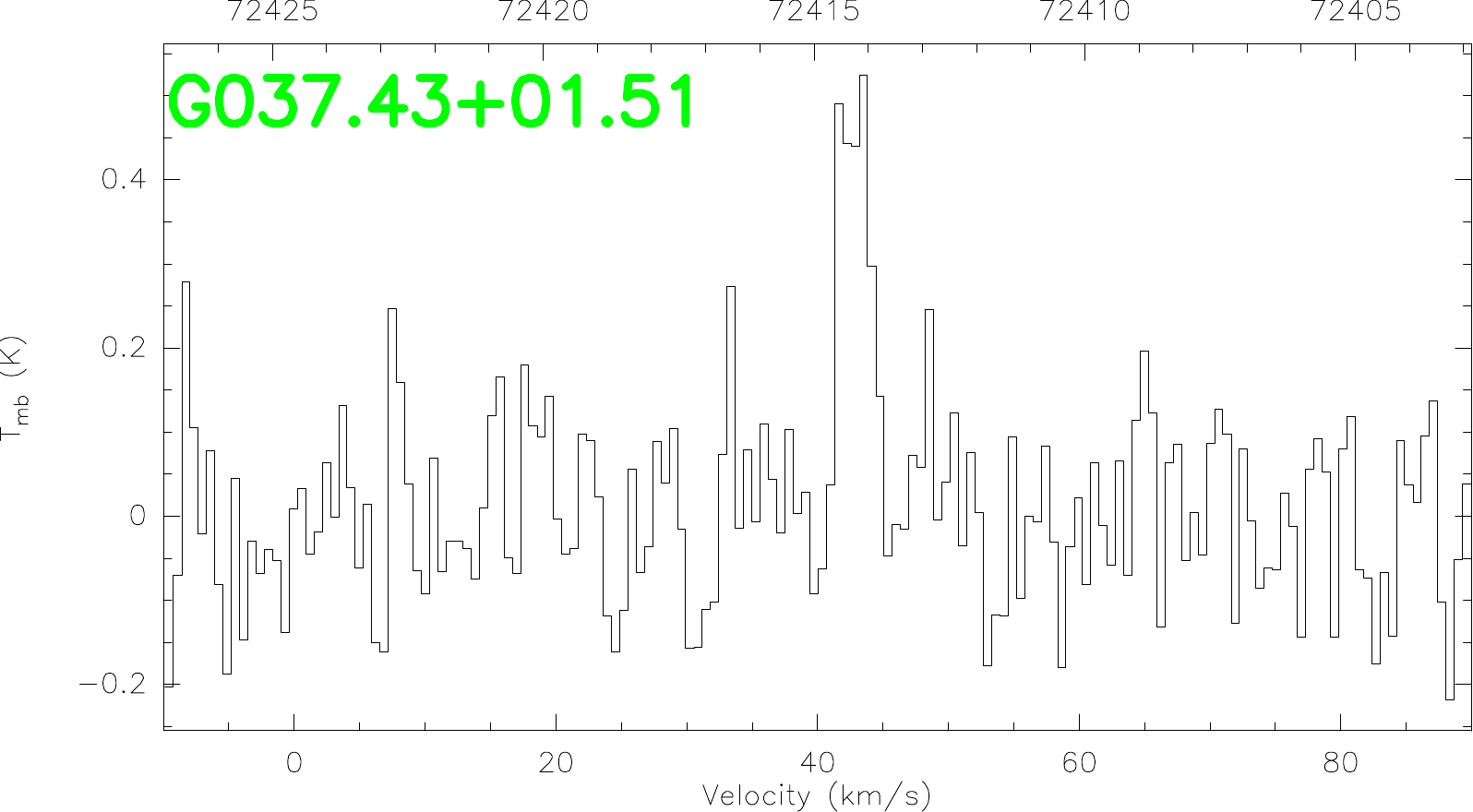}
\includegraphics[width=0.155\textwidth]{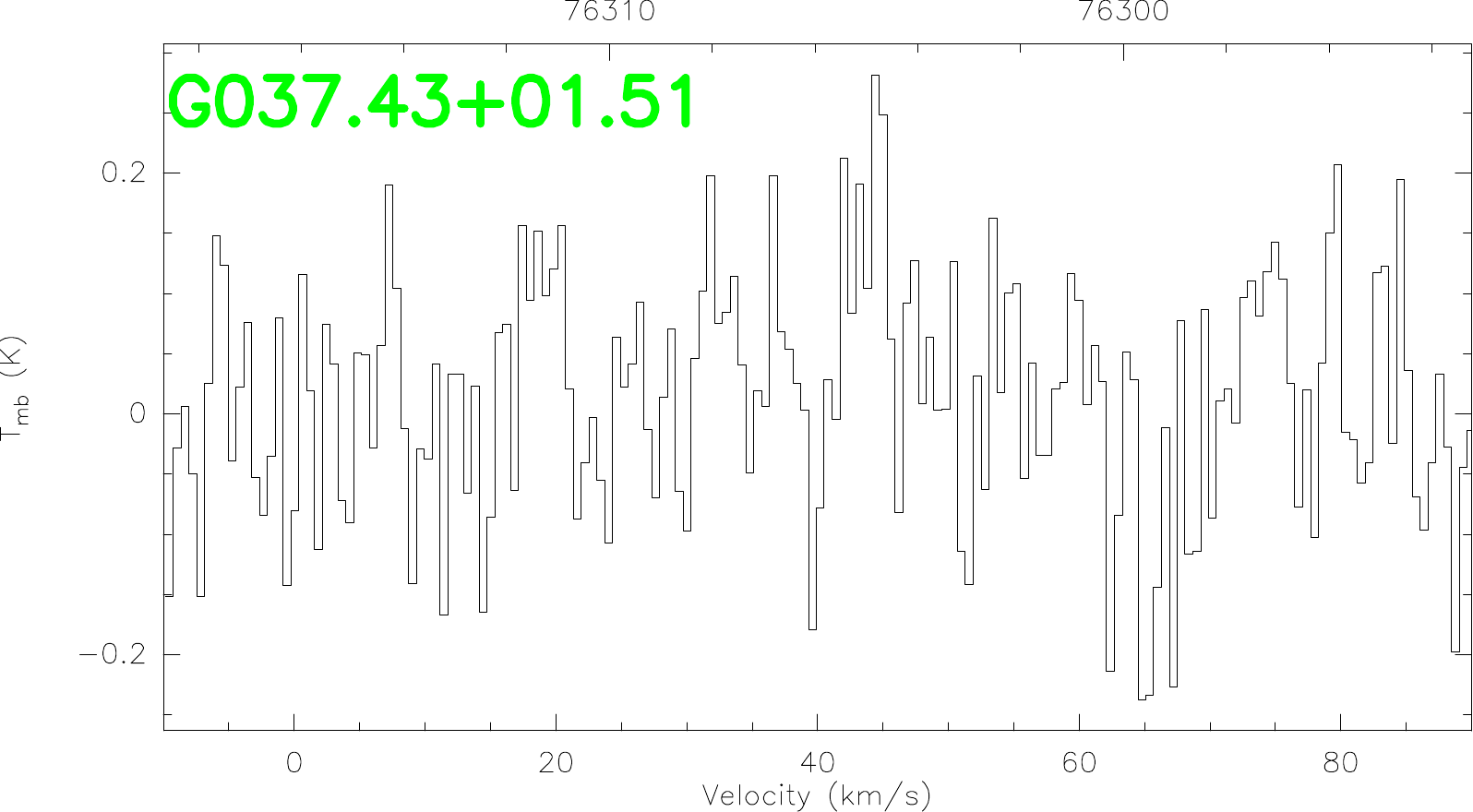}
\includegraphics[width=0.155\textwidth]{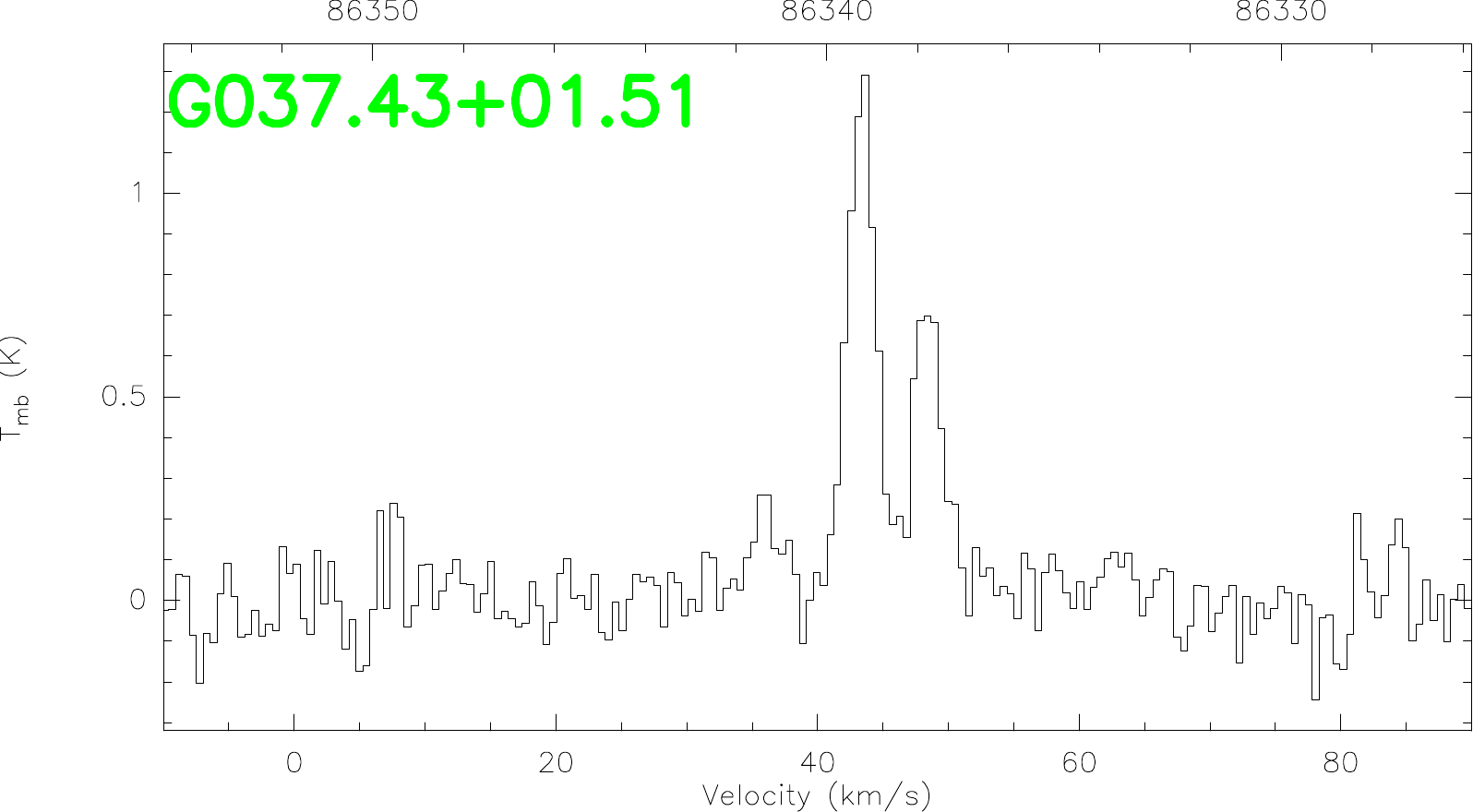}
\includegraphics[width=0.155\textwidth]{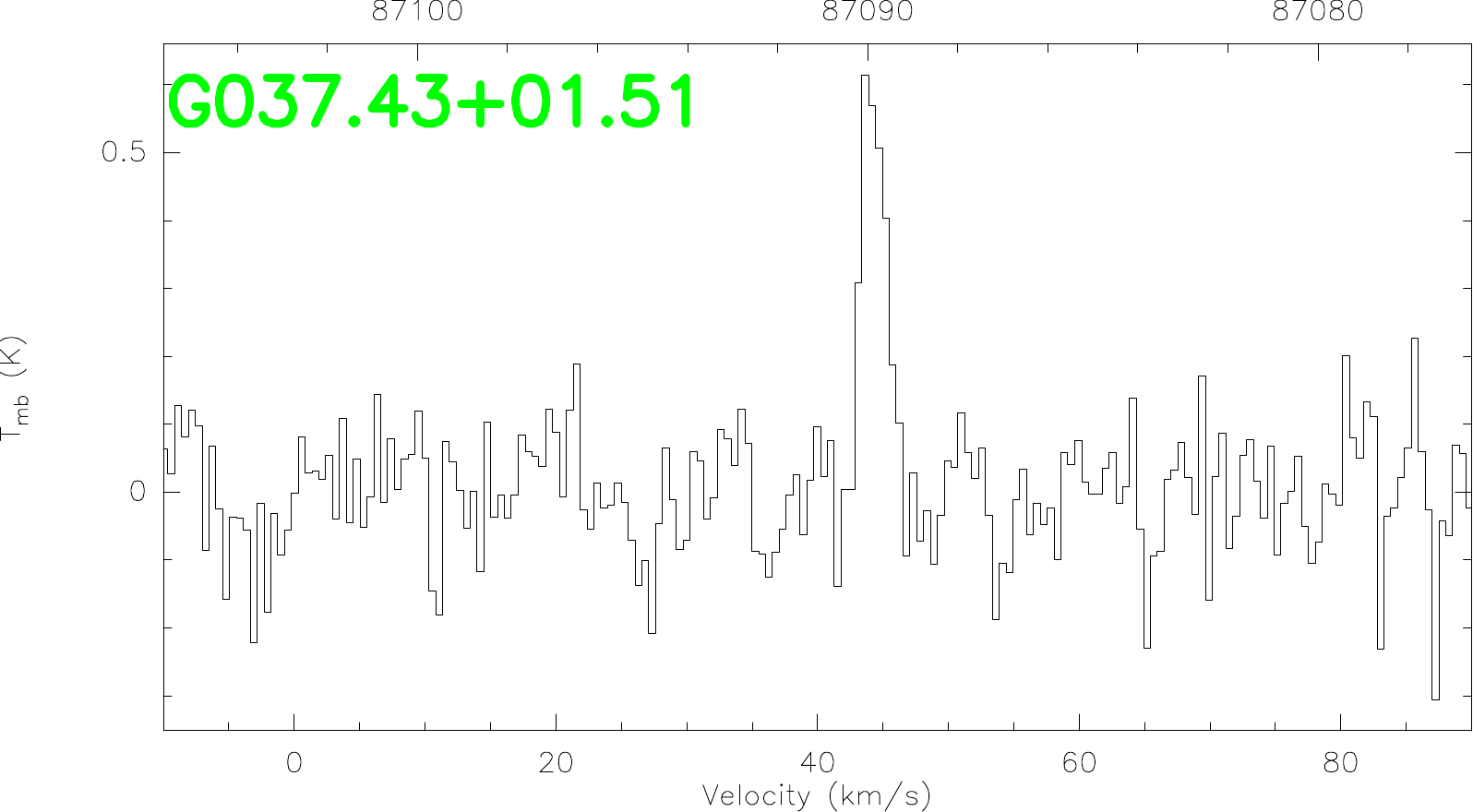}
\includegraphics[width=0.155\textwidth]{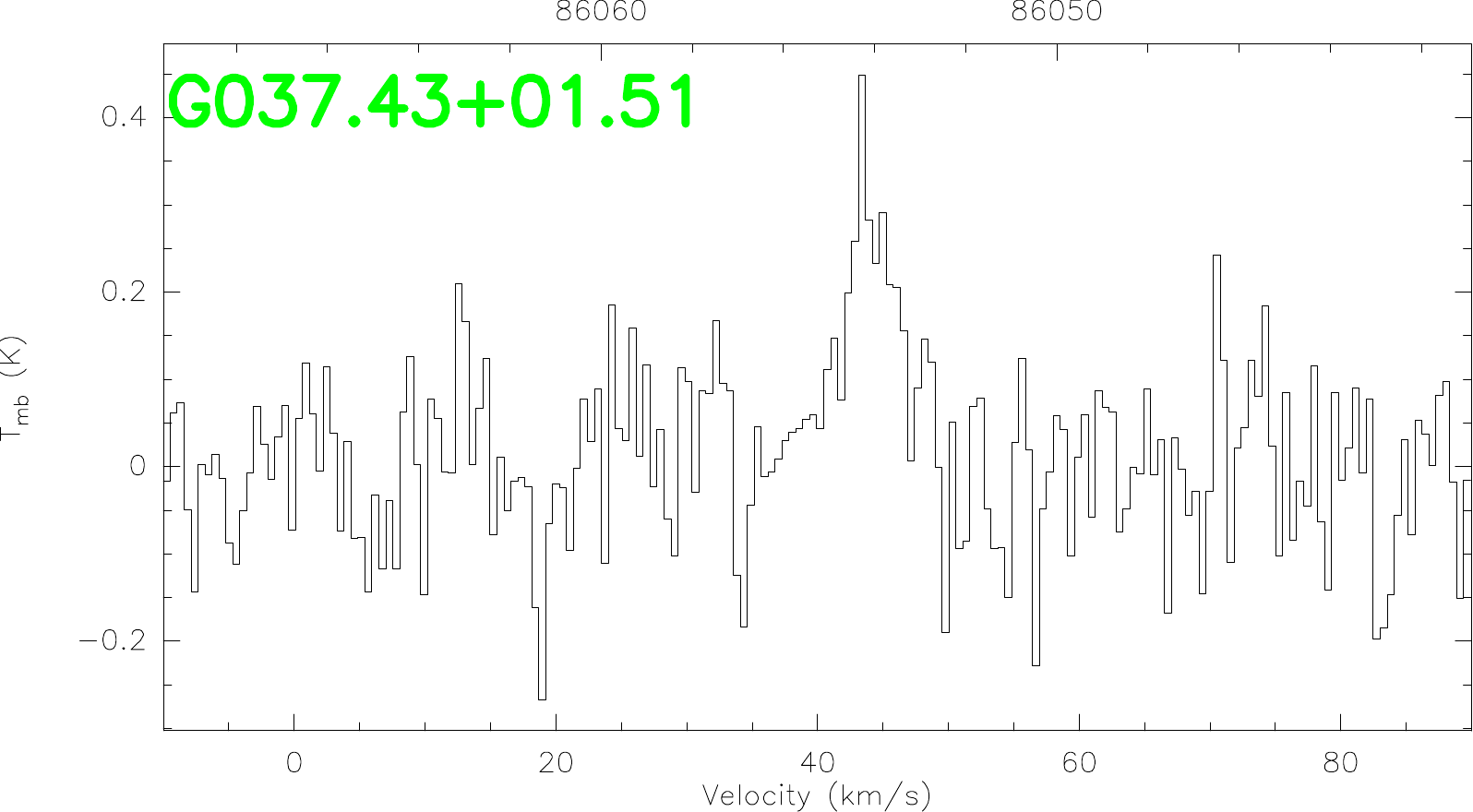}
\includegraphics[width=0.155\textwidth]{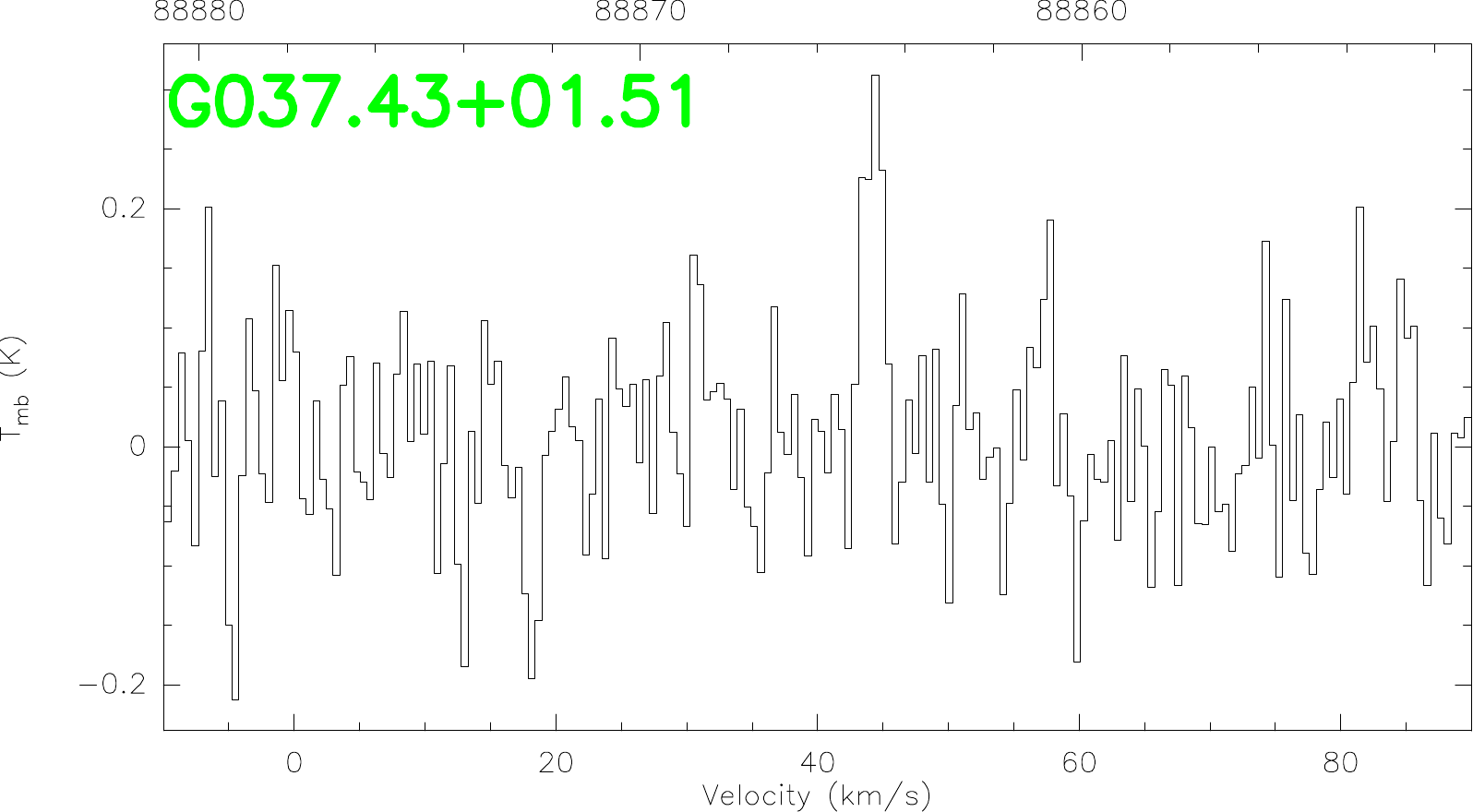}\\

\includegraphics[width=0.155\textwidth]{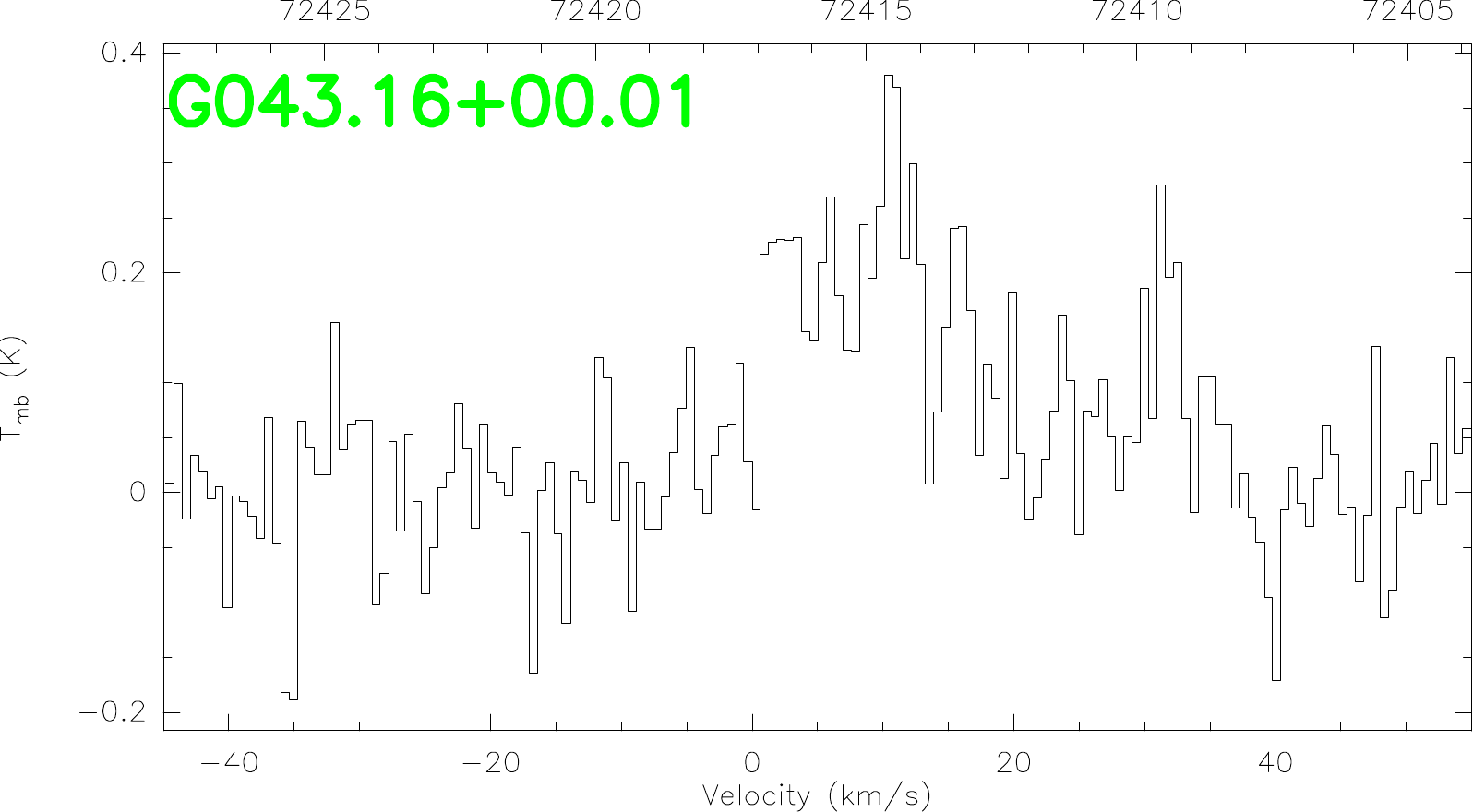}
\includegraphics[width=0.155\textwidth]{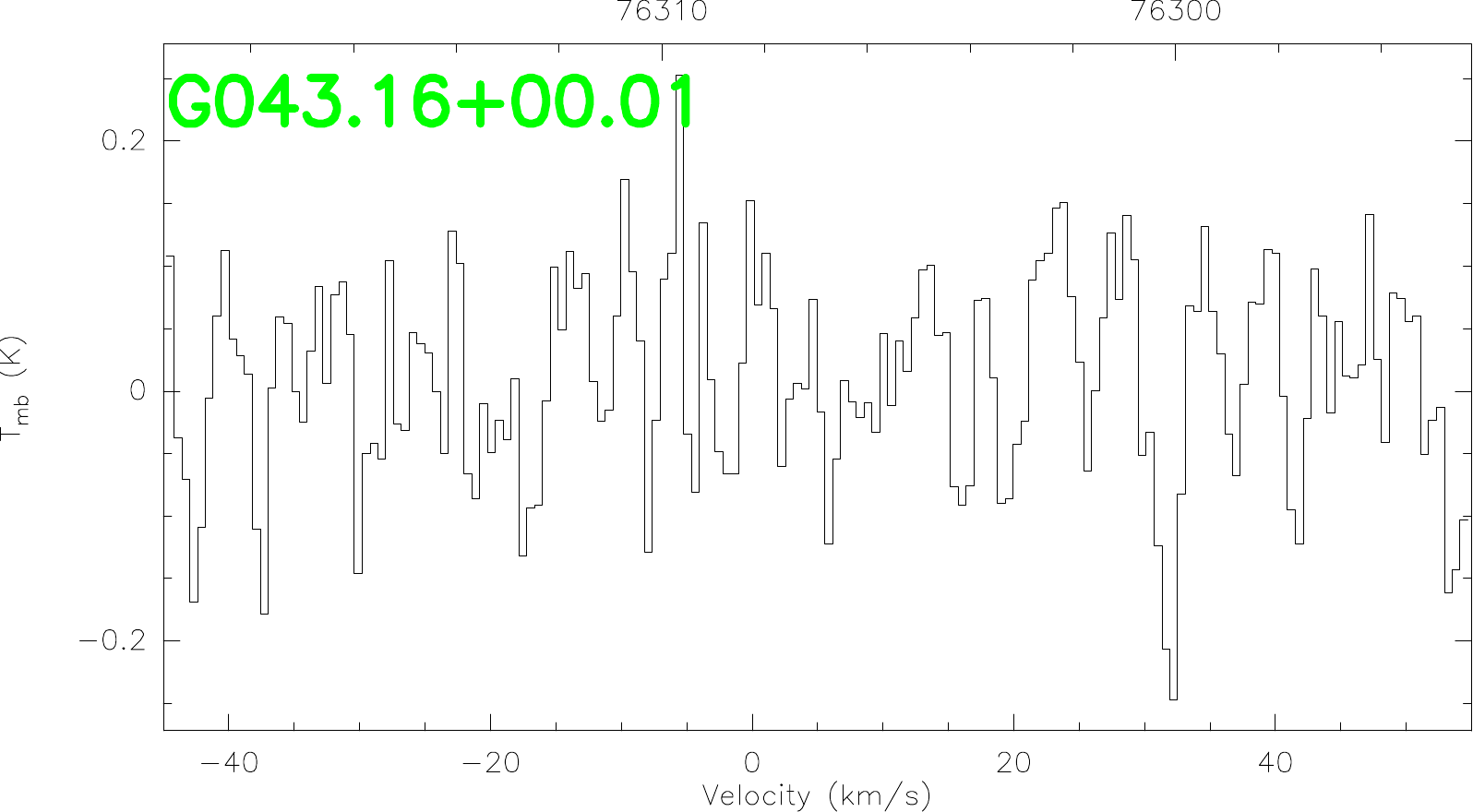}
\includegraphics[width=0.155\textwidth]{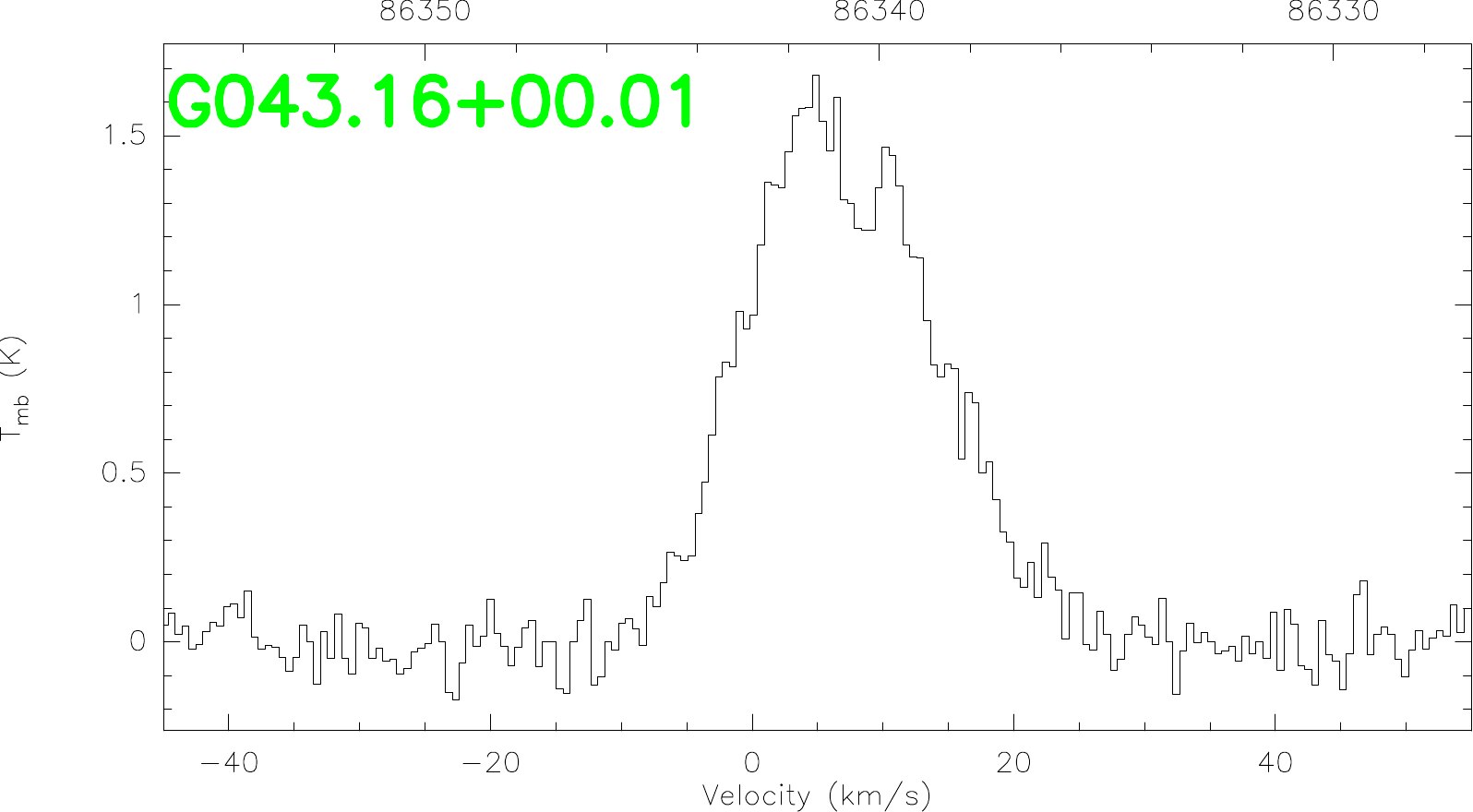}
\includegraphics[width=0.155\textwidth]{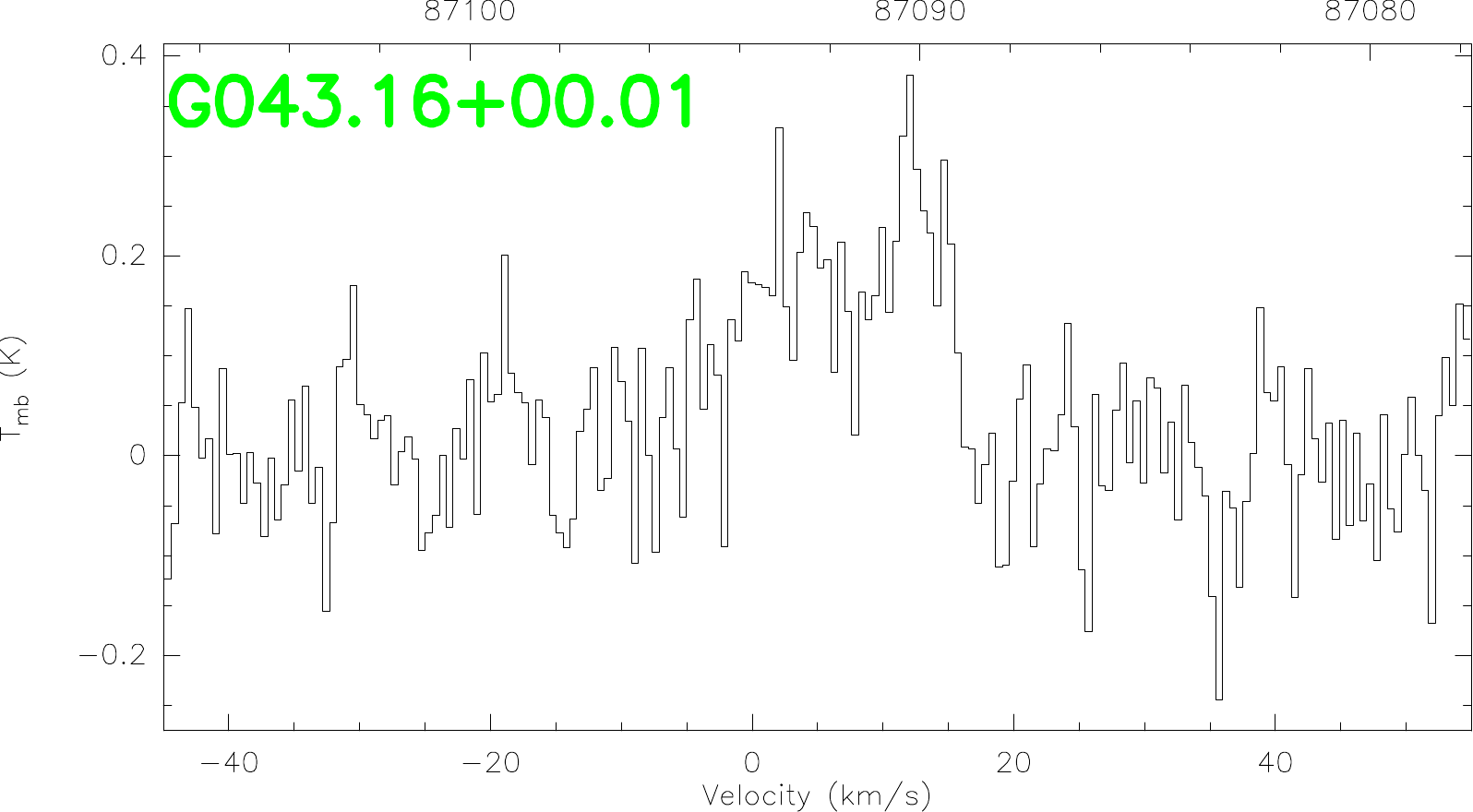}
\includegraphics[width=0.155\textwidth]{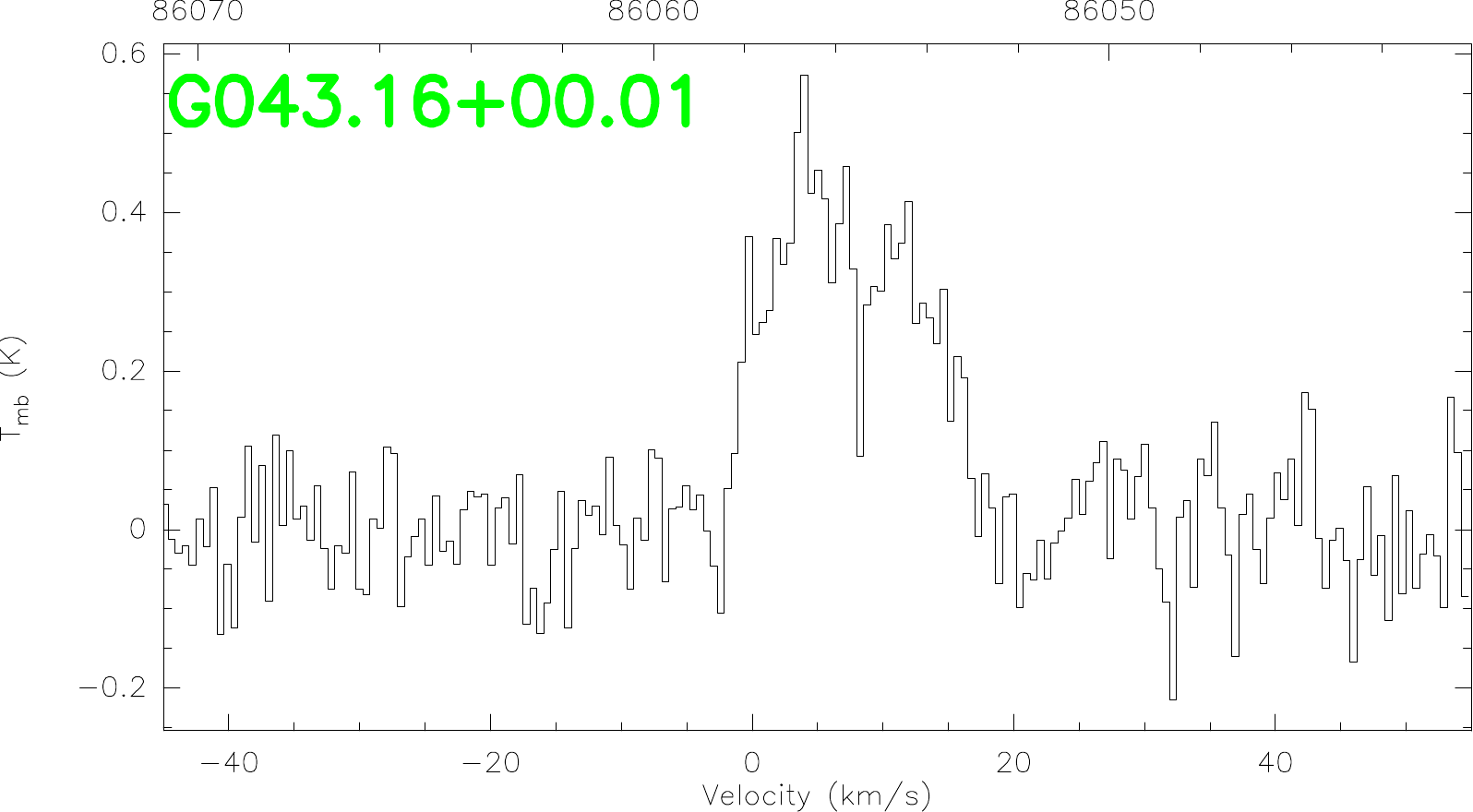}
\includegraphics[width=0.155\textwidth]{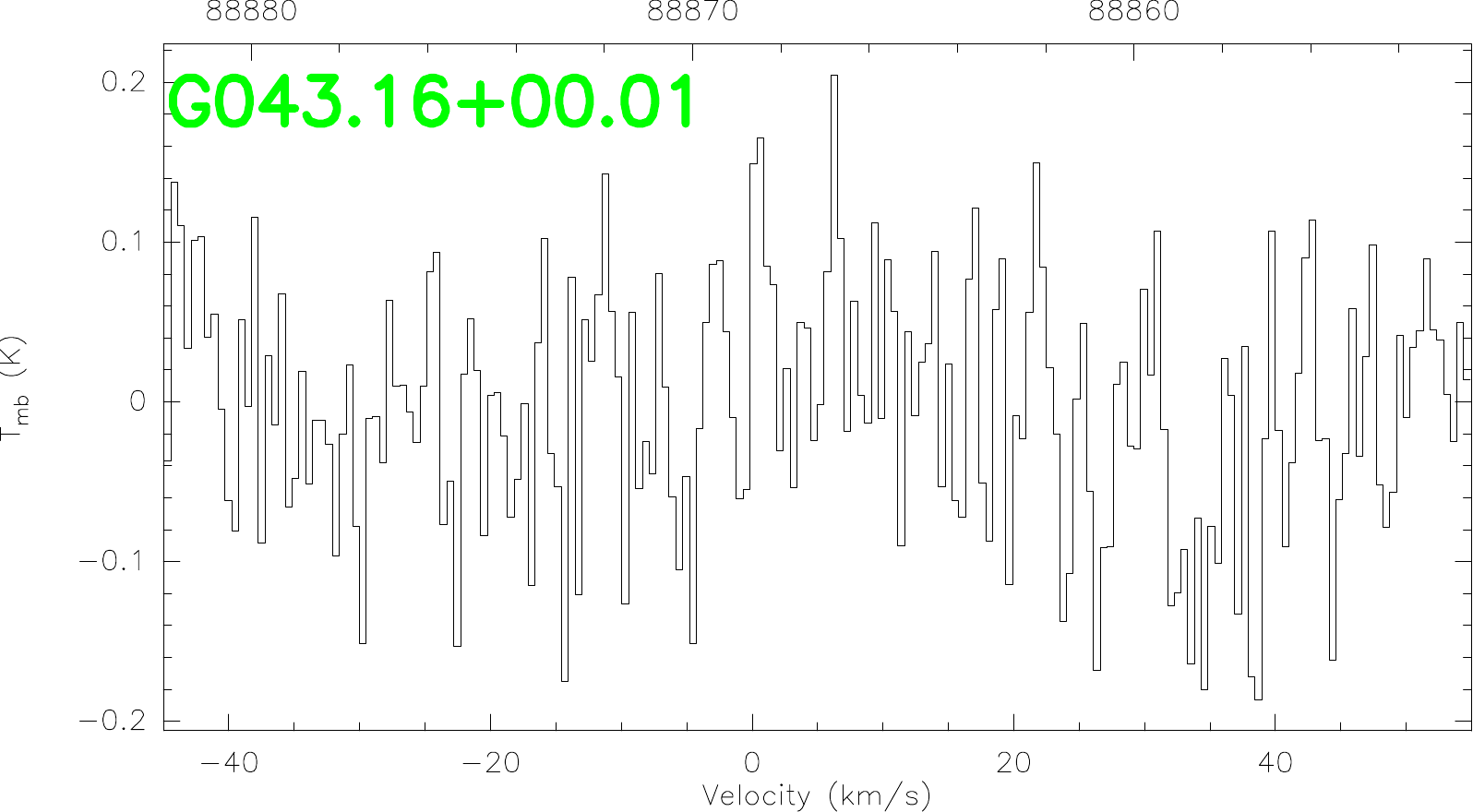}\\

\includegraphics[width=0.155\textwidth]{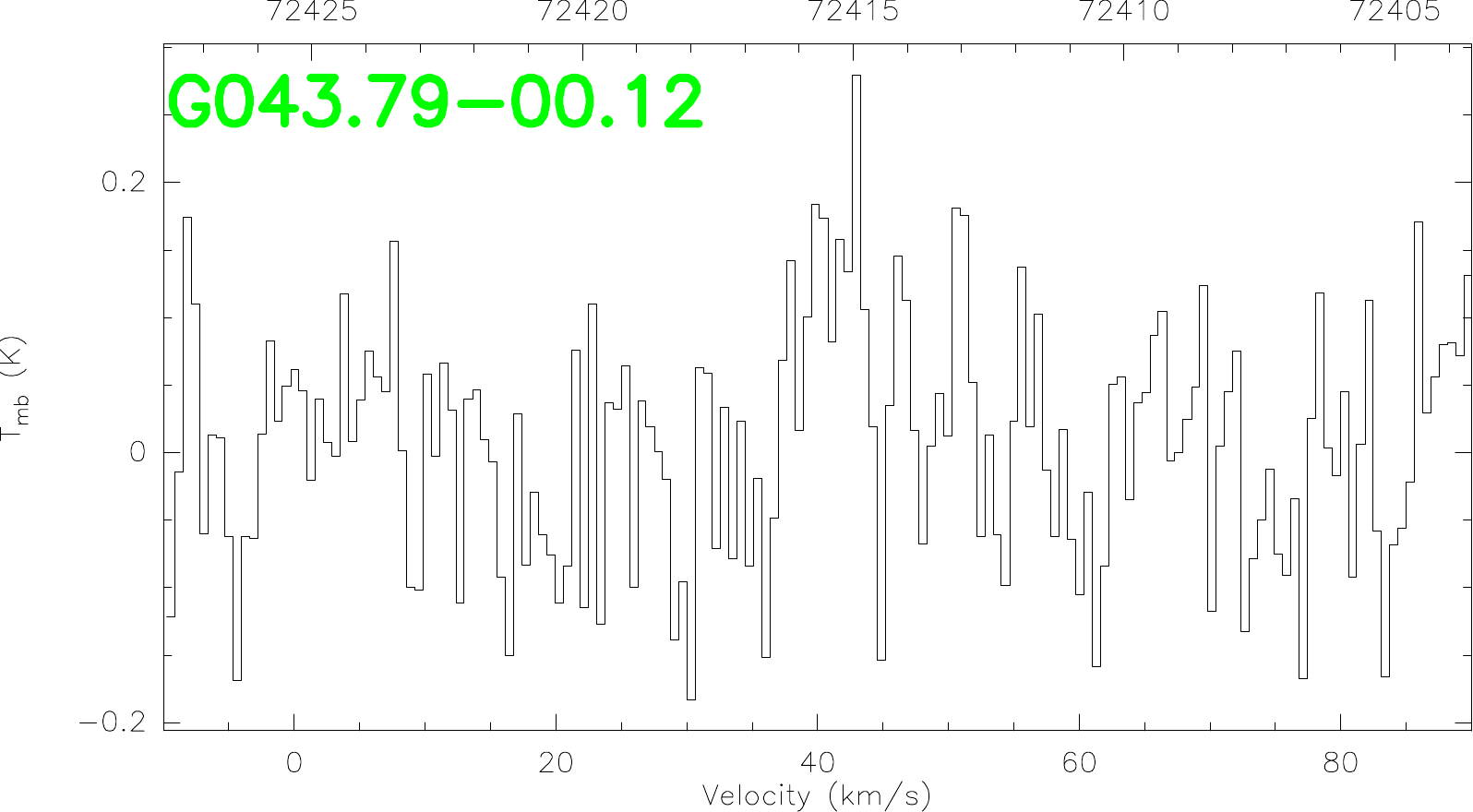}
\includegraphics[width=0.155\textwidth]{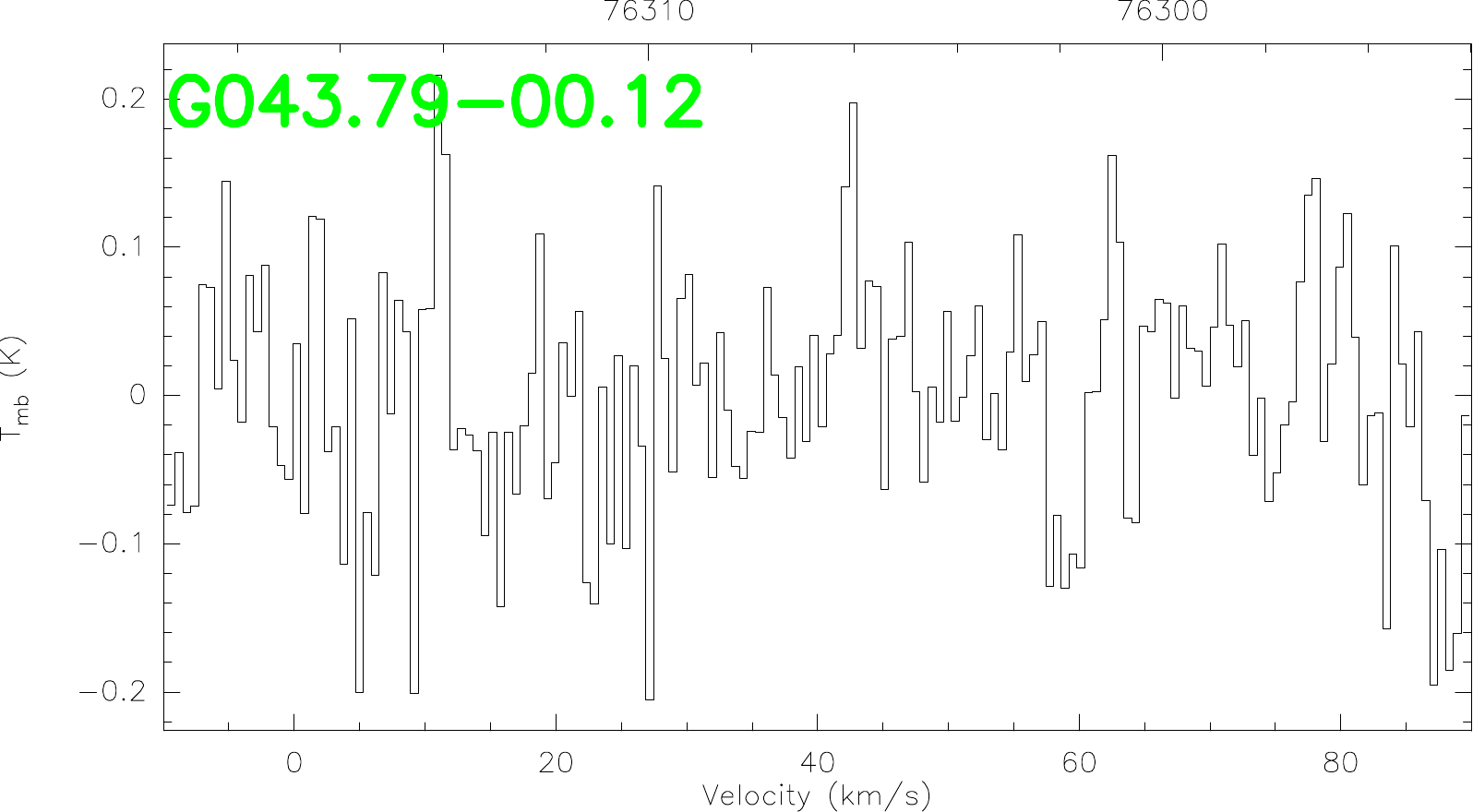}
\includegraphics[width=0.155\textwidth]{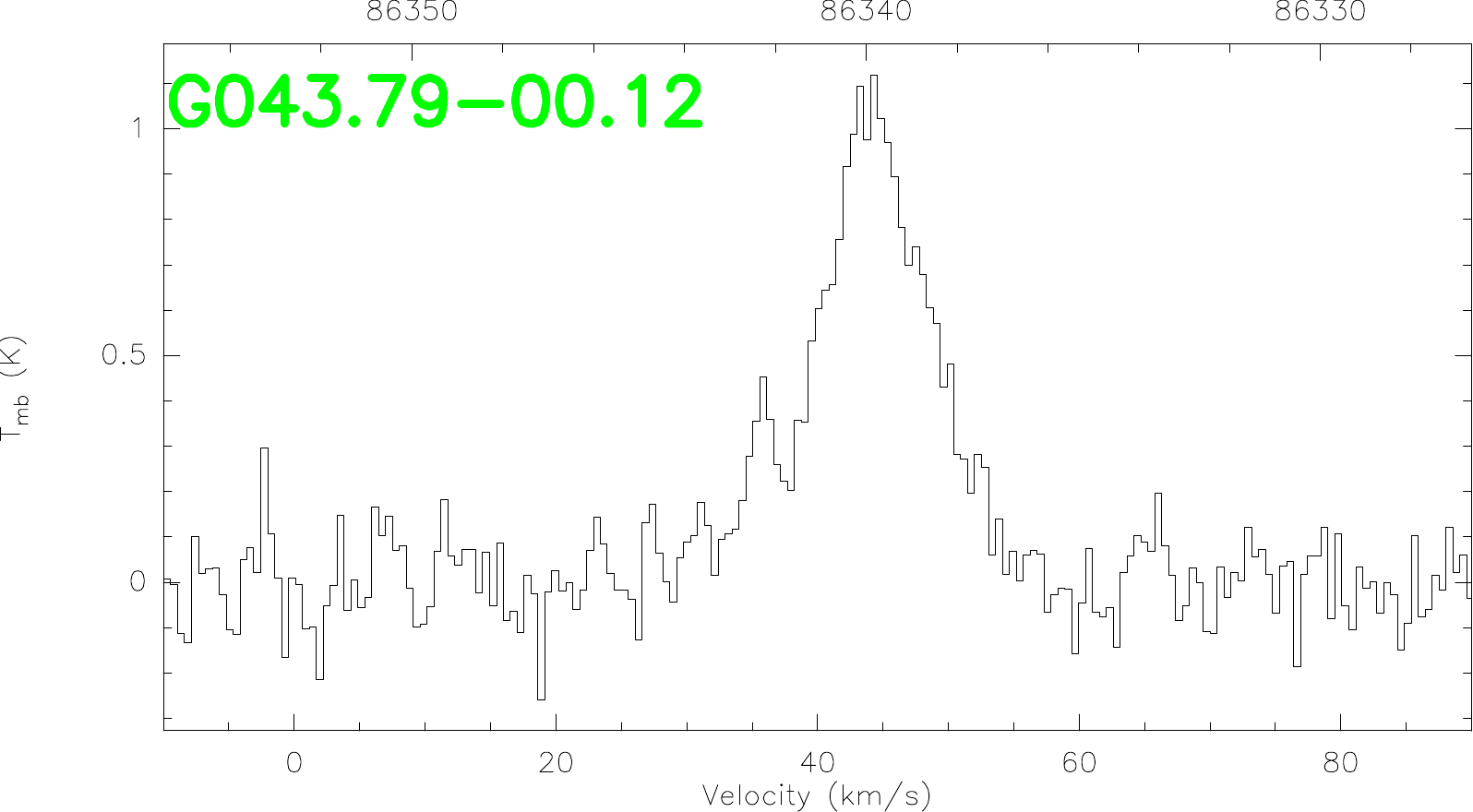}
\includegraphics[width=0.155\textwidth]{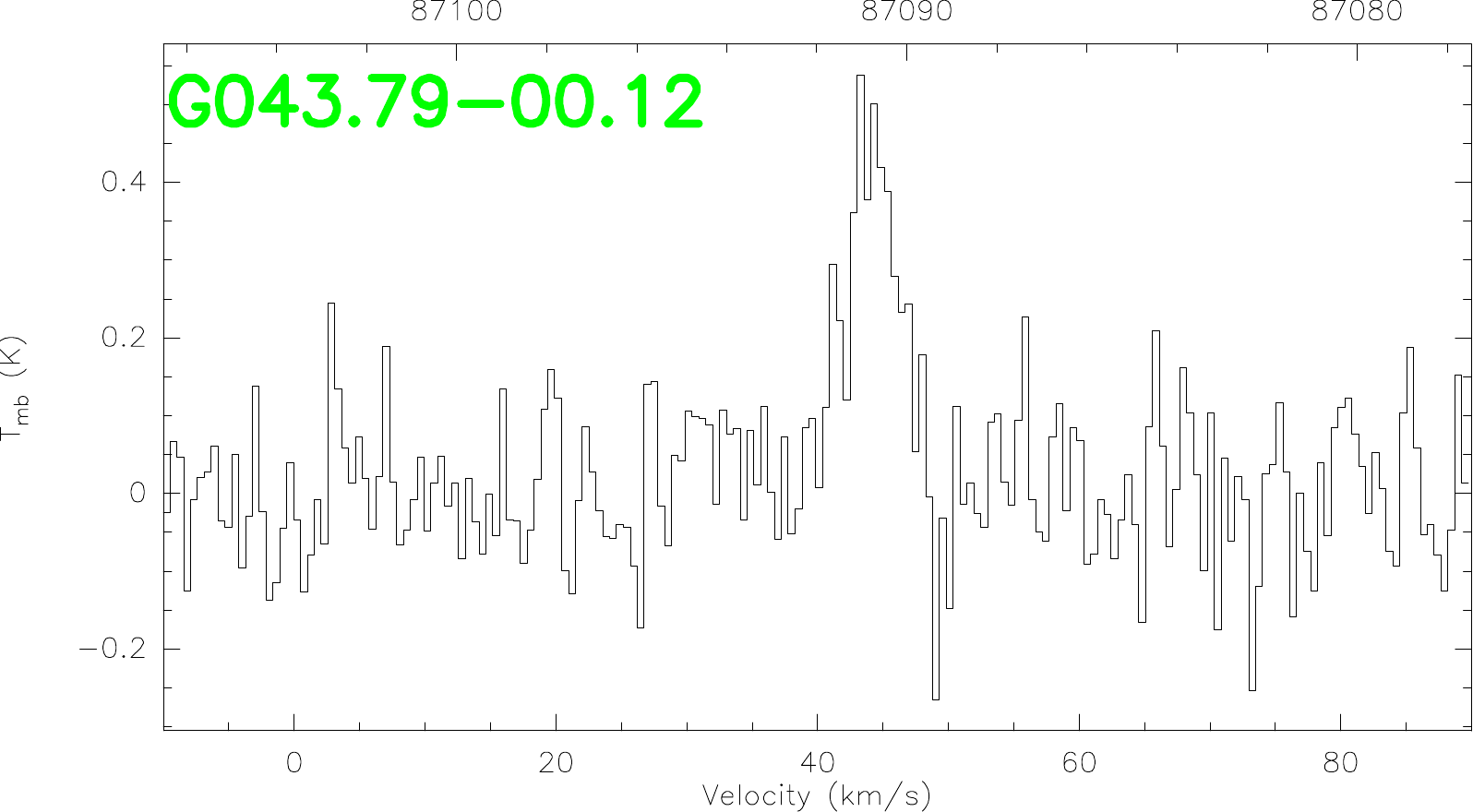}
\includegraphics[width=0.155\textwidth]{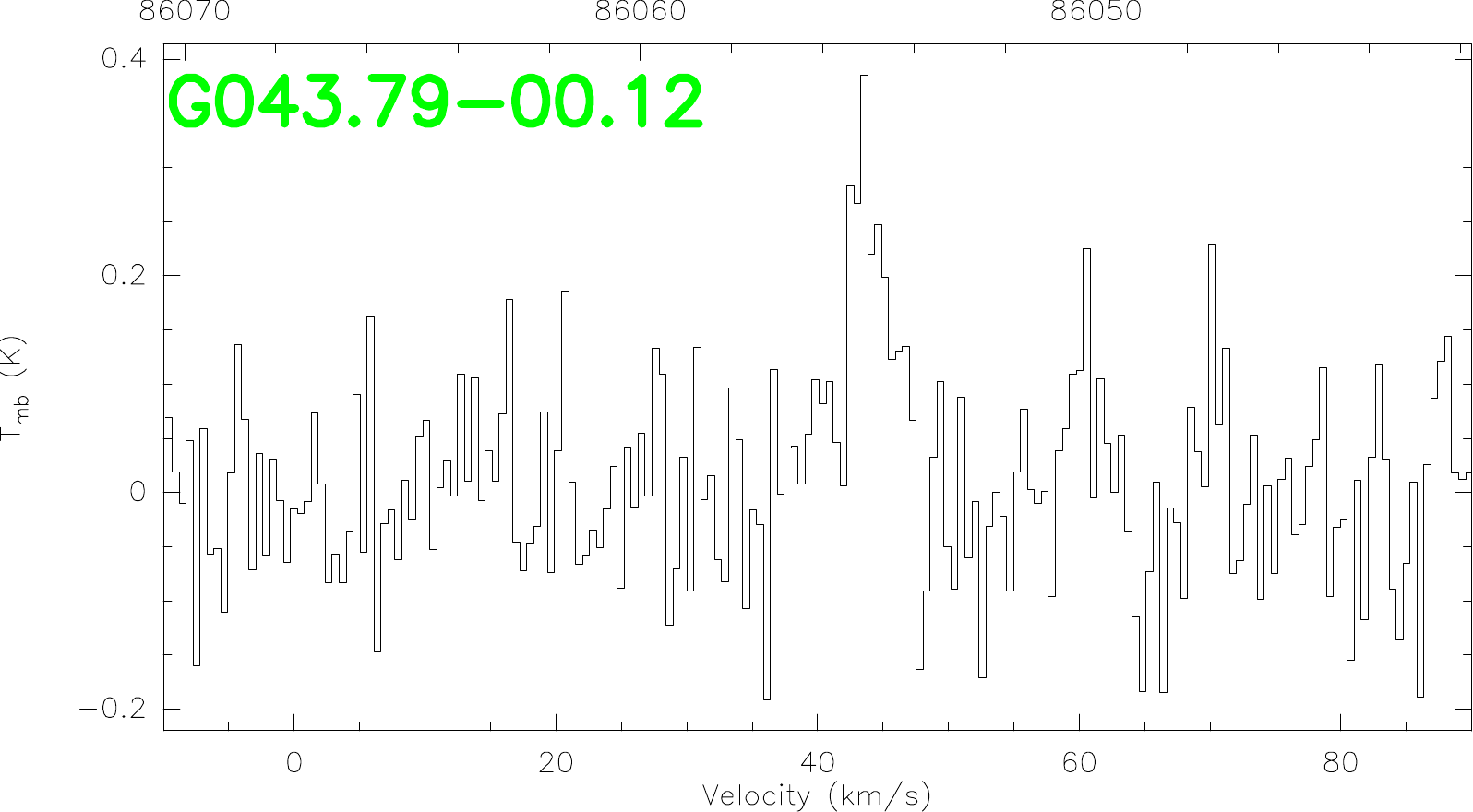}
\includegraphics[width=0.155\textwidth]{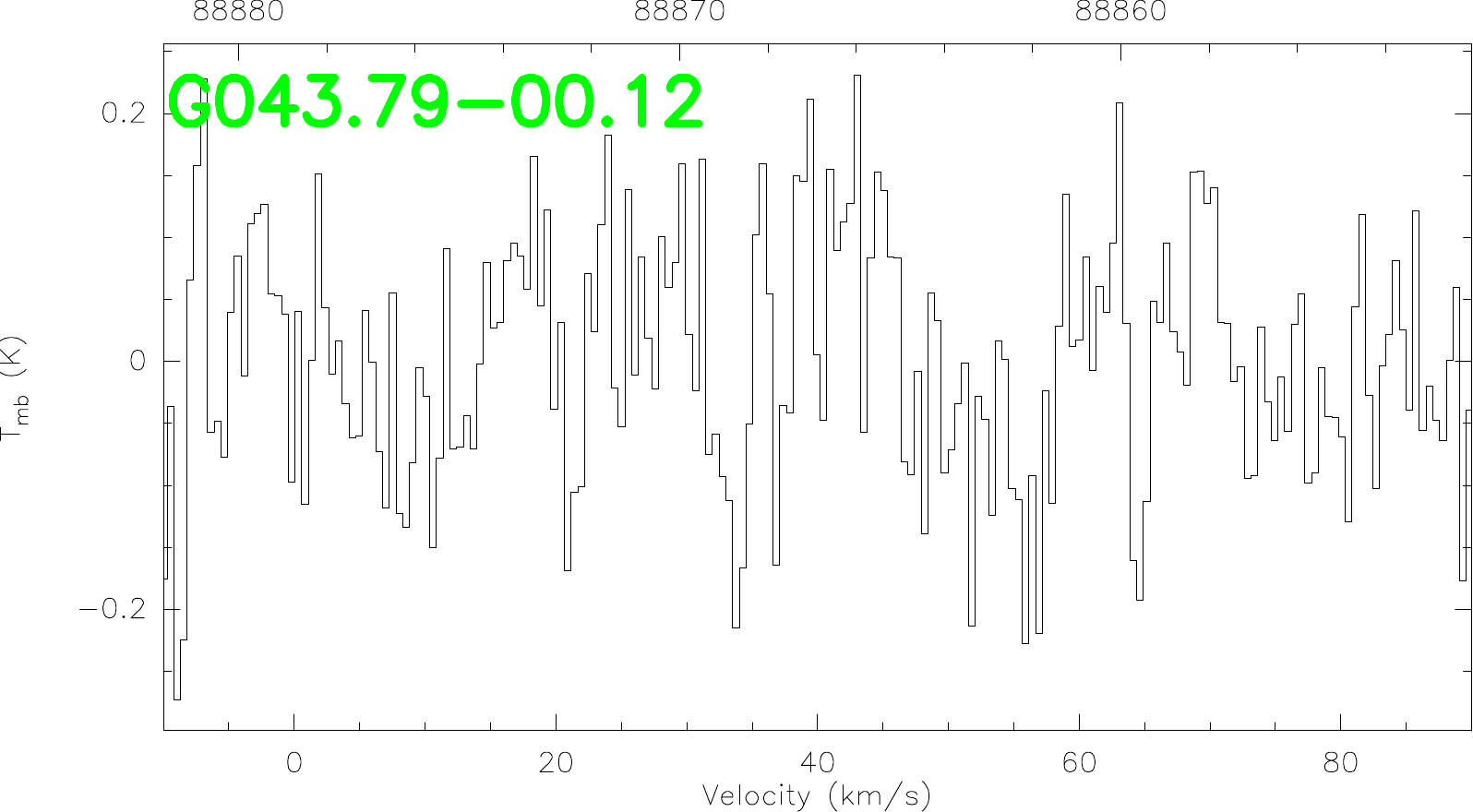}\\

\includegraphics[width=0.155\textwidth]{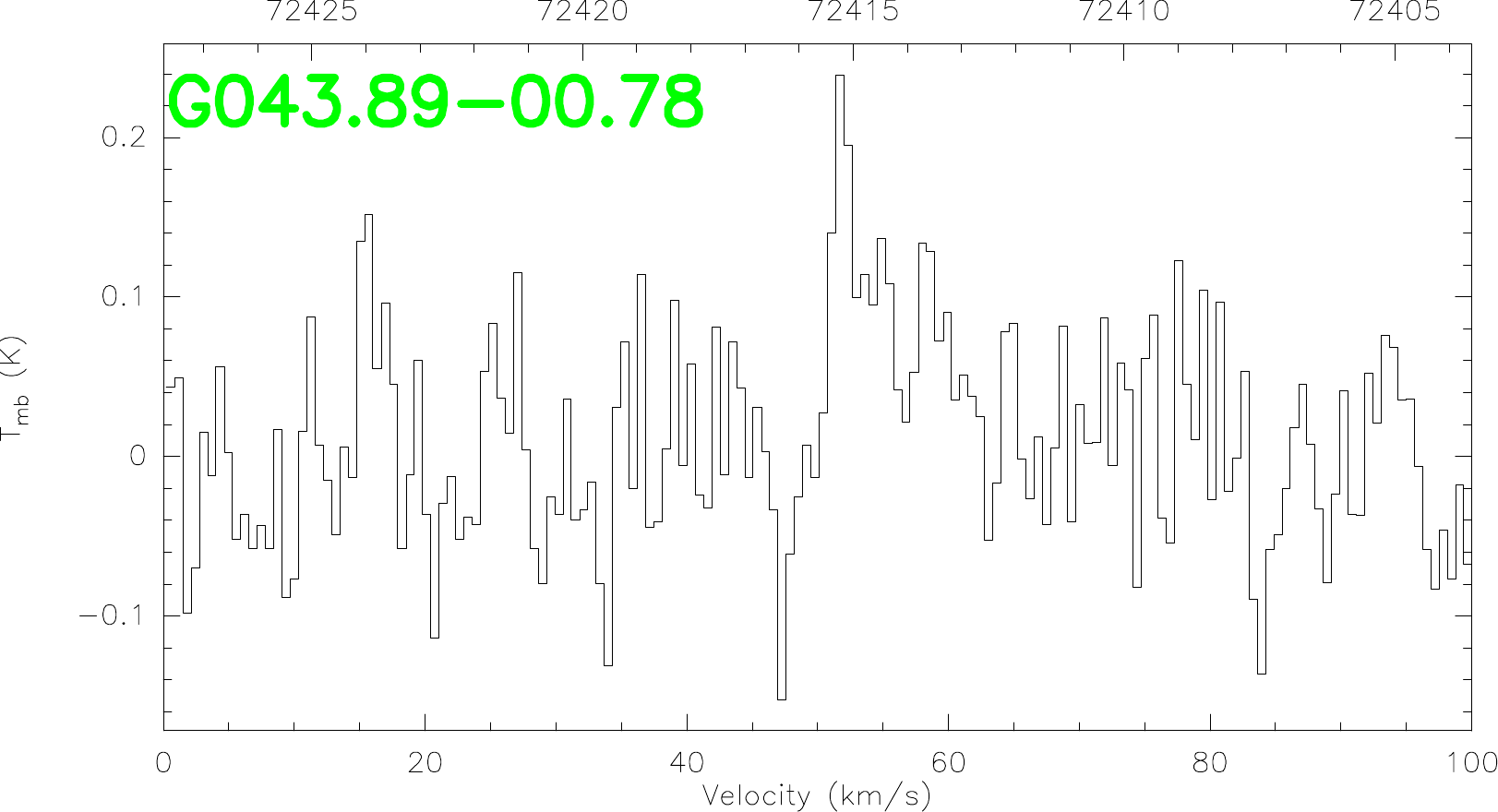}
\includegraphics[width=0.155\textwidth]{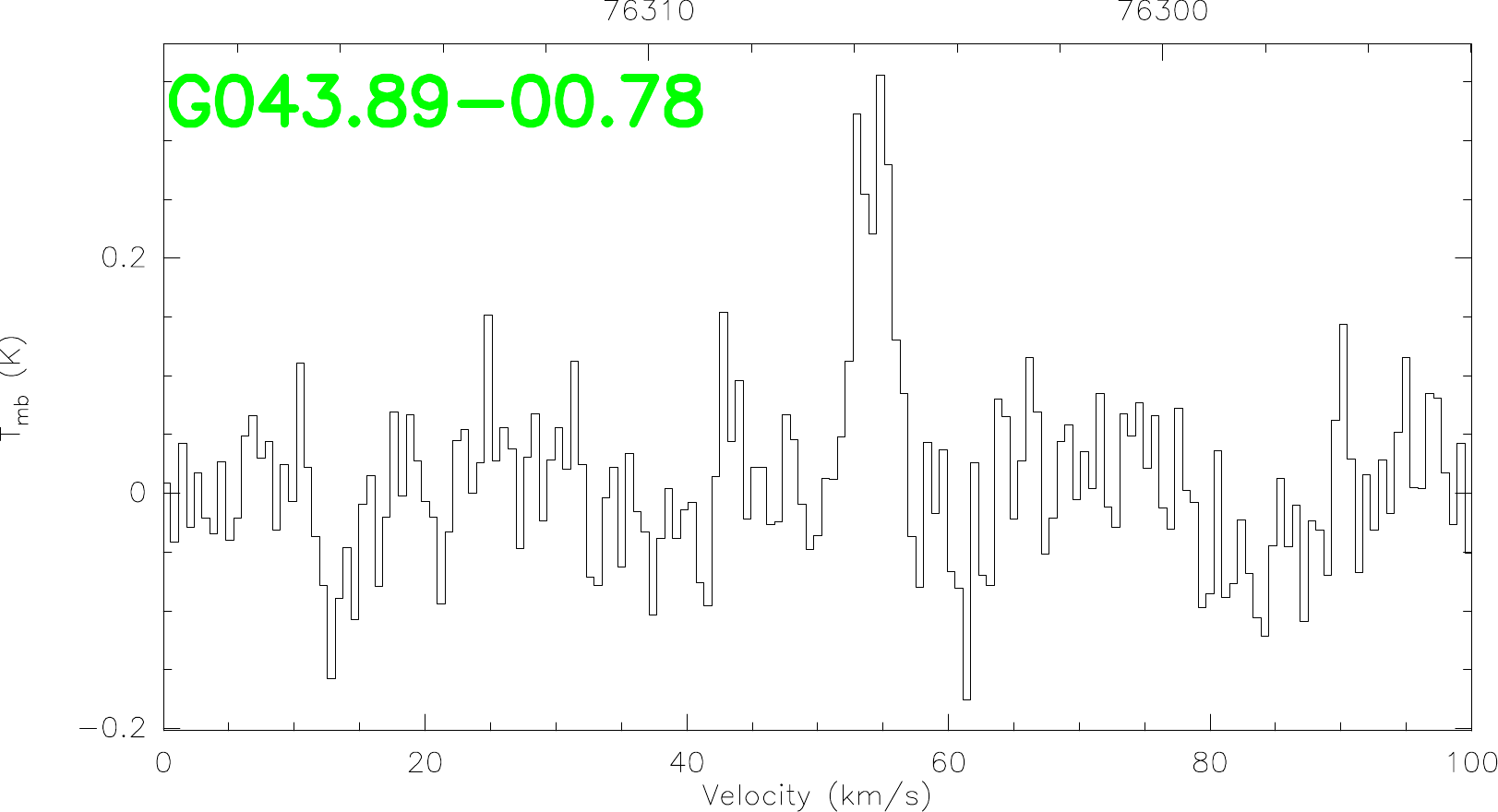}
\includegraphics[width=0.155\textwidth]{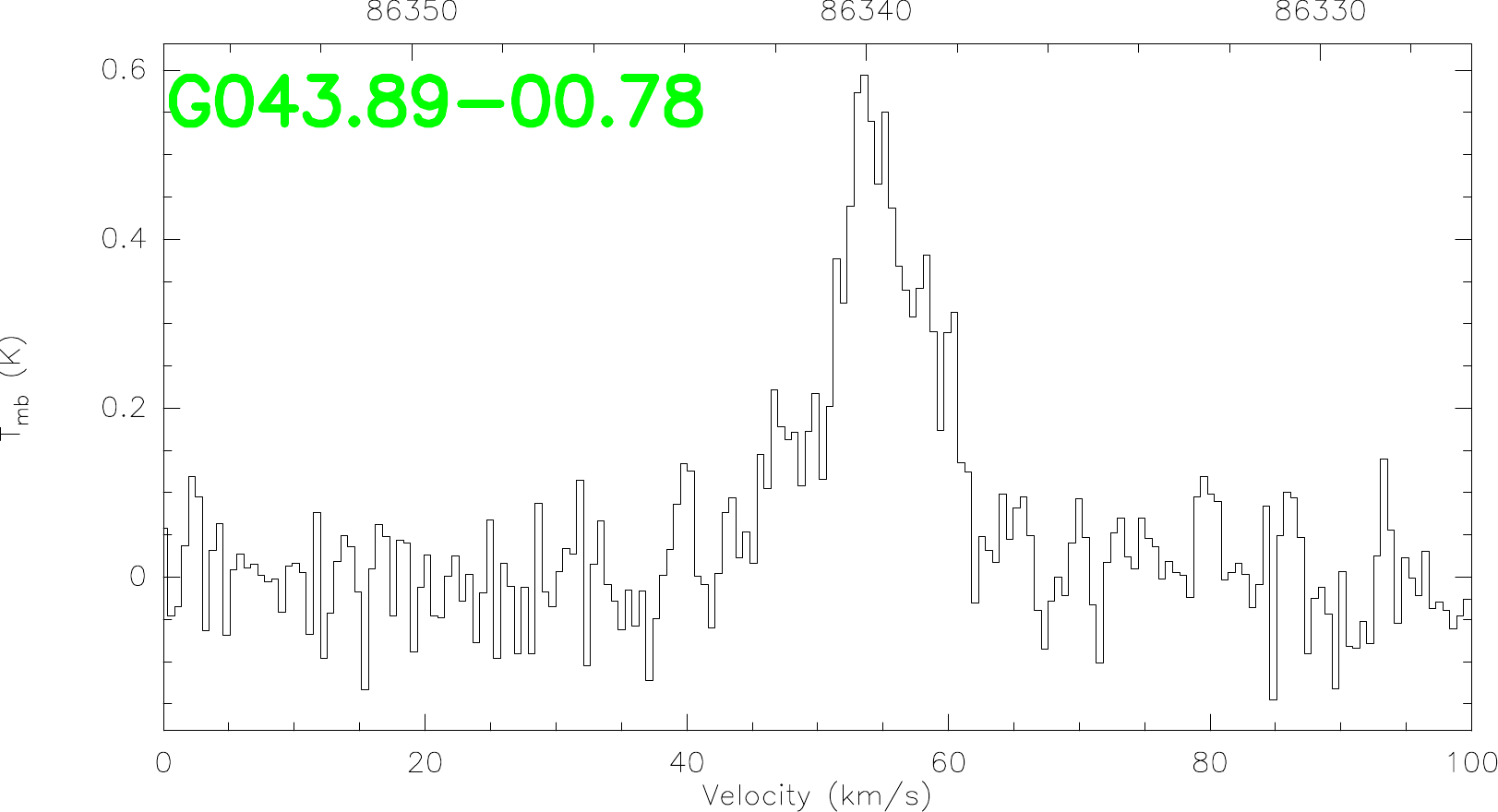}
\includegraphics[width=0.155\textwidth]{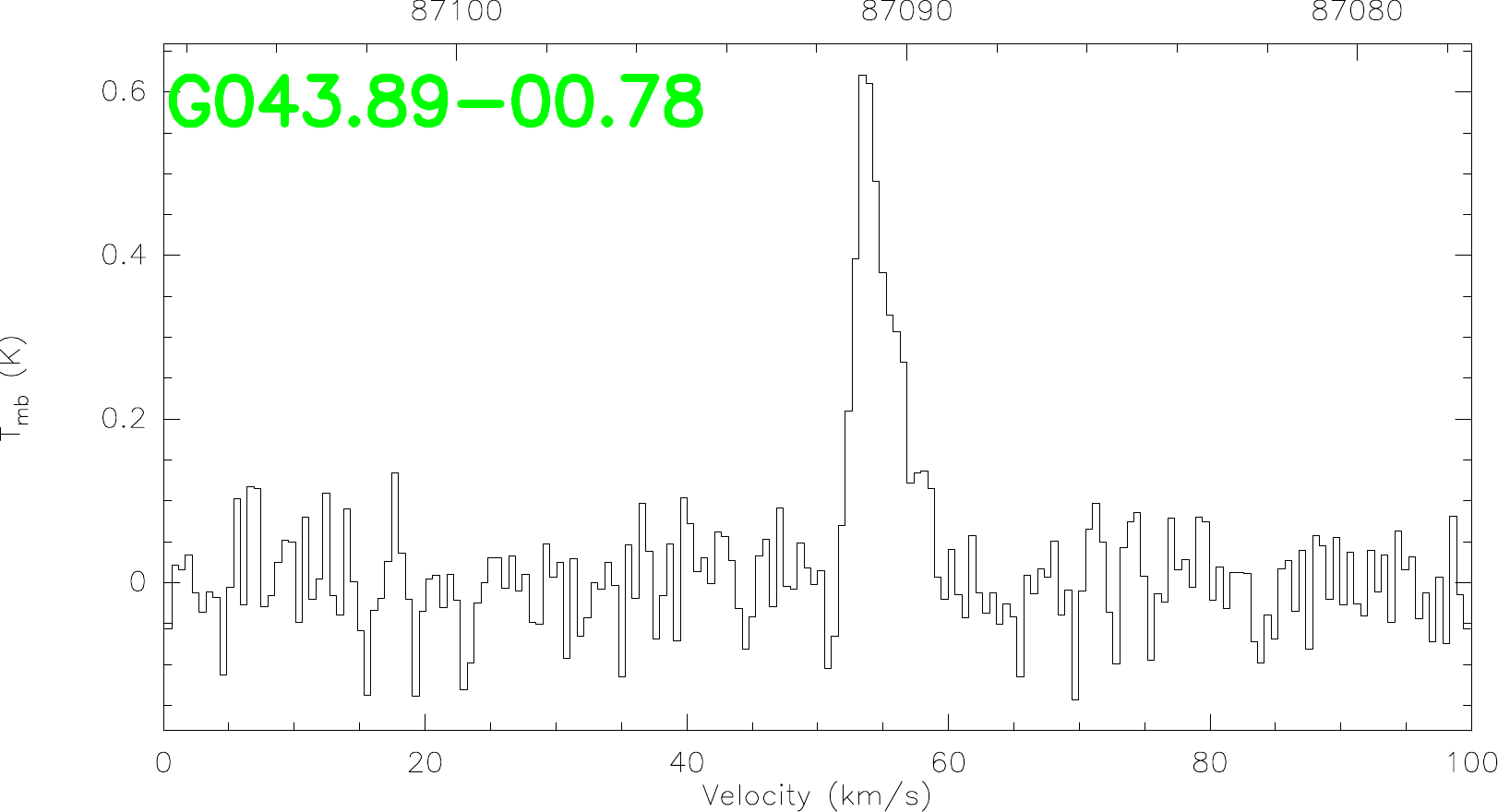}
\includegraphics[width=0.155\textwidth]{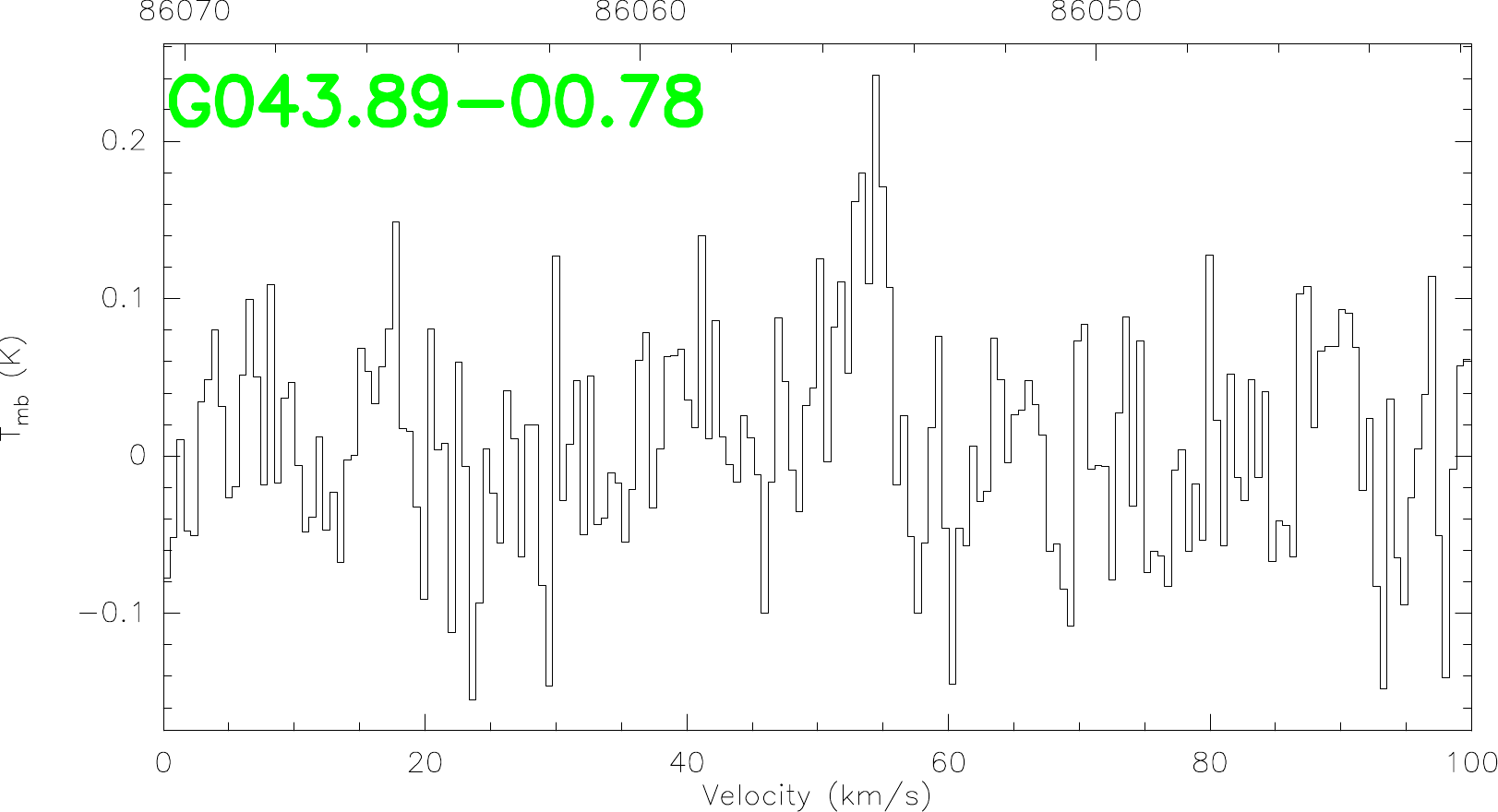}
\includegraphics[width=0.155\textwidth]{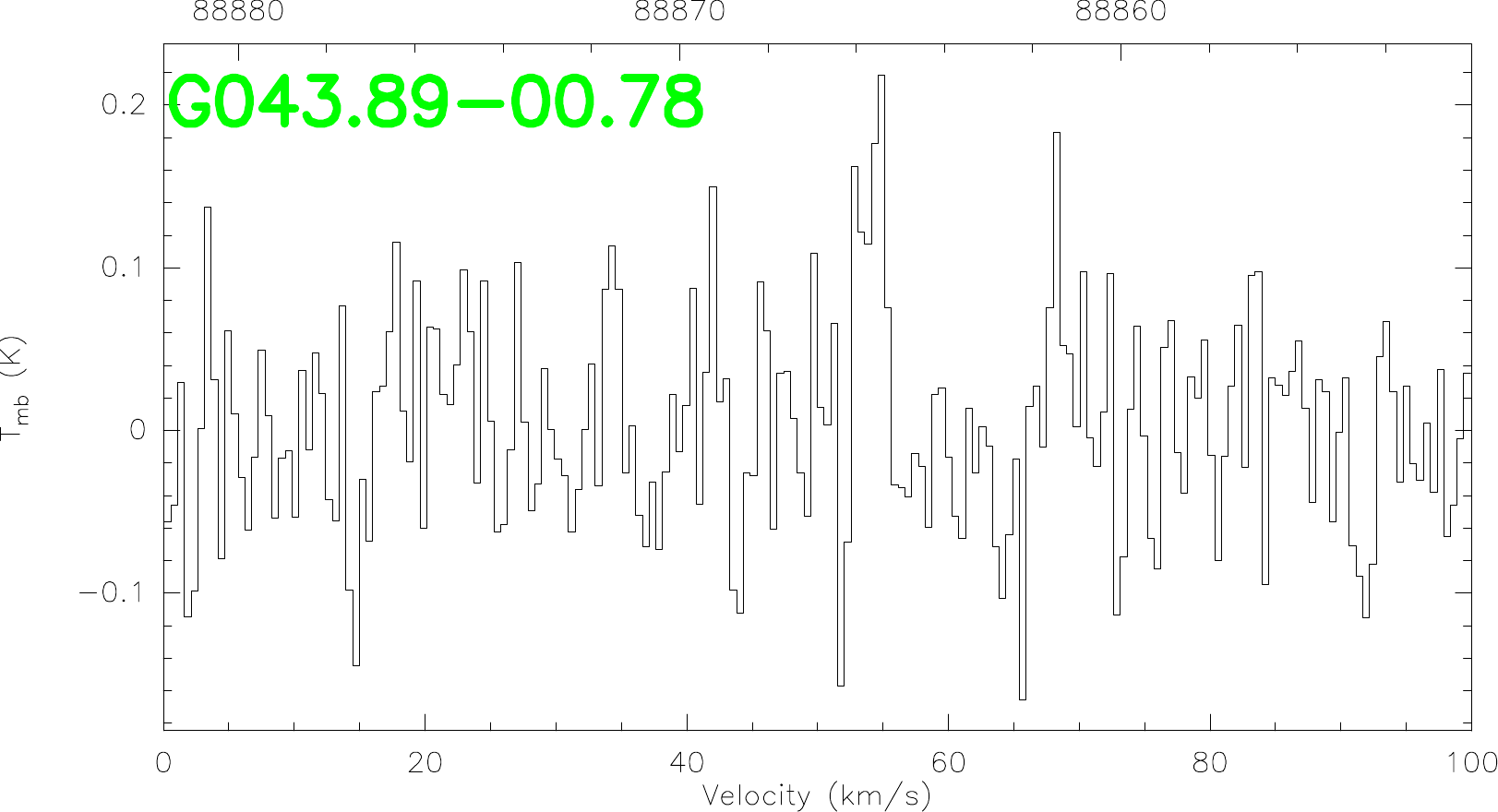}\\

\includegraphics[width=0.155\textwidth]{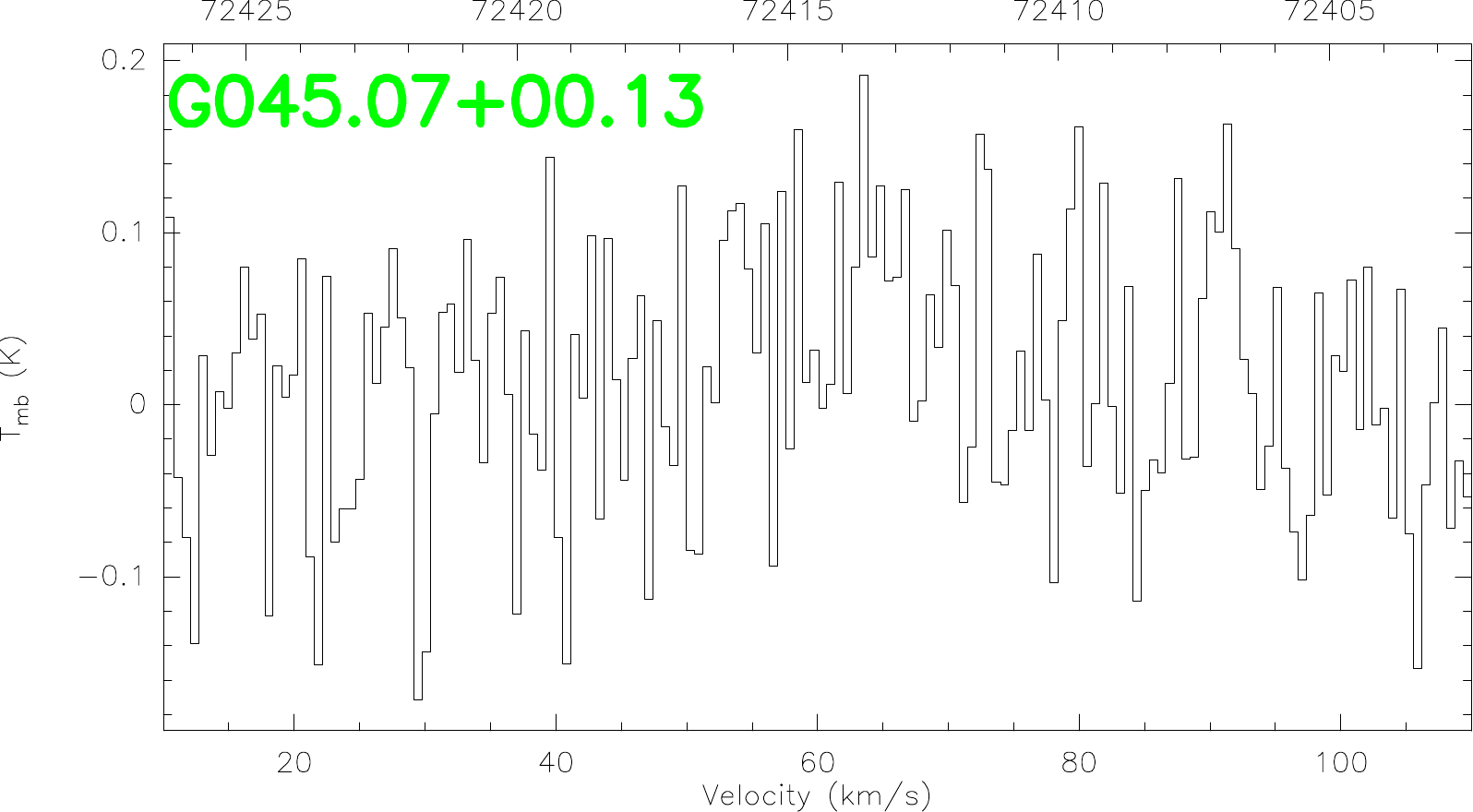}
\includegraphics[width=0.155\textwidth]{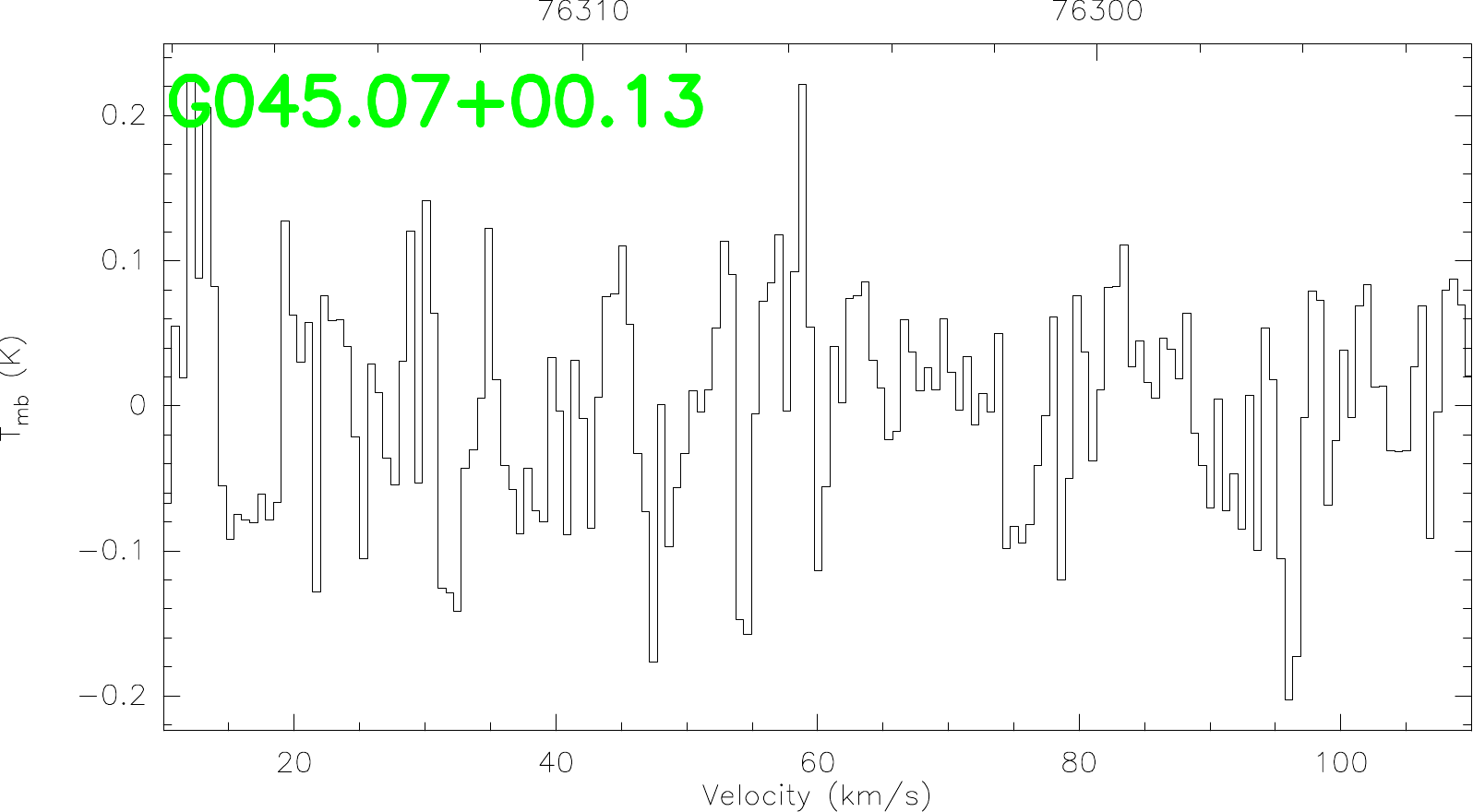}
\includegraphics[width=0.155\textwidth]{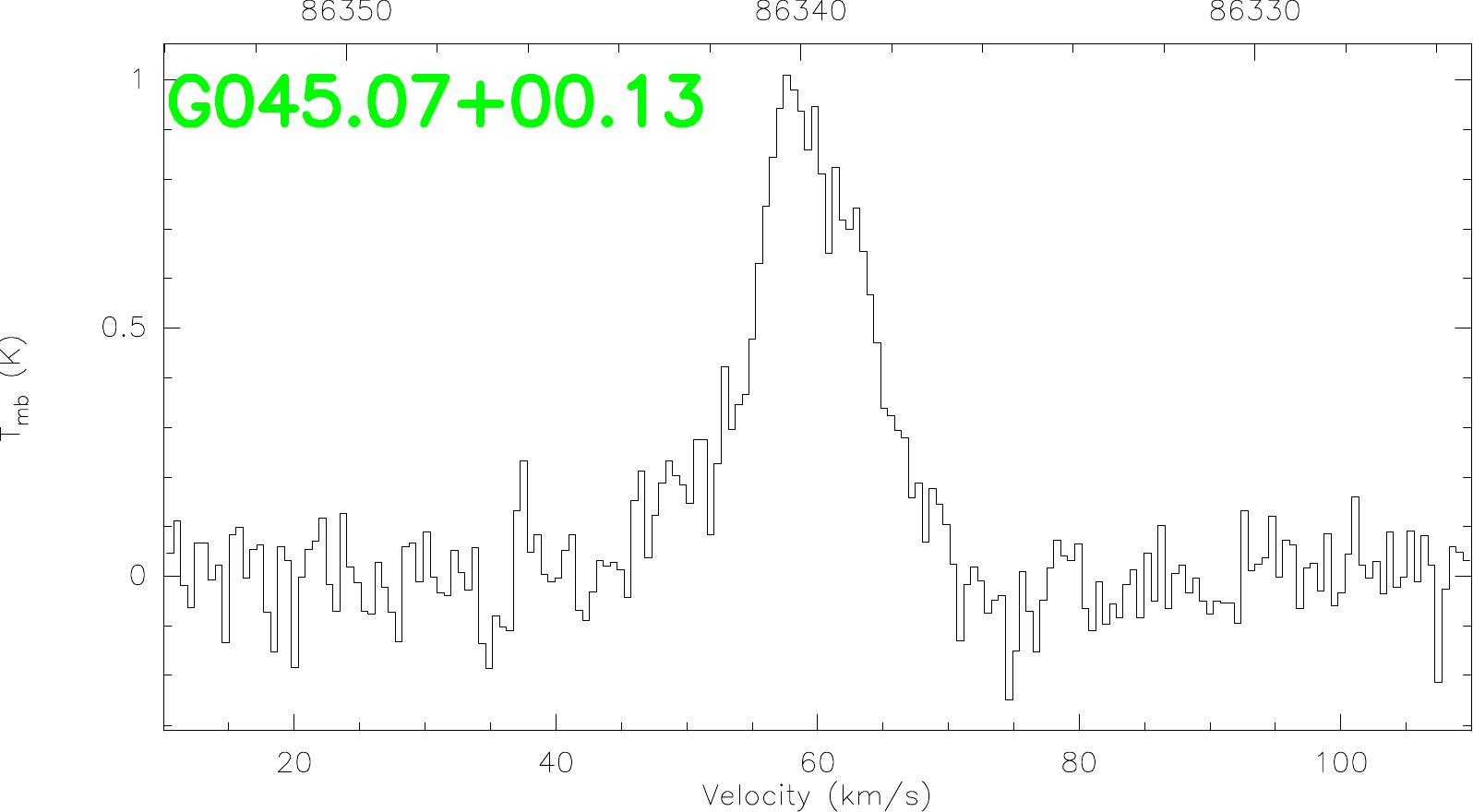}
\includegraphics[width=0.155\textwidth]{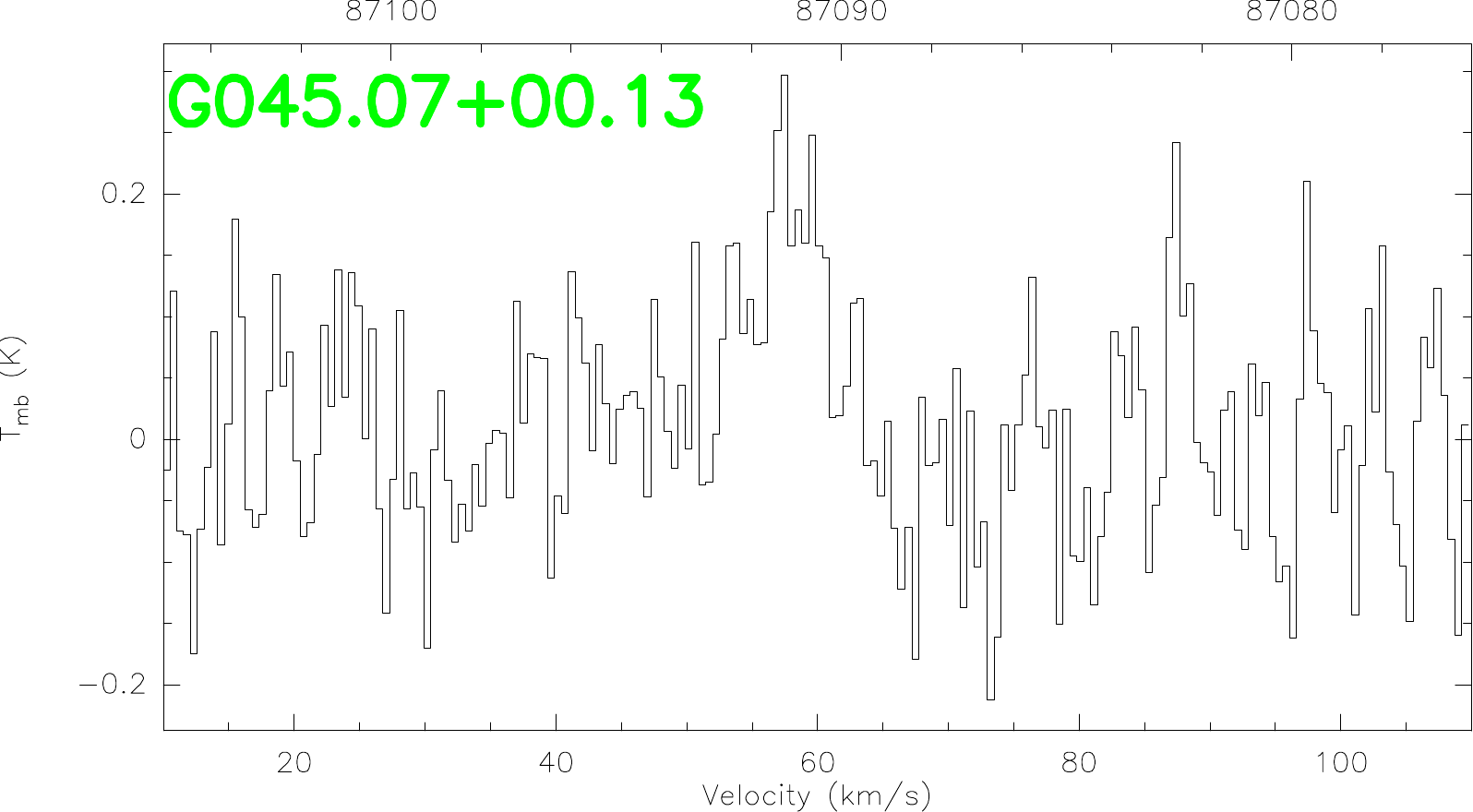}
\includegraphics[width=0.155\textwidth]{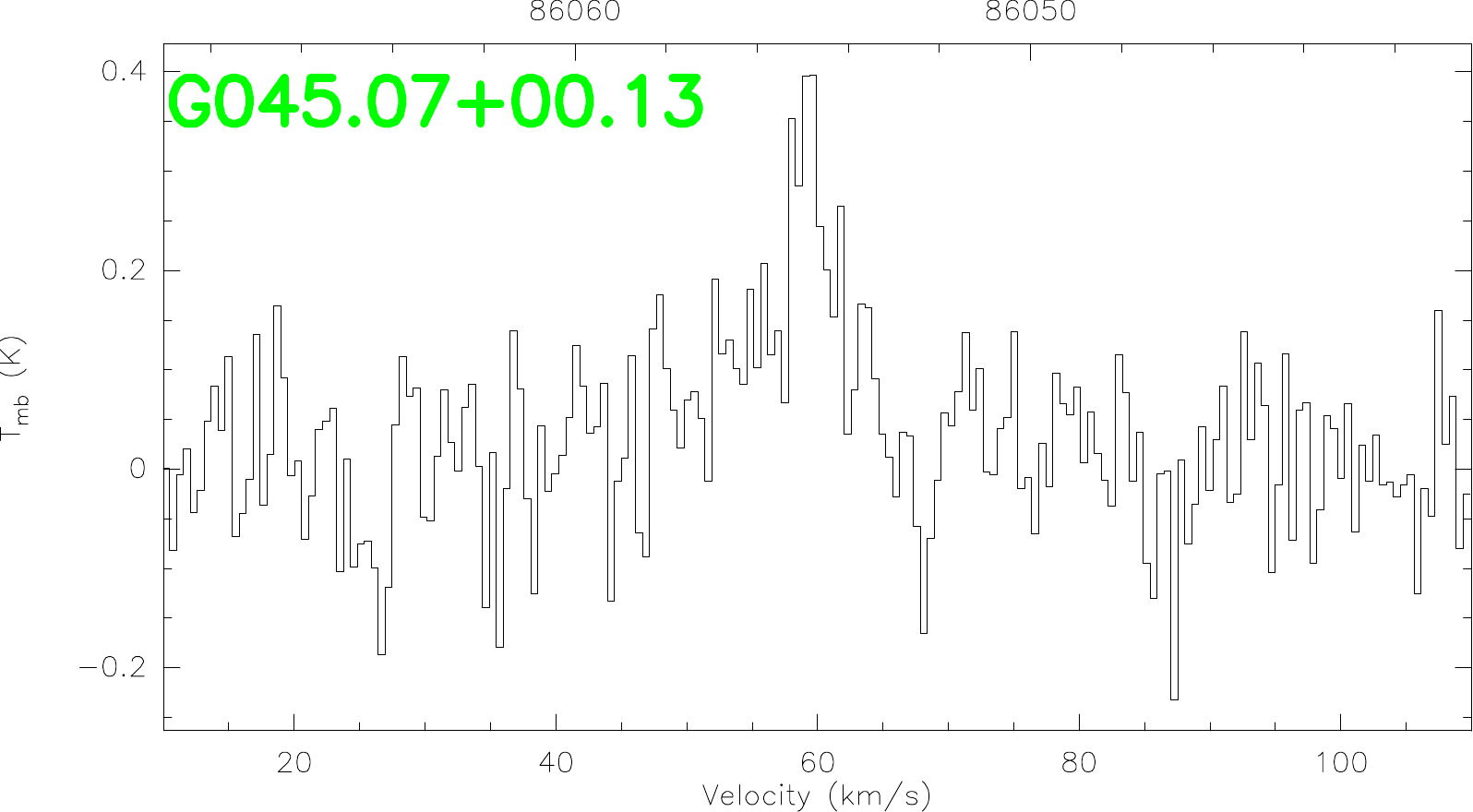}
\includegraphics[width=0.155\textwidth]{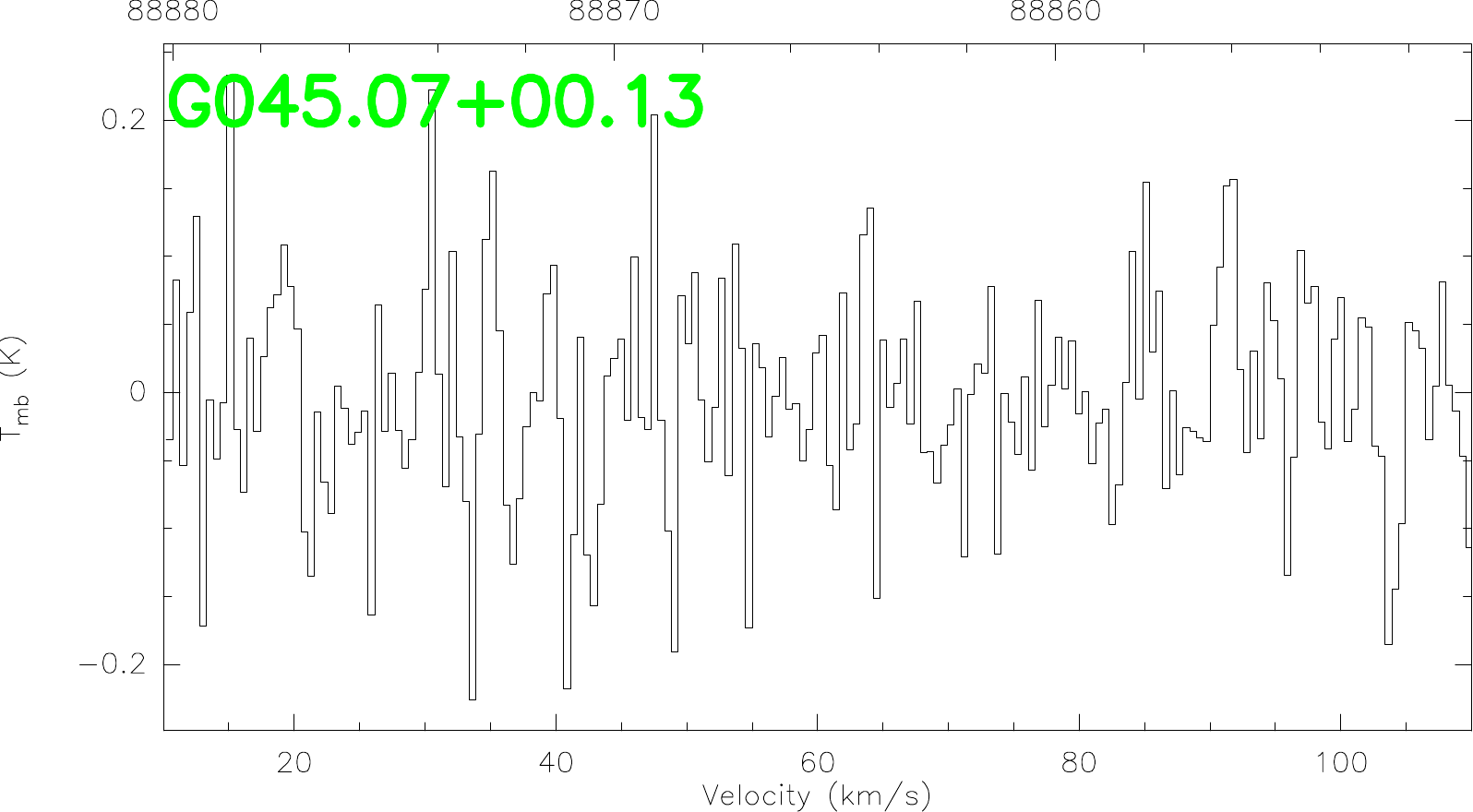}\\

\includegraphics[width=0.155\textwidth]{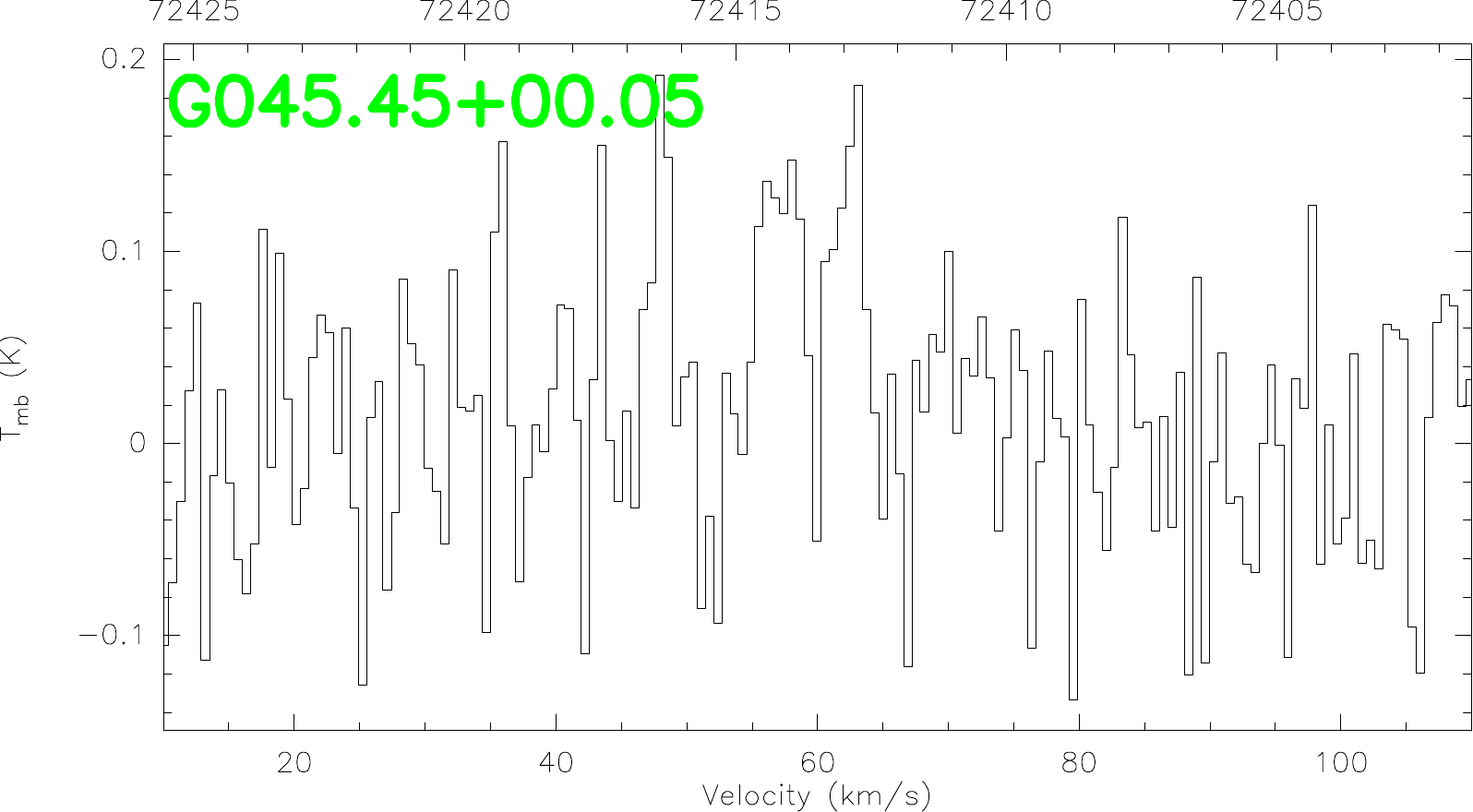}
\includegraphics[width=0.155\textwidth]{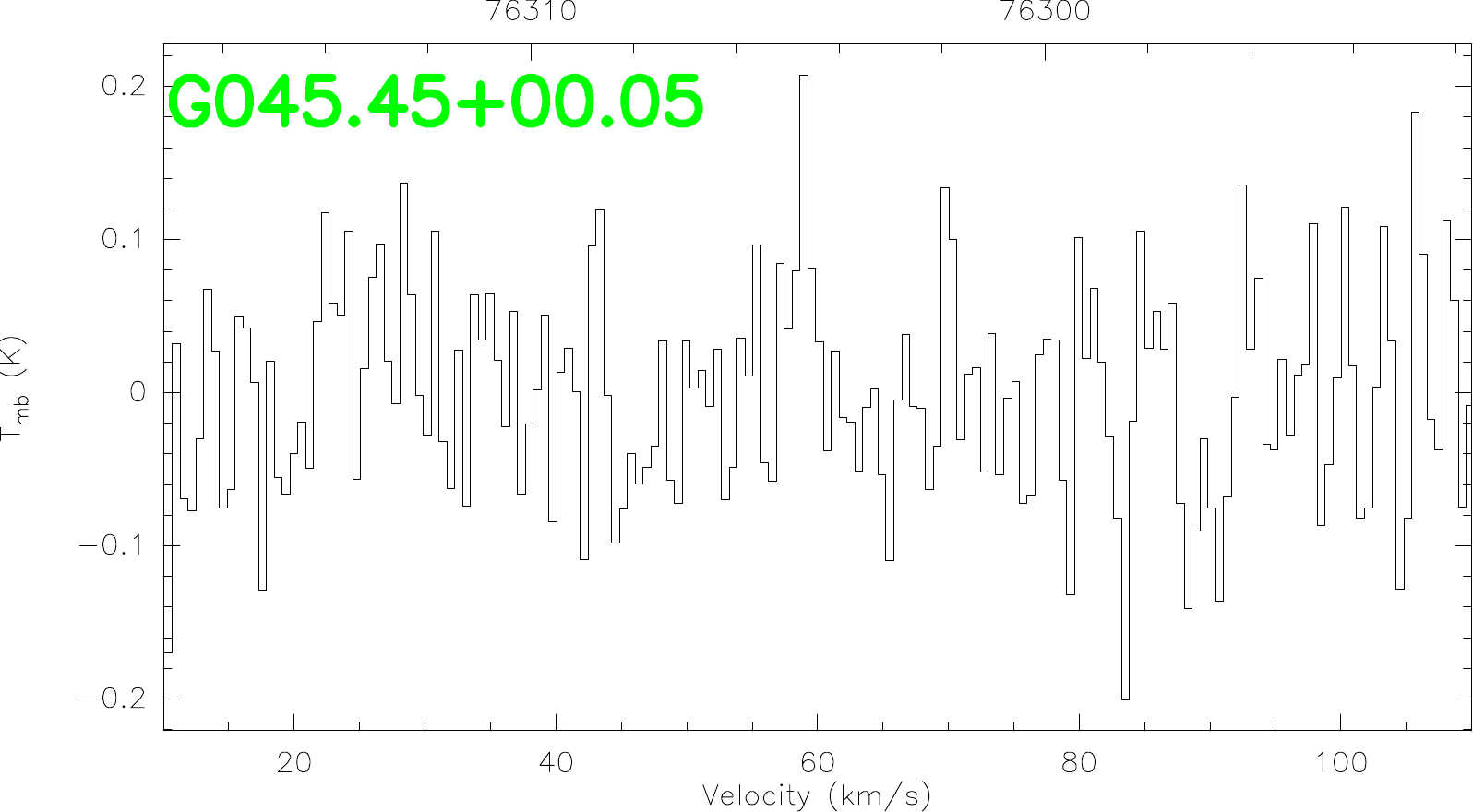}
\includegraphics[width=0.155\textwidth]{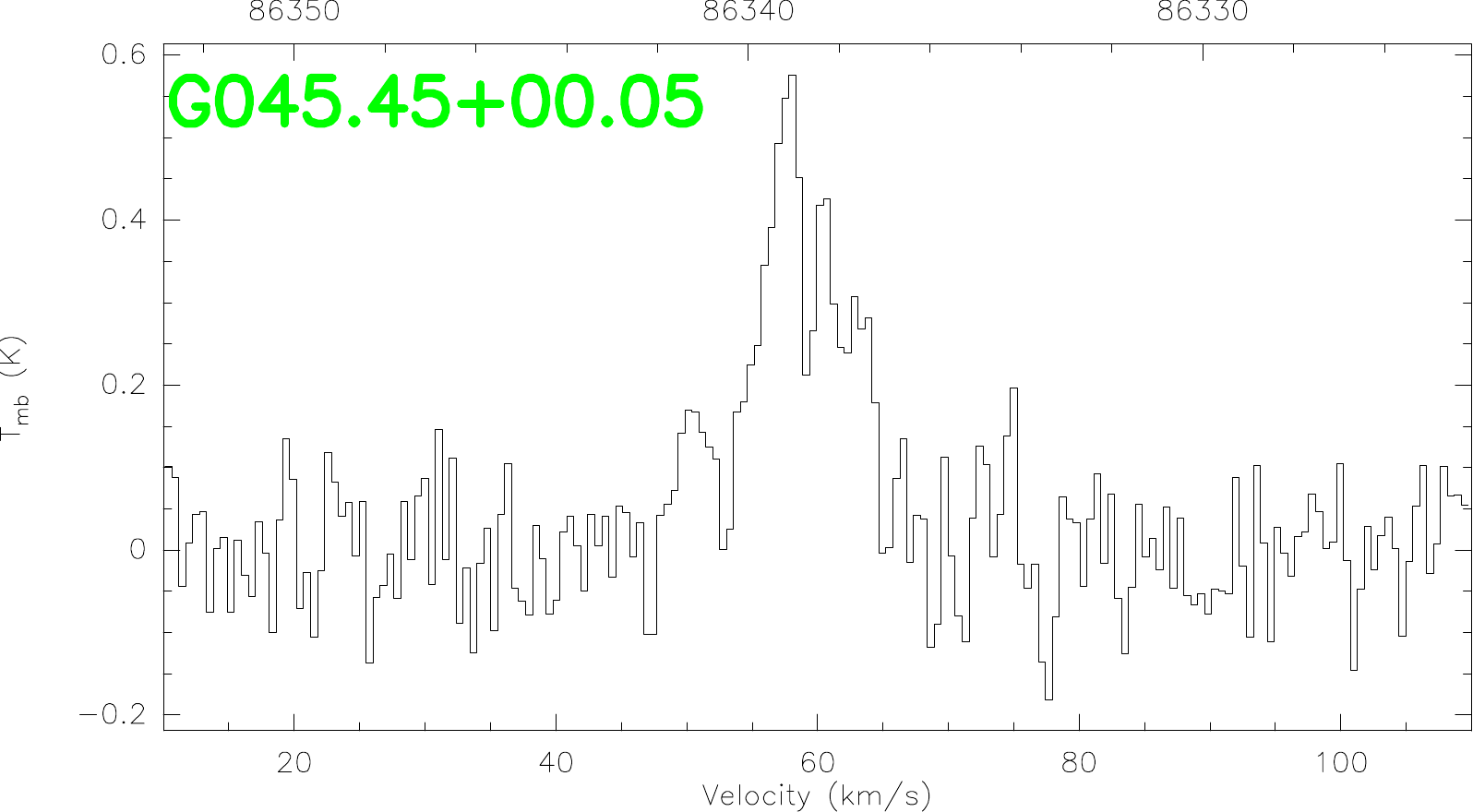}
\includegraphics[width=0.155\textwidth]{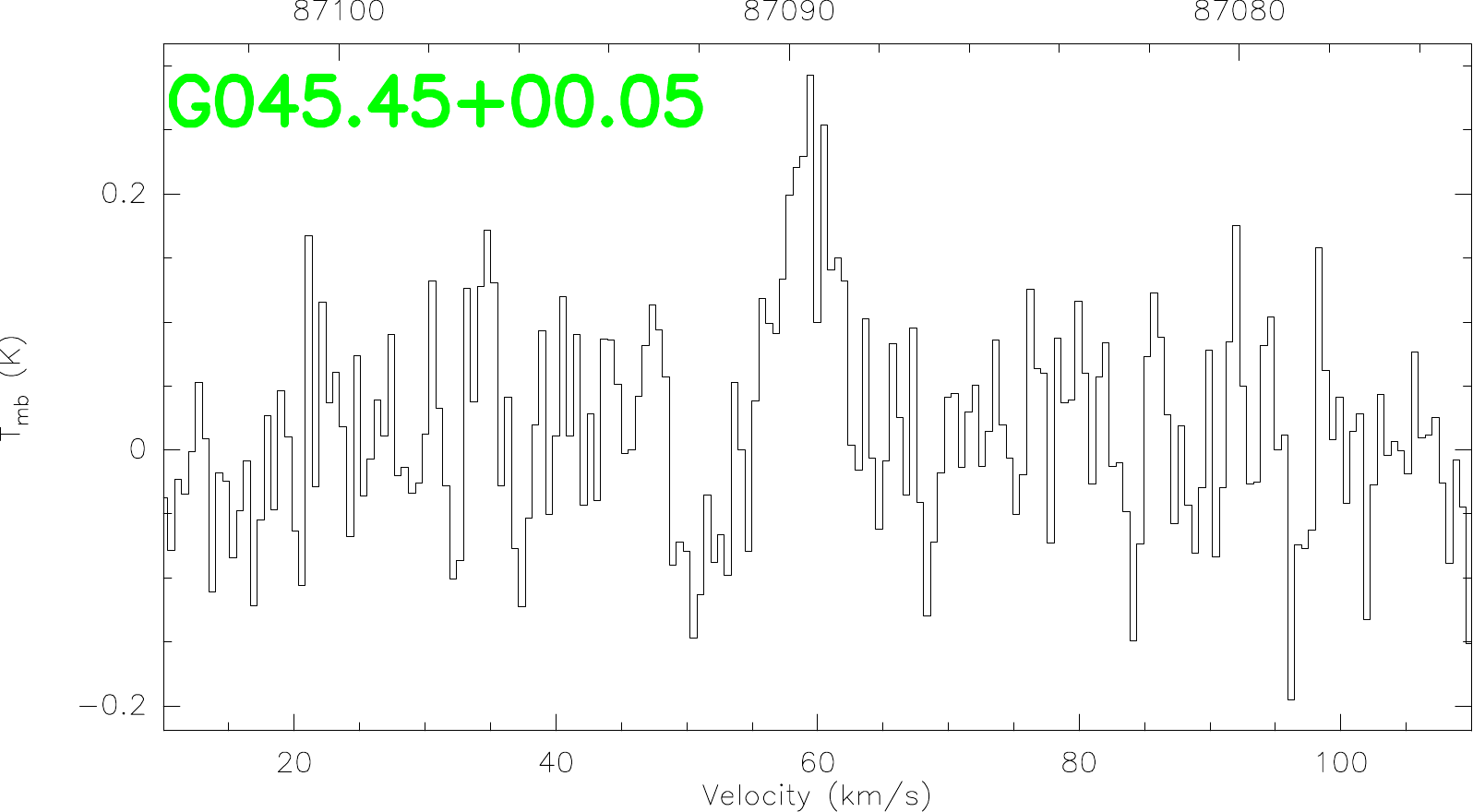}
\includegraphics[width=0.155\textwidth]{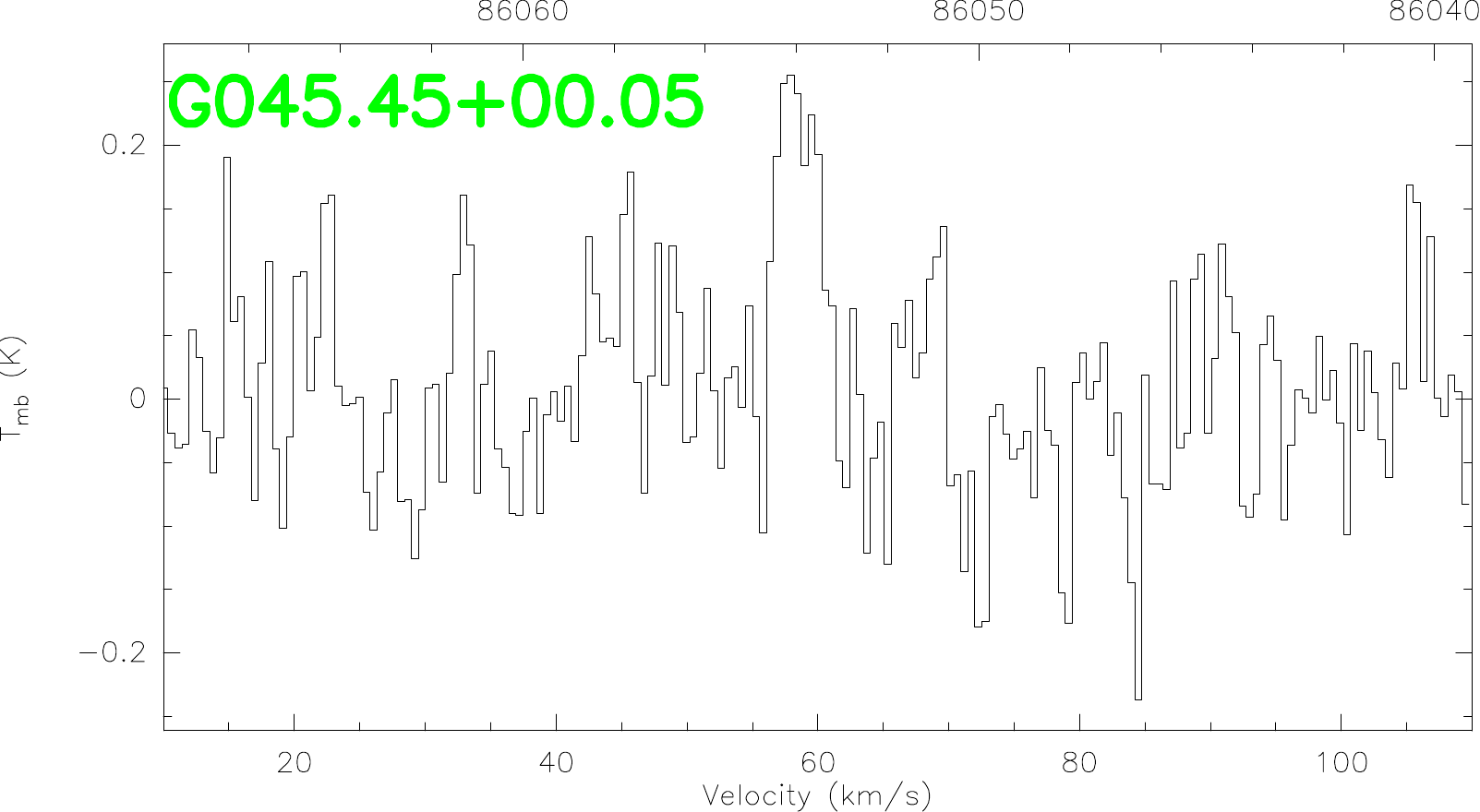}
\includegraphics[width=0.155\textwidth]{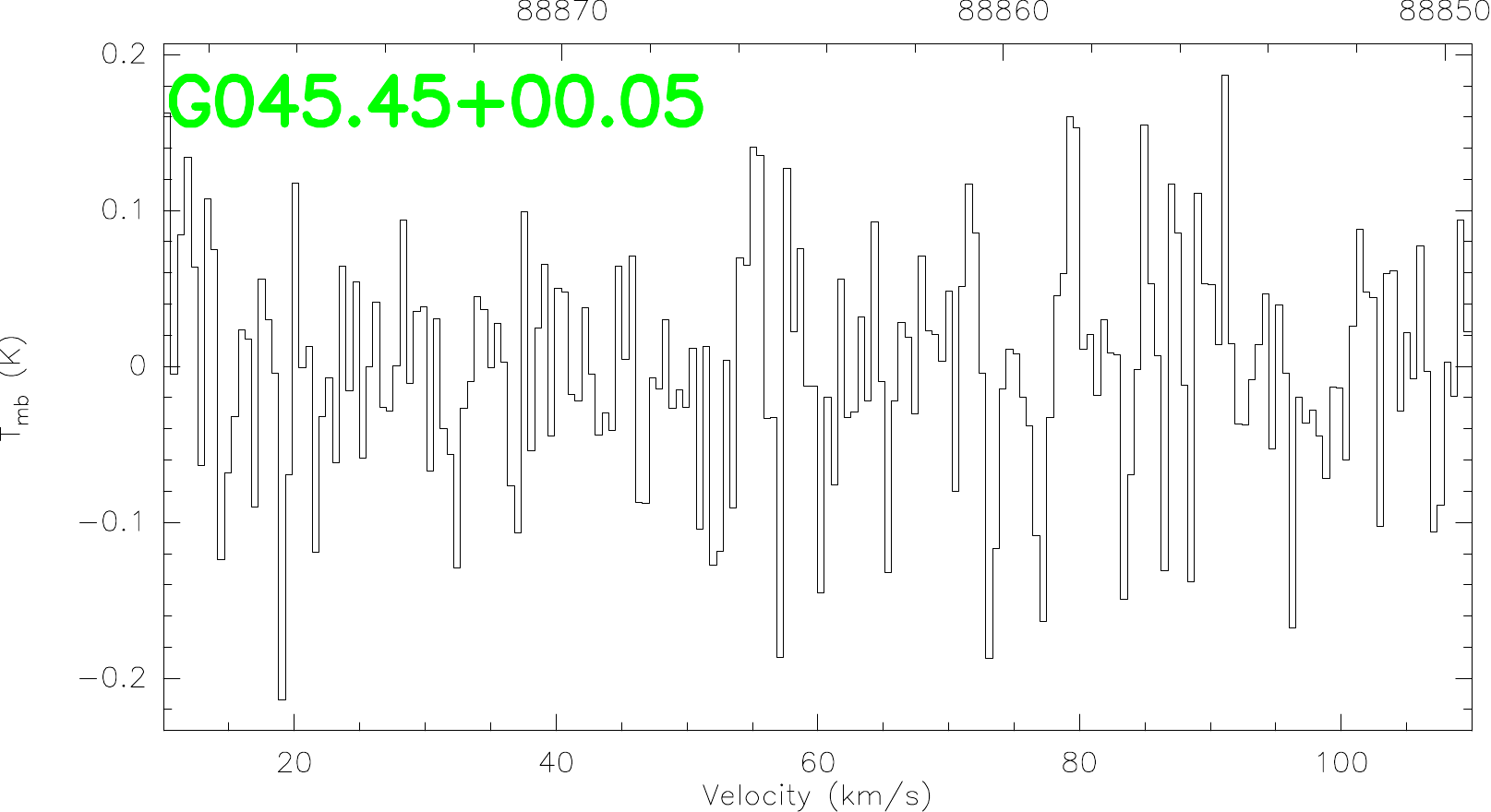}\\

\includegraphics[width=0.155\textwidth]{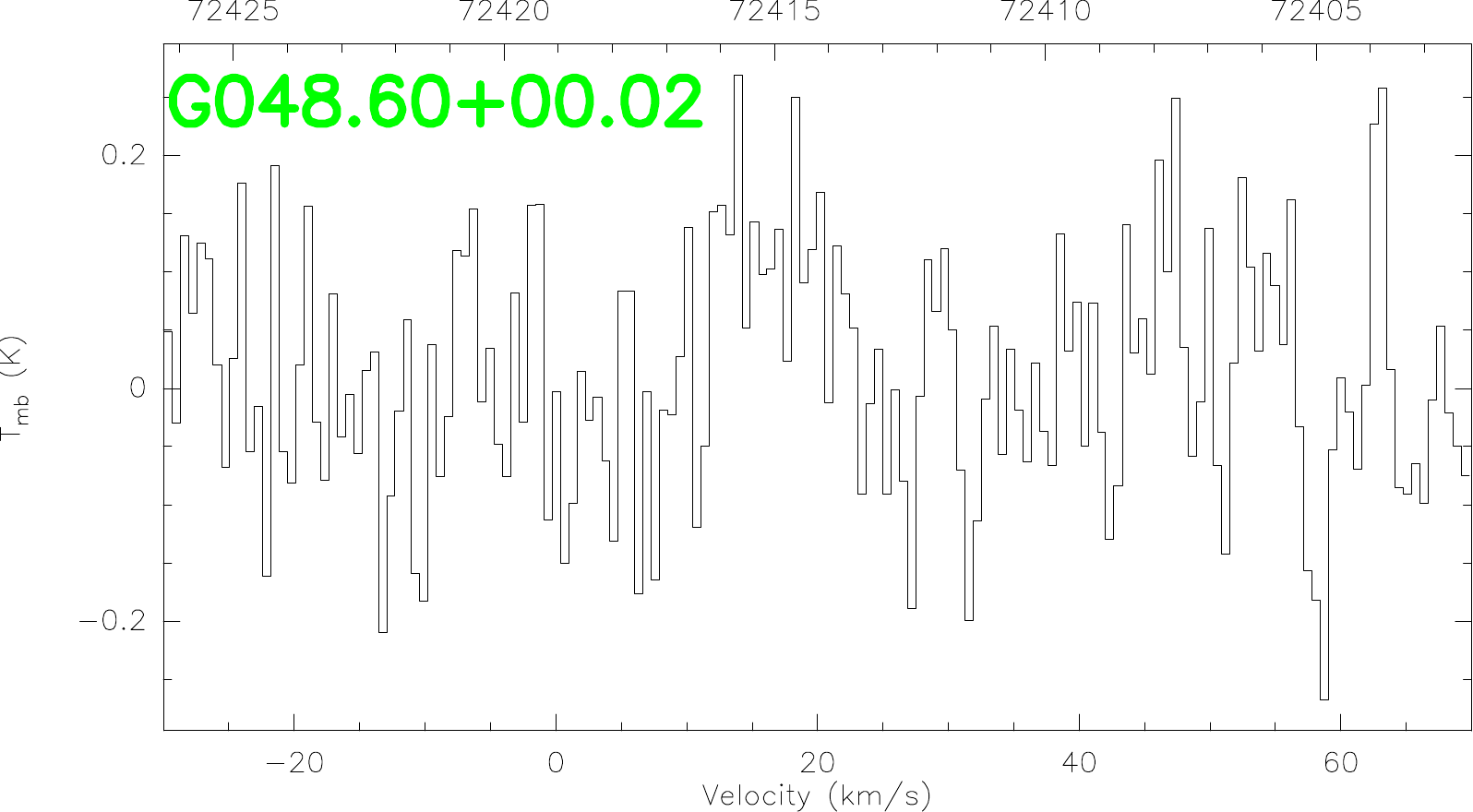}
\includegraphics[width=0.155\textwidth]{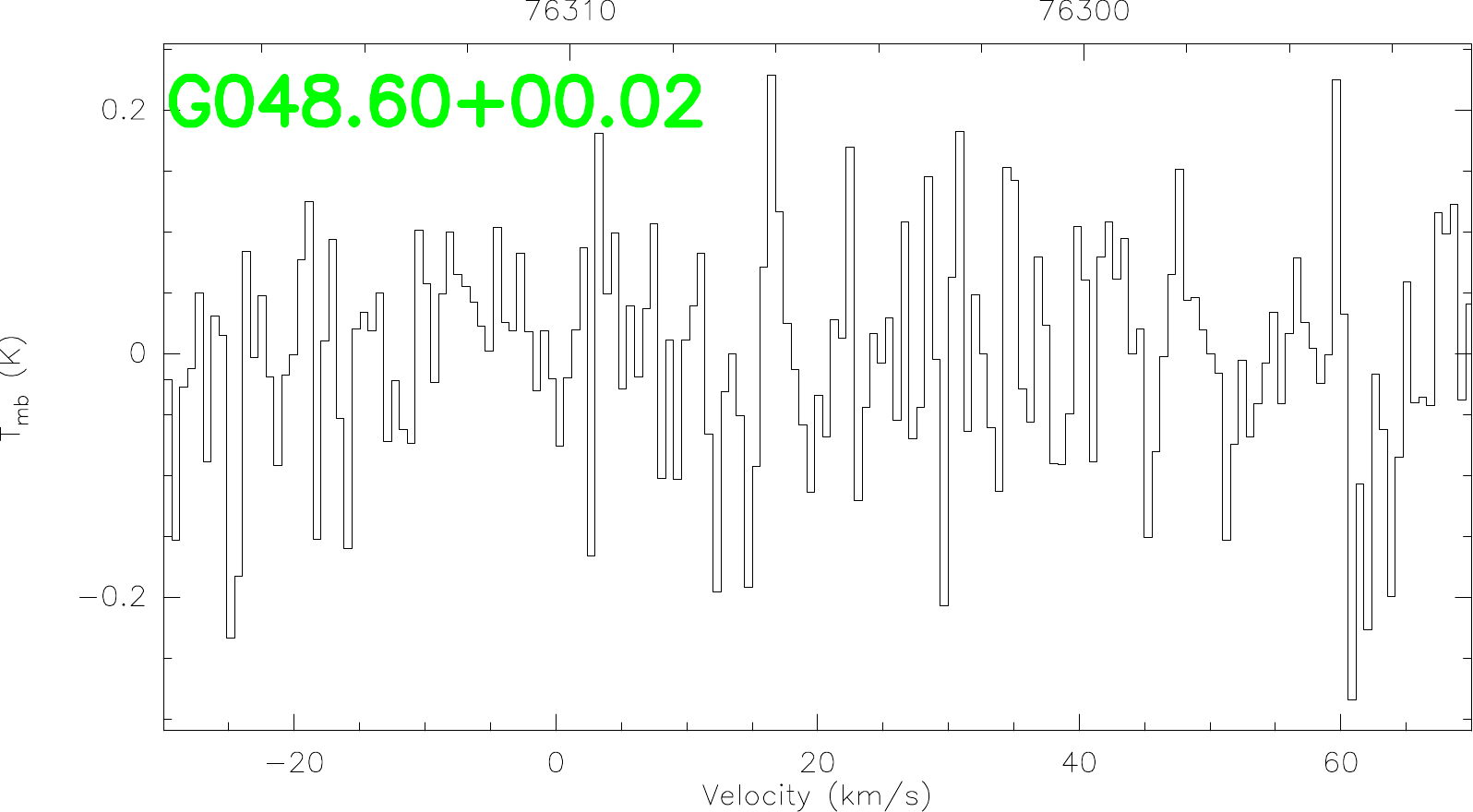}
\includegraphics[width=0.155\textwidth]{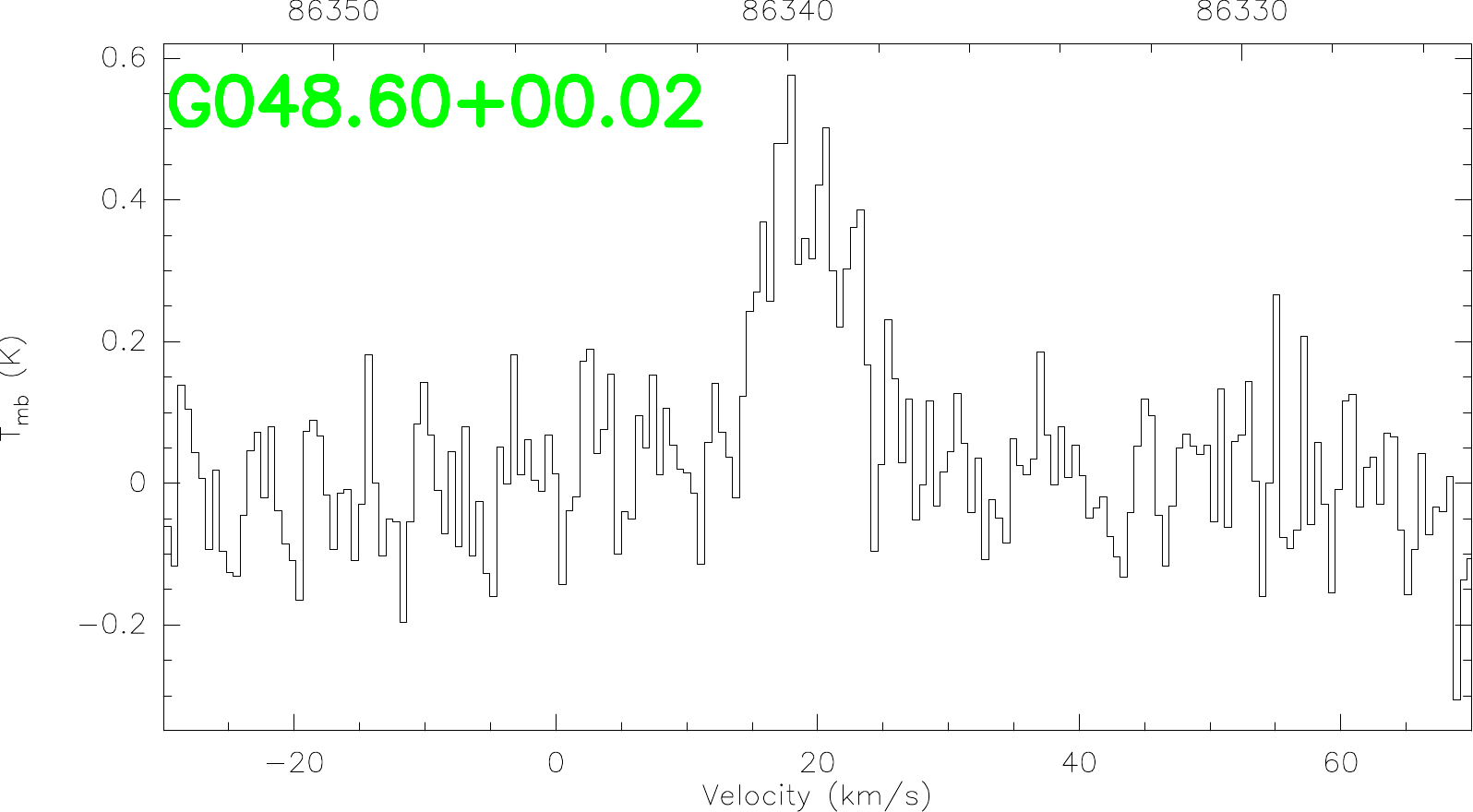}
\includegraphics[width=0.155\textwidth]{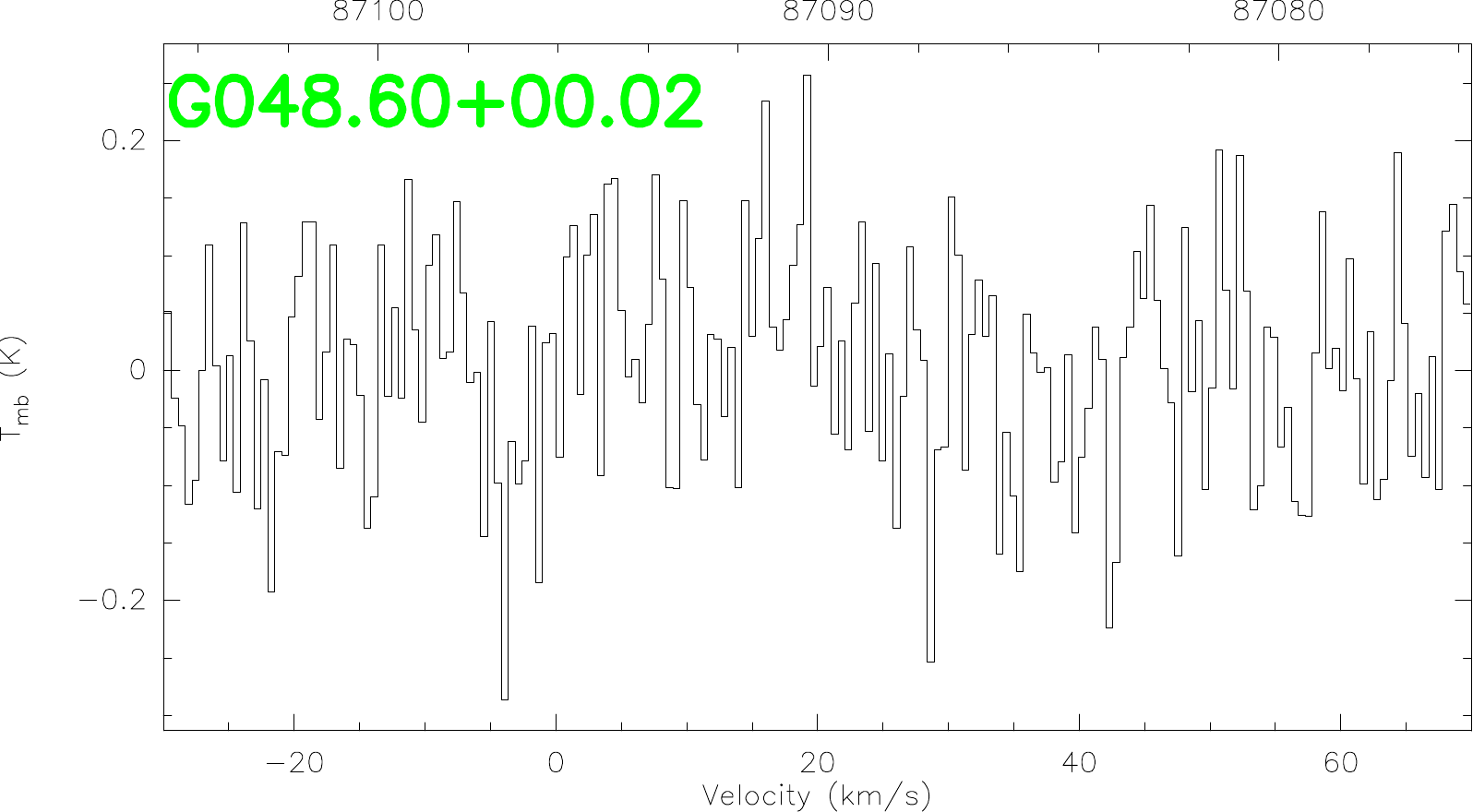}
\includegraphics[width=0.155\textwidth]{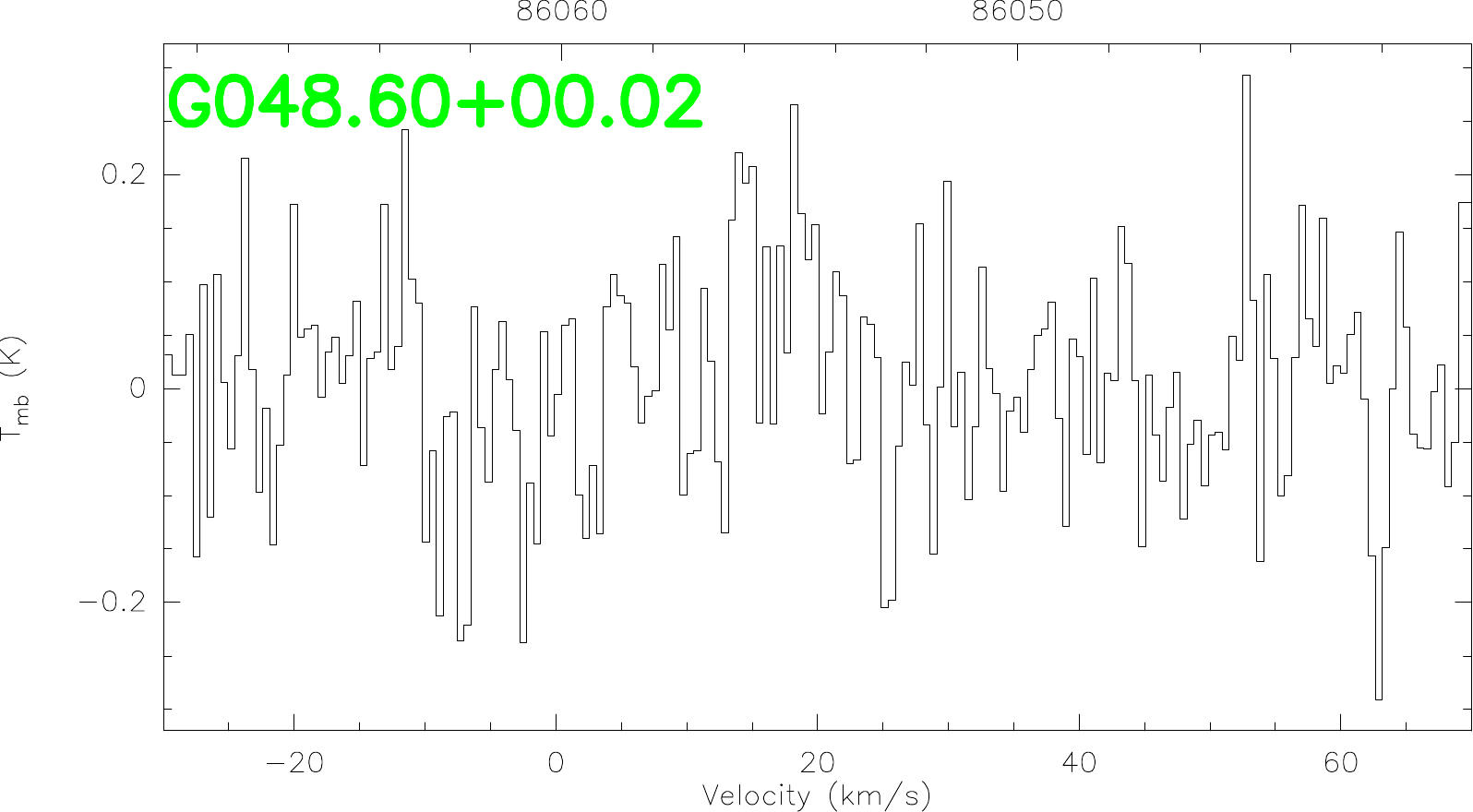}
\includegraphics[width=0.155\textwidth]{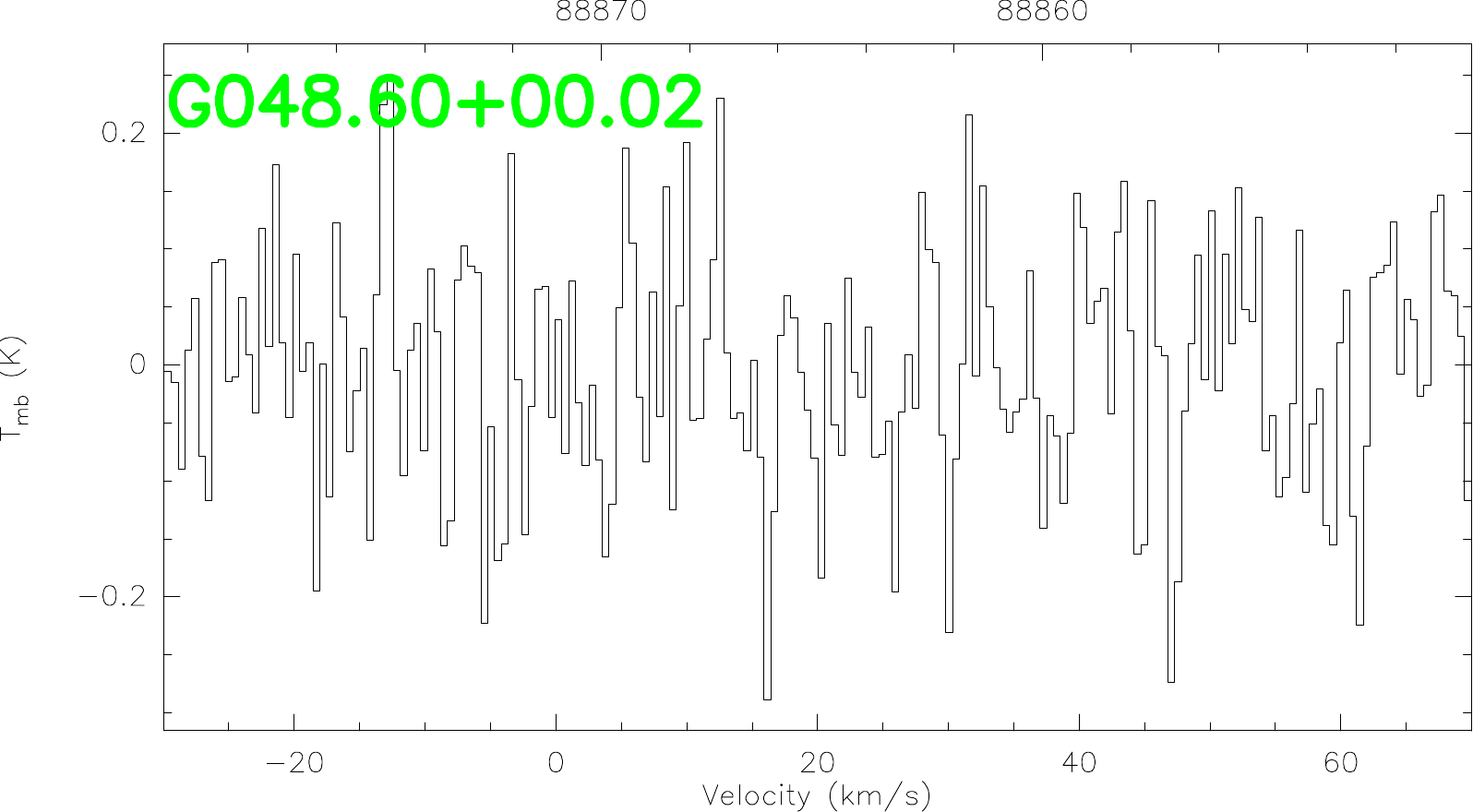}\\

\includegraphics[width=0.155\textwidth]{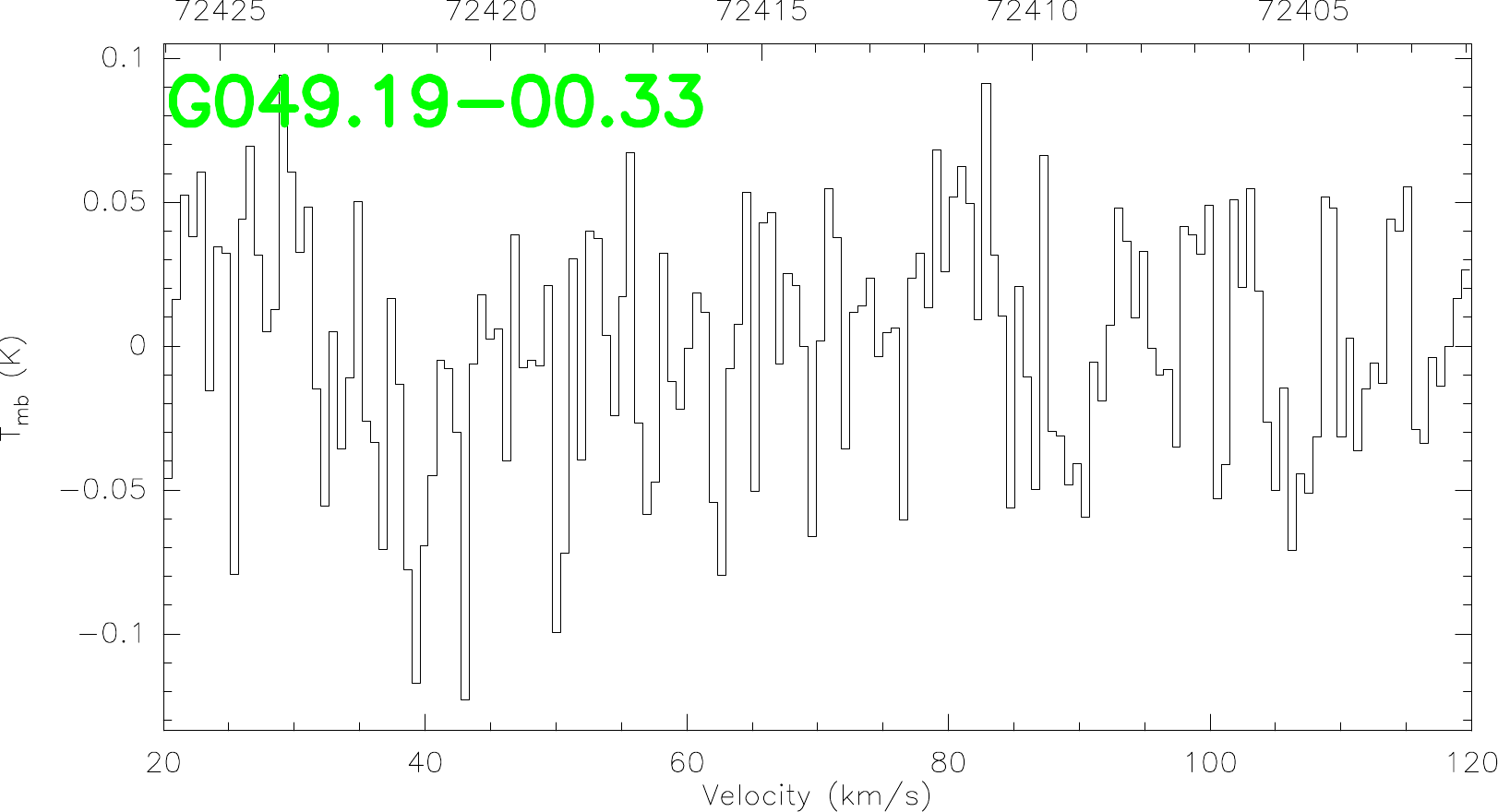}
\includegraphics[width=0.155\textwidth]{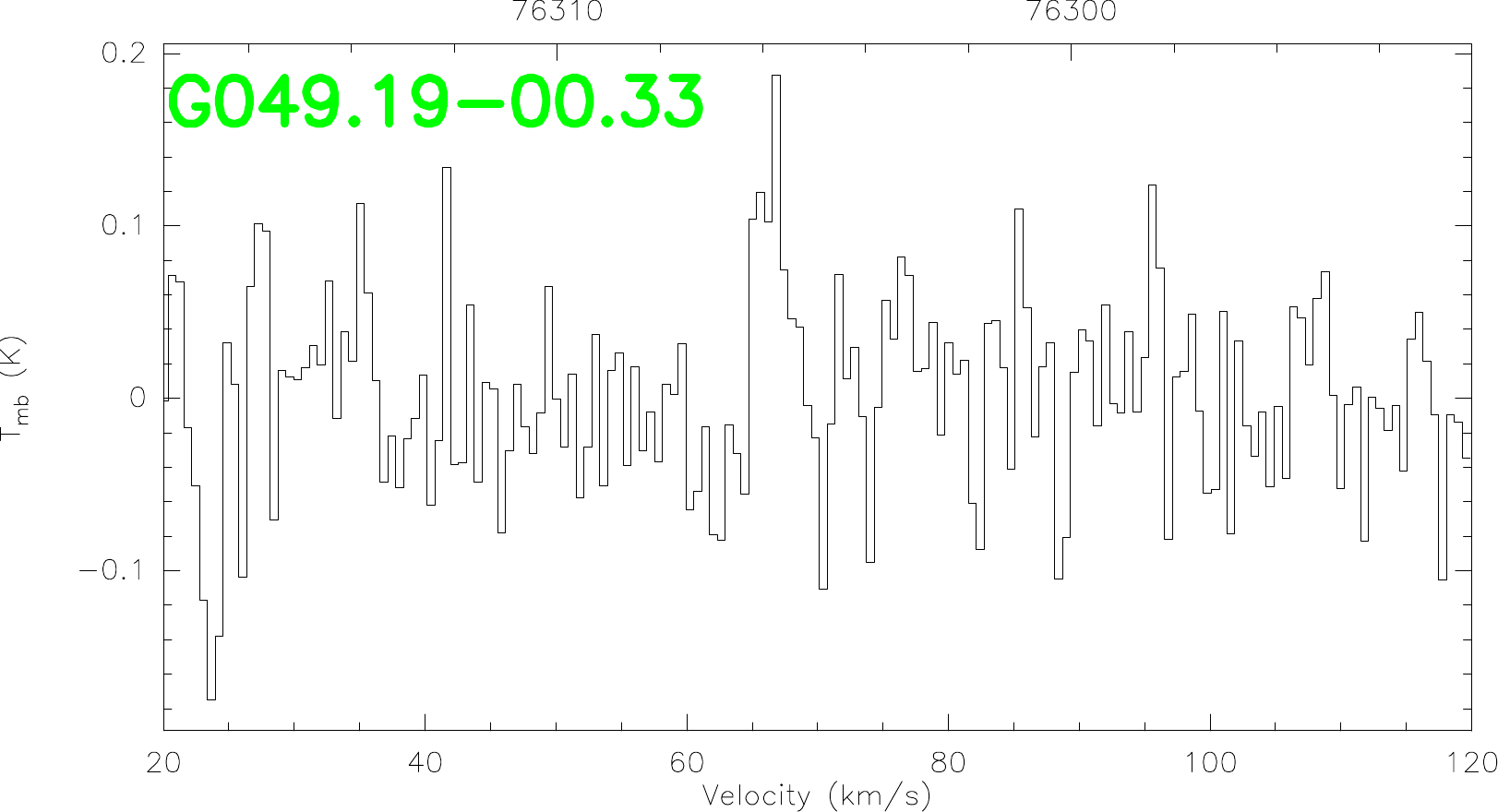}
\includegraphics[width=0.155\textwidth]{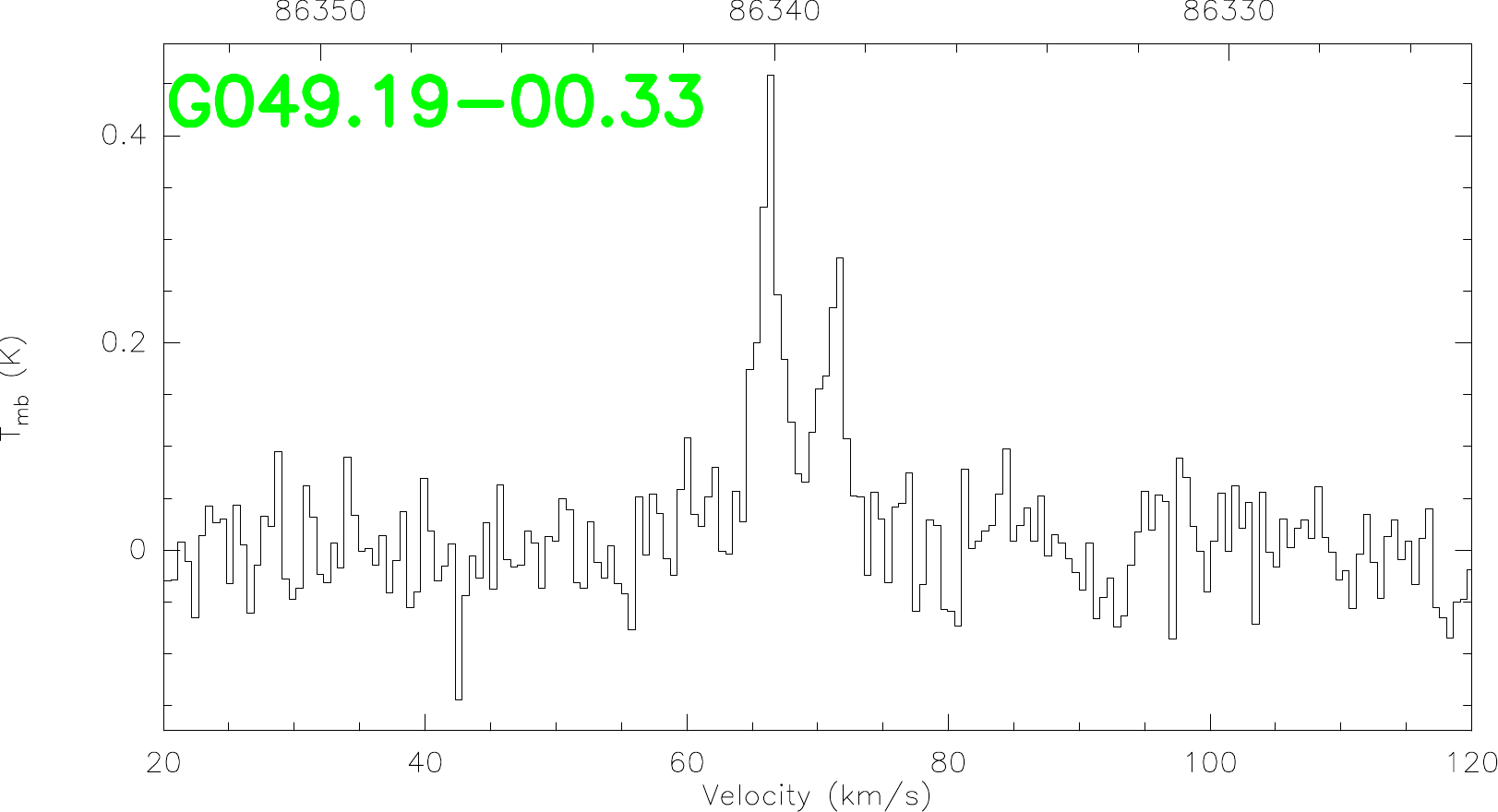}
\includegraphics[width=0.155\textwidth]{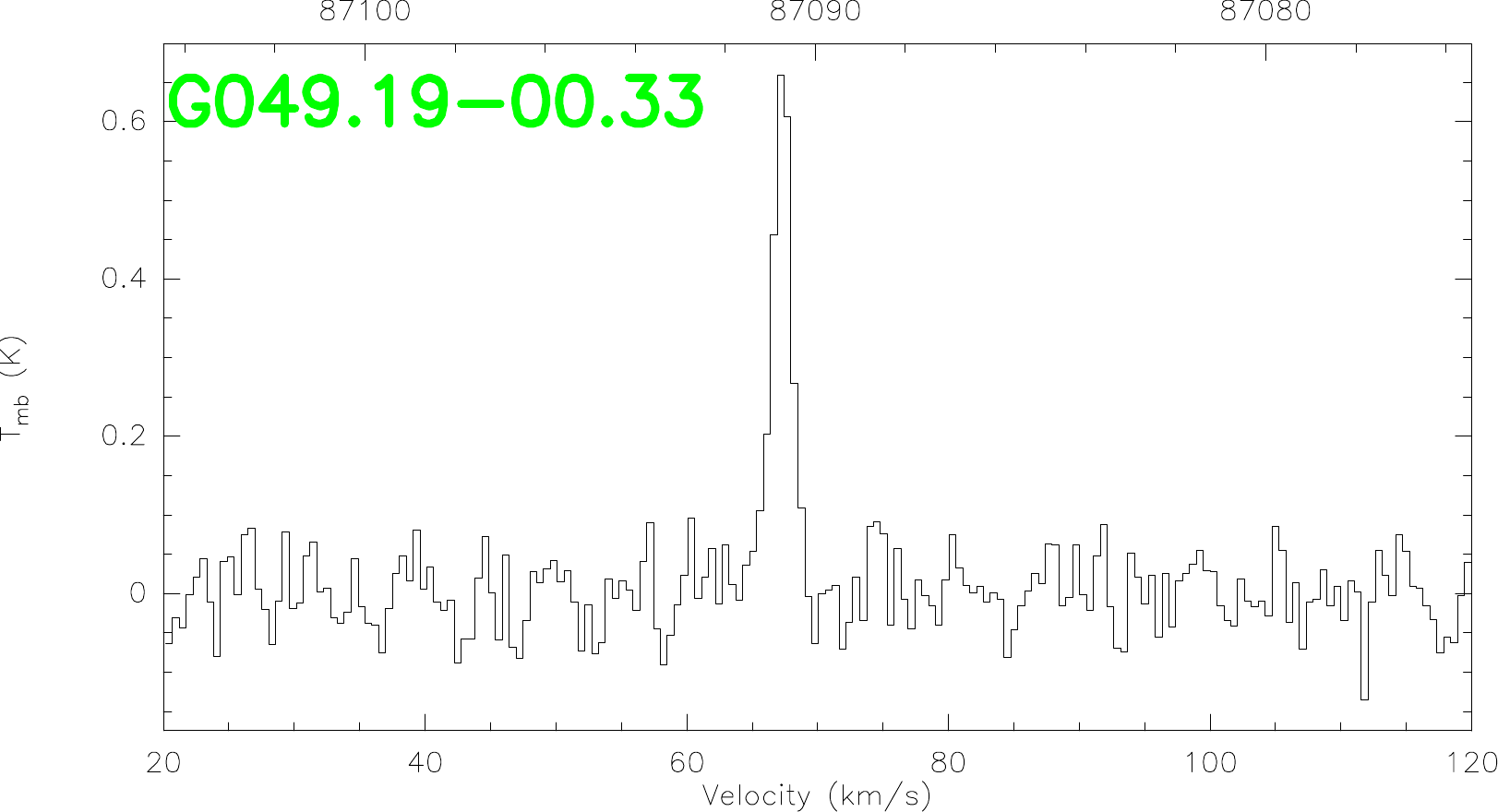}
\includegraphics[width=0.155\textwidth]{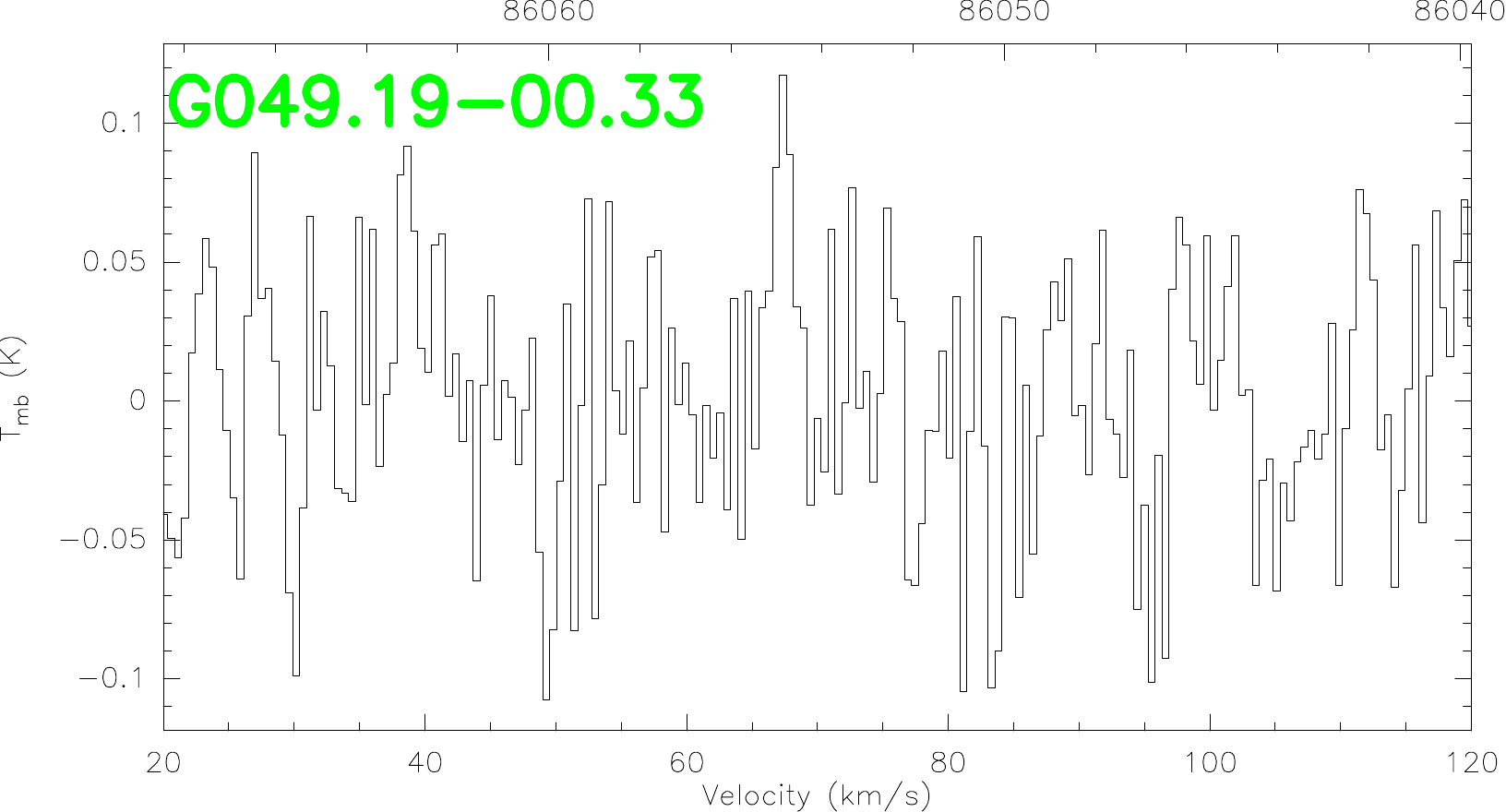}
\includegraphics[width=0.155\textwidth]{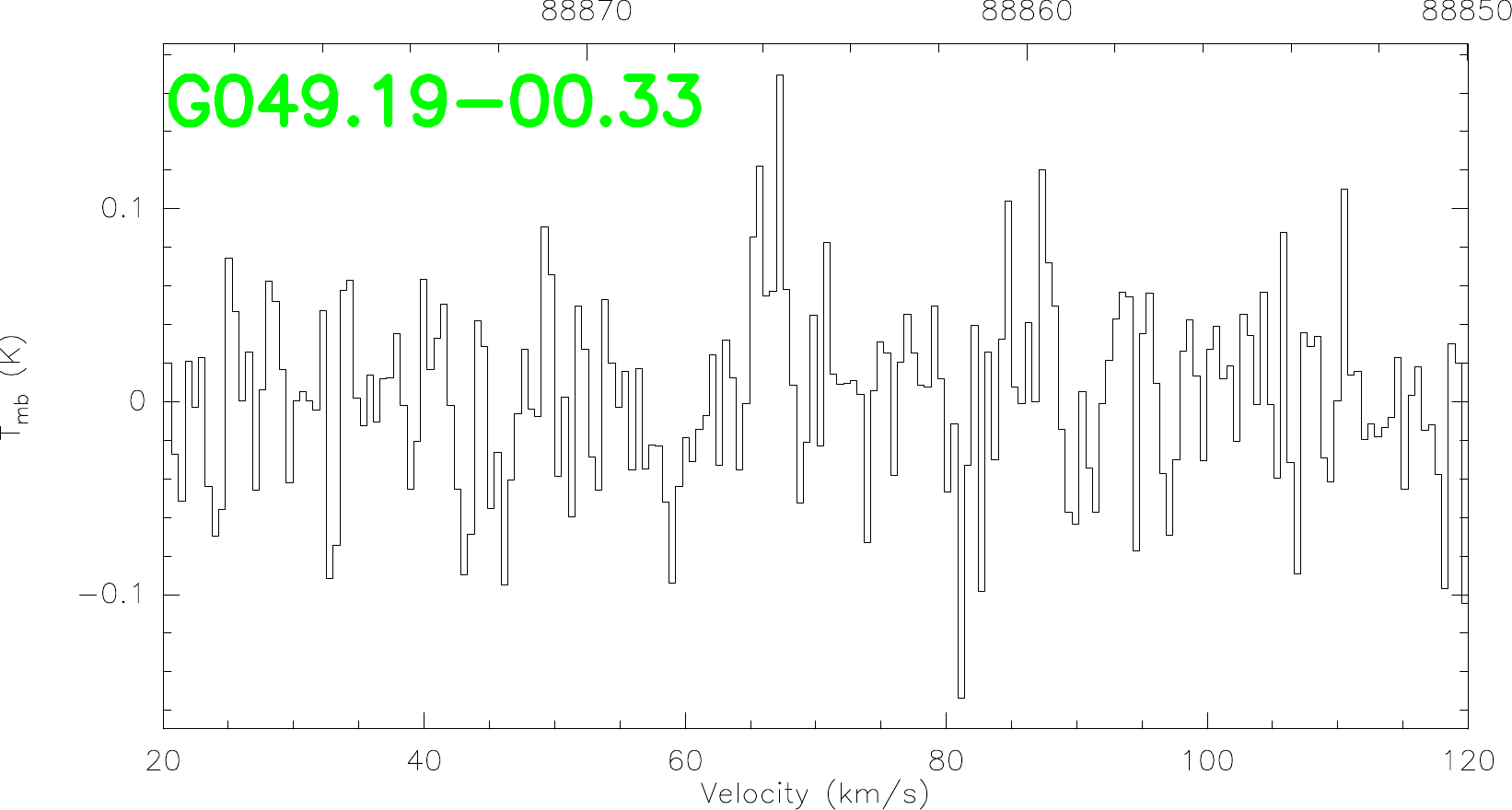}\\

\includegraphics[width=0.155\textwidth]{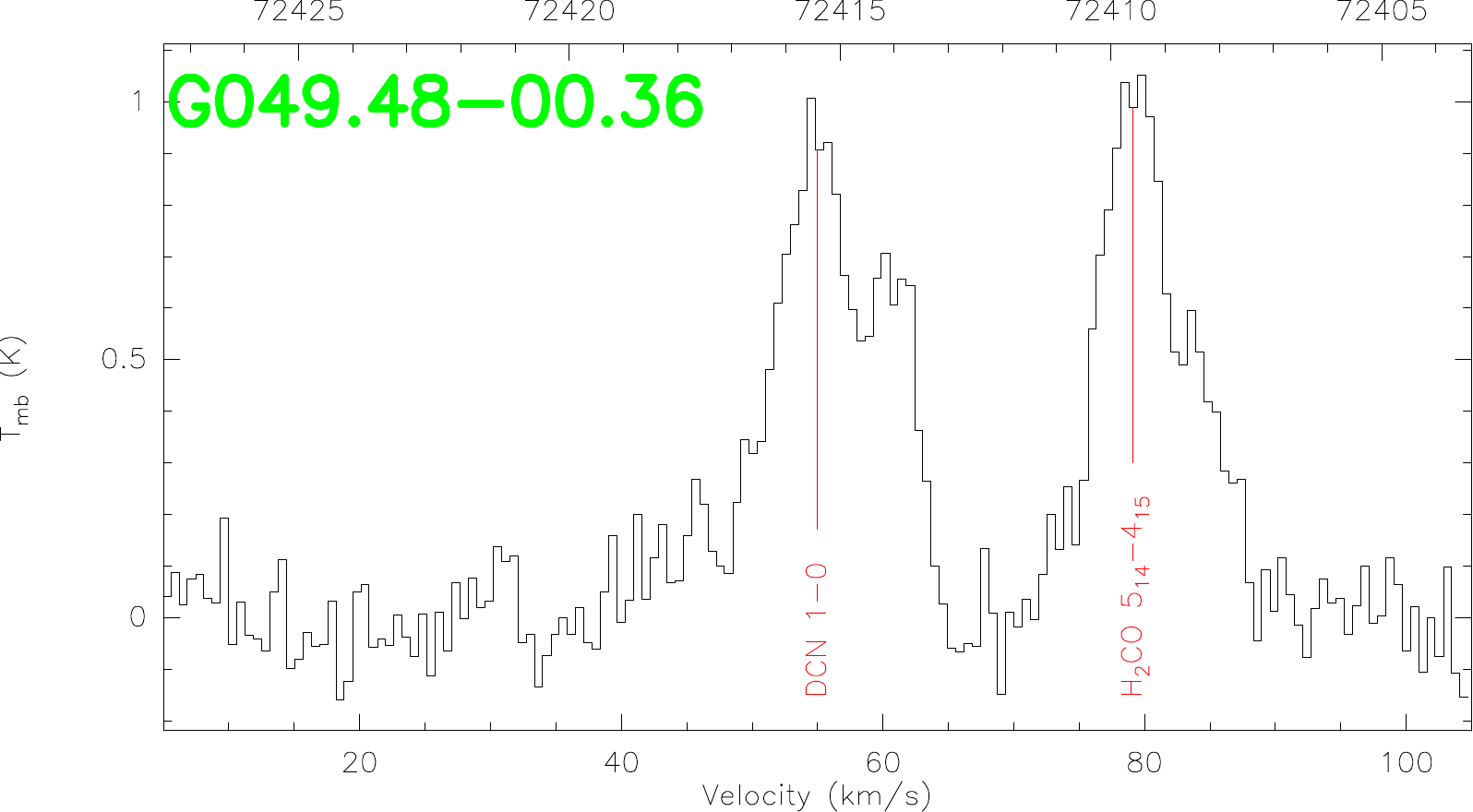}
\includegraphics[width=0.155\textwidth]{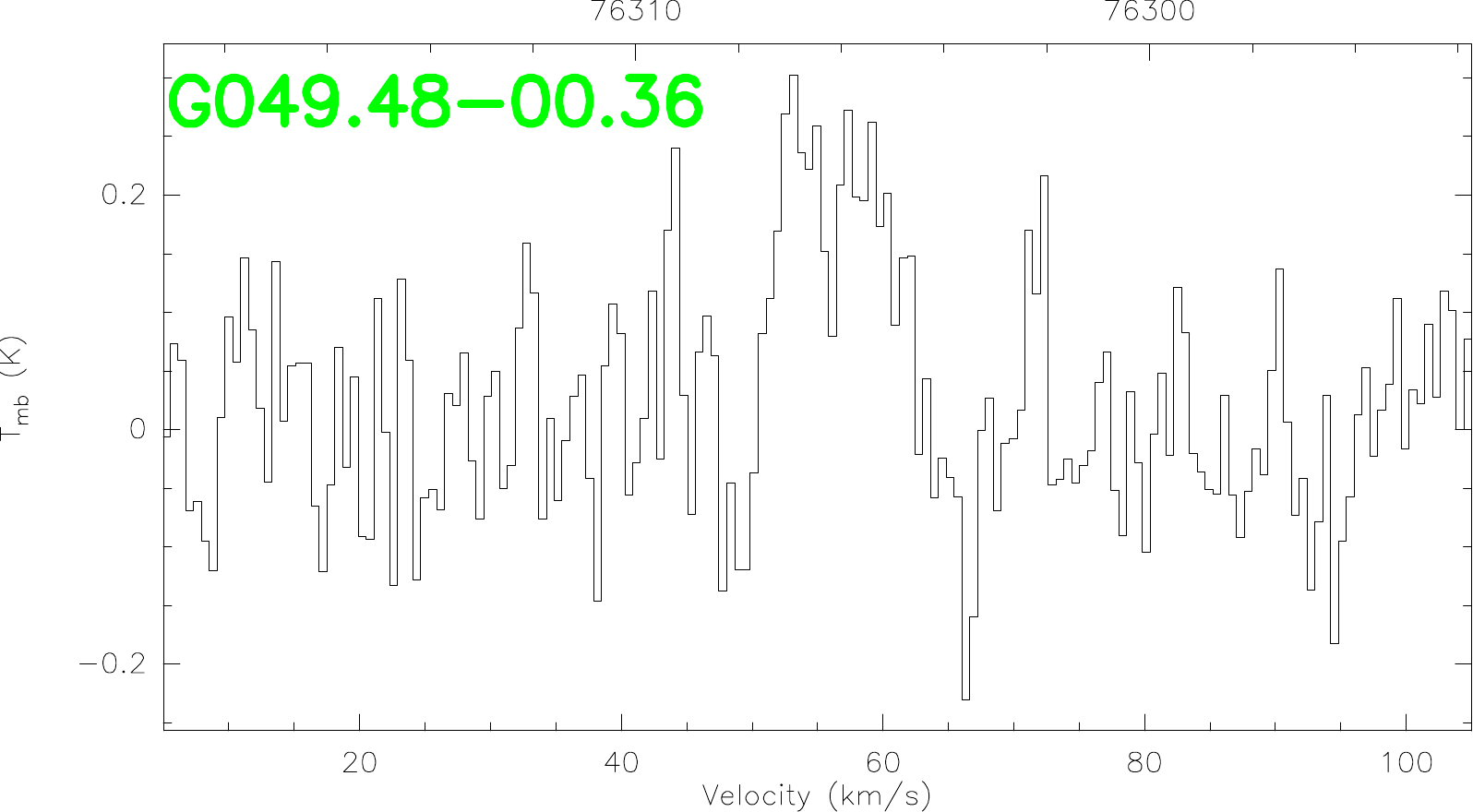}
\includegraphics[width=0.155\textwidth]{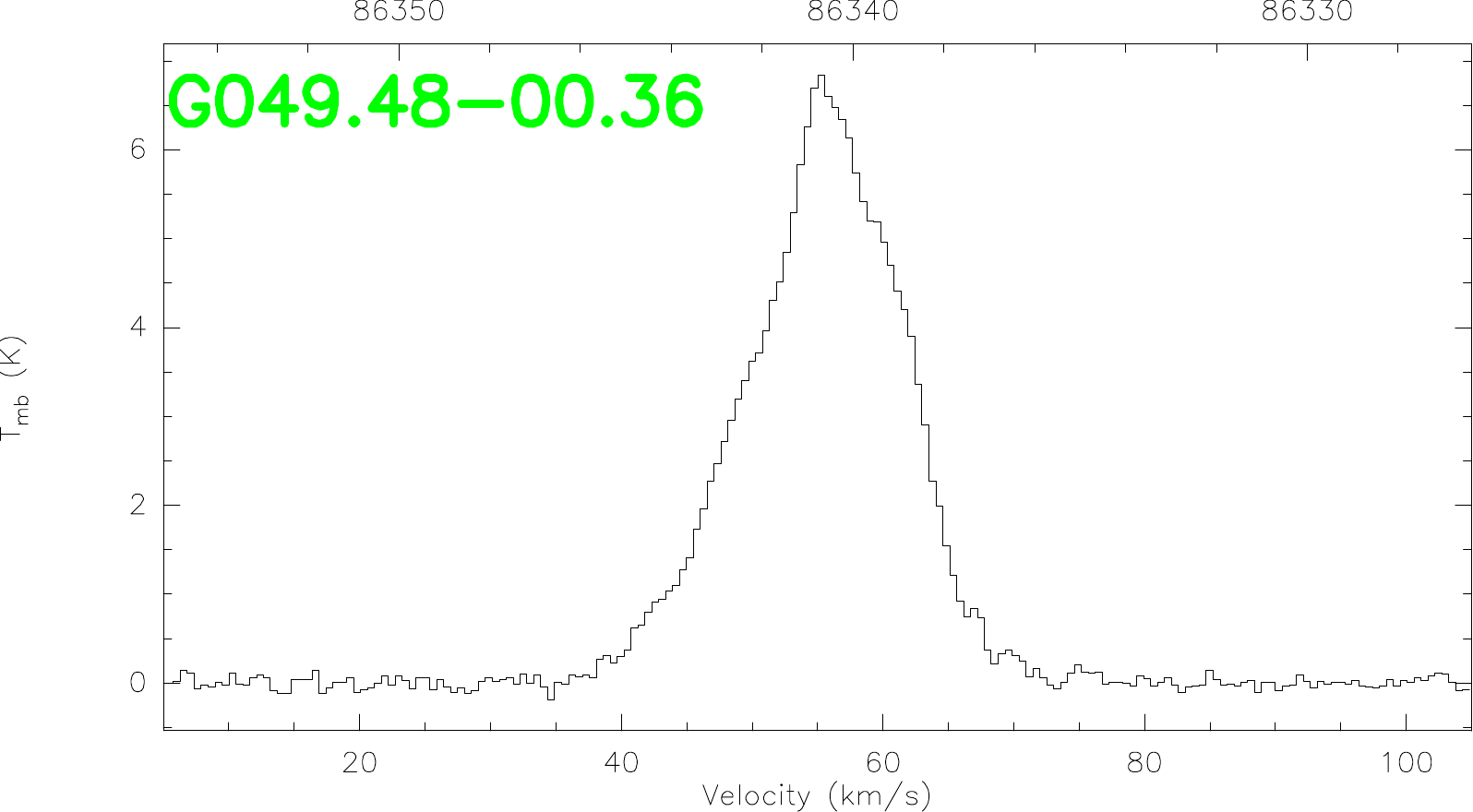}
\includegraphics[width=0.155\textwidth]{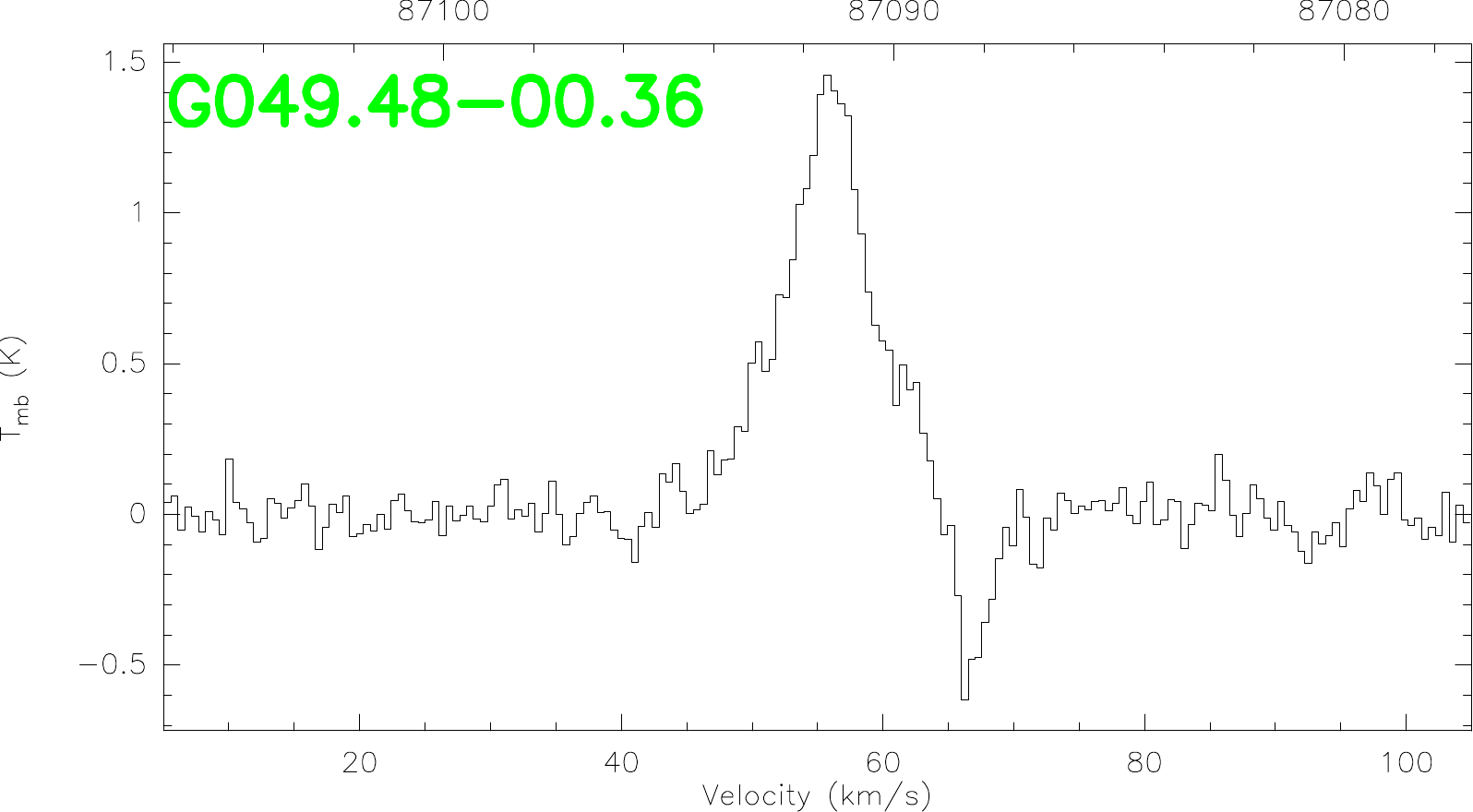}
\includegraphics[width=0.155\textwidth]{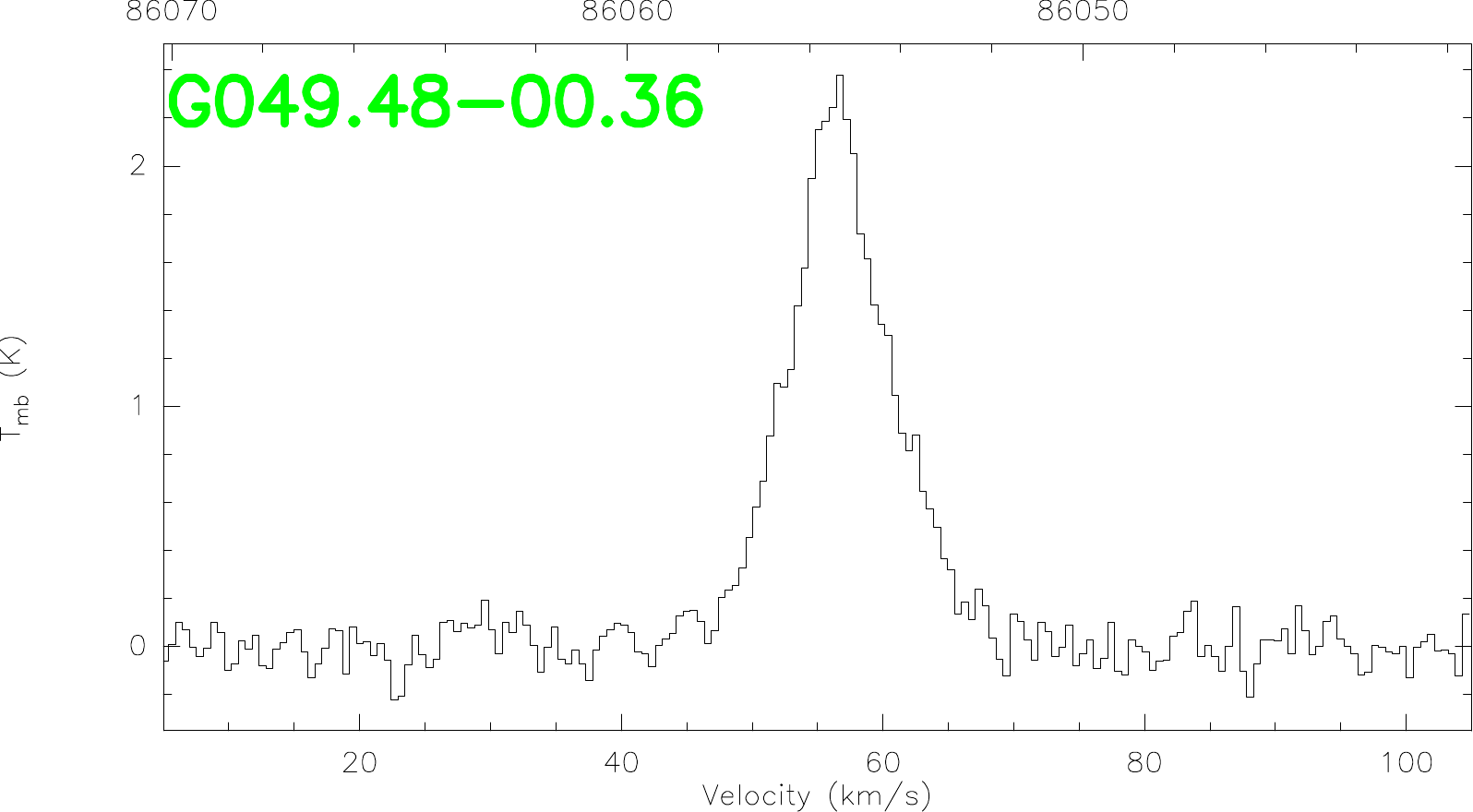}
\includegraphics[width=0.155\textwidth]{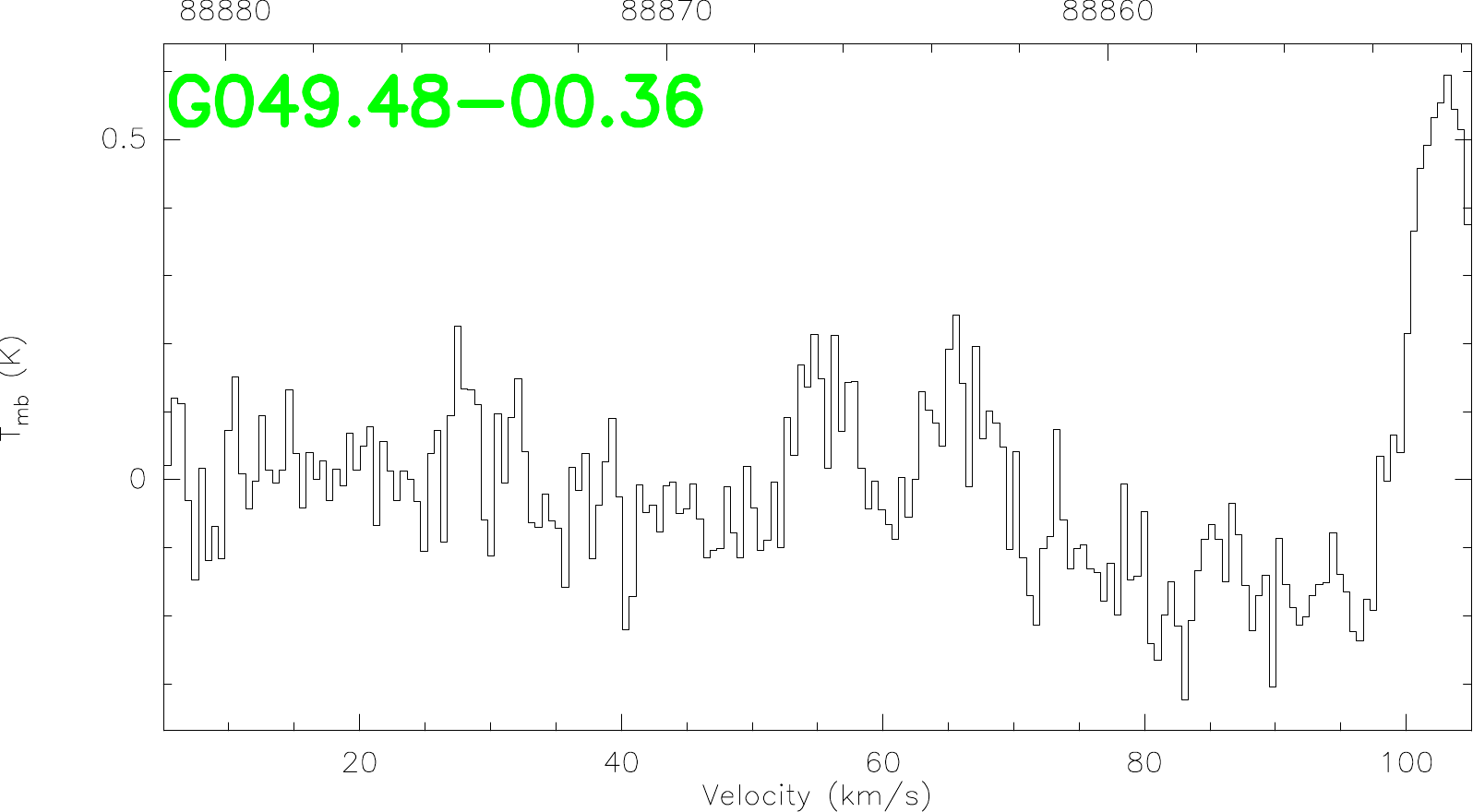}\\

\includegraphics[width=0.155\textwidth]{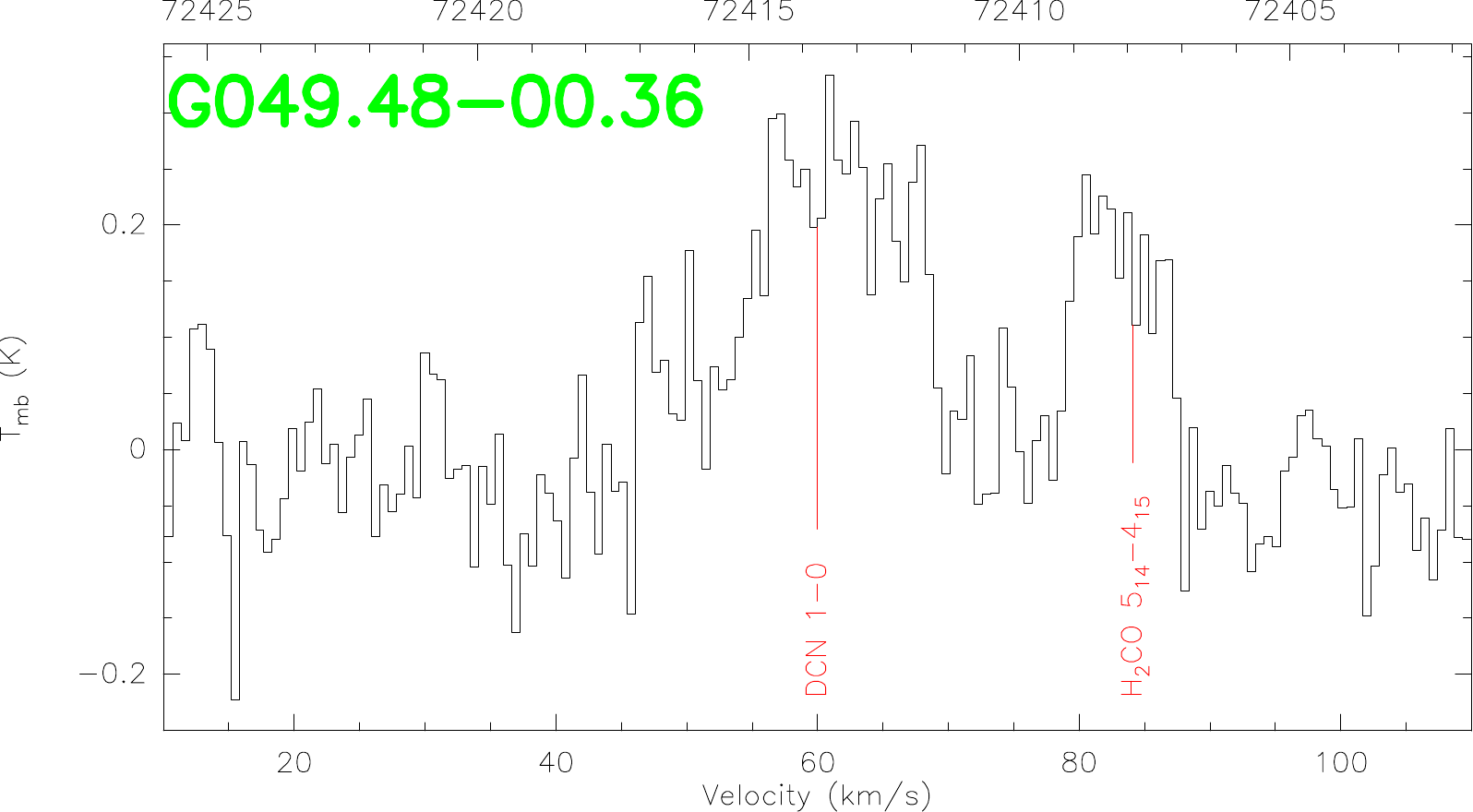}
\includegraphics[width=0.155\textwidth]{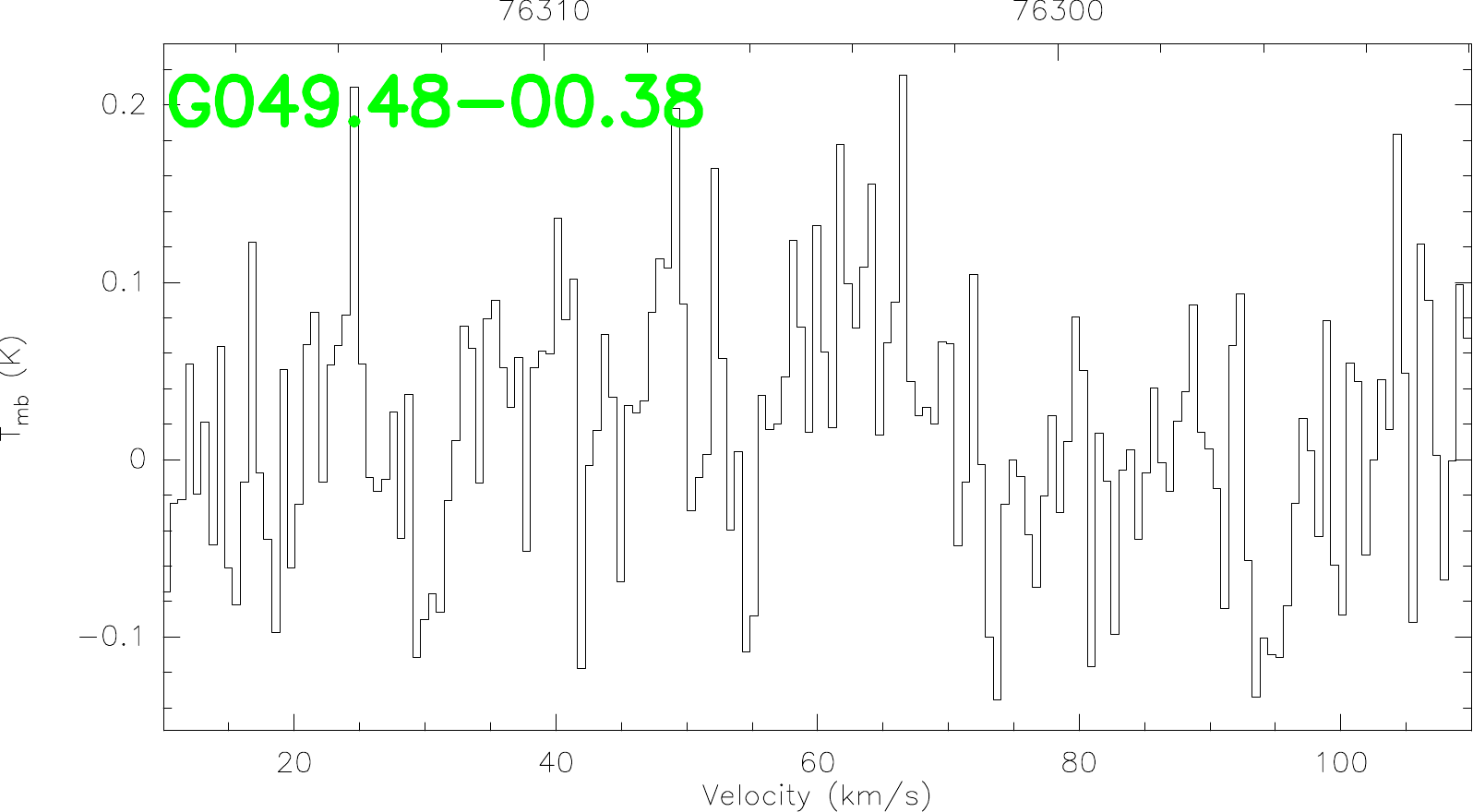}
\includegraphics[width=0.155\textwidth]{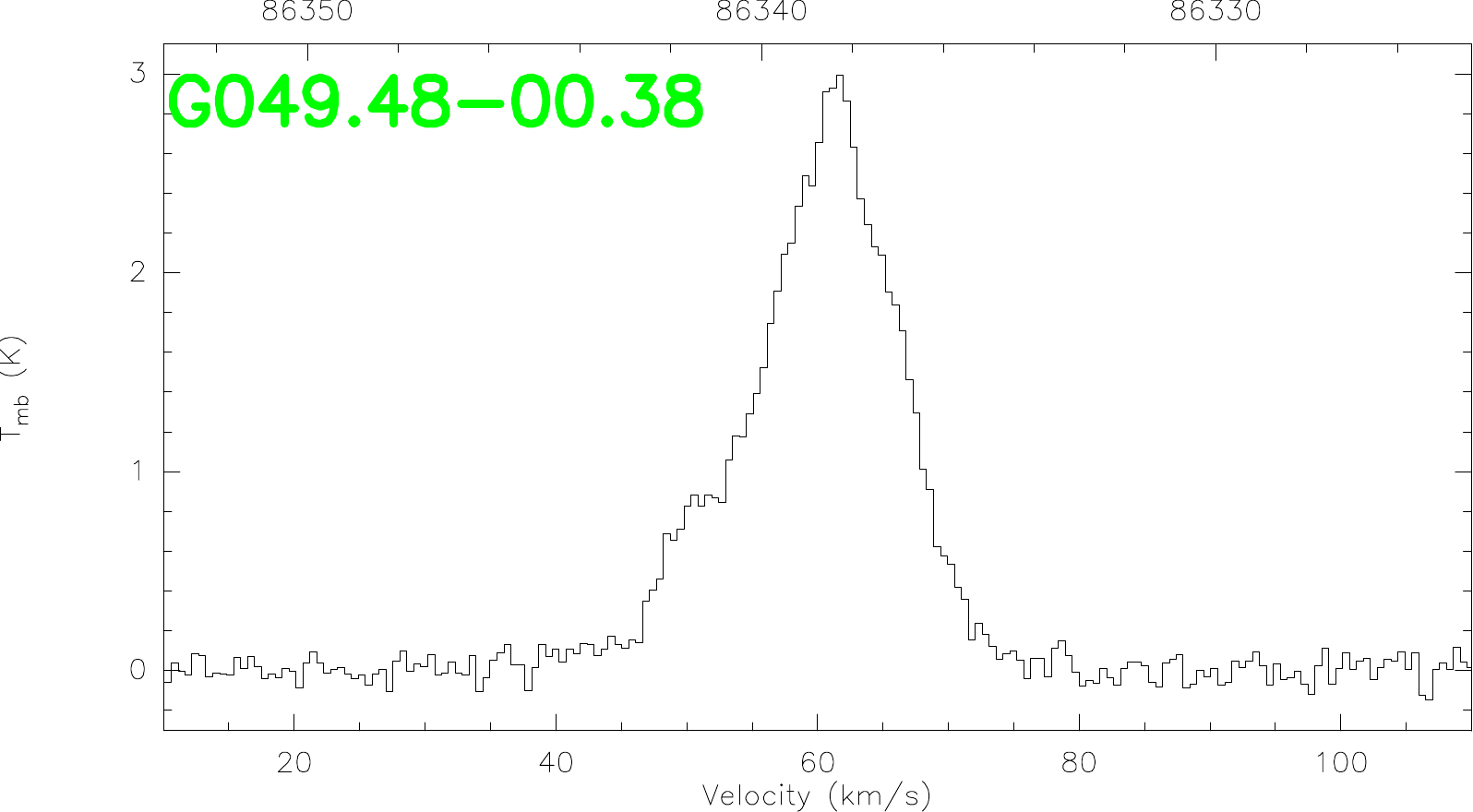}
\includegraphics[width=0.155\textwidth]{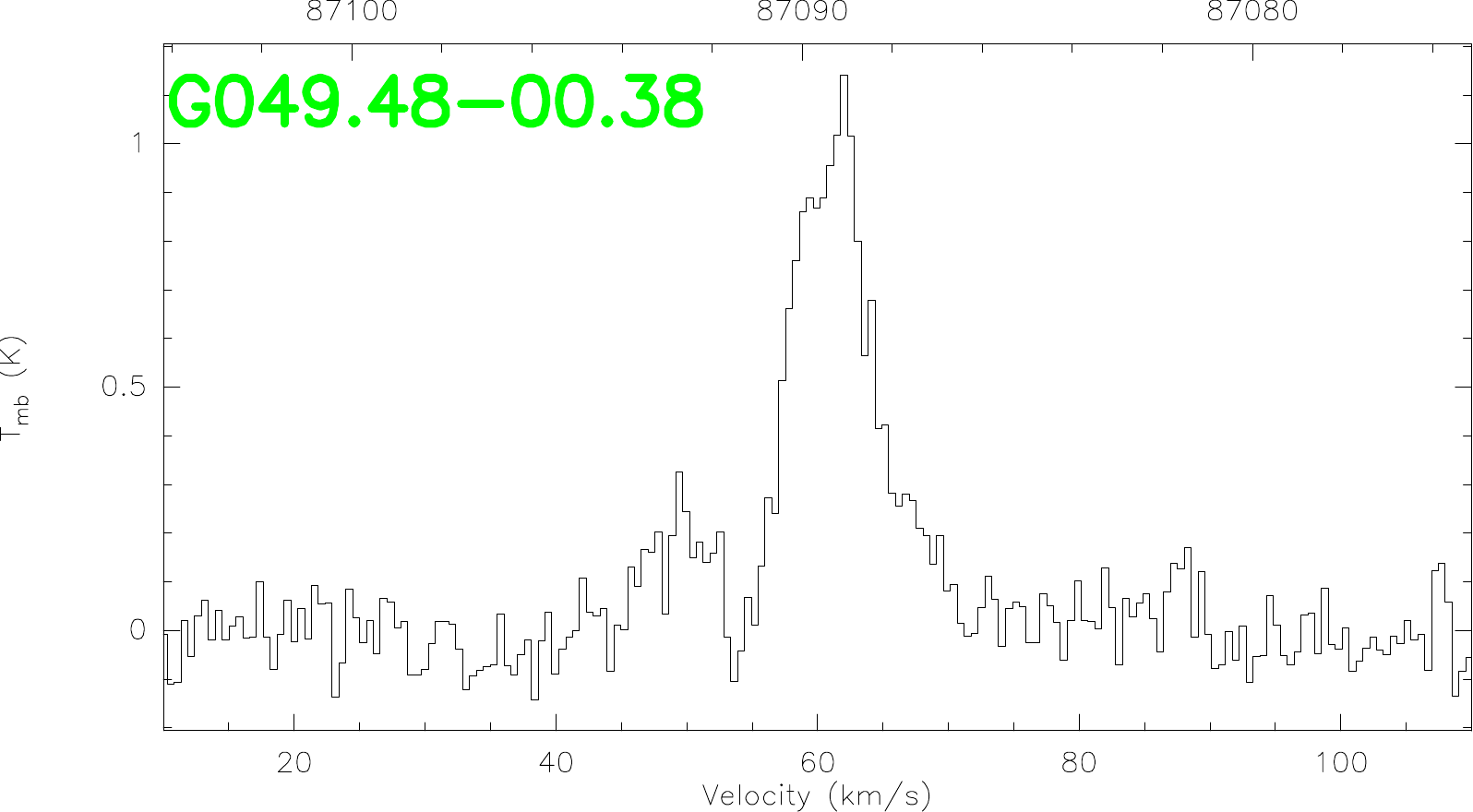}
\includegraphics[width=0.155\textwidth]{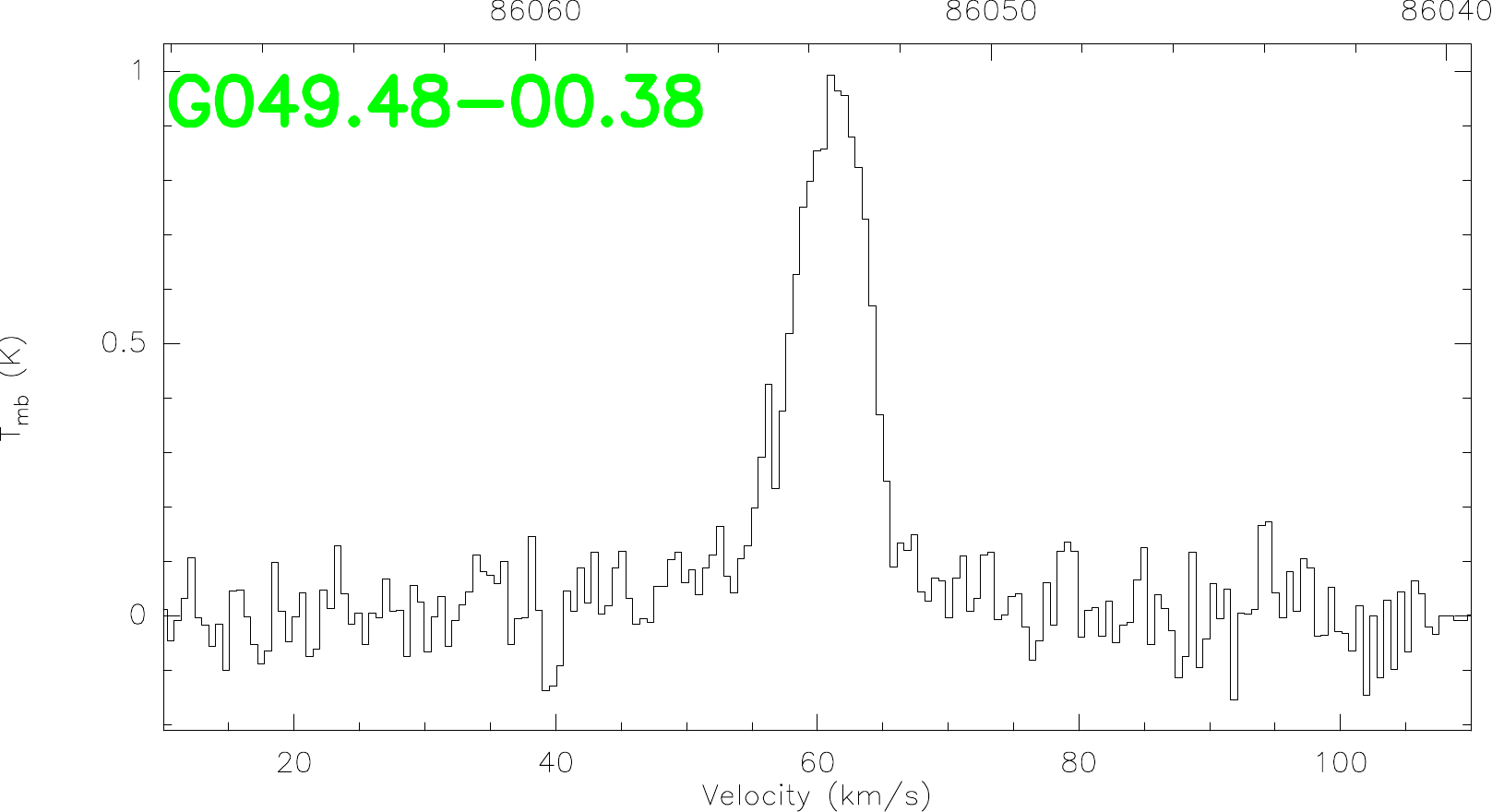}
\includegraphics[width=0.155\textwidth]{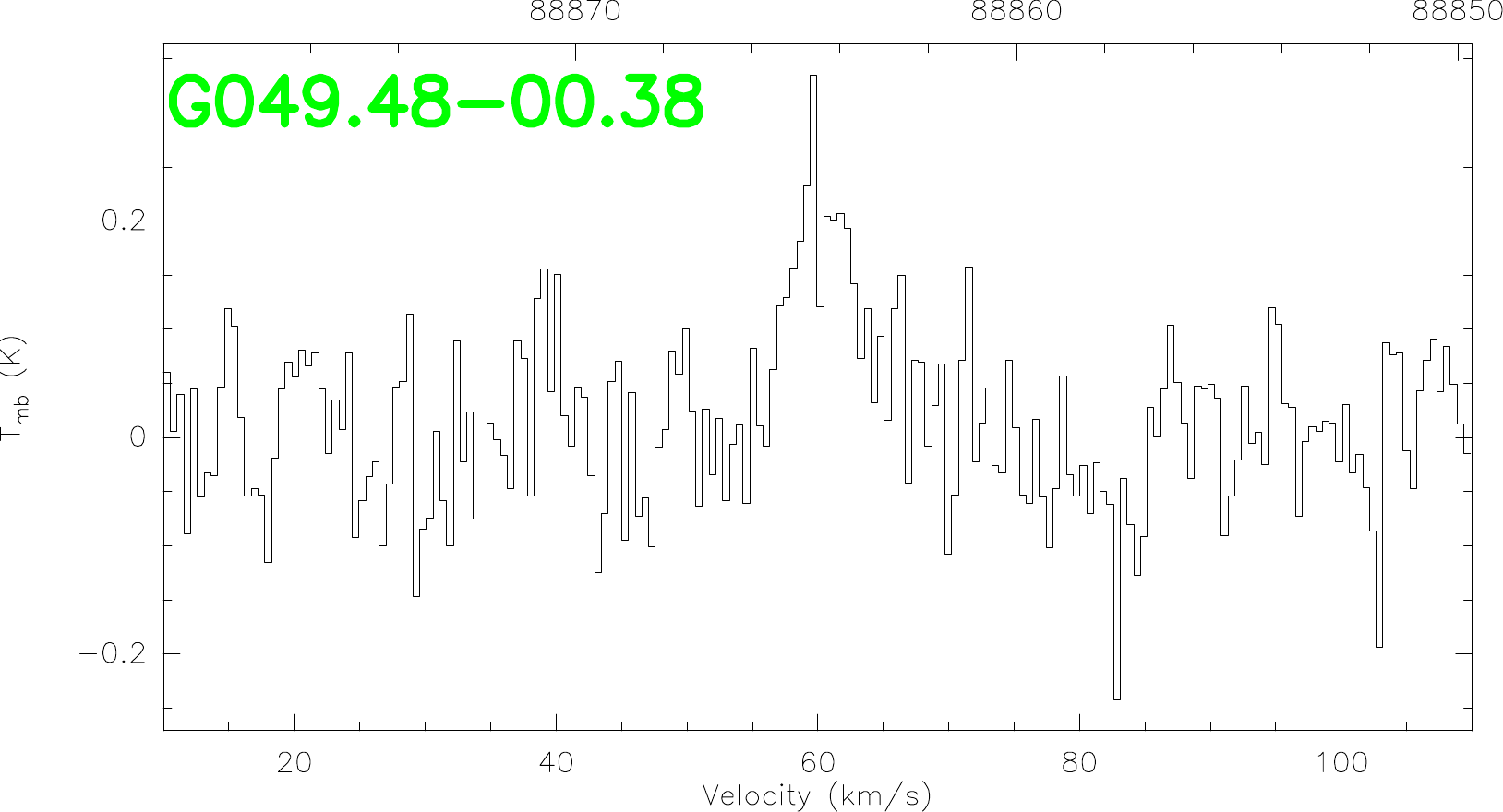}\\

\caption{Continued. Spectra of different molecular lines.}
\label{fig:multi_lines_page3}
\end{figure}

\begin{figure}[h]
\centering

\makebox[\colw][c]{DCN 1-0}
\makebox[\colw][c]{DNC 1-0}
\makebox[\colw][c]{H$^{13}$CN 1-0}
\makebox[\colw][c]{HN$^{13}$C 1-0}
\makebox[\colw][c]{HC$^{15}$N 1-0}
\makebox[\colw][c]{H$^{15}$NC 1-0}\\[0.2em]

\includegraphics[width=0.155\textwidth]{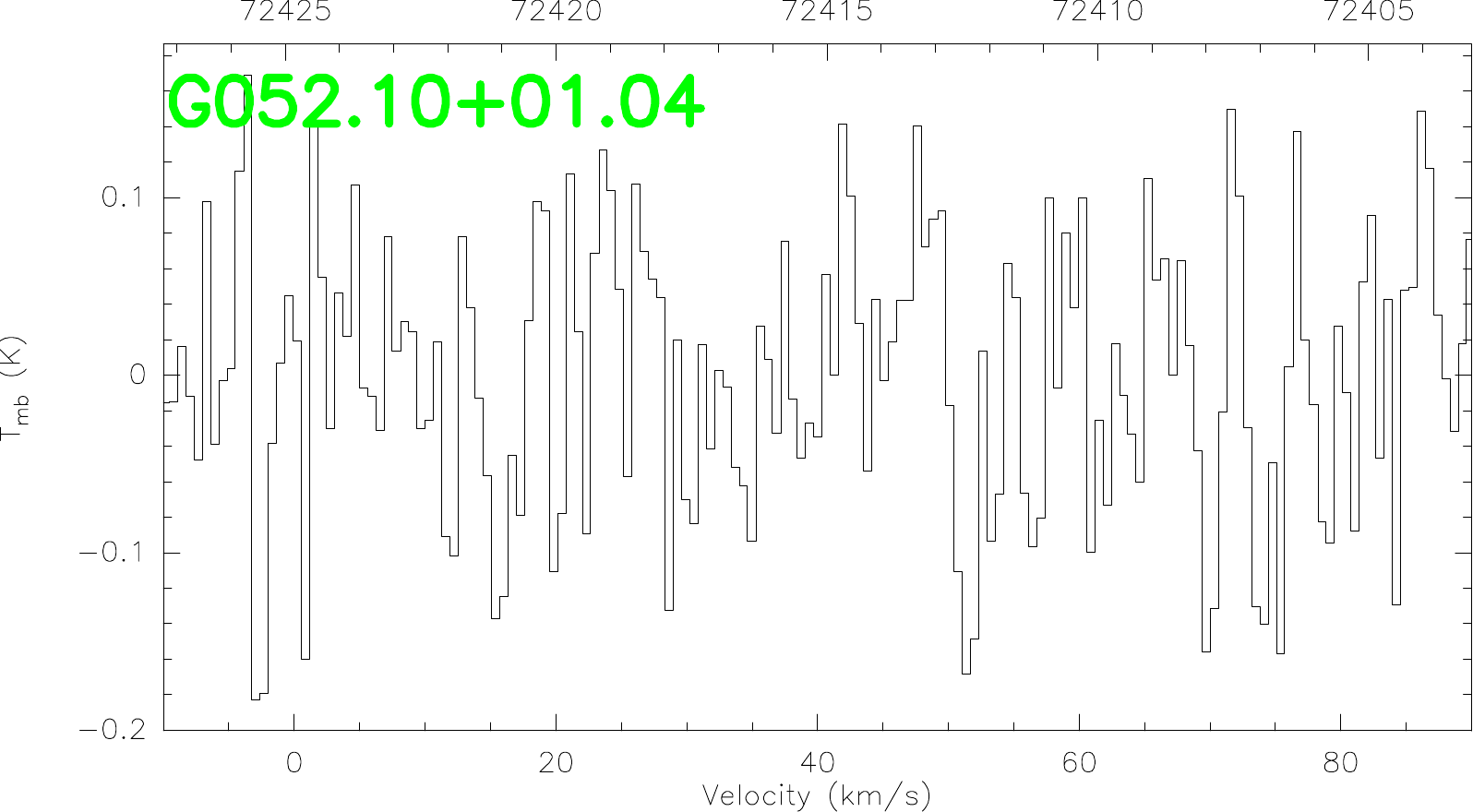}
\includegraphics[width=0.155\textwidth]{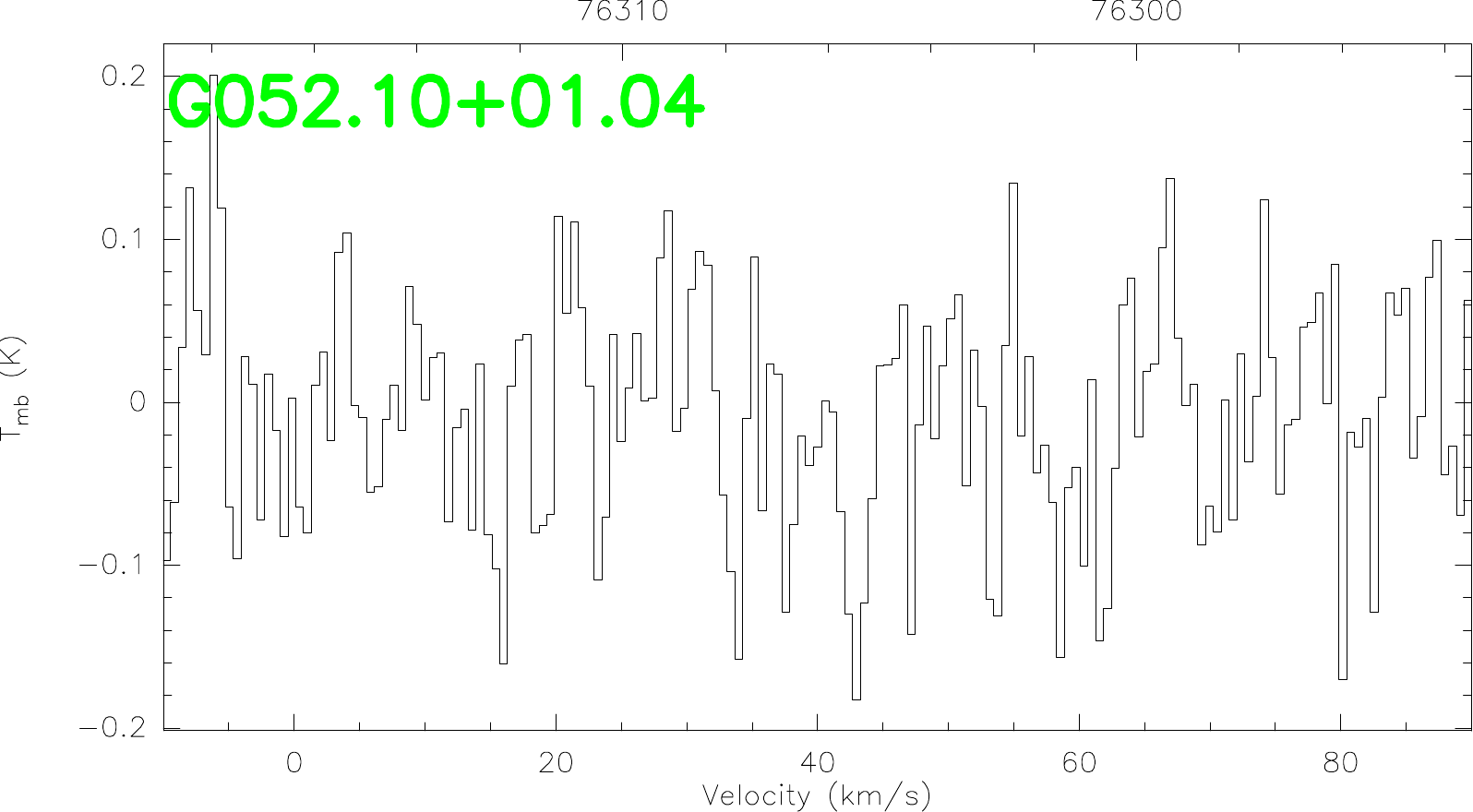}
\includegraphics[width=0.155\textwidth]{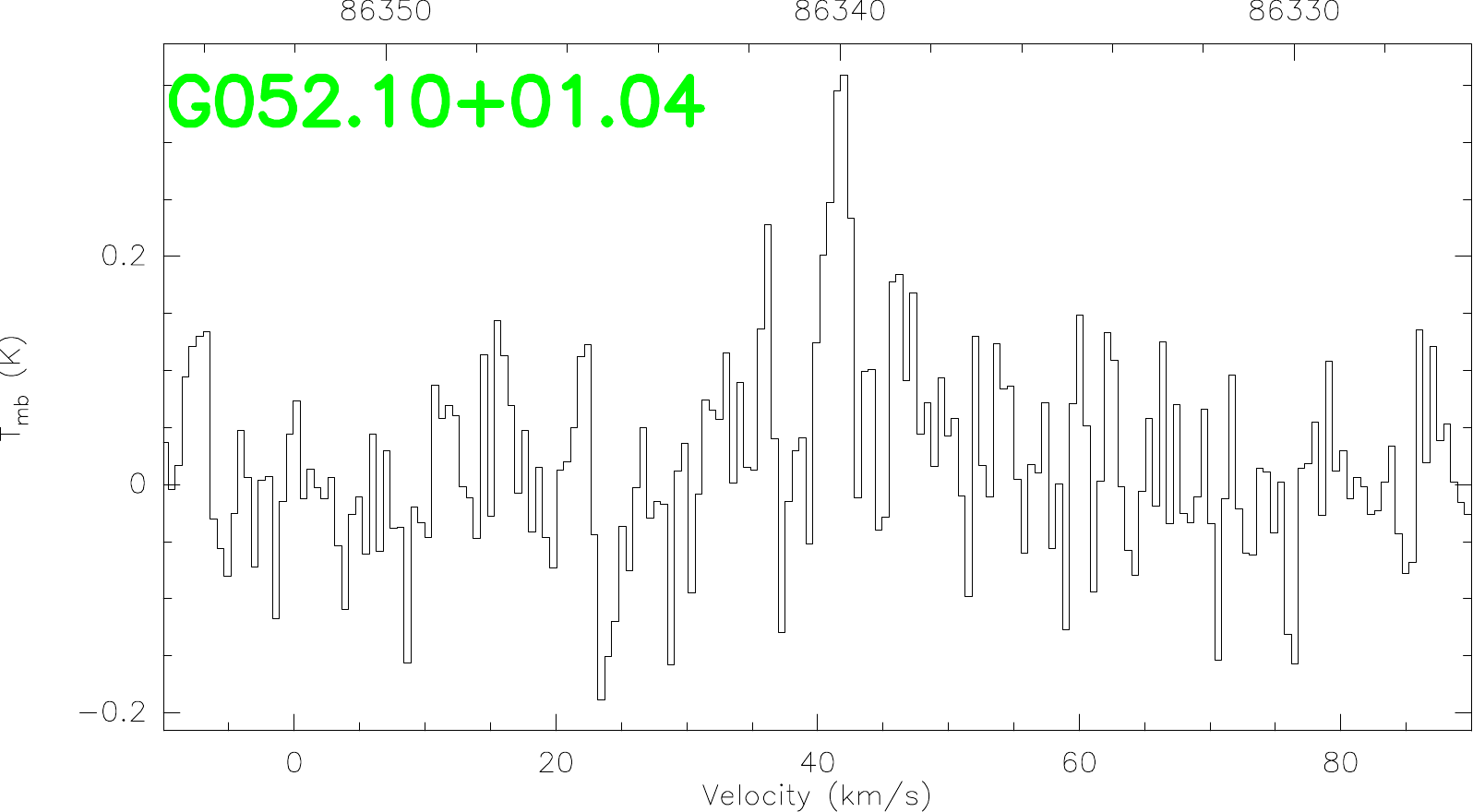}
\includegraphics[width=0.155\textwidth]{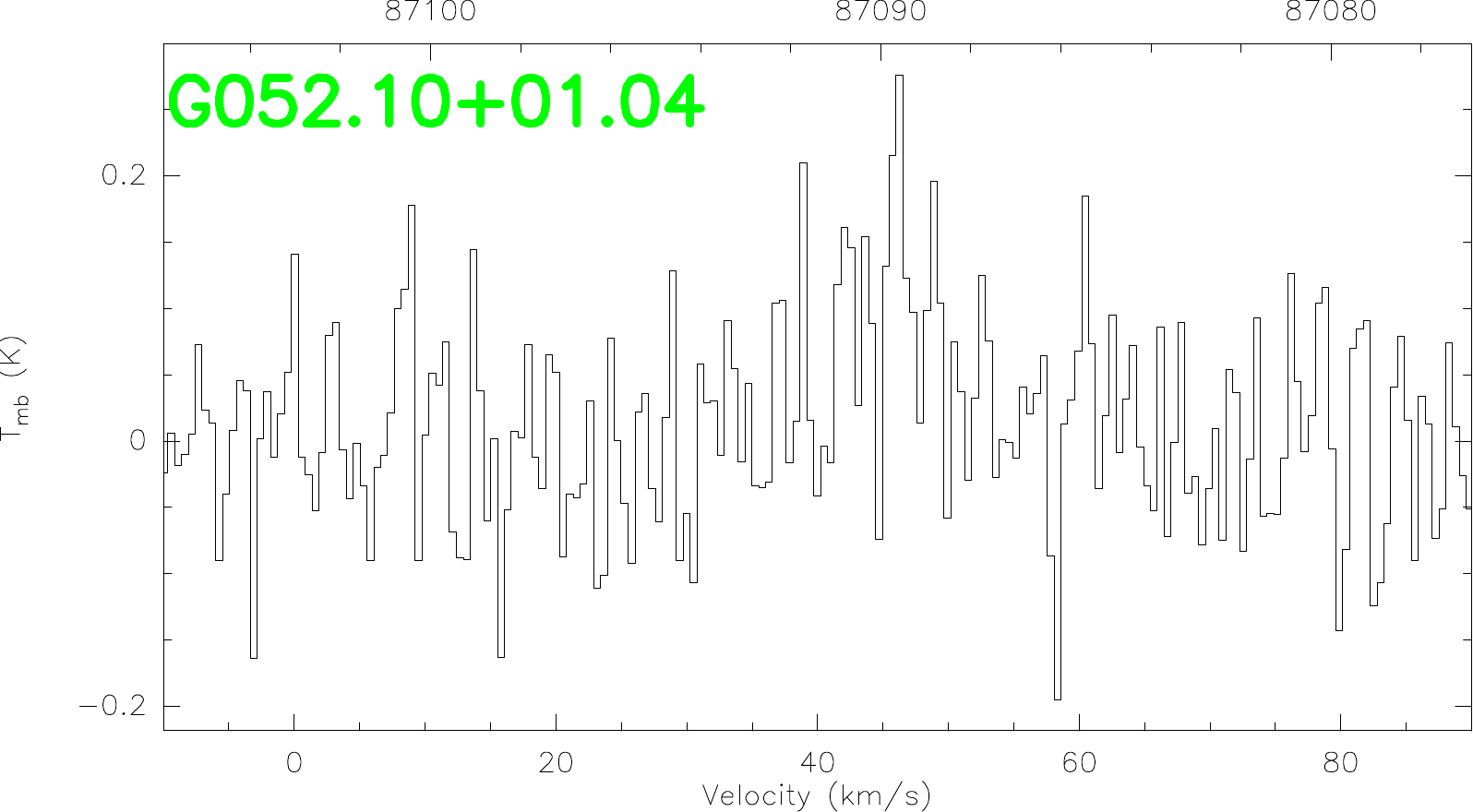}
\includegraphics[width=0.155\textwidth]{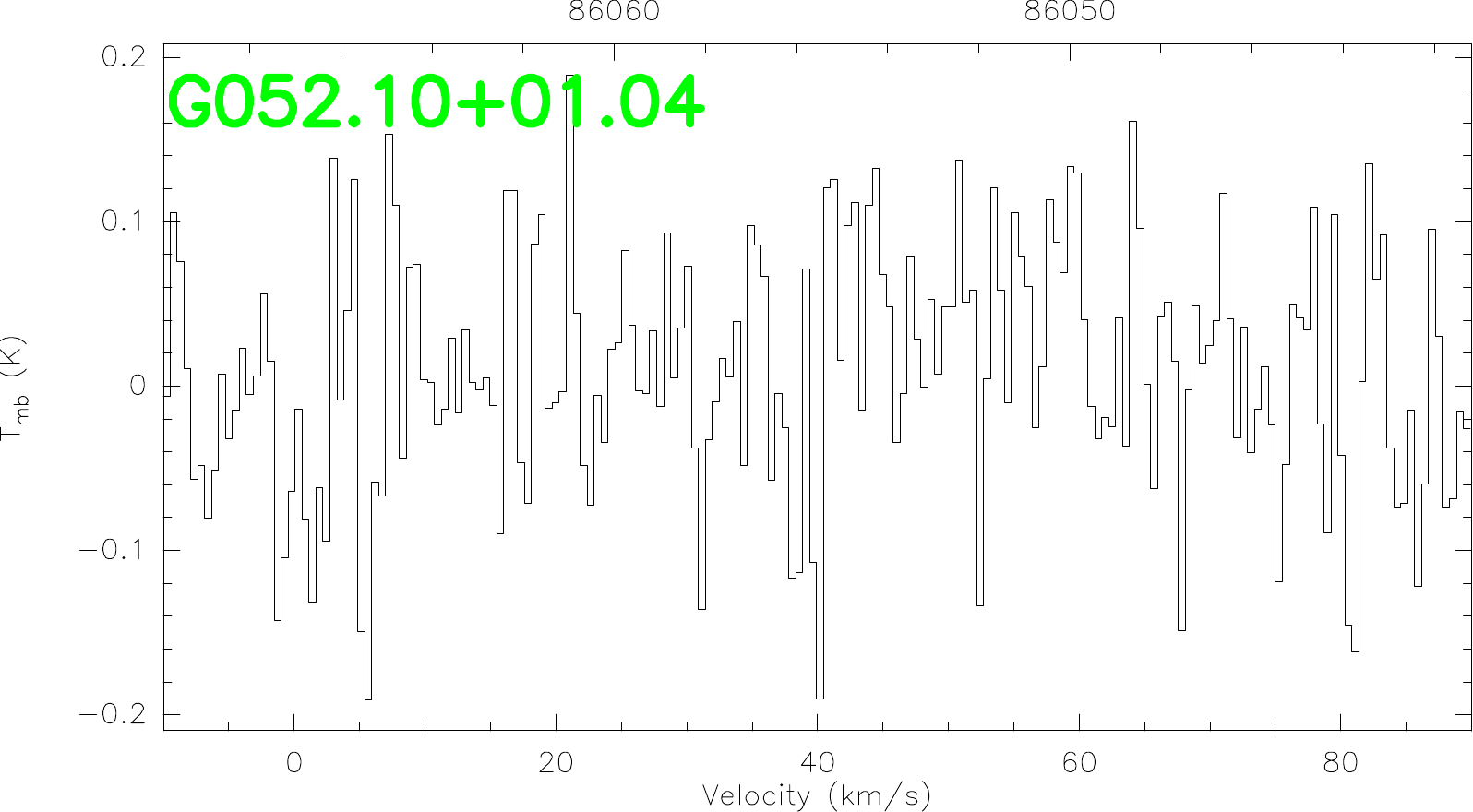}
\includegraphics[width=0.155\textwidth]{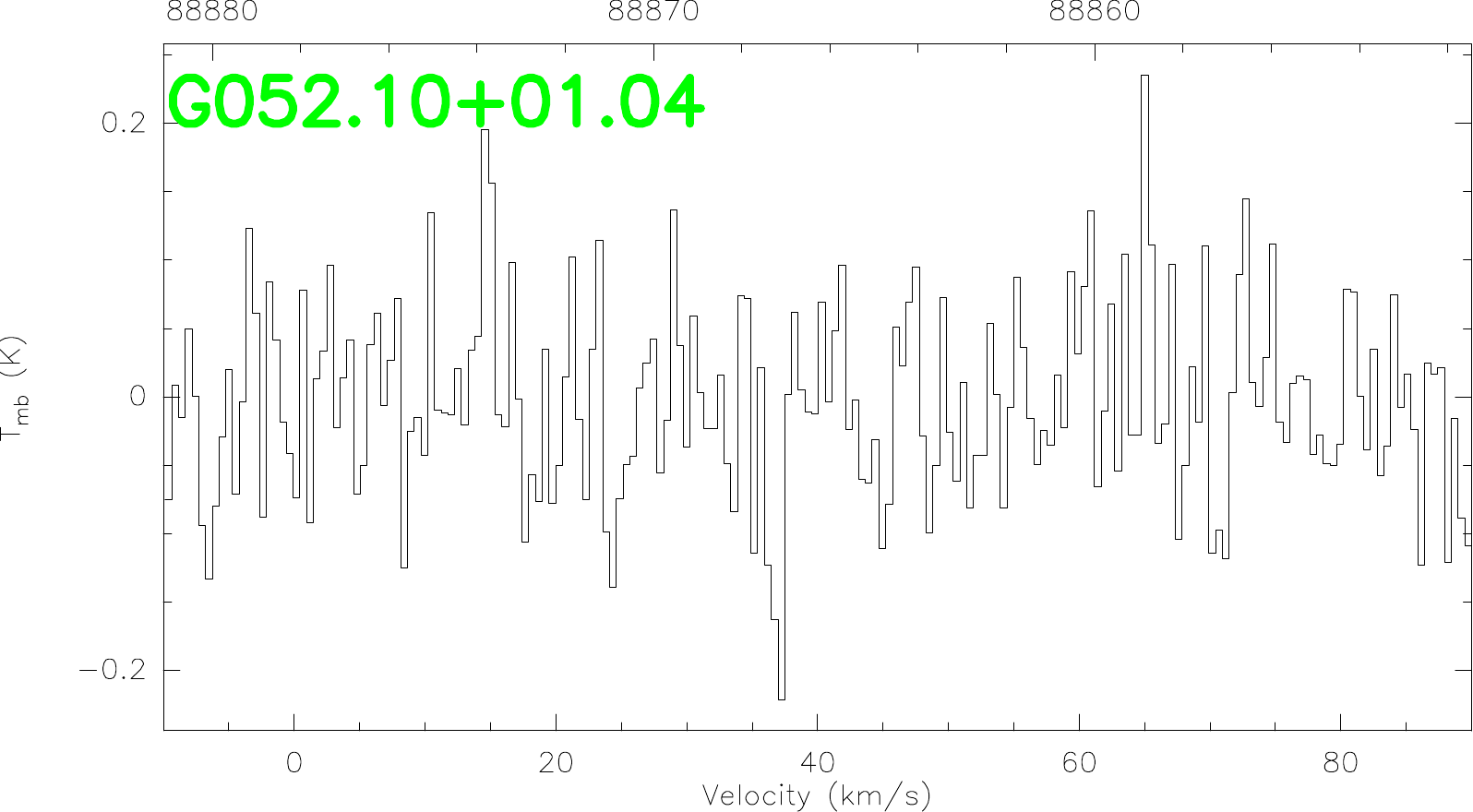}\\

\includegraphics[width=0.155\textwidth]{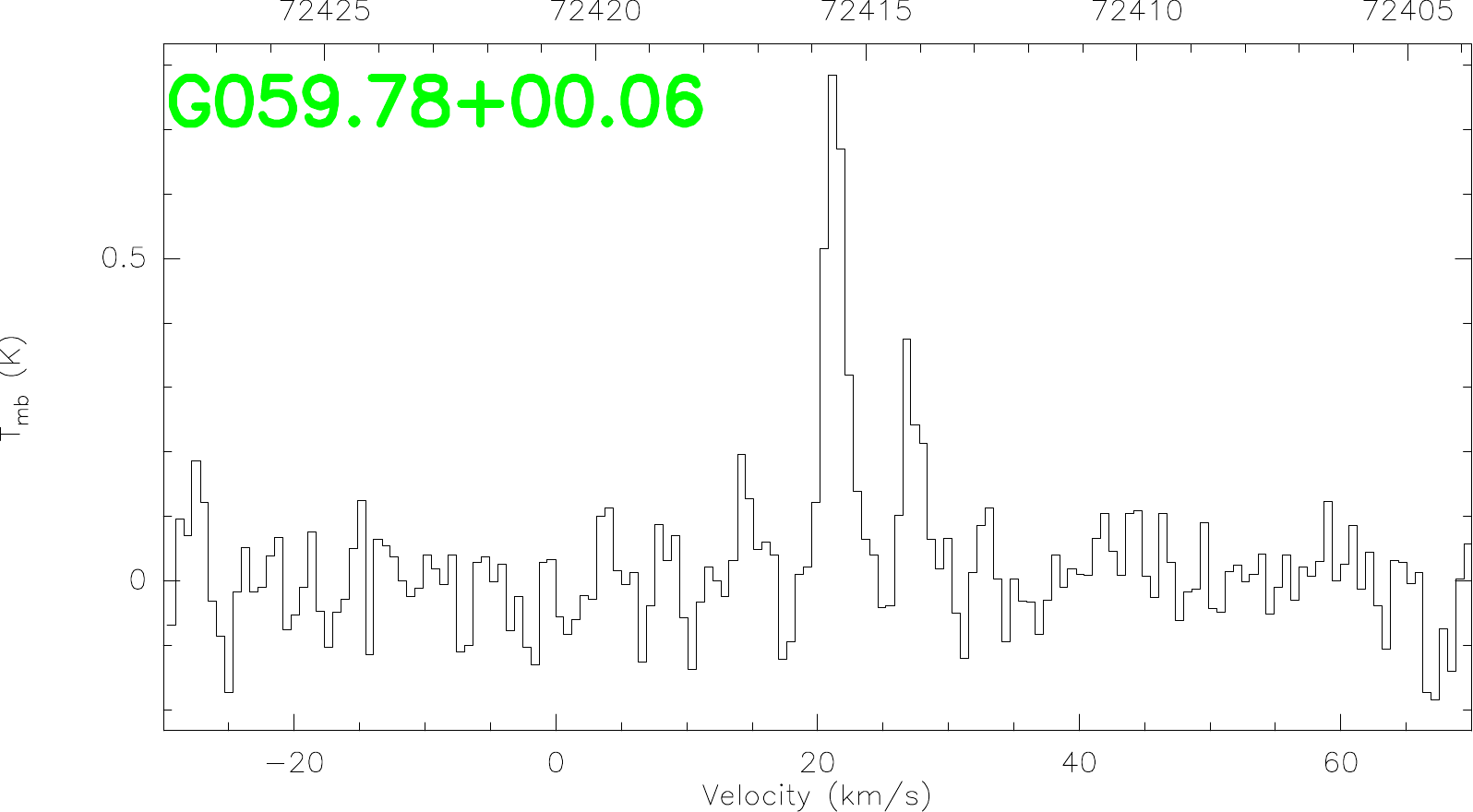}
\includegraphics[width=0.155\textwidth]{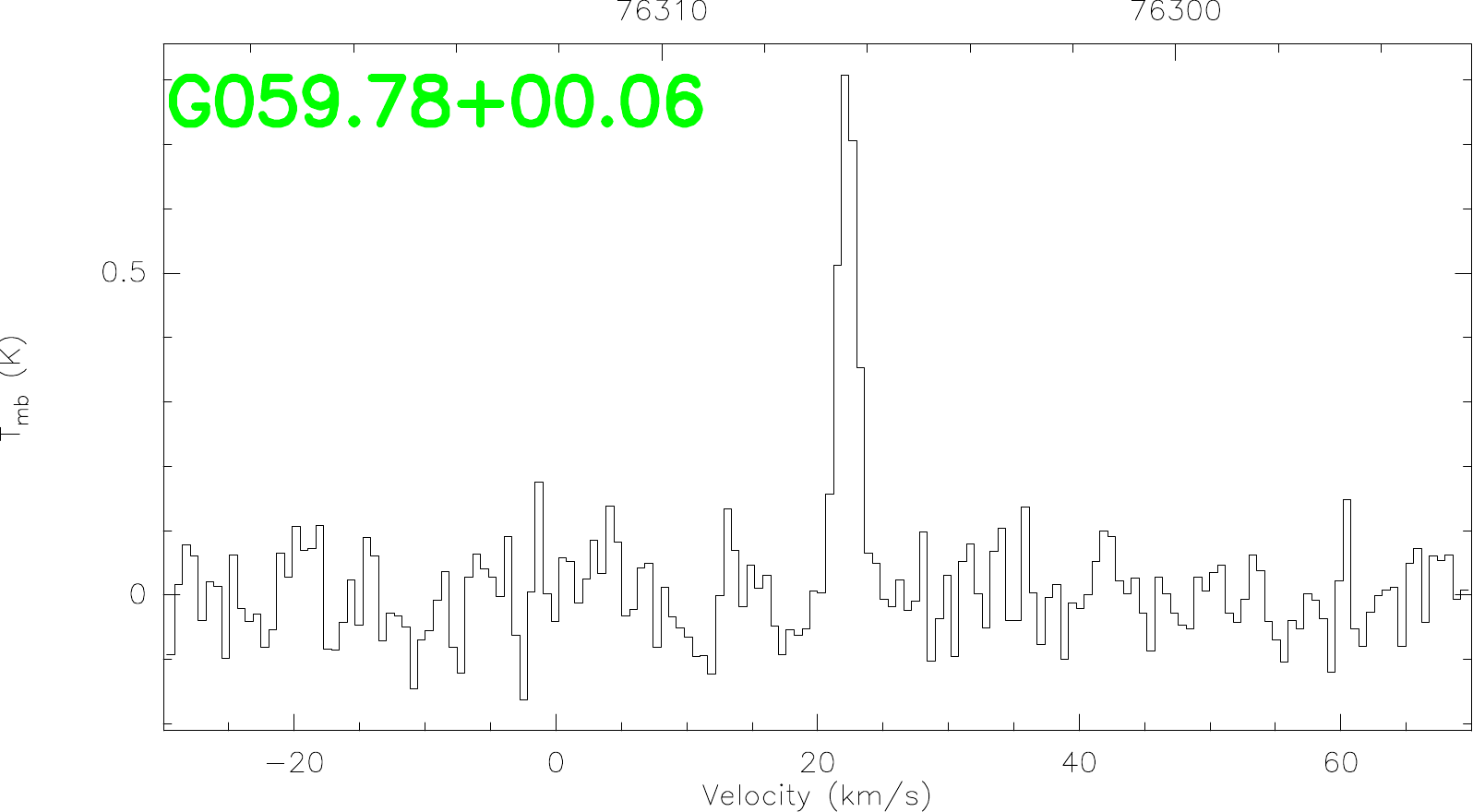}
\includegraphics[width=0.155\textwidth]{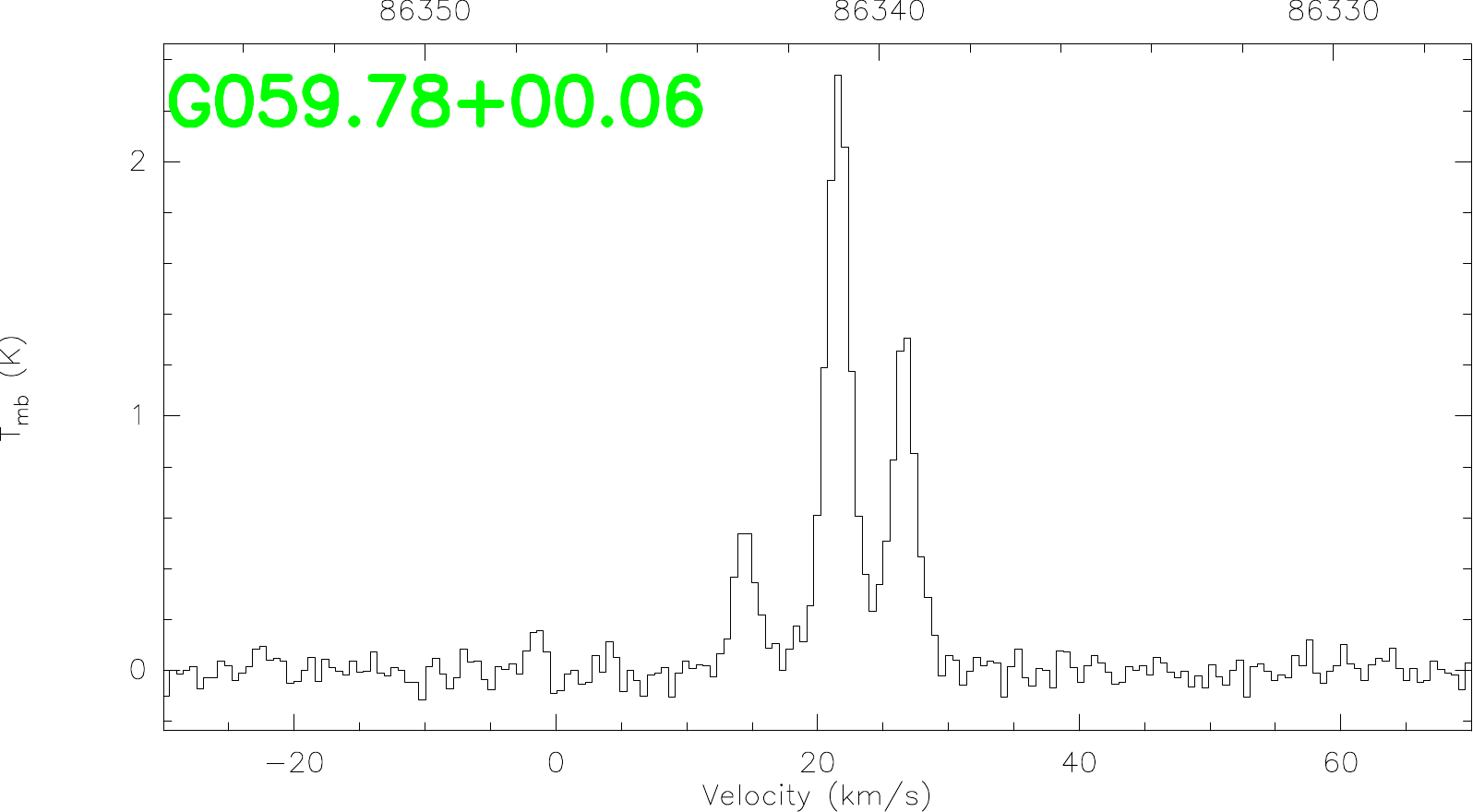}
\includegraphics[width=0.155\textwidth]{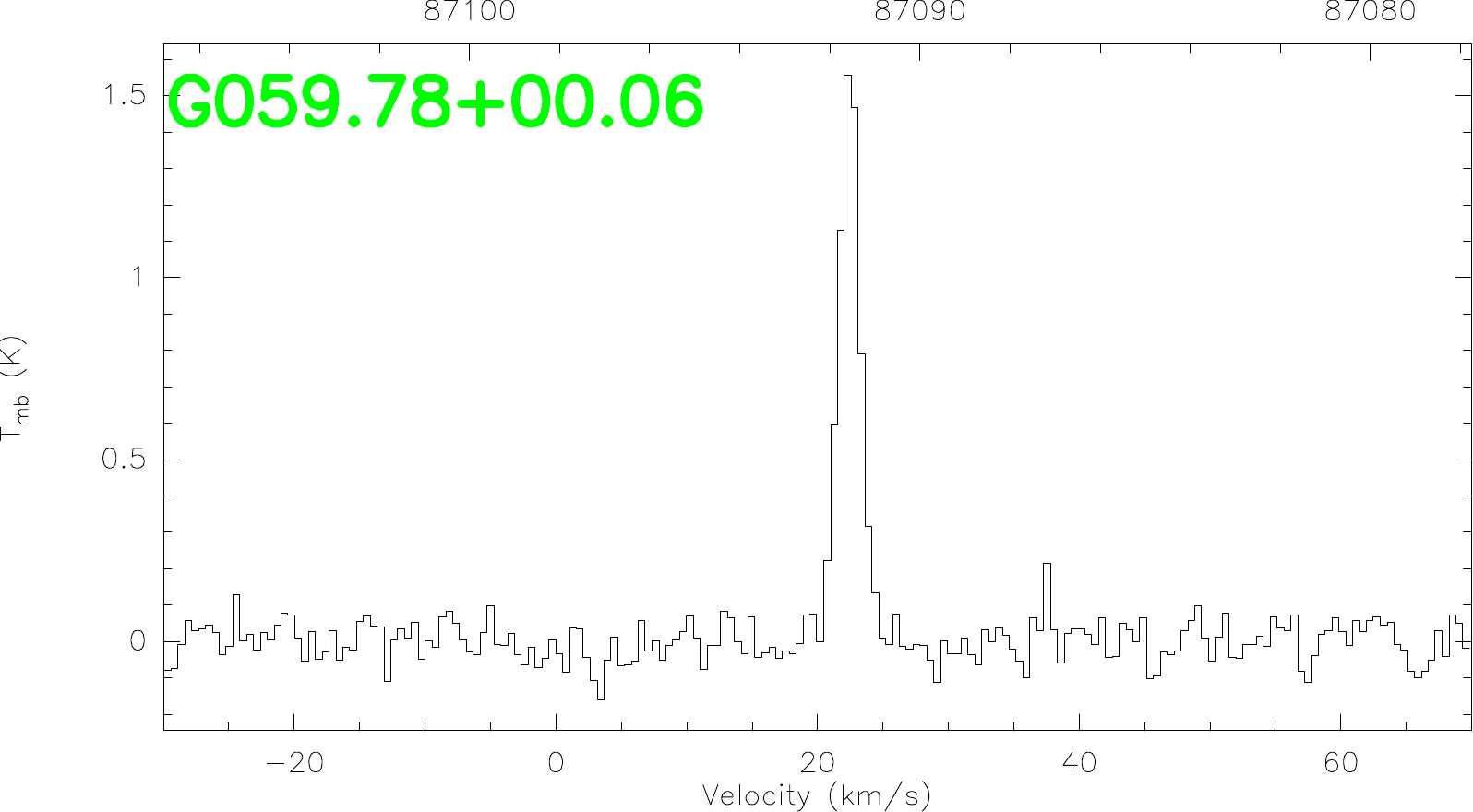}
\includegraphics[width=0.155\textwidth]{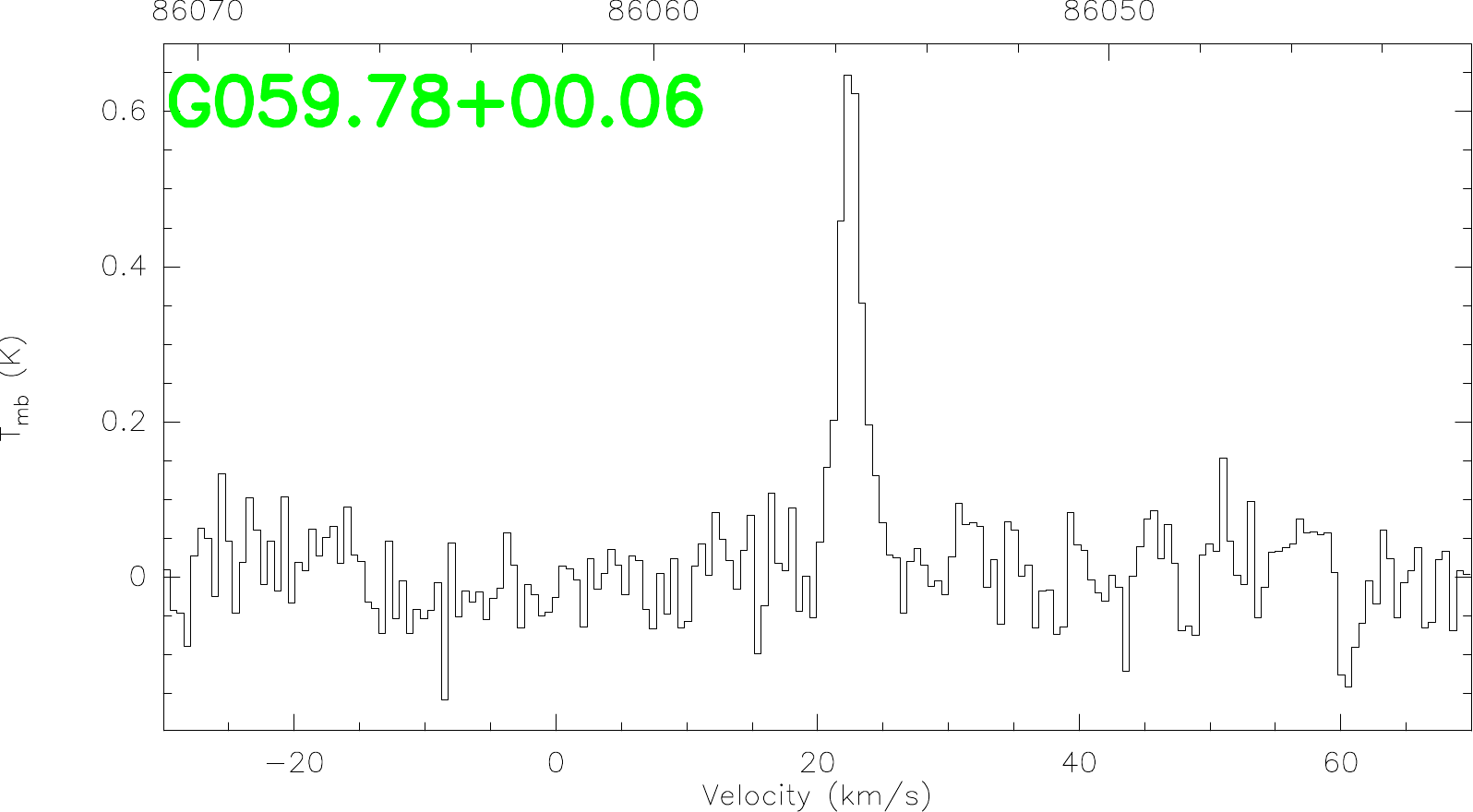}
\includegraphics[width=0.155\textwidth]{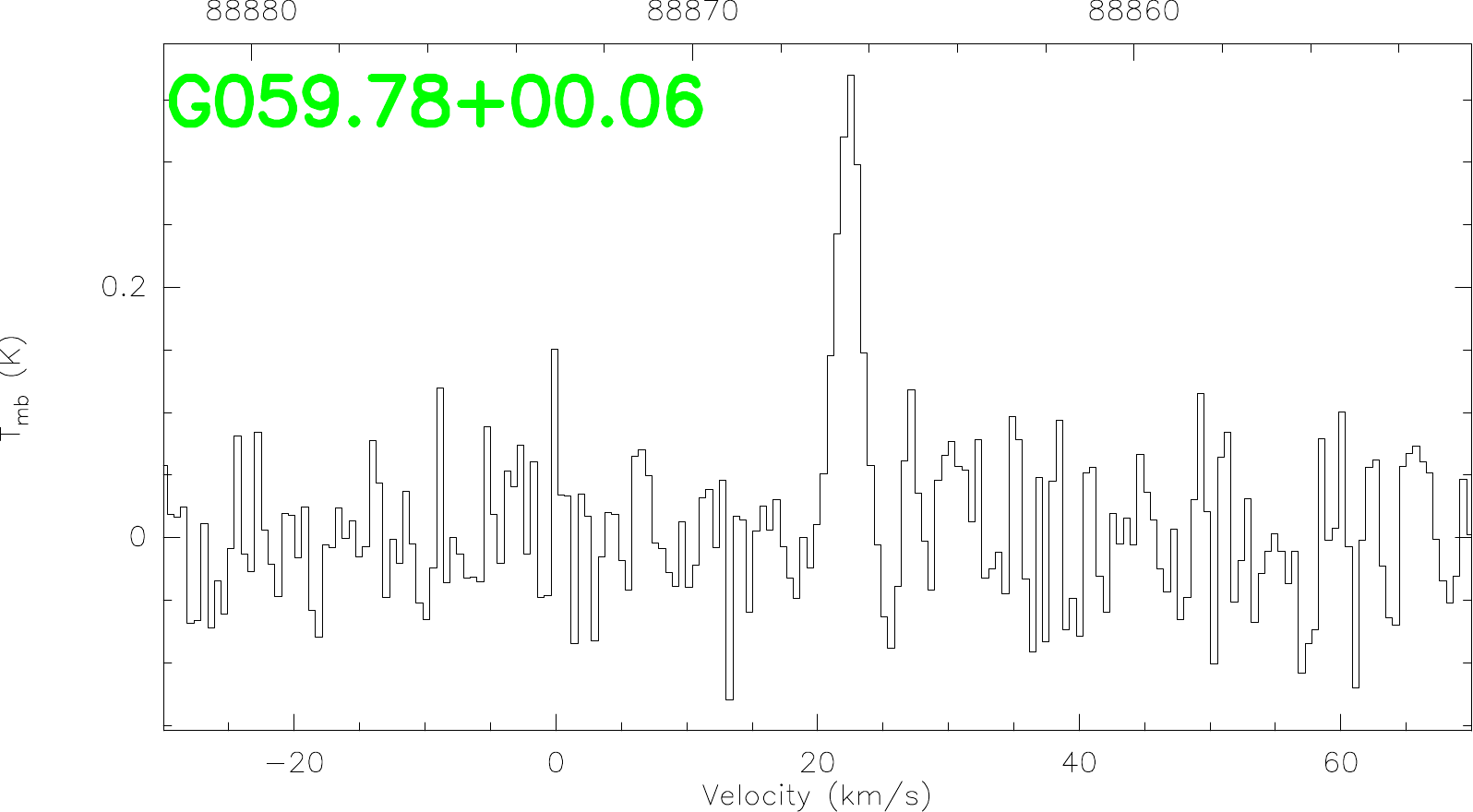}\\

\includegraphics[width=0.155\textwidth]{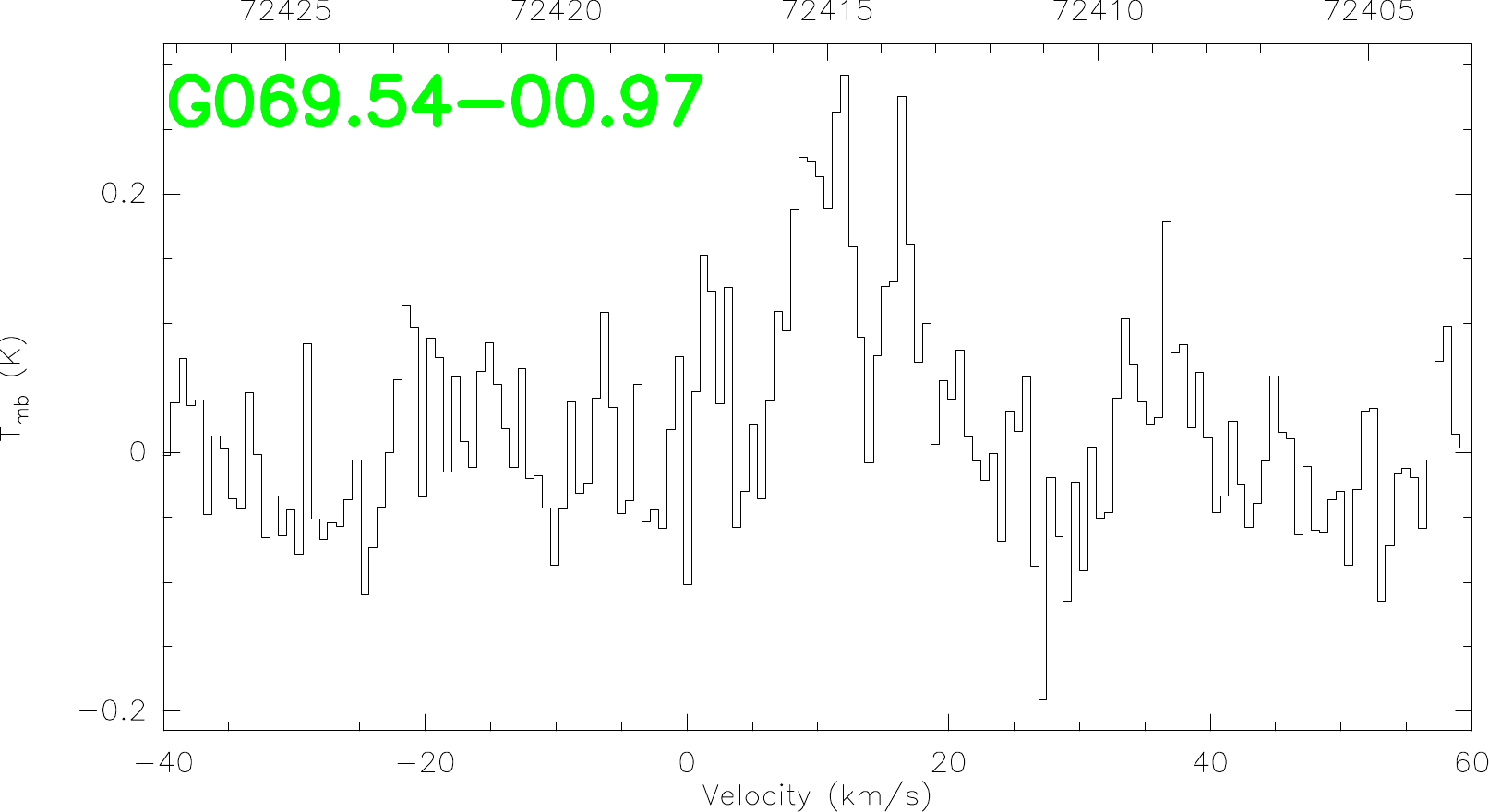}
\includegraphics[width=0.155\textwidth]{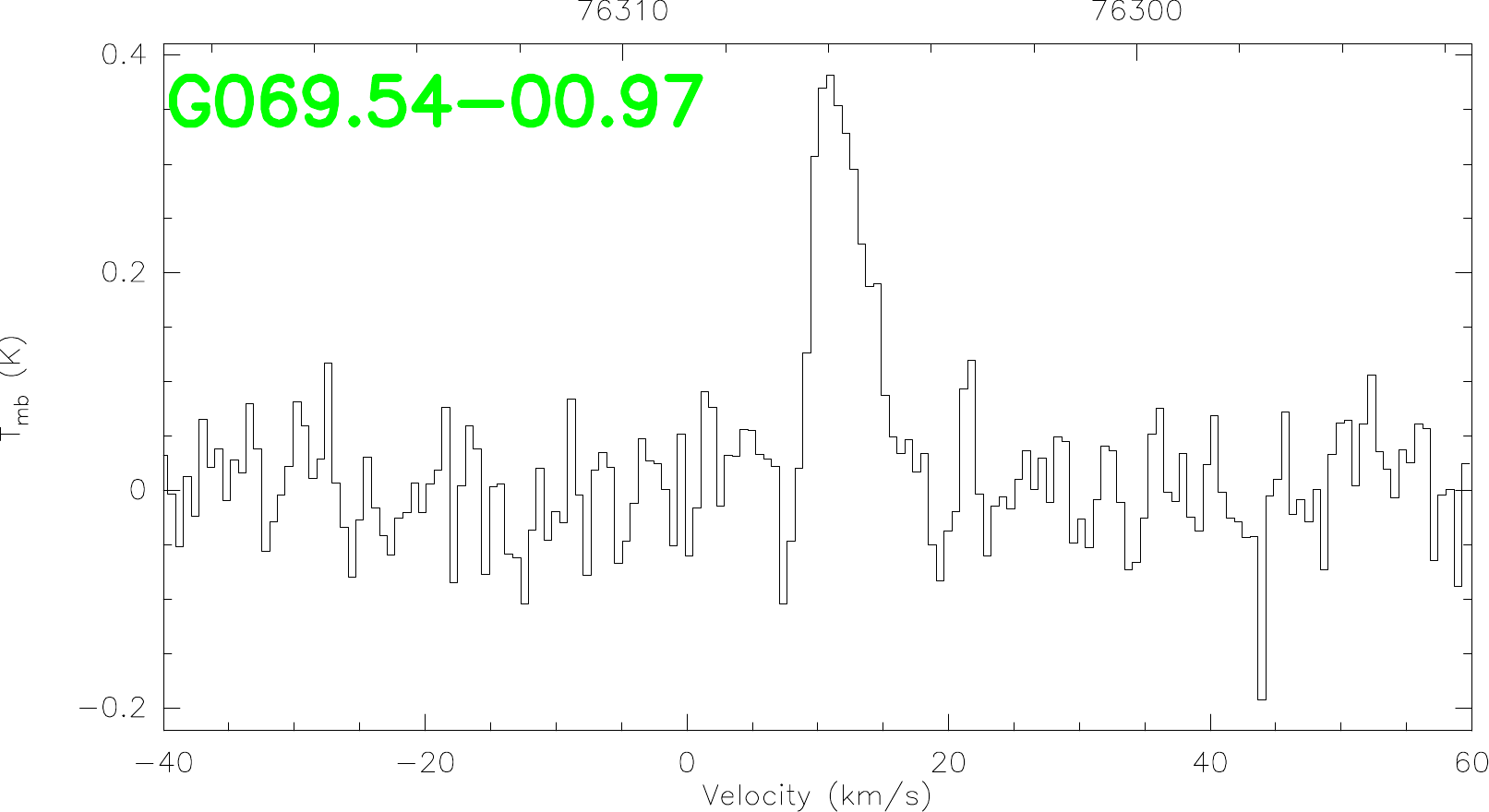}
\includegraphics[width=0.155\textwidth]{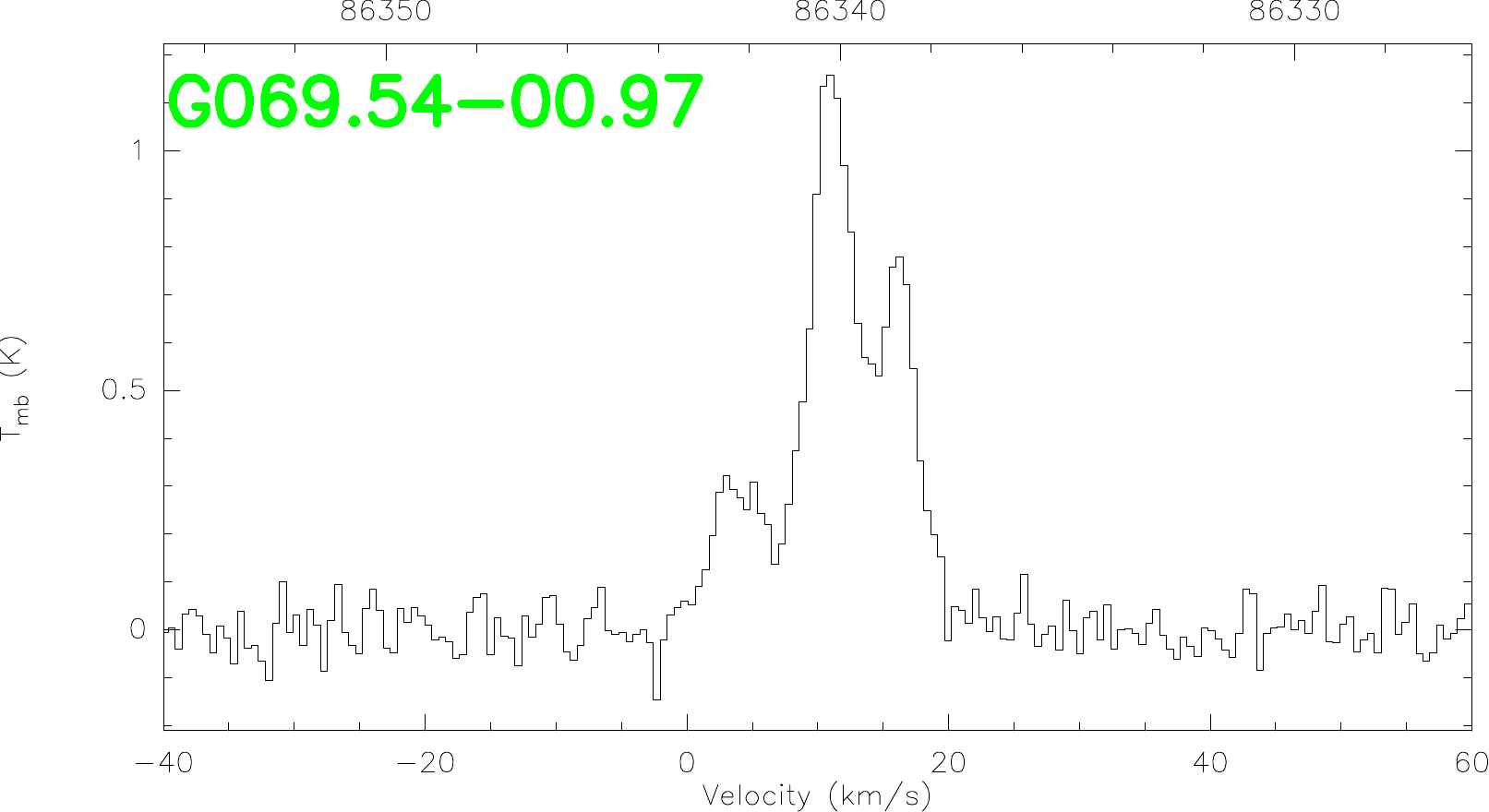}
\includegraphics[width=0.155\textwidth]{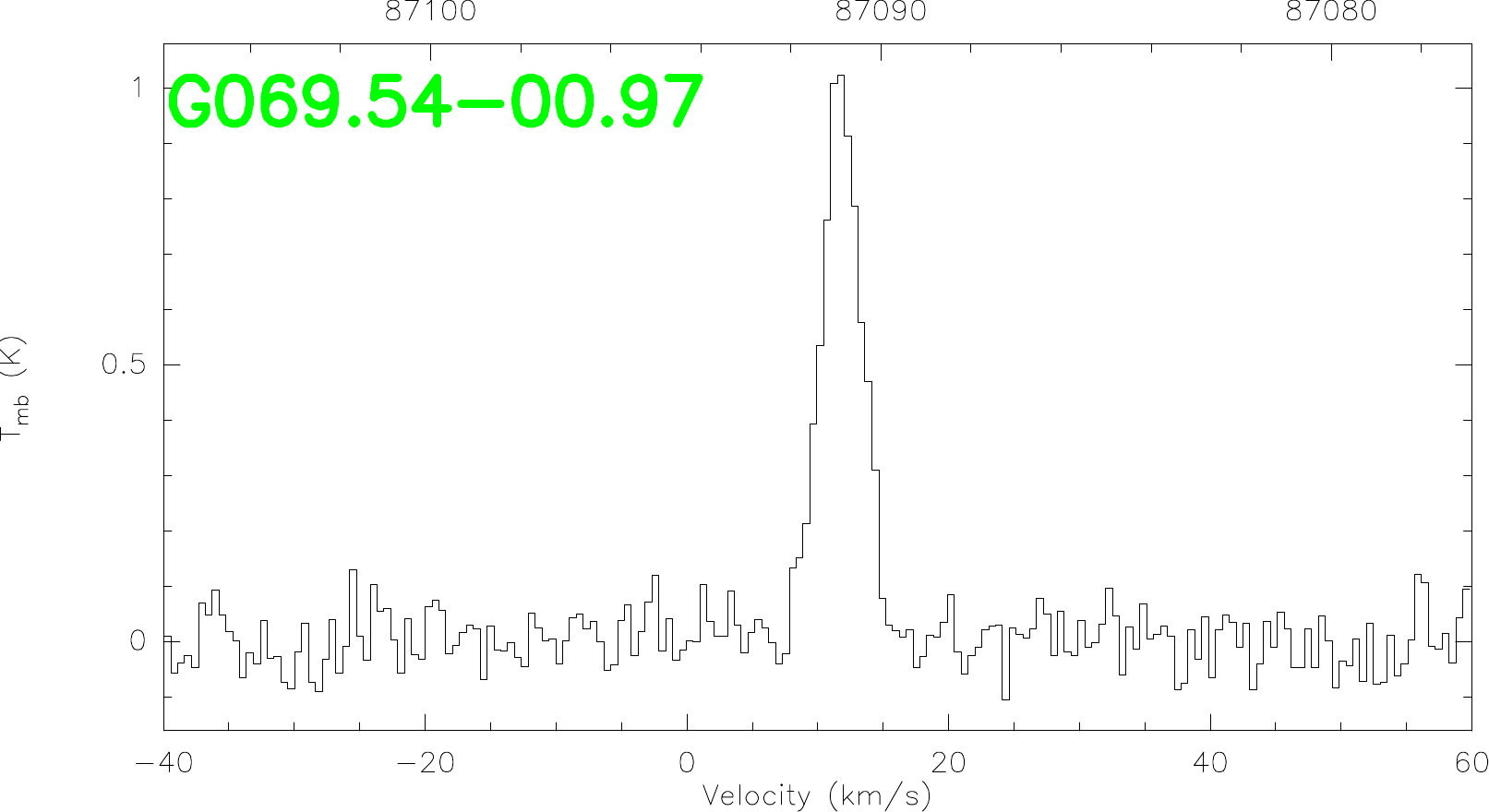}
\includegraphics[width=0.155\textwidth]{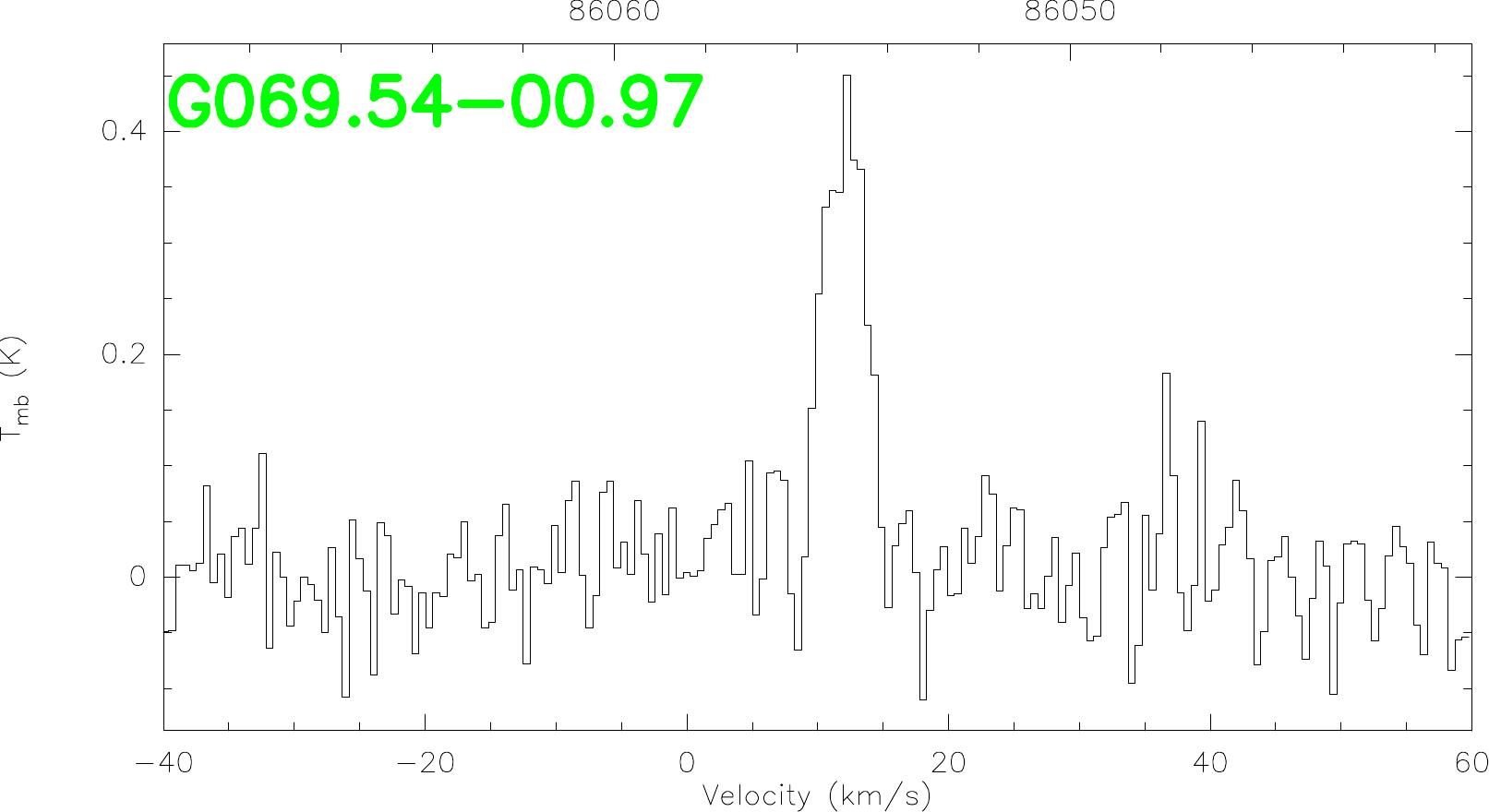}
\includegraphics[width=0.155\textwidth]{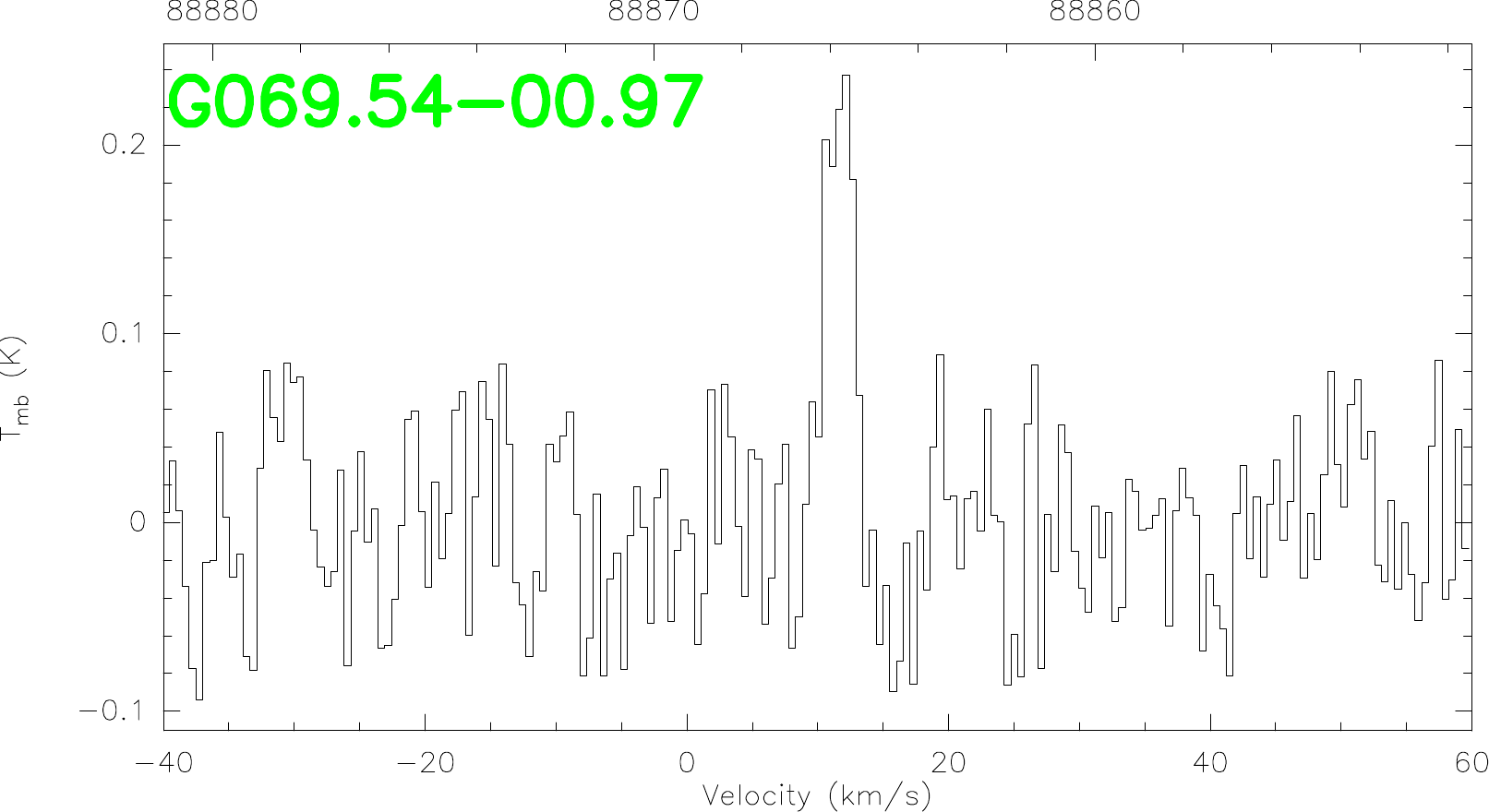}\\

\includegraphics[width=0.155\textwidth]{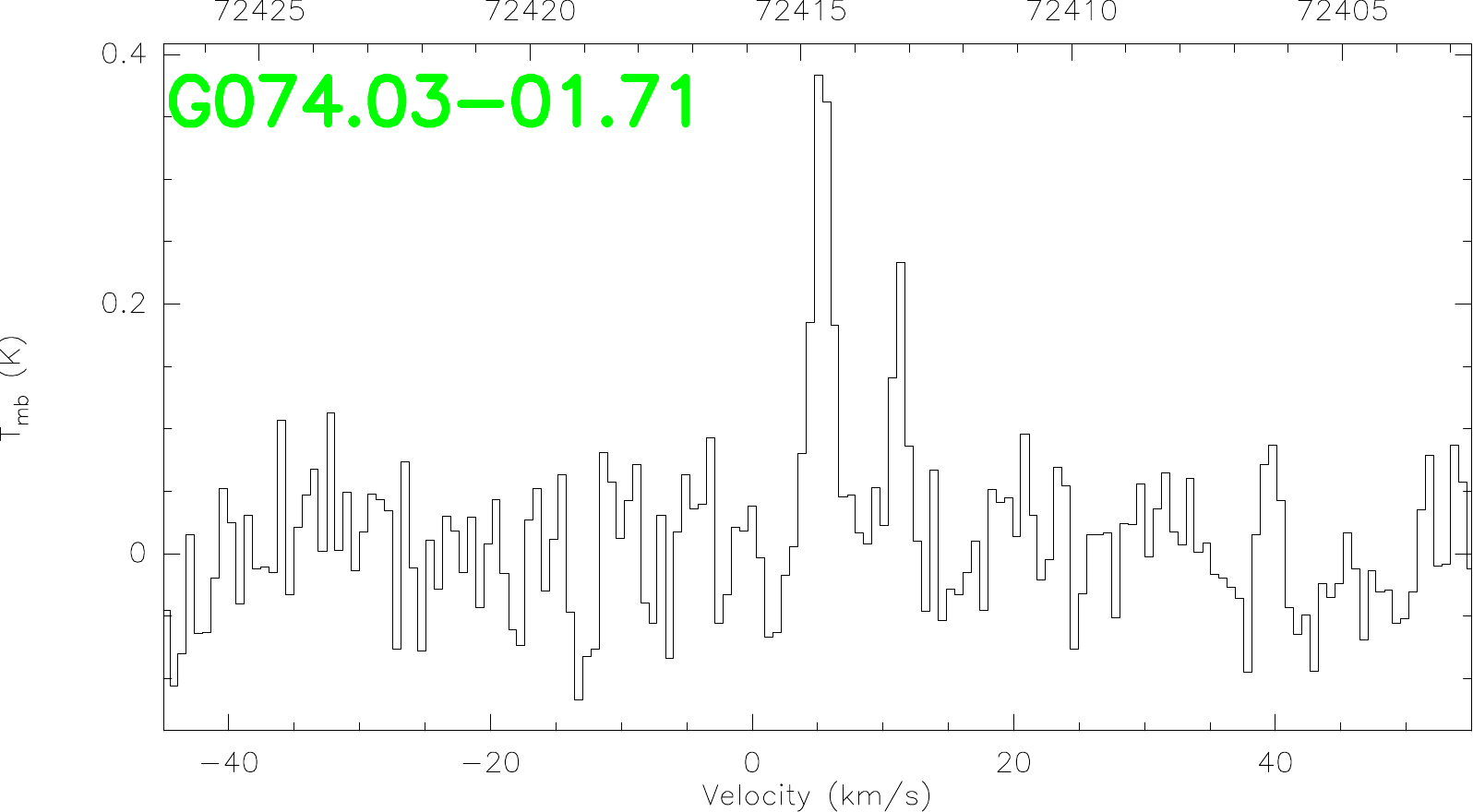}
\includegraphics[width=0.155\textwidth]{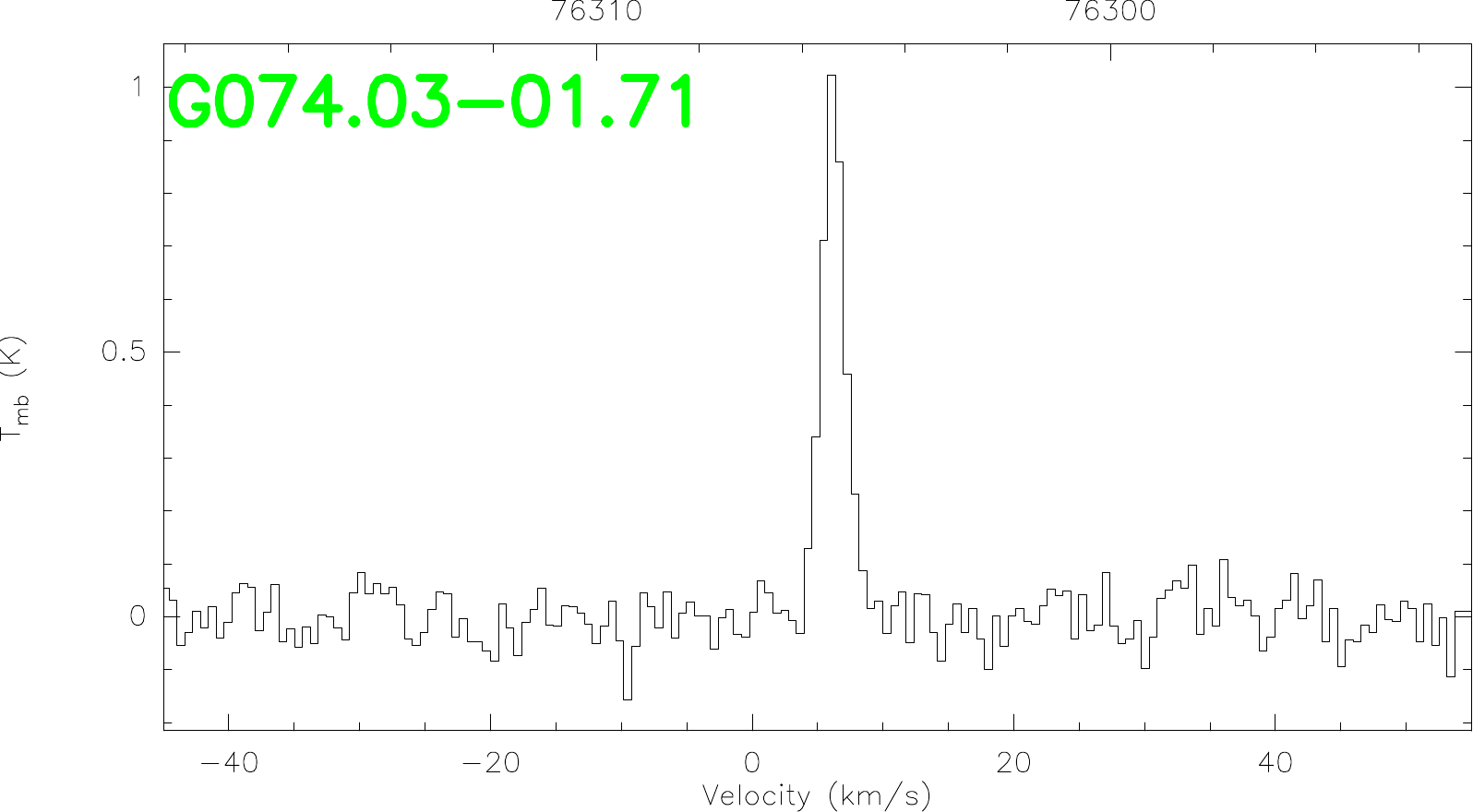}
\includegraphics[width=0.155\textwidth]{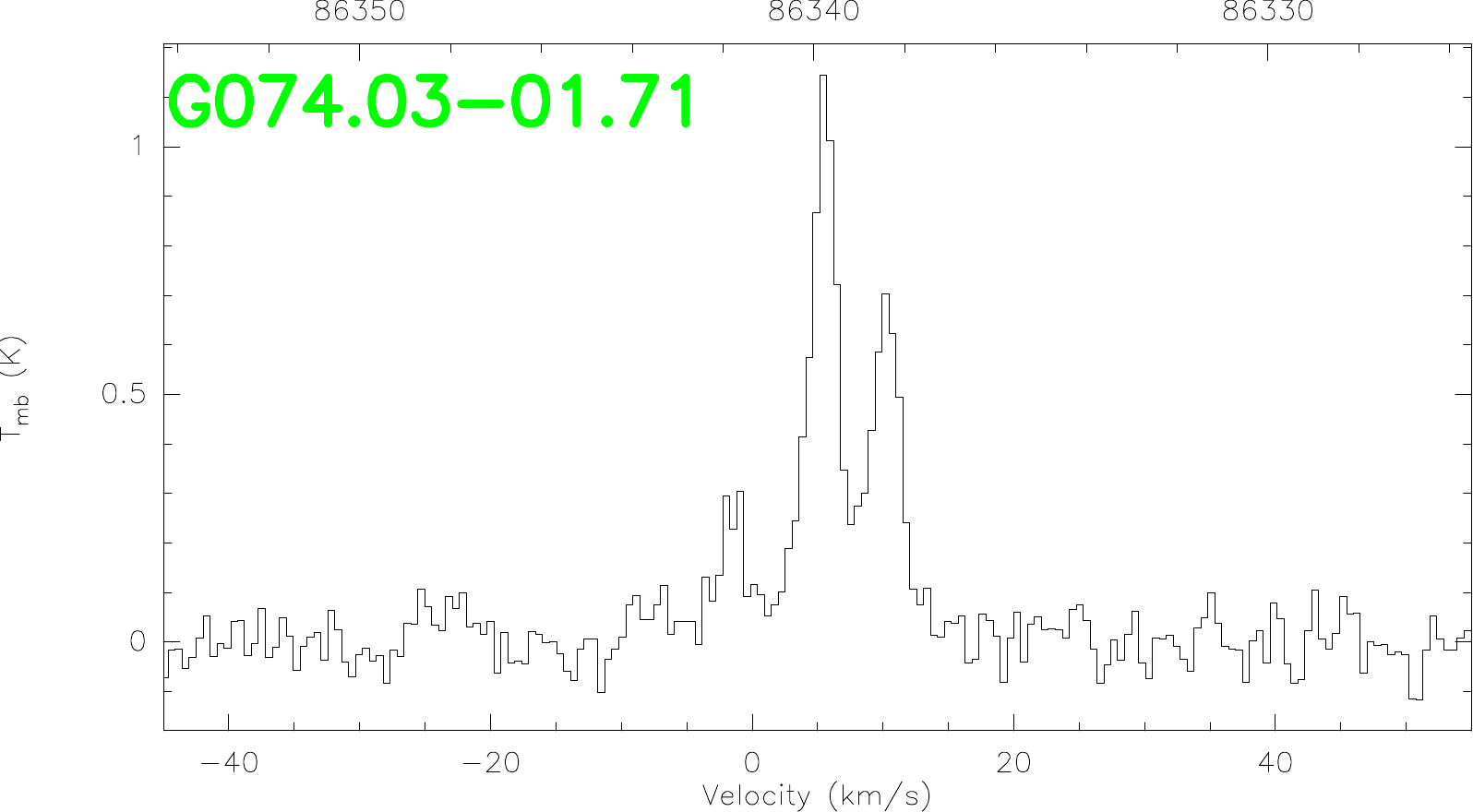}
\includegraphics[width=0.155\textwidth]{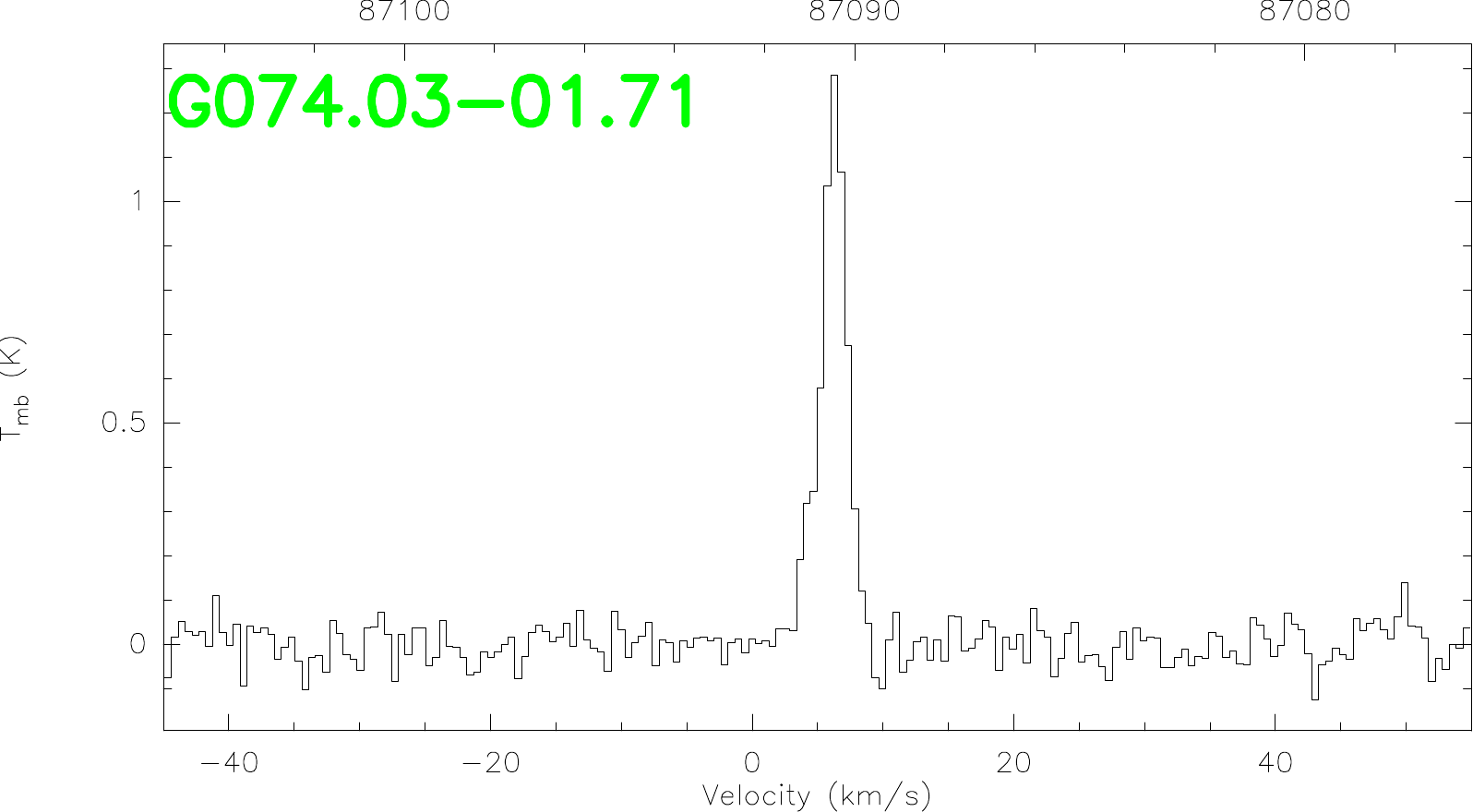}
\includegraphics[width=0.155\textwidth]{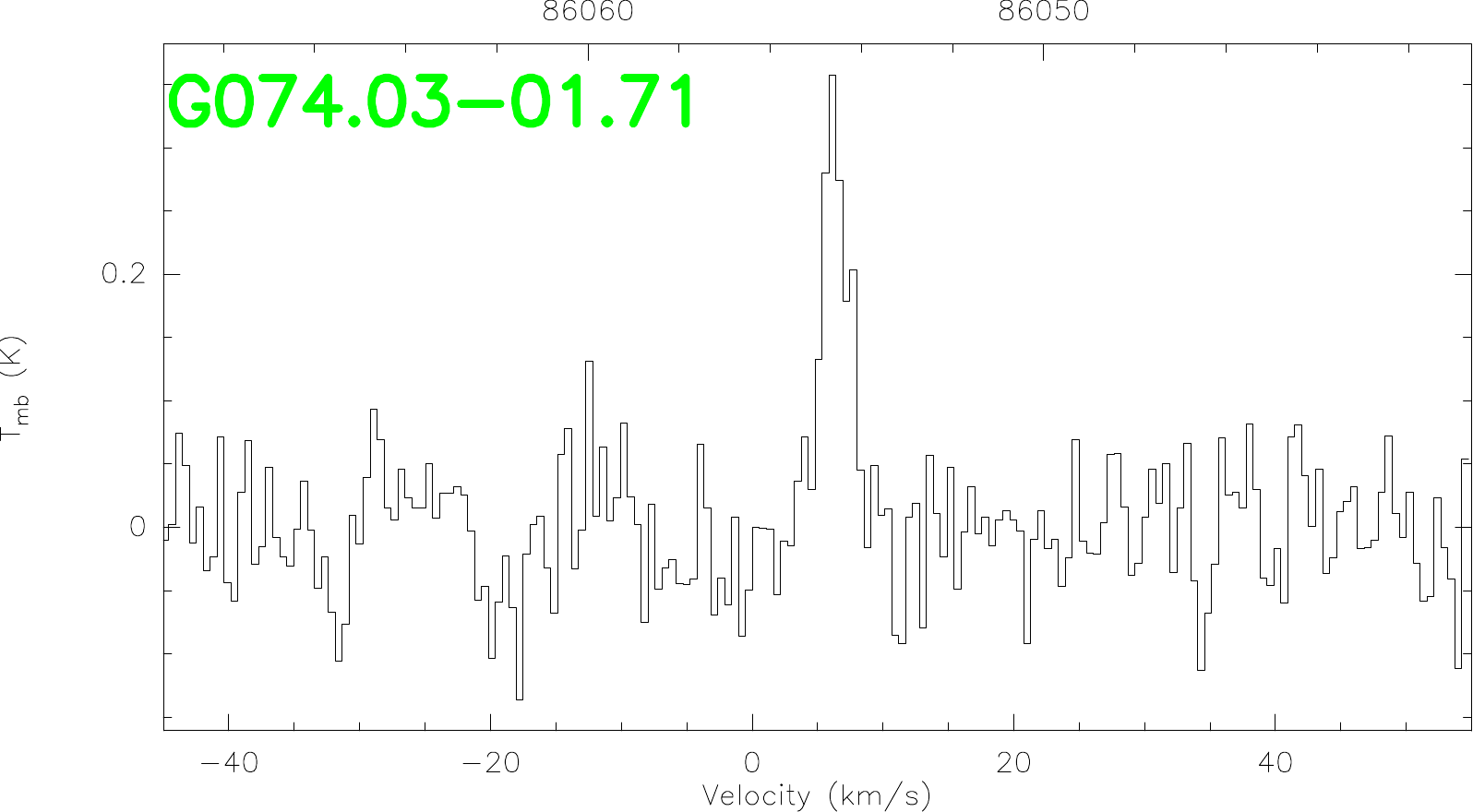}
\includegraphics[width=0.155\textwidth]{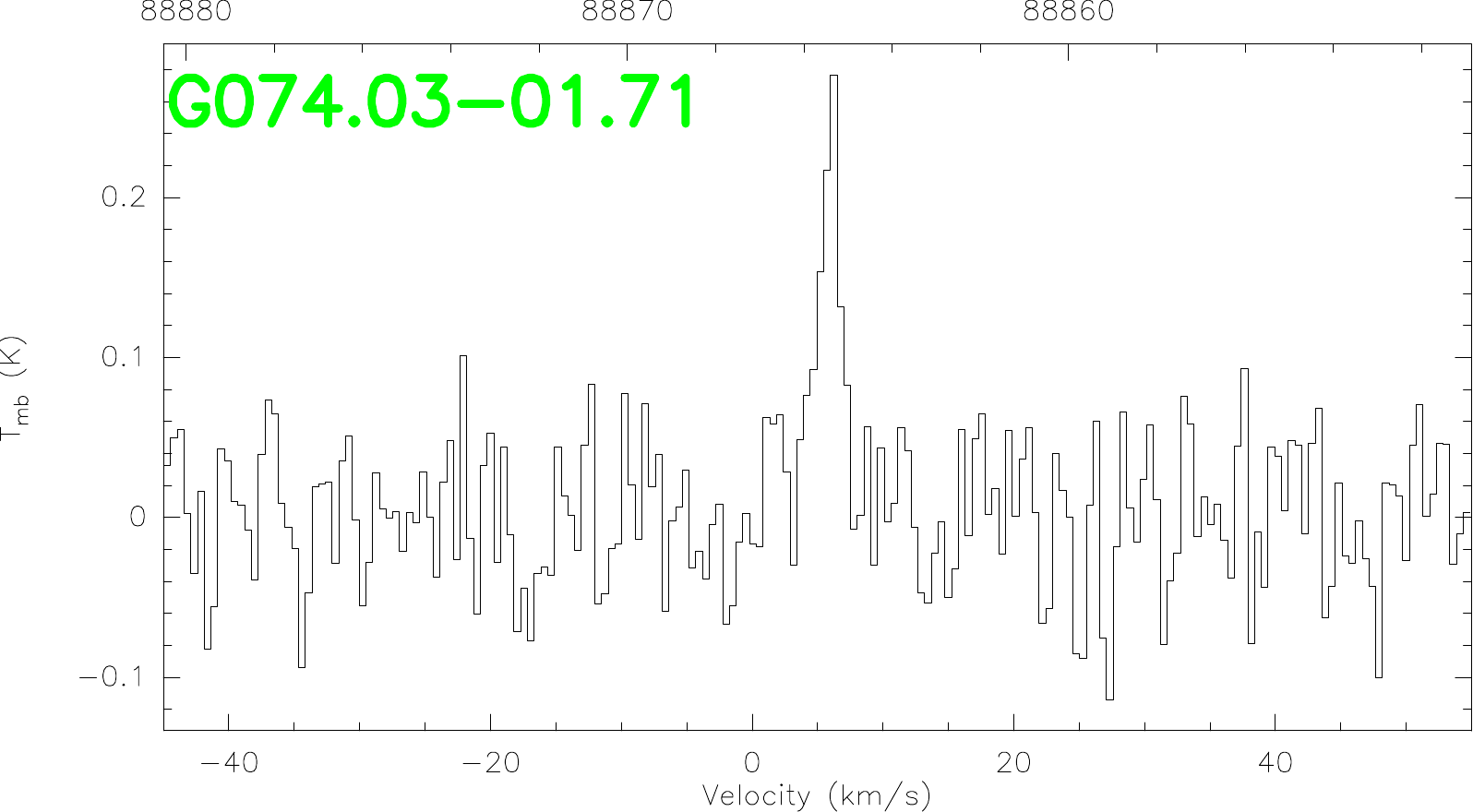}\\

\includegraphics[width=0.155\textwidth]{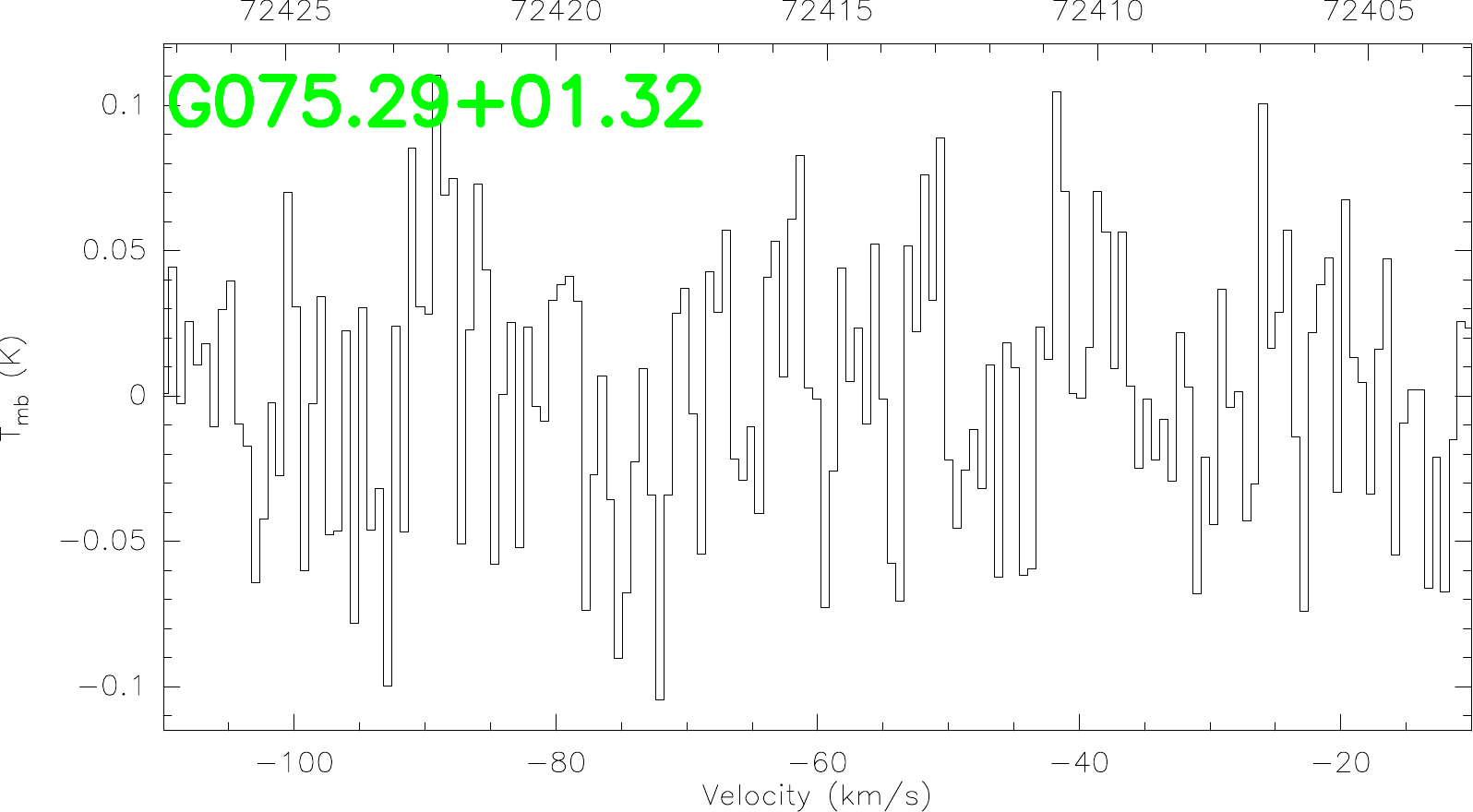}
\includegraphics[width=0.155\textwidth]{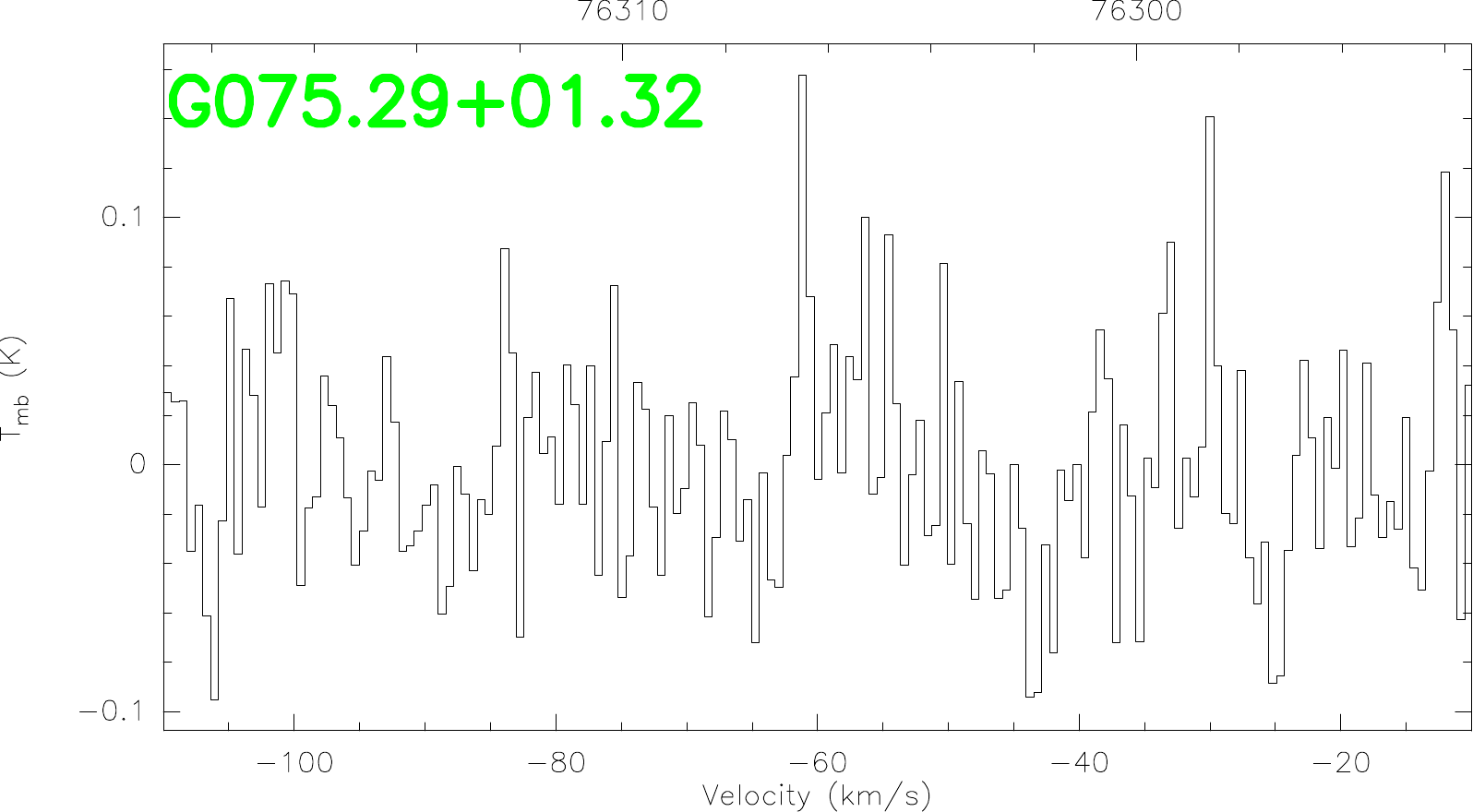}
\includegraphics[width=0.155\textwidth]{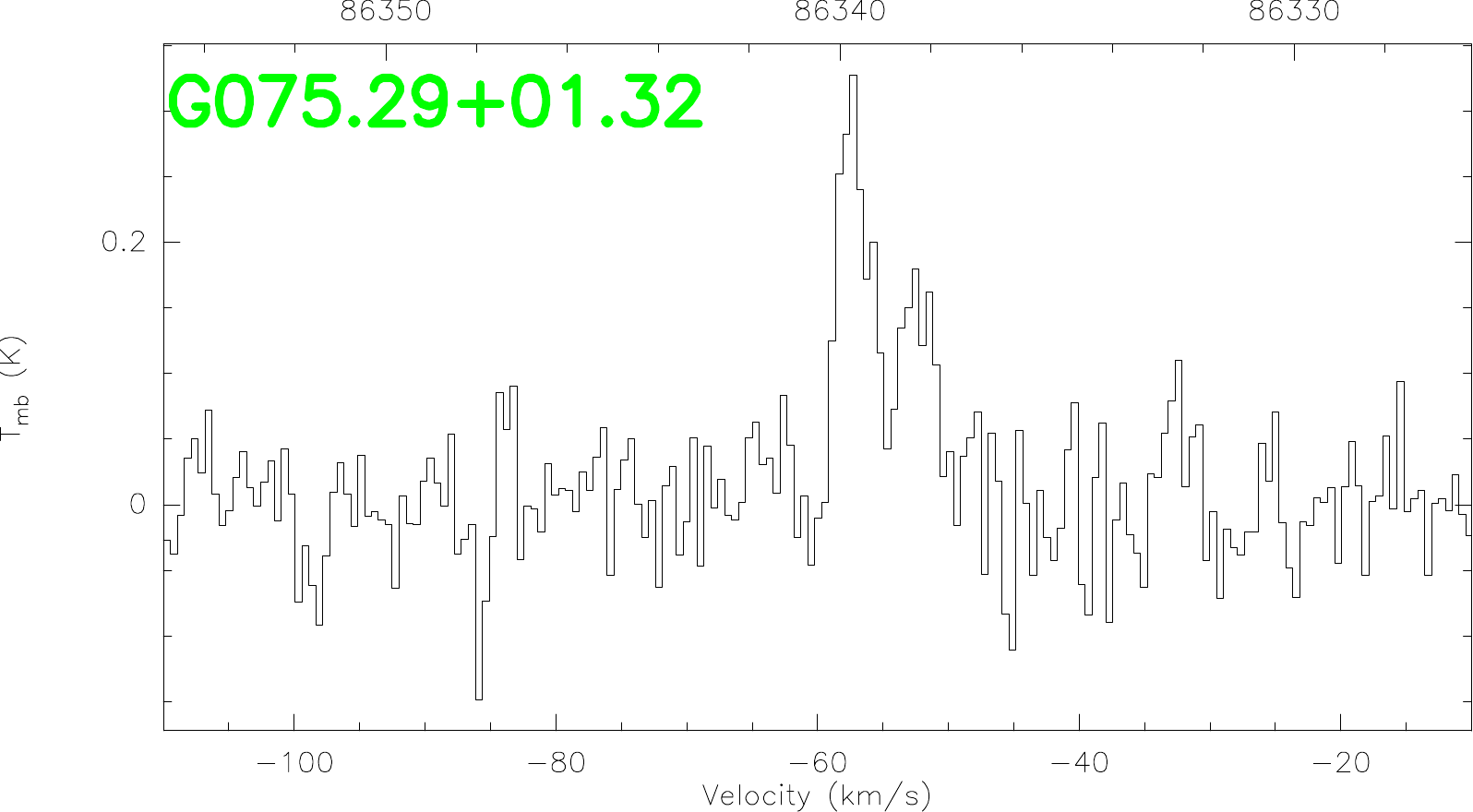}
\includegraphics[width=0.155\textwidth]{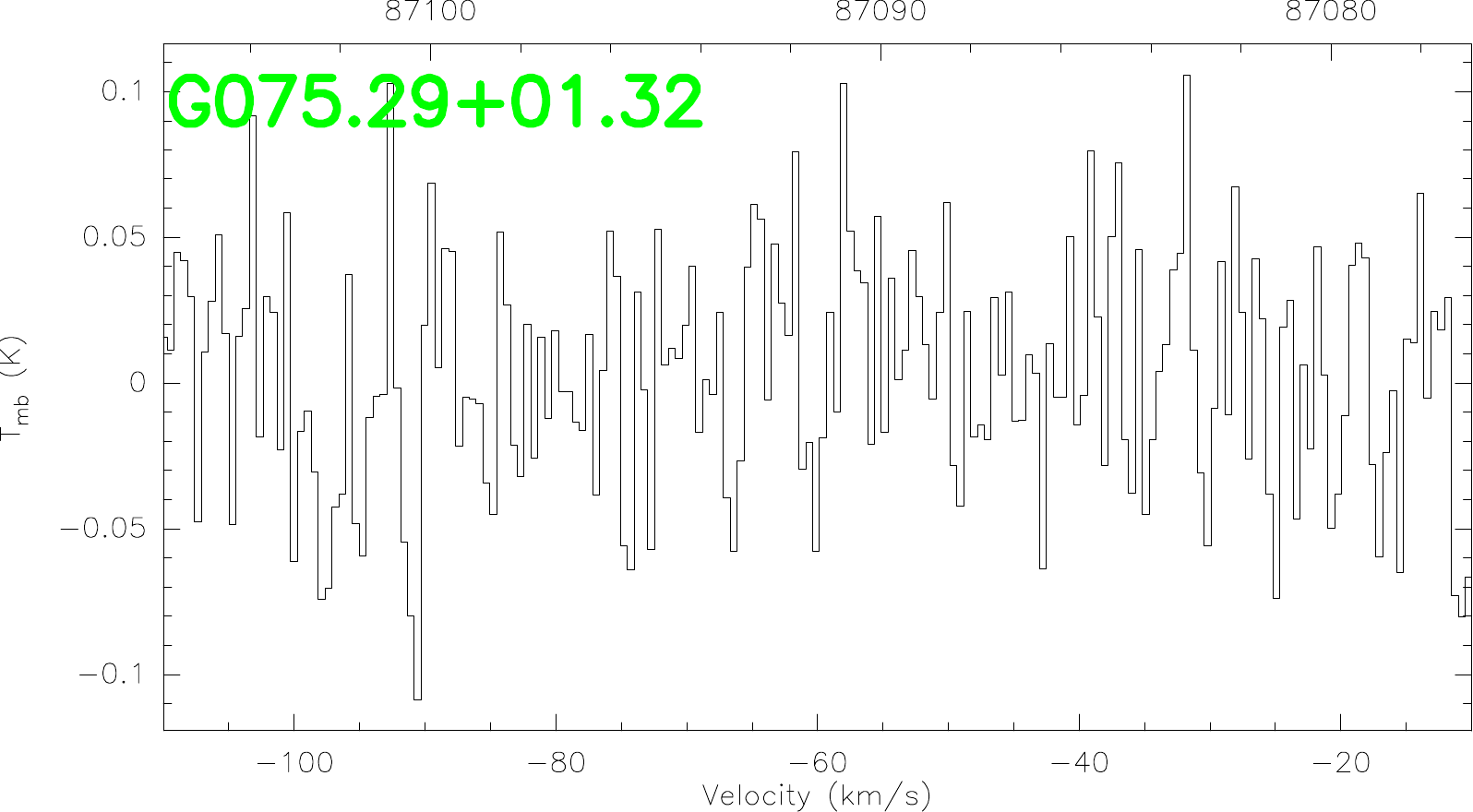}
\includegraphics[width=0.155\textwidth]{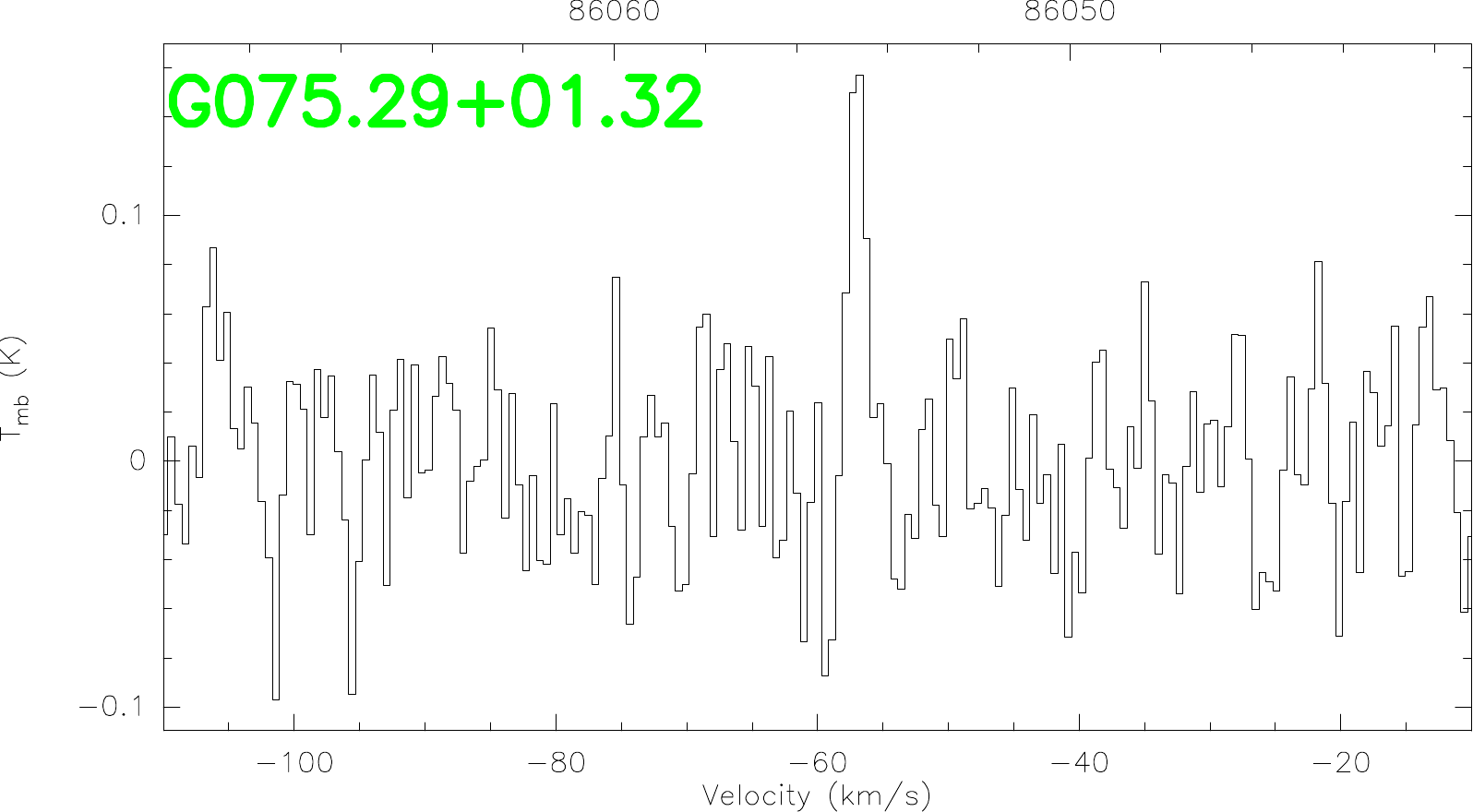}
\includegraphics[width=0.155\textwidth]{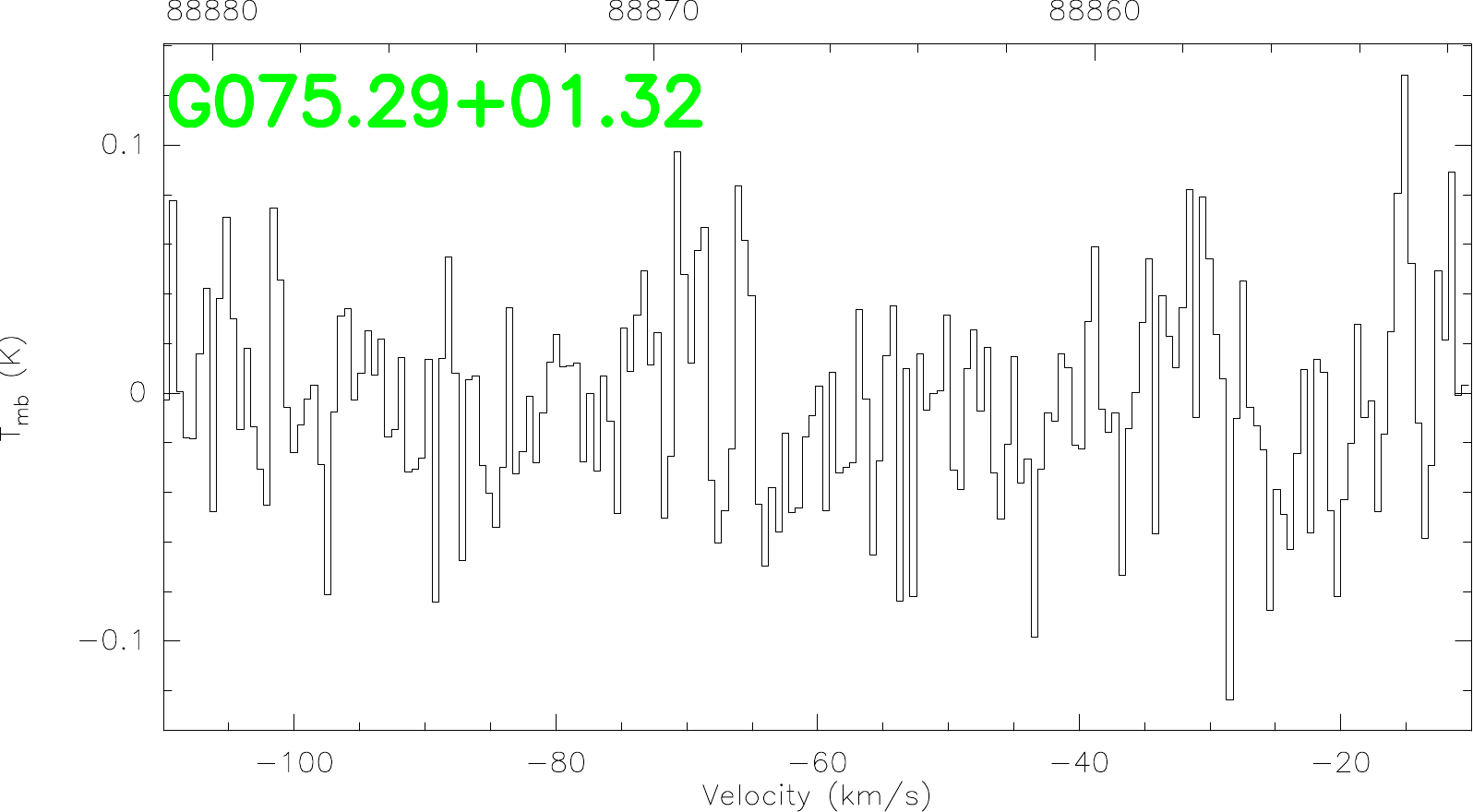}\\

\includegraphics[width=0.155\textwidth]{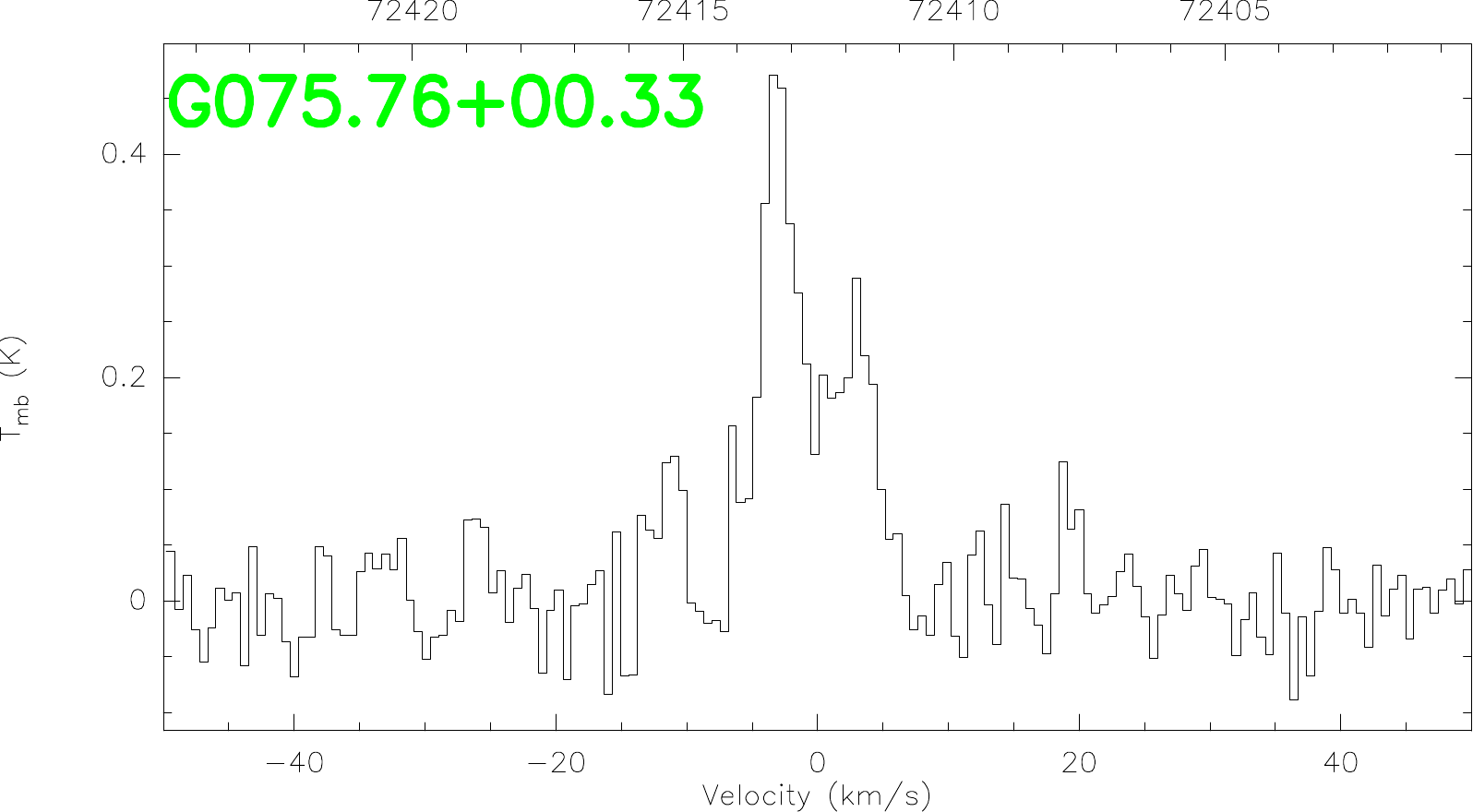}
\includegraphics[width=0.155\textwidth]{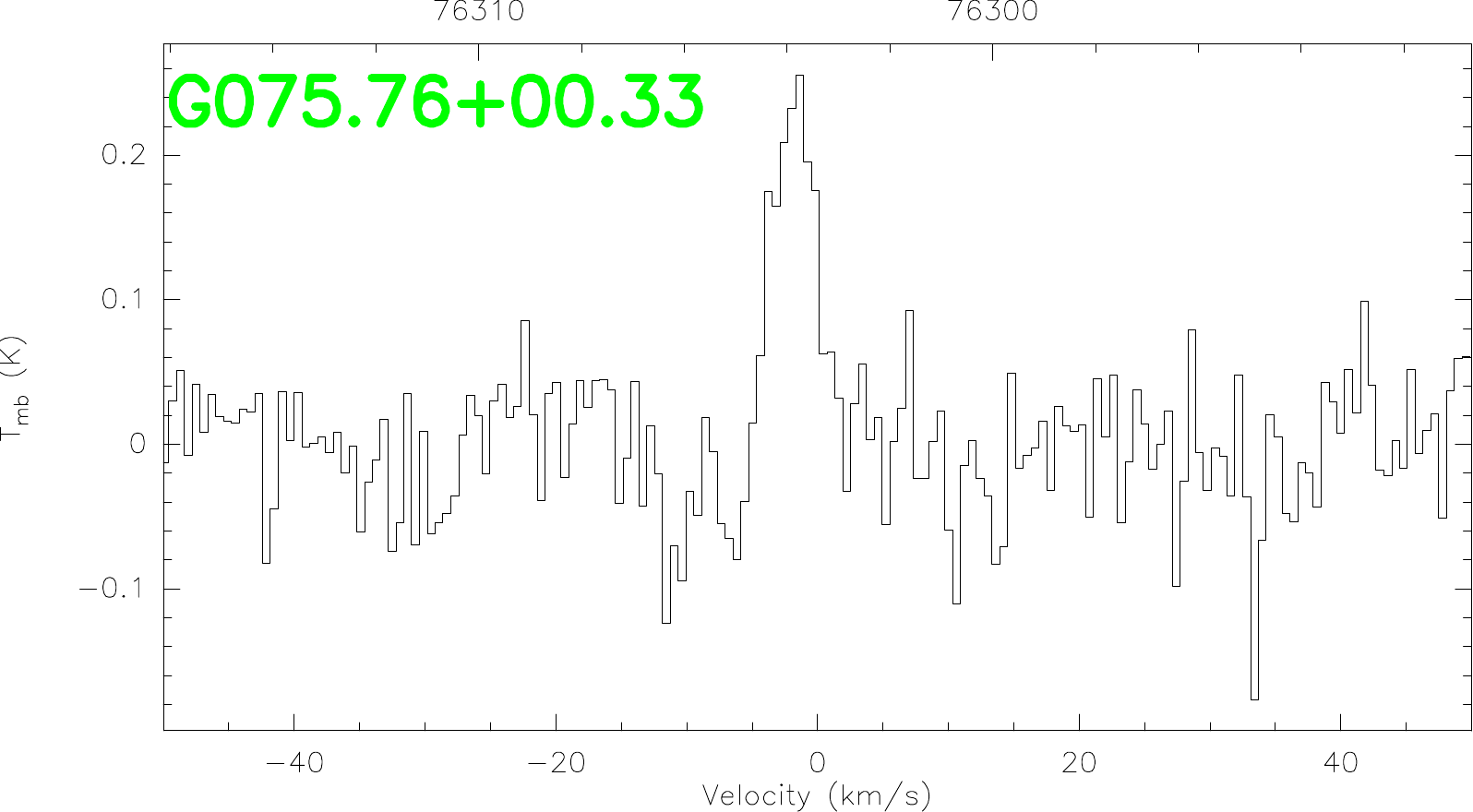}
\includegraphics[width=0.155\textwidth]{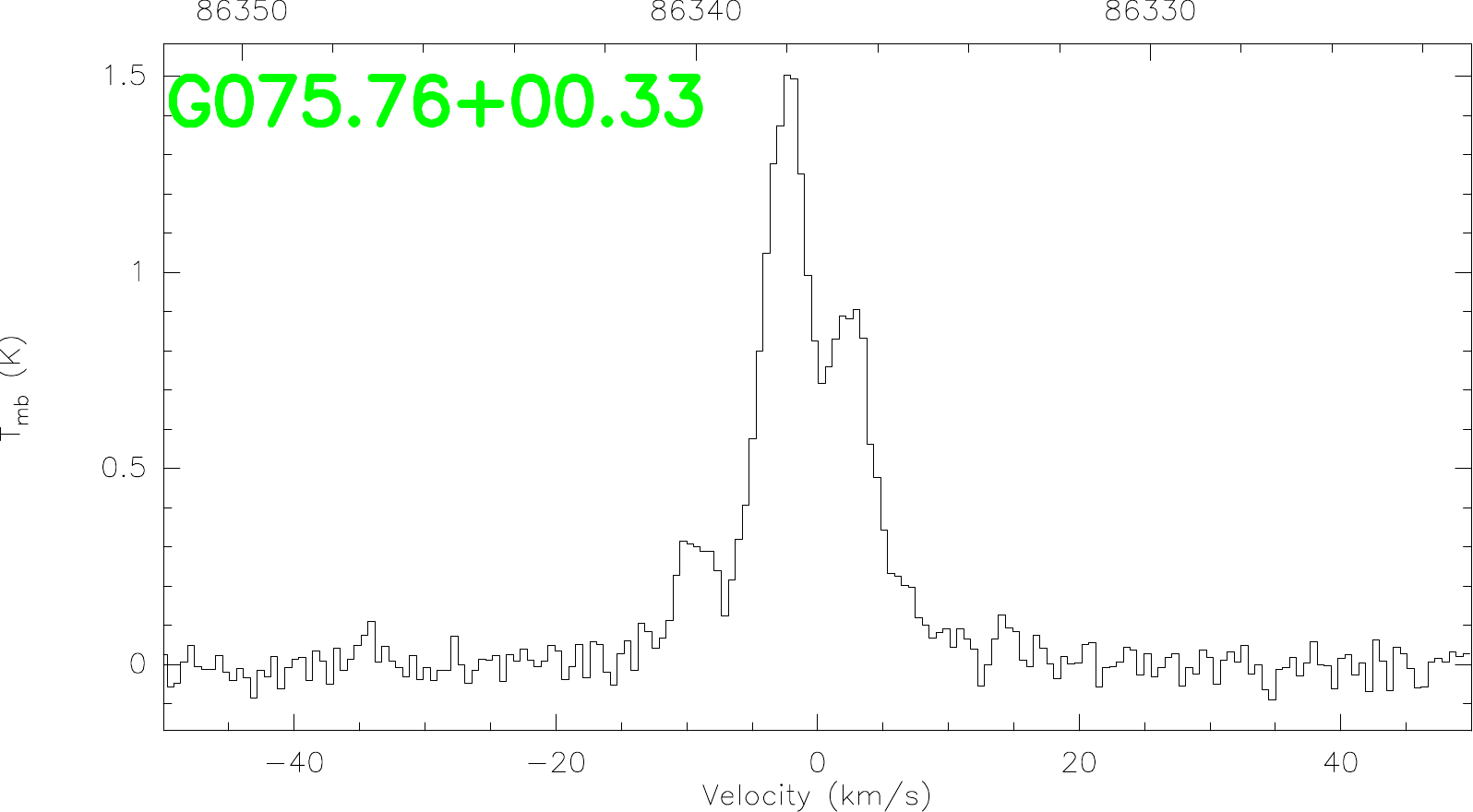}
\includegraphics[width=0.155\textwidth]{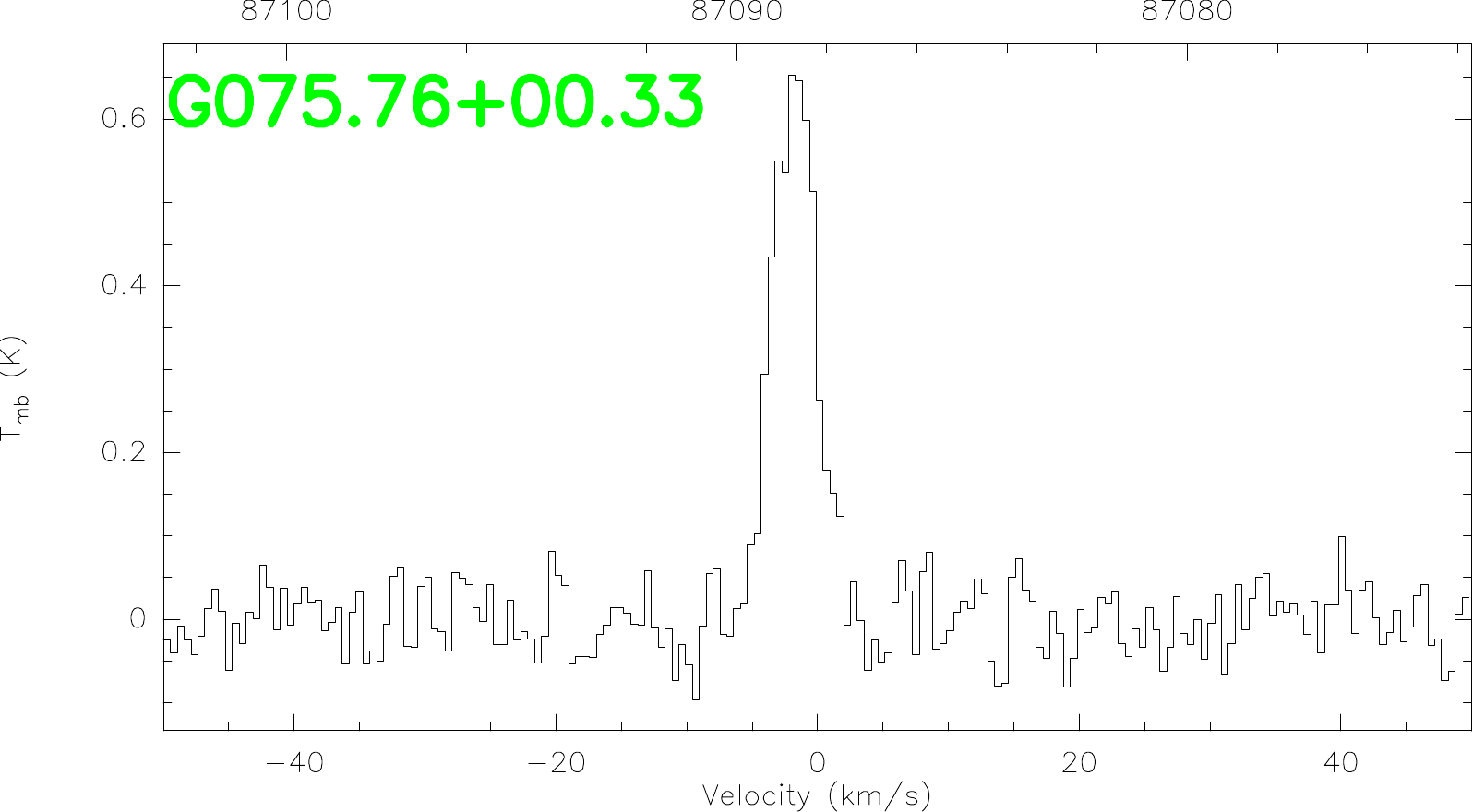}
\includegraphics[width=0.155\textwidth]{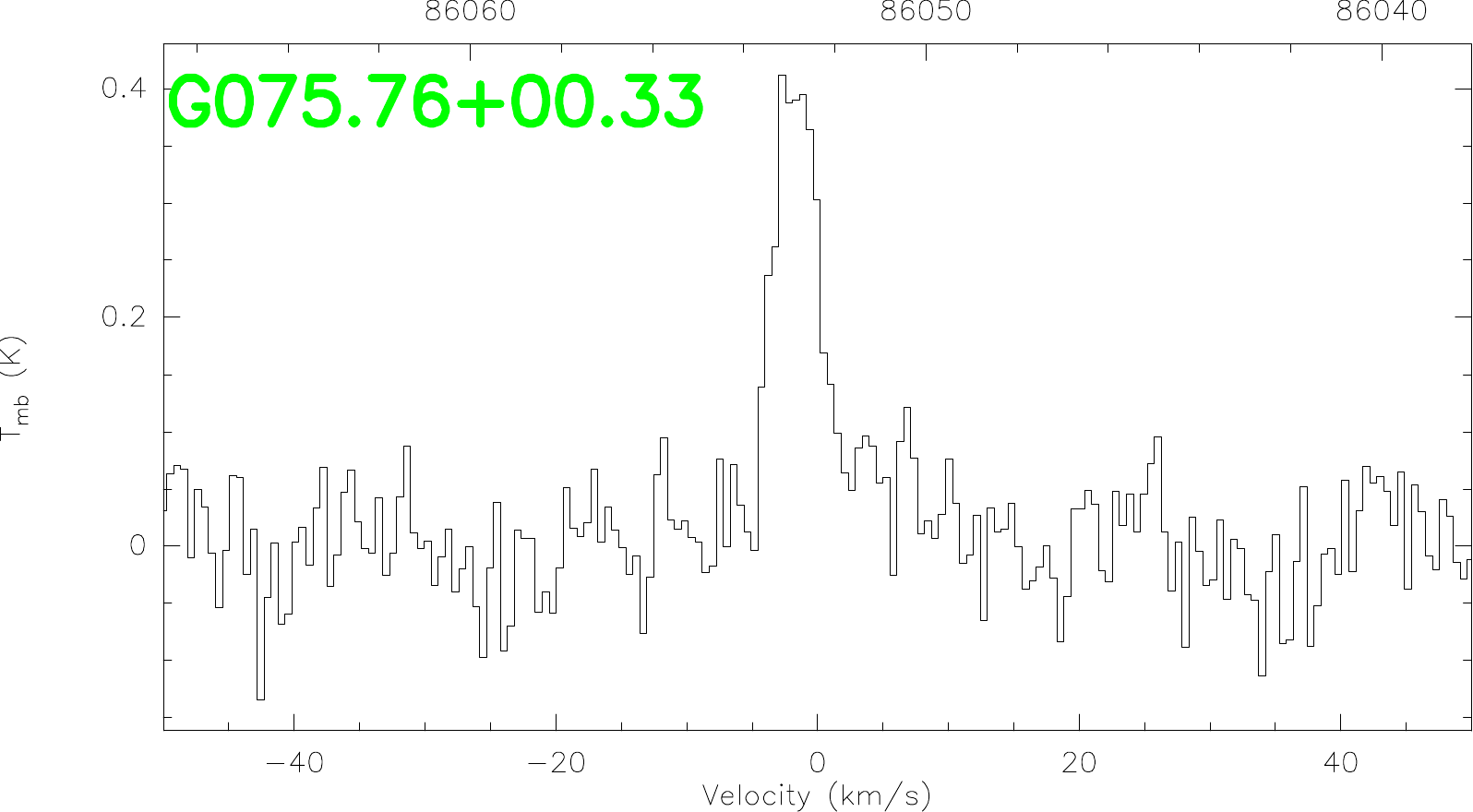}
\includegraphics[width=0.155\textwidth]{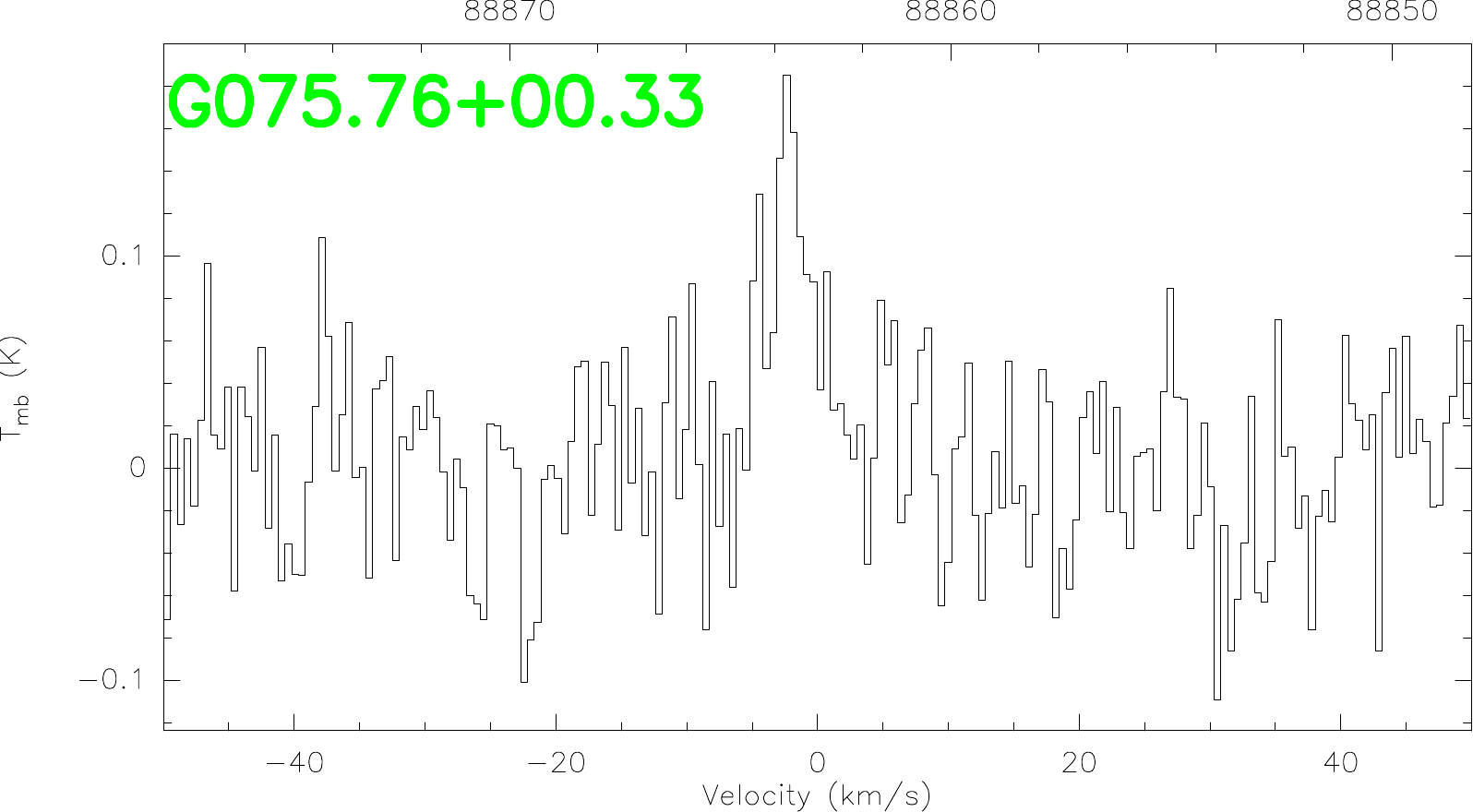}\\

\includegraphics[width=0.155\textwidth]{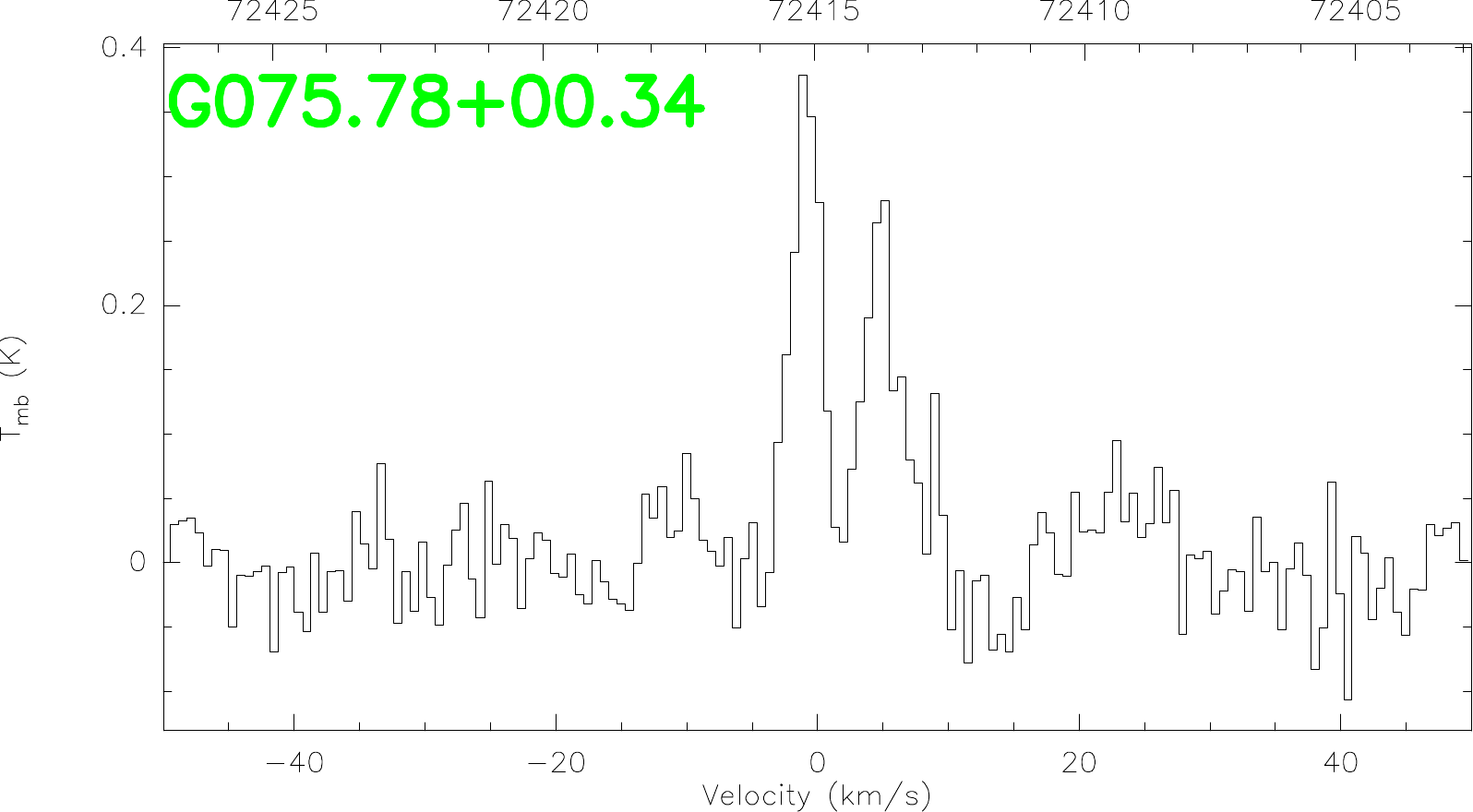}
\includegraphics[width=0.155\textwidth]{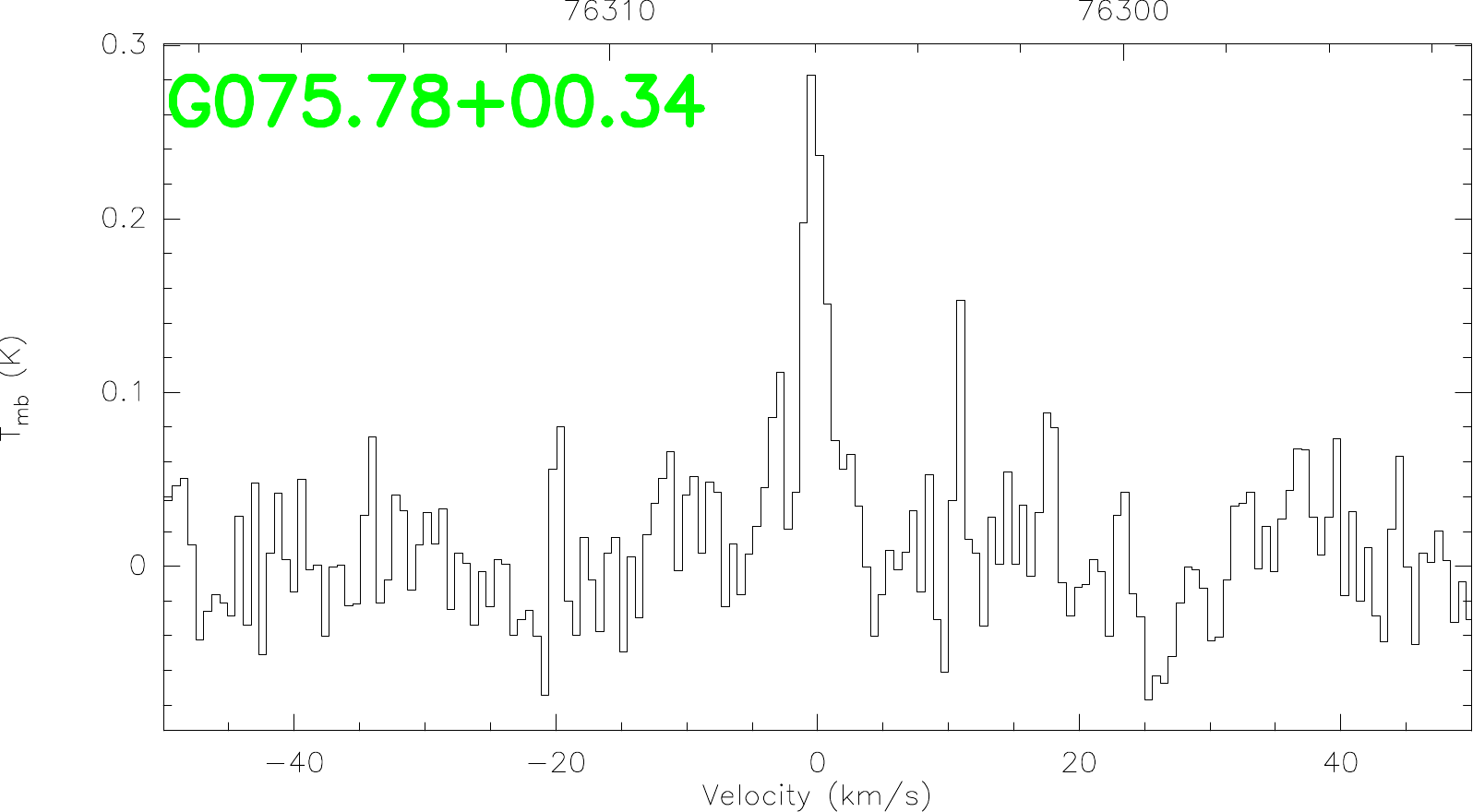}
\includegraphics[width=0.155\textwidth]{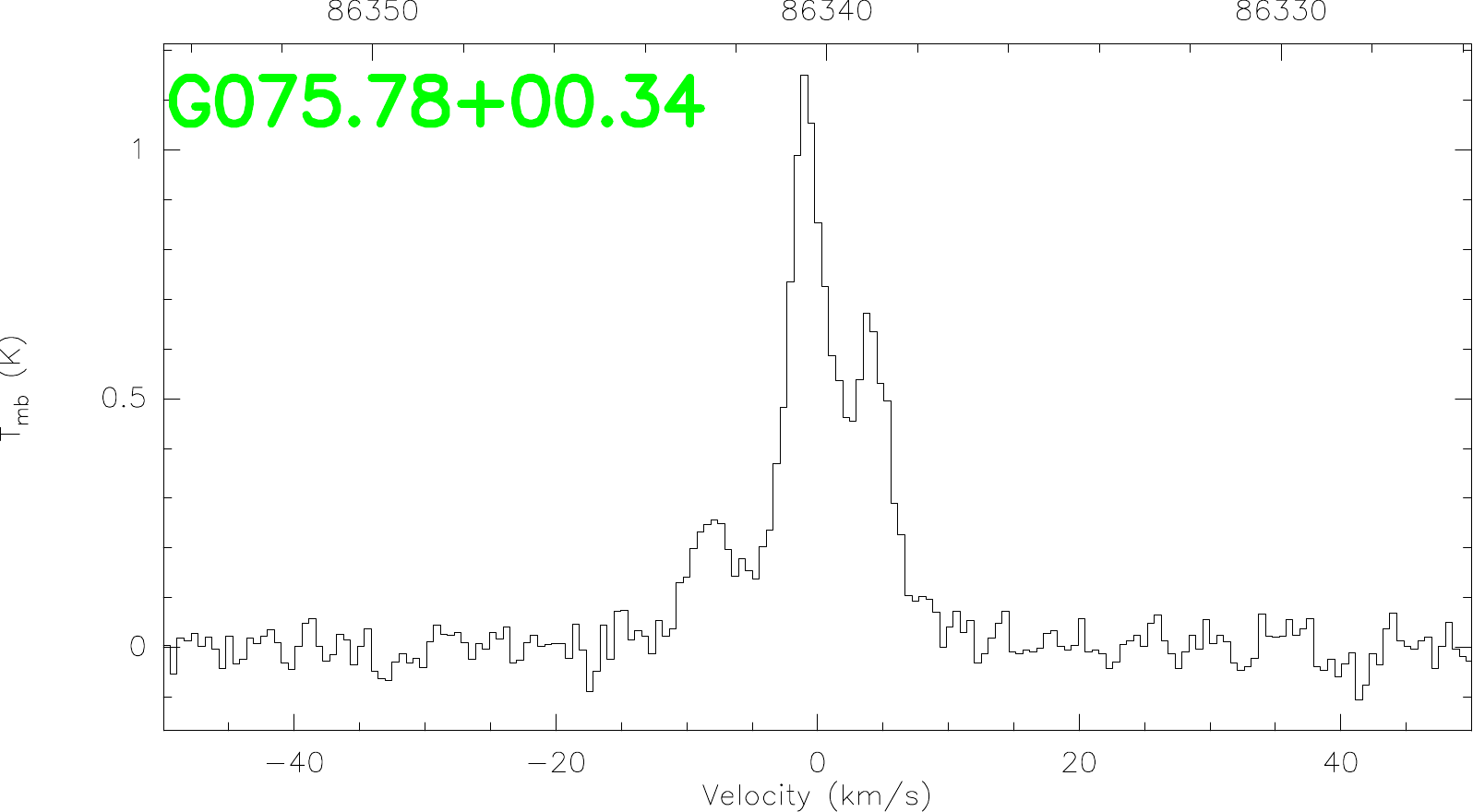}
\includegraphics[width=0.155\textwidth]{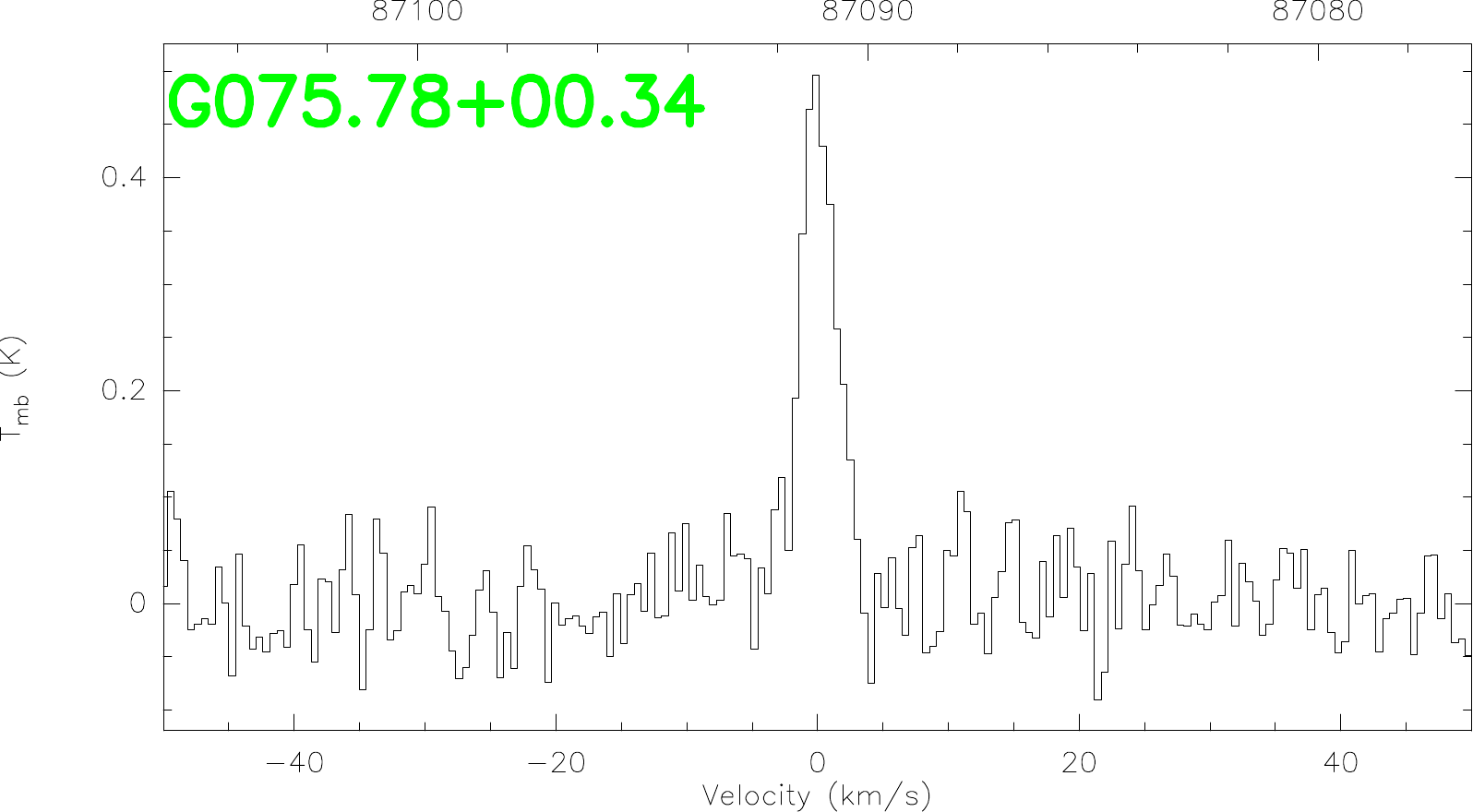}
\includegraphics[width=0.155\textwidth]{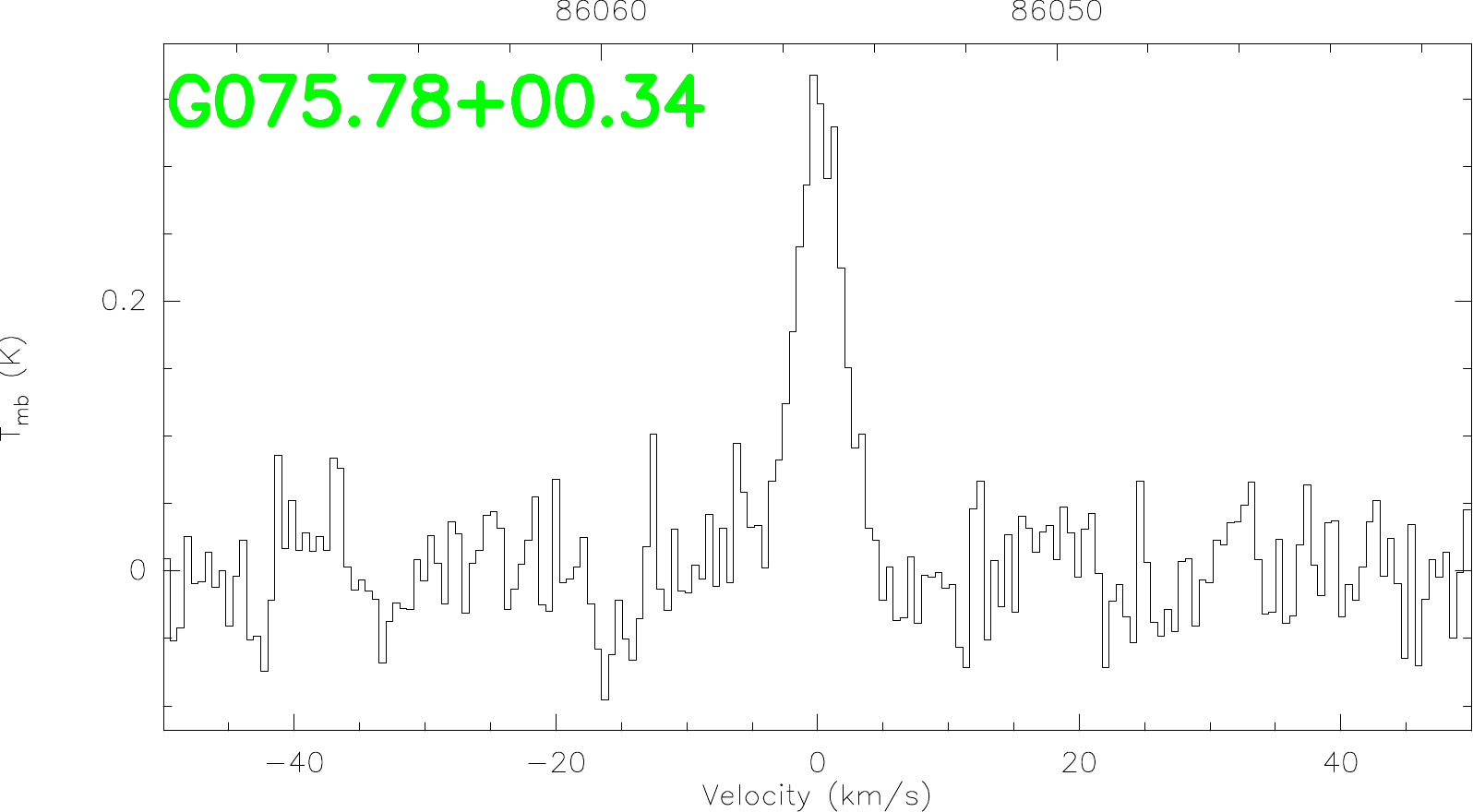}
\includegraphics[width=0.155\textwidth]{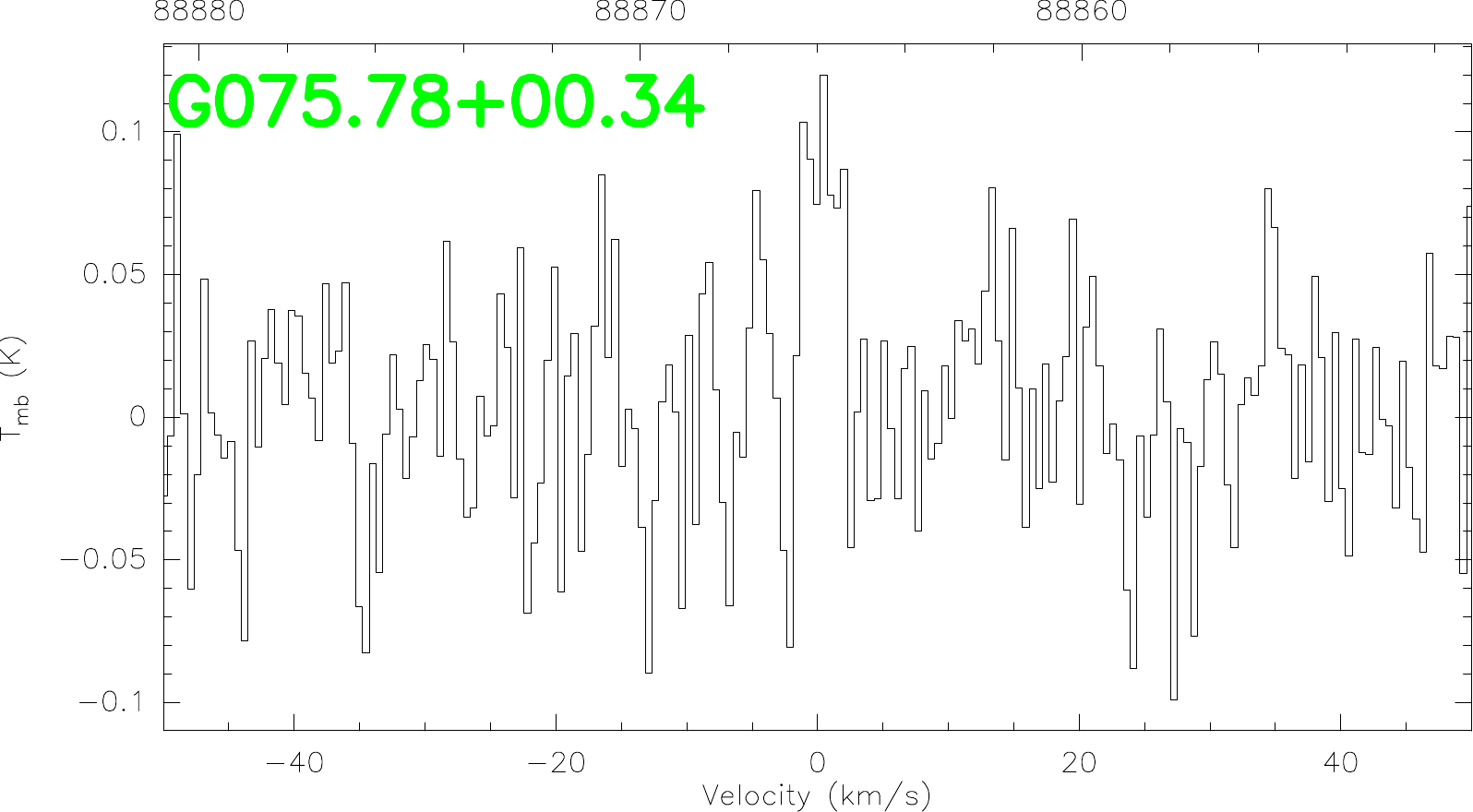}\\

\includegraphics[width=0.155\textwidth]{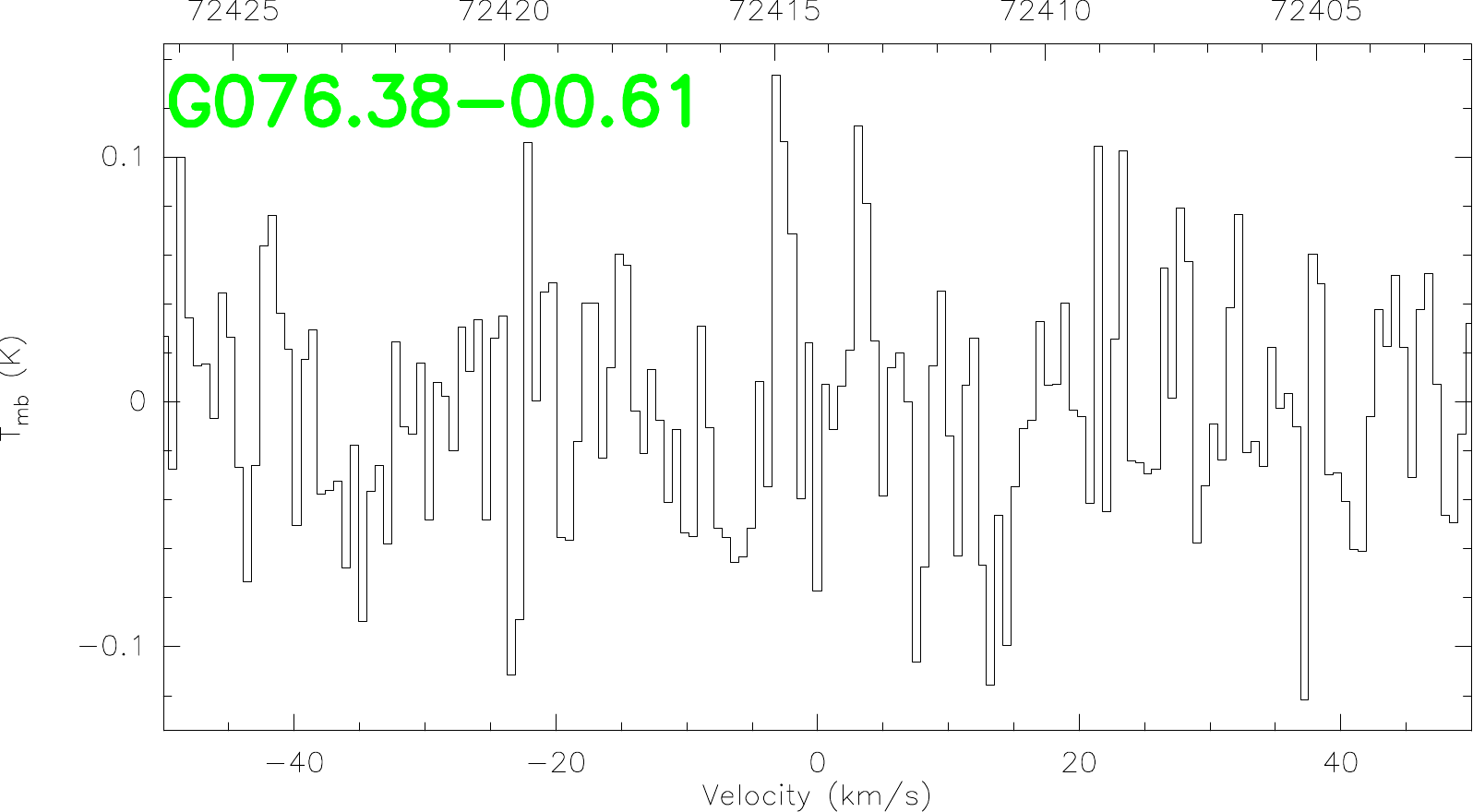}
\includegraphics[width=0.155\textwidth]{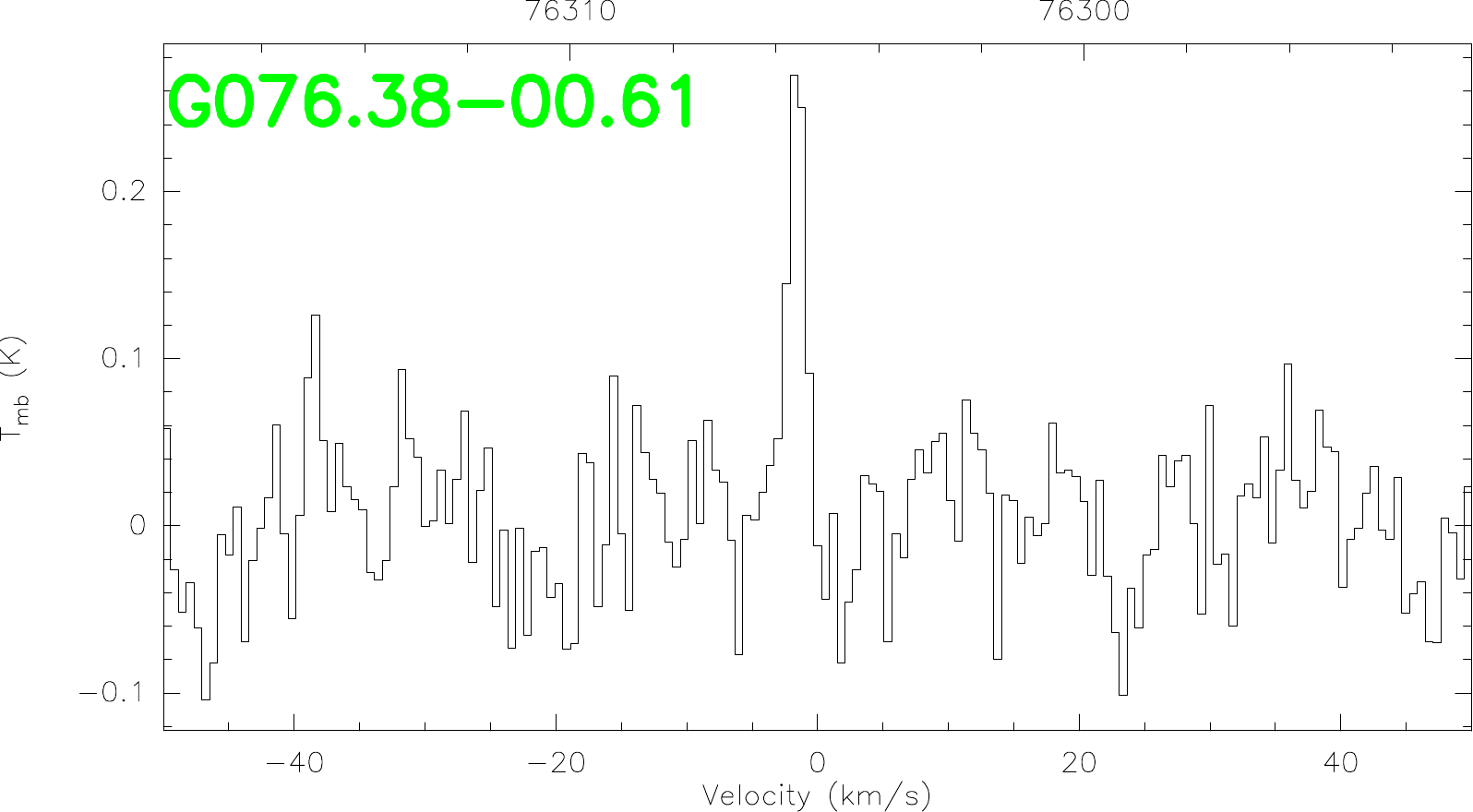}
\includegraphics[width=0.155\textwidth]{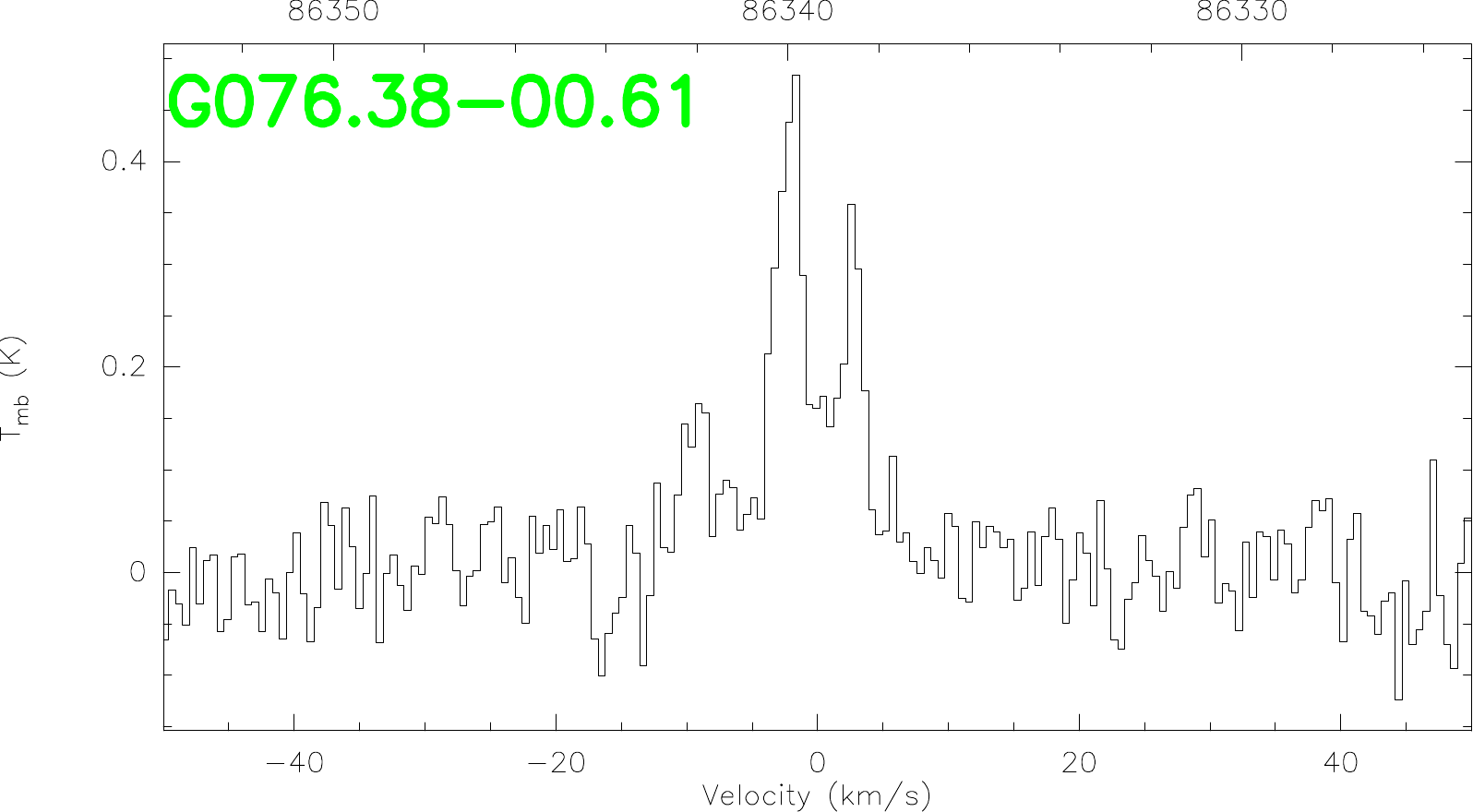}
\includegraphics[width=0.155\textwidth]{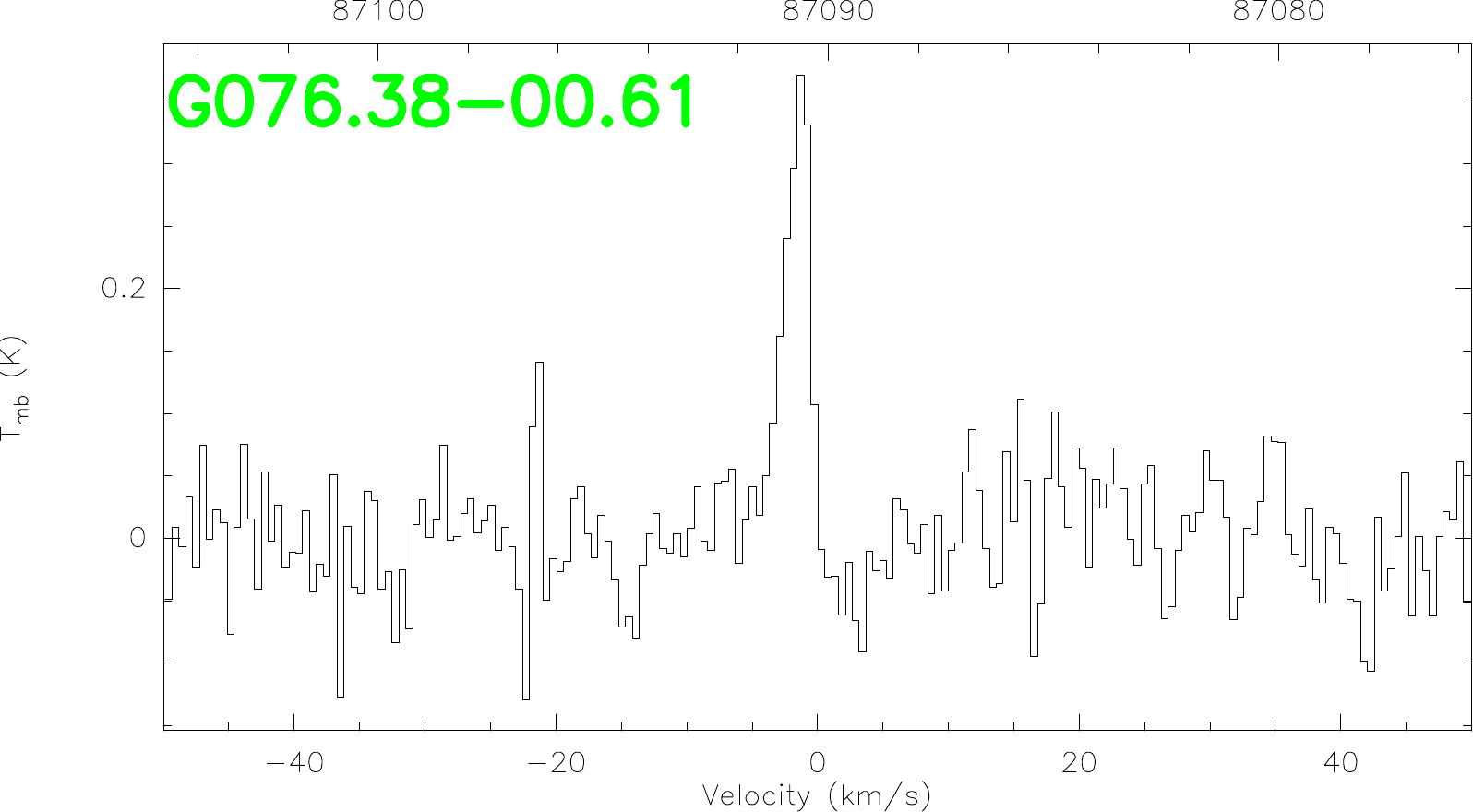}
\includegraphics[width=0.155\textwidth]{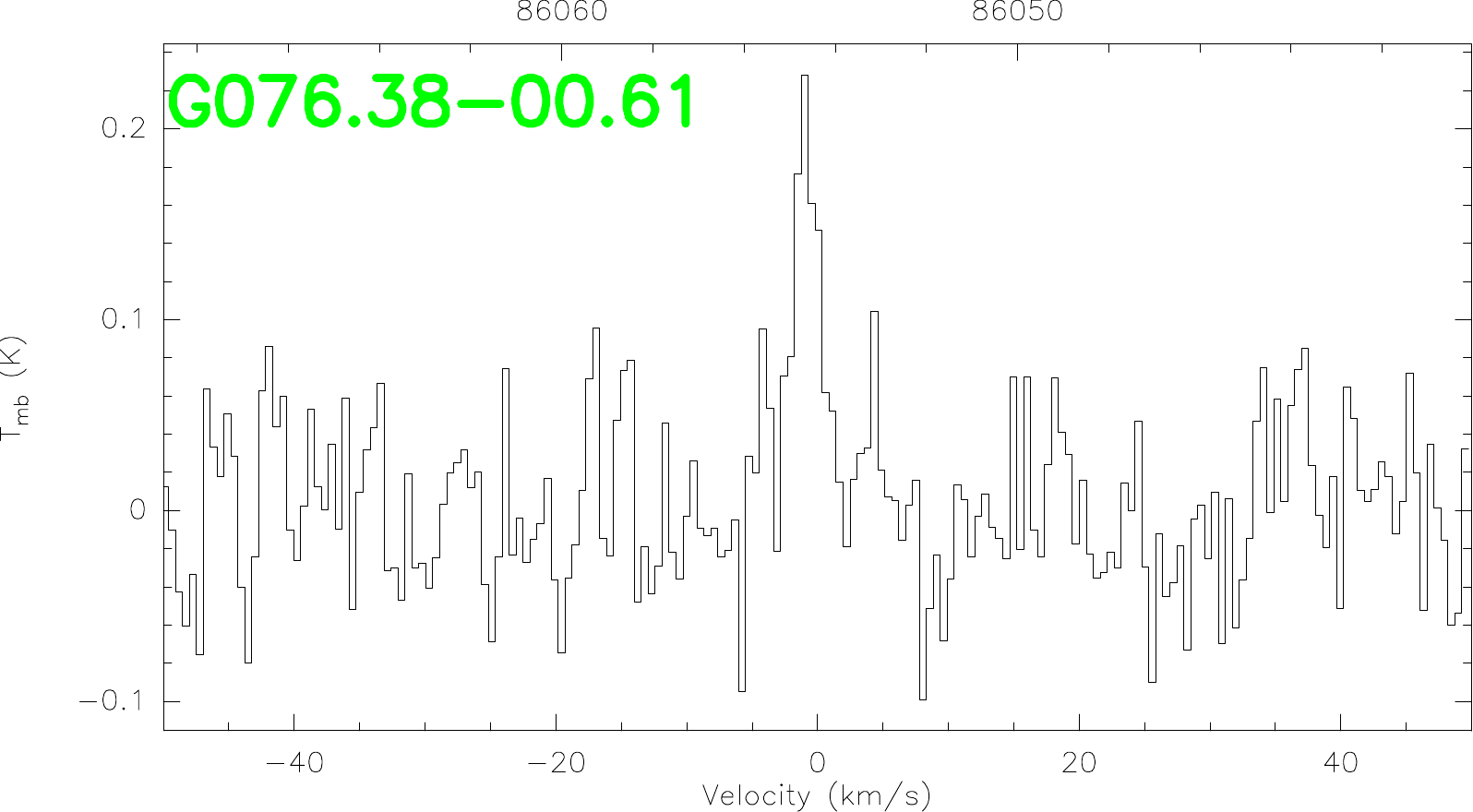}
\includegraphics[width=0.155\textwidth]{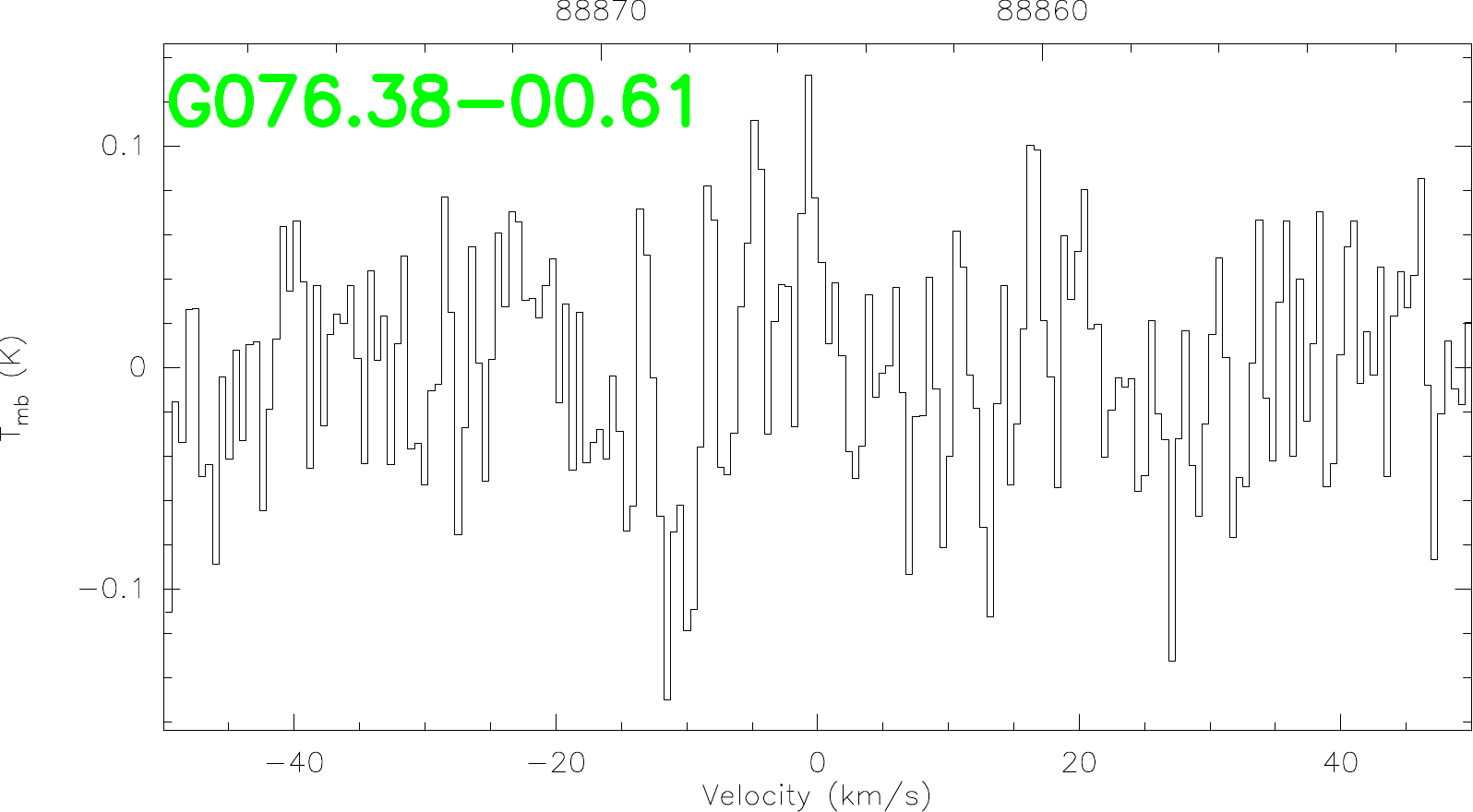}\\

\includegraphics[width=0.155\textwidth]{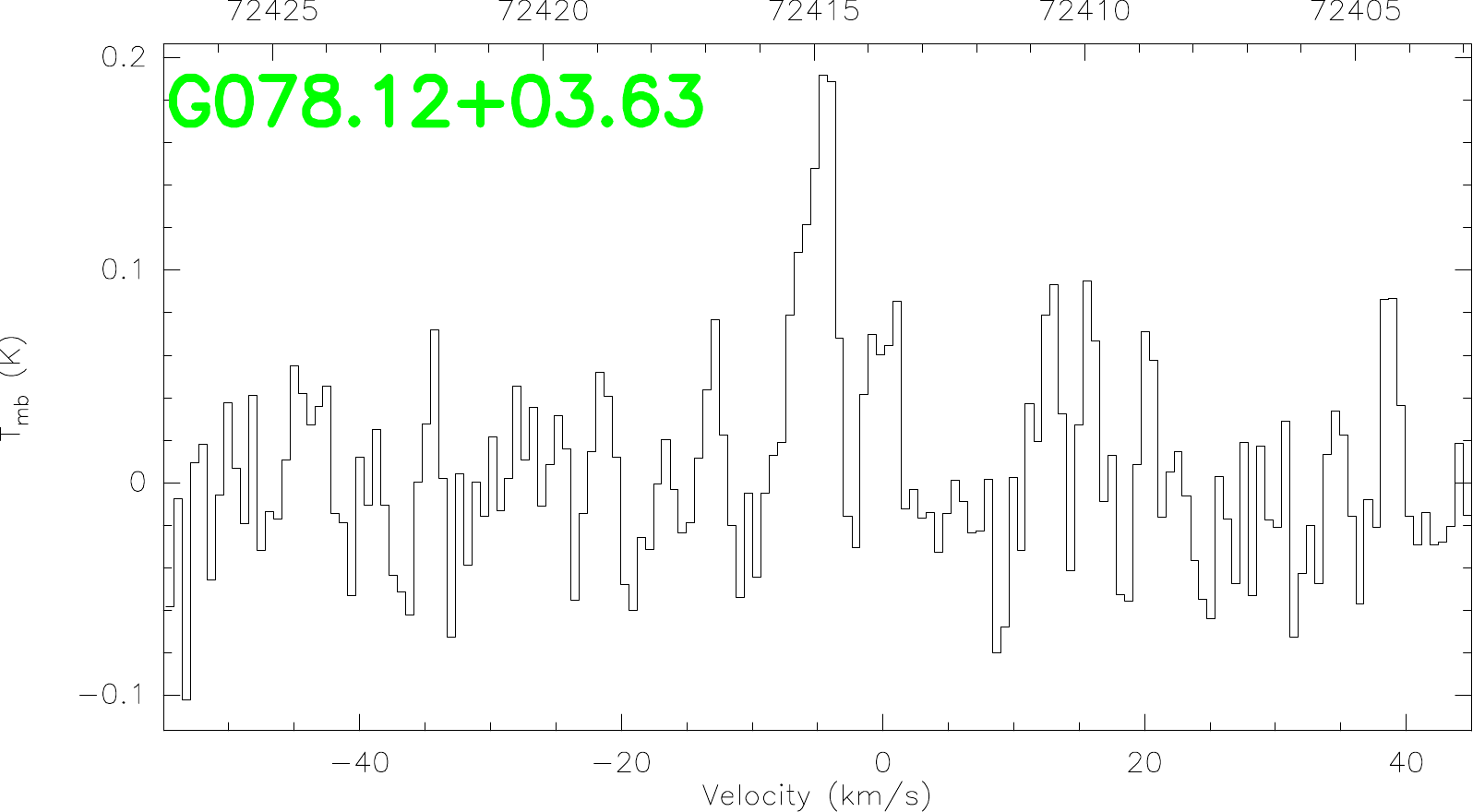}
\includegraphics[width=0.155\textwidth]{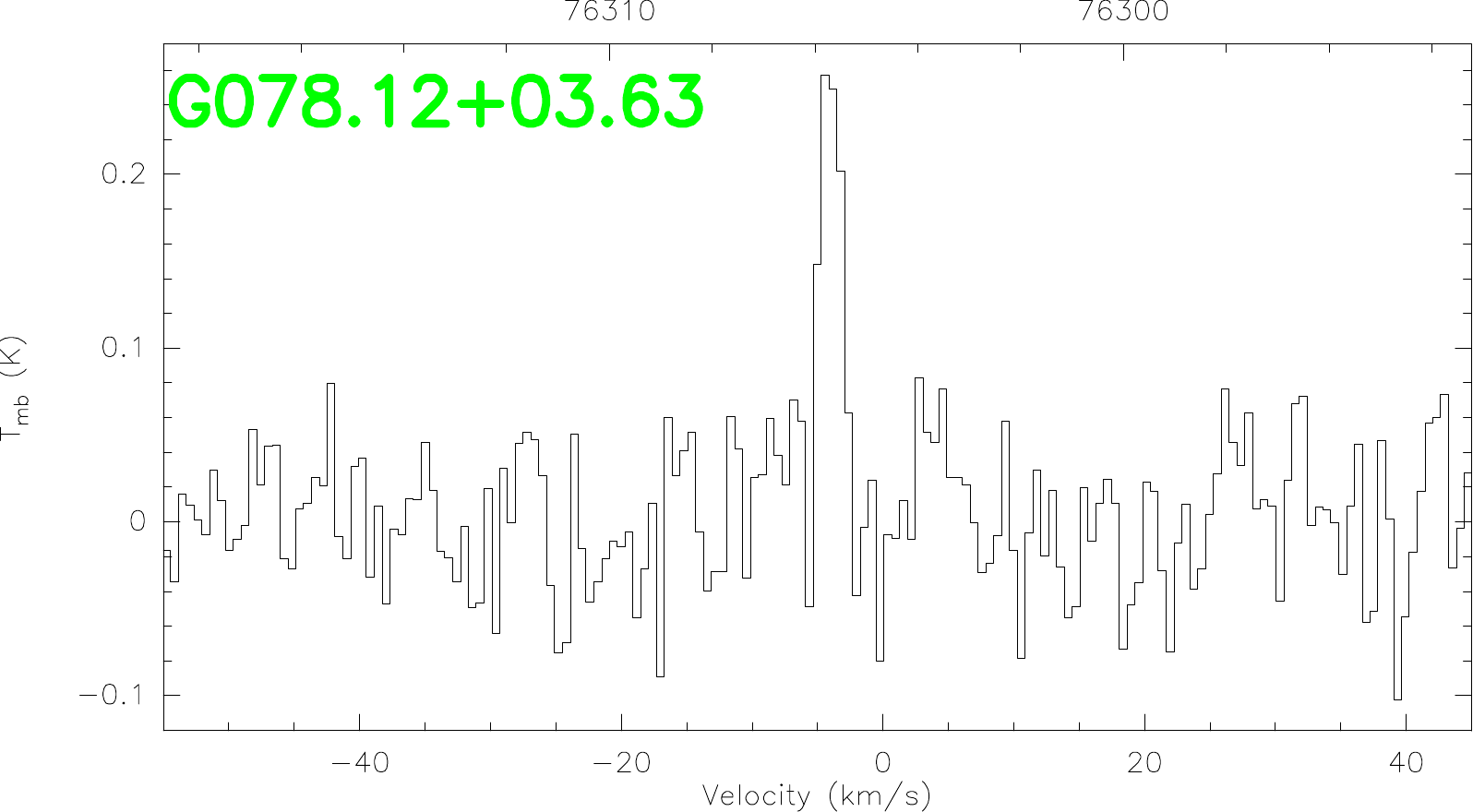}
\includegraphics[width=0.155\textwidth]{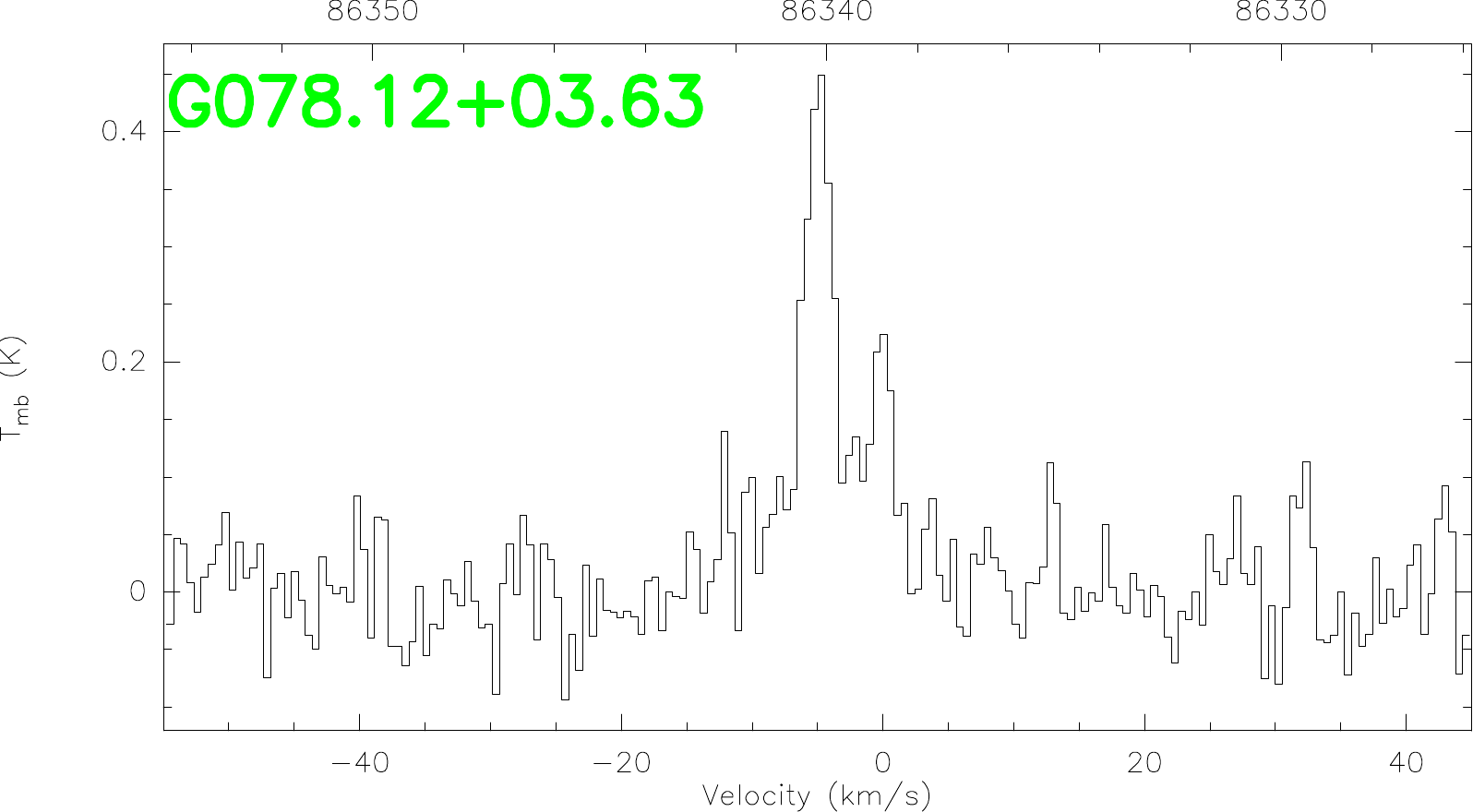}
\includegraphics[width=0.155\textwidth]{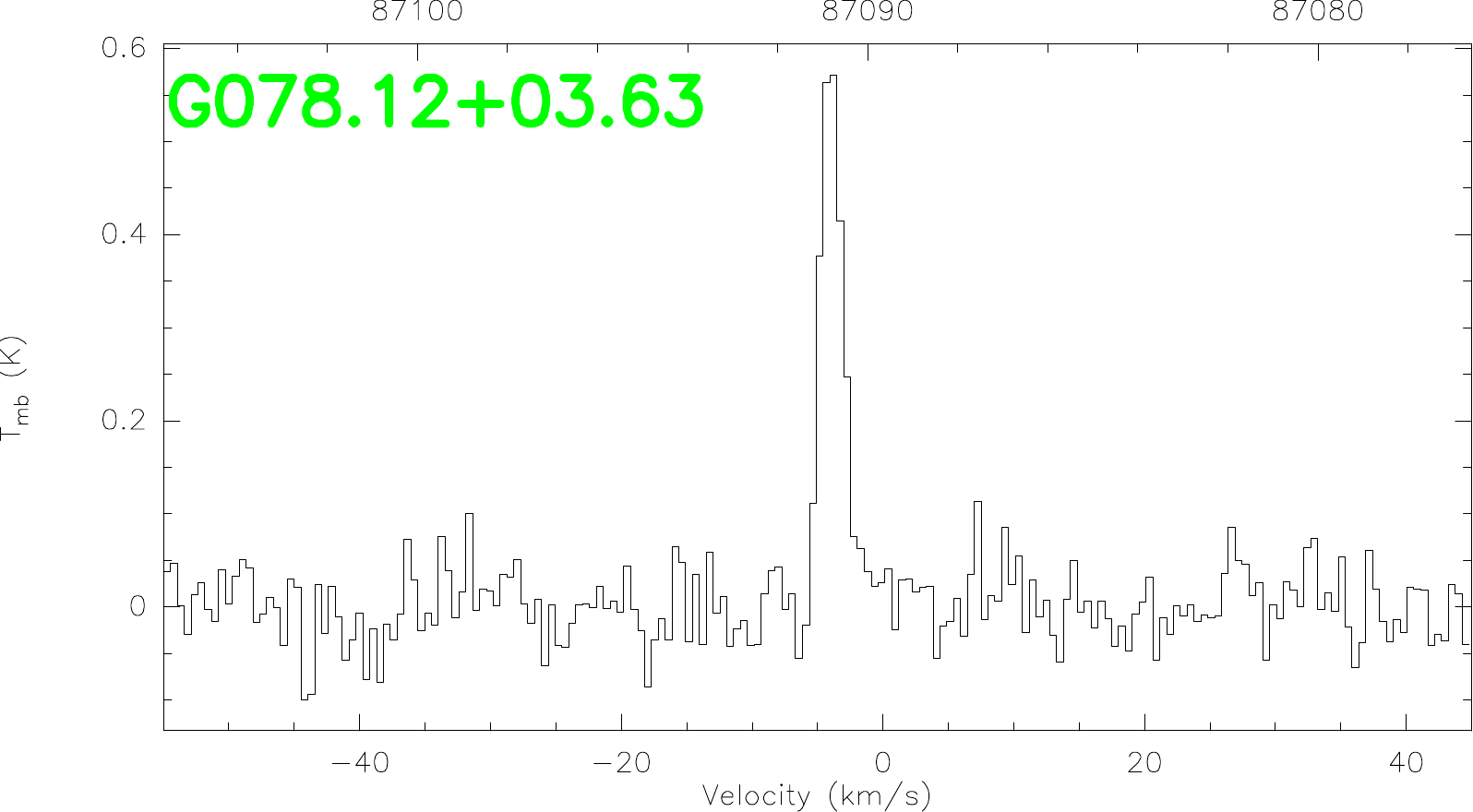}
\includegraphics[width=0.155\textwidth]{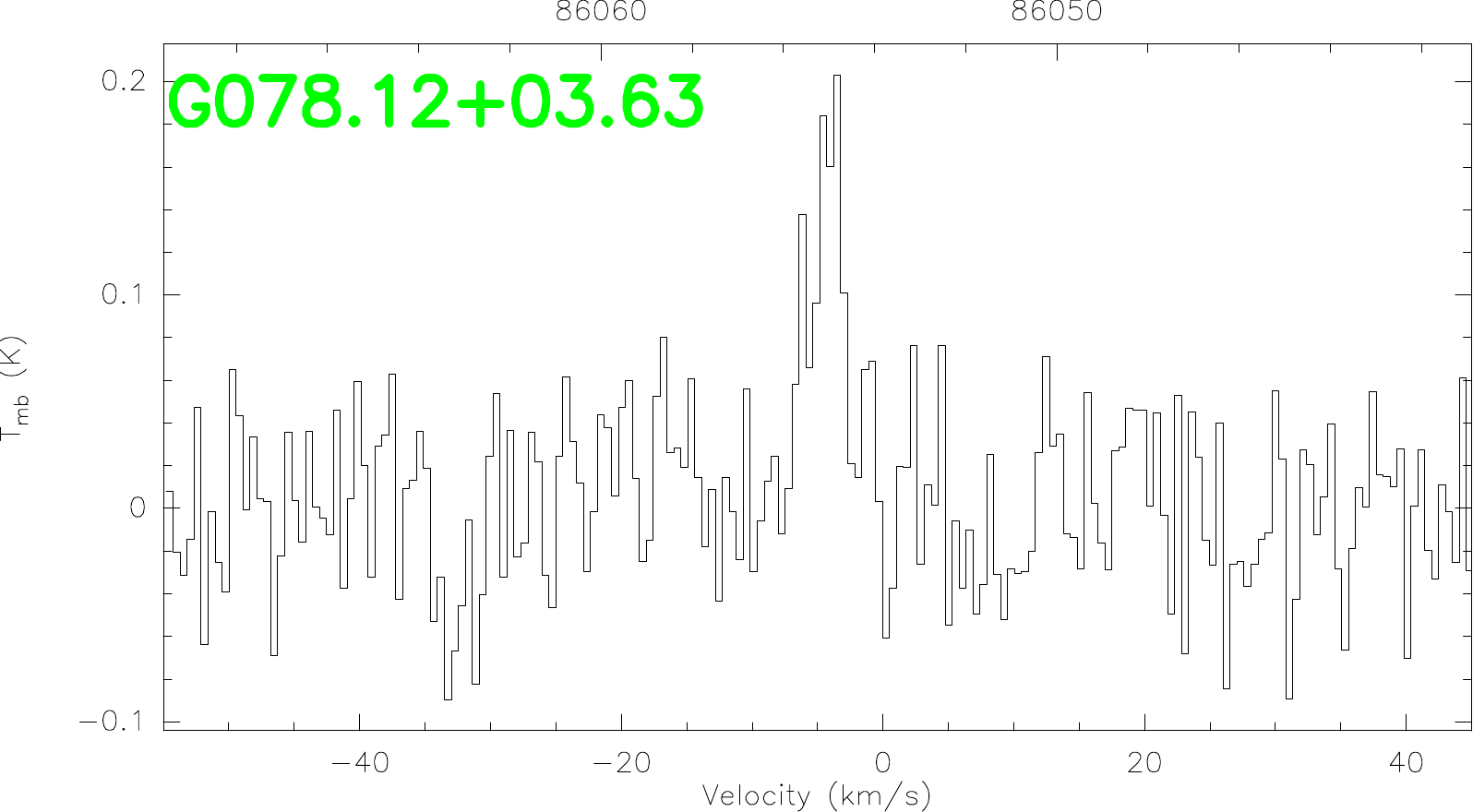}
\includegraphics[width=0.155\textwidth]{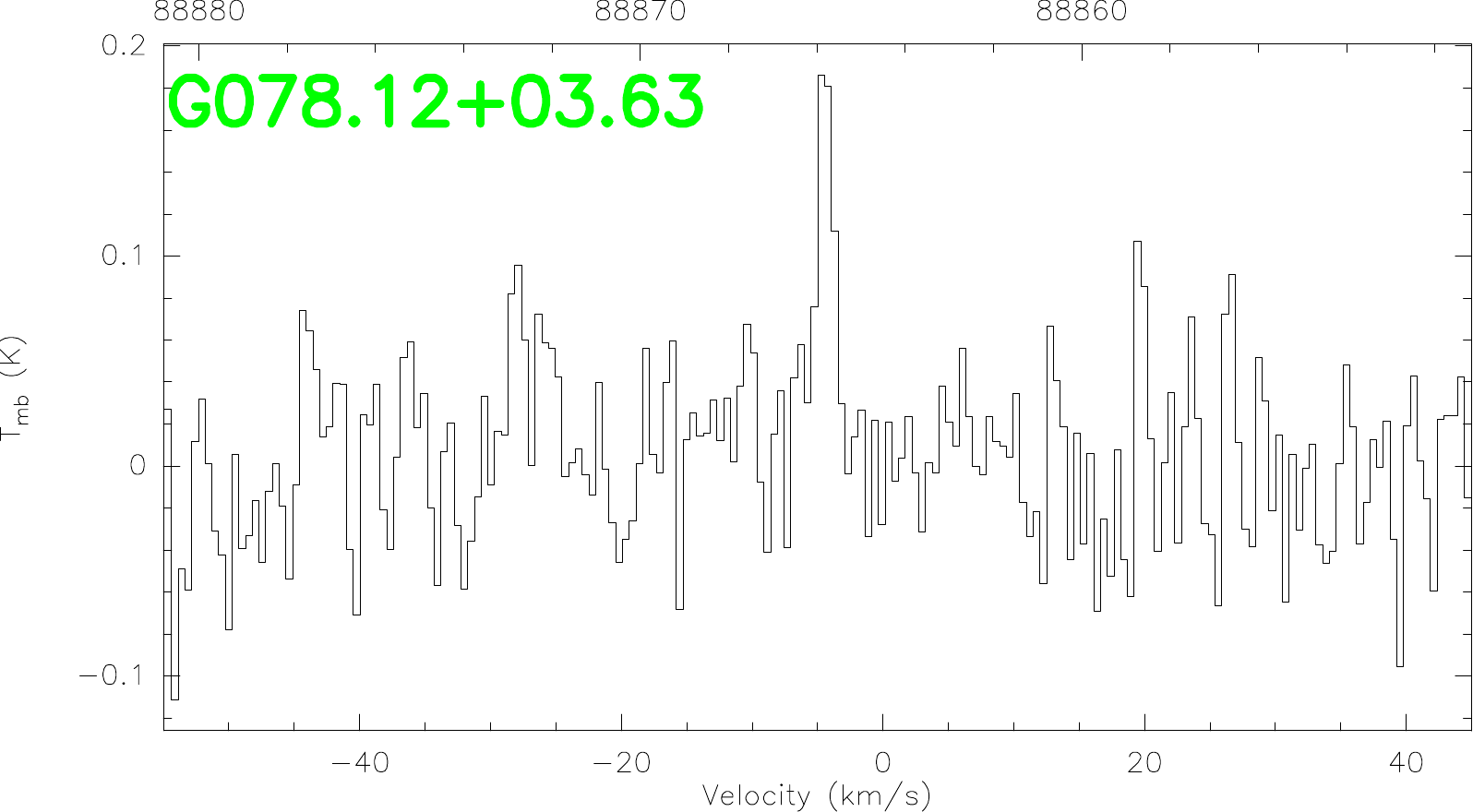}\\

\includegraphics[width=0.155\textwidth]{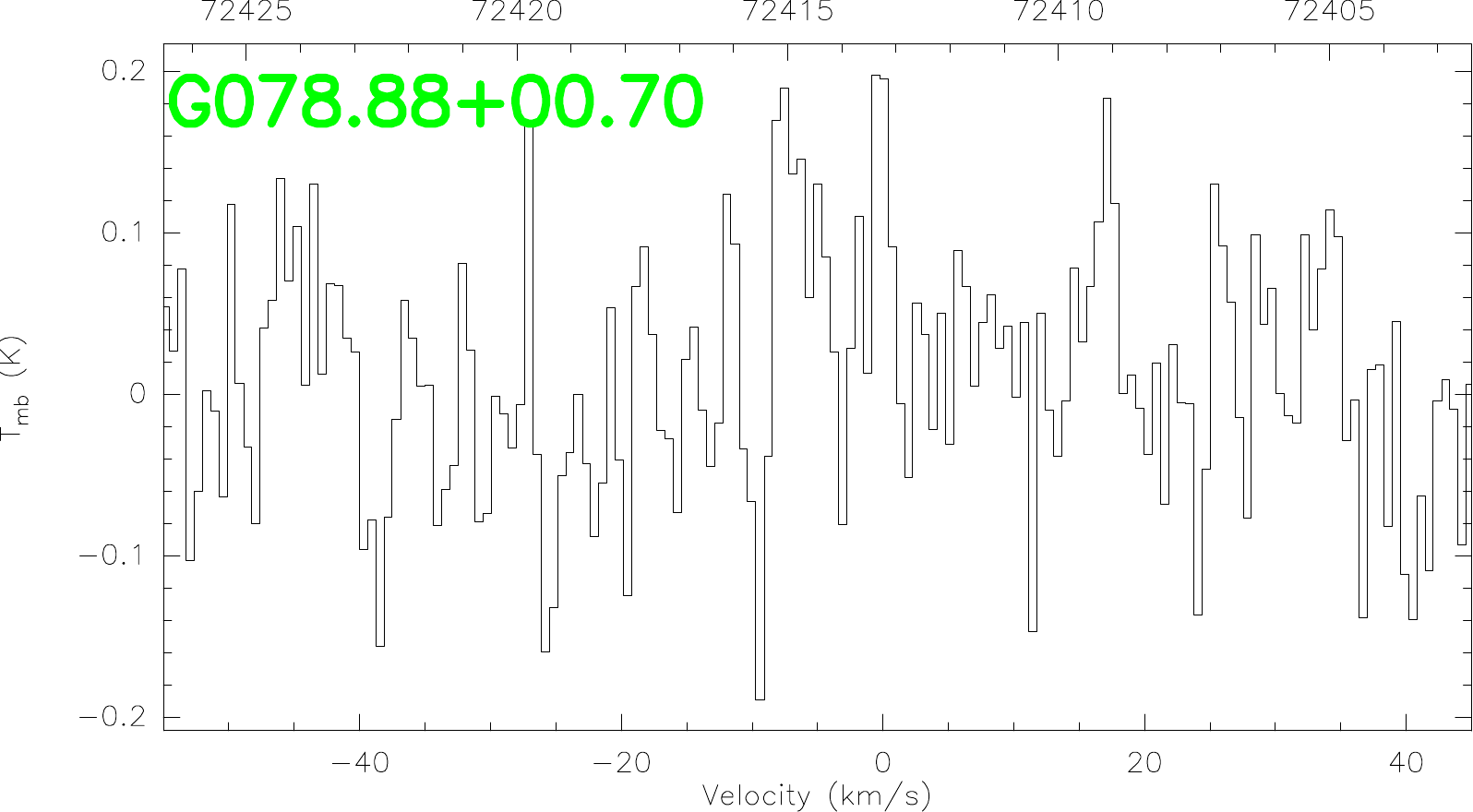}
\includegraphics[width=0.155\textwidth]{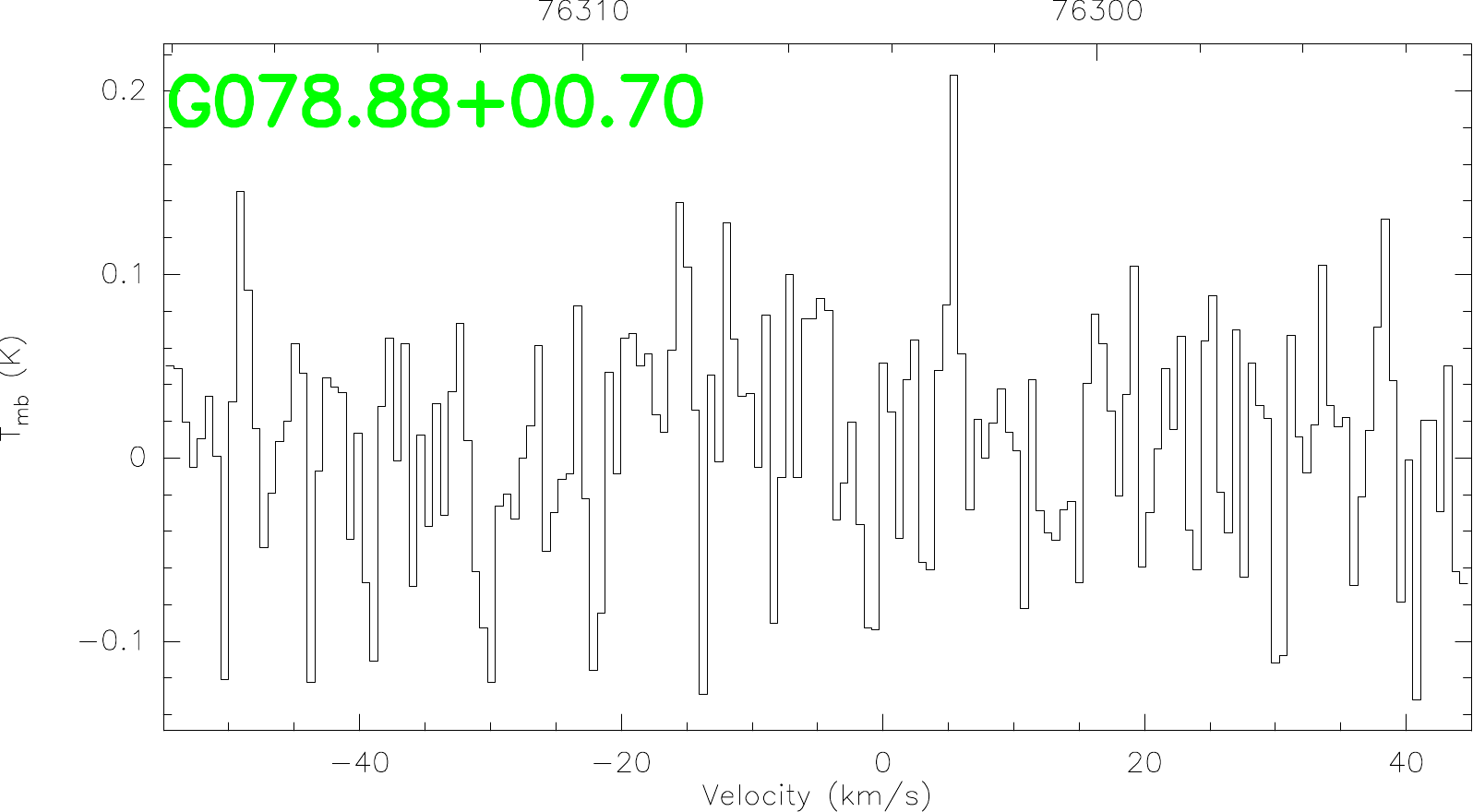}
\includegraphics[width=0.155\textwidth]{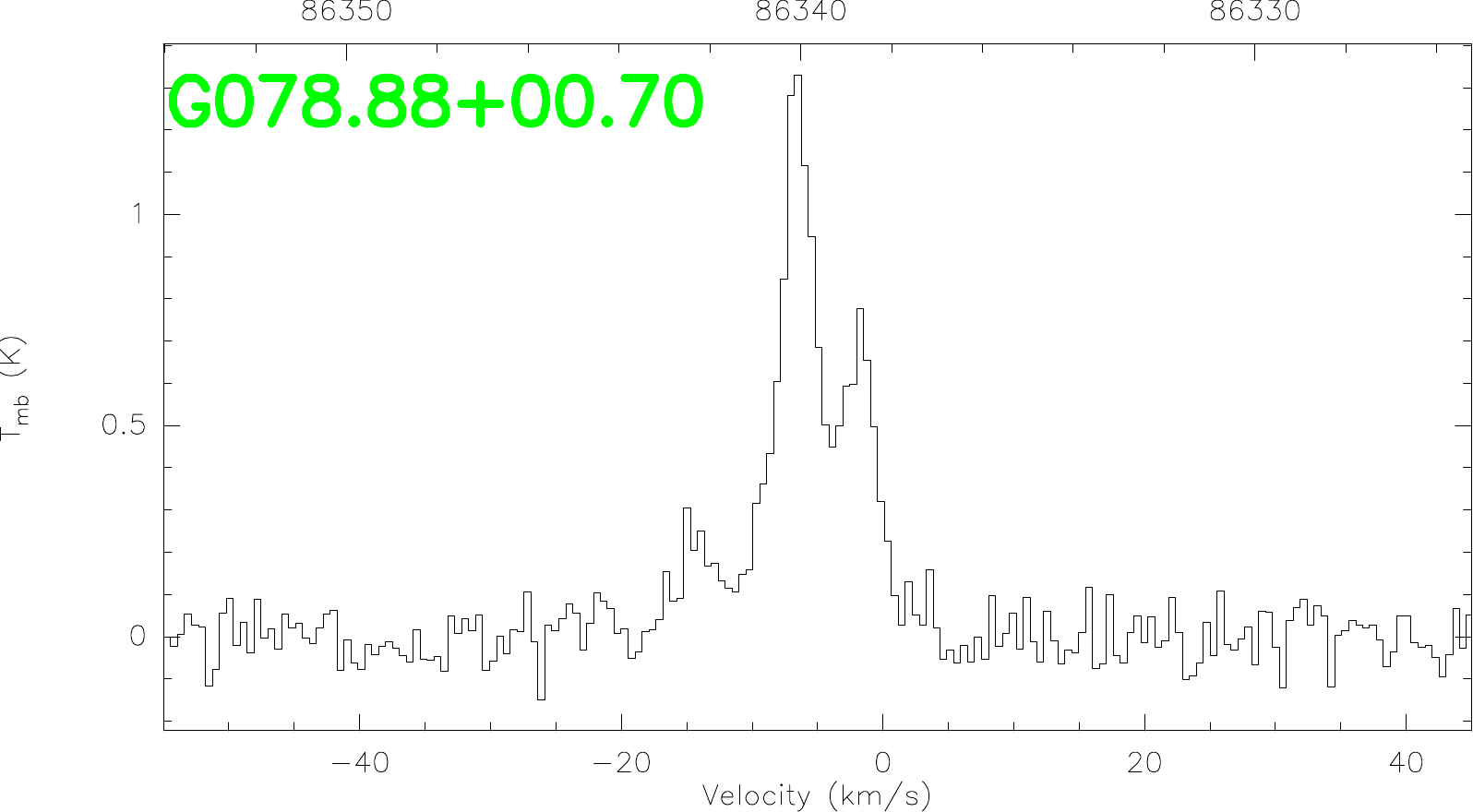}
\includegraphics[width=0.155\textwidth]{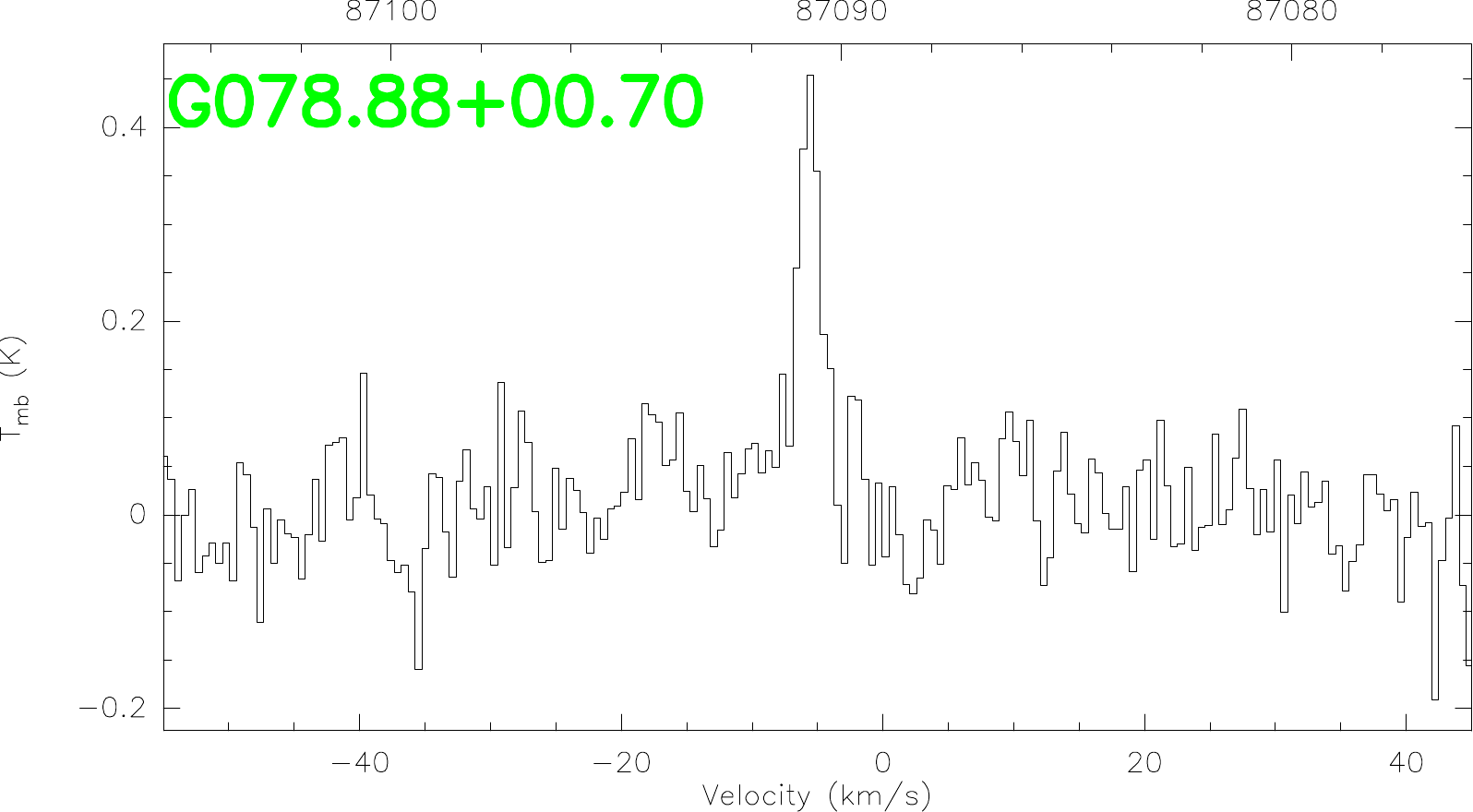}
\includegraphics[width=0.155\textwidth]{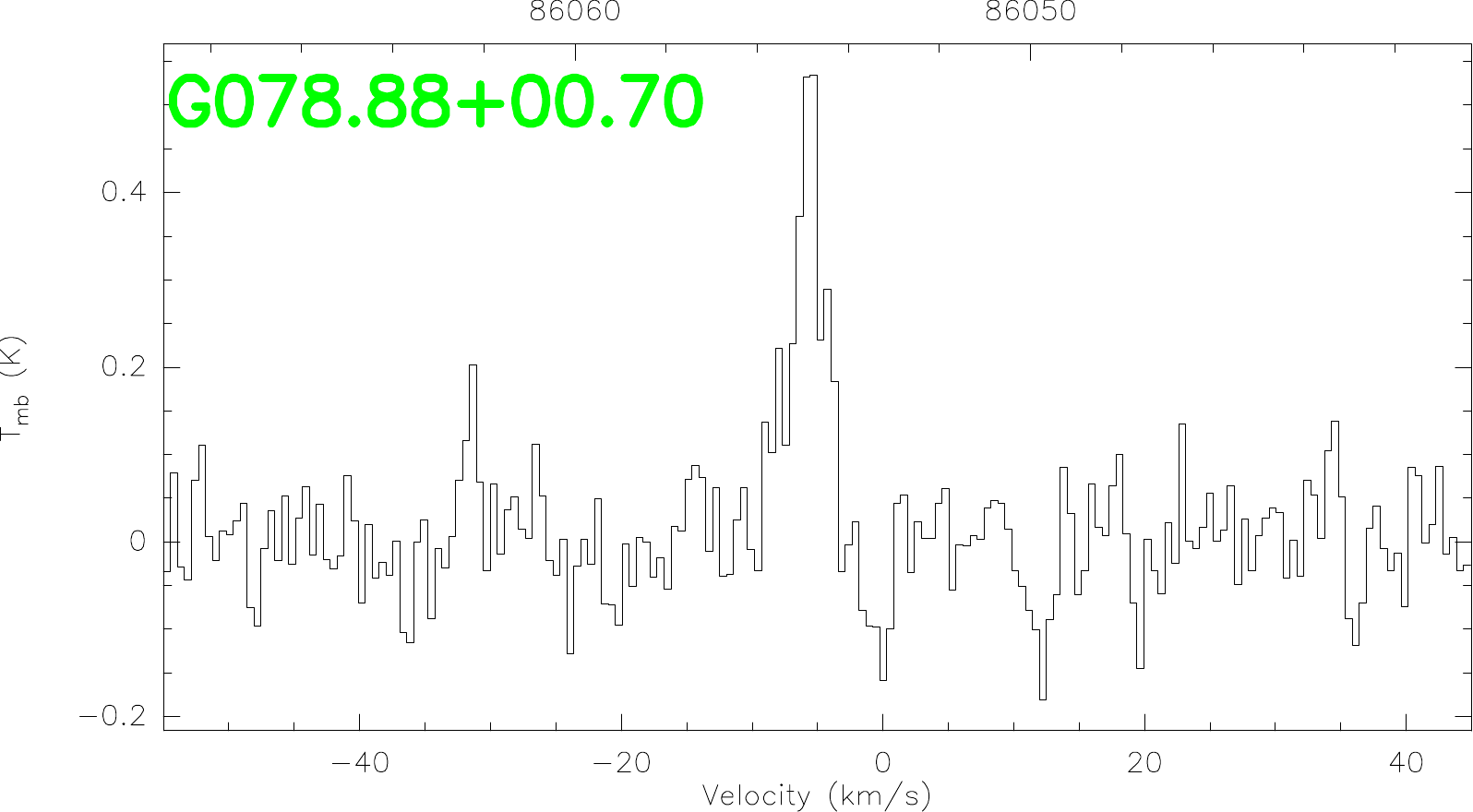}
\includegraphics[width=0.155\textwidth]{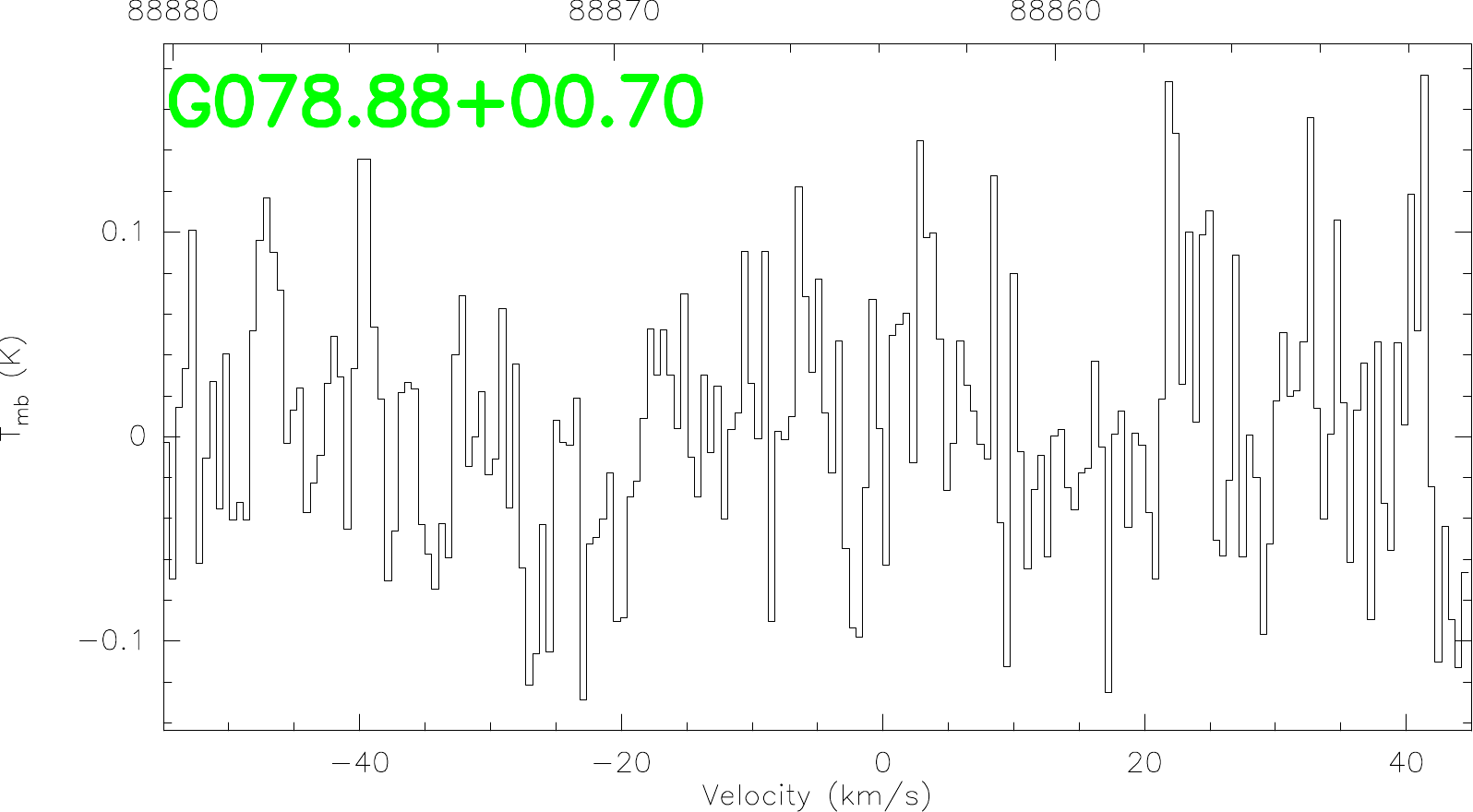}\\

\includegraphics[width=0.155\textwidth]{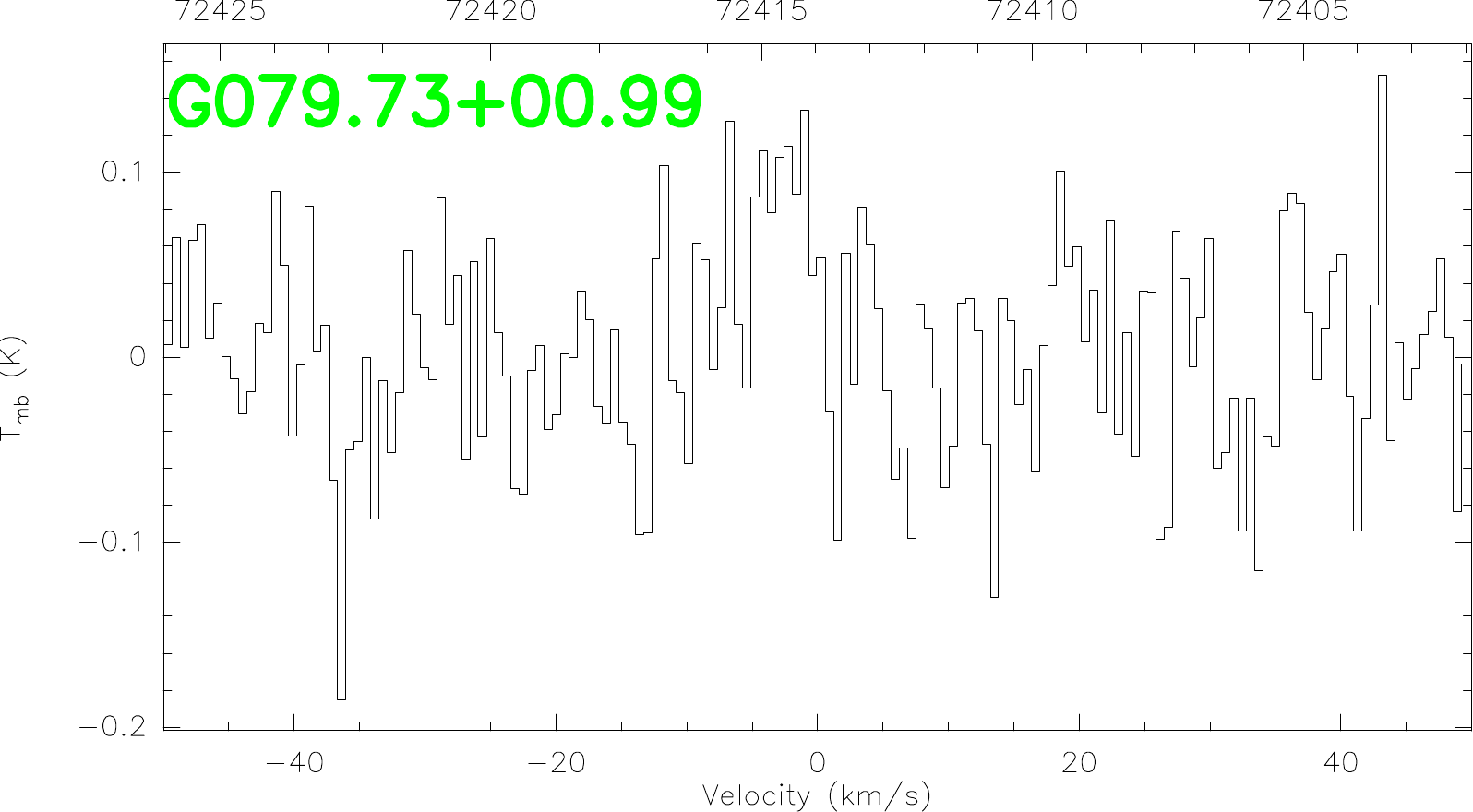}
\includegraphics[width=0.155\textwidth]{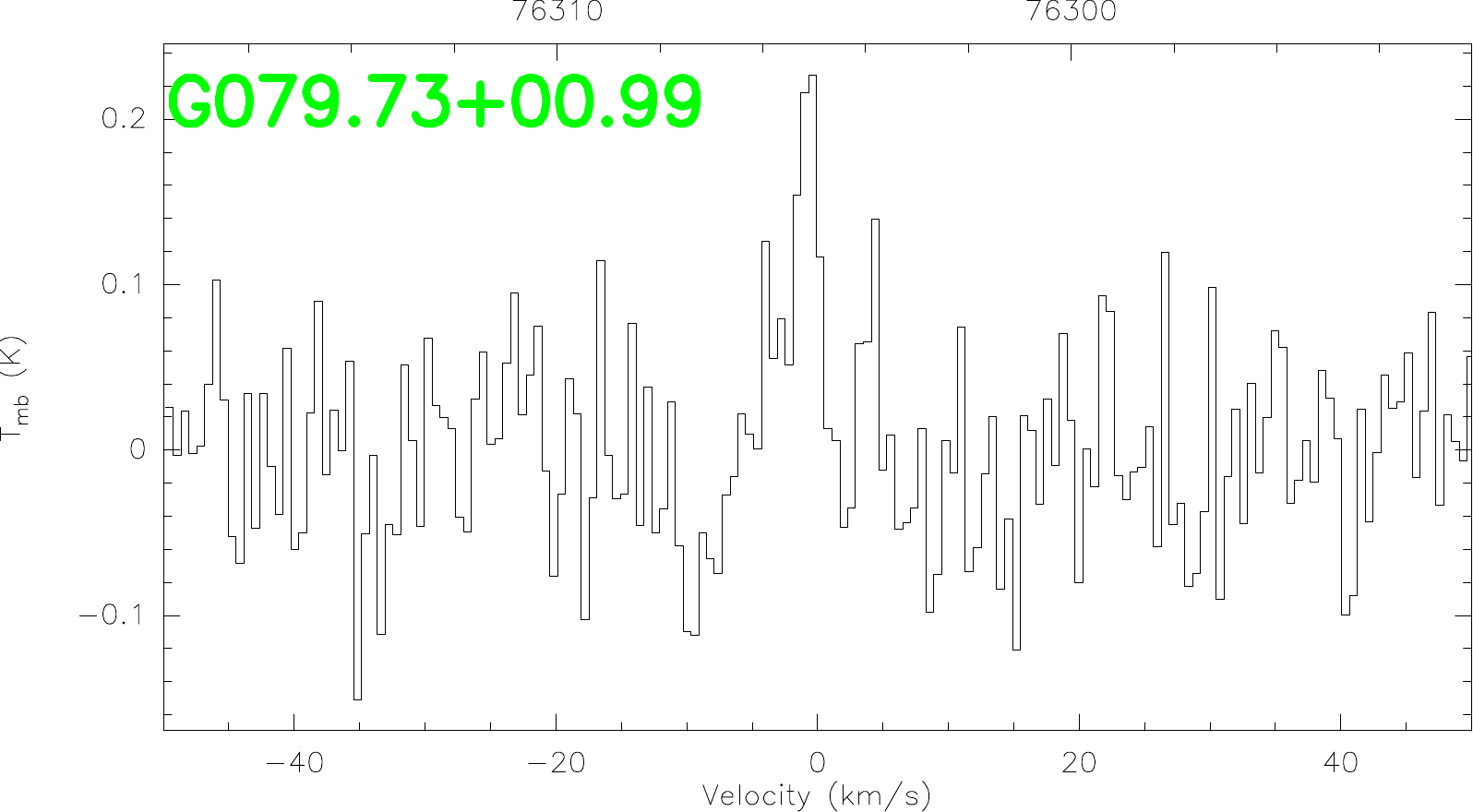}
\includegraphics[width=0.155\textwidth]{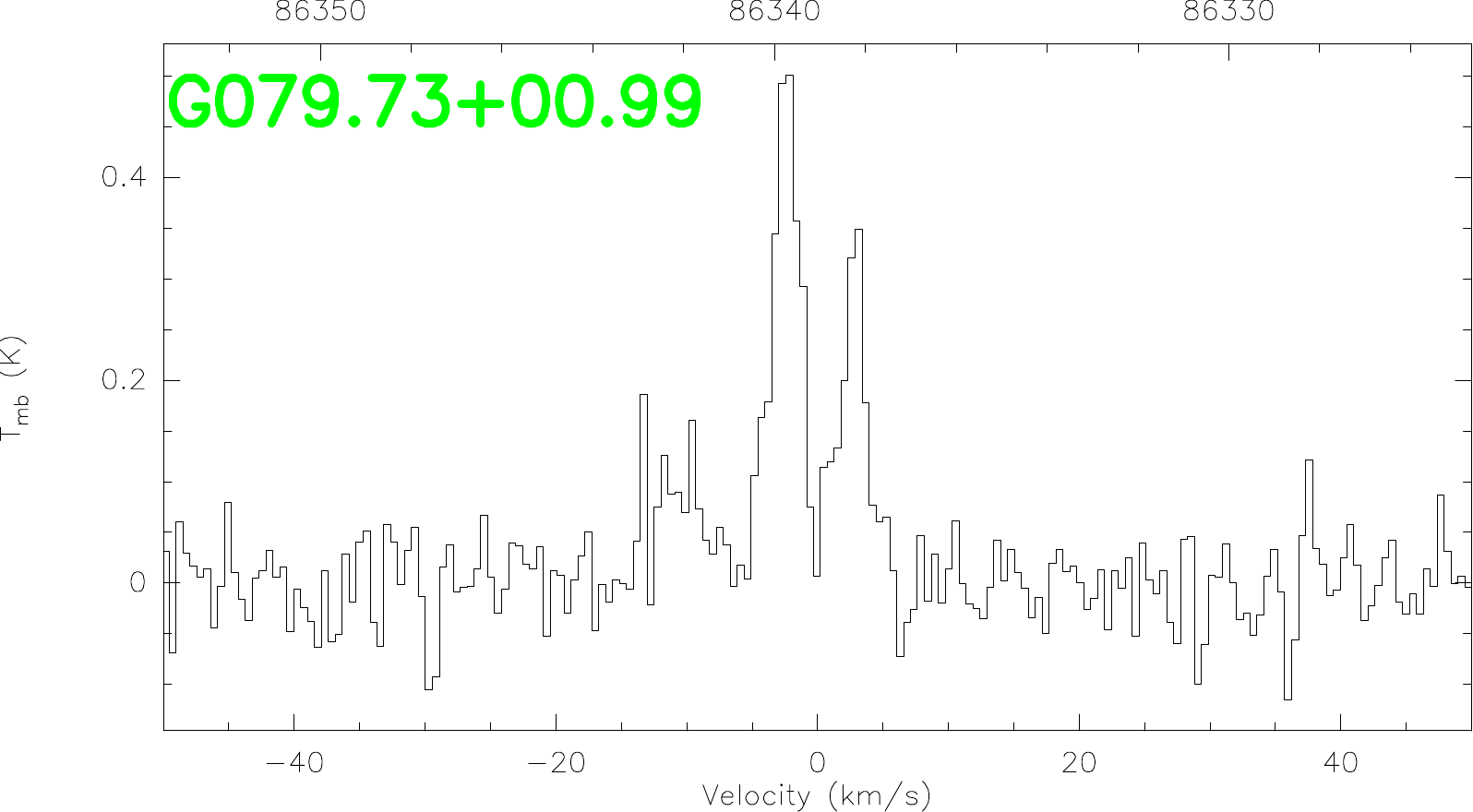}
\includegraphics[width=0.155\textwidth]{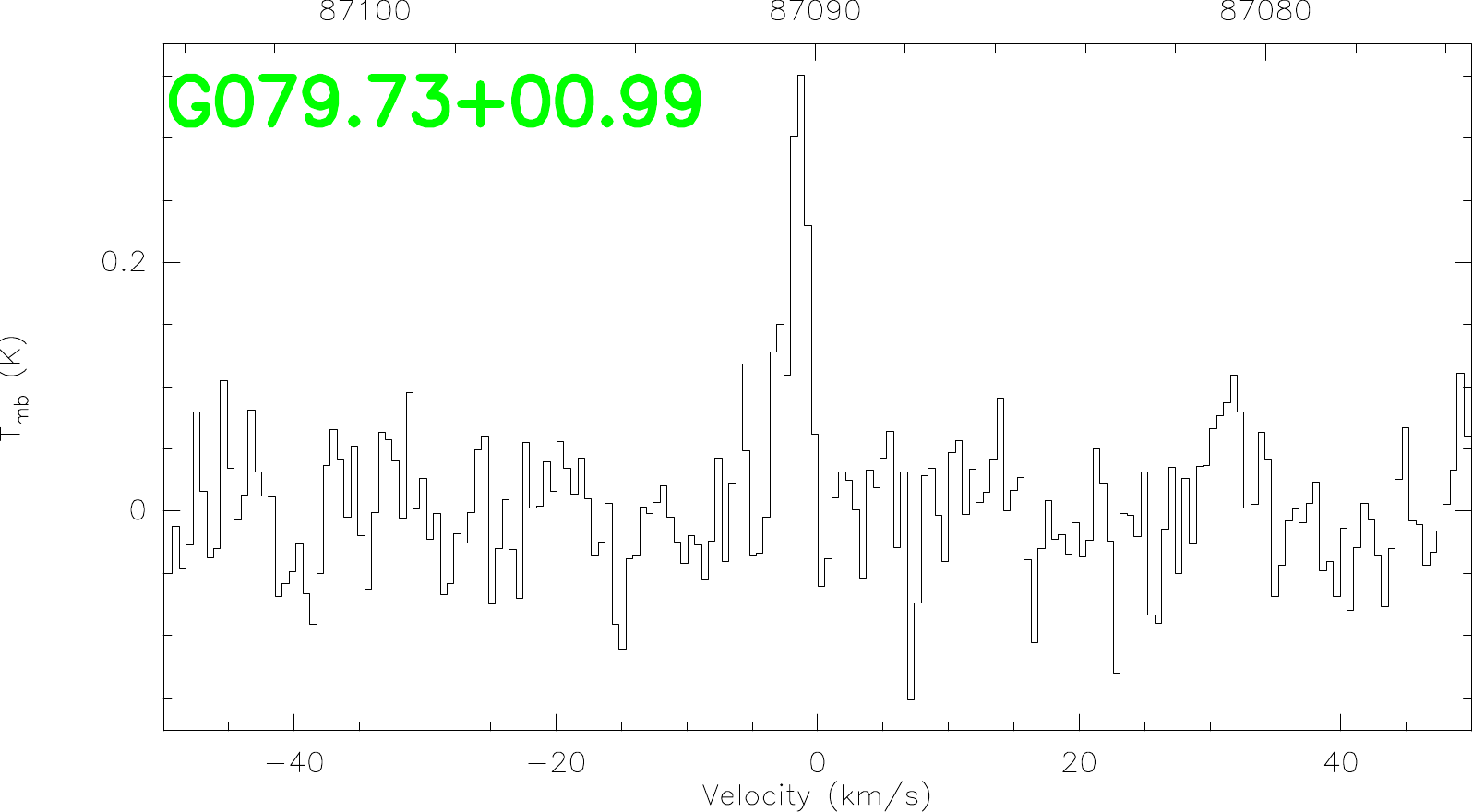}
\includegraphics[width=0.155\textwidth]{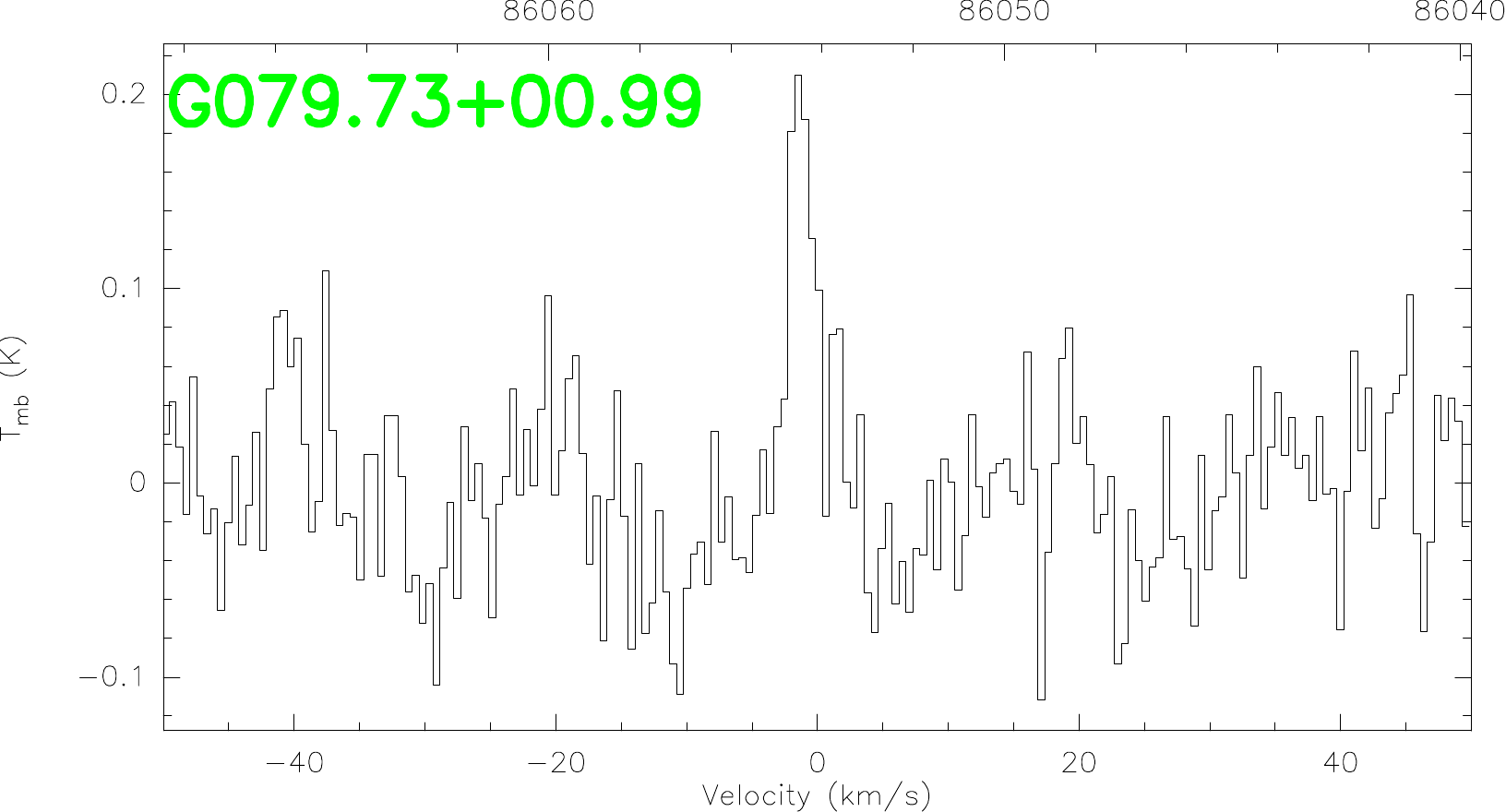}
\includegraphics[width=0.155\textwidth]{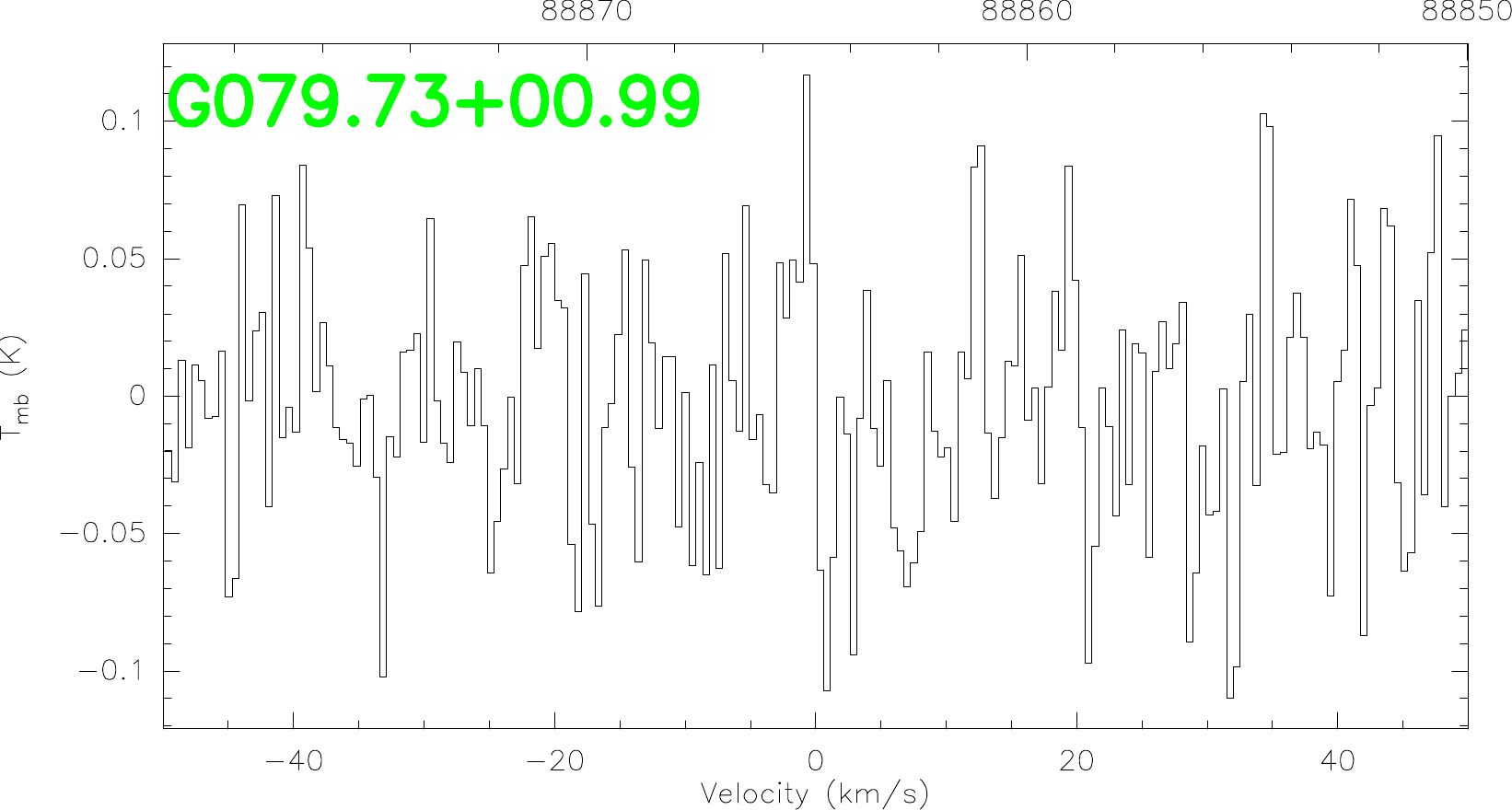}\\

\includegraphics[width=0.155\textwidth]{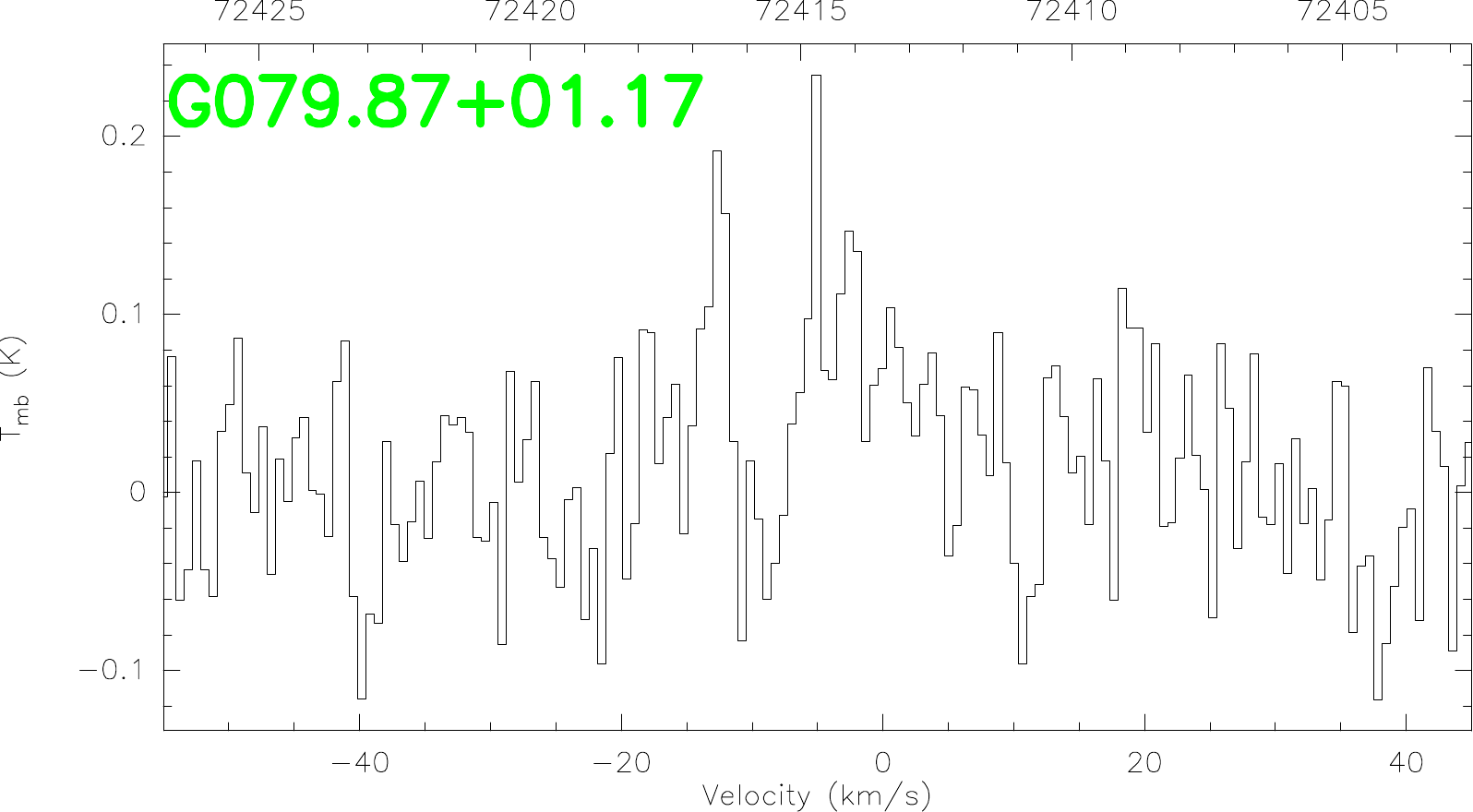}
\includegraphics[width=0.155\textwidth]{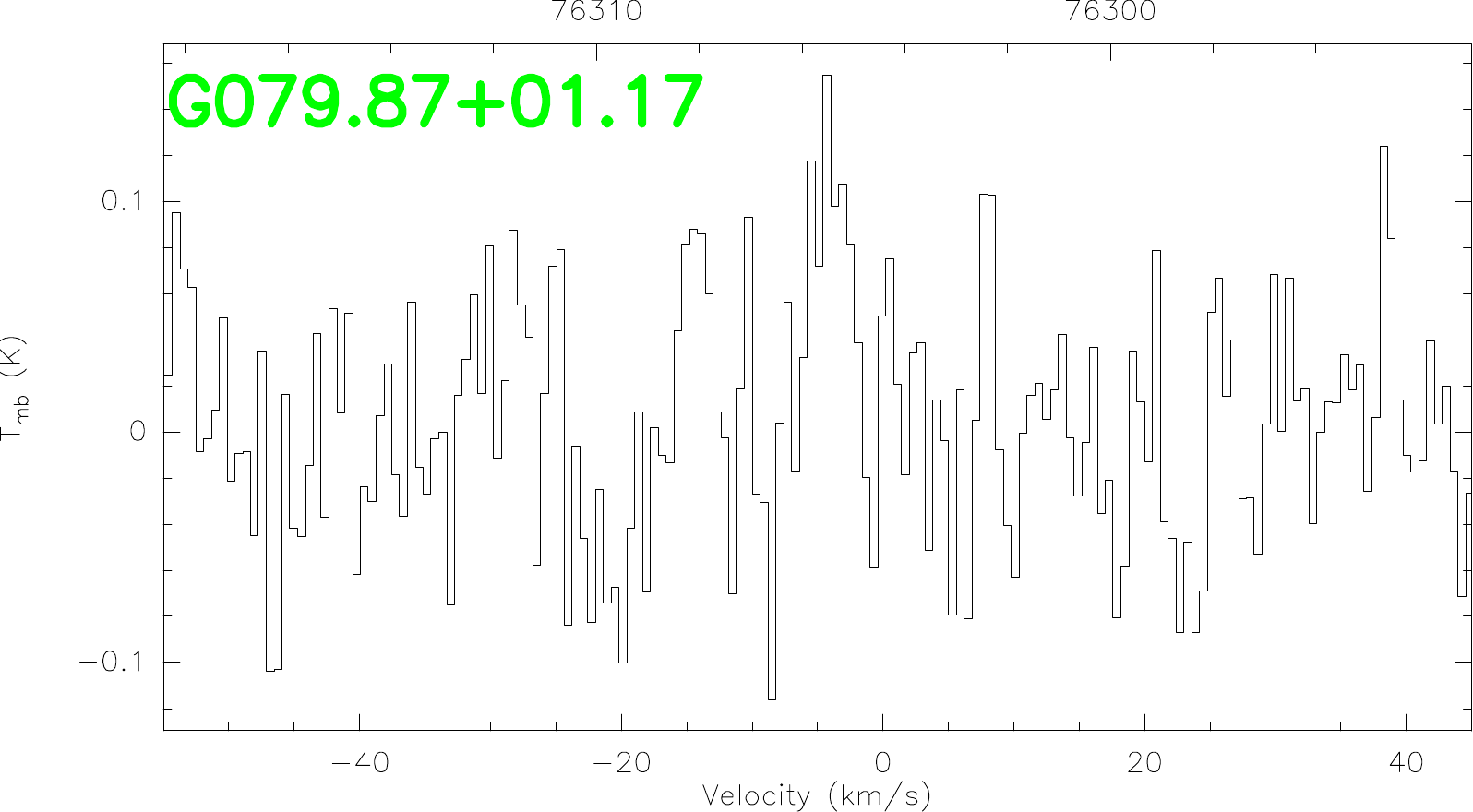}
\includegraphics[width=0.155\textwidth]{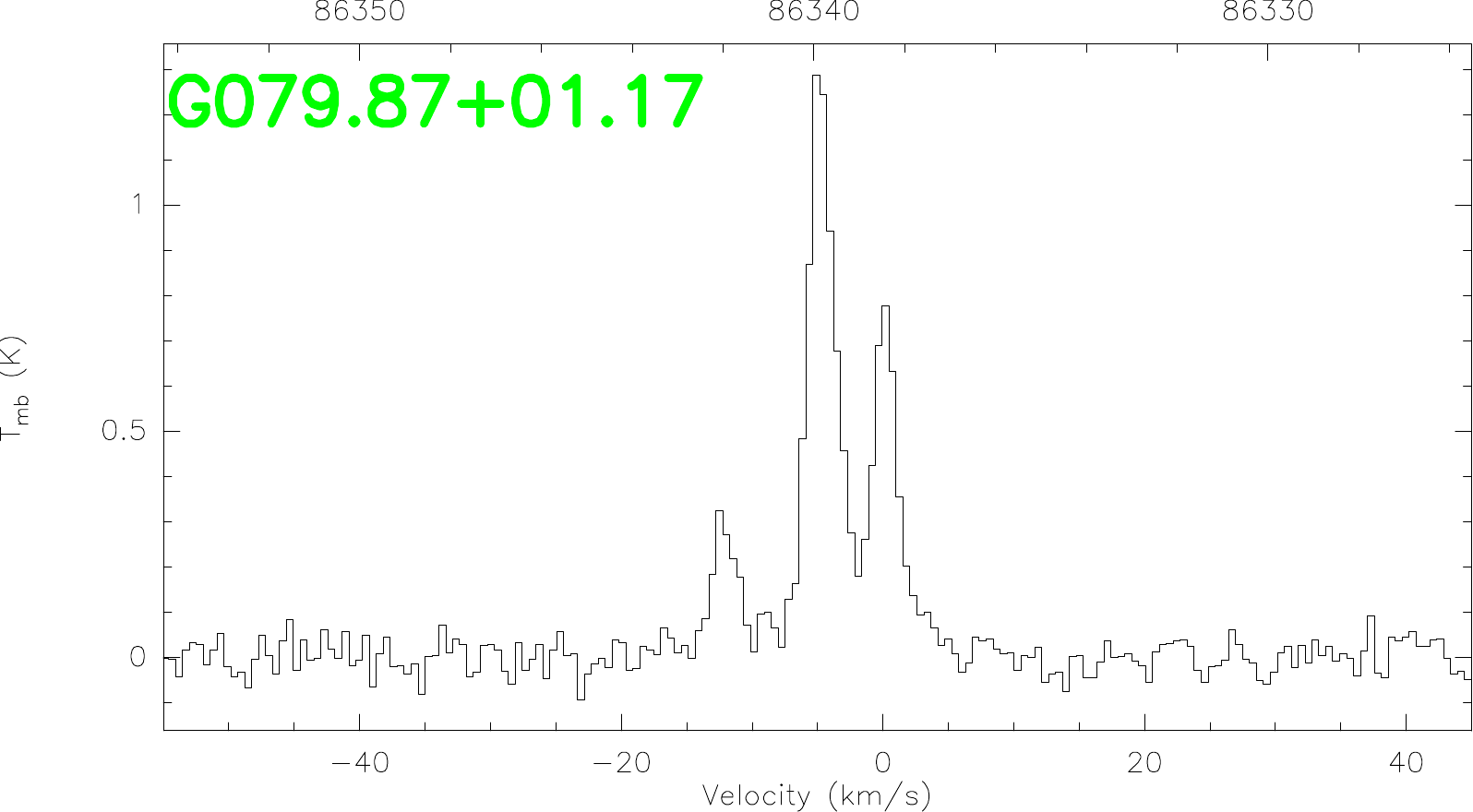}
\includegraphics[width=0.155\textwidth]{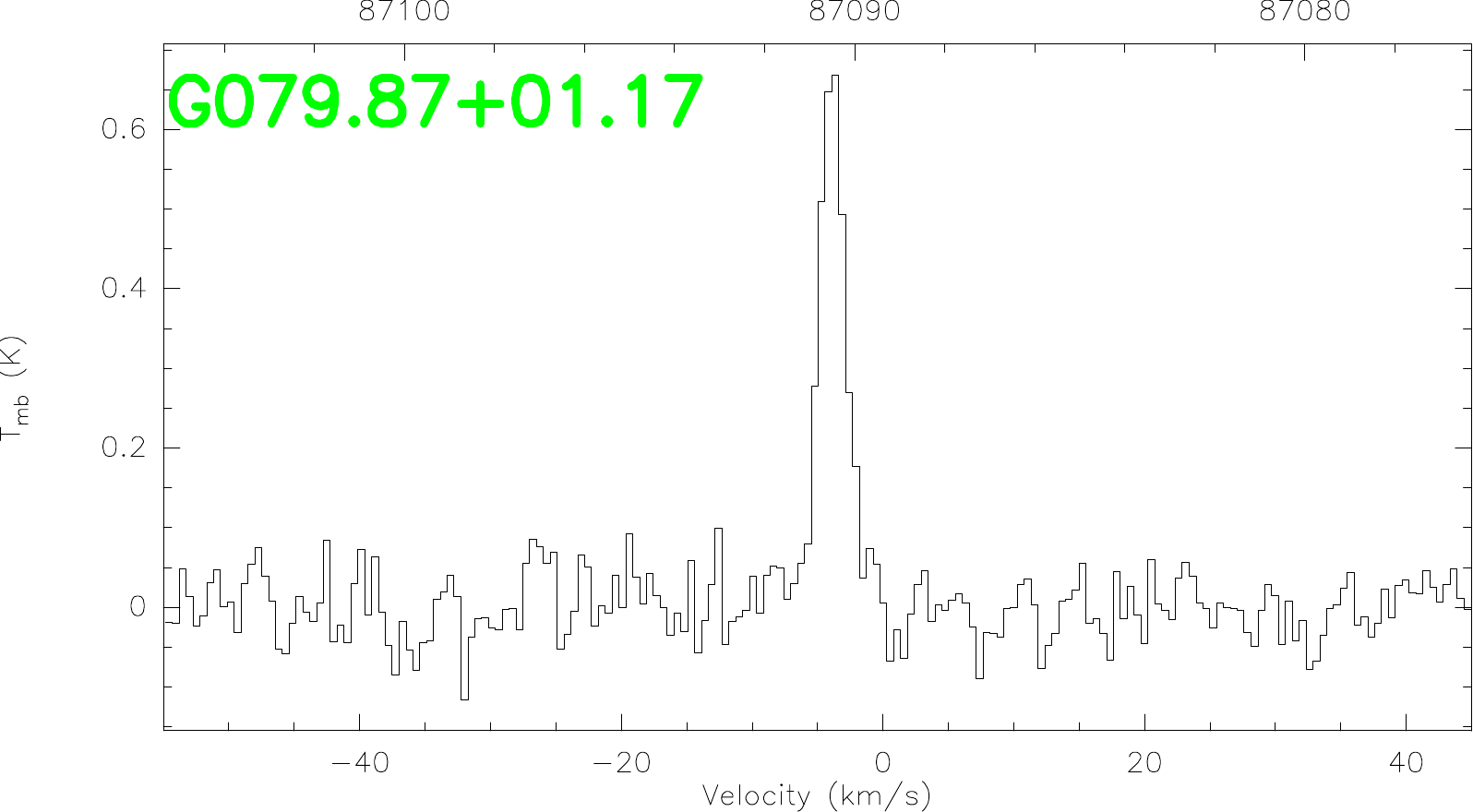}
\includegraphics[width=0.155\textwidth]{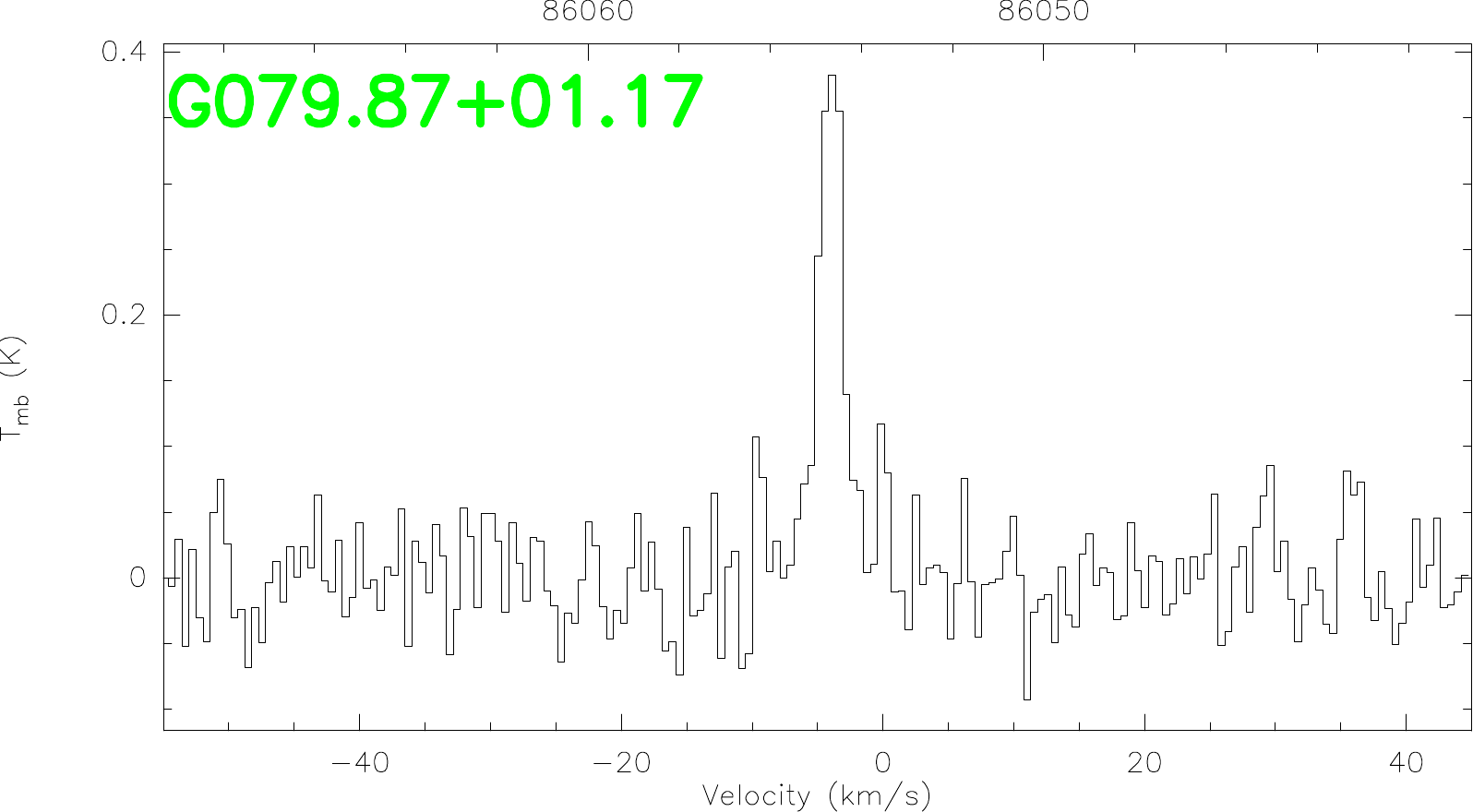}
\includegraphics[width=0.155\textwidth]{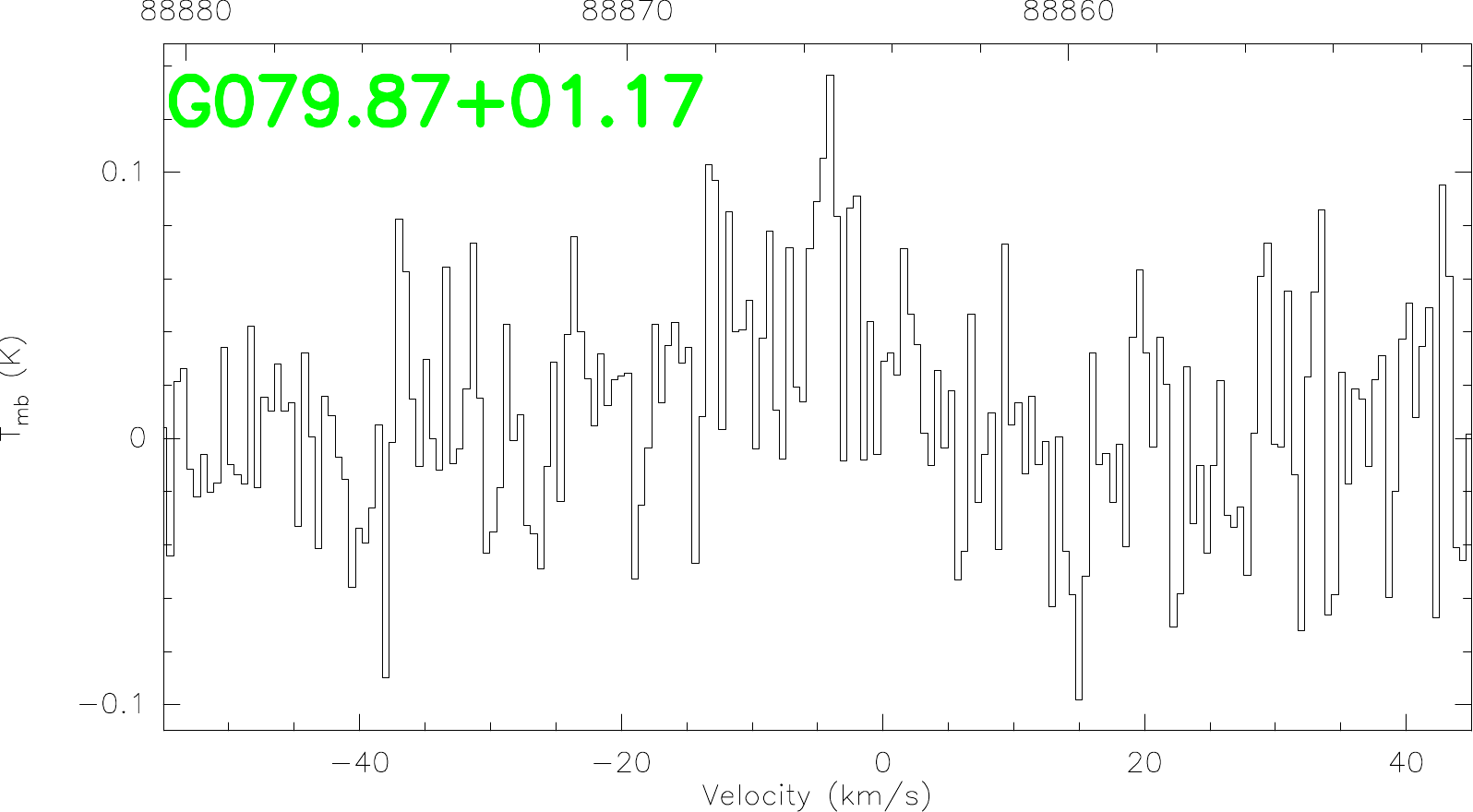}\\

\includegraphics[width=0.155\textwidth]{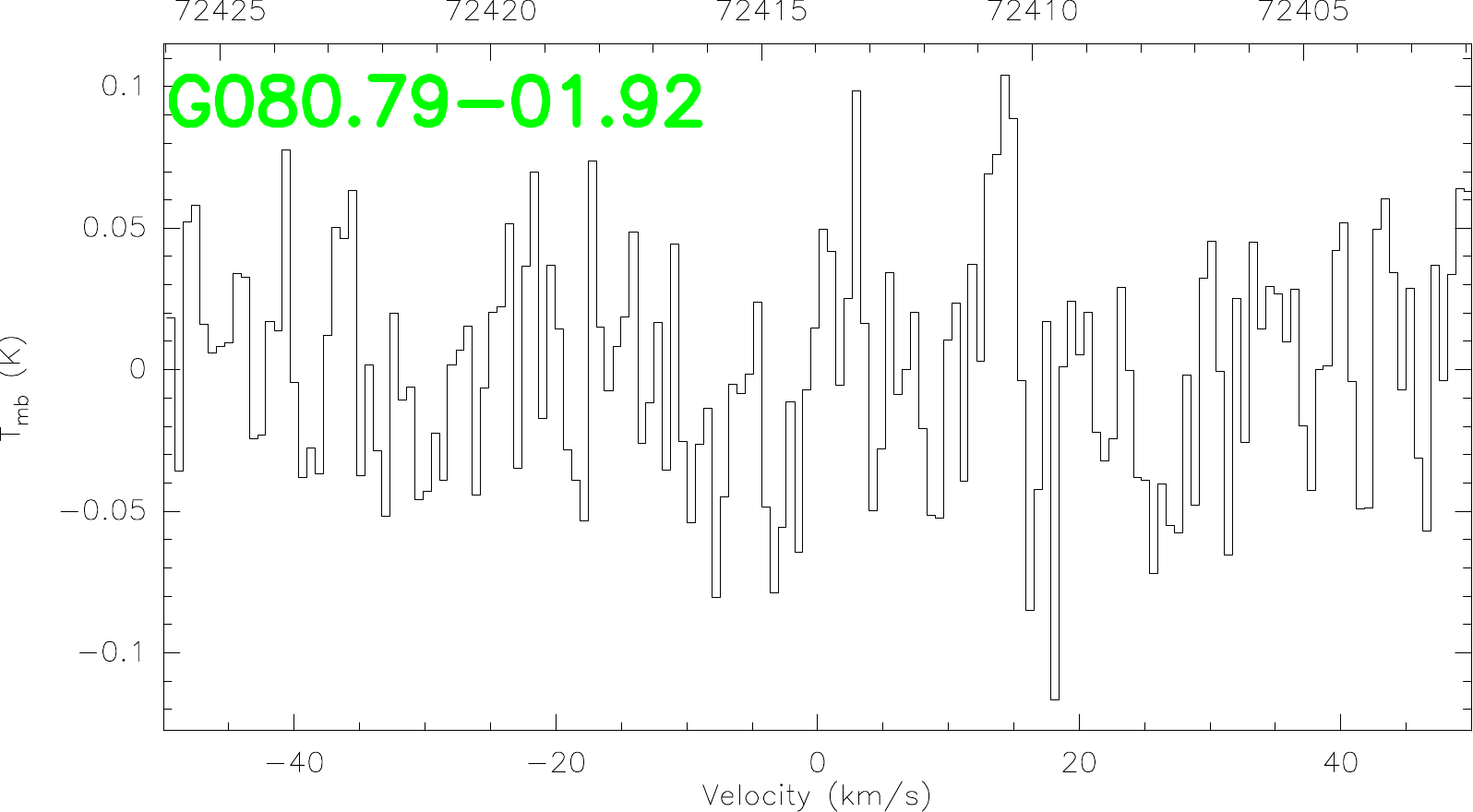}
\includegraphics[width=0.155\textwidth]{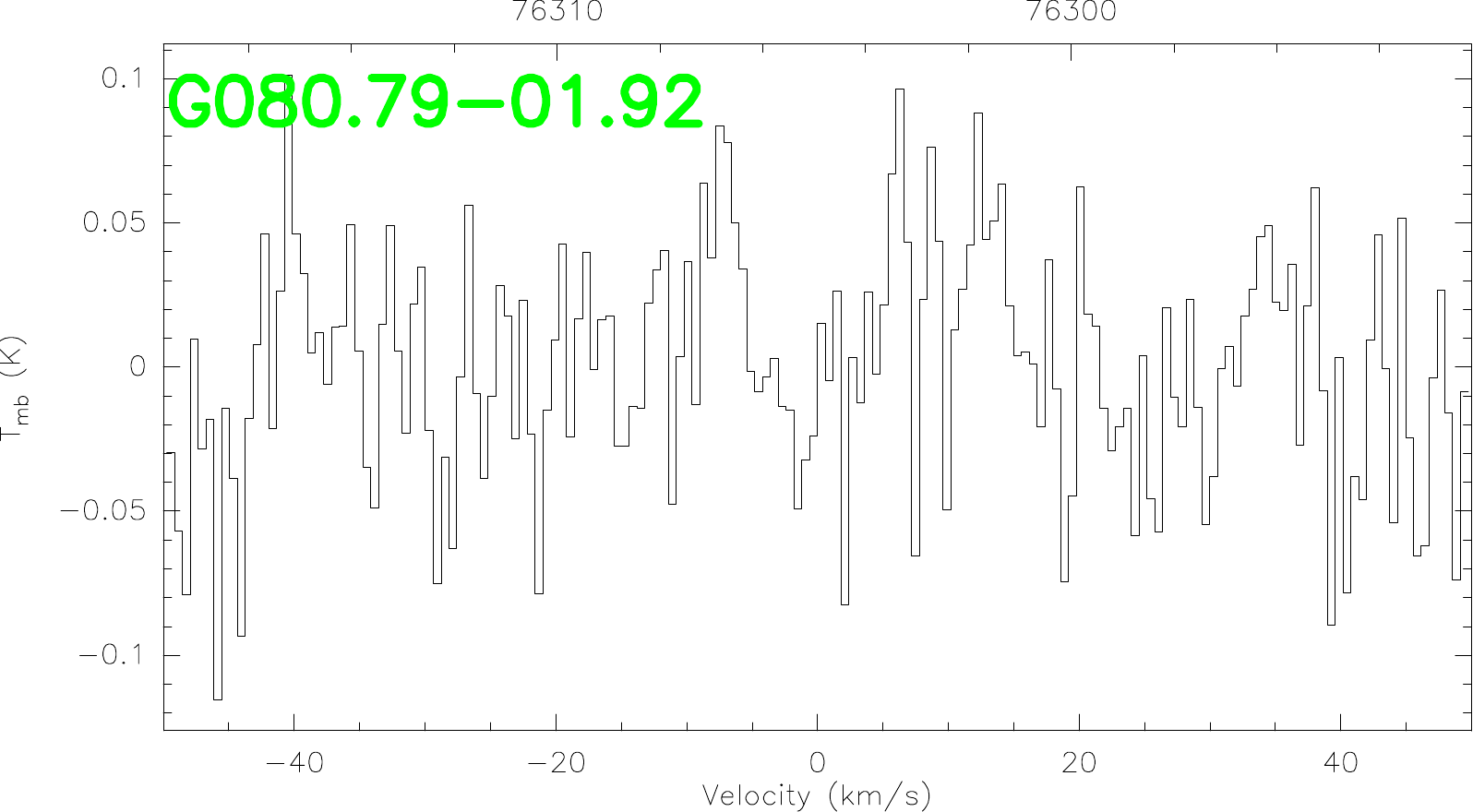}
\includegraphics[width=0.155\textwidth]{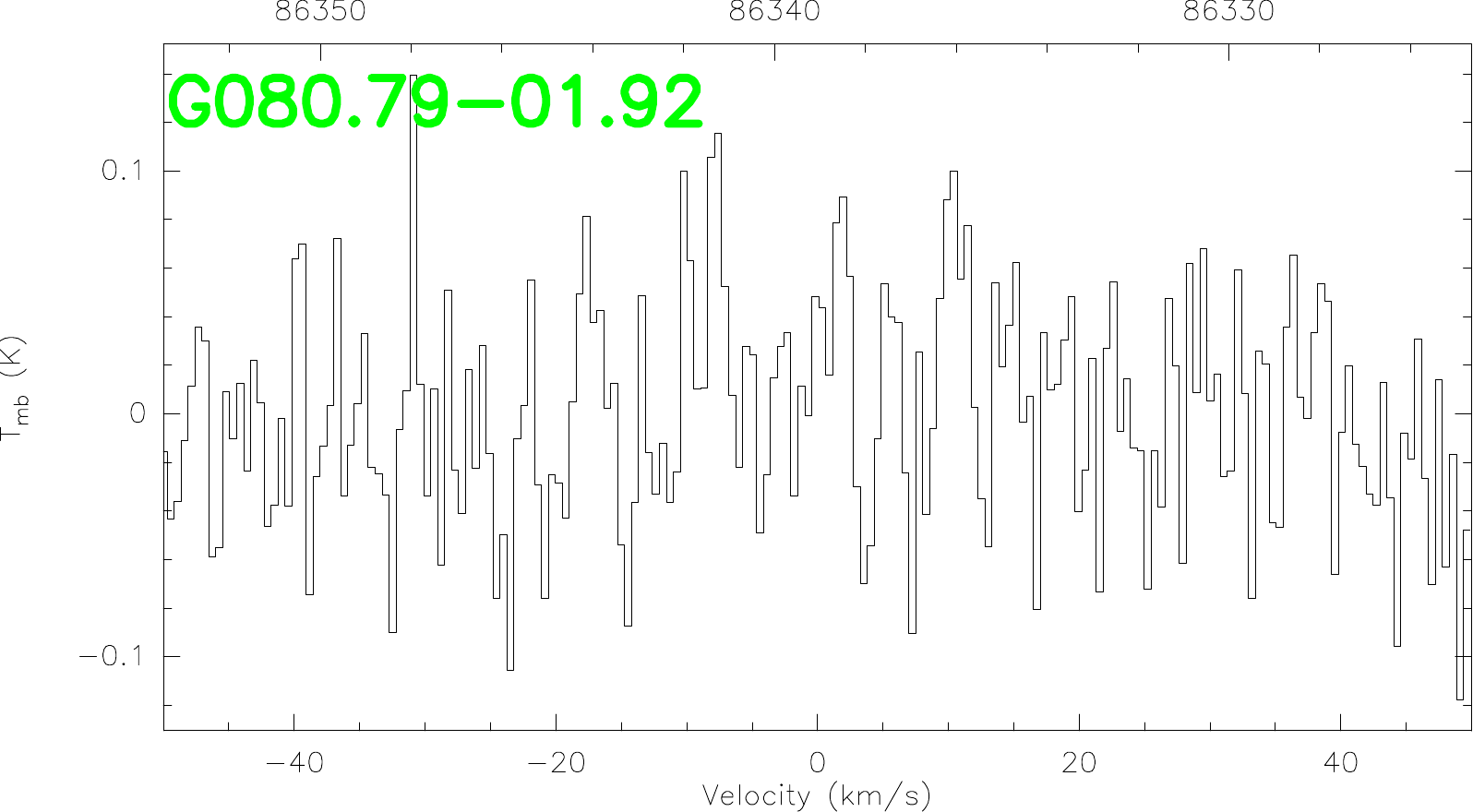}
\includegraphics[width=0.155\textwidth]{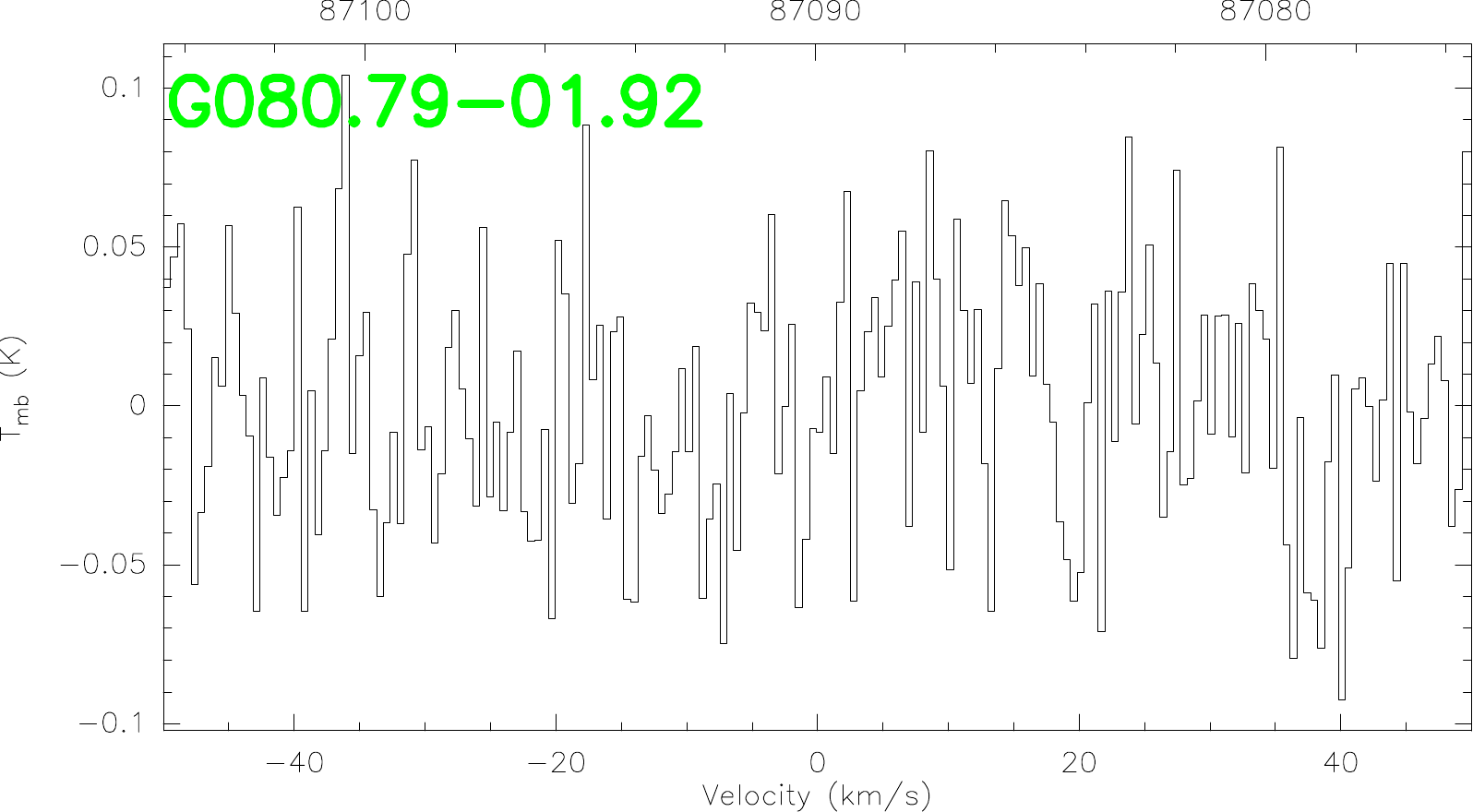}
\includegraphics[width=0.155\textwidth]{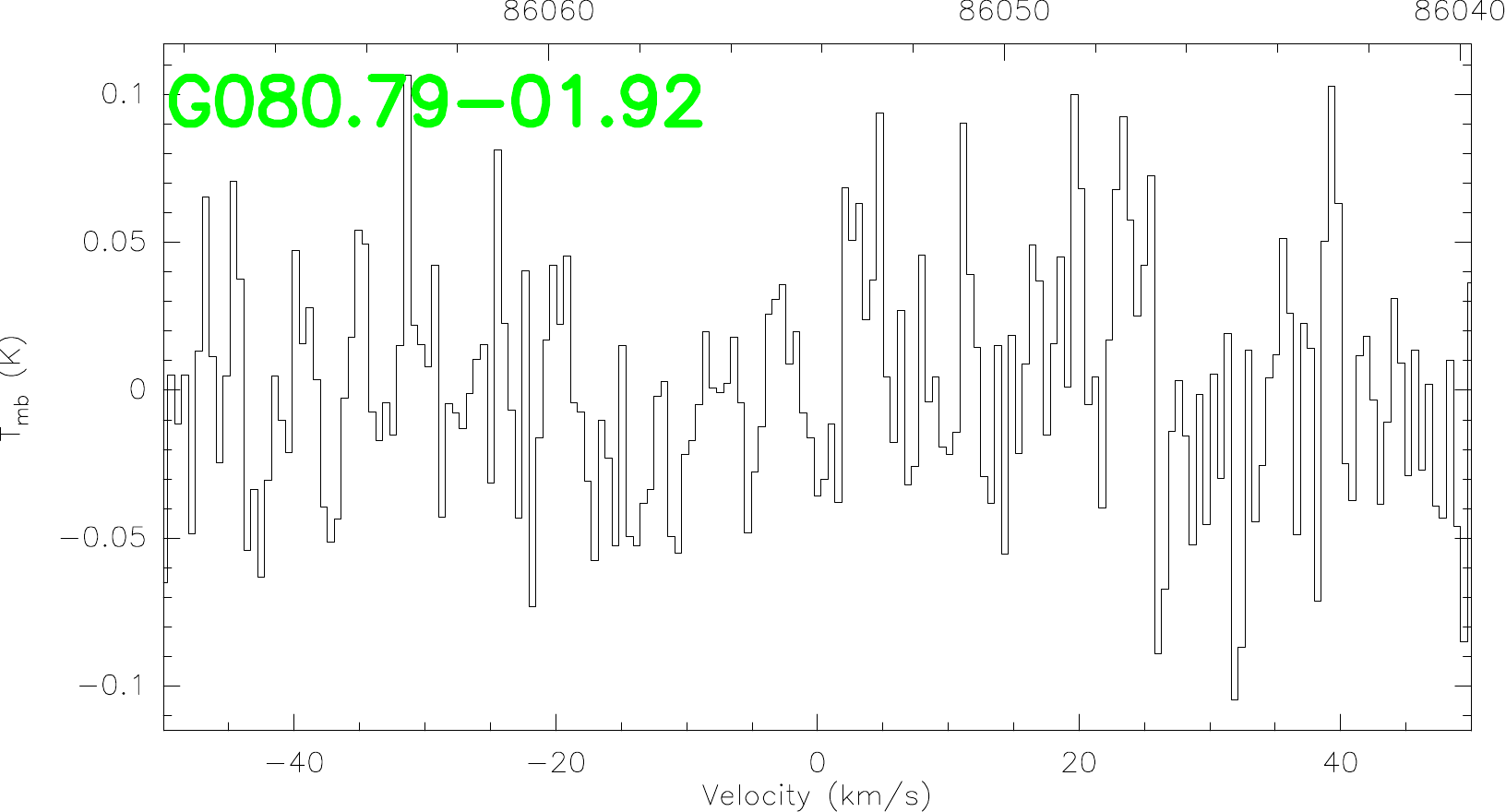}
\includegraphics[width=0.155\textwidth]{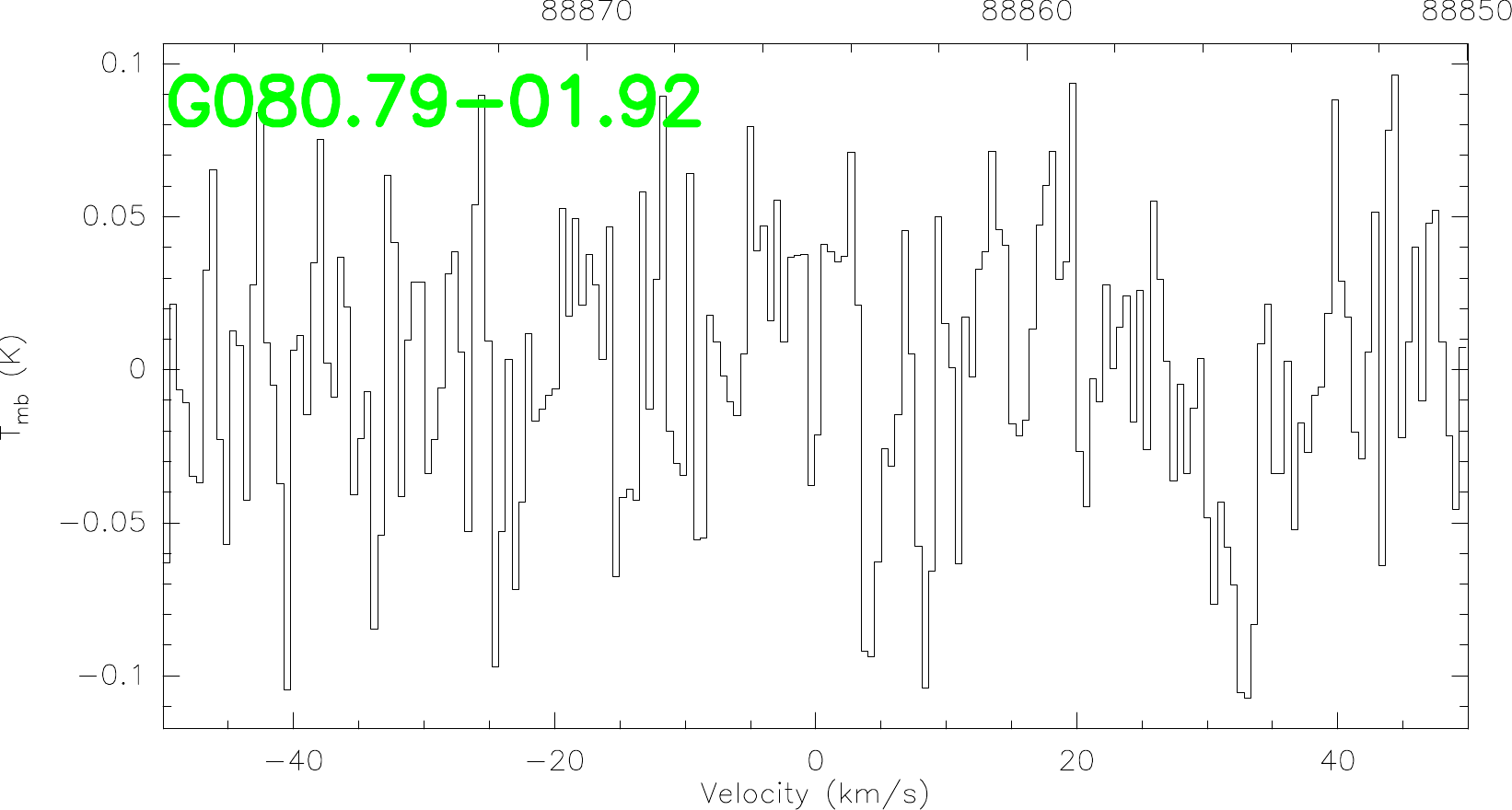}\\

\includegraphics[width=0.155\textwidth]{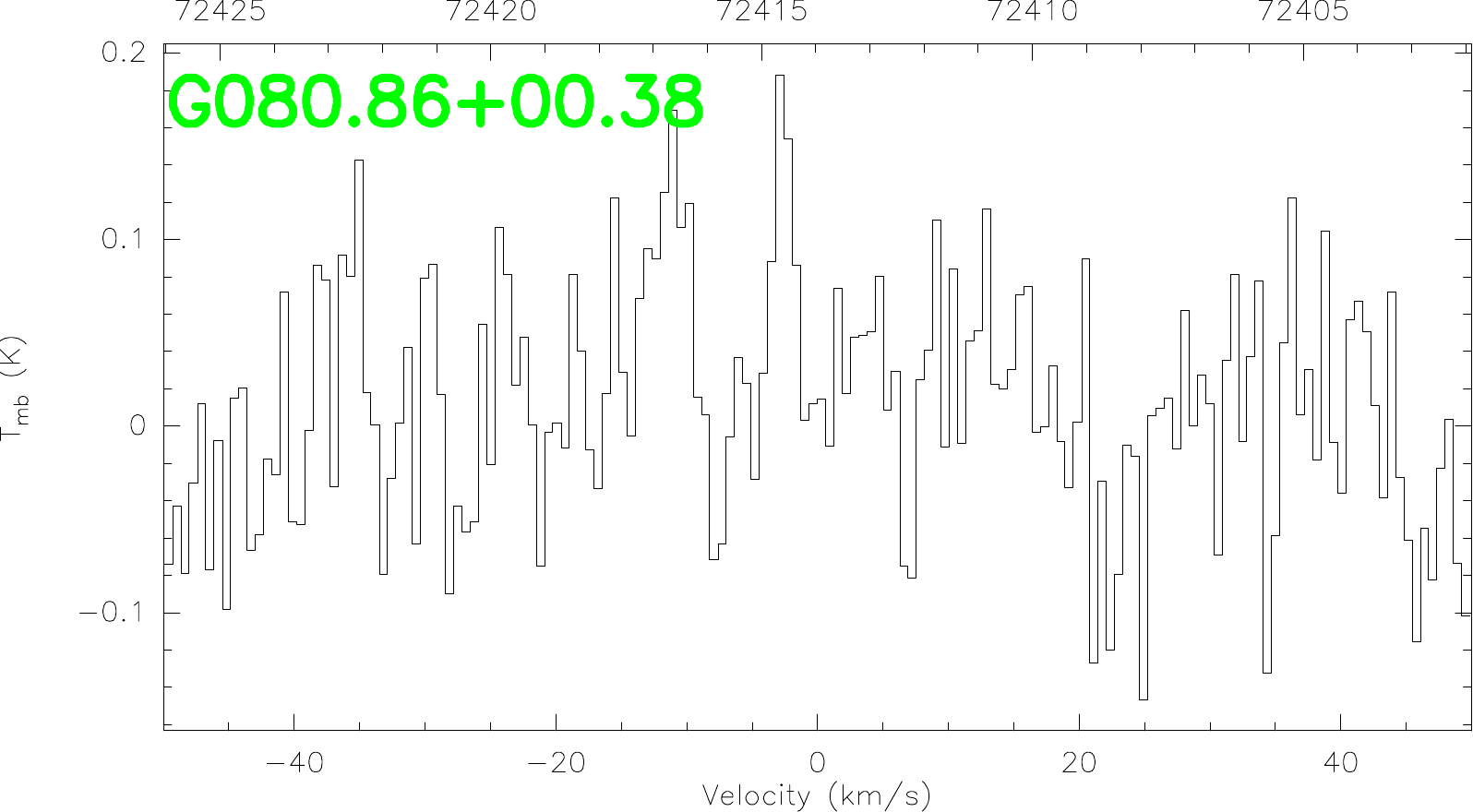}
\includegraphics[width=0.155\textwidth]{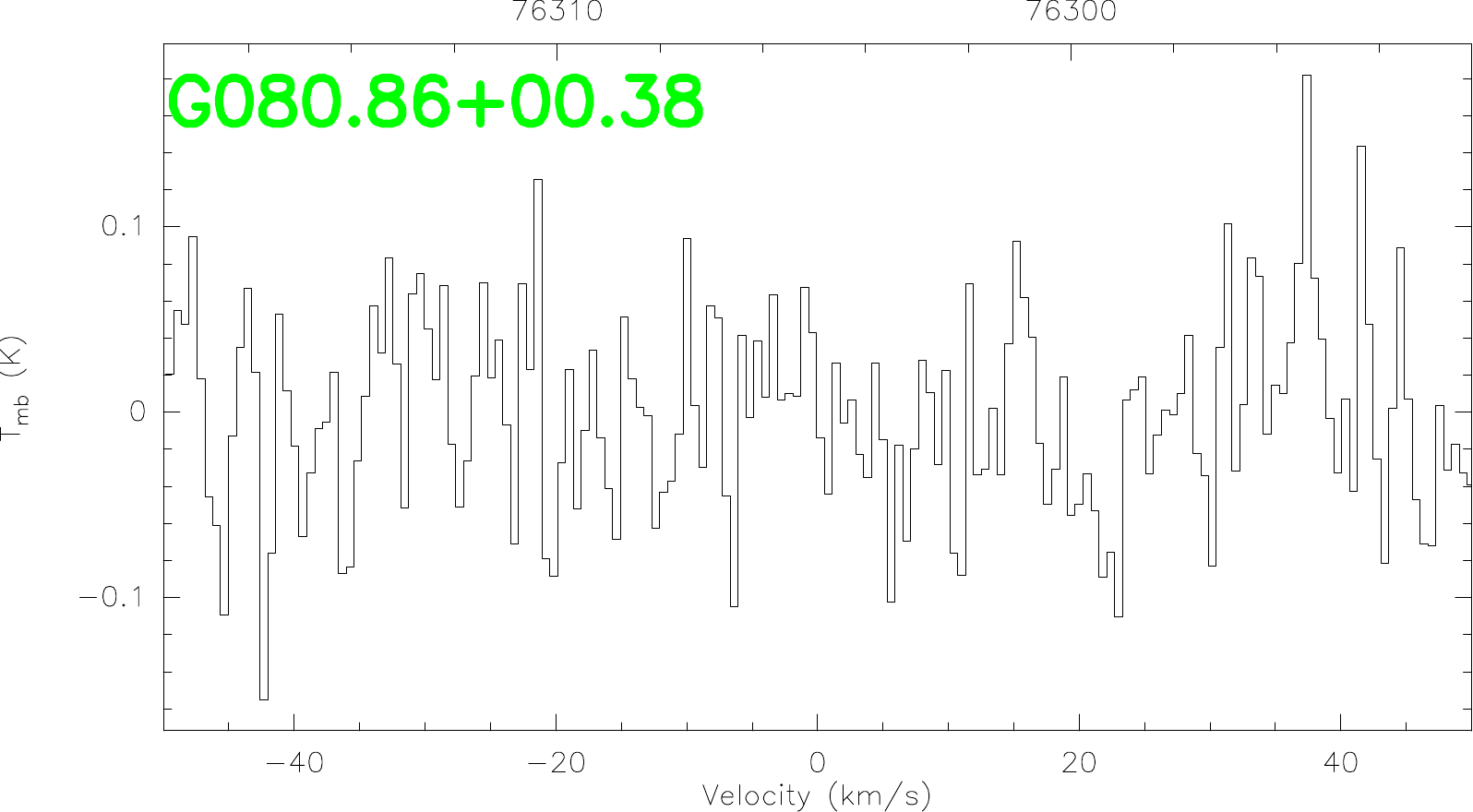}
\includegraphics[width=0.155\textwidth]{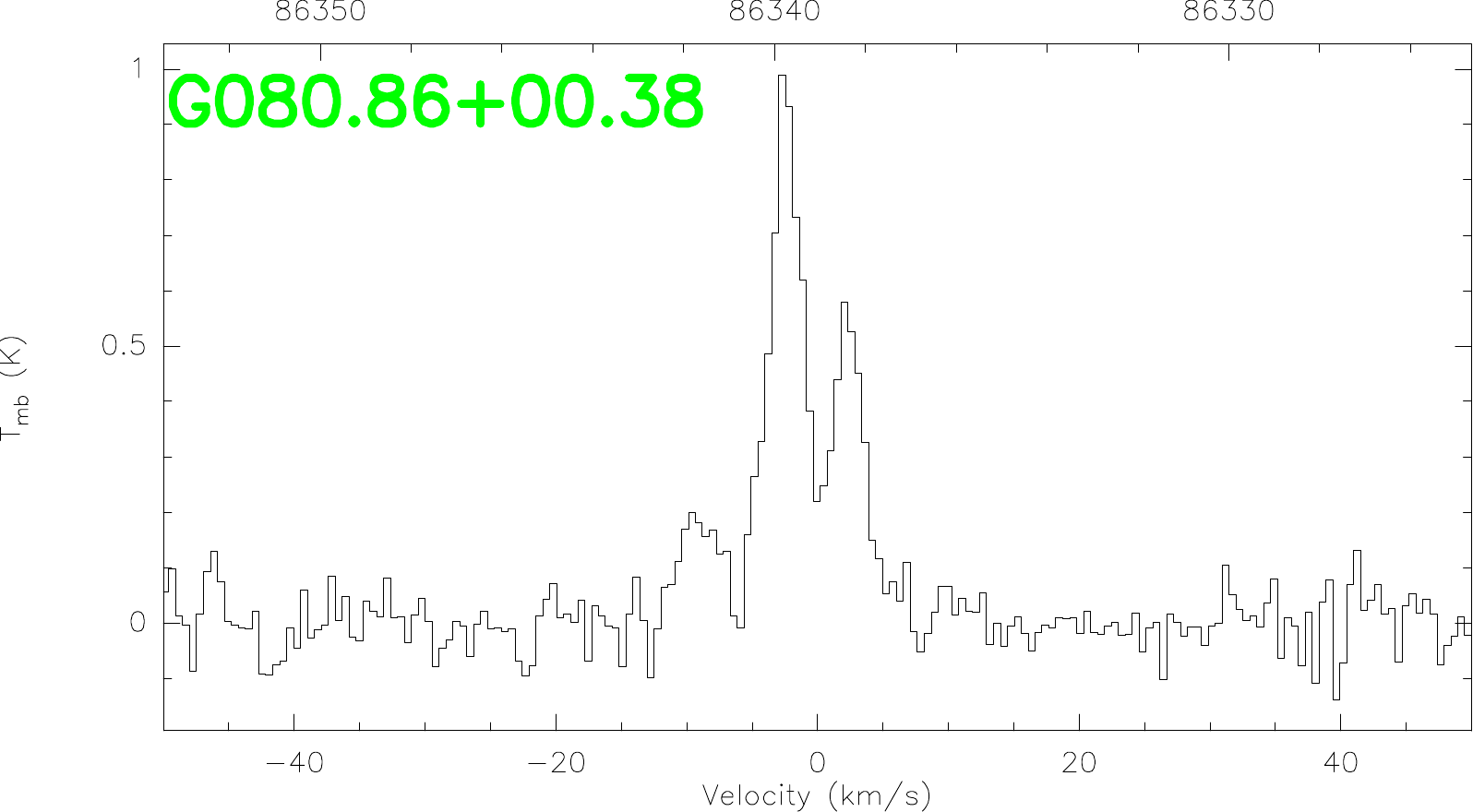}
\includegraphics[width=0.155\textwidth]{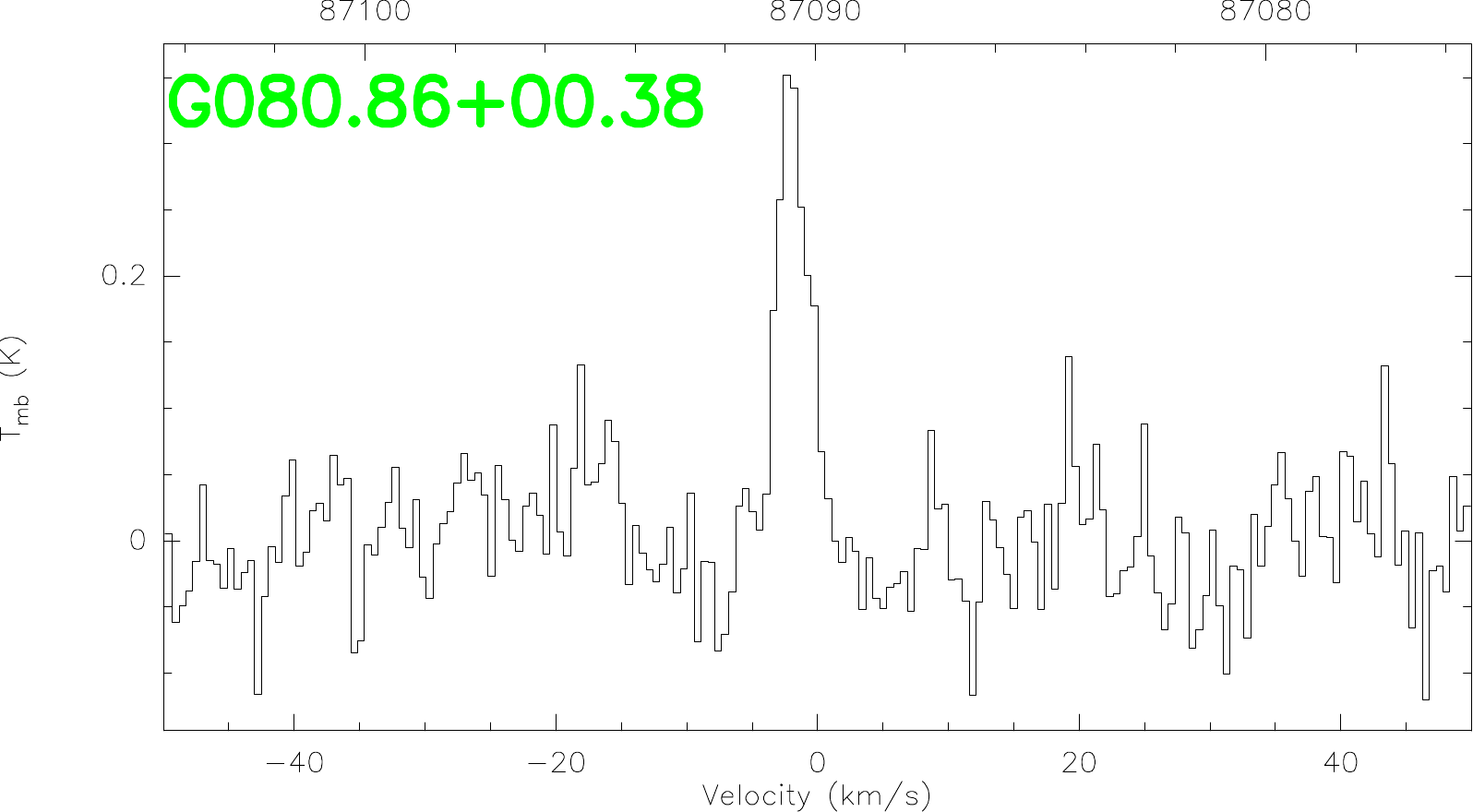}
\includegraphics[width=0.155\textwidth]{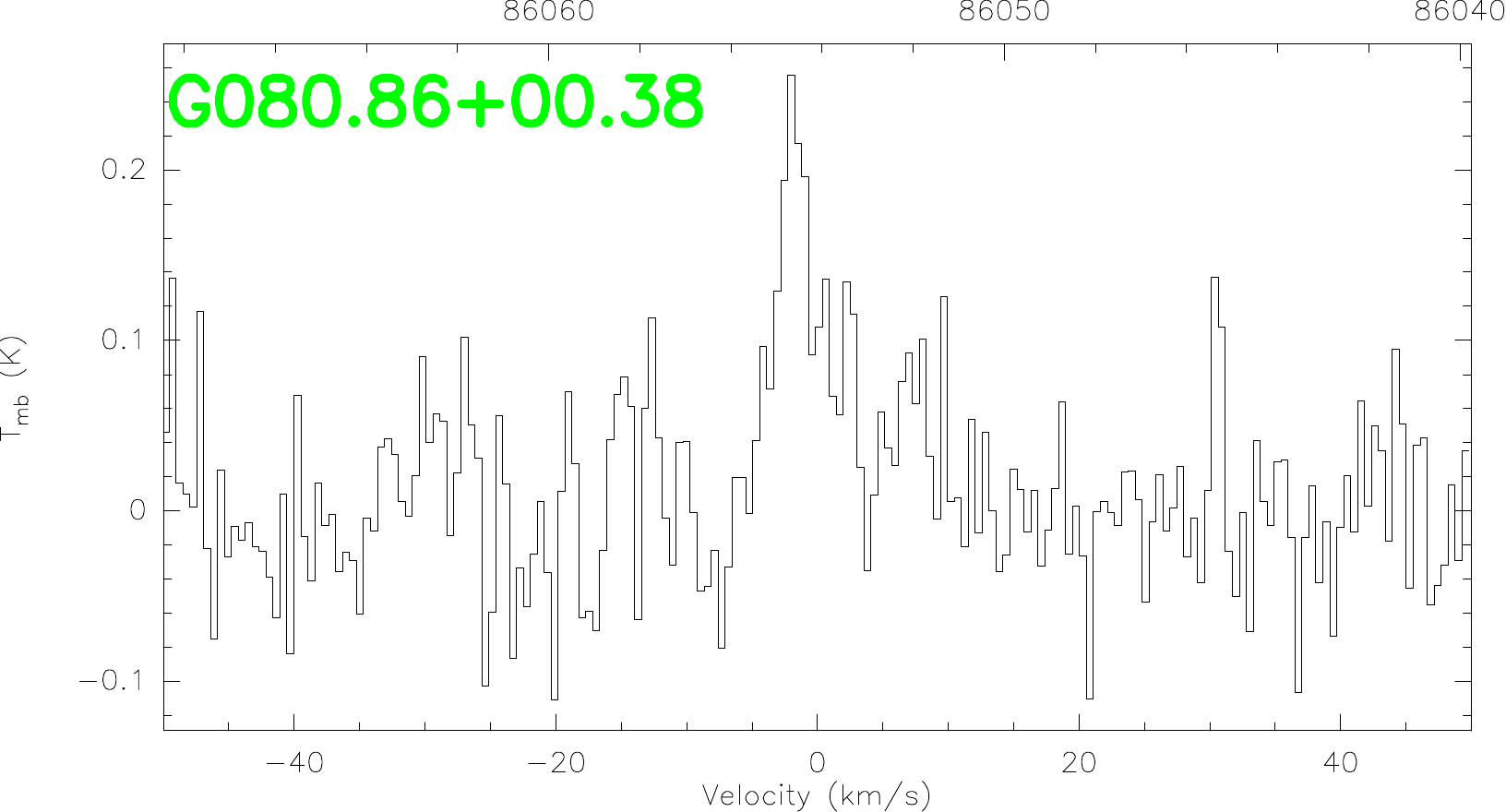}
\includegraphics[width=0.155\textwidth]{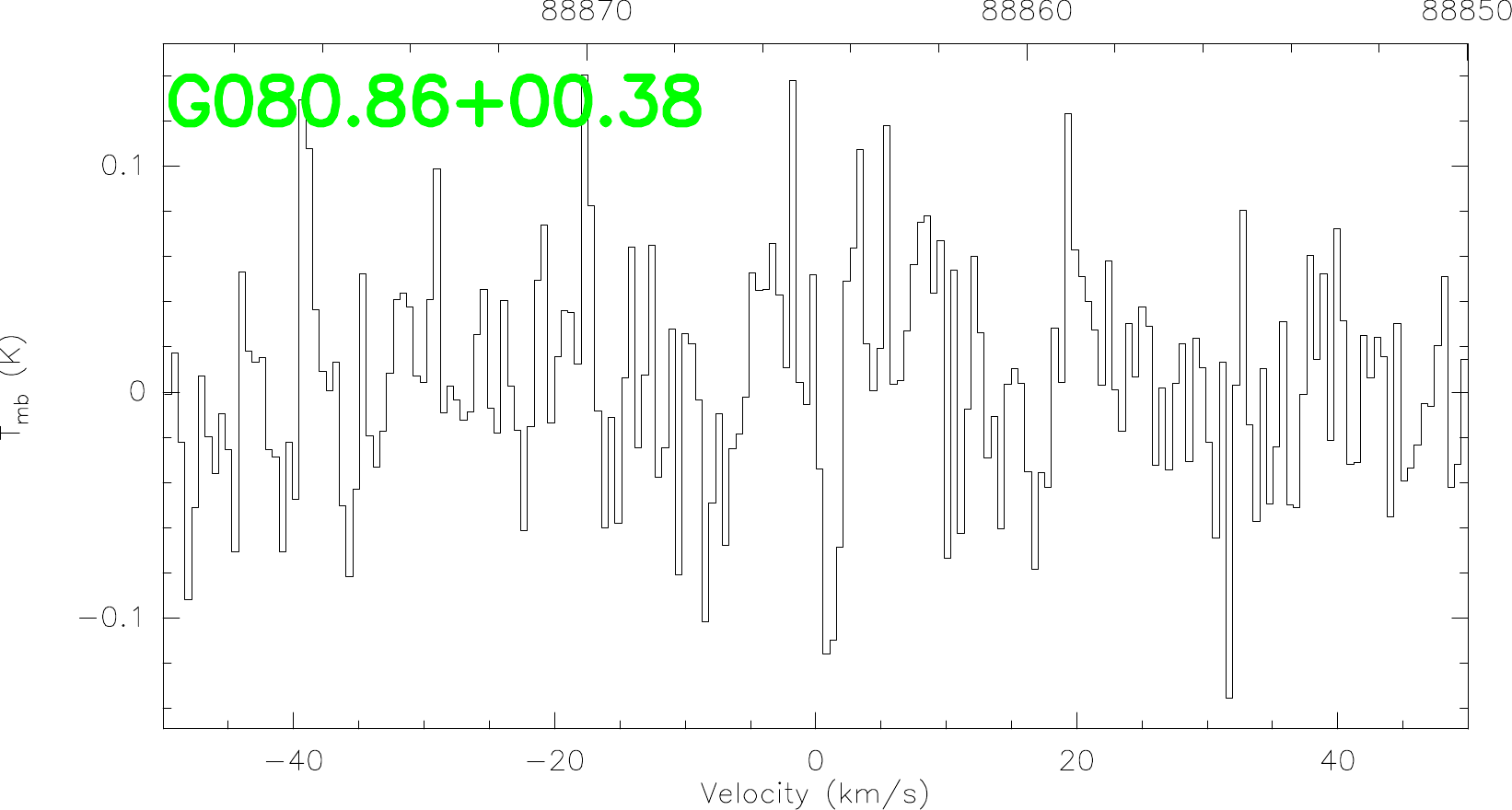}\\

\includegraphics[width=0.155\textwidth]{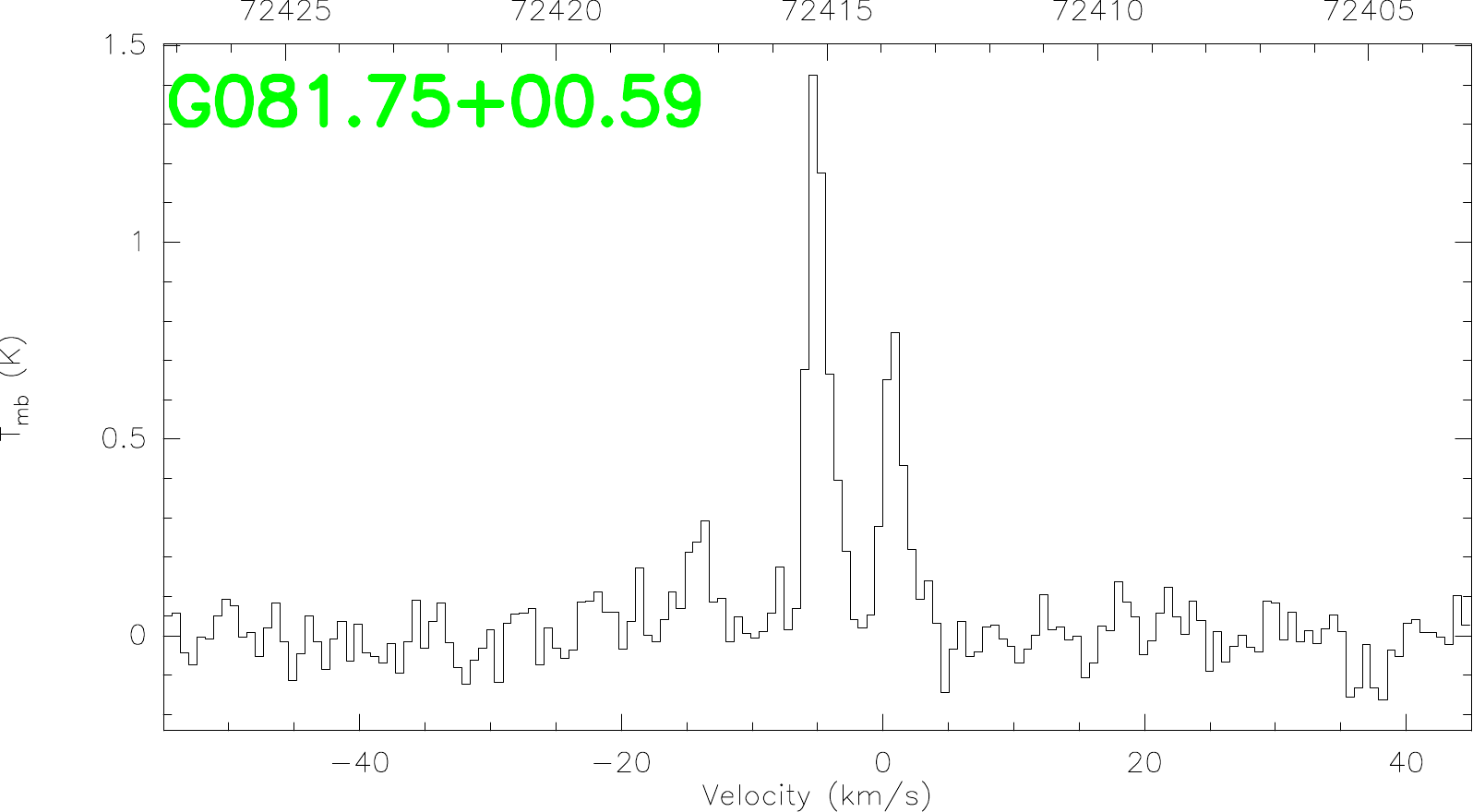}
\includegraphics[width=0.155\textwidth]{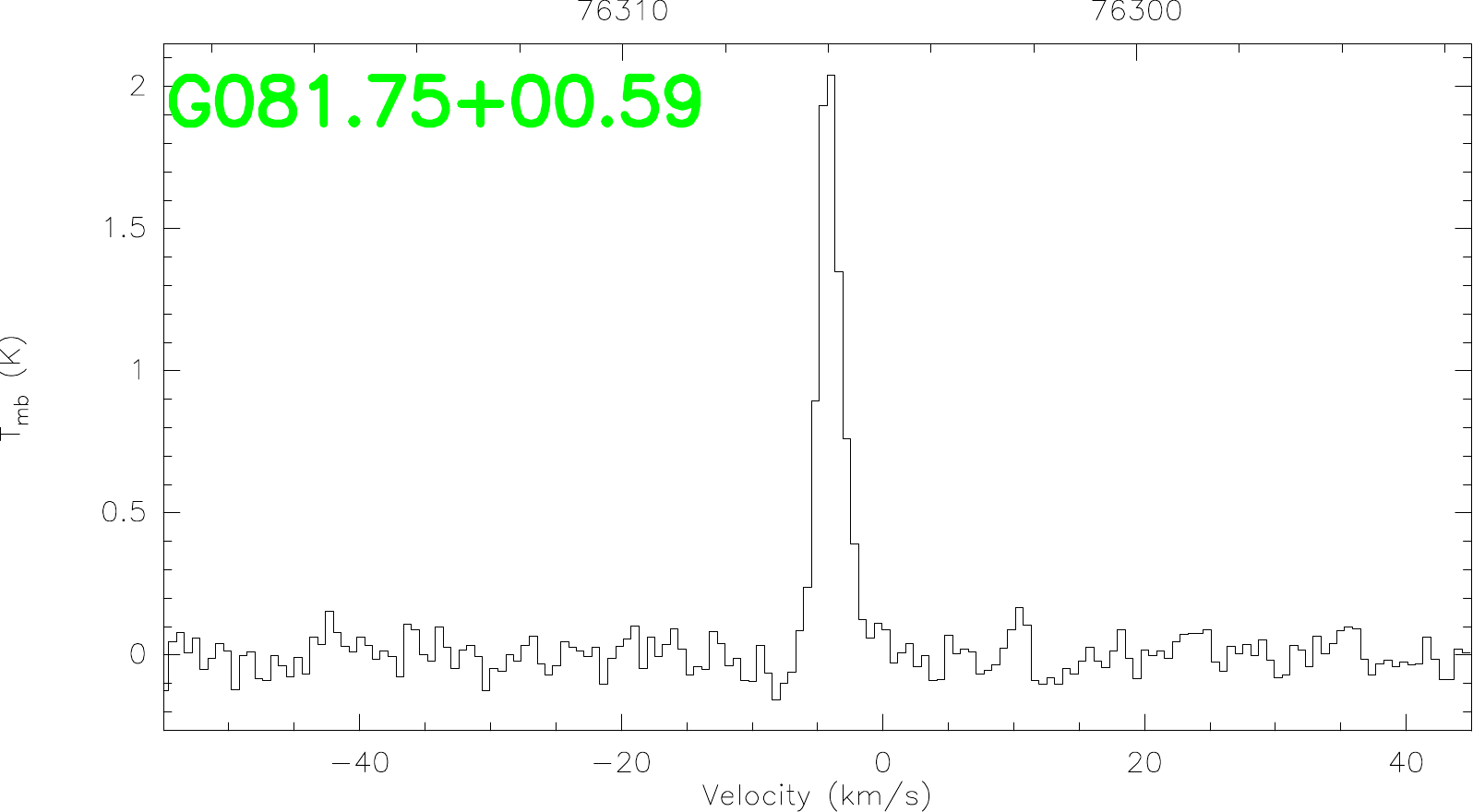}
\includegraphics[width=0.155\textwidth]{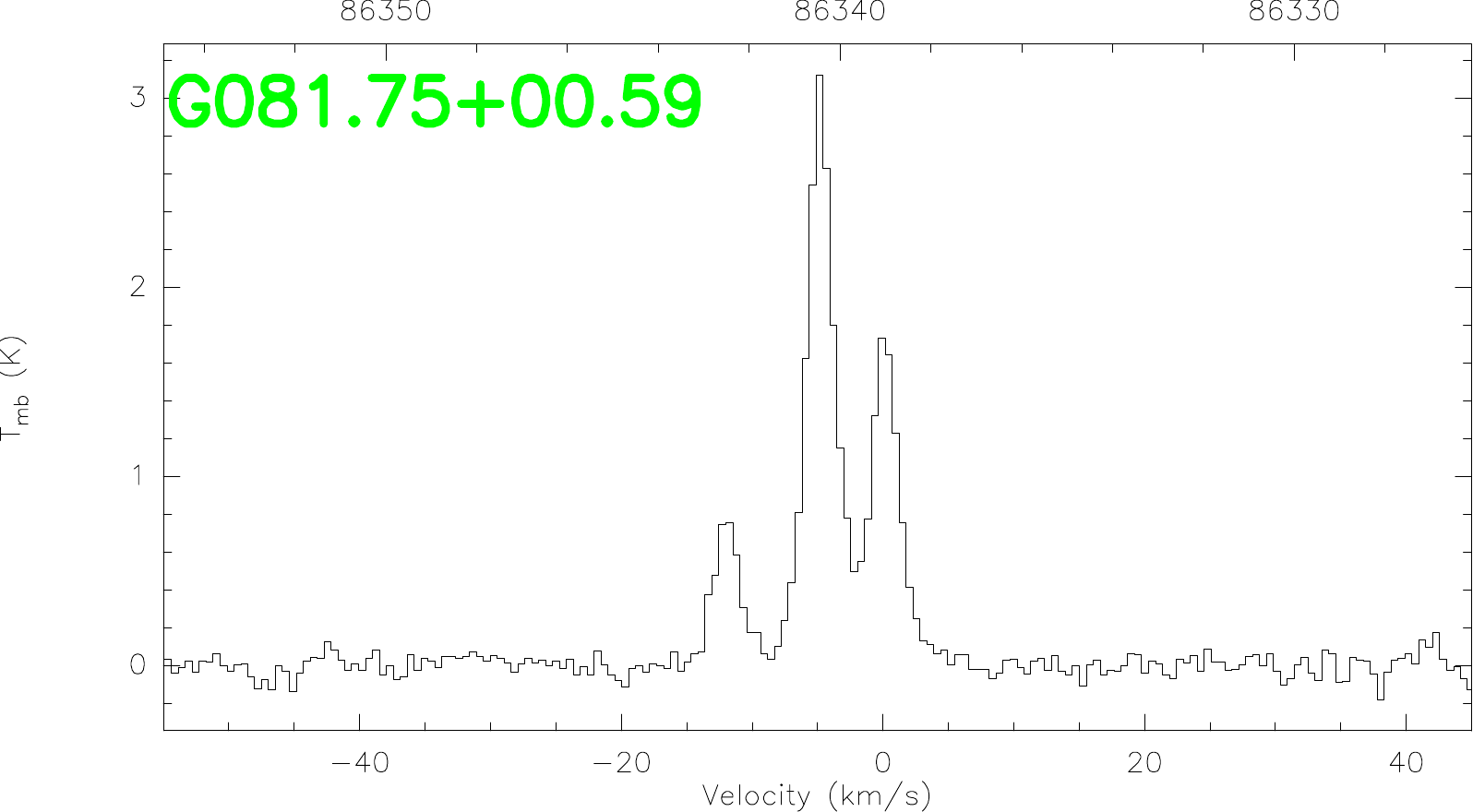}
\includegraphics[width=0.155\textwidth]{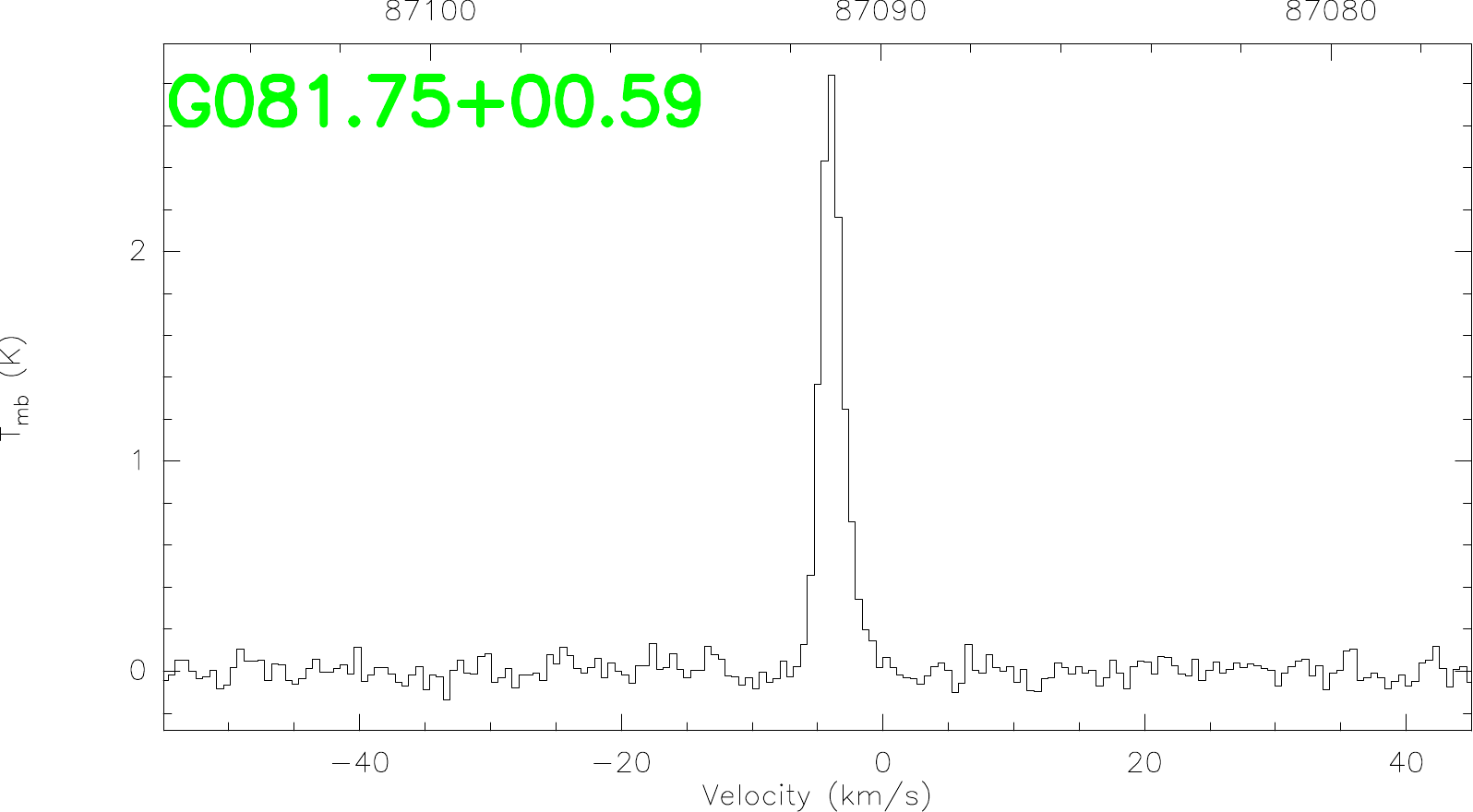}
\includegraphics[width=0.155\textwidth]{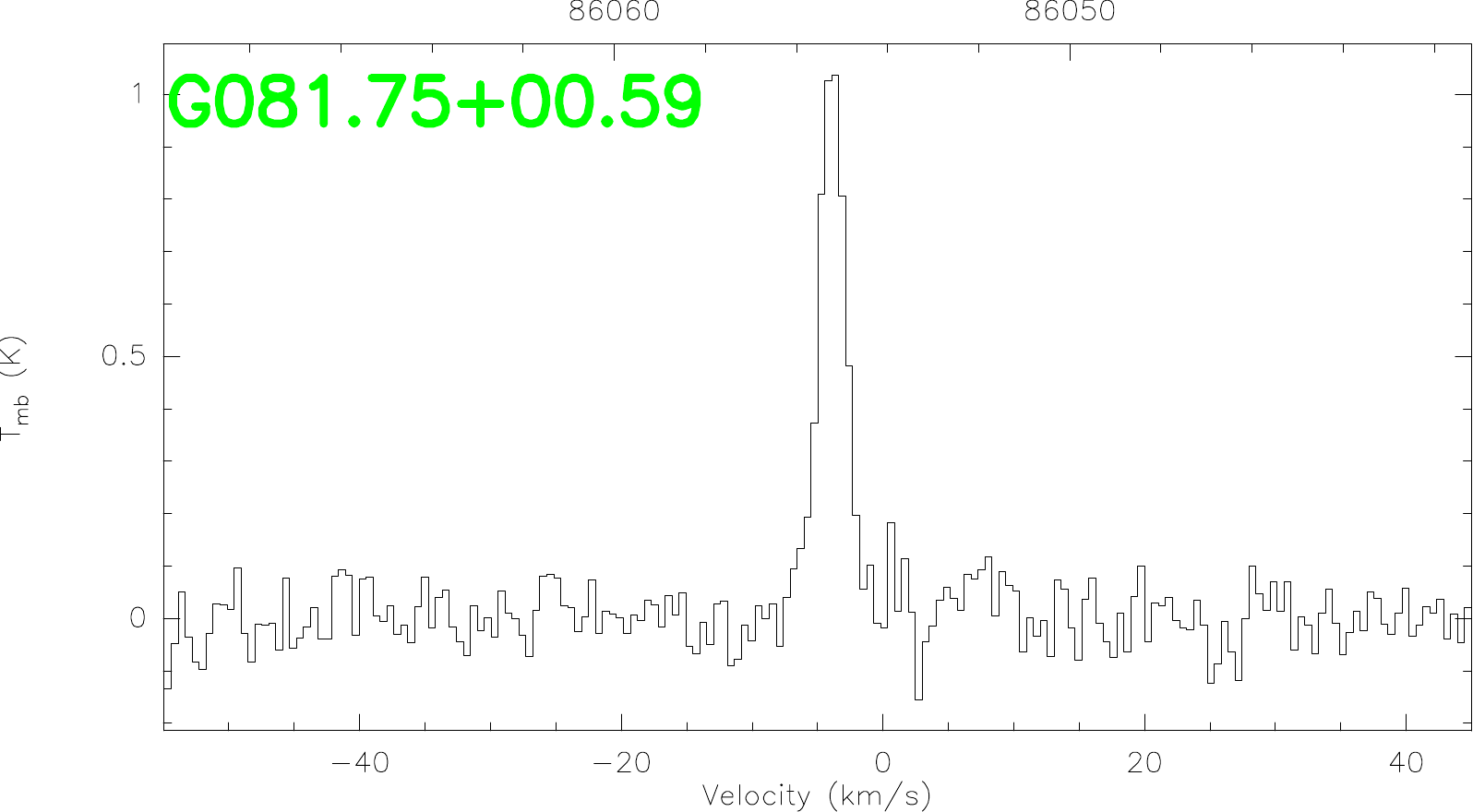}
\includegraphics[width=0.155\textwidth]{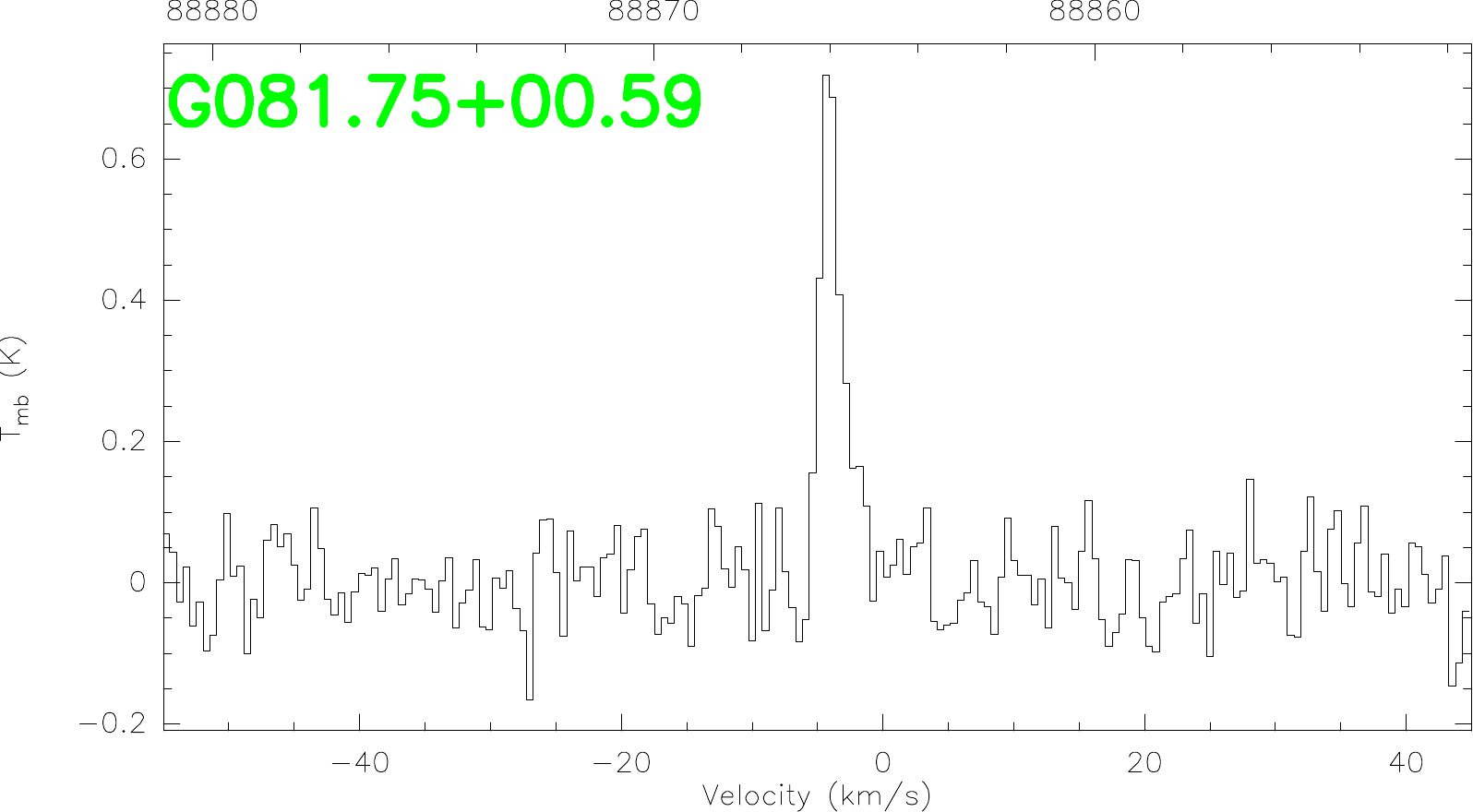}\\

\caption{Continued. Spectra of different molecular lines.}
\label{fig:multi_lines_page4}
\end{figure}

\begin{figure}[h]
\centering

\makebox[\colw][c]{DCN 1-0}
\makebox[\colw][c]{DNC 1-0}
\makebox[\colw][c]{H$^{13}$CN 1-0}
\makebox[\colw][c]{HN$^{13}$C 1-0}
\makebox[\colw][c]{HC$^{15}$N 1-0}
\makebox[\colw][c]{H$^{15}$NC 1-0}\\[0.2em]

\includegraphics[width=0.155\textwidth]{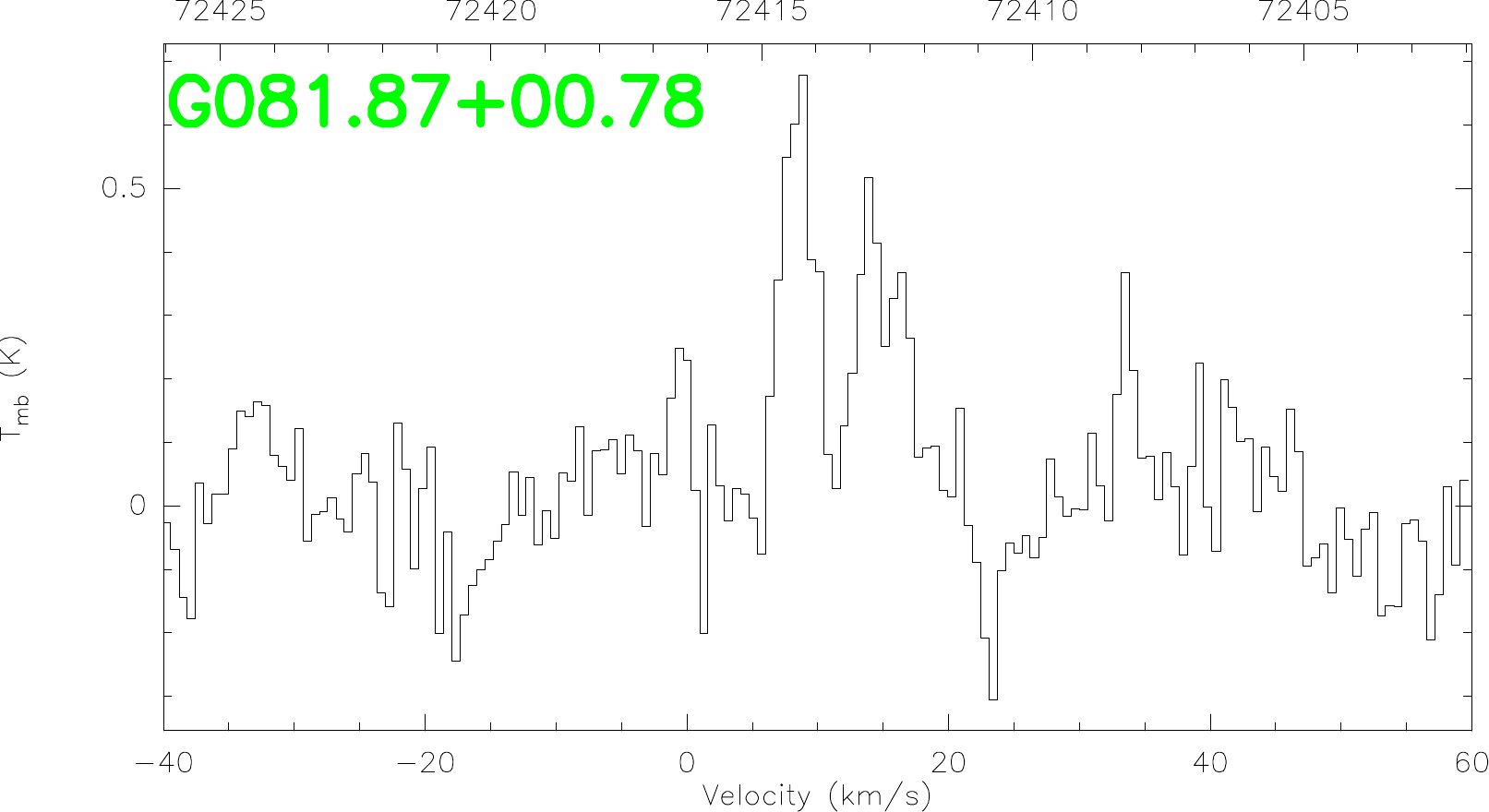}
\includegraphics[width=0.155\textwidth]{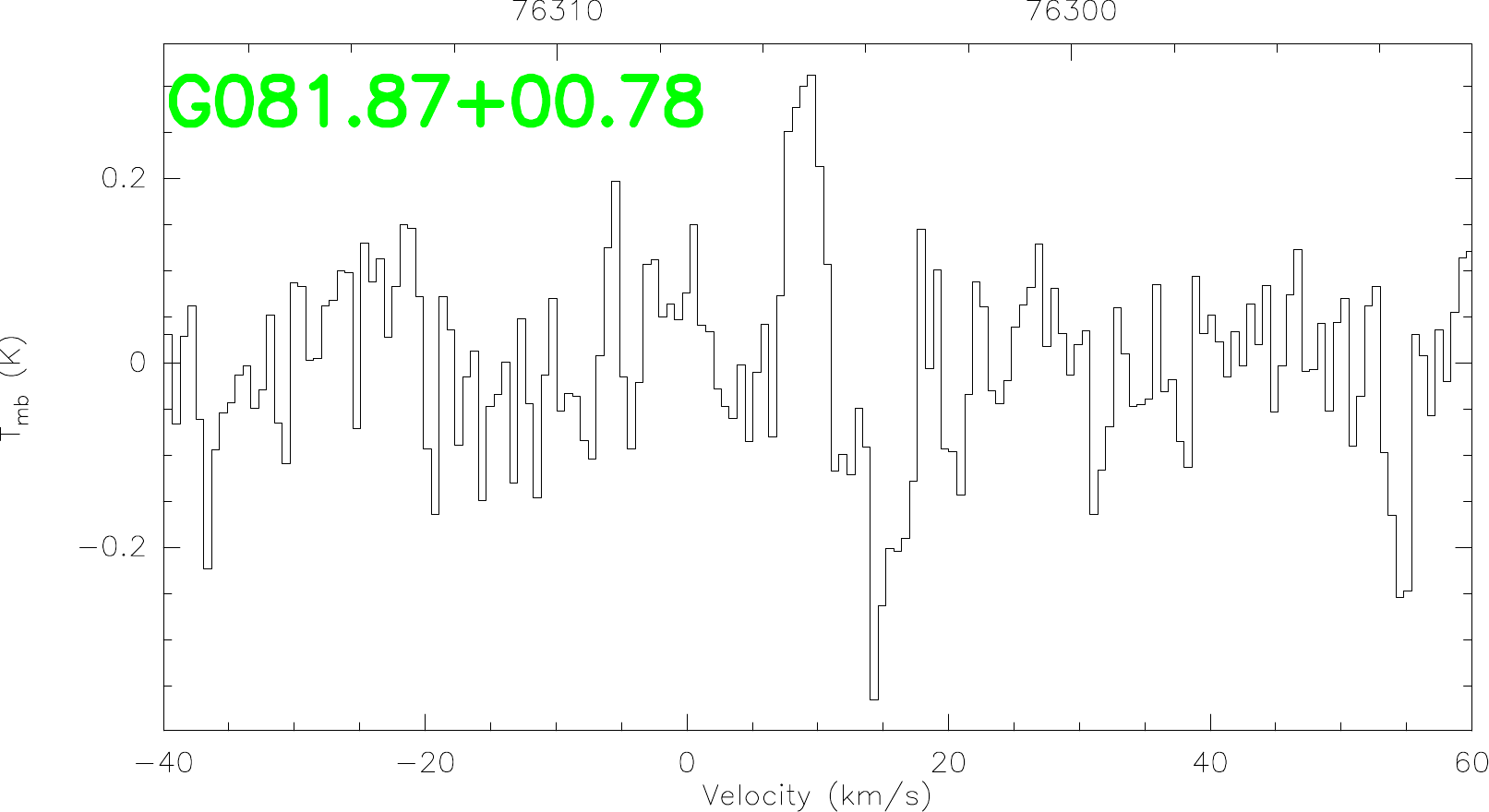}
\includegraphics[width=0.155\textwidth]{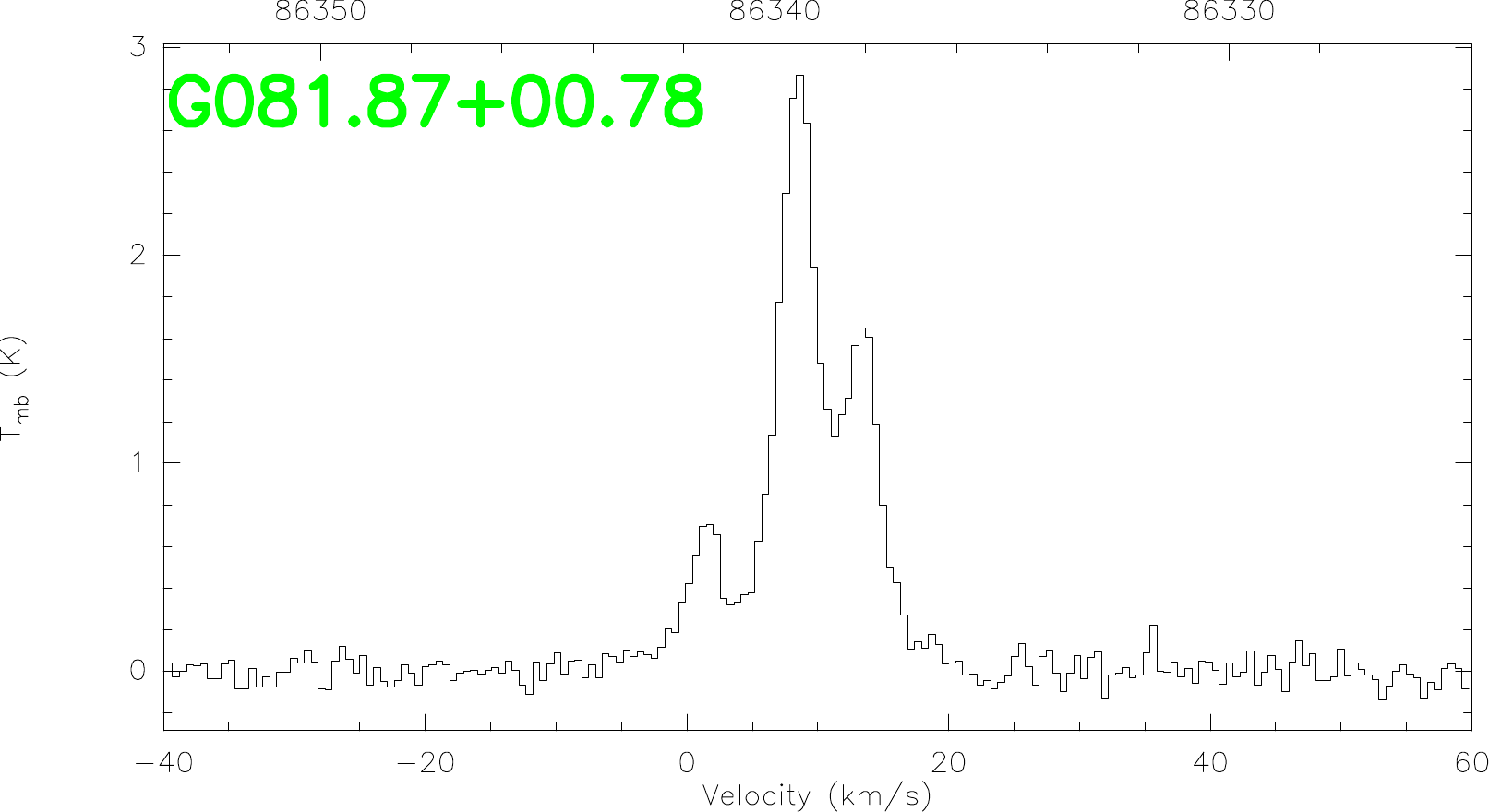}
\includegraphics[width=0.155\textwidth]{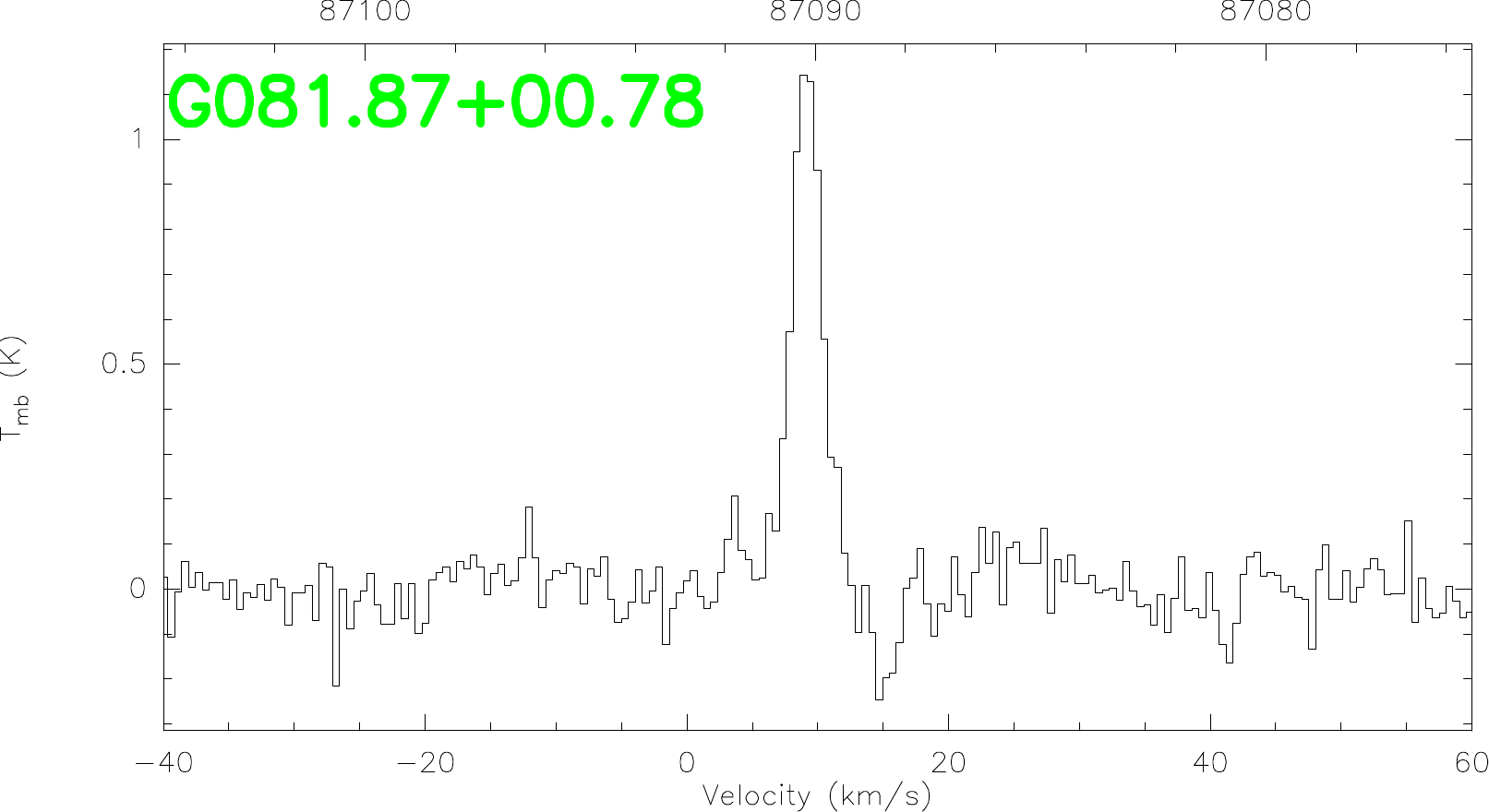}
\includegraphics[width=0.155\textwidth]{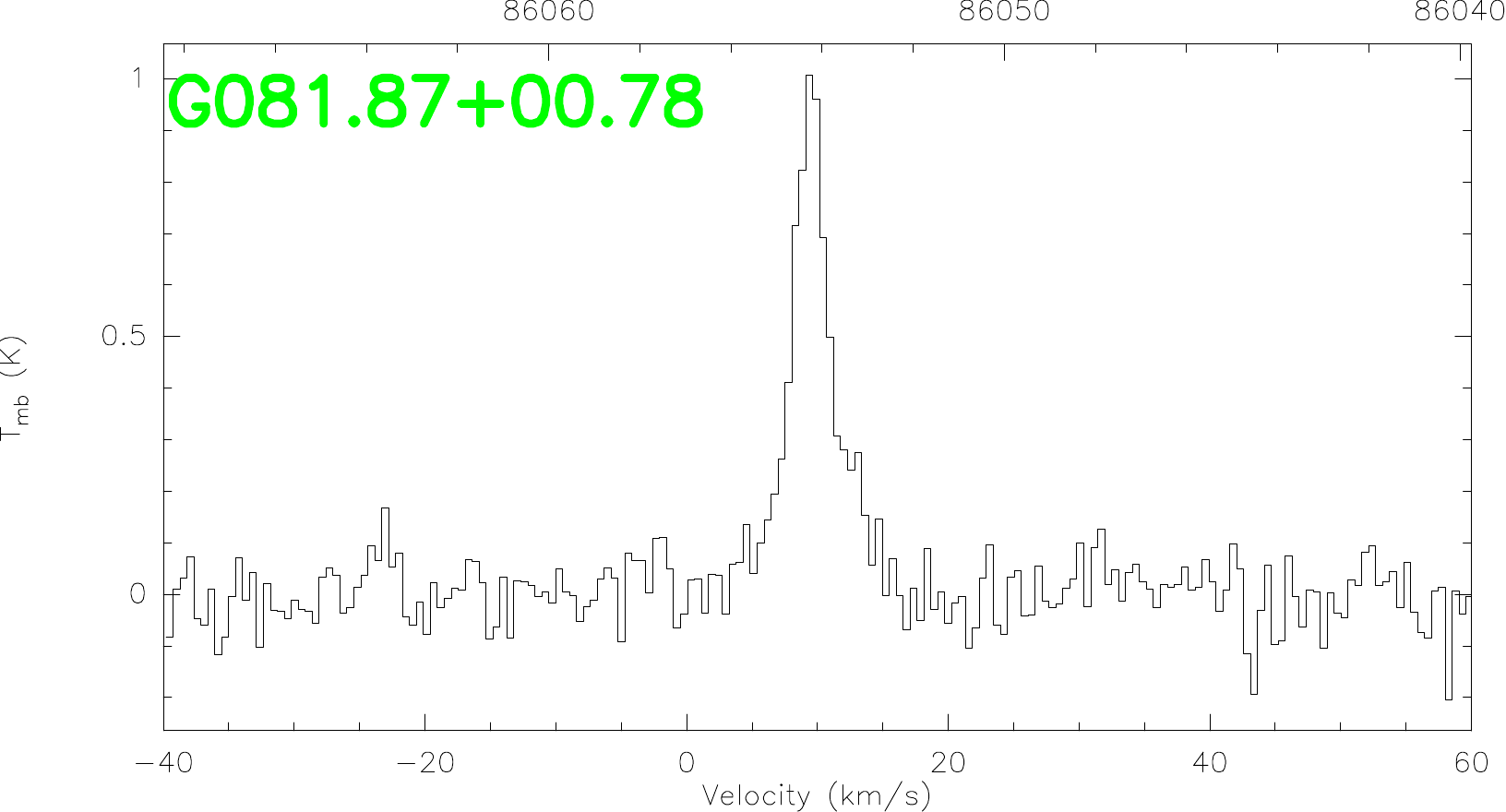}
\includegraphics[width=0.155\textwidth]{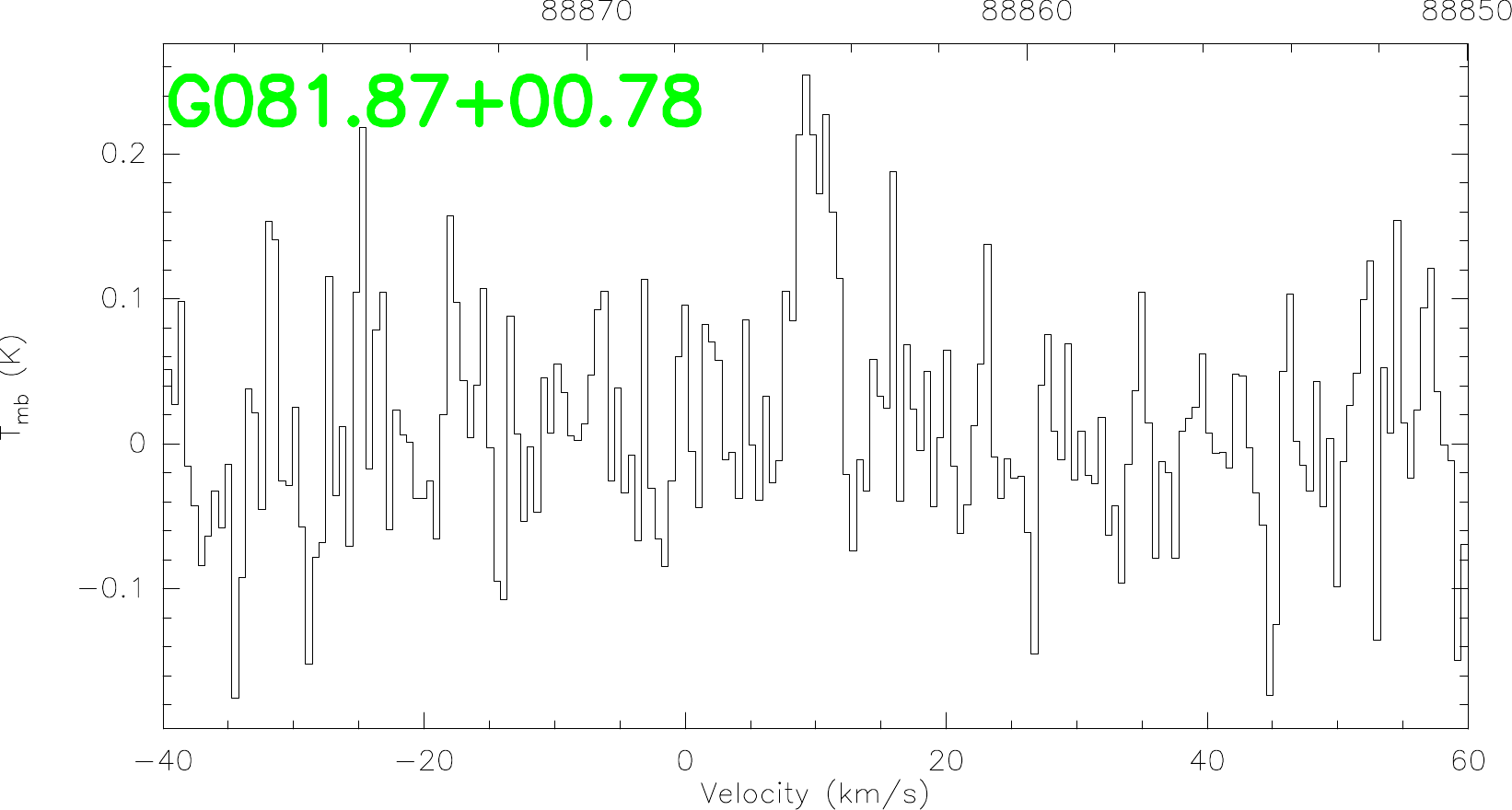}\\

\includegraphics[width=0.155\textwidth]{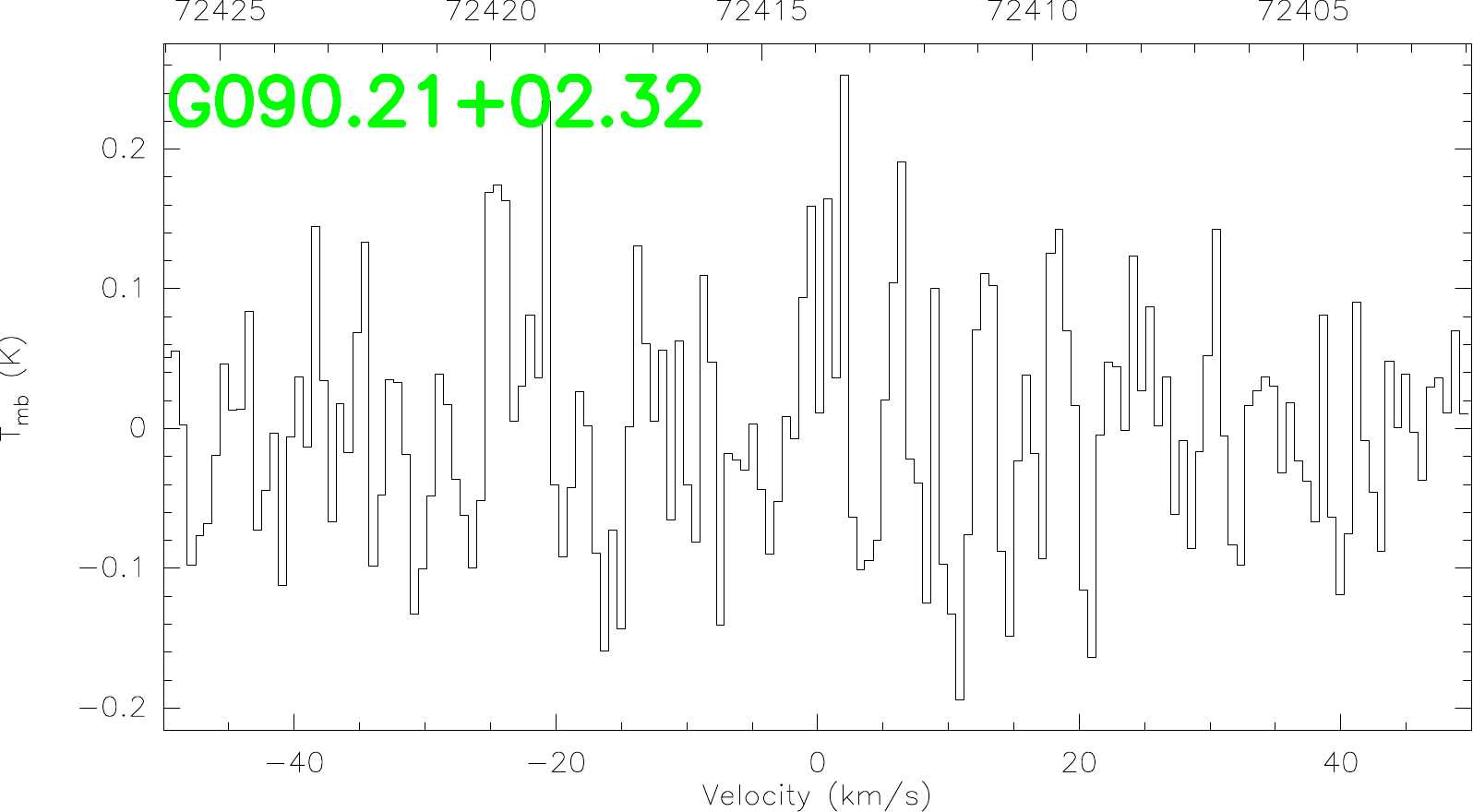}
\includegraphics[width=0.155\textwidth]{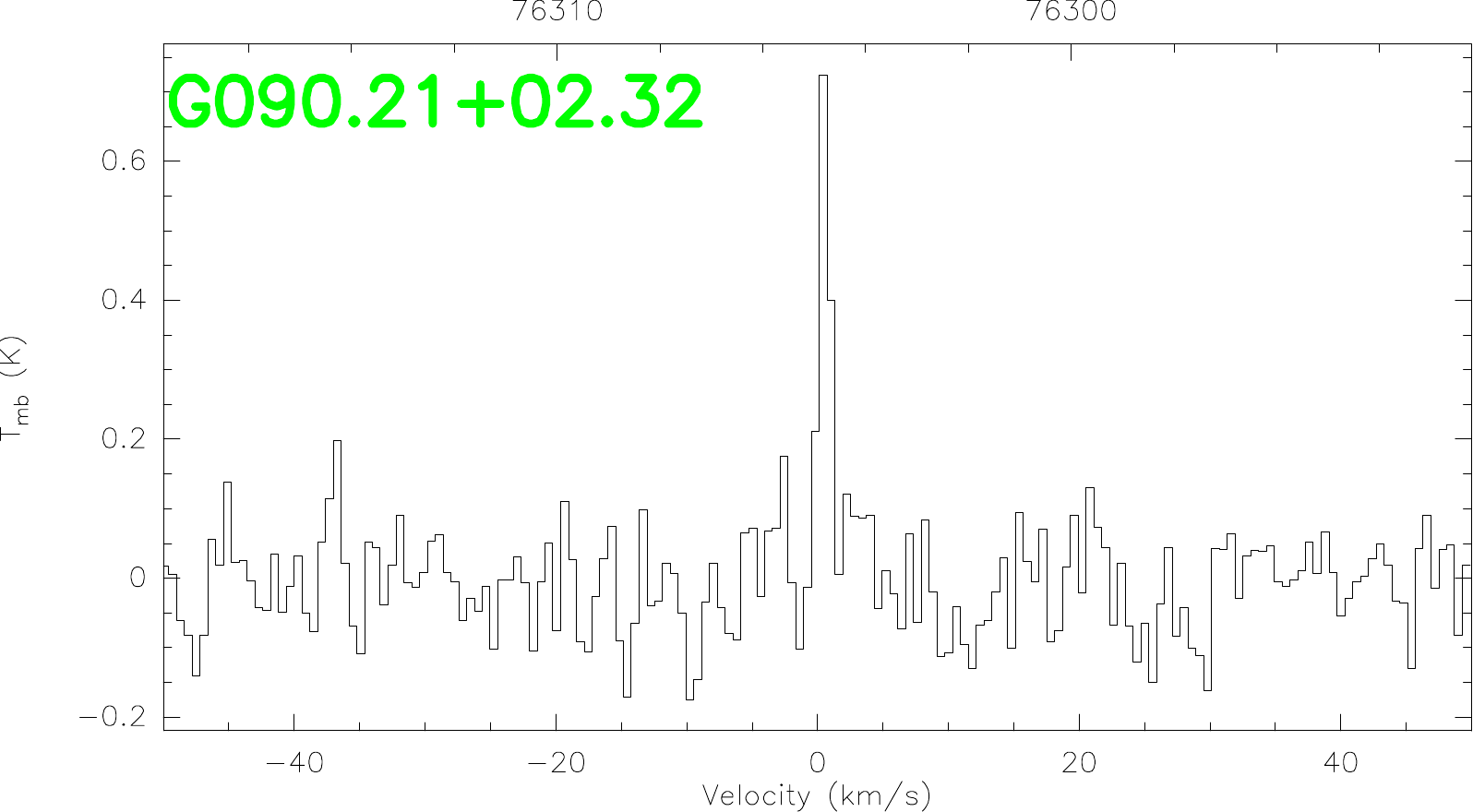}
\includegraphics[width=0.155\textwidth]{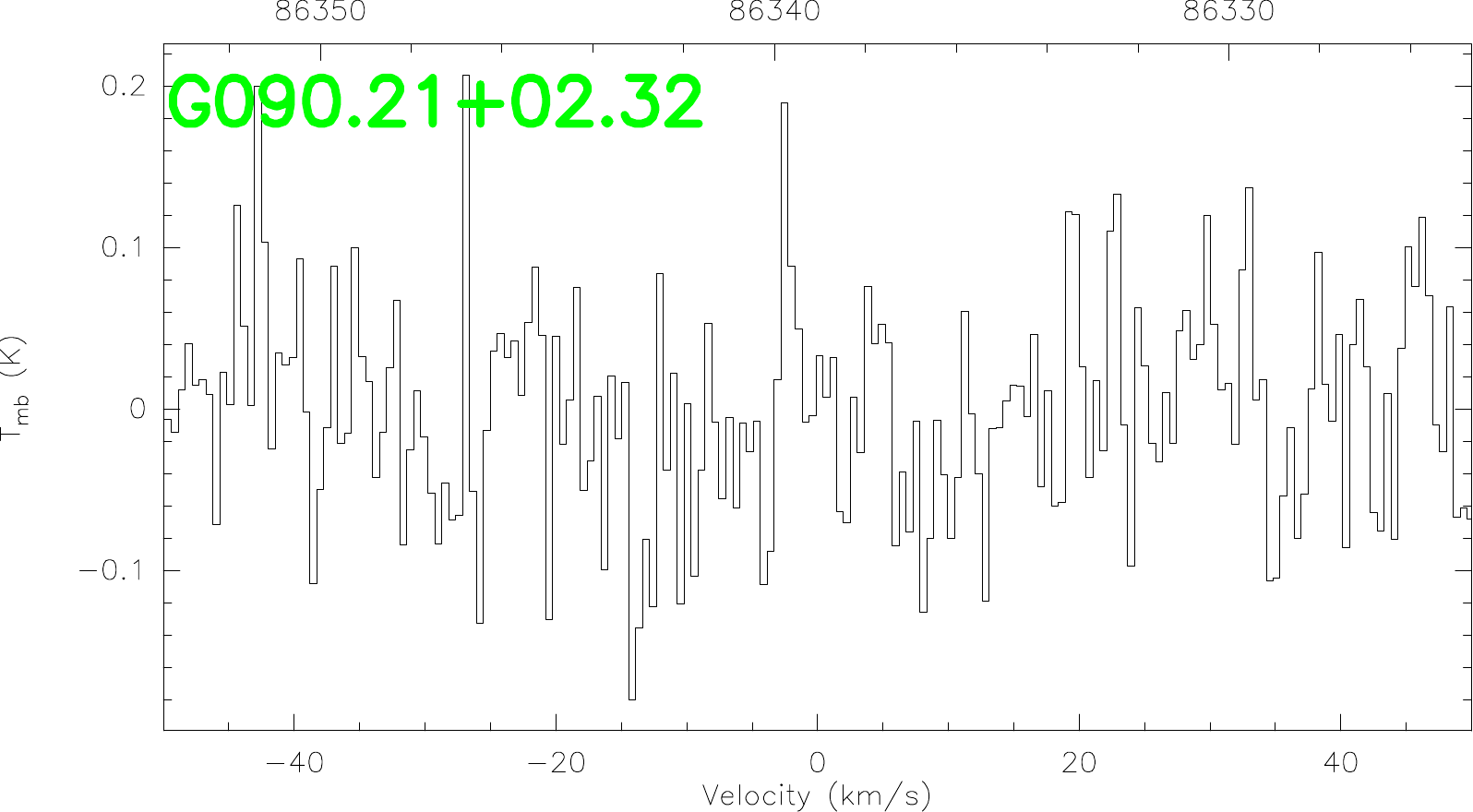}
\includegraphics[width=0.155\textwidth]{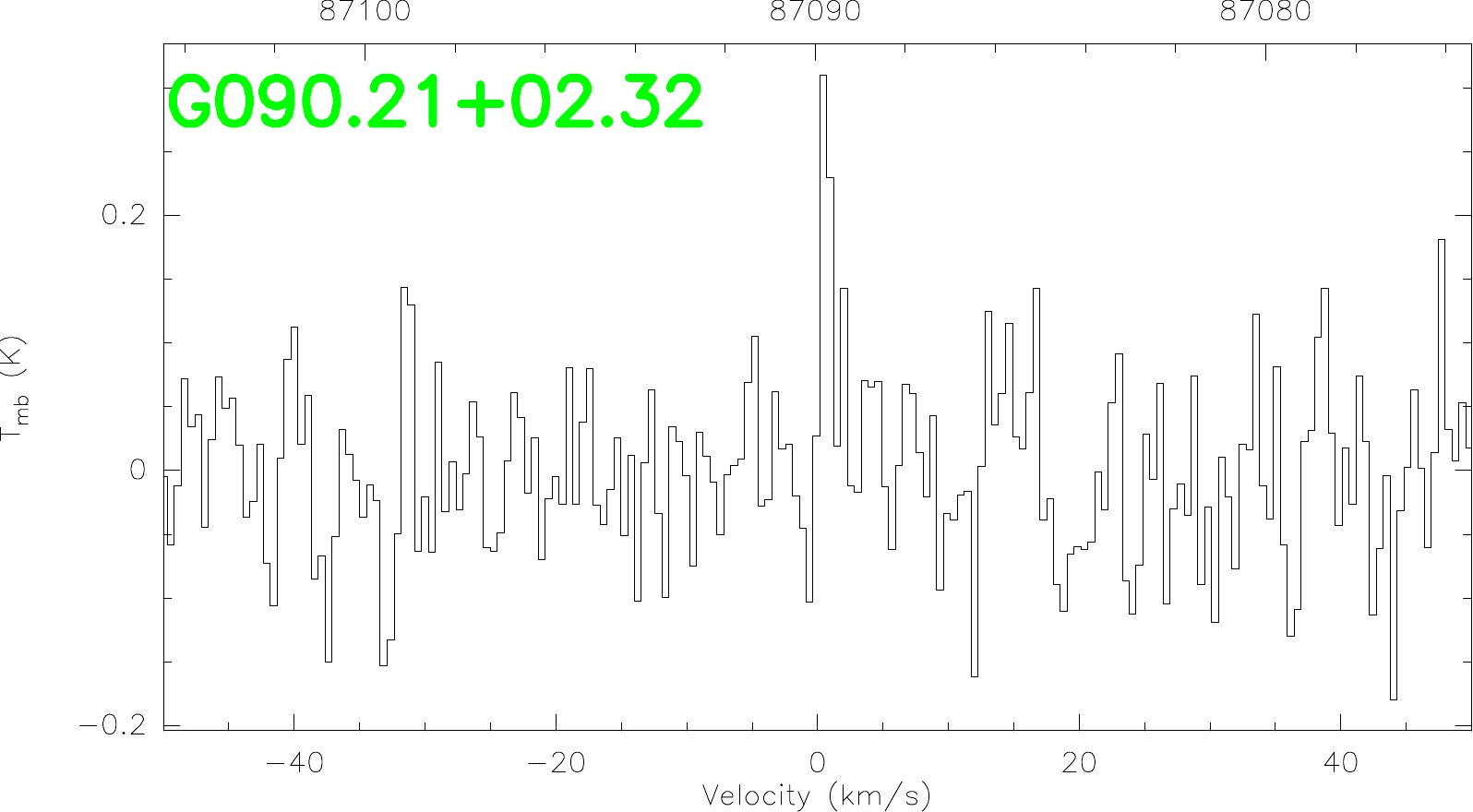}
\includegraphics[width=0.155\textwidth]{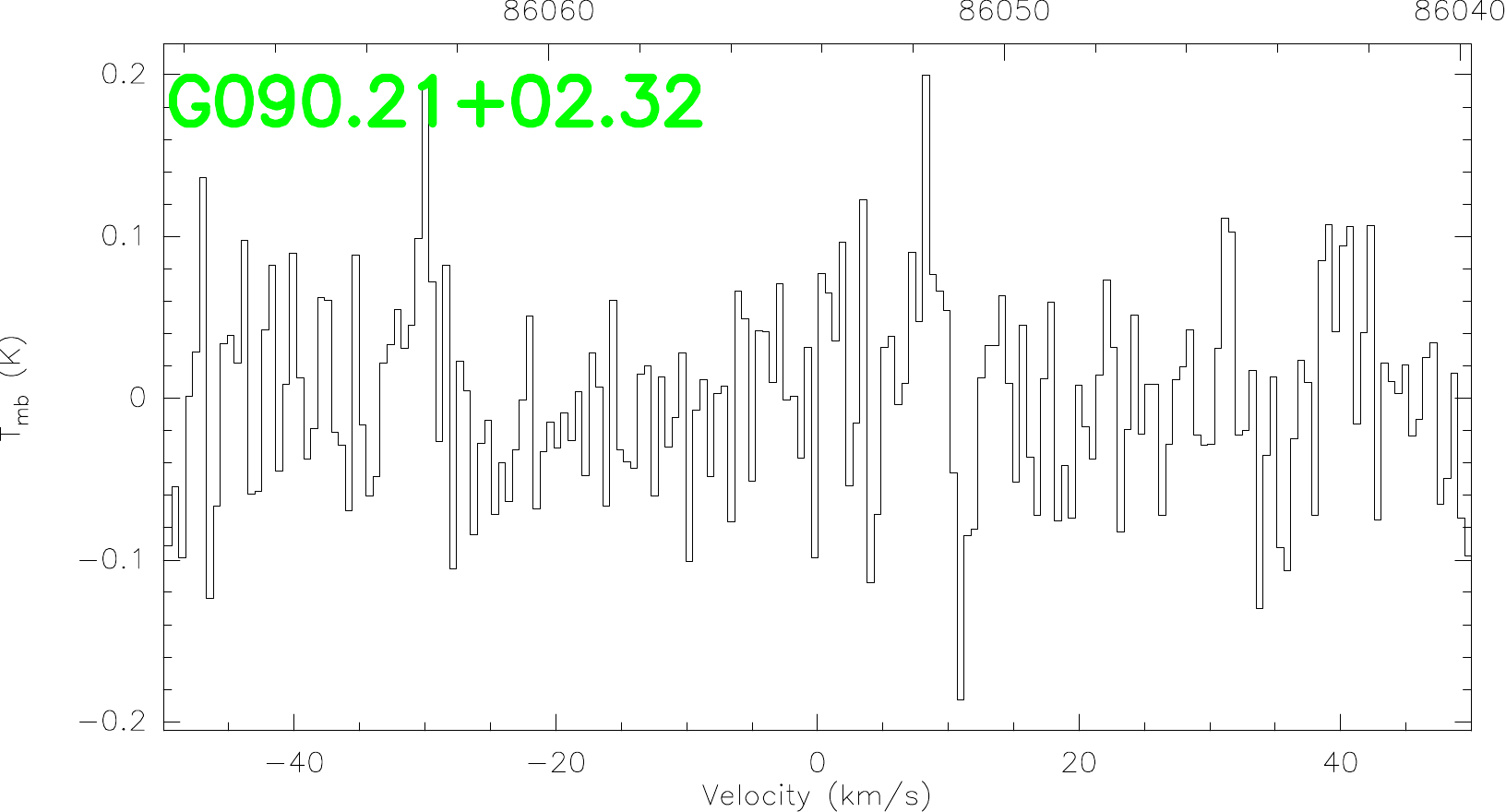}
\includegraphics[width=0.155\textwidth]{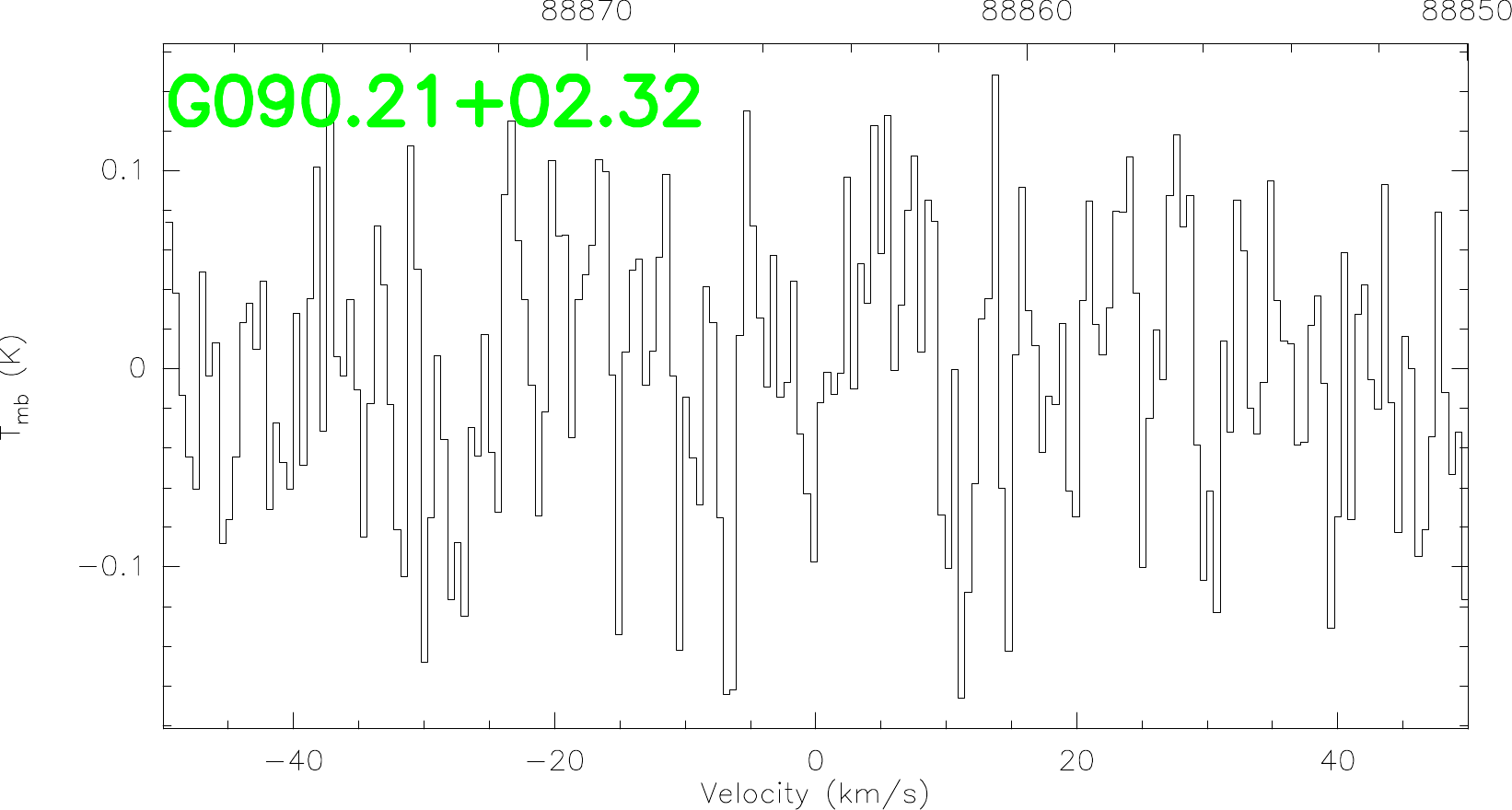}\\

\includegraphics[width=0.155\textwidth]{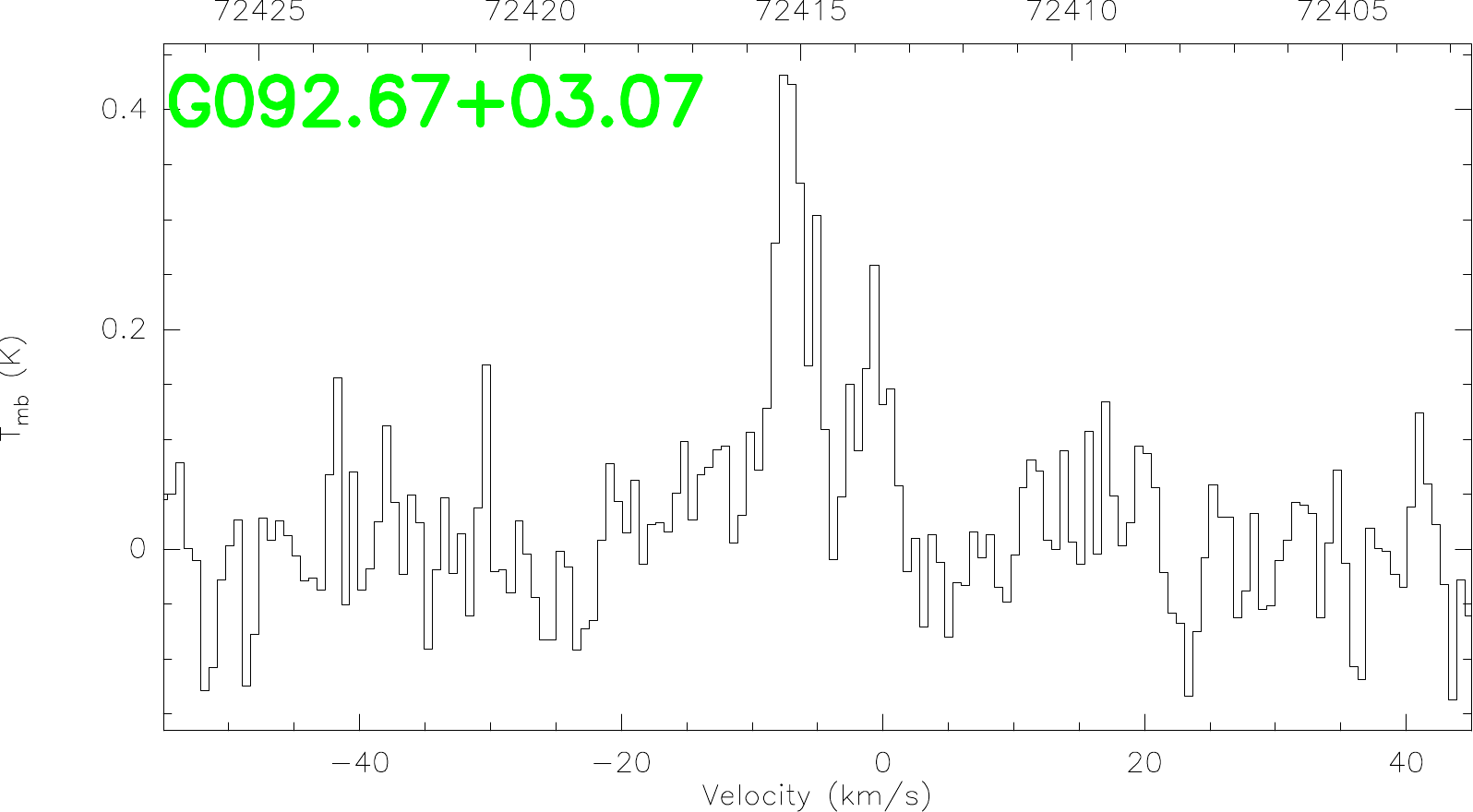}
\includegraphics[width=0.155\textwidth]{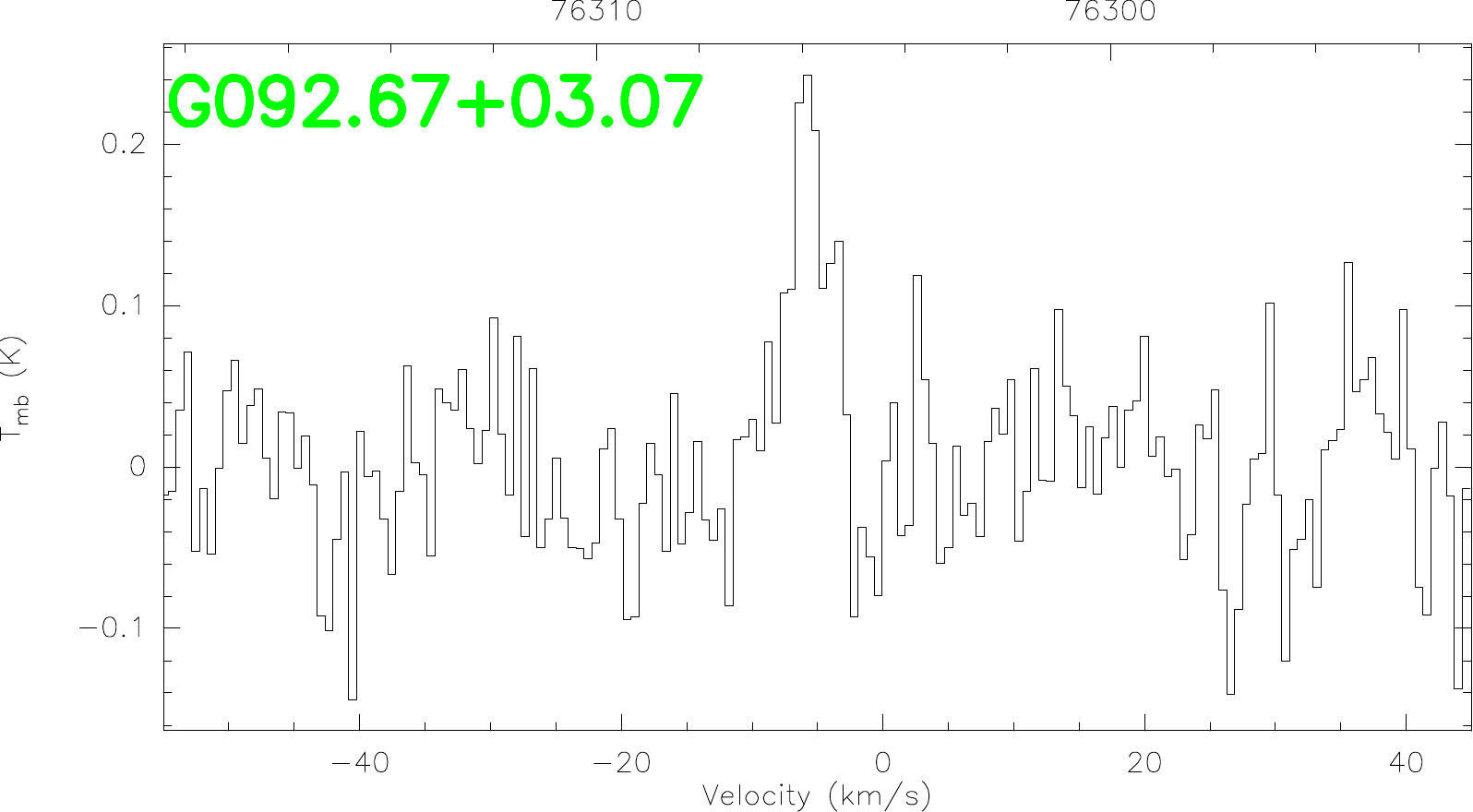}
\includegraphics[width=0.155\textwidth]{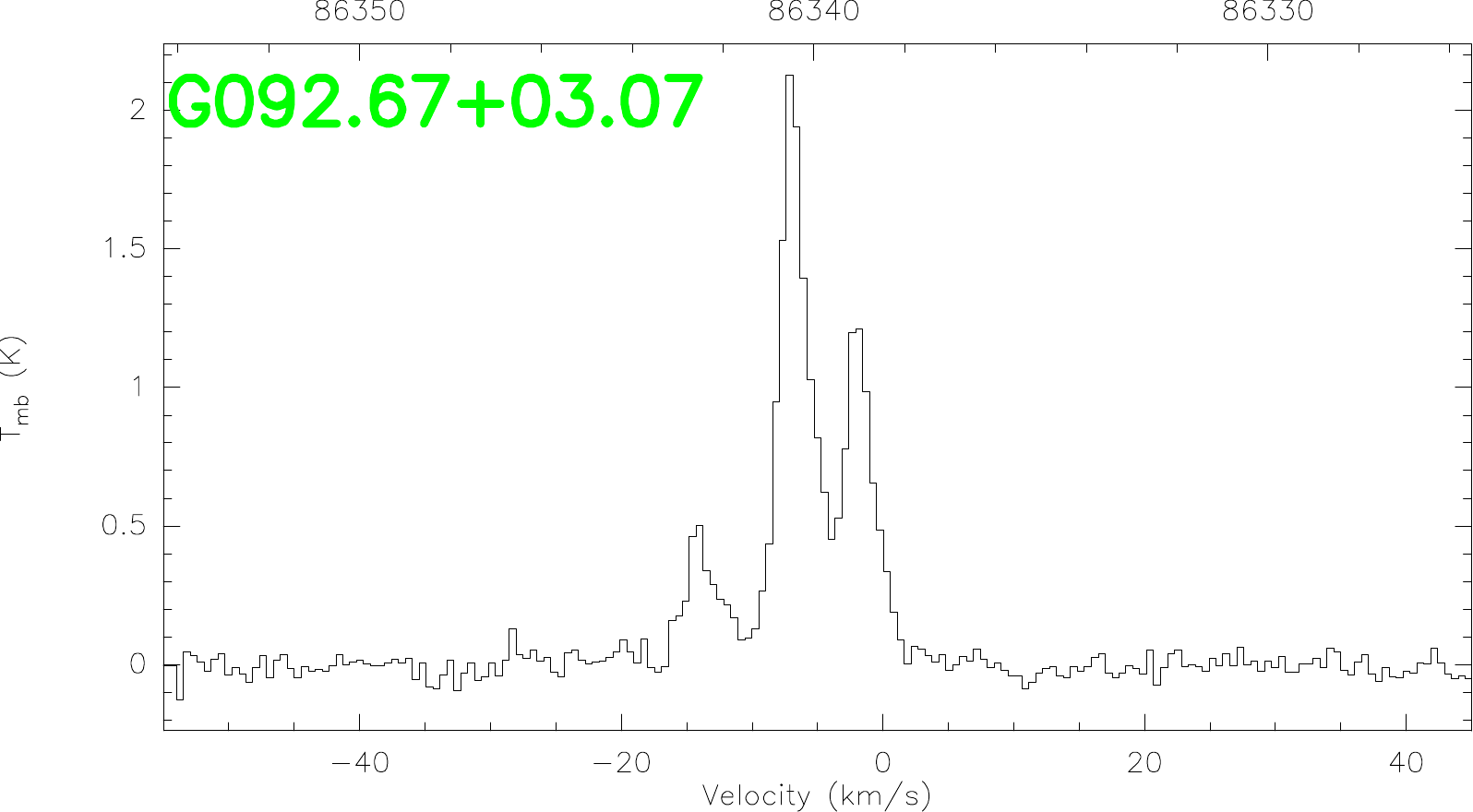}
\includegraphics[width=0.155\textwidth]{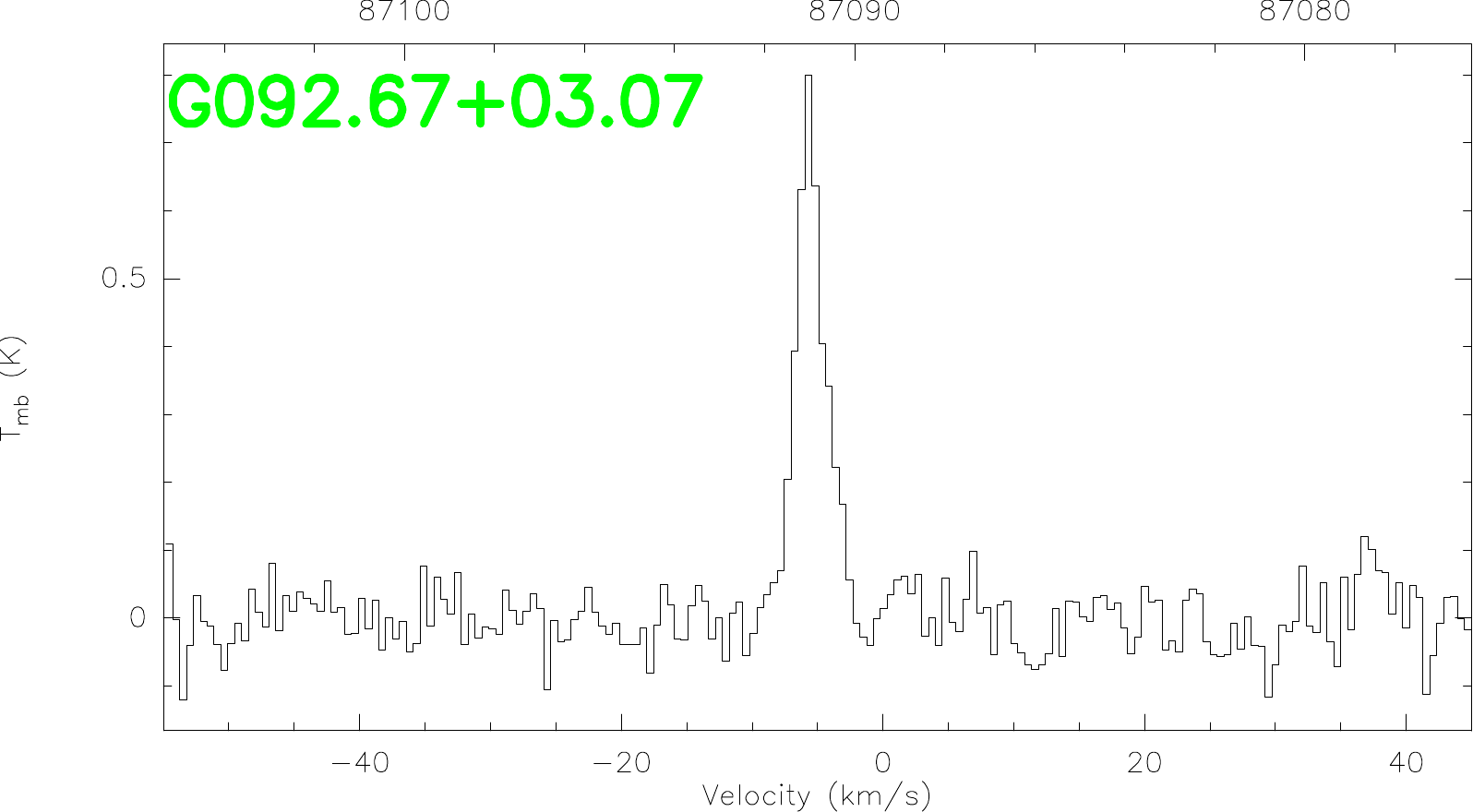}
\includegraphics[width=0.155\textwidth]{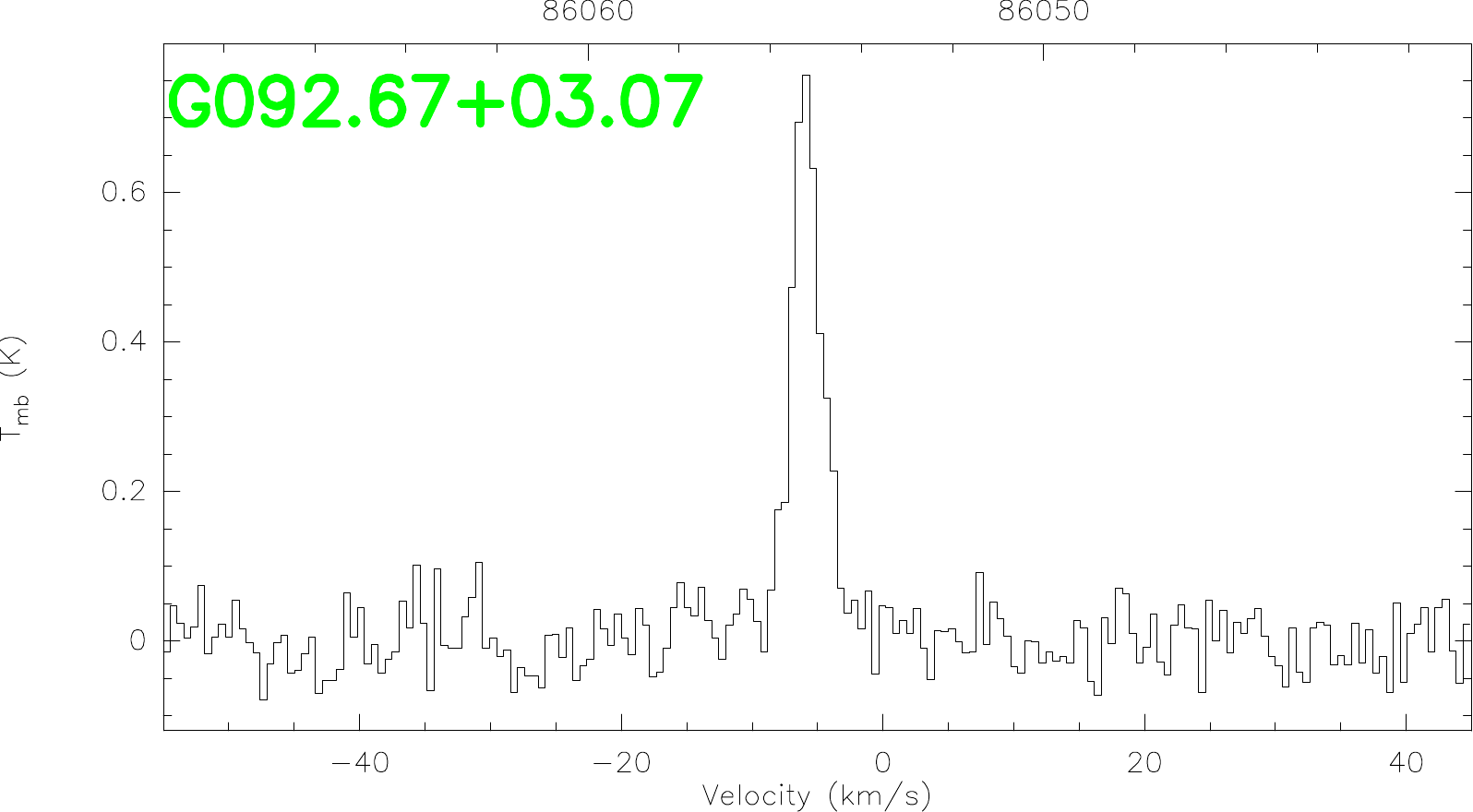}
\includegraphics[width=0.155\textwidth]{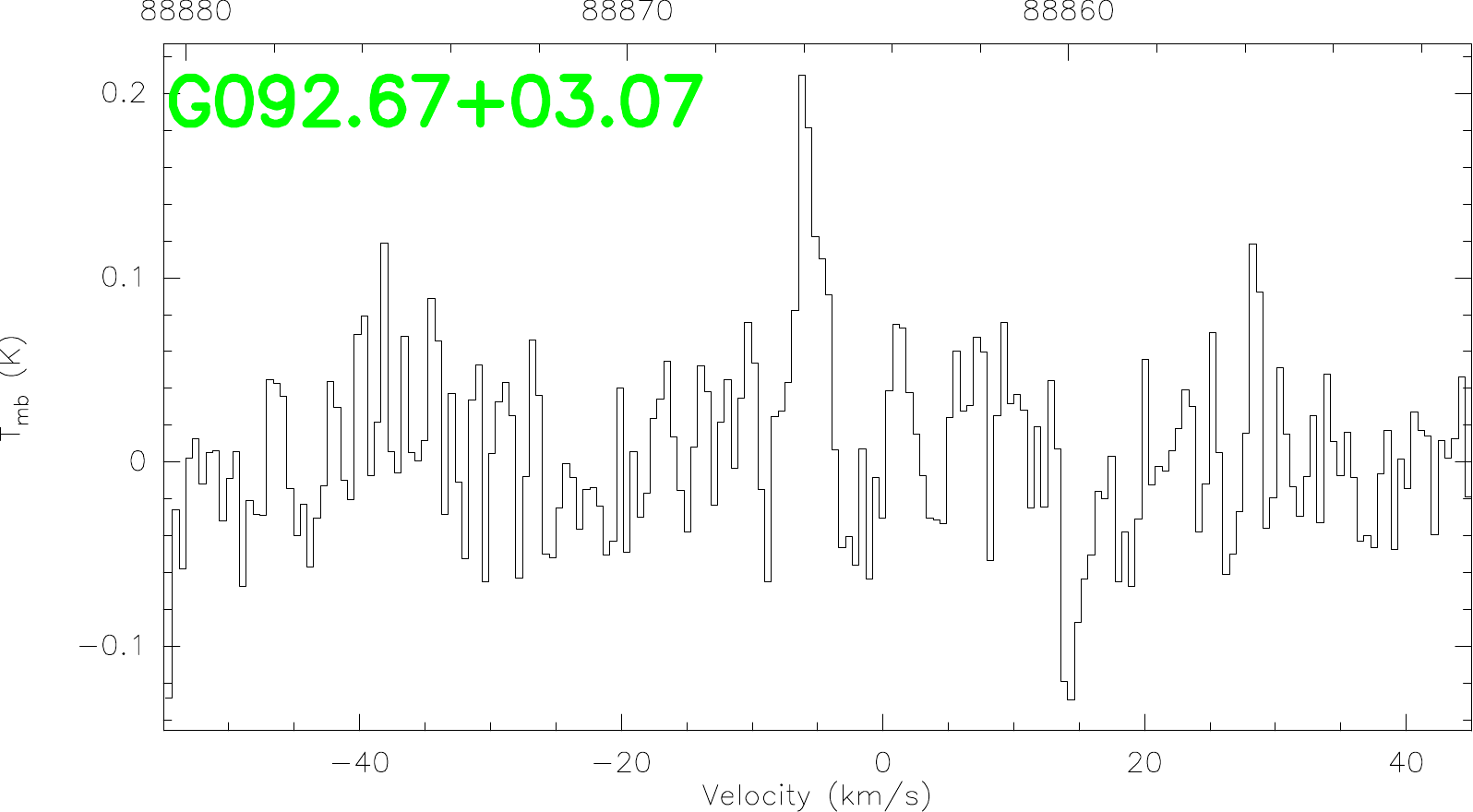}\\

\includegraphics[width=0.155\textwidth]{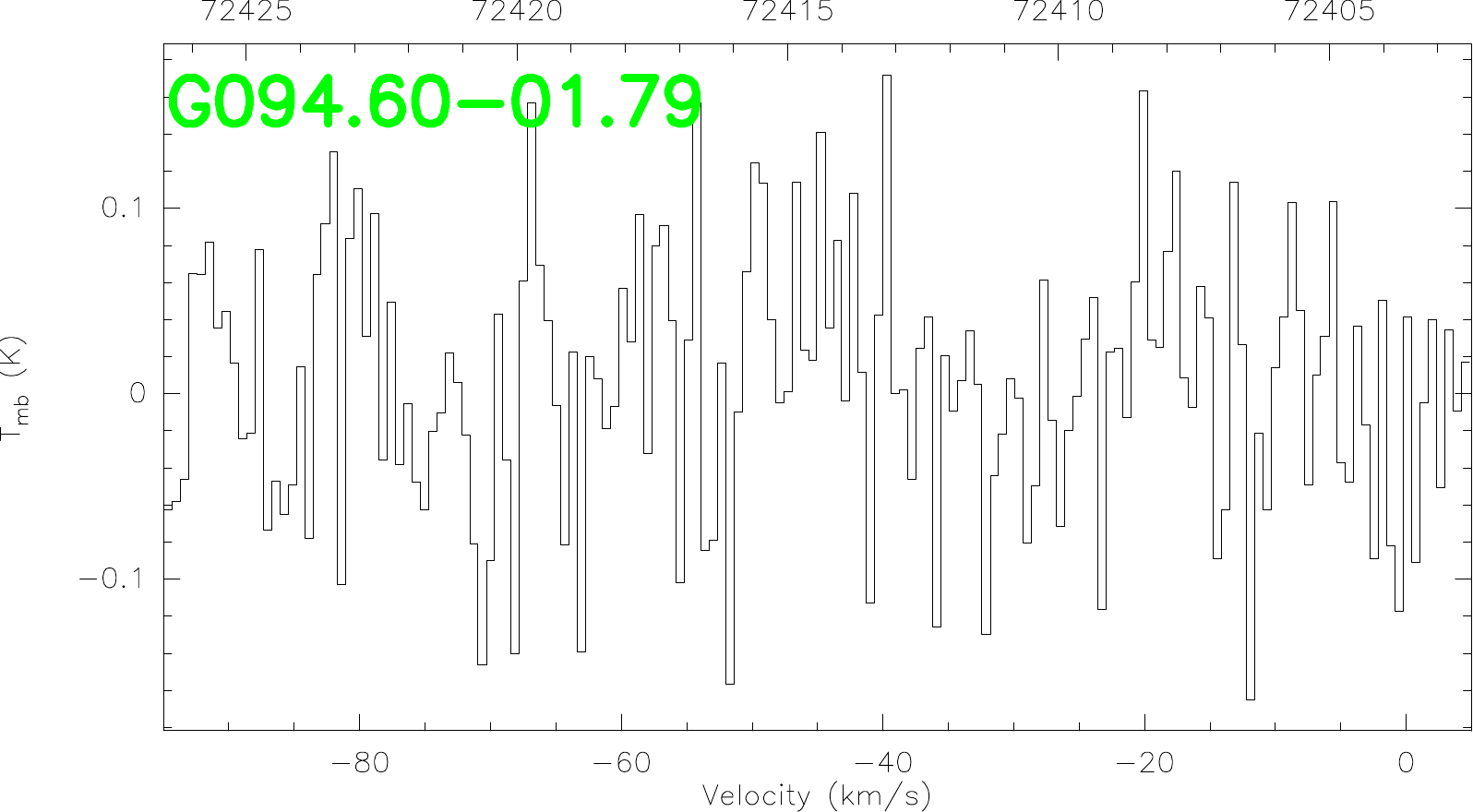}
\includegraphics[width=0.155\textwidth]{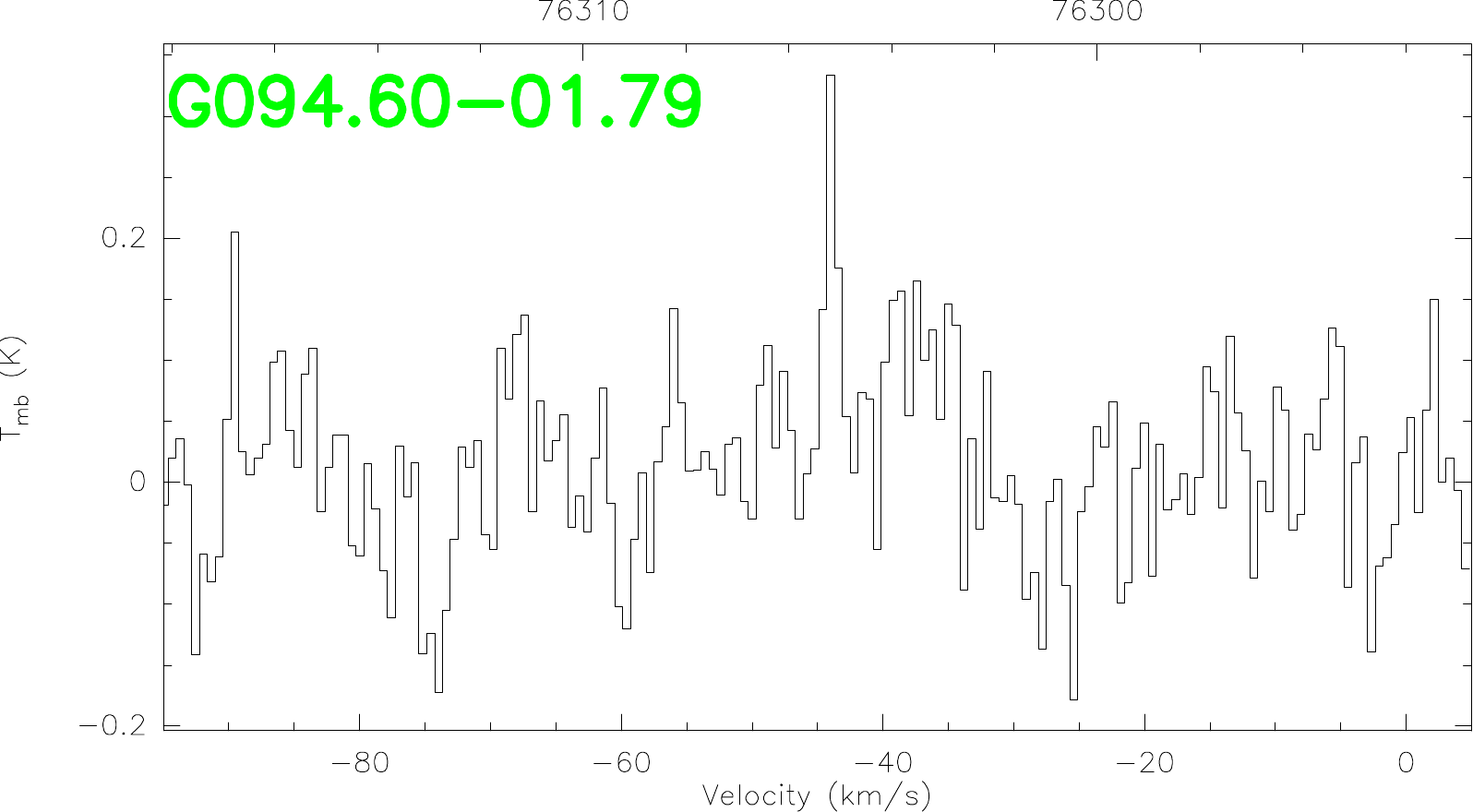}
\includegraphics[width=0.155\textwidth]{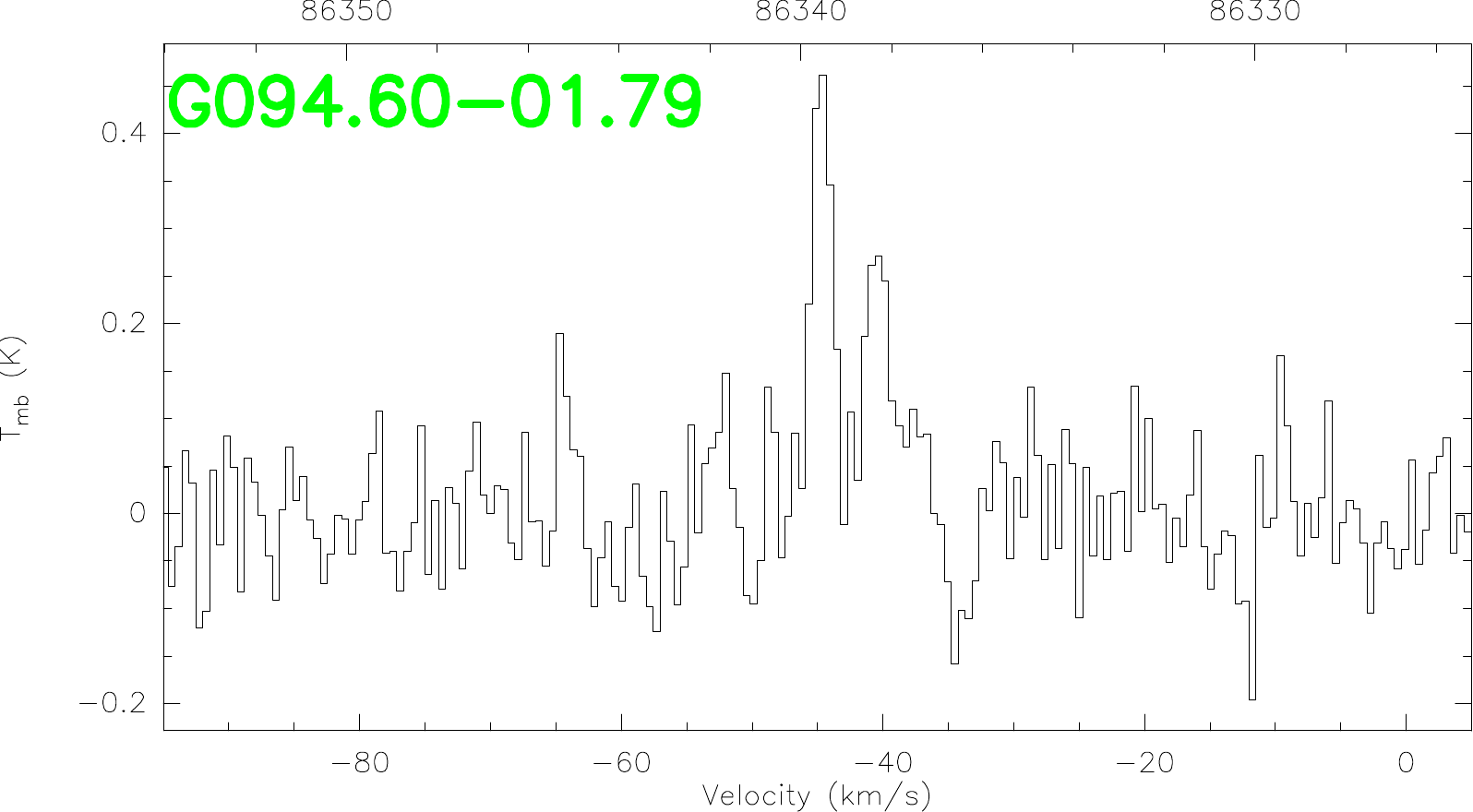}
\includegraphics[width=0.155\textwidth]{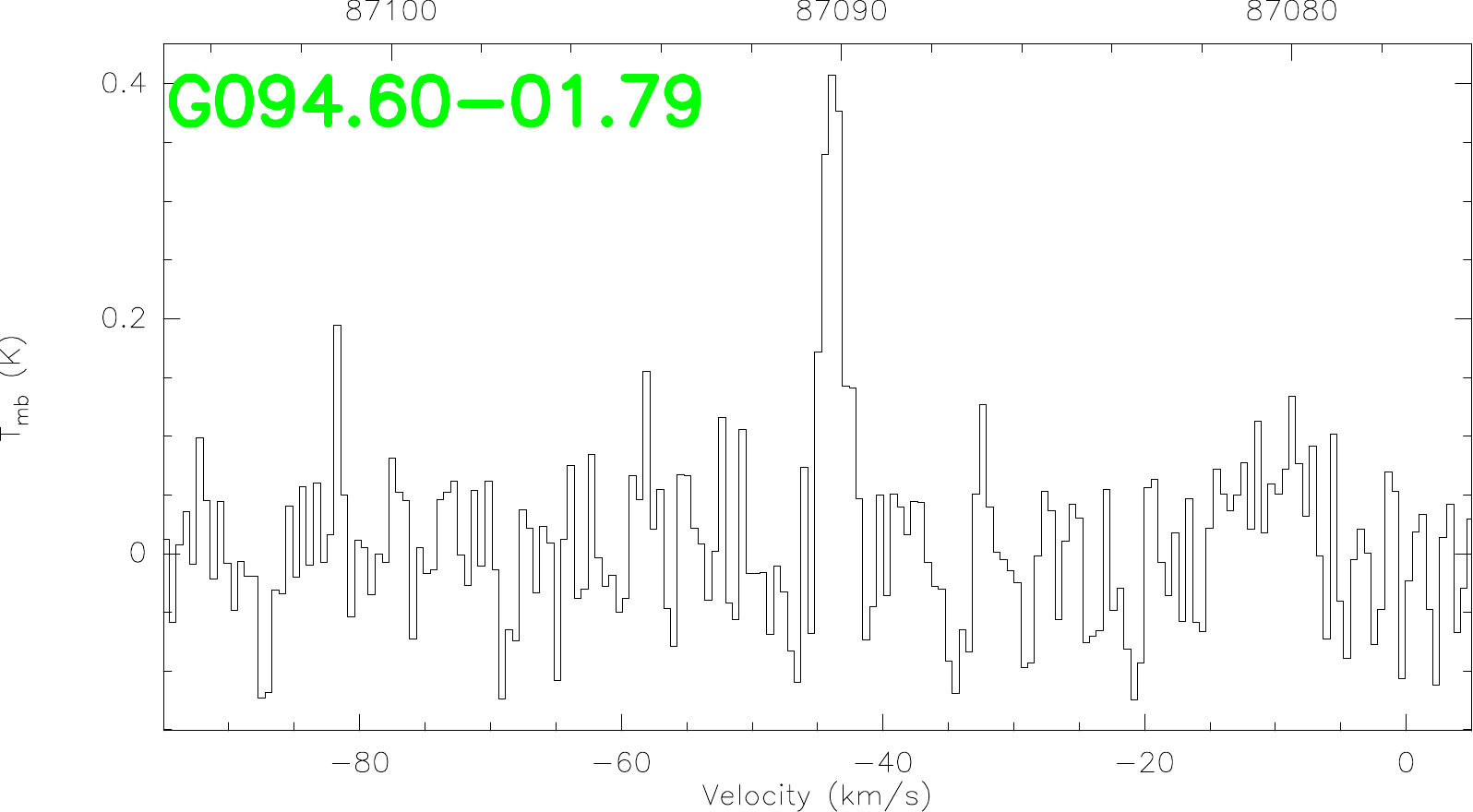}
\includegraphics[width=0.155\textwidth]{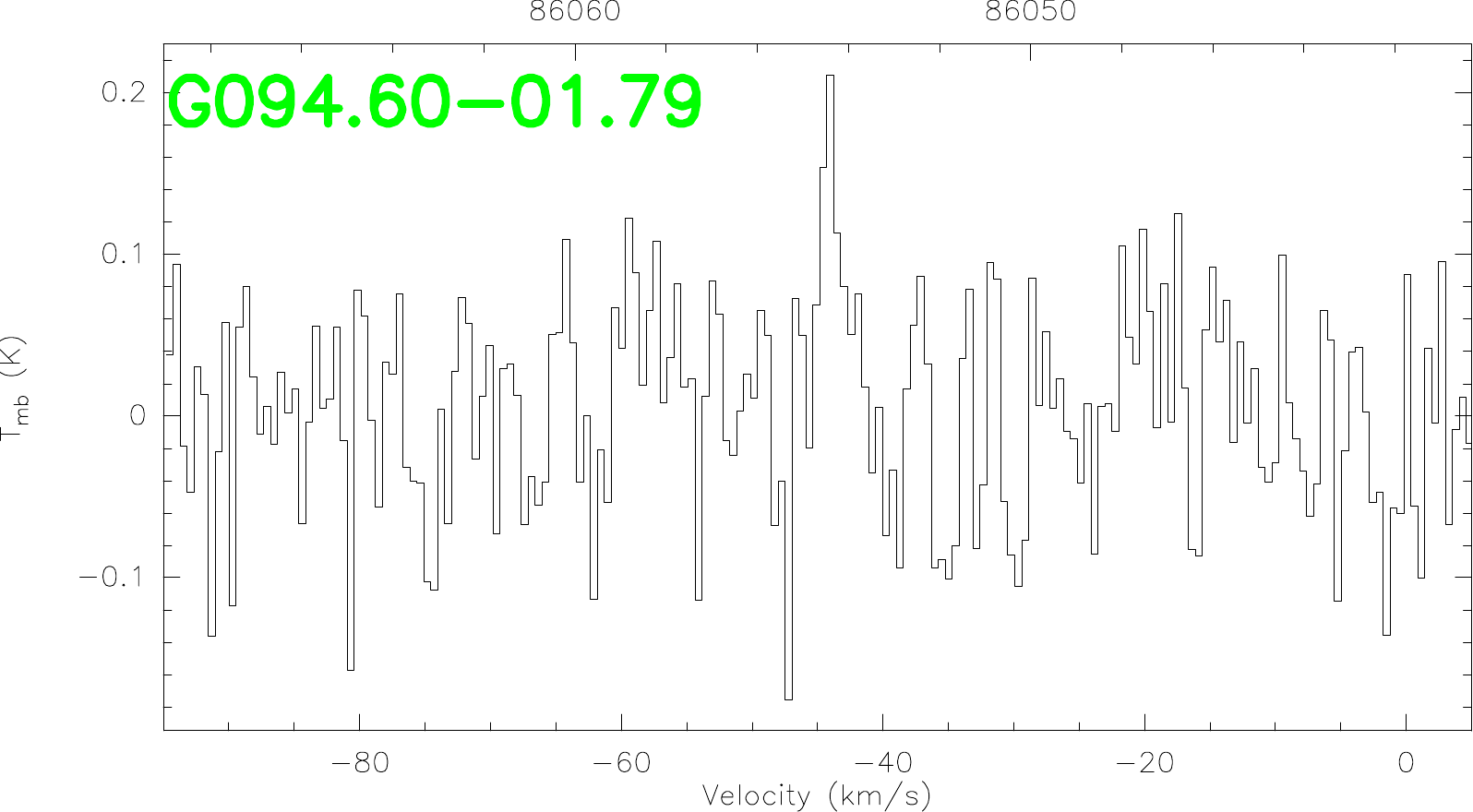}
\includegraphics[width=0.155\textwidth]{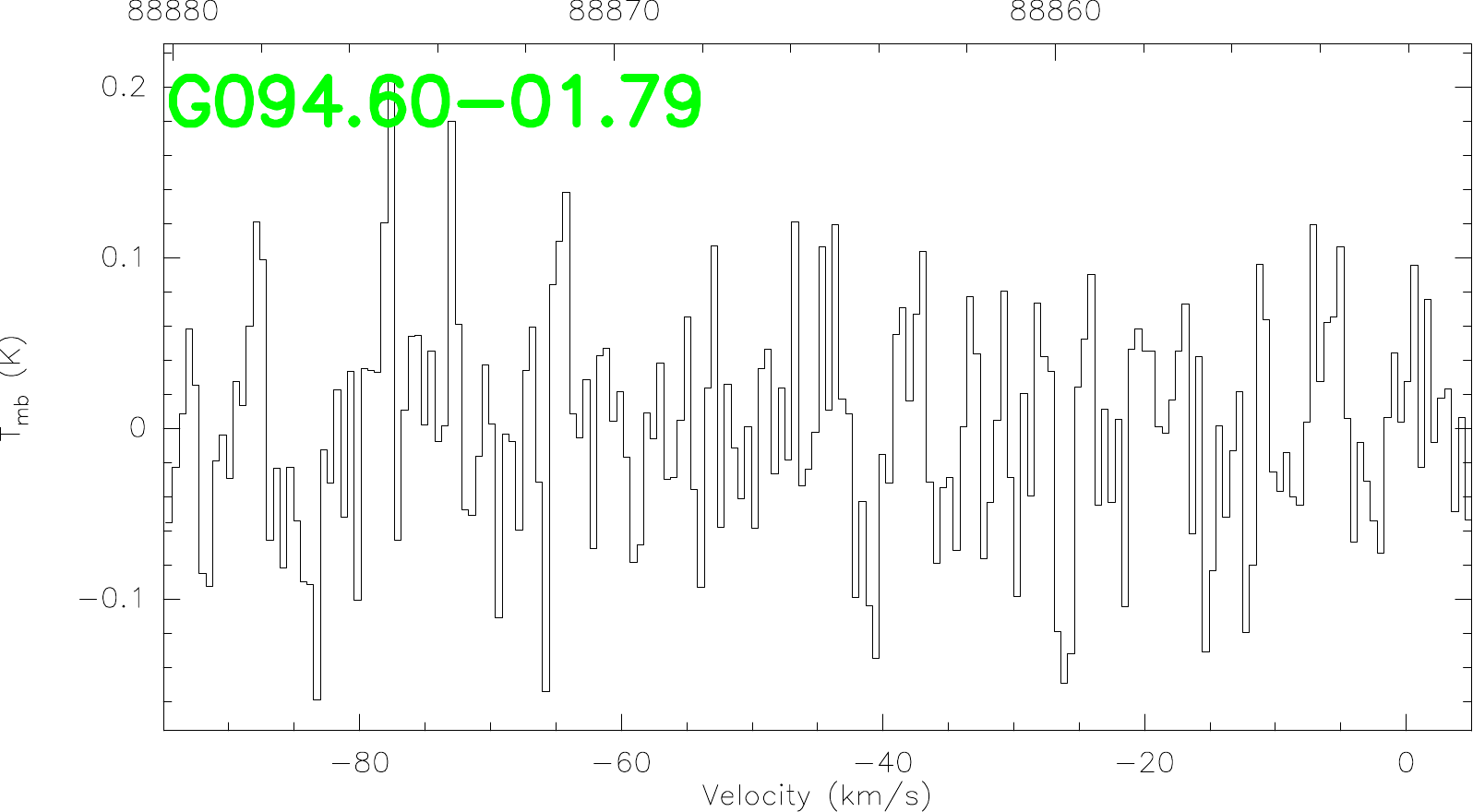}\\

\includegraphics[width=0.155\textwidth]{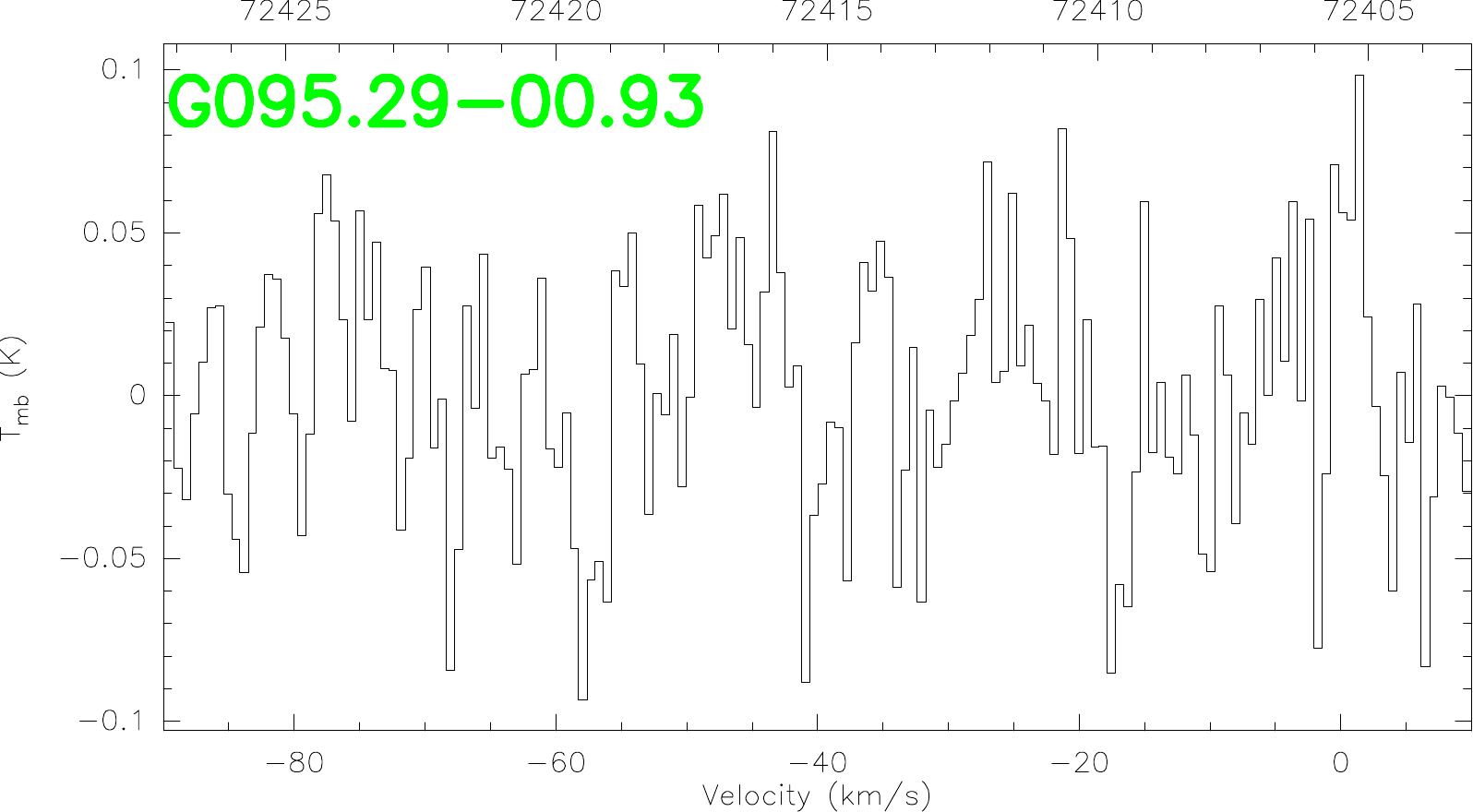}
\includegraphics[width=0.155\textwidth]{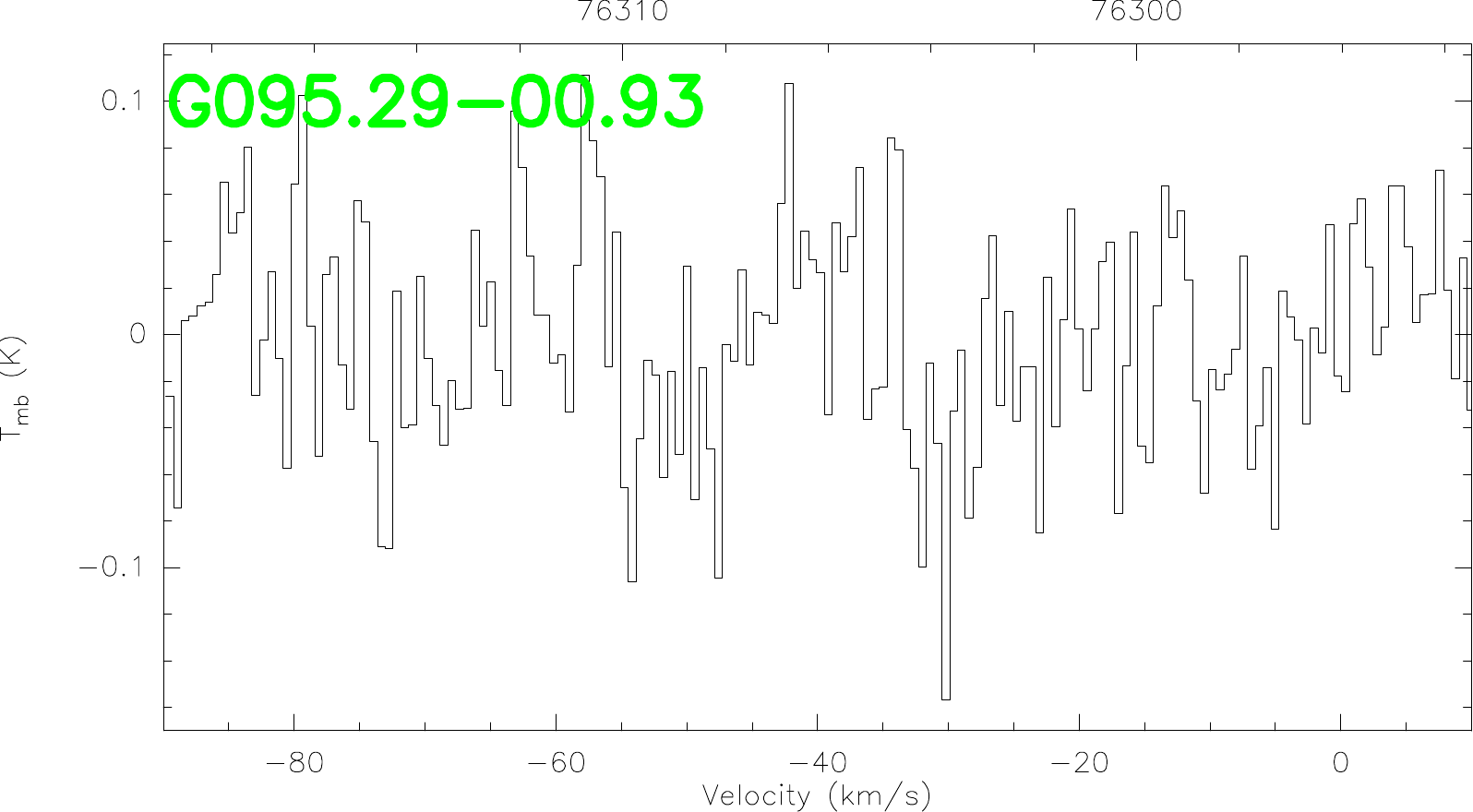}
\includegraphics[width=0.155\textwidth]{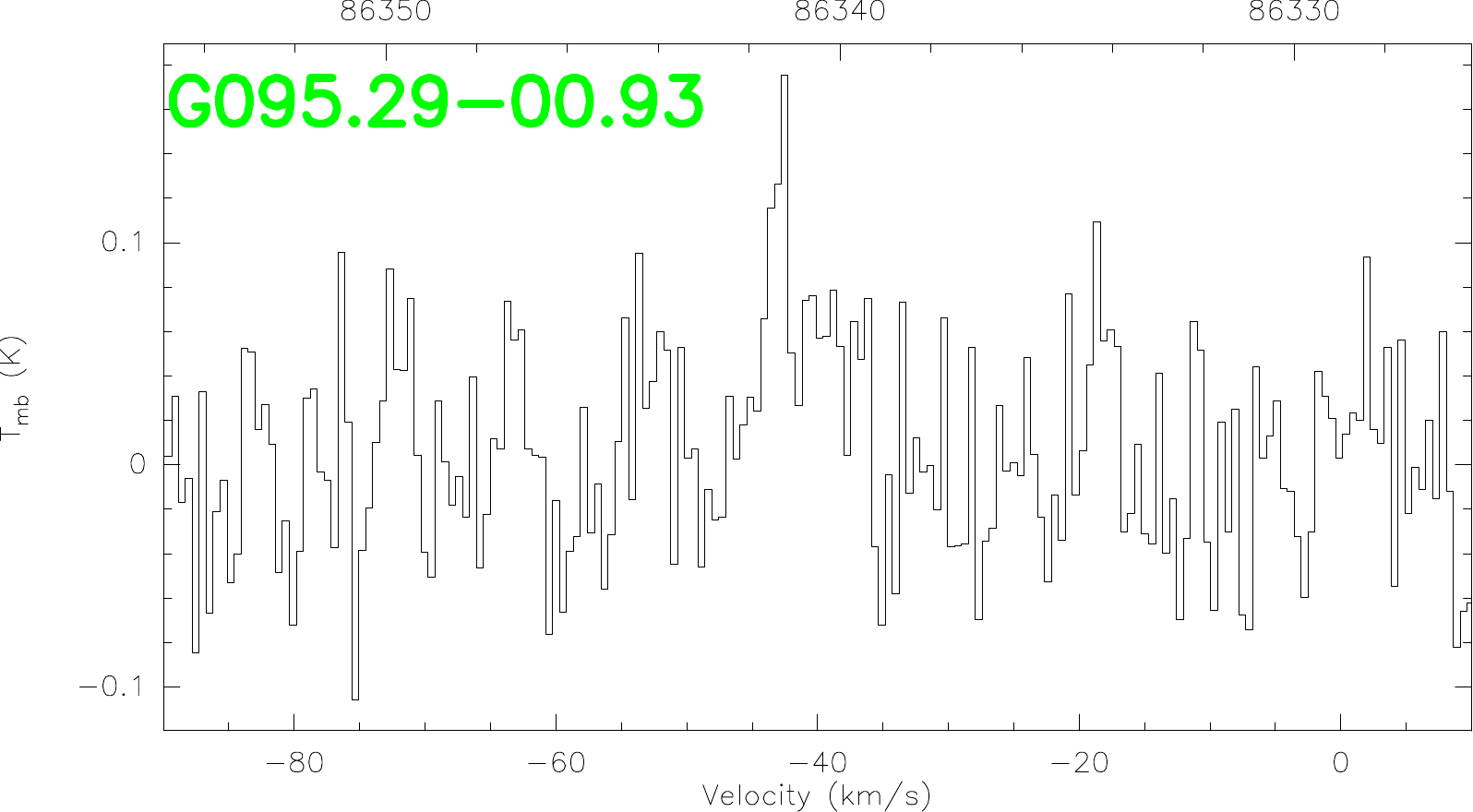}
\includegraphics[width=0.155\textwidth]{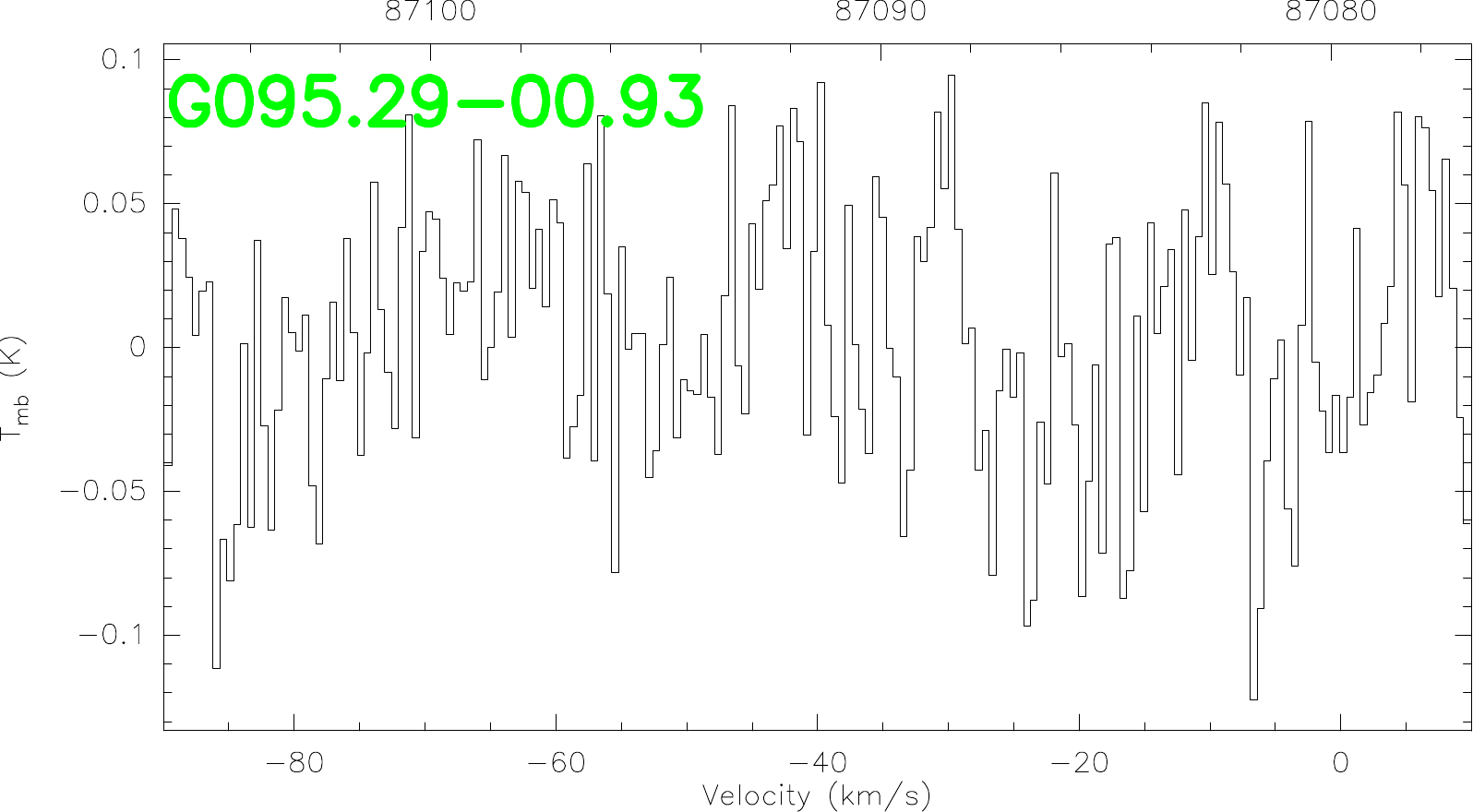}
\includegraphics[width=0.155\textwidth]{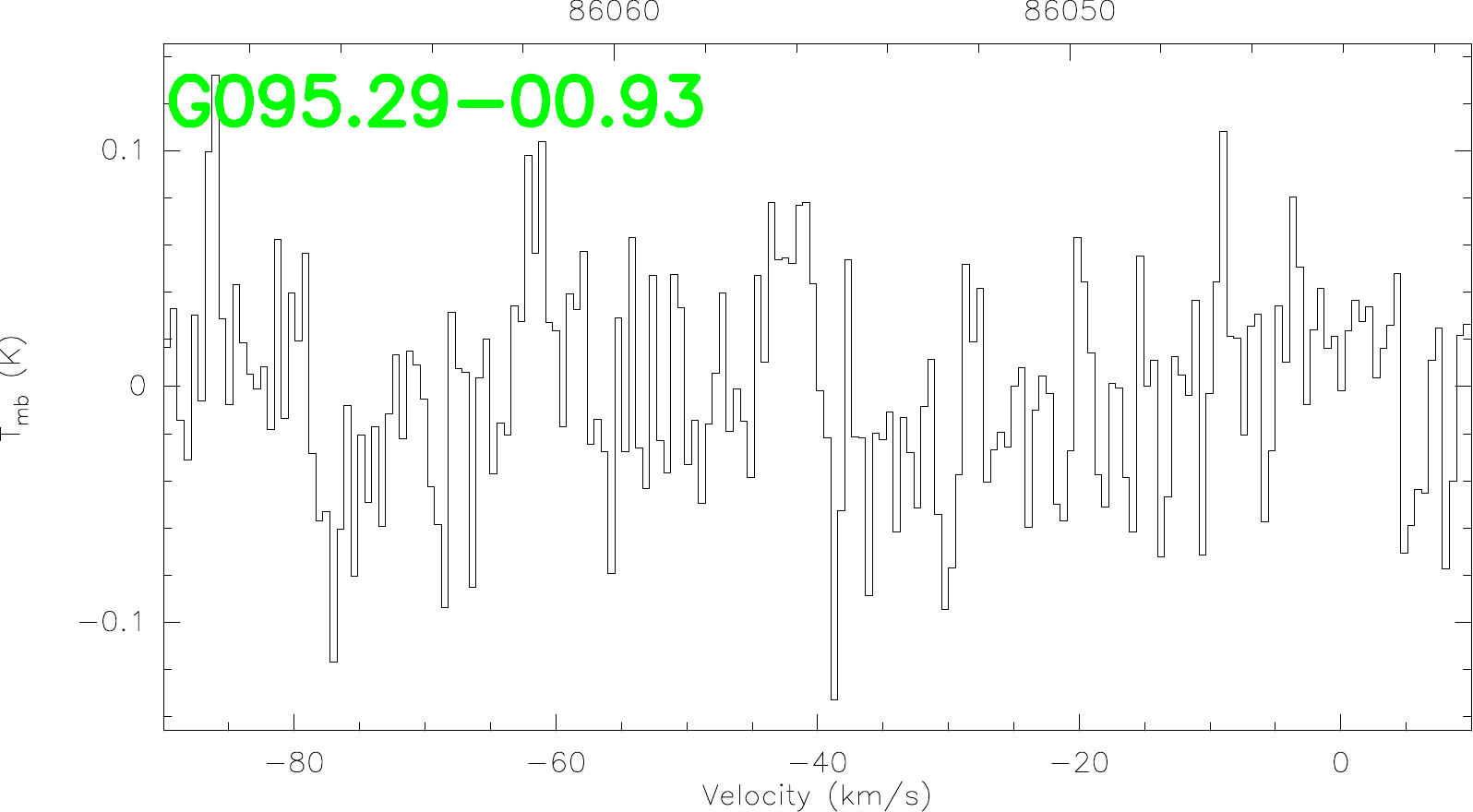}
\includegraphics[width=0.155\textwidth]{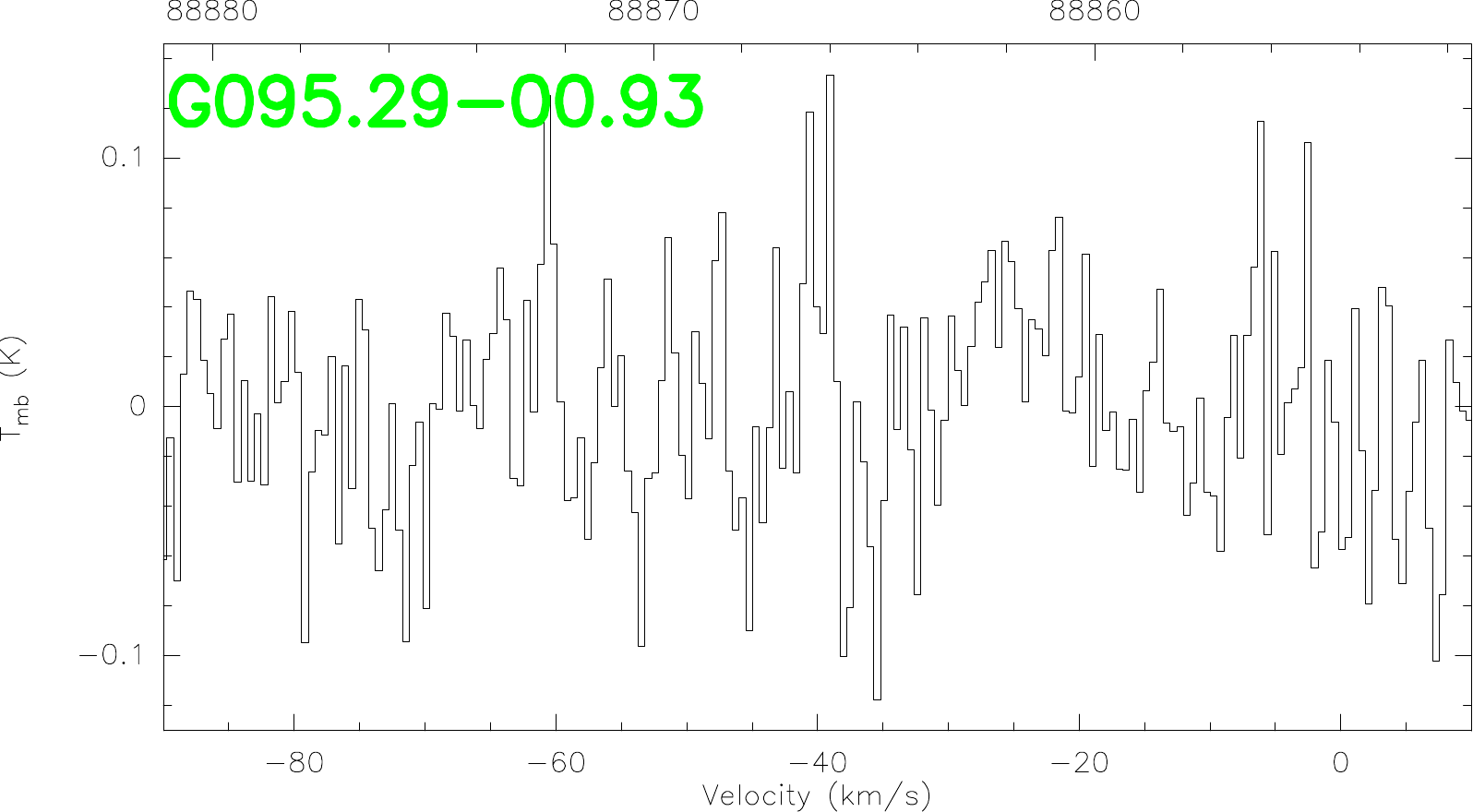}\\

\includegraphics[width=0.155\textwidth]{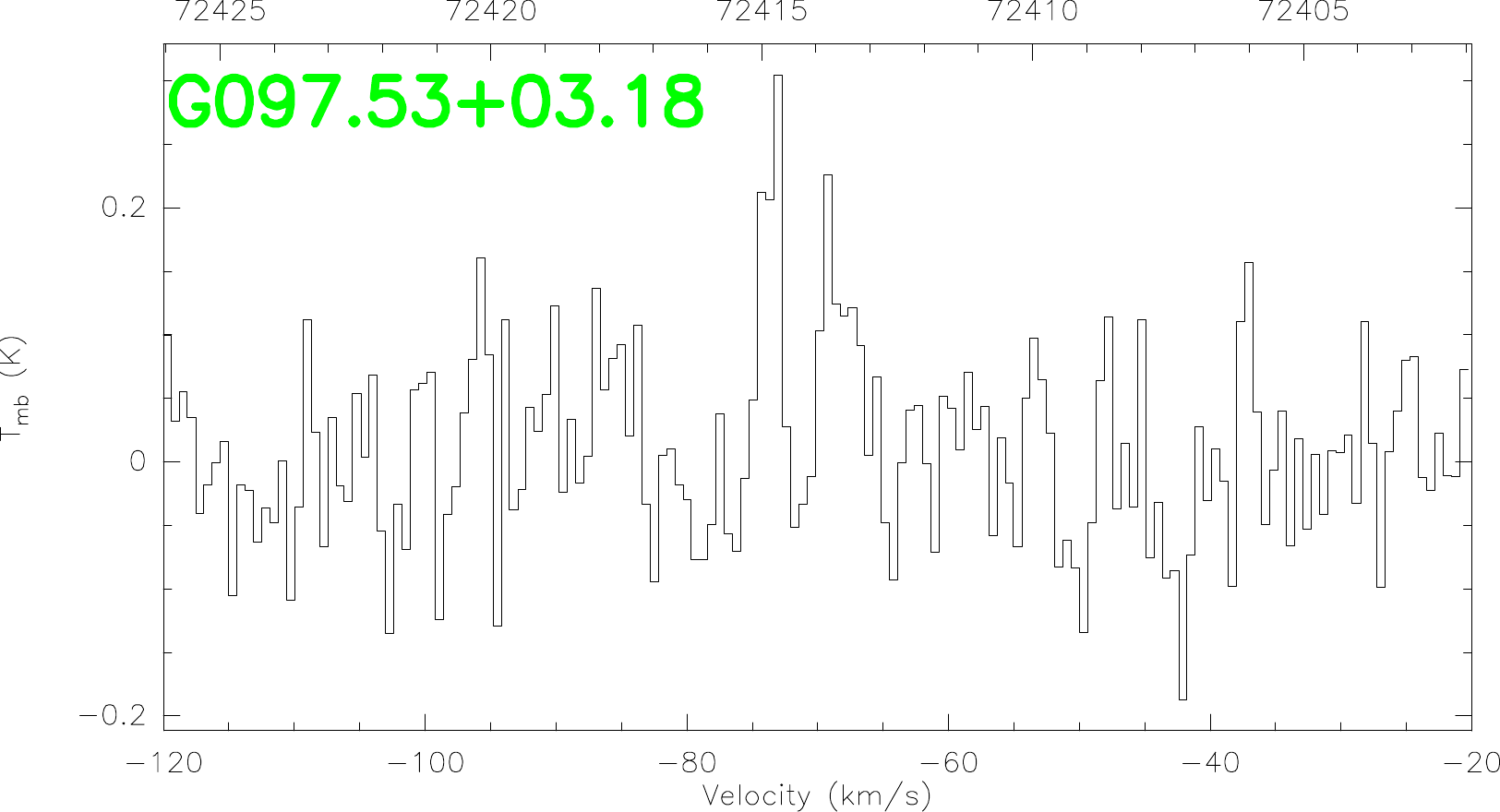}
\includegraphics[width=0.155\textwidth]{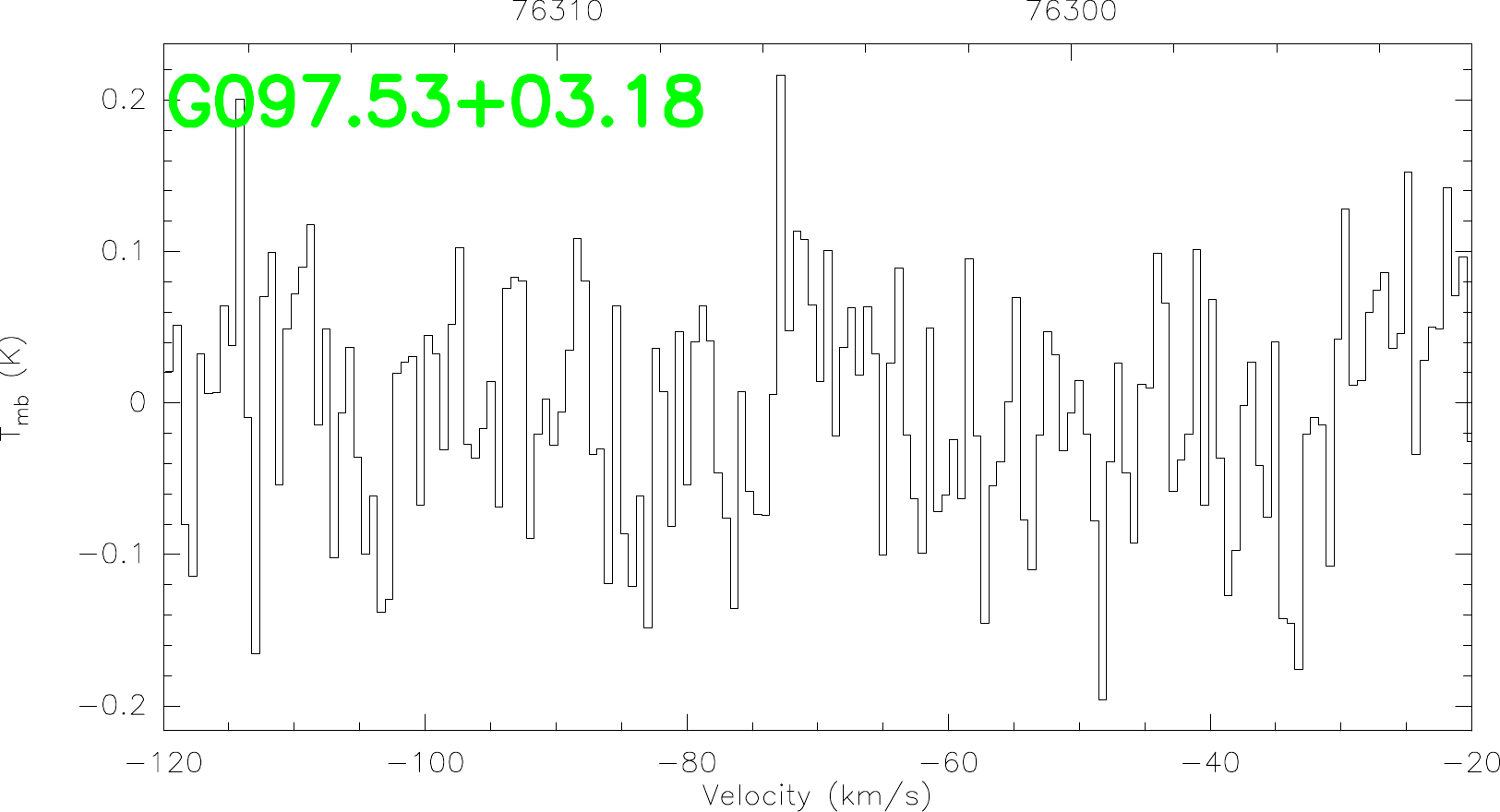}
\includegraphics[width=0.155\textwidth]{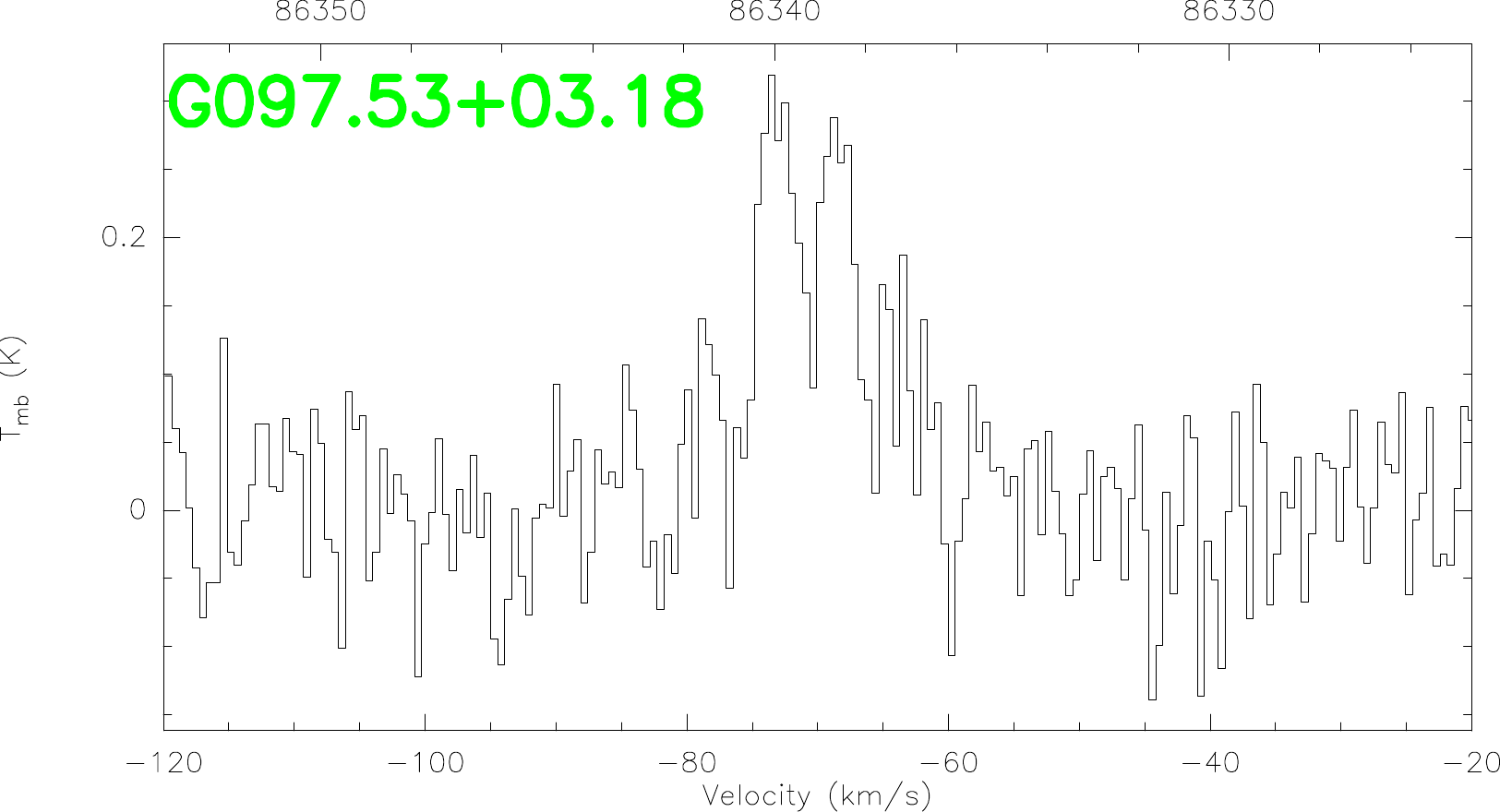}
\includegraphics[width=0.155\textwidth]{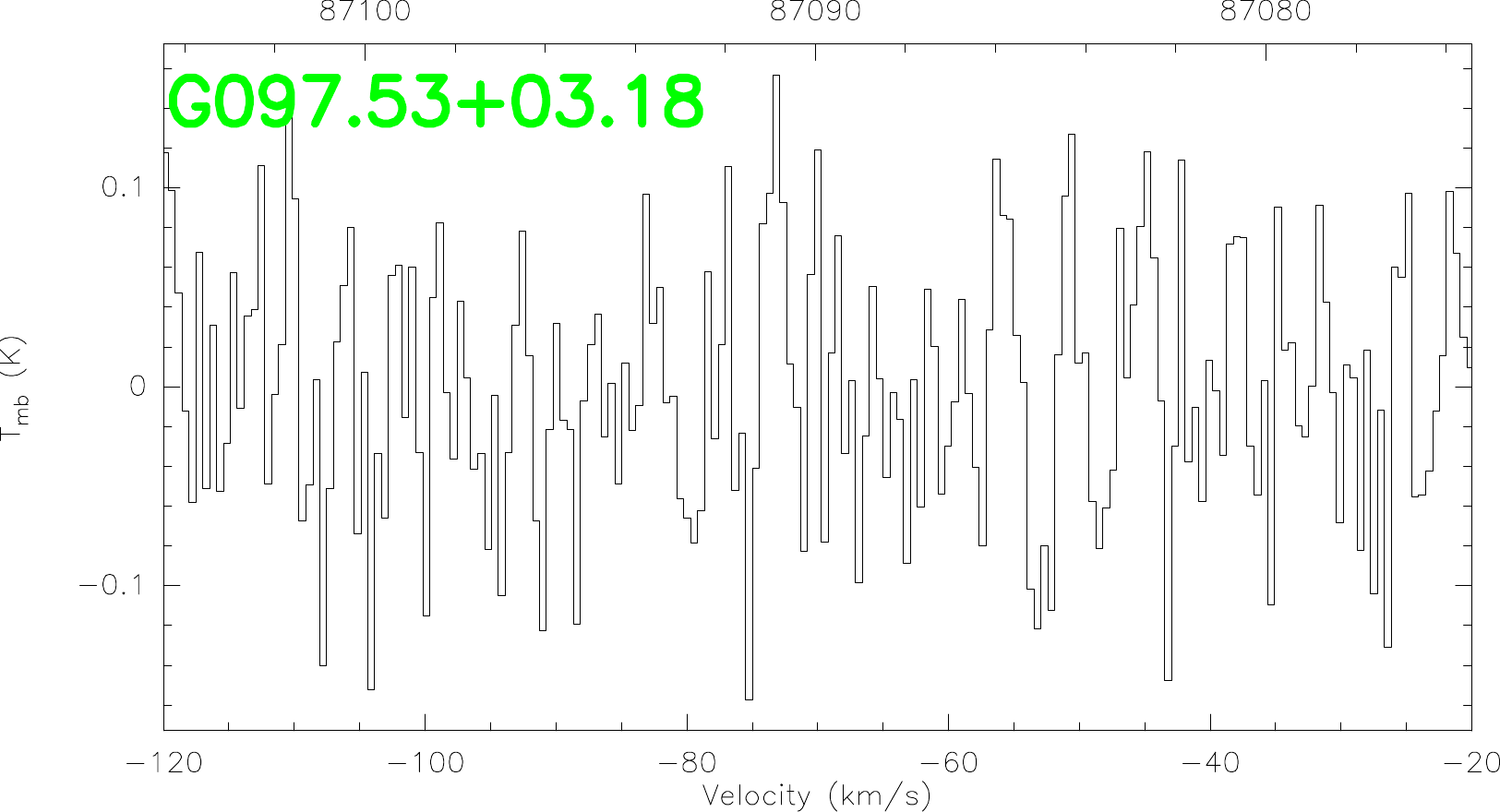}
\includegraphics[width=0.155\textwidth]{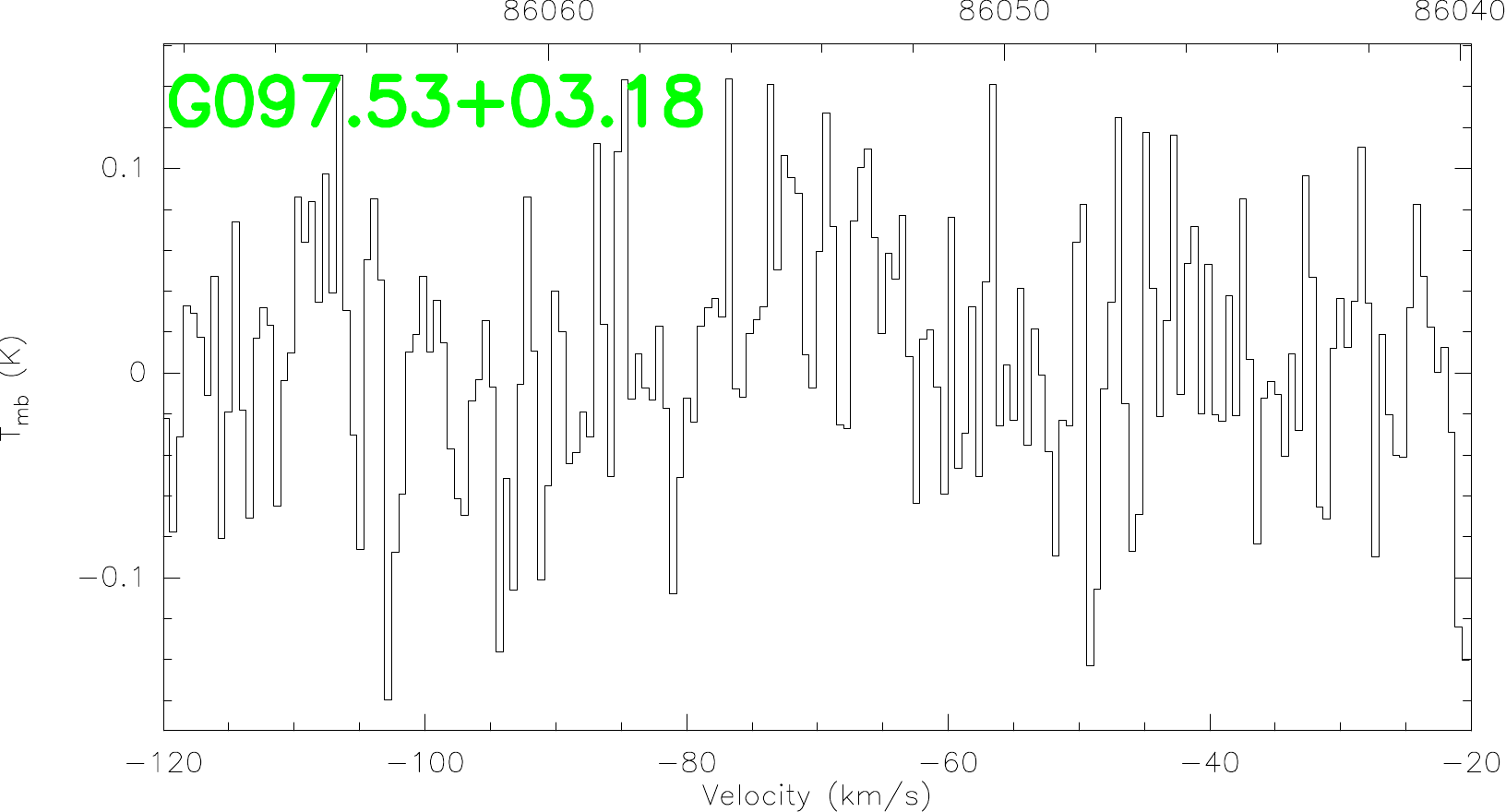}
\includegraphics[width=0.155\textwidth]{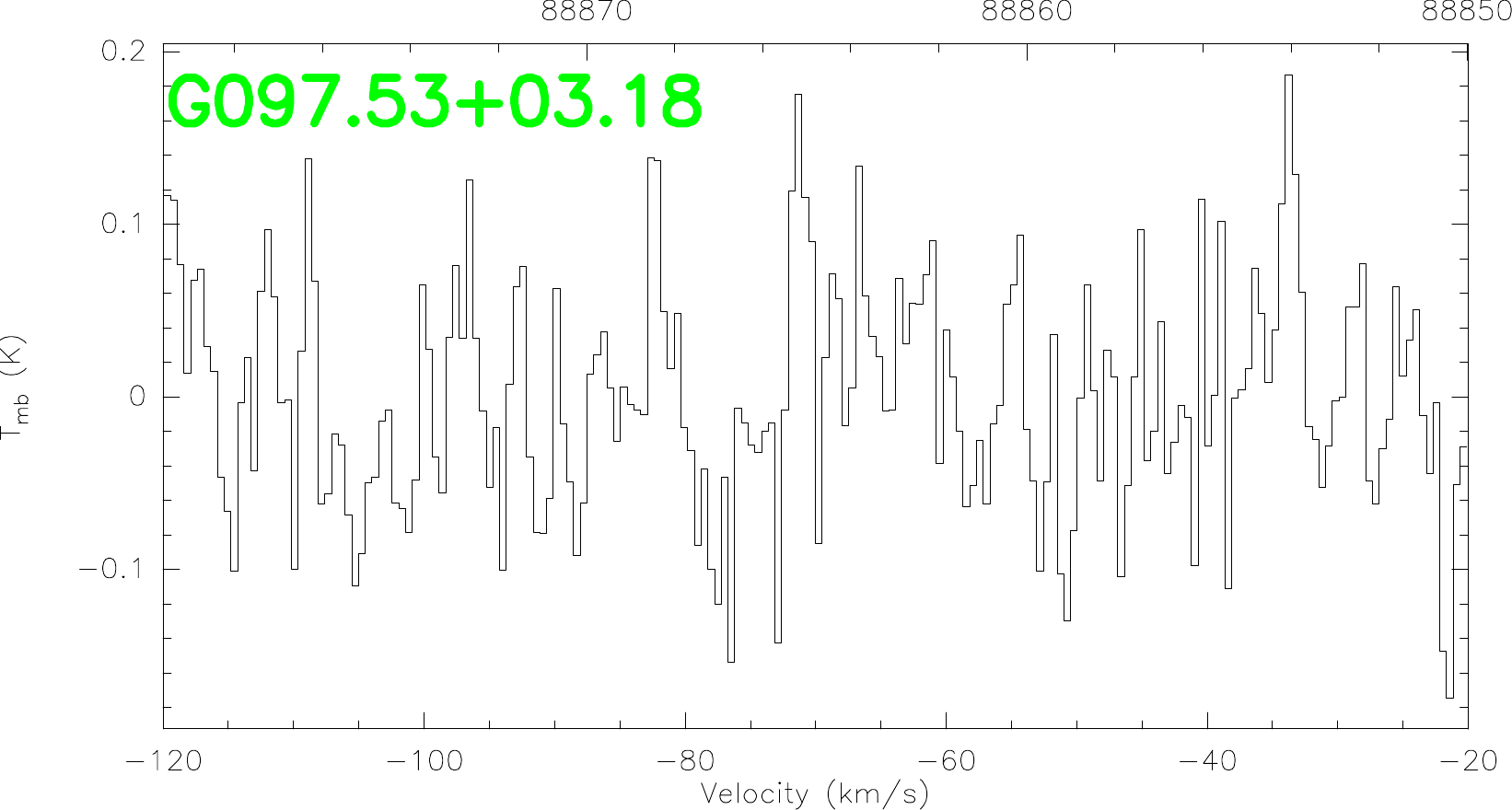}\\

\includegraphics[width=0.155\textwidth]{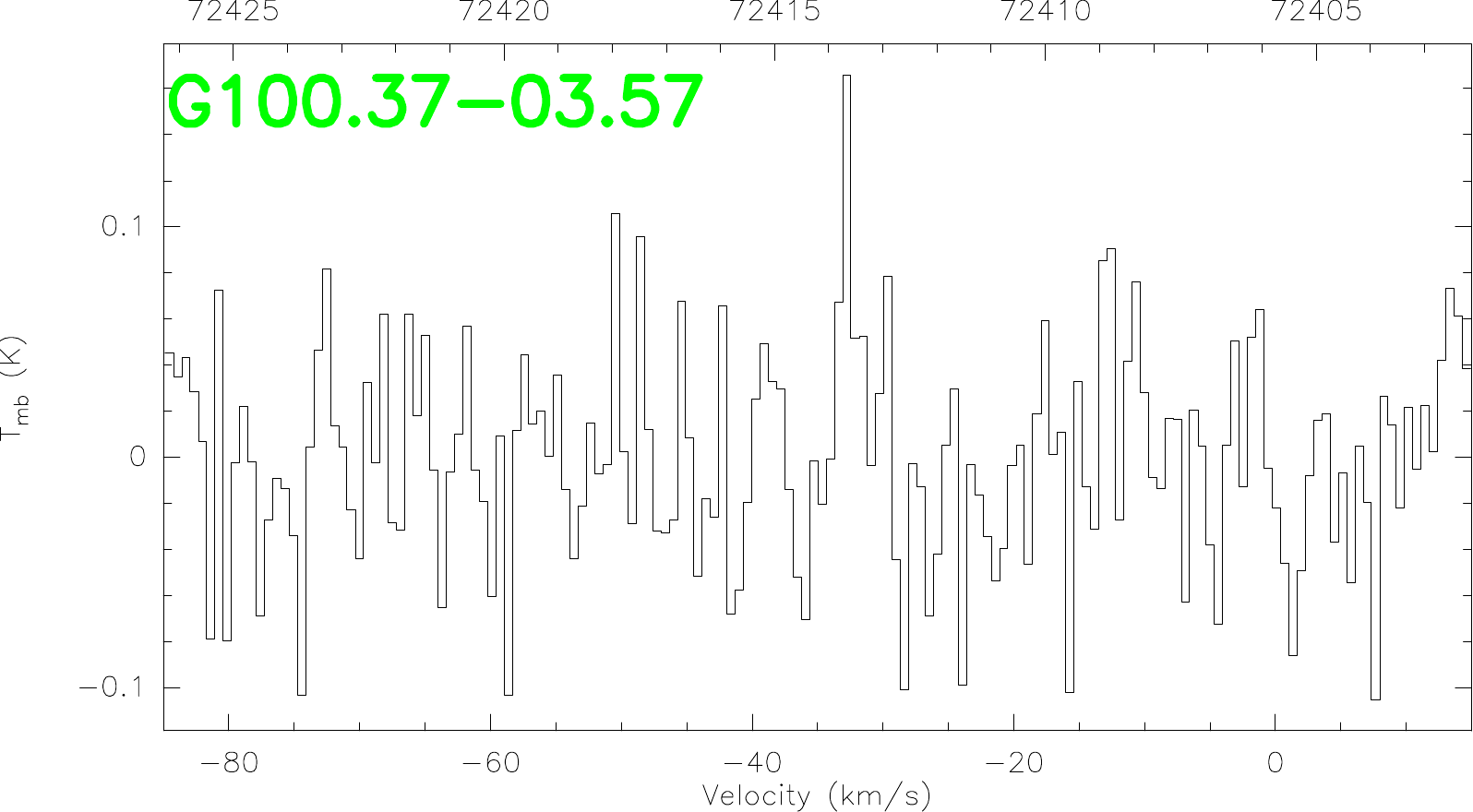}
\includegraphics[width=0.155\textwidth]{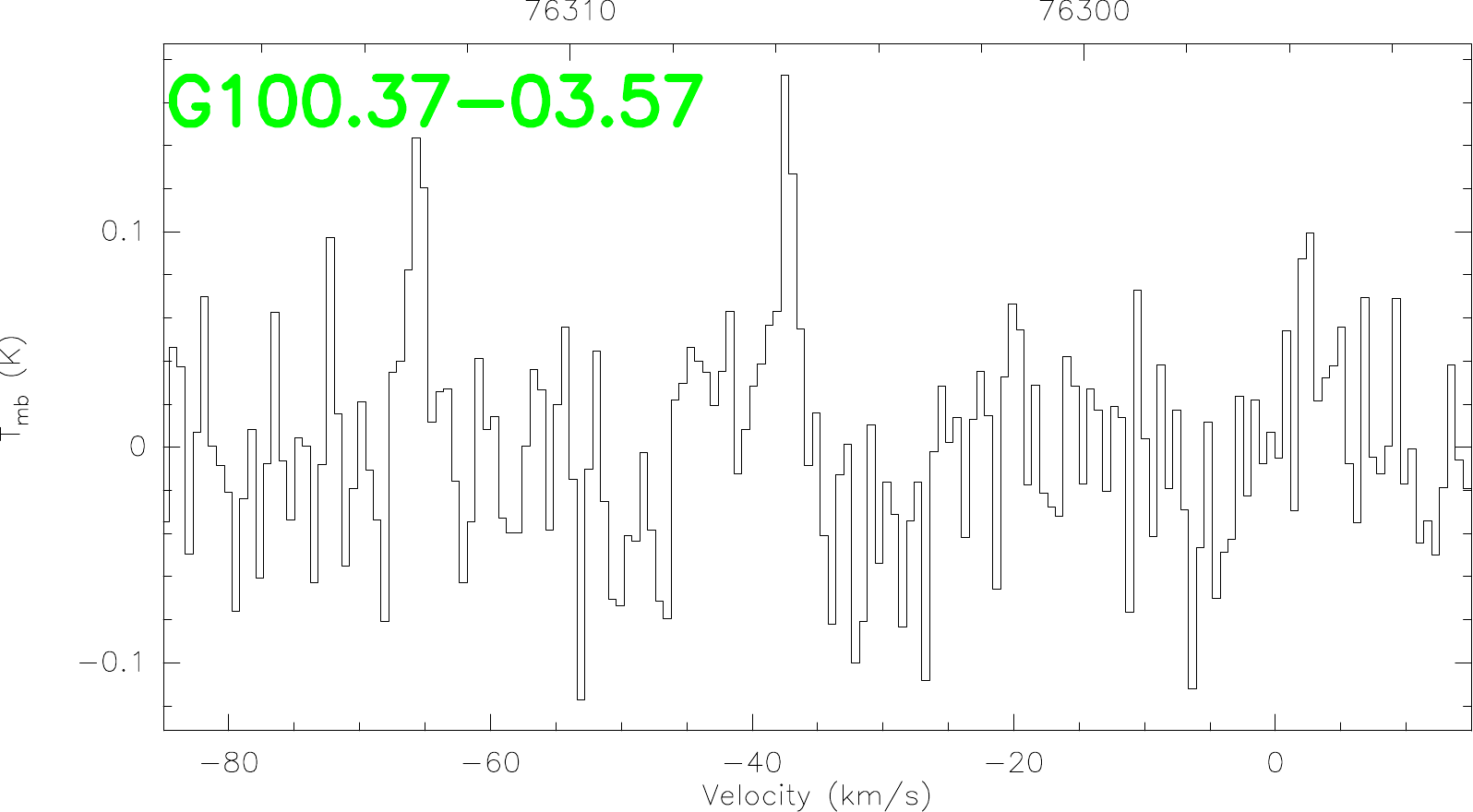}
\includegraphics[width=0.155\textwidth]{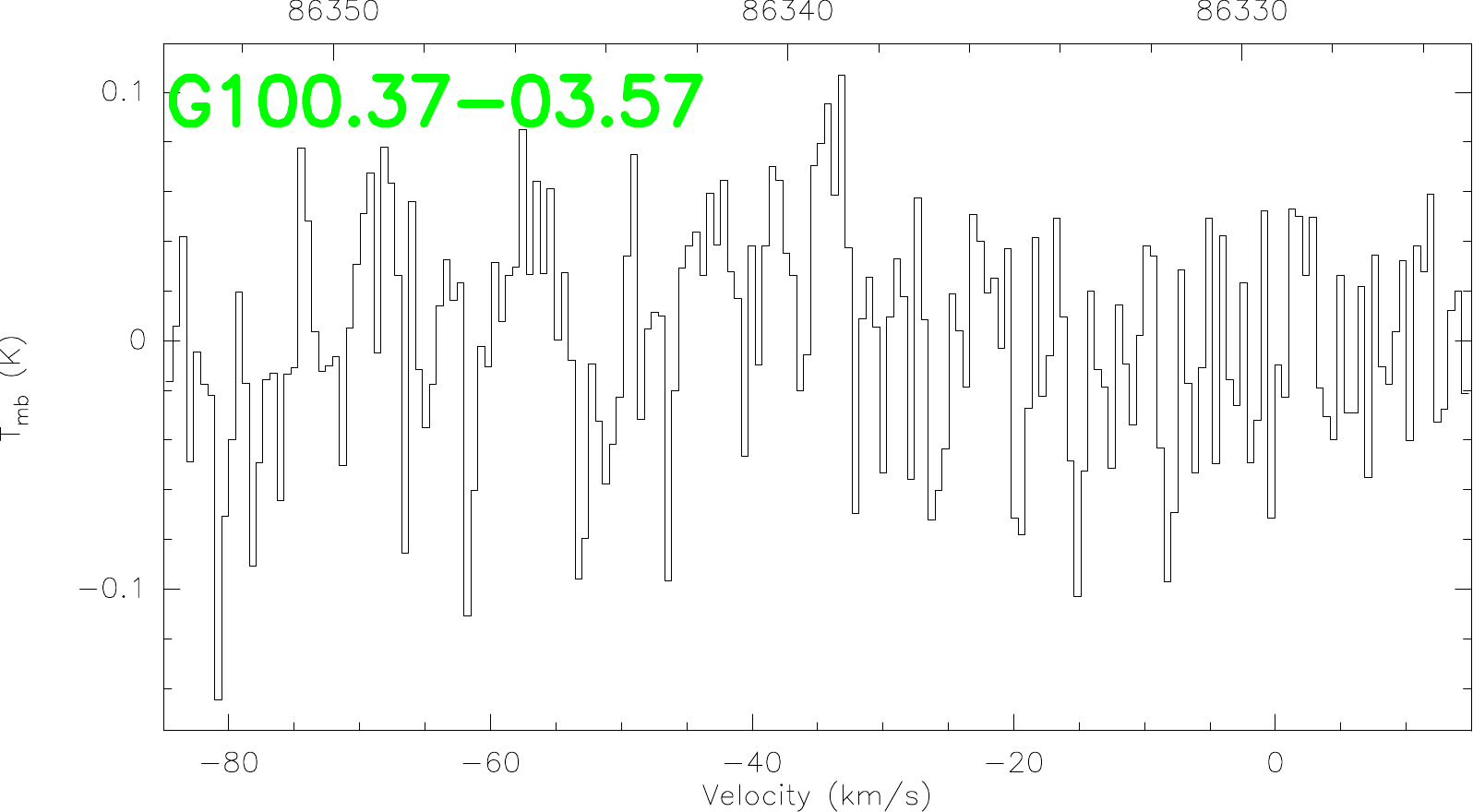}
\includegraphics[width=0.155\textwidth]{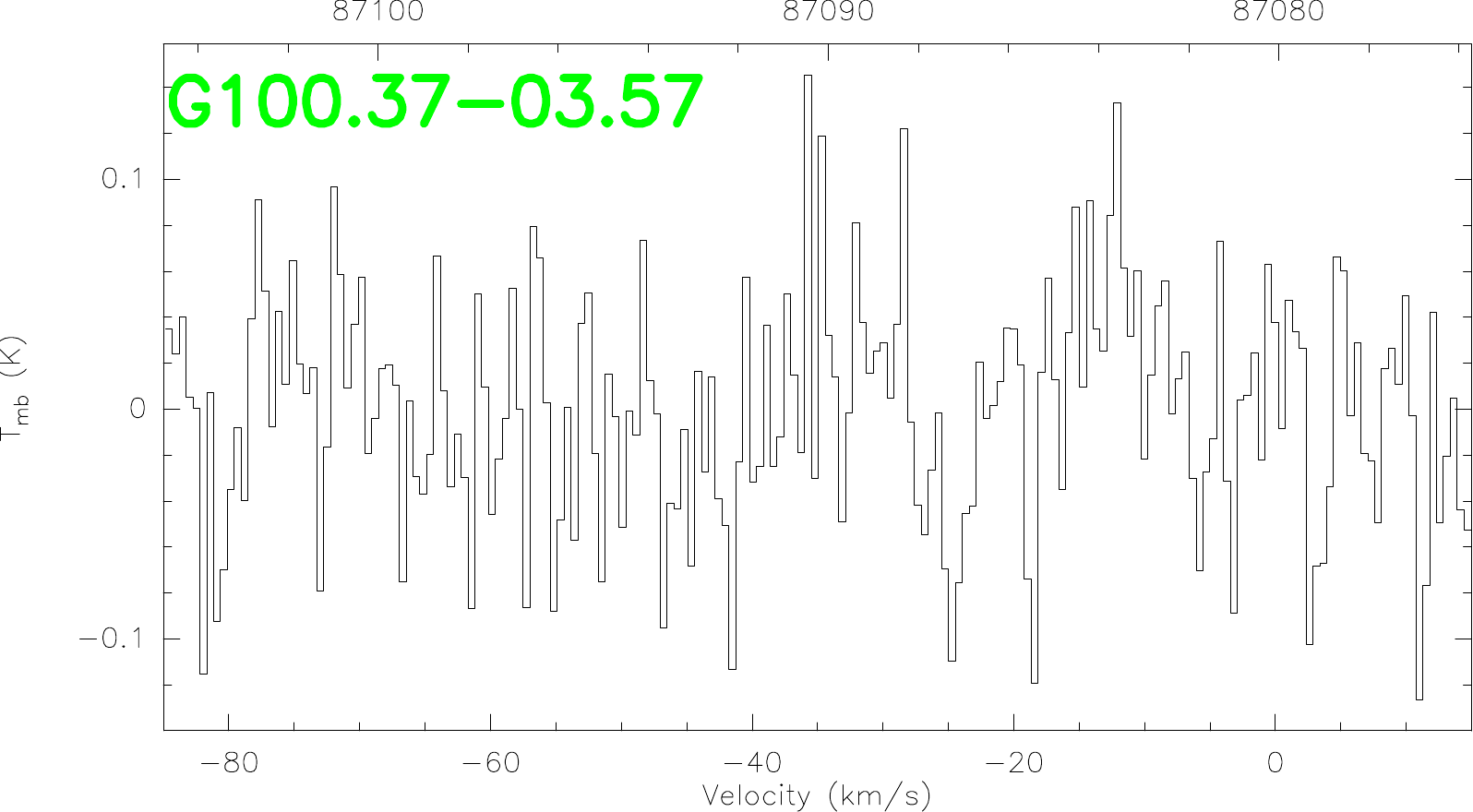}
\includegraphics[width=0.155\textwidth]{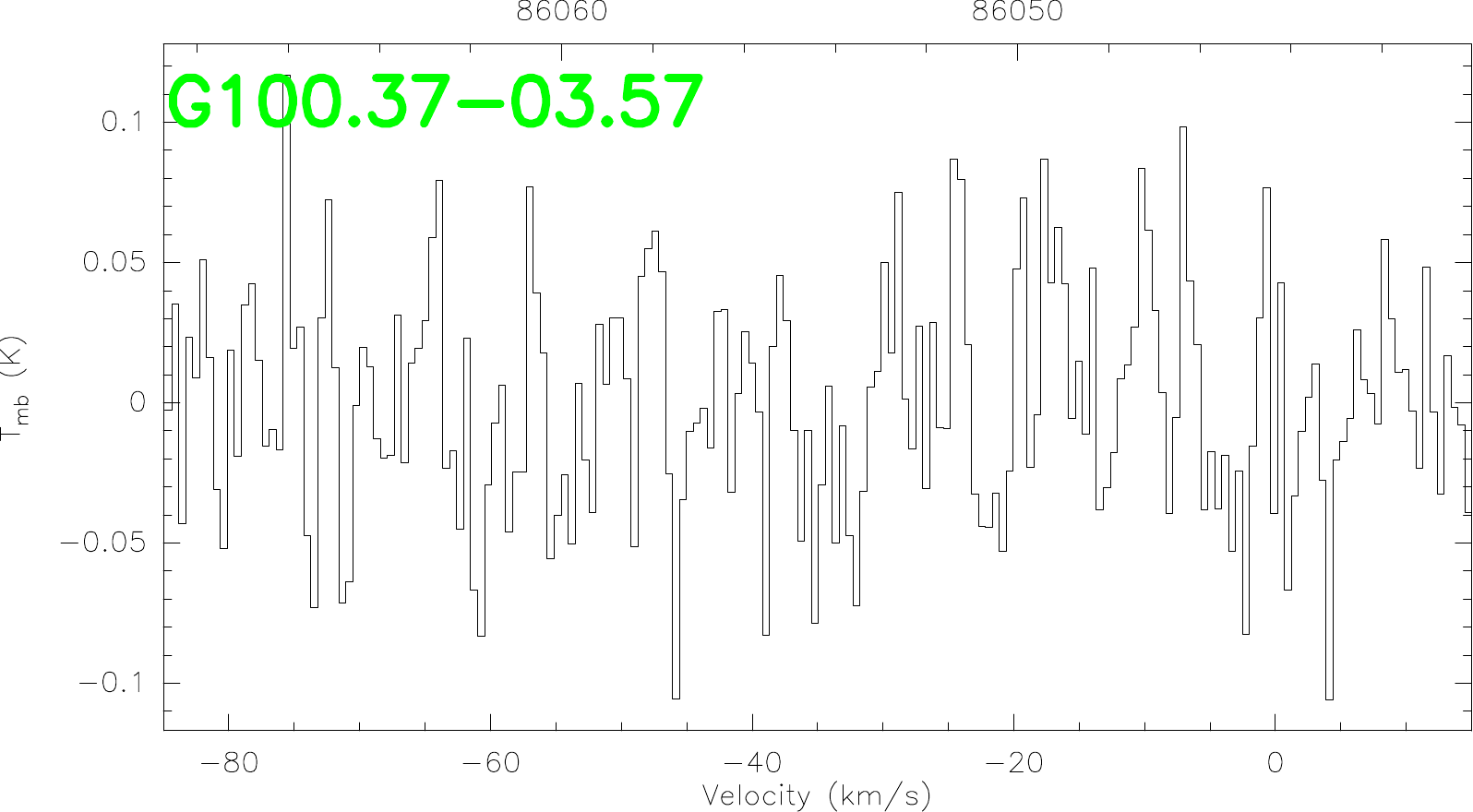}
\includegraphics[width=0.155\textwidth]{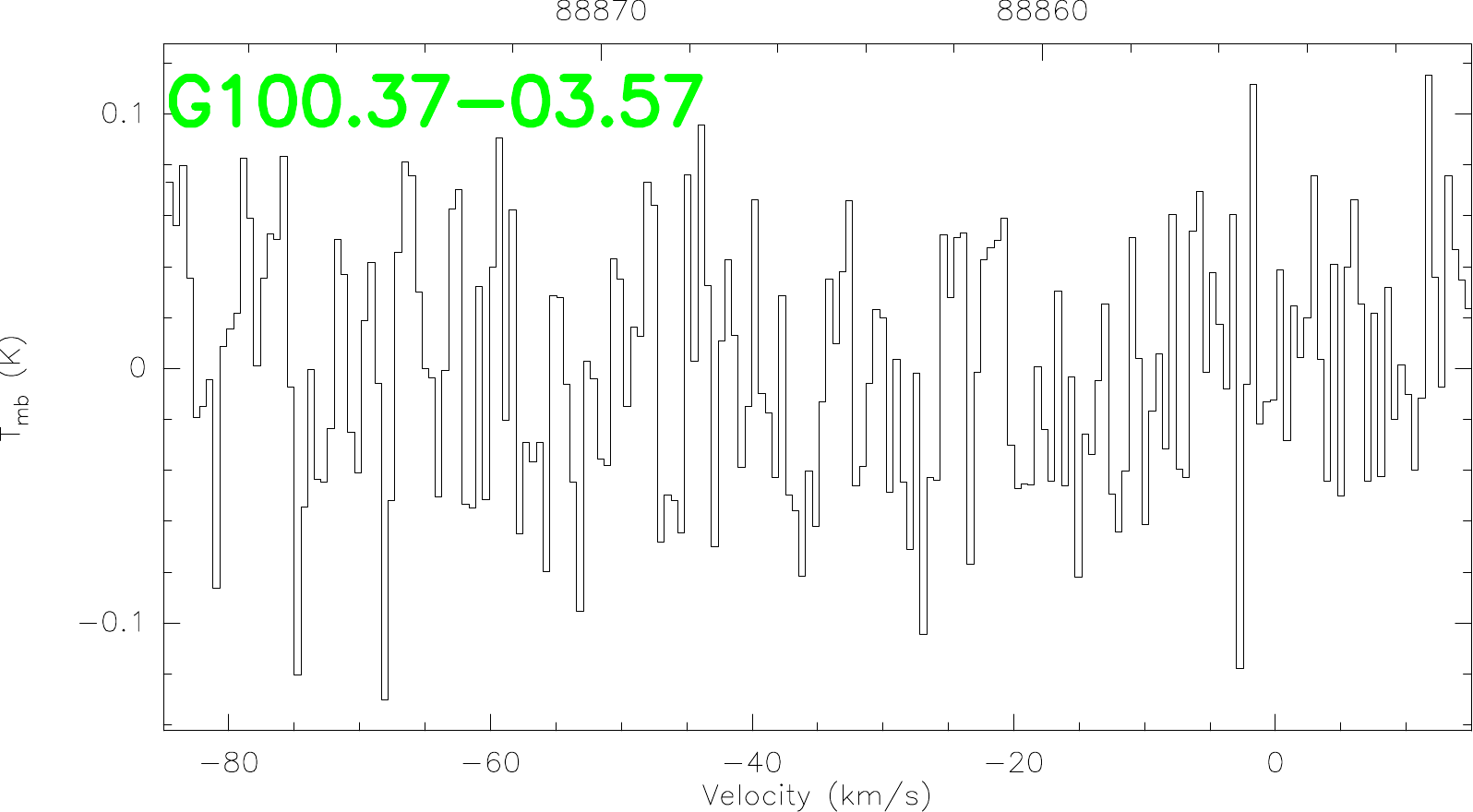}\\

\includegraphics[width=0.155\textwidth]{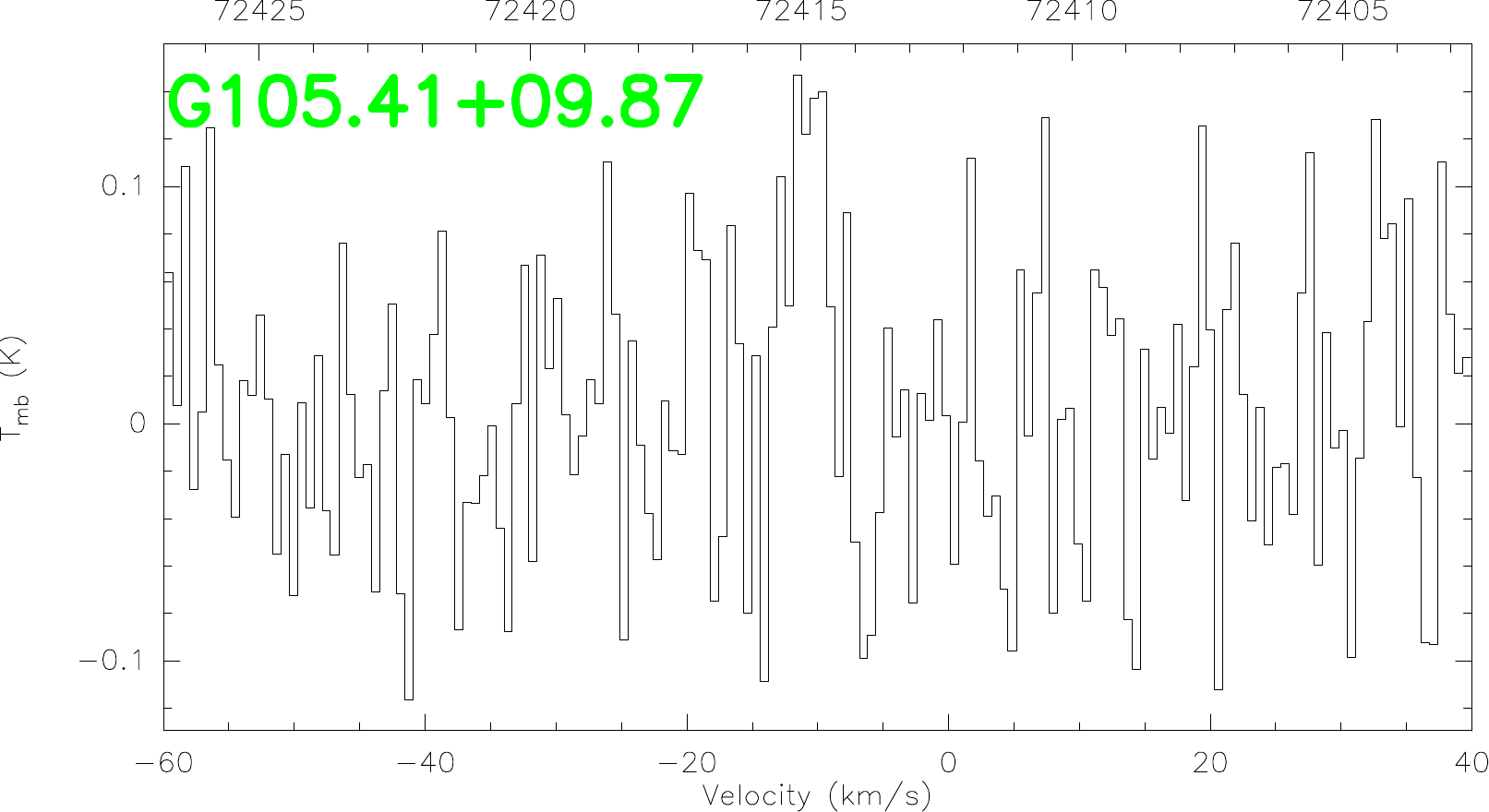}
\includegraphics[width=0.155\textwidth]{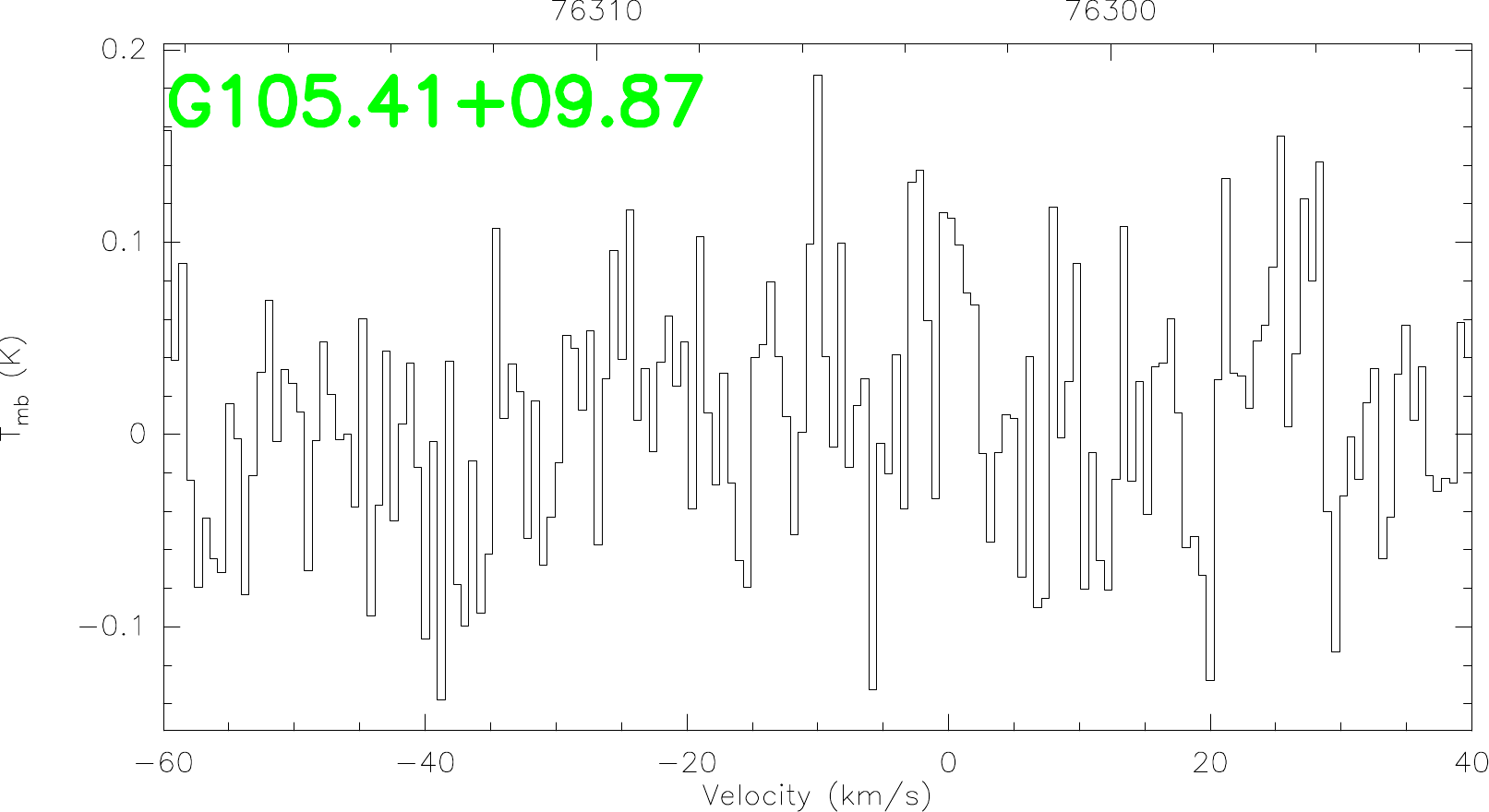}
\includegraphics[width=0.155\textwidth]{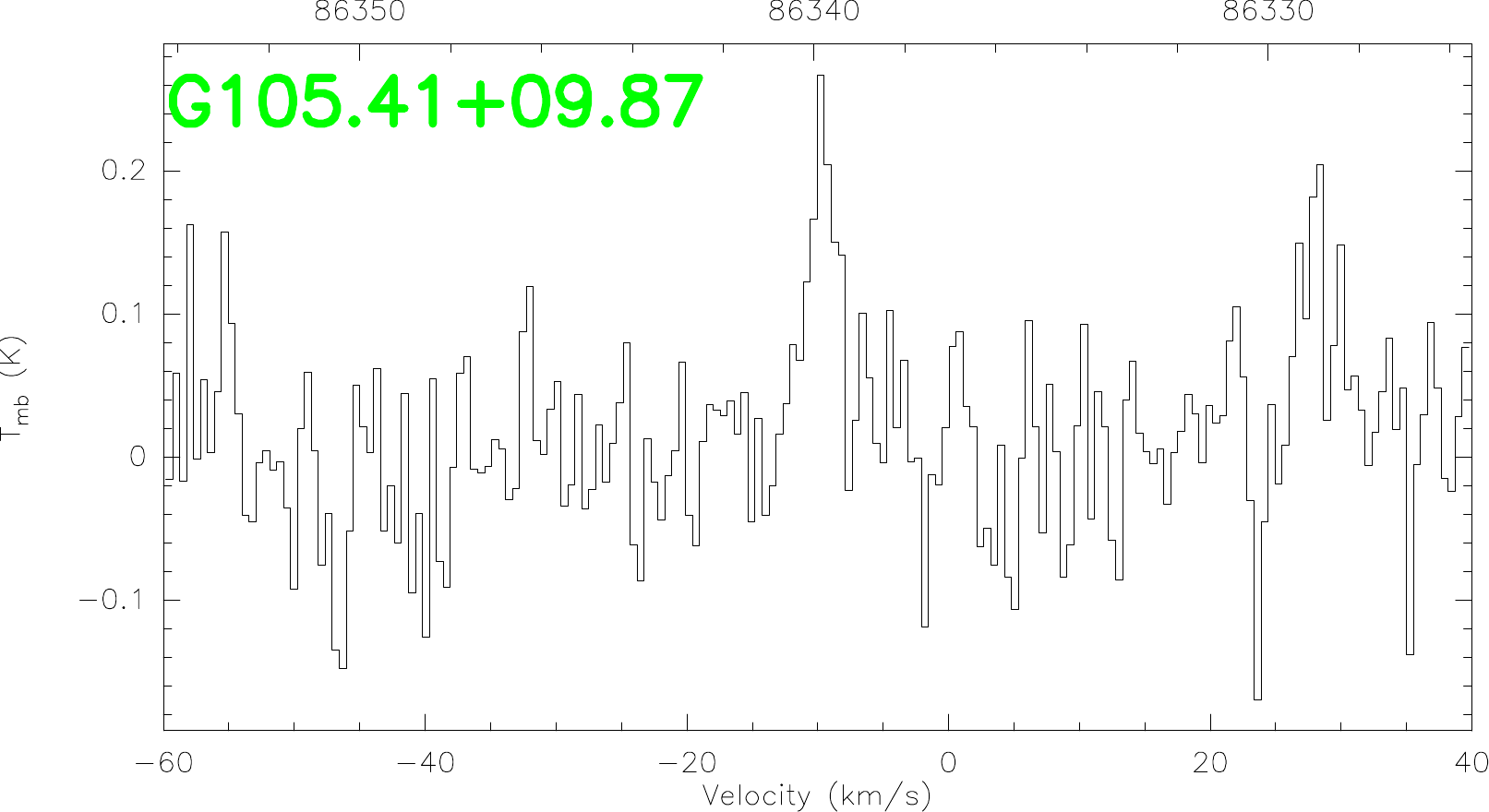}
\includegraphics[width=0.155\textwidth]{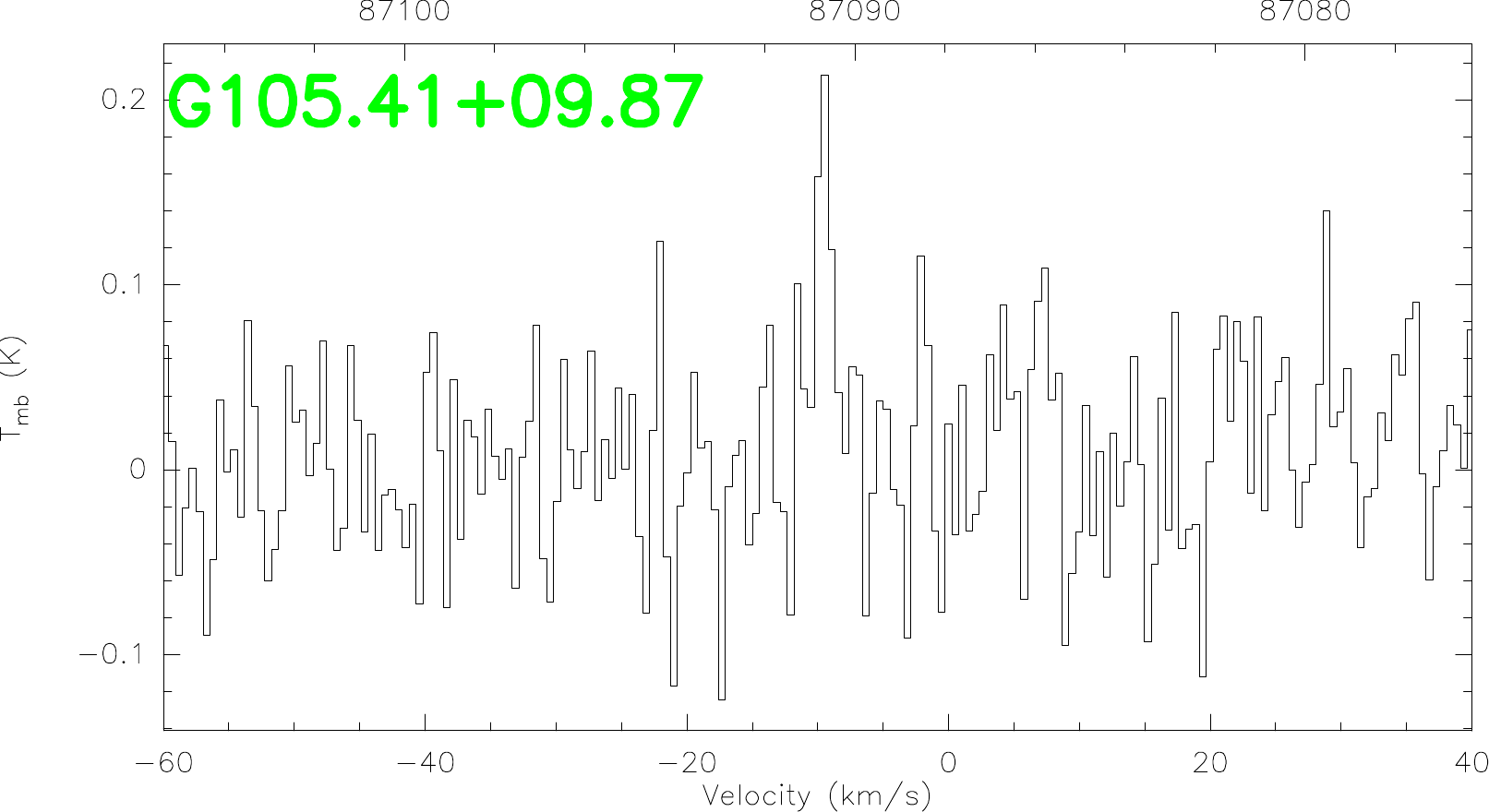}
\includegraphics[width=0.155\textwidth]{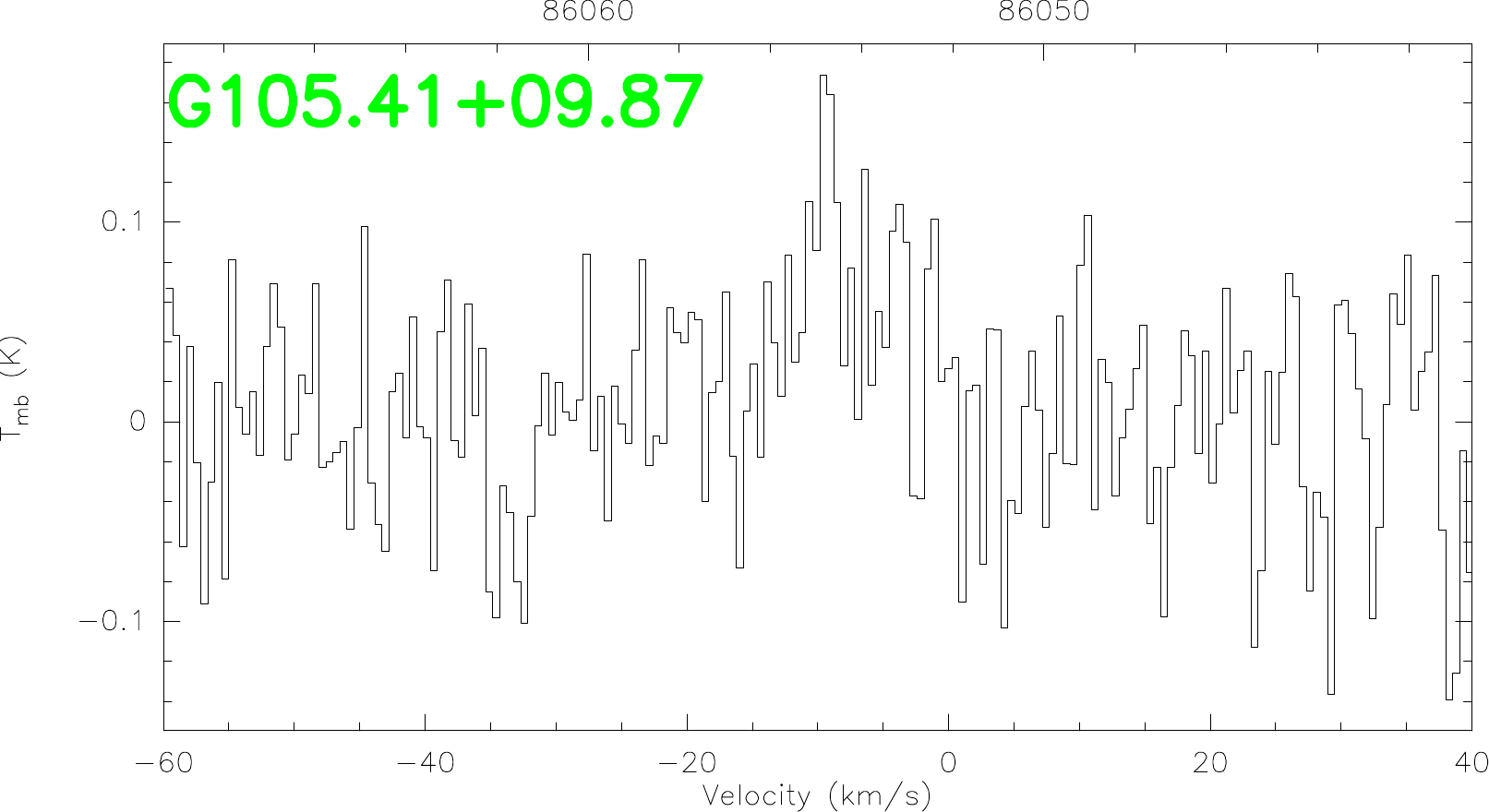}
\includegraphics[width=0.155\textwidth]{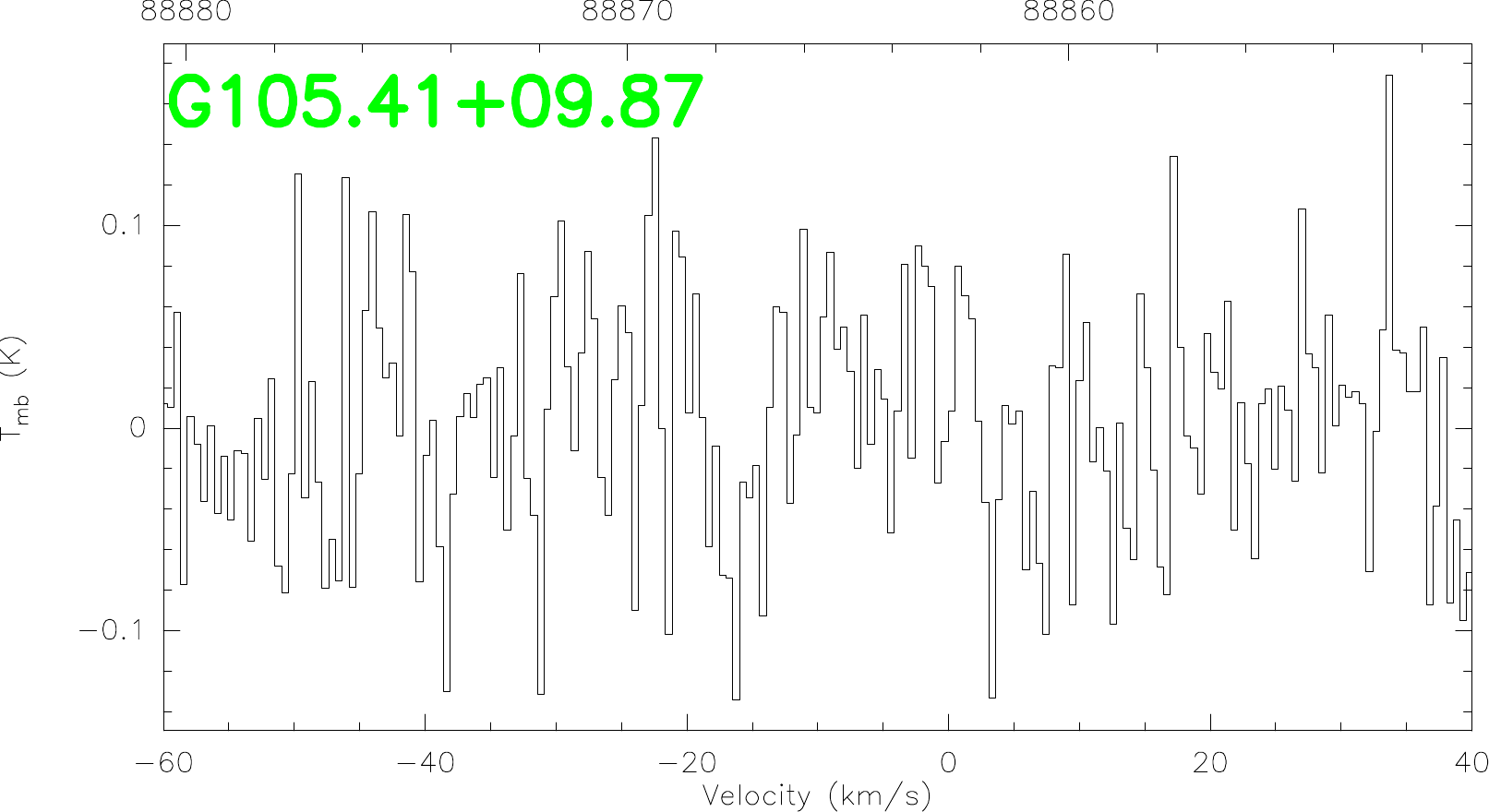}\\

\includegraphics[width=0.155\textwidth]{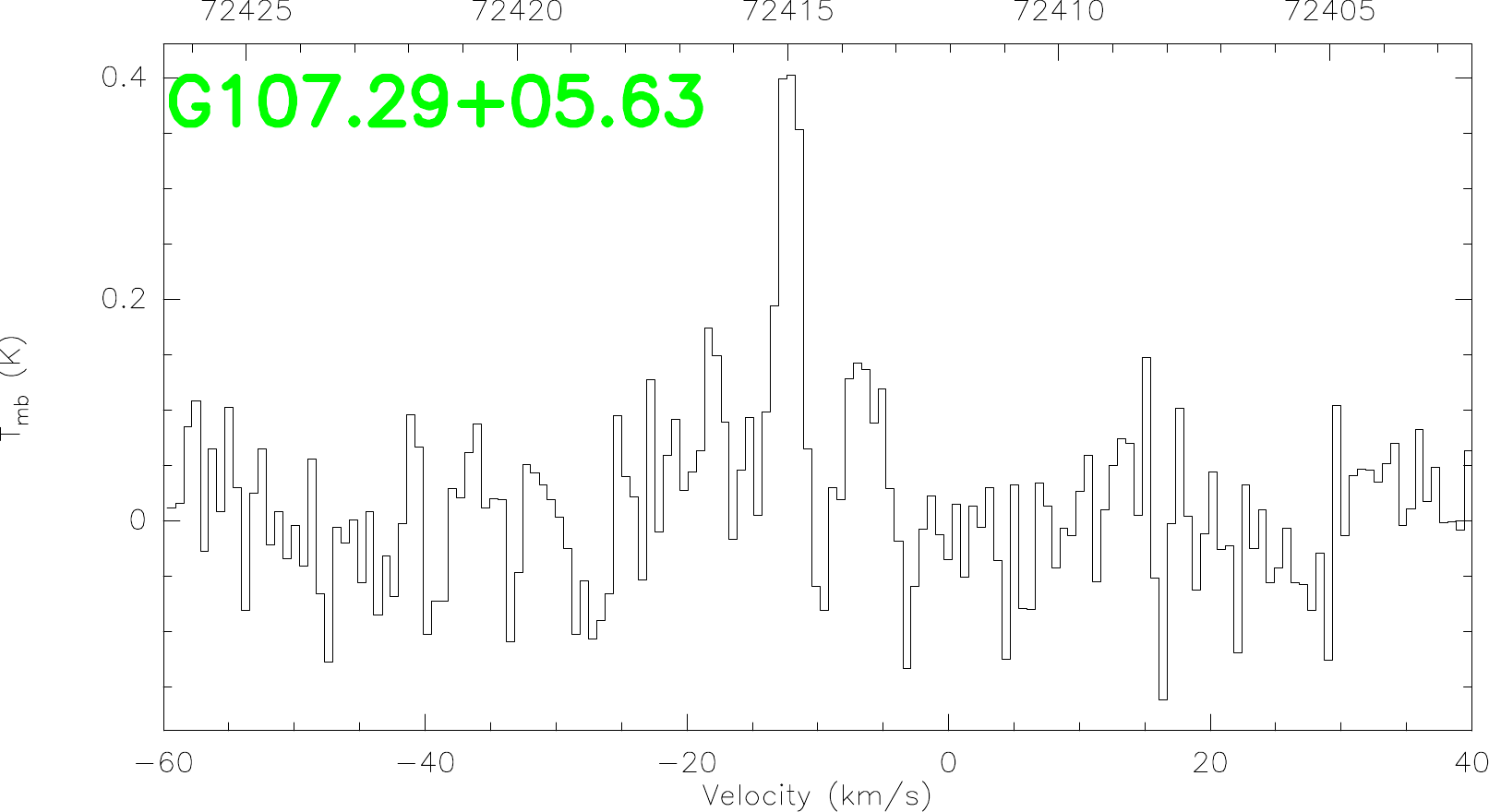}
\includegraphics[width=0.155\textwidth]{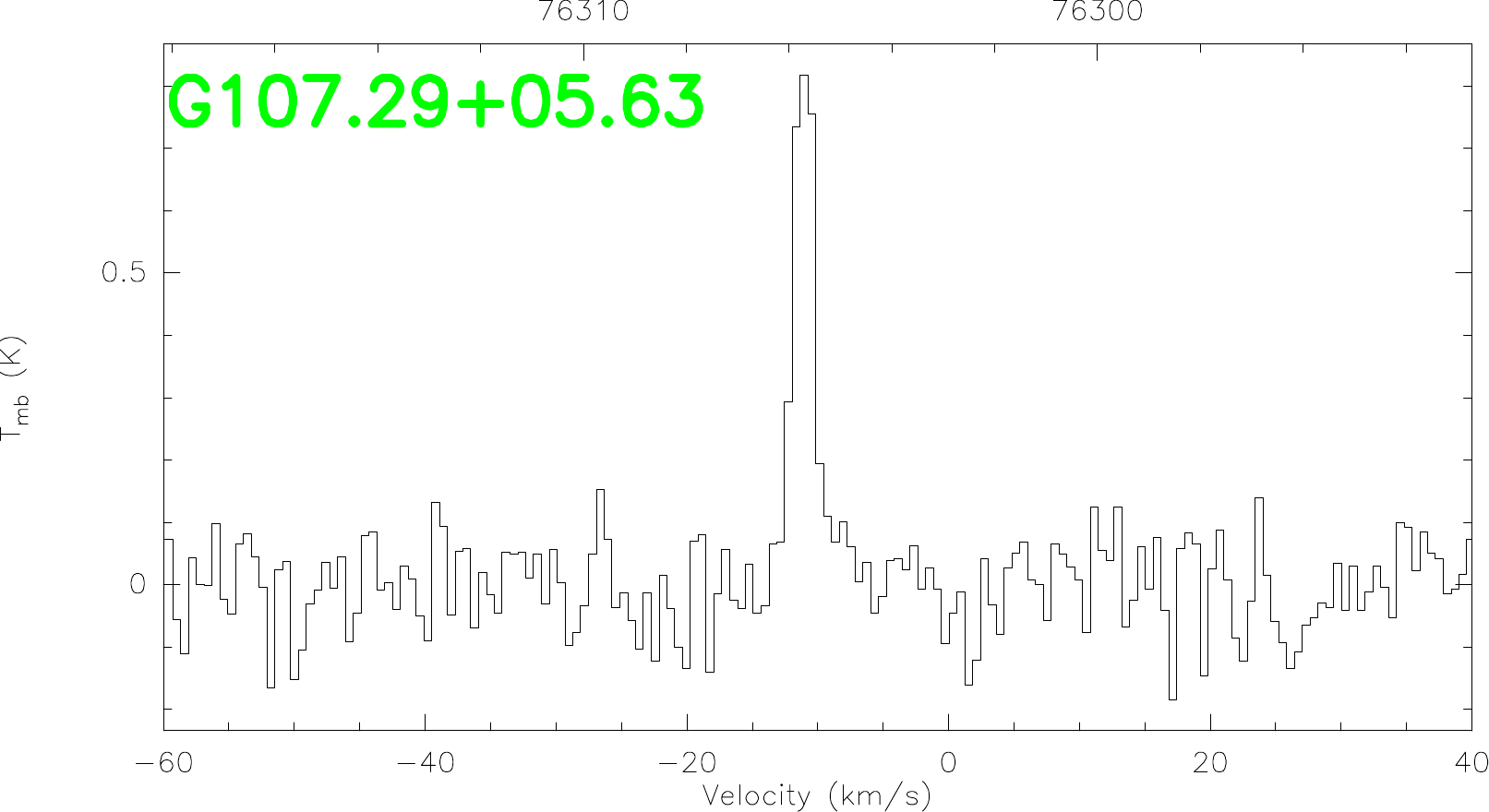}
\includegraphics[width=0.155\textwidth]{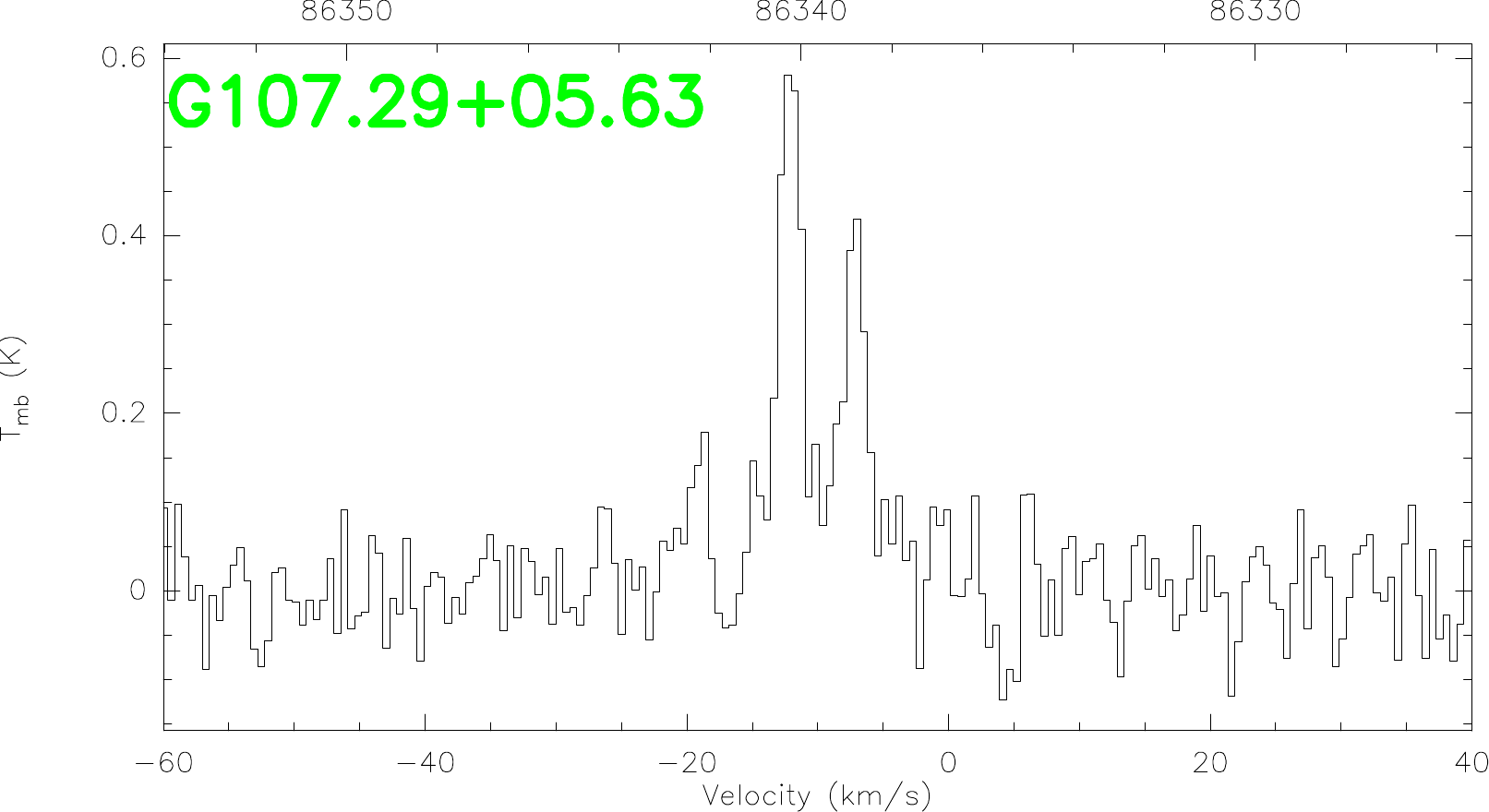}
\includegraphics[width=0.155\textwidth]{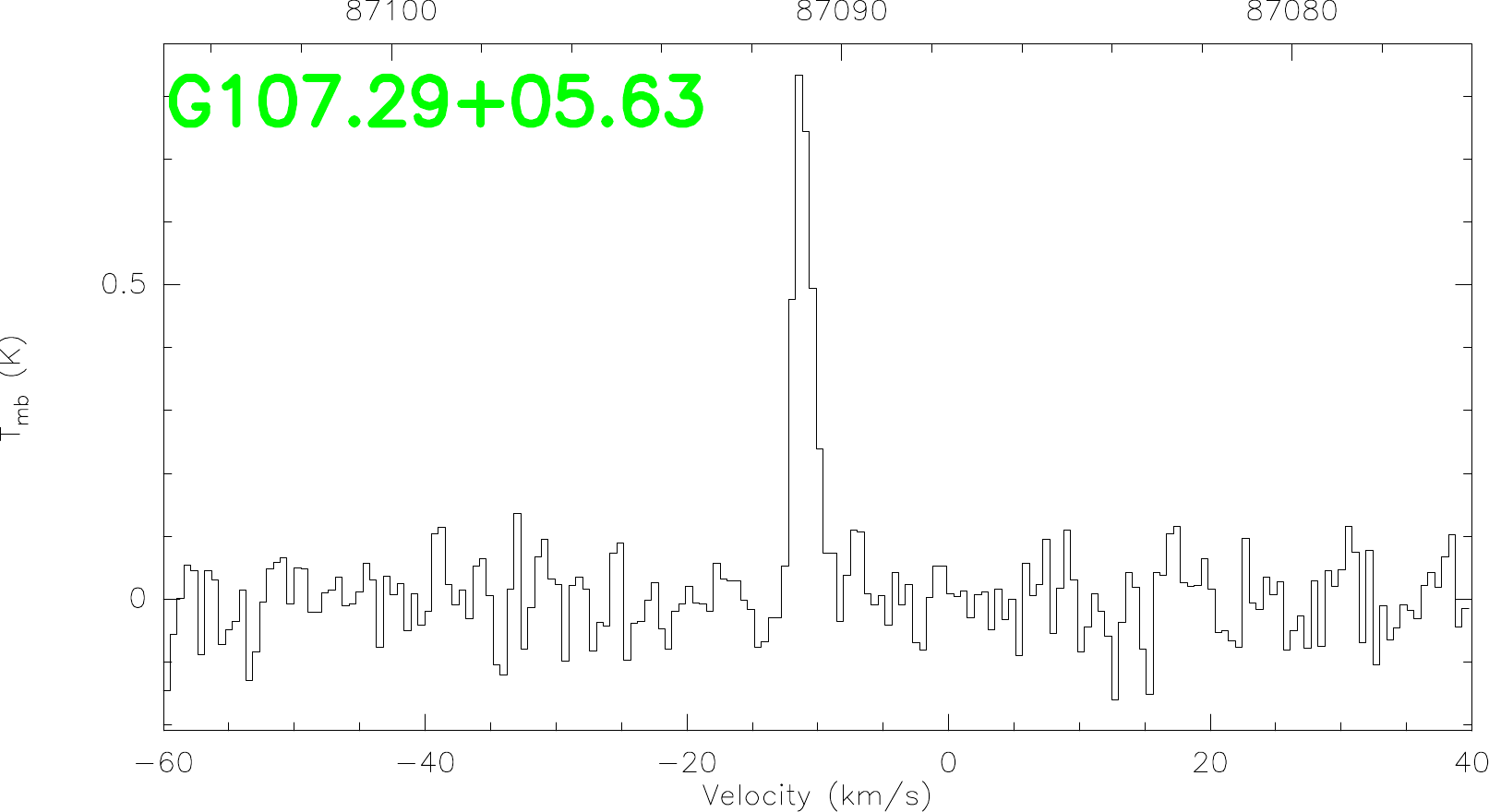}
\includegraphics[width=0.155\textwidth]{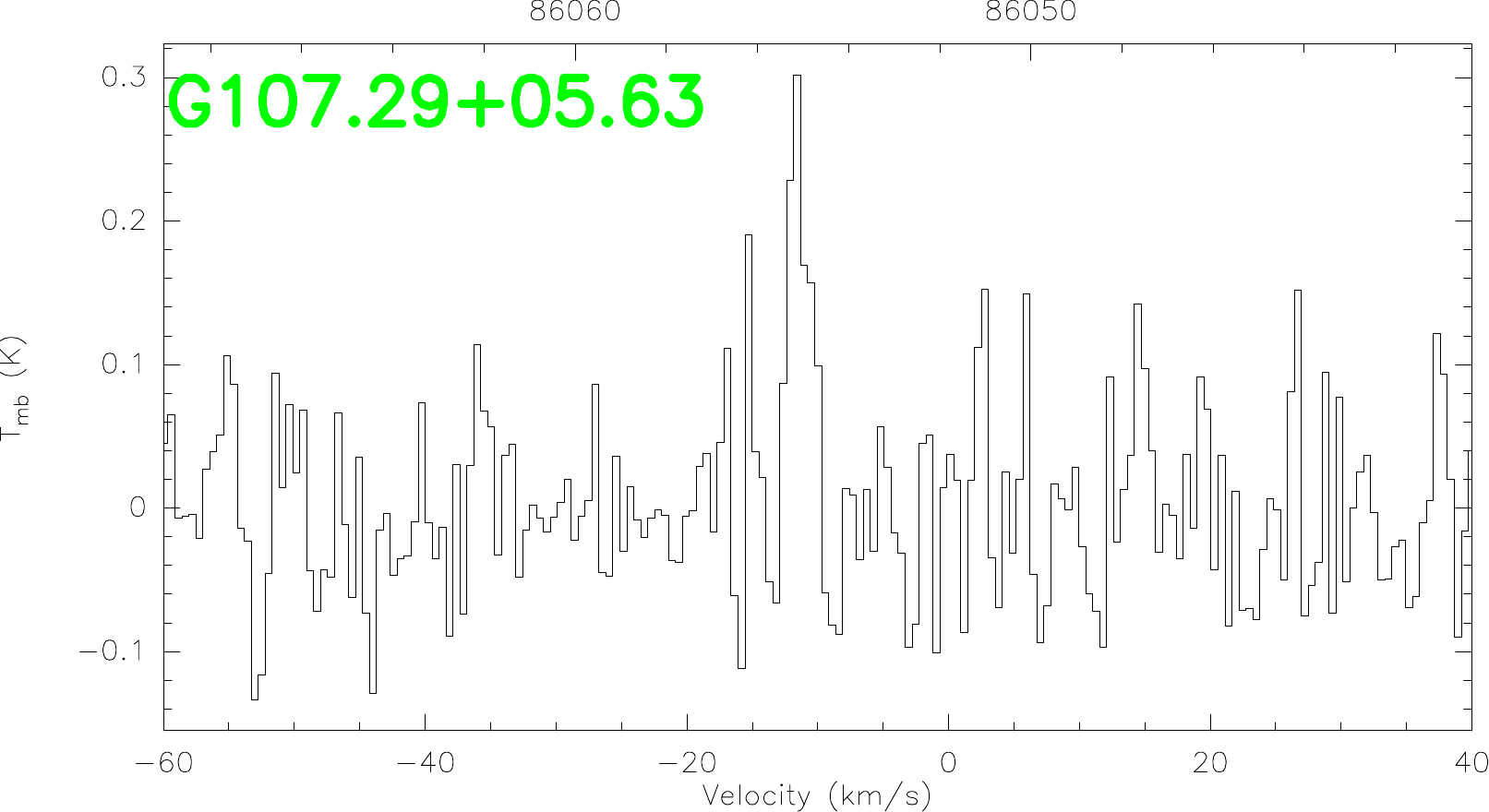}
\includegraphics[width=0.155\textwidth]{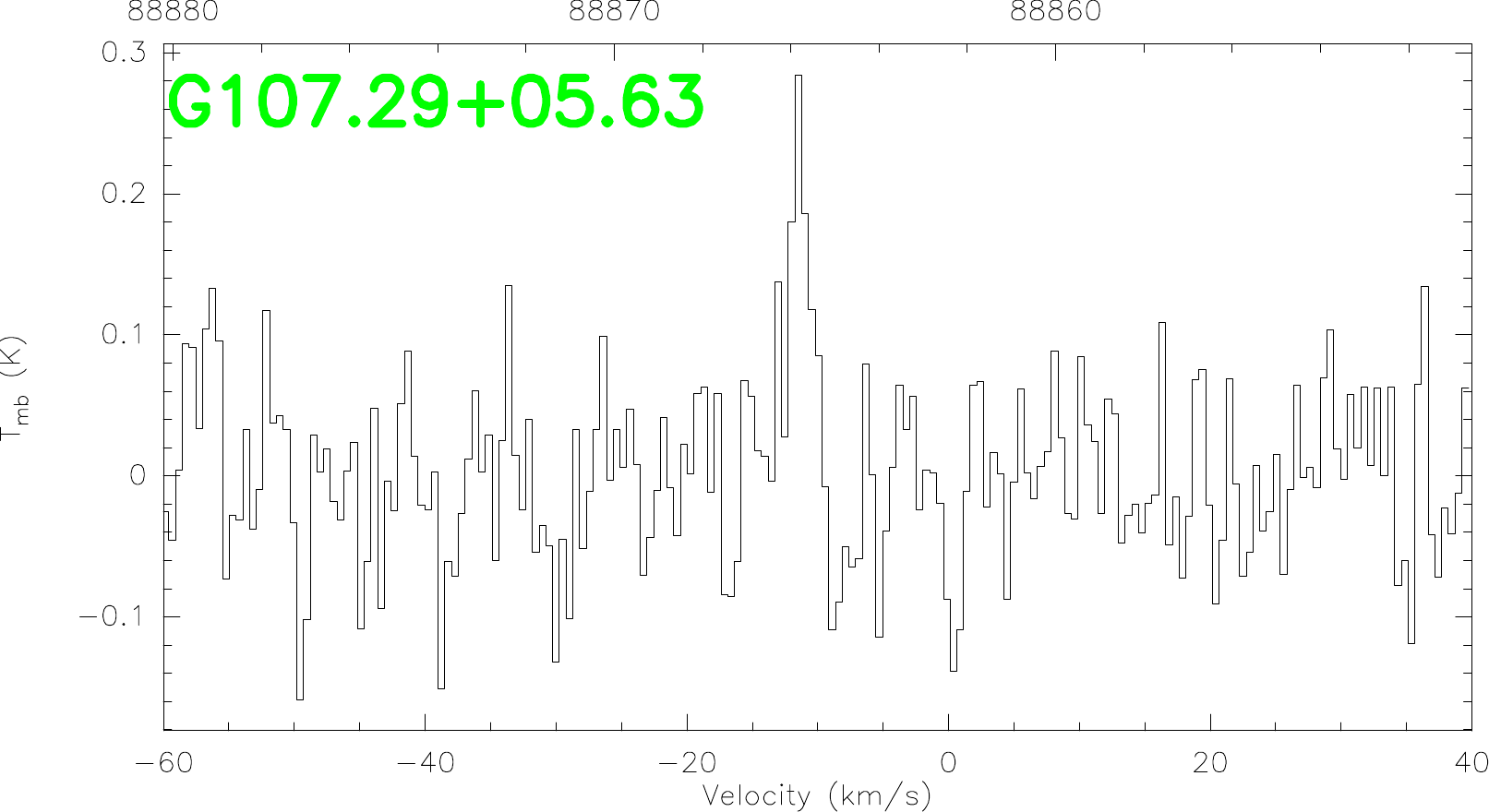}\\

\includegraphics[width=0.155\textwidth]{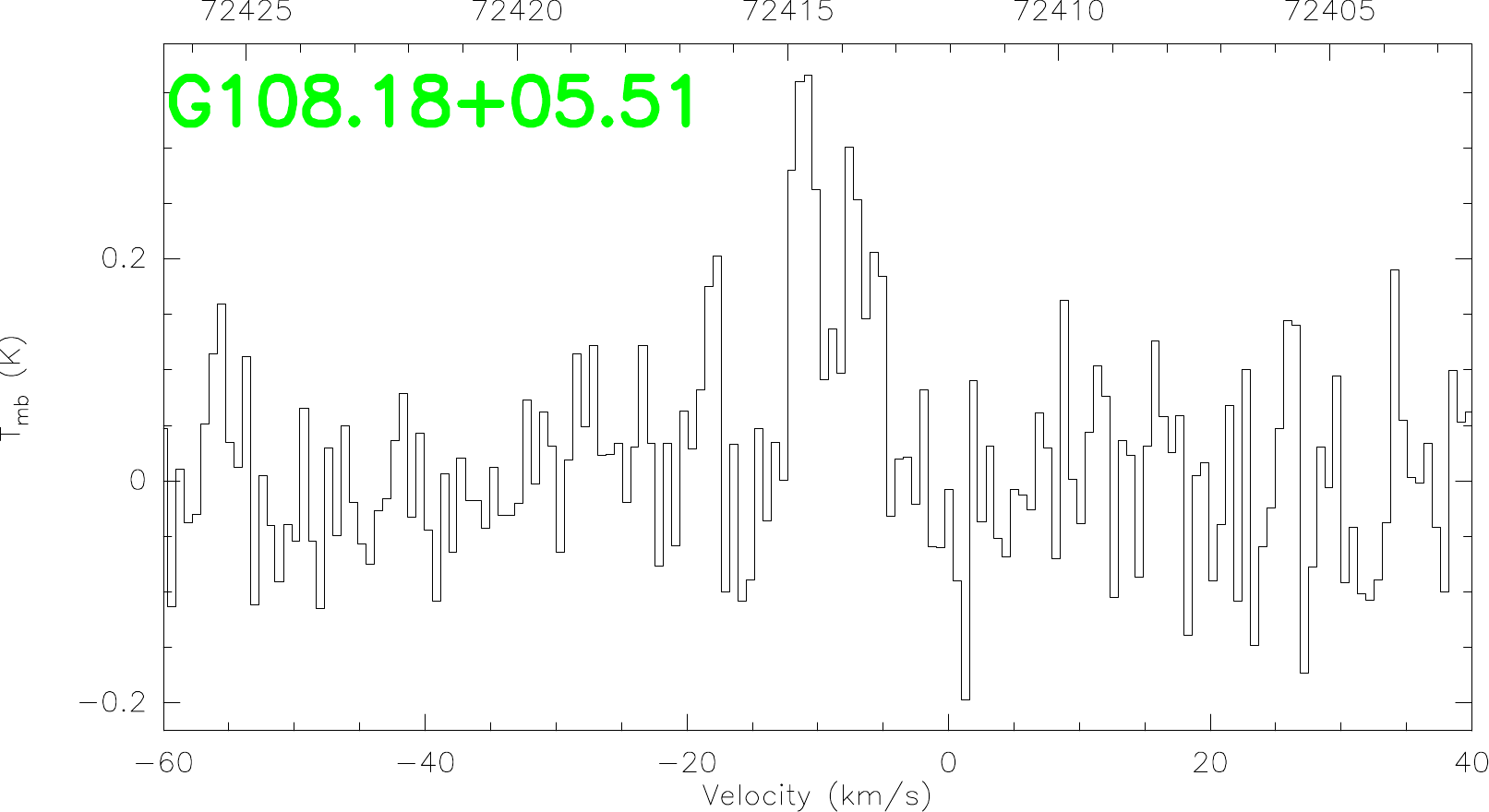}
\includegraphics[width=0.155\textwidth]{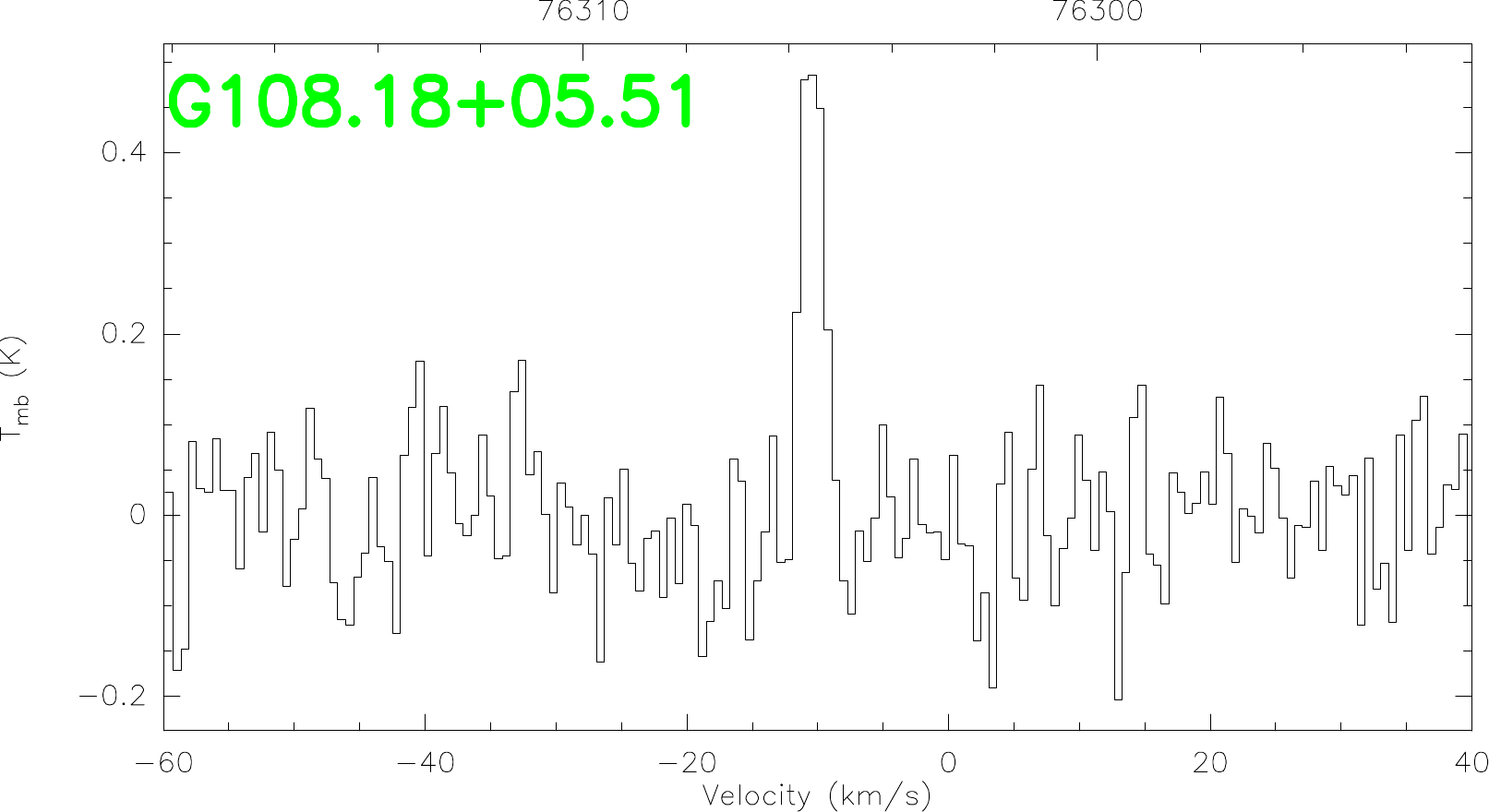}
\includegraphics[width=0.155\textwidth]{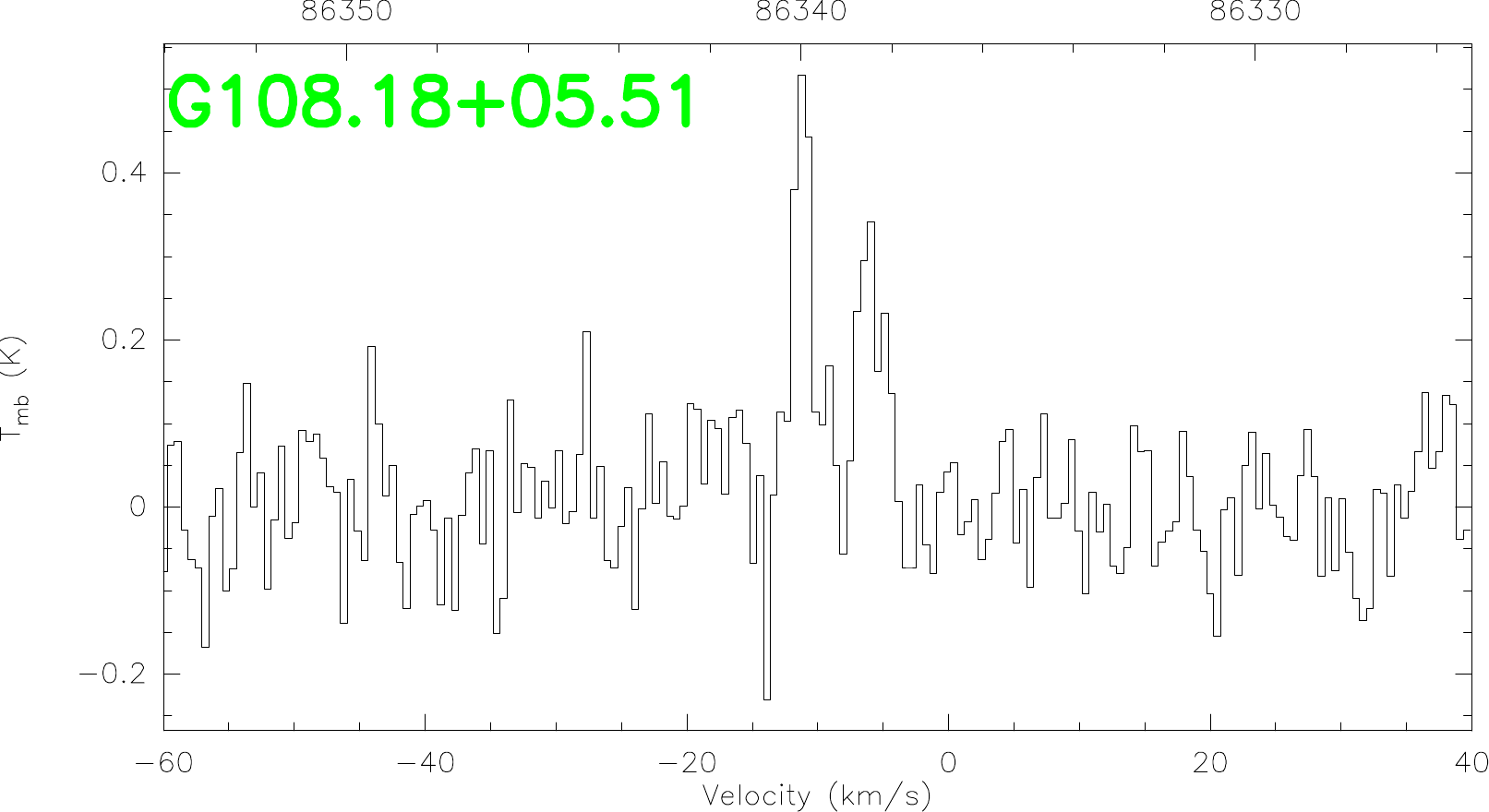}
\includegraphics[width=0.155\textwidth]{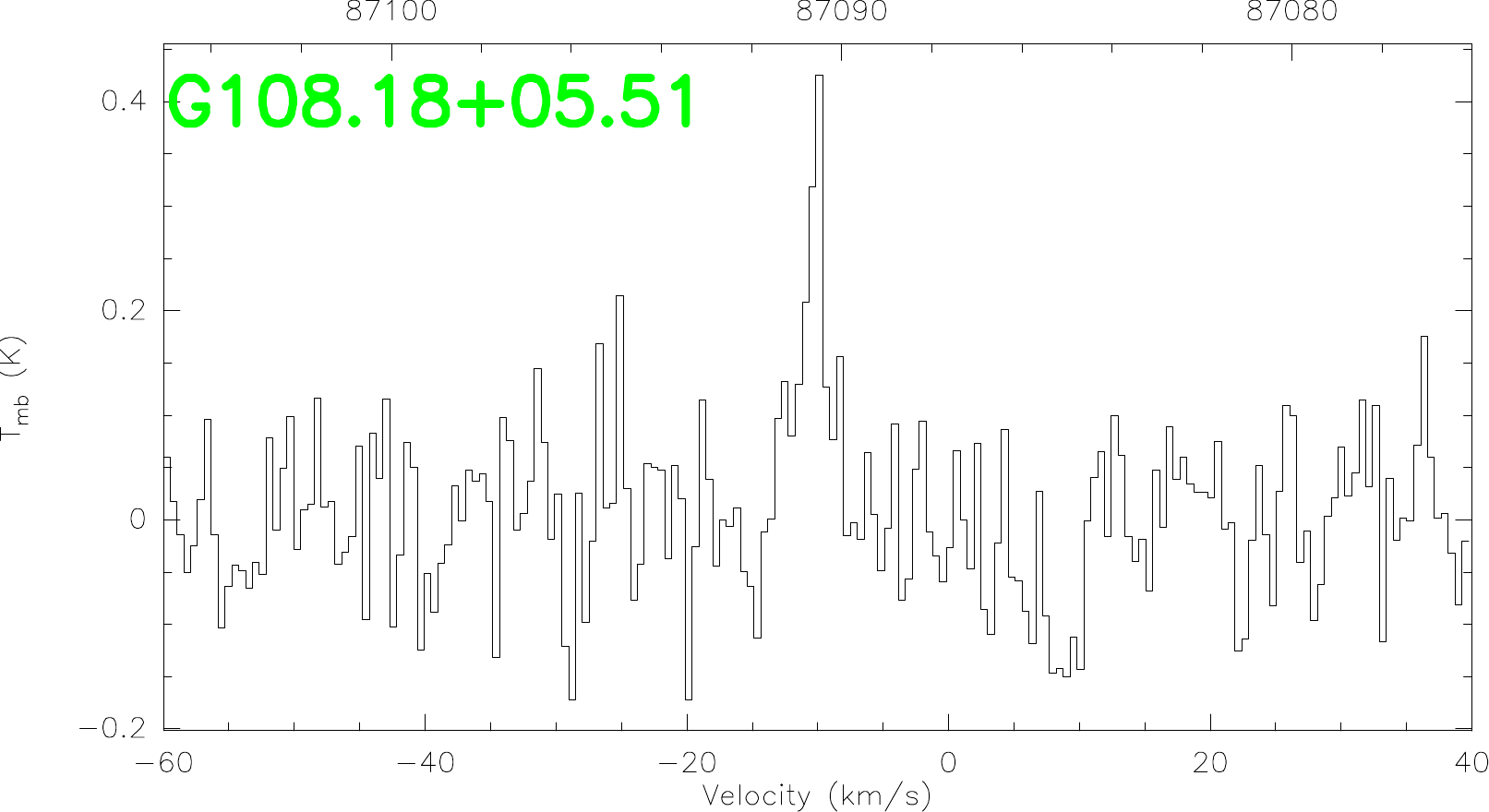}
\includegraphics[width=0.155\textwidth]{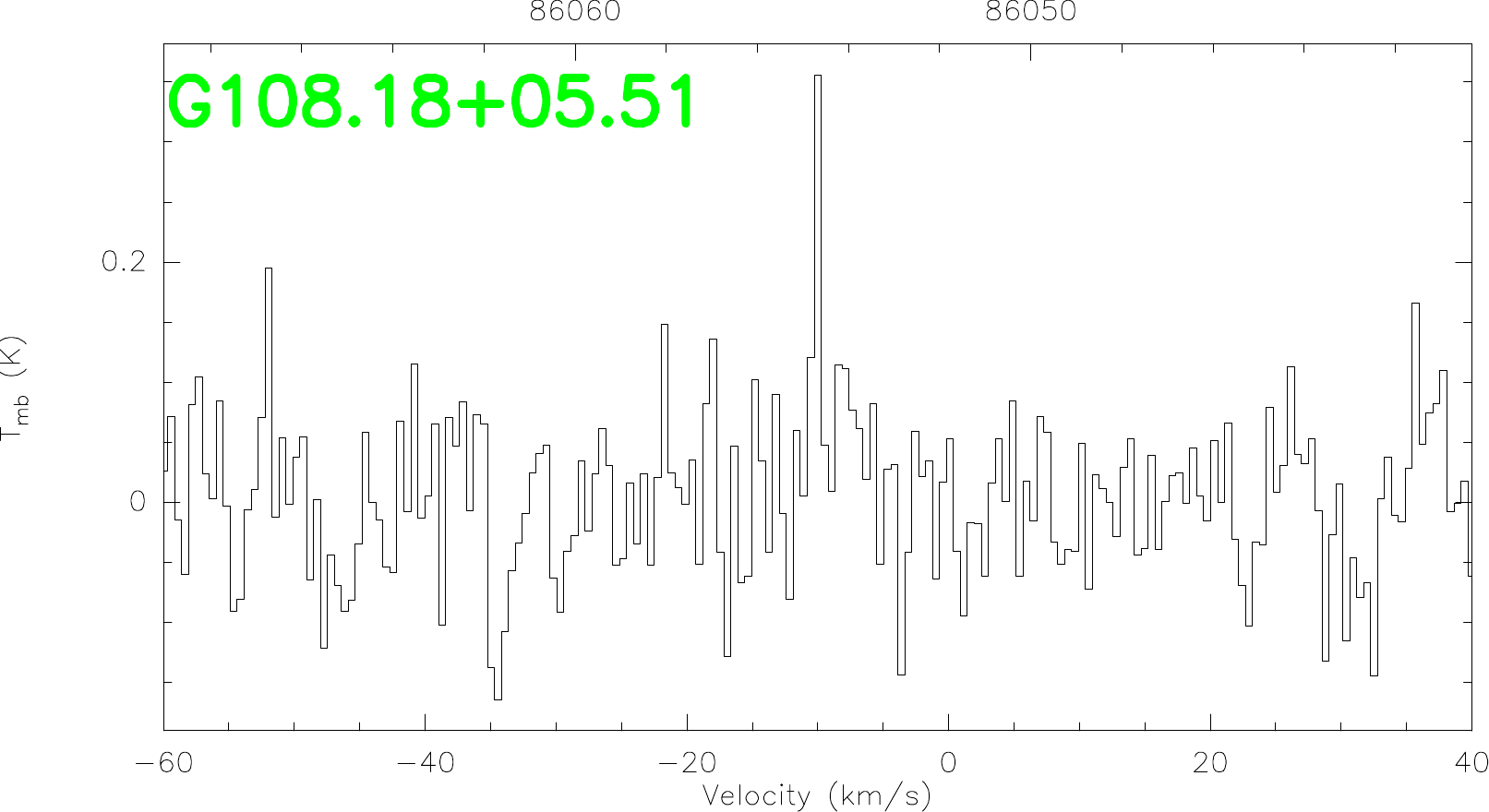}
\includegraphics[width=0.155\textwidth]{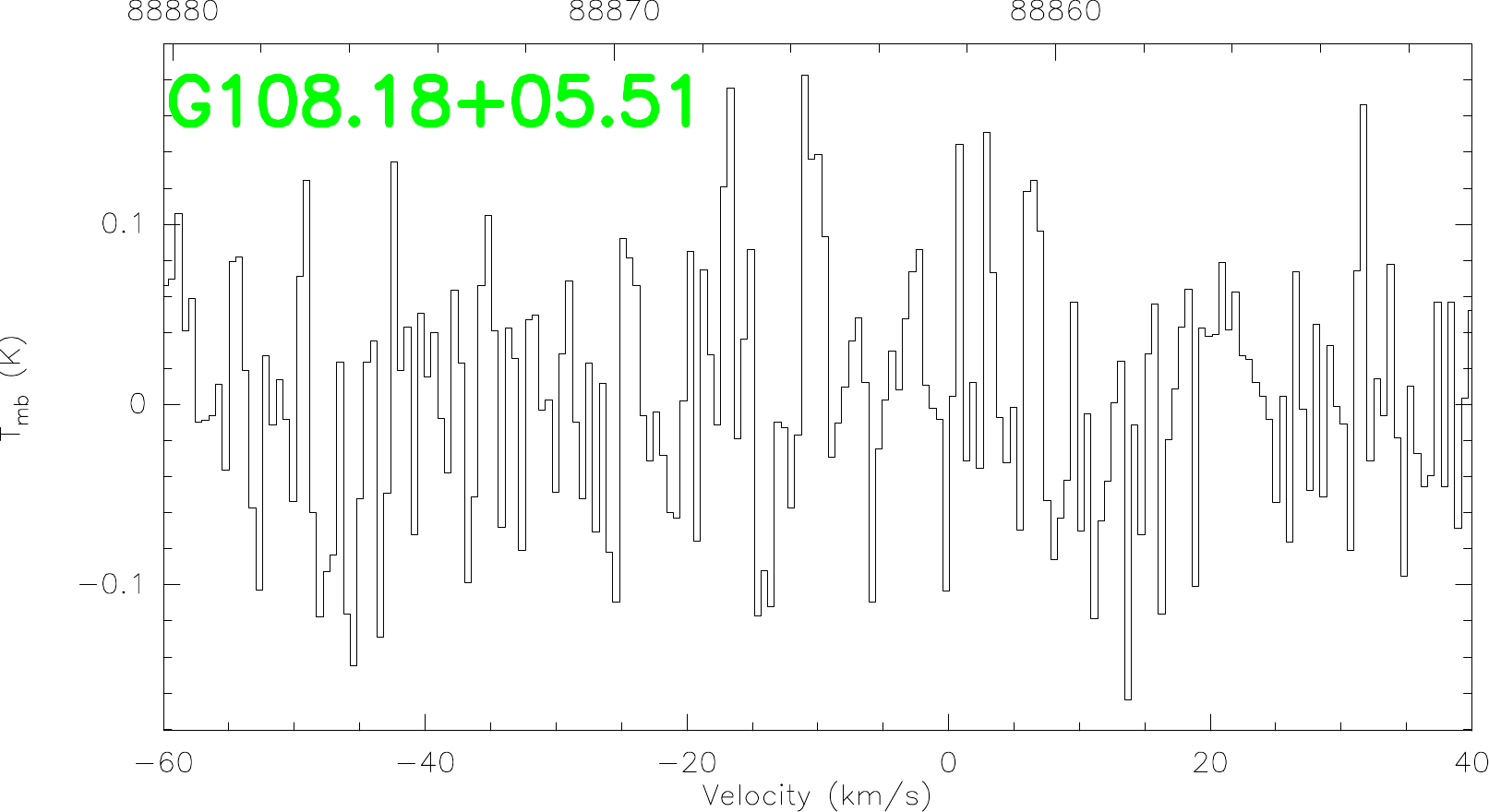}\\

\includegraphics[width=0.155\textwidth]{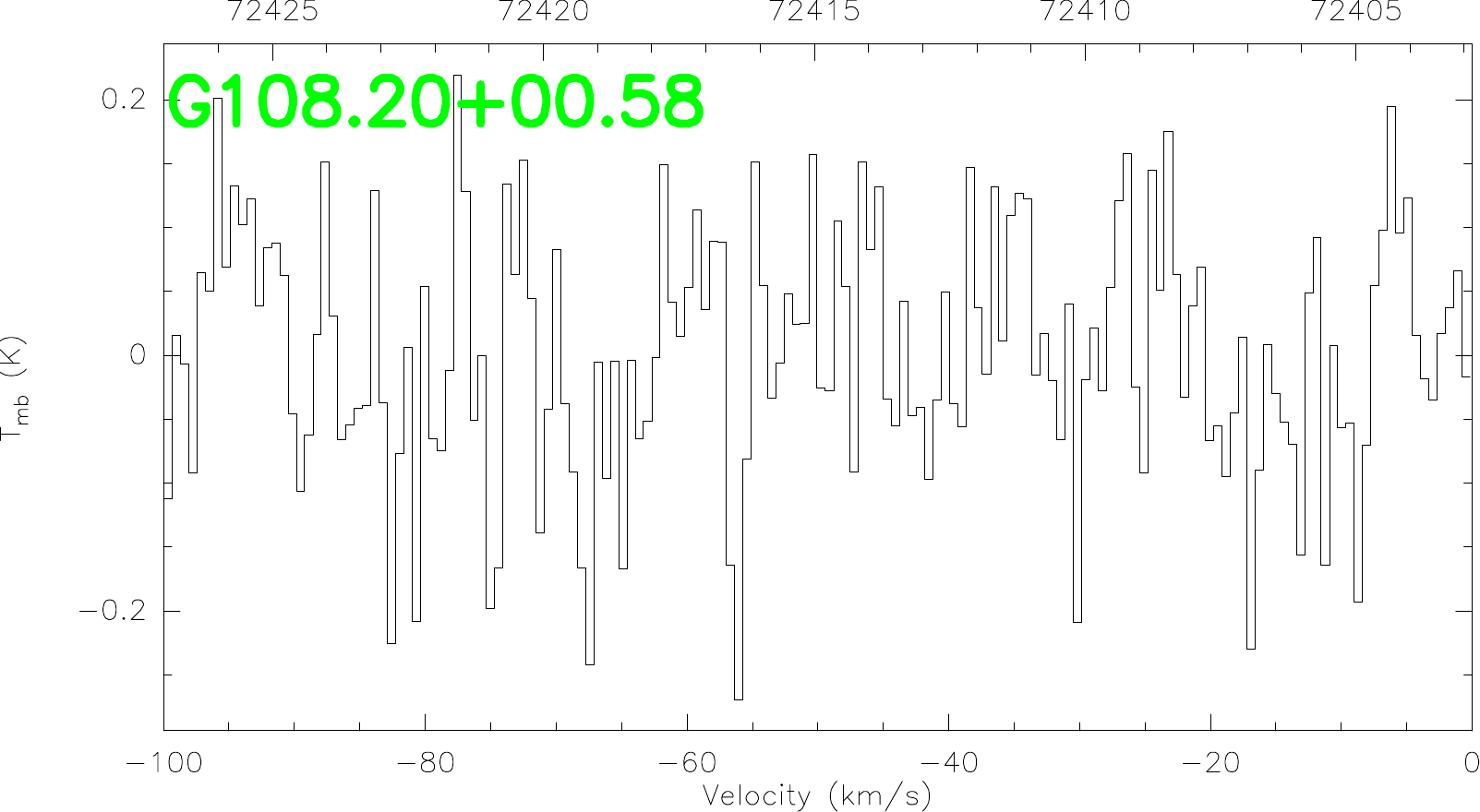}
\includegraphics[width=0.155\textwidth]{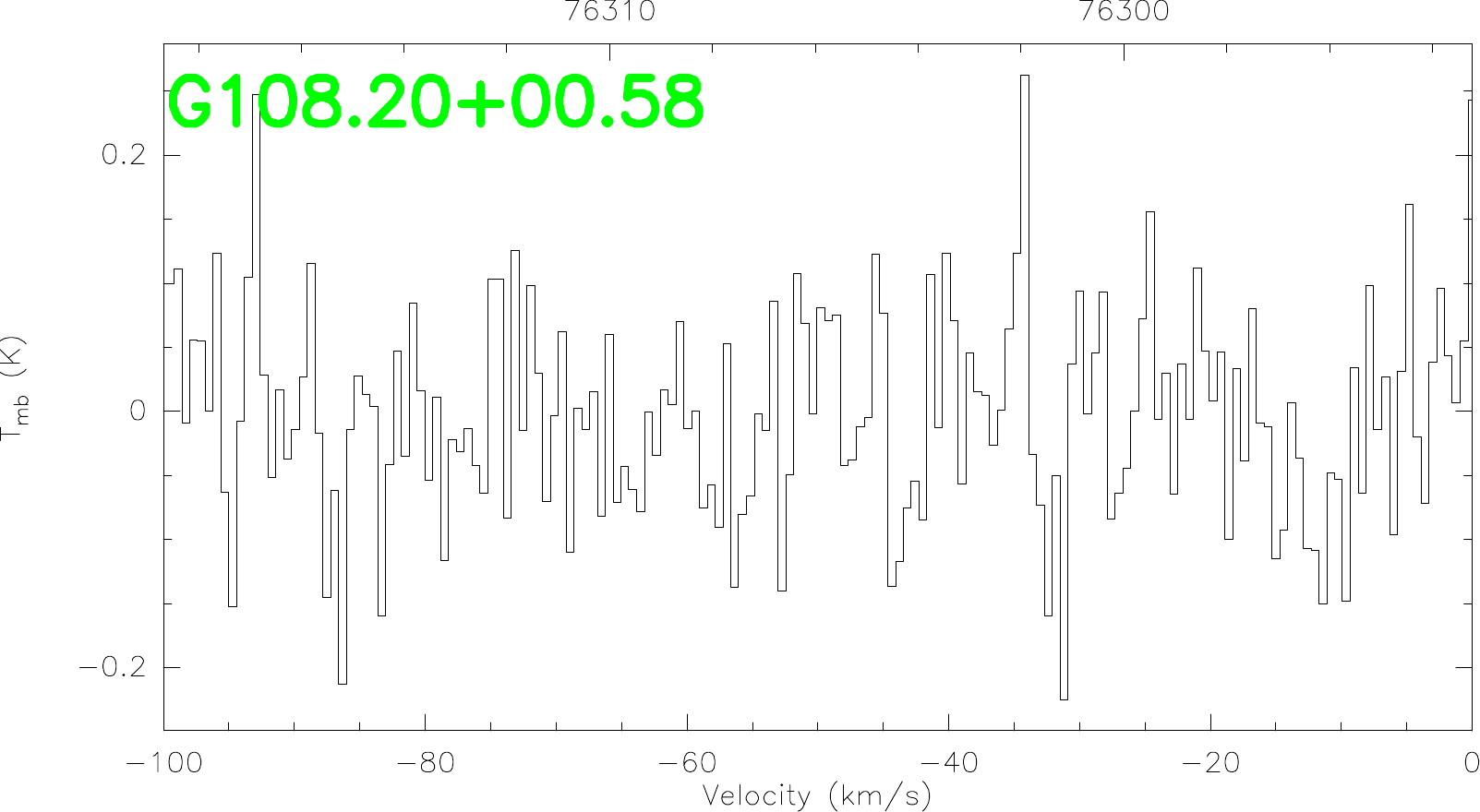}
\includegraphics[width=0.155\textwidth]{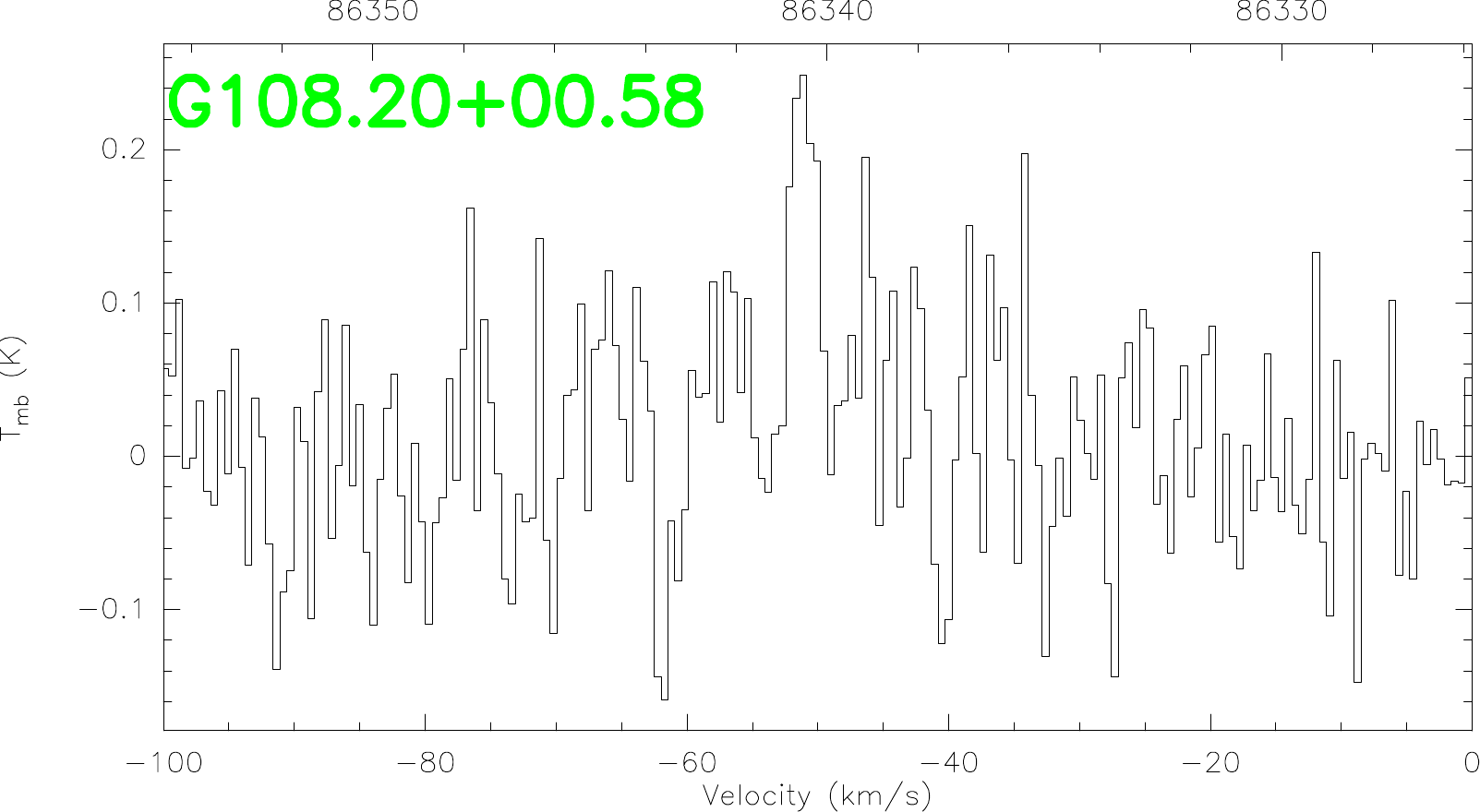}
\includegraphics[width=0.155\textwidth]{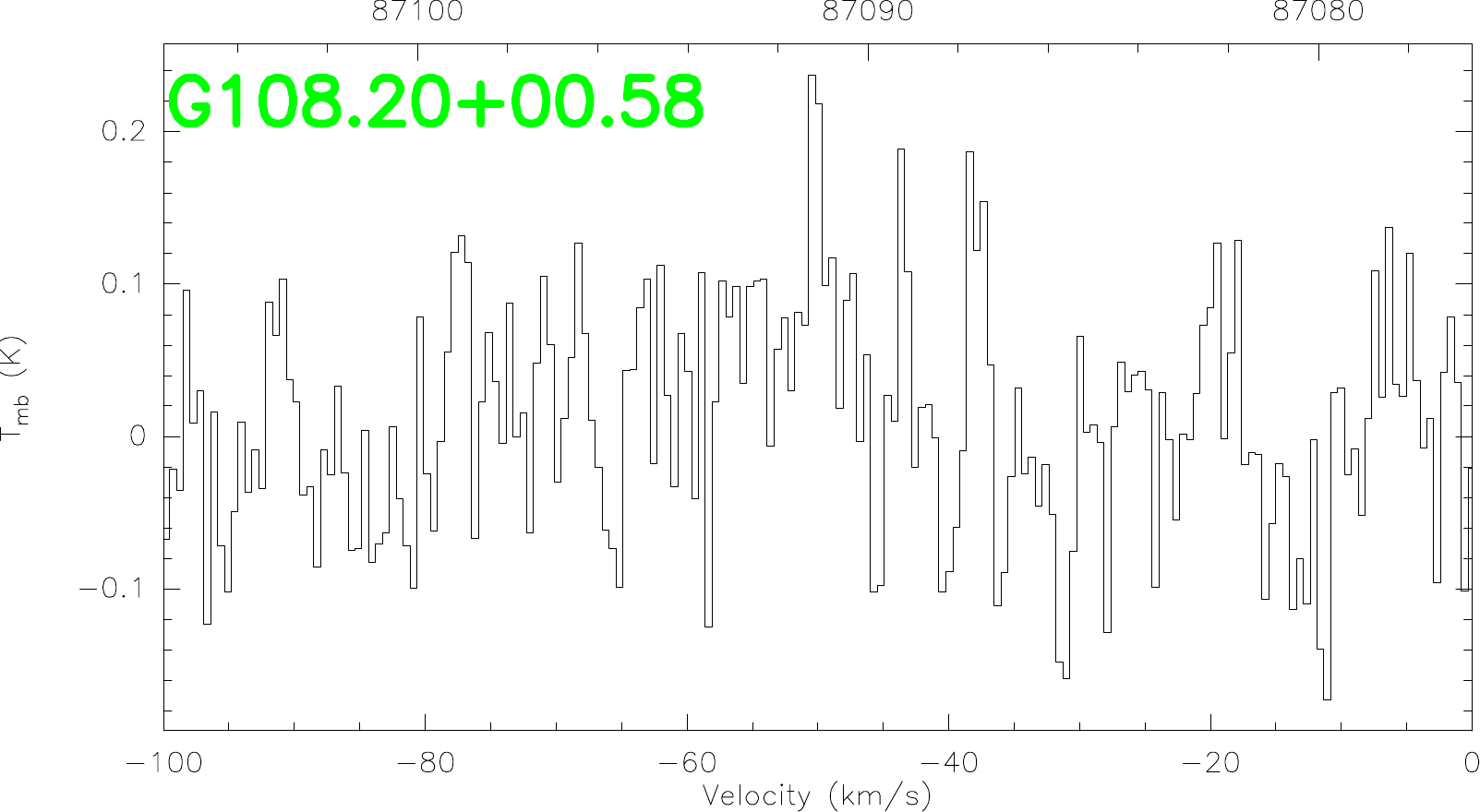}
\includegraphics[width=0.155\textwidth]{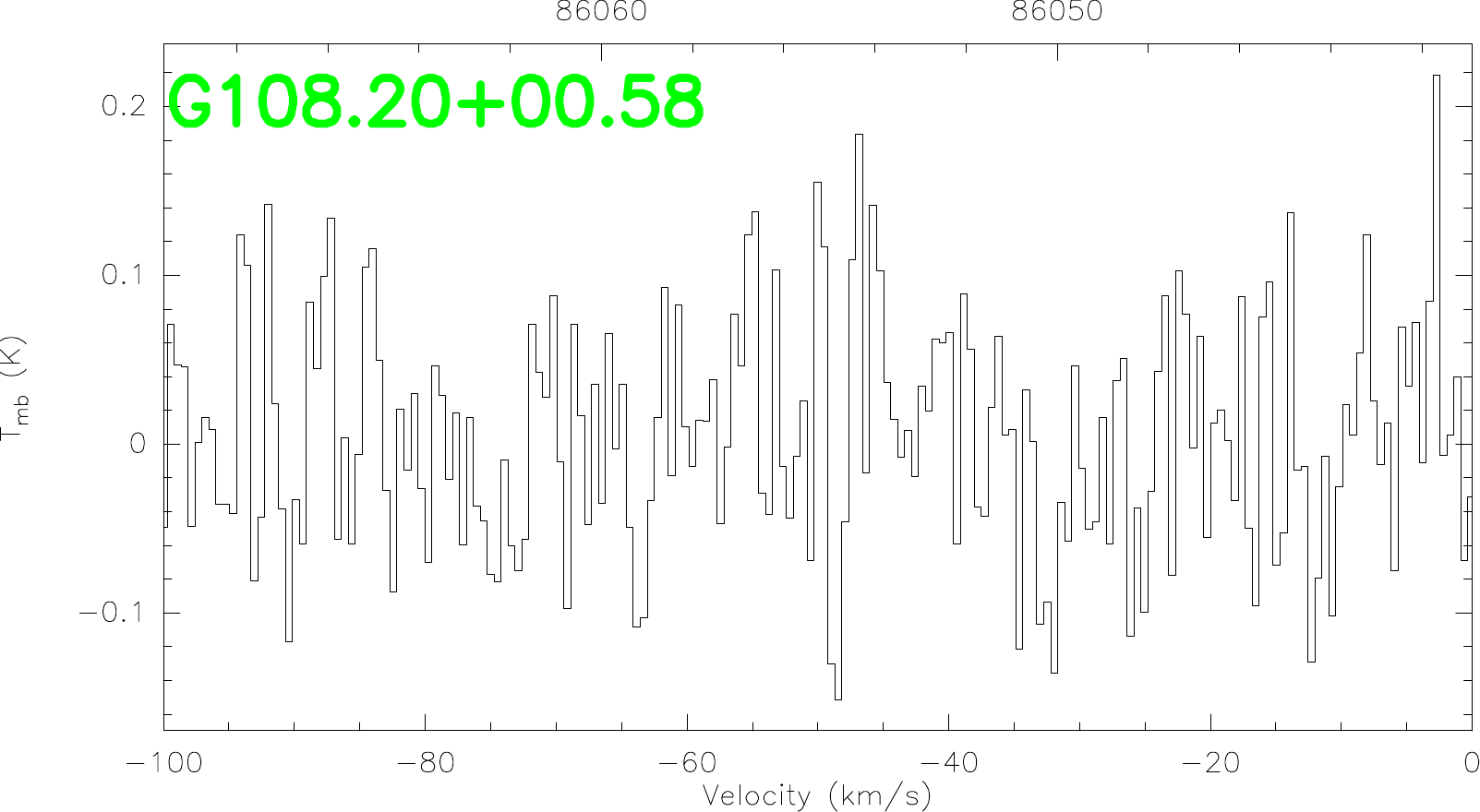}
\includegraphics[width=0.155\textwidth]{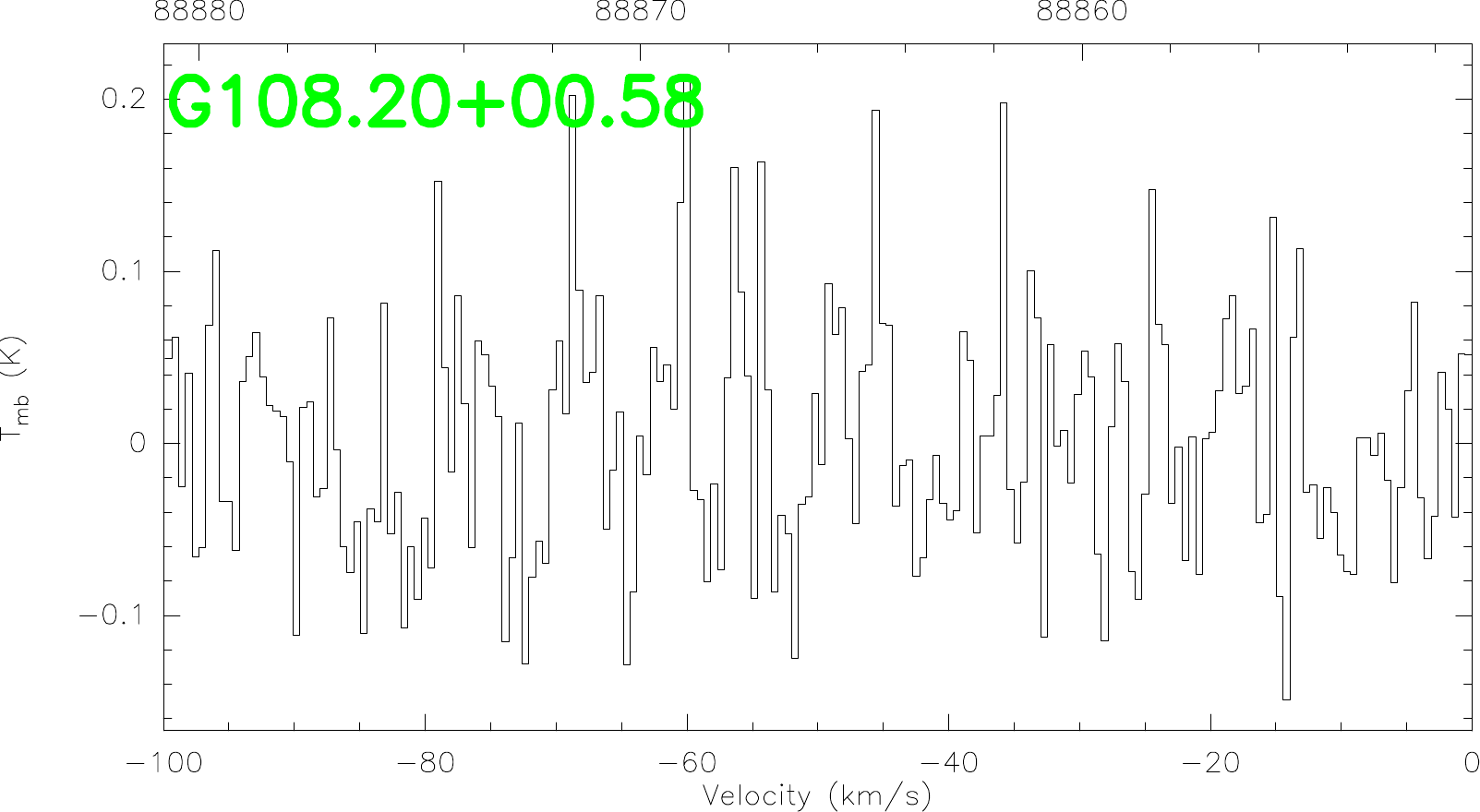}\\

\includegraphics[width=0.155\textwidth]{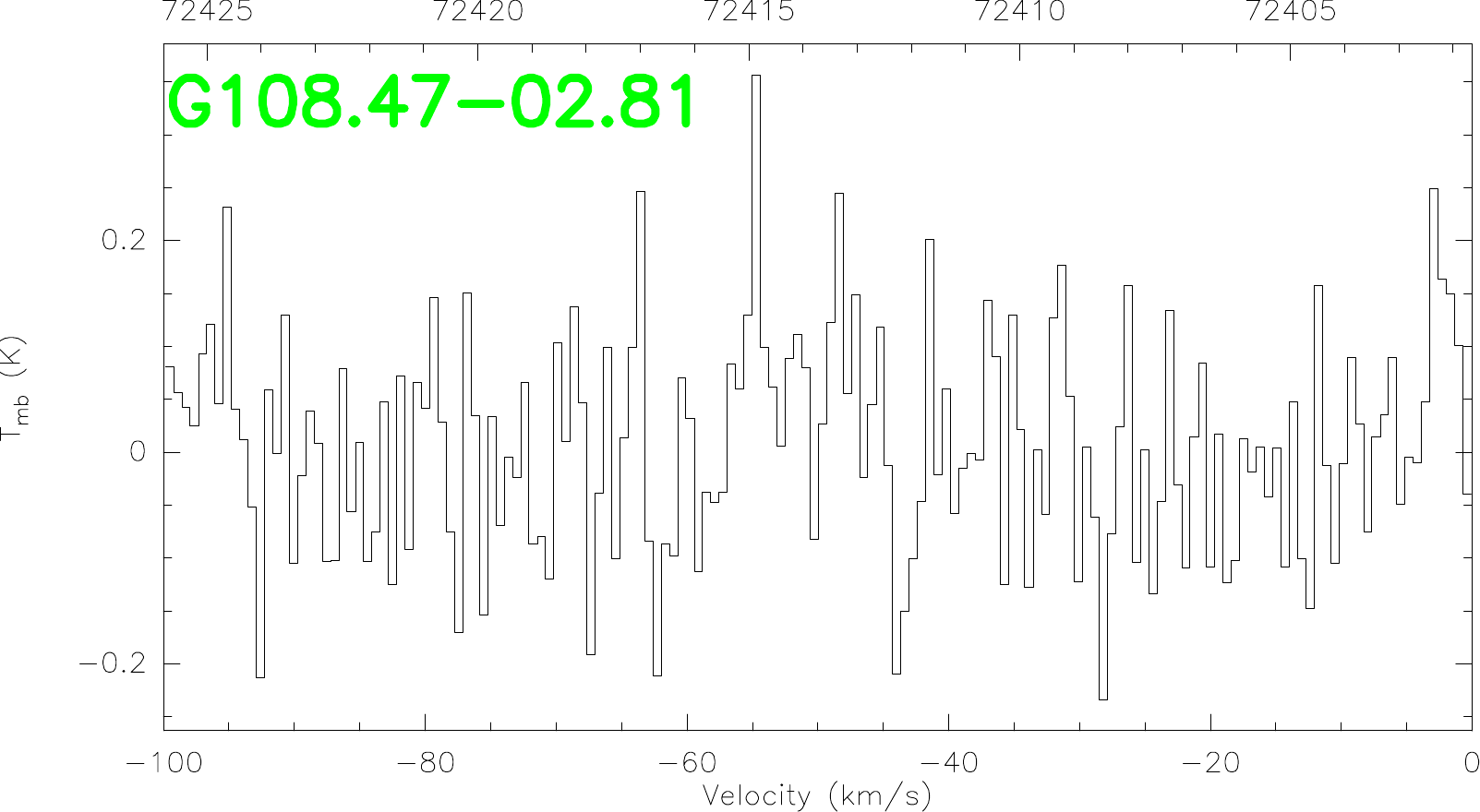}
\includegraphics[width=0.155\textwidth]{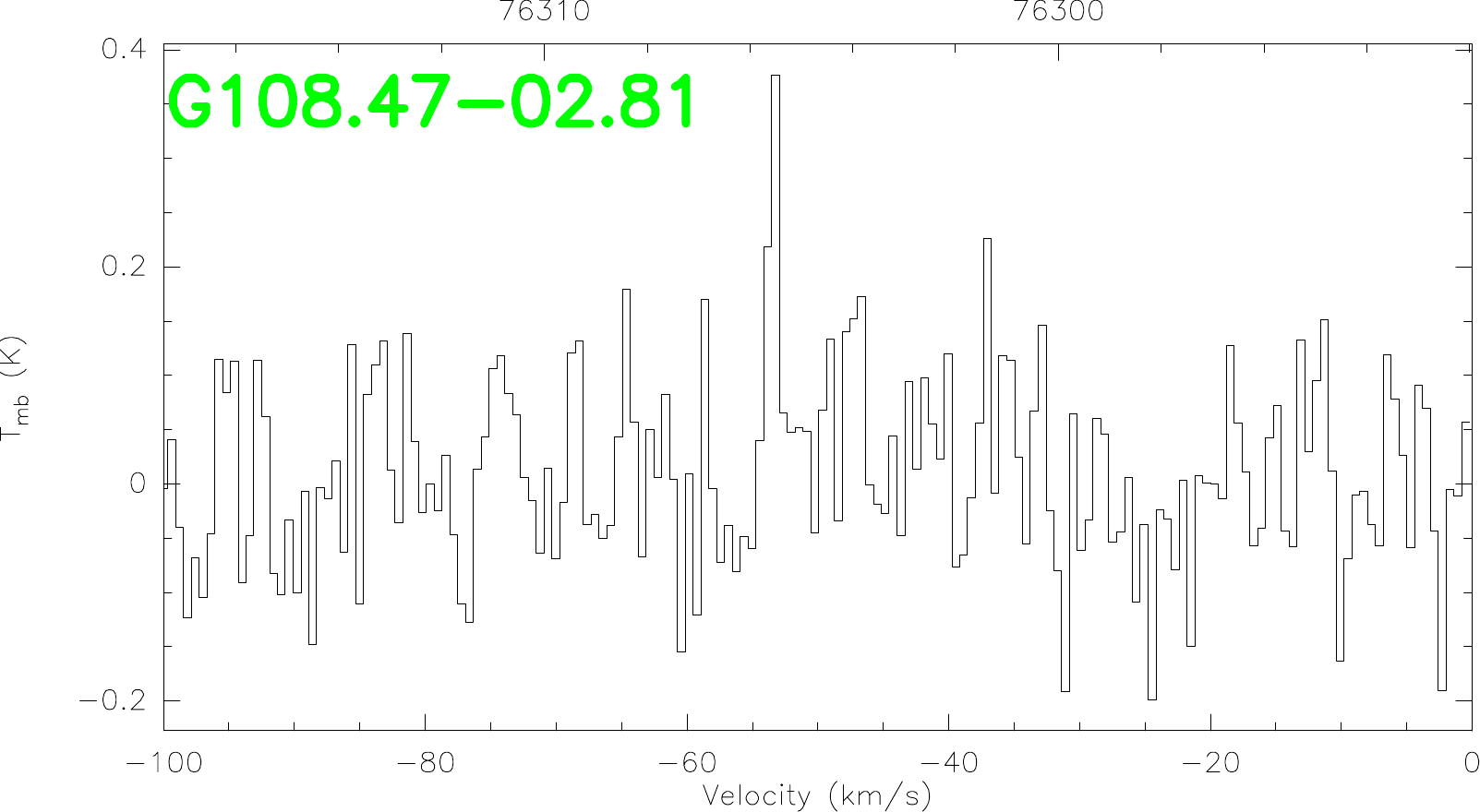}
\includegraphics[width=0.155\textwidth]{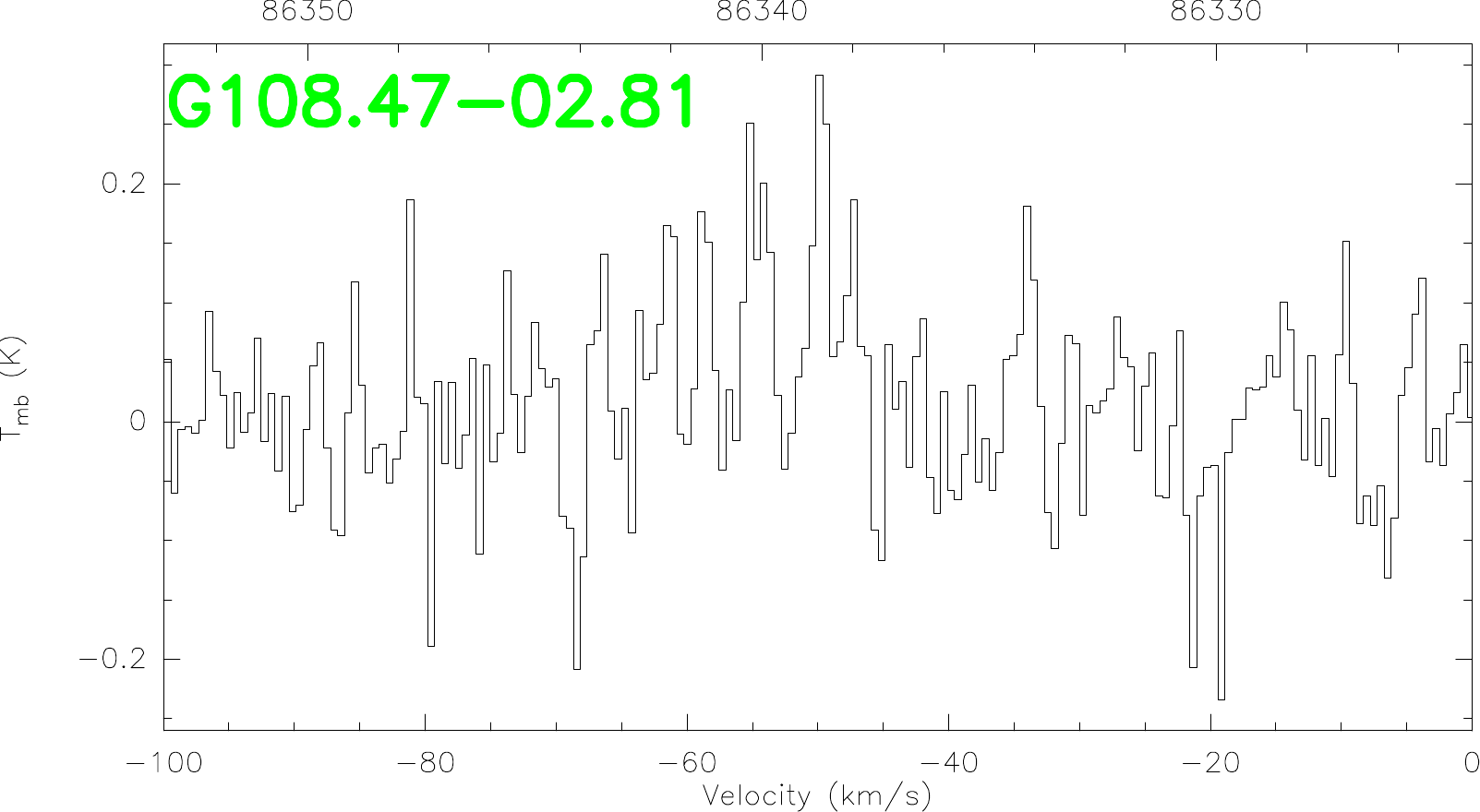}
\includegraphics[width=0.155\textwidth]{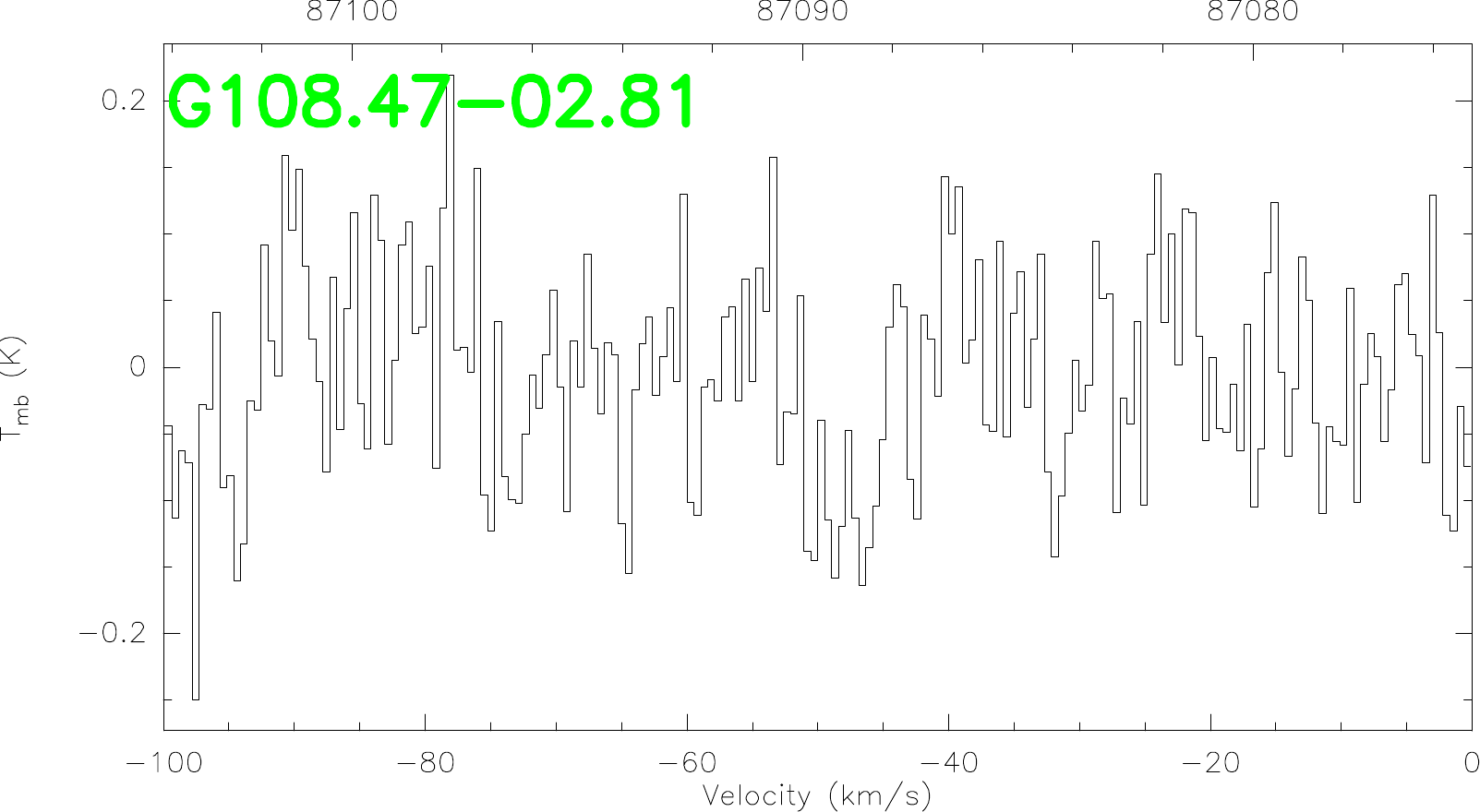}
\includegraphics[width=0.155\textwidth]{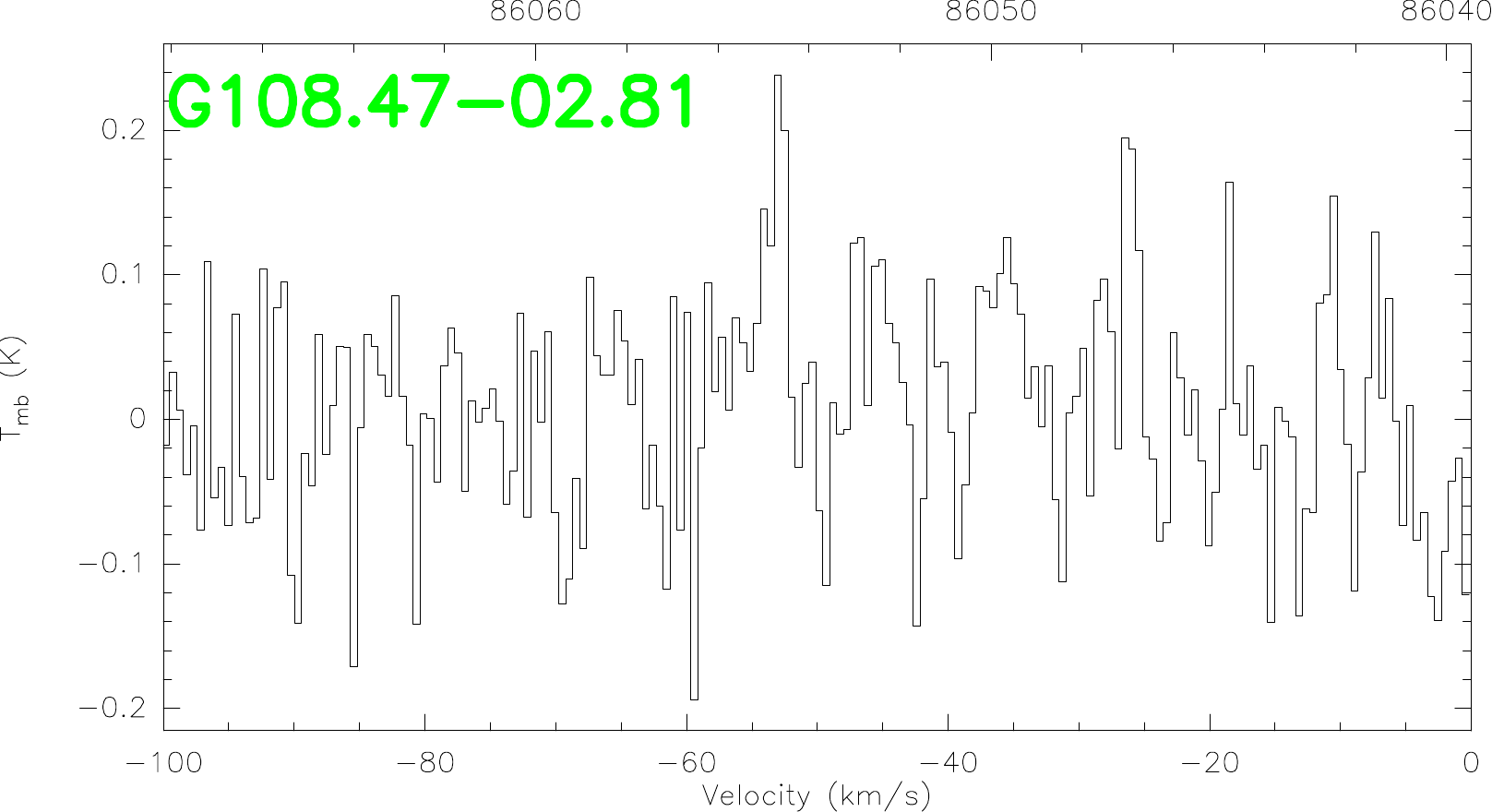}
\includegraphics[width=0.155\textwidth]{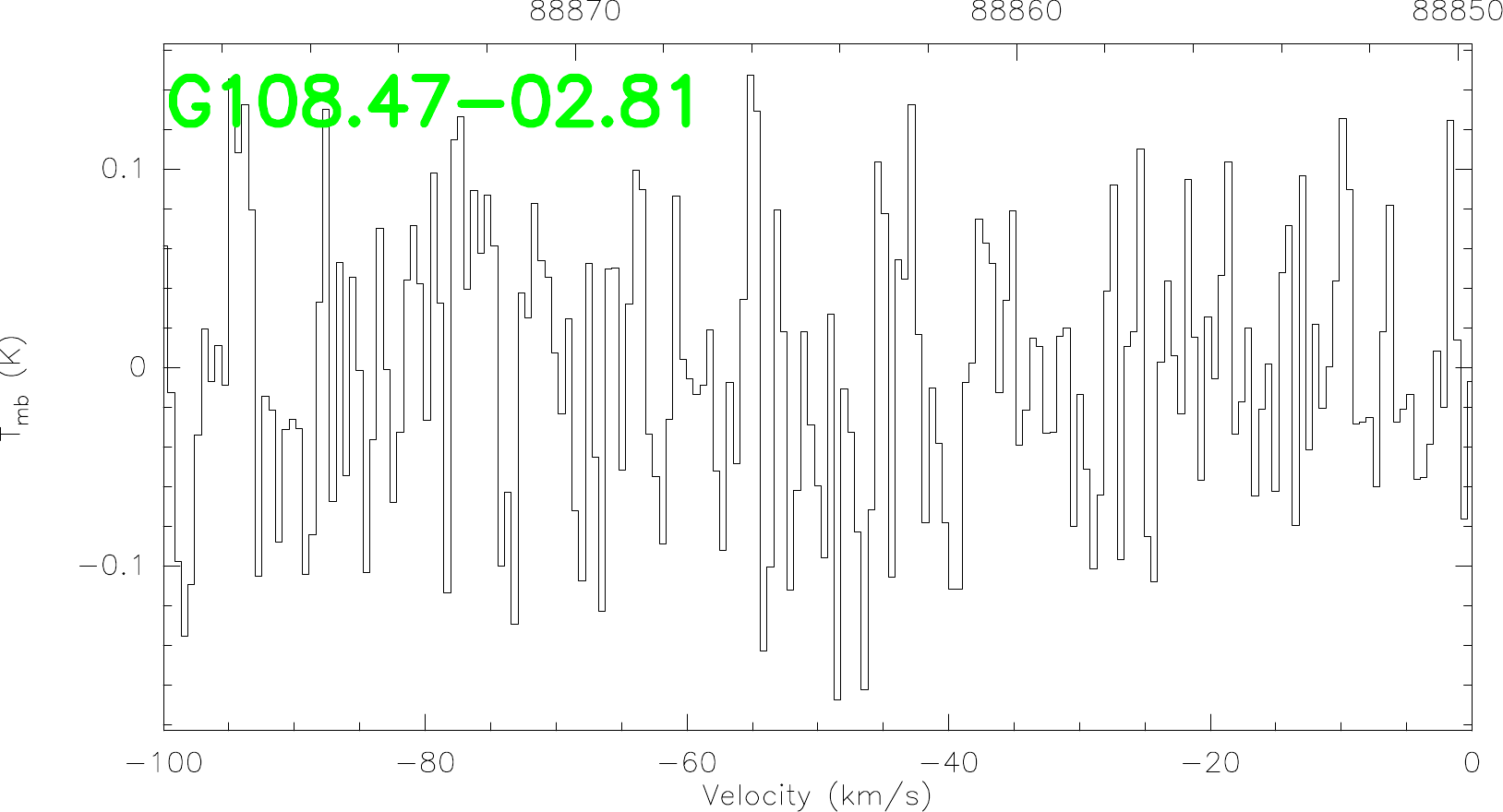}\\

\includegraphics[width=0.155\textwidth]{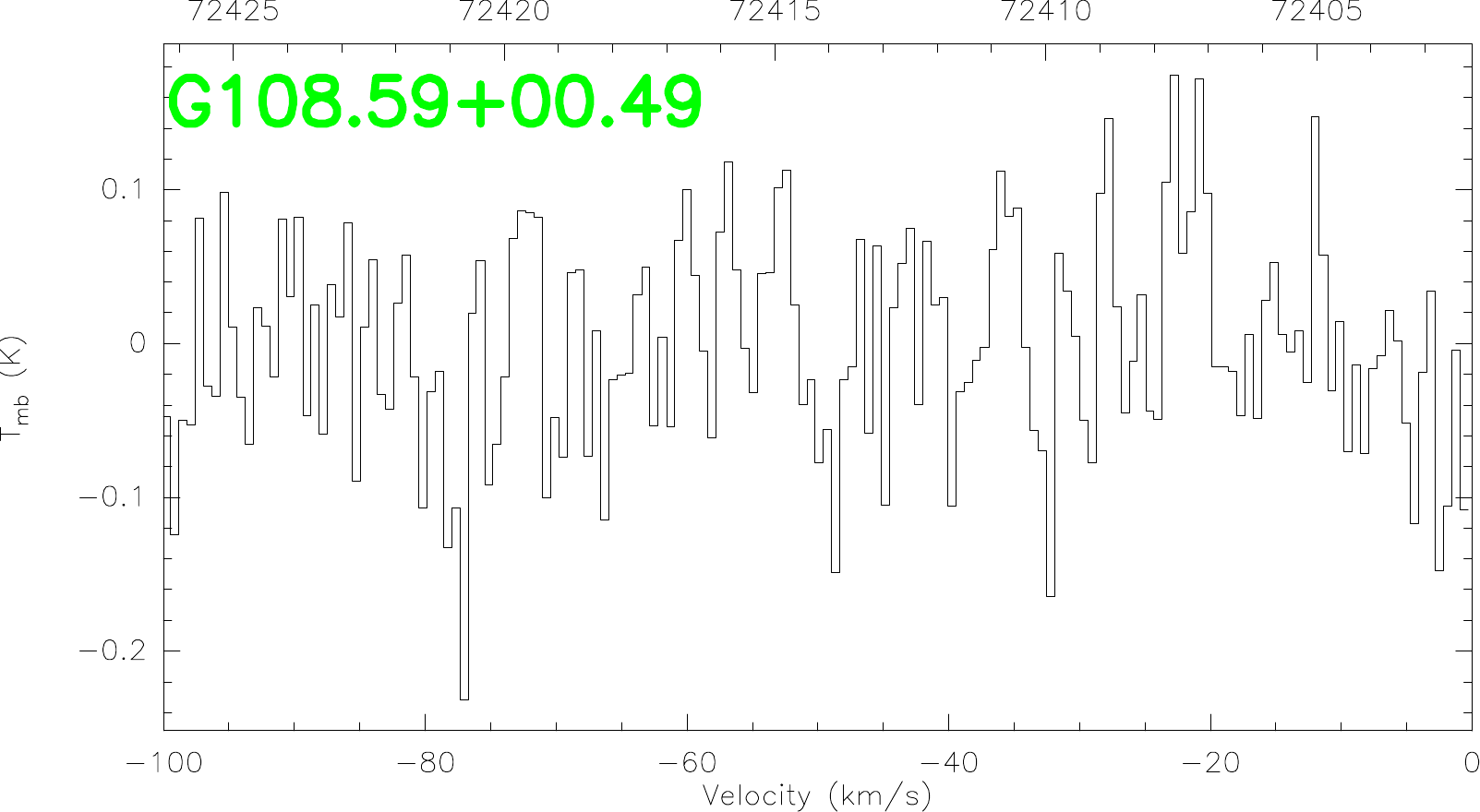}
\includegraphics[width=0.155\textwidth]{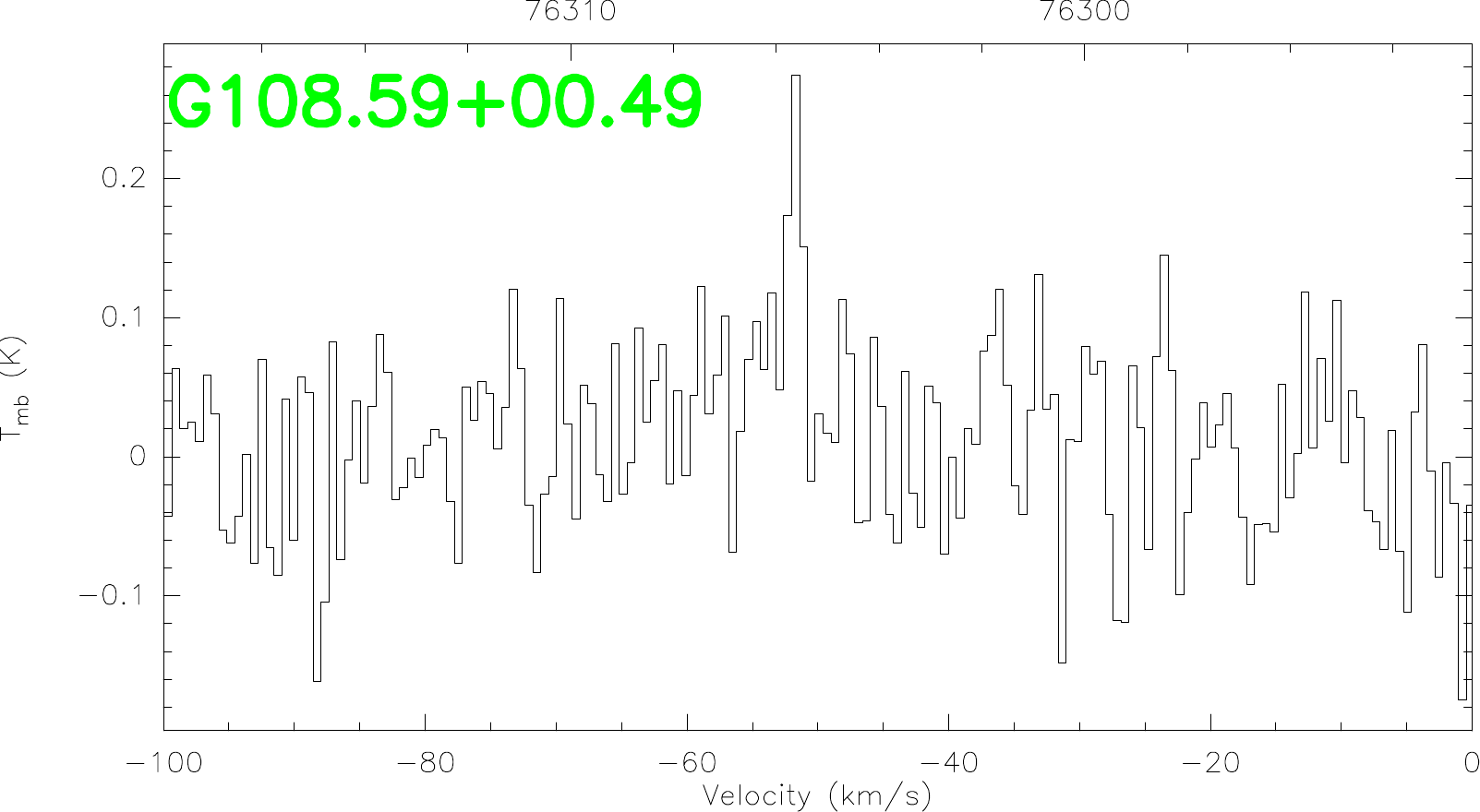}
\includegraphics[width=0.155\textwidth]{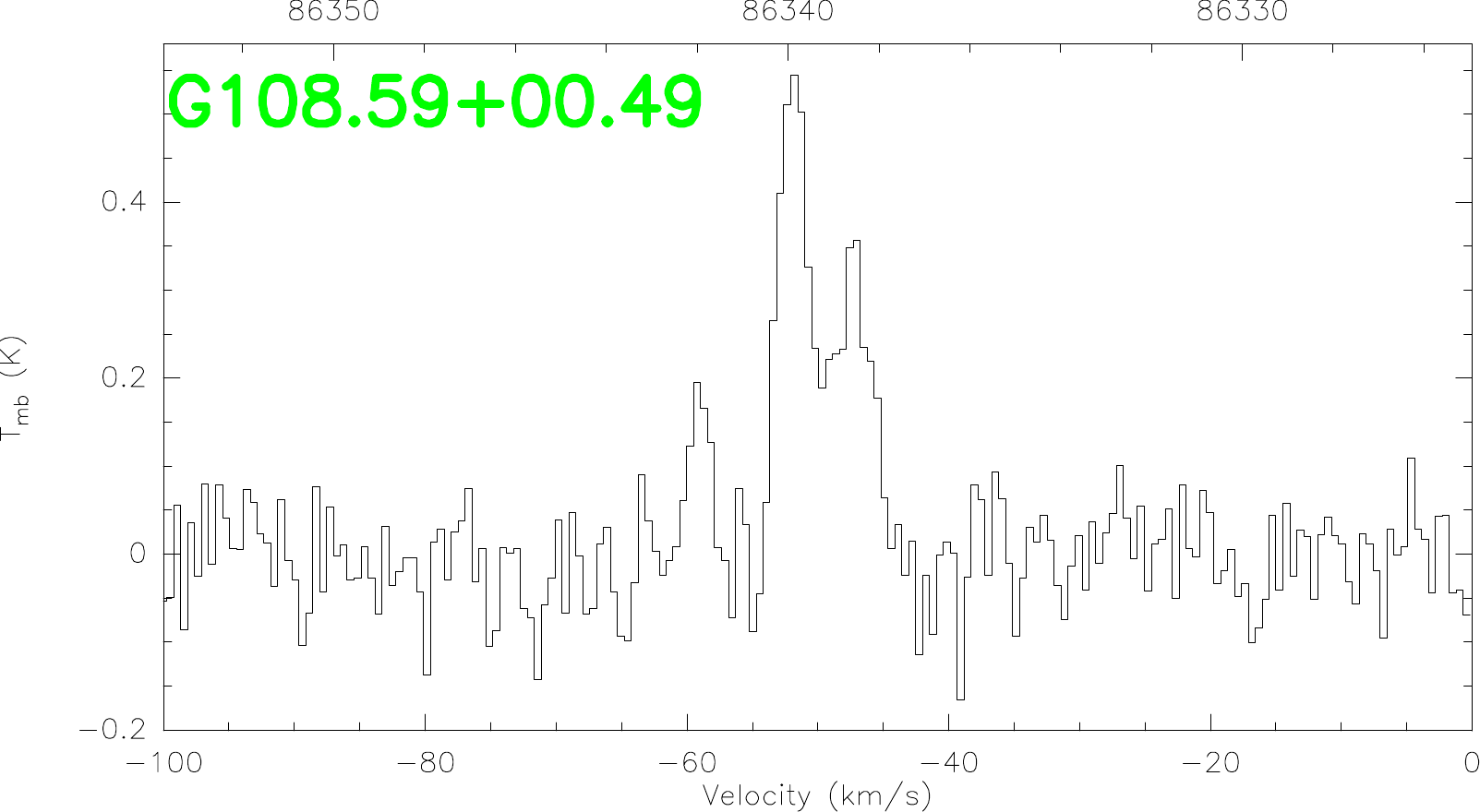}
\includegraphics[width=0.155\textwidth]{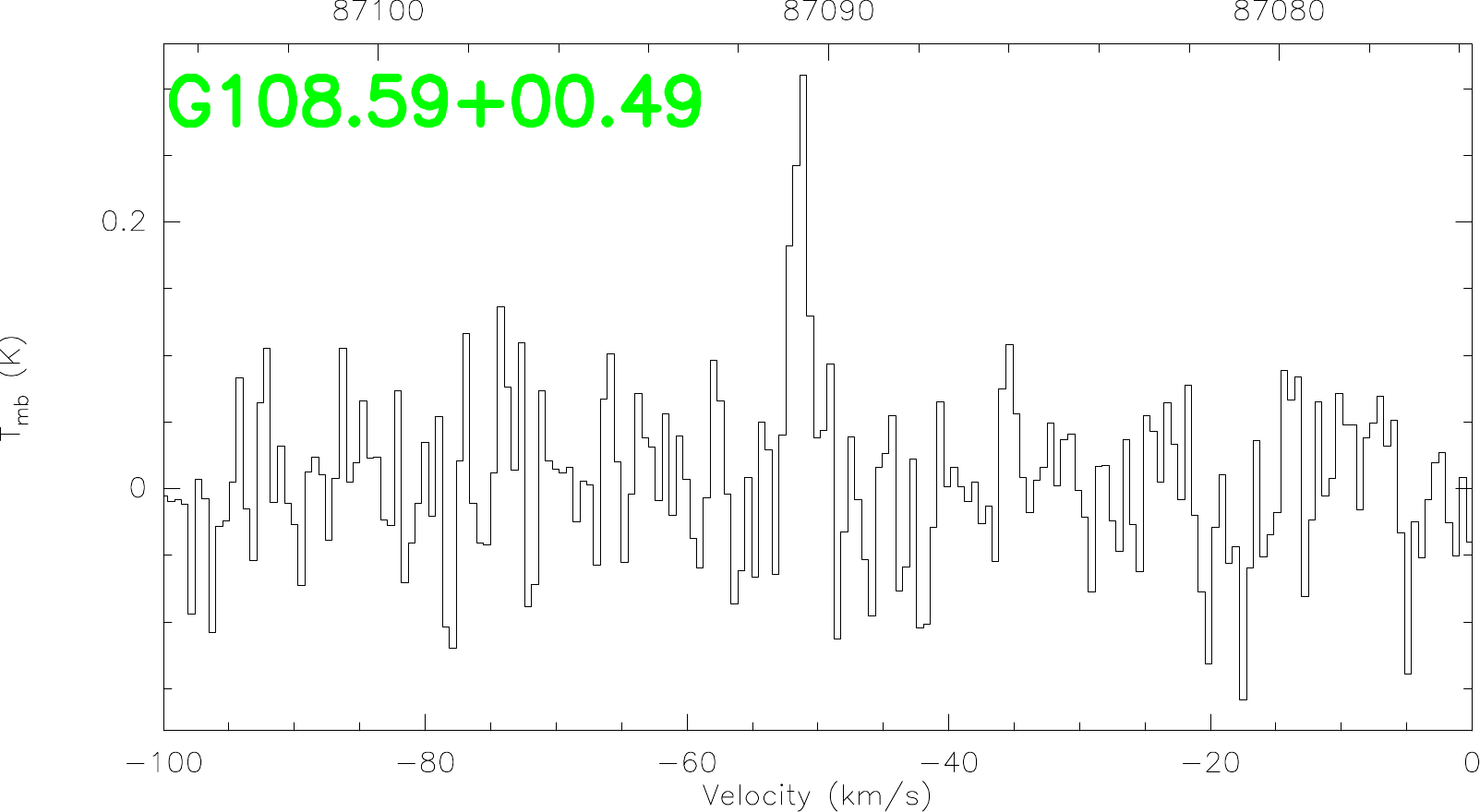}
\includegraphics[width=0.155\textwidth]{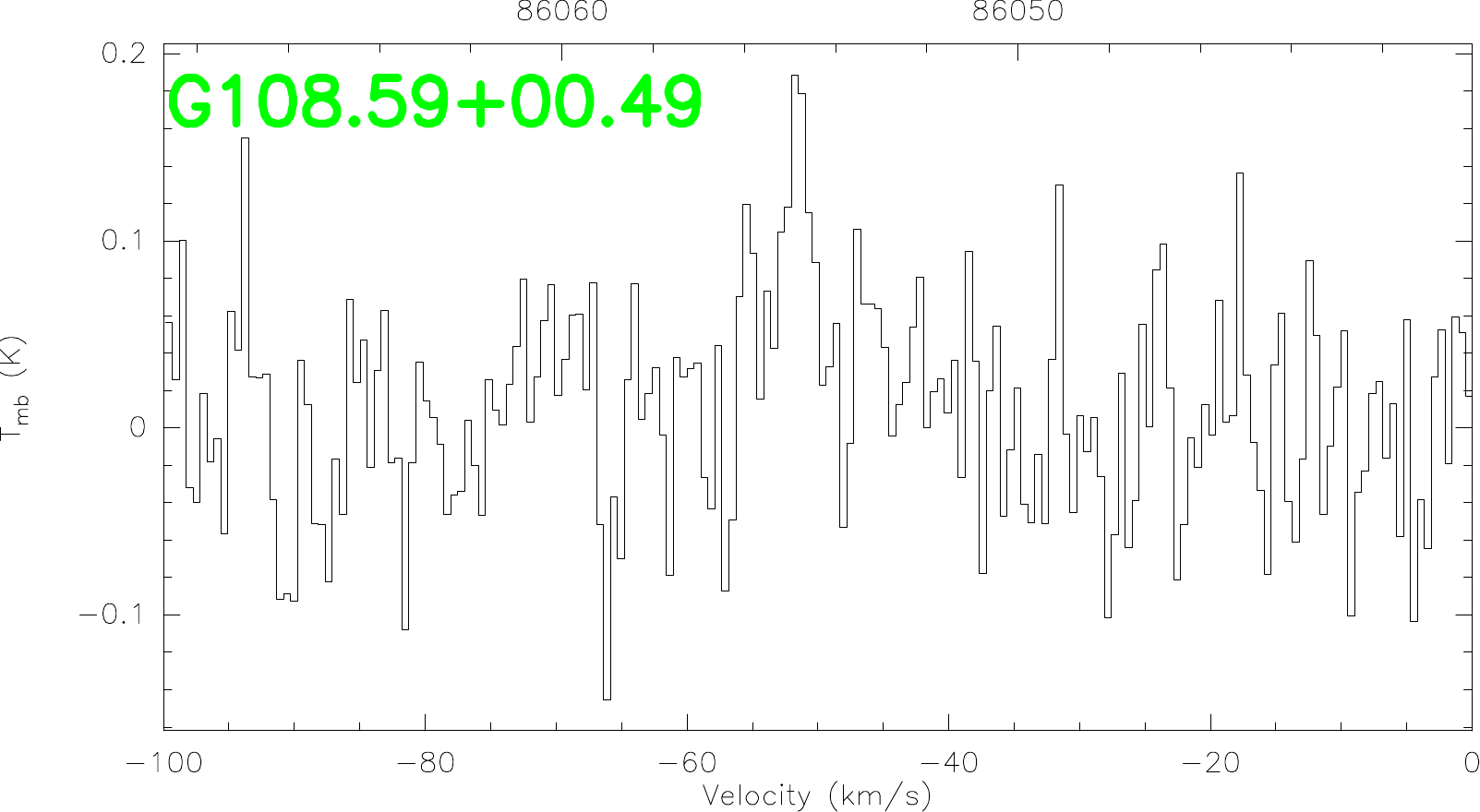}
\includegraphics[width=0.155\textwidth]{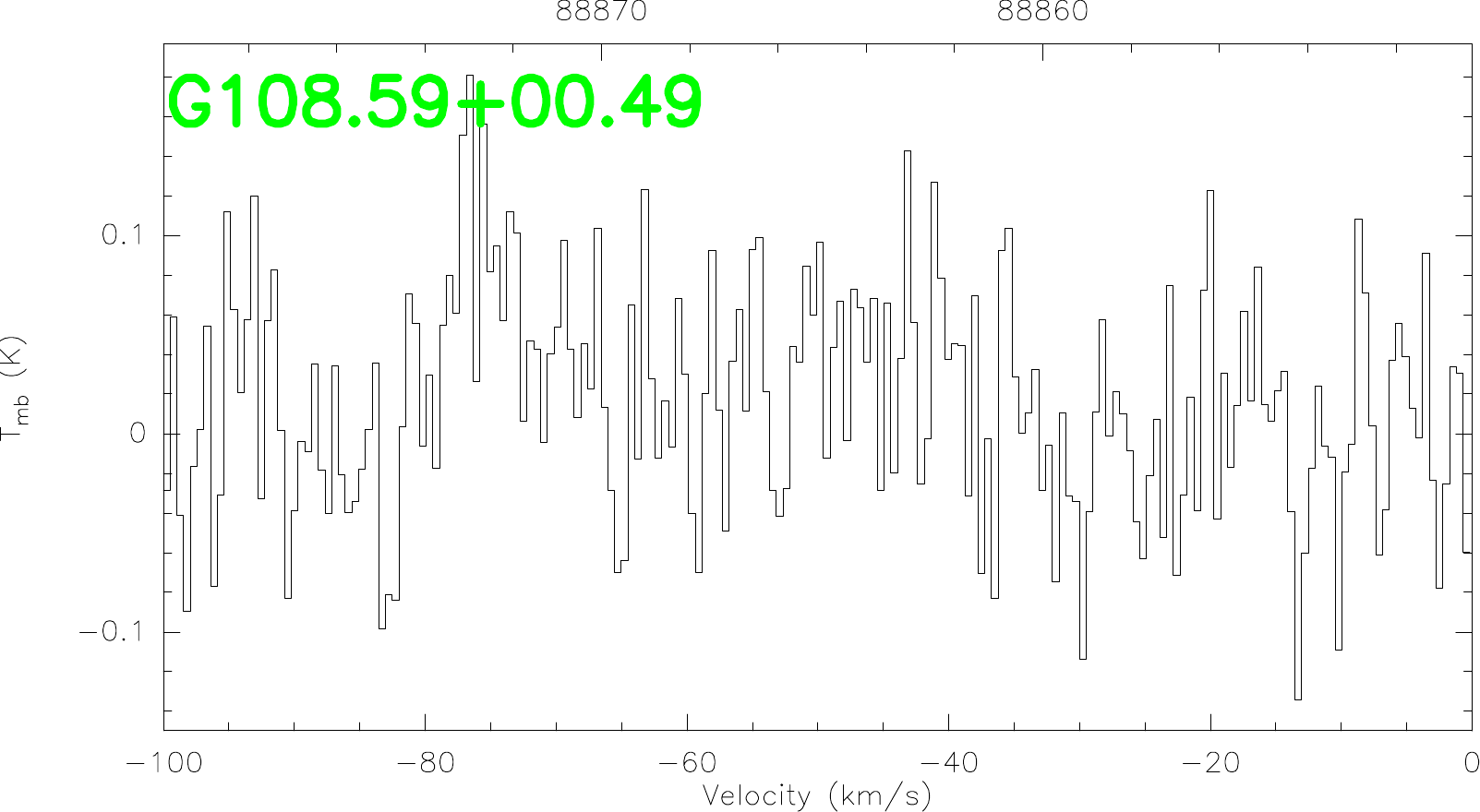}\\

\includegraphics[width=0.155\textwidth]{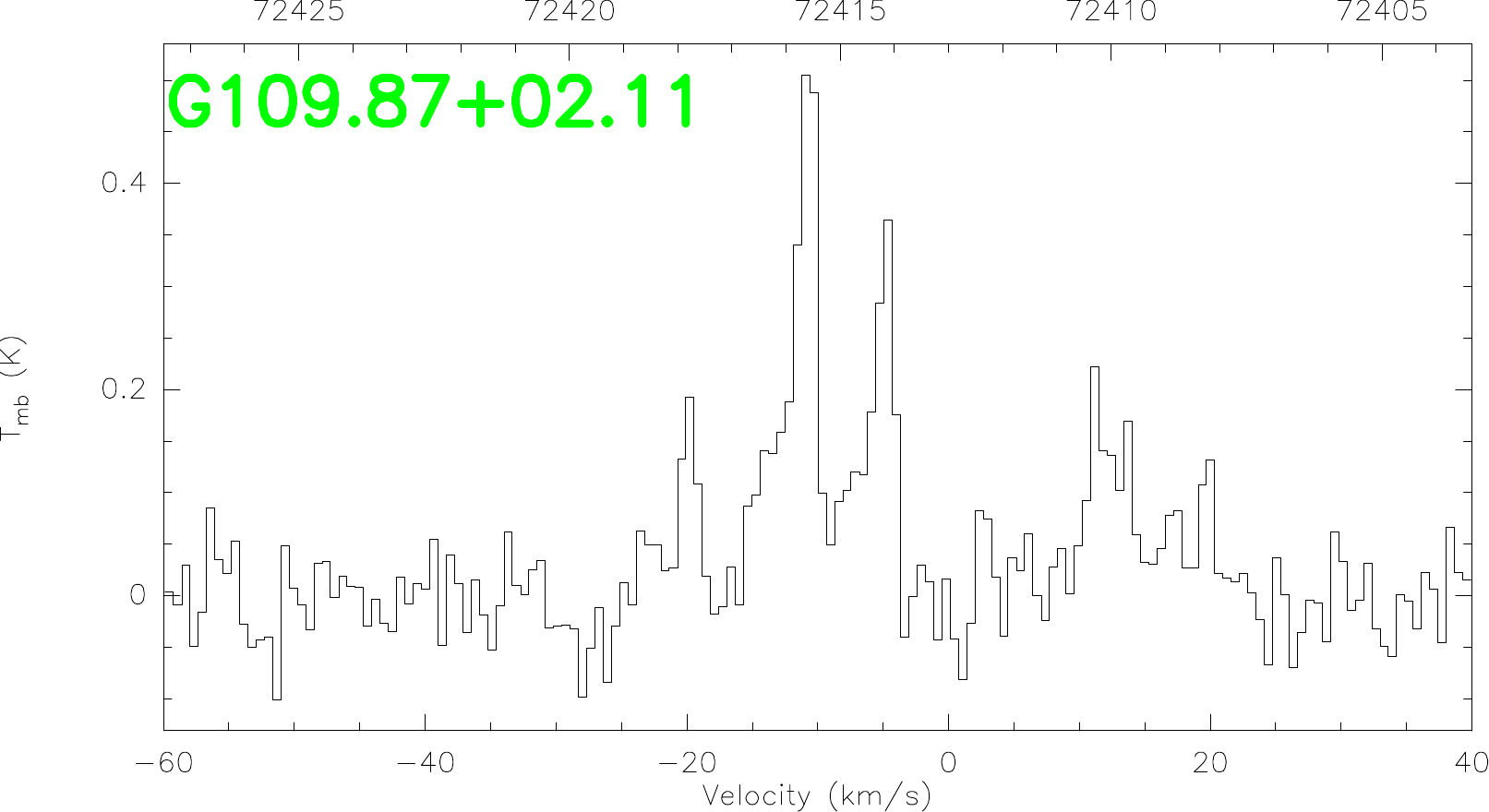}
\includegraphics[width=0.155\textwidth]{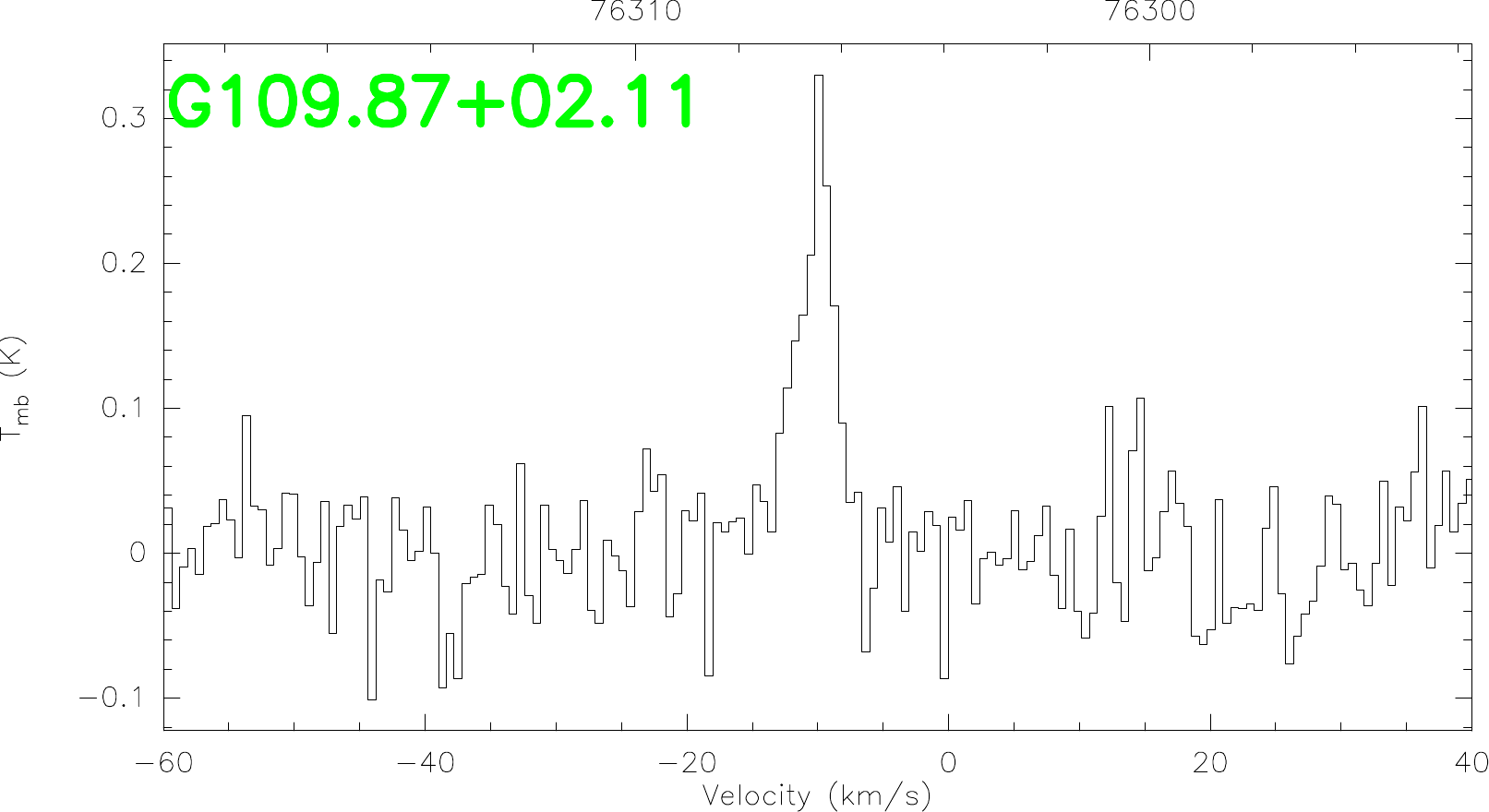}
\includegraphics[width=0.155\textwidth]{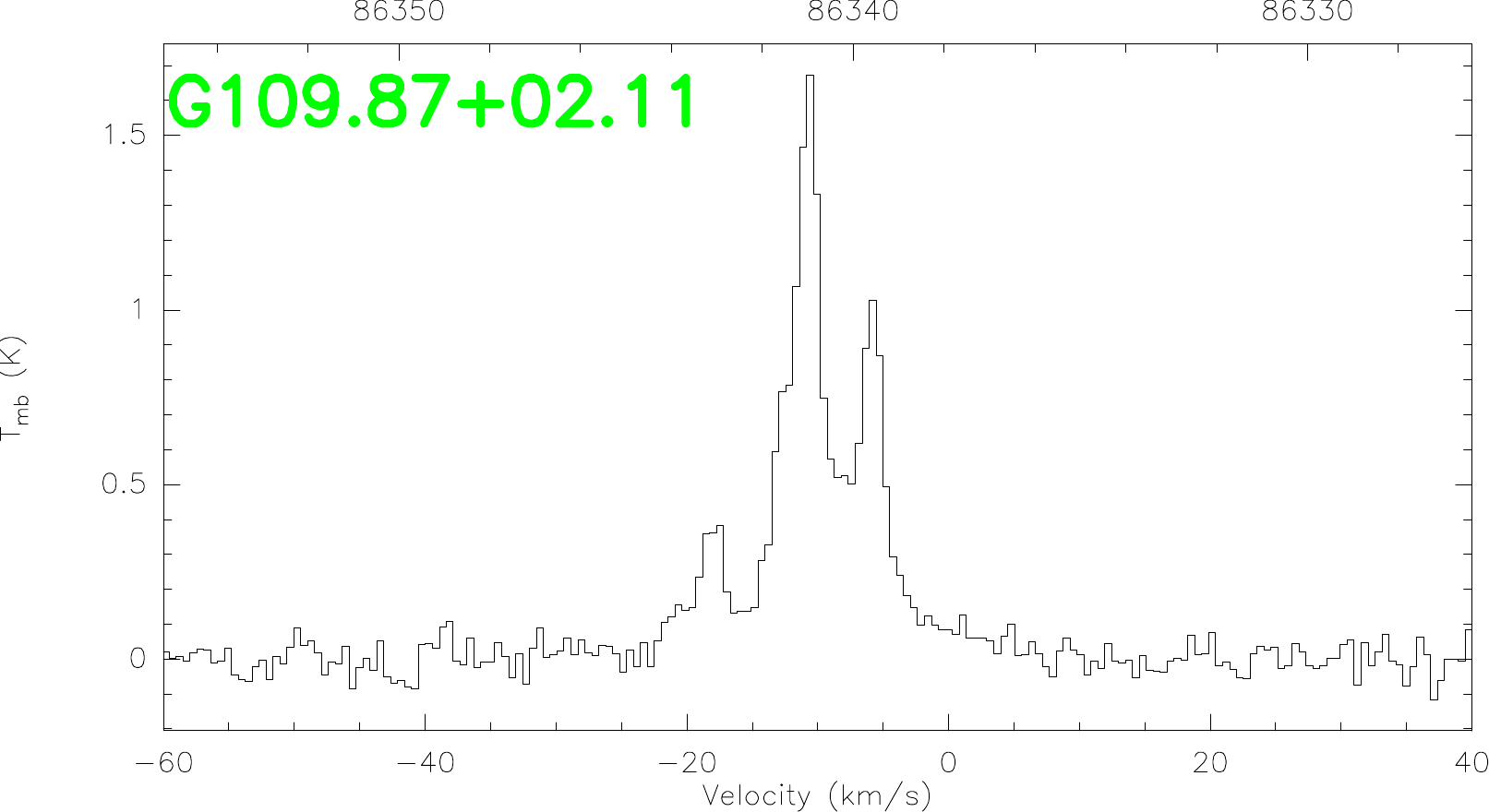}
\includegraphics[width=0.155\textwidth]{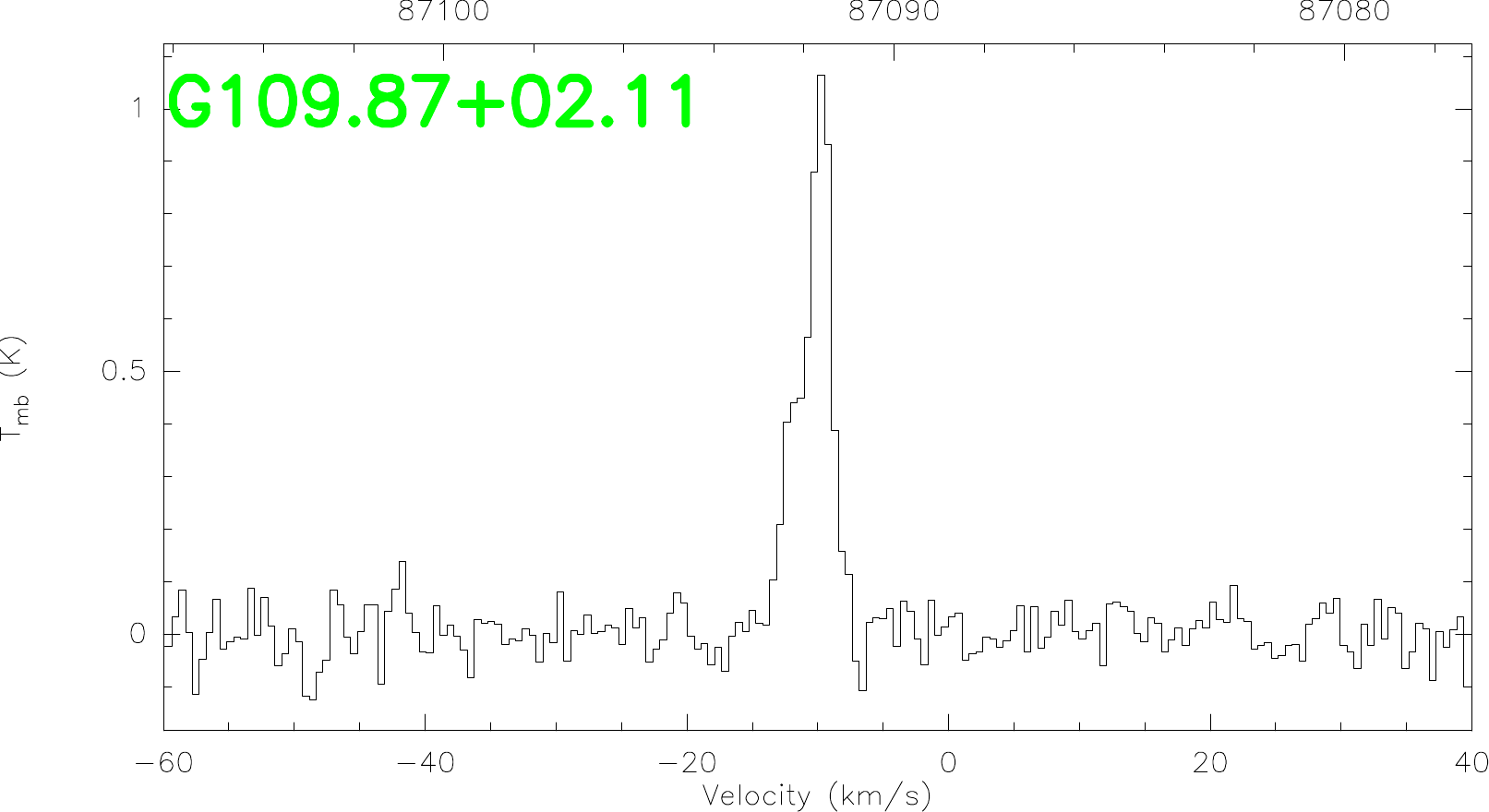}
\includegraphics[width=0.155\textwidth]{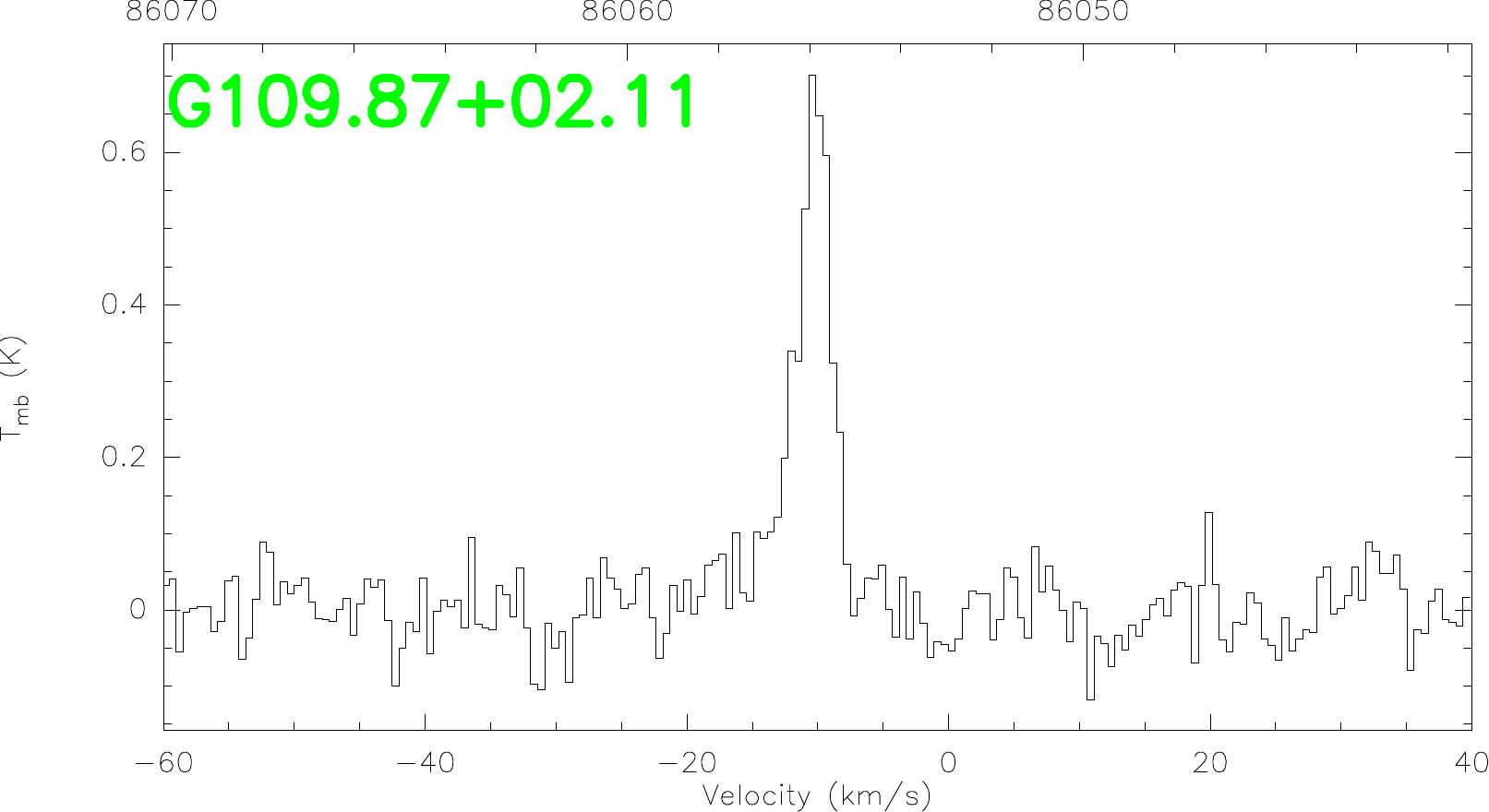}
\includegraphics[width=0.155\textwidth]{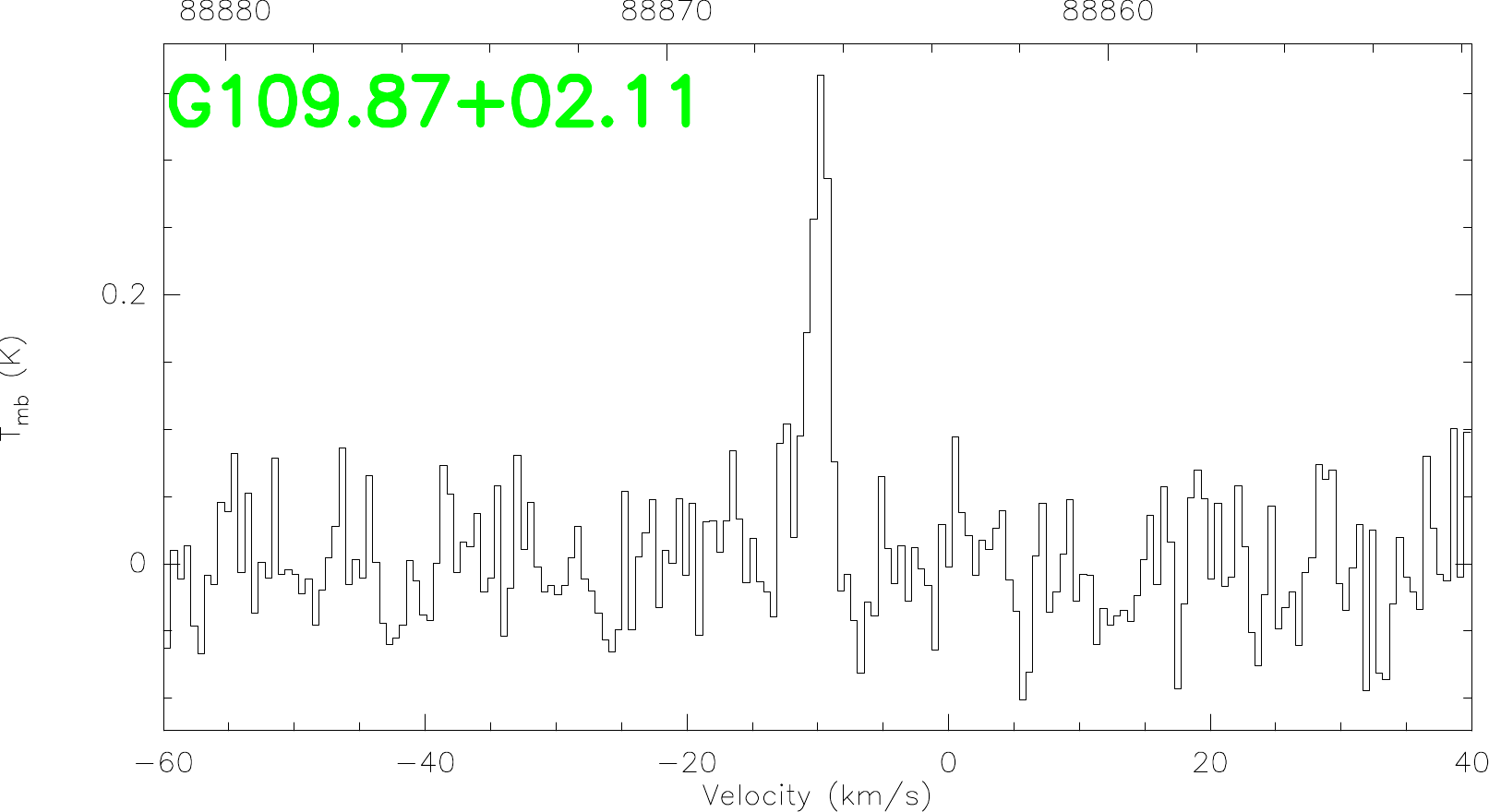}\\

\includegraphics[width=0.155\textwidth]{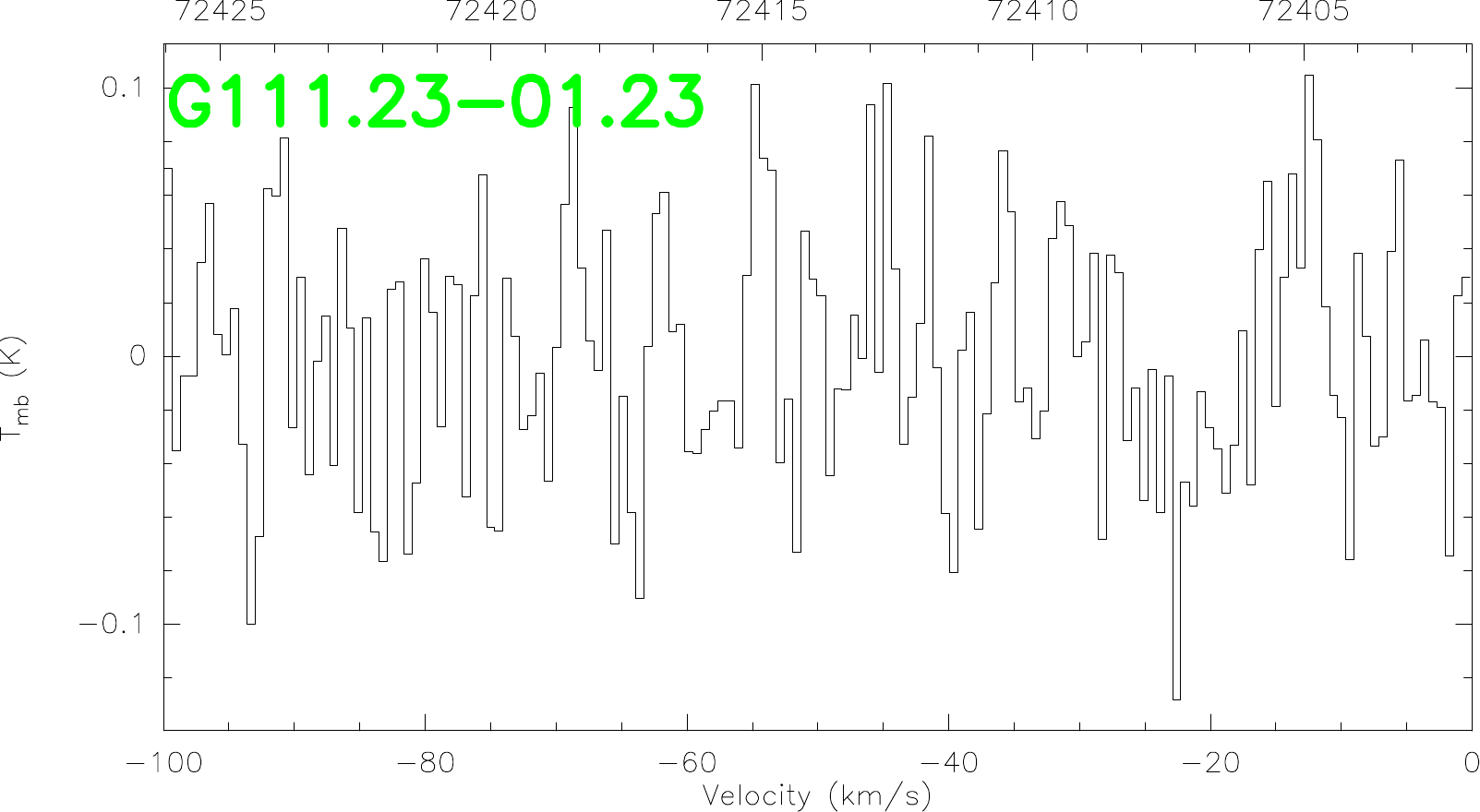}
\includegraphics[width=0.155\textwidth]{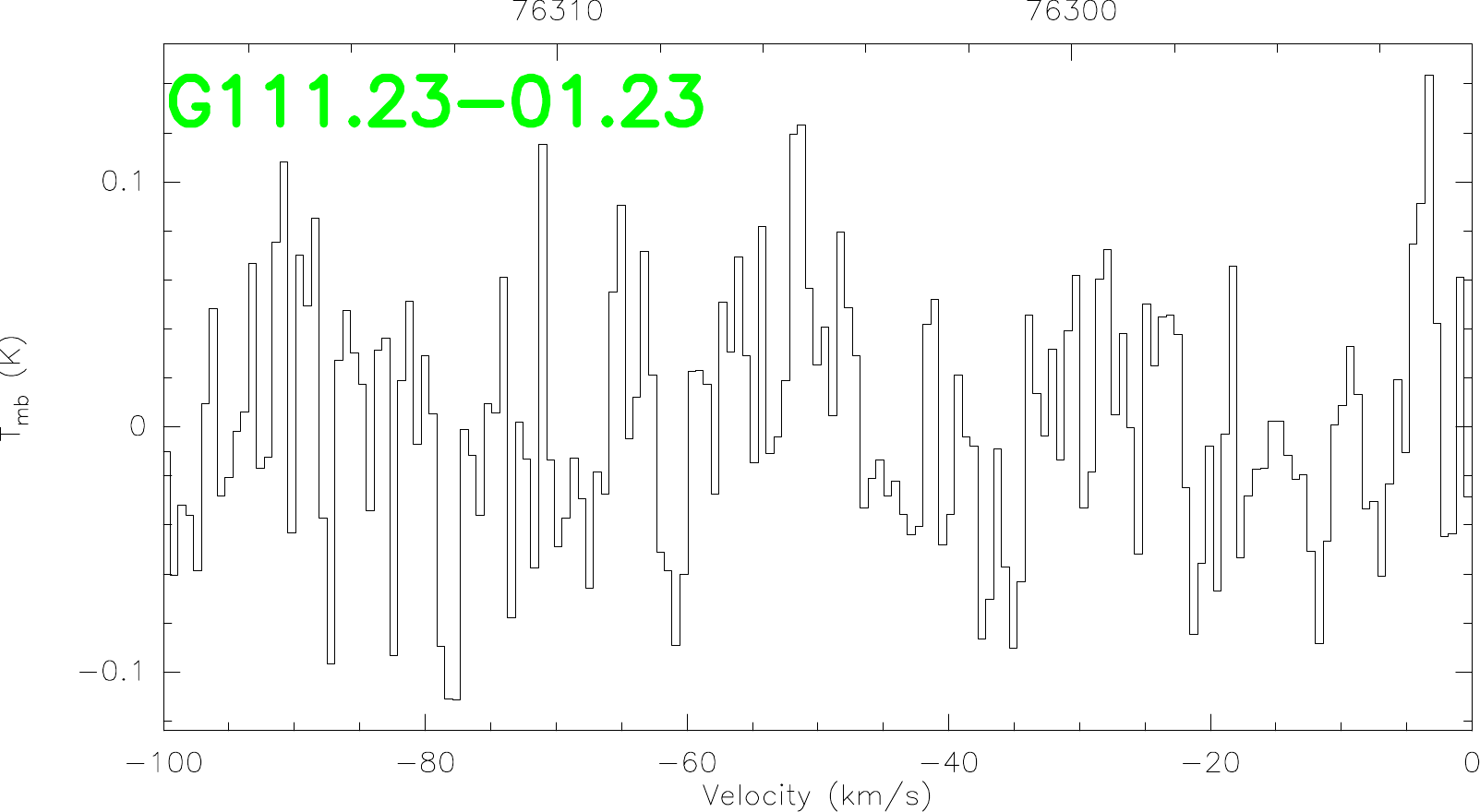}
\includegraphics[width=0.155\textwidth]{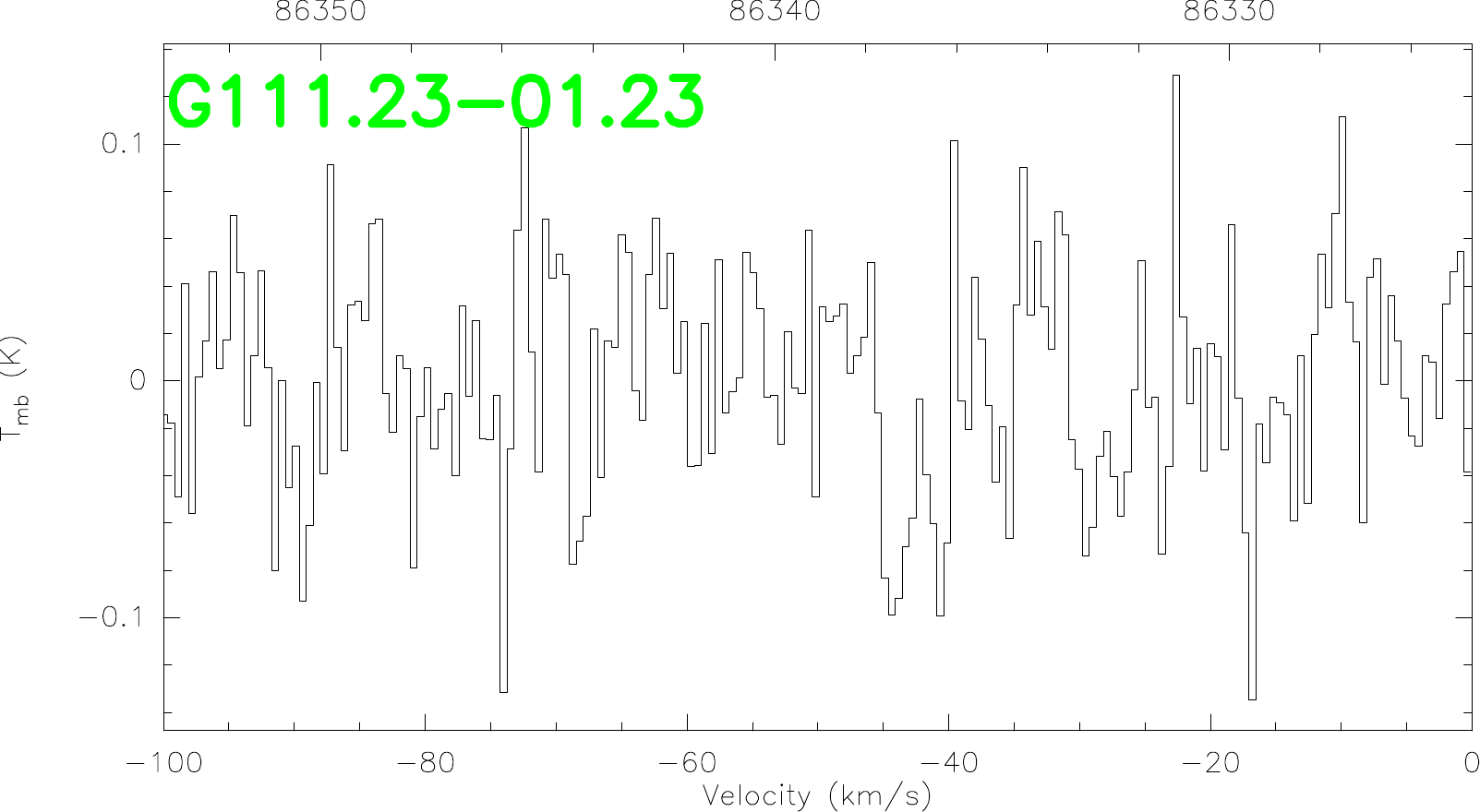}
\includegraphics[width=0.155\textwidth]{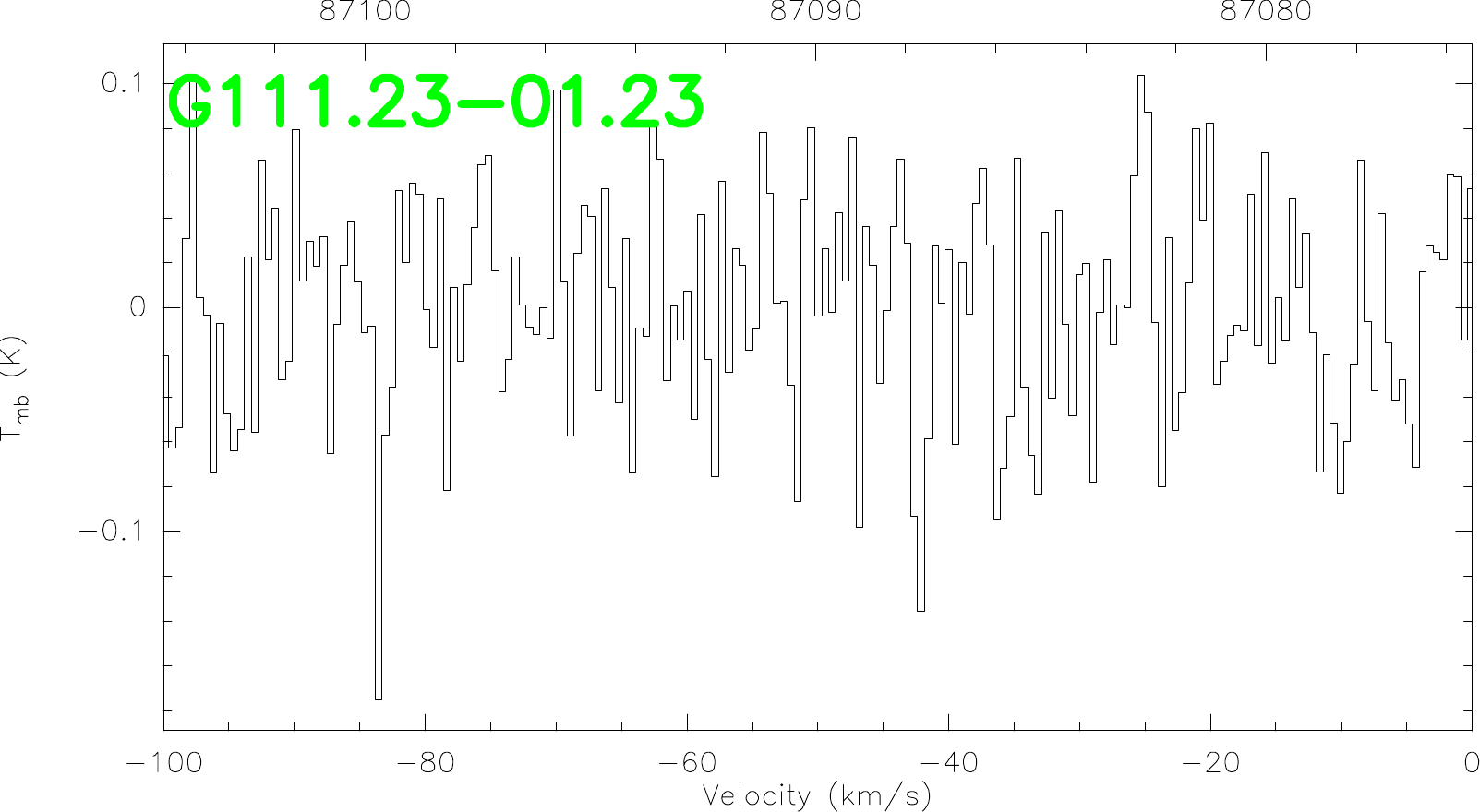}
\includegraphics[width=0.155\textwidth]{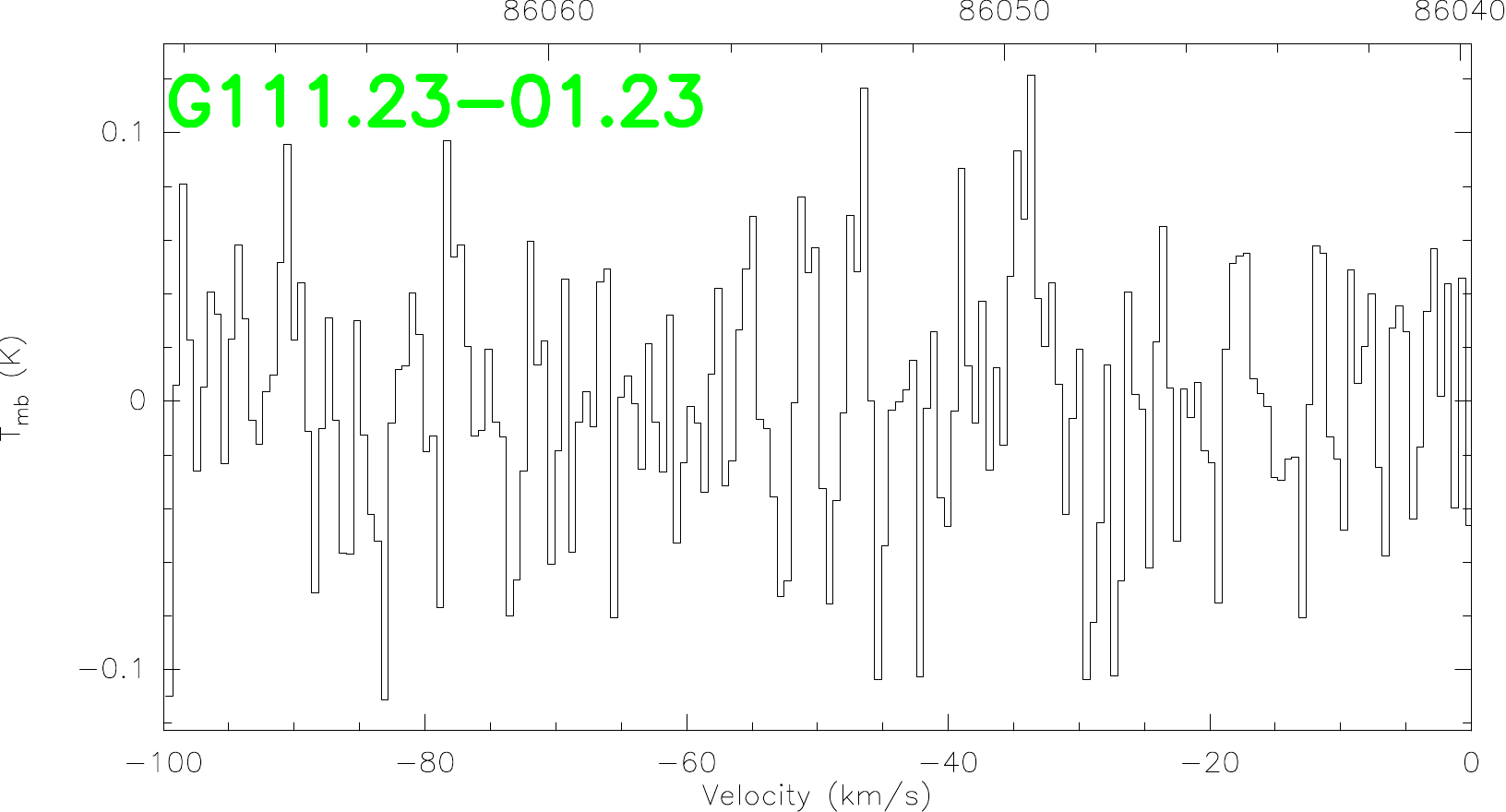}
\includegraphics[width=0.155\textwidth]{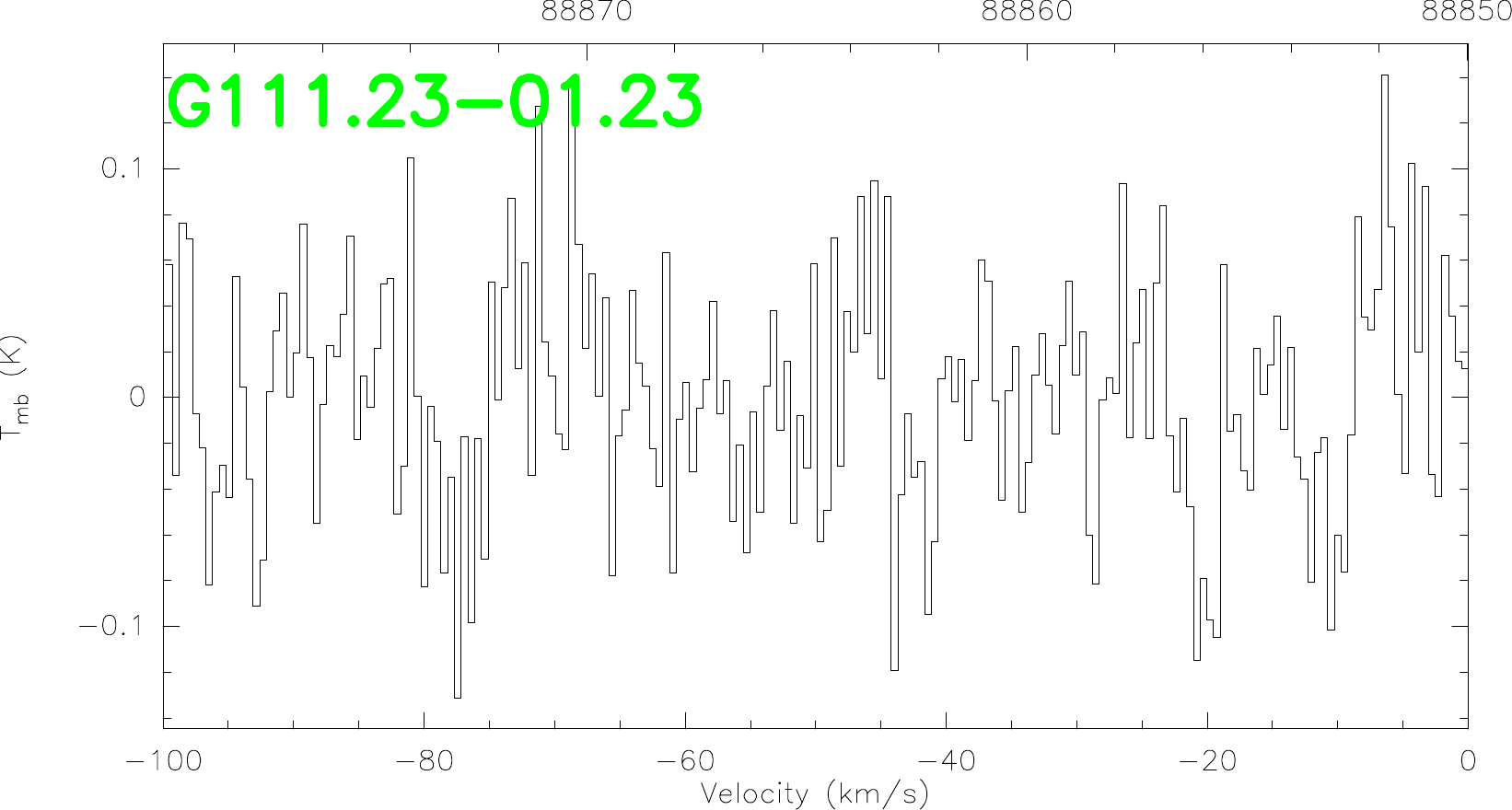}\\

\caption{Continued. Spectra of different molecular lines.}
\label{fig:multi_lines_page5}
\end{figure}

\begin{figure}[h]
\centering

\makebox[\colw][c]{DCN 1-0}
\makebox[\colw][c]{DNC 1-0}
\makebox[\colw][c]{H$^{13}$CN 1-0}
\makebox[\colw][c]{HN$^{13}$C 1-0}
\makebox[\colw][c]{HC$^{15}$N 1-0}
\makebox[\colw][c]{H$^{15}$NC 1-0}\\[0.2em]

\includegraphics[width=0.155\textwidth]{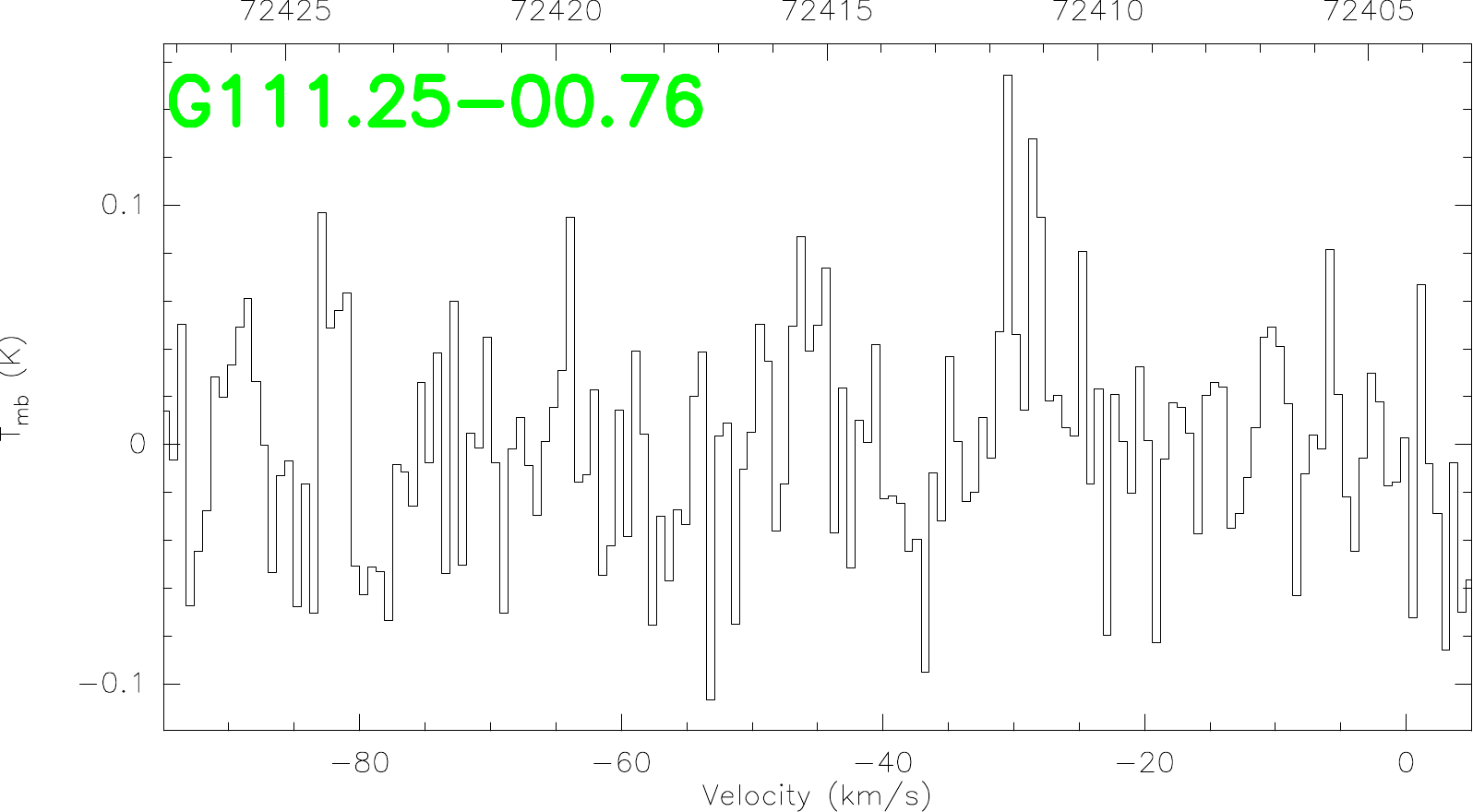}
\includegraphics[width=0.155\textwidth]{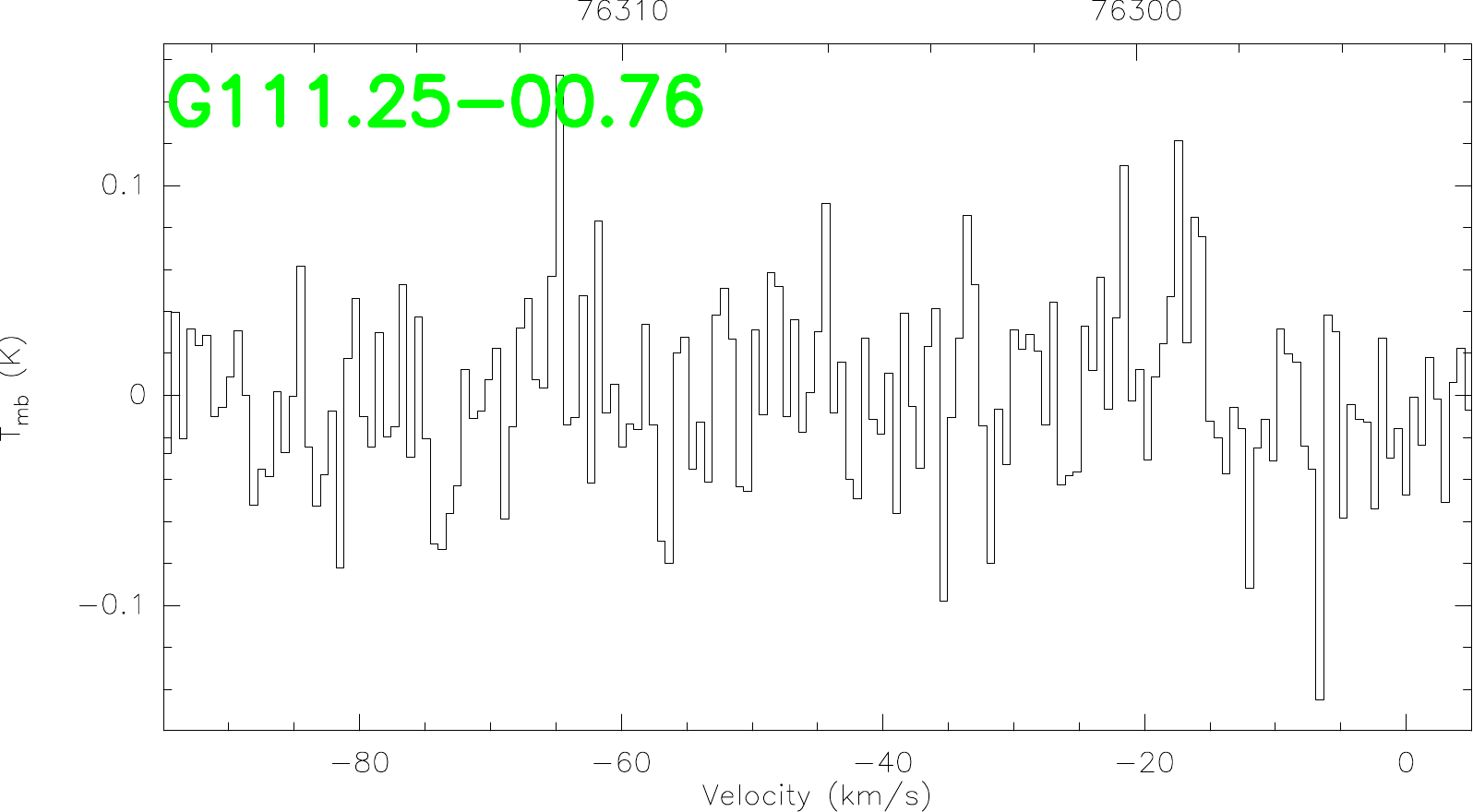}
\includegraphics[width=0.155\textwidth]{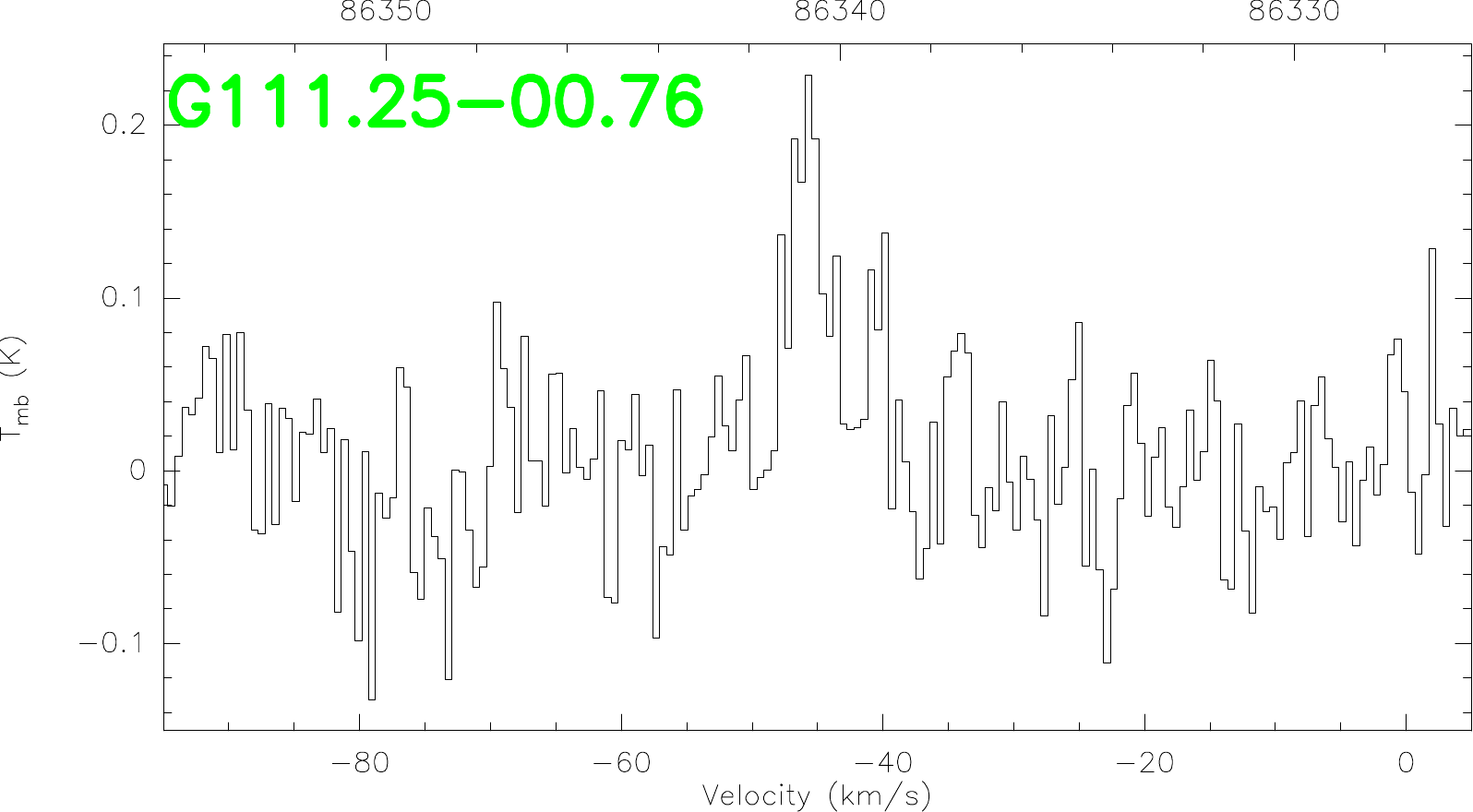}
\includegraphics[width=0.155\textwidth]{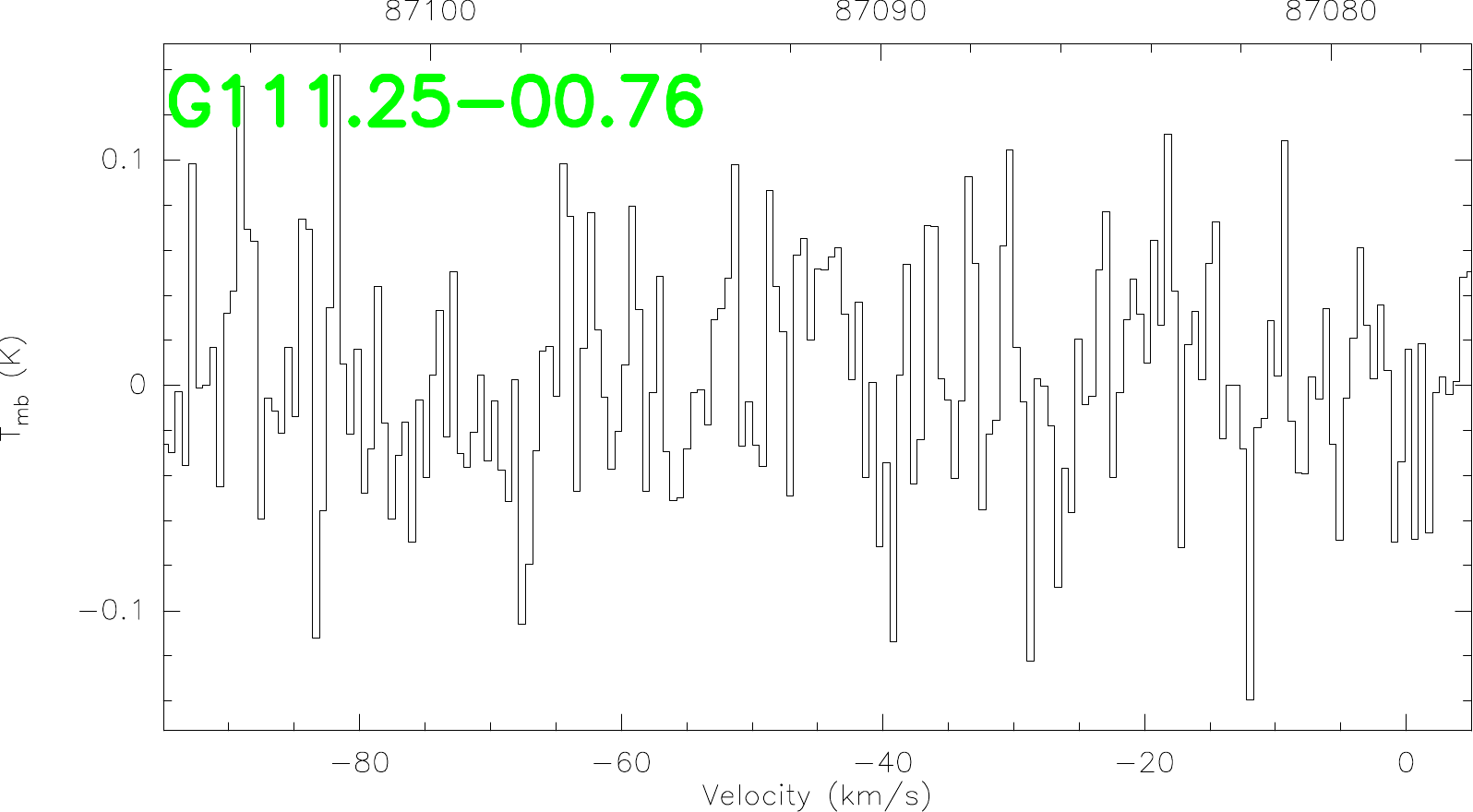}
\includegraphics[width=0.155\textwidth]{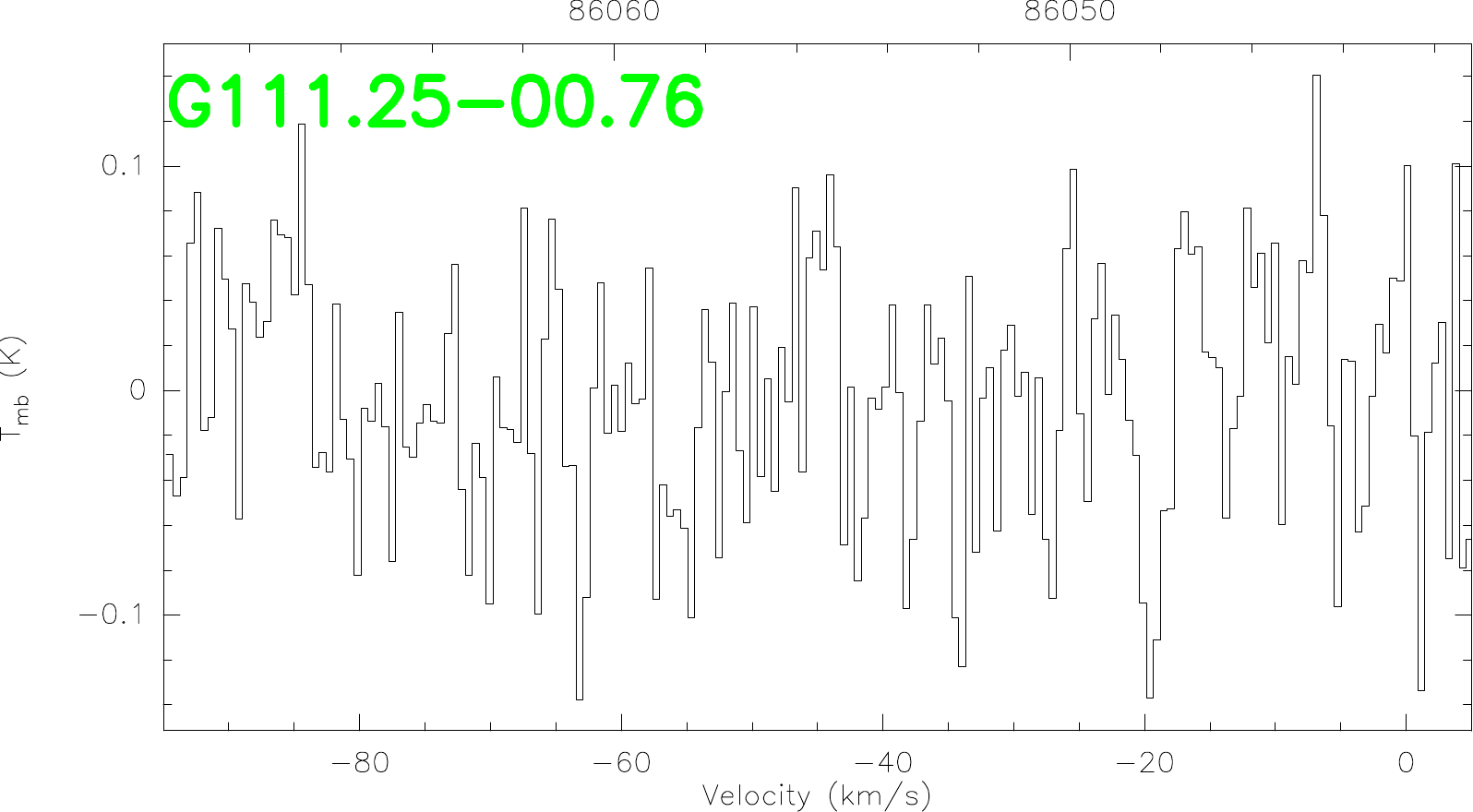}
\includegraphics[width=0.155\textwidth]{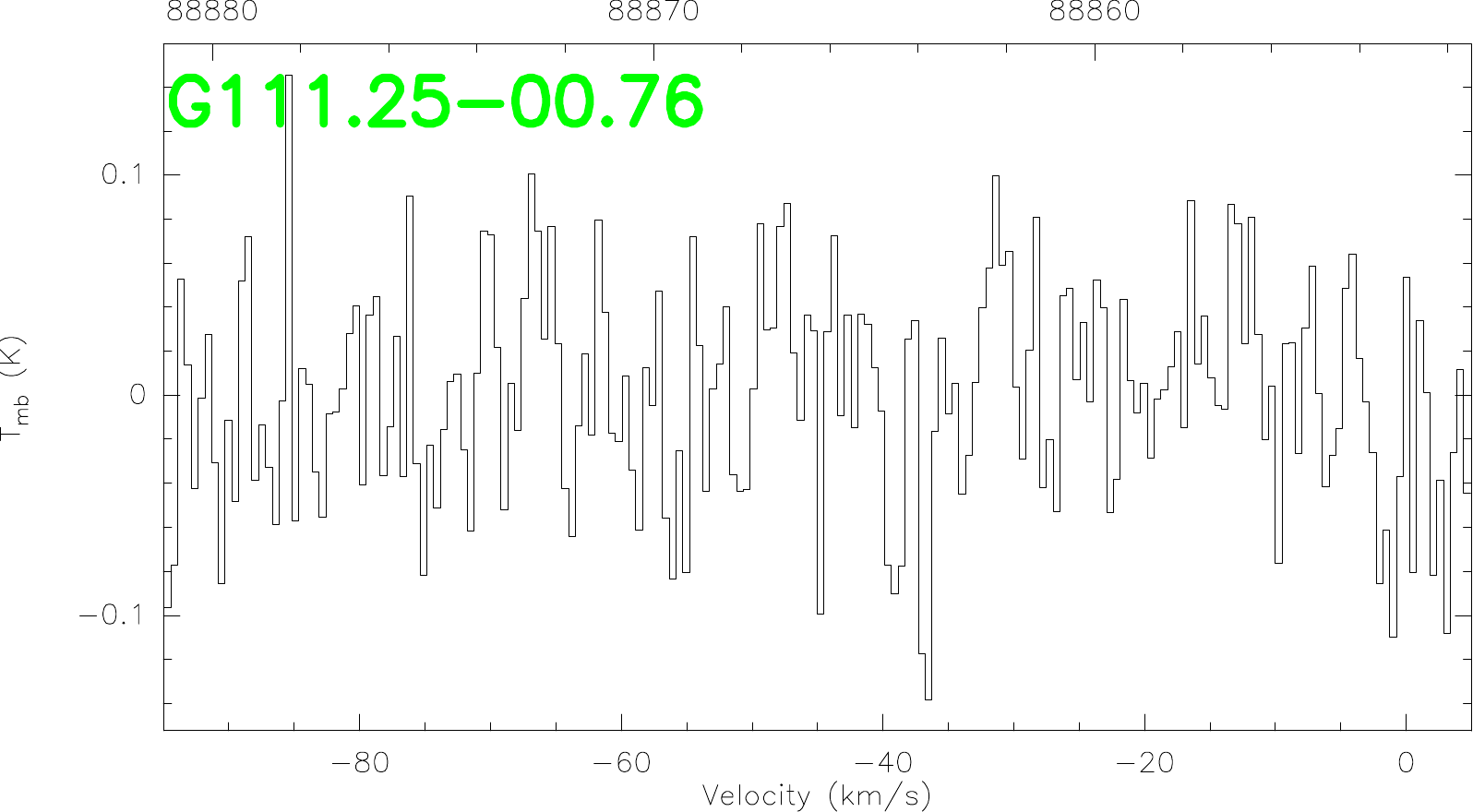}\\

\includegraphics[width=0.155\textwidth]{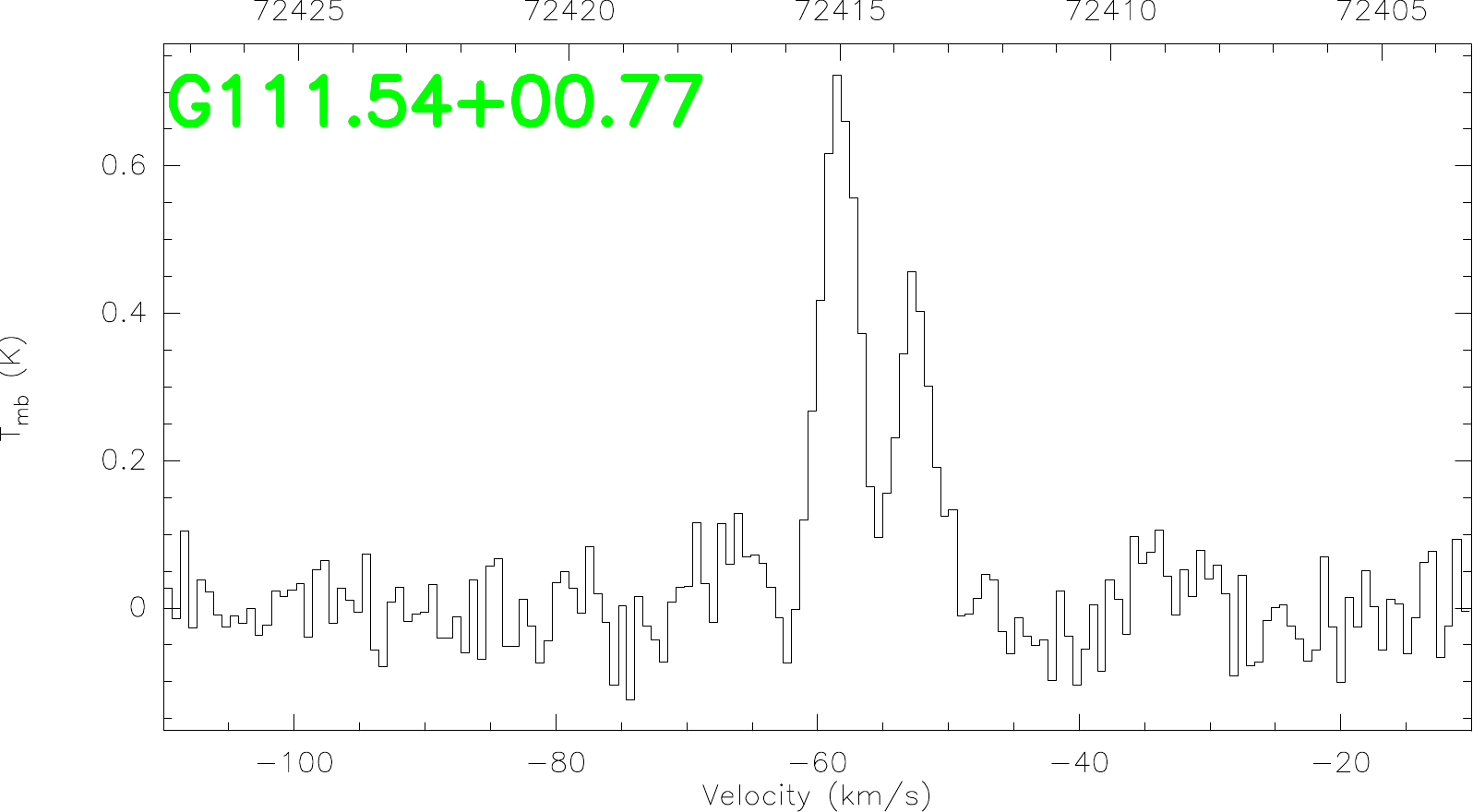}
\includegraphics[width=0.155\textwidth]{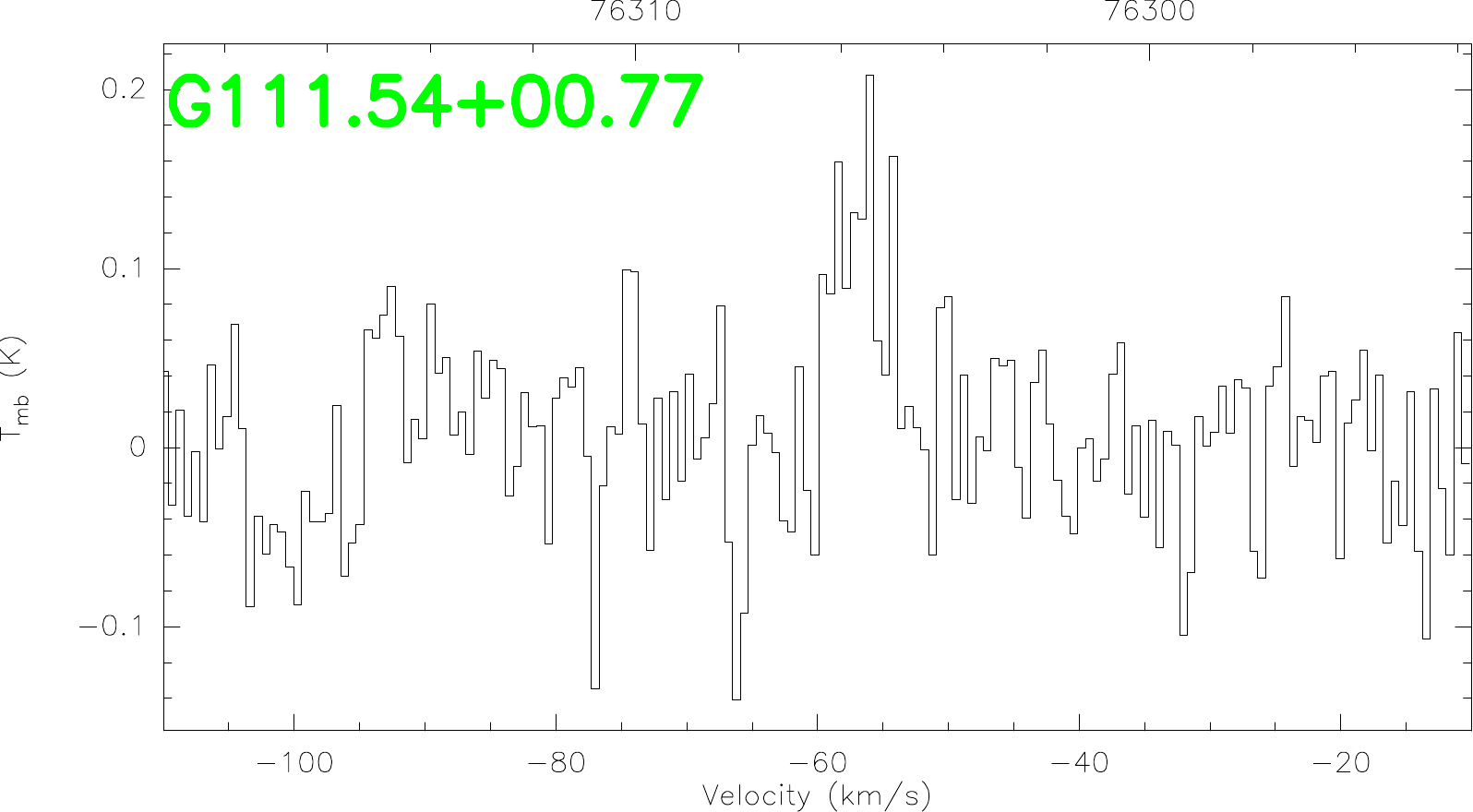}
\includegraphics[width=0.155\textwidth]{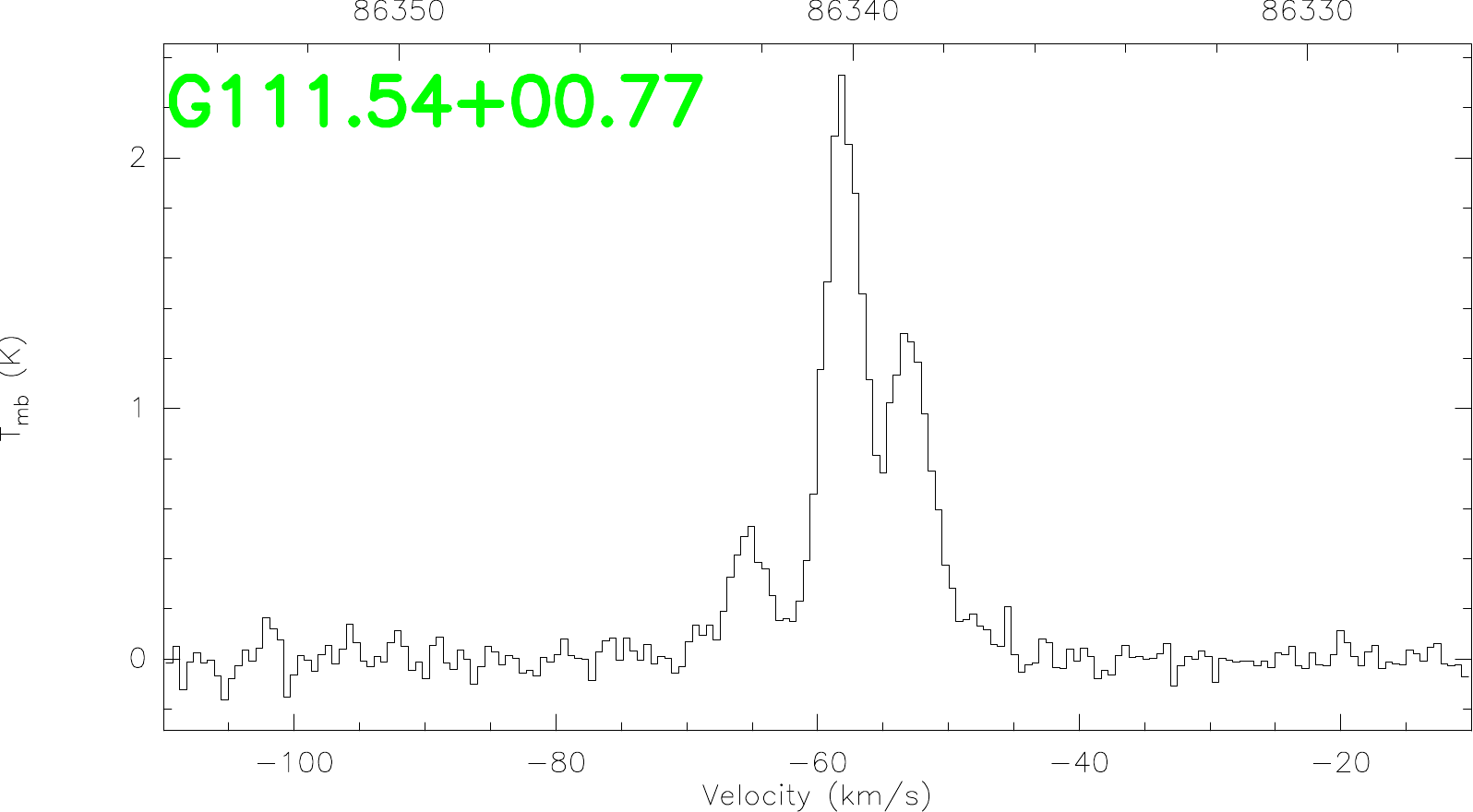}
\includegraphics[width=0.155\textwidth]{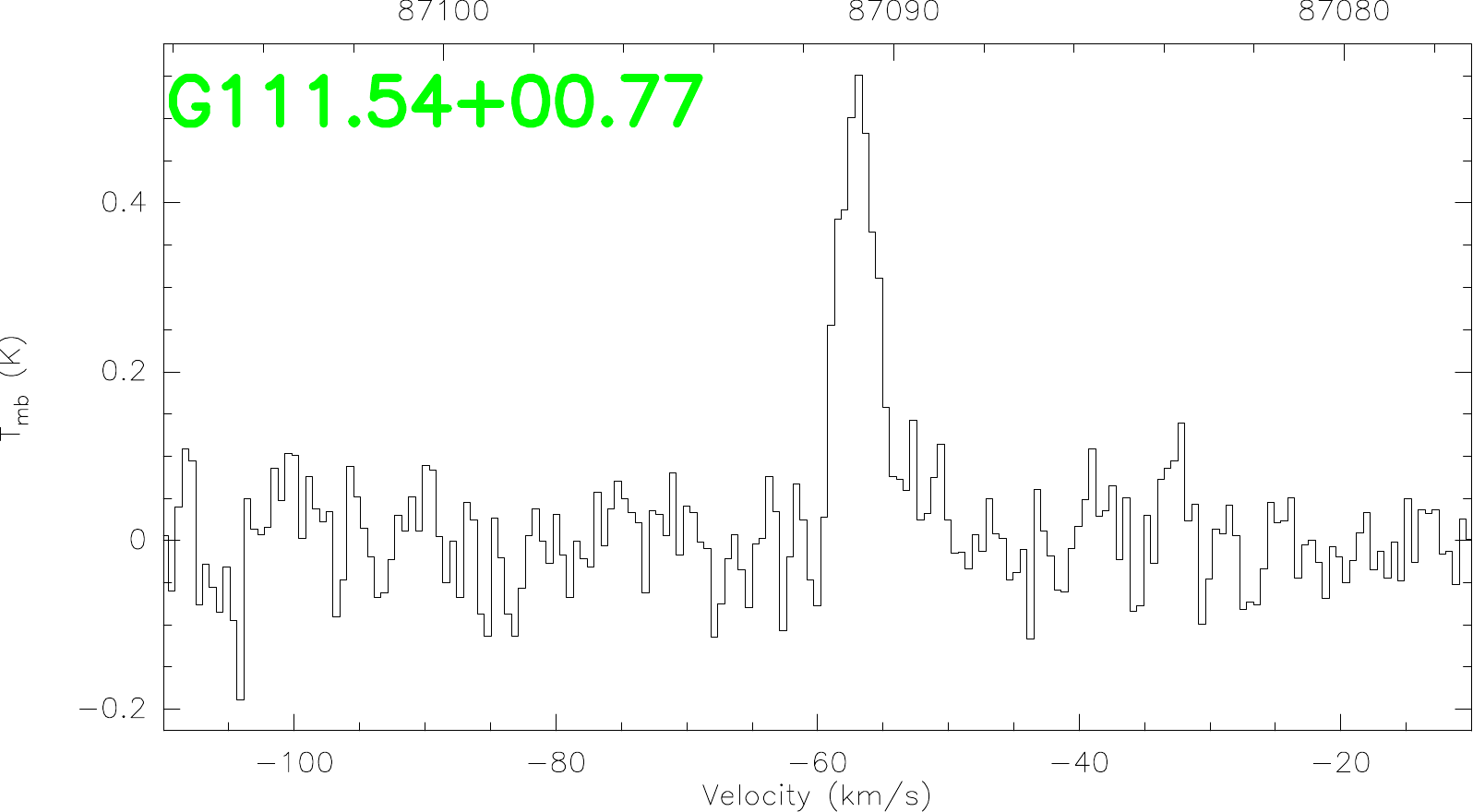}
\includegraphics[width=0.155\textwidth]{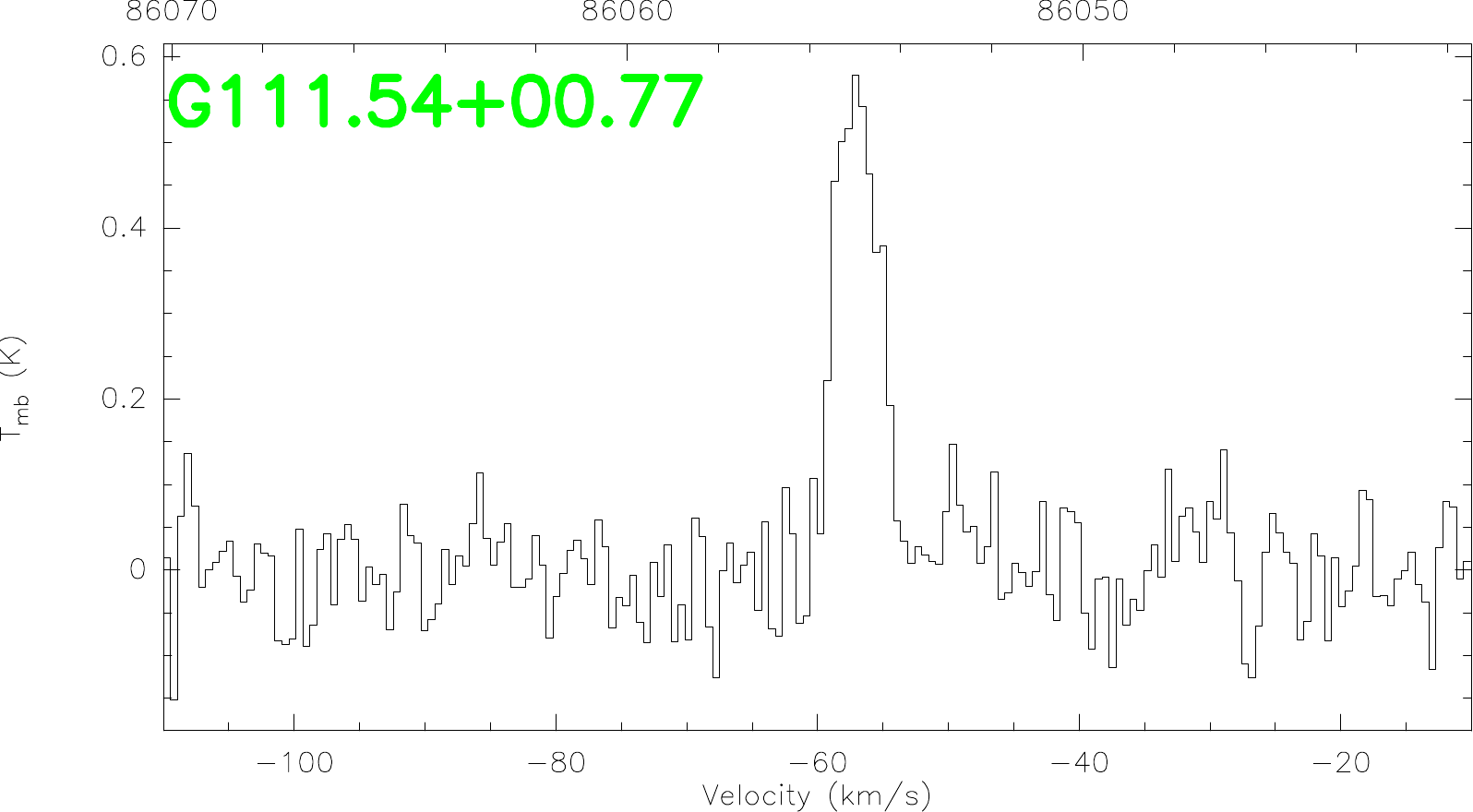}
\includegraphics[width=0.155\textwidth]{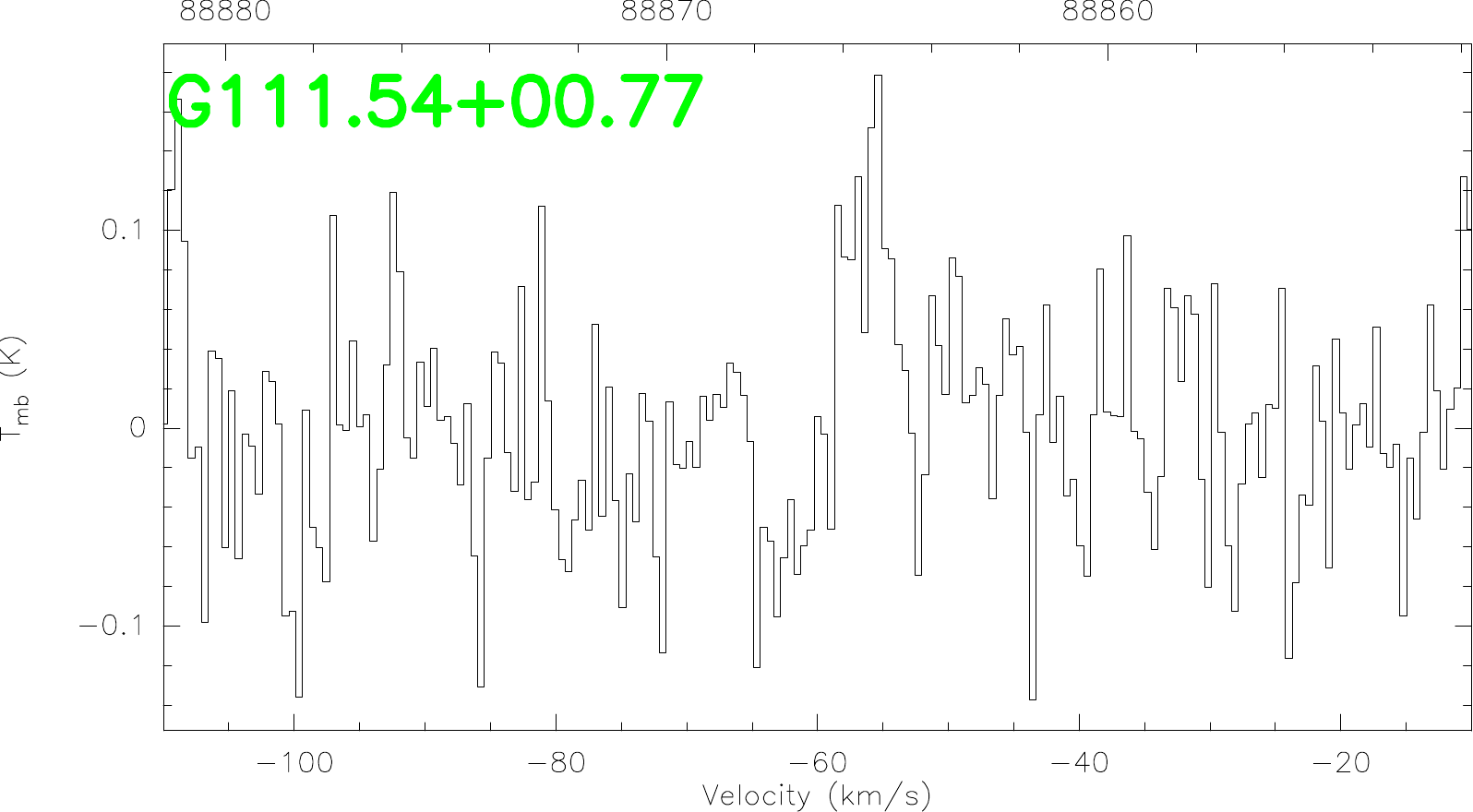}\\

\includegraphics[width=0.155\textwidth]{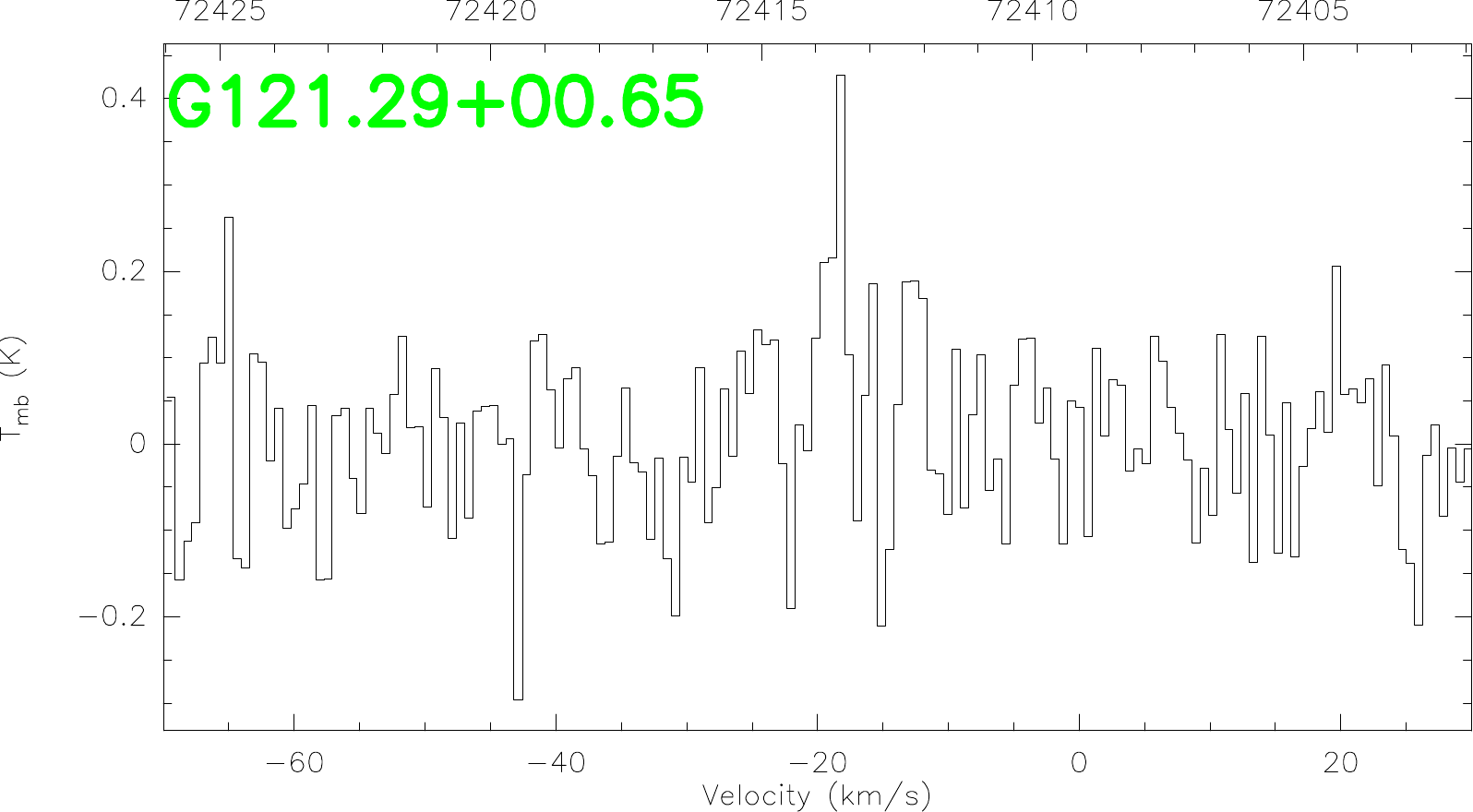}
\includegraphics[width=0.155\textwidth]{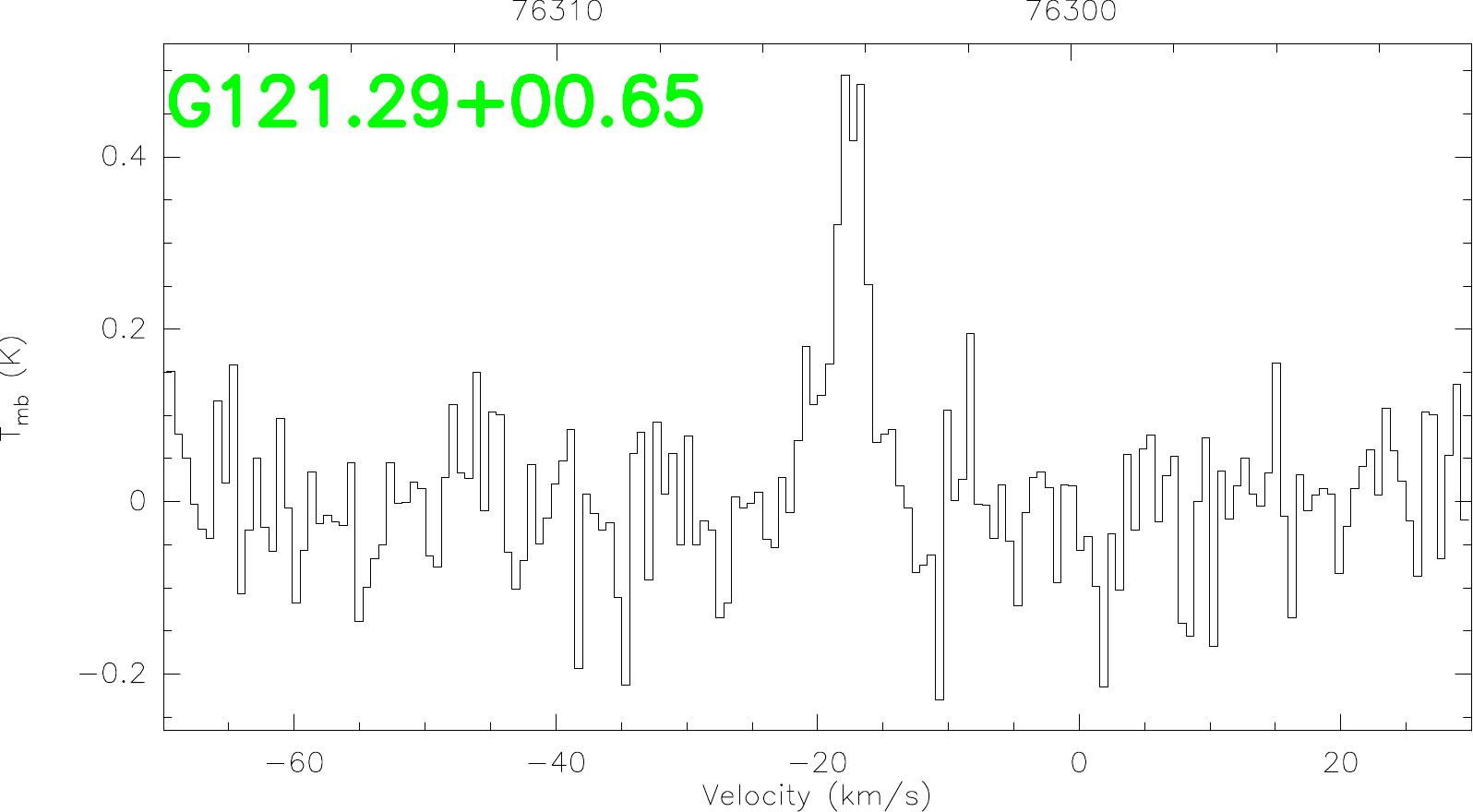}
\includegraphics[width=0.155\textwidth]{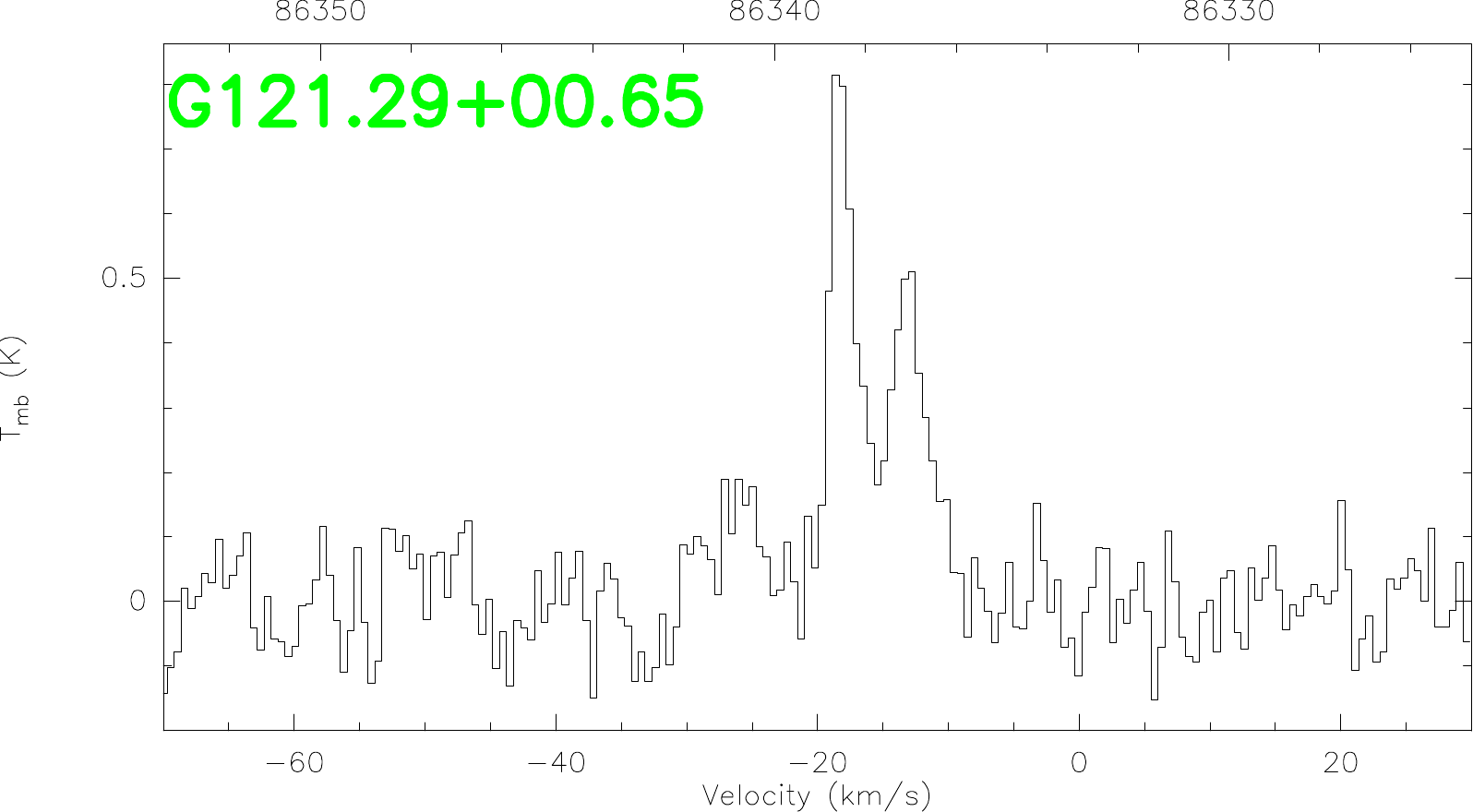}
\includegraphics[width=0.155\textwidth]{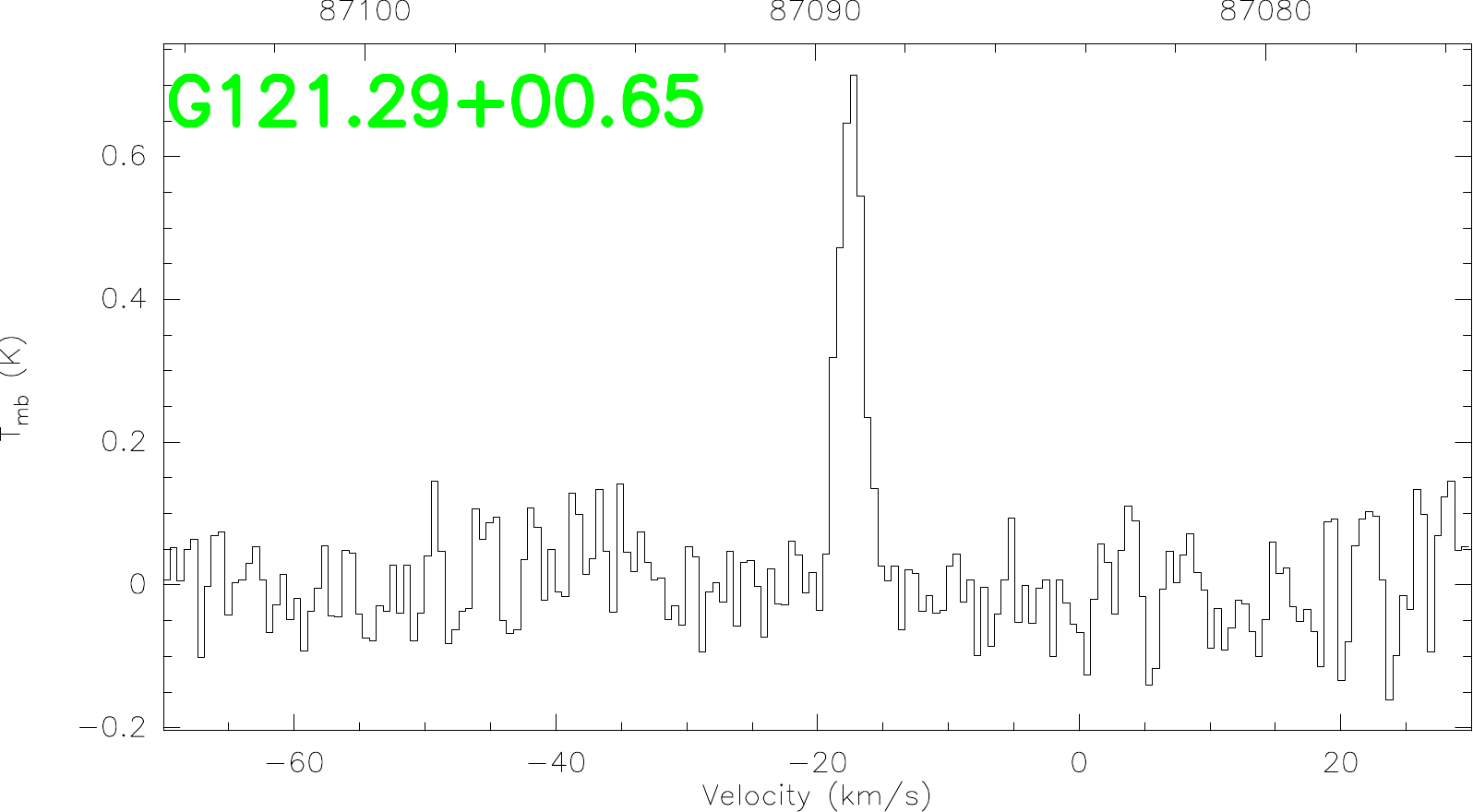}
\includegraphics[width=0.155\textwidth]{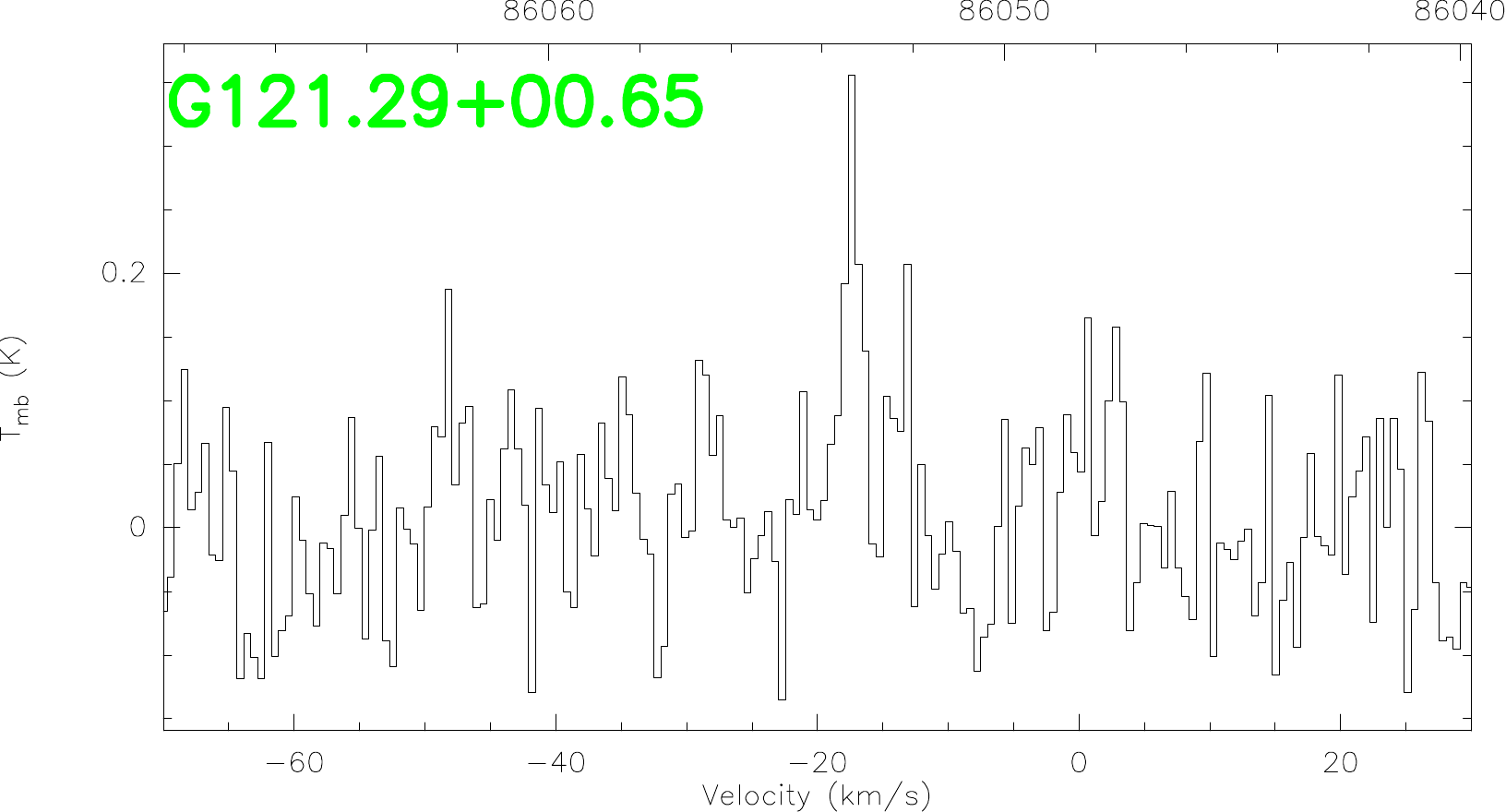}
\includegraphics[width=0.155\textwidth]{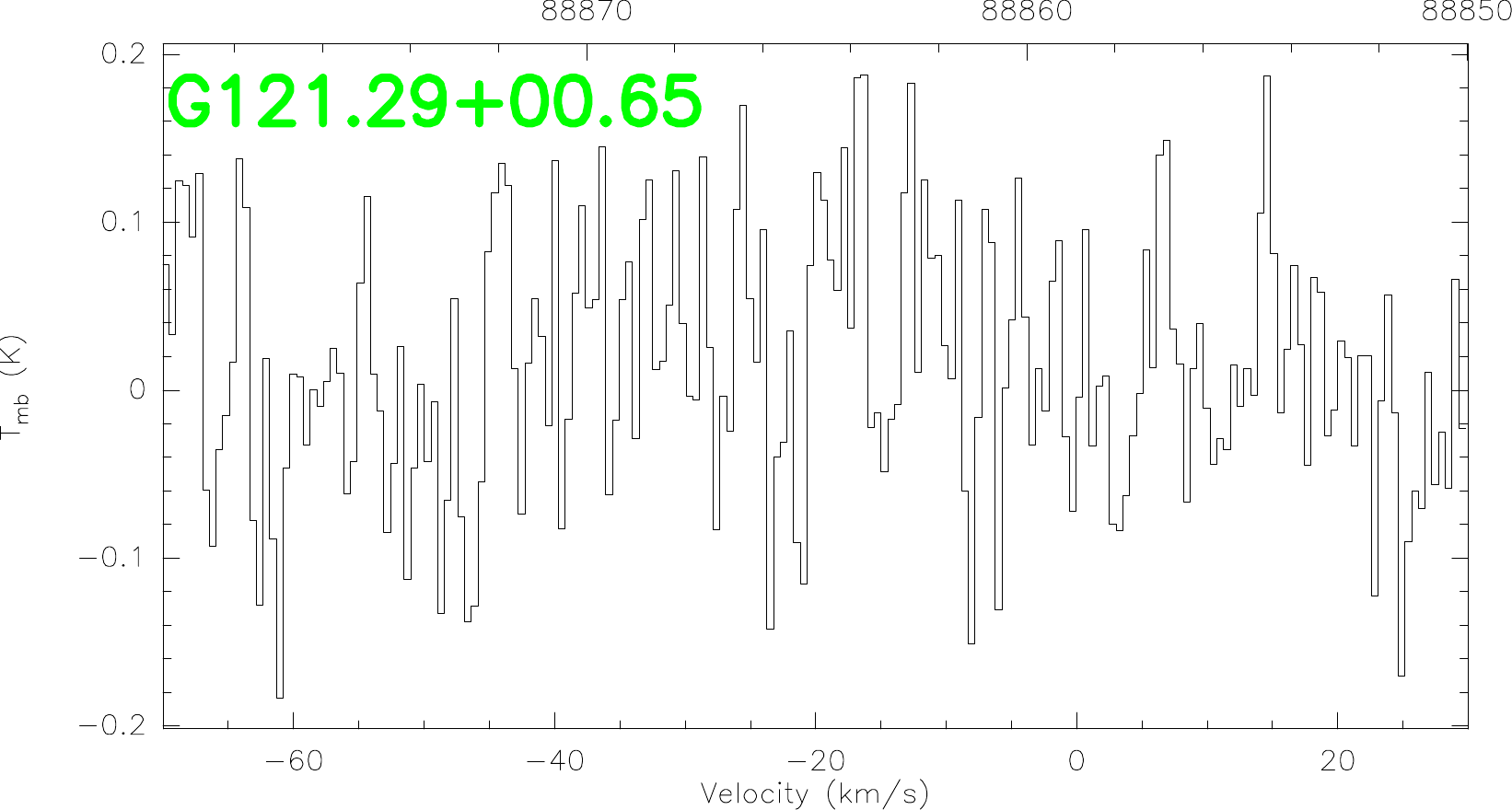}\\

\includegraphics[width=0.155\textwidth]{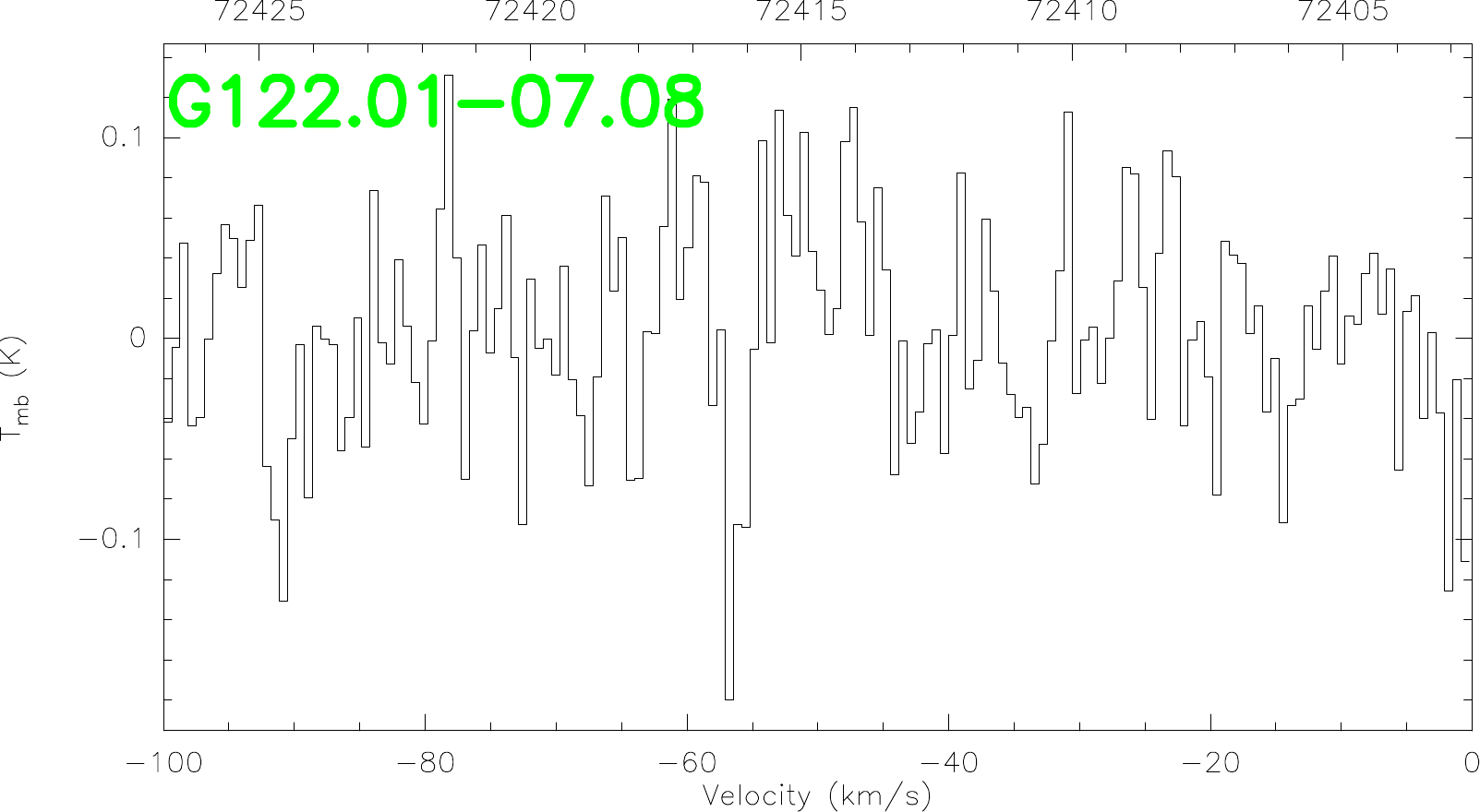}
\includegraphics[width=0.155\textwidth]{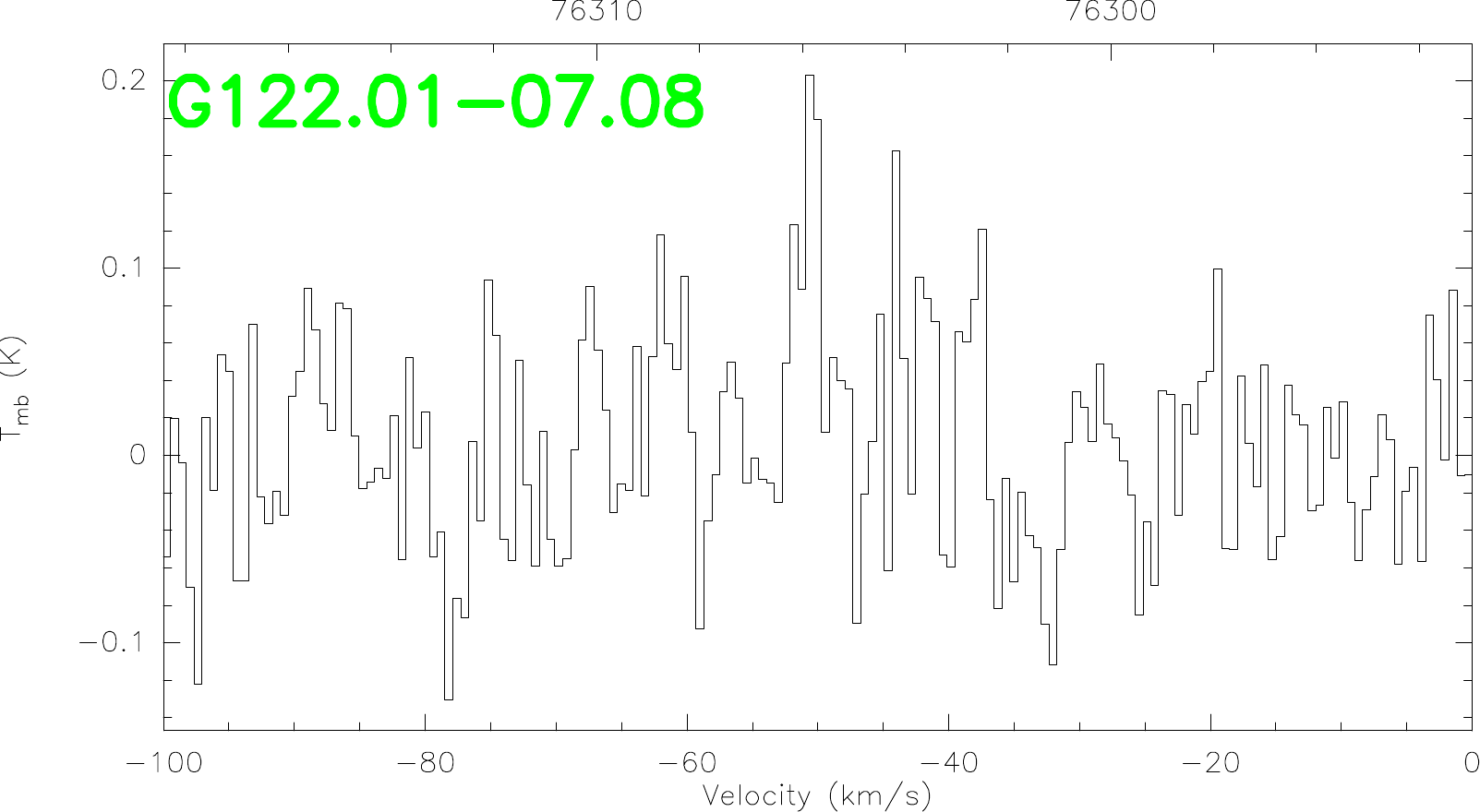}
\includegraphics[width=0.155\textwidth]{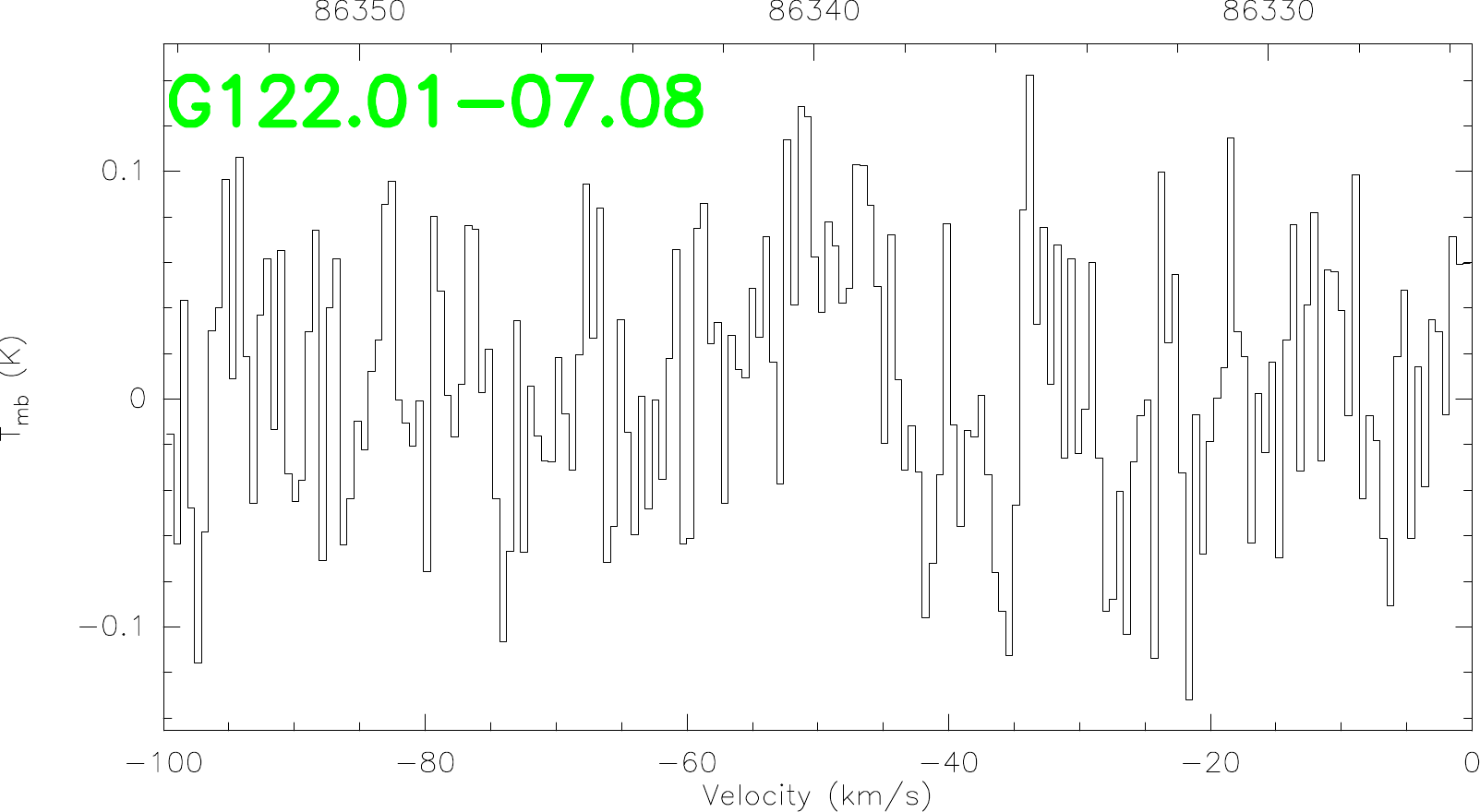}
\includegraphics[width=0.155\textwidth]{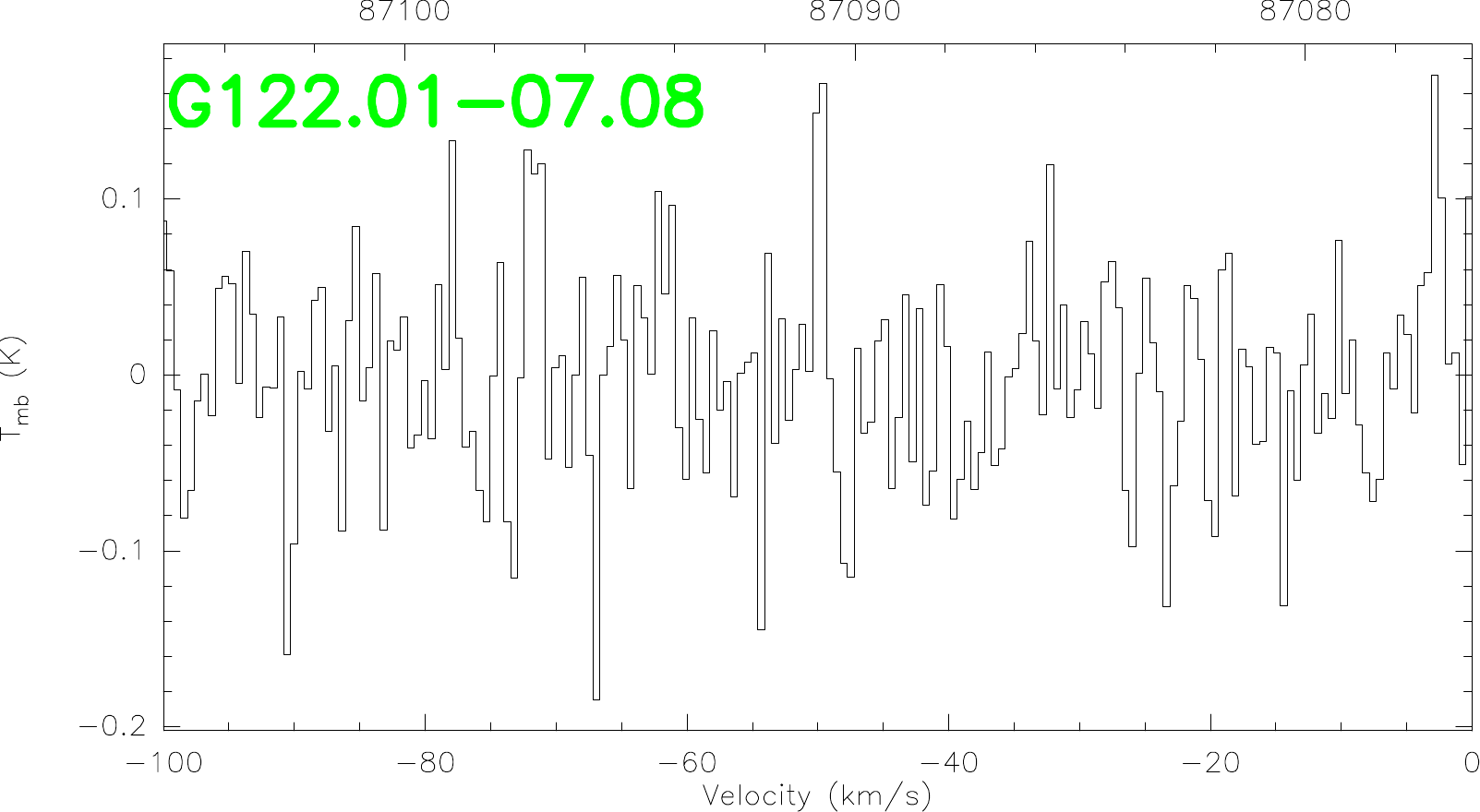}
\includegraphics[width=0.155\textwidth]{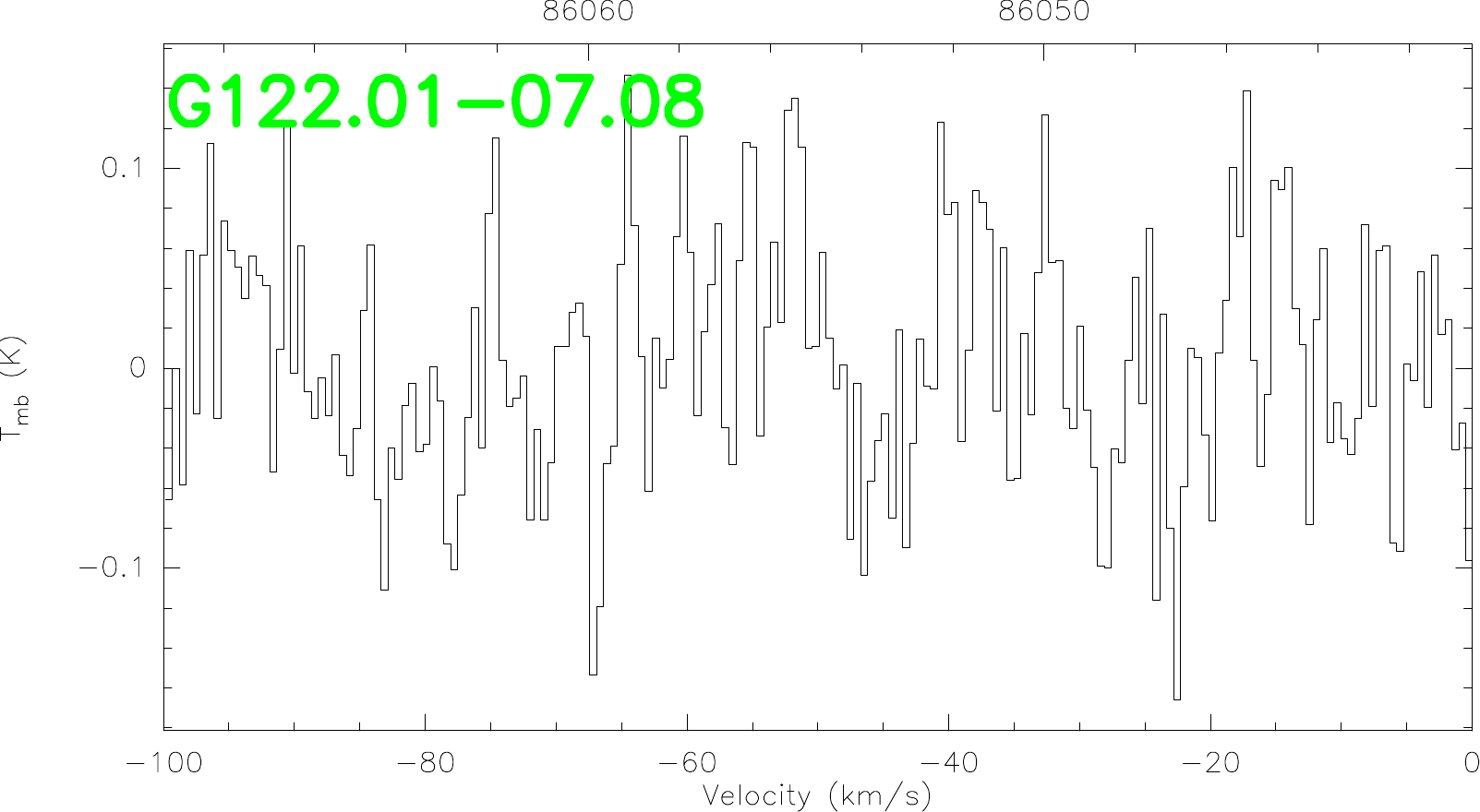}
\includegraphics[width=0.155\textwidth]{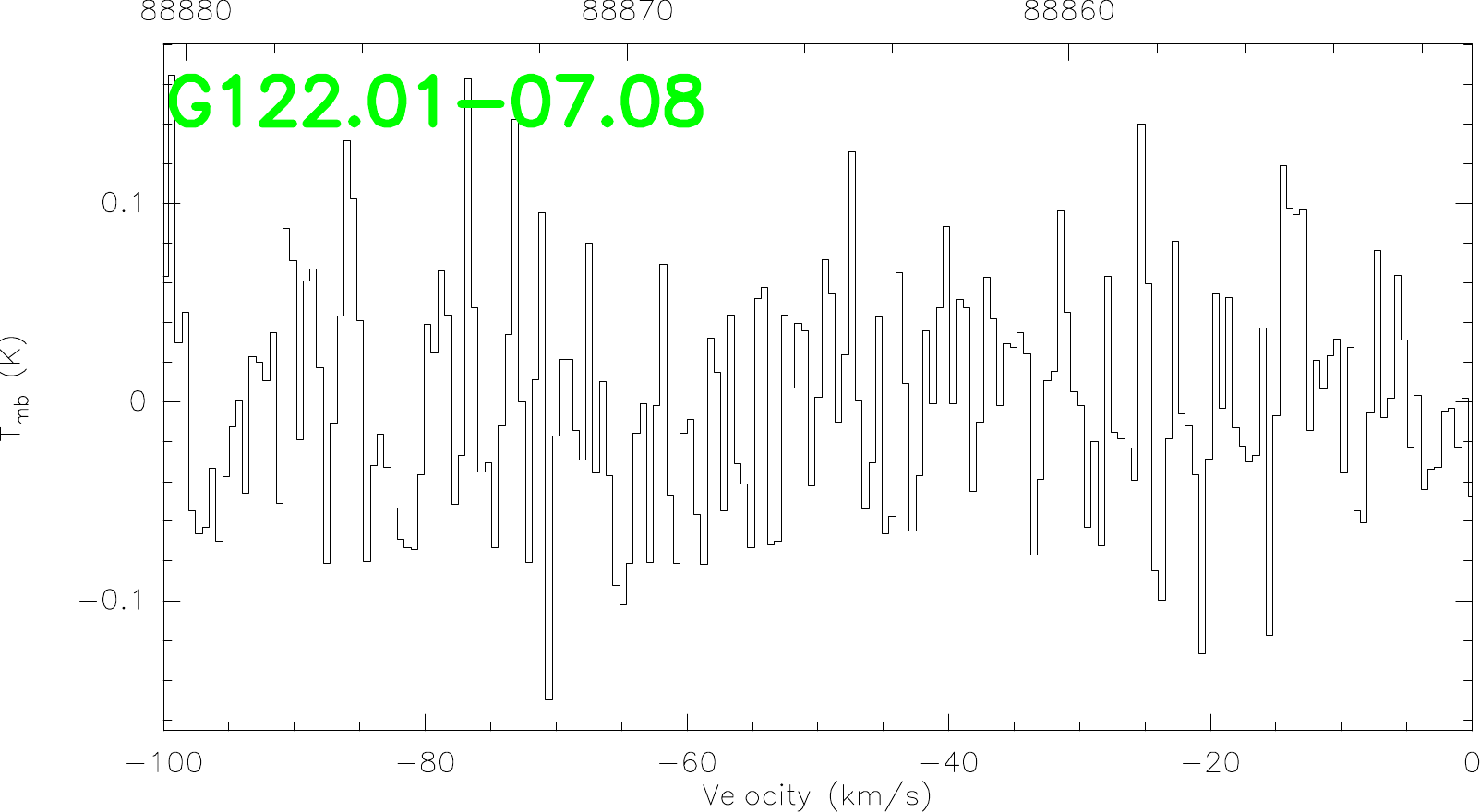}\\

\includegraphics[width=0.155\textwidth]{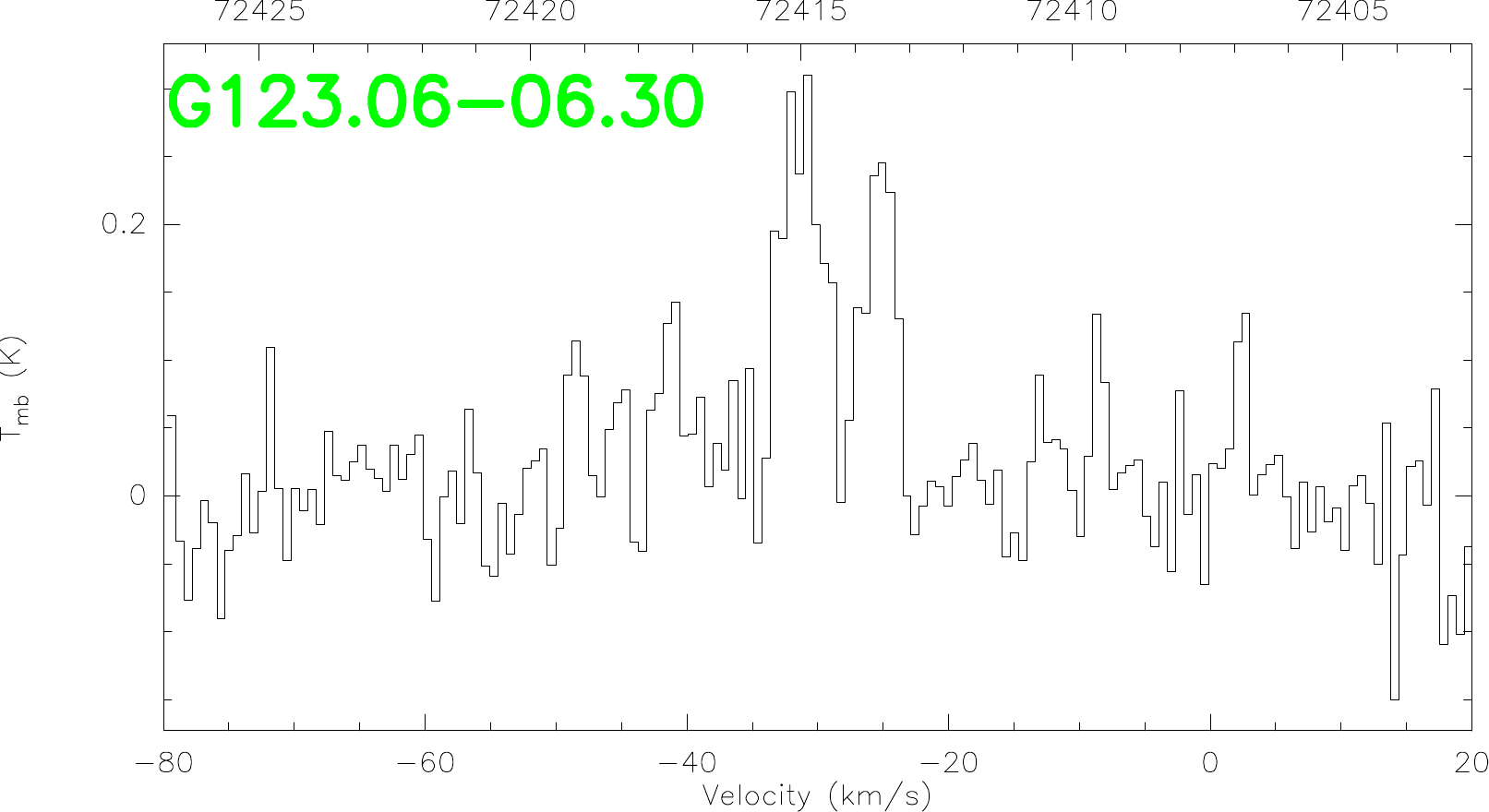}
\includegraphics[width=0.155\textwidth]{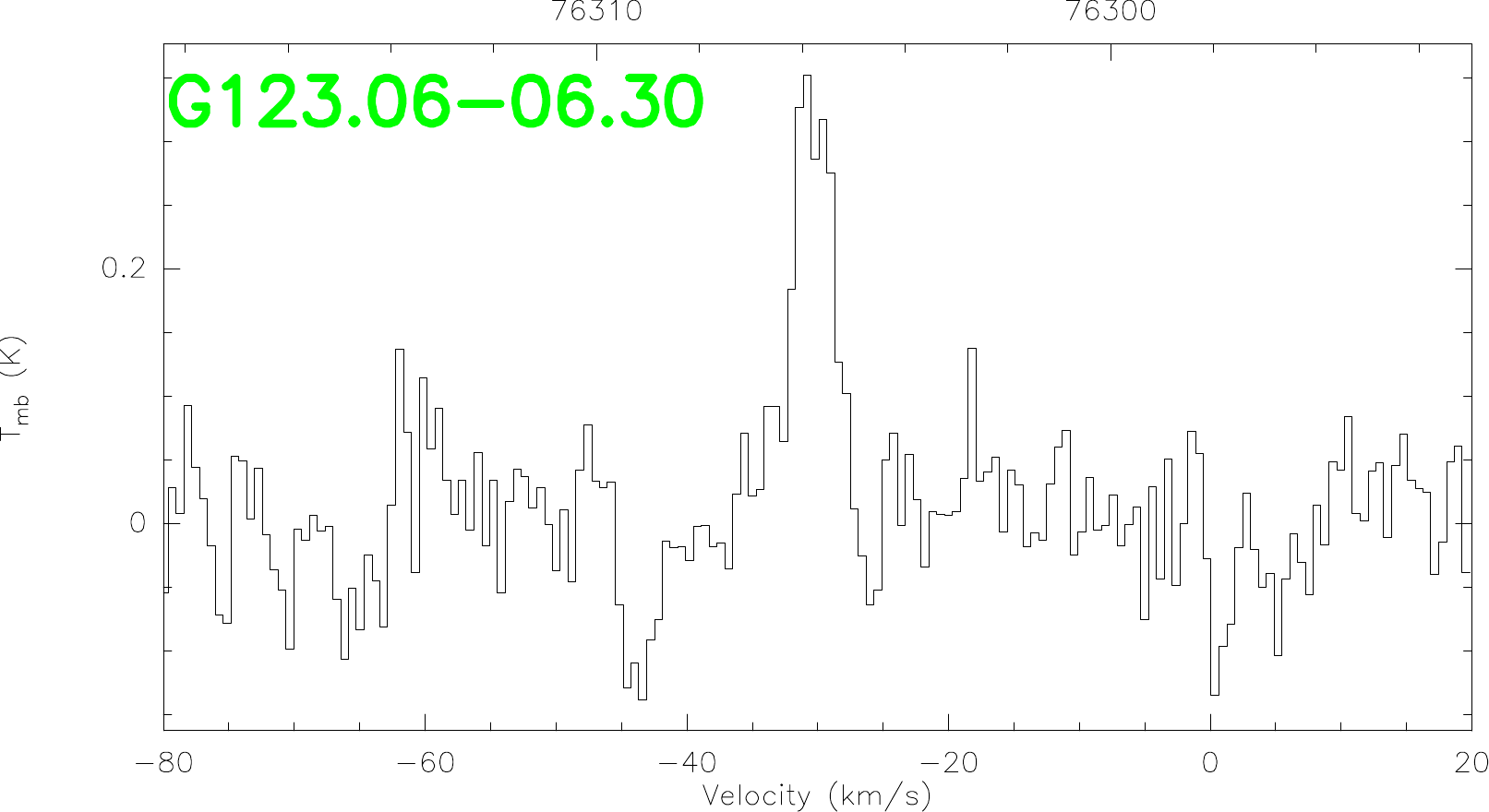}
\includegraphics[width=0.155\textwidth]{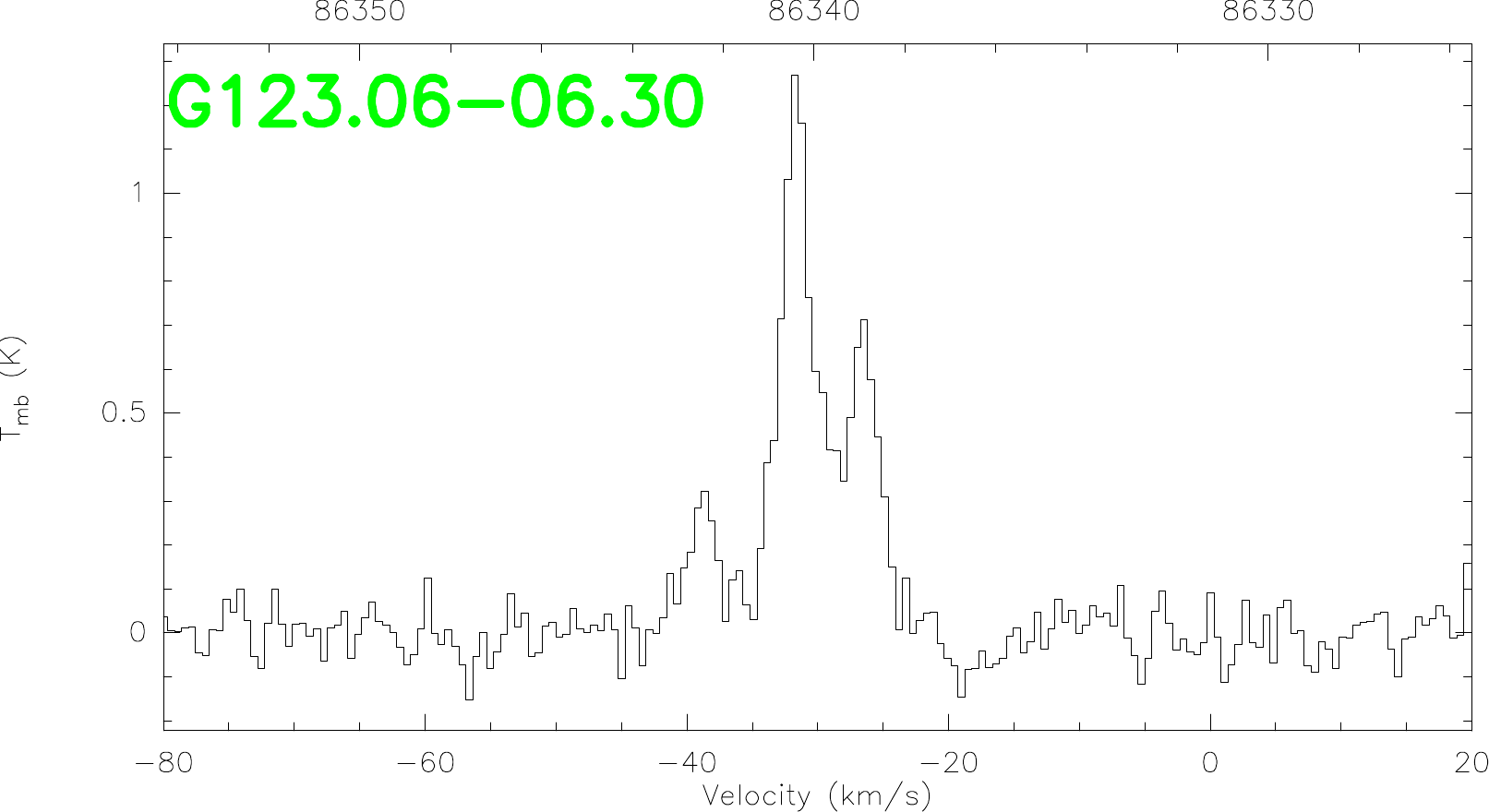}
\includegraphics[width=0.155\textwidth]{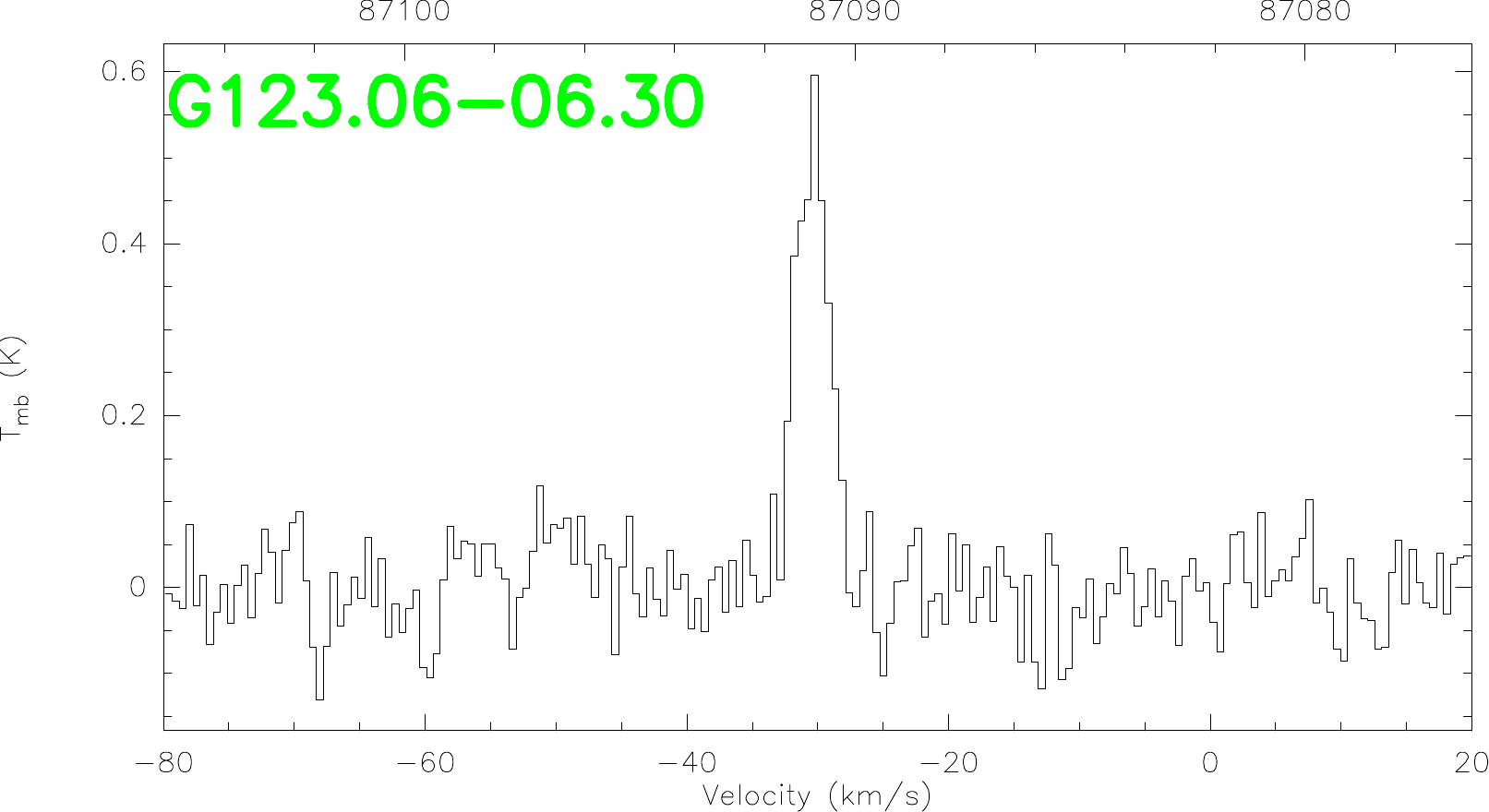}
\includegraphics[width=0.155\textwidth]{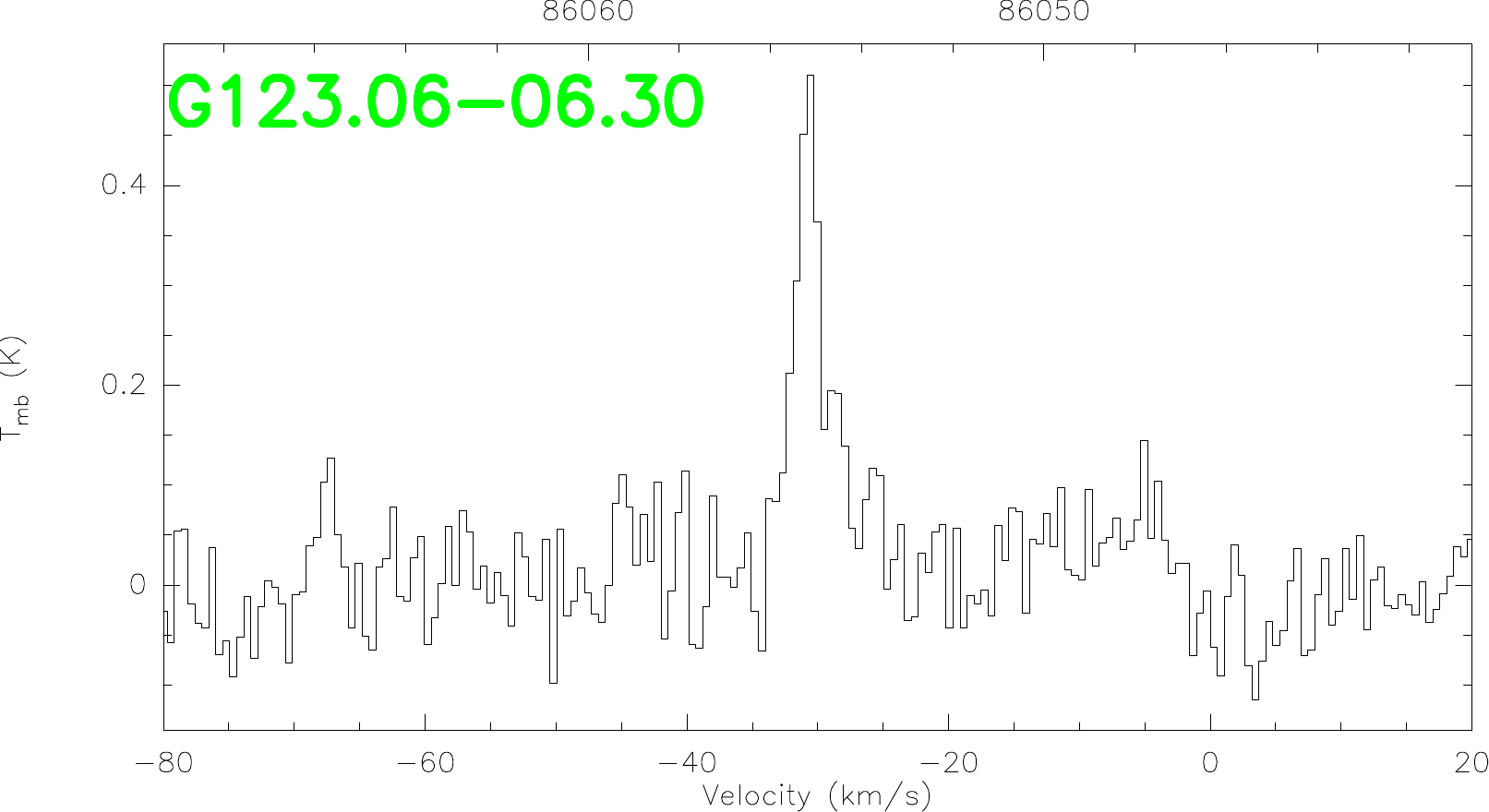}
\includegraphics[width=0.155\textwidth]{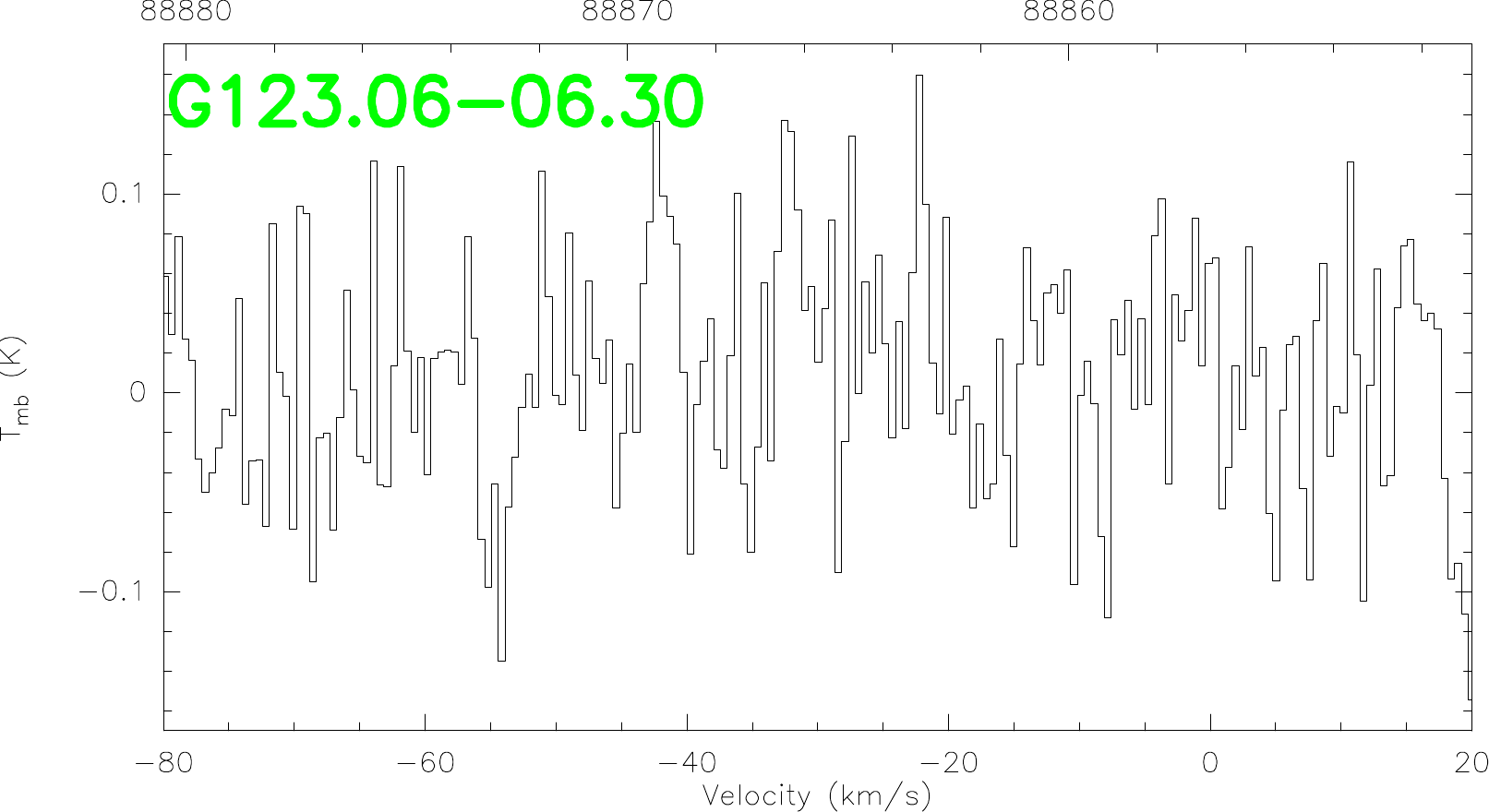}\\

\includegraphics[width=0.155\textwidth]{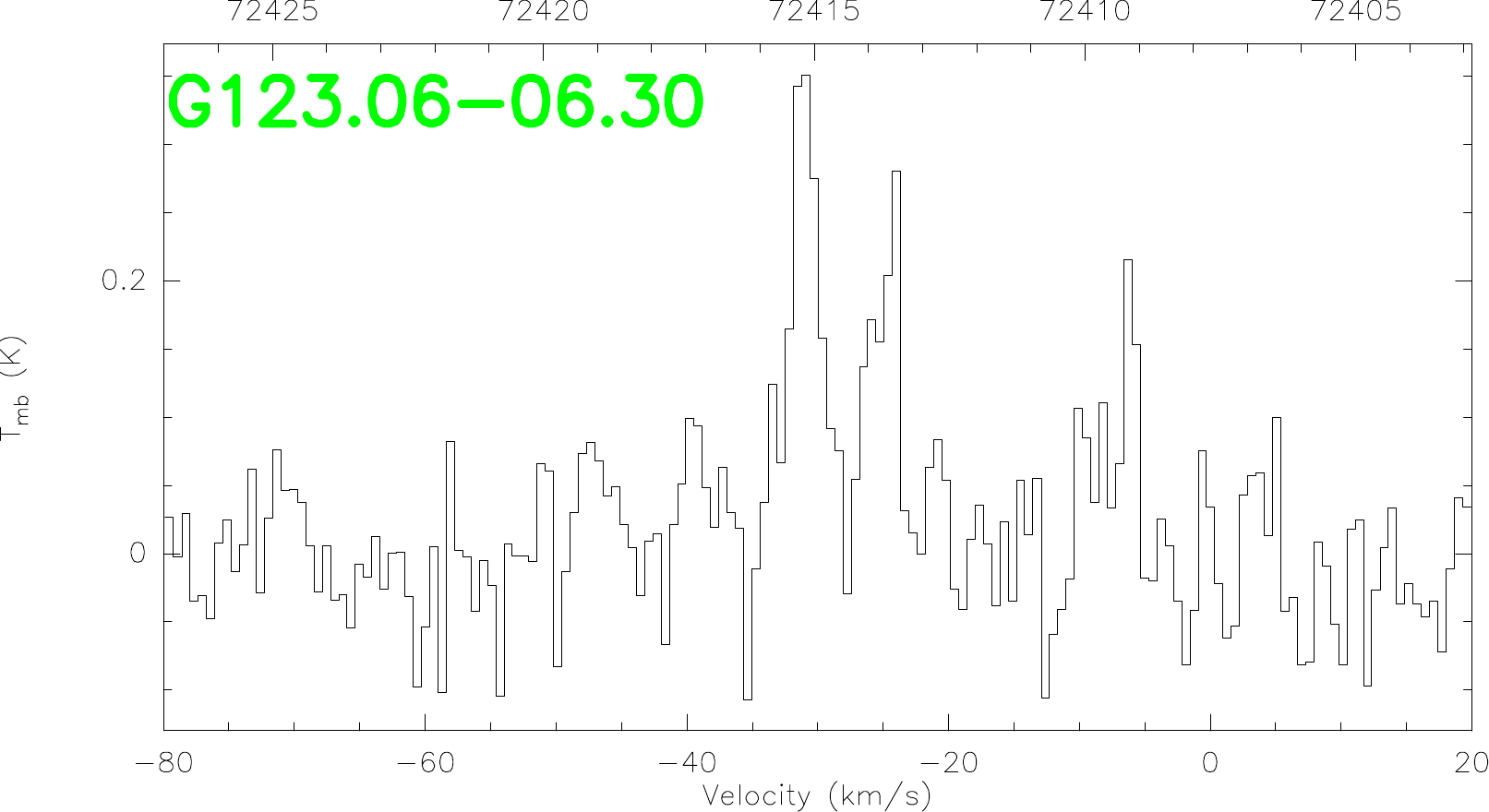}
\includegraphics[width=0.155\textwidth]{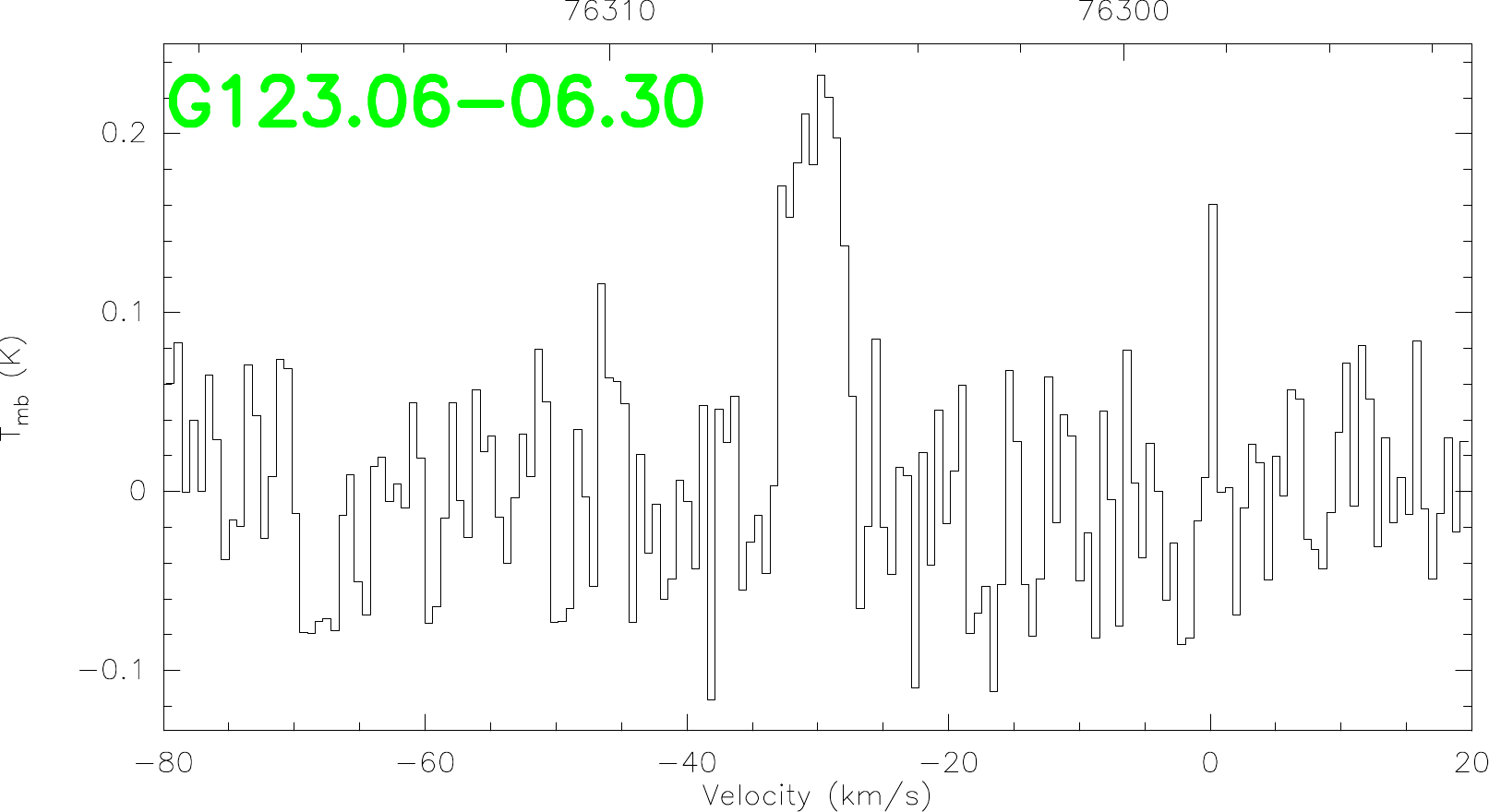}
\includegraphics[width=0.155\textwidth]{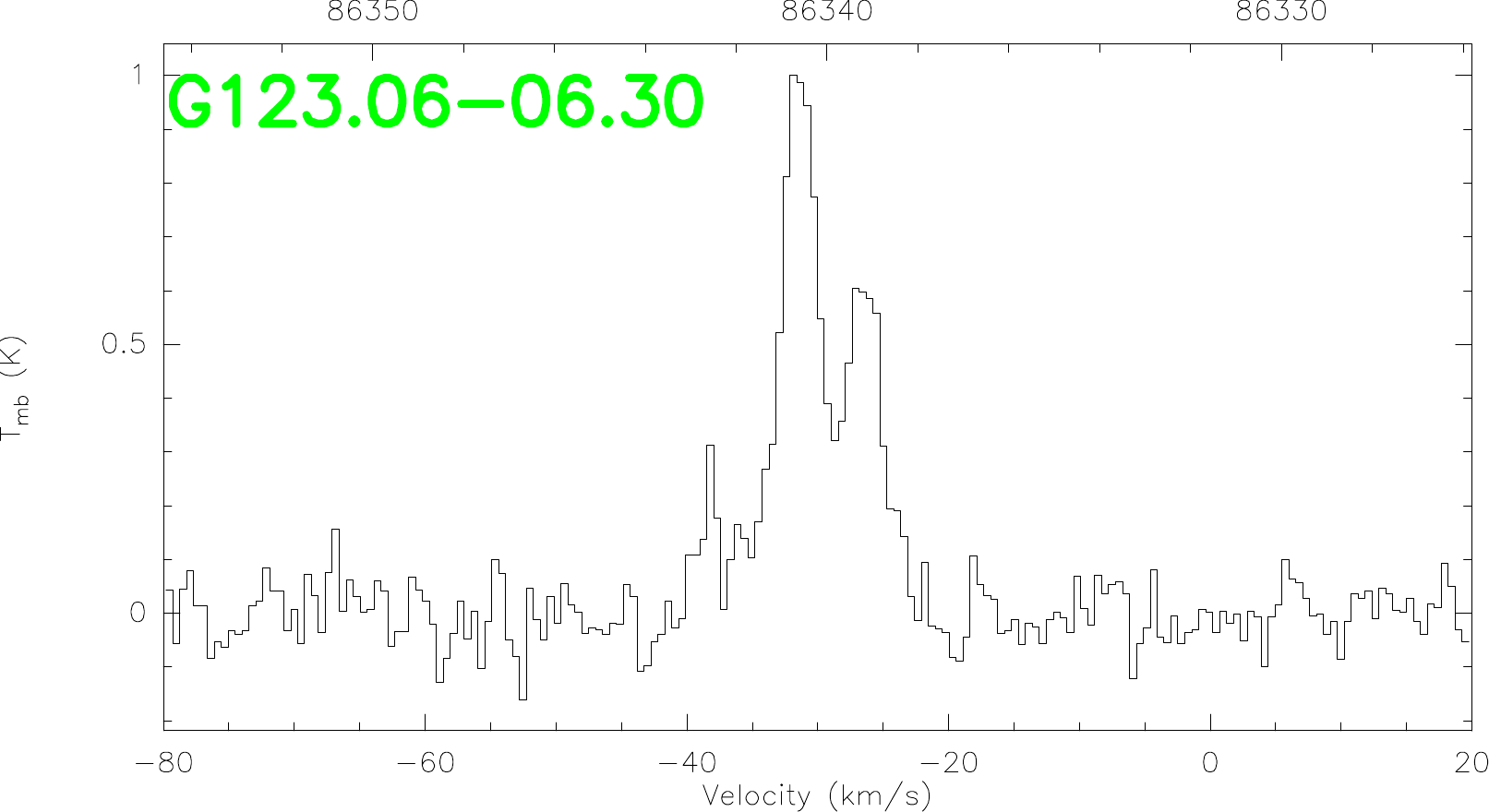}
\includegraphics[width=0.155\textwidth]{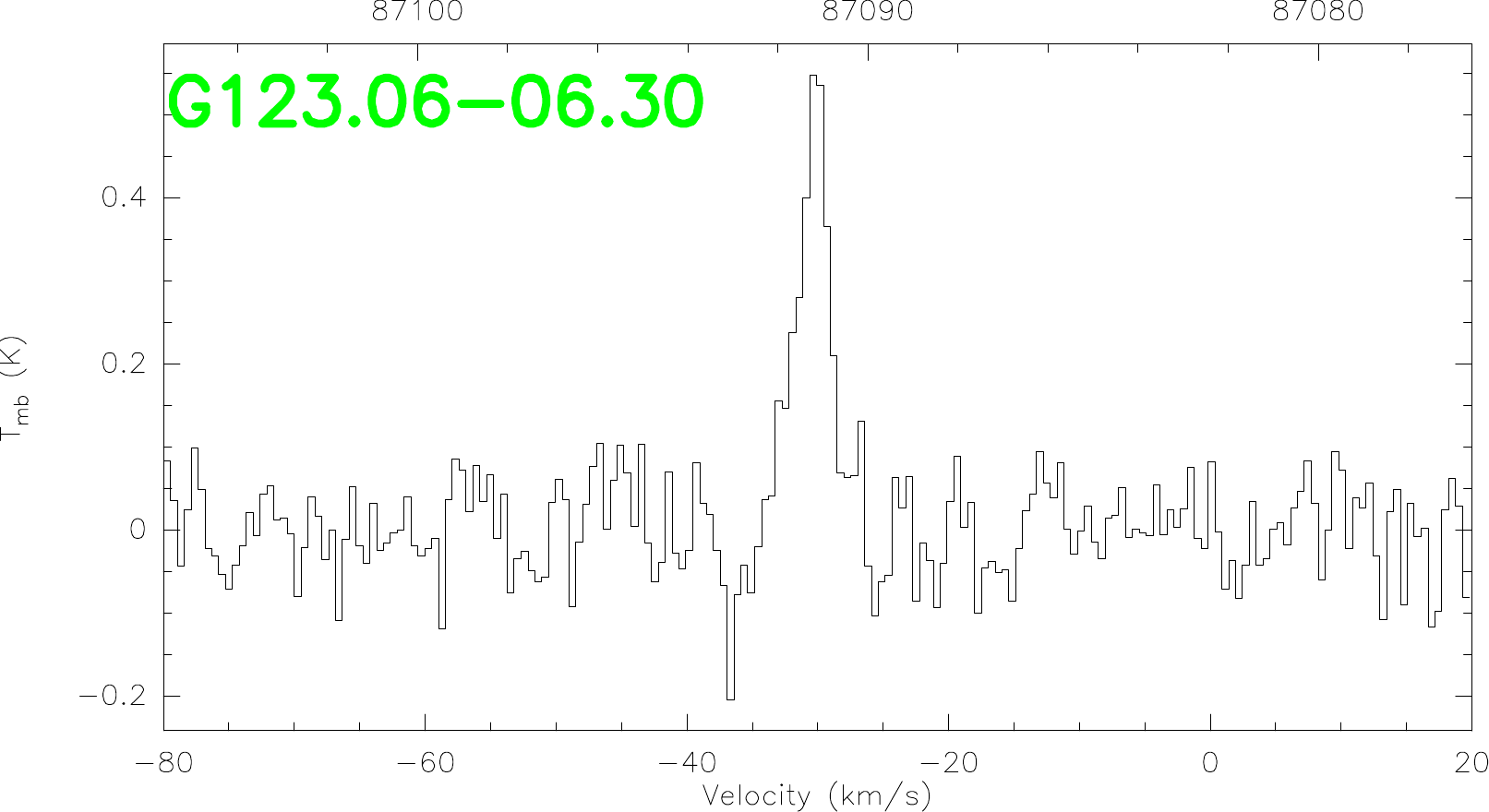}
\includegraphics[width=0.155\textwidth]{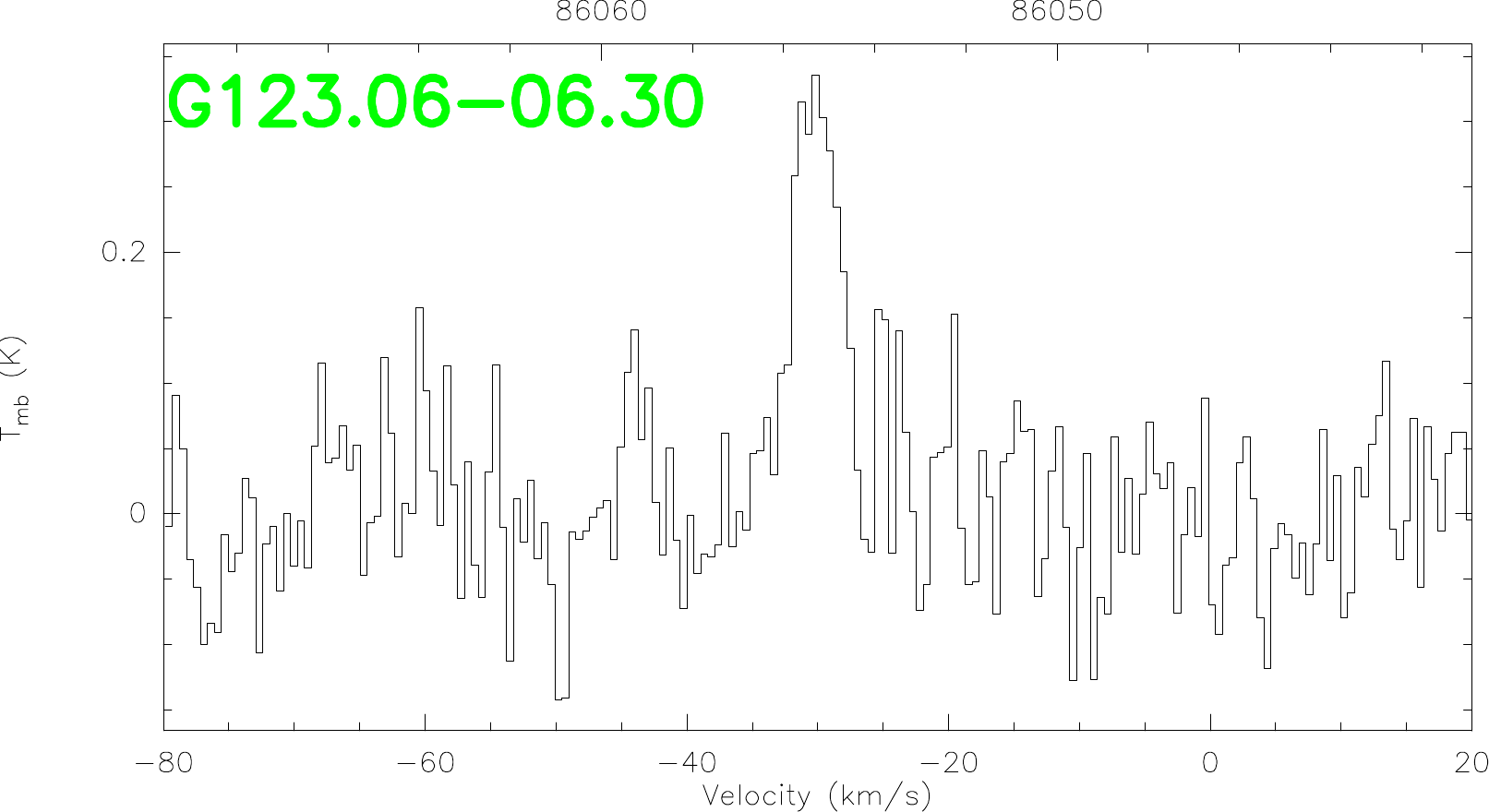}
\includegraphics[width=0.155\textwidth]{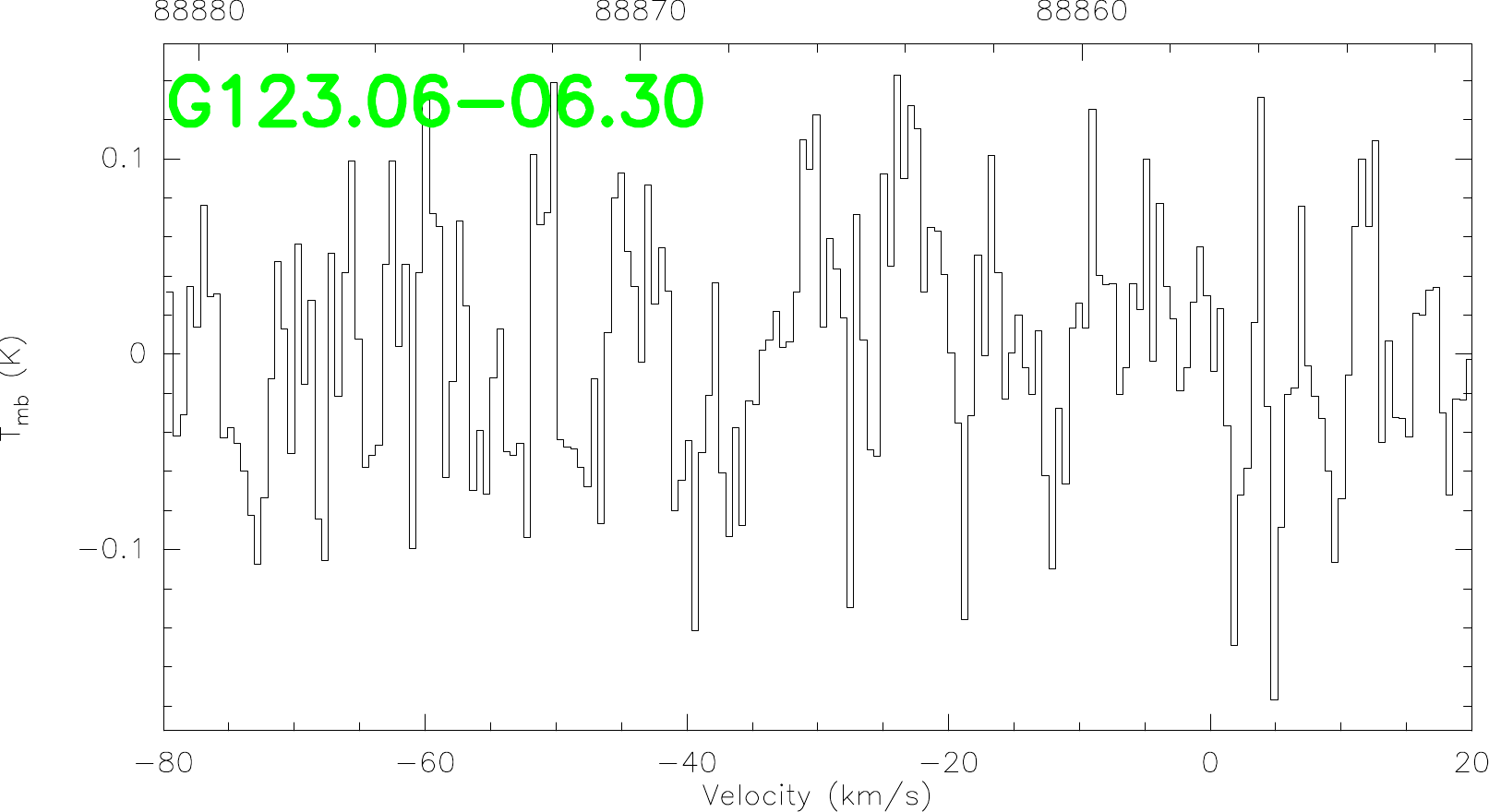}\\

\includegraphics[width=0.155\textwidth]{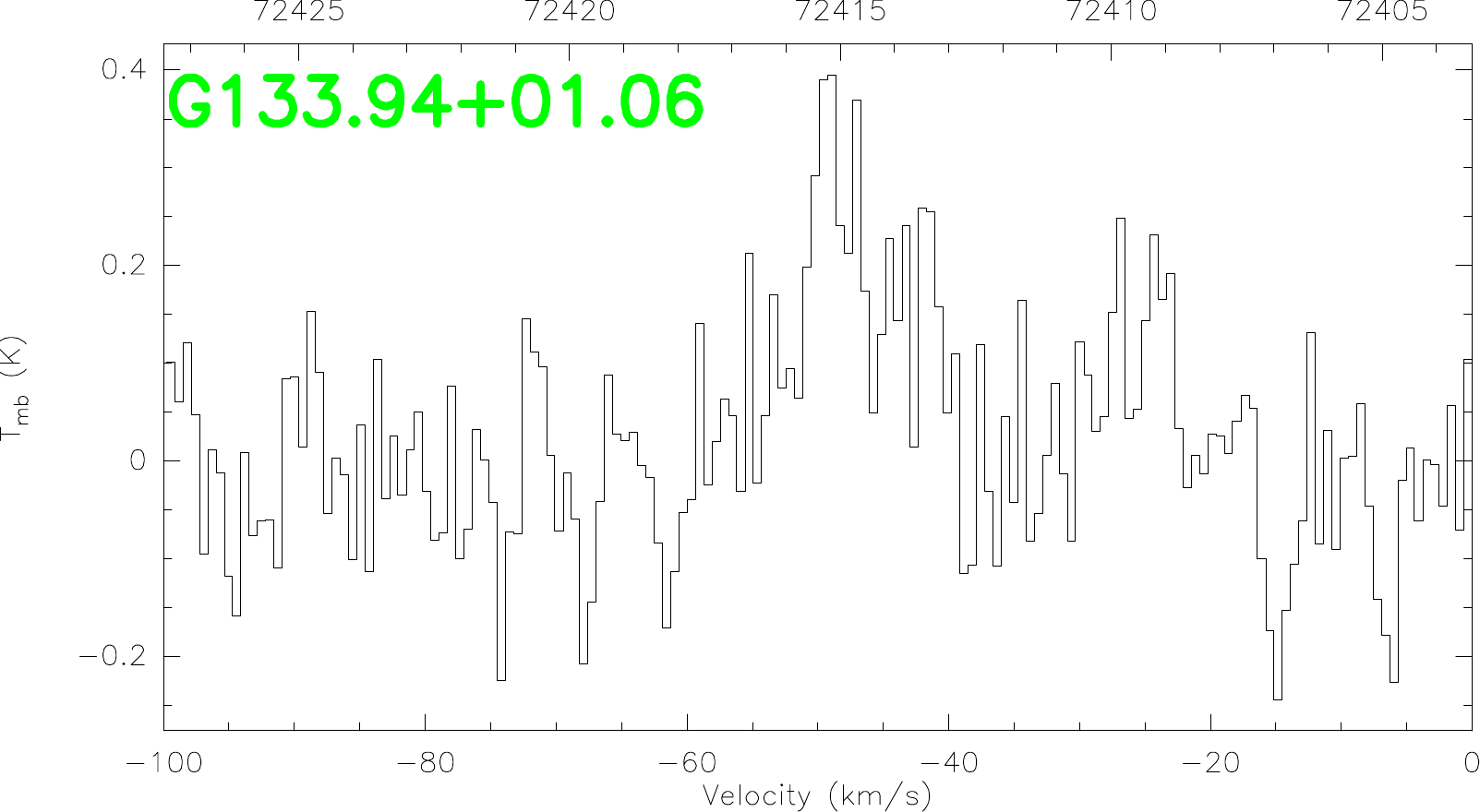}
\includegraphics[width=0.155\textwidth]{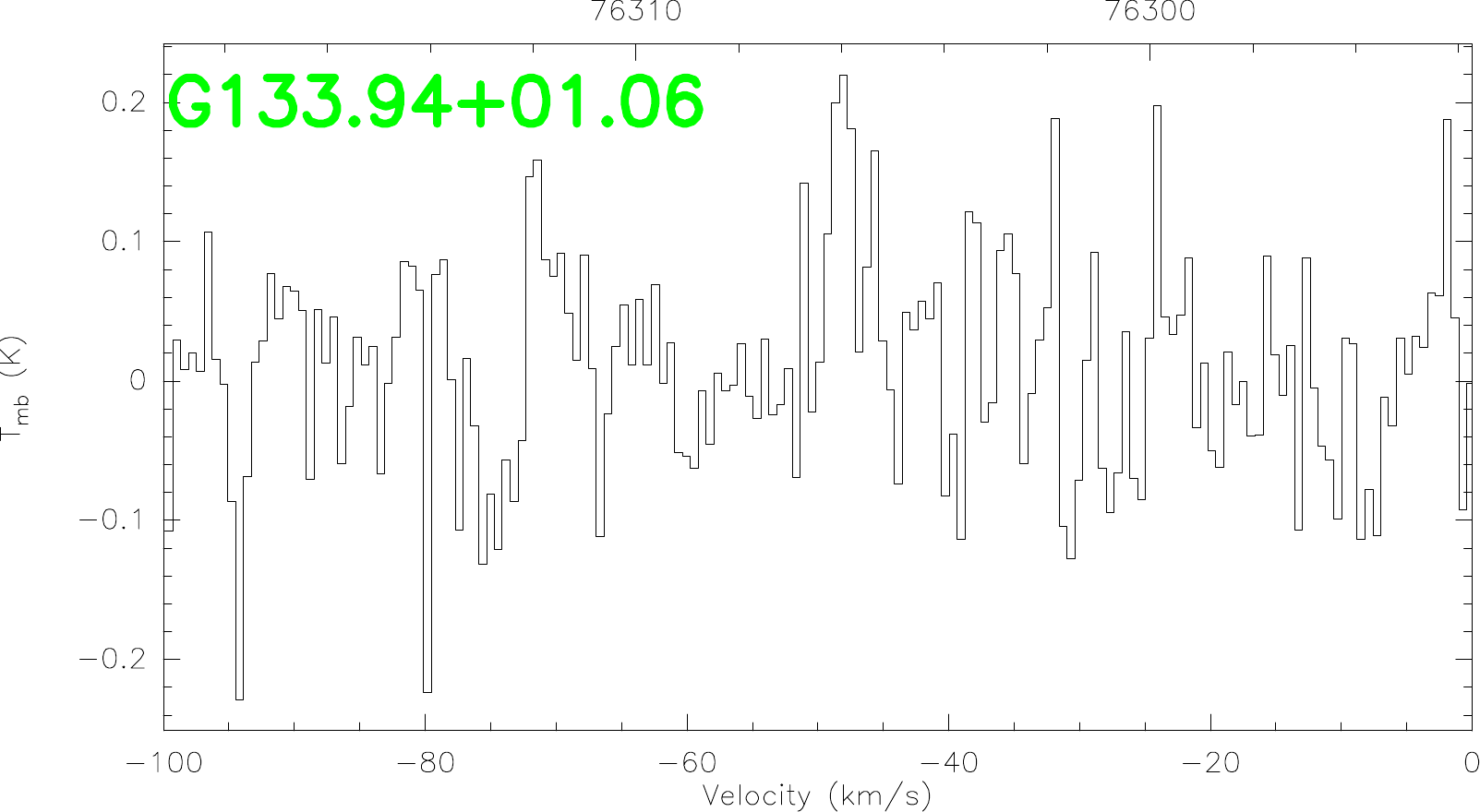}
\includegraphics[width=0.155\textwidth]{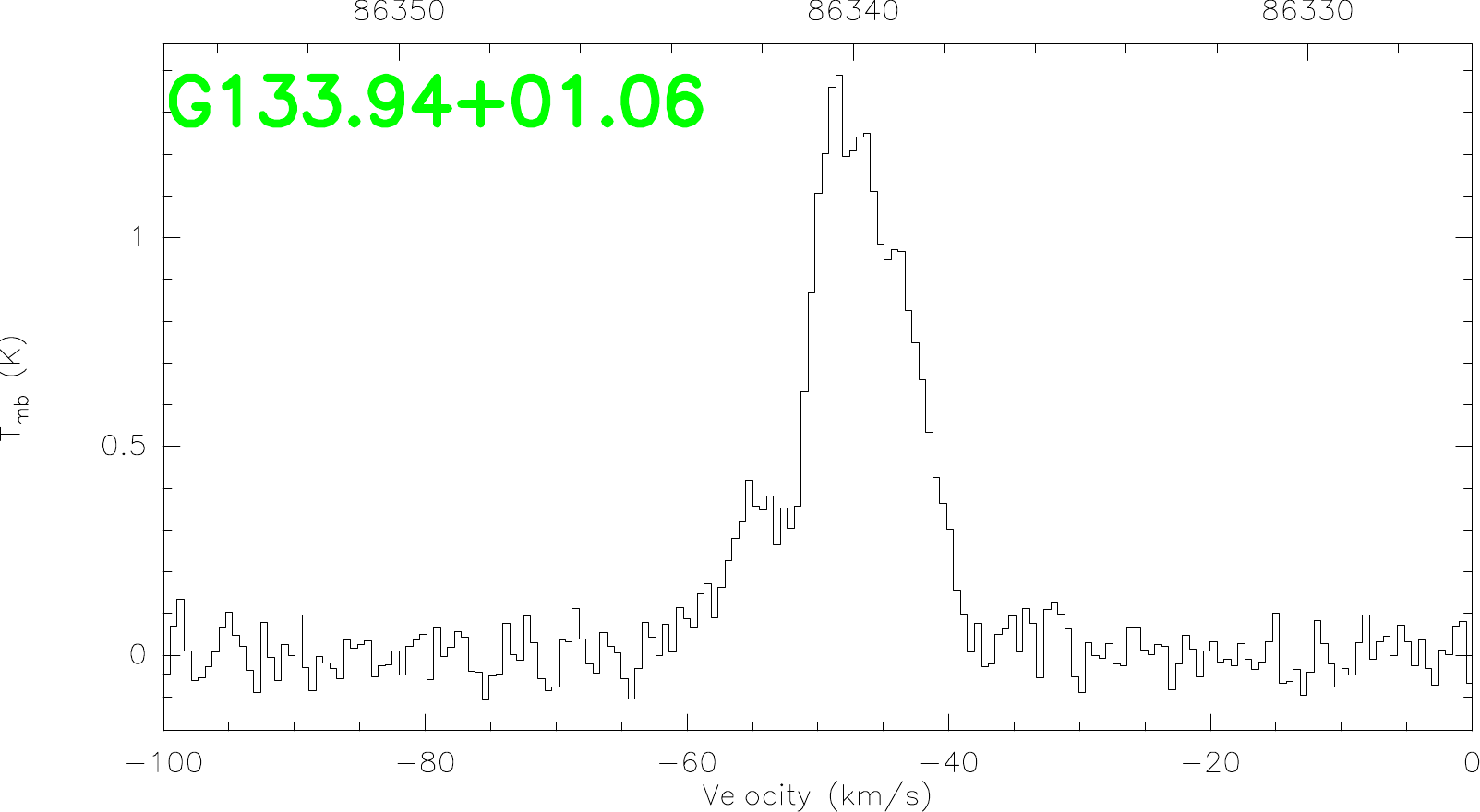}
\includegraphics[width=0.155\textwidth]{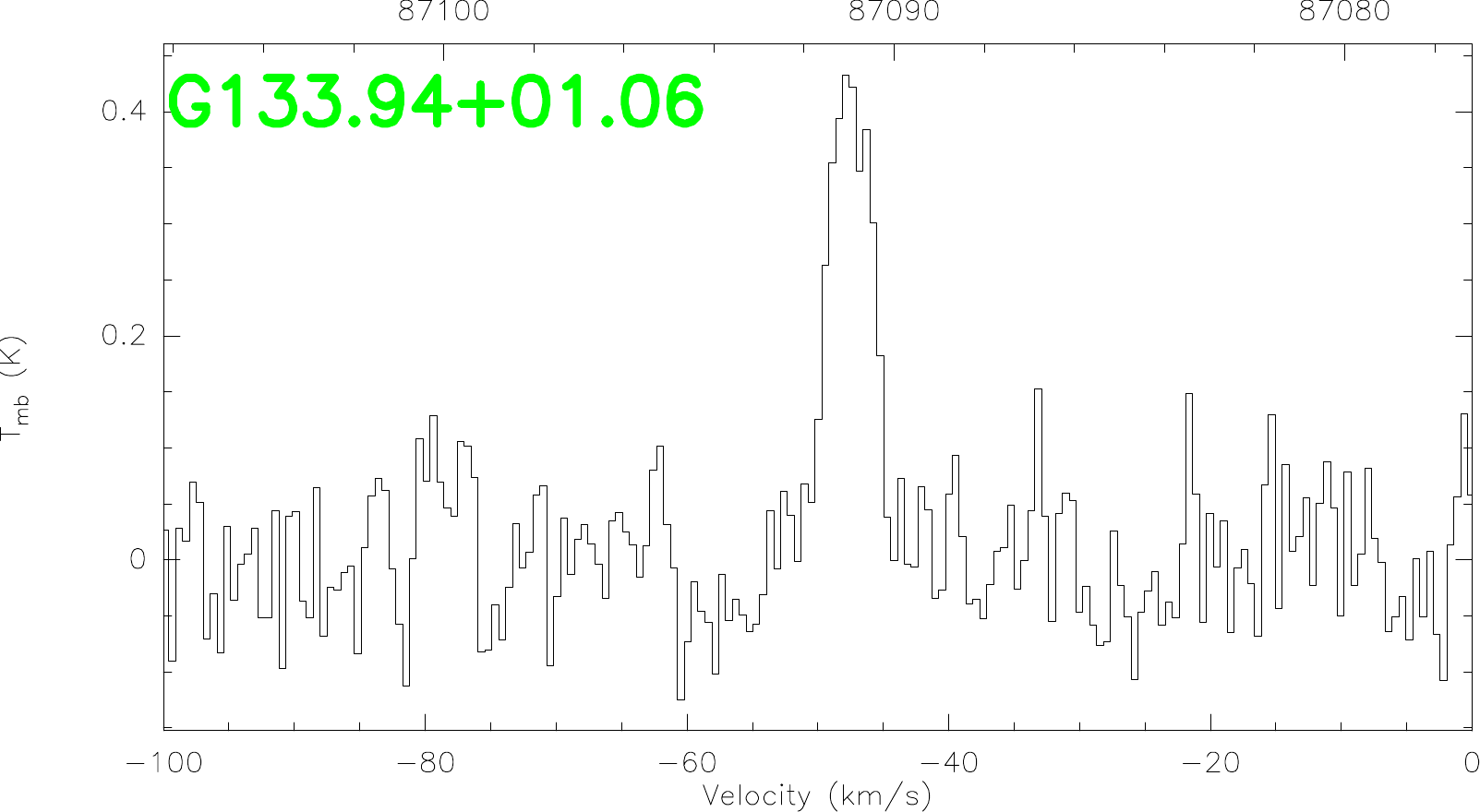}
\includegraphics[width=0.155\textwidth]{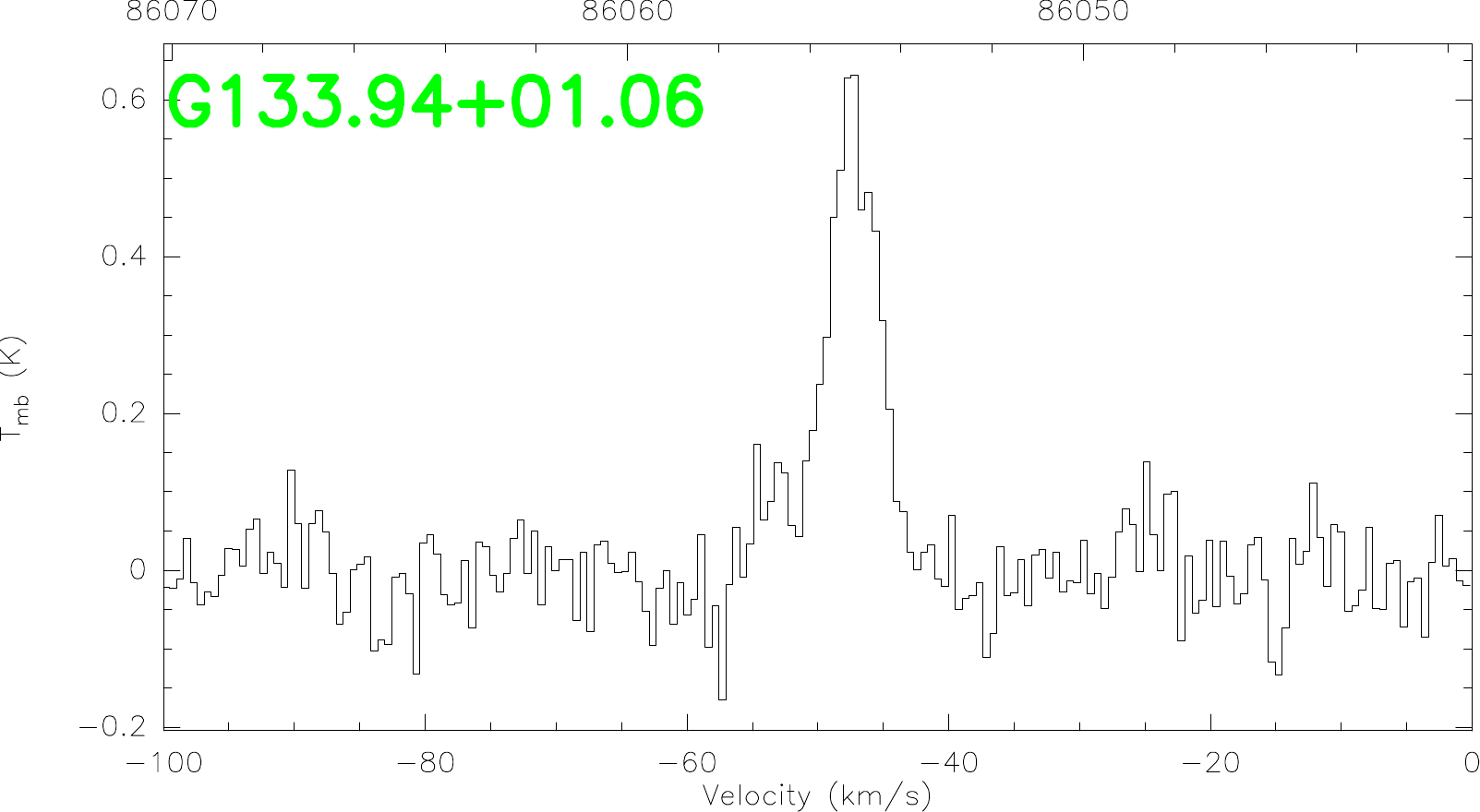}
\includegraphics[width=0.155\textwidth]{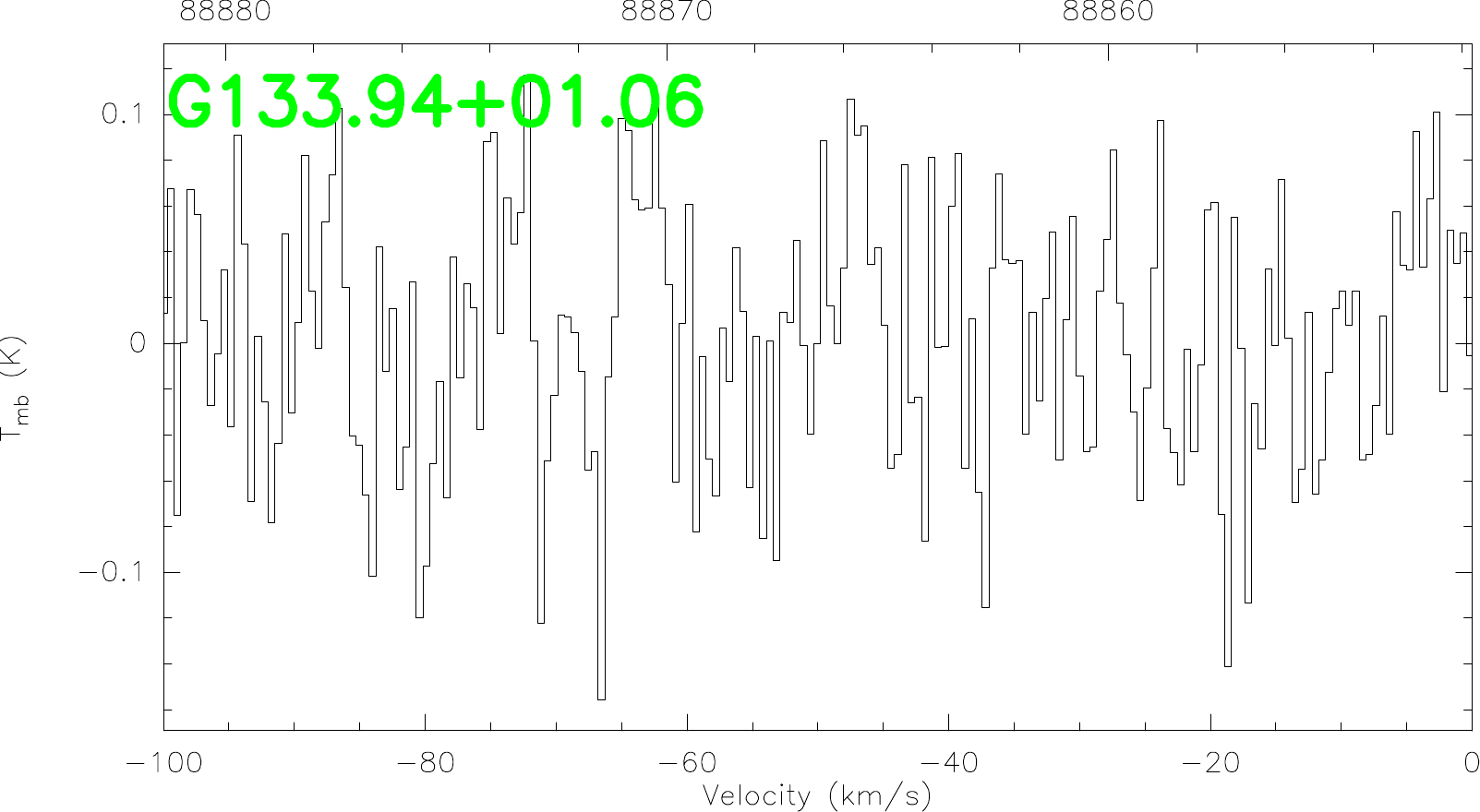}\\

\includegraphics[width=0.155\textwidth]{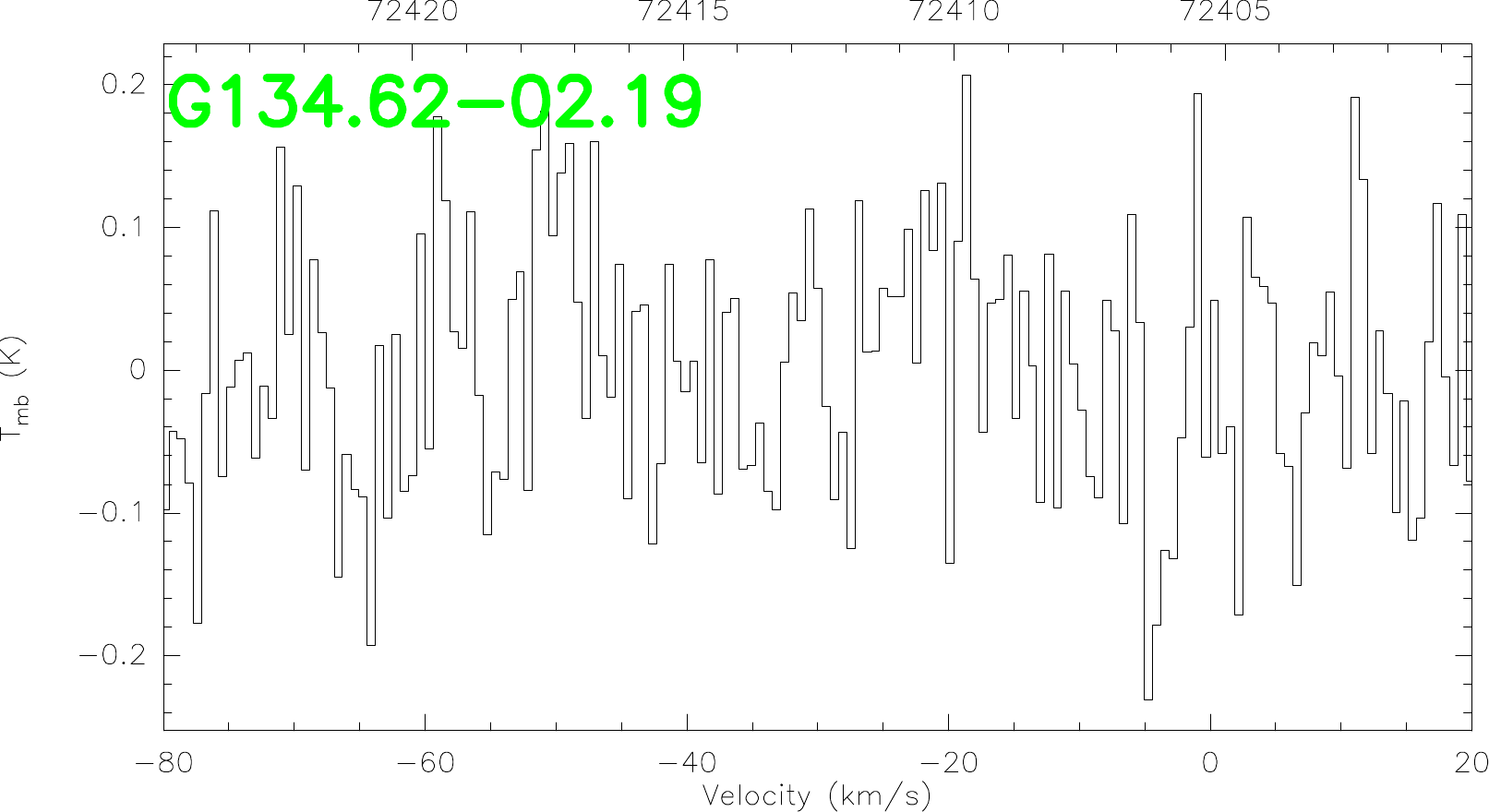}
\includegraphics[width=0.155\textwidth]{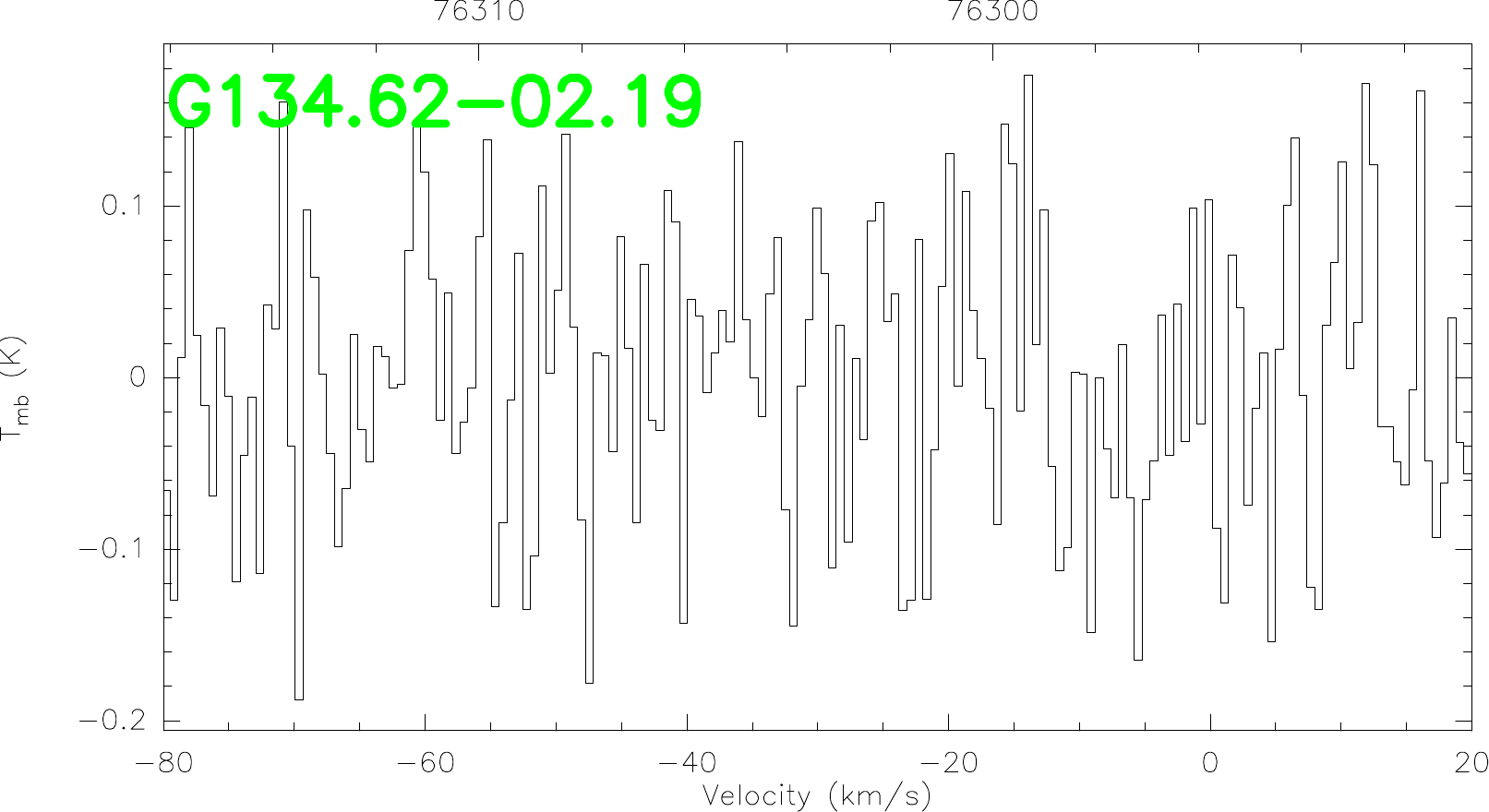}
\includegraphics[width=0.155\textwidth]{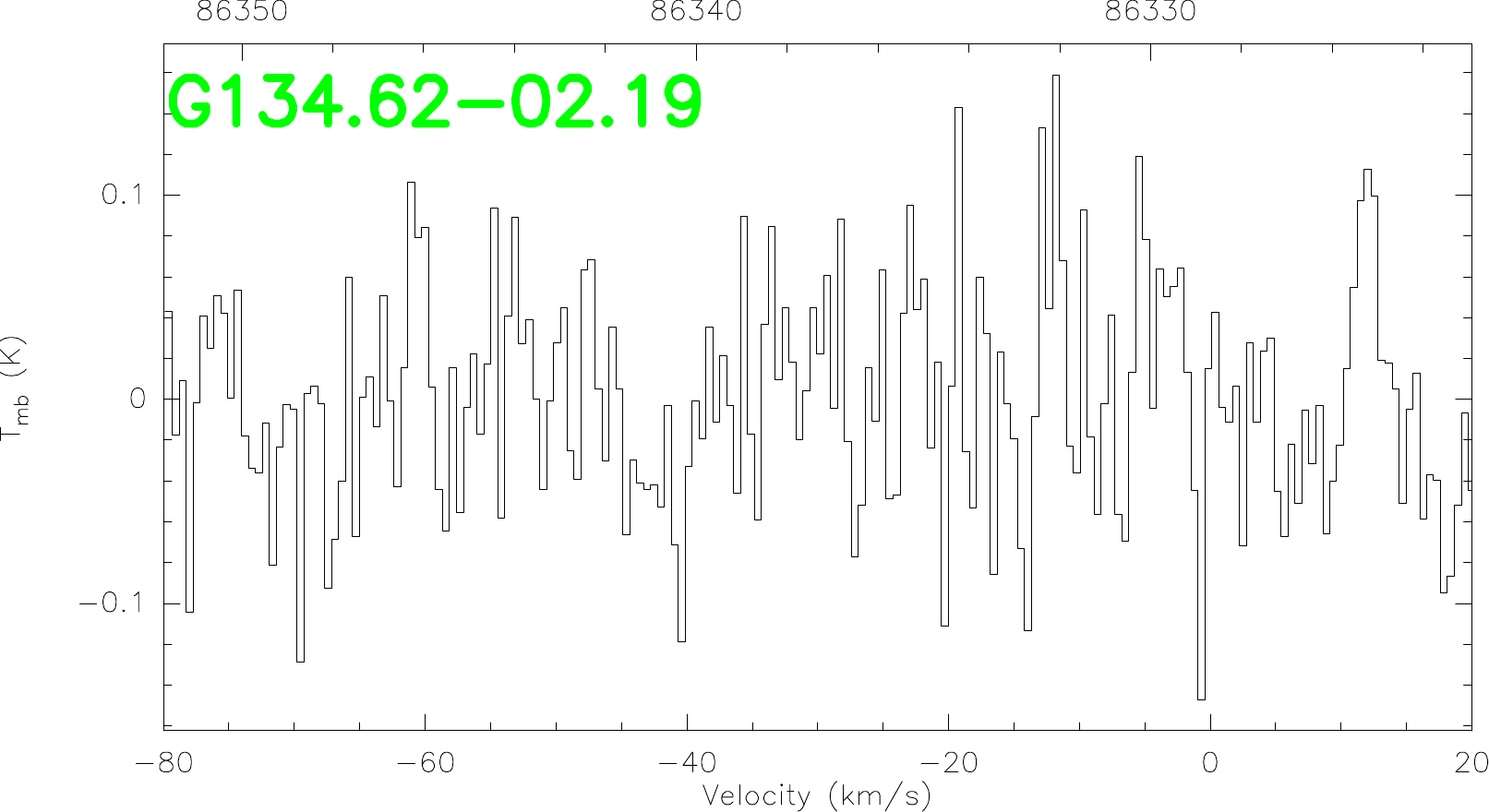}
\includegraphics[width=0.155\textwidth]{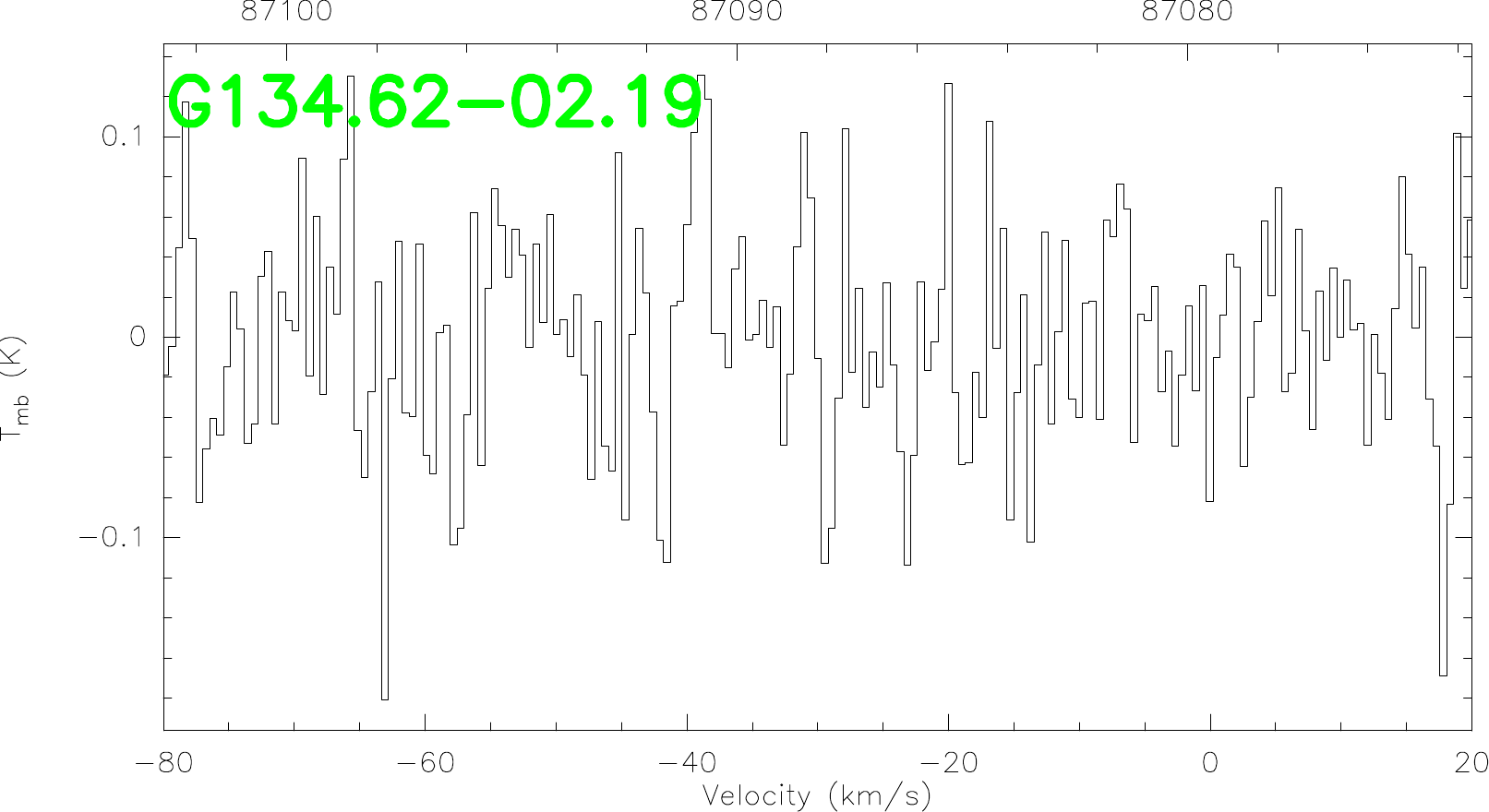}
\includegraphics[width=0.155\textwidth]{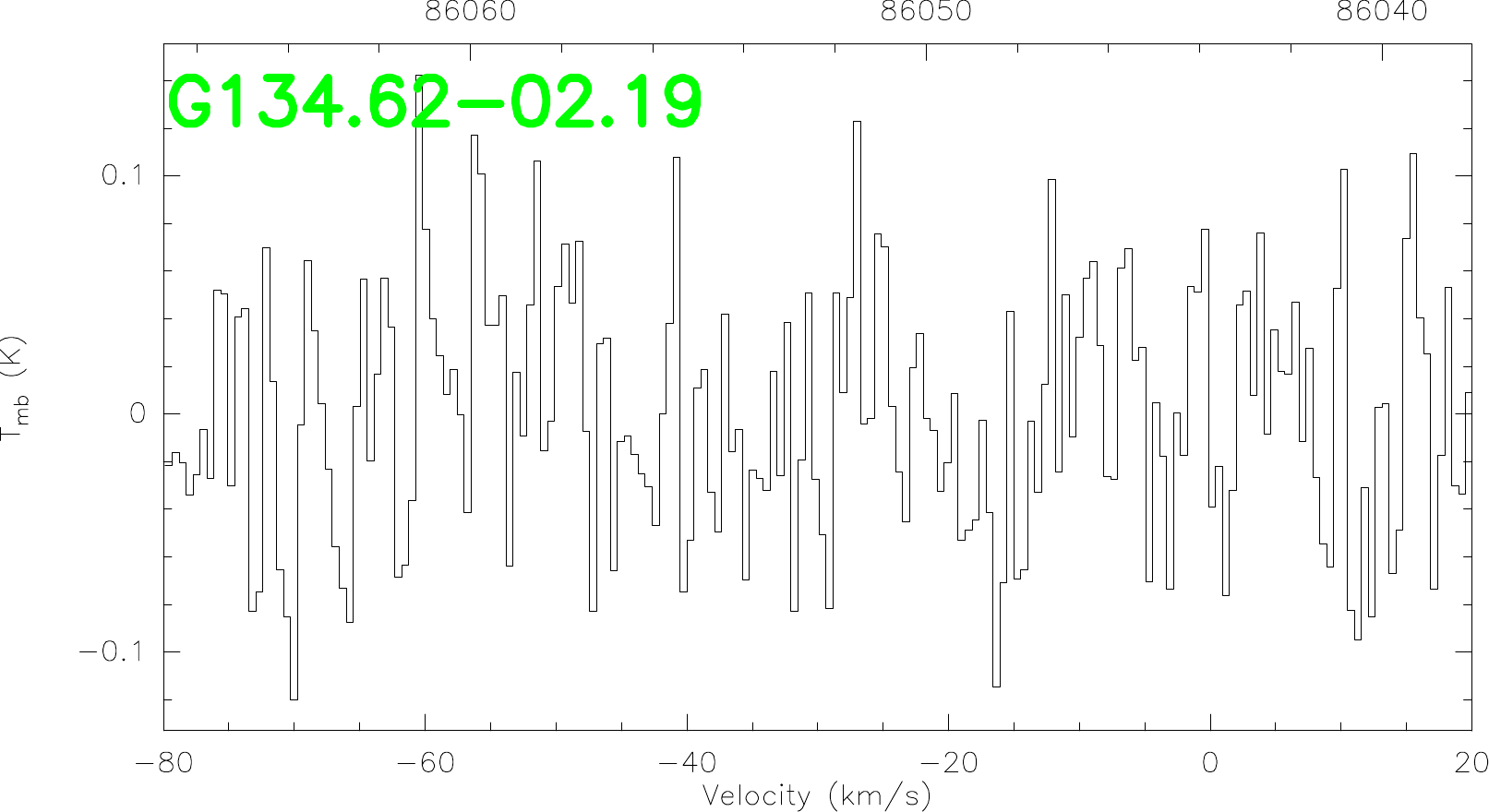}
\includegraphics[width=0.155\textwidth]{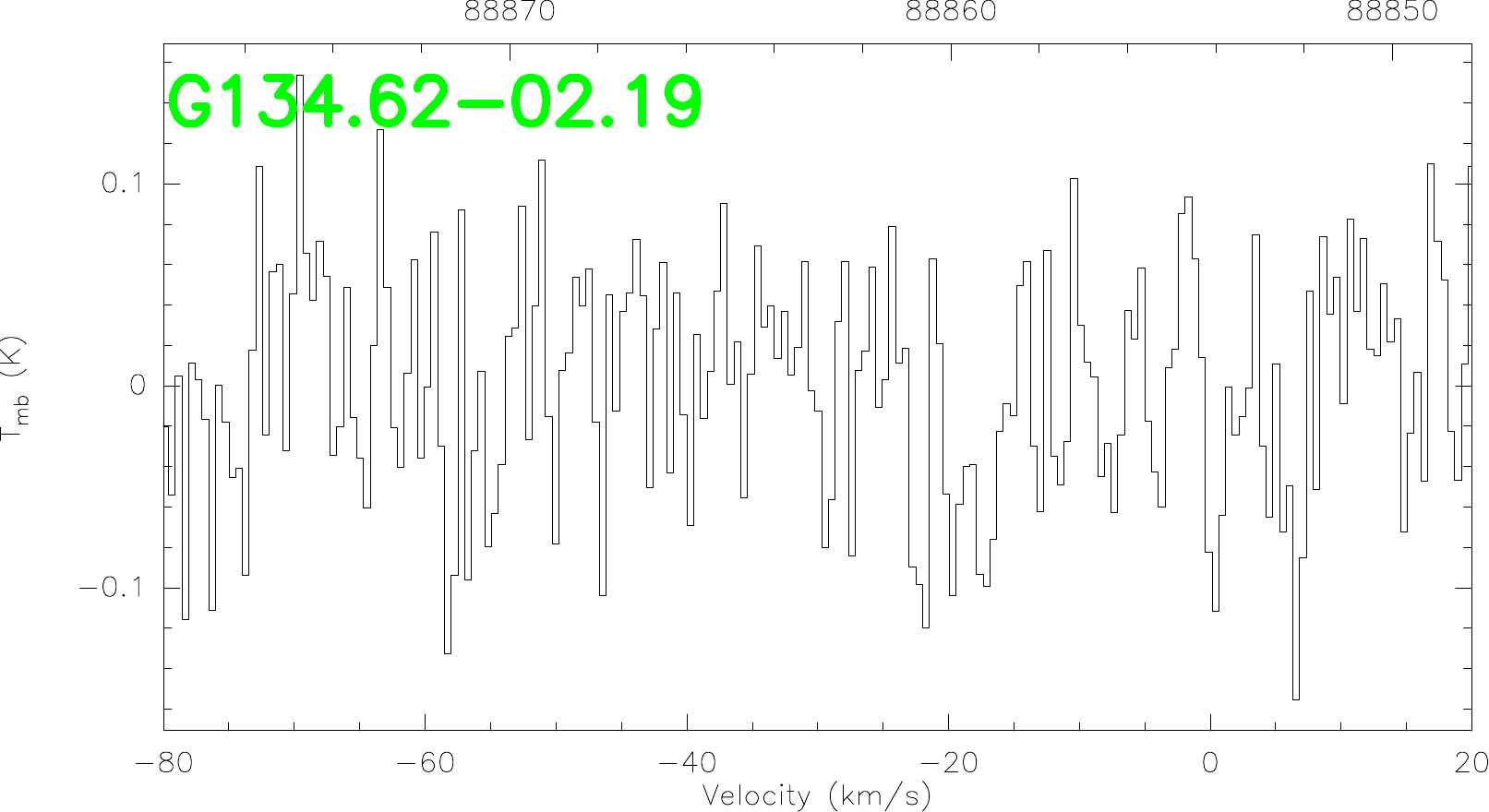}\\

\includegraphics[width=0.155\textwidth]{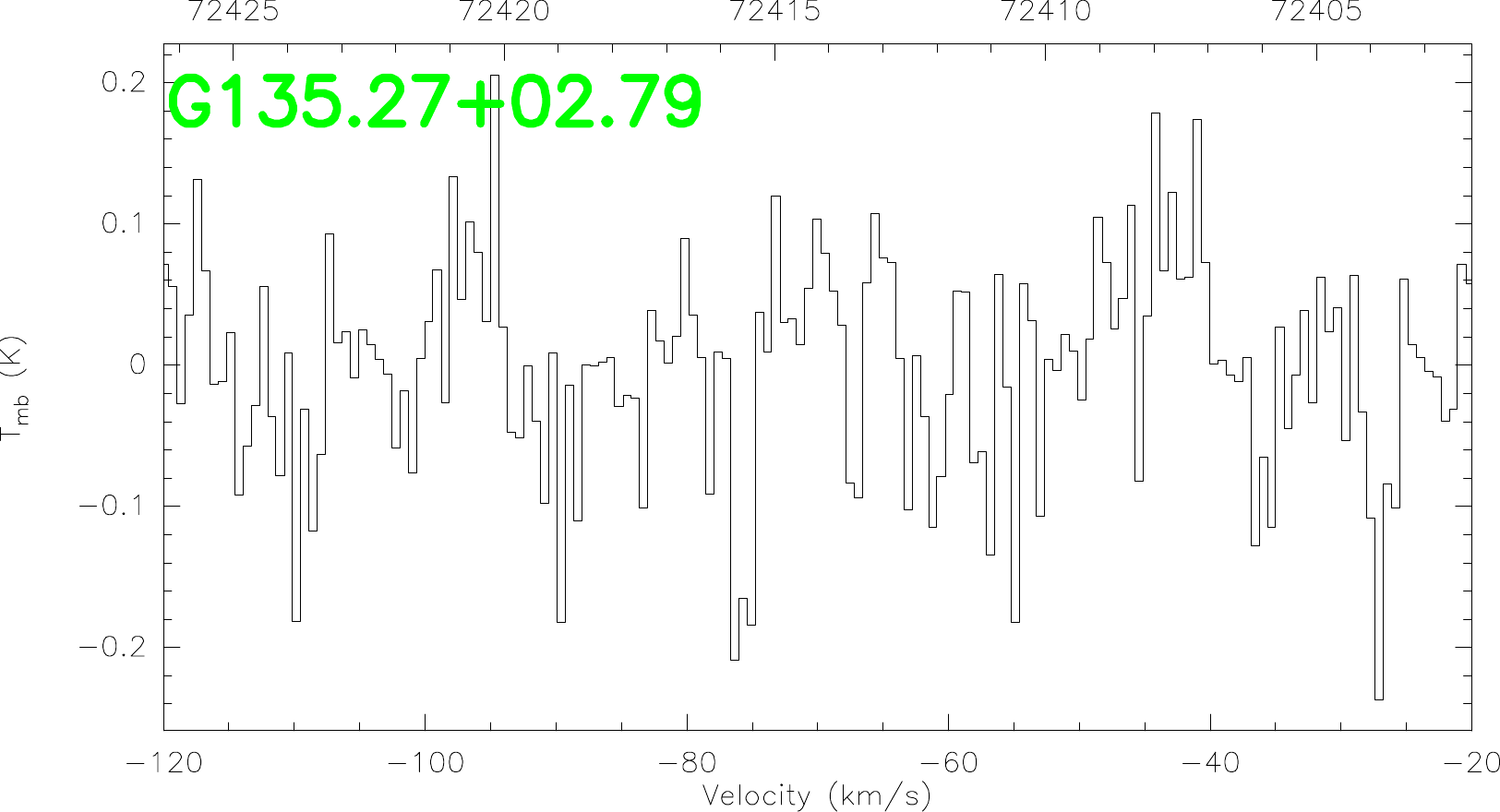}
\includegraphics[width=0.155\textwidth]{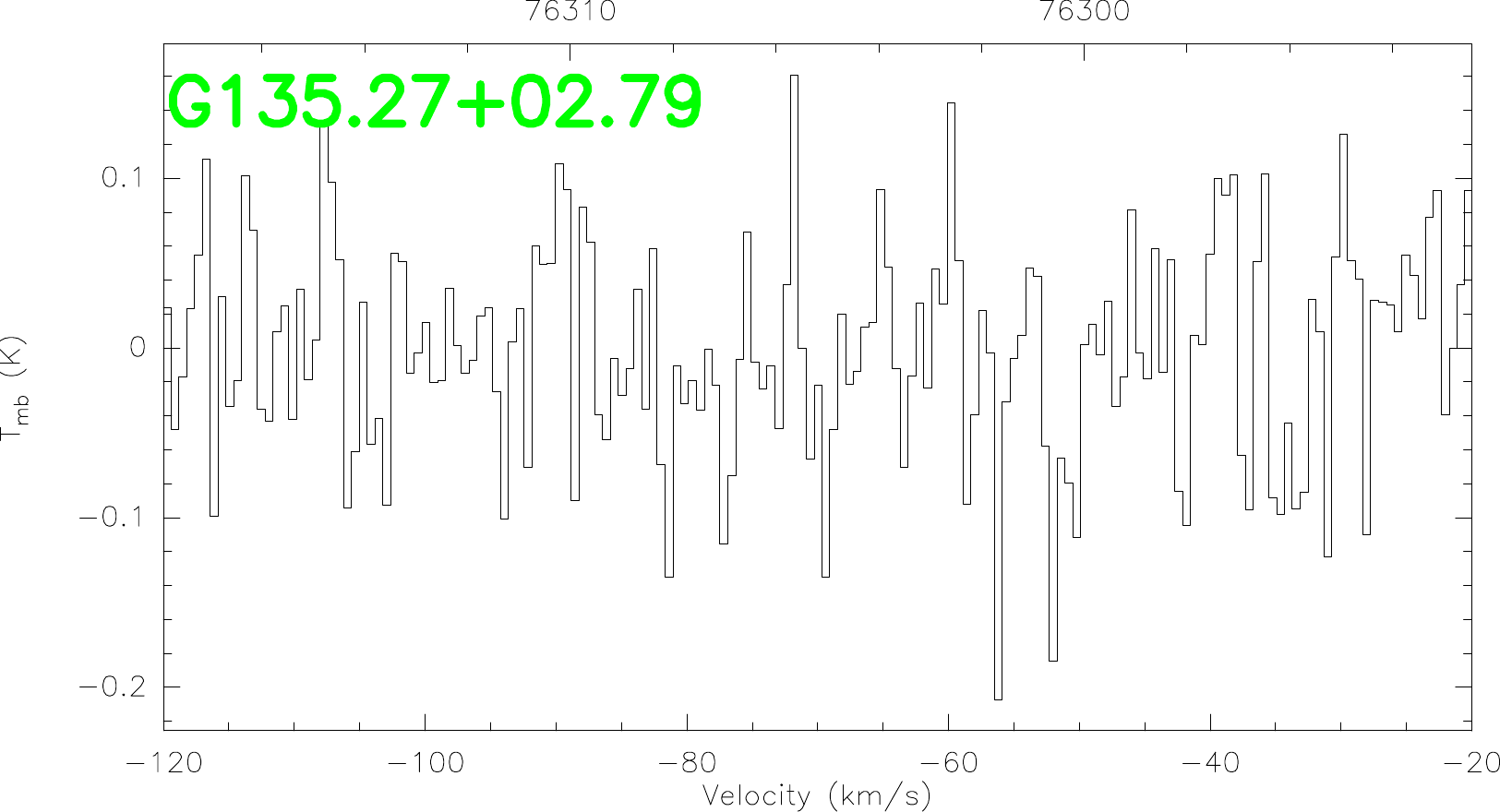}
\includegraphics[width=0.155\textwidth]{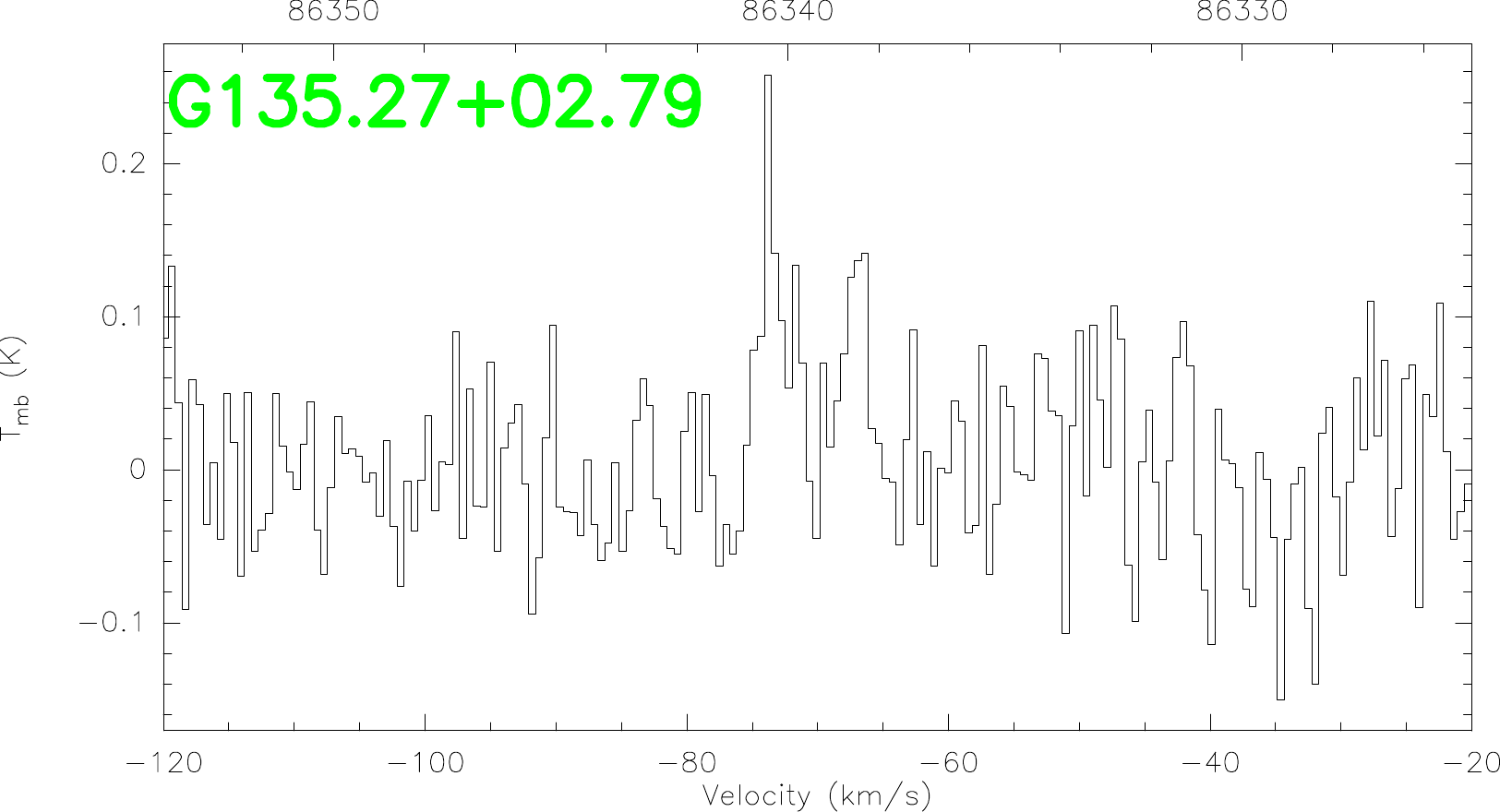}
\includegraphics[width=0.155\textwidth]{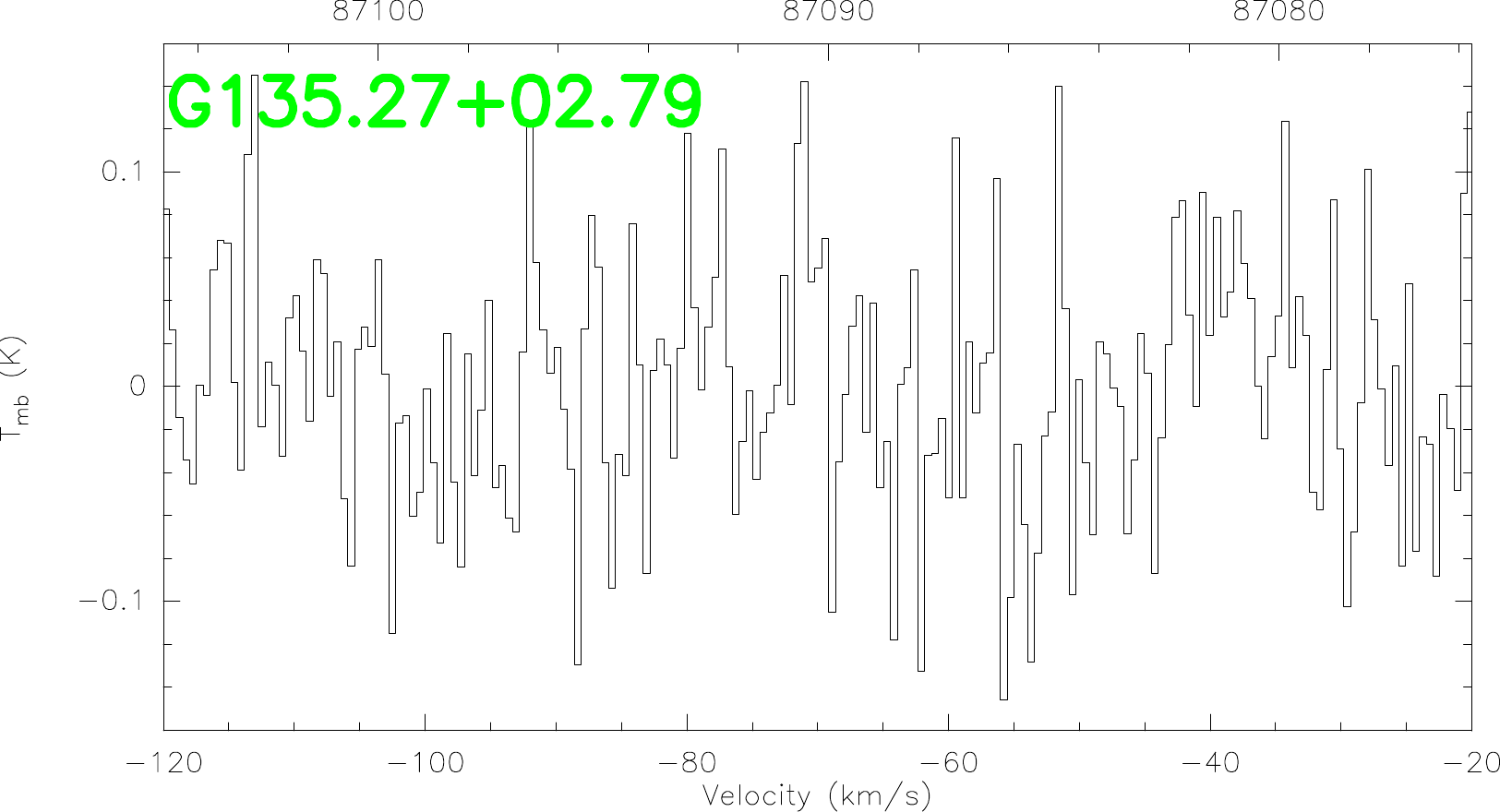}
\includegraphics[width=0.155\textwidth]{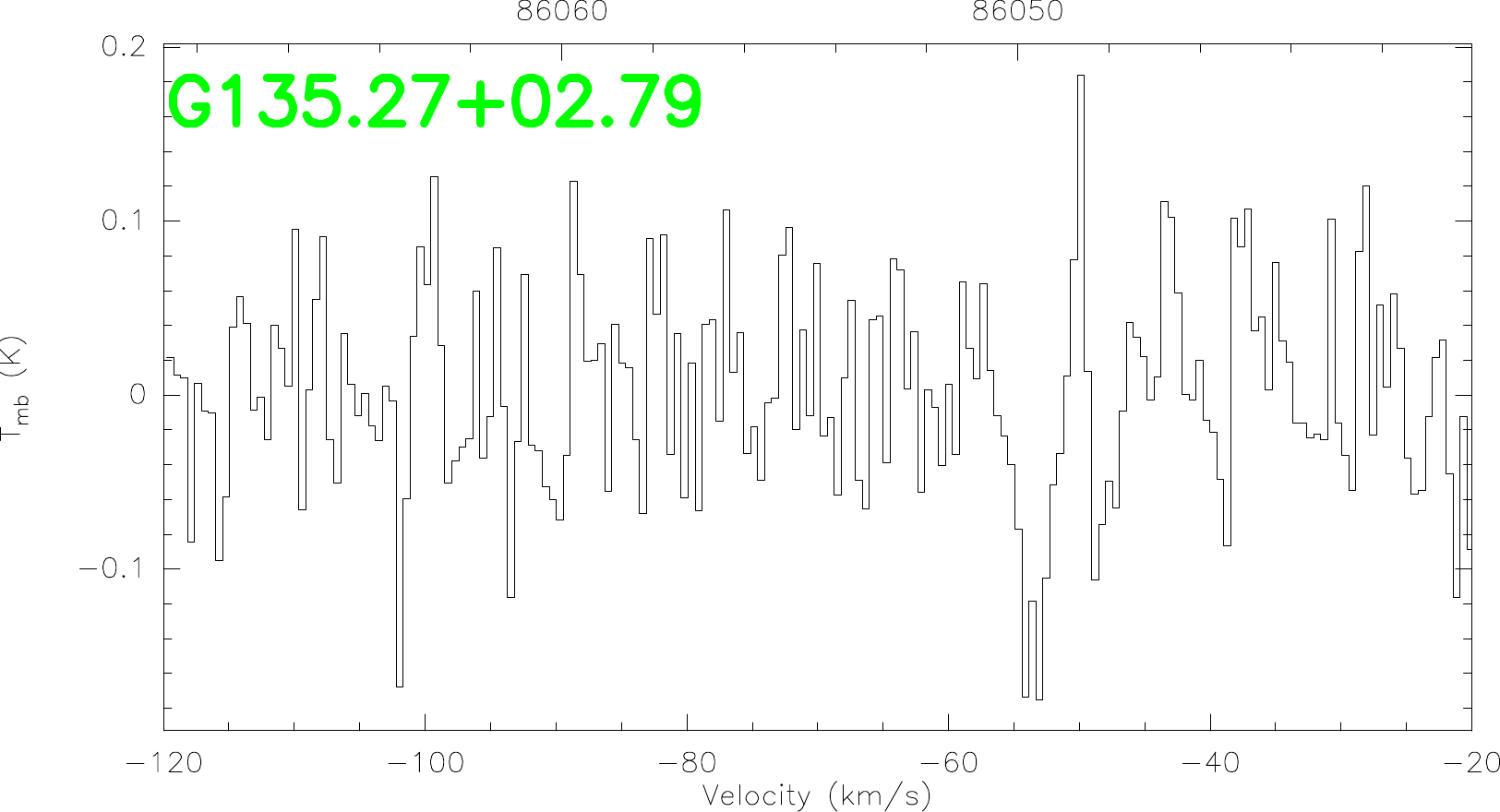}
\includegraphics[width=0.155\textwidth]{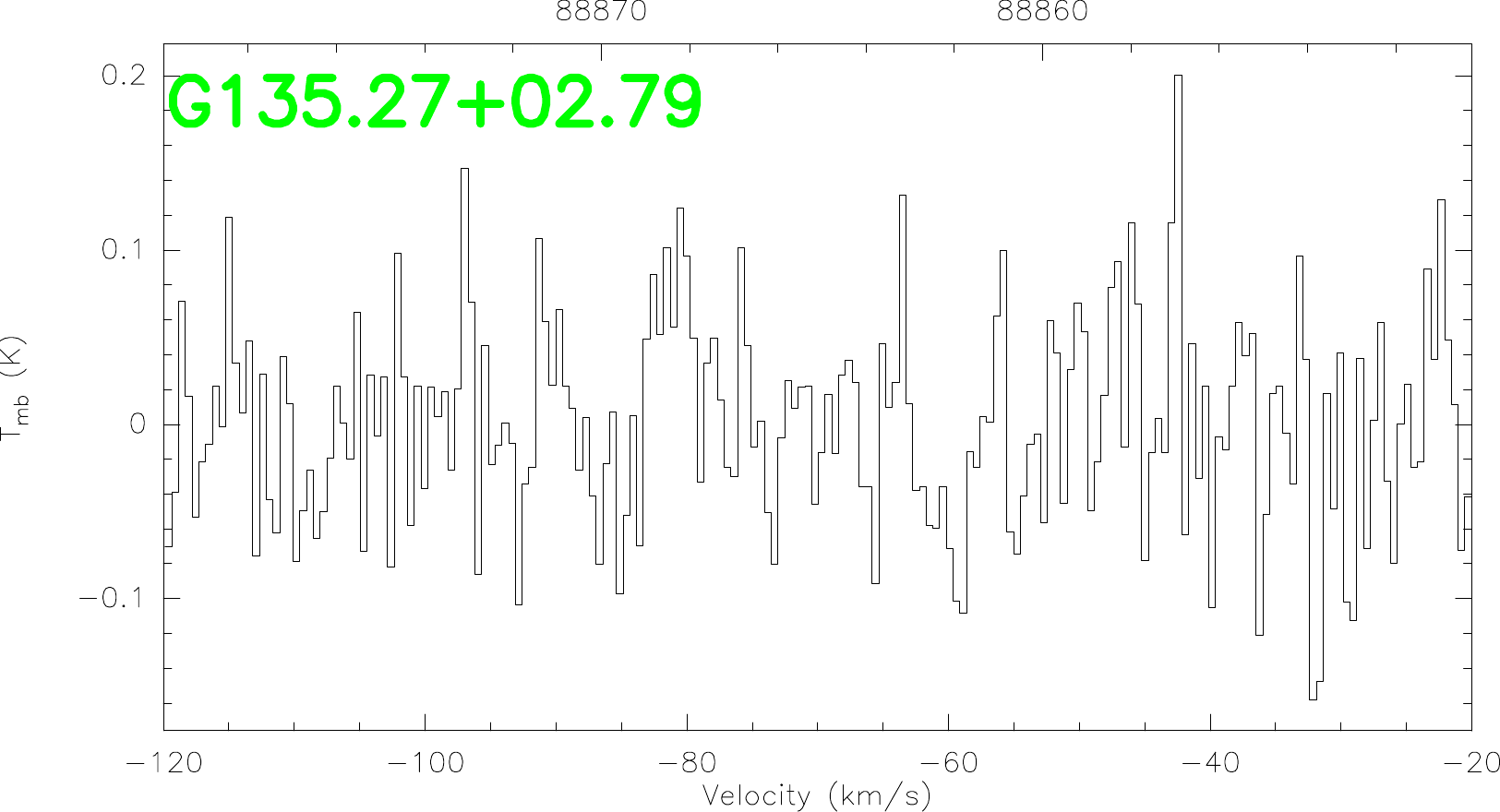}\\

\includegraphics[width=0.155\textwidth]{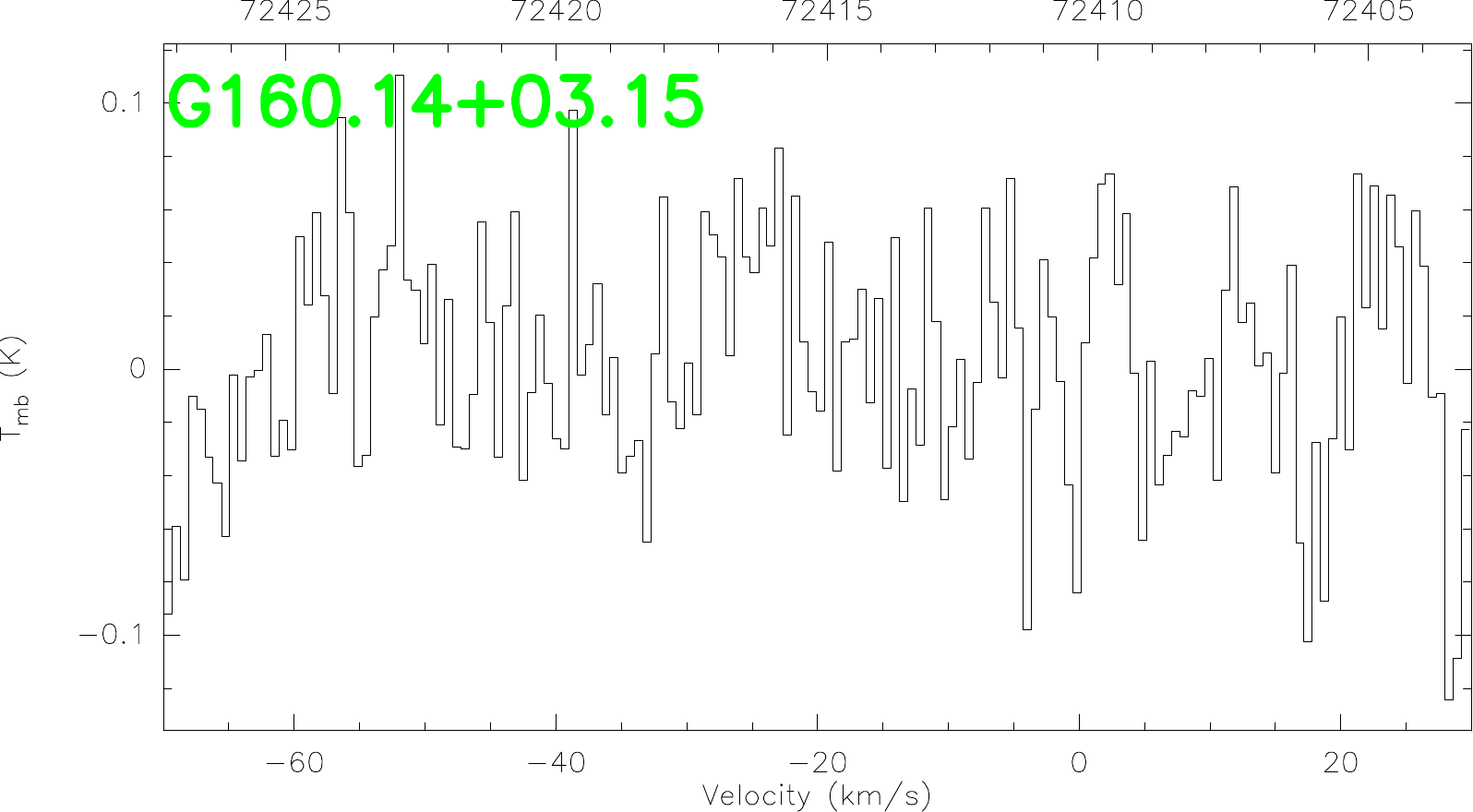}
\includegraphics[width=0.155\textwidth]{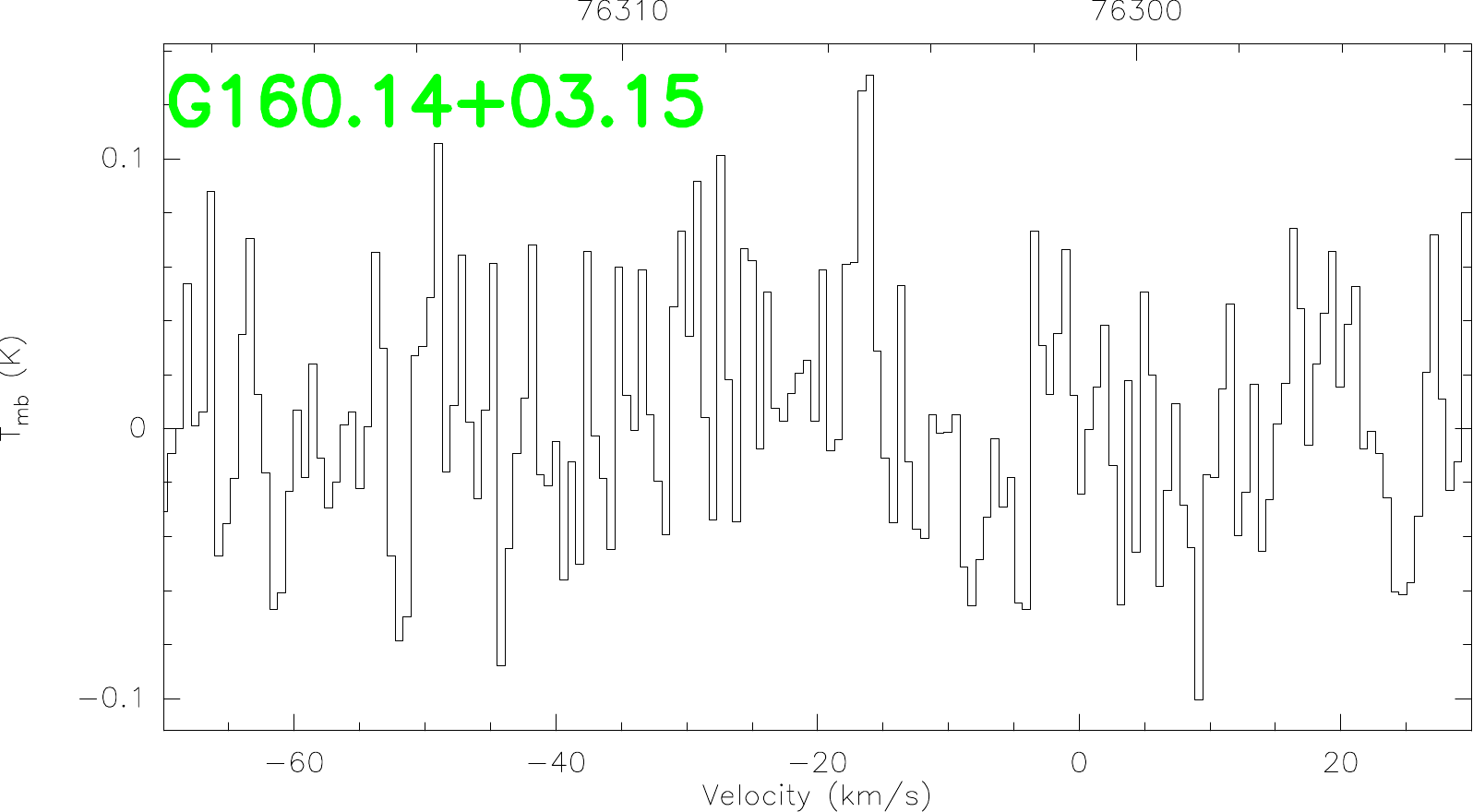}
\includegraphics[width=0.155\textwidth]{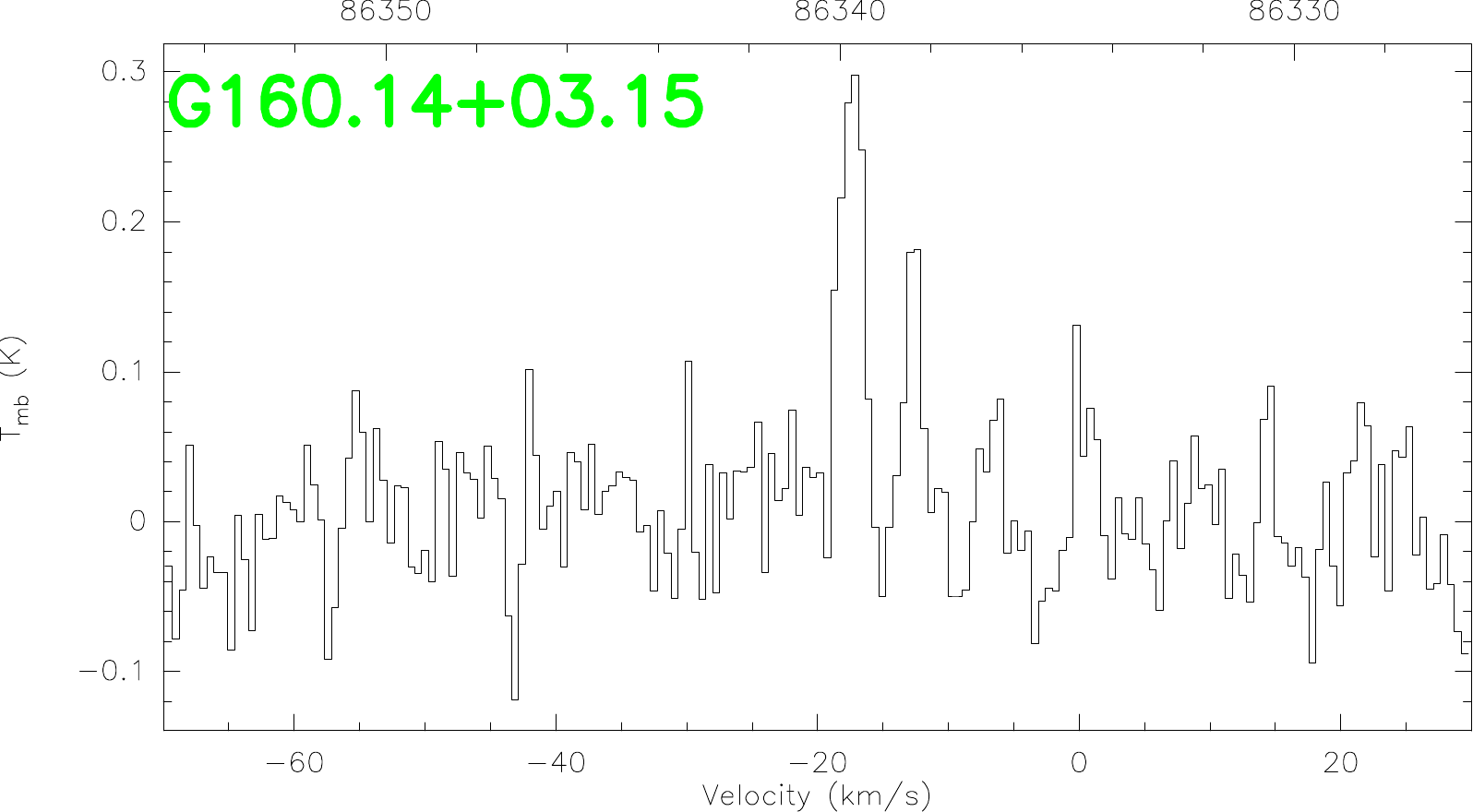}
\includegraphics[width=0.155\textwidth]{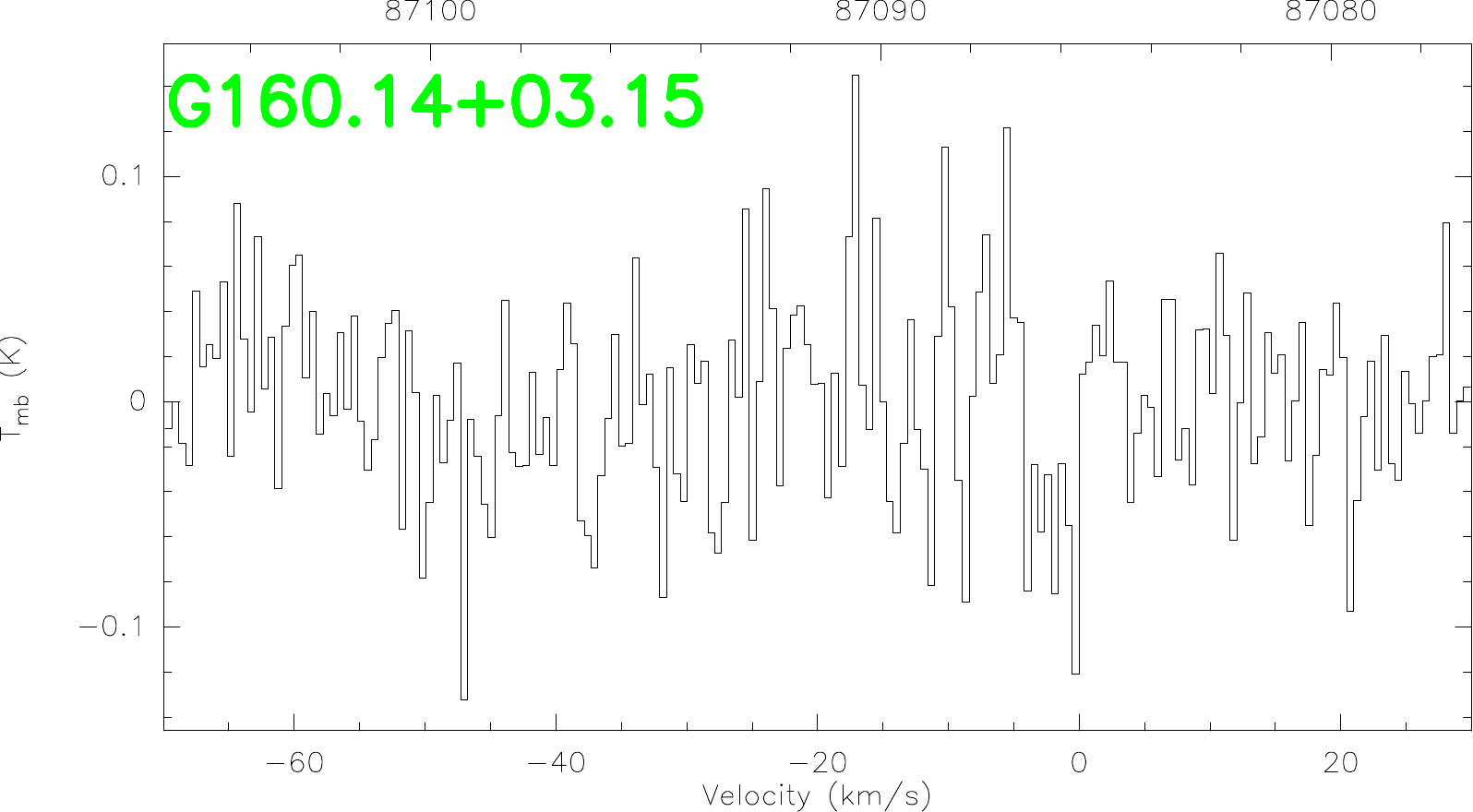}
\includegraphics[width=0.155\textwidth]{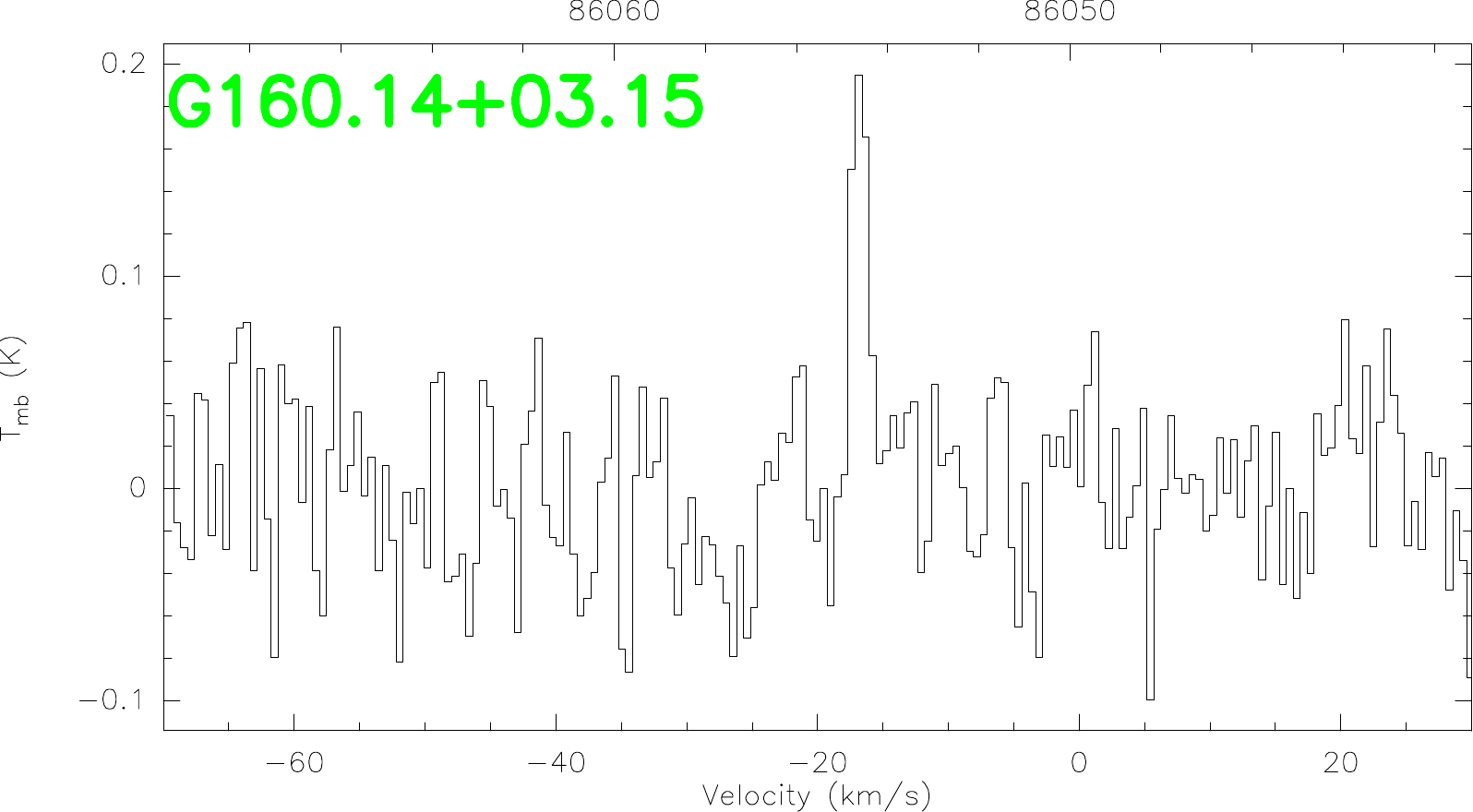}
\includegraphics[width=0.155\textwidth]{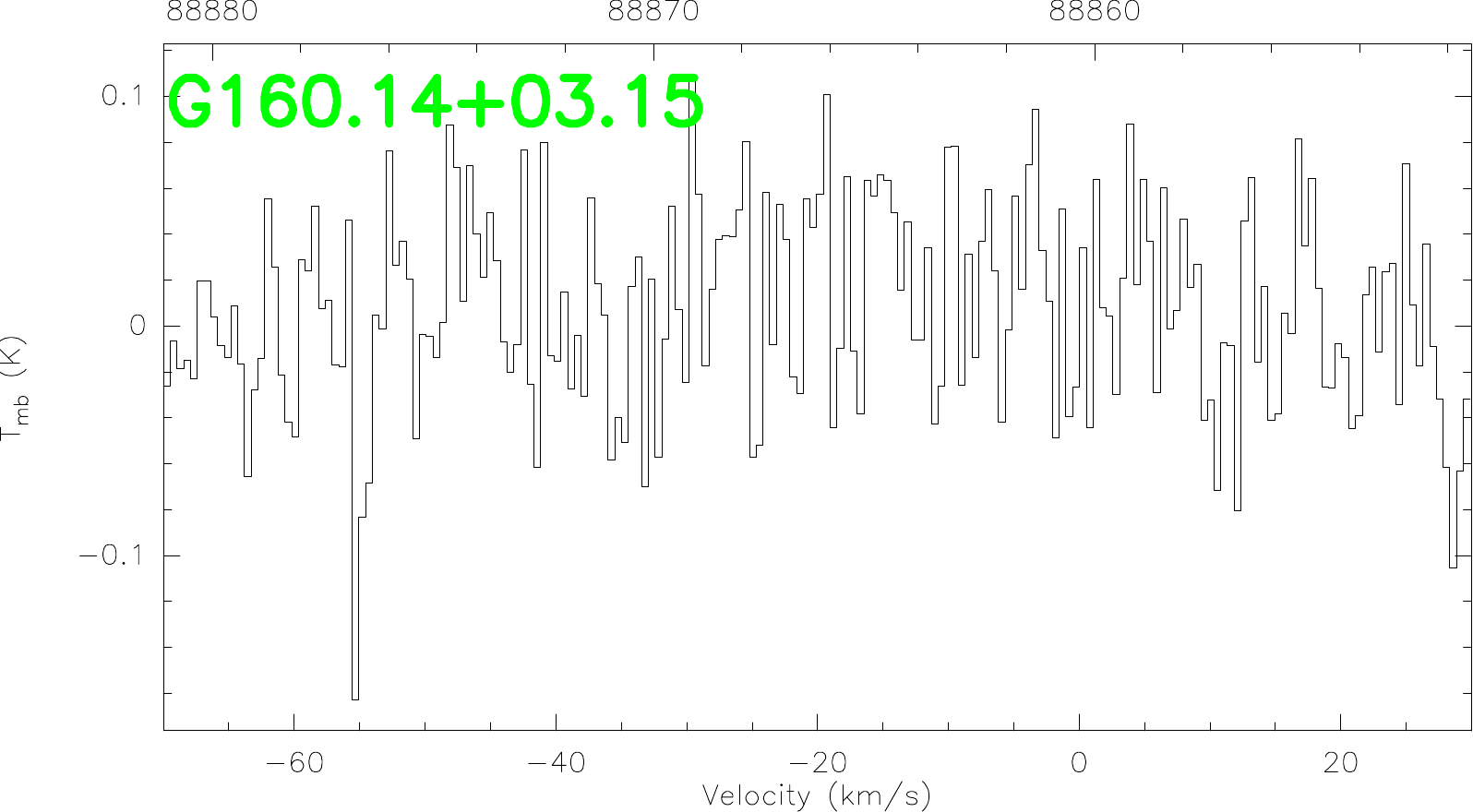}\\

\includegraphics[width=0.155\textwidth]{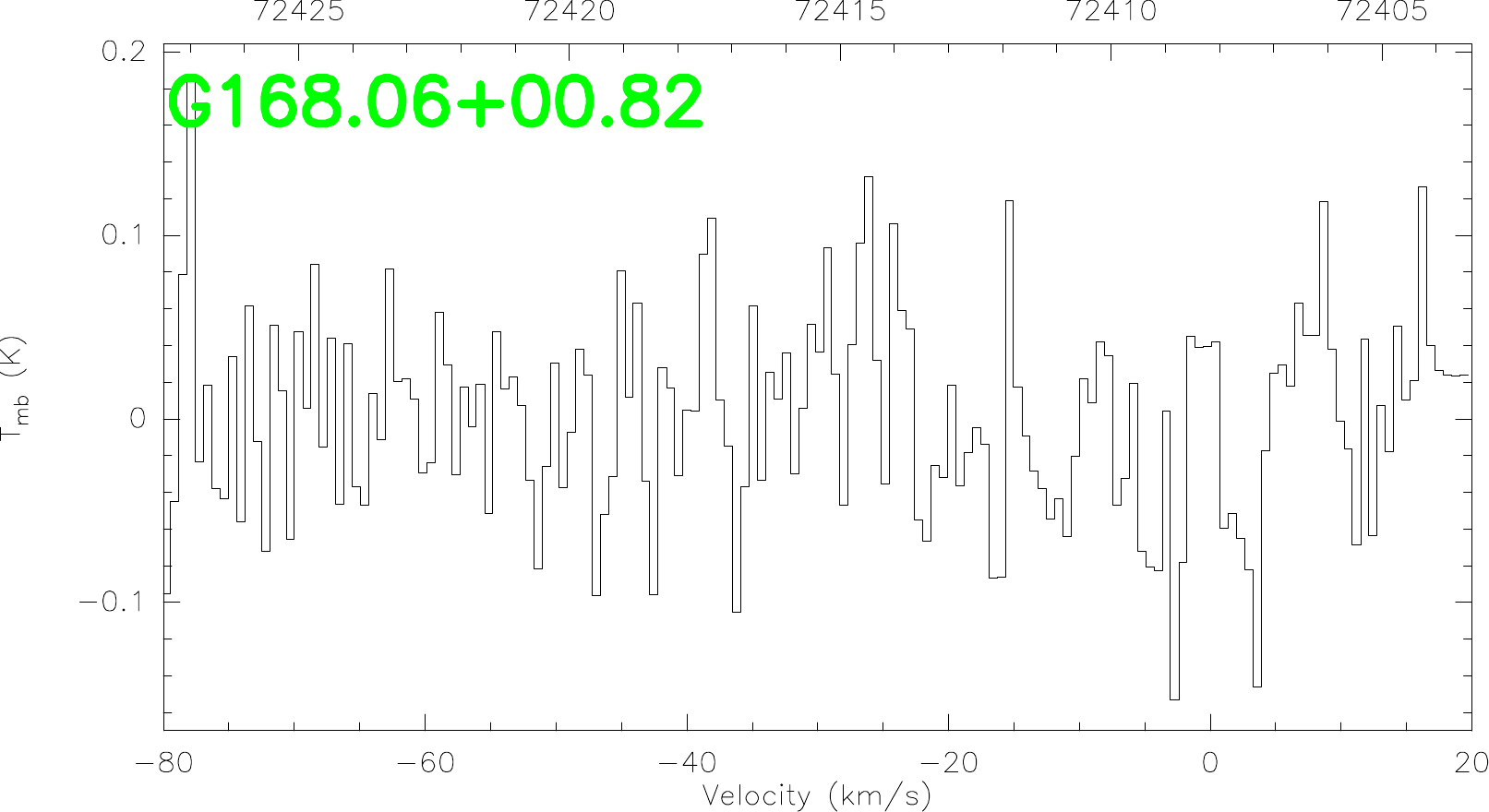}
\includegraphics[width=0.155\textwidth]{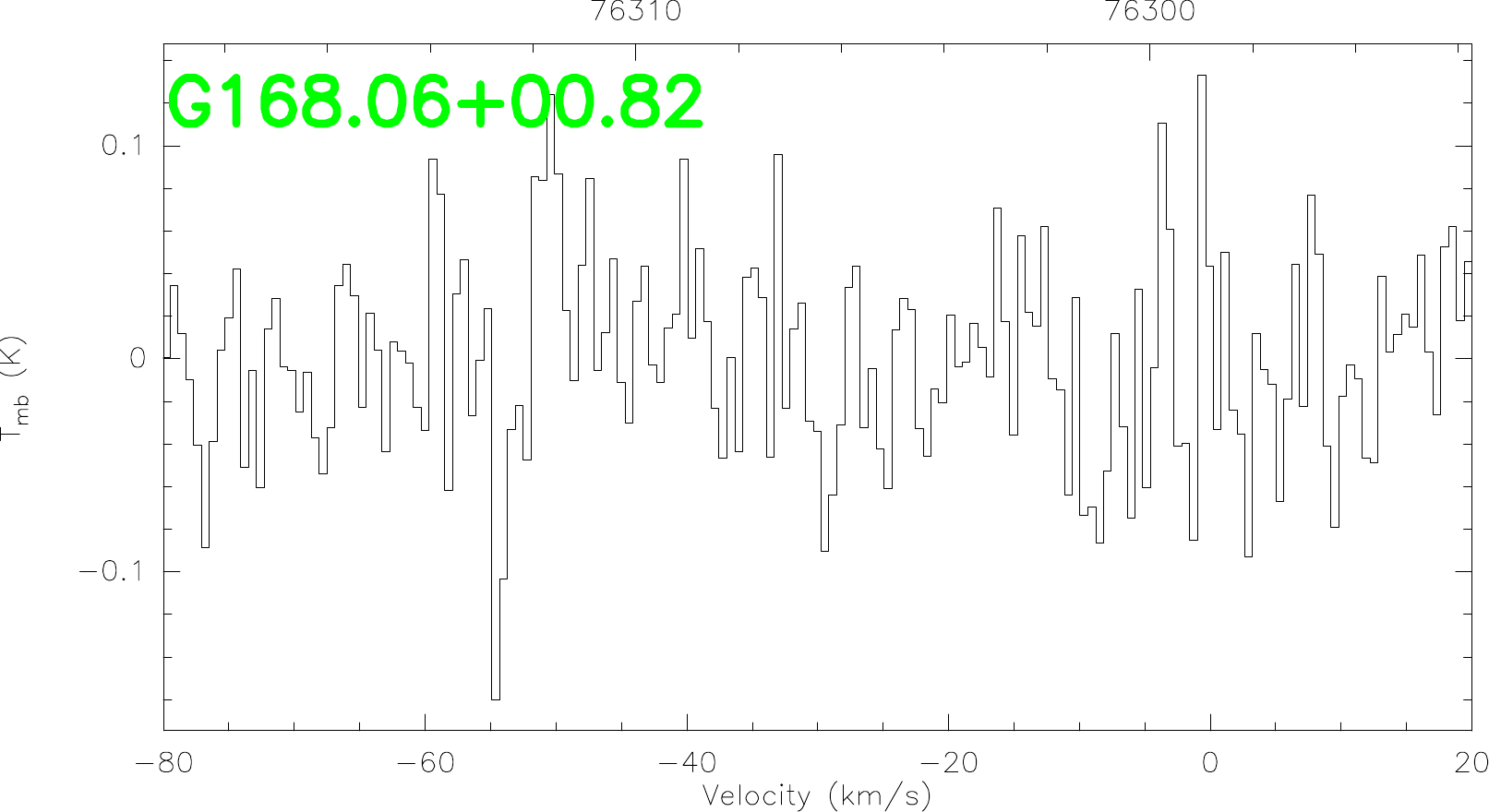}
\includegraphics[width=0.155\textwidth]{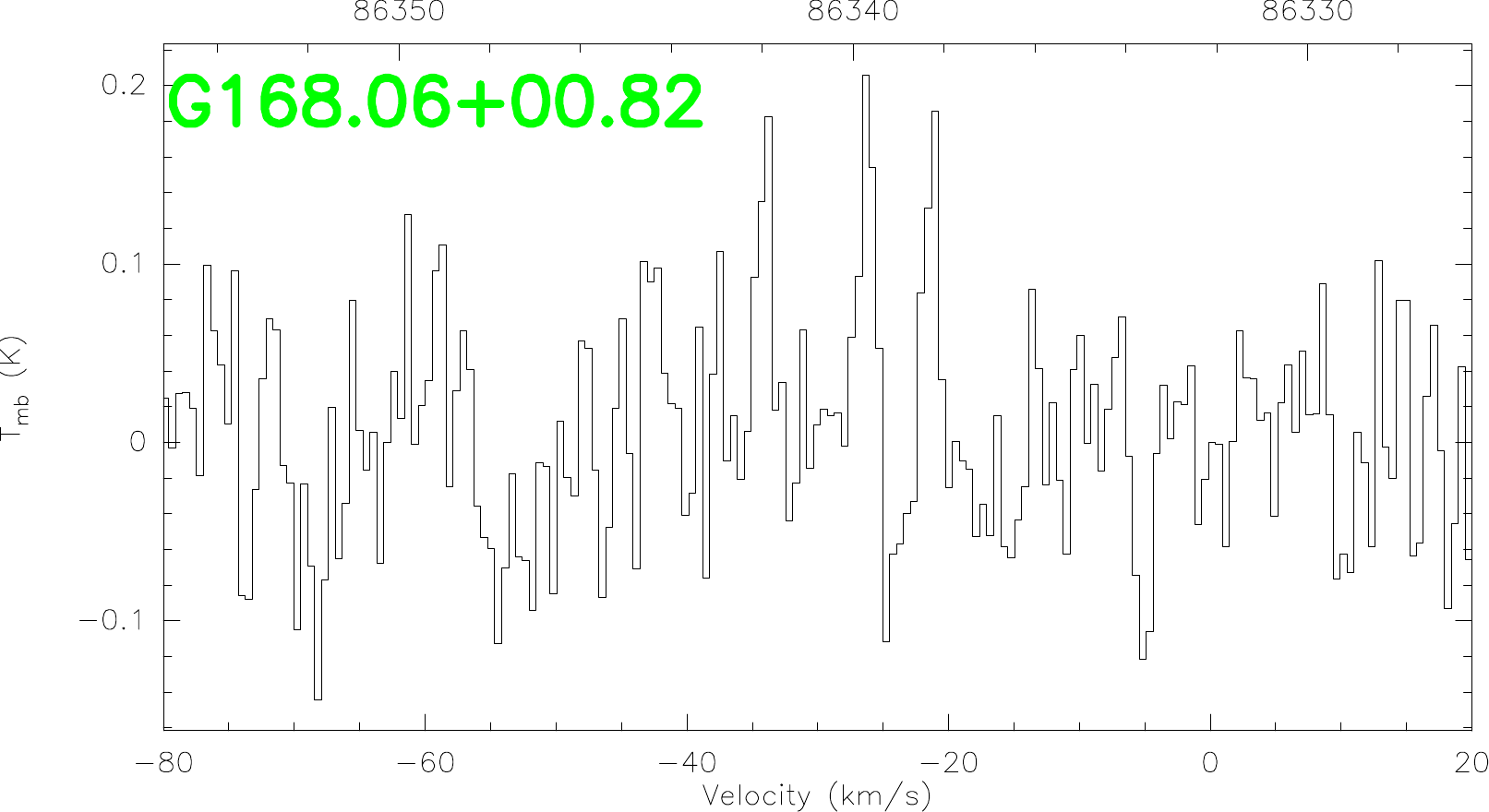}
\includegraphics[width=0.155\textwidth]{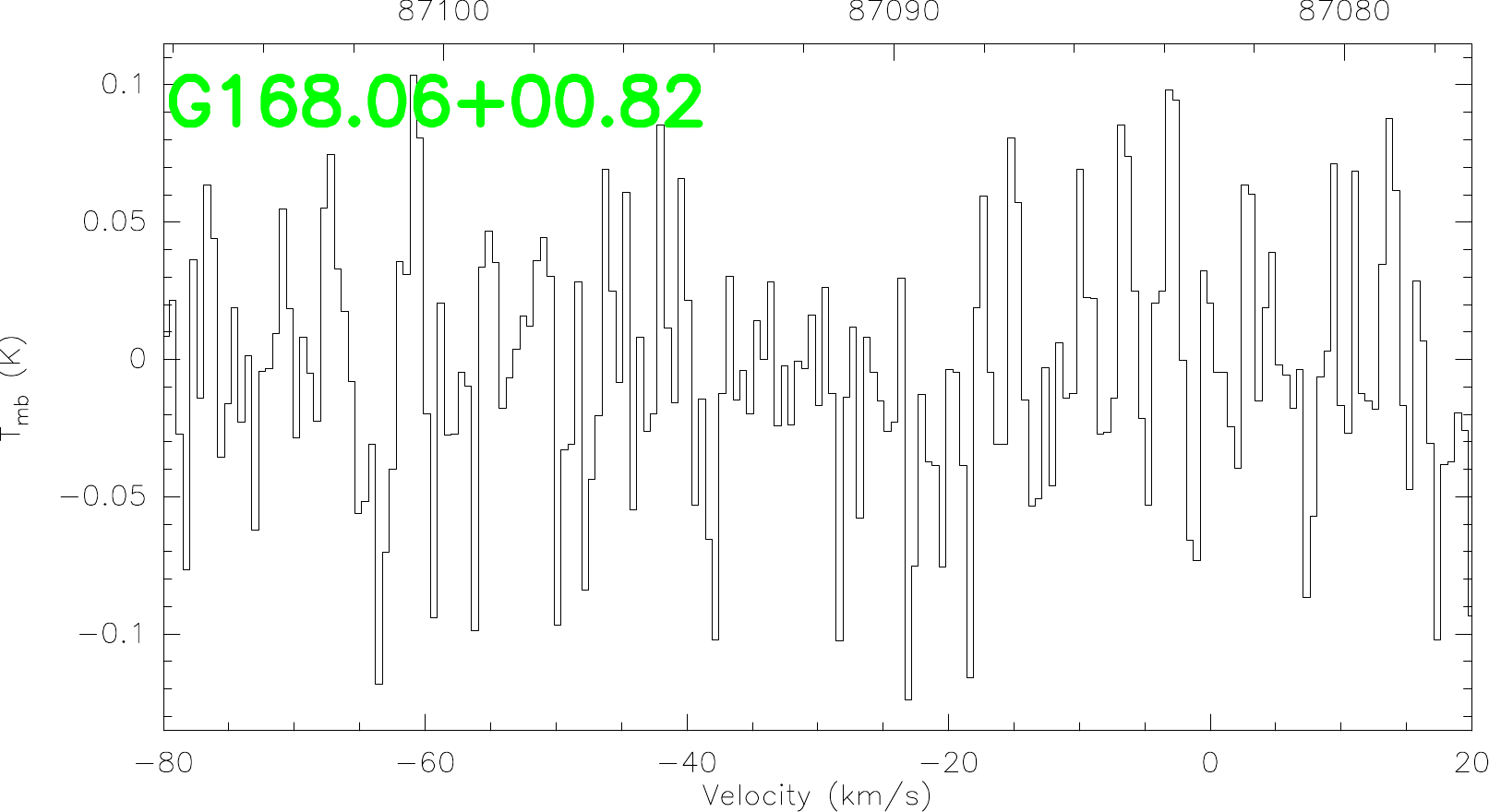}
\includegraphics[width=0.155\textwidth]{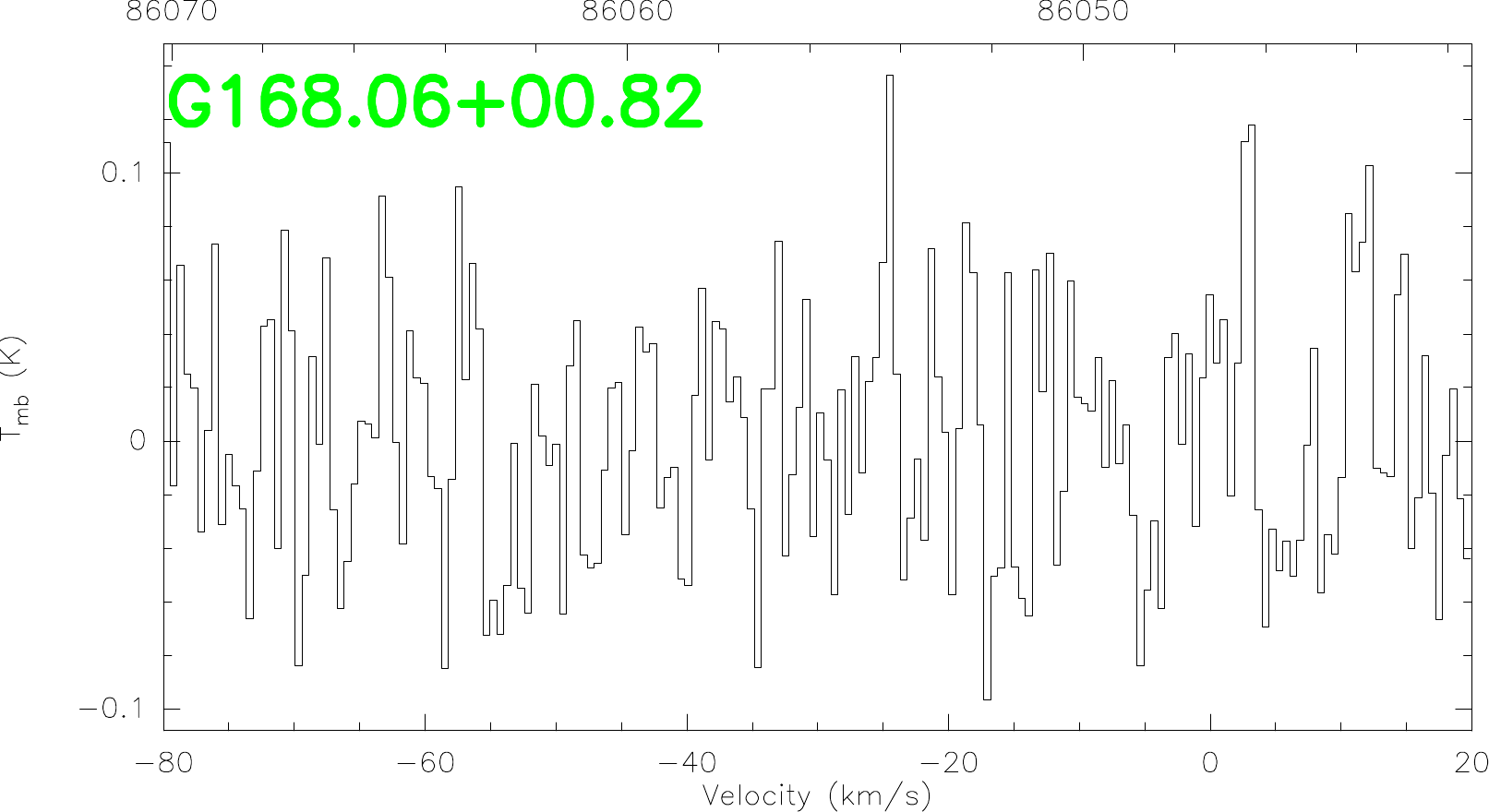}
\includegraphics[width=0.155\textwidth]{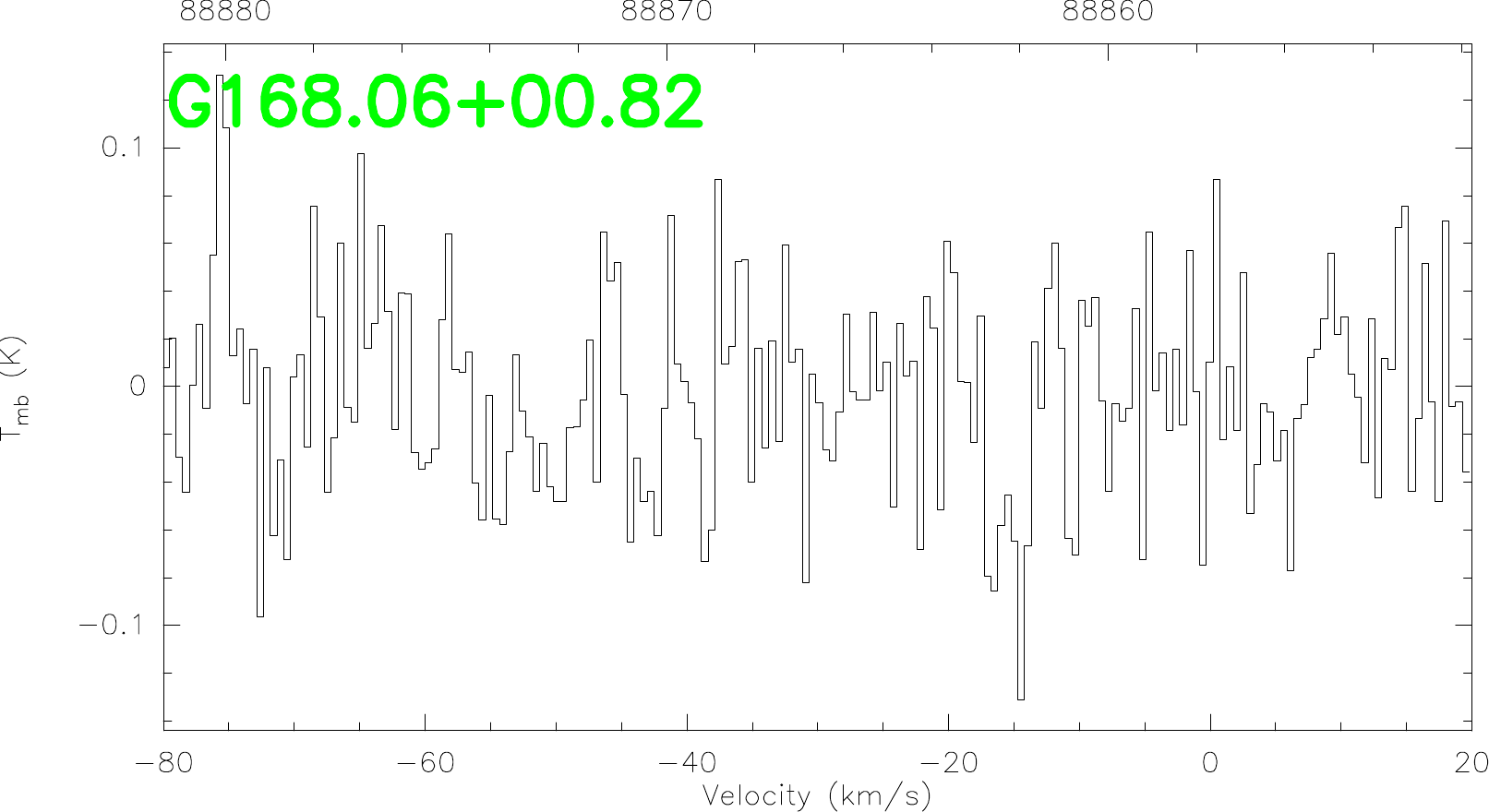}\\

\includegraphics[width=0.155\textwidth]{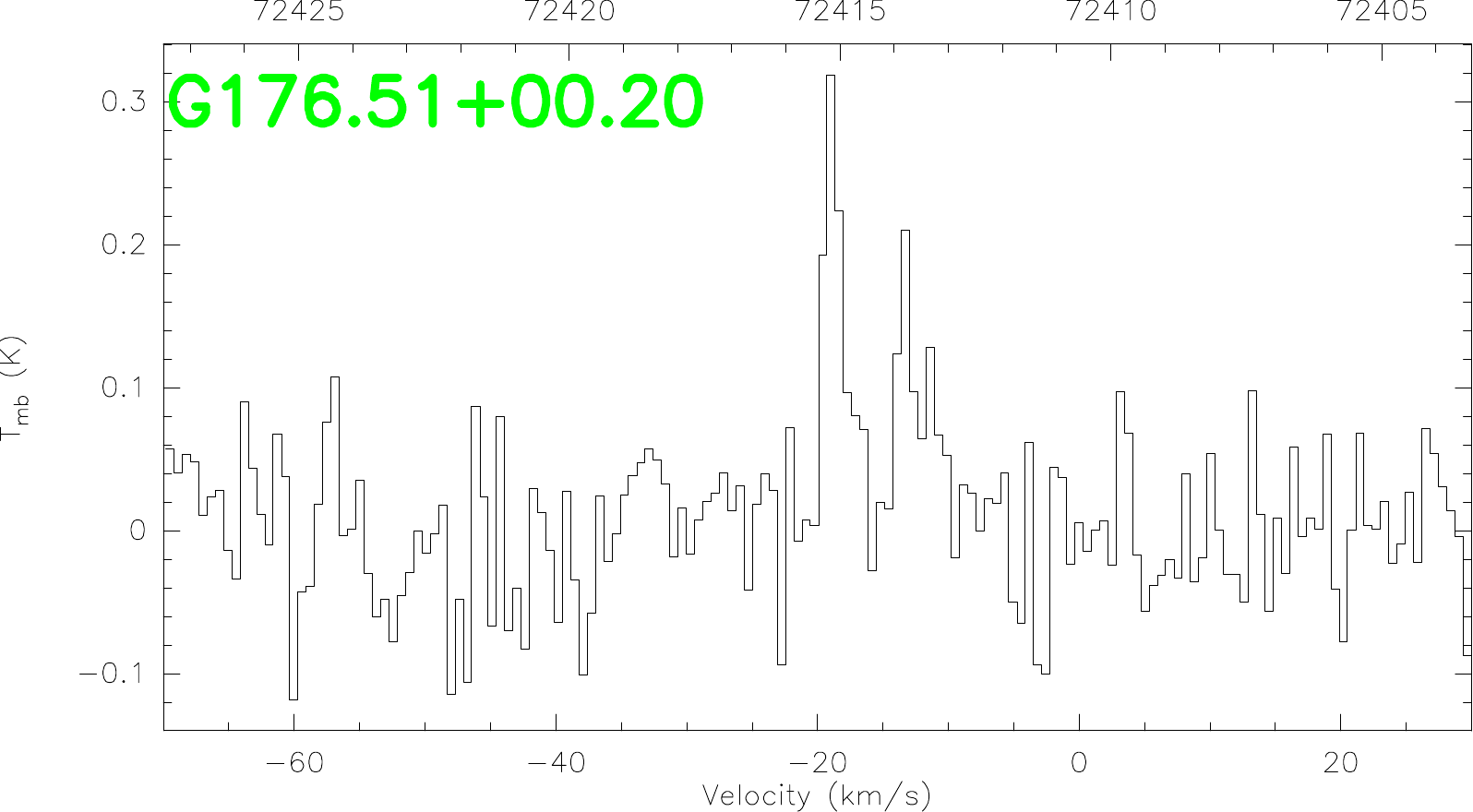}
\includegraphics[width=0.155\textwidth]{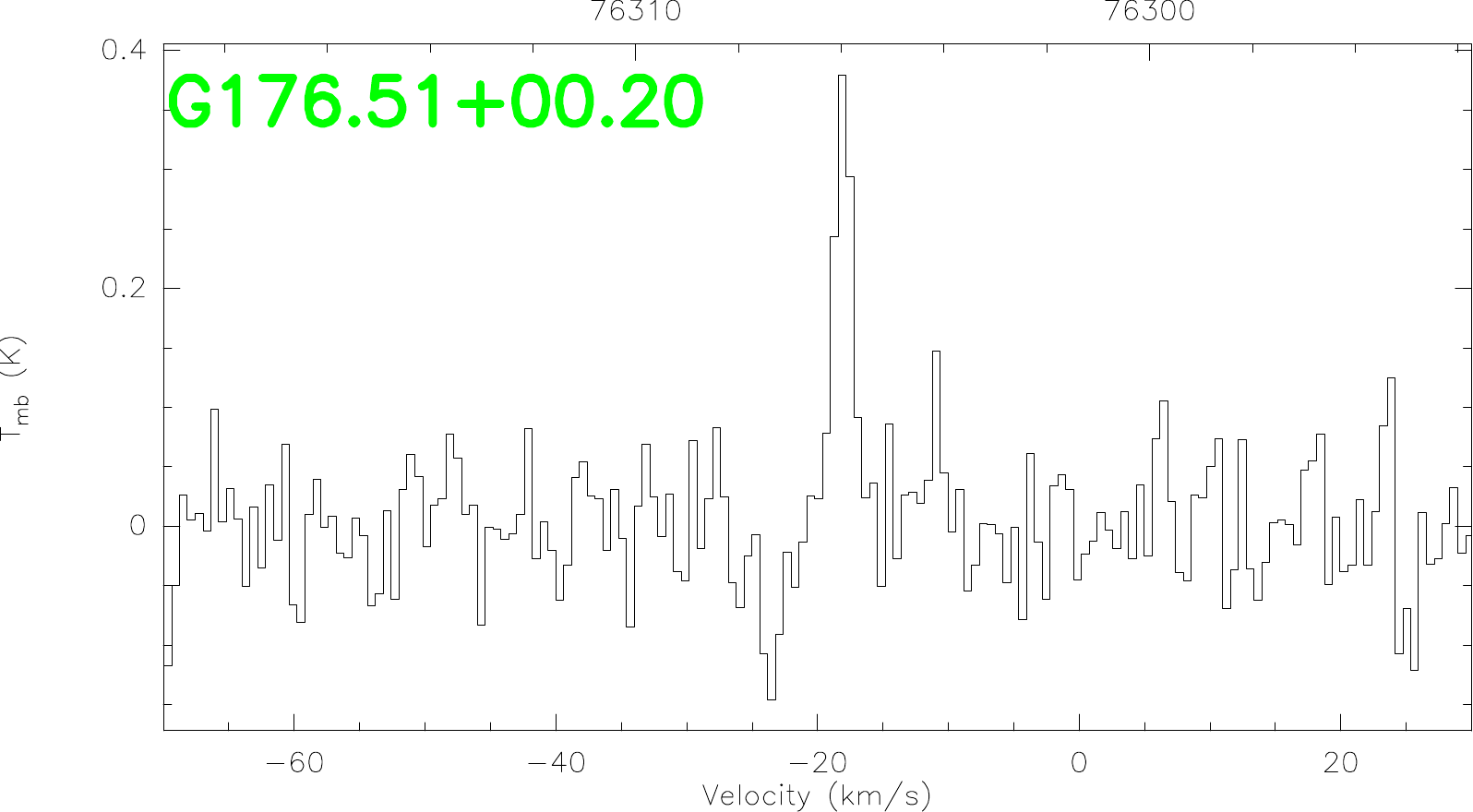}
\includegraphics[width=0.155\textwidth]{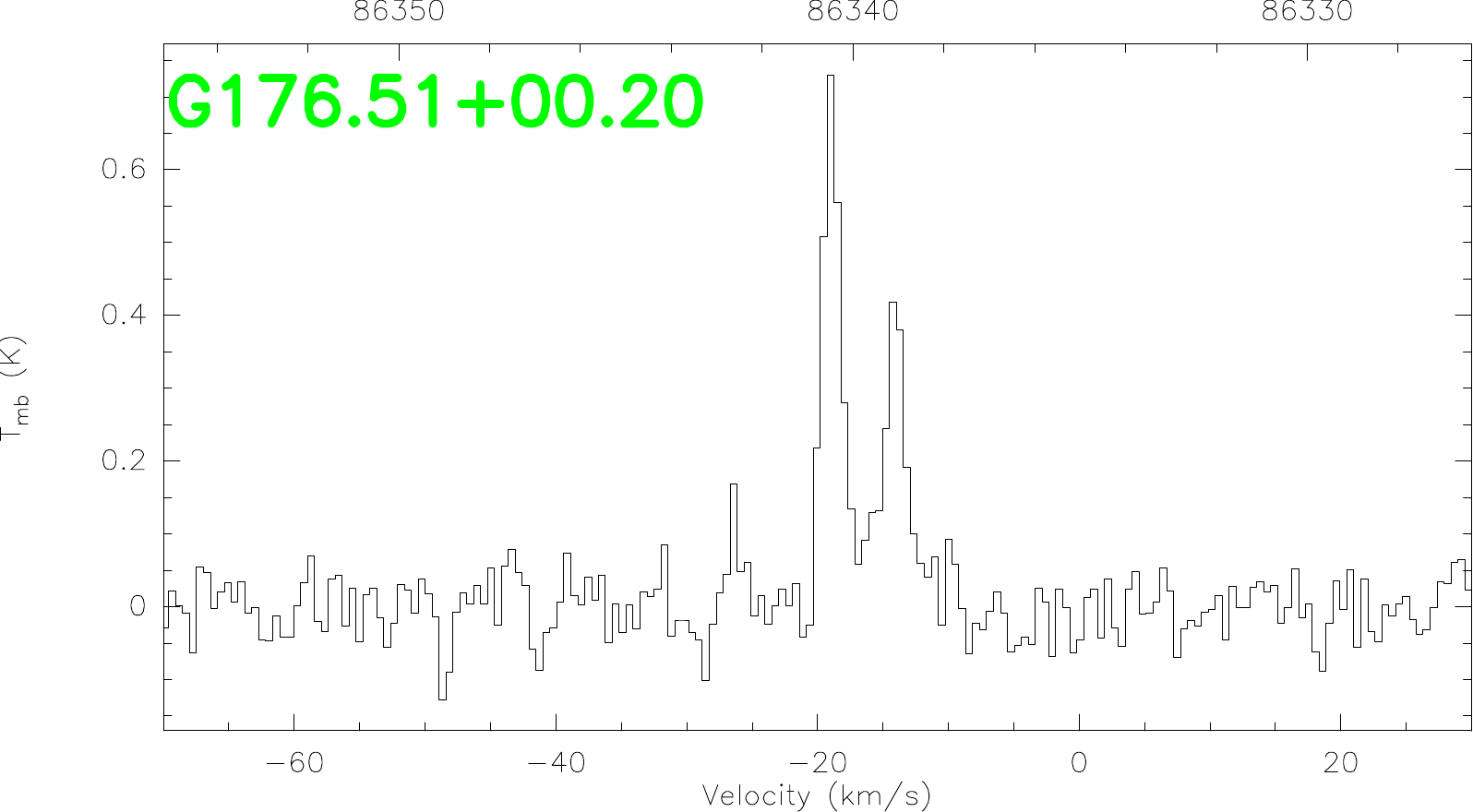}
\includegraphics[width=0.155\textwidth]{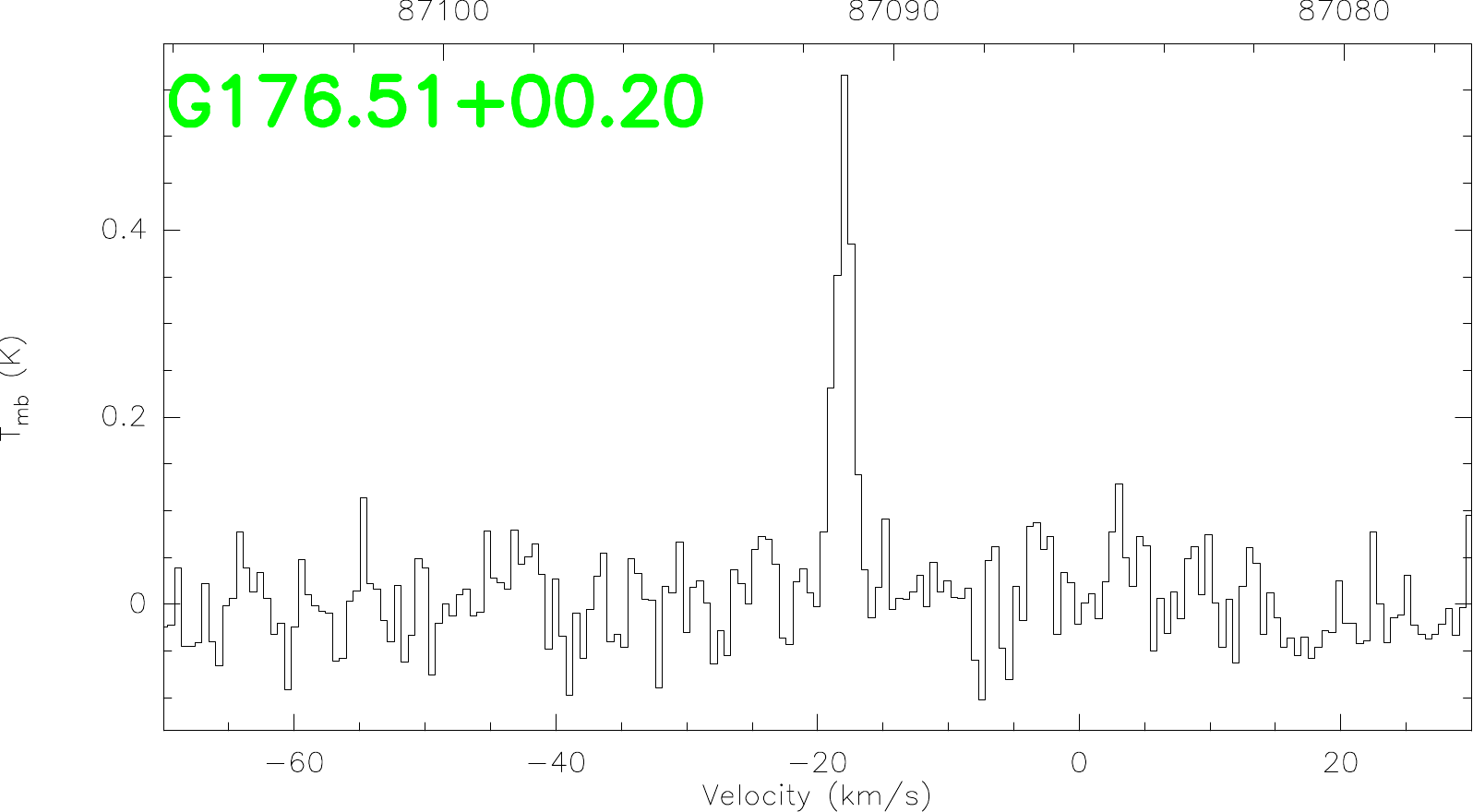}
\includegraphics[width=0.155\textwidth]{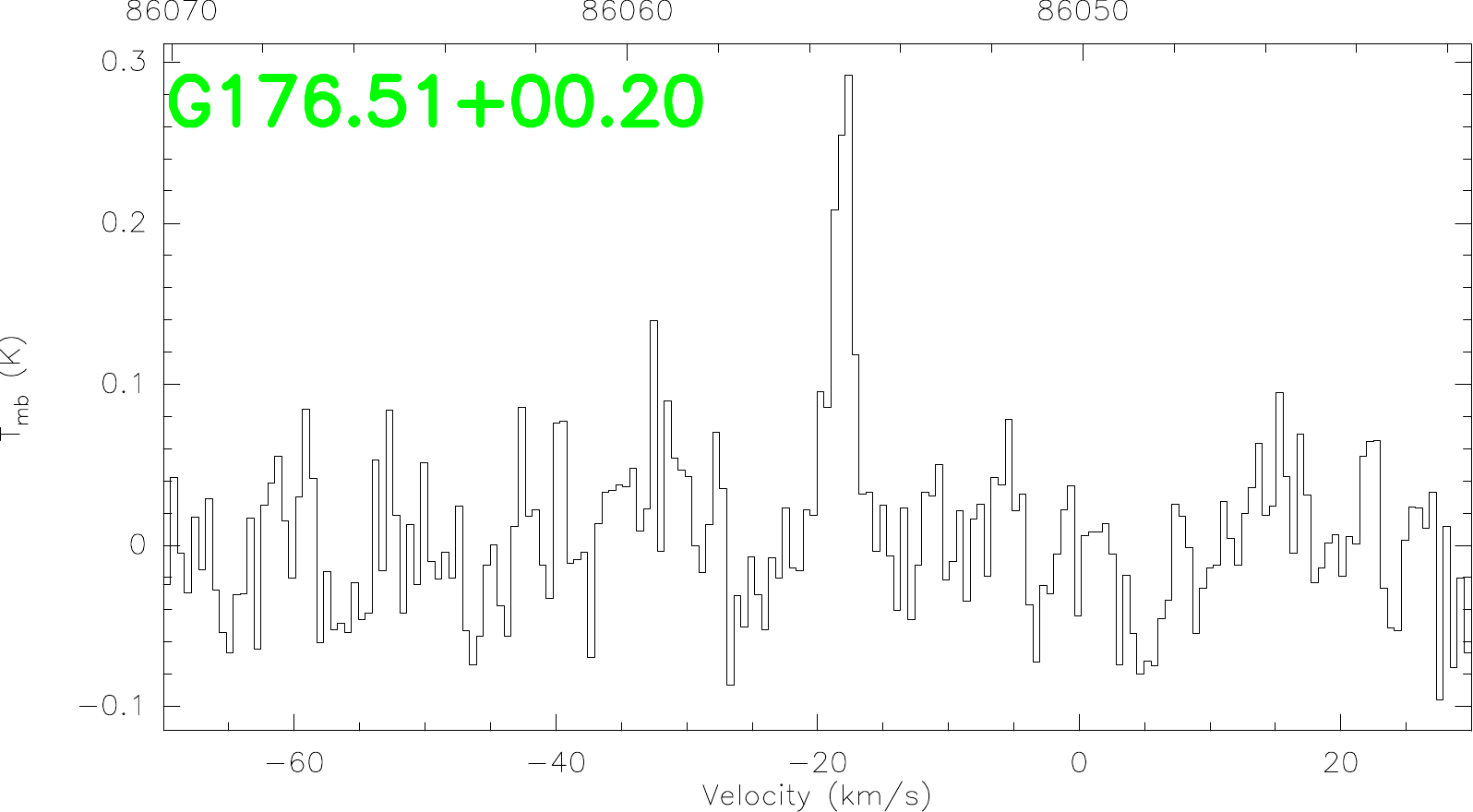}
\includegraphics[width=0.155\textwidth]{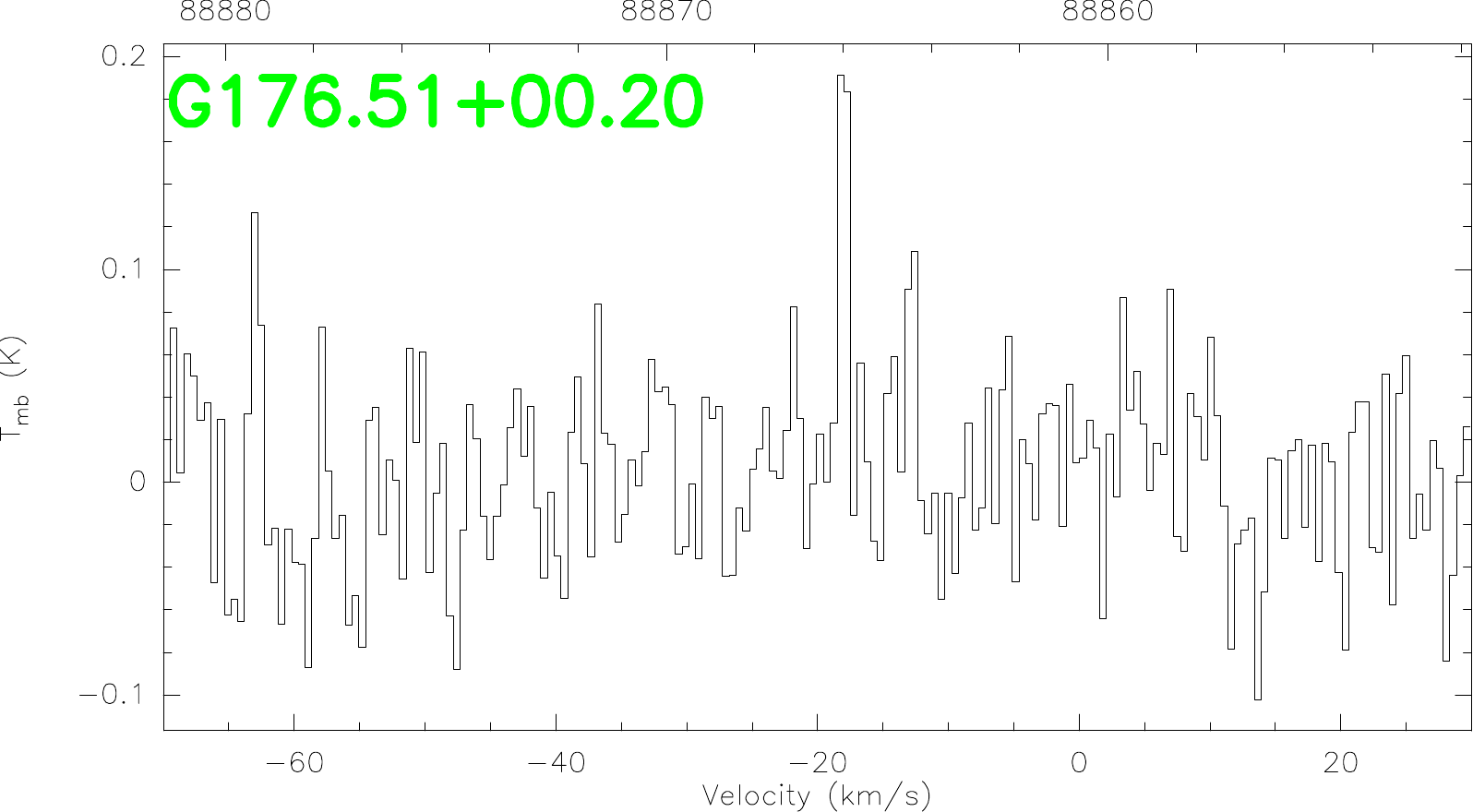}\\

\includegraphics[width=0.155\textwidth]{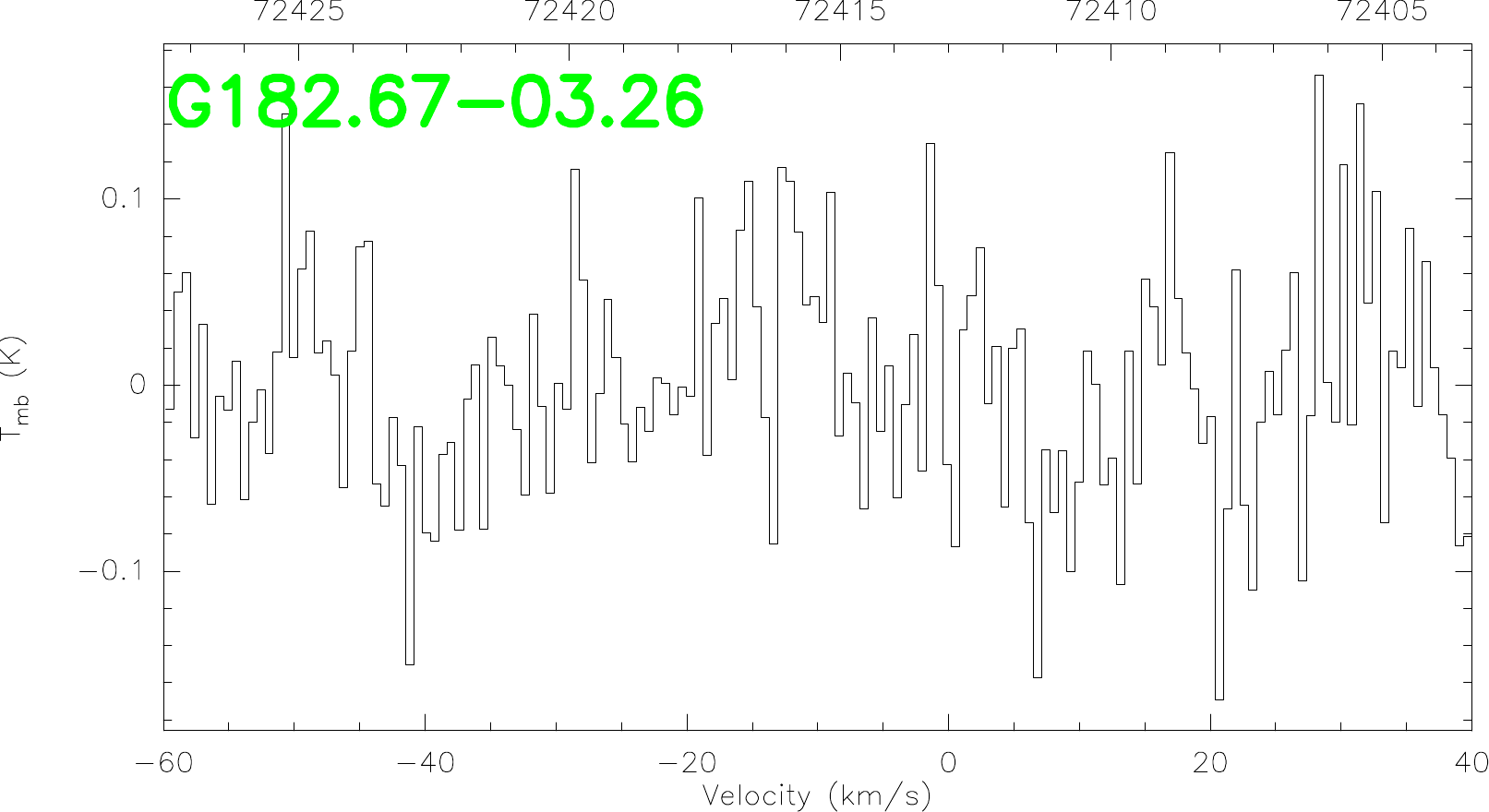}
\includegraphics[width=0.155\textwidth]{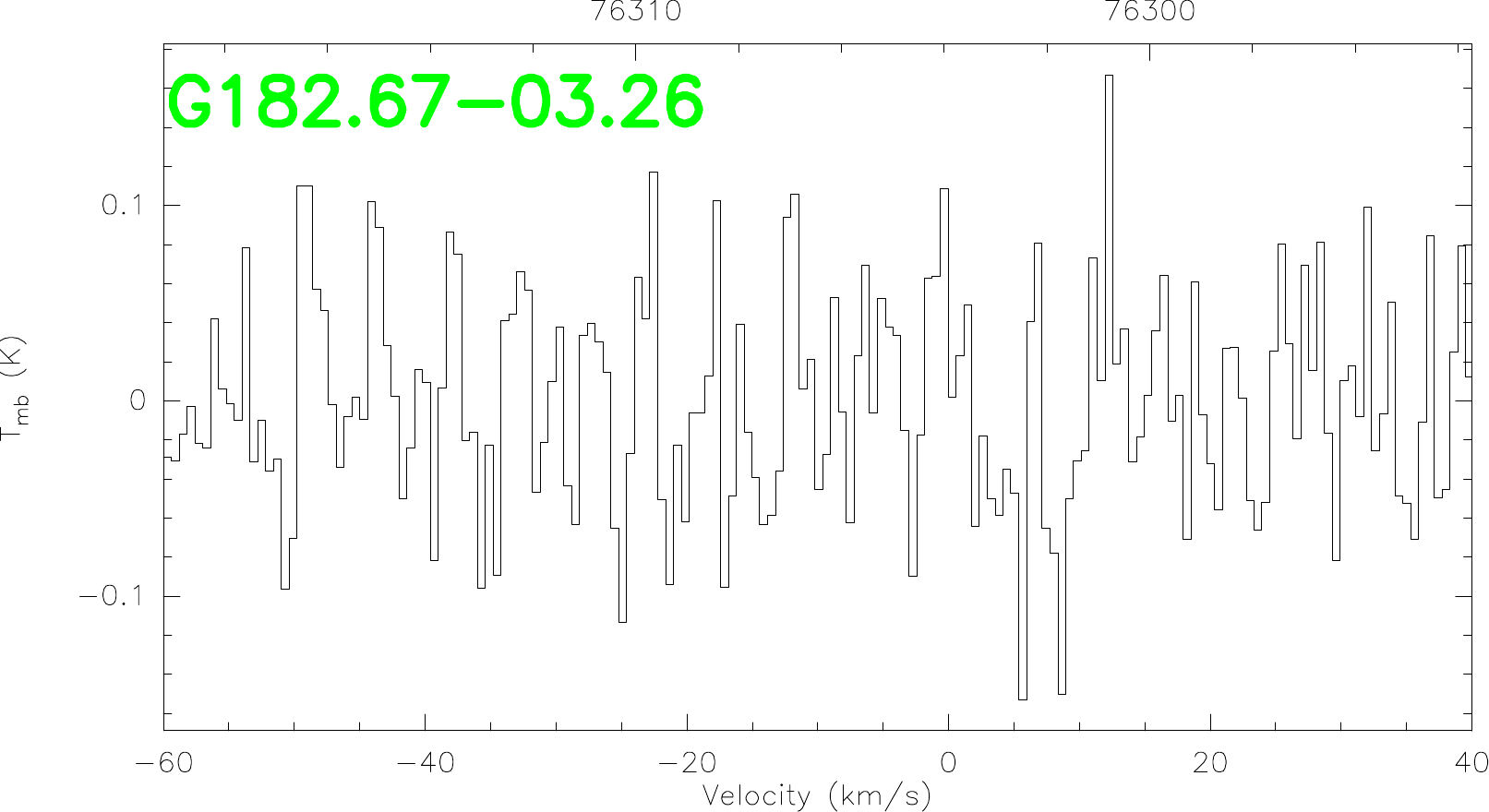}
\includegraphics[width=0.155\textwidth]{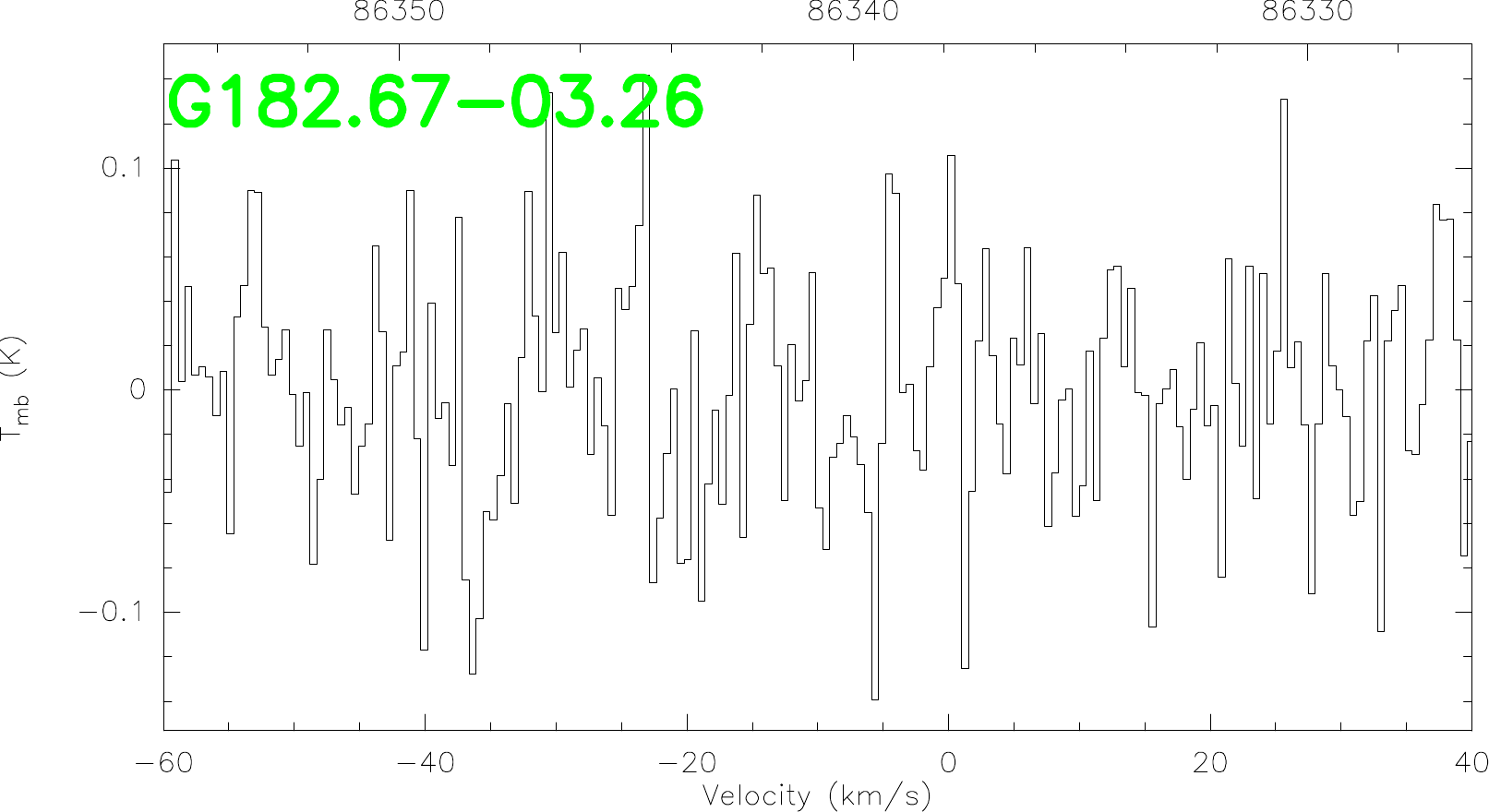}
\includegraphics[width=0.155\textwidth]{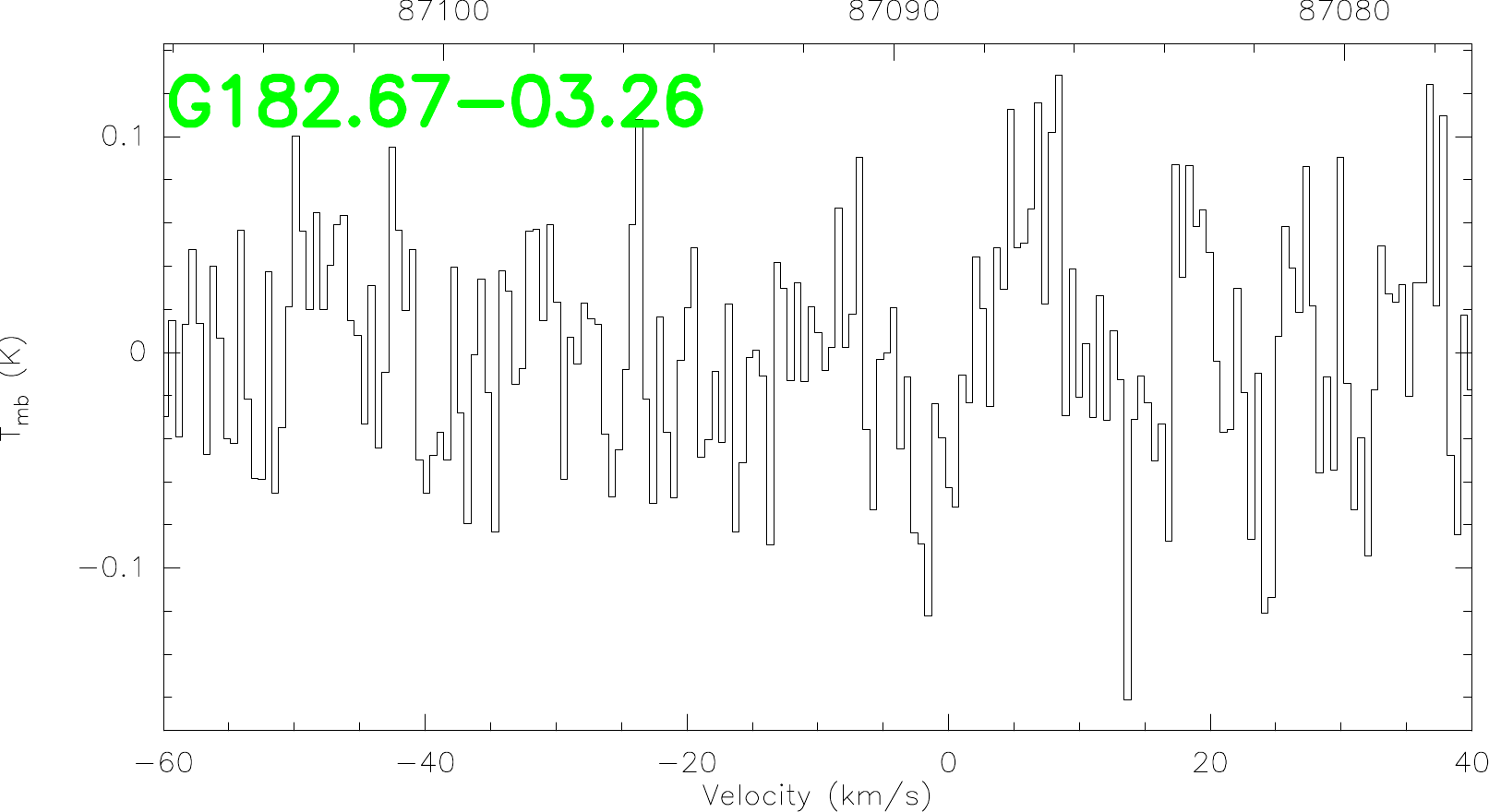}
\includegraphics[width=0.155\textwidth]{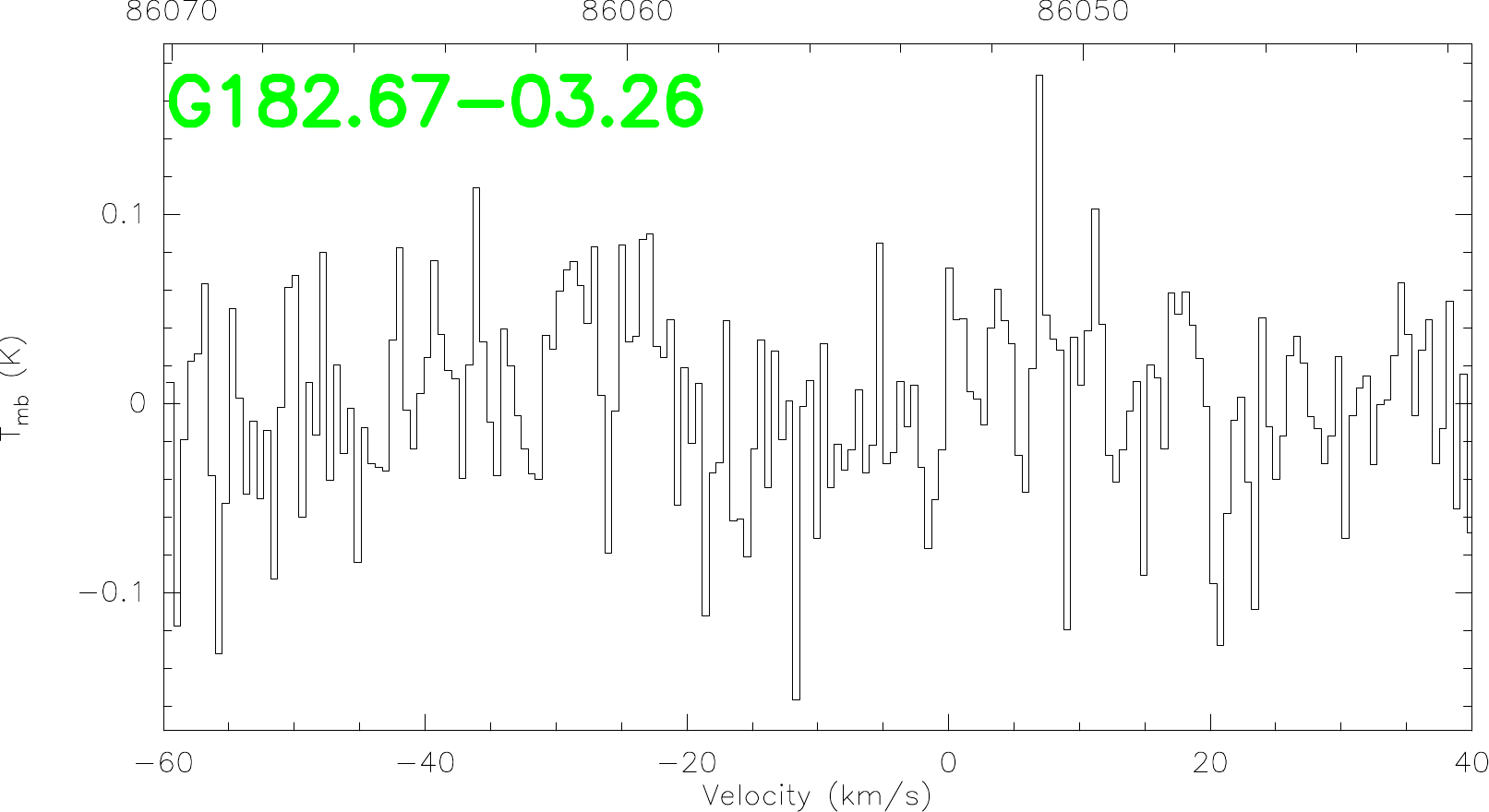}
\includegraphics[width=0.155\textwidth]{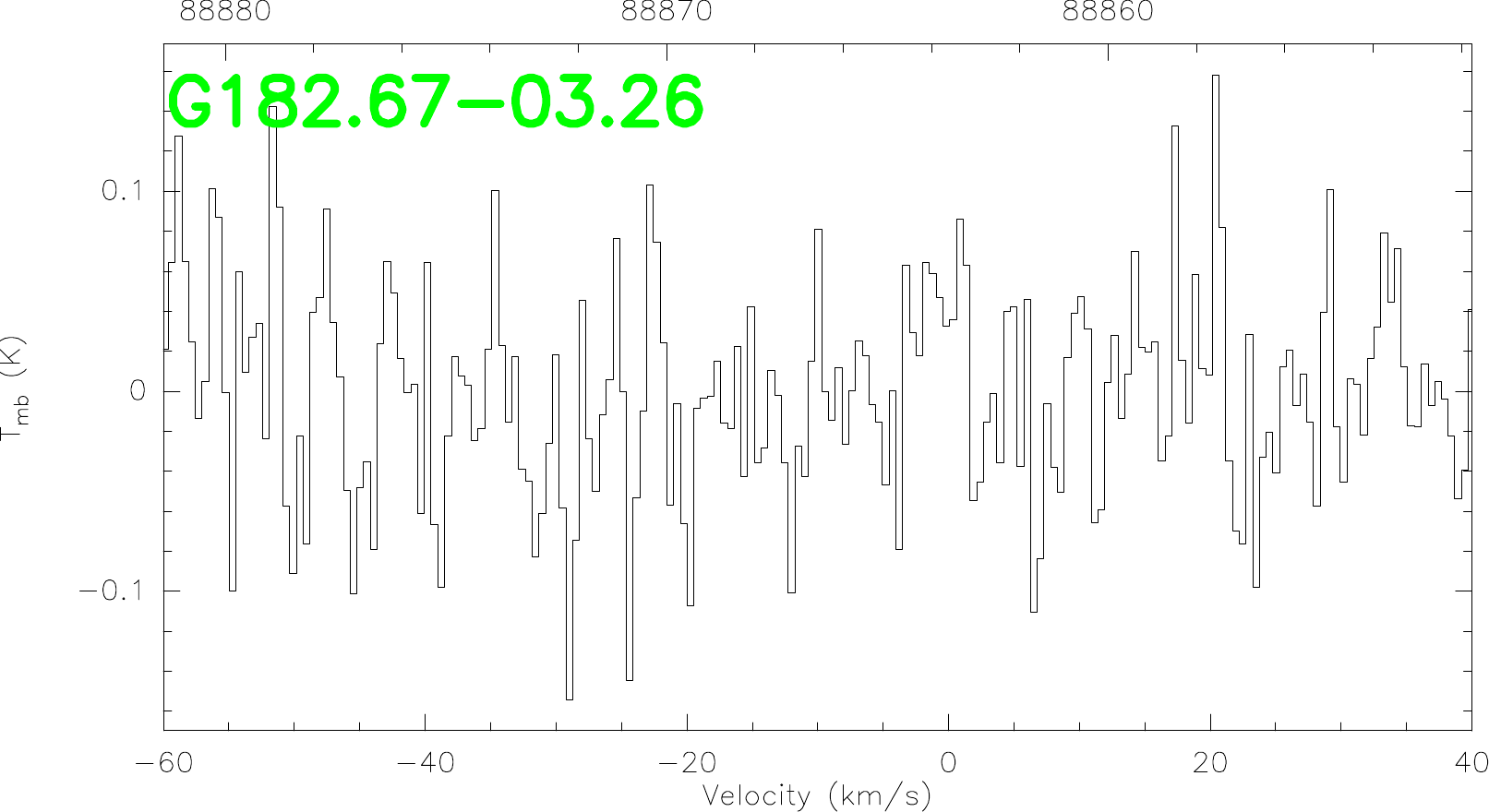}\\

\includegraphics[width=0.155\textwidth]{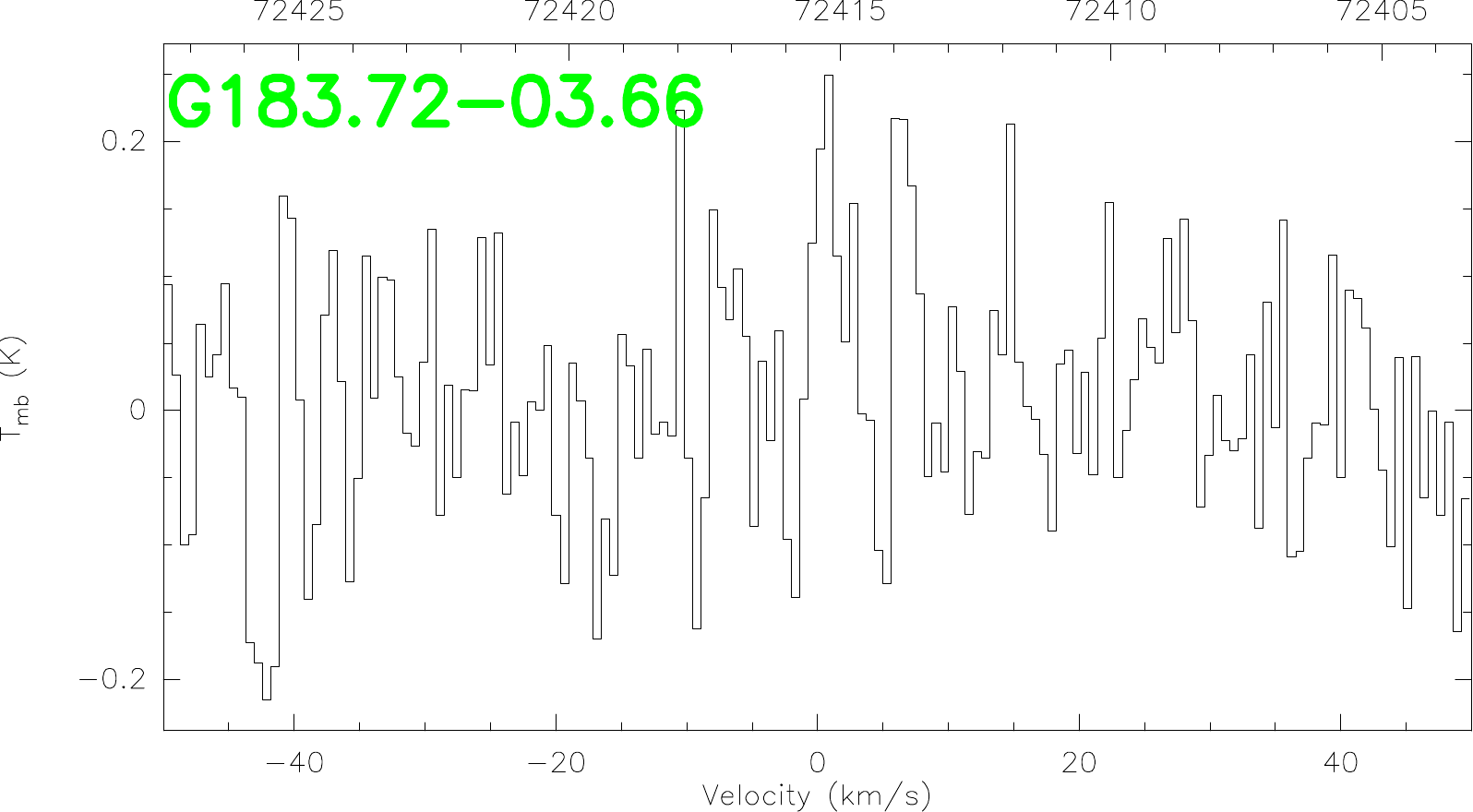}
\includegraphics[width=0.155\textwidth]{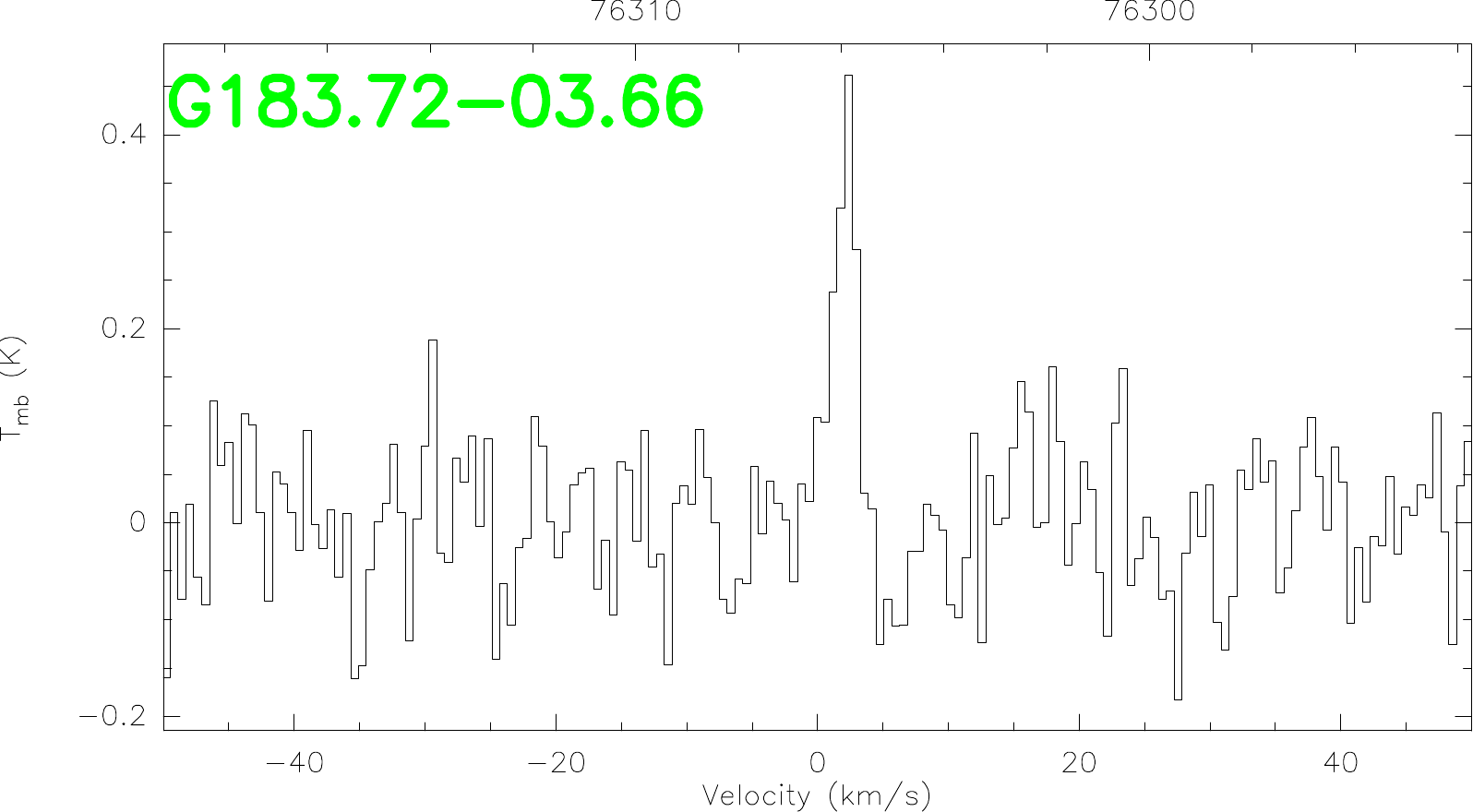}
\includegraphics[width=0.155\textwidth]{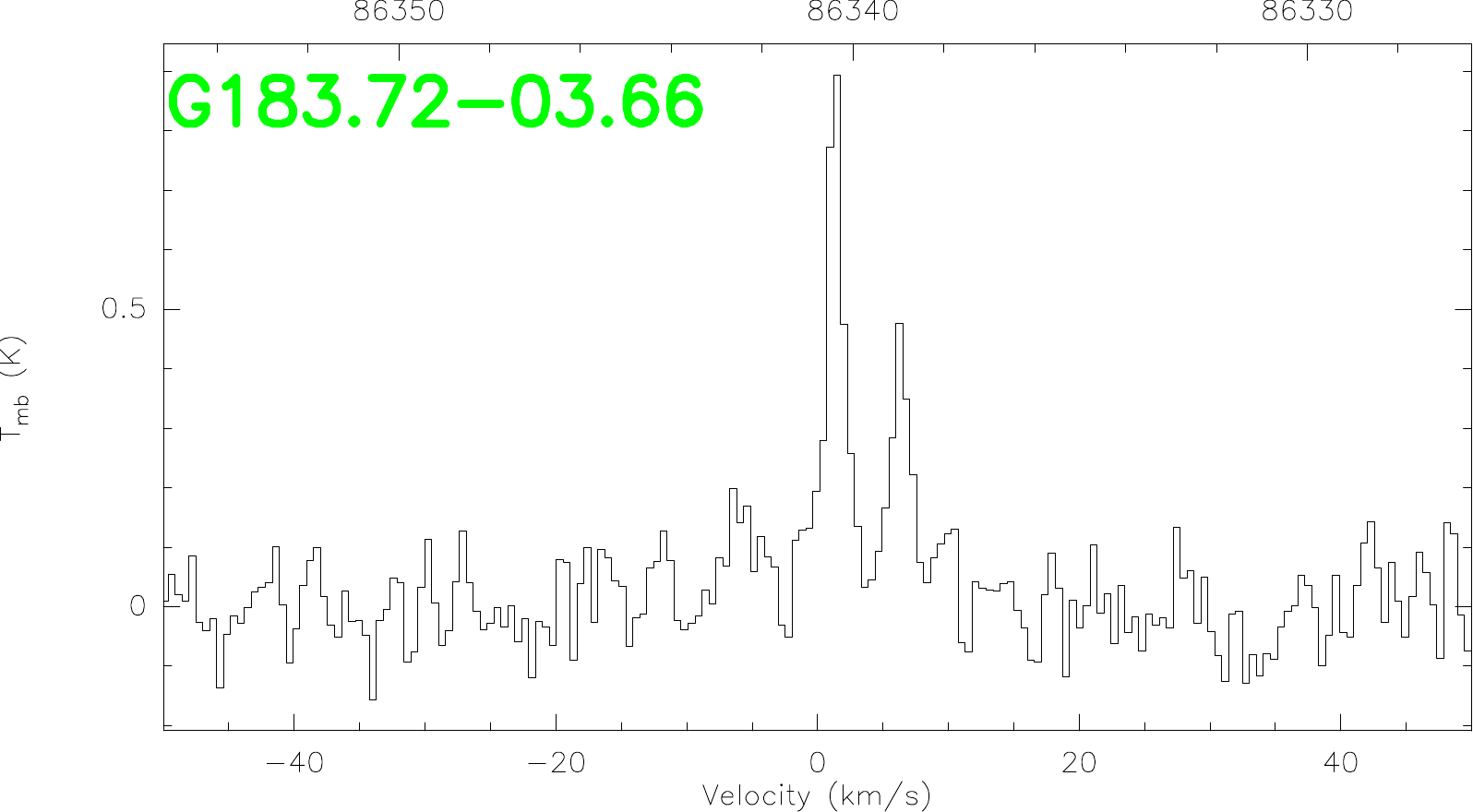}
\includegraphics[width=0.155\textwidth]{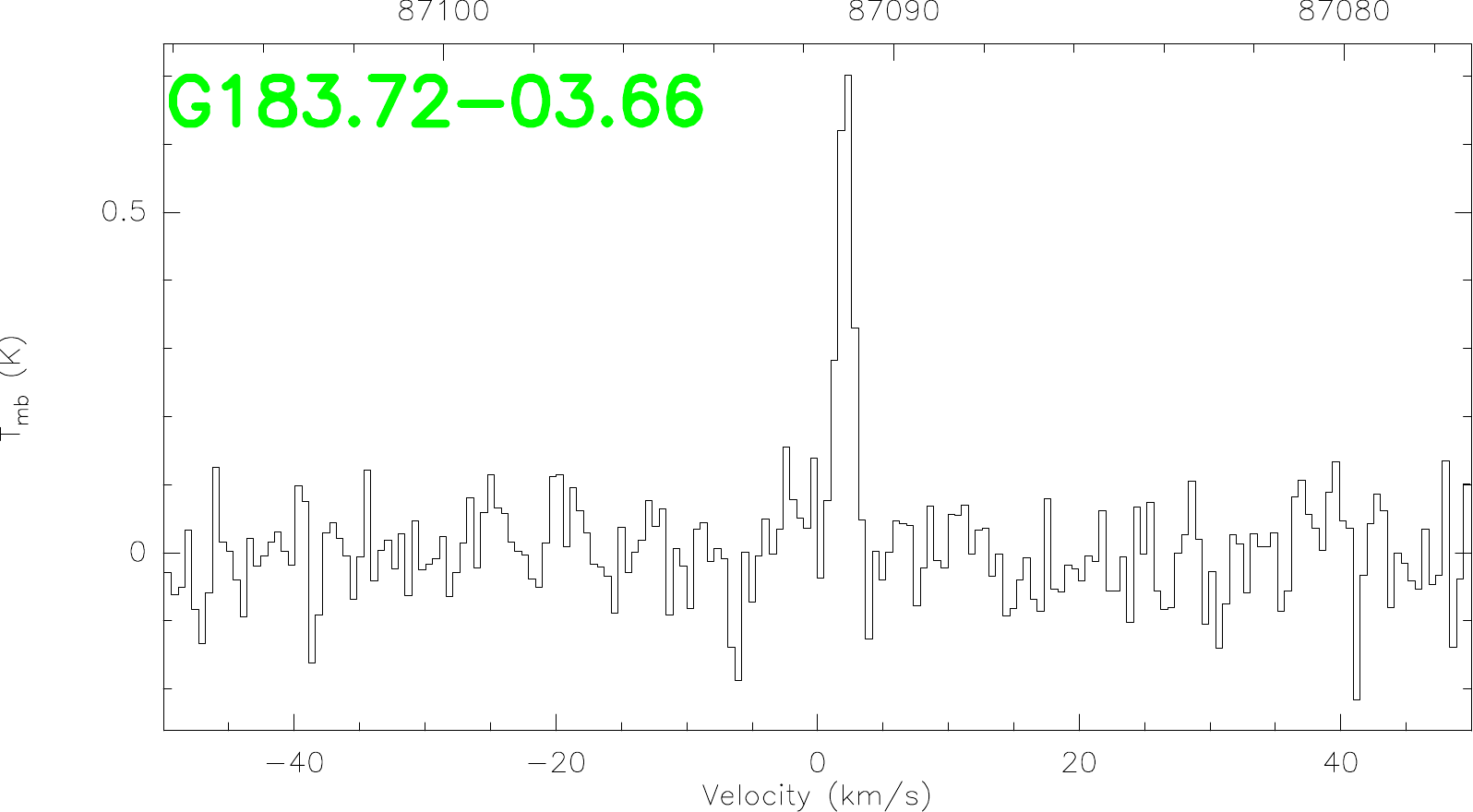}
\includegraphics[width=0.155\textwidth]{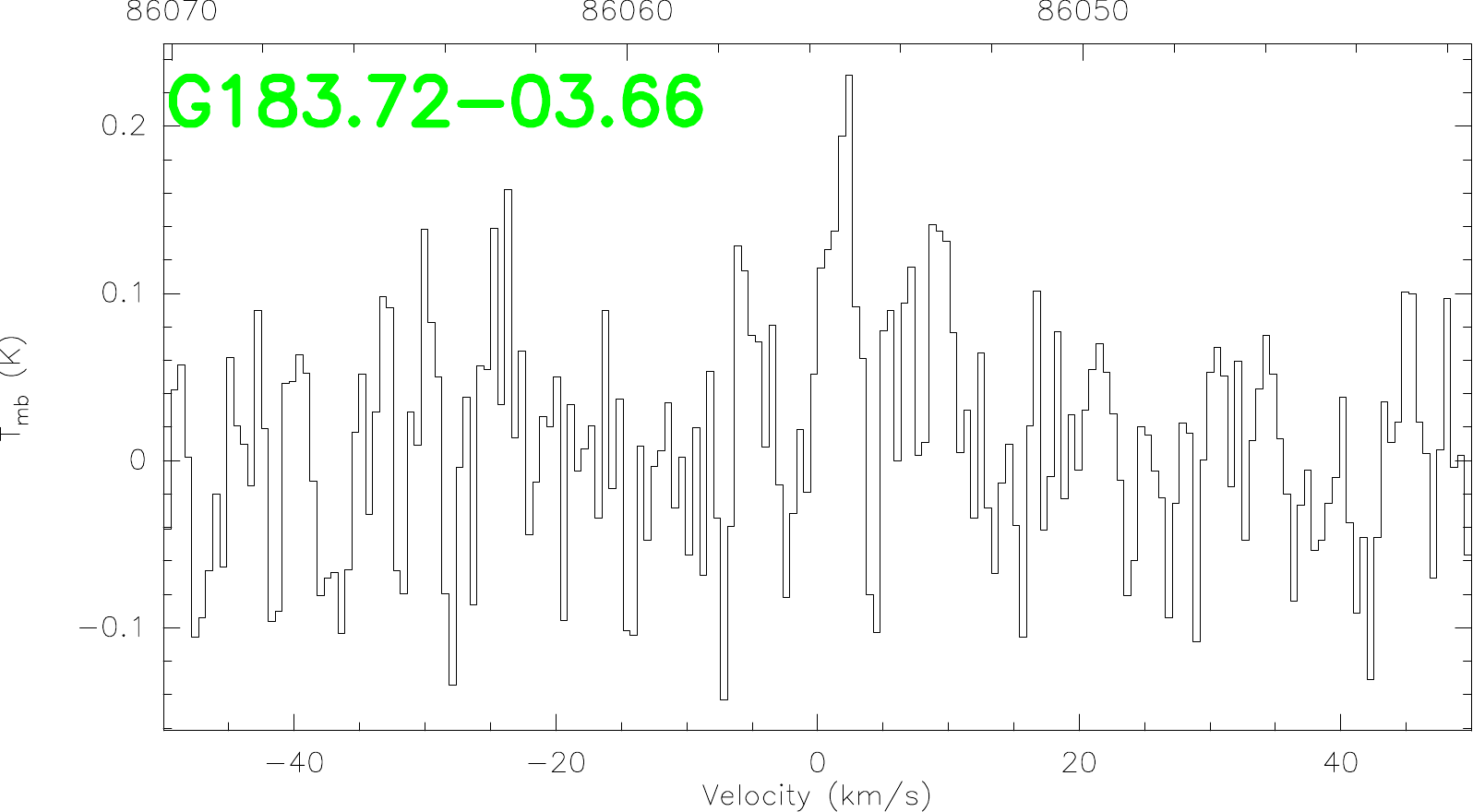}
\includegraphics[width=0.155\textwidth]{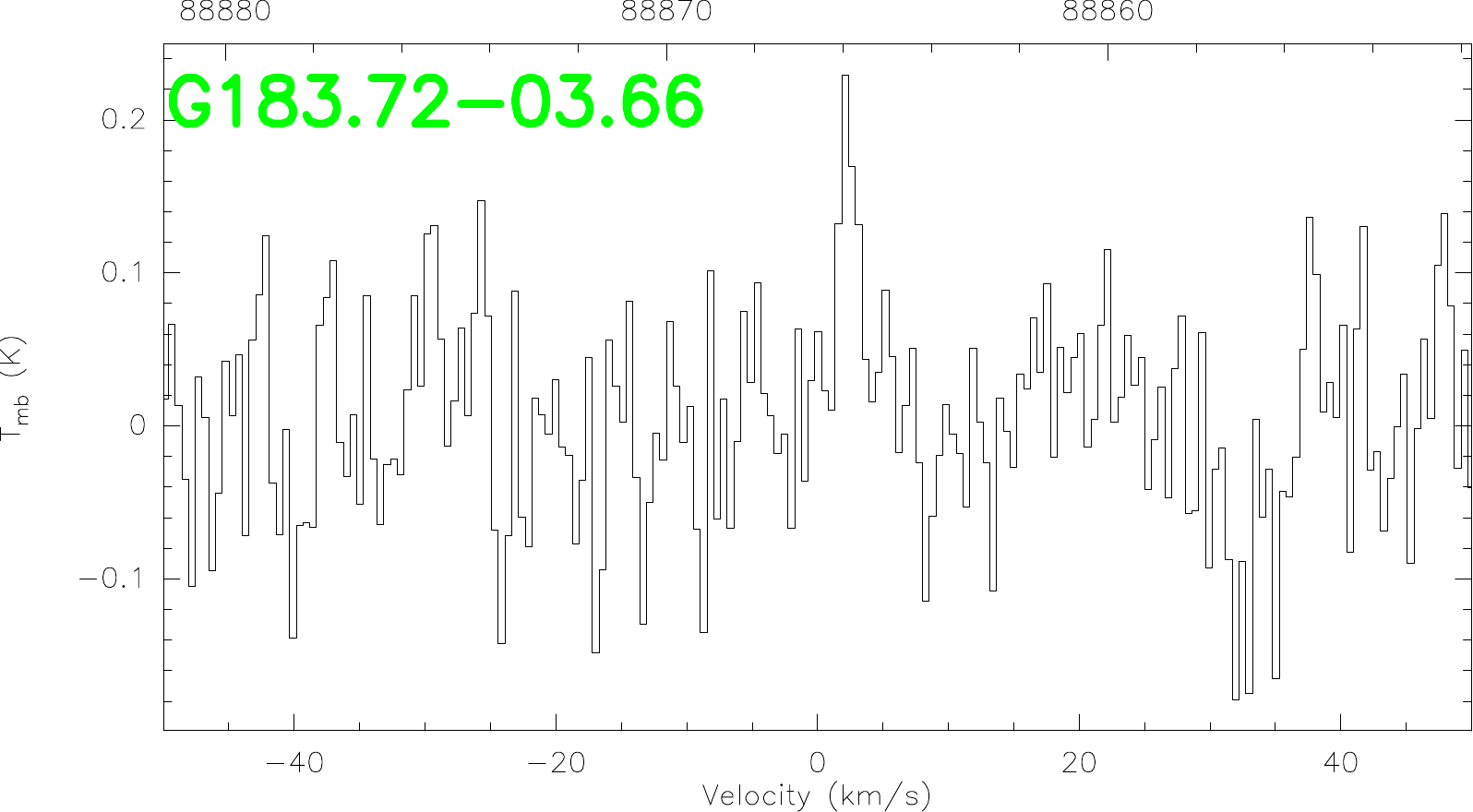}\\

\includegraphics[width=0.155\textwidth]{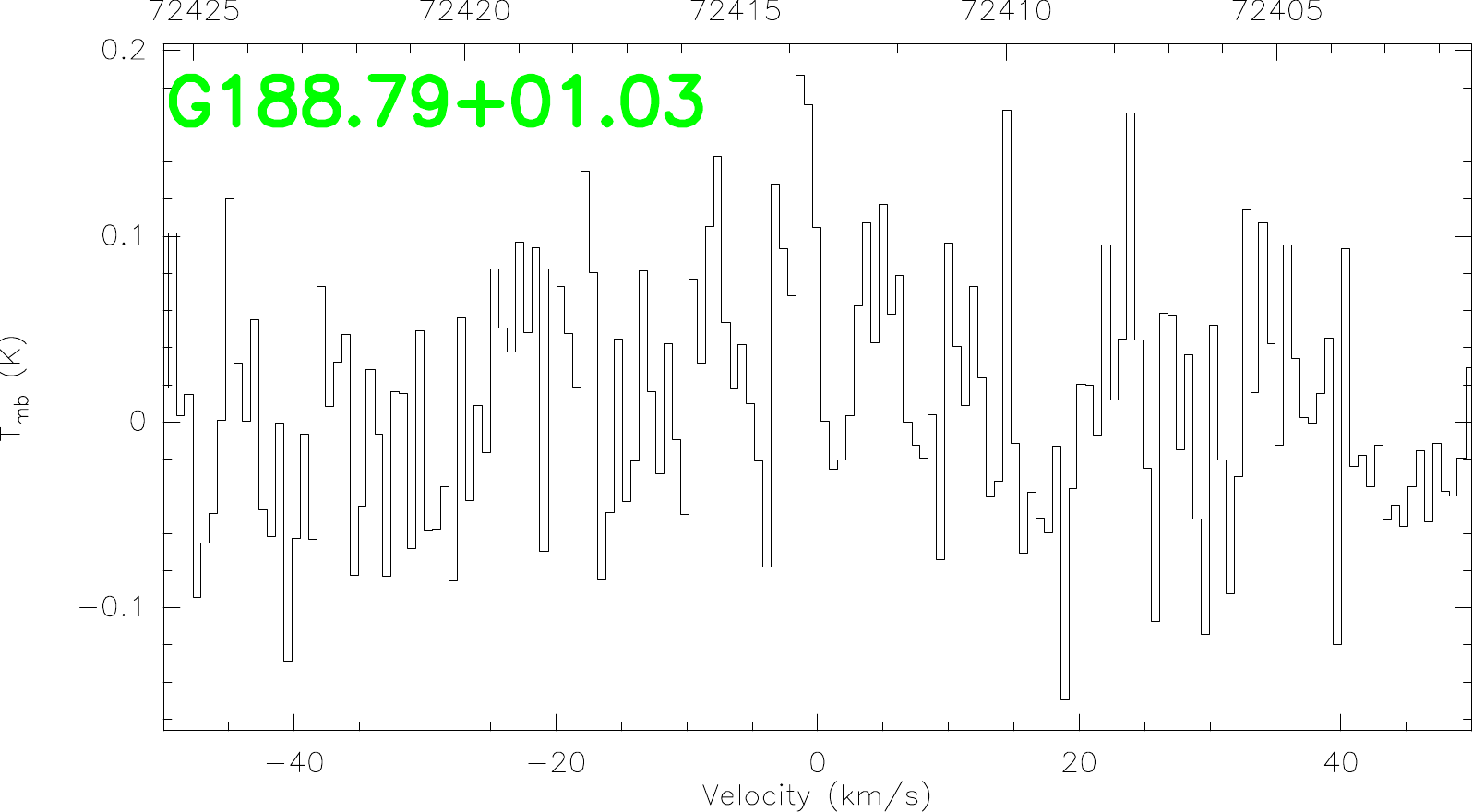}
\includegraphics[width=0.155\textwidth]{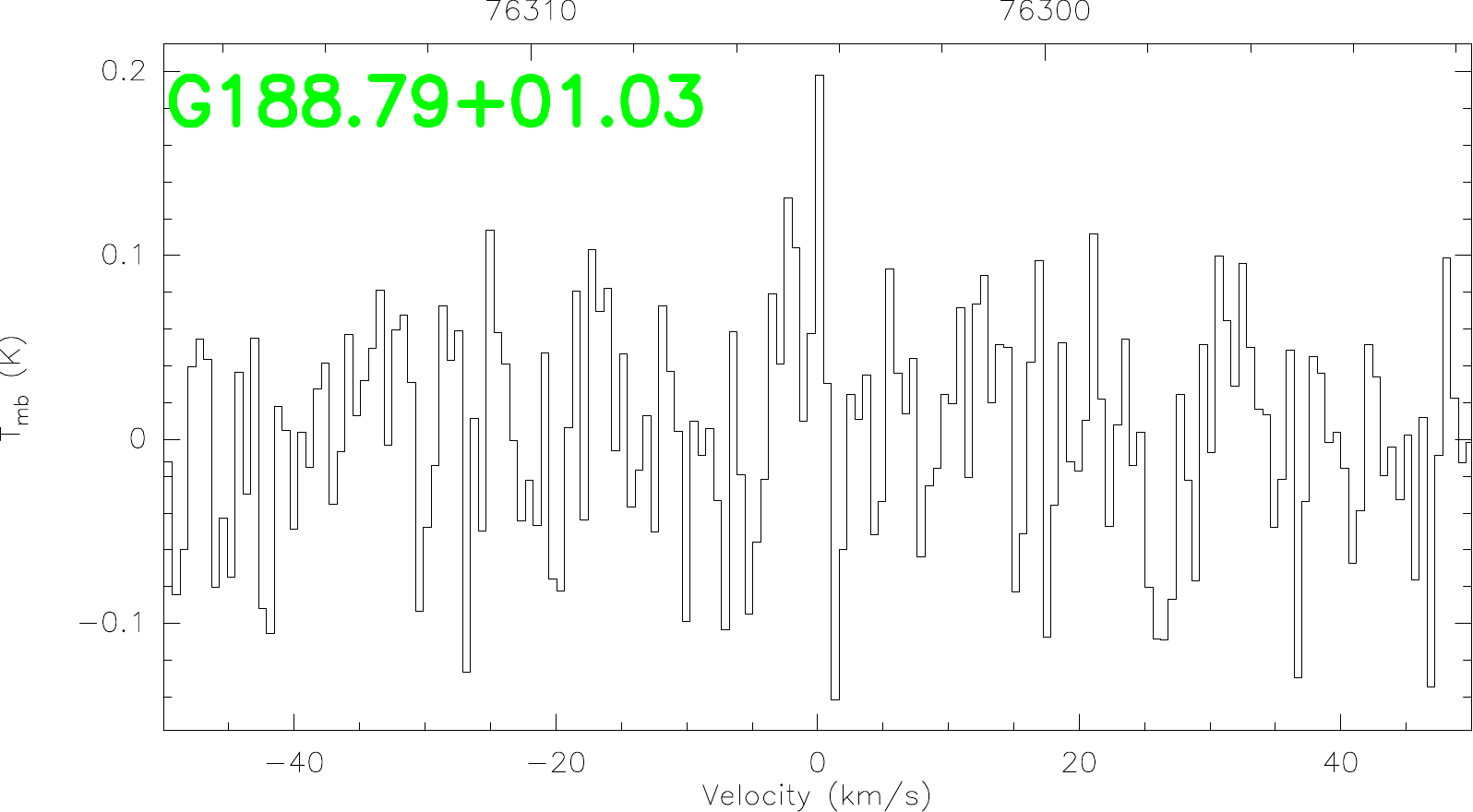}
\includegraphics[width=0.155\textwidth]{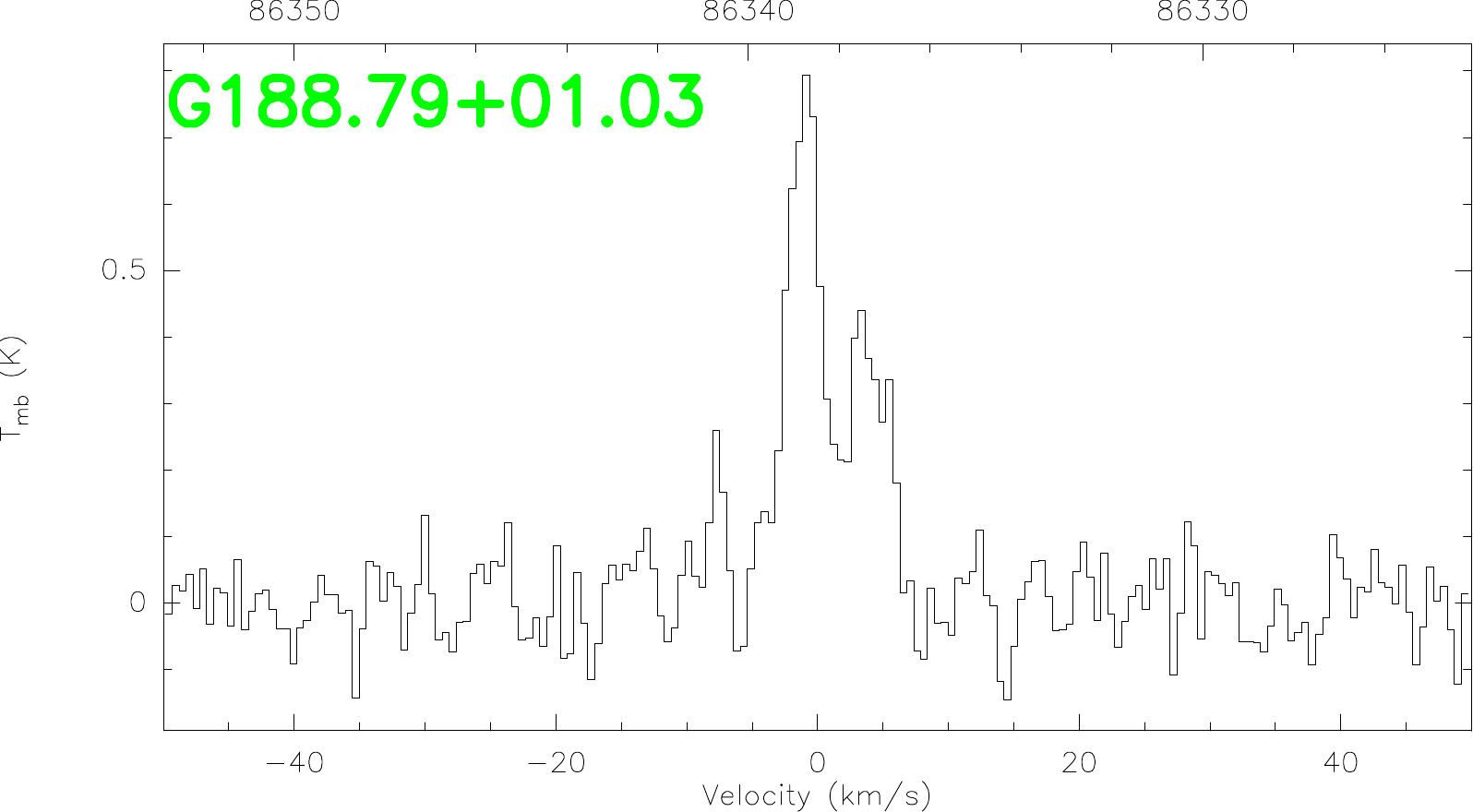}
\includegraphics[width=0.155\textwidth]{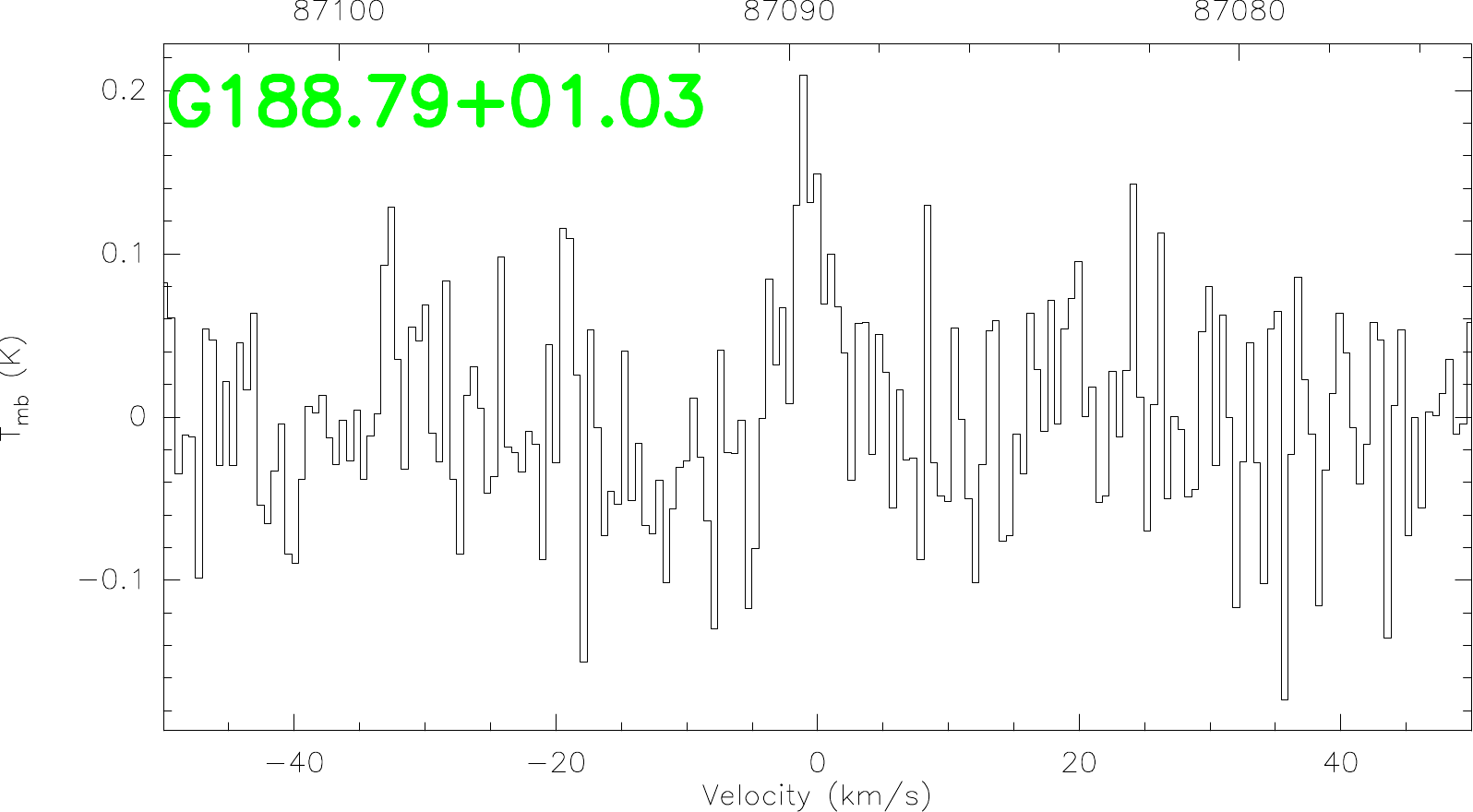}
\includegraphics[width=0.155\textwidth]{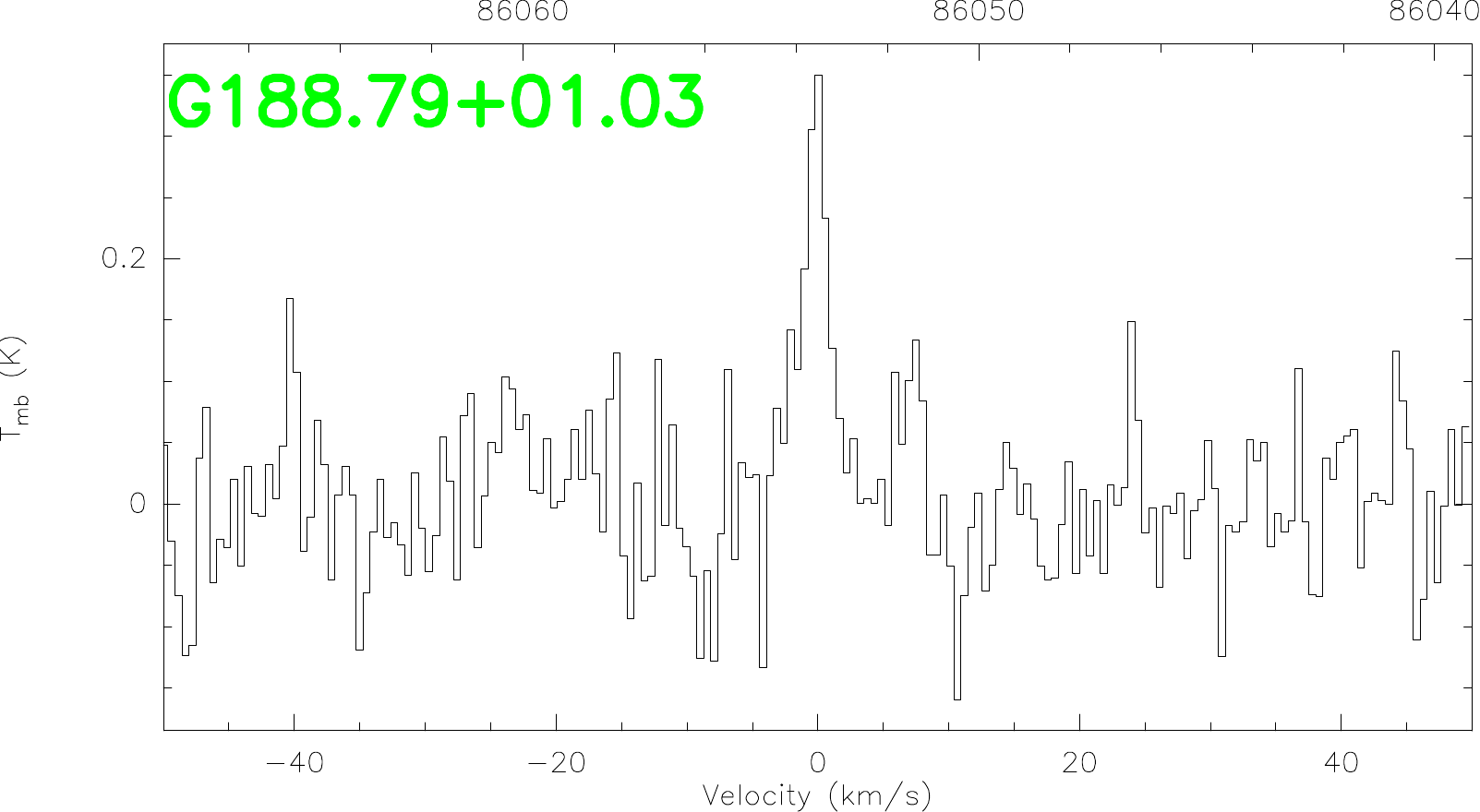}
\includegraphics[width=0.155\textwidth]{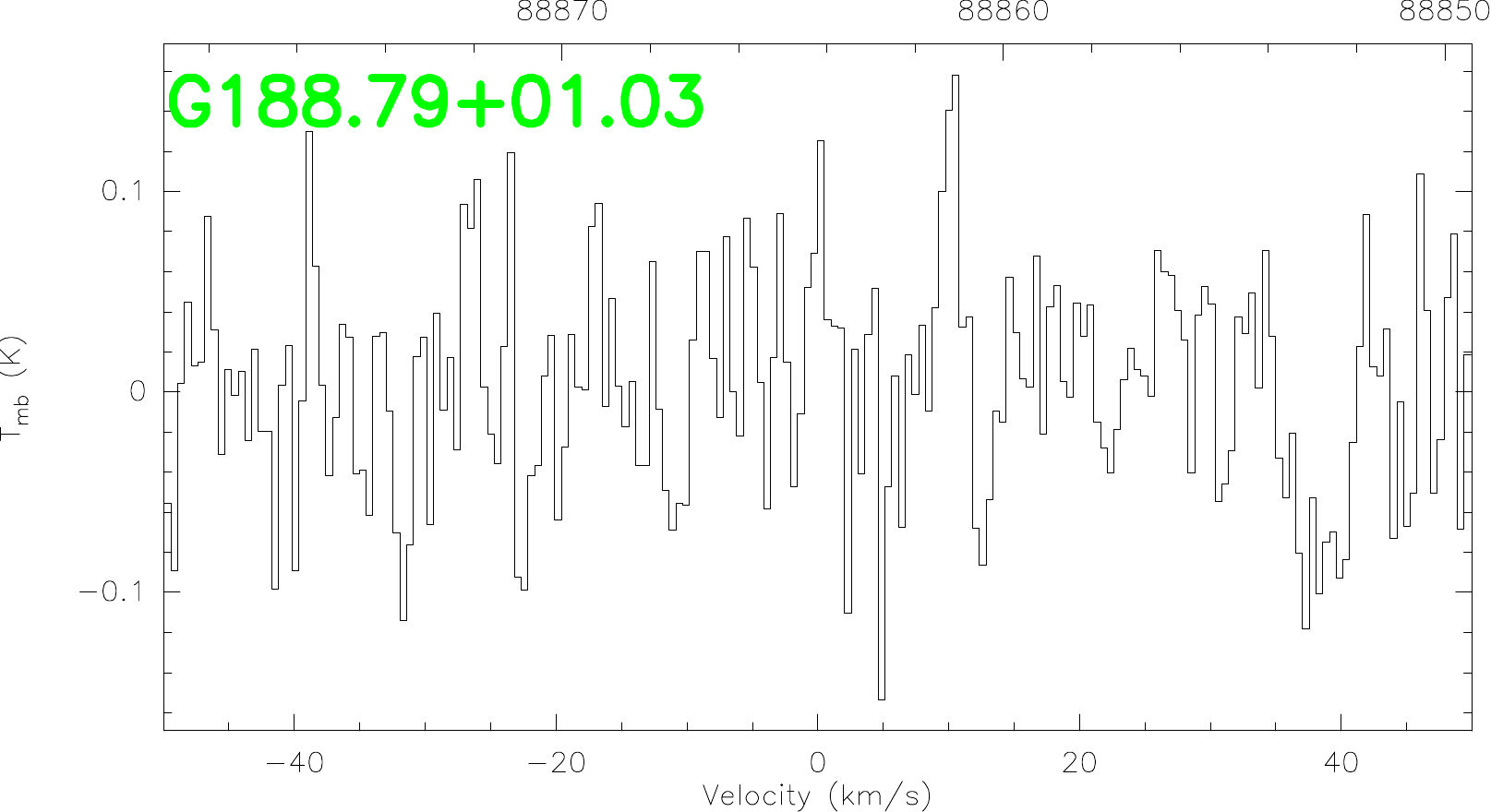}\\

\caption{Continued. Spectra of different molecular lines.}
\label{fig:multi_lines_page6}
\end{figure}

\begin{figure}[h]
\centering

\makebox[\colw][c]{DCN 1-0}
\makebox[\colw][c]{DNC 1-0}
\makebox[\colw][c]{H$^{13}$CN 1-0}
\makebox[\colw][c]{HN$^{13}$C 1-0}
\makebox[\colw][c]{HC$^{15}$N 1-0}
\makebox[\colw][c]{H$^{15}$NC 1-0}\\[0.2em]

\includegraphics[width=0.155\textwidth]{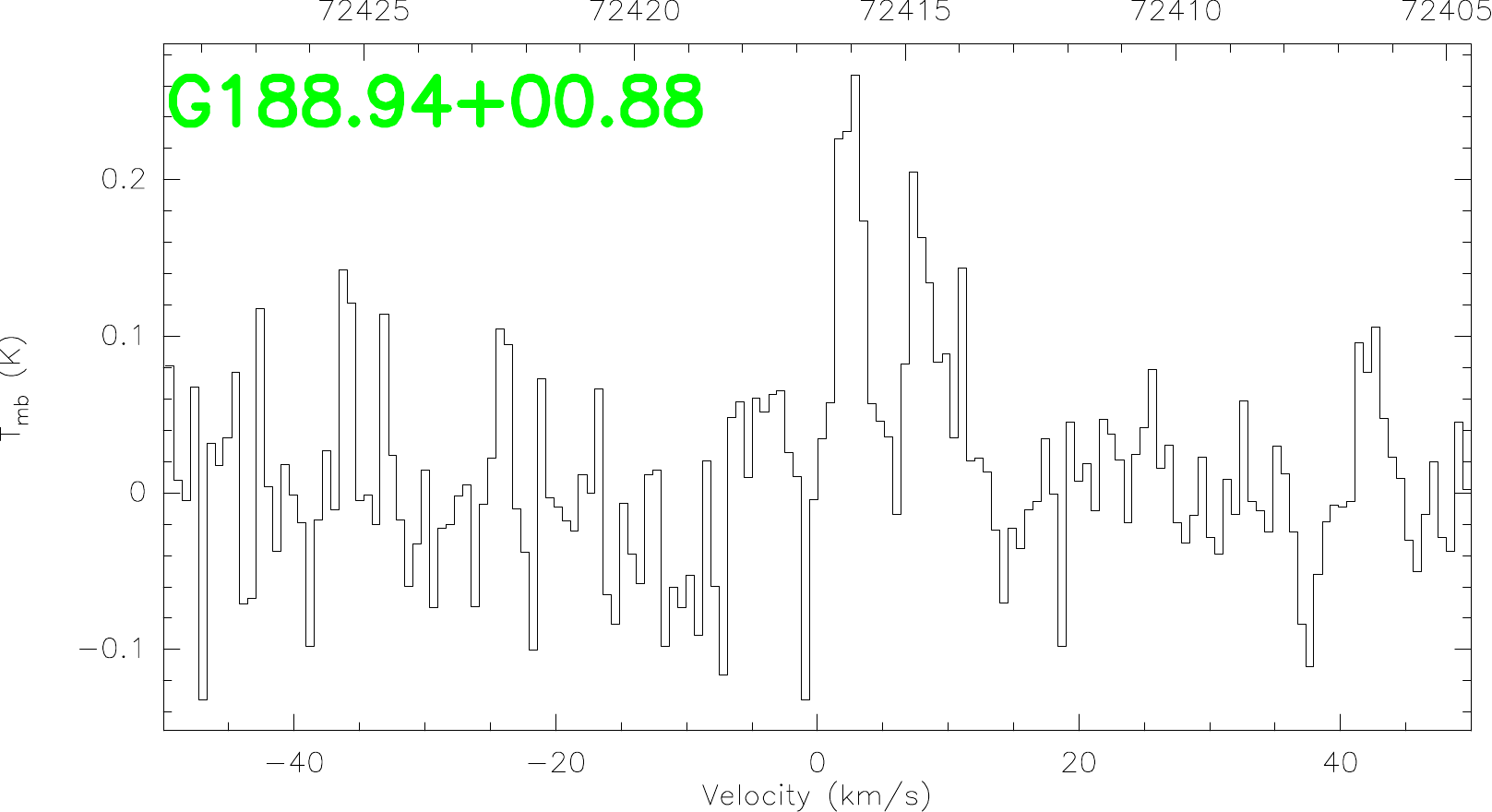}
\includegraphics[width=0.155\textwidth]{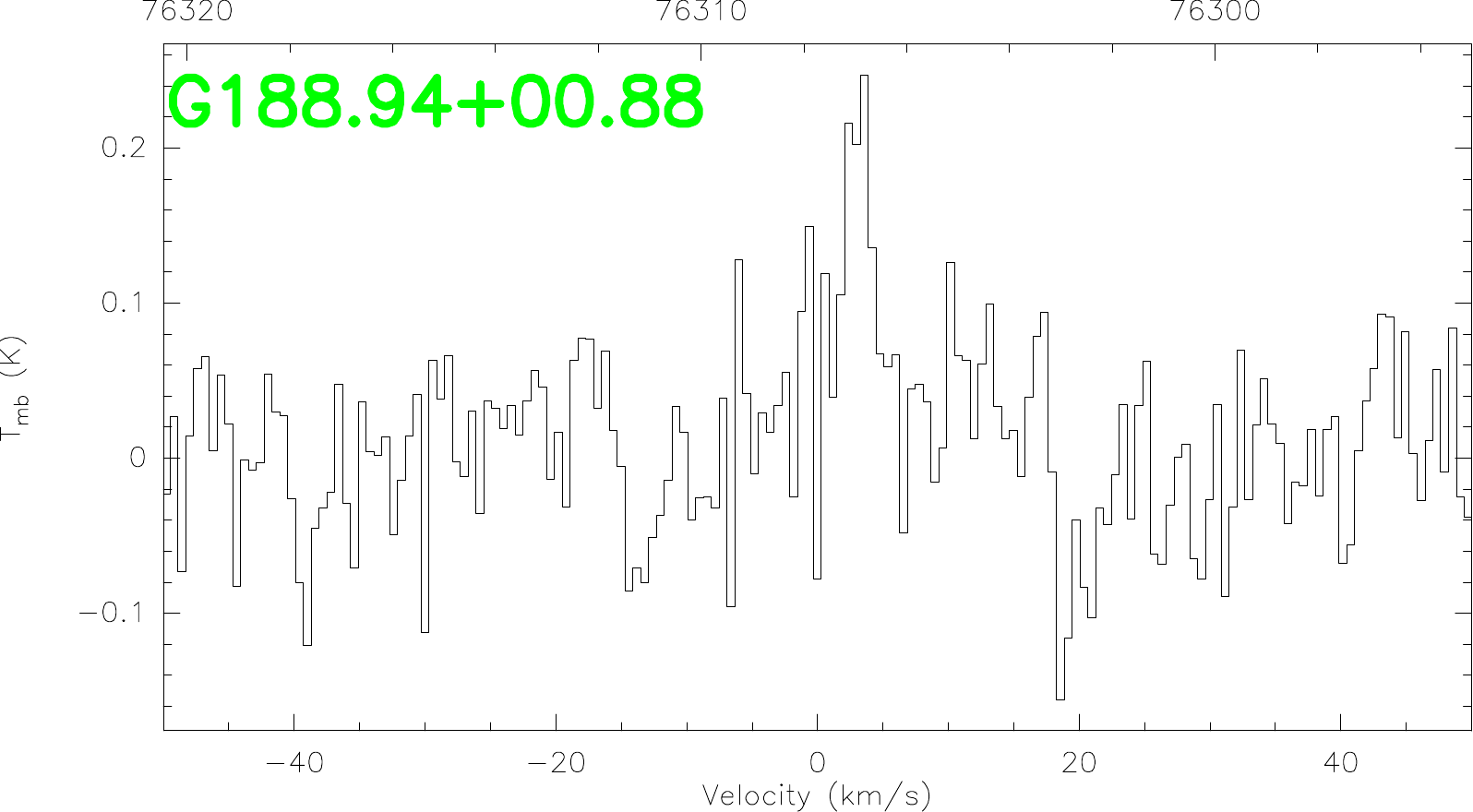}
\includegraphics[width=0.155\textwidth]{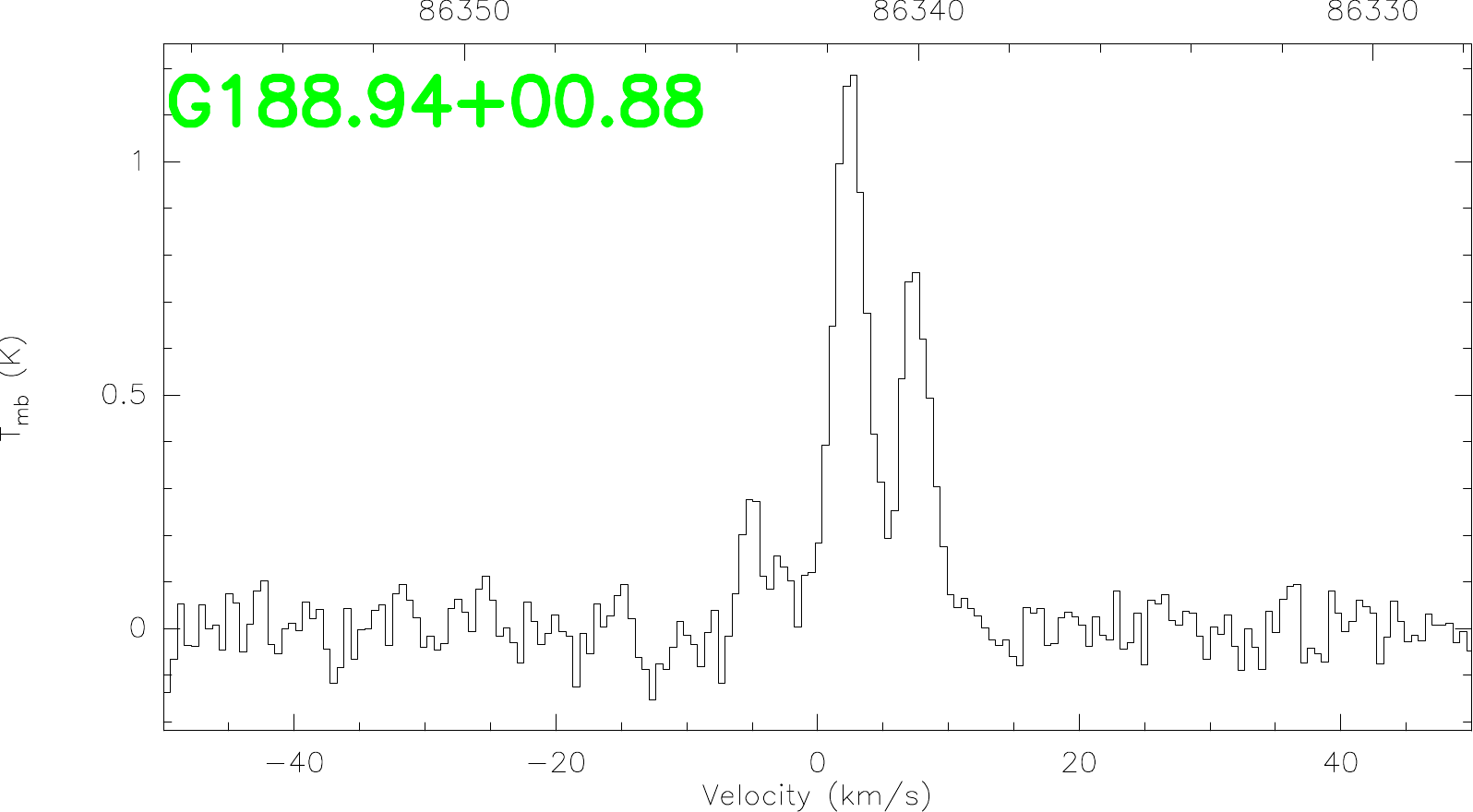}
\includegraphics[width=0.155\textwidth]{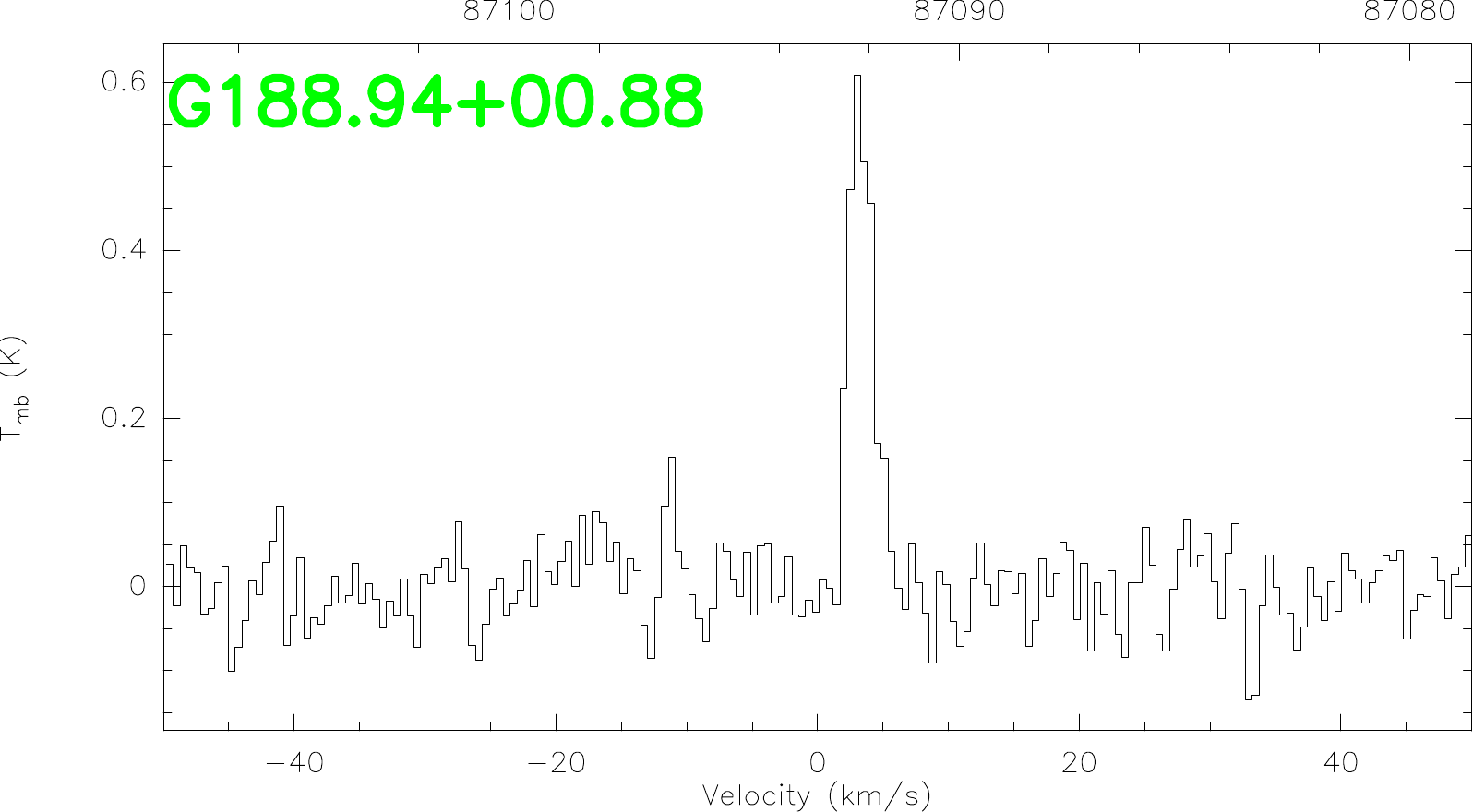}
\includegraphics[width=0.155\textwidth]{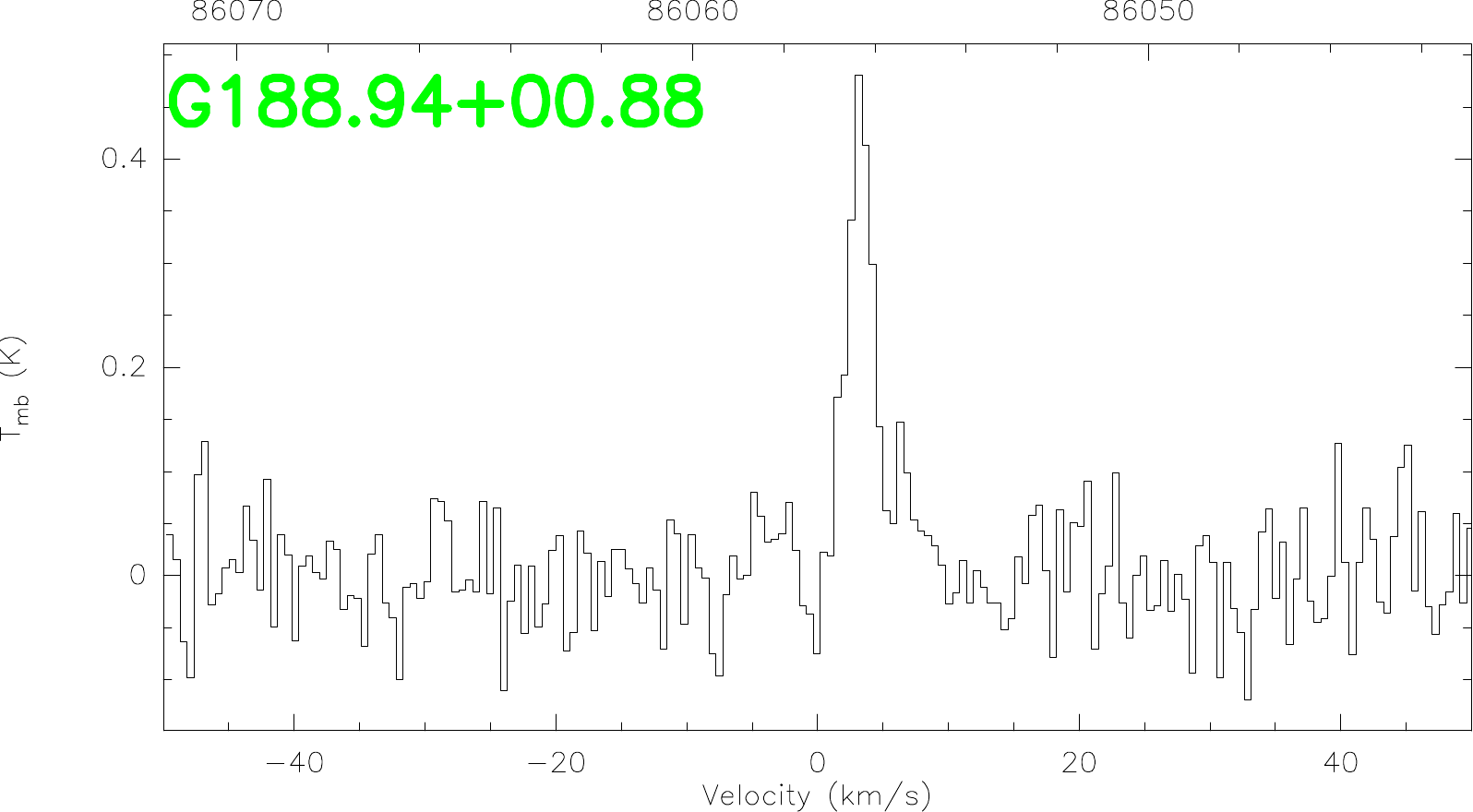}
\includegraphics[width=0.155\textwidth]{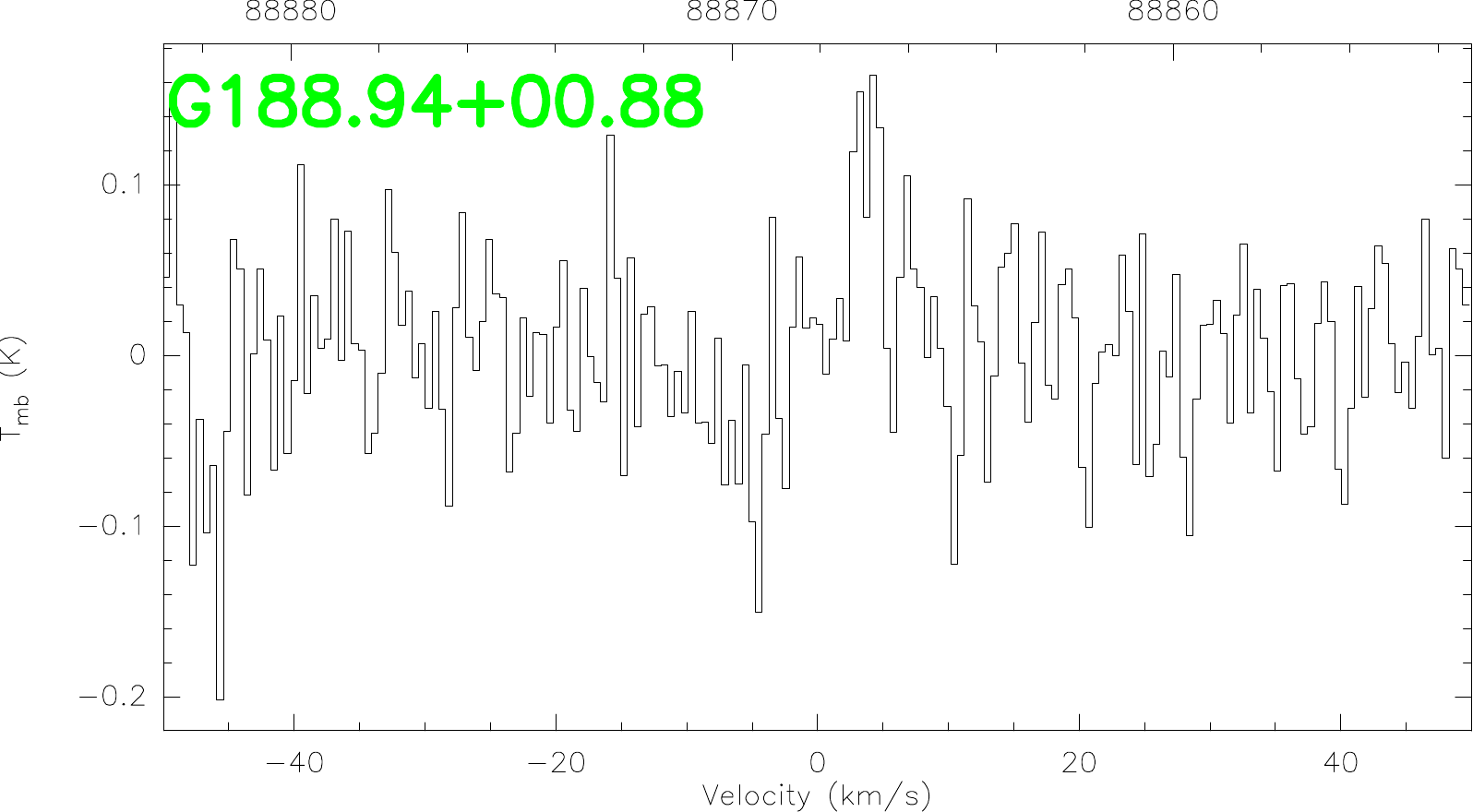}\\

\includegraphics[width=0.155\textwidth]{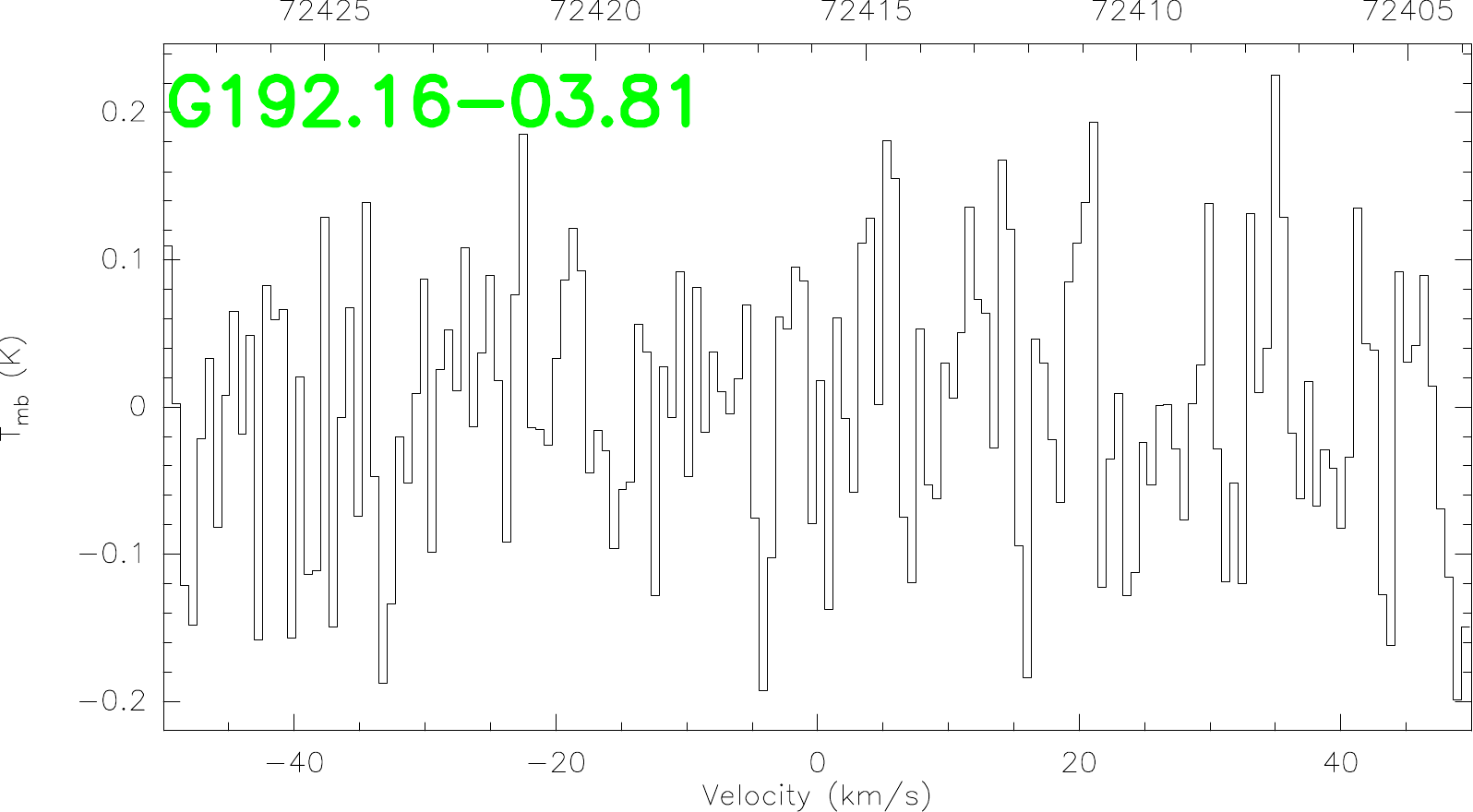}
\includegraphics[width=0.155\textwidth]{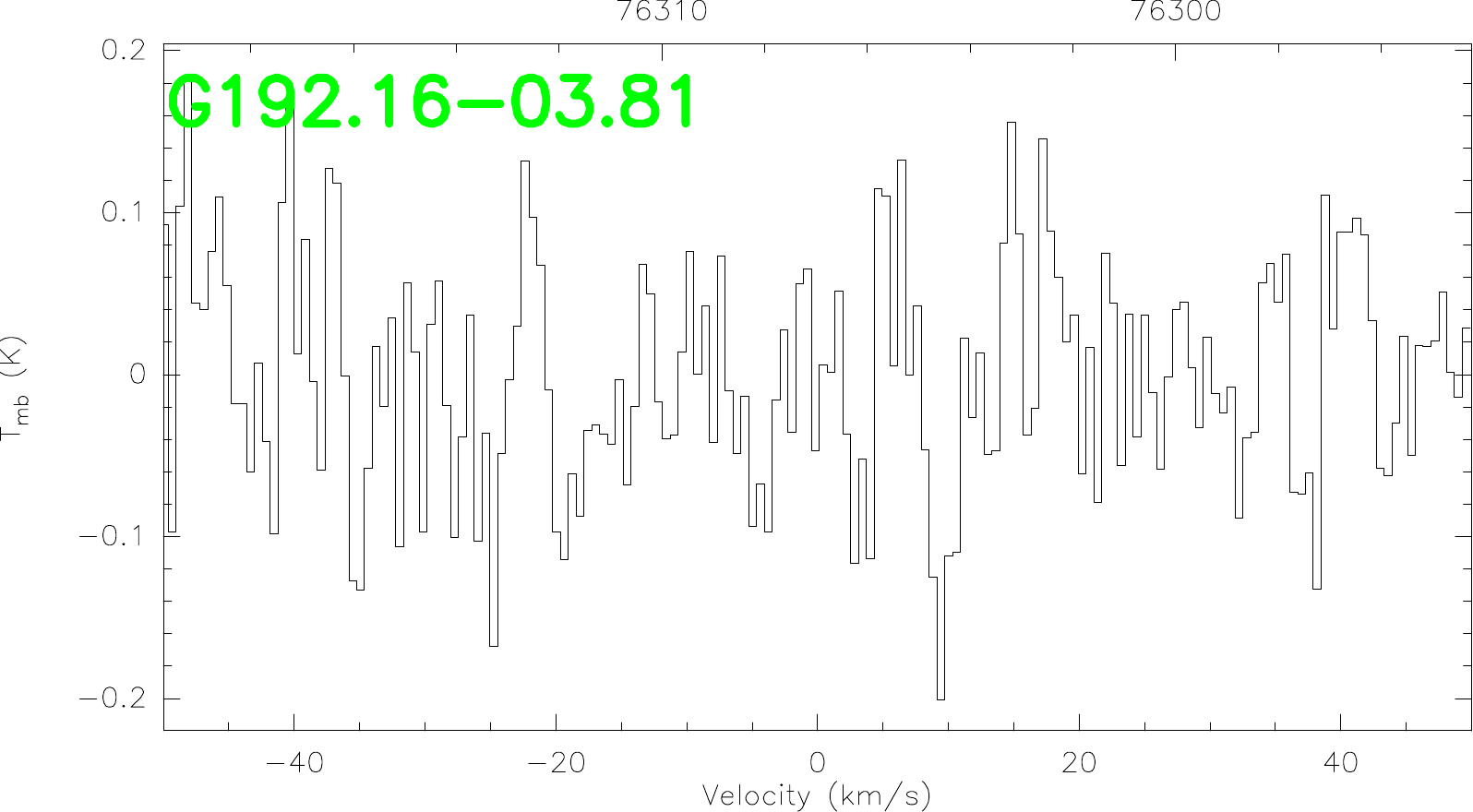}
\includegraphics[width=0.155\textwidth]{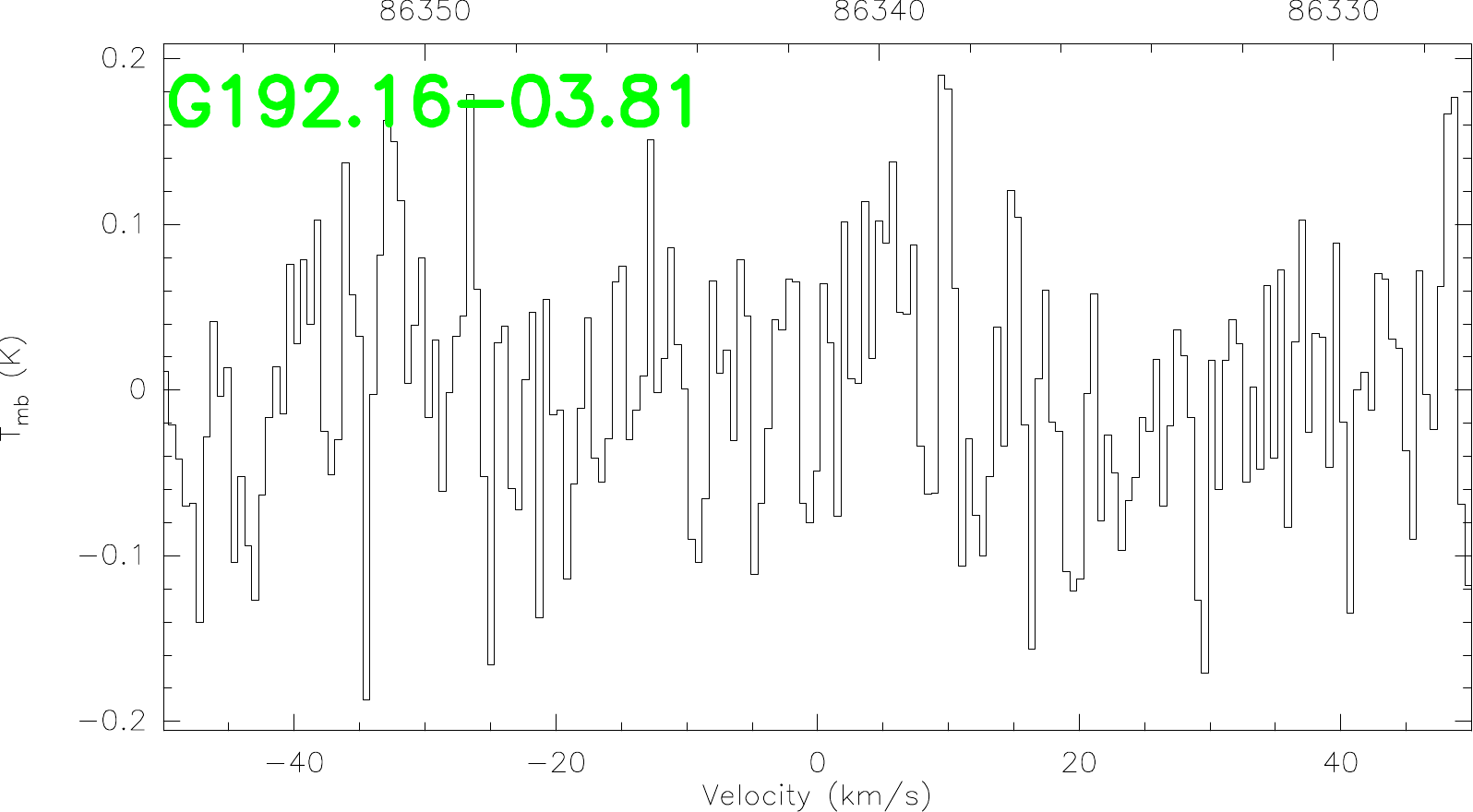}
\includegraphics[width=0.155\textwidth]{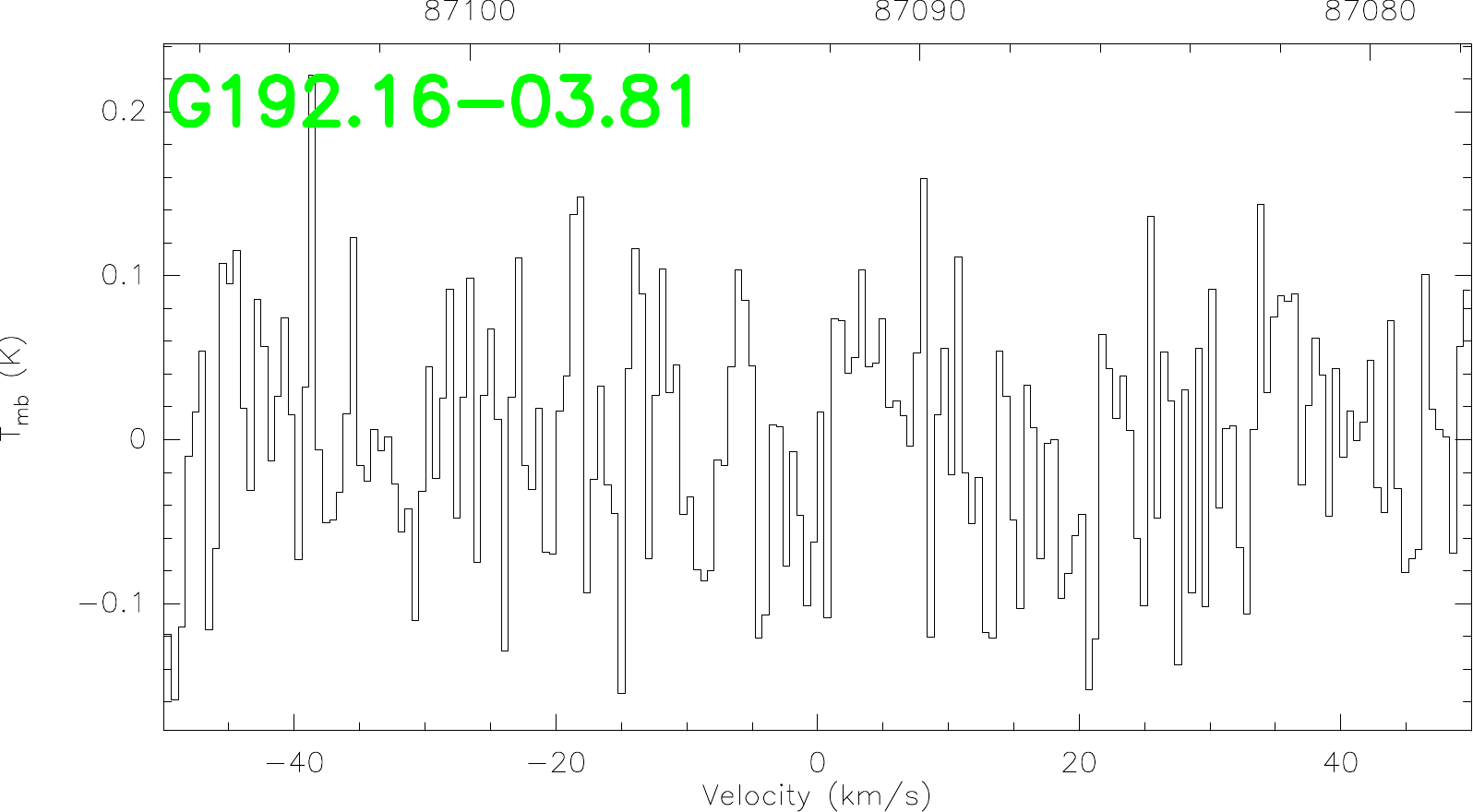}
\includegraphics[width=0.155\textwidth]{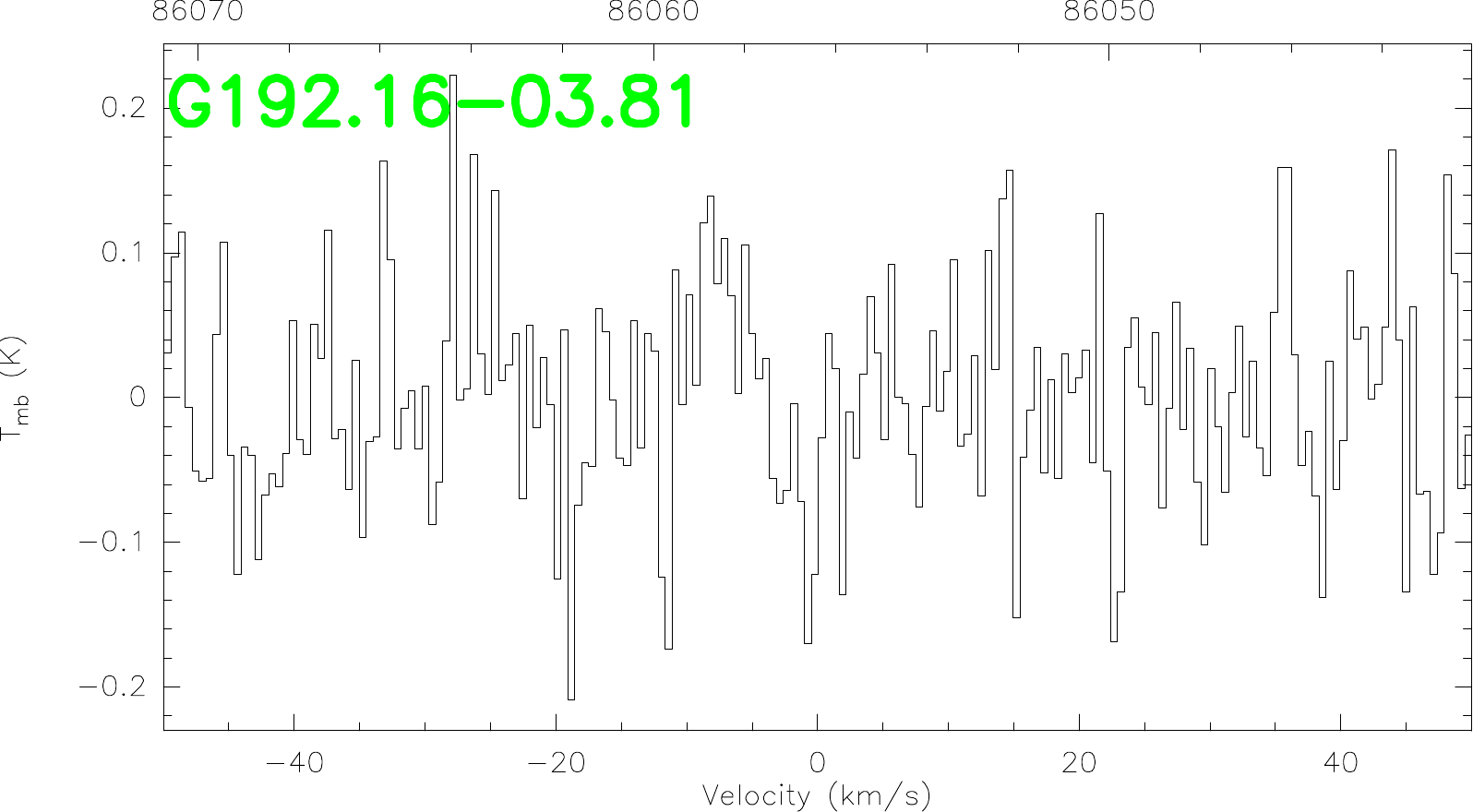}
\includegraphics[width=0.155\textwidth]{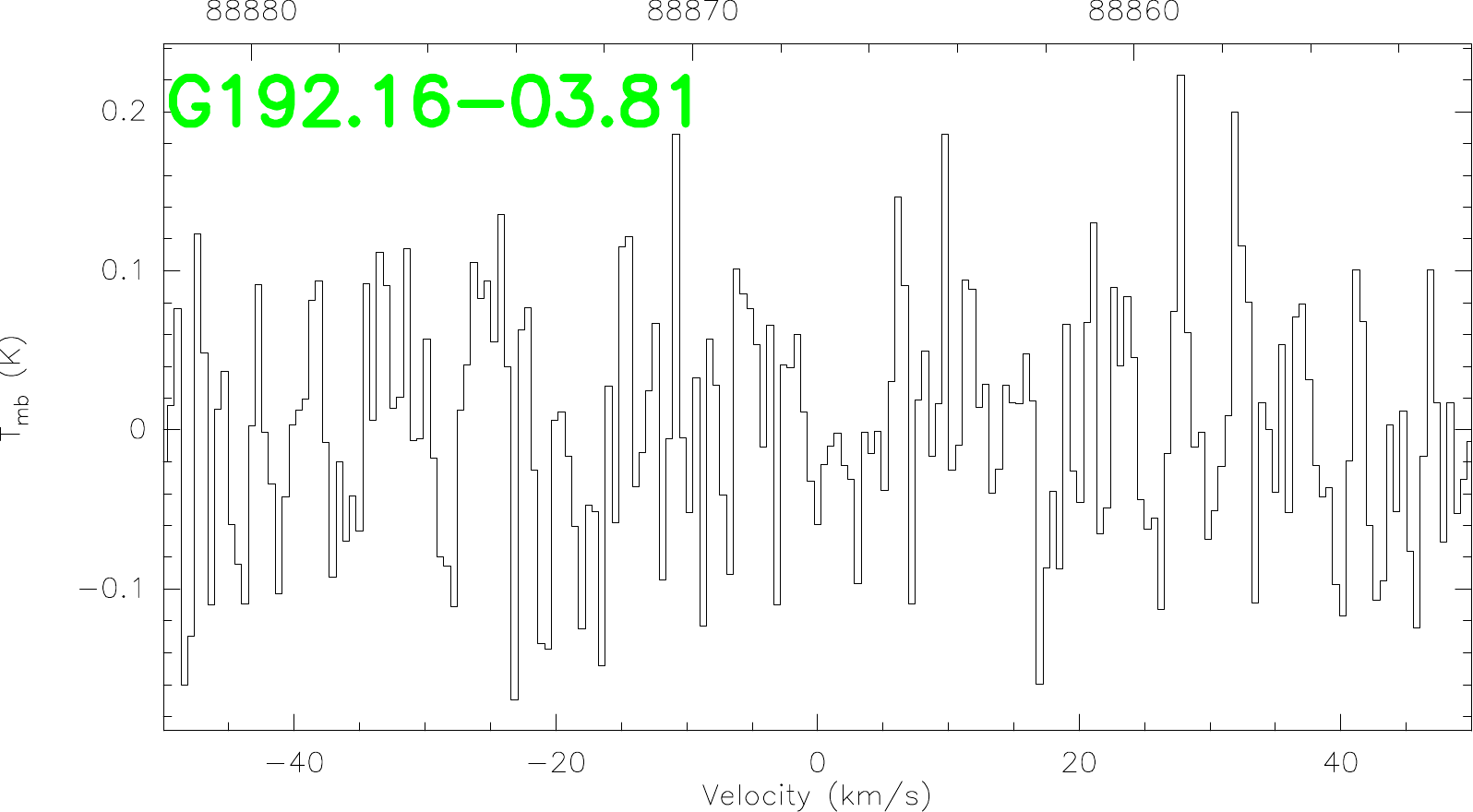}\\

\includegraphics[width=0.155\textwidth]{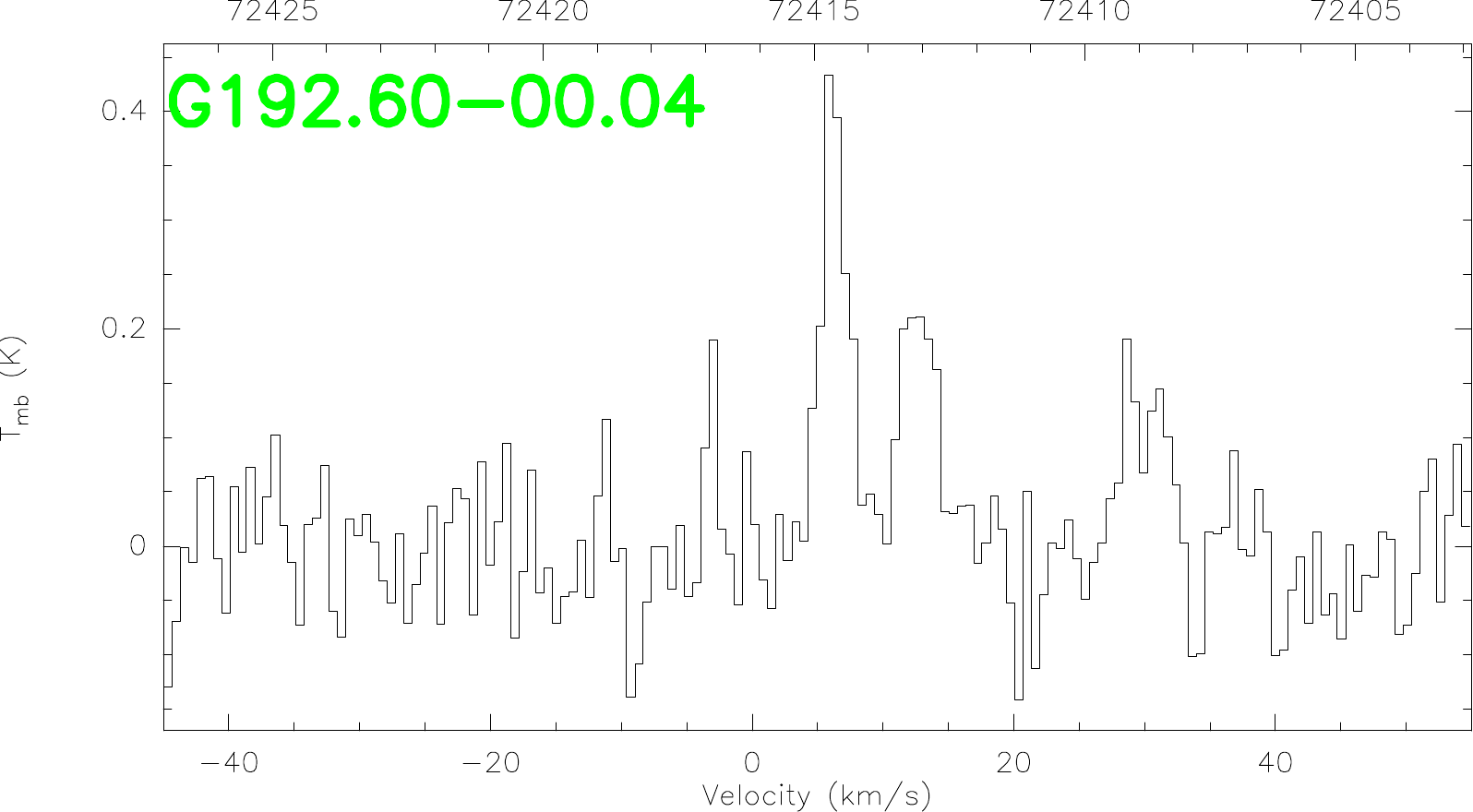}
\includegraphics[width=0.155\textwidth]{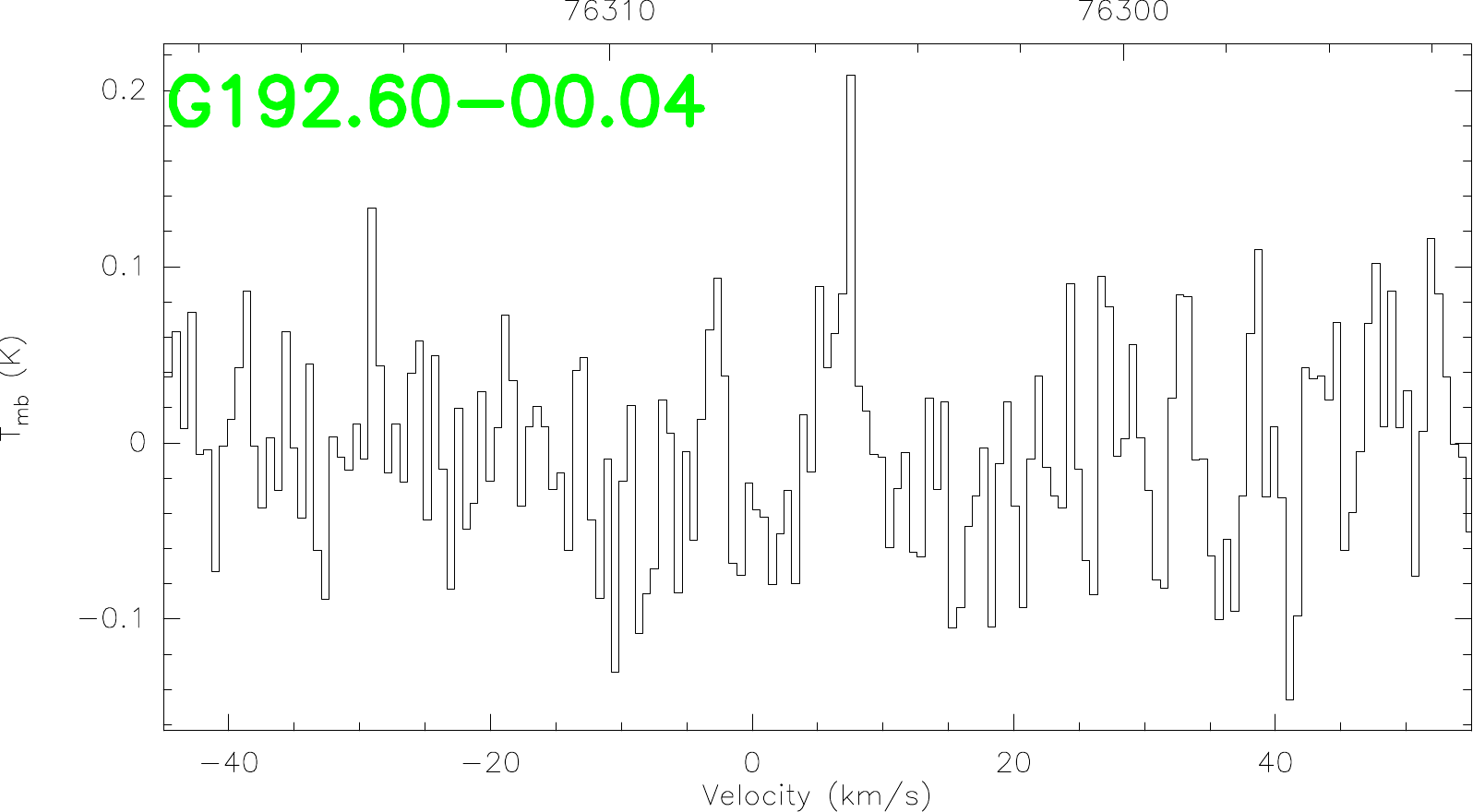}
\includegraphics[width=0.155\textwidth]{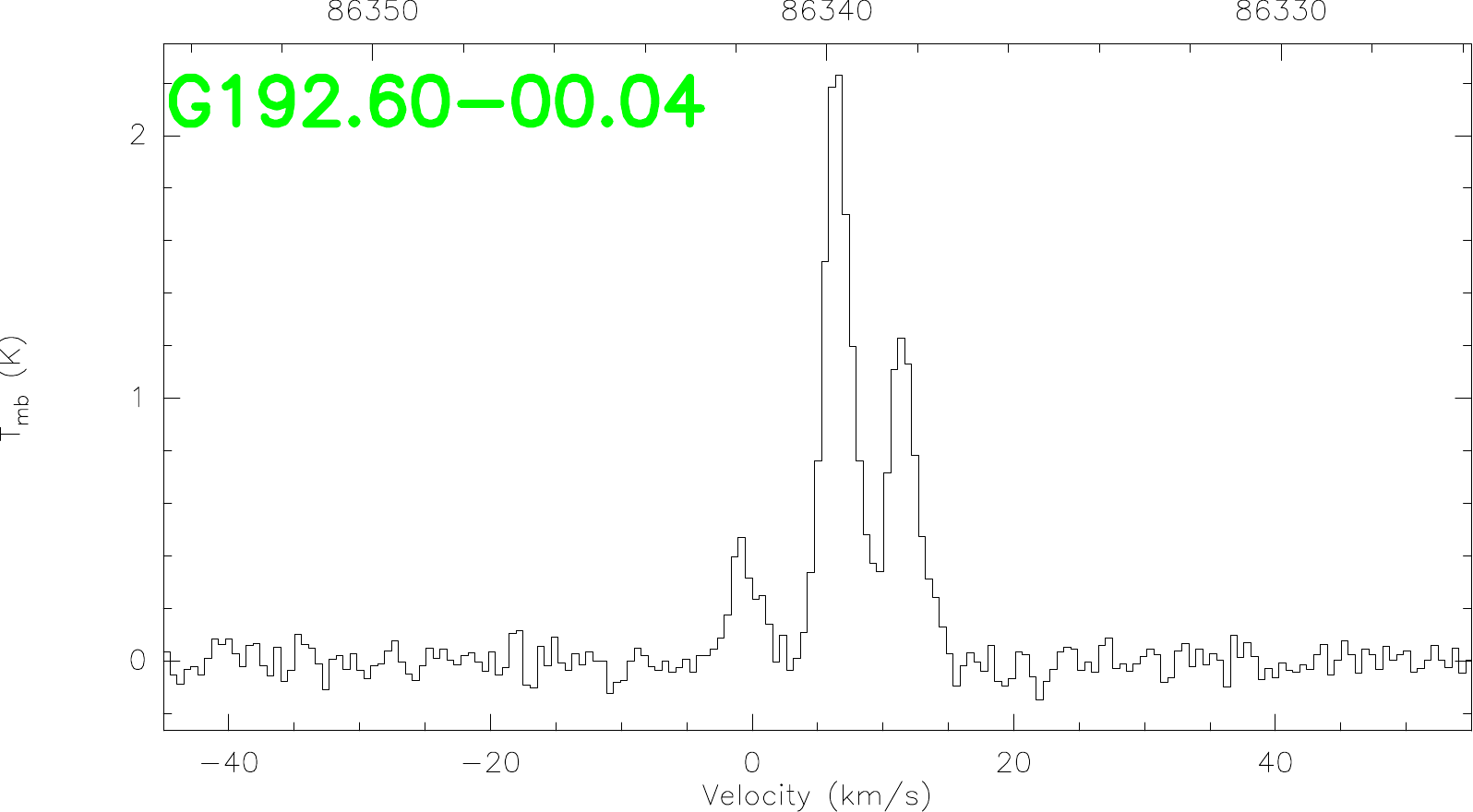}
\includegraphics[width=0.155\textwidth]{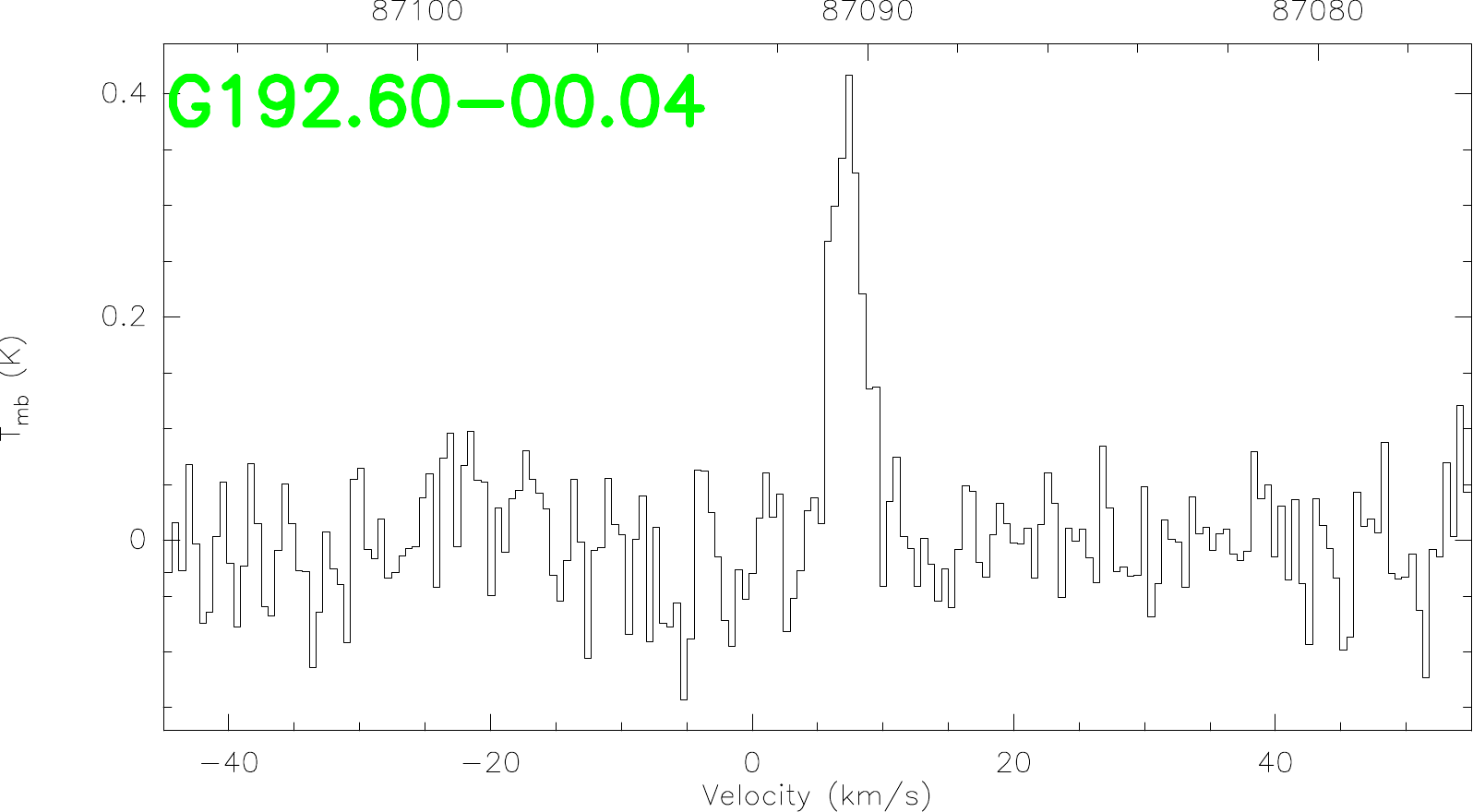}
\includegraphics[width=0.155\textwidth]{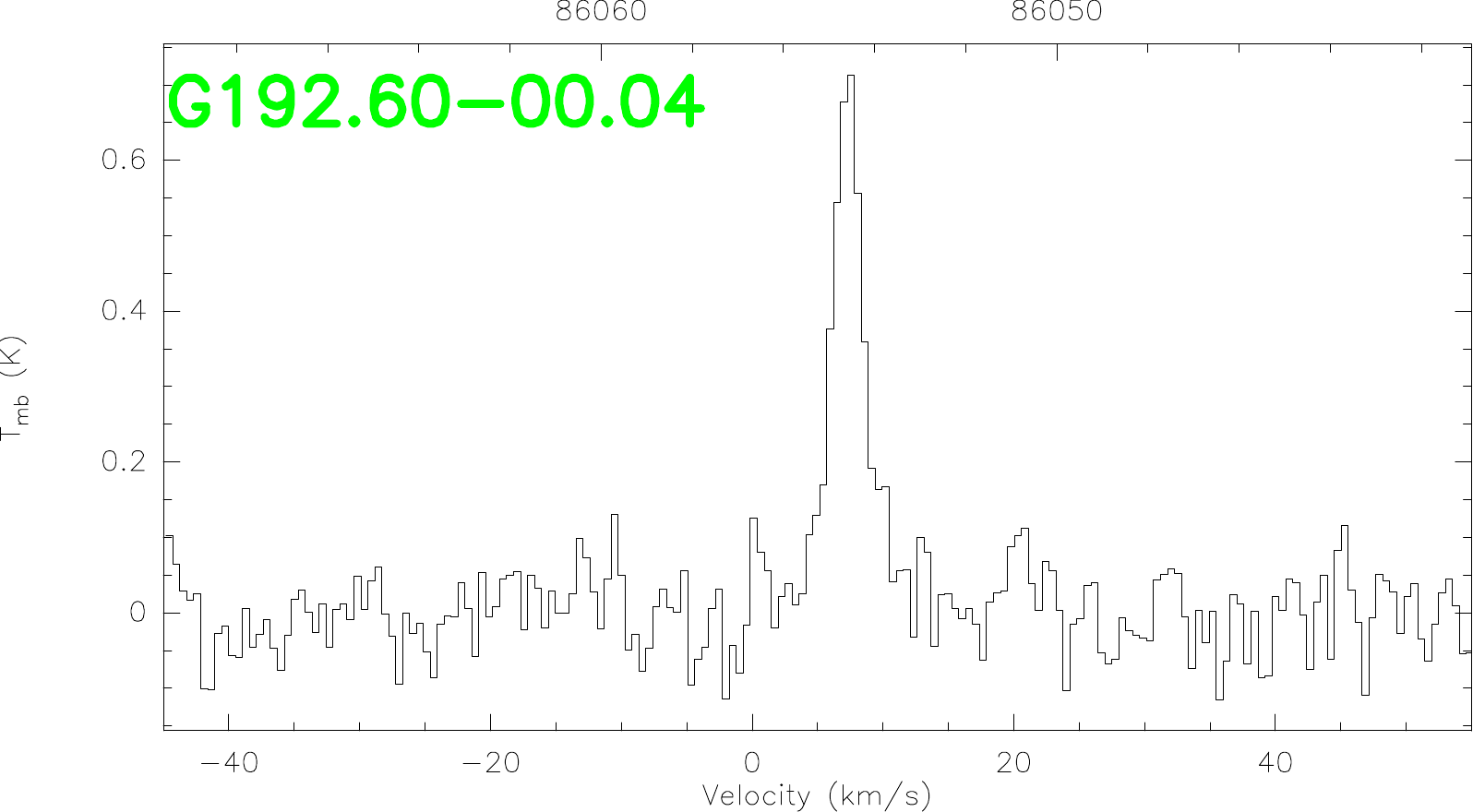}
\includegraphics[width=0.155\textwidth]{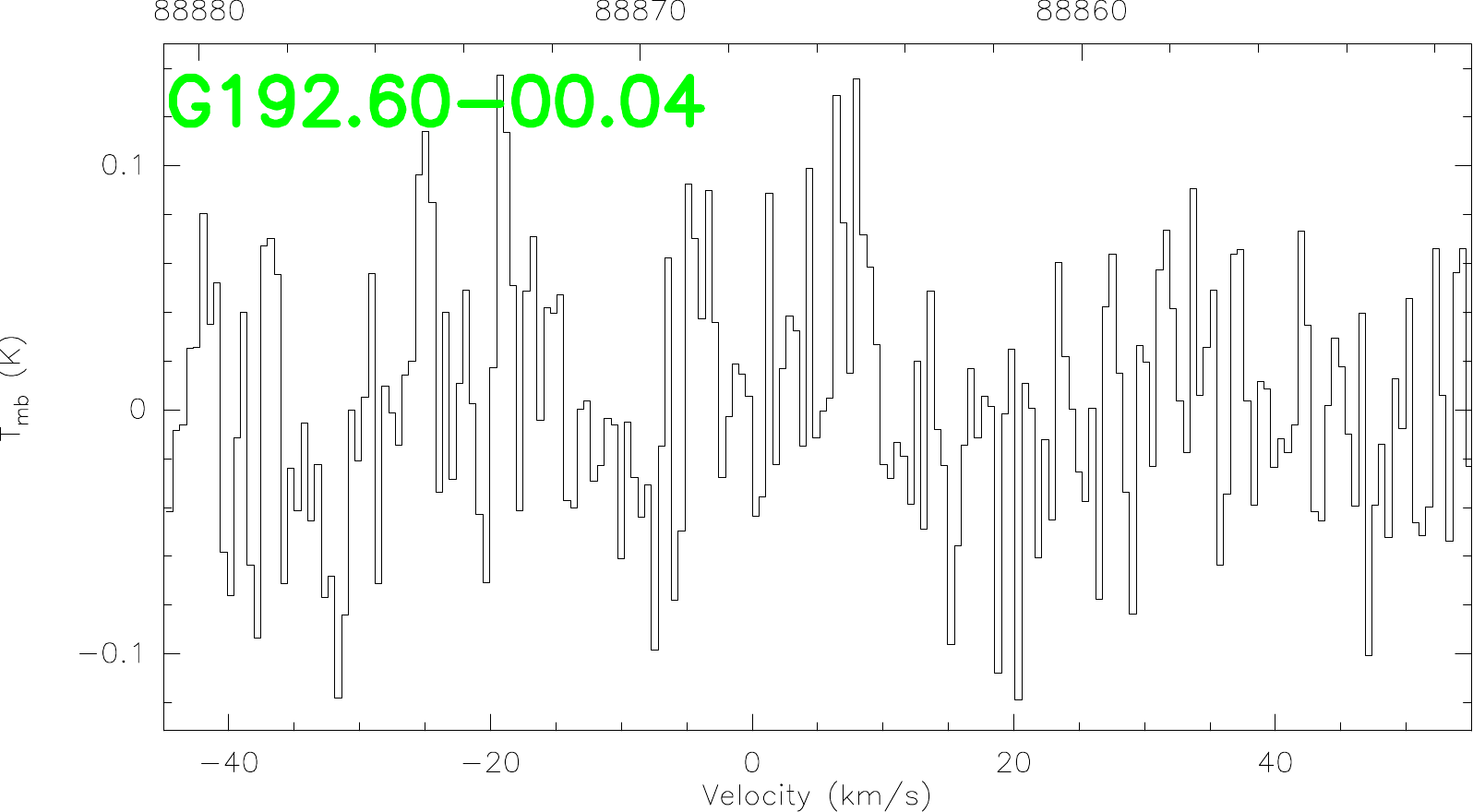}\\

\includegraphics[width=0.155\textwidth]{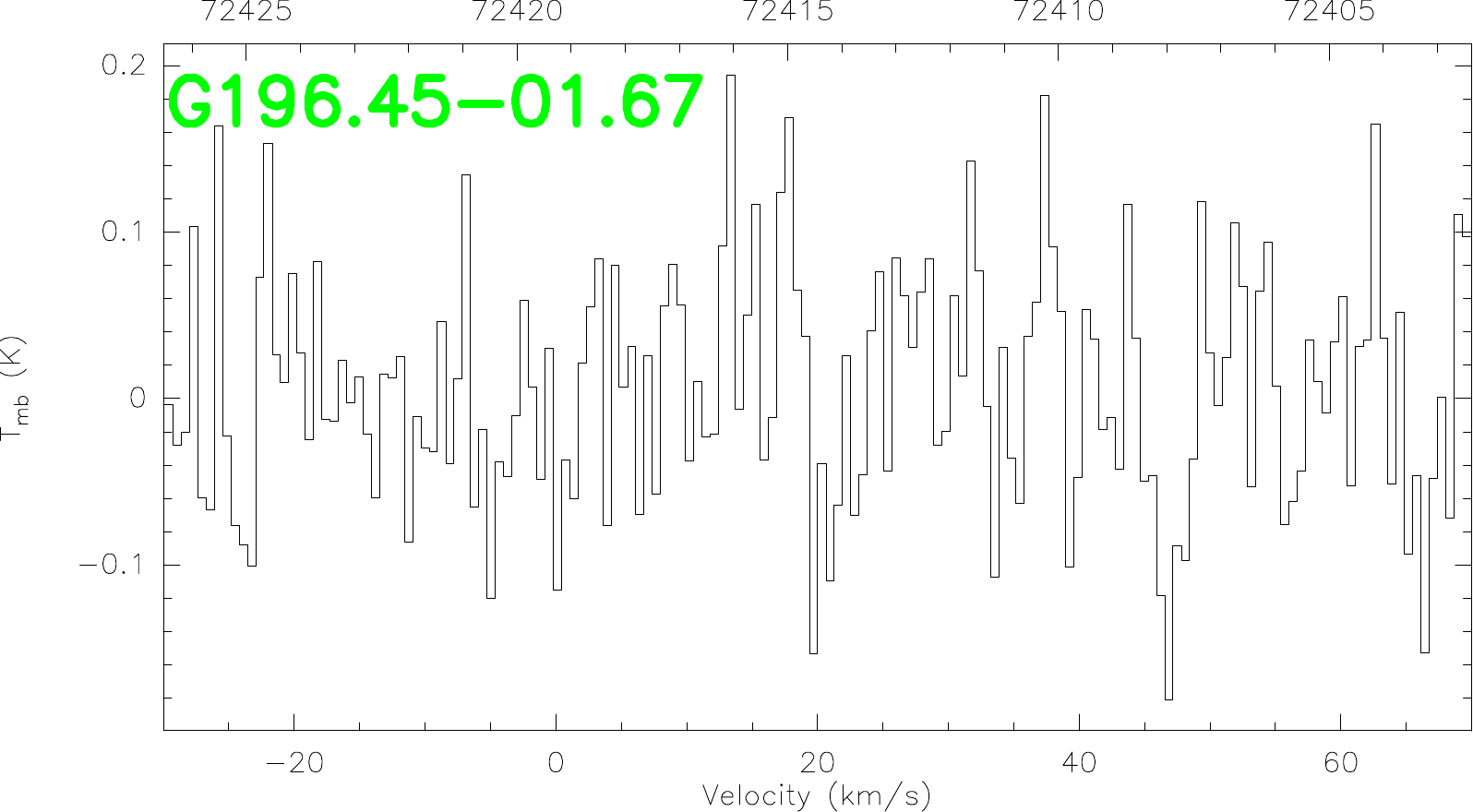}
\includegraphics[width=0.155\textwidth]{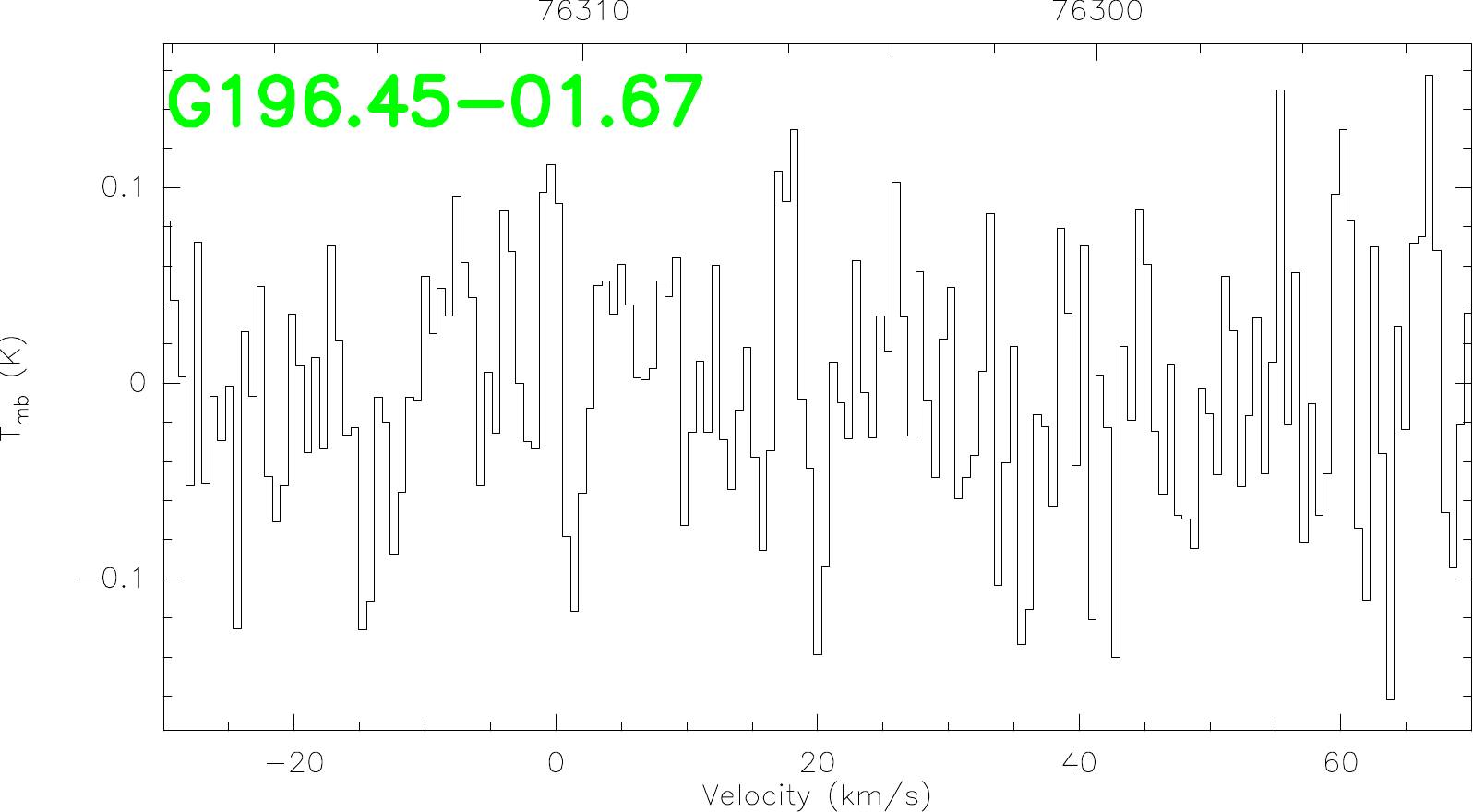}
\includegraphics[width=0.155\textwidth]{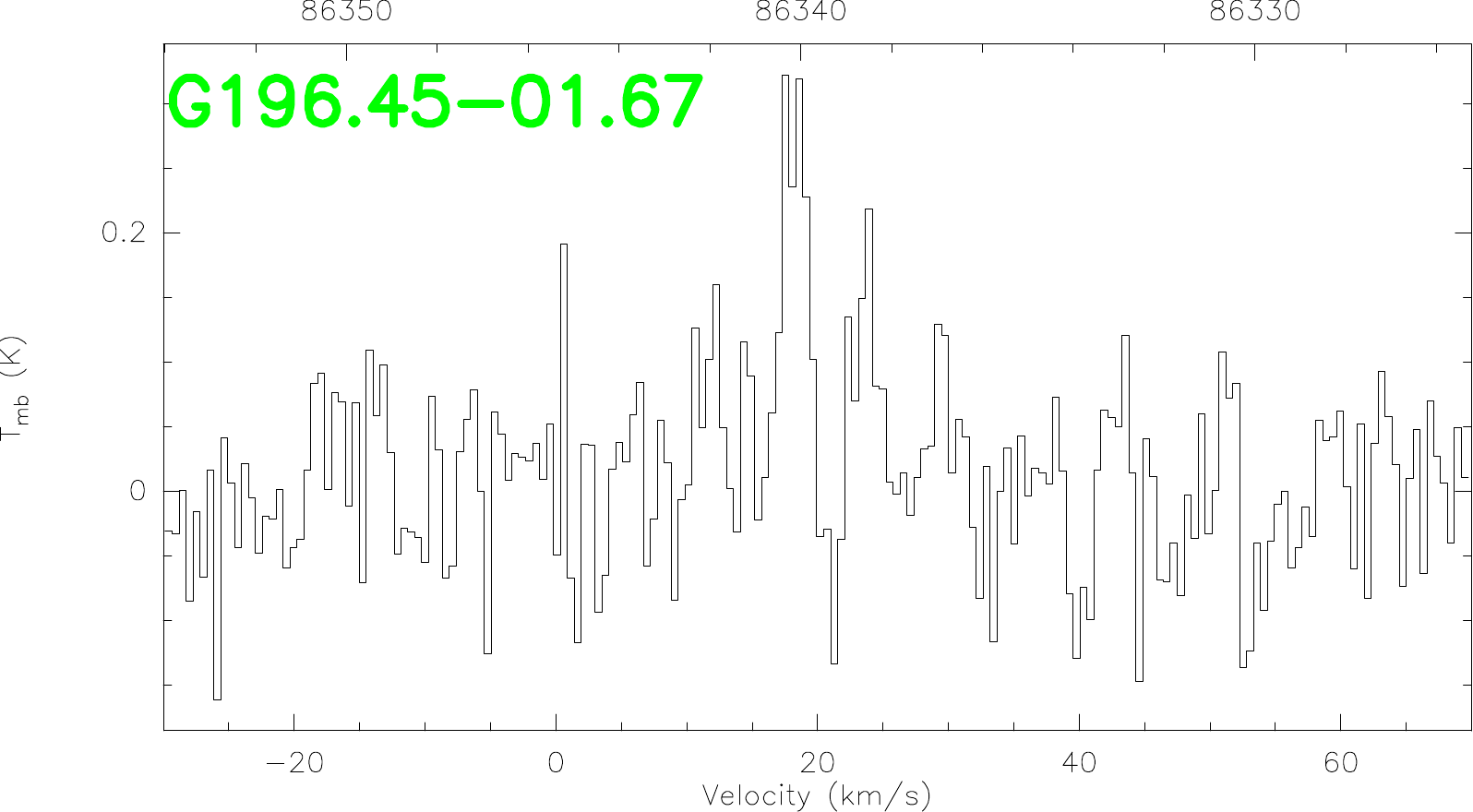}
\includegraphics[width=0.155\textwidth]{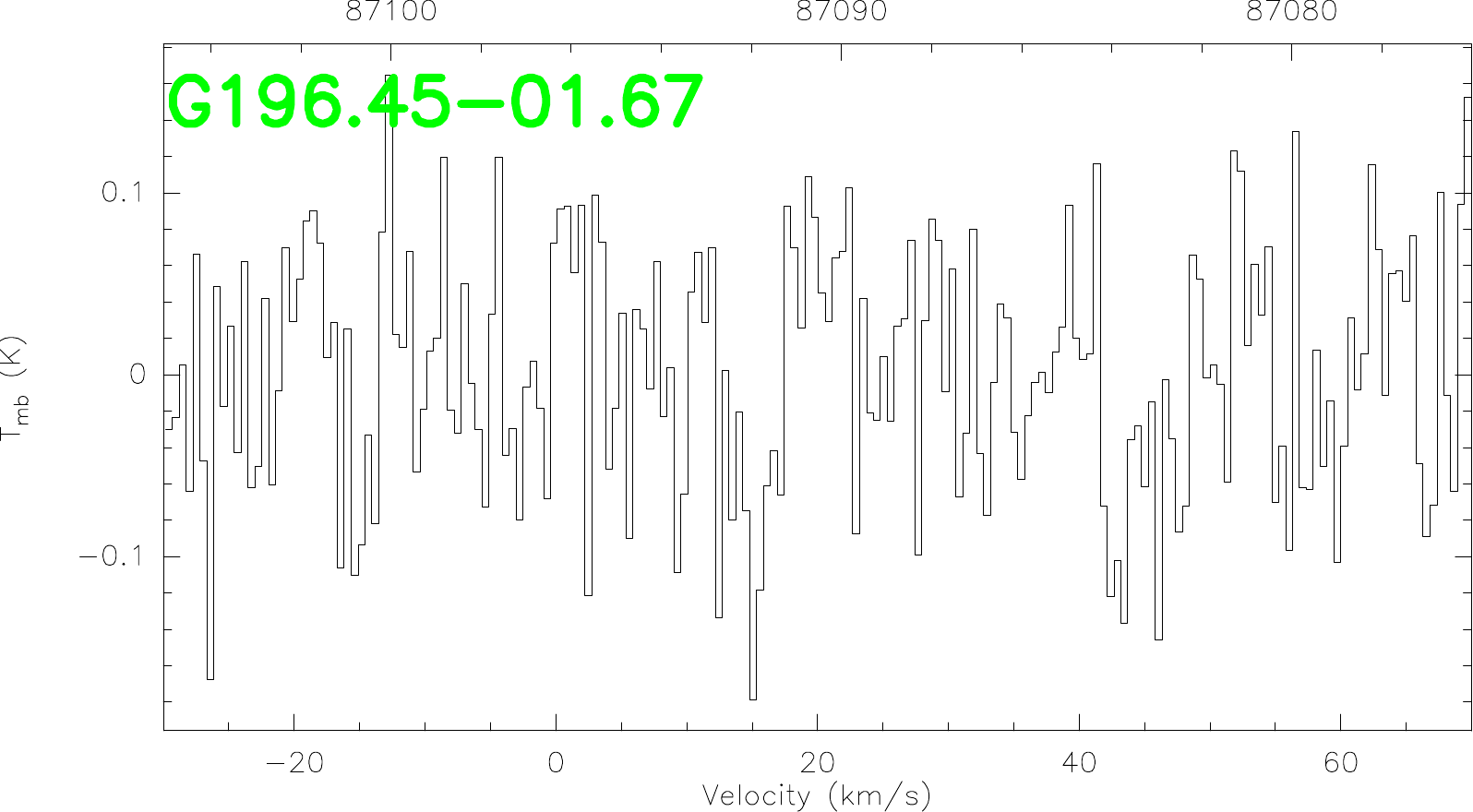}
\includegraphics[width=0.155\textwidth]{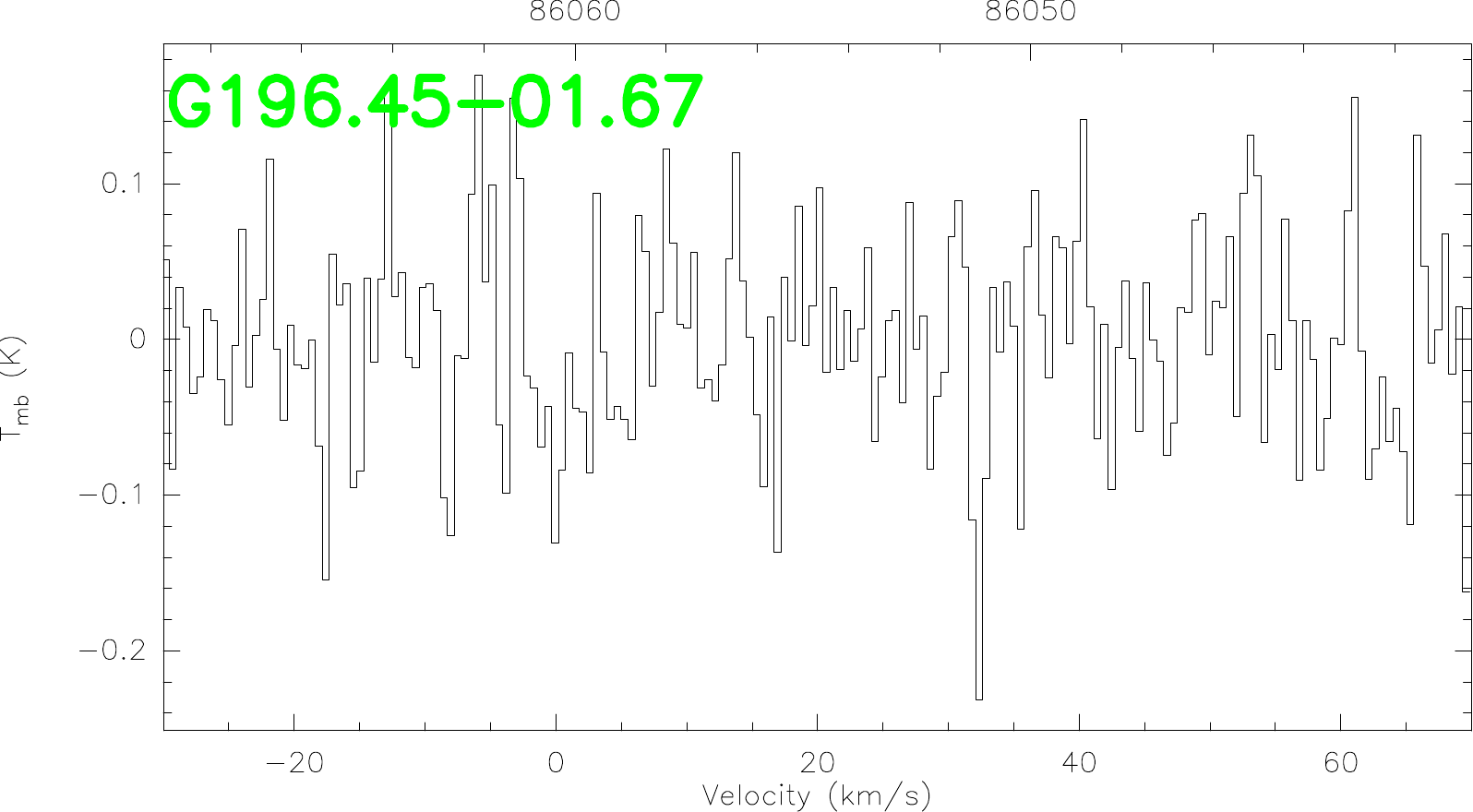}
\includegraphics[width=0.155\textwidth]{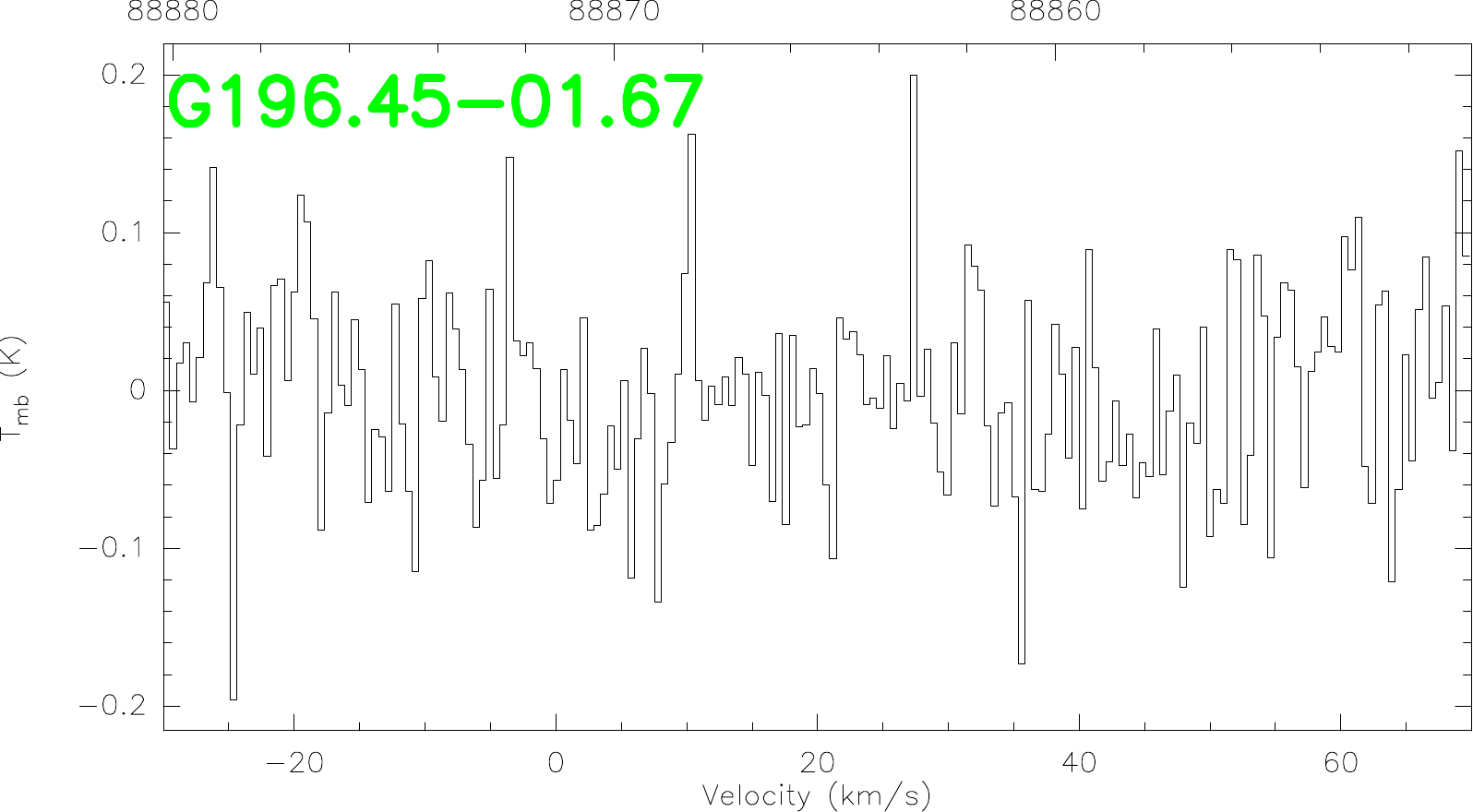}\\

\includegraphics[width=0.155\textwidth]{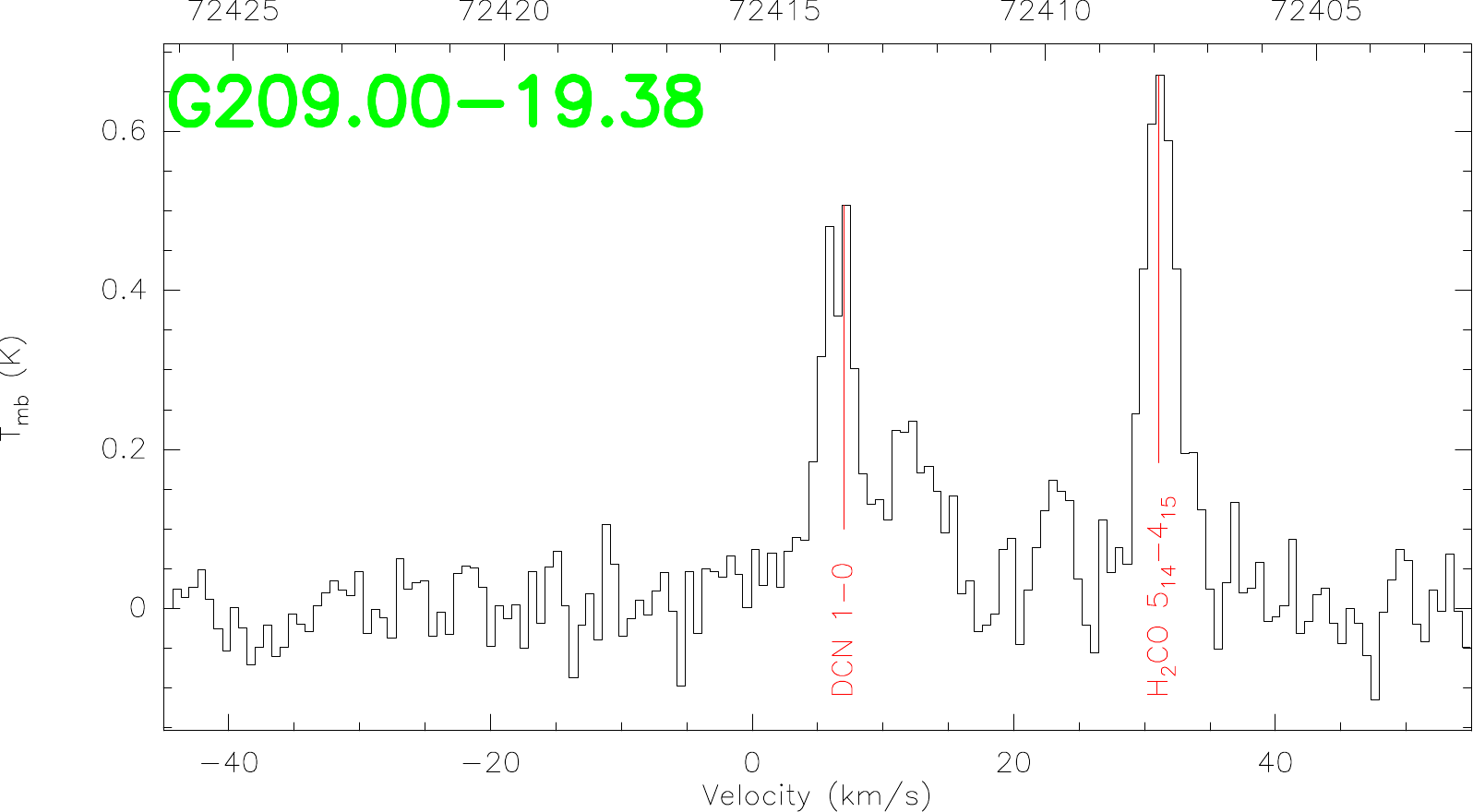}
\includegraphics[width=0.155\textwidth]{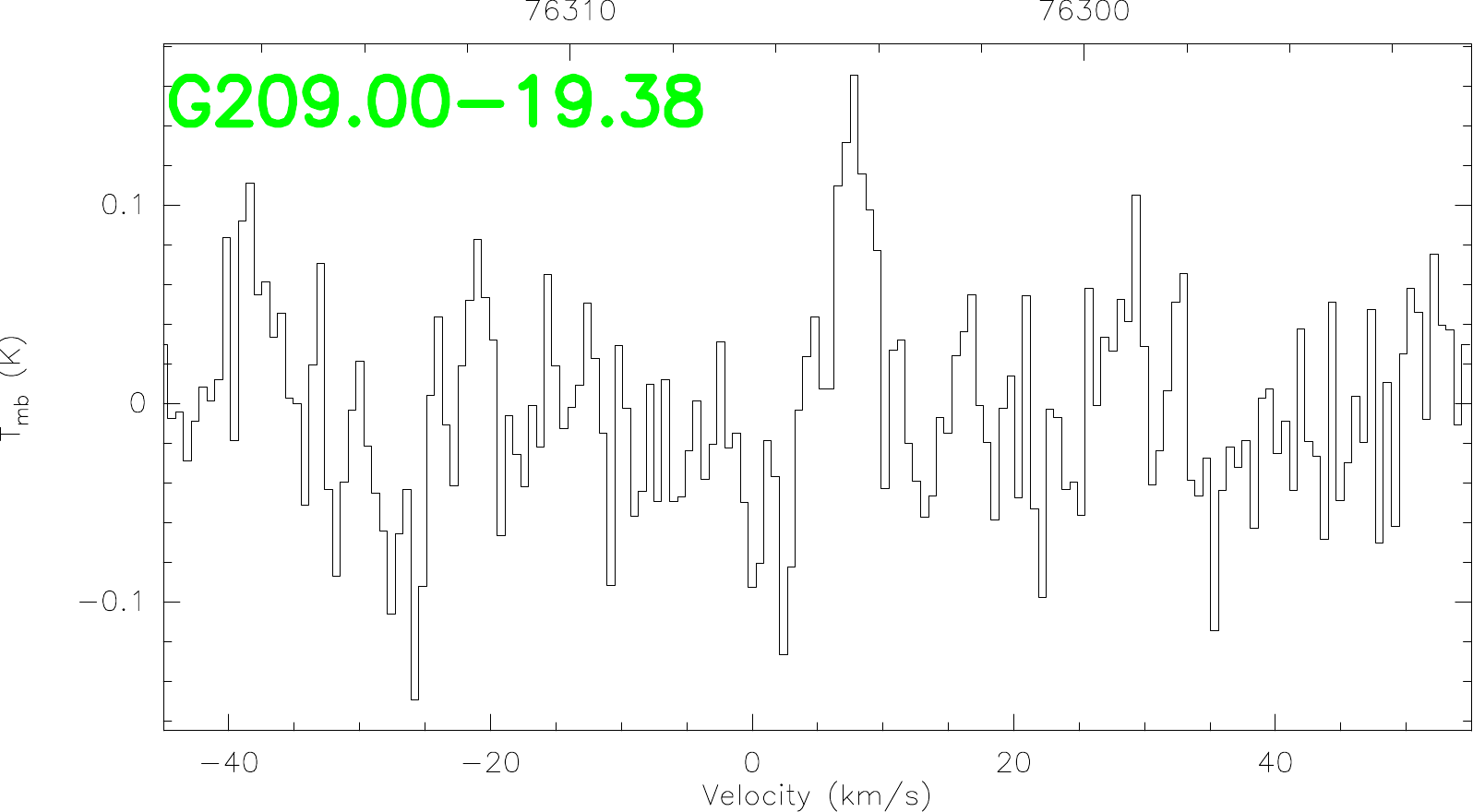}
\includegraphics[width=0.155\textwidth]{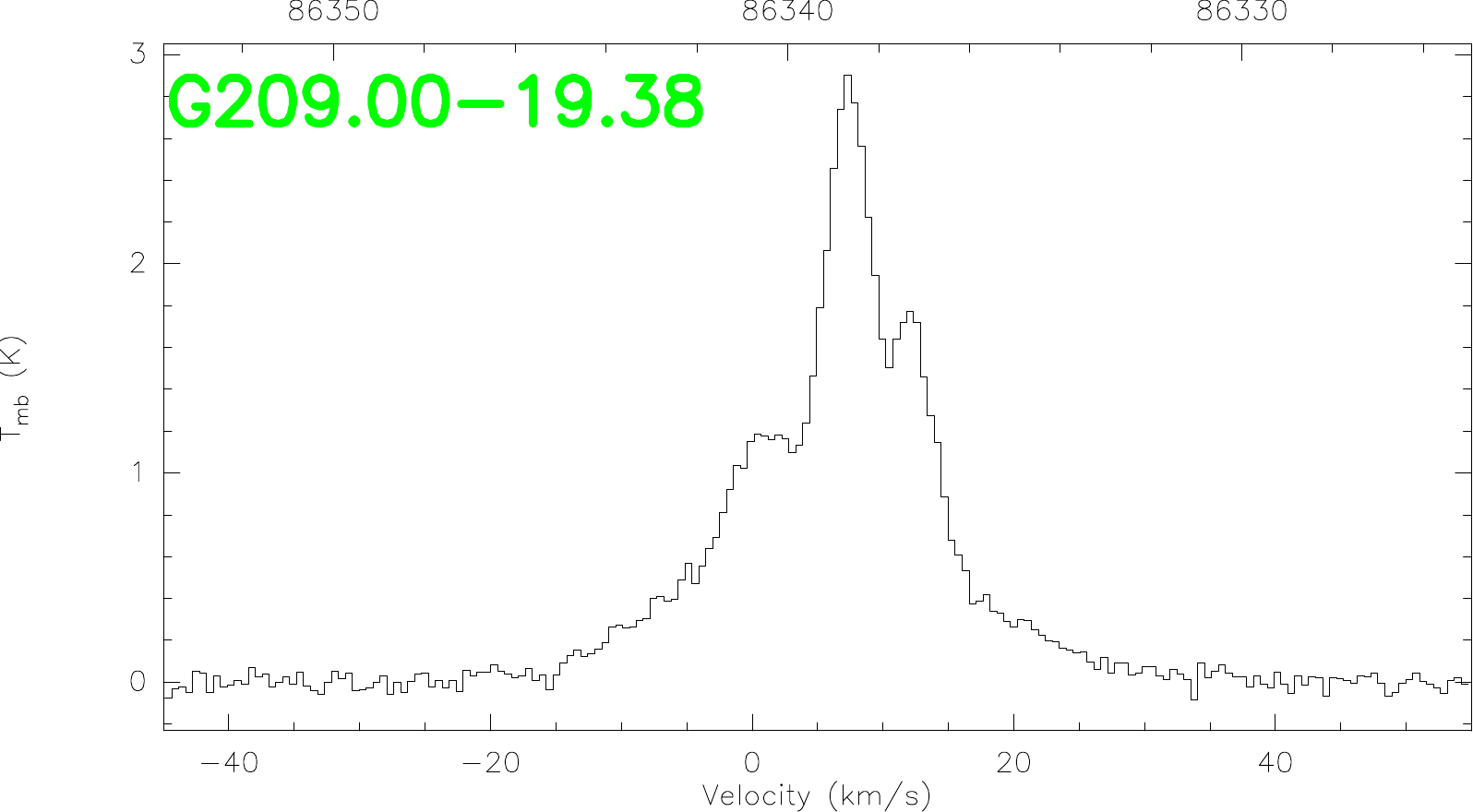}
\includegraphics[width=0.155\textwidth]{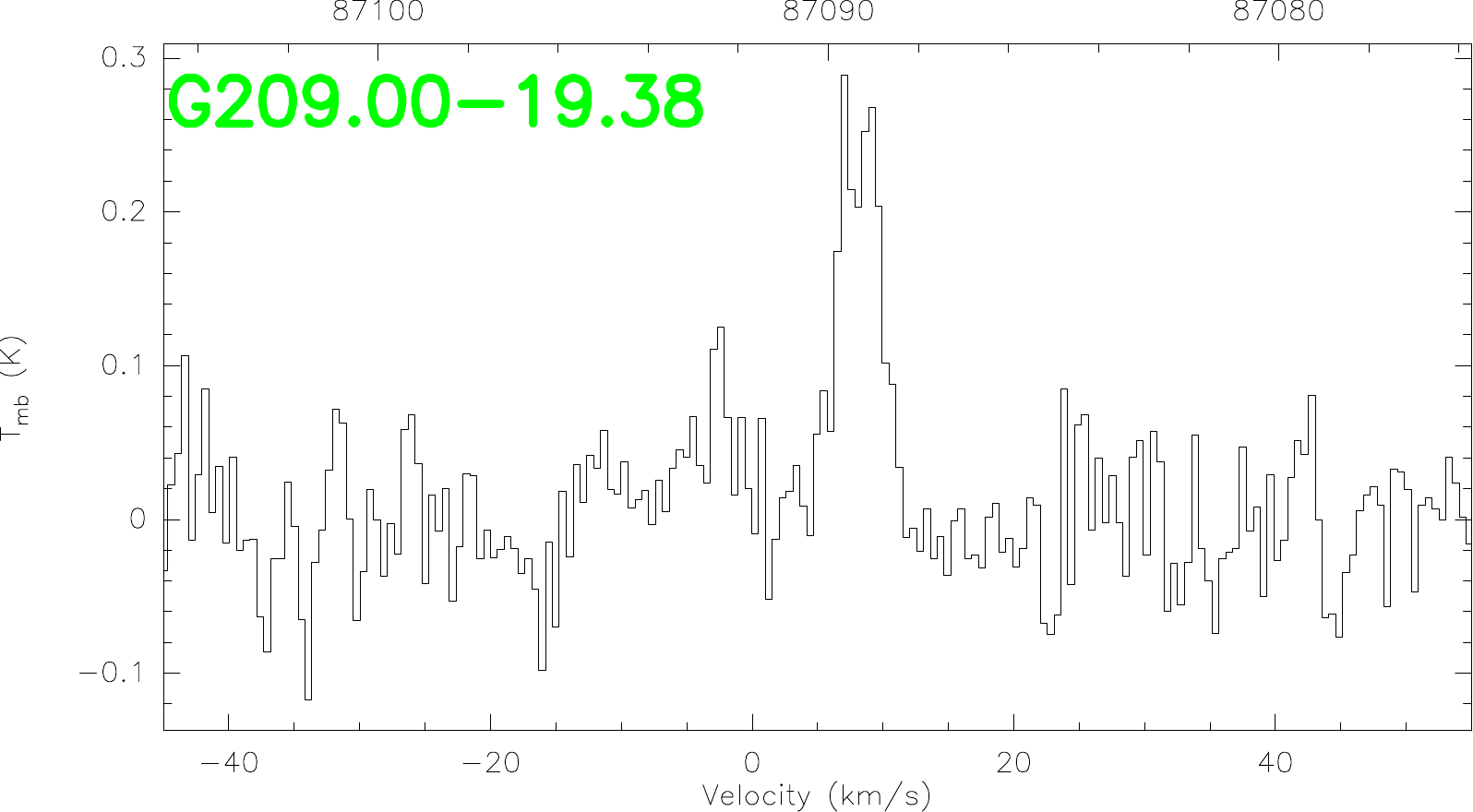}
\includegraphics[width=0.155\textwidth]{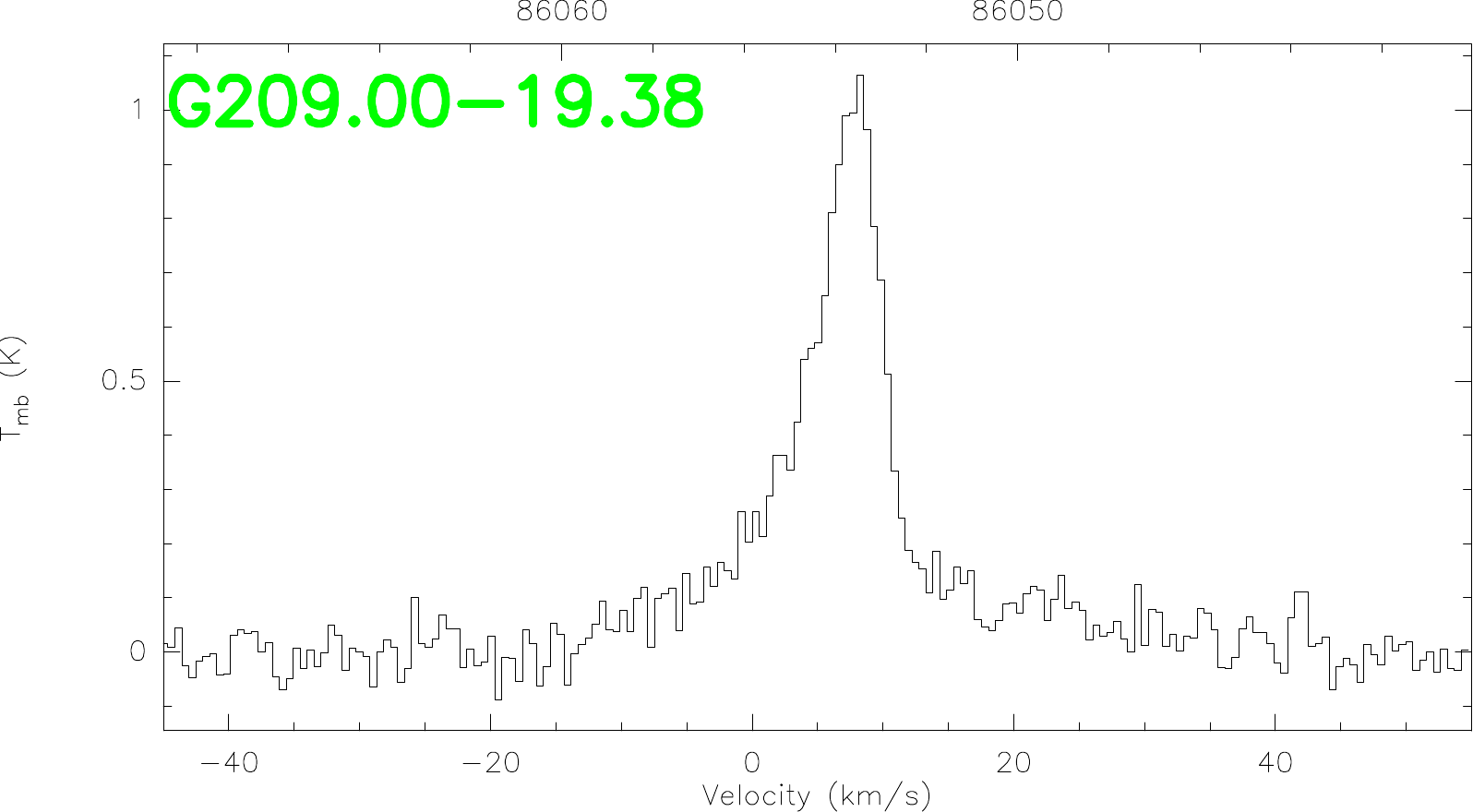}
\includegraphics[width=0.155\textwidth]{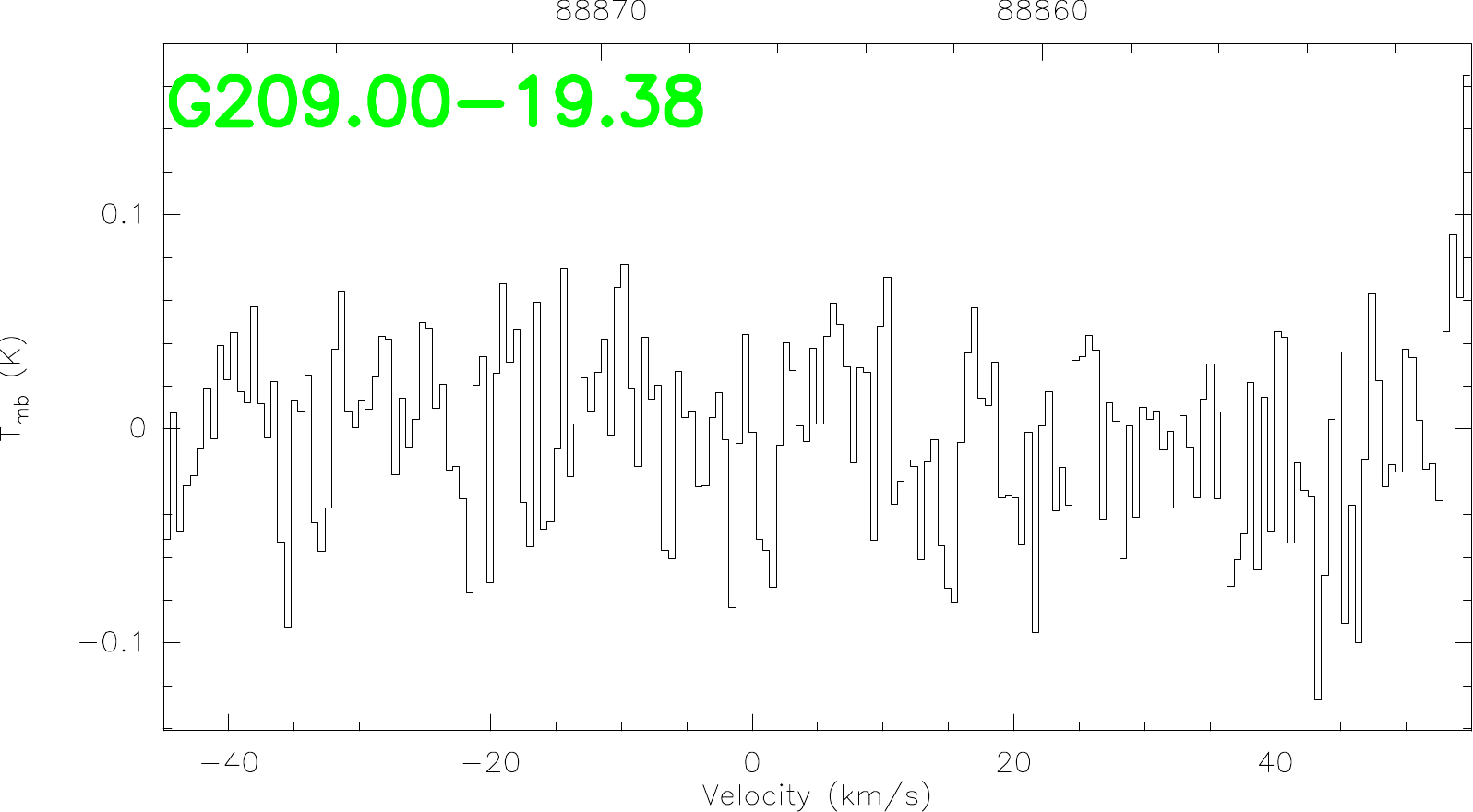}\\

\includegraphics[width=0.155\textwidth]{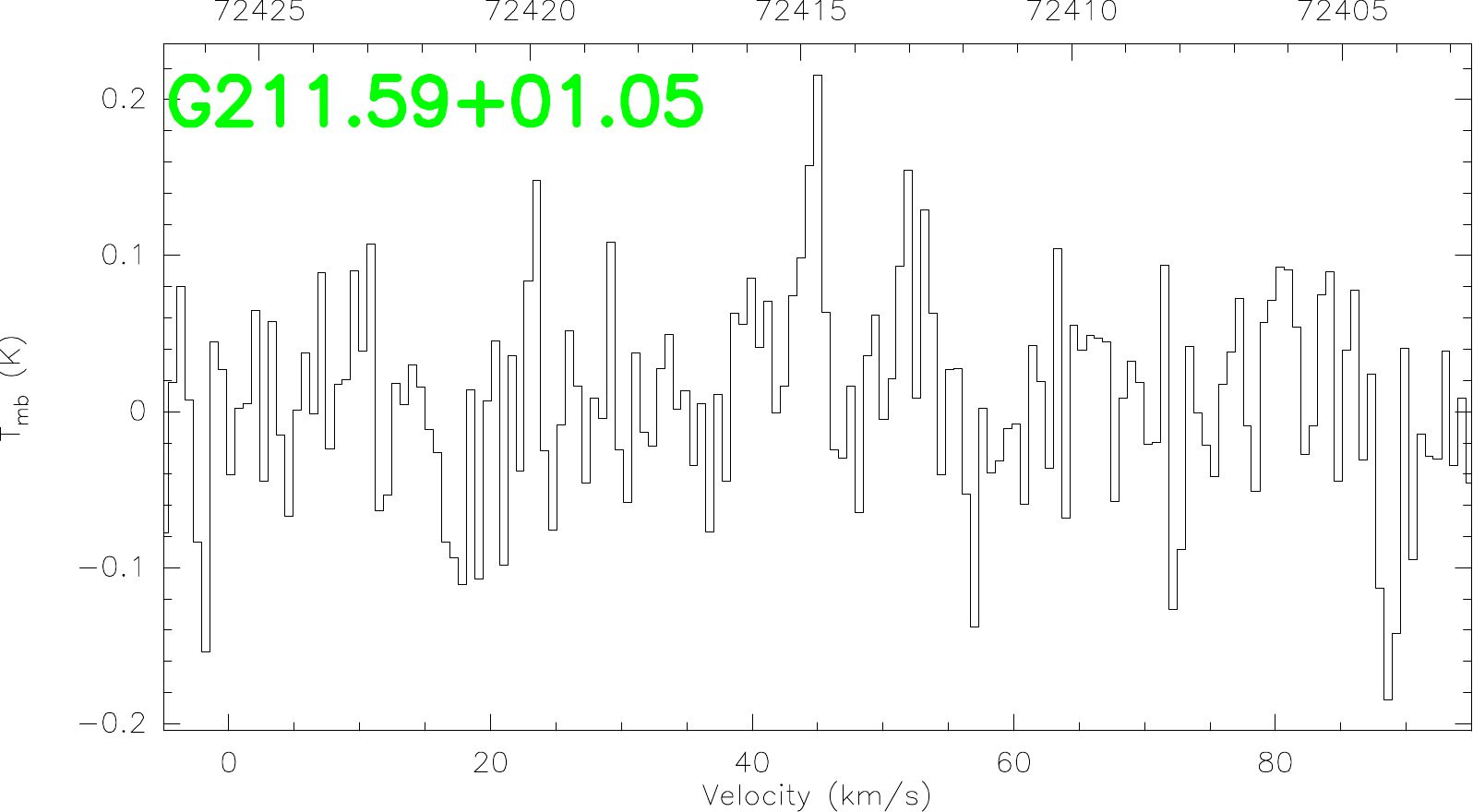}
\includegraphics[width=0.155\textwidth]{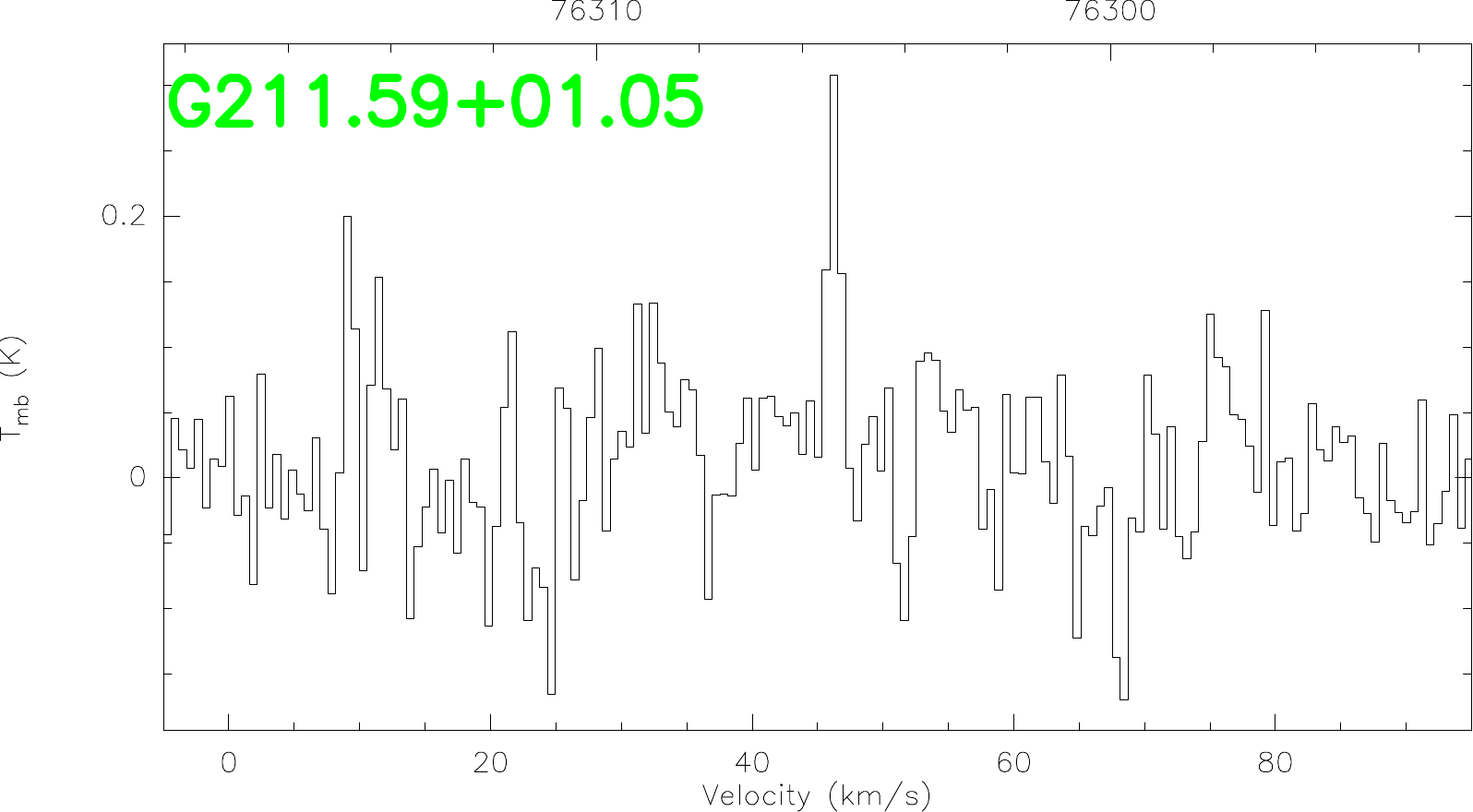}
\includegraphics[width=0.155\textwidth]{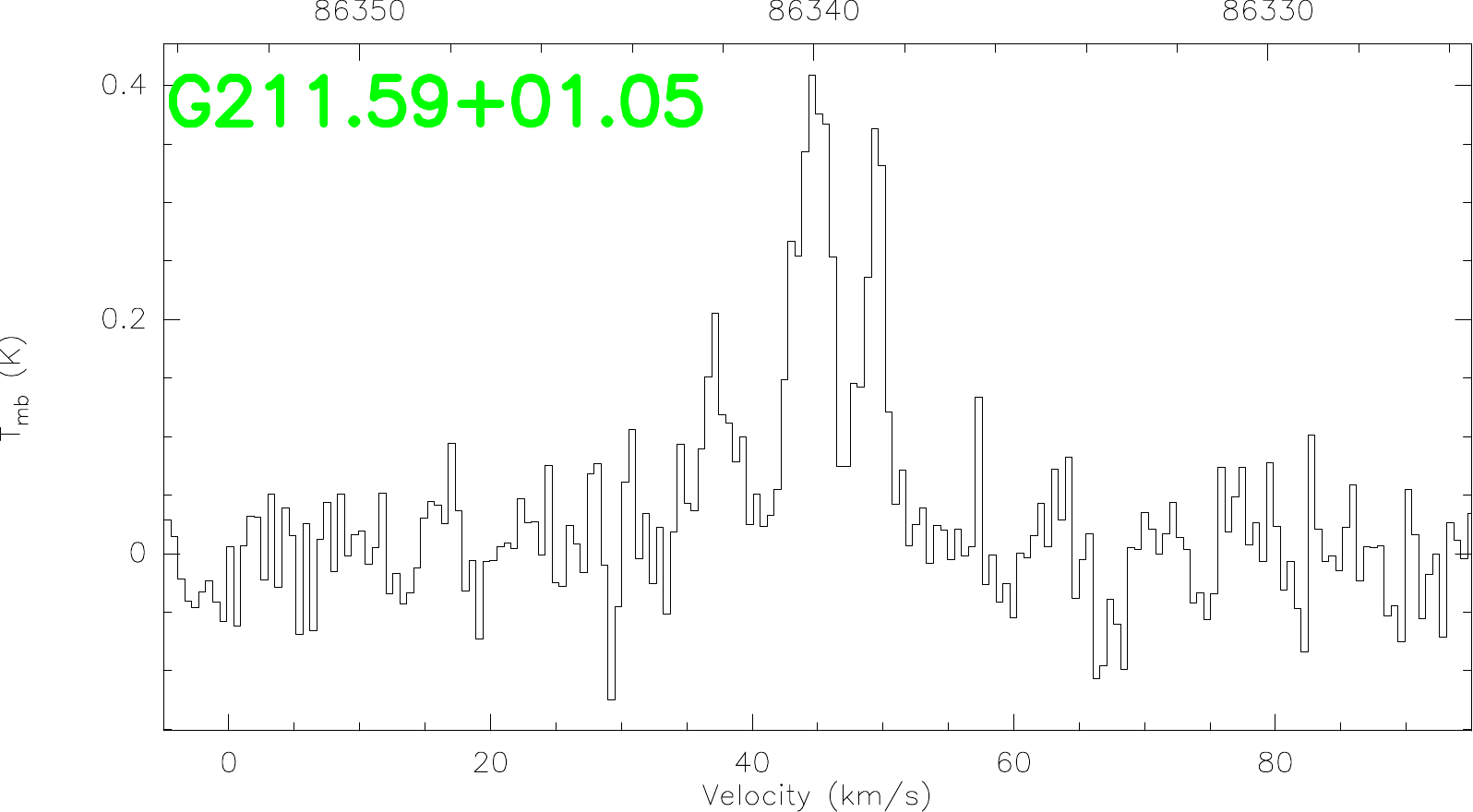}
\includegraphics[width=0.155\textwidth]{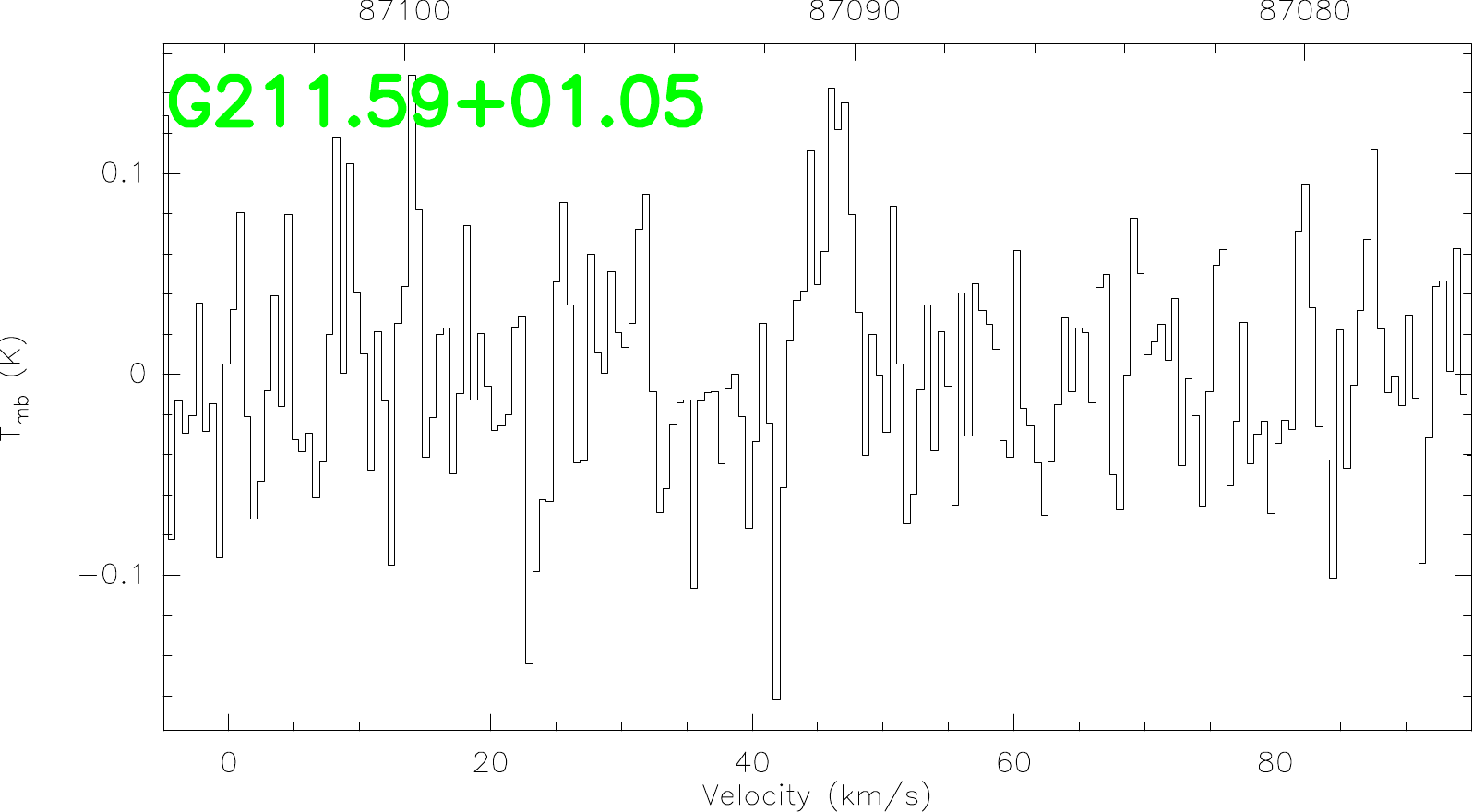}
\includegraphics[width=0.155\textwidth]{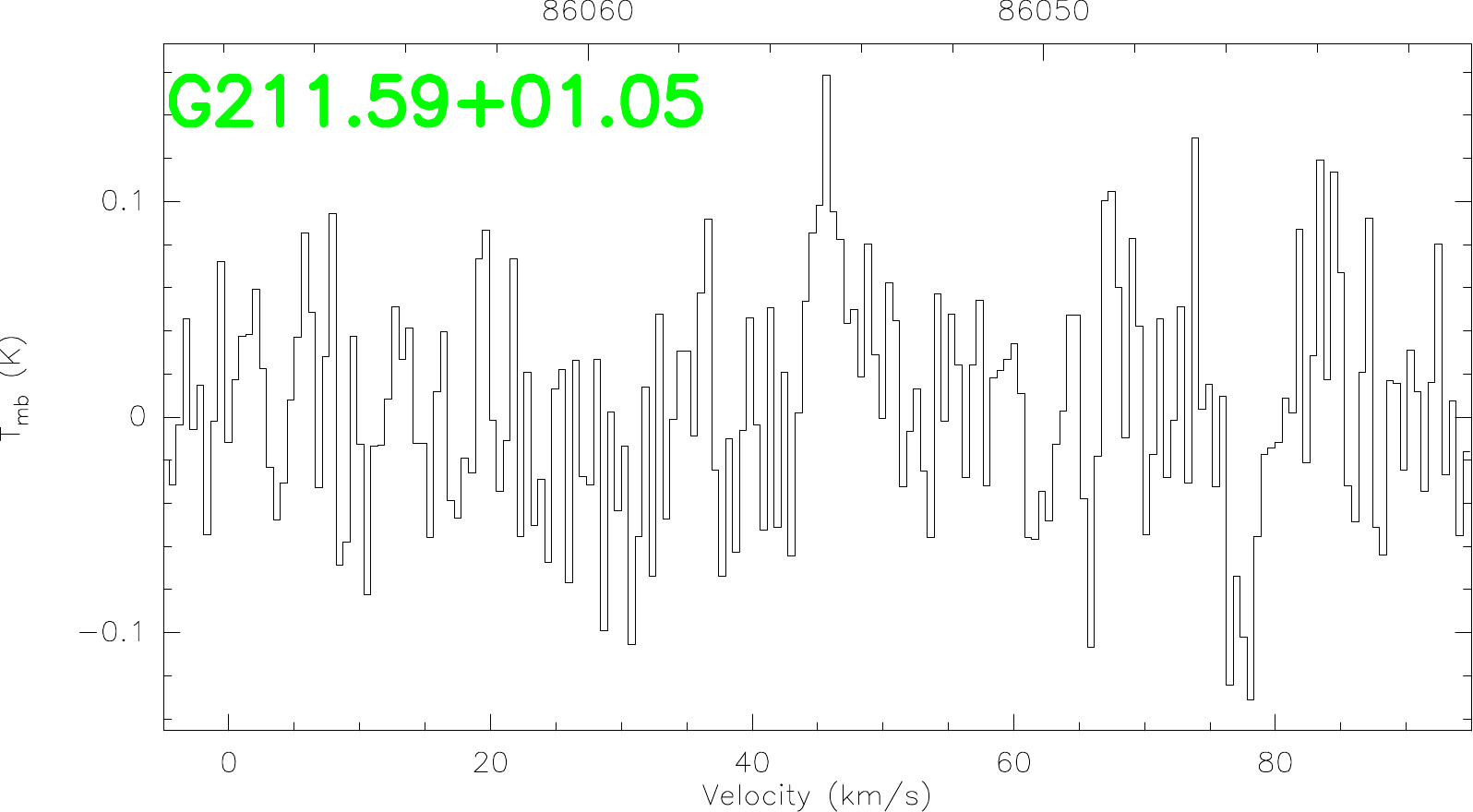}
\includegraphics[width=0.155\textwidth]{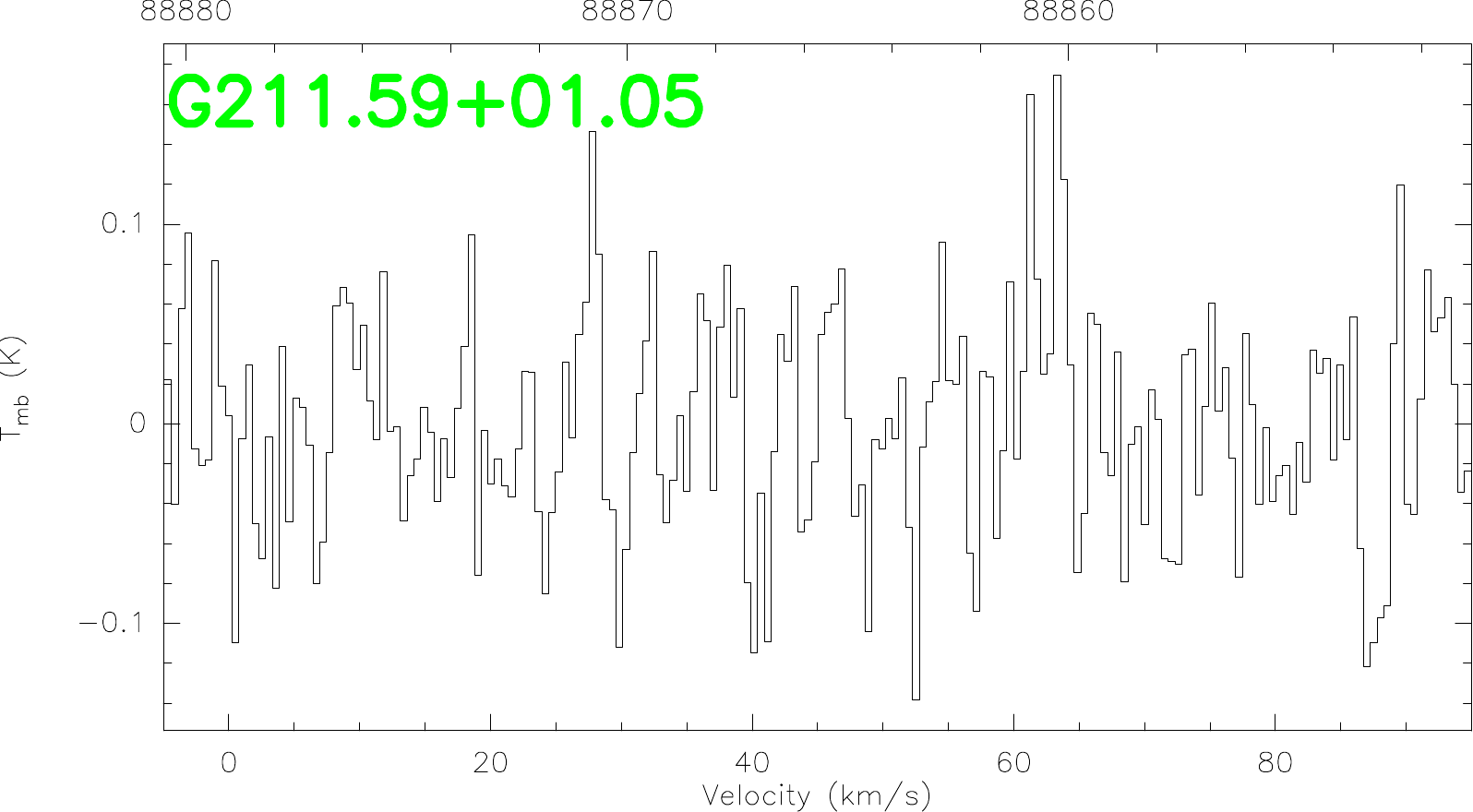}\\

\includegraphics[width=0.155\textwidth]{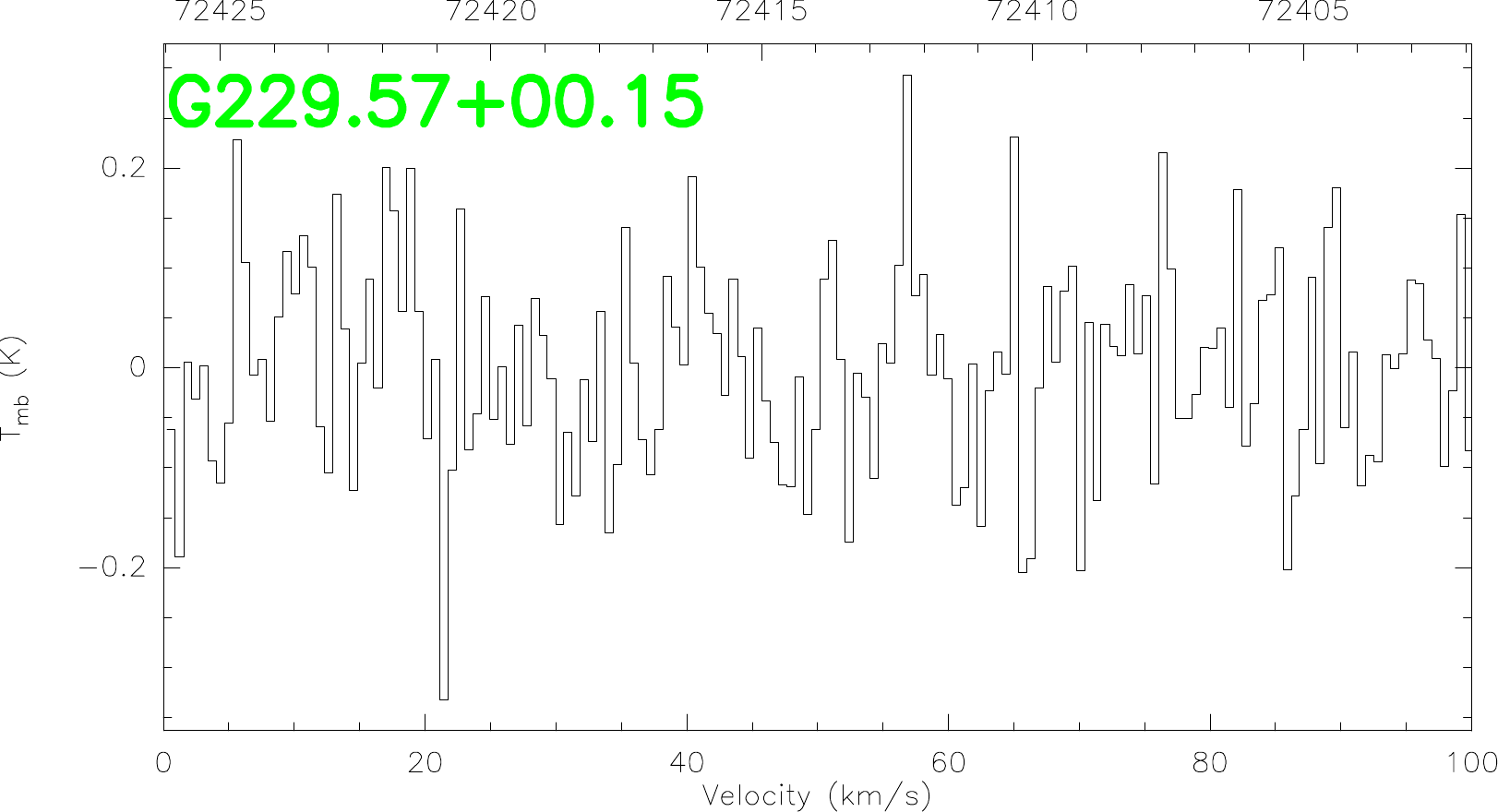}
\includegraphics[width=0.155\textwidth]{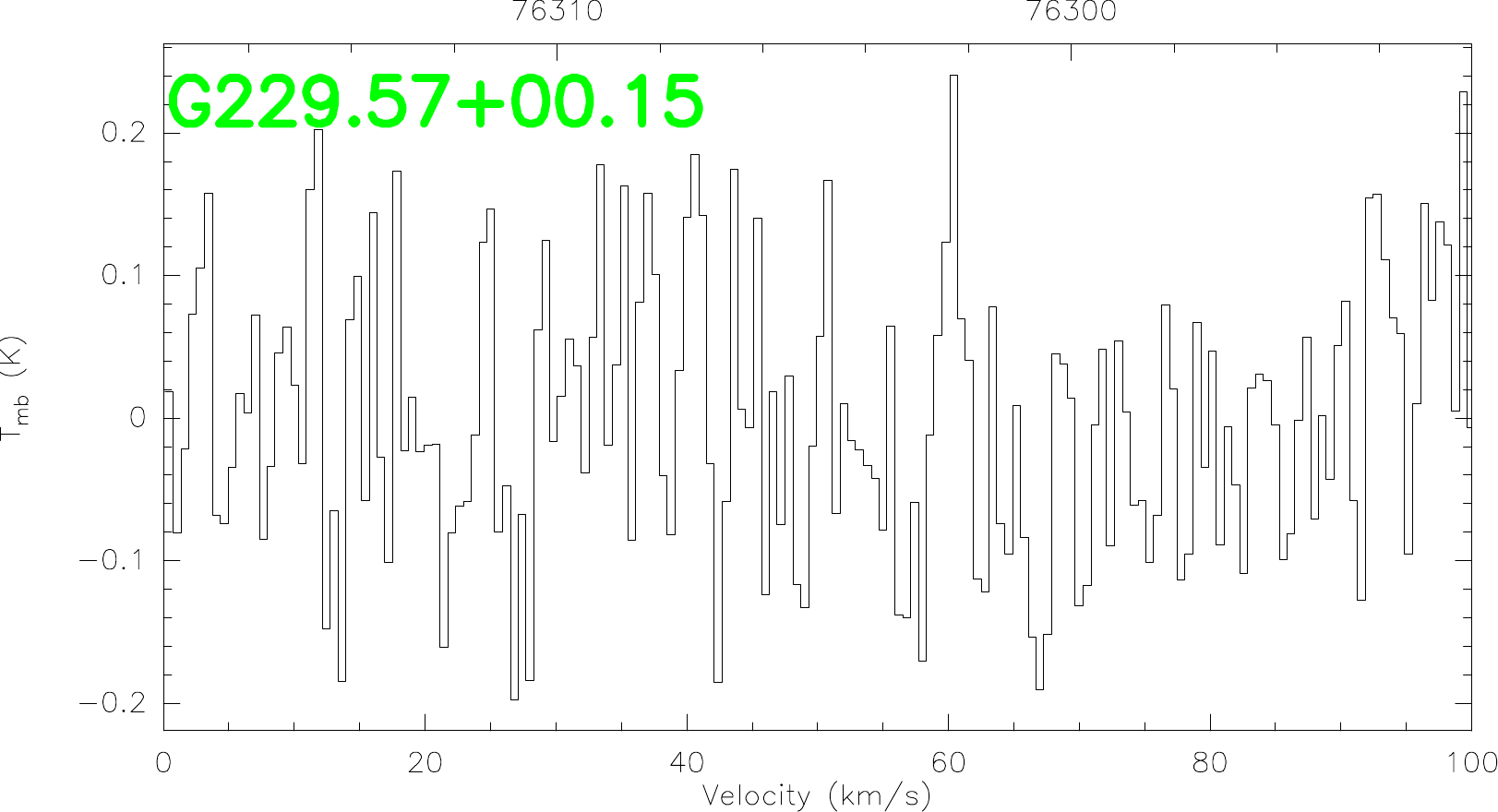}
\includegraphics[width=0.155\textwidth]{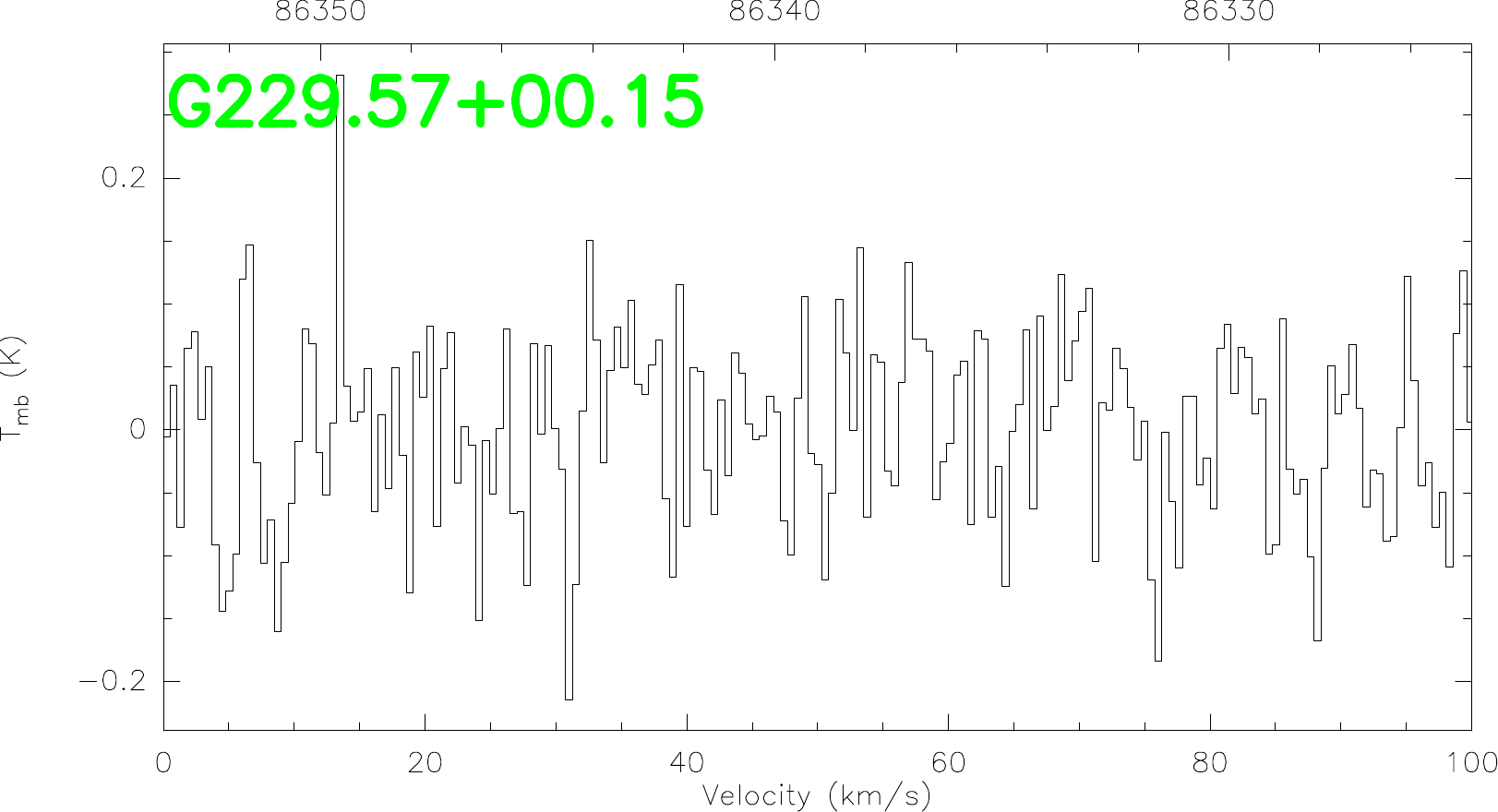}
\includegraphics[width=0.155\textwidth]{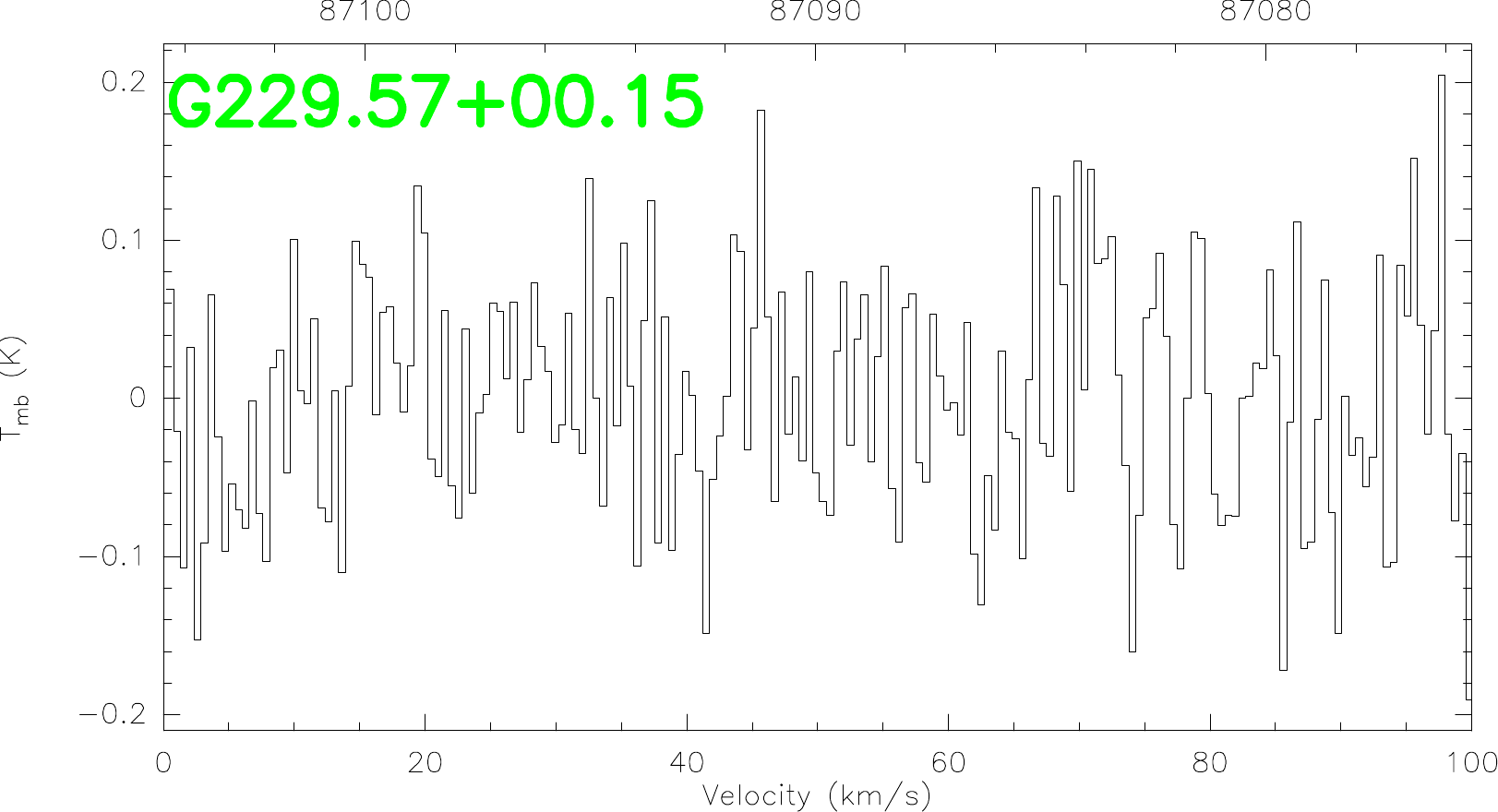}
\includegraphics[width=0.155\textwidth]{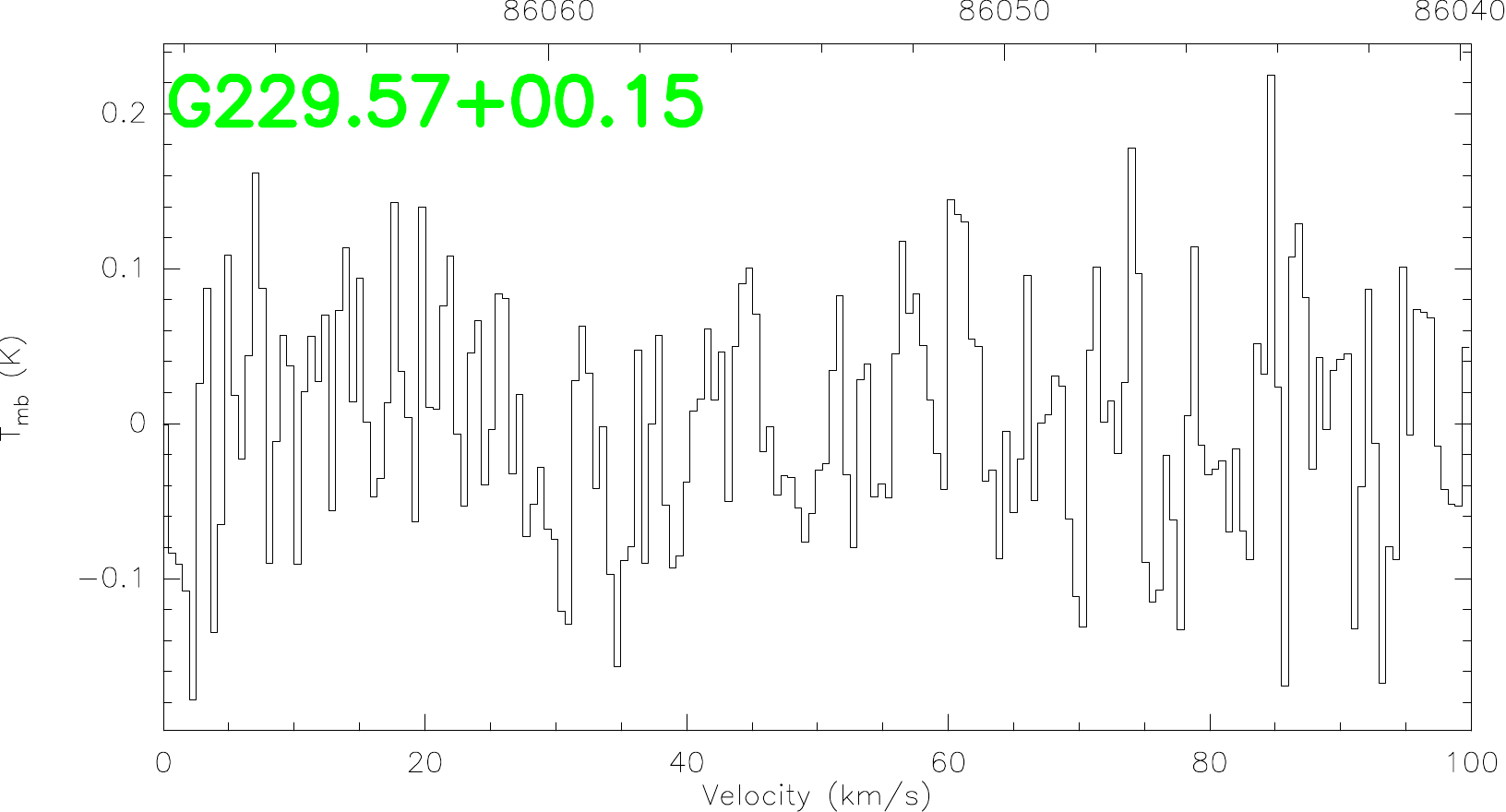}
\includegraphics[width=0.155\textwidth]{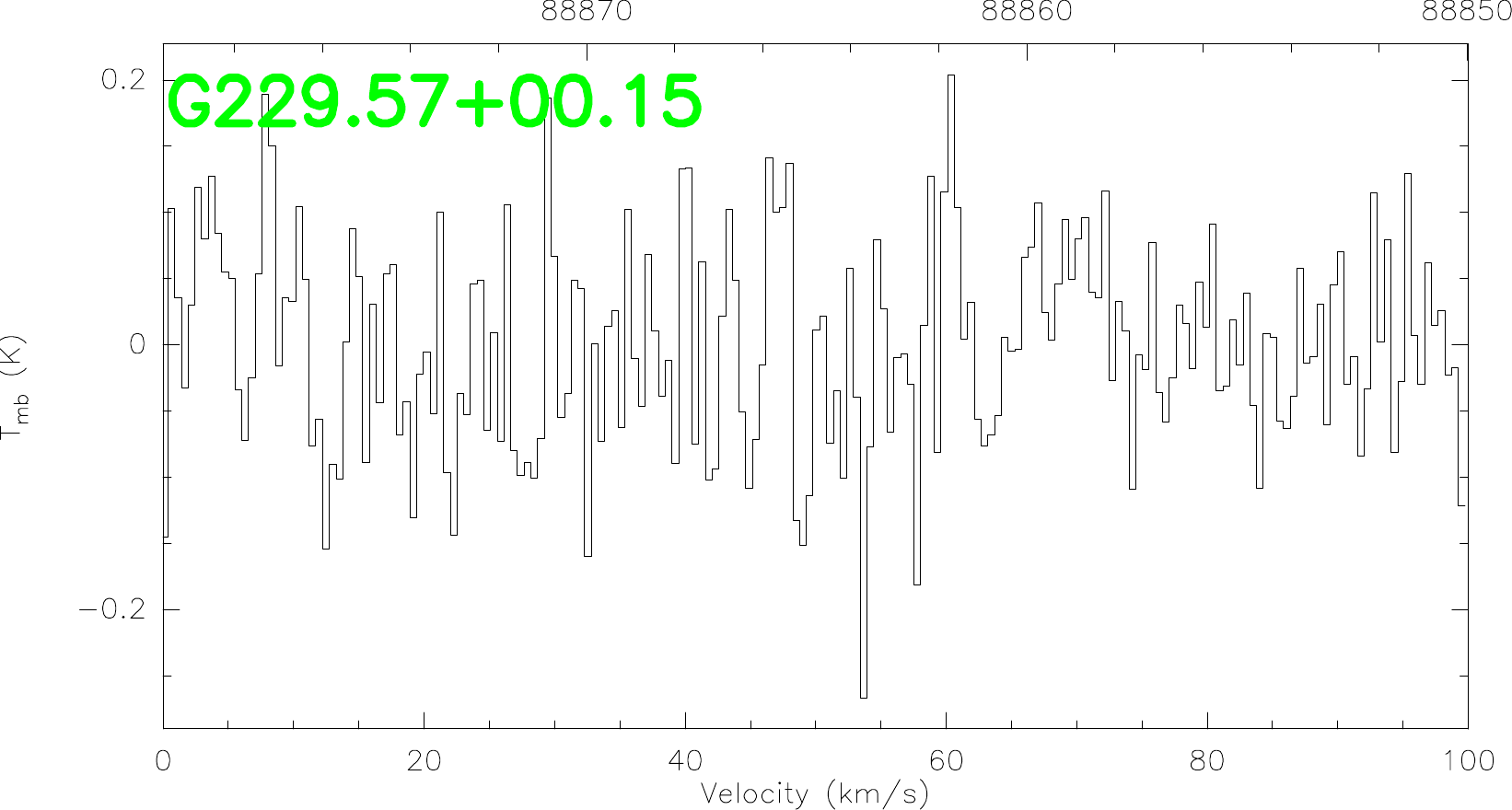}\\

\includegraphics[width=0.155\textwidth]{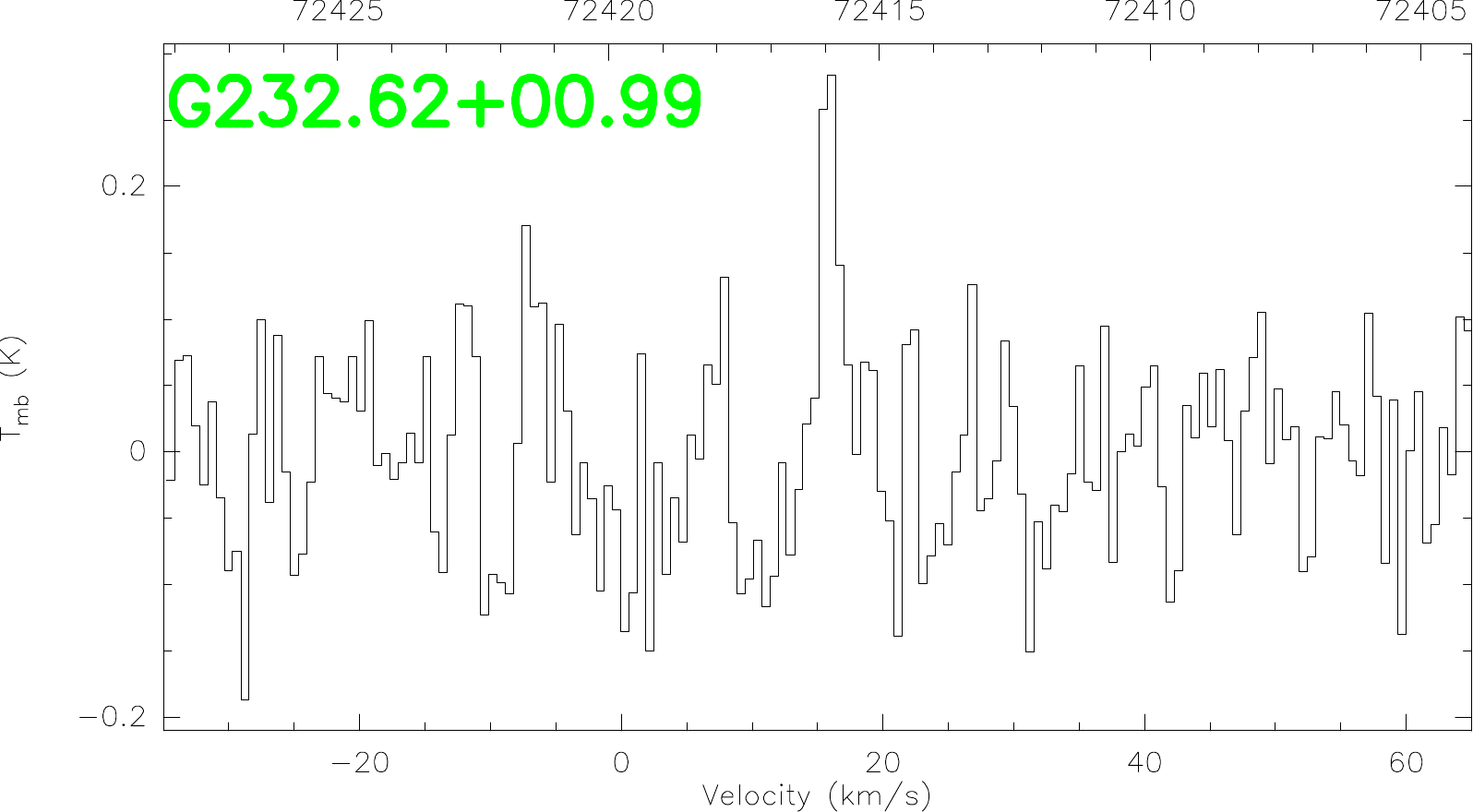}
\includegraphics[width=0.155\textwidth]{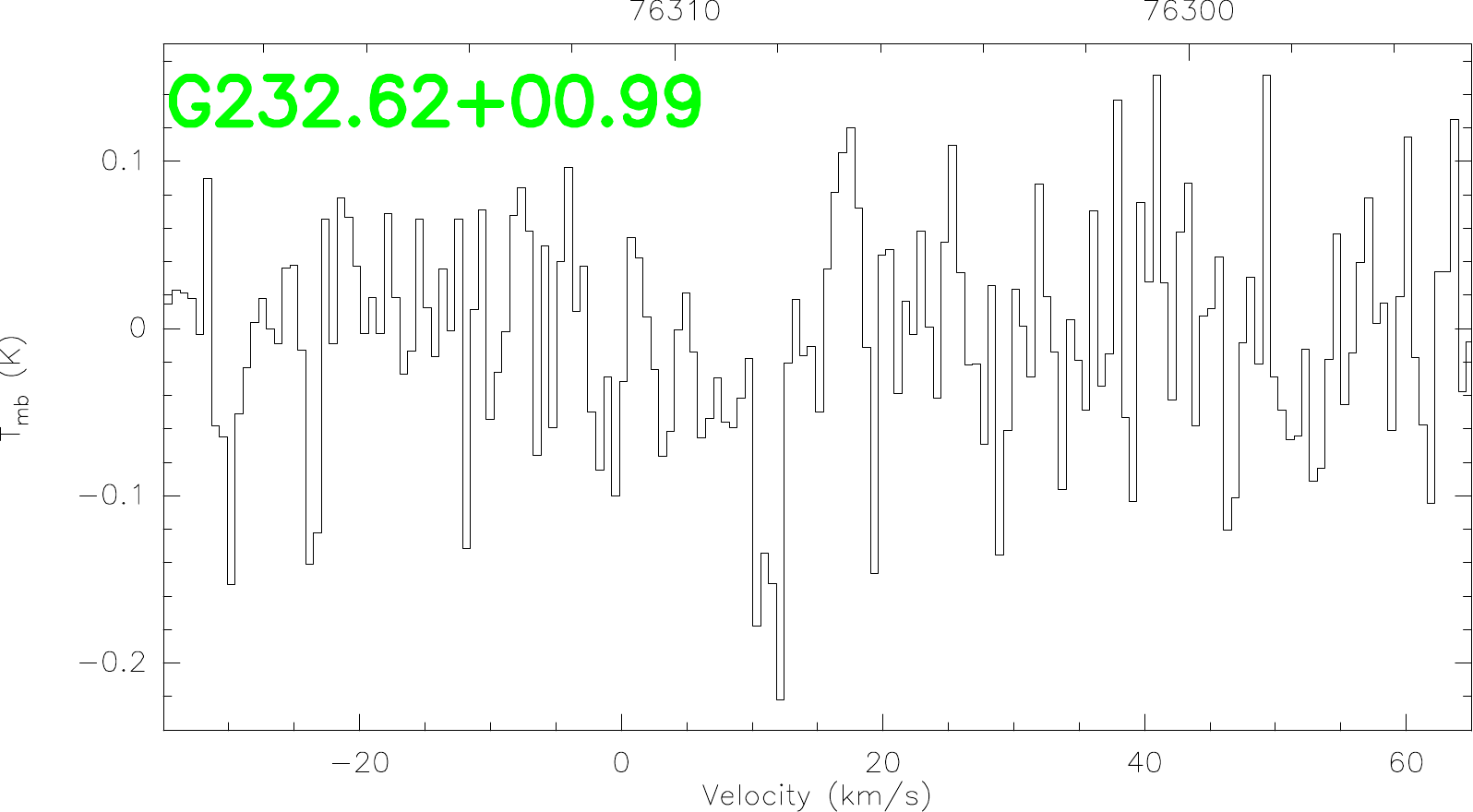}
\includegraphics[width=0.155\textwidth]{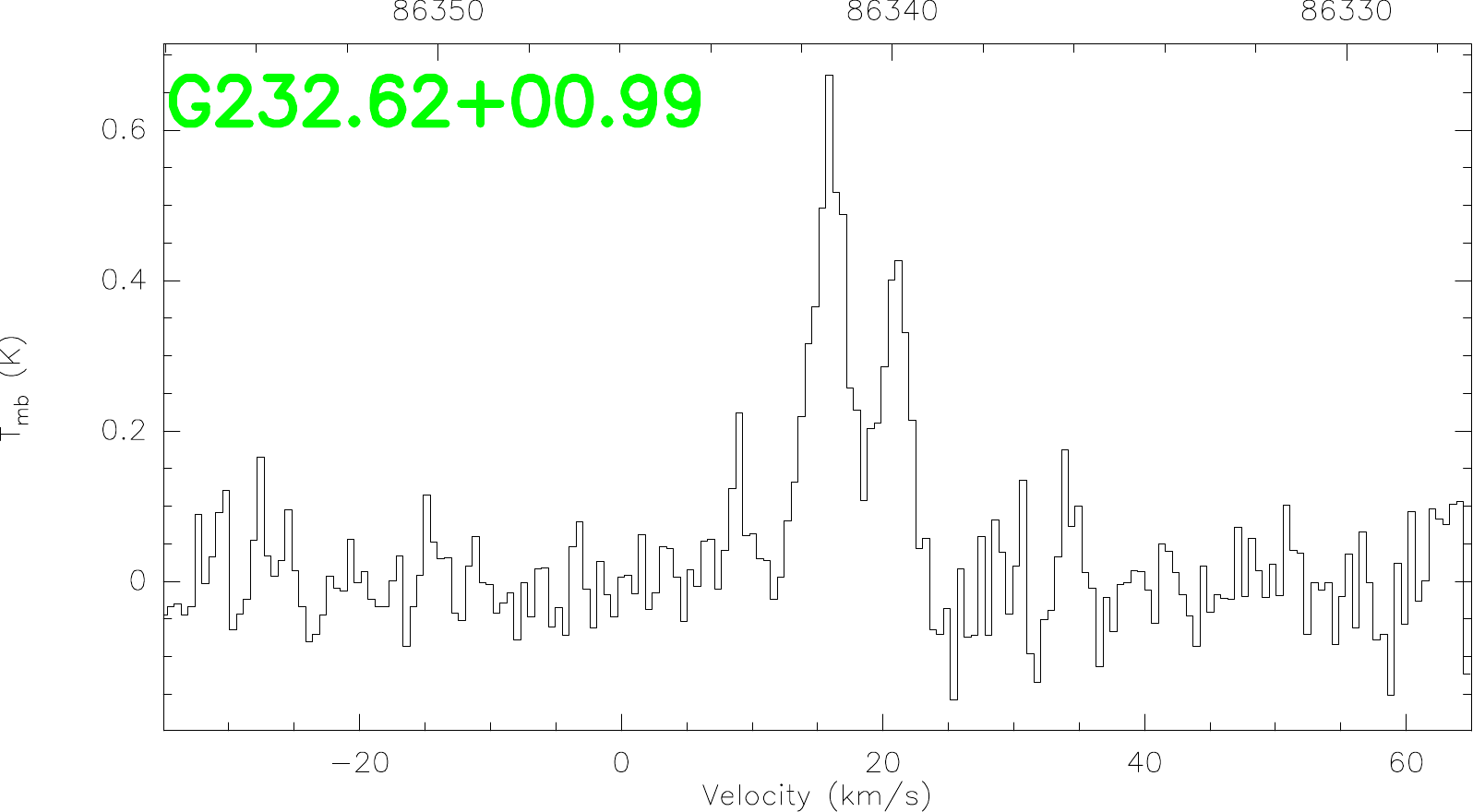}
\includegraphics[width=0.155\textwidth]{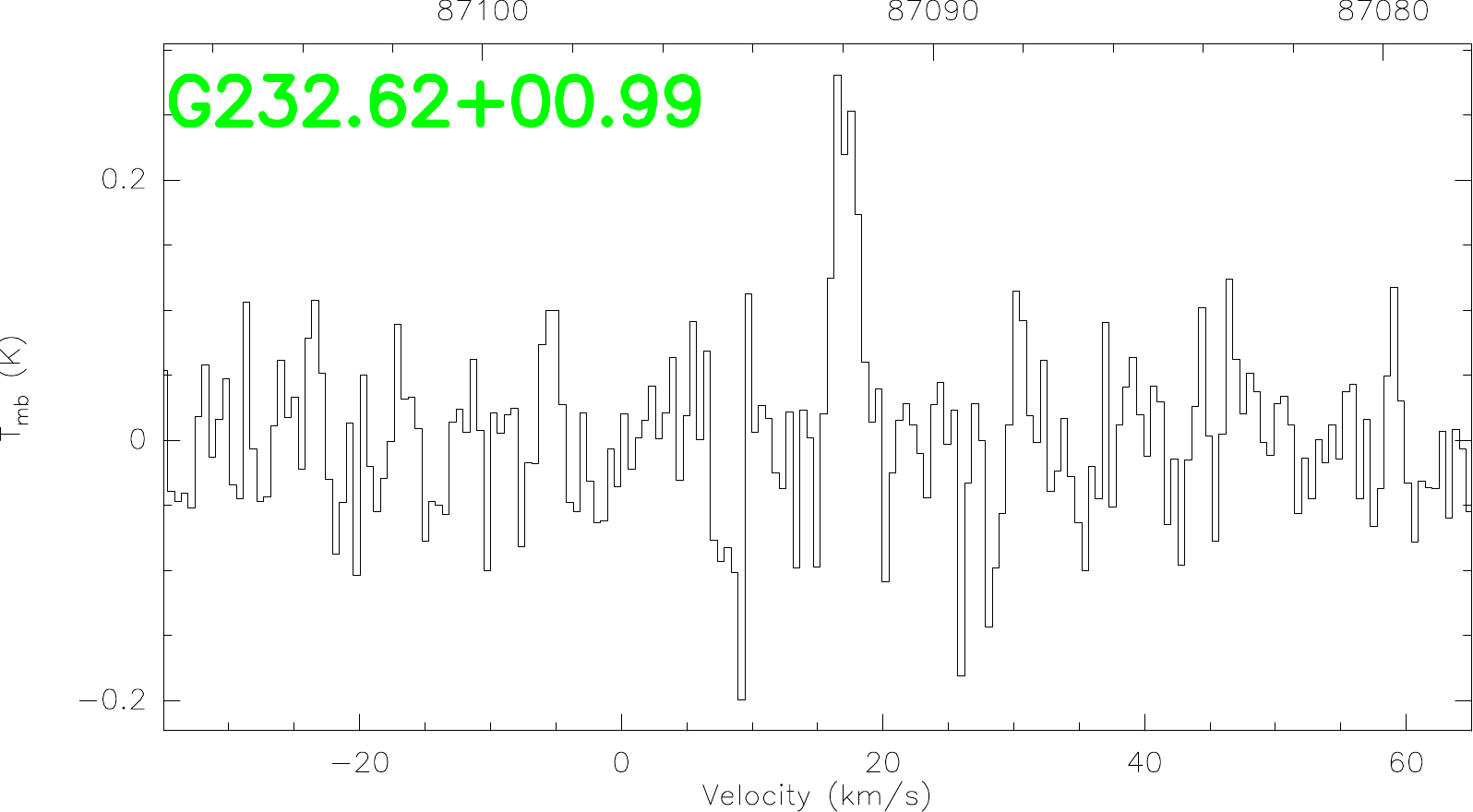}
\includegraphics[width=0.155\textwidth]{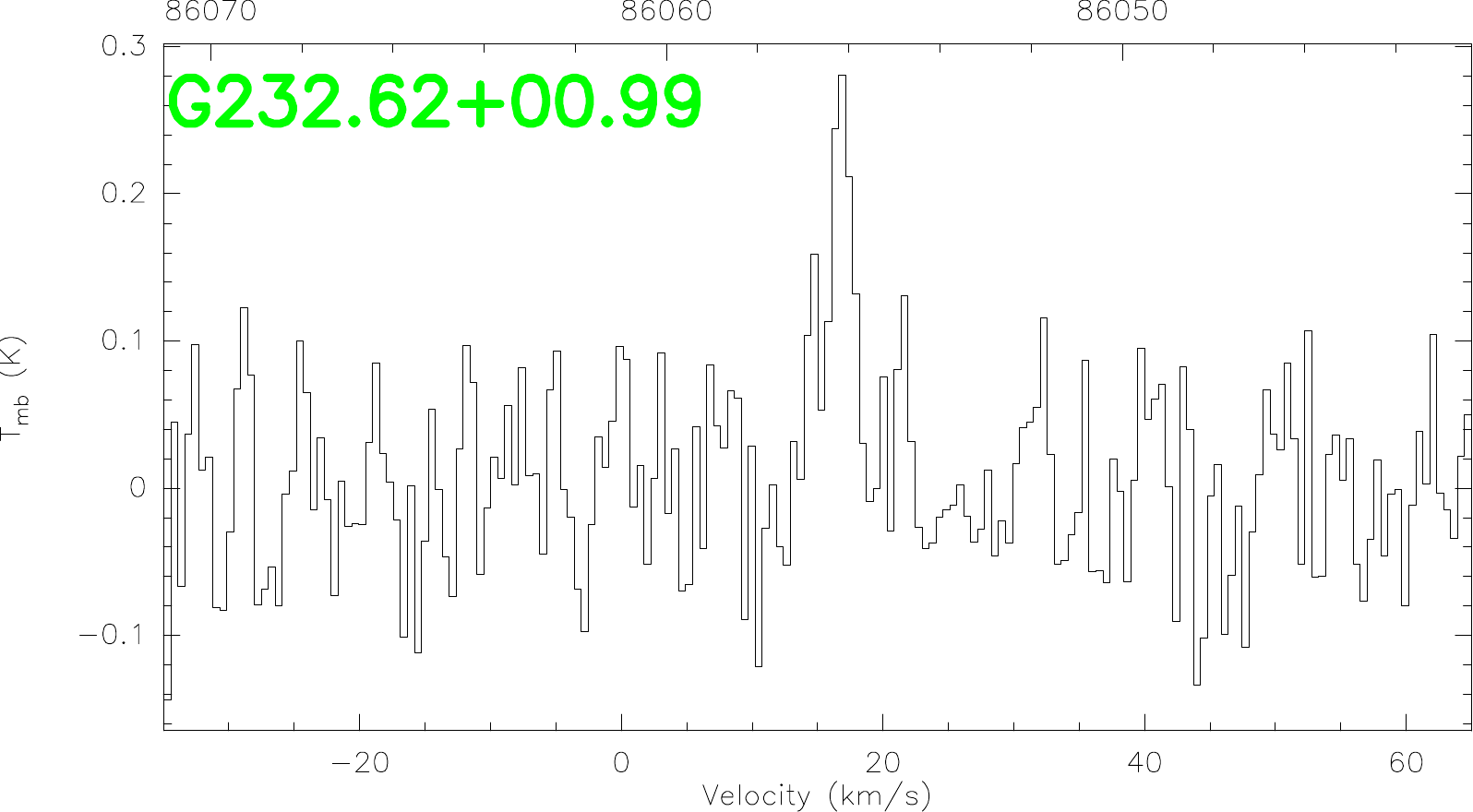}
\includegraphics[width=0.155\textwidth]{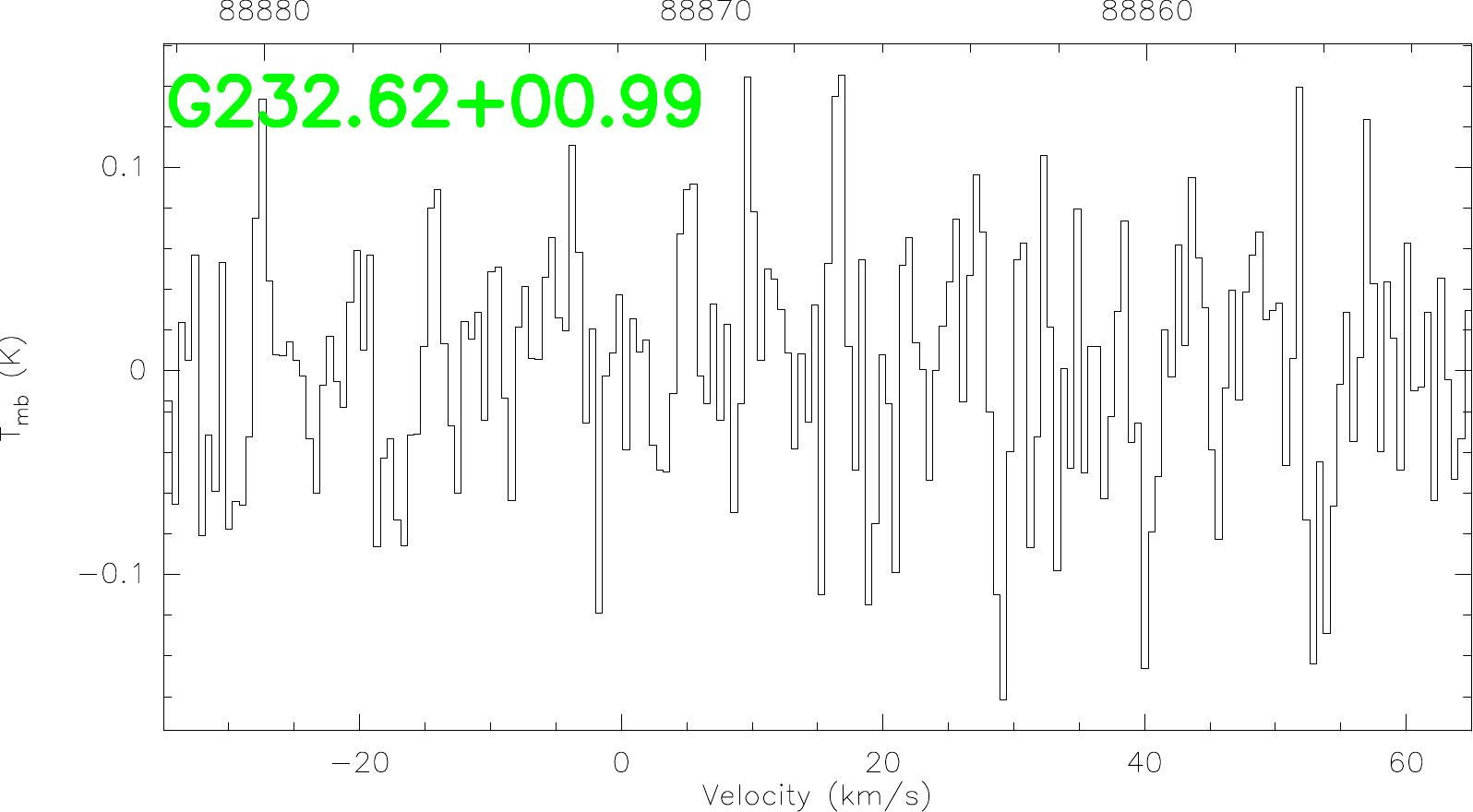}\\

\includegraphics[width=0.155\textwidth]{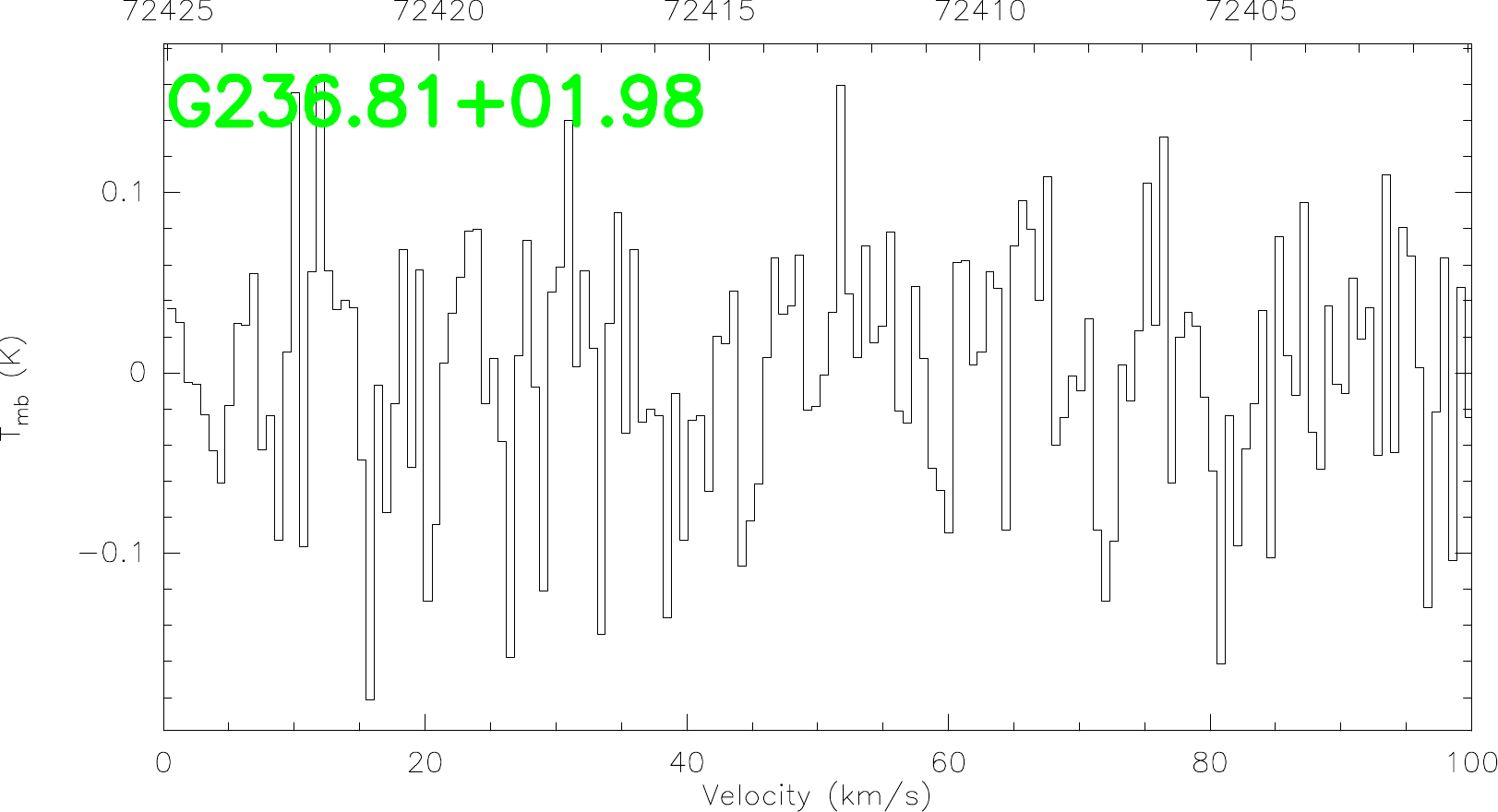}
\includegraphics[width=0.155\textwidth]{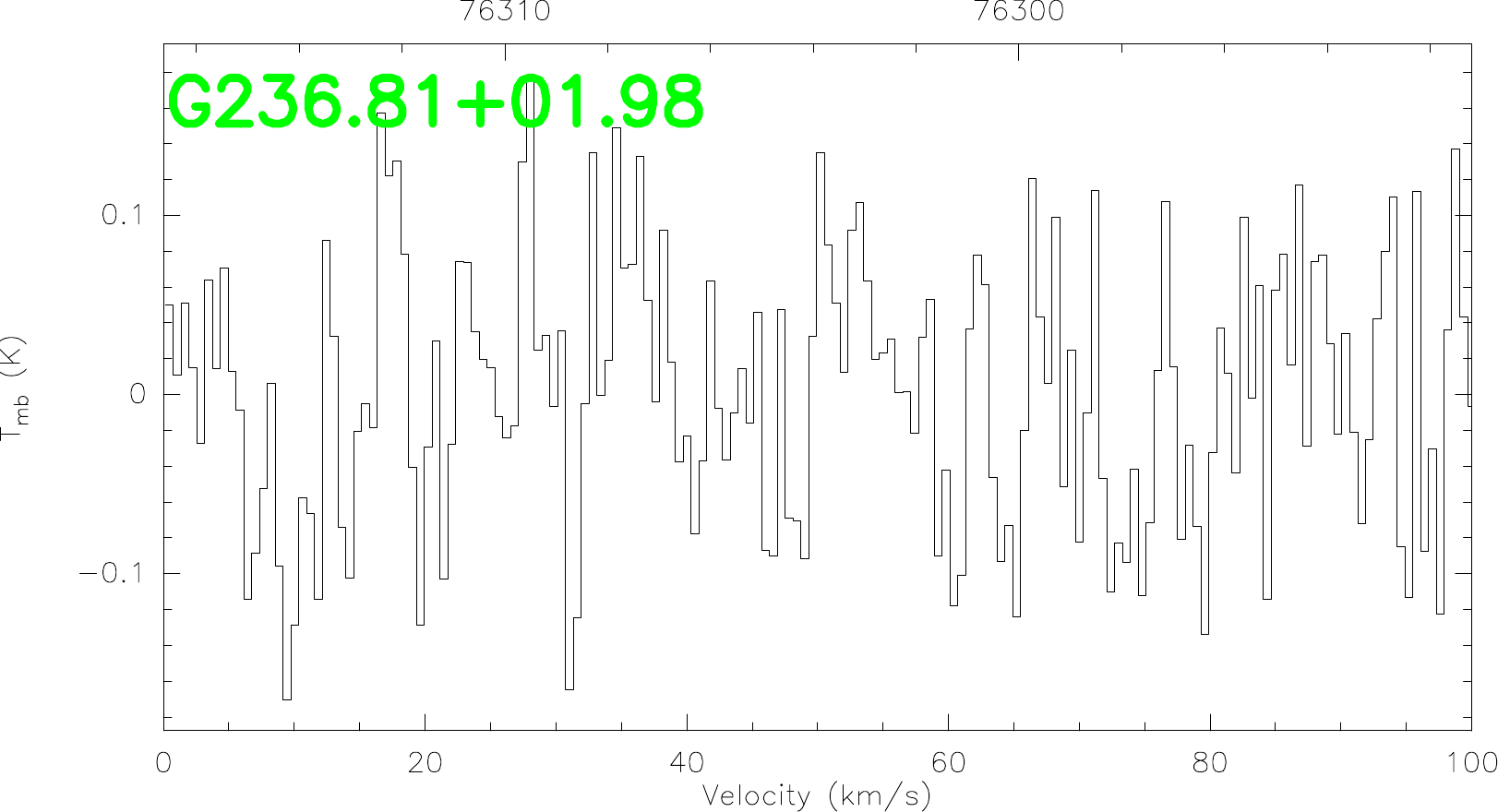}
\includegraphics[width=0.155\textwidth]{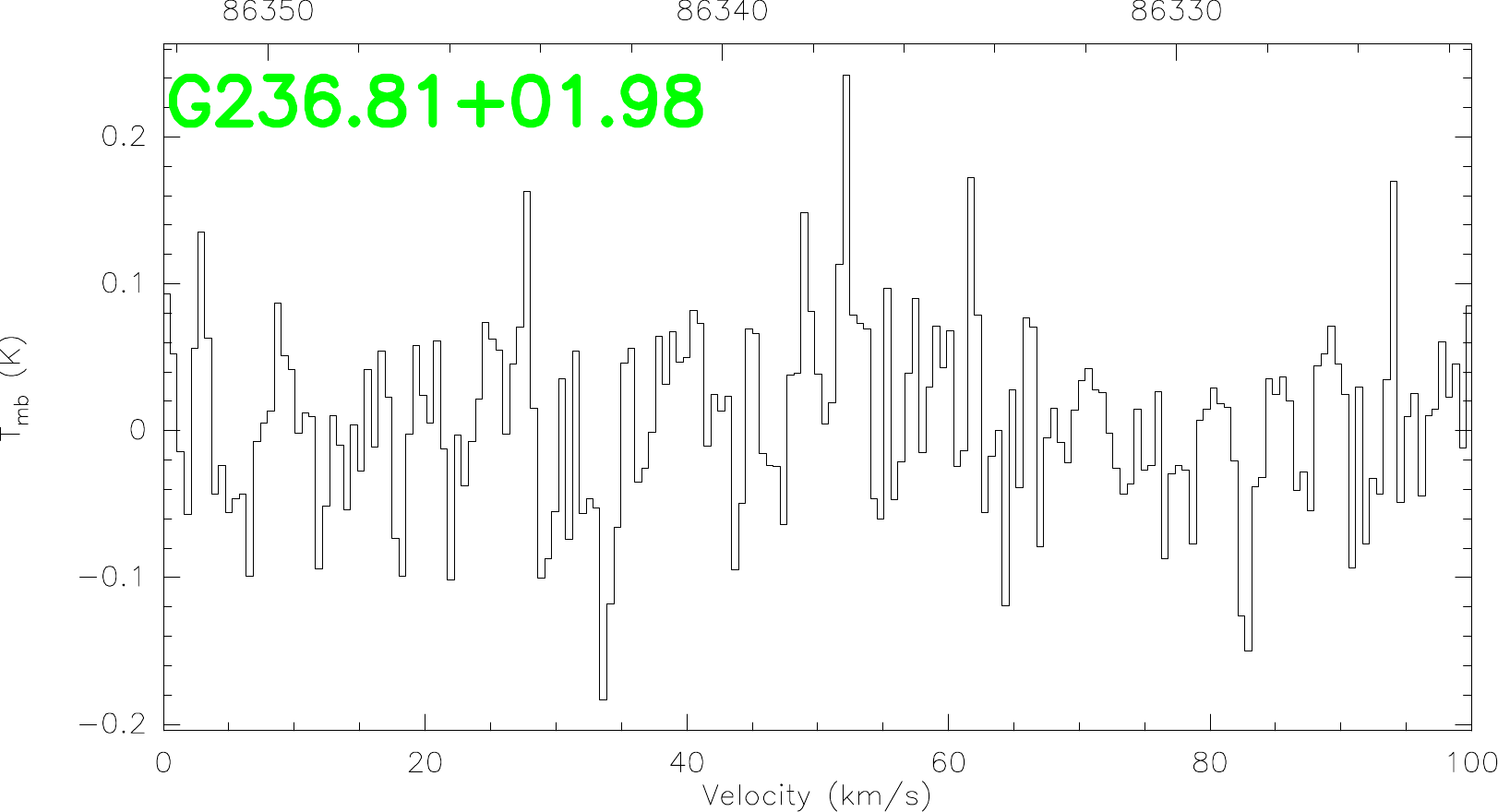}
\includegraphics[width=0.155\textwidth]{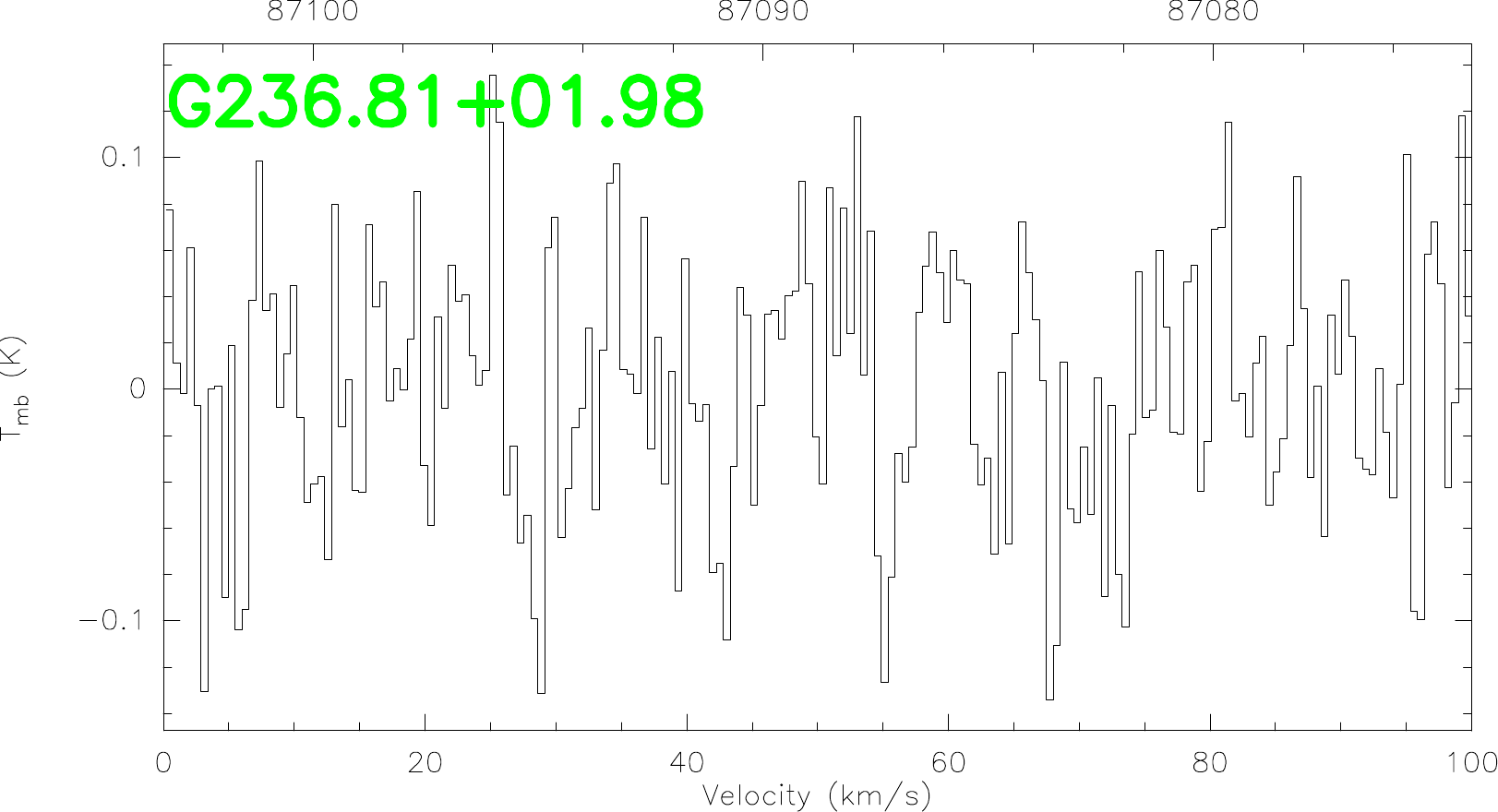}
\includegraphics[width=0.155\textwidth]{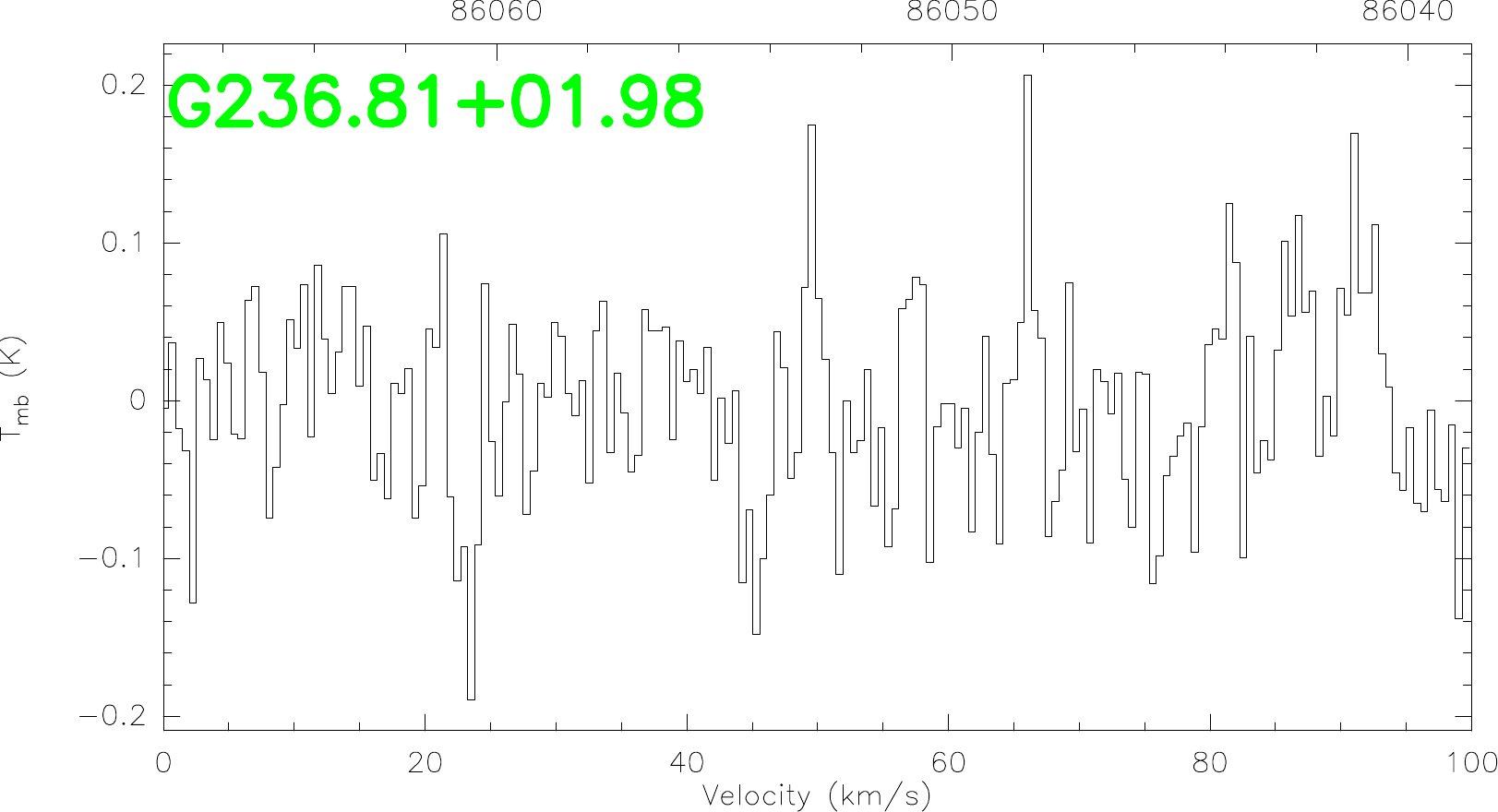}
\includegraphics[width=0.155\textwidth]{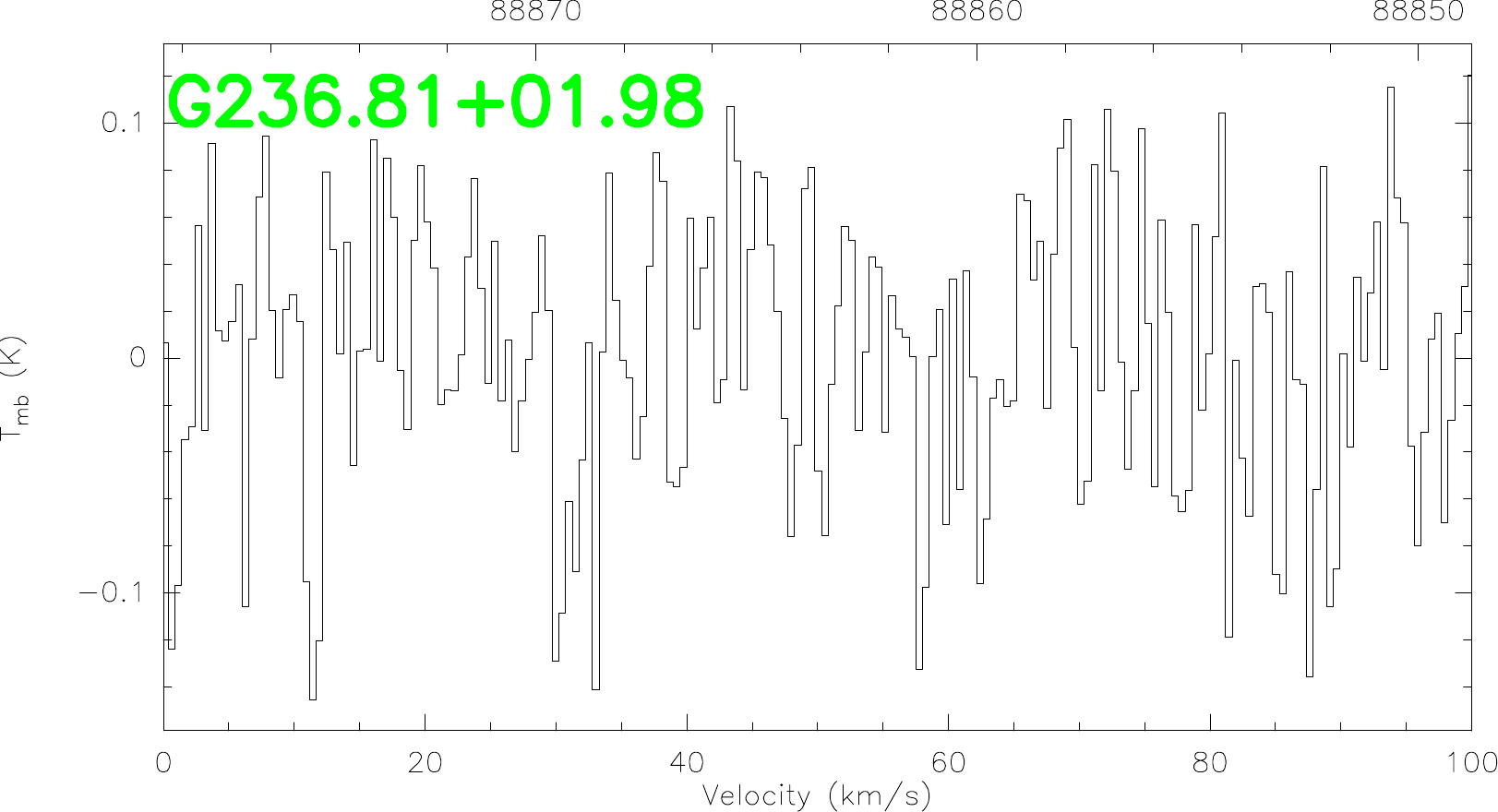}\\

\includegraphics[width=0.155\textwidth]{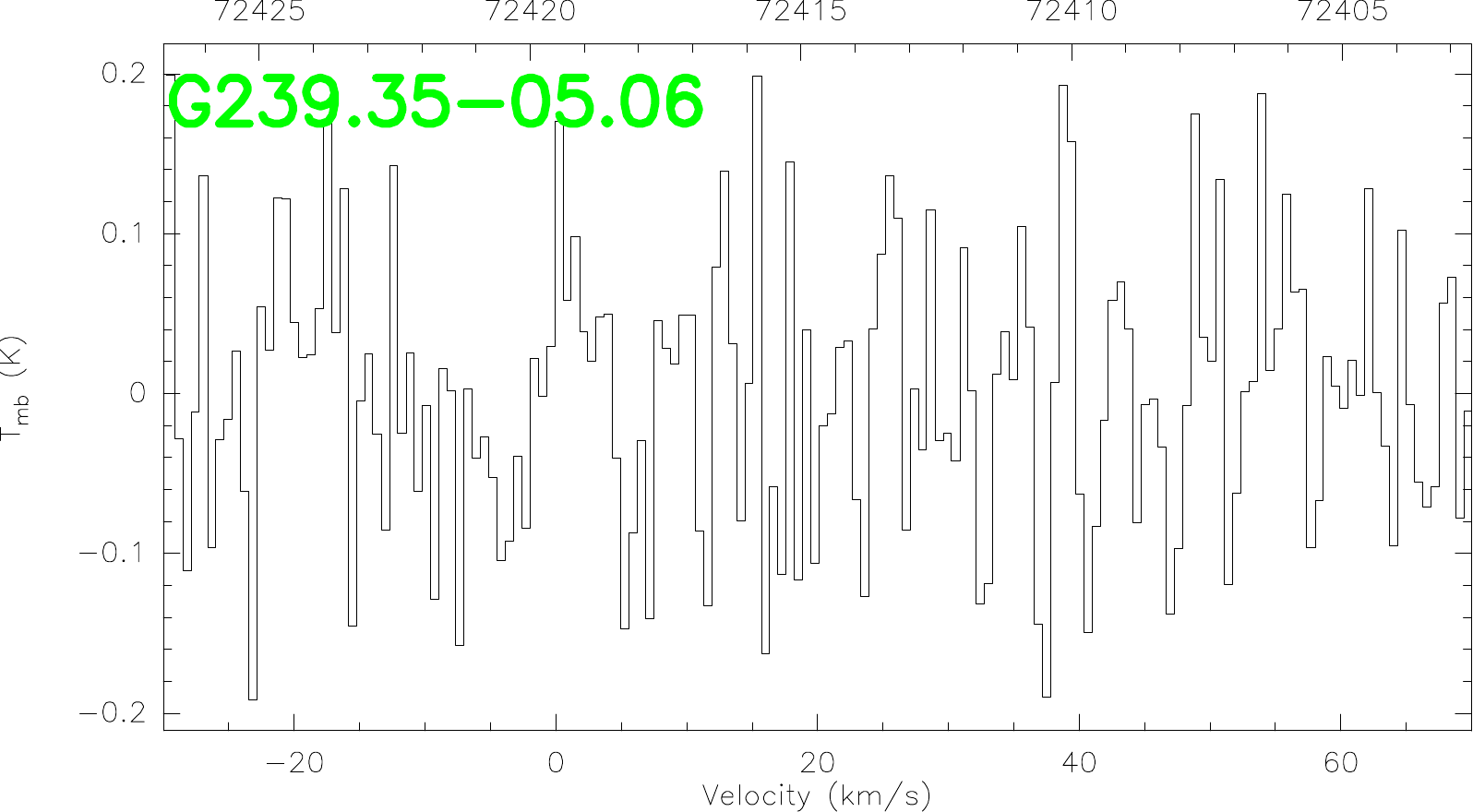}
\includegraphics[width=0.155\textwidth]{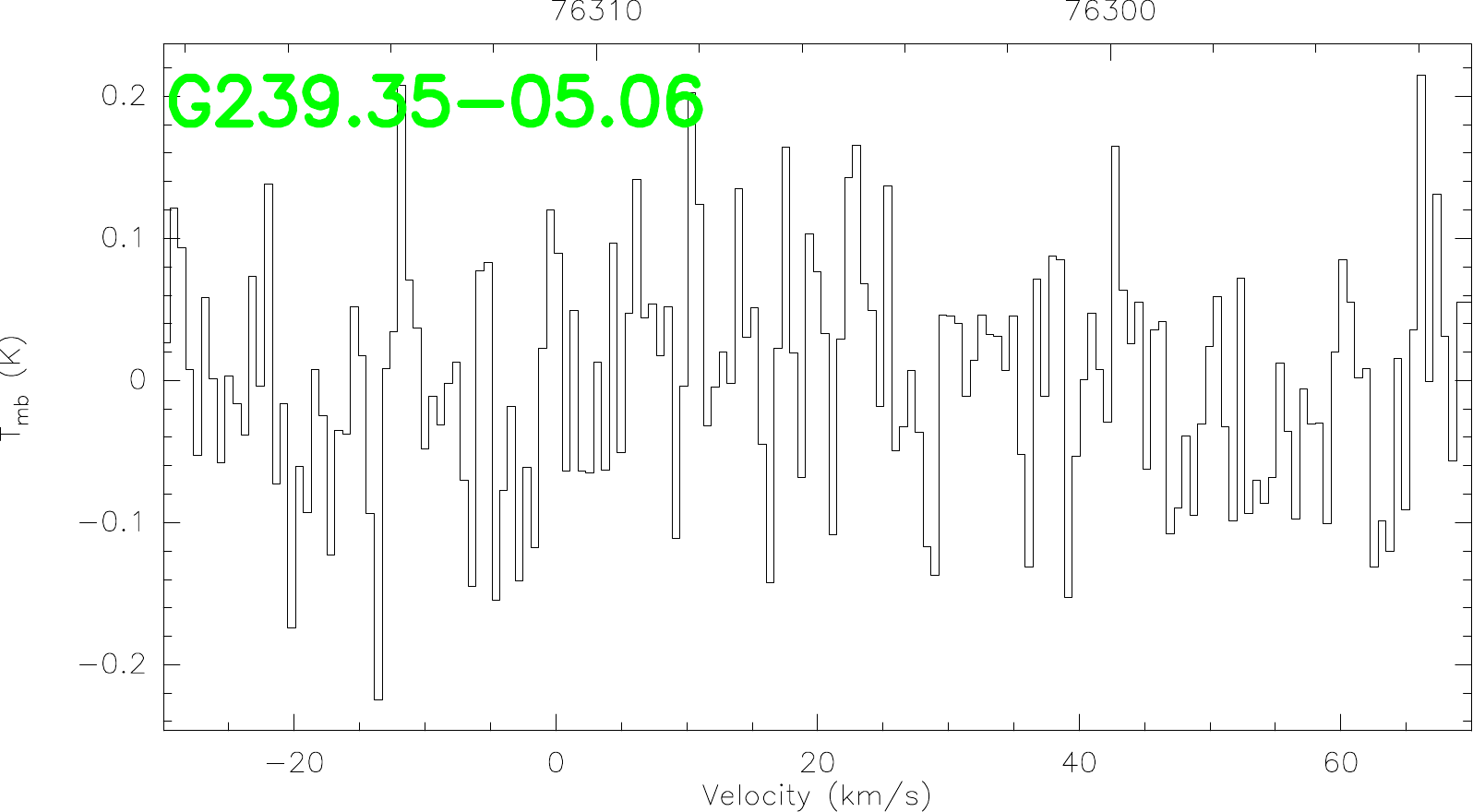}
\includegraphics[width=0.155\textwidth]{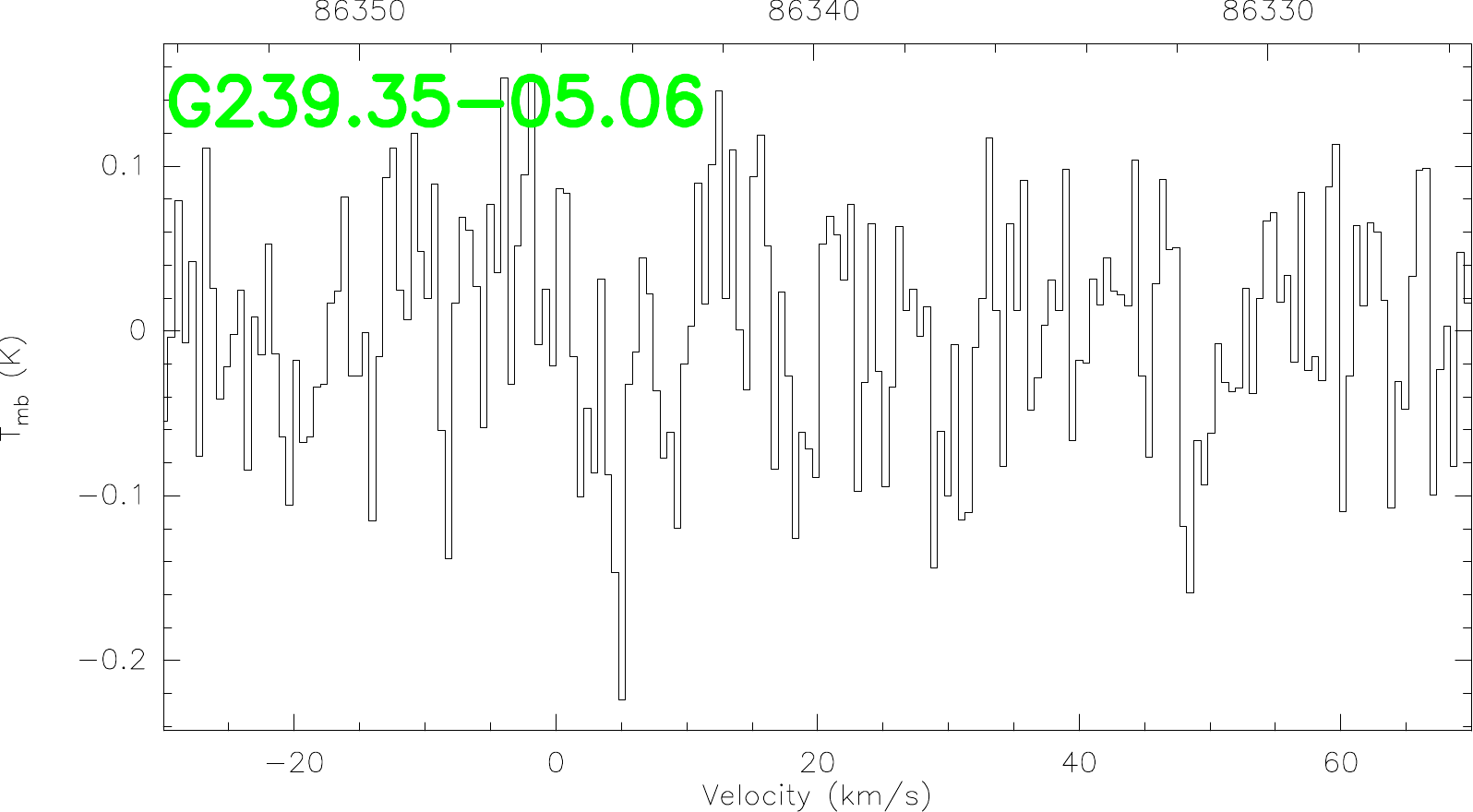}
\includegraphics[width=0.155\textwidth]{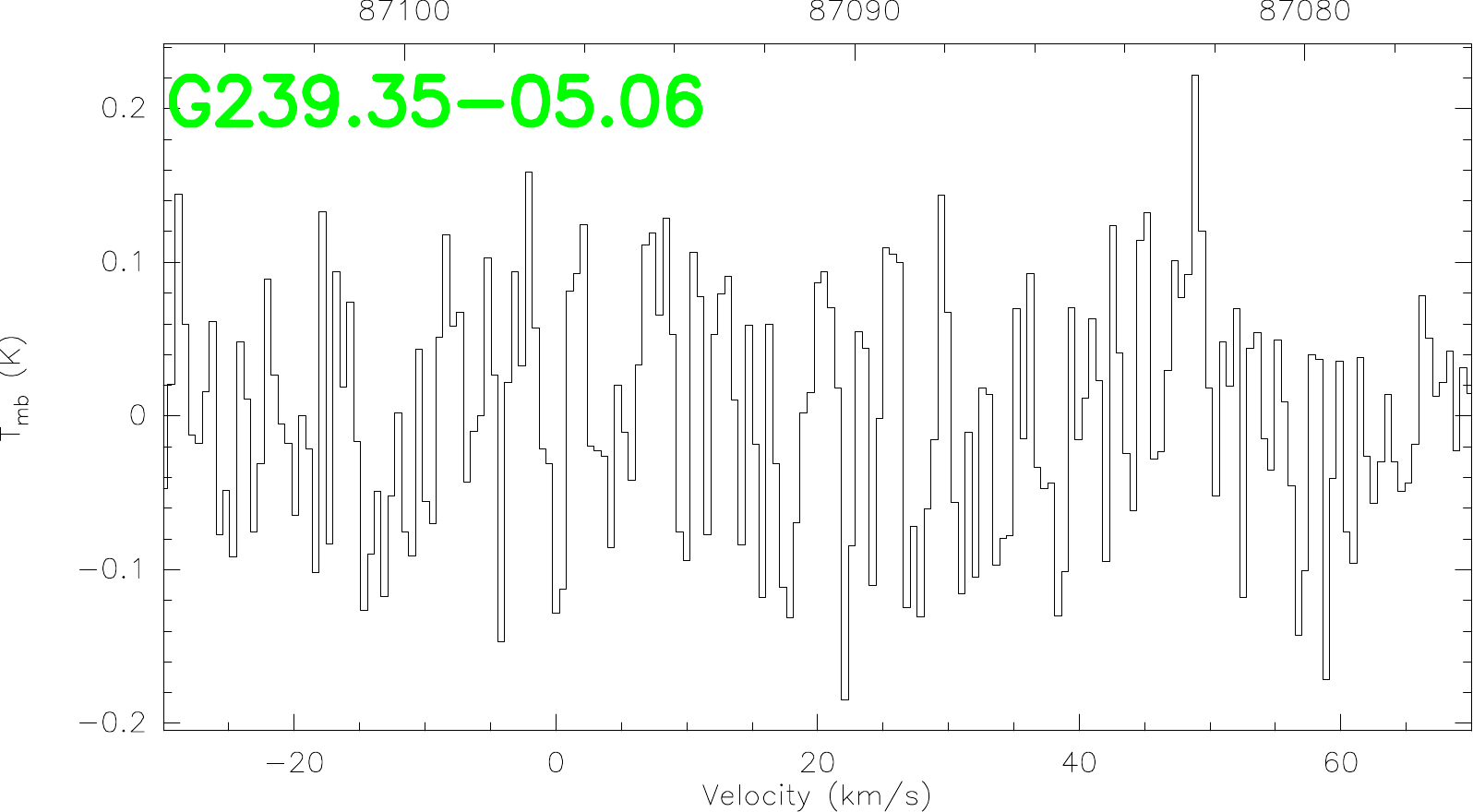}
\includegraphics[width=0.155\textwidth]{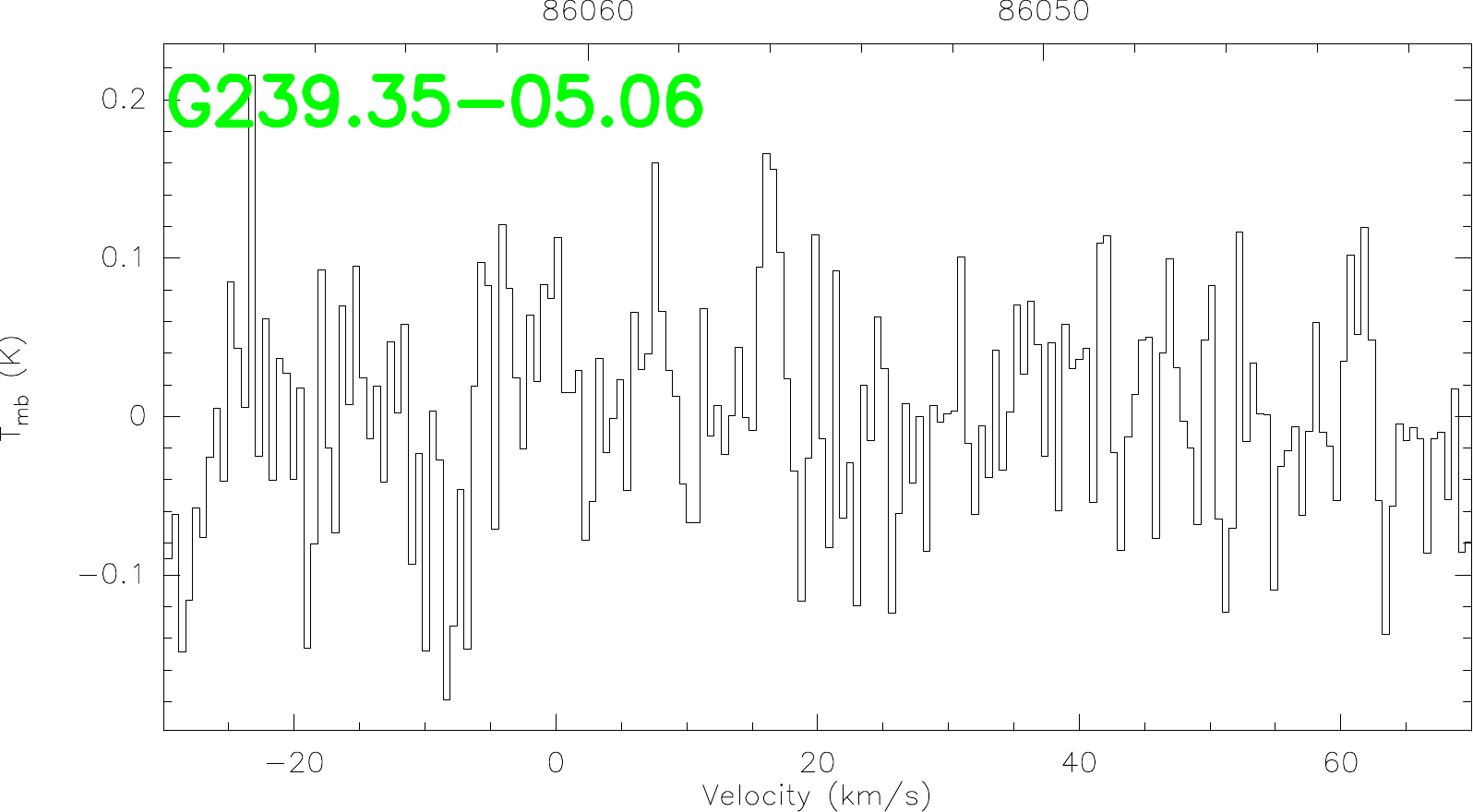}
\includegraphics[width=0.155\textwidth]{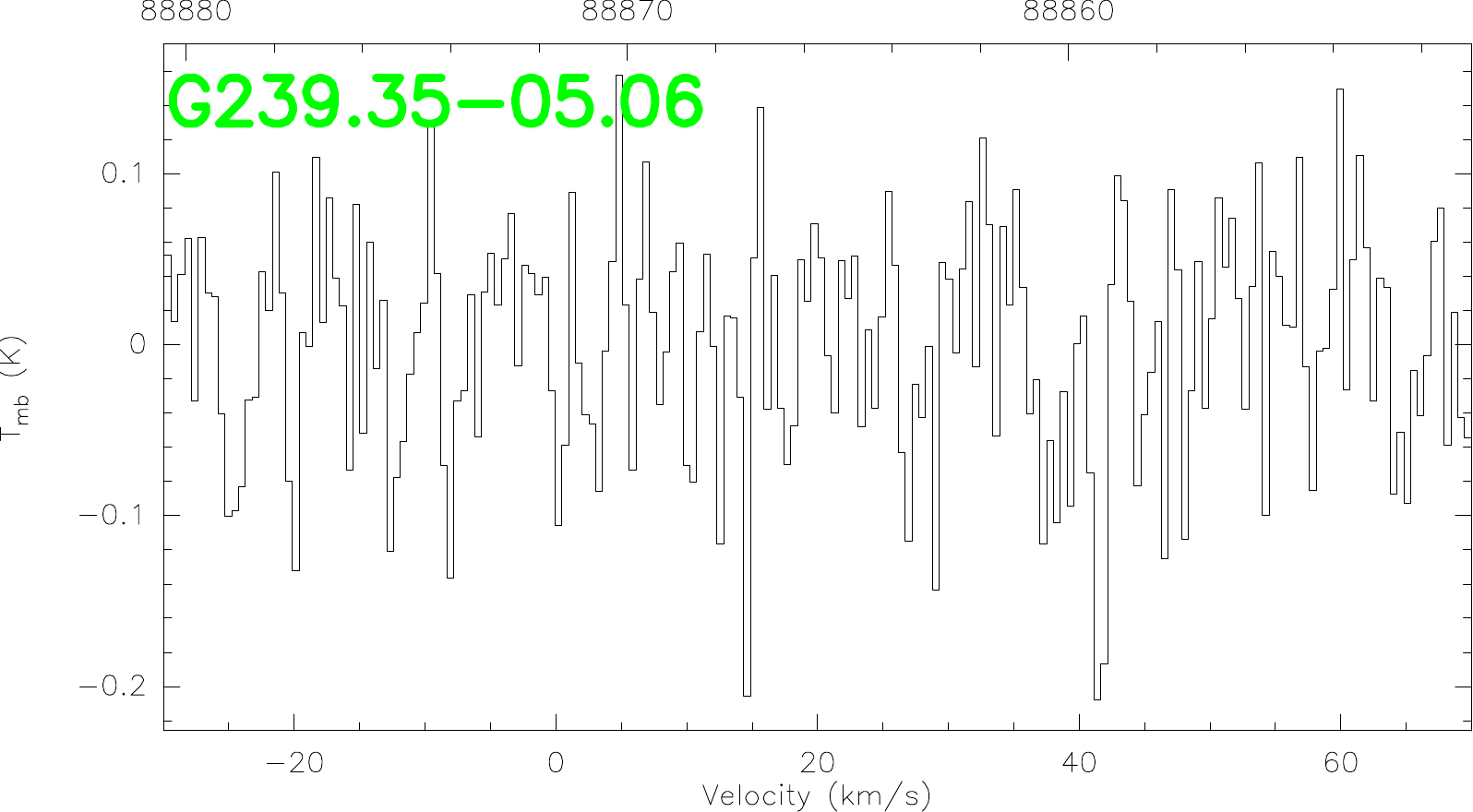}\\

\includegraphics[width=0.155\textwidth]{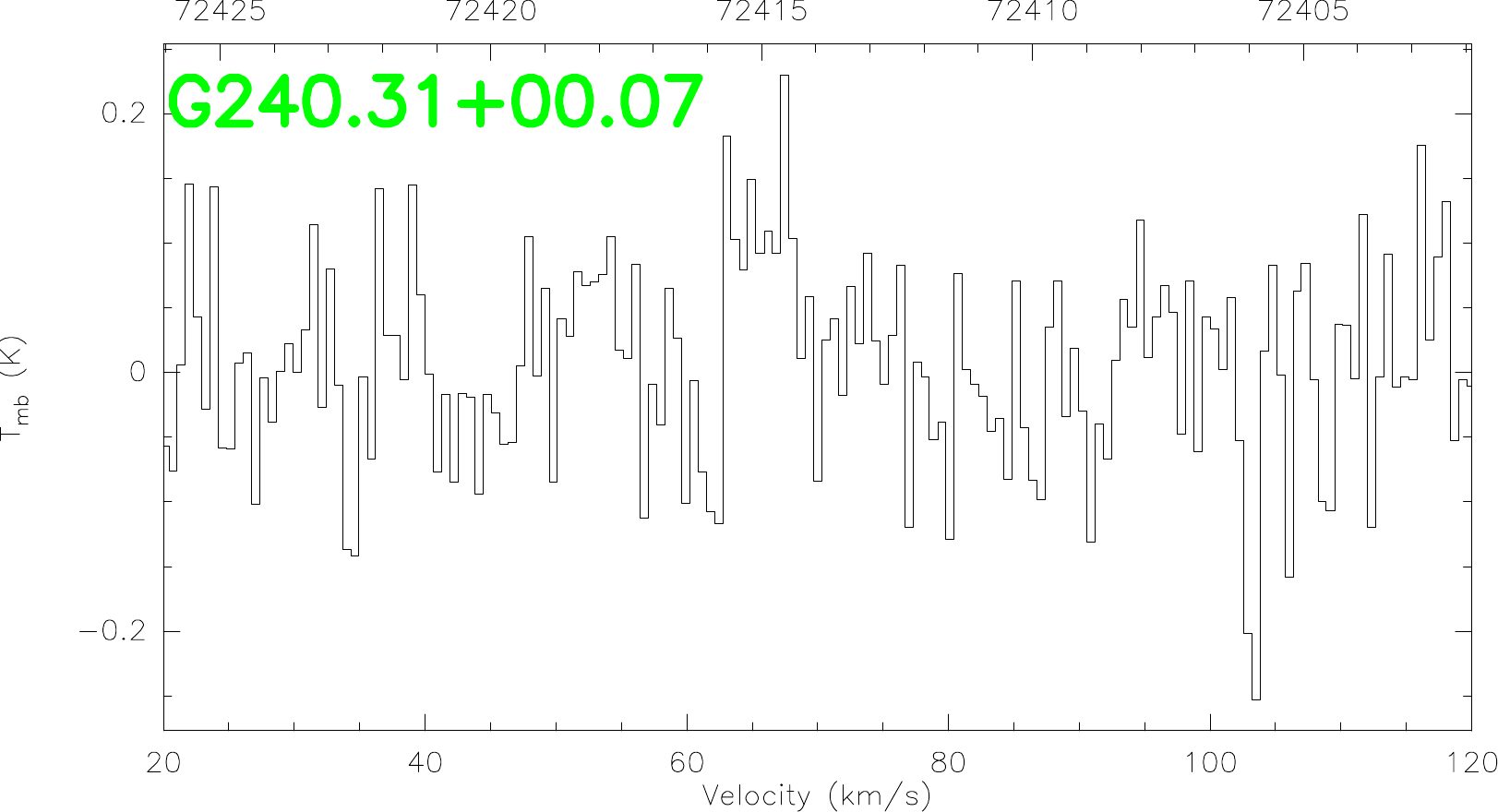}
\includegraphics[width=0.155\textwidth]{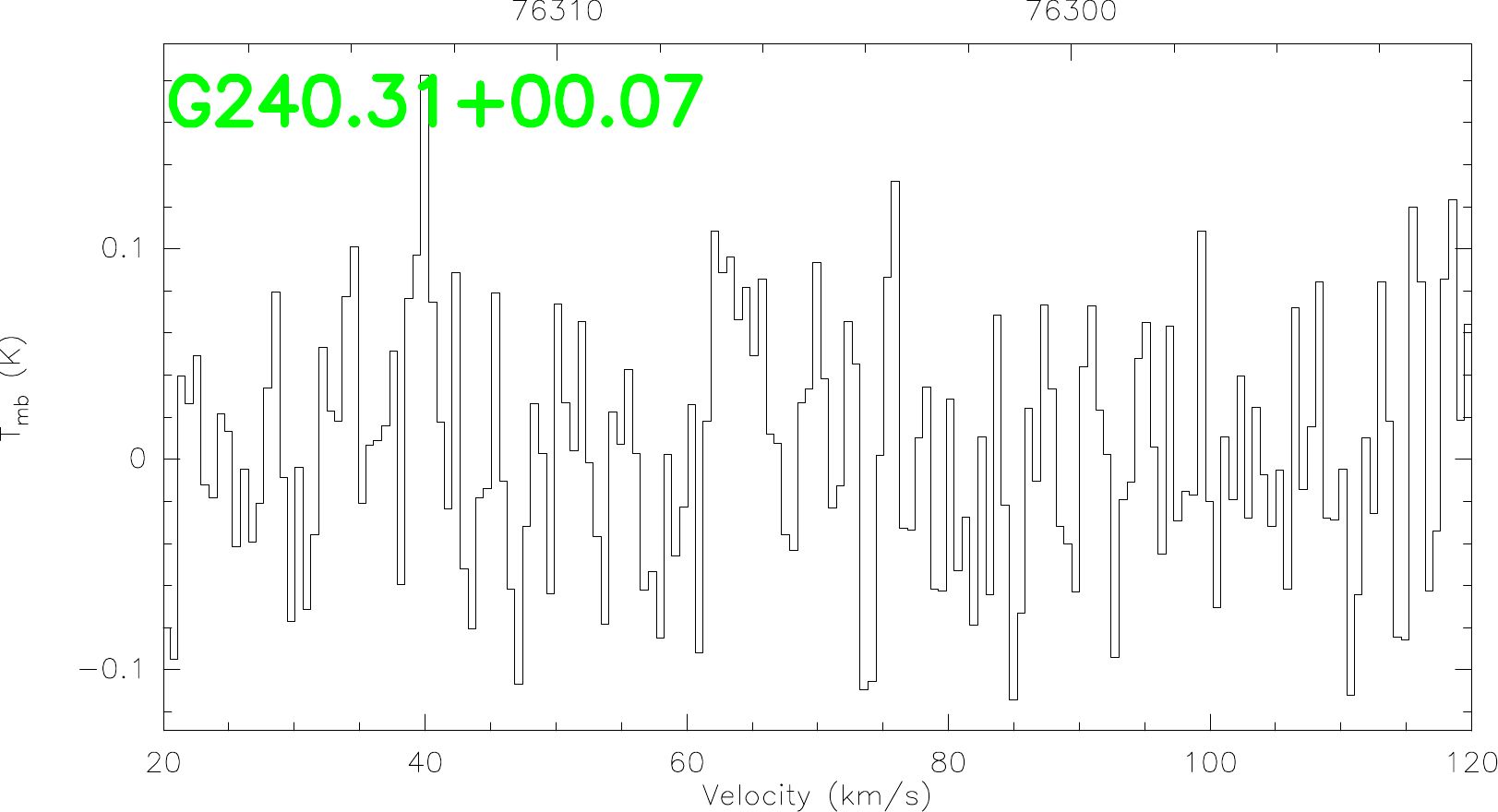}
\includegraphics[width=0.155\textwidth]{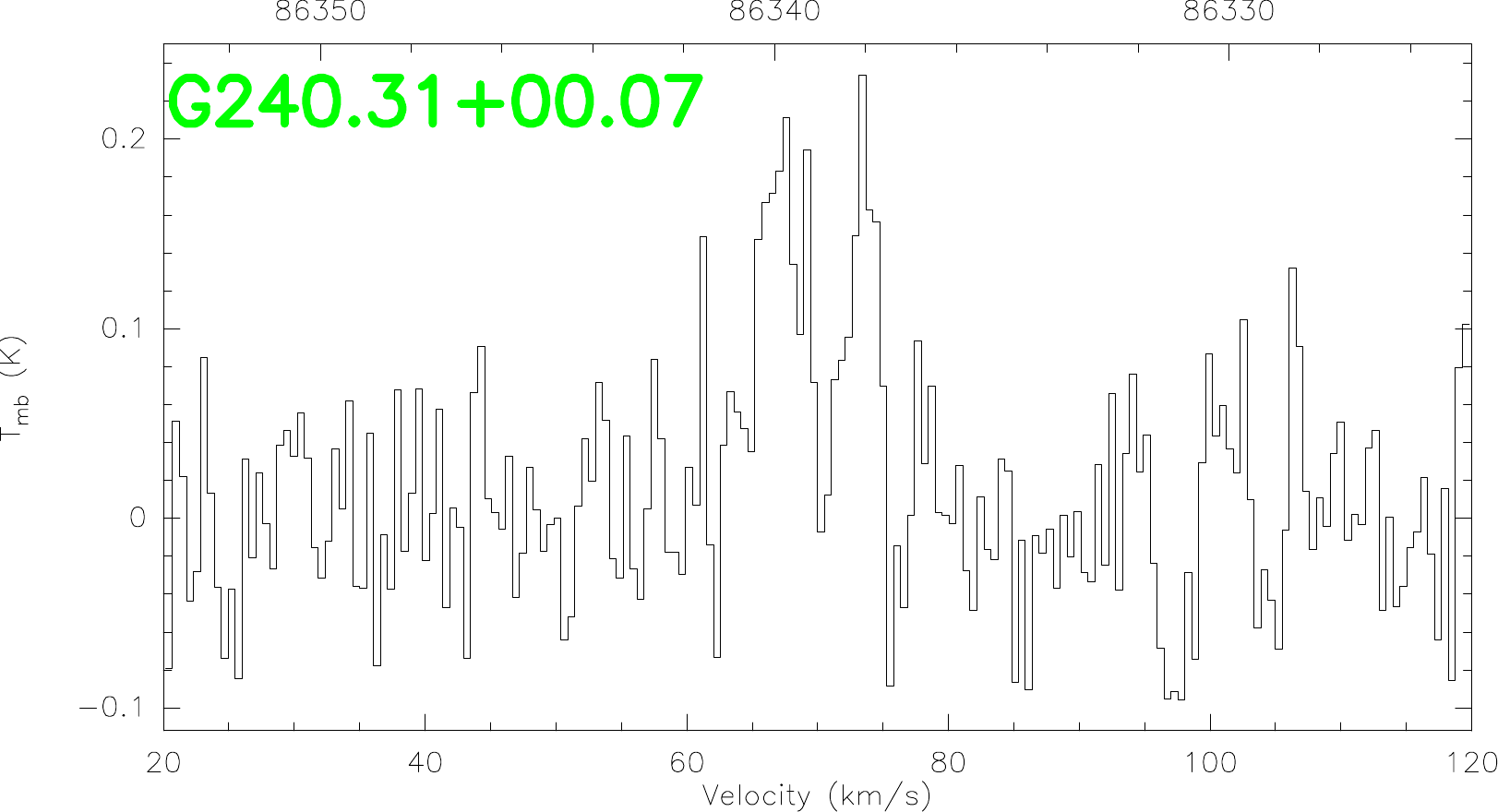}
\includegraphics[width=0.155\textwidth]{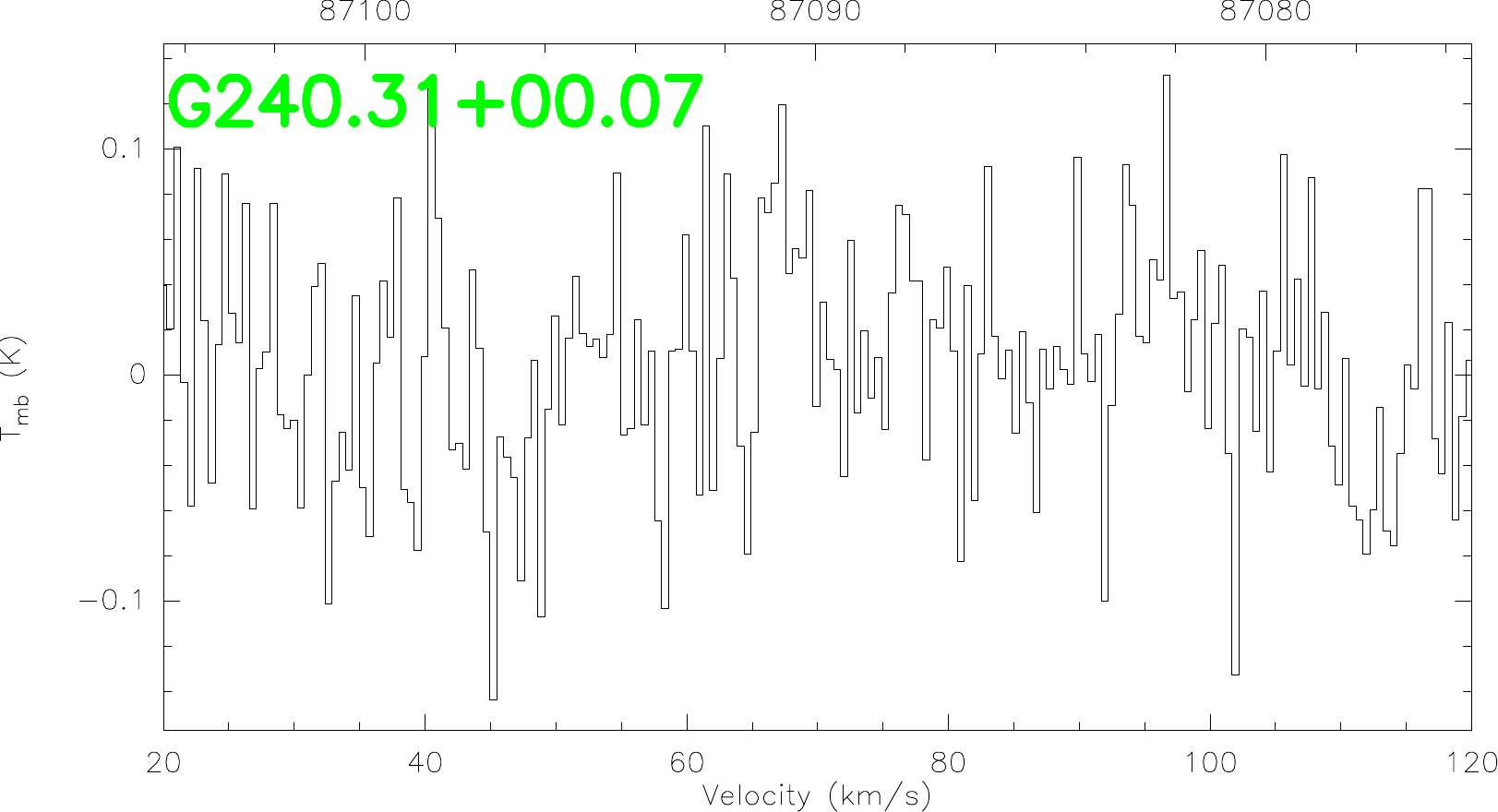}
\includegraphics[width=0.155\textwidth]{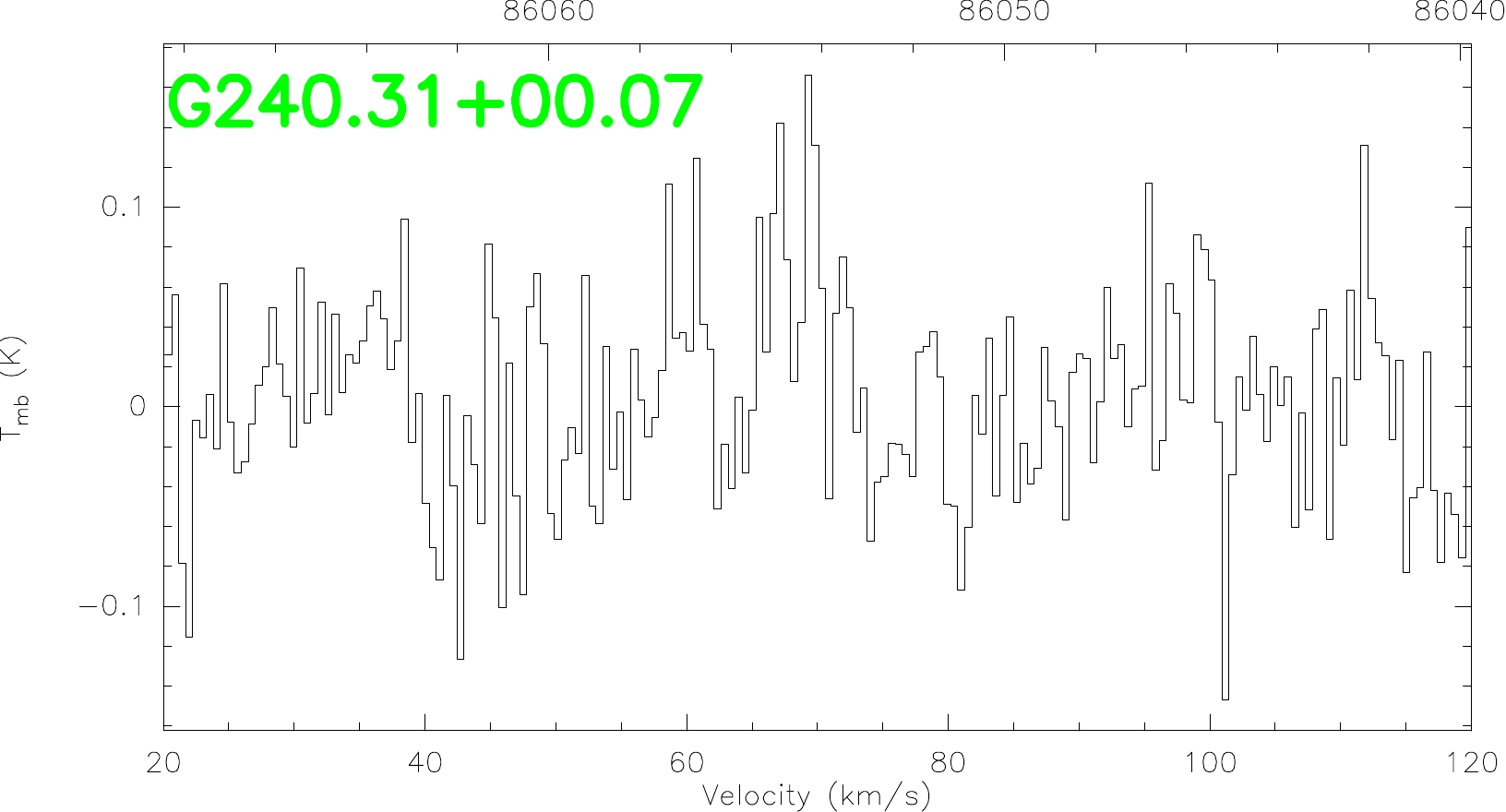}
\includegraphics[width=0.155\textwidth]{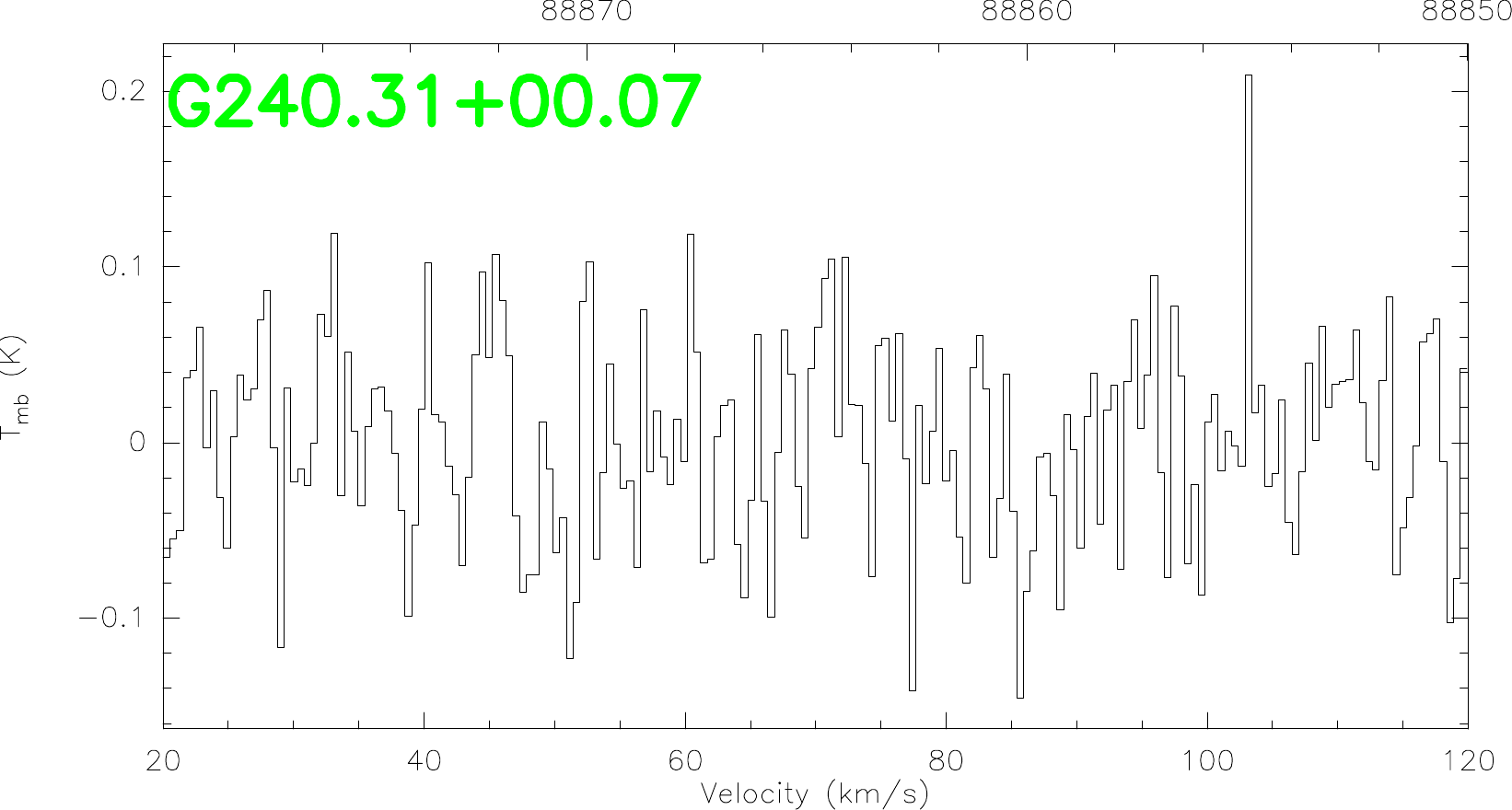}\\

\caption{Continued. Spectra of different molecular lines.}
\label{fig:multi_lines_page7}
\end{figure}

\clearpage

\section{Observed line parameters}
\label{app: line flux table}

Table \ref{app_tab: line flux table} presents the integrated intensities of the six observed transitions toward all 101 sources. 
Upper limits correspond to the 3$\sigma$ values estimated for non-detections,.

{\scriptsize
\begin{longtable}{lcccccccccccc}

\caption{Velocity-integrated intensities of the observed molecular transitions}
\label{app_tab: line flux table} \\

\hline
\hline
\multirow{3}{*}{Source} 
& \multicolumn{2}{c}{DCN 1-0} 
& \multicolumn{2}{c}{DNC 1-0} 
& \multicolumn{2}{c}{H$^{13}$CN 1-0} 
& \multicolumn{2}{c}{HN$^{13}$C 1-0} 
& \multicolumn{2}{c}{HC$^{15}$N 1-0} 
& \multicolumn{2}{c}{H$^{15}$NC 1-0} \\

\cmidrule(lr){2-3}
\cmidrule(lr){4-5}
\cmidrule(lr){6-7}
\cmidrule(lr){8-9}
\cmidrule(lr){10-11}
\cmidrule(lr){12-13}

& rms & $\int$$T_{\rm mb}{\rm dv}$
& rms & $\int$$T_{\rm mb}{\rm dv}$
& rms & $\int$$T_{\rm mb}{\rm dv}$
& rms & $\int$$T_{\rm mb}{\rm dv}$
& rms & $\int$$T_{\rm mb}{\rm dv}$
& rms & $\int$$T_{\rm mb}{\rm dv}$ \\

& (K) & (K $\cdot$ km\; s$^{-1}$)
& (K) & (K $\cdot$ km\; s$^{-1}$)
& (K) & (K $\cdot$ km\; s$^{-1}$)
& (K) & (K $\cdot$ km\; s$^{-1}$)
& (K) & (K $\cdot$ km\; s$^{-1}$)
& (K) & (K $\cdot$ km\; s$^{-1}$) \\

\hline
\endfirsthead

\caption{continued.}\\

\hline
\hline
\multirow{3}{*}{Source} 
& \multicolumn{2}{c}{DCN 1-0} 
& \multicolumn{2}{c}{DNC 1-0} 
& \multicolumn{2}{c}{H$^{13}$CN 1-0} 
& \multicolumn{2}{c}{HN$^{13}$C 1-0} 
& \multicolumn{2}{c}{HC$^{15}$N 1-0} 
& \multicolumn{2}{c}{H$^{15}$NC 1-0} \\

\cmidrule(lr){2-3}
\cmidrule(lr){4-5}
\cmidrule(lr){6-7}
\cmidrule(lr){8-9}
\cmidrule(lr){10-11}
\cmidrule(lr){12-13}

& rms & $\int$$T_{\rm mb}{\rm dv}$
& rms & $\int$$T_{\rm mb}{\rm dv}$
& rms & $\int$$T_{\rm mb}{\rm dv}$
& rms & $\int$$T_{\rm mb}{\rm dv}$
& rms & $\int$$T_{\rm mb}{\rm dv}$
& rms & $\int$$T_{\rm mb}{\rm dv}$ \\

& (K) & (K $\cdot$ km\; s$^{-1}$)
& (K) & (K $\cdot$ km\; s$^{-1}$)
& (K) & (K $\cdot$ km\; s$^{-1}$)
& (K) & (K $\cdot$ km\; s$^{-1}$)
& (K) & (K $\cdot$ km\; s$^{-1}$)
& (K) & (K $\cdot$ km\; s$^{-1}$) \\

\hline
\endhead

\hline
\multicolumn{13}{r}{\emph{Continued on next page}} \\
\endfoot

\hline
\endlastfoot

G000.67-00.03  & 0.16 & 29.10$\pm$0.71 & 0.04 & 2.08$\pm$0.30 & 0.12 & 33.91$\pm$0.95 & 0.03 & 6.40$\pm$0.75 & 0.03 & 7.39$\pm$0.72 & 0.03 & $\leq$0.06 \\ 
G005.88-00.39  & 0.13 & 6.85$\pm$0.51 & 0.02 & 3.70$\pm$0.18 & 0.05 & 84.38$\pm$0.29 & 0.01 & 20.32$\pm$0.15 & 0.01 & 12.06$\pm$0.19 & 0.01 & 3.38$\pm$0.13 \\ 
G009.62+00.19  & 0.08 & 2.77$\pm$0.32 & 0.02 & 0.72$\pm$0.14 & 0.06 & 32.30$\pm$0.34 & 0.01 & 6.49$\pm$0.15 & 0.01 & 5.07$\pm$0.17 & 0.01 & 0.53$\pm$0.10 \\ 
G010.47+00.02  & 0.22 & 2.85$\pm$0.83 & 0.02 & 2.26$\pm$0.20 & 0.05 & 60.44$\pm$0.27 & 0.01 & 12.62$\pm$0.18 & 0.01 & 18.00$\pm$0.16 & 0.01 & 1.40$\pm$0.16 \\ 
G010.62-00.38  & 0.09 & 6.70$\pm$0.34 & 0.02 & 2.78$\pm$0.24 & 0.11 & 66.49$\pm$0.52 & 0.01 & 16.62$\pm$0.17 & 0.01 & 9.38$\pm$0.18 & 0.01 & 2.29$\pm$0.17 \\ 
G011.49-01.48  & 0.07 & 1.61$\pm$0.23 & 0.02 & 0.79$\pm$0.11 & 0.04 & 5.22$\pm$0.14 & 0.01 & 1.09$\pm$0.07 & 0.01 & 1.14$\pm$0.09 & 0.01 & 0.24$\pm$0.06 \\ 
G011.91-00.61  & 0.08 & 1.82$\pm$0.22 & 0.02 & 1.97$\pm$0.11 & 0.05 & 19.90$\pm$0.25 & 0.01 & 5.91$\pm$0.13 & 0.01 & 2.65$\pm$0.19 & 0.01 & 0.74$\pm$0.11 \\ 
G012.02-00.03  & 0.07 & $\leq$0.23 & 0.02 & $\leq$0.03 & 0.07 & 2.33$\pm$0.25 & 0.01 & 1.07$\pm$0.13 & 0.01 & $\leq$0.02 & 0.02 & $\leq$0.02 \\ 
G012.68-00.18  & 0.10 & $\leq$0.34 & 0.03 & $\leq$0.03 & 0.09 & 9.37$\pm$0.38 & 0.02 & 2.42$\pm$0.14 & 0.02 & 2.21$\pm$0.24 & 0.02 & $\leq$0.02 \\ 
G012.80-00.20  & 0.07 & 2.96$\pm$0.27 & 0.02 & 1.36$\pm$0.20 & 0.08 & 15.50$\pm$0.27 & 0.02 & 6.12$\pm$0.20 & 0.02 & 1.04$\pm$0.12 & 0.02 & $\leq$0.03 \\ 
G012.88+00.48  & 0.09 & 3.45$\pm$0.33 & 0.02 & 1.52$\pm$0.21 & 0.08 & 18.71$\pm$0.31 & 0.02 & 5.81$\pm$0.16 & 0.02 & 3.71$\pm$0.18 & 0.02 & 0.61$\pm$0.19 \\ 
G012.90-00.24  & 0.08 & 1.11$\pm$0.32 & 0.02 & 2.28$\pm$0.16 & 0.09 & 4.34$\pm$0.30 & 0.02 & 3.96$\pm$0.16 & 0.02 & 0.77$\pm$0.18 & 0.02 & $\leq$0.03 \\ 
G012.90-00.26  & 0.12 & 3.37$\pm$0.44 & 0.03 & 2.54$\pm$0.20 & 0.09 & 20.80$\pm$0.39 & 0.02 & 7.63$\pm$0.23 & 0.02 & 3.57$\pm$0.29 & 0.02 & 1.16$\pm$0.24 \\ 
G013.87+00.28  & 0.07 & $\leq$0.23 & 0.02 & $\leq$0.03 & 0.06 & 6.32$\pm$0.24 & 0.02 & 1.28$\pm$0.14 & 0.02 & 0.65$\pm$0.11 & 0.02 & $\leq$0.02 \\ 
G014.33-00.64  & 0.11 & 4.55$\pm$0.41 & 0.02 & 3.49$\pm$0.18 & 0.07 & 24.24$\pm$0.33 & 0.01 & 7.23$\pm$0.12 & 0.01 & 4.00$\pm$0.14 & 0.02 & 1.04$\pm$0.11 \\ 
G014.63-00.57  & 0.08 & 2.48$\pm$0.30 & 0.02 & 2.44$\pm$0.15 & 0.06 & 7.63$\pm$0.19 & 0.01 & 4.13$\pm$0.09 & 0.01 & 1.47$\pm$0.09 & 0.01 & 0.95$\pm$0.07 \\ 
G015.03-00.67  & 0.10 & $\leq$0.34 & 0.03 & $\leq$0.05 & 0.10 & 8.03$\pm$0.31 & 0.02 & 1.53$\pm$0.22 & 0.02 & 1.43$\pm$0.16 & 0.02 & $\leq$0.03 \\ 
G016.58-00.05  & 0.10 & 1.76$\pm$0.28 & 0.03 & 1.96$\pm$0.21 & 0.09 & 12.66$\pm$0.33 & 0.02 & 4.10$\pm$0.18 & 0.02 & 1.82$\pm$0.27 & 0.02 & $\leq$0.03 \\ 
G023.00-00.41  & 0.08 & 1.25$\pm$0.29 & 0.02 & 1.45$\pm$0.15 & 0.06 & 15.38$\pm$0.27 & 0.01 & 7.96$\pm$0.15 & 0.02 & 2.57$\pm$0.24 & 0.01 & 1.05$\pm$0.13 \\ 
G023.43-00.18  & 0.06 & 0.77$\pm$0.18 & 0.02 & 1.01$\pm$0.12 & 0.05 & 6.74$\pm$0.21 & 0.01 & 2.07$\pm$0.12 & 0.01 & 1.23$\pm$0.15 & 0.01 & 0.27$\pm$0.08 \\ 
G023.65-00.12  & 0.08 & $\leq$0.26 & 0.02 & $\leq$0.03 & 0.06 & 1.87$\pm$0.20 & 0.01 & $\leq$0.02 & 0.01 & 0.41$\pm$0.11 & 0.01 & $\leq$0.02 \\ 
G023.70-00.19  & 0.07 & $\leq$0.25 & 0.02 & $\leq$0.03 & 0.05 & 2.41$\pm$0.21 & 0.01 & 1.24$\pm$0.10 & 0.01 & 0.45$\pm$0.09 & 0.01 & $\leq$0.01 \\ 
G025.70+00.04  & 0.06 & 0.65$\pm$0.18 & 0.02 & 0.63$\pm$0.11 & 0.04 & 5.45$\pm$0.17 & 0.01 & 2.29$\pm$0.08 & 0.01 & 0.65$\pm$0.10 & 0.01 & 0.35$\pm$0.09 \\ 
G027.36-00.16  & 0.06 & 2.38$\pm$0.23 & 0.02 & 2.23$\pm$0.14 & 0.06 & 9.26$\pm$0.28 & 0.01 & 4.51$\pm$0.11 & 0.01 & 1.27$\pm$0.14 & 0.01 & 0.62$\pm$0.20 \\ 
G028.86+00.06  & 0.10 & $\leq$0.33 & 0.02 & $\leq$0.03 & 0.07 & 8.02$\pm$0.26 & 0.02 & 1.53$\pm$0.14 & 0.01 & 1.04$\pm$0.18 & 0.01 & 0.21$\pm$0.05 \\ 
G029.86-00.04  & 0.08 & $\leq$0.26 & 0.02 & $\leq$0.03 & 0.05 & 3.89$\pm$0.18 & 0.01 & 1.59$\pm$0.10 & 0.01 & 0.54$\pm$0.12 & 0.01 & $\leq$0.02 \\ 
G029.95-00.01  & 0.10 & 2.12$\pm$0.32 & 0.03 & $\leq$0.04 & 0.06 & 26.07$\pm$0.27 & 0.02 & 4.75$\pm$0.17 & 0.02 & 3.87$\pm$0.21 & 0.02 & 0.87$\pm$0.16 \\ 
G031.28+00.06  & 0.06 & 2.14$\pm$0.22 & 0.02 & 0.97$\pm$0.15 & 0.05 & 17.70$\pm$0.17 & 0.01 & 5.25$\pm$0.11 & 0.01 & 2.61$\pm$0.12 & 0.01 & 0.67$\pm$0.08 \\ 
G031.58+00.07  & 0.09 & 1.33$\pm$0.28 & 0.02 & 0.82$\pm$0.18 & 0.07 & 9.45$\pm$0.29 & 0.02 & 3.60$\pm$0.14 & 0.01 & 1.51$\pm$0.13 & 0.02 & 0.83$\pm$0.18 \\ 
G032.04+00.05  & 0.06 & 1.61$\pm$0.21 & 0.02 & 1.28$\pm$0.13 & 0.05 & 14.24$\pm$0.21 & 0.01 & 4.38$\pm$0.12 & 0.01 & 1.81$\pm$0.13 & 0.01 & 0.63$\pm$0.11 \\ 
G033.64-00.22  & 0.07 & $\leq$0.22 & 0.02 & $\leq$0.03 & 0.05 & 3.00$\pm$0.16 & 0.01 & 0.90$\pm$0.10 & 0.01 & 0.41$\pm$0.10 & 0.01 & $\leq$0.02 \\ 
G034.39+00.22  & 0.06 & 2.85$\pm$0.23 & 0.02 & 3.00$\pm$0.13 & 0.06 & 12.78$\pm$0.26 & 0.01 & 5.36$\pm$0.10 & 0.02 & 1.32$\pm$0.14 & 0.01 & 1.00$\pm$0.15 \\ 
G035.02+00.34  & 0.07 & 1.73$\pm$0.25 & 0.02 & 0.61$\pm$0.15 & 0.07 & 10.72$\pm$0.27 & 0.02 & 2.49$\pm$0.13 & 0.02 & 1.83$\pm$0.16 & 0.02 & 0.50$\pm$0.12 \\ 
G035.19-00.74  & 0.08 & 3.30$\pm$0.20 & 0.02 & 1.78$\pm$0.13 & 0.07 & 19.00$\pm$0.29 & 0.02 & 6.77$\pm$0.15 & 0.02 & 3.15$\pm$0.18 & 0.01 & 1.04$\pm$0.11 \\ 
G035.20-01.73  & 0.05 & 1.53$\pm$0.14 & 0.01 & 0.46$\pm$0.13 & 0.04 & 8.99$\pm$0.17 & 0.01 & 1.79$\pm$0.10 & 0.01 & 1.51$\pm$0.08 & 0.01 & $\leq$0.02 \\ 
G037.43+01.51  & 0.10 & 1.63$\pm$0.27 & 0.03 & 0.73$\pm$0.20 & 0.09 & 6.66$\pm$0.29 & 0.02 & 1.43$\pm$0.13 & 0.02 & 1.61$\pm$0.23 & 0.02 & 0.56$\pm$0.10 \\ 
G043.16+00.01  & 0.08 & 3.30$\pm$0.28 & 0.02 & $\leq$0.07 & 0.07 & 27.65$\pm$0.35 & 0.02 & 3.80$\pm$0.32 & 0.02 & 6.14$\pm$0.29 & 0.02 & $\leq$0.05 \\ 
G043.79-00.12  & 0.08 & 0.92$\pm$0.22 & 0.02 & $\leq$0.04 & 0.09 & 11.60$\pm$0.37 & 0.02 & 2.30$\pm$0.20 & 0.02 & 1.17$\pm$0.18 & 0.02 & $\leq$0.03 \\ 
G043.89-00.78  & 0.06 & 1.11$\pm$0.19 & 0.02 & 1.13$\pm$0.12 & 0.06 & 5.27$\pm$0.20 & 0.01 & 2.08$\pm$0.12 & 0.01 & 0.66$\pm$0.13 & 0.01 & 0.42$\pm$0.08 \\ 
G045.07+00.13  & 0.08 & $\leq$0.28 & 0.02 & $\leq$0.04 & 0.08 & 10.40$\pm$0.30 & 0.02 & 1.57$\pm$0.24 & 0.02 & 2.24$\pm$0.27 & 0.02 & $\leq$0.04 \\ 
G045.45+00.05  & 0.07 & $\leq$0.25 & 0.02 & $\leq$0.03 & 0.07 & 4.47$\pm$0.21 & 0.02 & 1.10$\pm$0.16 & 0.02 & 0.99$\pm$0.15 & 0.02 & $\leq$0.03 \\ 
G048.60+00.02  & 0.10 & 1.27$\pm$0.33 & 0.03 & $\leq$0.03 & 0.09 & 3.76$\pm$0.26 & 0.02 & $\leq$0.03 & 0.02 & $\leq$0.03 & 0.02 & $\leq$0.03 \\ 
G049.19-00.33  & 0.04 & $\leq$0.13 & 0.01 & 0.38$\pm$0.09 & 0.04 & 1.87$\pm$0.13 & 0.01 & 1.27$\pm$0.07 & 0.01 & 0.21$\pm$0.06 & 0.01 & 0.29$\pm$0.07 \\ 
G049.48-00.36  & 0.27 & 8.00$\pm$1.12 & 0.02 & 2.07$\pm$0.25 & 0.07 & 93.27$\pm$0.32 & 0.02 & 11.66$\pm$0.35 & 0.02 & 20.63$\pm$0.29 & 0.01 & 1.42$\pm$0.28 \\ 
G049.48-00.38  & 0.09 & 3.92$\pm$0.34 & 0.02 & $\leq$0.04 & 0.06 & 38.61$\pm$0.28 & 0.02 & 7.85$\pm$0.25 & 0.01 & 7.10$\pm$0.20 & 0.02 & 1.50$\pm$0.19 \\ 
G052.10+01.04  & 0.08 & $\leq$0.26 & 0.02 & $\leq$0.03 & 0.07 & 1.82$\pm$0.20 & 0.02 & $\leq$0.02 & 0.02 & $\leq$0.02 & 0.02 & $\leq$0.02 \\ 
G059.78+00.06  & 0.07 & 2.50$\pm$0.23 & 0.02 & 1.60$\pm$0.10 & 0.05 & 10.42$\pm$0.16 & 0.01 & 3.26$\pm$0.08 & 0.01 & 1.45$\pm$0.09 & 0.01 & 0.85$\pm$0.08 \\ 
G069.54-00.97  & 0.06 & 2.30$\pm$0.23 & 0.01 & 1.75$\pm$0.12 & 0.04 & 9.40$\pm$0.16 & 0.01 & 3.87$\pm$0.10 & 0.01 & 1.67$\pm$0.10 & 0.01 & 0.63$\pm$0.07 \\ 
G074.03-01.71  & 0.05 & 1.11$\pm$0.15 & 0.01 & 2.28$\pm$0.08 & 0.05 & 6.32$\pm$0.16 & 0.01 & 3.02$\pm$0.07 & 0.01 & 0.84$\pm$0.08 & 0.01 & 0.55$\pm$0.07 \\ 
G075.29+01.32  & 0.04 & $\leq$0.15 & 0.01 & $\leq$0.02 & 0.04 & 1.61$\pm$0.13 & 0.01 & $\leq$0.01 & 0.01 & 0.26$\pm$0.05 & 0.01 & $\leq$0.01 \\ 
G075.76+00.33  & 0.04 & 2.94$\pm$0.15 & 0.01 & 1.06$\pm$0.10 & 0.04 & 12.66$\pm$0.15 & 0.01 & 2.75$\pm$0.08 & 0.01 & 1.92$\pm$0.11 & 0.01 & 0.71$\pm$0.10 \\ 
G075.78+00.34  & 0.04 & 2.27$\pm$0.14 & 0.01 & 0.63$\pm$0.08 & 0.03 & 8.21$\pm$0.12 & 0.01 & 1.65$\pm$0.08 & 0.01 & 1.58$\pm$0.08 & 0.01 & 0.30$\pm$0.06 \\ 
G076.38-00.61  & 0.05 & $\leq$0.15 & 0.01 & 0.48$\pm$0.06 & 0.04 & 3.05$\pm$0.15 & 0.01 & 0.85$\pm$0.07 & 0.01 & 0.53$\pm$0.07 & 0.01 & $\leq$0.01 \\ 
G078.12+03.63  & 0.04 & 0.78$\pm$0.13 & 0.01 & 0.55$\pm$0.06 & 0.04 & 2.40$\pm$0.13 & 0.01 & 1.28$\pm$0.06 & 0.01 & 0.59$\pm$0.08 & 0.01 & 0.31$\pm$0.05 \\ 
G078.88+00.70  & 0.08 & $\leq$0.25 & 0.02 & $\leq$0.02 & 0.06 & 8.20$\pm$0.20 & 0.01 & 1.03$\pm$0.10 & 0.01 & 1.49$\pm$0.12 & 0.01 & $\leq$0.02 \\ 
G079.73+00.99  & 0.05 & 0.74$\pm$0.17 & 0.01 & 0.53$\pm$0.10 & 0.04 & 2.71$\pm$0.13 & 0.01 & 0.65$\pm$0.08 & 0.01 & 0.50$\pm$0.07 & 0.01 & $\leq$0.01 \\ 
G079.87+01.17  & 0.05 & 1.08$\pm$0.19 & 0.01 & 0.45$\pm$0.10 & 0.04 & 6.68$\pm$0.13 & 0.01 & 1.71$\pm$0.07 & 0.01 & 0.95$\pm$0.06 & 0.01 & $\leq$0.01 \\ 
G080.79-01.92  & 0.04 & $\leq$0.13 & 0.01 & $\leq$0.02 & 0.05 & $\leq$0.14 & 0.01 & $\leq$0.01 & 0.01 & $\leq$0.01 & 0.01 & $\leq$0.01 \\ 
G080.86+00.38  & 0.06 & 0.99$\pm$0.23 & 0.02 & $\leq$0.02 & 0.05 & 5.59$\pm$0.17 & 0.01 & 1.01$\pm$0.08 & 0.01 & 0.95$\pm$0.13 & 0.01 & $\leq$0.01 \\ 
G081.75+00.59  & 0.06 & 5.58$\pm$0.24 & 0.02 & 4.58$\pm$0.11 & 0.06 & 15.13$\pm$0.21 & 0.01 & 6.10$\pm$0.08 & 0.01 & 2.71$\pm$0.09 & 0.01 & 1.48$\pm$0.09 \\ 
G081.87+00.78  & 0.10 & 4.00$\pm$0.40 & 0.02 & 0.86$\pm$0.15 & 0.06 & 20.00$\pm$0.26 & 0.01 & 3.37$\pm$0.13 & 0.01 & 3.61$\pm$0.15 & 0.02 & 0.79$\pm$0.12 \\ 
G090.21+02.32  & 0.08 & $\leq$0.26 & 0.02 & 0.85$\pm$0.08 & 0.07 & $\leq$0.18 & 0.01 & 0.32$\pm$0.07 & 0.01 & $\leq$0.01 & 0.02 & $\leq$0.01 \\ 
G092.67+03.07  & 0.06 & 2.49$\pm$0.20 & 0.01 & 0.81$\pm$0.11 & 0.05 & 10.93$\pm$0.17 & 0.01 & 2.07$\pm$0.08 & 0.01 & 2.17$\pm$0.08 & 0.01 & 0.46$\pm$0.07 \\ 
G094.60-01.79  & 0.07 & $\leq$0.23 & 0.02 & 0.40$\pm$0.09 & 0.06 & 1.53$\pm$0.18 & 0.01 & 0.82$\pm$0.09 & 0.01 & 0.36$\pm$0.11 & 0.01 & $\leq$0.02 \\ 
G095.29-00.93  & 0.04 & $\leq$0.13 & 0.01 & $\leq$0.02 & 0.04 & 0.58$\pm$0.11 & 0.01 & $\leq$0.01 & 0.01 & $\leq$0.01 & 0.01 & $\leq$0.01 \\ 
G097.53+03.18  & 0.07 & 0.88$\pm$0.18 & 0.02 & $\leq$0.03 & 0.05 & 2.98$\pm$0.19 & 0.01 & $\leq$0.02 & 0.01 & $\leq$0.02 & 0.02 & $\leq$0.02 \\ 
G100.37-03.57  & 0.04 & $\leq$0.15 & 0.01 & 0.28$\pm$0.08 & 0.05 & $\leq$0.14 & 0.01 & $\leq$0.01 & 0.01 & $\leq$0.01 & 0.01 & $\leq$0.01 \\ 
G105.41+09.87  & 0.06 & 0.54$\pm$0.18 & 0.02 & $\leq$0.02 & 0.06 & 1.58$\pm$0.13 & 0.01 & 0.29$\pm$0.06 & 0.01 & 0.96$\pm$0.18 & 0.01 & $\leq$0.01 \\ 
G107.29+05.63  & 0.06 & 2.08$\pm$0.22 & 0.02 & 1.83$\pm$0.11 & 0.05 & 3.20$\pm$0.16 & 0.01 & 1.55$\pm$0.08 & 0.01 & 0.53$\pm$0.08 & 0.01 & 0.50$\pm$0.09 \\ 
G108.18+05.51  & 0.07 & 1.90$\pm$0.24 & 0.02 & 1.14$\pm$0.12 & 0.07 & 1.99$\pm$0.19 & 0.02 & 0.83$\pm$0.15 & 0.01 & $\leq$0.02 & 0.01 & $\leq$0.02 \\ 
G108.20+00.58  & 0.10 & $\leq$0.33 & 0.02 & $\leq$0.04 & 0.07 & 1.28$\pm$0.12 & 0.02 & $\leq$0.02 & 0.02 & $\leq$0.02 & 0.01 & $\leq$0.02 \\ 
G108.47-02.81  & 0.10 & $\leq$0.34 & 0.02 & 0.40$\pm$0.10 & 0.07 & 1.51$\pm$0.18 & 0.02 & $\leq$0.02 & 0.02 & 0.43$\pm$0.12 & 0.02 & $\leq$0.02 \\ 
G108.59+00.49  & 0.07 & $\leq$0.22 & 0.02 & 0.39$\pm$0.09 & 0.05 & 2.50$\pm$0.16 & 0.01 & 0.49$\pm$0.07 & 0.01 & 0.64$\pm$0.15 & 0.01 & $\leq$0.01 \\ 
G109.87+02.11  & 0.05 & 2.56$\pm$0.19 & 0.01 & 0.97$\pm$0.09 & 0.04 & 10.47$\pm$0.18 & 0.01 & 2.87$\pm$0.09 & 0.01 & 2.23$\pm$0.09 & 0.01 & 0.66$\pm$0.06 \\ 
G111.23-01.23  & 0.05 & $\leq$0.15 & 0.01 & $\leq$0.02 & 0.05 & $\leq$0.14 & 0.01 & $\leq$0.01 & 0.01 & $\leq$0.01 & 0.01 & $\leq$0.02 \\ 
G111.25-00.76  & 0.04 & $\leq$0.15 & 0.01 & $\leq$0.02 & 0.05 & 1.11$\pm$0.12 & 0.01 & $\leq$0.01 & 0.01 & $\leq$0.02 & 0.01 & $\leq$0.01 \\ 
G111.54+00.77  & 0.05 & 4.44$\pm$0.19 & 0.01 & 0.69$\pm$0.11 & 0.05 & 16.22$\pm$0.21 & 0.01 & 1.91$\pm$0.11 & 0.01 & 2.45$\pm$0.12 & 0.01 & 0.53$\pm$0.10 \\ 
G121.29+00.65  & 0.09 & 1.17$\pm$0.30 & 0.02 & 1.58$\pm$0.18 & 0.07 & 4.39$\pm$0.24 & 0.02 & 1.66$\pm$0.10 & 0.02 & 0.54$\pm$0.10 & 0.02 & $\leq$0.02 \\ 
G122.01-07.08  & 0.05 & $\leq$0.18 & 0.01 & 0.42$\pm$0.10 & 0.06 & 0.66$\pm$0.14 & 0.01 & $\leq$0.02 & 0.01 & $\leq$0.02 & 0.01 & $\leq$0.02 \\ 
G123.06-06.30  & 0.05 & 2.25$\pm$0.17 & 0.01 & 1.30$\pm$0.11 & 0.05 & 7.01$\pm$0.19 & 0.01 & 1.76$\pm$0.09 & 0.01 & 1.45$\pm$0.11 & 0.01 & $\leq$0.02 \\ 
G123.06-06.30  & 0.05 & 1.86$\pm$0.19 & 0.01 & 1.07$\pm$0.11 & 0.05 & 6.29$\pm$0.18 & 0.01 & 1.58$\pm$0.11 & 0.01 & 1.51$\pm$0.15 & 0.01 & $\leq$0.02 \\ 
G133.94+01.06  & 0.10 & 3.16$\pm$0.34 & 0.02 & $\leq$0.03 & 0.05 & 13.76$\pm$0.20 & 0.01 & 1.84$\pm$0.12 & 0.01 & 2.86$\pm$0.13 & 0.01 & $\leq$0.02 \\ 
G134.62-02.19  & 0.09 & $\leq$0.30 & 0.02 & $\leq$0.03 & 0.05 & $\leq$0.16 & 0.01 & $\leq$0.02 & 0.01 & $\leq$0.02 & 0.01 & $\leq$0.02 \\ 
G135.27+02.79  & 0.08 & $\leq$0.25 & 0.02 & $\leq$0.02 & 0.05 & 0.73$\pm$0.16 & 0.01 & $\leq$0.02 & 0.01 & $\leq$0.02 & 0.01 & $\leq$0.02 \\ 
G160.14+03.15  & 0.04 & $\leq$0.15 & 0.01 & $\leq$0.02 & 0.04 & 1.14$\pm$0.10 & 0.01 & $\leq$0.01 & 0.01 & 0.32$\pm$0.05 & 0.01 & $\leq$0.01 \\ 
G168.06+00.82  & 0.06 & $\leq$0.18 & 0.01 & $\leq$0.02 & 0.06 & 0.53$\pm$0.17 & 0.01 & $\leq$0.01 & 0.01 & $\leq$0.01 & 0.01 & $\leq$0.01 \\ 
G176.51+00.20  & 0.05 & 1.08$\pm$0.14 & 0.01 & 0.68$\pm$0.07 & 0.04 & 2.21$\pm$0.13 & 0.01 & 0.93$\pm$0.06 & 0.01 & 0.59$\pm$0.07 & 0.01 & 0.20$\pm$0.04 \\ 
G182.67-03.26  & 0.06 & $\leq$0.20 & 0.02 & $\leq$0.02 & 0.05 & $\leq$0.16 & 0.01 & $\leq$0.02 & 0.01 & $\leq$0.01 & 0.01 & $\leq$0.02 \\ 
G183.72-03.66  & 0.09 & $\leq$0.29 & 0.02 & 0.89$\pm$0.12 & 0.06 & 3.39$\pm$0.20 & 0.01 & 1.09$\pm$0.08 & 0.01 & 0.51$\pm$0.10 & 0.02 & 0.39$\pm$0.09 \\ 
G188.79+01.03  & 0.07 & $\leq$0.24 & 0.02 & $\leq$0.02 & 0.05 & 4.38$\pm$0.17 & 0.01 & 0.56$\pm$0.12 & 0.01 & 0.84$\pm$0.11 & 0.01 & $\leq$0.02 \\ 
G188.94+00.88  & 0.05 & 1.62$\pm$0.10 & 0.01 & 0.70$\pm$0.12 & 0.05 & 6.65$\pm$0.19 & 0.01 & 1.39$\pm$0.07 & 0.01 & 1.16$\pm$0.09 & 0.01 & 0.35$\pm$0.08 \\ 
G192.16-03.81  & 0.09 & $\leq$0.30 & 0.02 & $\leq$0.03 & 0.07 & $\leq$0.23 & 0.02 & $\leq$0.02 & 0.02 & $\leq$0.02 & 0.02 & $\leq$0.02 \\ 
G192.60-00.04  & 0.06 & 2.04$\pm$0.21 & 0.02 & 0.22$\pm$0.07 & 0.05 & 10.54$\pm$0.18 & 0.01 & 1.20$\pm$0.09 & 0.01 & 2.16$\pm$0.10 & 0.01 & $\leq$0.01 \\ 
G196.45-01.67  & 0.07 & $\leq$0.24 & 0.02 & $\leq$0.02 & 0.06 & 1.45$\pm$0.19 & 0.02 & $\leq$0.02 & 0.02 & $\leq$0.02 & 0.01 & $\leq$0.02 \\ 
G209.00-19.38  & 0.12 & 2.05$\pm$0.42 & 0.01 & 0.51$\pm$0.10 & 0.04 & 35.68$\pm$0.19 & 0.01 & 1.11$\pm$0.09 & 0.01 & 7.41$\pm$0.19 & 0.01 & $\leq$0.01 \\ 
G211.59+01.05  & 0.06 & 0.85$\pm$0.19 & 0.02 & 0.38$\pm$0.08 & 0.04 & 2.84$\pm$0.14 & 0.01 & 0.41$\pm$0.09 & 0.01 & 0.37$\pm$0.10 & 0.01 & $\leq$0.02 \\ 
G229.57+00.15  & 0.10 & $\leq$0.34 & 0.03 & $\leq$0.04 & 0.08 & $\leq$0.23 & 0.02 & $\leq$0.02 & 0.02 & $\leq$0.02 & 0.02 & $\leq$0.02 \\ 
G232.62+00.99  & 0.07 & 0.63$\pm$0.15 & 0.02 & $\leq$0.02 & 0.06 & 3.47$\pm$0.19 & 0.01 & 0.61$\pm$0.08 & 0.01 & 0.67$\pm$0.12 & 0.01 & $\leq$0.01 \\ 
G236.81+01.98  & 0.07 & $\leq$0.22 & 0.02 & $\leq$0.03 & 0.06 & 0.64$\pm$0.12 & 0.01 & $\leq$0.02 & 0.01 & $\leq$0.02 & 0.01 & $\leq$0.02 \\ 
G239.35-05.06  & 0.09 & $\leq$0.29 & 0.02 & $\leq$0.03 & 0.07 & $\leq$0.21 & 0.02 & $\leq$0.02 & 0.02 & $\leq$0.02 & 0.02 & $\leq$0.02 \\ 
G240.31+00.07  & 0.07 & 0.70$\pm$0.16 & 0.02 & $\leq$0.02 & 0.05 & 1.48$\pm$0.13 & 0.01 & $\leq$0.02 & 0.01 & 0.49$\pm$0.11 & 0.01 & $\leq$0.02 \\ 

\end{longtable}
}

\section{Derived column densities}
\label{app: column densities table}

Table \ref{app_tab: column densities} lists the molecular column densities of DCN, DNC, H$^{13}$CN, HN$^{13}$C, HC$^{15}$N and H$^{15}$NC for the full source sample. 

\begin{longtable}{lccccccc}
\caption{Molecular column densities}\label{app_tab: column densities}\\
\hline\hline
Source & $N_{\rm tot}$(DCN) & $N_{\rm tot}$(DNC) & $N_{\rm tot}$(H$^{13}$CN) &
$N_{\rm tot}$(HN$^{13}$C) & $N_{\rm tot}$(HC$^{15}$N) & $N_{\rm tot}$(H$^{15}$NC) \\
 & ($\times$10$^{13}$ cm$^{-2}$) & ($\times$10$^{13}$ cm$^{-2}$) & ($\times$10$^{13}$ cm$^{-2}$) &
($\times$10$^{13}$ cm$^{-2}$) & ($\times$10$^{13}$ cm$^{-2}$) & ($\times$10$^{13}$ cm$^{-2}$) \\
\hline
\endfirsthead

\caption{continued.}\\
\hline\hline
Source & $N_{\rm tot}$(DCN) & $N_{\rm tot}$(DNC) & $N_{\rm tot}$(H$^{13}$CN) &
$N_{\rm tot}$(HN$^{13}$C) & $N_{\rm tot}$(HC$^{15}$N) & $N_{\rm tot}$(H$^{15}$NC) \\
 & ($\times$10$^{13}$ cm$^{-2}$) & ($\times$10$^{13}$ cm$^{-2}$) & ($\times$10$^{13}$ cm$^{-2}$) &
($\times$10$^{13}$ cm$^{-2}$) & ($\times$10$^{13}$ cm$^{-2}$) & ($\times$10$^{13}$ cm$^{-2}$) \\
\hline
\endhead

\hline
\multicolumn{8}{r}{\emph{Continued on next page}}\\
\endfoot

\hline

\endlastfoot

G000.67-00.03  & 16.16$\pm$0.39 & 1.01$\pm$0.14 & 13.61$\pm$0.38 & 2.42$\pm$0.28 & 2.98$\pm$0.29 & $\leq$0.37 \\ 
G005.88-00.39  & 5.35$\pm$0.40 & 2.52$\pm$0.12 & 47.34$\pm$0.17 & 10.75$\pm$0.08 & 6.81$\pm$0.10 & 2.19$\pm$0.08 \\ 
G009.62+00.19  & 1.68$\pm$0.19 & 0.38$\pm$0.07 & 14.12$\pm$0.15 & 2.67$\pm$0.06 & 2.23$\pm$0.08 & 0.27$\pm$0.05 \\ 
G010.47+00.02  & 1.58$\pm$0.46 & 1.09$\pm$0.10 & 24.25$\pm$0.11 & 4.77$\pm$0.07 & 7.27$\pm$0.07 & 0.65$\pm$0.08 \\ 
G010.62-00.38  & 4.62$\pm$0.24 & 1.67$\pm$0.15 & 32.94$\pm$0.26 & 7.76$\pm$0.08 & 4.67$\pm$0.09 & 1.32$\pm$0.10 \\ 
G011.49-01.48  & 0.75$\pm$0.11 & 0.32$\pm$0.05 & 1.76$\pm$0.05 & 0.35$\pm$0.02 & 0.39$\pm$0.03 & 0.09$\pm$0.02 \\ 
G011.91-00.61  & 0.83$\pm$0.10 & 0.78$\pm$0.05 & 6.59$\pm$0.08 & 1.84$\pm$0.04 & 0.88$\pm$0.06 & 0.28$\pm$0.04 \\ 
G012.02-00.03  & $\leq$0.39 & $\leq$0.13 & 0.93$\pm$0.10 & 0.40$\pm$0.05 & $\leq$0.10 & $\leq$0.13 \\ 
G012.68-00.18  & $\leq$0.56 & $\leq$0.16 & 3.76$\pm$0.15 & 0.91$\pm$0.05 & 0.89$\pm$0.10 & $\leq$0.15 \\ 
G012.80-00.20  & 2.13$\pm$0.19 & 0.86$\pm$0.13 & 8.03$\pm$0.14 & 2.99$\pm$0.10 & 0.54$\pm$0.07 & $\leq$0.22 \\ 
G012.88+00.48  & 2.09$\pm$0.20 & 0.80$\pm$0.11 & 8.16$\pm$0.13 & 2.39$\pm$0.07 & 1.63$\pm$0.08 & 0.31$\pm$0.10 \\ 
G012.90-00.24  & 0.57$\pm$0.16 & 1.02$\pm$0.07 & 1.61$\pm$0.11 & 1.39$\pm$0.06 & 0.29$\pm$0.07 & $\leq$0.16 \\ 
G012.90-00.26  & 1.95$\pm$0.25 & 1.28$\pm$0.10 & 9.05$\pm$0.16 & 3.01$\pm$0.09 & 1.50$\pm$0.12 & 0.56$\pm$0.12 \\ 
G013.87+00.28  & $\leq$0.36 & $\leq$0.13 & 2.42$\pm$0.09 & 0.46$\pm$0.05 & 0.25$\pm$0.04 & $\leq$0.12 \\ 
G014.33-00.64  & 2.41$\pm$0.22 & 1.61$\pm$0.08 & 9.29$\pm$0.13 & 2.61$\pm$0.04 & 1.54$\pm$0.06 & 0.46$\pm$0.05 \\ 
G014.63-00.57  & 1.45$\pm$0.18 & 1.24$\pm$0.07 & 3.22$\pm$0.08 & 1.64$\pm$0.04 & 0.62$\pm$0.04 & 0.47$\pm$0.04 \\ 
G015.03-00.67  & $\leq$0.86 & $\leq$0.36 & 4.88$\pm$0.19 & 0.87$\pm$0.13 & 0.87$\pm$0.10 & $\leq$0.24 \\ 
G016.58-00.05  & 0.81$\pm$0.13 & 0.78$\pm$0.08 & 4.21$\pm$0.11 & 1.29$\pm$0.06 & 0.61$\pm$0.09 & $\leq$0.15 \\ 
G023.00-00.41  & 0.63$\pm$0.15 & 0.64$\pm$0.07 & 5.66$\pm$0.10 & 2.76$\pm$0.05 & 0.95$\pm$0.09 & 0.45$\pm$0.05 \\ 
G023.43-00.18  & 0.38$\pm$0.09 & 0.44$\pm$0.05 & 2.43$\pm$0.08 & 0.70$\pm$0.04 & 0.45$\pm$0.05 & 0.11$\pm$0.03 \\ 
G023.65-00.12  & $\leq$0.43 & $\leq$0.15 & 0.75$\pm$0.08 & $\leq$0.09 & 0.16$\pm$0.04 & $\leq$0.11 \\ 
G023.70-00.19  & $\leq$0.41 & $\leq$0.14 & 0.97$\pm$0.09 & 0.47$\pm$0.04 & 0.18$\pm$0.04 & $\leq$0.09 \\ 
G025.70+00.04  & 0.27$\pm$0.07 & 0.23$\pm$0.04 & 1.62$\pm$0.05 & 0.64$\pm$0.02 & 0.20$\pm$0.03 & 0.12$\pm$0.03 \\ 
G027.36-00.16  & 1.14$\pm$0.11 & 0.94$\pm$0.06 & 3.24$\pm$0.10 & 1.49$\pm$0.04 & 0.45$\pm$0.05 & 0.25$\pm$0.08 \\ 
G028.86+00.06  & $\leq$0.62 & $\leq$0.17 & 3.61$\pm$0.12 & 0.65$\pm$0.06 & 0.47$\pm$0.08 & 0.11$\pm$0.03 \\ 
G029.86-00.04  & $\leq$0.43 & $\leq$0.17 & 1.56$\pm$0.07 & 0.60$\pm$0.04 & 0.22$\pm$0.05 & $\leq$0.09 \\ 
G029.95-00.01  & 0.90$\pm$0.14 & $\leq$0.16 & 8.05$\pm$0.08 & 1.38$\pm$0.05 & 1.20$\pm$0.07 & 0.31$\pm$0.06 \\ 
G031.28+00.06  & 1.43$\pm$0.15 & 0.57$\pm$0.09 & 8.54$\pm$0.08 & 2.39$\pm$0.05 & 1.26$\pm$0.06 & 0.37$\pm$0.04 \\ 
G031.58+00.07  & 0.72$\pm$0.15 & 0.39$\pm$0.09 & 3.70$\pm$0.11 & 1.33$\pm$0.05 & 0.60$\pm$0.05 & 0.37$\pm$0.08 \\ 
G032.04+00.05  & 0.89$\pm$0.12 & 0.62$\pm$0.06 & 5.70$\pm$0.08 & 1.65$\pm$0.05 & 0.73$\pm$0.05 & 0.29$\pm$0.05 \\ 
G033.64-00.22  & $\leq$0.37 & $\leq$0.15 & 1.21$\pm$0.06 & 0.34$\pm$0.04 & 0.17$\pm$0.04 & $\leq$0.11 \\ 
G034.39+00.22  & 1.74$\pm$0.14 & 1.60$\pm$0.07 & 5.63$\pm$0.11 & 2.23$\pm$0.04 & 0.59$\pm$0.06 & 0.51$\pm$0.07 \\ 
G035.02+00.34  & 0.96$\pm$0.14 & 0.30$\pm$0.07 & 4.30$\pm$0.11 & 0.94$\pm$0.05 & 0.74$\pm$0.06 & 0.23$\pm$0.06 \\ 
G035.19-00.74  & 1.92$\pm$0.12 & 0.90$\pm$0.07 & 7.97$\pm$0.12 & 2.68$\pm$0.06 & 1.33$\pm$0.07 & 0.51$\pm$0.05 \\ 
G035.20-01.73  & 0.85$\pm$0.08 & 0.22$\pm$0.06 & 3.61$\pm$0.07 & 0.68$\pm$0.04 & 0.61$\pm$0.03 & $\leq$0.09 \\ 
G037.43+01.51  & 2.25$\pm$0.37 & 0.87$\pm$0.24 & 6.57$\pm$0.29 & 1.32$\pm$0.12 & 1.59$\pm$0.22 & 0.63$\pm$0.11 \\ 
G043.16+00.01  & 1.83$\pm$0.15 & $\leq$0.36 & 11.09$\pm$0.14 & 1.44$\pm$0.12 & 2.48$\pm$0.12 & $\leq$0.32 \\ 
G043.79-00.12  & 0.51$\pm$0.12 & $\leq$0.19 & 4.66$\pm$0.15 & 0.87$\pm$0.08 & 0.47$\pm$0.07 & $\leq$0.21 \\ 
G043.89-00.78  & 0.62$\pm$0.10 & 0.54$\pm$0.06 & 2.11$\pm$0.08 & 0.79$\pm$0.04 & 0.27$\pm$0.05 & 0.19$\pm$0.04 \\ 
G045.07+00.13  & $\leq$0.47 & $\leq$0.22 & 4.17$\pm$0.12 & 0.59$\pm$0.09 & 0.91$\pm$0.11 & $\leq$0.21 \\ 
G045.45+00.05  & $\leq$0.41 & $\leq$0.16 & 1.79$\pm$0.08 & 0.41$\pm$0.06 & 0.40$\pm$0.06 & $\leq$0.17 \\ 
G048.60+00.02  & 0.71$\pm$0.18 & $\leq$0.18 & 1.51$\pm$0.10 & $\leq$0.14 & $\leq$0.16 & $\leq$0.19 \\ 
G049.19-00.33  & $\leq$0.28 & 0.23$\pm$0.05 & 0.92$\pm$0.06 & 0.59$\pm$0.03 & 0.10$\pm$0.03 & 0.17$\pm$0.04 \\ 
G049.48-00.36  & 5.50$\pm$0.77 & 1.24$\pm$0.15 & 46.15$\pm$0.16 & 5.44$\pm$0.16 & 10.27$\pm$0.14 & 0.81$\pm$0.16 \\ 
G049.48-00.38  & 3.07$\pm$0.27 & $<$0.31 & 21.70$\pm$0.16 & 4.16$\pm$0.13 & 4.02$\pm$0.11 & 0.98$\pm$0.13 \\ 
G052.10+01.04  & $\leq$0.44 & $\leq$0.14 & 0.73$\pm$0.08 & $\leq$0.11 & $\leq$0.11 & $\leq$0.13 \\ 
G059.78+00.06  & 2.00$\pm$0.19 & 1.11$\pm$0.07 & 5.97$\pm$0.09 & 1.76$\pm$0.04 & 0.84$\pm$0.05 & 0.56$\pm$0.05 \\ 
G069.54-00.97  & 1.16$\pm$0.12 & 0.77$\pm$0.05 & 3.43$\pm$0.06 & 1.33$\pm$0.03 & 0.61$\pm$0.04 & 0.27$\pm$0.03 \\ 
G074.03-01.71  & 0.62$\pm$0.09 & 1.11$\pm$0.04 & 2.55$\pm$0.06 & 1.15$\pm$0.03 & 0.34$\pm$0.03 & 0.26$\pm$0.03 \\ 
G075.29+01.32  & $\leq$0.25 & $\leq$0.09 & 0.65$\pm$0.05 & $\leq$0.06 & 0.11$\pm$0.02 & $\leq$0.08 \\ 
G075.76+00.33  & 2.10$\pm$0.11 & 0.66$\pm$0.06 & 6.51$\pm$0.08 & 1.33$\pm$0.04 & 0.99$\pm$0.06 & 0.42$\pm$0.06 \\ 
G075.78+00.34  & 1.54$\pm$0.10 & 0.37$\pm$0.05 & 4.03$\pm$0.06 & 0.76$\pm$0.04 & 0.78$\pm$0.04 & 0.17$\pm$0.03 \\ 
G076.38-00.61  & $\leq$0.39 & 0.37$\pm$0.05 & 1.90$\pm$0.09 & 0.50$\pm$0.04 & 0.33$\pm$0.05 & $\leq$0.11 \\ 
G078.12+03.63  & 0.42$\pm$0.07 & 0.26$\pm$0.03 & 0.94$\pm$0.05 & 0.47$\pm$0.02 & 0.23$\pm$0.03 & 0.14$\pm$0.02 \\ 
G078.88+00.70  & $\leq$0.35 & $\leq$0.09 & 2.76$\pm$0.07 & 0.33$\pm$0.03 & 0.50$\pm$0.04 & $\leq$0.08 \\ 
G079.73+00.99  & 0.37$\pm$0.09 & 0.23$\pm$0.05 & 0.98$\pm$0.05 & 0.22$\pm$0.03 & 0.18$\pm$0.03 & $\leq$0.05 \\ 
G079.87+01.17  & 0.49$\pm$0.09 & 0.18$\pm$0.04 & 2.23$\pm$0.04 & 0.54$\pm$0.02 & 0.32$\pm$0.02 & $\leq$0.05 \\ 
G080.79-01.92  & $\leq$0.22 & $\leq$0.08 & $\leq$0.17 & $\leq$0.06 & $\leq$0.06 & $\leq$0.08 \\ 
G080.86+00.38  & 0.54$\pm$0.12 & $\leq$0.10 & 2.21$\pm$0.07 & 0.38$\pm$0.03 & 0.38$\pm$0.05 & $\leq$0.08 \\ 
G081.75+00.59  & 2.95$\pm$0.13 & 2.11$\pm$0.05 & 5.79$\pm$0.08 & 2.20$\pm$0.03 & 1.04$\pm$0.03 & 0.66$\pm$0.04 \\ 
G081.87+00.78  & 2.06$\pm$0.20 & 0.39$\pm$0.07 & 7.46$\pm$0.10 & 1.18$\pm$0.04 & 1.35$\pm$0.05 & 0.34$\pm$0.05 \\ 
G090.21+02.32  & $\leq$0.43 & 0.41$\pm$0.04 & $\leq$0.22 & 0.12$\pm$0.03 & $\leq$0.05 & $\leq$0.07 \\ 
G092.67+03.07  & 1.58$\pm$0.13 & 0.45$\pm$0.06 & 4.98$\pm$0.08 & 0.89$\pm$0.03 & 0.99$\pm$0.03 & 0.24$\pm$0.04 \\ 
G094.60-01.79  & $\leq$0.39 & 0.19$\pm$0.04 & 0.61$\pm$0.07 & 0.31$\pm$0.03 & 0.14$\pm$0.05 & $\leq$0.09 \\ 
G095.29-00.93  & $\leq$0.22 & $\leq$0.09 & 0.23$\pm$0.05 & $\leq$0.07 & $\leq$0.07 & $\leq$0.08 \\ 
G097.53+03.18  & 0.49$\pm$0.10 & $\leq$0.15 & 1.20$\pm$0.08 & $\leq$0.09 & $\leq$0.10 & $\leq$0.12 \\ 
G100.37-03.57  & $\leq$0.24 & 0.14$\pm$0.04 & $\leq$0.17 & $\leq$0.07 & $\leq$0.06 & $\leq$0.09 \\ 
G105.41+09.87  & 0.30$\pm$0.10 & $\leq$0.08 & 0.63$\pm$0.05 & 0.11$\pm$0.02 & 0.39$\pm$0.07 & $\leq$0.07 \\ 
G107.29+05.63  & 1.16$\pm$0.12 & 0.88$\pm$0.05 & 1.29$\pm$0.06 & 0.59$\pm$0.03 & 0.21$\pm$0.03 & 0.23$\pm$0.04 \\ 
G108.18+05.51  & 1.05$\pm$0.13 & 0.55$\pm$0.06 & 0.80$\pm$0.08 & 0.32$\pm$0.06 & $\leq$0.08 & $\leq$0.11 \\ 
G108.20+00.58  & $\leq$0.56 & $\leq$0.18 & 0.51$\pm$0.05 & $\leq$0.11 & $\leq$0.11 & $\leq$0.13 \\ 
G108.47-02.81  & $\leq$0.56 & 0.19$\pm$0.05 & 0.61$\pm$0.07 & $\leq$0.12 & 0.18$\pm$0.05 & $\leq$0.13 \\ 
G108.59+00.49  & $\leq$0.37 & 0.19$\pm$0.04 & 1.00$\pm$0.07 & 0.19$\pm$0.03 & 0.26$\pm$0.06 & $\leq$0.08 \\ 
G109.87+02.11  & 1.53$\pm$0.11 & 0.51$\pm$0.05 & 4.61$\pm$0.06 & 1.17$\pm$0.04 & 0.97$\pm$0.04 & 0.33$\pm$0.03 \\ 
G111.23-01.23  & $\leq$0.26 & $\leq$0.10 & $\leq$0.17 & $\leq$0.07 & $\leq$0.07 & $\leq$0.10 \\ 
G111.25-00.76  & $\leq$0.25 & $\leq$0.09 & 0.44$\pm$0.05 & $\leq$0.07 & $\leq$0.09 & $\leq$0.09 \\ 
G111.54+00.77  & 3.12$\pm$0.13 & 0.42$\pm$0.07 & 8.19$\pm$0.11 & 0.91$\pm$0.05 & 1.24$\pm$0.06 & 0.31$\pm$0.06 \\ 
G121.29+00.65  & 0.51$\pm$0.13 & 0.61$\pm$0.07 & 1.40$\pm$0.08 & 0.50$\pm$0.03 & 0.17$\pm$0.03 & $\leq$0.09 \\ 
G122.01-07.08  & $\leq$0.30 & 0.21$\pm$0.05 & 0.26$\pm$0.06 & $\leq$0.09 & $\leq$0.09 & $\leq$0.11 \\ 
G123.06-06.30  & 1.05$\pm$0.08 & 0.53$\pm$0.05 & 2.38$\pm$0.06 & 0.56$\pm$0.03 & 0.49$\pm$0.04 & $\leq$0.09 \\ 
G123.06-06.30  & 1.15$\pm$0.11 & 0.57$\pm$0.06 & 2.78$\pm$0.08 & 0.66$\pm$0.05 & 0.67$\pm$0.06 & $\leq$0.12 \\ 
G133.94+01.06  & 1.98$\pm$0.22 & $\leq$0.19 & 6.23$\pm$0.09 & 0.78$\pm$0.05 & 1.30$\pm$0.06 & $\leq$0.13 \\ 
G134.62-02.19  & $\leq$0.49 & $\leq$0.16 & $\leq$0.20 & $\leq$0.08 & $\leq$0.08 & $\leq$0.11 \\ 
G135.27+02.79  & $\leq$0.42 & $\leq$0.13 & 0.29$\pm$0.06 & $\leq$0.08 & $\leq$0.10 & $\leq$0.11 \\ 
G160.14+03.15  & $\leq$0.25 & $\leq$0.09 & 0.46$\pm$0.04 & $\leq$0.06 & 0.13$\pm$0.02 & $\leq$0.08 \\ 
G168.06+00.82  & $\leq$0.31 & $\leq$0.09 & 0.21$\pm$0.07 & $\leq$0.07 & $\leq$0.07 & $\leq$0.08 \\ 
G176.51+00.20  & 0.86$\pm$0.11 & 0.47$\pm$0.05 & 1.27$\pm$0.07 & 0.50$\pm$0.03 & 0.34$\pm$0.04 & 0.13$\pm$0.03 \\ 
G182.67-03.26  & $\leq$0.34 & $\leq$0.11 & $\leq$0.19 & $\leq$0.08 & $\leq$0.08 & $\leq$0.10 \\ 
G183.72-03.66  & $\leq$0.46 & 0.42$\pm$0.06 & 1.30$\pm$0.08 & 0.40$\pm$0.03 & 0.20$\pm$0.04 & 0.17$\pm$0.04 \\ 
G188.79+01.03  & $\leq$0.40 & $\leq$0.13 & 1.76$\pm$0.07 & 0.21$\pm$0.05 & 0.34$\pm$0.04 & $\leq$0.10 \\ 
G188.94+00.88  & 0.86$\pm$0.05 & 0.32$\pm$0.06 & 2.54$\pm$0.07 & 0.50$\pm$0.03 & 0.45$\pm$0.03 & 0.16$\pm$0.03 \\ 
G192.16-03.81  & $\leq$0.50 & $\leq$0.14 & $\leq$0.27 & $\leq$0.10 & $\leq$0.12 & $\leq$0.14 \\ 
G192.60-00.04  & 1.53$\pm$0.16 & 0.15$\pm$0.05 & 5.69$\pm$0.10 & 0.61$\pm$0.04 & 1.17$\pm$0.06 & $\leq$0.12 \\ 
G196.45-01.67  & $\leq$0.40 & $\leq$0.13 & 0.58$\pm$0.08 & $\leq$0.10 & $\leq$0.11 & $\leq$0.12 \\ 
G209.00-19.38  & 1.29$\pm$0.27 & 0.28$\pm$0.05 & 16.16$\pm$0.09 & 0.47$\pm$0.04 & 3.38$\pm$0.09 & $\leq$0.09 \\ 
G211.59+01.05  & 0.47$\pm$0.11 & 0.18$\pm$0.04 & 1.14$\pm$0.06 & 0.15$\pm$0.03 & 0.15$\pm$0.04 & $\leq$0.10 \\ 
G229.57+00.15  & $\leq$0.57 & $\leq$0.19 & $\leq$0.27 & $\leq$0.11 & $\leq$0.12 & $\leq$0.14 \\ 
G232.62+00.99  & 0.35$\pm$0.08 & $\leq$0.11 & 1.39$\pm$0.08 & 0.23$\pm$0.03 & 0.27$\pm$0.05 & $\leq$0.08 \\ 
G236.81+01.98  & $\leq$0.37 & $\leq$0.15 & 0.25$\pm$0.05 & $\leq$0.08 & $\leq$0.09 & $\leq$0.11 \\ 
G239.35-05.06  & $\leq$0.48 & $\leq$0.16 & $\leq$0.25 & $\leq$0.11 & $\leq$0.11 & $\leq$0.13 \\ 
G240.31+00.07  & 0.39$\pm$0.09 & $\leq$0.12 & 0.59$\pm$0.05 & $\leq$0.08 & 0.20$\pm$0.04 & $\leq$0.11 \\ 

\end{longtable}

\section{Derived abundance ratios}
\label{app: ratios}

Table \ref{app_tab: ratios} lists the abundance ratios and deuterium fractionation derived from the column densities in Appendix \ref{app: column densities table}.

{\fontsize{7.5pt}{9pt}\selectfont
\begin{longtable}{lccccccc}
\caption{Abundance ratios and deuterium fractionation}
\label{app_tab: ratios} \\

\hline
\hline
\multirow{2}{*}{Source}
& \multirow{2}{*}{[DCN]/[DNC]}
& \multicolumn{3}{c}{Ratios from $^{13}$C isotopologues}
& \multicolumn{3}{c}{Ratios from $^{15}$N isotopologues} \\

\cmidrule(lr){3-5} \cmidrule(lr){6-8}

&
& [HCN]/[HNC]
& $D_{\rm frac}({\rm HCN})$
& $D_{\rm frac}({\rm HNC})$
& [HCN]/[HNC]
& $D_{\rm frac}({\rm HCN})$
& $D_{\rm frac}({\rm HNC})$ \\

\hline
\endfirsthead

\multicolumn{8}{c}
{{\tablename\ \thetable{} -- continued}} \\

\hline
\hline
\multirow{2}{*}{Source}
& \multirow{2}{*}{[DCN]/[DNC]}
& \multicolumn{3}{c}{Ratios from $^{13}$C isotopologues}
& \multicolumn{3}{c}{Ratios from $^{15}$N isotopologues} \\

\cmidrule(lr){3-5} \cmidrule(lr){6-8}

&
& [HCN]/[HNC]
& $D_{\rm frac}({\rm HCN})$
& $D_{\rm frac}({\rm HNC})$
& [HCN]/[HNC]
& $D_{\rm frac}({\rm HCN})$
& $D_{\rm frac}({\rm HNC})$ \\

\hline
\endhead

\hline
\multicolumn{8}{r}{\emph{Continued on next page}} \\
\endfoot

\hline
\multicolumn{8}{@{}l}{\footnotesize Notes.}\\
\multicolumn{8}{@{}l}{
\makebox[0pt][l]{
\parbox{0.88\linewidth}{
\footnotesize \textquotesingle...\textquotesingle{} indicates that neither of the two transitions was detected.
}}}\\

\endlastfoot

G000.67-00.03 & 16.06$\pm$2.33 & 5.62$\pm$0.68 & 0.058$\pm$0.002 & 0.020$\pm$0.004 & $\geq$7.99  & 0.037$\pm$0.004 & $\geq$0.019  \\ 
G005.88-00.39 & 2.12$\pm$0.19 & 4.40$\pm$0.04 & 0.0028$\pm$0.0002 & 0.0059$\pm$0.0003 & 3.10$\pm$0.13 & 0.0034$\pm$0.0003 & 0.0050$\pm$0.0003 \\ 
G009.62+00.19 & 4.41$\pm$1.00 & 5.28$\pm$0.13 & 0.0037$\pm$0.0004 & 0.0045$\pm$0.0009 & 8.34$\pm$1.68 & 0.0049$\pm$0.0006 & 0.0092$\pm$0.0026 \\ 
G010.47+00.02 & 1.45$\pm$0.44 & 5.08$\pm$0.08 & 0.0026$\pm$0.0008 & 0.0091$\pm$0.0008 & 11.16$\pm$1.31 & 0.0014$\pm$0.0004 & 0.011$\pm$0.002 \\ 
G010.62-00.38 & 2.77$\pm$0.28 & 4.24$\pm$0.05 & 0.0043$\pm$0.0002 & 0.0066$\pm$0.0006 & 3.55$\pm$0.27 & 0.0065$\pm$0.0004 & 0.0084$\pm$0.0010 \\ 
G011.49-01.48 & 2.33$\pm$0.48 & 5.08$\pm$0.37 & 0.0090$\pm$0.0013 & 0.020$\pm$0.003 & 4.21$\pm$1.05 & 0.0093$\pm$0.0015 & 0.017$\pm$0.005 \\ 
G011.91-00.61 & 1.06$\pm$0.14 & 3.57$\pm$0.09 & 0.0032$\pm$0.0004 & 0.011$\pm$0.001 & 3.10$\pm$0.51 & 0.0050$\pm$0.0007 & 0.015$\pm$0.002 \\ 
G012.02-00.03  & ... & 2.32$\pm$0.37 & $\leq$0.015  & $\leq$0.012  & ... & ... & ... \\ 
G012.68-00.18  & ... & 4.11$\pm$0.29 & $\leq$0.0035  & $\leq$0.0042  & $\geq$5.90  & $\leq$0.0031  & ... \\ 
G012.80-00.20 & 2.49$\pm$0.44 & 2.69$\pm$0.10 & 0.0065$\pm$0.0006 & 0.0070$\pm$0.0011 & $\geq$2.41  & 0.024$\pm$0.004 & $\geq$0.023  \\ 
G012.88+00.48 & 2.61$\pm$0.44 & 3.42$\pm$0.11 & 0.0061$\pm$0.0006 & 0.0079$\pm$0.0011 & 5.27$\pm$1.70 & 0.0079$\pm$0.0008 & 0.016$\pm$0.005 \\ 
G012.90-00.24 & 0.56$\pm$0.17 & 1.16$\pm$0.09 & 0.0083$\pm$0.0025 & 0.017$\pm$0.001 & $\geq$1.85  & 0.0097$\pm$0.0036 & $\geq$0.032  \\ 
G012.90-00.26 & 1.52$\pm$0.23 & 3.01$\pm$0.11 & 0.0051$\pm$0.0007 & 0.010$\pm$0.001 & 2.67$\pm$0.59 & 0.0067$\pm$0.0010 & 0.012$\pm$0.003 \\ 
G013.87+00.28  & ... & 5.23$\pm$0.60 & $\leq$0.0041  & $\leq$0.0078  & $\geq$2.05  & $\leq$0.0076  & ... \\ 
G014.33-00.64 & 1.50$\pm$0.15 & 3.56$\pm$0.08 & 0.0054$\pm$0.0005 & 0.013$\pm$0.001 & 3.35$\pm$0.38 & 0.0072$\pm$0.0007 & 0.016$\pm$0.002 \\ 
G014.63-00.57 & 1.17$\pm$0.16 & 1.96$\pm$0.07 & 0.010$\pm$0.001 & 0.017$\pm$0.001 & 1.34$\pm$0.14 & 0.011$\pm$0.002 & 0.013$\pm$0.001 \\ 
G015.03-00.67  & ... & 5.58$\pm$0.84 & $\leq$0.0040  & $\leq$0.0092  & $\geq$3.59  & $\leq$0.0062  & ... \\ 
G016.58-00.05 & 1.03$\pm$0.20 & 3.28$\pm$0.17 & 0.0050$\pm$0.0008 & 0.016$\pm$0.002 & $\geq$4.03  & 0.0062$\pm$0.0013 & $\geq$0.024  \\ 
G023.00-00.41 & 0.99$\pm$0.25 & 2.05$\pm$0.05 & 0.0031$\pm$0.0007 & 0.0064$\pm$0.0007 & 2.13$\pm$0.32 & 0.0034$\pm$0.0008 & 0.0073$\pm$0.0012 \\ 
G023.43-00.18 & 0.88$\pm$0.23 & 3.46$\pm$0.23 & 0.0047$\pm$0.0011 & 0.019$\pm$0.003 & 3.89$\pm$1.19 & 0.0048$\pm$0.0013 & 0.021$\pm$0.007 \\ 
G023.65-00.12  & ... & $\geq$8.62  & $\leq$0.014  & ... & $\geq$1.50  & $\leq$0.013  & ... \\ 
G023.70-00.19  & ... & 2.06$\pm$0.25 & $\leq$0.013  & $\leq$0.0090  & $\geq$2.12  & $\leq$0.013  & ... \\ 
G025.70+00.04 & 1.18$\pm$0.38 & 2.52$\pm$0.12 & 0.0044$\pm$0.0012 & 0.0094$\pm$0.0017 & 1.62$\pm$0.49 & 0.0072$\pm$0.0022 & 0.0098$\pm$0.0031 \\ 
G027.36-00.16 & 1.22$\pm$0.14 & 2.18$\pm$0.08 & 0.010$\pm$0.001 & 0.018$\pm$0.001 & 1.79$\pm$0.60 & 0.015$\pm$0.002 & 0.022$\pm$0.007 \\ 
G028.86+00.06  & ... & 5.55$\pm$0.54 & $\leq$0.0049  & $\leq$0.0073  & 4.25$\pm$1.30 & $\leq$0.0085  & $\leq$0.0097  \\ 
G029.86-00.04  & ... & 2.59$\pm$0.21 & $\leq$0.0078  & $\leq$0.0077  & $\geq$2.31  & $\leq$0.011  & ... \\ 
G029.95-00.01 & $\geq$5.71  & 5.82$\pm$0.21 & 0.0030$\pm$0.0005 & $\leq$0.0031  & 3.83$\pm$0.71 & 0.0038$\pm$0.0006 & $\leq$0.0025  \\ 
G031.28+00.06 & 2.53$\pm$0.46 & 3.58$\pm$0.08 & 0.0043$\pm$0.0004 & 0.0060$\pm$0.0009 & 3.40$\pm$0.42 & 0.0047$\pm$0.0005 & 0.0064$\pm$0.0012 \\ 
G031.58+00.07 & 1.86$\pm$0.57 & 2.78$\pm$0.14 & 0.0051$\pm$0.0011 & 0.0076$\pm$0.0017 & 1.59$\pm$0.38 & 0.0049$\pm$0.0011 & 0.0042$\pm$0.0013 \\ 
G032.04+00.05 & 1.44$\pm$0.24 & 3.45$\pm$0.11 & 0.0041$\pm$0.0005 & 0.0099$\pm$0.0010 & 2.48$\pm$0.45 & 0.0053$\pm$0.0008 & 0.0090$\pm$0.0018 \\ 
G033.64-00.22  & ... & 3.55$\pm$0.45 & $\leq$0.0081  & $\leq$0.012  & $\geq$1.53  & $\leq$0.012  & ... \\ 
G034.39+00.22 & 1.09$\pm$0.10 & 2.53$\pm$0.07 & 0.0066$\pm$0.0006 & 0.015$\pm$0.001 & 1.15$\pm$0.21 & 0.0086$\pm$0.0012 & 0.0091$\pm$0.0014 \\ 
G035.02+00.34 & 3.23$\pm$0.90 & 4.58$\pm$0.27 & 0.0050$\pm$0.0007 & 0.0070$\pm$0.0017 & 3.17$\pm$0.82 & 0.0056$\pm$0.0009 & 0.0055$\pm$0.0019 \\ 
G035.19-00.74 & 2.13$\pm$0.20 & 2.98$\pm$0.08 & 0.0053$\pm$0.0003 & 0.0074$\pm$0.0006 & 2.62$\pm$0.32 & 0.0073$\pm$0.0006 & 0.0090$\pm$0.0012 \\ 
G035.20-01.73 & 3.81$\pm$1.09 & 5.34$\pm$0.31 & 0.0056$\pm$0.0005 & 0.0078$\pm$0.0022 & $\geq$6.57  & 0.0053$\pm$0.0006 & $\geq$0.0091  \\ 
G037.43+01.51 & 2.58$\pm$0.82 & 4.98$\pm$0.51 & 0.0074$\pm$0.0013 & 0.014$\pm$0.004 & 2.51$\pm$0.57 & 0.0068$\pm$0.0015 & 0.0066$\pm$0.0021 \\ 
G043.16+00.01 & $\geq$5.04  & 7.72$\pm$0.66 & 0.0033$\pm$0.0003 & $\leq$0.0051  & $\geq$7.81  & 0.0048$\pm$0.0005 & ... \\ 
G043.79-00.12 & $\geq$2.65  & 5.35$\pm$0.51 & 0.0026$\pm$0.0006 & $\leq$0.0053  & $\geq$2.27  & 0.0060$\pm$0.0017 & ... \\ 
G043.89-00.78 & 1.13$\pm$0.22 & 2.69$\pm$0.18 & 0.0067$\pm$0.0011 & 0.016$\pm$0.002 & 1.37$\pm$0.38 & 0.011$\pm$0.003 & 0.014$\pm$0.003 \\ 
G045.07+00.13  & ... & 7.03$\pm$1.11 & $\leq$0.0025  & $\leq$0.0086  & $\geq$4.24  & $\leq$0.0025  & ... \\ 
G045.45+00.05  & ... & 4.33$\pm$0.67 & $\leq$0.0051  & $\leq$0.0088  & $\geq$2.41  & $\leq$0.0049  & ... \\ 
G048.60+00.02 & $\geq$3.97  & $\geq$10.82  & 0.0090$\pm$0.0024 & ... & ... & $\geq$0.019  & ... \\ 
G049.19-00.33 & $\leq$1.22  & 1.56$\pm$0.13 & $\leq$0.0068  & 0.0088$\pm$0.0021 & 0.63$\pm$0.24 & $\leq$0.013  & 0.0066$\pm$0.0022 \\ 
G049.48-00.36 & 4.43$\pm$0.82 & 8.49$\pm$0.26 & 0.0027$\pm$0.0004 & 0.0052$\pm$0.0006 & 12.62$\pm$2.52 & 0.0038$\pm$0.0005 & 0.011$\pm$0.003 \\ 
G049.48-00.38 & $>$10.01  & 5.22$\pm$0.17 & 0.0032$\pm$0.0003 & $<$0.0017  & 4.12$\pm$0.54 & 0.0070$\pm$0.0006 & $<$0.0029  \\
G052.10+01.04  & ... & $\geq$6.57  & $\leq$0.013  & ... & ... & ... & ... \\ 
G059.78+00.06 & 1.80$\pm$0.20 & 3.39$\pm$0.10 & 0.0069$\pm$0.0007 & 0.013$\pm$0.001 & 1.49$\pm$0.17 & 0.010$\pm$0.001 & 0.0086$\pm$0.0010 \\ 
G069.54-00.97 & 1.50$\pm$0.18 & 2.58$\pm$0.08 & 0.0067$\pm$0.0007 & 0.012$\pm$0.001 & 2.29$\pm$0.29 & 0.011$\pm$0.001 & 0.017$\pm$0.002 \\ 
G074.03-01.71 & 0.56$\pm$0.08 & 2.22$\pm$0.08 & 0.0048$\pm$0.0007 & 0.019$\pm$0.001 & 1.33$\pm$0.22 & 0.0080$\pm$0.0014 & 0.019$\pm$0.003 \\ 
G075.29+01.32  & ... & $\geq$11.12  & $\leq$0.0061  & ... & $\geq$1.41  & $\leq$0.0091  & ... \\ 
G075.76+00.33 & 3.20$\pm$0.33 & 4.88$\pm$0.16 & 0.0063$\pm$0.0003 & 0.0096$\pm$0.0009 & 2.35$\pm$0.35 & 0.0074$\pm$0.0006 & 0.0054$\pm$0.0009 \\ 
G075.78+00.34 & 4.13$\pm$0.58 & 5.28$\pm$0.27 & 0.0074$\pm$0.0005 & 0.0095$\pm$0.0013 & 4.58$\pm$0.92 & 0.0086$\pm$0.0007 & 0.0096$\pm$0.0022 \\ 
G076.38-00.61 & $\leq$1.07  & 3.79$\pm$0.37 & $\leq$0.0040  & 0.014$\pm$0.002 & $\geq$2.88  & $\leq$0.0052  & $\geq$0.014  \\ 
G078.12+03.63 & 1.63$\pm$0.32 & 2.00$\pm$0.14 & 0.0087$\pm$0.0015 & 0.011$\pm$0.001 & 1.63$\pm$0.35 & 0.010$\pm$0.002 & 0.010$\pm$0.002 \\ 
G078.88+00.70  & ... & 8.46$\pm$0.85 & $\leq$0.0024  & $\leq$0.0052  & $\geq$6.01  & $\leq$0.0030  & ... \\ 
G079.73+00.99 & 1.60$\pm$0.49 & 4.42$\pm$0.57 & 0.0073$\pm$0.0017 & 0.020$\pm$0.005 & $\geq$3.29  & 0.0089$\pm$0.0025 & $\geq$0.018  \\ 
G079.87+01.17 & 2.75$\pm$0.77 & 4.15$\pm$0.18 & 0.0043$\pm$0.0008 & 0.0065$\pm$0.0014 & $\geq$5.96  & 0.0068$\pm$0.0013 & $\geq$0.015  \\ 
G080.79-01.92  & ... & ... & ... & ... & ... & ... & ... \\ 
G080.86+00.38 & $\geq$5.46  & 5.88$\pm$0.51 & 0.0047$\pm$0.0011 & $\leq$0.0051  & $\geq$4.81  & 0.0063$\pm$0.0017 & ... \\ 
G081.75+00.59 & 1.40$\pm$0.07 & 2.63$\pm$0.05 & 0.0098$\pm$0.0004 & 0.0186$\pm$0.0005 & 1.59$\pm$0.11 & 0.011$\pm$0.001 & 0.012$\pm$0.001 \\ 
G081.87+00.78 & 5.31$\pm$1.09 & 6.30$\pm$0.25 & 0.0053$\pm$0.0005 & 0.0063$\pm$0.0012 & 3.97$\pm$0.61 & 0.0081$\pm$0.0009 & 0.0061$\pm$0.0014 \\ 
G090.21+02.32 & $\leq$1.05  & $\leq$1.86  & ... & 0.066$\pm$0.015 & ... & ... & $\geq$0.028  \\ 
G092.67+03.07 & 3.53$\pm$0.55 & 5.60$\pm$0.23 & 0.0060$\pm$0.0005 & 0.0095$\pm$0.0013 & 4.08$\pm$0.62 & 0.0083$\pm$0.0007 & 0.0096$\pm$0.0019 \\ 
G094.60-01.79 & $\leq$1.98  & 1.98$\pm$0.31 & $\leq$0.011  & 0.011$\pm$0.003 & $\geq$1.58  & $\leq$0.011  & $\geq$0.0089  \\ 
G095.29-00.93  & ... & $\geq$3.54  & $\leq$0.016  & ... & ... & ... & ... \\ 
G097.53+03.18 & $\geq$3.26  & $\geq$12.71  & 0.0061$\pm$0.0013 & ... & ... & $\geq$0.019  & ... \\ 
G100.37-03.57 & $\leq$1.79  & ... & ... & $\geq$0.033  & ... & ... & $\geq$0.0062  \\ 
G105.41+09.87 & $\geq$3.69  & 5.84$\pm$1.38 & 0.0089$\pm$0.0030 & $\leq$0.014  & $\geq$5.55  & 0.0033$\pm$0.0012 & ... \\ 
G107.29+05.63 & 1.31$\pm$0.16 & 2.19$\pm$0.16 & 0.017$\pm$0.002 & 0.028$\pm$0.002 & 0.91$\pm$0.22 & 0.023$\pm$0.004 & 0.016$\pm$0.003 \\ 
G108.18+05.51 & 1.91$\pm$0.31 & 2.53$\pm$0.51 & 0.025$\pm$0.004 & 0.033$\pm$0.007 & ... & $\geq$0.053  & $\geq$0.023  \\ 
G108.20+00.58  & ... & $\geq$4.84  & $\leq$0.018  & ... & ... & ... & ... \\ 
G108.47-02.81 & $\leq$2.91  & $\geq$5.23  & $\leq$0.016  & $\geq$0.028  & $\geq$1.37  & $\leq$0.013  & $\geq$0.0061  \\ 
G108.59+00.49 & $\leq$1.93  & 5.40$\pm$0.88 & $\leq$0.0065  & 0.018$\pm$0.005 & $\geq$3.44  & $\leq$0.0059  & $\geq$0.011  \\ 
G109.87+02.11 & 3.03$\pm$0.35 & 3.93$\pm$0.13 & 0.0063$\pm$0.0005 & 0.0081$\pm$0.0008 & 2.94$\pm$0.30 & 0.0097$\pm$0.0008 & 0.0094$\pm$0.0012 \\ 
G111.23-01.23  & ... & ... & ... & ... & ... & ... & ... \\ 
G111.25-00.76  & ... & $\geq$6.14  & $\leq$0.0095  & ... & ... & ... & ... \\ 
G111.54+00.77 & 7.37$\pm$1.20 & 9.00$\pm$0.53 & 0.0066$\pm$0.0003 & 0.0081$\pm$0.0014 & 4.00$\pm$0.79 & 0.0097$\pm$0.0006 & 0.0053$\pm$0.0013 \\ 
G121.29+00.65 & 0.85$\pm$0.24 & 2.80$\pm$0.23 & 0.0067$\pm$0.0018 & 0.022$\pm$0.003 & $\geq$1.91  & 0.0097$\pm$0.0031 & $\geq$0.022  \\ 
G122.01-07.08 & $\leq$1.46  & $\geq$3.04  & $\leq$0.020  & $\geq$0.041  & ... & ... & $\geq$0.0079  \\ 
G123.06-06.30 & 1.98$\pm$0.23 & 4.21$\pm$0.24 & 0.0074$\pm$0.0006 & 0.016$\pm$0.002 & $\geq$5.73  & 0.0087$\pm$0.0009 & $\geq$0.025  \\ 
G123.06-06.30 & 2.01$\pm$0.29 & 4.22$\pm$0.32 & 0.0071$\pm$0.0007 & 0.015$\pm$0.002 & $\geq$5.59  & 0.0070$\pm$0.0010 & $\geq$0.019  \\ 
G133.94+01.06 & $\geq$10.48  & 7.94$\pm$0.53 & 0.0055$\pm$0.0006 & $\leq$0.0042  & $\geq$9.92  & 0.0087$\pm$0.0010 & ... \\ 
G134.62-02.19  & ... & ... & ... & ... & ... & ... & ... \\ 
G135.27+02.79  & ... & $\geq$3.45  & $\leq$0.020  & ... & ... & ... & ... \\ 
G160.14+03.15  & ... & $\geq$7.22  & $\leq$0.0079  & ... & $\geq$1.59  & $\leq$0.0069  & ... \\ 
G168.06+00.82  & ... & $\geq$3.25  & $\leq$0.017  & ... & ... & ... & ... \\ 
G176.51+00.20 & 1.82$\pm$0.31 & 2.52$\pm$0.22 & 0.012$\pm$0.002 & 0.017$\pm$0.002 & 2.54$\pm$0.59 & 0.013$\pm$0.002 & 0.018$\pm$0.004 \\ 
G182.67-03.26  & ... & ... & ... & ... & ... & ... & ... \\ 
G183.72-03.66 & $\leq$1.12  & 3.30$\pm$0.32 & $\leq$0.0060  & 0.018$\pm$0.003 & 1.14$\pm$0.36 & $\leq$0.0074  & 0.0076$\pm$0.0021 \\ 
G188.79+01.03  & ... & 8.28$\pm$1.78 & $\leq$0.0037  & $\leq$0.0098  & $\geq$3.23  & $\leq$0.0047  & ... \\ 
G188.94+00.88 & 2.65$\pm$0.48 & 5.09$\pm$0.30 & 0.0056$\pm$0.0004 & 0.011$\pm$0.002 & 2.85$\pm$0.65 & 0.0063$\pm$0.0006 & 0.0068$\pm$0.0019 \\ 
G192.16-03.81  & ... & ... & ... & ... & ... & ... & ... \\ 
G192.60-00.04 & 10.46$\pm$3.56 & 9.29$\pm$0.69 & 0.0046$\pm$0.0005 & 0.0041$\pm$0.0014 & $\geq$9.86  & 0.0048$\pm$0.0006 & $\geq$0.0045  \\ 
G196.45-01.67  & ... & $\geq$5.92  & $\leq$0.0093  & ... & ... & ... & ... \\ 
G209.00-19.38 & 4.63$\pm$1.32 & 34.25$\pm$2.69 & 0.0015$\pm$0.0003 & 0.011$\pm$0.002 & $\geq$37.83  & 0.0015$\pm$0.0003 & $\geq$0.012  \\ 
G211.59+01.05 & 2.55$\pm$0.77 & 7.38$\pm$1.70 & 0.0060$\pm$0.0014 & 0.018$\pm$0.005 & $\geq$1.48  & 0.010$\pm$0.004 & $\geq$0.0058  \\ 
G229.57+00.15  & ... & ... & ... & ... & ... & ... & ... \\ 
G232.62+00.99 & $\geq$3.29  & 6.04$\pm$0.89 & 0.0044$\pm$0.0011 & $\leq$0.0081  & $\geq$3.24  & 0.0053$\pm$0.0016 & ... \\ 
G236.81+01.98  & ... & $\geq$3.14  & $\leq$0.024  & ... & ... & ... & ... \\ 
G239.35-05.06  & ... & ... & ... & ... & ... & ... & ... \\ 
G240.31+00.07 & $\geq$3.35  & $\geq$7.89  & 0.010$\pm$0.003 & ... & $\geq$1.89  & 0.0074$\pm$0.0023 & ... \\ 

\end{longtable}
}

\section{Molecular linewidths}
\label{app: FWHM table}

Table \ref{app_tab: FWHM} lists the FWHMs of the six observed transitions toward all 101 sources. The FWHMs of H$^{13}$CN 1-0 and DCN 1-0 were derived using hyperfine-structure (HFS) fitting, while DNC 1-0, HN$^{13}$C 1-0, HC$^{15}$N 1-0 and H$^{15}$NC 1-0 were obtained from single-component Gaussian fitting.

\begin{longtable}{lccccccc}
\caption{Molecular FWHM}\label{app_tab: FWHM}\\
\hline\hline
Source & DCN 1-0 & DNC 1-0 & H$^{13}$CN 1-0 & HN$^{13}$C 1-0 & HC$^{15}$N 1-0 & H$^{15}$NC 1-0 \\
 & (km\; s$^{-1}$) & (km\; s$^{-1}$) & (km\; s$^{-1}$) & (km\; s$^{-1}$) & (km\; s$^{-1}$) & (km\; s$^{-1}$) \\
\hline
\endfirsthead

\caption{continued.}\\
\hline\hline
Source & DCN 1-0 & DNC 1-0 & H$^{13}$CN 1-0 & HN$^{13}$C 1-0 & HC$^{15}$N 1-0 & H$^{15}$NC 1-0 \\
 & (km\; s$^{-1}$) & (km\; s$^{-1}$) & (km\; s$^{-1}$) & (km\; s$^{-1}$) & (km\; s$^{-1}$) & (km\; s$^{-1}$) \\
\hline
\endhead

\hline
\multicolumn{8}{r}{\emph{Continued on next page}}\\
\endfoot

\hline
\multicolumn{8}{@{}l}{\footnotesize Notes.}\\
\multicolumn{8}{@{}l}{
\makebox[0pt][l]{
\parbox{0.88\linewidth}{
\footnotesize \textquotesingle...\textquotesingle{} indicates that neither of the two transitions was detected.
}}}\\
\endlastfoot

G000.67-00.03  & ... & 5.3$\pm$0.9  & ... & 7.4$\pm$0.9 & 11.1$\pm$0.9 & 6.6$\pm$1.6 \\ 
G005.88-00.39 & 3.5$\pm$0.3 & 3.1$\pm$0.2 & 4.6$\pm$0.1 & 4.00$\pm$0.04 & 4.9$\pm$0.1 & 4.0$\pm$0.2 \\ 
G009.62+00.19 & 4.6$\pm$1.2 & 2.1$\pm$0.5 & 6.3$\pm$0.2 & 4.4$\pm$0.1 & 5.9$\pm$0.3 & 2.3$\pm$0.6 \\ 
G010.47+00.02 & 5.6$\pm$1.6 & 6.5$\pm$0.8 & 12.0$\pm$0.1 & 6.7$\pm$0.1 & 9.3$\pm$0.1 & 6.0$\pm$0.8 \\ 
G010.62-00.38 & 5.4$\pm$0.3 & 8.2$\pm$0.8 & 5.9$\pm$0.1 & 6.4$\pm$0.1 & 5.9$\pm$0.1 & 5.8$\pm$0.5 \\ 
G011.49-01.48 & 1.9$\pm$0.2 & 2.2$\pm$0.4 & 2.4$\pm$0.1 & 1.9$\pm$0.2 & 2.9$\pm$0.3 & 1.2$\pm$0.3 \\ 
G011.91-00.61 & 3.3$\pm$0.5 & 2.4$\pm$0.2 & 6.2$\pm$0.3 & 3.3$\pm$0.1 & 7.2$\pm$0.7 & 2.2$\pm$0.4 \\ 
G012.02-00.03  & ...  & ... & 4.8$\pm$1.2 & 3.5$\pm$0.4  & ...  & ... \\ 
G012.68-00.18  & ...  & ... & 3.5$\pm$0.3 & 2.6$\pm$0.2 & 8.2$\pm$1.0  & ... \\ 
G012.80-00.20 & 3.9$\pm$0.4 & 4.2$\pm$0.9 & 3.1$\pm$0.1 & 3.7$\pm$0.1 & 2.0$\pm$0.3 & 3.5$\pm$2.6 \\ 
G012.88+00.48 & 3.2$\pm$0.4 & 5.2$\pm$1.0 & 3.7$\pm$0.1 & 3.4$\pm$0.1 & 5.0$\pm$0.3 & 3.0$\pm$1.4 \\ 
G012.90-00.24 & 2.4$\pm$0.5 & 2.3$\pm$0.2 & 3.5$\pm$0.3 & 2.5$\pm$0.1 & 2.9$\pm$0.9  & ... \\ 
G012.90-00.26 & 2.8$\pm$0.4 & 3.4$\pm$0.3 & 4.3$\pm$0.2 & 3.9$\pm$0.1 & 6.7$\pm$0.6 & 4.8$\pm$0.9 \\ 
G013.87+00.28  & ...  & ... & 3.3$\pm$0.2 & 3.1$\pm$0.4 & 1.9$\pm$0.4  & ... \\ 
G014.33-00.64 & 2.6$\pm$0.3 & 2.8$\pm$0.2 & 3.4$\pm$0.1 & 2.9$\pm$0.1 & 3.8$\pm$0.2 & 2.3$\pm$0.3 \\ 
G014.63-00.57 & 2.2$\pm$0.4 & 2.8$\pm$0.2 & 2.5$\pm$0.1 & 2.2$\pm$0.1 & 2.8$\pm$0.2 & 1.9$\pm$0.2 \\ 
G015.03-00.67  & ...  & ... & 3.4$\pm$0.1 & 3.0$\pm$0.6 & 3.3$\pm$0.4  & ... \\ 
G016.58-00.05 & 2.4$\pm$0.5 & 4.6$\pm$0.5 & 3.7$\pm$0.1 & 3.3$\pm$0.2 & 6.0$\pm$1.2  & ... \\ 
G023.00-00.41 & 3.9$\pm$0.8 & 3.0$\pm$0.3 & 4.9$\pm$0.2 & 3.8$\pm$0.1 & 8.5$\pm$1.0 & 4.0$\pm$0.5 \\ 
G023.43-00.18 & 3.4$\pm$1.3 & 4.0$\pm$0.5 & 5.9$\pm$0.6 & 4.8$\pm$0.3 & 8.9$\pm$1.2 & 2.4$\pm$0.8 \\ 
G023.65-00.12  & ...  & ... & 12.5$\pm$3.2  & ... & 3.1$\pm$0.8  & ... \\ 
G023.70-00.19  & ...  & ... & 3.3$\pm$0.3 & 3.2$\pm$0.3 & 2.9$\pm$0.6  & ... \\ 
G025.70+00.04 & 2.7$\pm$1.0 & 2.8$\pm$0.5 & 3.6$\pm$0.1 & 2.4$\pm$0.1 & 4.0$\pm$0.7 & 2.2$\pm$0.9 \\ 
G027.36-00.16 & 3.1$\pm$0.3 & 2.9$\pm$0.2 & 4.4$\pm$0.2 & 3.6$\pm$0.1 & 5.5$\pm$0.8 & 3.9$\pm$2.2 \\ 
G028.86+00.06  & ...  & ... & 3.5$\pm$0.2 & 3.2$\pm$0.4 & 4.5$\pm$1.1 & 0.9$\pm$0.2 \\ 
G029.86-00.04  & ...  & ... & 3.1$\pm$0.1 & 3.2$\pm$0.2 & 3.7$\pm$1.1  & ... \\ 
G029.95-00.01 & 3.5$\pm$1.0  & ... & 4.1$\pm$0.1 & 4.0$\pm$0.2 & 4.8$\pm$0.3 & 2.9$\pm$0.7 \\ 
G031.28+00.06 & 3.2$\pm$0.3 & 3.7$\pm$0.7 & 4.7$\pm$0.1 & 3.7$\pm$0.1 & 4.5$\pm$0.2 & 1.9$\pm$0.3 \\ 
G031.58+00.07 & 2.3$\pm$0.5 & 2.6$\pm$0.9 & 3.5$\pm$0.1 & 3.3$\pm$0.2 & 3.2$\pm$0.3 & 3.7$\pm$1.2 \\ 
G032.04+00.05 & 5.0$\pm$0.9 & 3.7$\pm$0.4 & 4.5$\pm$0.2 & 3.7$\pm$0.1 & 5.3$\pm$0.4 & 3.3$\pm$0.7 \\ 
G033.64-00.22  & ...  & ... & 2.6$\pm$0.1 & 2.9$\pm$0.4 & 2.7$\pm$0.9  & ... \\ 
G034.39+00.22 & 2.8$\pm$0.4 & 2.6$\pm$0.1 & 3.4$\pm$0.1 & 2.7$\pm$0.1 & 3.7$\pm$0.5 & 2.7$\pm$0.6 \\ 
G035.02+00.34 & 3.5$\pm$0.5 & 3.7$\pm$0.9 & 3.8$\pm$0.1 & 3.3$\pm$0.2 & 3.5$\pm$0.4 & 2.3$\pm$0.7 \\ 
G035.19-00.74 & 3.4$\pm$0.3 & 3.1$\pm$0.2 & 4.4$\pm$0.1 & 3.5$\pm$0.1 & 4.7$\pm$0.3 & 2.6$\pm$0.4 \\ 
G035.20-01.73 & 4.5$\pm$0.5 & 2.1$\pm$0.9 & 4.1$\pm$0.1 & 4.5$\pm$0.3 & 4.4$\pm$0.3 & 2.1$\pm$1.1 \\ 
G037.43+01.51 & 2.5$\pm$0.3 & 3.2$\pm$1.0 & 2.5$\pm$0.1 & 2.1$\pm$0.2 & 5.0$\pm$0.9 & 1.7$\pm$0.3 \\ 
G043.16+00.01 & 10.8$\pm$1.8  & ... & 12.8$\pm$0.5 & 16.4$\pm$1.5 & 13.7$\pm$0.7  & ... \\ 
G043.79-00.12 & 4.1$\pm$1.1  & ... & 6.0$\pm$0.3 & 4.7$\pm$0.5 & 3.7$\pm$0.7  & ... \\ 
G043.89-00.78 & 3.5$\pm$0.5 & 3.2$\pm$0.3 & 4.6$\pm$0.3 & 3.5$\pm$0.2 & 3.3$\pm$0.8 & 2.0$\pm$0.4 \\ 
G045.07+00.13  & ...  & ... & 5.7$\pm$0.4 & 7.0$\pm$1.3 & 8.1$\pm$1.3  & ... \\ 
G045.45+00.05  & ...  & ... & 4.2$\pm$0.3 & 4.5$\pm$0.7 & 3.4$\pm$0.5  & ... \\ 
G048.60+00.02 & 3.6$\pm$1.1  & ... & 4.6$\pm$0.5  & ...  & ...  & ... \\ 
G049.19-00.33  & ... & 2.3$\pm$0.5 & 2.5$\pm$0.2 & 1.8$\pm$0.1 & 1.8$\pm$0.6 & 2.5$\pm$0.5 \\ 
G049.48-00.36 & 5.4$\pm$0.9 & 8.1$\pm$0.9 & 9.4$\pm$0.1 & 8.3$\pm$0.3 & 9.2$\pm$0.2 & 5.7$\pm$1.6 \\ 
G049.48-00.38 & 6.1$\pm$1.2  & ... & 8.1$\pm$0.1 & 7.3$\pm$0.3 & 7.0$\pm$0.2 & 6.3$\pm$1.0 \\ 
G052.10+01.04  & ...  & ... & 2.3$\pm$0.3  & ...  & ...  & ... \\ 
G059.78+00.06 & 1.8$\pm$0.1 & 1.8$\pm$0.1 & 2.19$\pm$0.04 & 1.9$\pm$0.1 & 2.1$\pm$0.2 & 2.2$\pm$0.2 \\ 
G069.54-00.97 & 3.5$\pm$0.5 & 4.3$\pm$0.3 & 3.7$\pm$0.1 & 3.6$\pm$0.1 & 3.7$\pm$0.2 & 2.4$\pm$0.3 \\ 
G074.03-01.71 & 1.6$\pm$0.2 & 2.1$\pm$0.1 & 2.6$\pm$0.1 & 2.3$\pm$0.1 & 2.4$\pm$0.3 & 2.1$\pm$0.3 \\ 
G075.29+01.32  & ...  & ... & 2.8$\pm$0.2  & ... & 1.4$\pm$0.3  & ... \\ 
G075.76+00.33 & 3.5$\pm$0.2 & 3.8$\pm$0.4 & 4.0$\pm$0.1 & 3.9$\pm$0.1 & 4.3$\pm$0.3 & 4.6$\pm$0.7 \\ 
G075.78+00.34 & 2.7$\pm$0.2 & 2.2$\pm$0.4 & 3.6$\pm$0.1 & 3.2$\pm$0.2 & 4.3$\pm$0.3 & 2.6$\pm$0.4 \\ 
G076.38-00.61  & ... & 1.6$\pm$0.2 & 2.8$\pm$0.2 & 2.2$\pm$0.2 & 2.4$\pm$0.4 & 1.3$\pm$0.6 \\ 
G078.12+03.63 & 2.8$\pm$0.3 & 1.8$\pm$0.2 & 2.8$\pm$0.2 & 2.0$\pm$0.1 & 3.3$\pm$0.5 & 1.5$\pm$0.3 \\ 
G078.88+00.70  & ...  & ... & 3.3$\pm$0.1 & 2.3$\pm$0.3 & 2.9$\pm$0.3  & ... \\ 
G079.73+00.99 & 2.9$\pm$0.7 & 2.2$\pm$0.6 & 2.4$\pm$0.1 & 1.8$\pm$0.3 & 2.2$\pm$0.4  & ... \\ 
G079.87+01.17 & 2.2$\pm$0.4 & 3.1$\pm$0.7 & 2.45$\pm$0.05 & 2.4$\pm$0.1 & 2.2$\pm$0.2  & ... \\ 
G080.79-01.92  & ...  & ...  & ...  & ...  & ...  & ... \\ 
G080.86+00.38 & 1.2$\pm$0.2  & ... & 2.9$\pm$0.1 & 2.8$\pm$0.3 & 4.7$\pm$0.8  & ... \\ 
G081.75+00.59 & 1.8$\pm$0.1 & 2.1$\pm$0.1 & 2.48$\pm$0.03 & 2.07$\pm$0.03 & 2.4$\pm$0.1 & 1.9$\pm$0.1 \\ 
G081.87+00.78 & 2.8$\pm$0.3 & 2.4$\pm$0.4 & 3.44$\pm$0.04 & 2.7$\pm$0.1 & 3.7$\pm$0.2 & 3.0$\pm$0.4 \\ 
G090.21+02.32  & ... & 1.1$\pm$0.1  & ... & 0.8$\pm$0.3  & ...  & ... \\ 
G092.67+03.07 & 3.2$\pm$0.3 & 3.4$\pm$0.5 & 2.76$\pm$0.03 & 2.7$\pm$0.1 & 2.8$\pm$0.1 & 2.3$\pm$0.4 \\ 
G094.60-01.79  & ... & 1.2$\pm$0.3 & 2.0$\pm$0.2 & 1.8$\pm$0.2 & 1.8$\pm$0.8  & ... \\ 
G095.29-00.93  & ...  & ... & 3.2$\pm$0.6  & ...  & ...  & ... \\ 
G097.53+03.18 & 2.1$\pm$0.4  & ... & 6.1$\pm$7.1  & ...  & ...  & ... \\ 
G100.37-03.57  & ... & 1.6$\pm$0.6  & ...  & ...  & ...  & ... \\ 
G105.41+09.87 & 1.4$\pm$0.6  & ... & 2.6$\pm$0.5 & 1.3$\pm$0.4 & 9.5$\pm$2.1  & ... \\ 
G107.29+05.63 & 2.1$\pm$0.2 & 1.9$\pm$0.1 & 2.2$\pm$0.1 & 1.7$\pm$0.1 & 1.8$\pm$0.3 & 1.9$\pm$0.4 \\ 
G108.18+05.51 & 3.3$\pm$0.6 & 1.9$\pm$0.2 & 1.9$\pm$0.2 & 2.4$\pm$0.7  & ...  & ... \\ 
G108.20+00.58  & ...  & ... & 2.2$\pm$0.4 & 3.8$\pm$2.3  & ...  & ... \\ 
G108.47-02.81  & ... & 0.9$\pm$0.3 & 1.3$\pm$0.3  & ... & 1.9$\pm$0.7  & ... \\ 
G108.59+00.49  & ... & 1.4$\pm$0.4 & 2.7$\pm$0.1 & 1.5$\pm$0.3 & 4.3$\pm$1.5  & ... \\ 
G109.87+02.11 & 1.9$\pm$0.3 & 3.4$\pm$0.4 & 3.2$\pm$0.1 & 2.9$\pm$0.1 & 3.2$\pm$0.2 & 1.8$\pm$0.2 \\ 
G111.23-01.23  & ...  & ...  & ...  & ...  & ...  & ... \\ 
G111.25-00.76  & ...  & ... & 2.9$\pm$0.3  & ...  & ...  & ... \\ 
G111.54+00.77 & 3.2$\pm$0.1 & 4.4$\pm$0.7 & 3.38$\pm$0.05 & 3.3$\pm$0.2 & 3.9$\pm$0.2 & 3.6$\pm$0.7 \\ 
G121.29+00.65 & 1.5$\pm$0.3 & 3.1$\pm$0.5 & 2.7$\pm$0.1 & 2.2$\pm$0.1 & 1.6$\pm$0.4  & ... \\ 
G122.01-07.08  & ... & 2.2$\pm$0.6 & 1.2$\pm$0.3  & ...  & ...  & ... \\ 
G123.06-06.30 & 2.9$\pm$0.3 & 3.5$\pm$0.3 & 3.1$\pm$0.1 & 3.1$\pm$0.2 & 3.2$\pm$0.3  & ... \\ 
G123.06-06.30 & 2.6$\pm$0.3 & 4.1$\pm$0.4 & 3.2$\pm$0.1 & 3.0$\pm$0.3 & 4.3$\pm$0.5  & ... \\ 
G133.94+01.06 & 4.3$\pm$0.5  & ... & 4.9$\pm$0.1 & 3.9$\pm$0.3 & 4.6$\pm$0.3  & ... \\ 
G134.62-02.19  & ...  & ...  & ...  & ...  & ...  & ... \\ 
G135.27+02.79  & ...  & ... & 2.6$\pm$0.6  & ...  & ...  & ... \\ 
G160.14+03.15  & ...  & ... & 1.8$\pm$0.2  & ... & 1.4$\pm$0.2  & ... \\ 
G168.06+00.82  & ...  & ... & 0.9$\pm$3.6  & ...  & ...  & ... \\ 
G176.51+00.20 & 1.9$\pm$0.4 & 1.7$\pm$0.2 & 1.7$\pm$0.1 & 1.7$\pm$0.1 & 2.0$\pm$0.3 & 0.7$\pm$0.4 \\ 
G182.67-03.26  & ...  & ...  & ...  & ...  & ...  & ... \\ 
G183.72-03.66  & ... & 2.0$\pm$0.3 & 1.8$\pm$0.1 & 1.4$\pm$0.1 & 2.3$\pm$0.5 & 1.7$\pm$0.5 \\ 
G188.79+01.03  & ...  & ... & 3.0$\pm$0.1 & 3.4$\pm$0.9 & 2.6$\pm$0.4  & ... \\ 
G188.94+00.88 & 2.7$\pm$0.4 & 3.0$\pm$0.7 & 2.7$\pm$0.1 & 2.2$\pm$0.1 & 2.4$\pm$0.2 & 2.3$\pm$0.5 \\ 
G192.16-03.81  & ...  & ...  & ...  & ...  & ...  & ... \\ 
G192.60-00.04 & 2.5$\pm$0.3 & 1.0$\pm$0.4 & 2.47$\pm$0.04 & 2.8$\pm$0.2 & 3.0$\pm$0.2  & ... \\ 
G196.45-01.67  & ...  & ... & 1.9$\pm$0.3  & ...  & ...  & ... \\ 
G209.00-19.38 & 6.5$\pm$0.2 & 2.9$\pm$0.7 & 9.8$\pm$0.2 & 3.9$\pm$0.3 & 7.8$\pm$0.3  & ... \\ 
G211.59+01.05 & 2.4$\pm$0.7 & 1.2$\pm$0.2 & 2.6$\pm$0.2 & 3.0$\pm$0.8 & 2.7$\pm$1.0  & ... \\ 
G229.57+00.15  & ...  & ...  & ...  & ...  & ...  & ... \\ 
G232.62+00.99 & 1.1$\pm$0.2  & ... & 2.9$\pm$0.1 & 2.0$\pm$0.3 & 2.6$\pm$0.7  & ... \\ 
G236.81+01.98  & ...  & ... & 0.9$\pm$0.5  & ...  & ...  & ... \\ 
G239.35-05.06  & ...  & ...  & ...  & ...  & ...  & ... \\ 
G240.31+00.07 & 4.0$\pm$1.0  & ... & 3.0$\pm$0.5  & ... & 4.8$\pm$1.0  & ... \\

\end{longtable}

\end{document}